# Mu2e Conceptual Design Report

March 2012

Fermi National Accelerator Laboratory
Batavia, IL 60510
www.fnal.gov

Contacts: R. Ray – Project Manager (rray@fnal.gov)
D. Glenzinski – Deputy Project Manager (douglasg@fnal.gov)

Fermilab

## DISCLAIMER

# The Mu2e Project Team

L. Bartoszek
*Bartoszek Engineering, Aurora IL*

J. Miller
*Boston University, Boston Massachusetts*

V. Polychronakos, P. Yamin
*Brookhaven National Laboratory, Upton, New York*

C. Cheng, B. Echenard, J. Hanson, D.G. Hitlin, F. Porter, R.Y. Zhu
*California Institute of Technology, Pasadena, California*

S. Oh, C. Wang
*Duke University, Durham, North Carolina*

G. Ambrosio, N. Andreev, D. Arnold, K. Biery, R. Bossert, M. Bowden, J. Brandt, M. Buehler, A. Burov, R. Carcagno, R. Coleman, J. Dey, G. Deuerling, S. Dixon, B. Drendel, D. Evbota, I. Fang, S. Feher, G. Gallo, K. Genser, D. Glenzinski, S. Hansen, E. Huedem, W. Jaskierny, R. Jedziniak, C. Johnstone, V.S. Kashikhin, V.V. Kashikhin, P. Kasper, M. Kim, A. Klebaner, I. Kourbanis, J. Kowalkowski, K. Krempetz, R. Kwarciany, T. Lackowski, M. Lamm, M. Larwill, T. Leveling, C. Lindenmeyer, M. Lopes, W. Masayoshi, L. Michelotti, J. Morgan, A. Mukherjee, V. Nagaslaev, T. Nicol, J. Niehoff, B. Norris, D. Orris, R. Ostojic, T. Page, T. Peterson, A. Pla-Dalmau, E. Prebys, P. Prieto, V. Pronskikh, R. Rabehl, R.E. Ray, R. Rechenmacher, R. Rivera, P. Rubinov, V.L. Rusu, C. Sylvester, Z. Tang, M. Tartaglia, J. Theilacker, J. Tompkins, G. Van Zandbergen, R. Wagner, R. Walton, R. Wands, S. Werkema, R. Wielgos, J. Wu, M. Xiao, R. Yamada
*Fermi National Accelerator Laboratory, Batavia, Illinois*

M. Cordelli, S. Giovannella, F. Happacher, A. Luca, S. Miscetti, G. Pileggi, B. Ponzio,A. Saputi, I. Sarra
*Laboratori Nazionali di Frascati, Italy*

E.V. Hungerford
*University of Houston, Texas*

D.N. Brown, C. Grace, J.K. Chang, S.S. Fung, D. Gnani, Yu.G. Kolomensky, T. Ma, H. von der Lippe, J.P. Walder
*Lawrence Berkeley National Laboratory and University of California, Berkeley*

L. De Lorenzis, F. Grancagnolo, A. Maffezzoli, A. Miccoli, G. Zavarise,
*Instituto Nazionale di Fisica Nucleare, Lecce, Italy*

T. Ito, K. Yarritu
*Los Alamos National Laboratory, Los Alamos New Mexico*

R. Carosi, F. Puccinelli
*Instituto Nazionale di Fisica Nucleare, Pisa, Italy*

M. Corcoran, J. Orduna
*Rice University, Houston, Texas*

C. Densham, J. Odell, M. Woodward
*Rutherford Appleton Laboratory, Didcot, United Kingdom*

G. Pauletta
*The University of Udine and INFN/Trieste Udine, Italy*

C.E. Dukes, R. Ehrlich, M. Frank, E. Frlež, S. Goadhouse, C. Group, E. Ho,
Y. Oksuzian, D. Počanić, J. Steward
*University of Virginia, Charlottesville, Virginia*

K.R. Lynch, J.L. Popp
*York College of the City University of New York*

# The Mu2e Collaboration

E. Barnes, R.M. Carey, V. Khalatian, J.P. Miller, B.L. Roberts
*Boston University, Boston Massachusetts*

W.J. Marciano, V. Polychronakos, Y. Semertzidis, P. Yamin
*Brookhaven National Laboratory, Upton, New York*

C. Cheng, B. Echenard, D.G. Hitlin, F. Porter, R.Y. Zhu
*California Institute of Technology, Pasadena, California*

E. Heckmaier, T.I. Kang, D. Kocen, G. Lim, W. Molzon, Z. You
*University of California, Irvine California*

A. Artikov, J. Budagov, Y. Davydov, V. Glagolev, A. Simonenko, I. Suslov, V. Tereshchenko
*Joint Institute for Nuclear Research, Dubna, Russia*

S. Oh, C. Wang
*Duke University, Durham, North Carolina*

G. Ambrosio, R.H. Bernstein, R.M. Coleman, F. DeJongh, N. Evans, S. Feher, M. Fischler,
A. Gaponenko, D.A. Glenzinski, J.A. Johnstone, P. Kasper, R.K. Kutschke, M.J. Lamm,
V. Logashenko, M. McAteer, L. Michelotti, A. Mukherjee, S. Nagaitsev, V. Nagaslaev,
D.V. Neuffer, A.J. Norman, R. Ostojic, C.C. Polly, M. Popovic, E.J. Prebys, V. Pronskikh, R.E. Ray,
P. Rubinov, V.L. Rusu, P. Shanahan, R.S. Tschirhart, S. Werkema, H.B. White Jr., K. Yonehara
*Fermi National Accelerator Laboratory, Batavia, Illinois*

P. Ciambrone, M. Cordelli, S. Giovannella, F. Happacher, A. Luca,
M. Martini, S. Miscetti, A. Saputi, I. Sarra, G. Venanzoni
*Laboratori Nazionali di Frascati, Italy*

D. Alexander, X. Huang, E. Hungerford, K. Lau
*University of Houston, Texas*

P.T. Debevec, G.D. Gollin
*University of Illinois, Urbana-Champaign, Illinois*

D.N. Brown, Yu.G, Kolomensky, M.J. Lee
*Lawrence Berkeley National Laboratory and University of California, Berkeley*

M. Cascella, L. De Lorenzis, F. Grancagnolo, A. L'Erario, A. Maffezzoli,
G. Onorato, G.M. Piacentino, S. Rella, G. Tassielli, G. Zavarise
*Instituto Nazionale di Fisica Nucleare, Lecce, Italy*

J. Kozminski
*Lewis University, Romeoville, Illinois*

M. Cooper, T. Ito, R. McCrady, J. Ramsey, K. Yarritu
*Los Alamos National Laboratory, Los Alamos New Mexico*

D.M. Kawall, K.S. Kumar
*University of Massachusetts, Amherst*

R. Djilkibaev, V. Matushko
*Institute for Nuclear Research, Moscow, Russia*

R.J. Abrams, C.M. Ankenbrandt, R.P. Johnson, S.A. Kahn,
T.J. Roberts, C. Yoshikawa
*Muons Inc., Batavia, Illinois*

D. Hedin, A. Dychkant, A. Yurkewicz
*Northern Illinois University, DeKalb, Illinois*

A.L. de Gouvea
*Northwestern University, Evanston, Illinois*

D.M. Asner, R. Bonicalzi, J.E. Fast, G. Warren, L.S. Wood
*Pacific Northwest National Laboratory, Richland, Washington*

F. Cervelli, R. Carosi, S. Di Falco, G. Gallucci, T. Lomtadze, L. Ristori, C. Vannini
*Instituto Nazionale di Fisica Nucleare, Pisa, Italy*

V. Biliyar, A. Chandra, M.D. Corcoran
*Rice University, Houston, Texas*

D. Cauz, G. Pauletta
*The University of Udine and INFN/Trieste Udine, Italy*

C.E. Dukes, R. Ehrlich, M. Frank, E. Frlež, S. Goadhouse, C. Group, R.J. Hirosky, P.Q. Hung,
Y. Oksuzian, K.D. Paschke, D. Počanić
*University of Virginia, Charlottesville, Virginia*

D.W. Hertzog, P. Kammel
*University of Washington, Seattle*

K.R. Lynch, J.L. Popp
*York College of the City University of New York*

This page intentionally left blank

# Table of Contents

# 1 Executive Summary

## 1.1 Introduction

Fermi National Accelerator Laboratory and the Mu2e Collaboration, composed of about 135 scientists and engineers from 26 Universities and Laboratories around the world, have collaborated to create this conceptual design for a new facility to study charged lepton flavor violation using the existing Department of Energy investment in the Fermilab accelerator complex.

Mu2e proposes to measure the ratio of the rate of the neutrinoless, coherent conversion of muons into electrons in the field of a nucleus, relative to the rate of ordinary muon capture on the nucleus:

$$R_{\mu e} = \frac{\mu^- + A(Z,N) \to e^- + A(Z,N)}{\mu^- + A(Z,N) \to \nu_\mu + A(Z-1,N)}.$$

The conversion process is an example of charged lepton flavor violation (CLFV), a process that has never been observed experimentally. The significant motivation behind the search for muon-to-electron conversion is discussed in Chapter 3. The current best experimental limit on muon-to-electron conversion, $R_{\mu e} < 7 \times 10^{-13}$ (90% CL), is from the SINDRUM II experiment [1]. With $3.6 \times 10^{20}$ delivered protons Mu2e will probe four orders of magnitude beyond the SINDRUM II sensitivity, measuring $R_{\mu e}$ with a single event sensitivity of $5.4 \times 10^{-17}$. Observation of this process would provide unambiguous evidence for physics beyond the Standard Model and can help to illuminate discoveries made at the LHC or point to new physics beyond the reach of the LHC.

The conversion of a muon to an electron in the field of a nucleus occurs coherently, resulting in a monoenergetic electron near the muon rest energy that recoils off of the nucleus in a two-body interaction. This distinctive signature has several experimental advantages including the near-absence of background from accidentals and the suppression of background electrons near the conversion energy from muon decays.

At the proposed Mu2e sensitivity there are a number of processes that can mimic a muon-to-electron conversion signal. Controlling these potential backgrounds drives the overall design of Mu2e. These backgrounds result principally from five sources:

1. Intrinsic processes that scale with beam intensity and include muon decay in orbit (DIO) and radiative muon capture (RMC).
2. Processes that are delayed because of particles that spiral slowly down the muon beamline, such as antiprotons.

3. Prompt processes where the detected electron is nearly coincident in time with the arrival of a beam particle at the muon stopping target.
4. Electrons that are initiated by cosmic rays.
5. Events that result from reconstruction errors induced by additional activity in the detector from conventional processes.

A general description of these backgrounds can be found in Section 3.2 and a detailed description combined with estimates of background rates in Mu2e can be found in Section 3.5.

## 1.2 Scope

To achieve the sensitivity goal described above a high intensity, low energy muon beam coupled with a detector capable of efficiently identifying 105 MeV electrons while minimizing background from conventional processes will be required. The muon beam is created by an 8 GeV, pulsed beam of protons striking a production target. The scope of work required to meet the scientific and technical objectives of Mu2e is listed below.

- Modify the accelerator complex to transfer 8 GeV protons from the Fermilab Booster to the detector while the 120 GeV neutrino program is operating. To accomplish this the existing Recycler and Debuncher Rings will be modified to re-bunch batches of protons from the Booster and then slow extract beam to the Mu2e detector.
- Design and construct a new beamline from the Debuncher Ring to the Mu2e detector. The beamline includes an *extinction insert* that removes residual out-of-time protons.
- Design and construct the Mu2e superconducting solenoid system (Figure 1.1) consisting of a *Production Solenoid* that contains the target for the primary proton beam, an S-shaped *Transport Solenoid* that serves as a magnetic channel for pions and muons of the correct charge and momentum range and a *Detector Solenoid* that houses the muon stopping target and the detector elements.
- Design and construct the Mu2e detector (Figure 1.1) consisting of a tracker, a calorimeter, a stopping target monitor, a cosmic ray veto, an extinction monitor and the electronics, trigger and data acquisition required to read out, select and store the data. The tracker accurately measures the trajectory of charged particles, the calorimeter provides independent measurements of energy, position and time, the stopping target monitor measures the characteristic X-ray spectrum from the formation of muonic atoms, the cosmic ray veto identifies cosmic ray muons traversing the detector region that can cause backgrounds and the extinction monitor detects scattered protons from the stopping target to determine the fraction of out-of-time beam.

- Design and construct a facility to house the Mu2e detector and the associated infrastructure (see Figure 1.2). This includes an underground detector enclosure and a surface building to house necessary equipment and infrastructure that can be accessed while beam is being delivered to the detector.

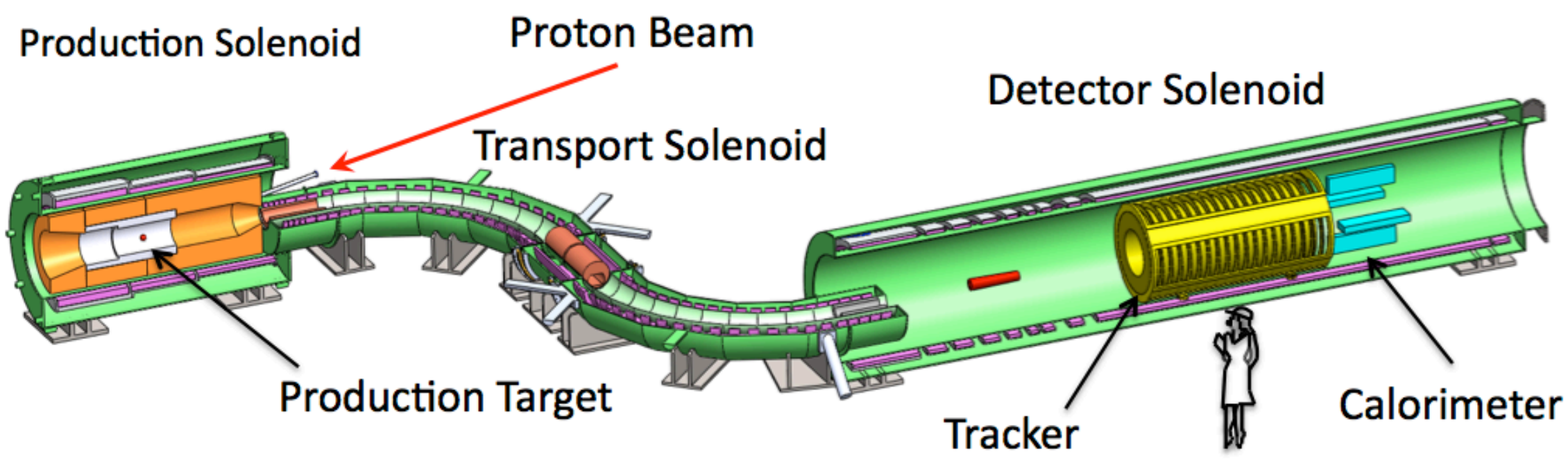

Figure 1.1. The Mu2e Detector. The cosmic ray veto, surrounding the Detector Solenoid is not shown.

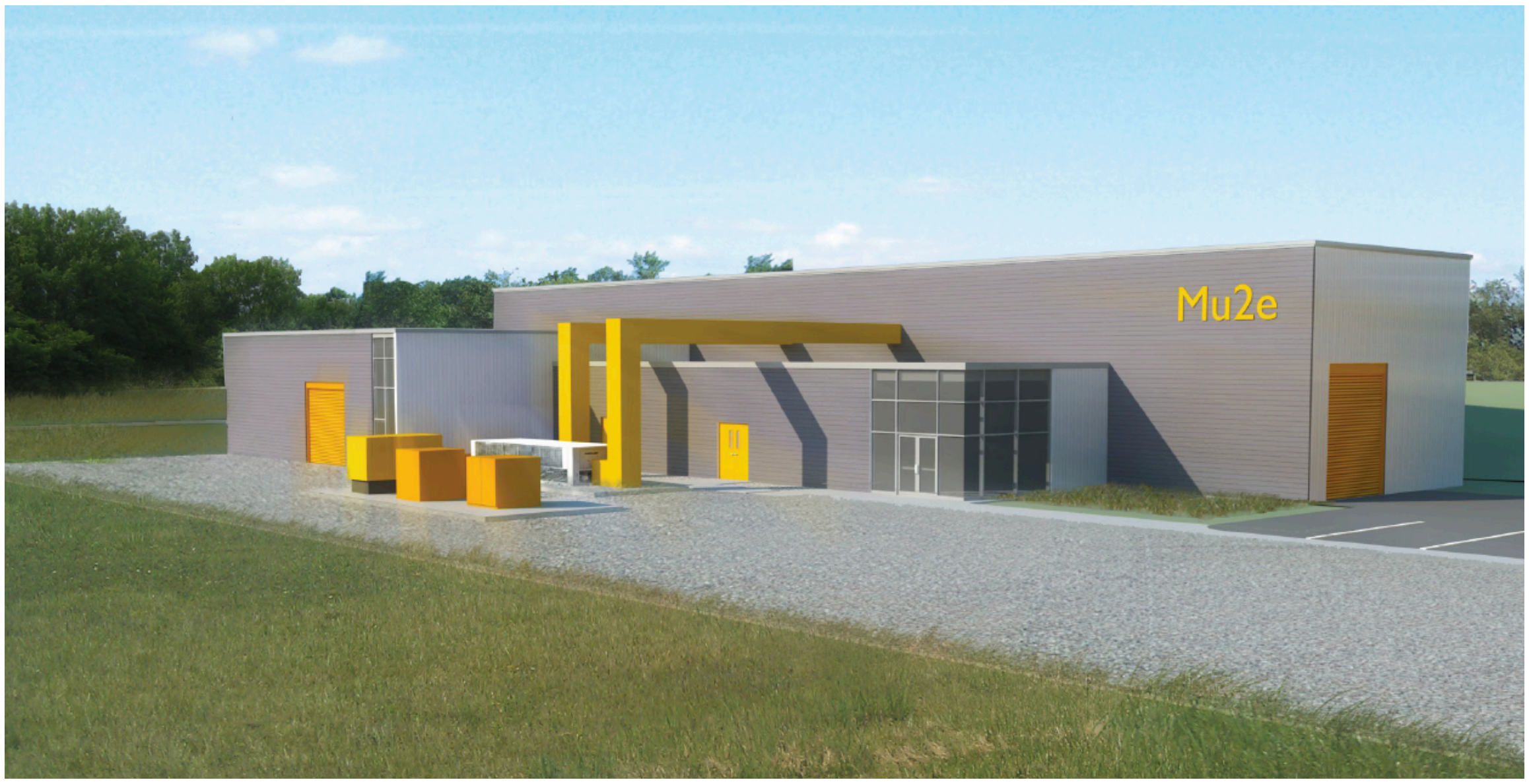

Figure 1.2. Depiction of the above-grade portion of the Mu2e facility.

Mu2e is integrated into Fermilab's overall science program that includes many experiments that use the same machines and facilities, though often in different ways. Because of the overlapping needs of several experimental programs, the scope of work described above will be accomplished through a variety of mechanisms. The NOvA and g-2 Projects both include upgrades to the Recycler Ring that will be used by Mu2e. In addition, there is infrastructure required by both Mu2e and g-2 that will be funded as common Accelerator Improvement Projects (AIPs) and General Plant Projects (GPPs). These common projects will be managed by Fermilab to ensure completion on a time scale consistent with the Lab's overall program plan and to guarantee that the needs of the overall program are satisfied. The scope of the Mu2e Project and the common projects will be described in Chapter 2.

## 1.3 Cost and Schedule

The total project cost for Mu2e is in the range \$208M - \$287M. This includes the base cost plus contingency and overhead, escalated to actual year dollars. There is very little scope range associated with the Mu2e. A schedule range of 60 - 84 months is proposed for the range of the construction project from CD3a – CD4.

## 1.4 Acquisition Strategy

The acquisition strategy relies on the Fermi Research Alliance (FRA), the Department of Energy Managing and Operating (M&O) contractor for Fermi National Accelerator Laboratory (Fermilab), to directly manage the Mu2e acquisition. The design, fabrication, assembly, installation, testing and commissioning for the Mu2e Project will be performed by the Mu2e Project scientific and technical staff provided by Fermilab and the various Mu2e collaborating institutions. Much of the subcontracted work to be performed for Mu2e consists of hardware fabrication and conventional facilities construction.

## 1.5 References

[1] W. Bertl et al., Eur. Phys. J. **C47**, 337 (2006).

# 2 Project Overview

## 2.1 Project Mission

The primary mission of the Mu2e Project is to design and construct a facility that will enable the most sensitive search ever made for the coherent conversion of muons into electrons in the field of a nucleus. Mu2e will measure the ratio of muon conversions to conventional muon captures:

$$R_{\mu e} = \frac{\mu^- + A(Z,N) \rightarrow e^- + A(Z,N)}{\mu^- + A(Z,N) \rightarrow \nu_\mu + A(Z-1,N)}$$

with a single event sensitivity of $5.4 \times 10^{-17}$ (90% C.L.). Mu2e will be 10,000 times more sensitive to this process than previous experiments. Observation of this process would provide unambiguous evidence for physics beyond the Standard Model and can help to illuminate discoveries made at the LHC or point to new physics beyond the reach of the LHC.

To achieve this significant leap in sensitivity, Mu2e requires an intense low energy muon beam and a state-of-the-art detector capable of precision measurements in the presence of high rates.

## 2.2 Scope Required to Satisfy the Mission Requirements

A conceptual design has been developed for the Mu2e Project that meets the Mission Requirements described in Section 2.1. The scope includes the following:

- A proton beam that can produce an intense secondary muon beam with a structure that allows time for the muons to decay before the next pulse arrives.
- A pion capture and muon transport system that efficiently captures charged pions and transports negatively charged decay muons to a target where they can be stopped. The momentum spectrum of the transported muon beam must be low enough to ensure that a significant fraction of the muons can be brought to rest in a thin target.
- A detector that is capable of efficiently and accurately identifying and analyzing conversion electrons with momenta near 105 MeV/c while rejecting backgrounds from conventional processes.
- A detector hall facility to house the experimental apparatus.

Mu2e is one of three new experiments at Fermilab that require modifications to the existing accelerator complex. Mu2e, g-2 and NOvA all require modifications to the Recycler Ring. Mu2e and g-2 both require modifications to the Debuncher Ring and to the transfer lines that will carry beam from the Recycler to the Debuncher. Additionally, Mu2e and g-2 both require a new beamline to transport beam from the Debuncher Ring to their experimental halls. NOvA is scheduled to begin taking beam in 2013, g-2 in 2016 and Mu2e in 2020. Modifications to the accelerator complex will be made by all 3 Projects. The scope of the modifications depends on the exact needs of the experiment and the time at which they are needed. Additionally, some of the modifications that are required by both Mu2e and g-2 will be accomplished by a set of common projects. The set of common projects required by both Mu2e and g-2 include one Accelerator Improvement Project (AIP) and two General Plant Projects (GPPs). By taking advantage of numerous synergies between Mu2e and g-2, Fermilab can support two unique, world-class muon experiments for less than the cost of the individual programs executed independently.

The full Mu2e scope of work required to execute the Mu2e experiment will be described in the sections that follow. The funding source will be identified in each case. The majority of this Conceptual Design Report will be confined to the scope of work included in the Mu2e Project, but it is important for Fermilab and the funding agencies to understand the full scope of the work required.

### *2.2.1* Primary Proton Beam

Mu2e requires a high intensity, pulsed proton beam to produce an intense beam of low energy muons with the time structure required by the experiment. As shown in Figure 2.1, batches of protons from the Booster will be transported to the Recycler Ring where they will be re-bunched by a new RF system. The re-bunched beam will be kicked out of the Recycler into existing transfer lines that will deliver the protons to the Debuncher Ring. A resonant extraction system in the Debuncher will slow extract protons to the Mu2e detector through a new external beamline.

***Recycler Ring Modifications***

Mu2e, g-2 and NOvA all require the ability to transport protons from the Booster to the Recycler Ring. Currently, the MI-8 line connects the Booster to the Main Injector, which resides in the same beamline enclosure as the Recycler Ring. Part of the scope of the NOvA Project is to connect MI-8 to the Recycler Ring and to construct and install a kicker system to inject Booster batches into the Recycler. Mu2e and g-2 will also use this new injection system.

Both Mu2e and g-2 require the ability to re-bunch beam in the Recycler Ring. A new 2.5 MHz RF system will divide batches of protons from the Booster into four smaller bunches that will be transferred one-at-a-time to the existing P1 line. A new connection is required from the Recycler Ring to the P1 line, which currently connects to the Main Injector. A new extraction kicker is also required. The RF system, Recycler to P1 connection and the extraction kicker are part of the g-2 Project scope.

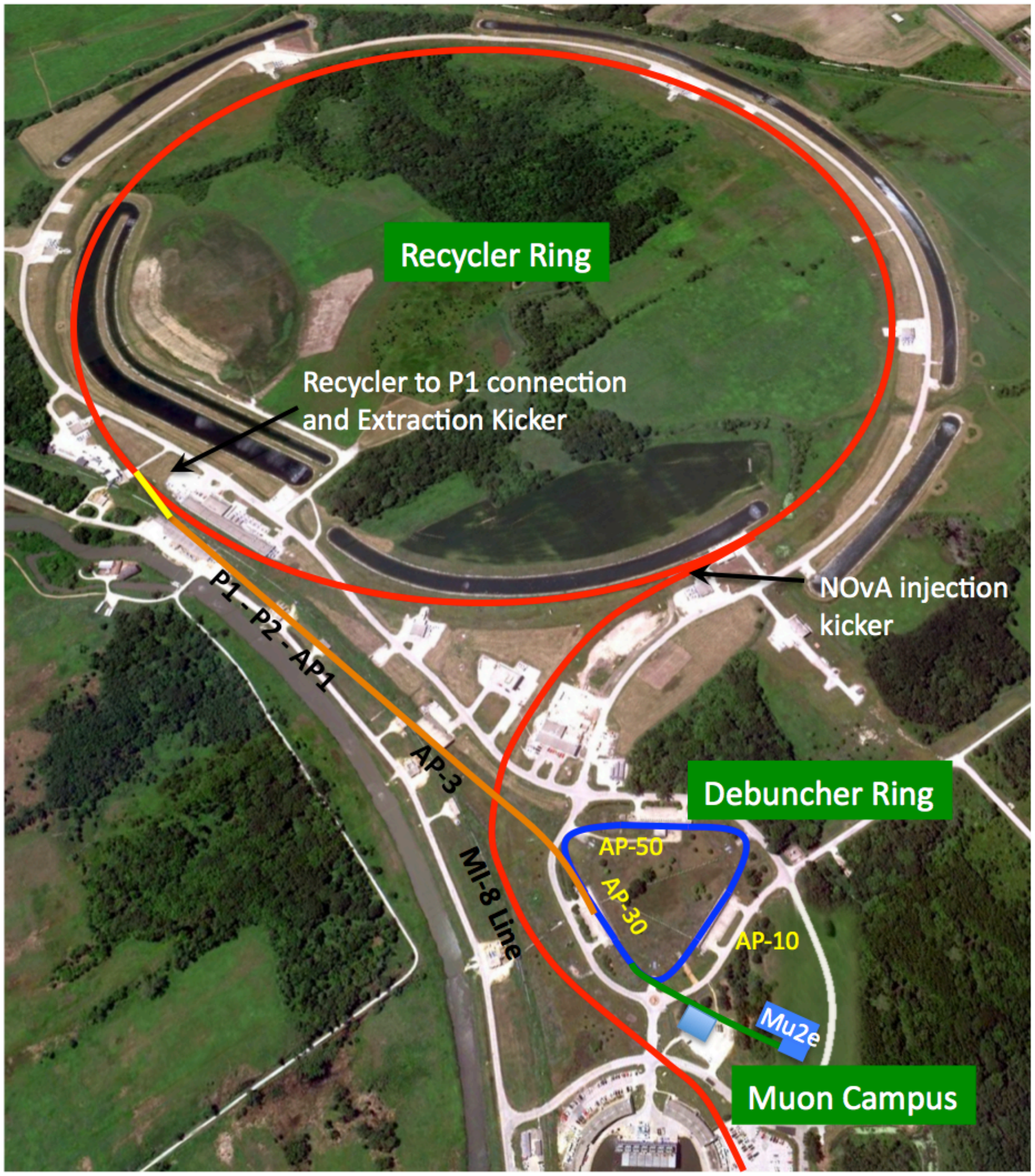

Figure 2.1. Layout of the Mu2e facility (lower right) relative to the accelerator complex that will provide proton beam to the detector. Protons are transported from the Booster through the MI-8 beamline to the Recycler Ring where they will circulate while they are re-bunched by a new 2.5 MHz RF system. The reformatted bunches are kicked into the P1 line and transported to the Debuncher Ring where they are slow extracted to the Mu2e detector through a new external beamline.

### *Transfer Lines and Debuncher Injection*

Proton bunches formed in the Recycler Ring will be kicked into the P1 line and transported to the Debuncher Ring through a series of existing transfer lines. For Mu2e running, protons will traverse the P2, AP1 and AP3 lines before being injected into the Debuncher Ring by a new Debuncher injection kicker. The proton bunches will be captured in the Debuncher by a new 2.4 MHz RF system consisting of RF modules that are identical to the RF modules needed for the Recycler Ring. For g-2 running, protons will traverse the P2 and AP1 lines before intersecting a production target at AP0. Muons from the target will be collected and transported through the AP3 line and injected into the Debuncher Ring using the same Debuncher injection kicker required for Mu2e. A proton abort system is required in the Debuncher for Mu2e should it become necessary to dump proton bunches for operational reasons and to clean up the ring at the end of a resonant extraction cycle. The abort system will also be used by g-2 to eliminate residual protons circulating with their muon beam. After traversing the Debuncher Ring several times the protons lag behind the muons and can be kicked into the abort. Stochastic cooling tanks and other equipment used for antiproton production will be removed from the Debuncher Ring to open up the beam aperture as much as possible. Because these modifications are required for both Mu2e and g-2, they will be funded as an Accelerator Improvement Project (AIP) and managed by the Accelerator Division Muon Department.

### *Debuncher Modifications exclusive to Mu2e*

Mu2e requires slow extracted proton beam to be delivered to the Mu2e detector. A new resonant extraction system is required that delivers narrow microbunches to the detector that are separated by the revolution period of the Debuncher Ring (~1.7 μs). The resonant extraction system consists of sextupoles, quadrupoles, an RF knockout device and an electrostatic extraction septum along with the controls and instrumentation necessary to operate and control the resonant extraction process. Internal shielding at loss points in the beamline tunnel and Debuncher Ring are also required for Mu2e operation. This scope is all funded as part of the Mu2e Project.

### *Muon Campus Beamline Enclosure*

The final run from the Debuncher Ring to both the Mu2e and g-2 detectors requires a new external beamline. The beamline enclosure scope includes excavation, the concrete enclosure itself, utilities, lighting and the above grade berm. g-2 requires only the upstream portion of the beamline and their schedule calls for the beamline to be in place several years before the full beamline is needed for Mu2e. However, it is most efficient from a cost and management perspective to construct the entire enclosure as part of a single construction package. This avoids paying twice for mobilization, eliminates the overhead associated with multiple bid processes and avoids potential interface issues. The Beamline Enclosure is funded as a General Plant Project (GPP) and will be managed by Fermilab's Facilities Engineering Services Section (FESS).

***Muon Campus External Beamline***

Most of the components required for the Muon Campus External Beamline will be recycled from the Accumulator Ring, which is not used by either Mu2e or g-2. This results in a significant cost reduction compared to fabrication of new components. Mu2e will transport 8 GeV protons through the entire external beamline while g-2 will transport 3 GeV muons through just the upstream section. g-2 requires extra quadrupoles for stronger focusing in the upstream section of the external beamline to compensate for the larger transverse size of the lower energy muon beam. The upstream section of the beamline will be funded by the g-2 Project and the Mu2e Project will fund the downstream section. The downstream section required for Mu2e will be assembled and commissioned several years after the upstream section is assembled for g-2.

Mu2e requires well-defined pulses of protons separated by about 1.7 μs with very little residual beam between pulses. An *extinction system*, in the form of a high frequency *AC dipole* (see Chapter 5), will be constructed to suppress unwanted beam between successive microbunches. The extinction system is funded as part of the Mu2e Project.

### *2.2.2* Superconducting Solenoids

The 8 GeV protons provided by the Fermilab accelerator complex will be used to produce a high intensity, low energy muon beam. A significant fraction of the muons will be brought to rest in a stopping target made from a series of thin foils. Electrons that emerge from the stopping foils will be accurately momentum analyzed. These tasks will all take place in the evacuated inner bore of a series of superconducting magnets.

The *Production Solenoid* is a high field magnet with a graded solenoidal field varying smoothly from 4.6 Tesla to 2.5 Tesla. The gradient will be formed by 3 axial coils with a decreasing number of windings, made of aluminum stabilized NbTi. The solenoid is approximately 4 m long with an inner bore diameter of approximately 1.5 m. The Production Solenoid will be used to capture pions and the muons into which they decay. 8 GeV protons striking a production target near the center of the Production Solenoid initiate this process. A heat and radiation shield, constructed from bronze, will line the inside of the Production Solenoid to limit the heat load in the cold mass from secondaries produced in the production target and to limit radiation damage to the superconducting cable.

The S-shaped *Transport Solenoid* consists of a set of superconducting solenoids and toroids that form a magnetic channel that efficiently transmits low energy negatively charged muons from the Production Solenoid. The Transport Solenoid consists of five distinct regions: a 1 m long straight section, a 90° curved section, a second straight section about 2 m long, a second 90° curved section that brings the beam back to its original direction, and a third straight section of 1 m length. The major radius of the two curved sections is about 3 m and the resulting total magnetic length of the Transport Solenoid along its axis is about 13 m. The inner warm bore of the Transport Solenoid cryostat has a diameter of about 0.5 m. A relatively tight specification exists on the fields in the three straight sections where negative field gradients must be maintained at all places to eliminate backgrounds from late arriving particles. High energy negatively charged particles, positively charged particles and line-of-sight neutral particles will nearly all be eliminated by the two 90° bends combined with a series of absorbers and collimators.

The *Detector Solenoid* is nearly 11 m long with a clear bore diameter of about 2 m. The Detector Solenoid will provide a graded field for the muon stopping target located upstream and a uniform field downstream for the detector elements that analyze conversion electrons.

The solenoids are the cost and schedule driver for the project. The Production and Detector Solenoids will be designed and constructed in industry, likely by different vendors. The relatively unique Transport Solenoid will be designed and fabricated at Fermilab, though many of the components (superconducting cable, cryostats, etc.) will be procured from industry. The make-buy decisions are based on the similarity of the Production and Detector Solenoids to other solenoids fabricated in industry and to the limited availability of resources at Fermilab. The superconducting cable and the tooling required for fabricating the solenoids are long-lead items that must be procured early.

Significant infrastructure is required to support the operation of the solenoids. This includes power, quench protection, cryogens (liquid nitrogen and liquid helium), control and safety systems as well as mechanical supports to resist the significant magnetic forces on the magnets.

***Muon Campus Cryoplant***

Both Mu2e and g-2 require liquid helium to cool superconducting magnets. Rather than construct independent cryogenic facilities for each experiment, a common facility will be constructed to service both. Existing Tevatron compressors will drive compressed He from the Tevatron ring to a low bay attached to the g-2 detector hall (MC1). The low bay will contain 3 recycled Tevatron satellite refrigerators that can handle the dynamic loads of both experiments simultaneously. Cold lines will run from the refrigerators to each experimental detector hall. The Muon Campus Cryo facility will be funded as a General Plant Project (GPP).

### *2.2.3* The Mu2e Detector

The Mu2e detector contains components that operate in the evacuated warm bore of the Detector Solenoid as well as components outside of the Detector Solenoid. It must accurately and redundantly measure the energy of 105 MeV electrons that signal muon conversions in aluminum while eliminating backgrounds. This requires good momentum resolution and particle ID in the presence of high rates. The detector will consist of a tracker and a calorimeter that will provide redundant energy/momentum, timing, and trajectory measurements. A cosmic ray veto surrounds the Detector Solenoid to tag incoming muons and a stopping target monitor is located outside the downstream end of the detector solenoid to detect X-rays from the formation of muonic atoms.

The Mu2e tracker is designed to accurately measure the helical trajectory of electrons in a uniform magnetic field in order to determine their momenta. The accuracy with which the trajectory can be determined is limited by multiple scattering in the tracker. To meet these requirements the alternative selected for the Mu2e tracker is a low mass array of straw drift tubes aligned transverse to the axis of the Detector Solenoid. The Mu2e straw tubes are 5 mm in diameter, constructed from two layers of 6 μm thick Mylar and filled with an 80-20 mixture of Argon-$CO_2$ gas. Approximately 21,600 Mylar tubes will either be procured from a vendor or assembled by Mu2e. Each straw will have a wire inserted along its axis that will be securely terminated on a plate attached to a gas manifold at each end. The straws will then be accurately assembled into planes and tensioned to prevent sagging. One or two assembly factories will be established for these tasks. The tracker electronics will be located on or near the straws, inside of the Detector Solenoid vacuum. This requires low power electronics and a cooling system to remove the heat. The tracker electronics will be a mix of custom and commercial parts.

High rates of hits in the tracker may cause pattern recognition errors that add tails to the resolution function and result in background. Accidental hits can combine with or obscure hits from lower energy particles to leave behind a set of hits that might reconstruct to a trajectory consistent with a higher energy conversion electron. Extrapolating the fitted trajectory to the downstream calorimeter and comparing the calculated intercept with the measured position in the calorimeter may help to identify backgrounds that result from reconstruction errors. The calorimeter may also be used in a hardware, software or firmware trigger. The proposed calorimeter consists of approximately 1936 LYSO crystals located downstream of the tracker and arranged in four vanes. Each crystal is $3 \times 3 \times 11$ cm$^3$ and will be equipped with two Avalanche Photodiodes (APDs) for operation in the magnetic field of the Detector Solenoid. The crystals will have to be cooled to increase the light output.

Cosmic ray muons are a known source of potential background for experiments like Mu2e. Suppression of this background requires a large area cosmic ray veto surrounding the active detectors. Because of the large area required, the selected detector technology must be relatively inexpensive. The preferred alternative is a detector constructed from extruded scintillator with embedded wavelength shifting fibers read out into photodetectors. The scintillator will be produced at the Fermilab scintillator extrusion facility and shipped to an assembly factory where fibers will be inserted, silicon photomultipliers (SiPMs) attached and the counters packaged into self-contained modules. After testing, the modules will be shipped to Fermilab for installation.

The rate of muon captures at the muon stopping target must be determined as a normalization for muon conversions, should they be observed. A Muon Stopping Target Monitor consisting of a germanium crystal will measure the muon stopping rate during the live time of the experiment by measuring the characteristic X-ray spectrum from the formation of muonic atoms. The capture rate is derived from the stopping rate.

### *2.2.4* Conventional Facilities

The conventional facilities for the Mu2e Project include the site preparation, Mu2e surface building and the underground enclosure to house the Mu2e detector. Routing of utilities from nearby locations and installation of new transformers to power the facility are included in the scope of the conventional facilities work. Together the conventional facilities comprise approximately 25,000 ft$^2$ of new construction space.

The optimized length of the external beamline required for Mu2e results in a location for the Mu2e Detector Hall that intersects Kautz Road. As a result, it is necessary to re-route Kautz Road to the northwest of the Mu2e facility. This will be combined with some required utility work for g-2 and funded as a General Plant Project (GPP) so that it can be executed in a timely fashion and coordinated with other work in the area. This will reduce the overall cost. The Mu2e conventional facilities are shown in Figure 2.2.

#### *Sustainable Design and LEED*

The Mu2e Project is not required to meet the Leadership in Energy and Environmental Design (LEED)-Gold certification due to the function and operation of the facility. Specifically, the facility will not be occupied on a regular basis. In lieu of LEED-Gold certification, the Project plans to utilize guiding principles and ASHRAE recommendations to meet sustainability goals. A paper describing this approach [1] will be submitted to DOE-SC.

### *2.2.5* Summary of Scope Required to Satisfy the Mission Requirements

The scope of work required to produce an operational Mu2e experiment that satisfies the Mission requirements includes the scope of the Mu2e Project as well as the scope contained in several common projects that overlap with g-2. The common work is contained in one Accelerator Improvement Project (AIP) and two General Plant Projects (GPP). The AIP is for modifications and improvements to existing transfer lines and the Debuncher Ring. The GPPs are for the Muon Campus Beamline enclosure and the Muon Campus Cryogenic facility. Mu2e will also take advantage of functionality provided by the NOvA and g-2 Projects to inject, bunch and extract beam from the Recycler Ring. The operational scope required for Mu2e is summarized in Table 2.1. In each case the funding source is identified. The focus of this Conceptual Design Report will be on the items included in the Mu2e Project.

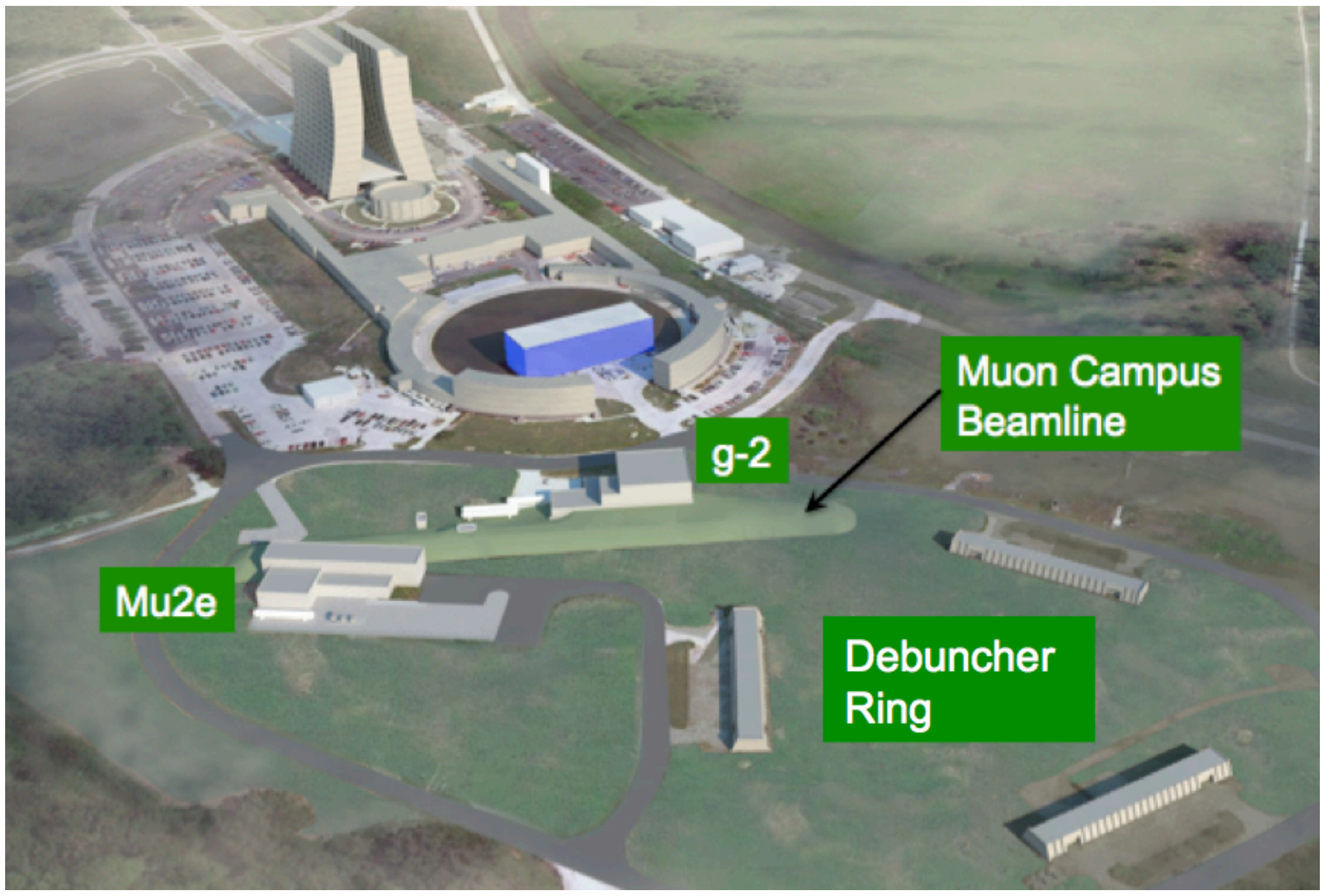

Figure 2.2. The Fermilab Muon Campus includes the detector halls for the Mu2e and g-2 experiments.

## 2.3 Project Organization

The Mu2e Project consists of nine subprojects coordinated by a central Project Office located at Fermilab. The subprojects, or Level 2 systems, are:

1. Project Management
2. Accelerator Systems
3. Conventional Construction
4. Solenoids
5. Muon Beamline
6. Tracker
7. Calorimeter
8. Cosmic Ray Veto
9. Trigger and DAQ.

| Item | Description | Funding Source |
|---|---|---|
| Recycler upgrades | • MI-8 to Recycler Connection<br>• Recycler injection kicker | NOvA |
| Recycler upgrades | • 2.5 MHz RF system<br>• Recycler to P1 connection<br>• Recycler extraction kicker<br>• MI-52 extension | g-2 |
| Transfer Line and Debuncher Modifications | • Beam transport from Recycler Ring to Debuncher<br>• Debuncher injection kicker<br>• Proton abort system<br>• Removal of Collider equipment | AIP |
| Mu2e specific Debuncher modifications | • 2.4 MHz RF system<br>• Resonant extraction system | Mu2e |
| Muon Campus Beamline Enclosure | Beamline tunnel to house external beamline elements for g-2 and Mu2e. | GPP |
| Muon Campus Site Preparation | • Site prep<br>• Diversion of Kautz Road<br>• General Utility work | GPP |
| Muon Campus Beamline (upstream) | Upstream elements, including increased quad density required for g-2 | g-2 |
| Muon Campus Beamline (downstream) | • Downstream elements required for Mu2e<br>• AC dipole extinction system<br>• Proton beam dump | Mu2e |
| Conventional Construction | • Surface Building<br>• Underground enclosure to house detector<br>• Utilities | Mu2e |
| Solenoids | • Production Solenoid<br>• Transport Solenoid<br>• Detector Solenoid<br>• Power<br>• Quench protection<br>• Cryo distribution | Mu2e |
| Muon Campus Cryo Plant | • Cryo refrigerators<br>• Warm lines for compressed Helium<br>• Cold lines to Mu2e and g-2 detector halls | GPP |
| Detector | • Tracker<br>• Calorimeter<br>• Cosmic Ray Veto<br>• Stopping Target Monitor<br>• Trigger and data acquisition system | Mu2e |

Table 2.1. The scope of work required to produce an operational Mu2e experiment. This includes the scope of the Mu2e Project as well as the scope contained in several common projects that overlap with g-2.

The Fermilab Project Office is headed by the Project Manager and assisted by a Deputy Project Manager and two Project Engineers. Project office support staff includes a Financial Manager, Project Controls Specialists, an ES&H Coordinator, a Risk Manager, a Quality Control Manager, a Configuration Control Manager and administrative support. Fermilab provides additional ES&H support and oversight. The Mu2e Project Office has developed overriding plans for project management, risk management, configuration control and quality assurance [2][3][4][5].

### *2.3.1* Work Breakdown Structure

The Mu2e Project has been organized into a Work Breakdown Structure (WBS). The WBS contains a complete definition of the Project's scope and forms the basis for planning, executing and controlling project activities. The Project WBS is shown in Figure 2.3 and Figure 2.4 down to level 3. Items are defined as specific deliverables (WBS 1.2 – 1.9) or Project Management (WBS 1.1).

1.1 *Project Management* – Project Office administrative and management activities that integrate across the entire project (management, regulatory compliance, quality assurance, safety, project controls, budget, risk management, etc.)

1.2 *Accelerator* – All phases of R&D, design, procurement, installation, integration and testing of the accelerator systems that are part of the Mu2e Project.

1.3 *Conventional Construction* - All phases of design, procurement, construction and integration of the conventional construction facilities including site preparation and access to utility systems.

1.4 *Solenoids* – All phases of R&D, design, procurement, installation, integration, testing and commissioning of the superconducting solenoid system and associated infrastructure including systems to distribute cryogens, power and quench protection.

1.5 *Muon Beamline* – All phases of R&D, design, procurement, installation, integration, testing and commissioning of the series of deliverables associated with the Muon Beamline system.

1.6 *Tracker* – All phases of R&D, design, procurement, assembly, installation, integration, testing and commissioning of the tracker, tracker electronics and associated support infrastructure.

1.7 *Calorimeter* – The project scope for the calorimeter includes procurement, testing and processing of 1/3 of the crystals as well as the R&D, design, construction and installation of the calibration system and the front end electronics. The rest of the calorimeter will be provided in-kind by INFN.

1.8 *Cosmic Ray Veto* - All phases of R&D, design, procurement, assembly, installation, integration, testing and commissioning of the cosmic ray veto, the veto electronics and associated support infrastructure.

1.9 *Trigger and DAQ* - All phases of R&D, design, procurement, assembly, installation, integration, testing and commissioning of the data acquisition system.

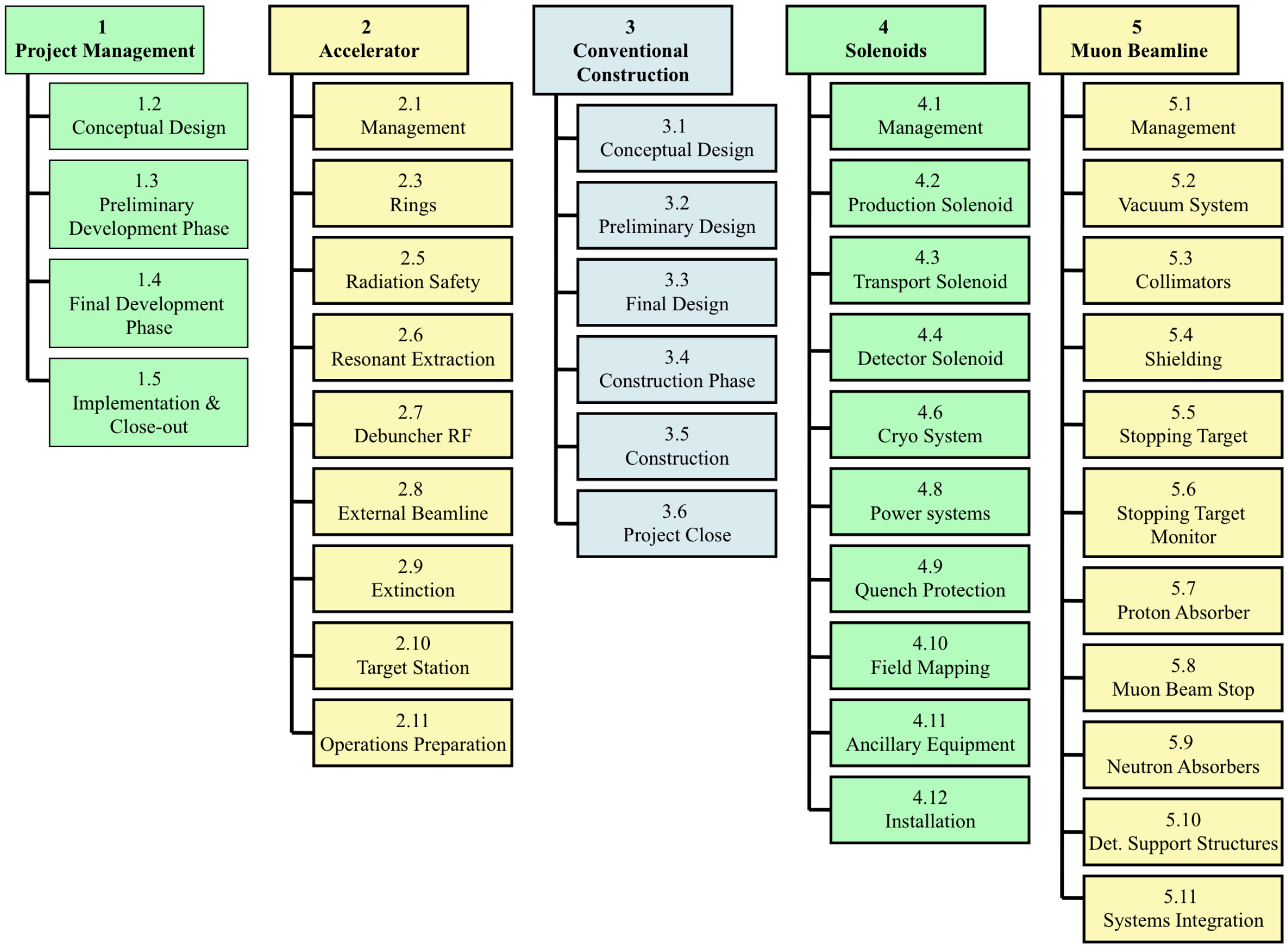

Figure 2.3. Mu2e Project WBS to Level 3 for Project Management, Accelerator, Conventional Construction, Solenoids and Muon Beamline.

## 2.4 Project Management

### *2.4.1* Project Controls

The Mu2e Project will be in full compliance with the DOE certified FRA Earned Value Management System (EVMS). The Earned Value Management System is used to monitor, analyze, and report project performance. Mu2e's EVMS implementation uses Primavera P6 scheduling software for the resource loaded cost and schedule, Cobra for escalation, burdening, and earned value reporting and analysis and Fermilab's Oracle Project Accounting system for tracking obligations and actual costs. The Fermilab EVMS description can be accessed through the Fermilab Office of Project Management website [6]. EVMS reporting is not required at the CD-1 stage of the Project, but the cost and schedule for CD-1 is being constructed with this eventual requirement in mind.

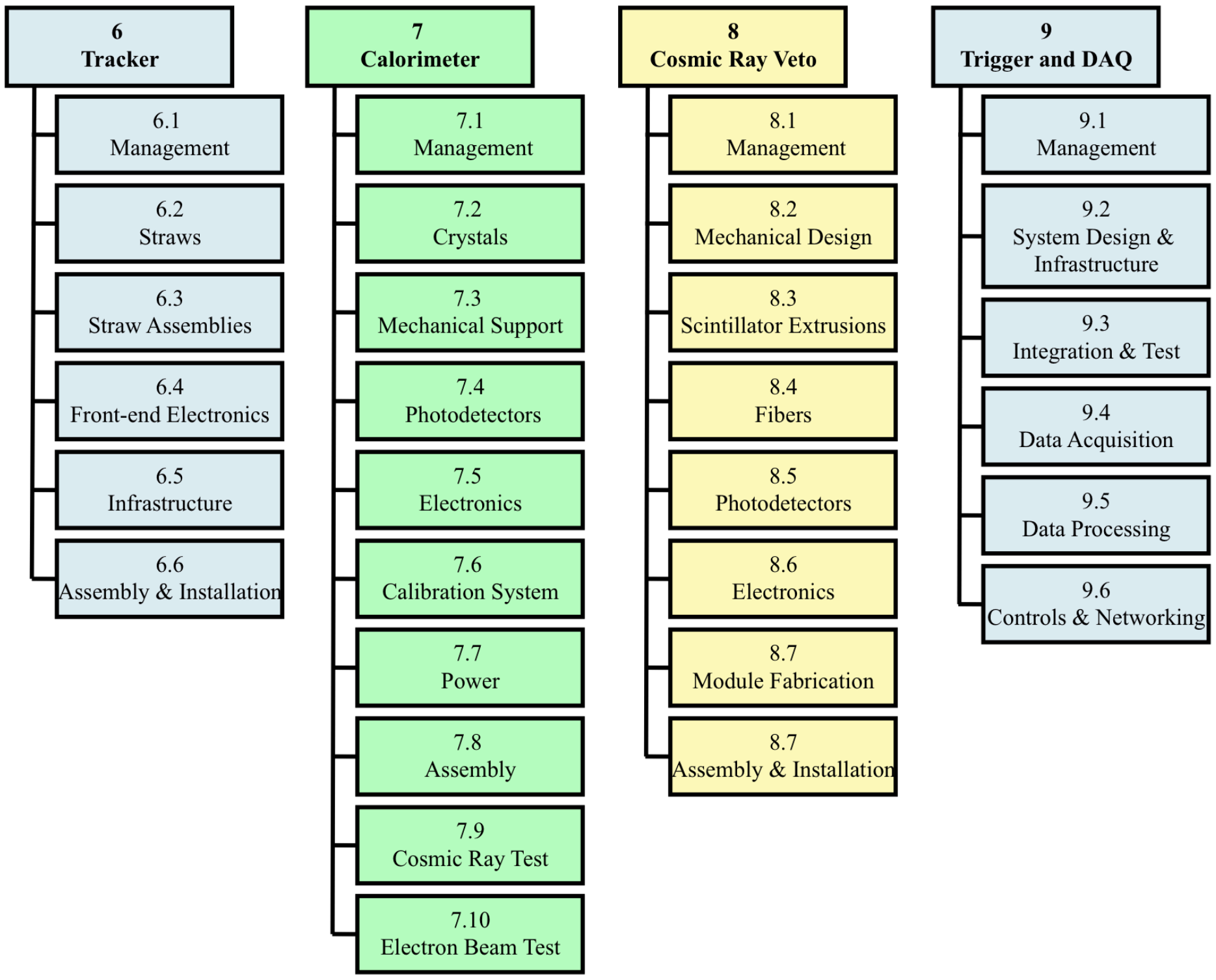

Figure 2.4. Mu2e Project WBS to Level 3 for the Tracker, Calorimeter, Cosmic Ray Veto and the Trigger and DAQ.

### *2.4.2* ES&H Management

The Laboratory Director has overall responsibility for establishing and maintaining Fermilab's ES&H policy. It is Fermilab's policy to integrate environment, safety and health protection into all aspects of work, utilizing the principles and core functions of the Integrated ES&H Management System and implemented through the appropriate lines of management. The Mu2e Project Manager reports to the Fermilab Director or his designee and is responsible for implementing Fermilab's ES&H policies into all aspects of the Mu2e Project. The draft Mu2e Project Management Plan [2] includes a section on Integrated Safety Management that describes how the Mu2e Project ES&H policies fit within the DOE approved Fermilab ES&H program.

The philosophy of Integrated Safety Management (ISM) will be incorporated into all work on Mu2e, including any work done on the Fermilab site by subcontractors and sub-tier contractors. Integrated Safety Management is a system for performing work safely and in an environmentally responsible manner. The term "integrated" is used to indicate that the ES&H management systems are normal and natural elements of doing work. The intent is to integrate the management of ES&H with the management of the other primary elements of work: quality, cost and schedule.

### *2.4.3* Quality Assurance

Quality Assurance and Quality Control (QA/QC) systems are designed, as part of the Quality Management Program, to ensure that the components of the Mu2e Project meet the design specifications and operate within the parameters mandated by the requirements of the Mu2e physics program. The Mu2e Project Manager is responsible for achieving performance goals. The Mu2e Quality Assurance Manager is responsible for ensuring that a quality system is established, implemented, and maintained in accordance with requirements. The Quality Assurance manager will provide oversight and support to the project participants to ensure a consistent quality program.

The QA/QC elements in place for the Mu2e Project draw heavily on the experience gained from similar projects in the past. Senior management recognizes that prompt identification and documentation of deficiencies, coupled with the identification and correction of the root causes, are key aspects of any effective QA/QC Program. The Project Manager endorses and promotes an environment in which all personnel are expected to identify nonconforming items or activities and potential areas for improvement.

### *2.4.4* Configuration Management

Configuration Management is a formalized process to manage proposed system changes and provide an audit trail to manage and maintain the evolution of system configurations. A Configuration Management Plan establishes a baseline, defines the rules for changing that baseline and records changes as they occur. The origin of changes and their status at any subsequent point should be readily identifiable.

The Mu2e Project uses several tools to achieve this objective, including a document control system that supports versioning and document signoff to "approve" a version, drawing management systems, and software control with a versioning and release system based on a software repository.

### *2.4.5* Risk Management

Project risk in Mu2e is mitigated through a structured and integrated process for identifying, evaluating, tracking, abating and managing risks in terms of three risk categories: cost, schedule and technical performance. A Risk Management Board, chaired by the Project Manager, meets regularly to identify risks and develop mitigation plans.

Any project faces both threats and opportunities and must strive to exploit the opportunities while ensuring that the threats do not derail the project. Numerous informal and formal approaches are used to identify threats and opportunities, assessing their likelihood and prioritizing them for possible mitigation or exploitation. The key to successful risk management is to implement a deliberate approach to accepting, preventing, mitigating or avoiding them. The Mu2e Project becomes aware of potential risks in many ways, notably during work planning, meetings and reviews as well as via lessons learned from others. Risk is managed during the planning and design phase by implementing appropriate actions, such as ensuring adequate contingency and schedule float, pursuing multiple parallel approaches and/or developing backup options. Every effort must be made to specify these actions in a manner that reduces the risk to an acceptably low level.

Risks that are identified will be managed as early as possible to assure that they do not delay the timely completion of the project or stress its budget in unexpected ways. The Mu2e Risk Management Plan [3] is under configuration management.

## 2.5 Conceptual Design and Alternatives Analysis

The Conceptual Design process is the exploration of concepts, specifications and designs for meeting the mission needs, and the development of alternatives that are technically viable, affordable and sustainable. The conceptual design provides sufficient detail to produce a more refined cost estimate range and to evaluate the merits of the project.

The starting point for the Mu2e conceptual design is the Project Mission described in Section 2.1 and the scientific proposal developed by the Mu2e Collaboration [7]. Based on the Mu2e sensitivity goal and an evaluation of all known sources of potential backgrounds, a set of physics requirements was developed. These are described in Section 3.6. A series of detailed requirements documents were developed to describe the functionality needed from the various detector and facility subsystems to satisfy the physics requirements.

A list of the Mu2e requirements documents appears in Table 2.2 along with the assigned document database number. Requirements documents are under configuration management and have been electronically signed by the relevant subsystem managers responsible for developing a conceptual design that satisfies the requirements. Signatories are automatically notified by the document database anytime a change is made to a requirements document. A modified document must be re-signed to become official.

Alternatives have been evaluated for satisfying the mission need as well as the set of requirements defined in the documents in Table 2.2. The goal of an alternatives analysis is to choose the most efficient, cost effective path to satisfy the requirements. Evaluation of alternatives may be made in terms of the three components of a project baseline: technical performance, cost and schedule.

| Topic | Document Database Number |
|---|---|
| Proton Beam | Mu2e-doc-1105 |
| Extinction | Mu2e-doc-1175 |
| Extinction Monitoring | Mu2e-doc-894 |
| Production Target | Mu2e-doc-887 |
| Heat and Radiation Shield | Mu2e-doc-1092 |
| Proton Beam Absorber | Mu2e-doc-948 |
| Conventional Facilities | Mu2e-doc-1088 |
| Production Solenoid | Mu2e-doc-945 |
| Transport Solenoid | Mu2e-doc-947 |
| Detector Solenoid | Mu2e-doc-946 |
| Cryoplant | Mu2e-doc-1509 |
| Cryo Distribution | Mu2e-doc-1244 |
| Quench Protection | Mu2e-doc-1238 |
| Solenoid Power System | Mu2e-doc-1237 |
| Magnetic Field Mapping | Mu2e-doc-1275 |
| Stopping Target | Mu2e-doc-1437 |
| Stopping Target Monitor | Mu2e-doc-1438 |
| Transport Solenoid Collimators | Mu2e-doc-1129 |
| Muon Beam Stop | Mu2e-doc-1351 |
| Vacuum System | Mu2e-doc-1481 |
| Proton Absorber | Mu2e-doc-1439 |
| Neutron Absorbers | Mu2e-doc-1371 |
| Muon Beamline Shielding | Mu2e-doc-1506 |
| Detector Support and Installation System | Mu2e-doc-1383 |
| Pbar Window | Mu2e-doc-941 |
| Tracker | Mu2e-doc-732 |
| Calorimeter | Mu2e-doc-864 |
| Cosmic Ray Veto | Mu2e-doc-944 |
| Calibration | Mu2e-doc-1182 |
| Trigger and DAQ | Mu2e-doc-1150 |

Table 2.2. List of Mu2e requirements documents.

Programmatic alternatives have been evaluated and are discussed in detail in the Mu2e Acquisition Strategy document [8]. Fermilab is the best choice among alternative sites for Mu2e because of the use of existing infrastructure and the ability to share common costs with g-2.

The Mu2e conceptual design consists of the set of technical alternatives selected as the result of the alternative analysis. The alternatives that have been considered for each subsystem are discussed in Chapters 5-12. In each case the relevant requirements are listed, the basis for the alternative selection is described and an assessment of how well the selected alternative satisfies the requirements is made. The conceptual design process is shown pictorially in Figure 2.5.

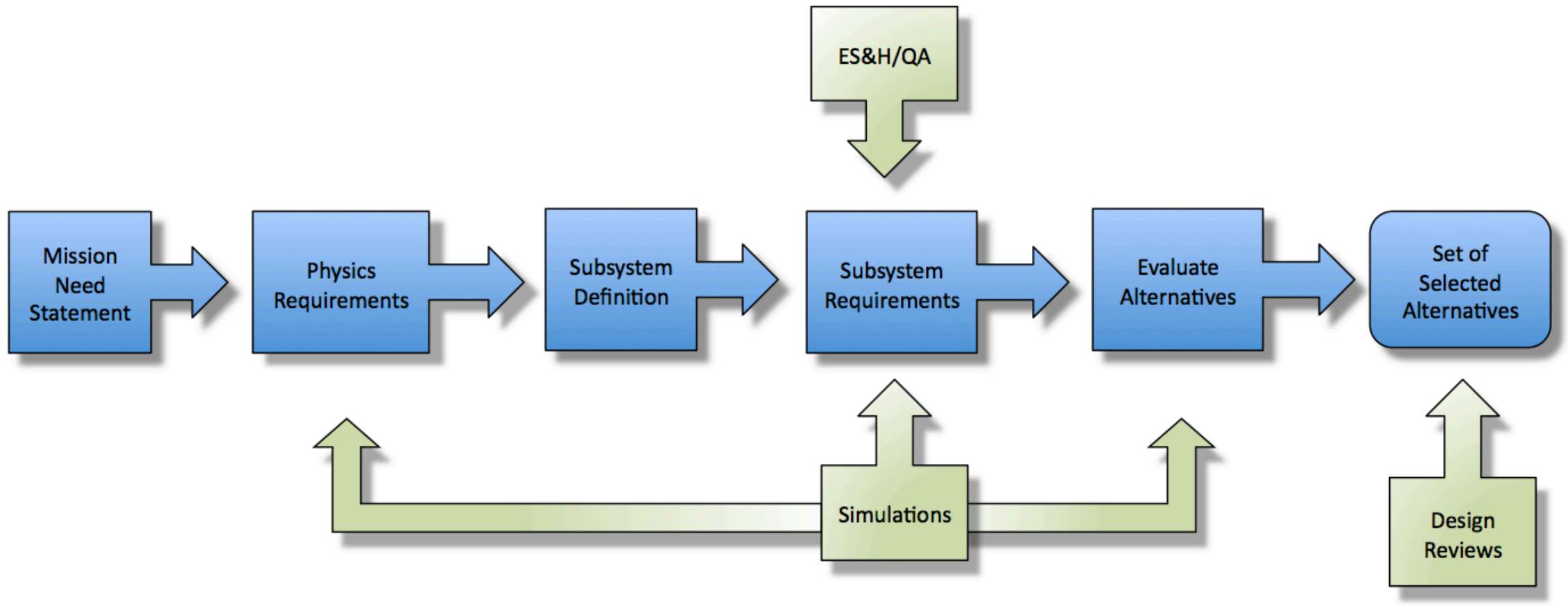

Figure 2.5. The Mu2e design process and flow-down of requirements.

The Mu2e schedule is determined in the early stages by the time required to navigate the CD process and in the later phases by the time required to construct and install the solenoids. This is true for any set of alternatives, so schedule has not played a significant role in alternative selection.

The technical performance of the Fermilab accelerator complex is well understood and the technical risks associated with the required accelerator modifications are relatively small. Evaluation of technical risks played a significant role in the selection of alternatives for several Mu2e subsystems including the tracker, the calorimeter crystals and the cosmic ray veto photodetectors.

## 2.6 Key Performance Parameters required to Obtain Expected Outcome

Project completion (CD-4) will be accomplished when the scope defined in the WBS dictionary has been completed and the apparatus has been demonstrated to be functioning by achieving Key Performance Parameters (KPP) listed below. The WBS dictionary is under change control. After achieving the Key Performance Parameters, the Project Manager will request acceptance and approval of CD-4. The Key Performance Parameters are those that demonstrate functionality of the system while achievement of beam parameters required for the experimental program will be obtained after routine tuning and operation of the accelerator complex.

Key Performance Parameters for CD-4:

- Beamline is ready for commissioning with beam, as demonstrated by acceptance testing of individual components.

- Superconducting solenoid system is capable of producing a low energy muon beam and analyzing 105 MeV electrons, as demonstrated by acceptance testing.

- Detector system is ready for commissioning with beam, as demonstrated by identifying tracks from cosmic rays.

The Key Performance Parameters will evolve as the project matures and will be finalized and approved as part of the Performance Baseline at CD-2.

## 2.7 Cost and Schedule

A preliminary high-level summary of the cost range for the Mu2e Project, at the second level of the Work Breakdown Structure, is shown in Table 2.3. The cost range was developed from an initial bottoms-up midpoint estimate of the base cost plus estimate uncertainty. The overall estimate uncertainty for the technical component of the Project (not counting the largely Level-of-Effort Project Management subsystem) is 36%. The cost range is applied on top of this estimate uncertainty. The cost range was developed around the midpoint estimate by evaluating the risks and opportunities catalogued in the Mu2e Risk and Opportunity Registry [9]. The total Project cost for Mu2e is in the range \$208M - \$287M. The cost range, broken down by Level 2 subsystem, is shown in Table 2.4. A preliminary Level 0 milestone schedule to construct Mu2e is shown in Table 2.5. A schedule range of 60 - 84 months is proposed for the range of the construction project from CD3a to Project completion. 18 months of programmatic float has been added to the estimated Project completion date to arrive at a CD-4 date.

### *2.7.1* Life Cycle Cost Assumptions

The life cycle costs for Mu2e are based on known costs for other operational experiments and facilities. The costs are then appropriately scaled to Mu2e [10]. The costs are in FY12 dollars.

***Accelerator and Beamline costs***

Mu2e will run simultaneously with the 120 GeV neutrino program at Fermilab, using booster batches that are available while the Main Injector is ramping. Therefore, the differential cost of running Mu2e is limited to the cost of running the antiproton facility and the Muon Campus External Beamline.

| WBS | L2 Cost Element | Base Estimate ($k) | Estimate Uncertainty ($k) | Estimate Uncertainty % | Total |
|---|---|---|---|---|---|
| 1 | Project Management | 20,491 | 83 | 0% | 20,574 |
| 2 | Accelerator | 35,362 | 9,939 | 33% | 45,301 |
| 3 | Conventional Construction | 18,535 | 5,979 | 33% | 24,514 |
| 4 | Solenoids | 71,713 | 26,520 | 39% | 98,233 |
| 5 | Muon Beamline | 10,944 | 3,118 | 33% | 14,062 |
| 6 | Tracker | 6,971 | 1,915 | 33% | 8,886 |
| 7 | Calorimeter | 4,294 | 1,270 | 26% | 5,391 |
| 8 | Cosmic Ray Veto | 4,418 | 1,359 | 33% | 5,777 |
| 9 | Trigger & DAQ | 4,941 | 1,461 | 32% | 6,402 |
| | **Total** | **$177,669** | **$51,644** | **32%** | **$229,313** |

Table 2.3 Level 2 midpoint cost estimate (base cost + estimate uncertainty) breakout for the Mu2e Project. Costs are fully burdened and escalated into actual year $k. Estimate uncertainty percentage is for the work remaining. Completed work has an estimate uncertainty of 0%.

| | | Total Cost | | Cost Range | |
|---|---|---|---|---|---|
| WBS | L2 Subsystem | Base + Estimate Uncertainty | Estimate Uncertainty % (on Remaining Work) | Low End | High End |
| 1 | Project Management | 20,574 | 0% | 16,696 | 30,699 |
| 2 | Accelerator | 45,301 | 33% | 42,779 | 56,607 |
| 3 | Conventional Construction | 24,514 | 33% | 21,197 | 30,165 |
| 4 | Solenoids | 98,233 | 39% | 95,933 | 111,458 |
| 5 | Muon Beamline | 14,062 | 33% | 13,582 | 21,152 |
| 6 | Tracker | 8,886 | 33% | 8,123 | 10,442 |
| 7 | Calorimeter | 5,391 | 30% | 0 | 10,700 |
| 8 | Cosmic Ray Veto | 5,777 | 33% | 3,669 | 8,009 |
| 9 | Trigger & DAQ | 6,402 | 32% | 6,128 | 7,551 |
| | **Total** | **$229,313** | **32%** | **$208,107** | **$286,783** |

Table 2.4. The Mu2e Project Cost Range, by level 2 subsystem. Costs are fully burdened and escalated into actual year $k. Estimate uncertainty percentage is for the work remaining. Completed work has an estimate uncertainty of 0%.

### *Solenoids*

Based on the expected operating currents of the planned power converters, the Mu2e solenoids would require about 300 kW of power. The solenoids require both liquid nitrogen and liquid helium. The liquid helium will be recycled while the liquid nitrogen will not. Based on estimates, we will consume about 60 liters/hr of liquid nitrogen.

Based on experience with CDF and D0, the Mu2e cryo system will need constant monitoring and oversight. Dealing with small problems before they turn into big problems prevents downtime, loss of data and money. We estimate that 2 people will be required to monitor the cryo system 24/7 during data taking. This corresponds to about 8.5 FTEs of technician effort totaling about $1.2M per year in FY12 dollars.

| Major Milestone Events | Preliminary Schedule |
|---|---|
| CD-0 (Approve Mission Need) | $1^{st}$ Qtr, FY10 (A) |
| CD-1 (Approve Alternative Selection and Cost Range) | $4^{th}$ Qtr, FY12 |
| CD-2 (Approve Performance Baseline) | $2^{d}$ Qtr, FY14 |
| CD-3a (Approve Start of Long-lead Procurement) | $2^{d}$ Qtr, FY14 |
| CD-3b (Approve Start of Construction) | $4^{th}$ Qtr, FY15 |
| Key Performance Parameters Satisfied | $4^{th}$ Qtr, FY19 |
| CD-4 (Includes 18 months of programmatic float) | $2^{d}$ Qtr, FY21 |

Table 2.5 Preliminary Level 0 Milestone Schedule. Note that satisfying the Key Performance Parameters is not a Level 0 milestone but is added to provide context for the CD-4 milestone.

***Detector Hall***

The power requirements for the Mu2e detector hall, not including the power required to power the solenoids, assumes a 15% duty factor on 500 kW and a cost of $0.06/kWH resulting in a cost of $40k/year. This includes a $0.015 kW/h charge for the social cost of carbon imposed by the DOE. We assume a cost of $5.61/sq. ft. per year for the cost of facility maintenance. This is the day-to-day cost to sustain a property in a state that allows one to realize the anticipated useful life of a fixed asset [11]. The total cost of facility maintenance for the detector hall (15,500 sq. ft.) is $87k/year.

The annual operating costs for Mu2e are listed in Table 2.6.

***Decontamination and Decommissioning Costs***

Decontamination and decommissioning of the Mu2e detector is relatively straightforward by Fermilab standards. The proton production target and the Production Solenoid heat shield will experience significant irradiation. Remote handling will be required to remove these devices and they will be stored at a safe location at Fermilab along with similarly irradiated materials.

Mu2e is a relatively unique experiment and much of the equipment is unique to this application. The scintillating crystals for the calorimeter could potentially be used for another experiment, though their short length may limit their applicability. Other equipment that could be reused includes power supplies, cryo equipment, vacuum pumps, photo detectors and scintillator.

| Item | $k per year |
|---|---|
| Accelerator and Beamline | $3800 |
| Power costs, detector hall | $40 |
| Maintenance cost | $87 |
| Phones and networking | $25 |
| Solenoid power | $158 |
| Liquid nitrogen | $60 |
| Cryo plant power | $158 |
| Cryo System operation and monitoring | $1200 |
| Chamber gas (argon/CO2) | $20 |
| Data storage | $17 |
| **Total** | **$5565** |

Table 2.6. Annual operating costs for Mu2e.

The base cost to install the Mu2e detector is about $10M in AY dollars (assume installation in 2019-2020). We assume that the cost to remove the Mu2e detector from the detector hall is equal to the cost to install it. In AY$, where we assume D&D in 2030, we estimate the base cost to be $12.8M. A 30% contingency added to this estimate would bring it up to $16.6M. This is consistent with the estimate to disassemble the BaBar detector of $9.4M (FY07) [12].

***Total Life Cycle Costs***

The total life cycle cost for Mu2e assumes a 4-year run to commission the accelerator and detector and accumulate data followed by two years of data analysis [10]. When the experiment is operating and taking data all of the costs included in Table 2.6 are applicable. When the experiment is not operating only the detector hall power, maintenance, phone, networking and data storage costs apply. The life cycle cost for Mu2e, including D&D, is shown in Table 2.7 and totals about $39M.

| Activity | Duration | Annual Cost ($k) | Total Cost ($k) |
|---|---|---|---|
| Data taking | 4 years | $3765 | $22,260 |
| Data analysis | 2 years | $169 | $338 |
| Data processing infrastructure | 1 time charge | | $130 |
| D&D | | | $16,600 |
| **Total** | | | **$39,328** |

Table 2.7. Life cycle costs for Mu2e assuming a 4 year run followed by 2 years of data analysis.

## 2.8 References

[1] R. Walton, "Mu2e Strategy for Sustainability," Mu2e-doc-2005.

[2] R. Ray, "Mu2e Project Management Plan," Mu2e-doc-508.

[3] R. Ray, "Mu2e Risk Management Plan," Mu2e-doc-461.

[4] R. Ray, "Mu2e Configuration Management Plan," Mu2e-doc-509.

[5] R. Ray, "Mu2e Quality Management Plan," Mu2e-doc-677.

[6] http://www.fnal.gov/directorate/OPMO/OPMOhome.htm.

[7] R. M. Carey et al., "Mu2e Proposal," Mu2e-doc-388.

[8] R. Ray et al., "Acquisition Strategy for the Muon to Electron Conversion Project," Mu2e-doc-1074.

[9] M. Dinnon, "Mu2e Risk Registry," Mu2e-doc-1463.

[10] R. Ray, "Life Cycle Costs for Mu2e," Mu2e-doc-524.

[11] The Whitestone Building Maintenance And Repair Cost Reference, 2008 -2009.

[12] W. Wisniewski, "D&D Phase Overview," http://www-conf.slac.stanford.edu/bfactory-decom/Talks/Wisniewski2.pdf

# 3 Muon to Electron Conversion

## 3.1 Physics Motivation

Before the discovery of neutrino oscillations, it was generally understood that lepton flavor changing processes were forbidden in the Standard Model (SM) and that the lepton flavor numbers $L_e$, $L_\mu$ and $L_\tau$ were conserved. This is because neutrinos were taken to be massless, which trivially allows one to diagonalize the mass matrices for the charged leptons and neutrinos simultaneously. However, after the discovery of neutrino oscillations, we knew that mixing among the lepton families occurs, giving rise to lepton flavor violating (LFV) processes. This will also be generically true for any model that includes a mechanism for generating neutrino masses. The rate at which LFV processes occur in the neutrino sector is constrained by the measured neutrino mixing parameters, but the rate at which charged lepton flavor violating (CLFV) occur is model dependent and can vary over many orders of magnitude. For example, in the minimal extension to the SM where neutrino mass is generated by introducing three right-handed SU(2) singlet fields and three new Yukawa couplings, the CLFV process $\mu^- N \rightarrow e^- N$ can only occur through loop diagrams whose amplitudes are proportional to $(\Delta m^2_{ij} / M^2_w)^2$ where $\Delta m^2_{ij}$ is the mass-squared difference between the $i^{th}$ and $j^{th}$ neutrino mass eigenstates. Because the neutrino mass differences are so small relative to $M_w$ the rates of CLFV decays in the modified SM are effectively zero (e.g. $< 10^{-50}$ for both $\mu^+ \rightarrow e^+ \gamma$ and $\mu^- N \rightarrow e^- N$). On the other hand, many New Physics (NP) models predict significant enhancements to CLFV rates and to the $\mu^- N \rightarrow e^- N$ process in particular. Many well-motivated physics models predict rates for CLFV processes that are within a few orders of magnitude of the current experimental bounds. These include the MSSM with right-handed neutrinos, SUSY with R-parity violation; models with leptoquarks, new gauge bosons, large extra-dimensions and a non-minimal Higgs sector [1]. The Mu2e experiment, with a single-event sensitivity of a few $10^{-17}$ for the ratio of $\mu^- N \rightarrow e^- N$ conversions to conventional muon captures, will have excellent discovery potential over a wide range of new physics models and could prove to be a powerful discriminant.

### *3.1.1* Charged Lepton Flavor Violation

There is an active global program to search for CLFV processes using rare decays of muons, taus, kaons, and *B*-mesons. The ratio of rates among various CLFV processes is model dependent and varies widely depending on the underlying physics responsible. Thus, it is important to pursue experiments sensitive to different processes in order to elucidate the mechanism responsible for CLFV effects. The most stringent limits come from the muon sector because of the relative “ease” with which an intense source of muons can be produced. Three rare muon processes stand out: $\mu^+ \rightarrow e^+ \gamma$ , $\mu^+ \rightarrow e^+e^+e^-$ and $\mu^- N \rightarrow e^- N$. Searches for these processes have yielded null results and set upper limits on the corresponding rates. The experimental limits (all at 90% CL) on the branching ratios are: for B($\mu^+ \to e^+ \gamma$ ) < 2.4 × $10^{-12}$ ([5]), B($\mu^+ \to e^+e^+e^-$)< 1.0 × $10^{-12}$ [3], and $R_{\mu e}$(Ti) ($\mu \to e$ conversion on gold) < 7 × $10^{-13}$ [19]. In this decade significant improvement is possible on all three modes.

The MEG experiment [5], operating at PSI, has already reached 2.4 × $10^{-12}$ and hopes to achieve ≥ $10^{-13}$ for the $\mu^+ \to e^+\gamma$ branching ratio, while the proposed COMET [6] experiment at JPARC and Mu2e at Fermilab will each reach sensitivities of $10^{-16}$ – $10^{-17}$ on $R_{\mu e}$(Al). It is important to note that these two processes have complementary sensitivity to new physics effects and the results from both are helpful in order to untangle the underlying physics. To illustrate this one can estimate the sensitivity of a given CLFV process in a model independent manner by adding lepton-flavor-violating effective operators to the Standard Model Lagrangian where Λ is the mass scale of new

$$L_{\mathrm{CLFV}} = \frac{m_\mu}{(1+\kappa)\Lambda^2}\bar{\mu}_R\sigma_{\mu\nu}e_L F^{\mu\nu} + \frac{\kappa}{(1+\kappa)}\bar{\mu}_L\gamma_\mu e_L\left(\sum_{q=u,d}\bar{q}_L\gamma^\mu q_L\right),$$

physics and κ is an arbitrary parameter controlling the relative contribution of the two terms [7]. Most new physics contributions are accounted for in these two classes of effective operators. If κ << 1, the first term, a dimension five magnetic-moment-type operator, is dominant. If κ >> 1, the second term, a four-fermion interaction-type operator, is dominant. Simply put, the first term arises from loops with an emitted photon. If the photon is real, one observes $\mu^+ \to e\,\gamma$. The second term includes contact terms and a variety of other processes not resulting in a photon. Therefore, the $\mu^- N \to e^- N$ and $\mu^+ \to e^+e^+e^-$ processes are sensitive to new physics regardless of the relative contributions of the first and second terms. The new physics scale, Λ, to which these two processes are sensitive as a function of κ is shown in Figure 3.1. The projected sensitivity of the MEG experiment will probe Λ values up to 2000 - 4000 TeV for κ << 1 scenarios, while having little sensitivity for the case that κ >> 1. The projected sensitivity of the Mu2e experiment will probe Λ values from 3000 to over 10000 TeV over *all* values of κ. It should be noted that these effective operators provide a good description of most of the new physics scenarios in which large CLFV effects might appear in $\mu^+ \to e^+\gamma$ and $\mu^- N \to e^- N$, and the conclusions regarding relative sensitivity are generically true. As demonstrated by Figure 3.1, a Mu2e experiment sensitive to rates in the range of $10^{-16}$ – $10^{-17}$ is interesting and important in all MEG scenarios. If MEG observes a signal, then Mu2e should also, and the ratio of measured rates can be used to simultaneously constrain Λ and κ (limiting which types of new physics models remain viable). On the other hand, a null result from MEG does not preclude a Mu2e discovery since the new physics may be dominated by interactions to which the $\mu^+ \to e\,\gamma$ process is blind. An example of the complementary nature of these two processes in the context of a specific model is provided in Figure 3.2, which depicts a scan of the parameter space of a Littlest Higgs Model with T-parity [8]. The different colored points refer to different choices for the structure of the mirror-lepton mixing matrix that gives rise to the CLFV effects. The combination of results from MEG and Mu2e would severely constrain the allowed parameter space of this model and could distinguish between the Littlest Higgs Model and the Minimal Supersymmetric models in a transparent way, as the correlations between the two CLFV processes are significantly different in the two models. Several other specific examples are discussed in Ref. [1].

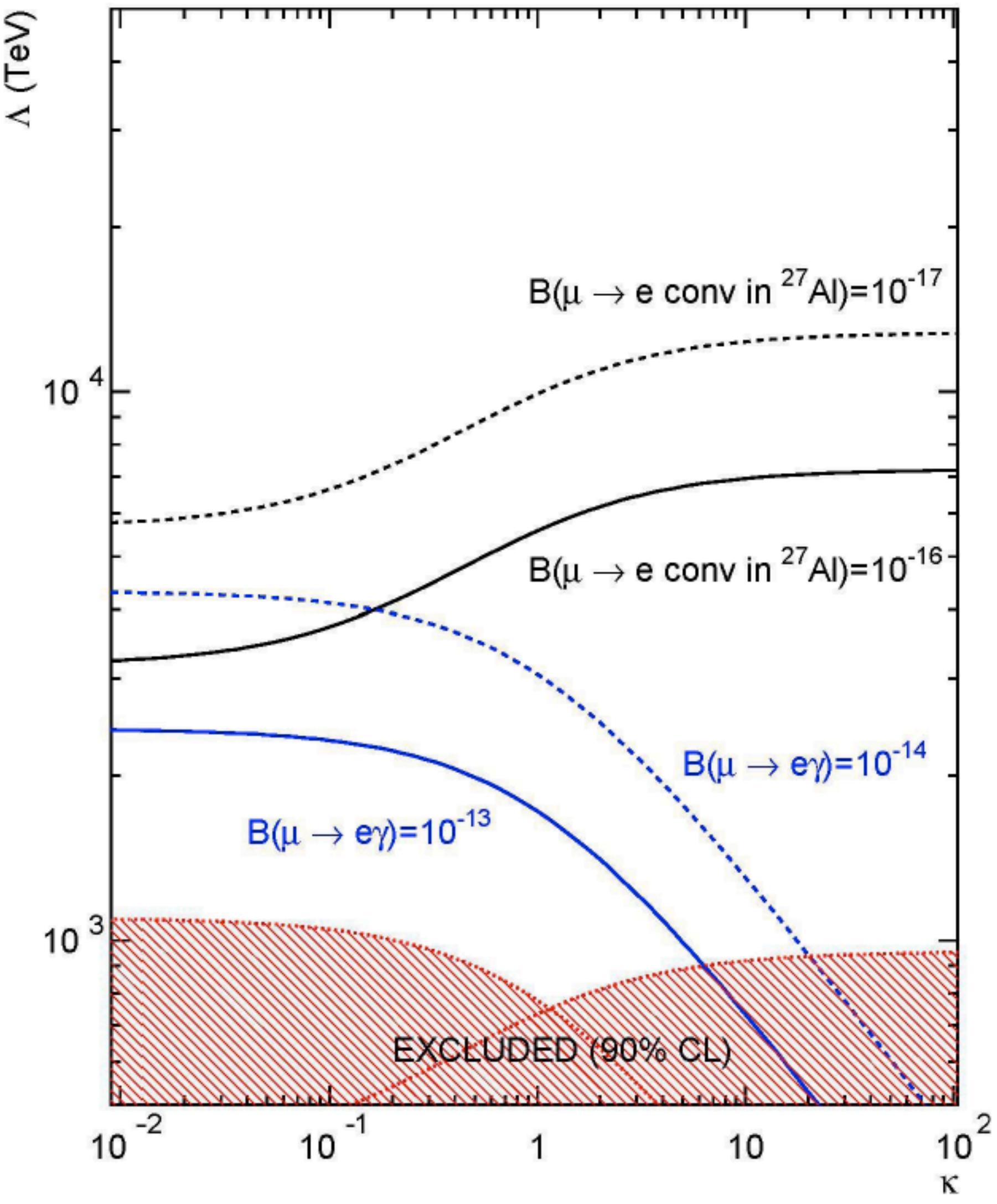

Figure 3.1. The sensitivity to the scale of new physics, Λ, as a function of κ, for a muon to electron conversion experiment with a sensitivity of $10^{-16} - 10^{-17}$ is compared to that for a muon-to-electron-gamma experiment with a sensitivity of $10^{-12} - 10^{-13}$. See the text for a definition of κ. The excluded region of parameter space, based on current experimental limits, is shaded.

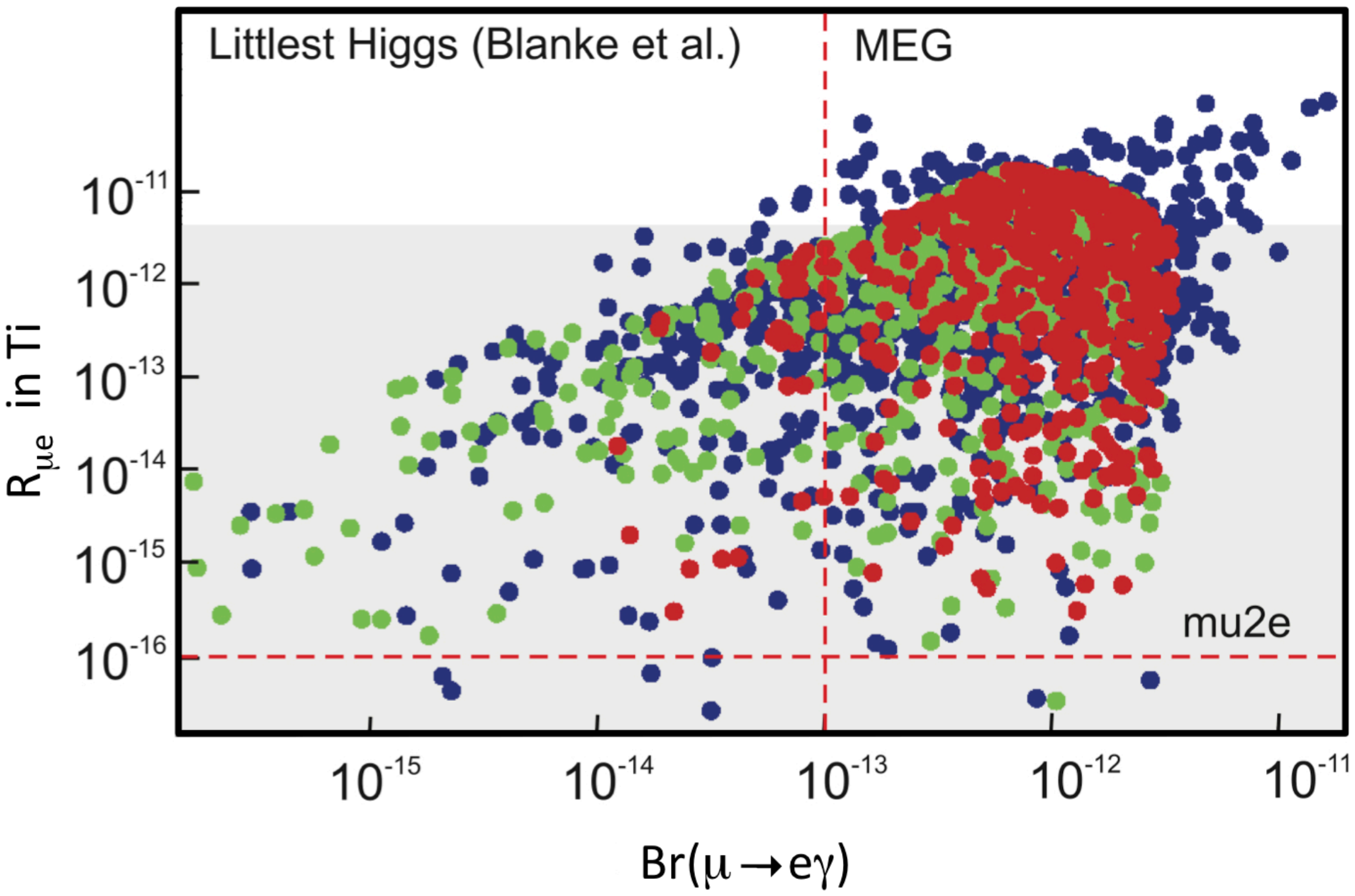

Figure 3.2. The predicted rate of muon to electron conversion in titanium is compared to the predicted branching ratio for μ → eγ in the context of the Littlest Higgs model with T-parity [8]. The red points assume that a PMNS-like matrix describes the mixing matrix of the mirror leptons, while the green points assume that a CKM-like Matrix describes the mixing. The blue points are a general scan of the parameters of the mirror lepton mixing matrix. The shaded region is the parameter space not excluded by current CLFV results in conversion experiments.

### *3.1.2* Charged Lepton Flavor Violation in the LHC Era

By the time the next generation of CLFV experiments reach their target sensitivities, the LHC experiments are expected to have analyzed many $fb^{-1}$ of data collected at center-of-mass energies of 7 TeV or higher, thus exploring physics at the TeV scale. The HEP community is anxious to learn whether the LHC data reveals the new phenomena predicted by many new physics models. The importance and impact of pursuing next generation CLFV experiments is independent of what the LHC data might reveal over the next several years. As discussed in the previous sections, these and other ultra-rare processes probe new physics scales that are orders of magnitude beyond the direct reach of the LHC, and thus may offer the only evidence of new physics phenomena should it lie at mass scales significantly above the TeV range. A more optimistic scenario would assume new physics discoveries at the LHC *do* occur and ask to what degree do measurements of CLFV processes complement the LHC experiments? The LHC experiments do not, aside from specially-tuned cases, have sensitivities to CLFV processes that approach that of next-generation $\mu^+ \to e^+\gamma$ and $\mu^- N \to e^- N$ experiments. Thus, Mu2e probes the underlying physics in a unique manner with a sensitivity that is significantly better than any other CFLV process can hope to accomplish on a similar timescale. Moreover, many of the new physics scenarios for which the LHC has discovery potential predict rates for $\mu^- N \rightarrow e^- N$ in the discovery range for Mu2e (i.e. larger than $10^{-16}$). Figure 3.2 shows an example in the context of a specific new physics model where the parameter space corresponds to those scenarios in which the LHC would discover new phenomena. As discussed above, this is a Littlest Higgs with T-parity model and it is clear that the information provided by the CFLV measurements would help pin down the viable parameter space. Another example is given in Figure 3.3, which shows the predicted $\mu^- N \rightarrow e^- N$ rate in titanium as a function of the universal gaugino mass at the GUT scale, $M_{1/2}$, in the context of an SO(10) SUSY GUT model [9]. The SUSY parameter space is explored for those scenarios for which the LHC has discovery sensitivity and the different color points correspond to different assumptions about the neutrino Yukawa couplings. Again, a measurement of the $\mu^- N \rightarrow e^- N$ rate would significantly restrict the viable parameter space in a manner that the LHC experiments are unable to do. We have only examined two specific models here but the results are representative of the power of muon-to-electron conversion. It is generally understood that an experiment sensitive to $\mu^- N \rightarrow e^- N$ rates at the level of $10^{-16}$ to $10^{-17}$ would have discovery potential that overlaps the parameter space to which the LHC is sensitive and would help constrain that parameter space in a manner complementary to what the LHC experiments can accomplish on their own [1][9].

## 3.2 Signal and Backgrounds for Muon Conversion Experiments

The conversion of a muon to an electron in the field of a nucleus is coherent: the muon recoils off the entire nucleus and the kinematics are those of two-body decay. The mass of a nucleus is large compared to the electron mass so the recoil terms are small. A conversion electron is therefore monoenergetic with energy slightly less than the muon rest mass (more detail is given below). The muon energy of 105.6 MeV is well above the maximum energy of the electron from muon decay (given by the Michel spectrum) at 52.8 MeV; hence, the vast majority of muon decays do not contribute background, subject to an important qualification discussed below. This distinctive signature has several experimental advantages including the near-absence of background from accidentals and the suppression of background electrons near the conversion energy from muon decays.

When a negatively charged muon stops in a target it rapidly cascades down to the 1S state [41]. Capture, decay or conversion of the muon takes place with a mean lifetime that has been measured in various materials and ranges from less than ~100 ns (high-Z nuclei) to over 2 μs (low-Z nuclei) [10]. Neutrinoless conversion of a muon will produce an electron with an energy that is slightly less than the rest mass of the muon and depends

on the target nucleus:

$$E_e = m_\mu c^2 - B_\mu(Z) - C(A),$$

where Z and A are the number of protons and nucleons in the nucleus, $B_\mu$ is the atomic binding energy of the muon and C(A) is the nuclear recoil energy. In the case of muonic aluminum, the energy of the conversion electron is 104.97 MeV and the muon lifetime is 864 ns [10]. An electron of this energy signals the conversion.

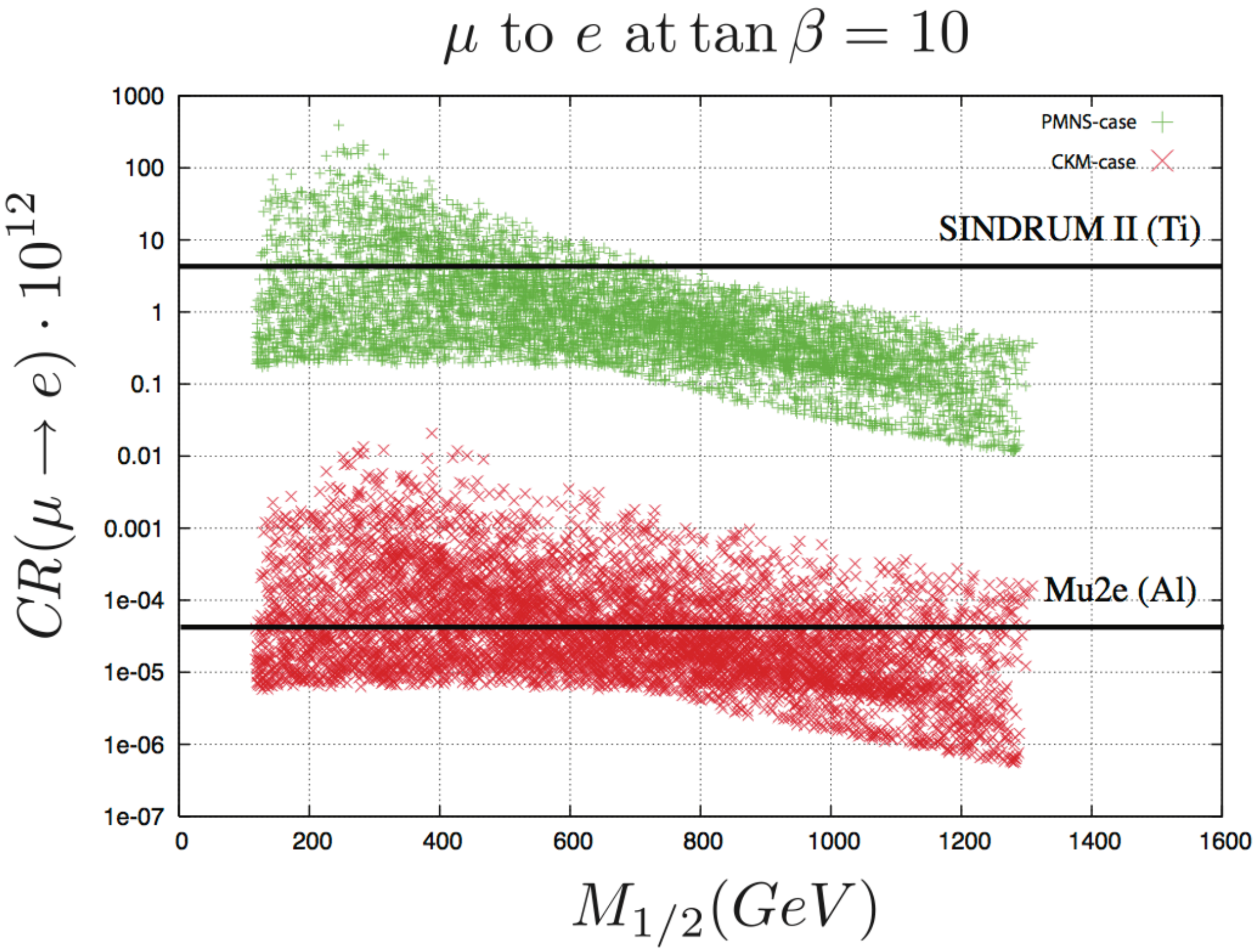

Figure 3.3. The predicted rate for muon-to-electron conversion in titanium for various scenarios in the context of the SUSY GUT model described in [9]. The SUSY parameter space explored corresponds to that for which the LHC has discovery sensitivity. The different colored points correspond to different assumptions about the neutrino Yukawa couplings. The horizontal lines represent the current limit [4] and the projected sensitivity of Mu2e.

At the proposed Mu2e sensitivity there are a number of processes that can mimic a muon-to-electron conversion signal. Controlling these potential backgrounds drives the overall design of Mu2e. These backgrounds result principally from five sources:

1. Intrinsic processes that scale with beam intensity; these include muon decay-in-orbit (DIO) and radiative muon capture (RMC).

2. Processes that are delayed because of particles that spiral slowly down the muon beam line, such as antiprotons.
3. Prompt processes where the detected electron is nearly coincident in time with the arrival of a beam particle at the muon stopping target (e.g. radiative pion capture, RPC).
4. Electrons or muons that are initiated by cosmic rays.
5. Events that result from reconstruction errors induced by additional activity in the detector from conventional processes.

A free muon decays according to the Michel spectrum with a peak probability at the maximum energy at about half the muon rest energy (52.8 MeV) and far from the 105 MeV conversion electron energy. If the muon is bound in atomic orbit, the outgoing electron can exchange momentum with the nucleus, resulting in an electron with energy (ignoring the neutrino mass) equal to that of a conversion electron, however with very small probability. At the kinematic limit of the bound decay, the two neutrinos carry away no momentum and the electron recoils against the nucleus, simulating the two-body final state of muon to electron conversion. The differential energy spectrum of electrons from muon decay-in-orbit falls rapidly near the endpoint, approximately as $(E_{endpoint} - E_e)^5$. The spectrum of electron energies that results from muon decays in orbit in aluminum, our target of choice, is illustrated in Figure 3.4 where the most prominent feature is the Michel peak. As described above, the nuclear recoil slightly distorts the Michel peak and gives rise to a small tail that extends out to the conversion energy. Because of the rapid decrease in the DIO rate as the electron energy approaches the endpoint, the background can be suppressed through adequate resolution on the electron momentum. To reduce the DIO background, the central part of the energy resolution function must be narrow and high energy tails must be suppressed. This depends on the intrinsic resolution of the tracker detector as well as the amount of material traversed by conversion electrons.

To date, there have been no experimental measurements of the DIO spectrum with sufficient sensitivity near the endpoint energy. The rate is very low and therefore far more stopped muons than in previous experiments are required to see it. The shape of the spectrum near the endpoint is dominated by phase space considerations that are generally understood but important corrections to account for nuclear effects must also be included. The veracity of these corrections is untested by experiment. However, a number of theoretical calculations of the DIO spectra of various nuclei have been done over the years, in particular a recent one by Czarnecki, et al. [24]. The uncertainty in the rate versus energy near the endpoint is estimated at less than 20%.

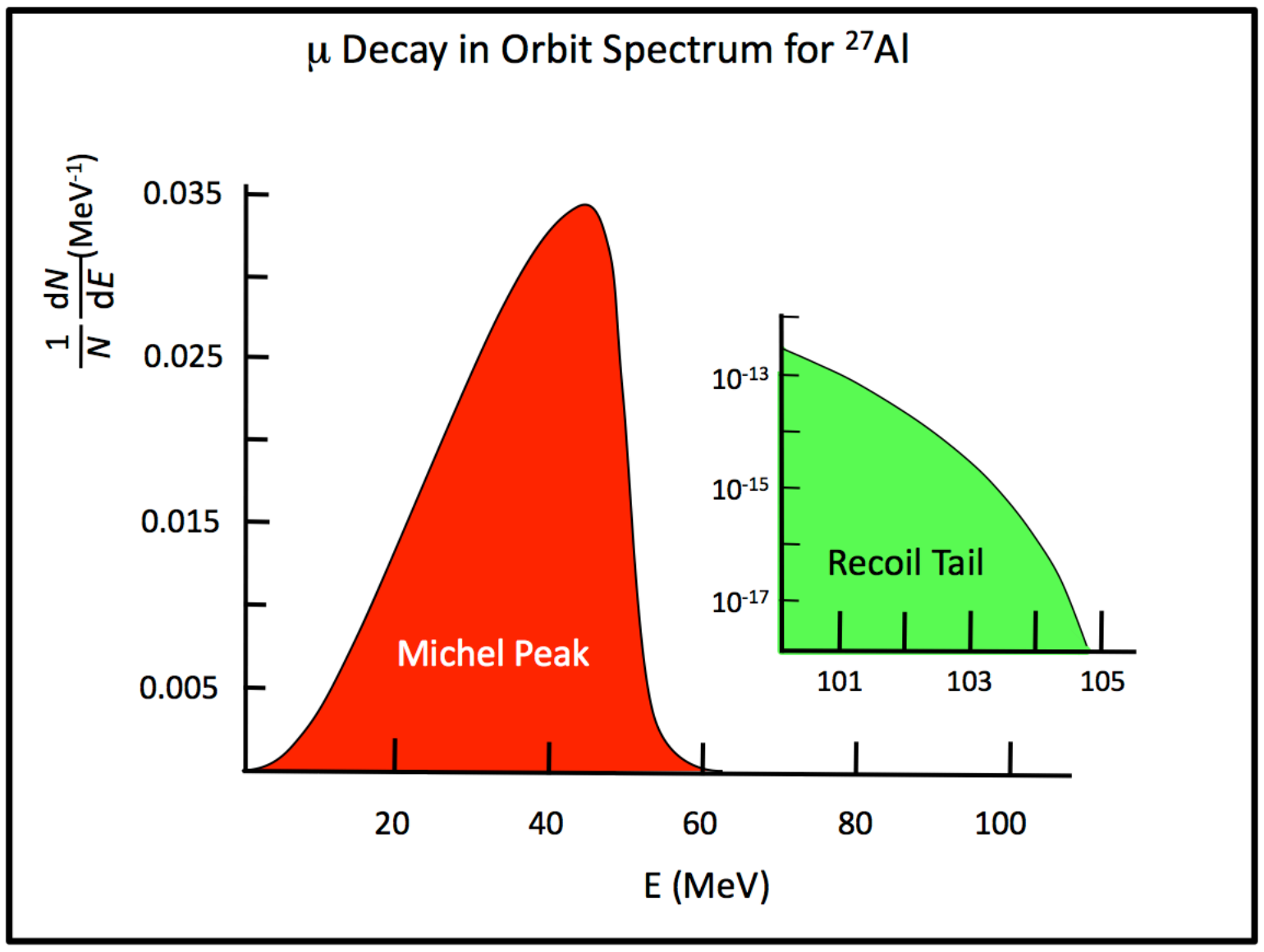

Figure 3.4. The electron energy spectrum for muon decay-in-orbit in aluminum. The recoiling nucleus results in a small tail (blown up on the right) that extends out to the conversion energy.

Radiative muon capture on the nucleus ($\mu^-$ Al $\rightarrow \gamma\, \nu$ Mg) is an intrinsic source of high energy photons that can convert to an electron-positron pair in the stopping target or other surrounding material, producing an electron near the conversion electron energy. Photons can also convert internally. These internal and external rates, by numerical accident, are approximately equal for the Mu2e stopping target configuration. Radiative muon capture can produce photons with an endpoint energy close to the conversion electron energy but shifted because of the difference in mass of the initial and final nuclear states. Ideally, the stopping target is chosen so that the minimum masses of daughter nuclei are all at least a couple of MeV/c$^2$ above the rest mass of the stopping target nucleus, in order to push the RMC photon energy below the conversion electron energy; for aluminum the RMC endpoint energy is 102.4 MeV, about 2.6 MeV below the conversion electron energy. The shape of the photon spectrum and the rate of radiative muon capture are not well known for medium mass nuclei and experiments have not had enough data to observe events near the kinematic endpoint. The electrons that result from photon conversions cannot exceed the RMC kinematic endpoint for the energy of the radiated photon, so the planned energy resolution of the conversion peak (on the order of 1 MeV FWHM including energy straggling and tracking uncertainties) can render this background negligible.

Most low-energy muon beams have large pion contaminations. Pions can produce background when they are captured in the stopping target or surrounding material and produce a high energy photon through radiative pion capture (RPC):

$$\pi^- N \to \gamma N^*$$

RPC occurs in 2.1% of pion captures for an aluminum target. The kinematic endpoint is near the pion rest mass energy with a broad distribution that peaks at about 110 MeV. If the photon then converts in the stopping material, one sees an electron-positron pair and in the case of an asymmetric conversion, the outgoing electron can be near the conversion energy. In addition, the photon can internally convert:

$$\pi^- N \to e^+ e^- N^*$$

and by numerical accident, the internal and external conversion rates are about equal. Thus electrons resulting from photon conversions, both internal and external, can produce background. RPC background can be suppressed with a pulsed proton beam: the search for conversion electrons is delayed until virtually all pions have decayed or annihilated in material. Beam electrons near the conversion energy that scatter in the target, along with the in-flight decay of a muon in the region of the stopping target are other examples of prompt backgrounds.

Cosmic rays (electrons, muons, photons) are a potential source of electrons near the conversion electron energy. If such electrons have trajectories that appear to originate in the stopping target they can fake a muon conversion. Identifying an incoming cosmic ray particle can reject these events. Passive shielding and veto counters around the spectrometer help to suppress this background. Note that this background scales with the experiment's live time rather than with beam intensity.

Track reconstruction can be affected by other activity in the detector, causing tails in the energy resolution response function that can move low-energy DIO electrons into the signal momentum window. Additional activity in the detector primarily originates from the muon beam, from multiple DIO electrons within a narrow time window, and from muon capture on a target nucleus that results in the emission of photons, neutrons and protons. The protons ejected from the nucleus following muon capture have a very small kinetic energy and are highly ionizing, so the large pulses they leave behind in tracking chambers can shadow hits from low energy electrons, potentially adding to the likelihood of reconstruction errors. Ejected neutrons can be captured on hydrogen or other atoms and produce low-energy photons. Low-momentum electrons can be created in the tracker by photons that undergo Compton scattering, photo-production, or pair production, and by delta-ray emission from electrons and protons. Because of the low mass of the tracker, these electrons can spiral a considerable distance through the detector before they range out, generating a substantial number of in-time hits. Electron-generated hits caused by neutron-generated photons are the most common and difficult to remove form of background activity. The rate of background activity scales linearly with beam intensity. The momentum resolution tails depend roughly linearly on the rate of additional detector activity. The scaling rates can be controlled through careful design of the detector and reconstruction software, and by using estimates of track reconstruction quality when selecting physics samples.

## 3.3 Previous Muon Conversion Experiments

Lagarrigue and Peyrou first searched for muon to electron conversion in 1952 [11]. Many other searches have been performed since ([12] - [19]). The techniques employed in the most recent experiments are particularly noteworthy and provide important input for more sensitive searches.

In 1988 a search for muon to electron conversion was performed at TRIUMF [18]. A 73 MeV/c muon beam was stopped in a titanium target at a rate of $10^6$ $\mu^-$/sec. A hexagonal time projection chamber located in a uniform 0.9 Tesla axial field was used to measure the energy of electrons. Scintillation counters were used to tag those electron candidates coincident with the arrival of a particle at the stopping target as prompt background. No events were observed with energies consistent with the muon-to-electron conversion hypothesis. However, nine events with momenta exceeding 106 MeV/c were observed. The source of these events was thought to be cosmic rays, a hypothesis that was later confirmed in a separate experiment that measured the cosmic ray induced background with the beam turned off. The limit from the TRIUMF search was $4.6 \times 10^{-12}$ (90% CL).

The 1993 SINDRUM II experiment, performed at PSI, focused negative muons with a momentum of 88 MeV/c and an intensity of $1.2 \times 10^7$ $\mu^-$/sec on a Titanium target [4]. During a 50 day run a total of $3 \times 10^{13}$ muons were stopped. The electron energy was measured with a spectrometer inside a superconducting solenoid with a 1 Tesla field. The spectrometer consisted of several cylindrical detectors surrounding the target on the beam axis. Two drift chambers provided the tracking while scintillation and Cerenkov hodoscopes were used for the timing of the track elements and electron identification. A scintillation beam counter in front of the target helped to recognize prompt background electrons produced by radiative capture of beam pions or beam electrons scattering off the target. The pion contamination was reduced by a factor of $10^6$ by passing the beam through a thin moderator that reduced the muon flux by 30%. The few surviving pions had very low momenta and a simulation showed that ~ 99.9% of them decayed before reaching the target. Electrons from radiative pion capture in the moderator could reach the target and scatter into the detector solid angle. This background was easily recognized since it was strongly peaked in the forward direction and had a characteristic time correlation with the cyclotron RF.

Figure 3.5 shows electron candidates from the SINDRUM II experiment on Titanium. The electron spectrum is well described by muon decay-in-orbit. No events were observed with energies consistent with the muon-to-electron conversion hypothesis resulting in a limit of $6.1 \times 10^{-13}$ (90% CL).

In 2000 SINDRUM II performed a new search for muon to electron conversion using a 53 MeV/c muon beam and a gold target [19]. During a 75-day live time $4.4 \times 10^{13}$ muons were stopped. The results are shown in Figure 3.6. The sample was divided into events occurring in the 10 nsec after the beam pulse (prompt) and the remaining 10 nsec. Forward prompt events have been removed from the top plot where the electron spectrum is well described by muon decays in orbit and there are no events observed in the signal region. One electron event, thought to be pion induced, was identified at higher energy. This interpretation is consistent with the data shown in the bottom plot of Figure 3.6 that contains only prompt forward events where we would expect an enhancement from the fast radiative pion capture process. A final limit on muon to electron conversion in gold was set at $7 \times 10^{-13}$ (90%).

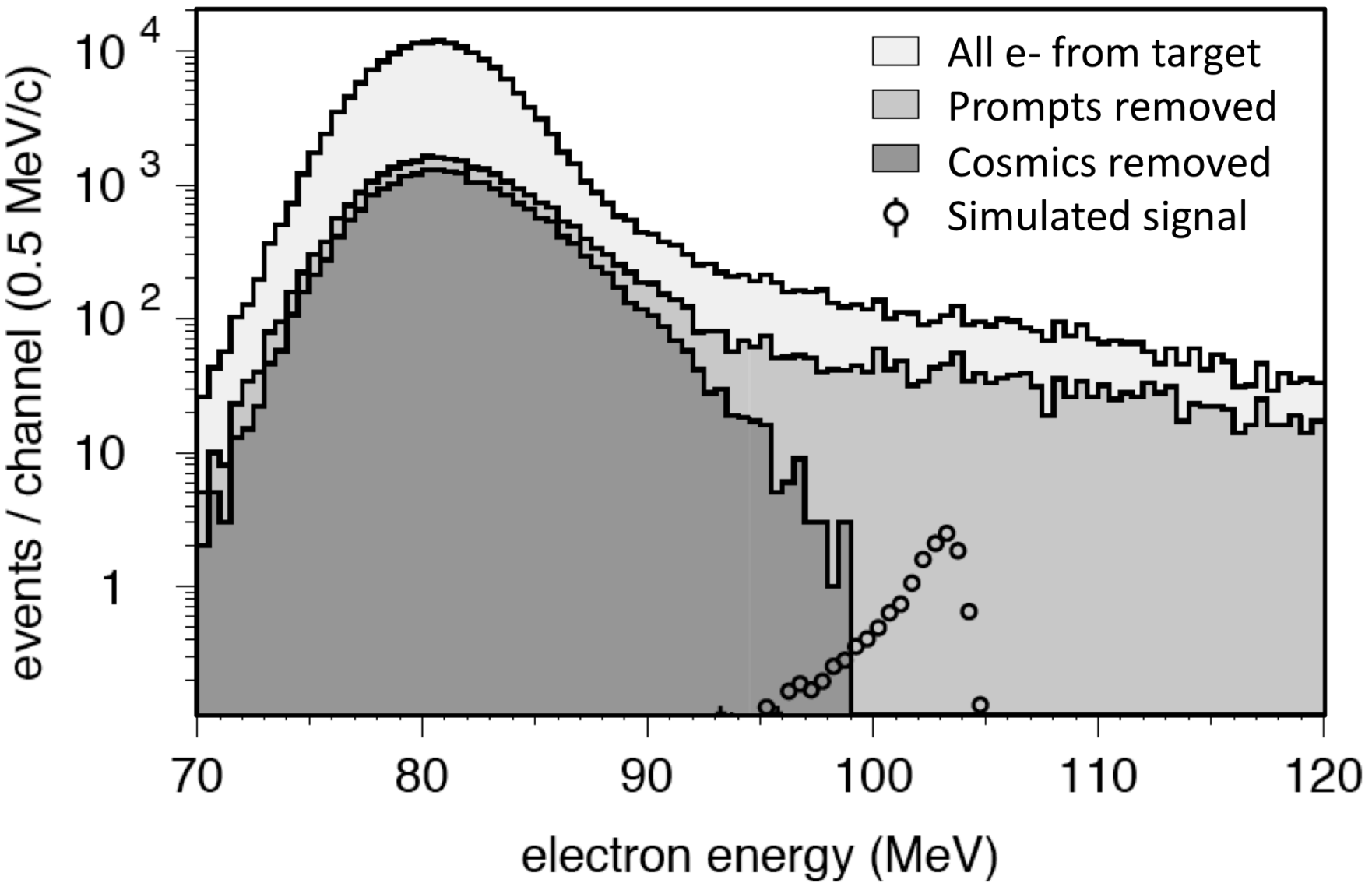

Figure 3.5. Electron energy spectrum from the 1993 SINDRUM II experiment using a Titanium target. Shown are all electrons from the target and the remaining events after suppression of prompt and cosmic ray backgrounds. A GEANT simulation of conversion electrons is superimposed on the data assuming $R_{\mu e} = 4 \times 10^{-12}$.

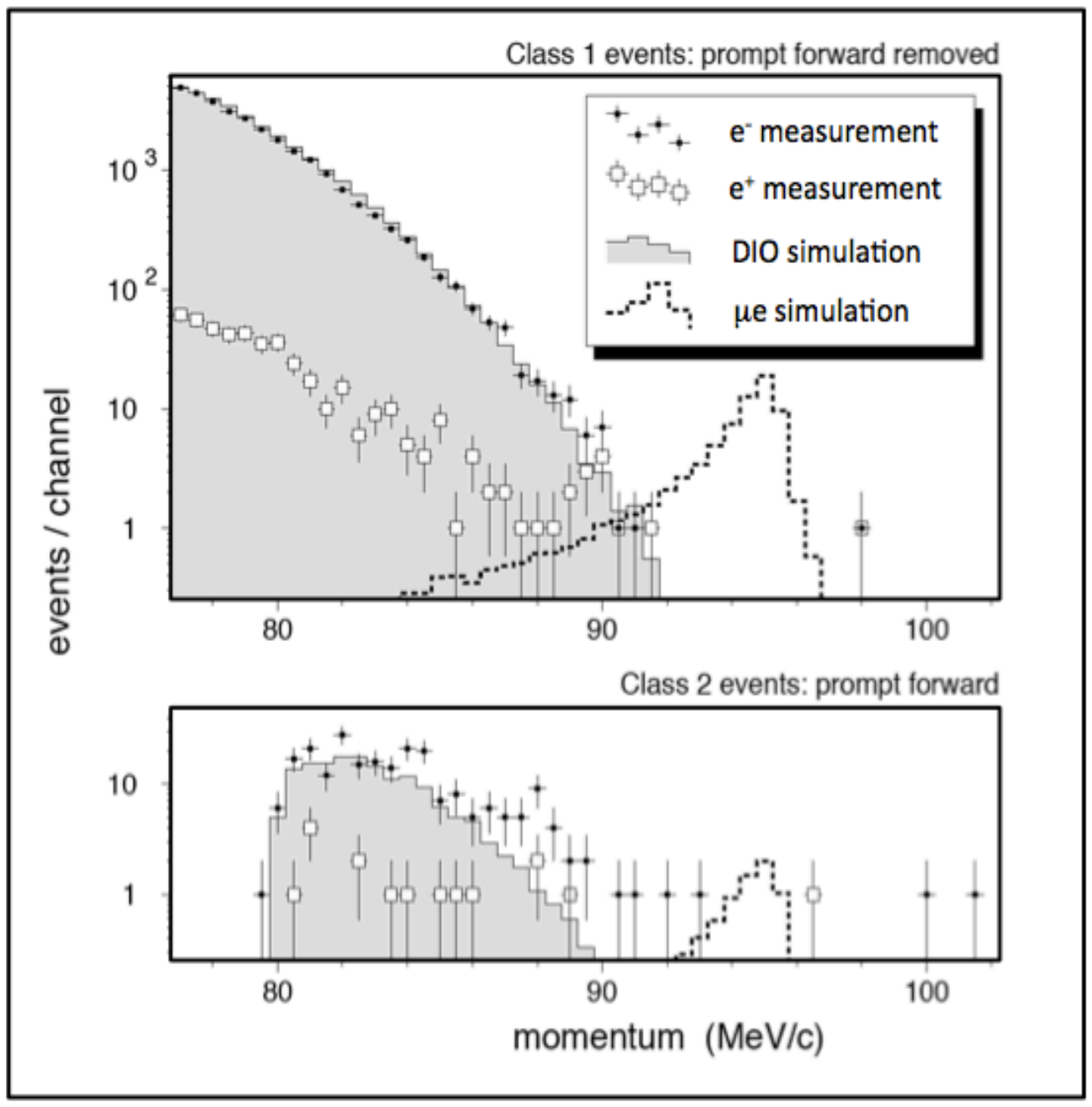

Figure 3.6. Momentum distribution for electrons from the SINDRUM II experiment using a gold target. Forward prompt events have been removed from the top plot while the bottom plot shows only prompt forward events where an enhancement from radiative pion captures would be expected. Note that the conversion energy for gold is 95.6 MeV.

## 3.4 Overview of Mu2e

Previous muon to electron conversion experiments have failed to observe events in the signal region, though events at higher energies have been observed that have been attributed to pion background and cosmic rays. Based on these results there would appear to be considerable room for improvement for an experiment with sufficient muon intensity, momentum resolution and rate capability so long as prompt backgrounds and cosmic rays are controlled. Mu2e proposes to improve on previous measurements by a factor of approximately 10,000 by deployment of a highly efficient solenoidal muon beam channel and a state-of-the-art detector combined with the power and flexibility of Fermilab's accelerator complex. The major improvements implemented for Mu2e that make this significant leap in sensitivity possible are discussed below. The Mu2e apparatus is shown in Figure 3.7.

An integrated array of superconducting solenoids forms a graded magnetic system that includes the *Production Solenoid*, the *Transport Solenoid* and the *Detector Solenoid*. The Production Solenoid contains the *production target* that intercepts an 8 GeV kinetic energy, high intensity, pulsed proton beam. The S-shaped Transport Solenoid transports low energy $\mu^-$ from the Production Solenoid to the Detector Solenoid and allows sufficient path length for a large fraction of the pions to decay to muons. Additionally, the Transport Solenoid attenuates nearly all high energy negatively charged particles, positively charged particles and line-of-sight neutral particles. The upstream section of the Detector Solenoid houses the muon stopping target and has a graded magnetic field. The graded field increases the acceptance for conversion electrons and plays a key role in rejecting certain backgrounds. The downstream section of the Detector Solenoid has a nearly uniform field (<1% non-uniformity) in the region occupied by the *tracker* and the *calorimeter*. The tracking detector is made from low mass straw tubes oriented transverse to the solenoid axis. The current Mu2e pattern recognition program is a based on a hit-level Monte Carlo with almost all known sources of accidental activity. This simulation, the pattern recognition, and the tracking algorithm are under intense and continuing development and numbers quoted in this CDR should be considered a snapshot of an ongoing process. A pattern recognition program has been developed for Mu2e. The track fitting algorithm is a Kalman Filter adapted from BaBar. We have chosen a set of cuts on the Kalman Filter reconstructed variables that will certainly evolve with further study; those cuts are reflected in the resolutions quoted below.

The momentum resolution is dominated by fluctuations in the energy lost in the target and proton absorber (Chapter 8), multiple scattering, and bremsstrahlung of the electron in the tracker. Energy loss in material creates the long low-side tail and the tracker resolution plays little role. Current estimates for the resolution are presented in Section 3.5.2. The calorimeter consists of LYSO crystals arranged in four vanes. All of these components will be discussed in detail later in this report.

To increase the sensitivity to muon-to-electron conversion by a factor of 10,000 the intensity of stopped muons will be increased to $1.3 \times 10^{10}$ per second. This significant increase in stopped muons is achieved by placing the production target in a graded solenoidal field that varies from 2.5 – 4.6 T. A proton beam enters the Production Solenoid moving in the direction of increasing field strength, opposite the outgoing muon beam direction and away from the detectors. A large fraction of the confined pions decay, producing muons. The graded field steadily increases the pitch of the muons, effectively accelerating them into the lower field of the Transport Solenoid that transports negatively charged muons within the desired momentum range to the stopping target.[1] The MuSIC R&D effort at Osaka University has recently validated this approach, demonstrating the principle of high muon yields from a target in a superconducting solenoid for the first time [23]. For the Mu2e system, using the QGSP-BERT model of particle production, the resulting efficiency is ~0.0016 stopped muons per incident proton.

[1] This overall scheme was first suggested by Djilkibaev, Lobashev and collaborators in an earlier proposal called MELC [20]. Proponents of the muon collider have subsequently adopted their ideas for muon collection in graded solenoids [21][22].

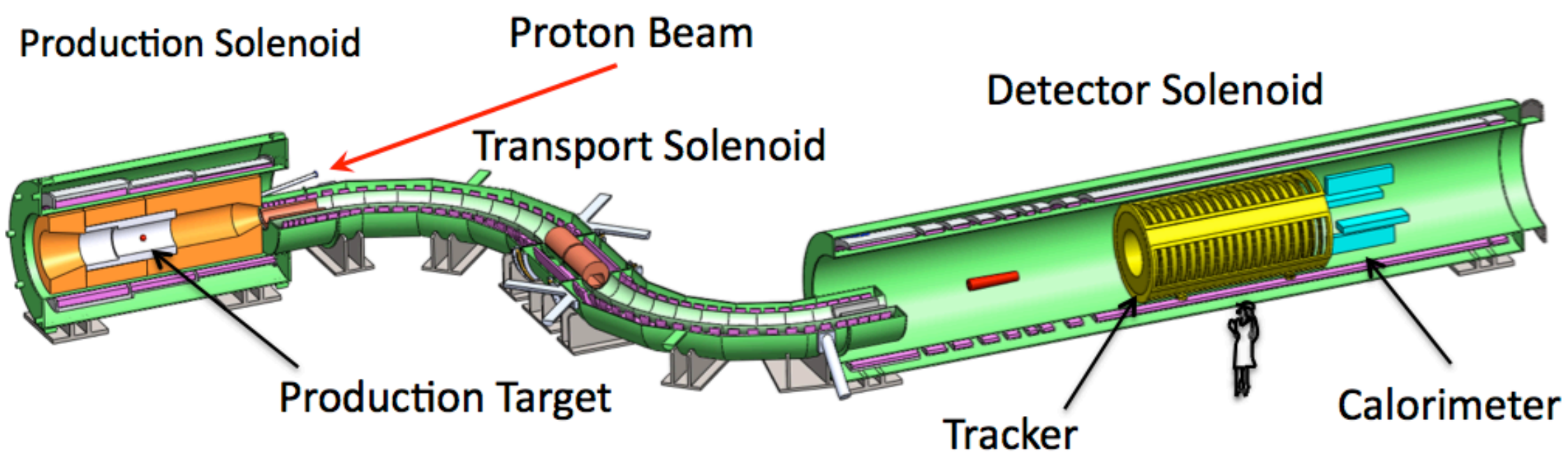

Figure 3.7. The proposed Mu2e apparatus. The Cosmic Ray Veto that surrounds the detector Solenoid and absorbers inside the Detector Solenoid are not shown.

We described the previous best experiment, SINDRUM-II, in Section 3.3. The SINDRUM method of using beam counters to tag and veto prompt backgrounds can no longer be used at the rates required for Mu2e. Those prompt backgrounds are dominated by radiative pion captures (or RPCs, described in Section 3.5.3) in their conversion stopping target. The relevant timescale was the pion lifetime of 21 nsec. The PSI beam of SINDRUM and SINDRUM-II was a continuous stream of short beam bursts every 20 nsec. Therefore the timescales are comparable and this process limited the experiment. If we use a pulsed beam with separation large compared to the pion lifetime, one can wait for pions to decay and thereby largely eliminate the radiative pion capture background. Since the muon lifetime in a stopping target like Al is long (864 nsec) the loss of muons is acceptable if the time between pulses is not much longer than the muon lifetime and one simply waits for the pions to decay. Mu2e will therefore search for conversion electrons between proton pulses during times when the flux of particles in the secondary muon beam is relatively low and after the RPC process has dropped by roughly $10^{11}$. Fermilab provides a nearly perfect ring for such an experiment. The Fermilab Debuncher, unused in the post-collider era, can be re-purposed. It will supply a single circulating bunch that will be slow-extracted, providing a pulsed beam to Mu2e every cyclotron period of 1695 nsec for 8 GeV protons. This circumference of 1695 nsec is about twice the muon lifetime in aluminum (so the loss in muons from their decay is acceptable), and the storage and extraction process can be made to have little or no beam between pulses. Figure 3.8 shows the beam structure and the delayed search window.

The muon stopping target will be located in a graded solenoidal field that varies smoothly from 2.0 to 1.0 Tesla. The active detector will be displaced downstream of the stopping target in a uniform field region. This configuration increases the acceptance for conversion electrons, suppresses backgrounds, and allows for a reduction of rates in the active detector.

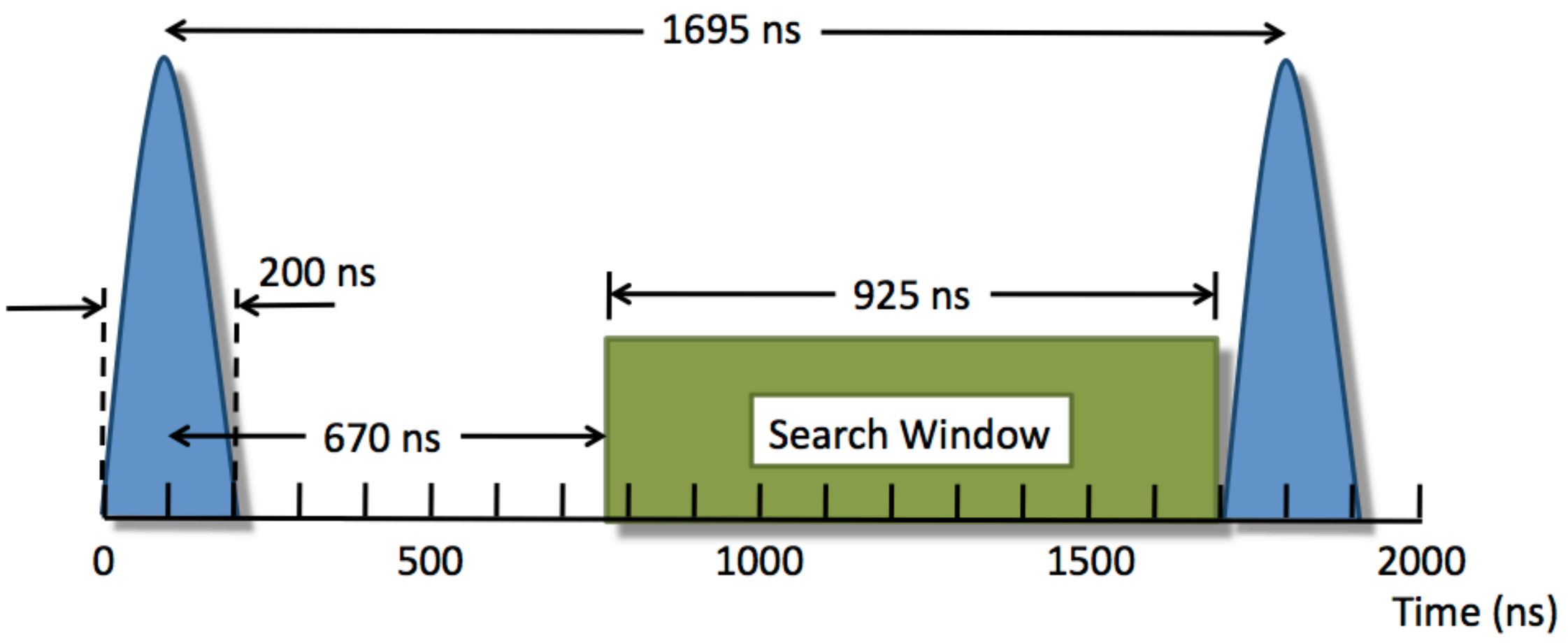

Figure 3.8. The Mu2e spill cycle for the proton beam and the delayed search window that allows for the effective elimination of prompt backgrounds when the number of protons between pulses is suppressed to the required level.

The 105 MeV conversion electrons (along with decay-in-orbit electrons from normal Michel decay) are produced isotropically in the stopping target. The tracker surrounds a central region with no instrumentation: the vast majority of electrons from Michel decay, all but a few parts in $10^{-16}$, have radii in the 1 Tesla field that are too small to intercept the tracker and are thus essentially invisible; the few remaining are a source of background we will discuss at length. Here, a final gradient field in the region of the stopping target and before the tracker plays a critical role. Conversion electrons, at 105 MeV, emitted at 90° ± 30° with respect to the solenoid axis ($p_t$ > 90 MeV/c) are projected forward and pitched by the gradient into helical trajectories with large radii that intercept the tracking detector. Electrons in this range that emerge from the target in the direction opposite the tracking detector (upstream) see an increasing magnetic field that reflects them back towards the detector. In addition to nearly doubling the geometric acceptance for conversion electrons, the graded field helps to reject background by shifting the transverse momentum of electrons passing through it. Conversion electrons within the acceptance of the tracker originate from the stopping target with transverse momenta > 90 MeV/c. The graded magnetic field shifts the transverse momentum of the conversion electrons into the range between 75 - 86 MeV/c by the time they reach the tracker. Electrons with a total momentum of 105 MeV/c that are generated upstream of the stopping target, at the entrance to the Detector Solenoid, cannot reach the tracking detector with more than 75 MeV/c of transverse momentum because of the effect of the graded field, eliminating many potential sources of background.

The active detector consists of low mass straw tubes oriented transverse to the Detector Solenoid, a calorimeter constructed from LYSO crystals arranged in 4 vanes and a cosmic ray veto that surrounds the Detector Solenoid (Figure 3.7). The detector is displaced downstream of the stopping target in order to:

- reduce the acceptance for neutrons and photons emitted from the stopping target and to allow space for absorbers to attenuate protons ejected by nuclei as part of the muon capture process in the stopping target.
- provide a region for the aforementioned gradient field to pitch conversion electrons into a region of good acceptance. This helps to reduce accidental activity in the detector. Beam particles entering the Detector Solenoid and, as stated, the vast majority of electrons from muon decay-in-orbit (the Michel peak in Figure 3.4) pass undisturbed through the evacuated center of the detector. The size and gradient of the field, the size of the central evacuated region, and their geometric locations, are therefore jointly designed to maximize acceptance for conversion electrons while also greatly reducing decay-in-orbit background.
- reduce activity in the detector from the remnant muon beam, about half the intensity entering the Detector Solenoid (stopping more of it would require more stopping material, yielding more accidental activity and smearing out the conversion peak because of energy loss straggling). The remnant muon beam enters a specially designed re-entrant absorber (Chapter 8); the beam absorber minimizes albedo that could increase accidental activity in the detectors.

## 3.5 Background and Sensitivity

The known processes that may create backgrounds for muon conversion experiments were discussed in general in Section 3.2. In this section we explicitly estimate the backgrounds expected by Mu2e. Eliminating potential backgrounds drives many of the design features of the Mu2e detector. We conclude with our current estimate of the experimental sensitivity.

### *3.5.1* Muon Decay-in-orbit

Muons that have been captured by a stopping target nucleus can decay-in-orbit (DIO) to an electron and two neutrinos and are a potential source of background. To date there have been no experimental measurements of the DIO spectrum with sufficient sensitivity to measure the part of the spectrum that is important for Mu2e, although the spectrum is calculable to better than 20% near the endpoint. The simulations for Mu2e use the recent calculation of Czarnecki et al. [24]. Integrating this spectrum over the relevant portion near the conversion energy tells us only a few $10^{-17}$ of all muon decays could fall into the Mu2e signal region. This size sets a scale for the ultimate sensitivity of the experiment, but the situation is more complicated. The fraction of events that will actually be observed in the signal region depends not only on the DIO spectrum but also on the energy resolution of the spectrometer as well as the mean energy loss and straggling in the detector. This will tell us where to set the lower edge in energy for our acceptance window. We next examine this in more detail.

The current version of the GEANT4 simulation of Mu2e has background hits from particles produced in the muon stopping process along with delta rays generated as particles propagate through the detector. Hits are modeled as 100 nsec long with hits overlapping if their time separation is < 100 nsec. The energy deposit is recorded. Time division is included with a location-dependent (position along the wire) resolution of approximately 8 cm. The hits, including background, are then given to a reconstruction algorithm based on the BaBar track software. The numbers we present for signals and backgrounds in conversion electrons, decay-in-orbit events, and radiative pion captures are based on full reconstruction and cuts are made on reconstructed quantities. The reconstruction algorithm is still in its infancy and we are actively studying improvements to the algorithm as well as using it to improve the detector design. While we believe the efficiency will improve by a factor of about two and the background from decay-in-orbit tails will greatly decrease, we present the algorithms and their results as they are today. We have performed a simulation of the experiment with the current resolutions and acceptance described above and the existing fitting algorithm convolved with the Czarnecki et al. spectrum and determined a DIO background of 0.22 ± 0.06 events. The signal window of 103.5 to 104.7 MeV/c reconstructed momentum was chosen to optimize the crude figure-of-merit of $S/\sqrt{B}$; a more sophisticated scheme (Mu2e-doc-1821 and 1856) is under development.[2]

The DIO shape and magnitude must be verified from the data. The rate observed in the Mu2e data, in both magnitude and shape, will be compared to simulations. The calibration from $\pi^+ \rightarrow e^+ \nu_e$ decays will be used to understand the detector response. The monochromatic electrons from $\pi^+ \rightarrow e^+ \nu_e$ decays will set the absolute calibration and the measured shape of the electron spectrum will be compared with simulations to verify that the resolution is fully understood. Once the resolution is understood it can be used to study the rapidly falling DIO spectrum near the endpoint. Data taken at different rates can be used to study the effects of accidental hits that might cause catastrophic reconstruction errors.

[2] We have doubled the acceptance to the ultimately expected value for some beam-related backgrounds, since we want to ensure the design for collimators, windows, etc. will sufficiently suppress backgrounds in the best (worst for backgrounds) possible case; an example will be found in Section 3.5.4 on antiprotons. However, for DIOs we have kept the estimate. The location of the lower momentum cut depends on the resolution and acceptance and since we think both will improve, we are making the most pessimistic case given our current knowledge.

Effects from energy loss and resolution are not shown in the pure calculation of Figure 3.9, but they can be seen in the full simulation of Figure 3.10, which shows the shape of the DIO spectrum with all known significant effects included, normalized to the expected run. Figure 3.11 gives the corresponding shape of the conversion peak.

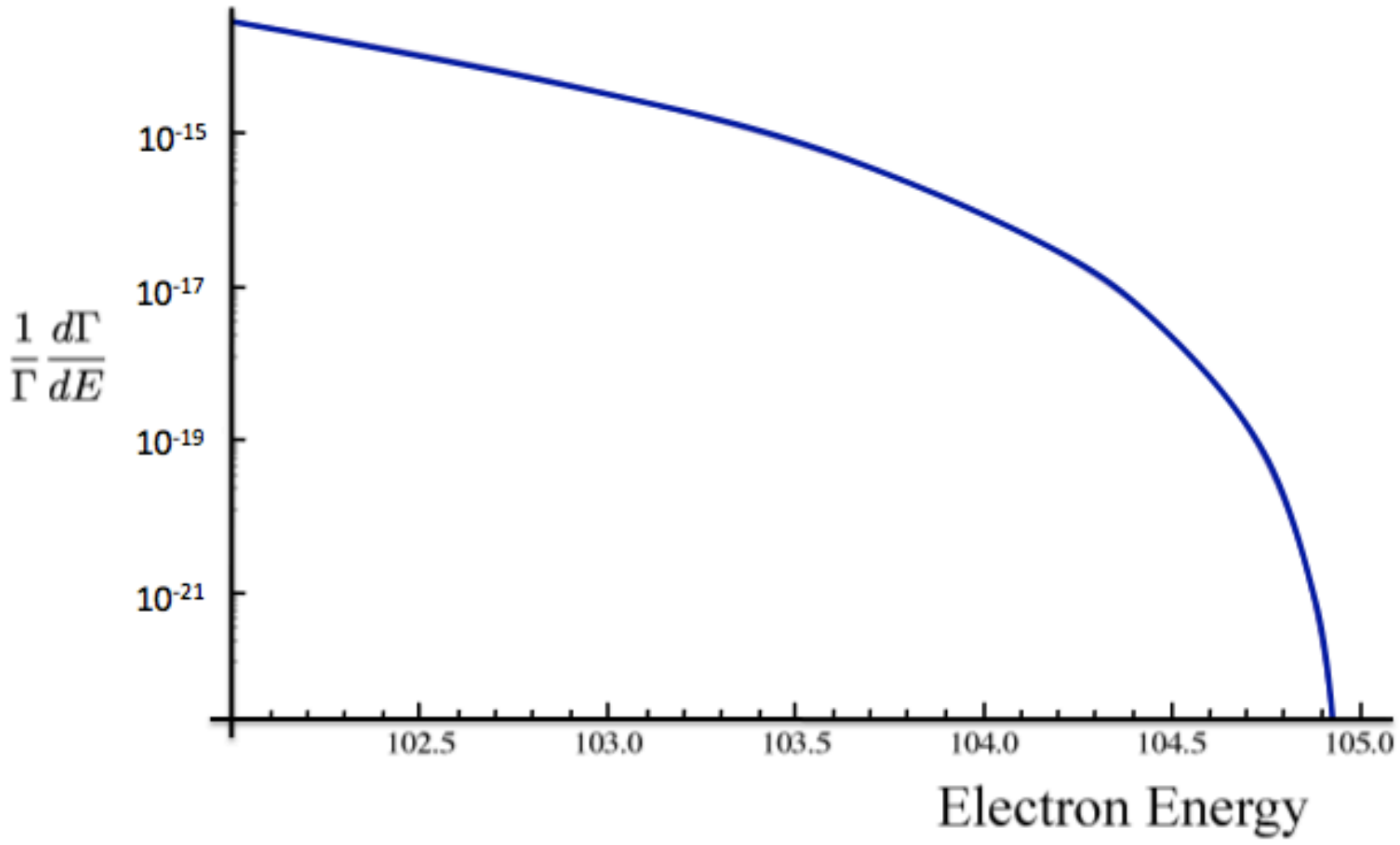

Figure 3.9. Muon decay-in-orbit rate (1/Γ)(dΓ/dE) (normalized to unity) spectrum near the endpoint energy from the fit given in Ref. [25]. Energy is in MeV.

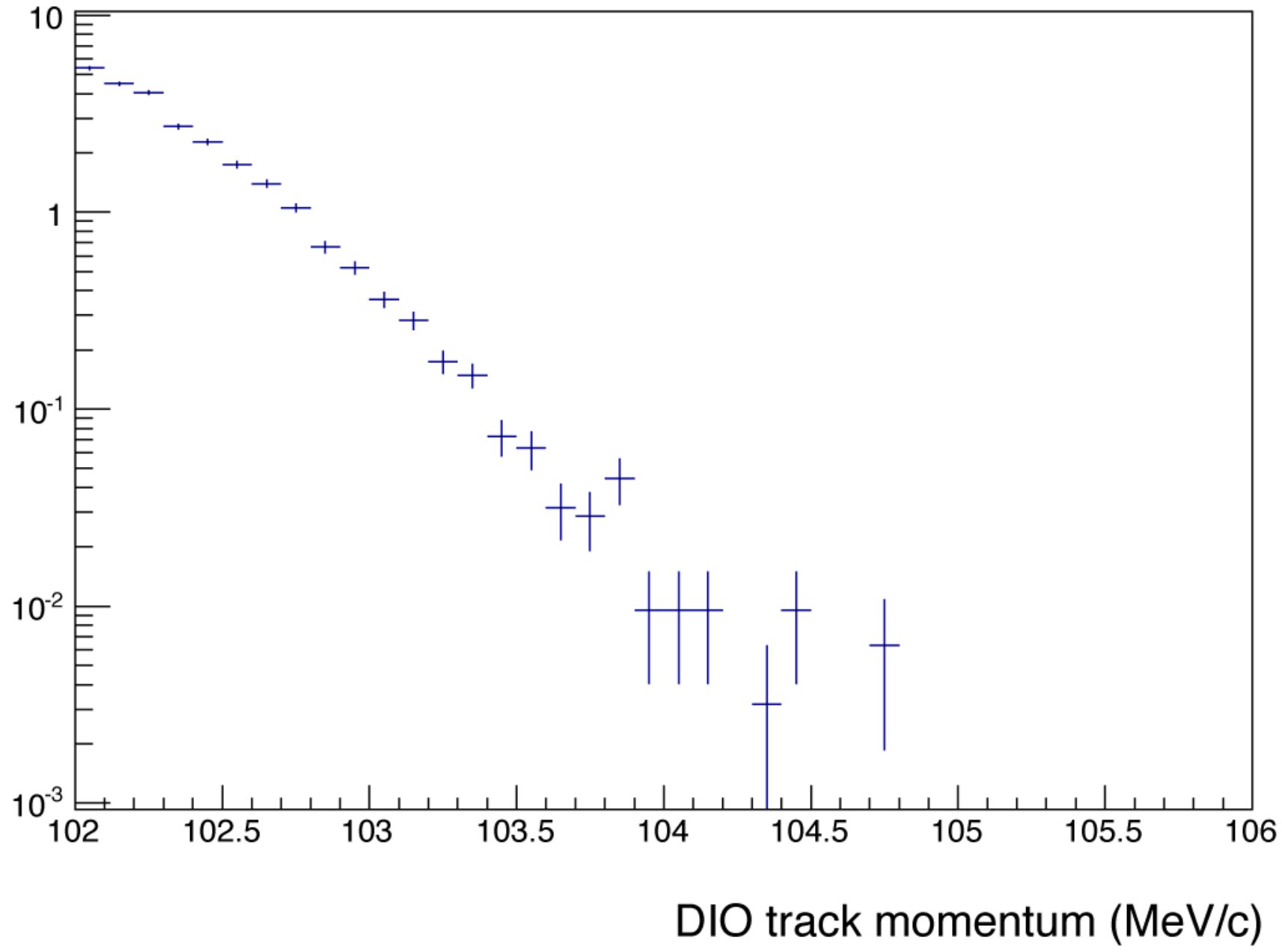

Figure 3.10. The DIO momentum spectrum with all known significant effects. The raw spectrum has been propagated through the simulation and the reconstructed momentum after all current cuts is shown. The plot is normalized for the experiment, and the momentum window is currently 103.5 to 104.7 MeV/c as shown in Figure 3.11.

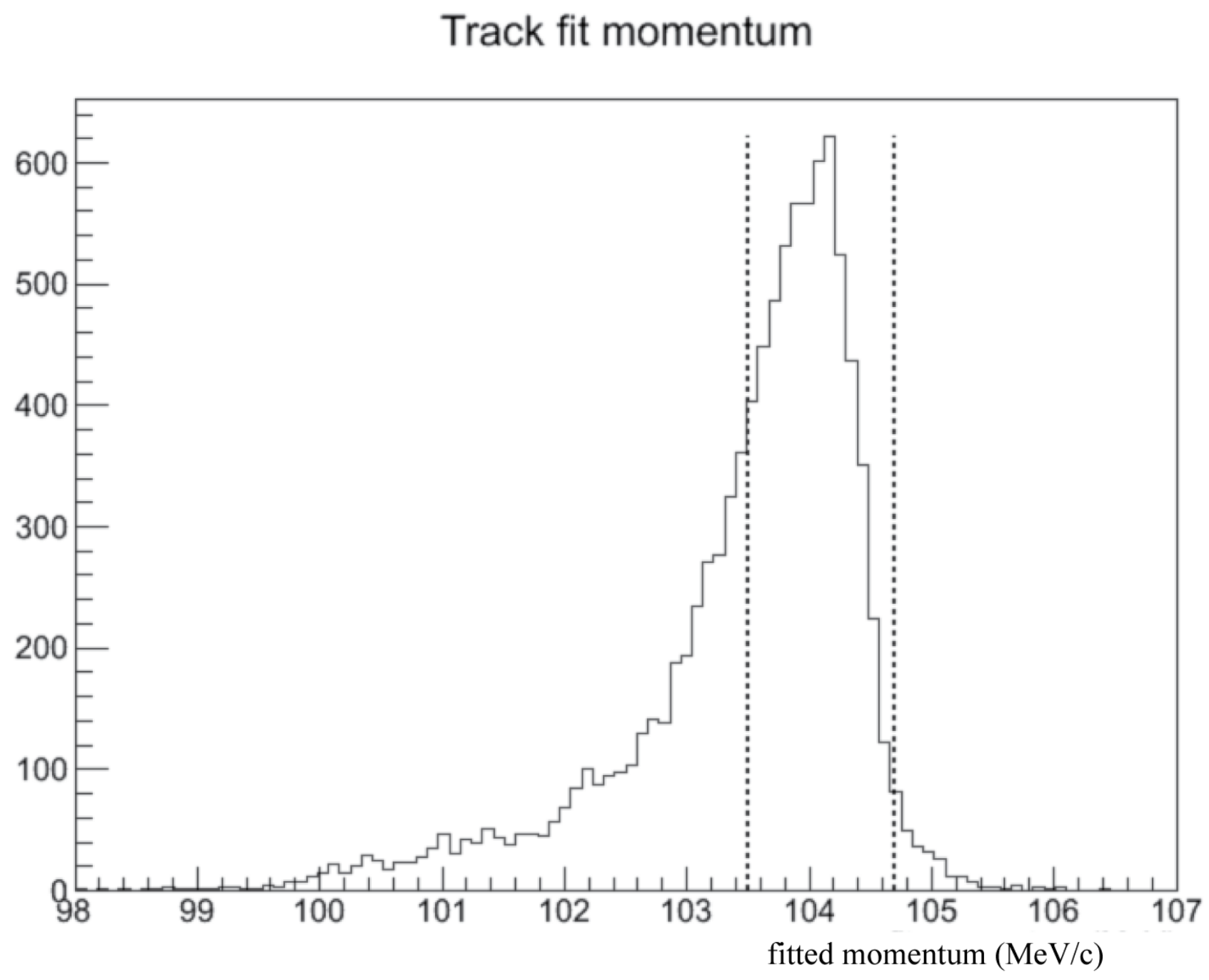

Figure 3.11 The Monte Carlo generated spectrum of conversion electrons in the Mu2e spectrometer after energy loss, straggling and detector resolution. The dashed lines indicate the cuts used for this CDR. Normalization is arbitrary.

### *3.5.2* Background due to Reconstruction Errors

Now that we have presented the signal and main resolution-related background (DIOs) we have a context for a description of the existing pattern recognition and fitting algorithms. The other backgrounds depend only weakly on reconstruction algorithms and resolution.

Activity in the tracker can lead to background in several ways. DIOs can be promoted into the signal region by hits that are obscured by nearby, extra hits that lead to failures of the reconstruction algorithm. Either case may result in the promotion of DIOs into the signal window. Extra hits can be produced by delta rays, Compton electrons, and photo-electrons, which we will refer to collectively as delta rays. Extra hits can also arise from multiple DIO events. When muons are captured, on average about 0.1 protons, two photons, and two neutrons are ejected (Mu2e-doc-1619 and [41]). Protons from muon capture on the stopping target are slow moving, and therefore highly ionizing, and can thus shadow other hits. The resulting large pulses can also result in cross-talk with neighboring straws. Most proton hits are identified and discarded using pulse height information, but hits that overlap in time and space with the protons are lost. Monte Carlo studies show that this shadowing results in <1% hit inefficiency, even on the straws with the highest hit rates. Neutrons can produce photons when they slow down and capture.

These photons, as well as those produced in the original capture, can produce low energy electrons that create noise. A typical signal event with the corresponding backgrounds within a 100 ns window around the conversion electron is shown in Figure 3.12. The total tracker occupancy is less than 1%.

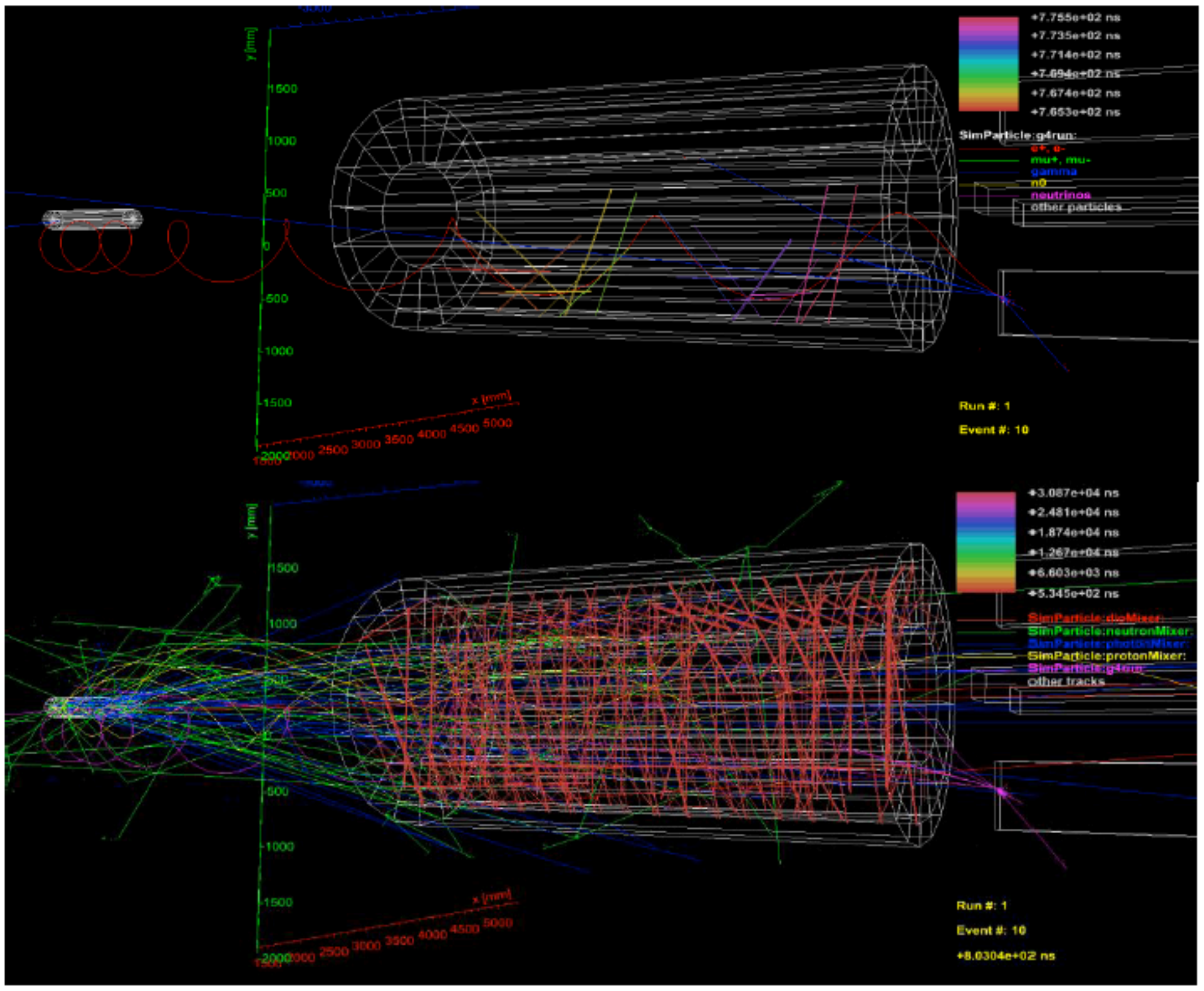

Figure 3.12. Display of a single conversion electron event. The top shows just the conversion electron, the bottom includes all the expected backgrounds within ±50 ns of the conversion.

Mu2e has developed an algorithm to find and reconstruct electron tracks in the presence of backgrounds, using just the tracker information. The track reconstruction uses the straw hit position, time, measured pulse height, and time division along the straws to separate signal hits from background. Clusters of hits produced by delta rays and other low-energy electrons are removed using a dedicated algorithm. Hits consistent in time and space coming from a helix are collected and fit using successively more accurate algorithms, culminating in a Kalman filter fit. Tracks are selected based on their measured parameters and fit quality. The reconstructed momentum resolution with our chosen cuts is shown in Figure 3.13. Details of the track reconstruction are presented in section Tracker chapter of this report, Section 9.4.3.

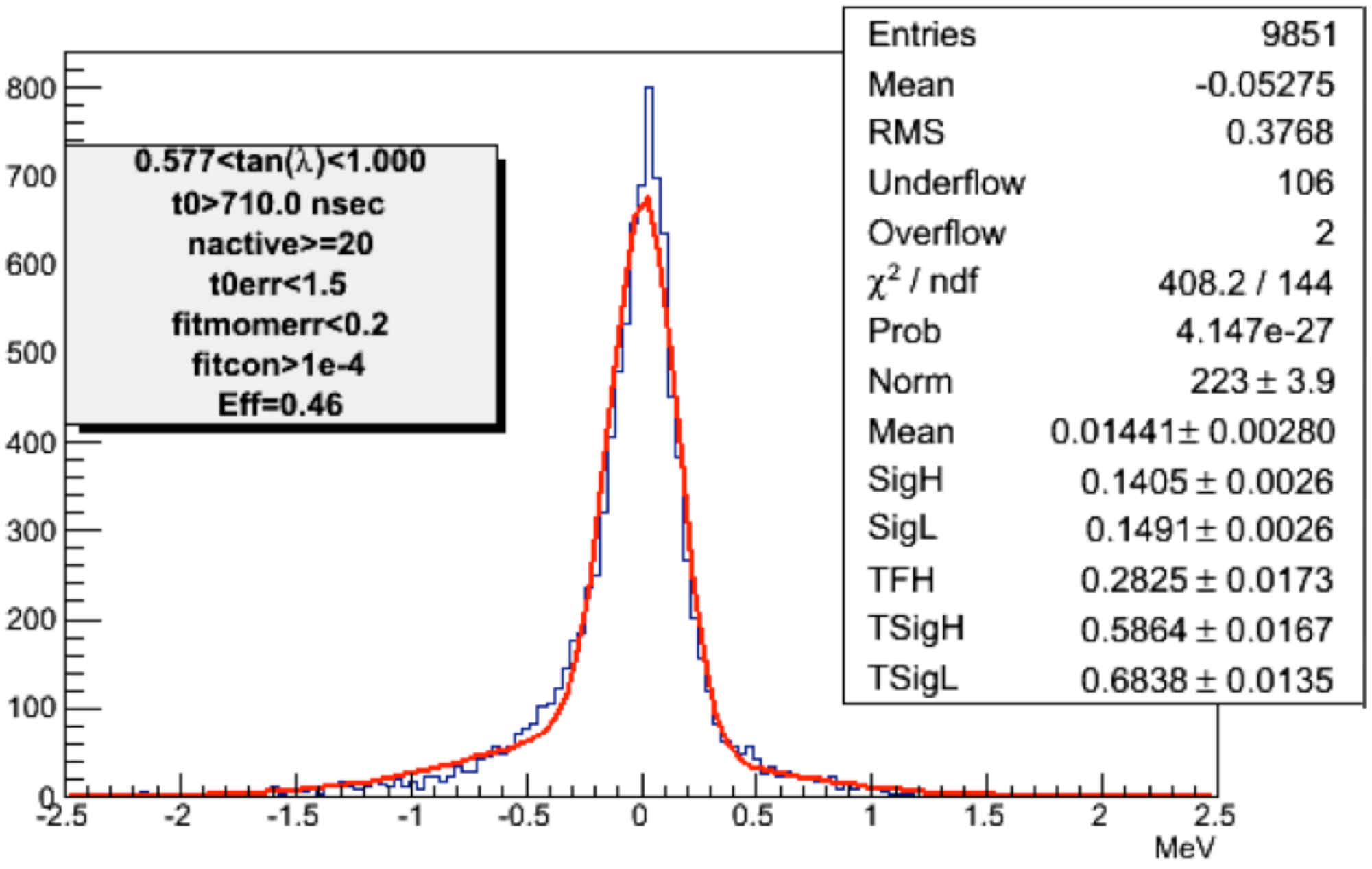

Figure 3.13. Reconstructed track momentum resolution with cuts used in the text. A double-sided Gaussian was fit to indicate the size and width of the tails, but the results in this CDR use the reconstructed momentum, not a *post-facto* smearing.

We have discussed various cuts to minimize background; Figure 3.15 shows the overall acceptance of 5.25%. The main cuts are:

- Events in the live gate, which leaves 51%. The live gate begins ~700 ns after the proton pulse and ends when the next proton pulse arrives. This is required to remove pions that can produce radiative pion capture, discussed in Section 3.5.3. Note that in Figure 3.15, the ratio, 0.5839, differs from 0.51. This is due to the bias introduced by the cuts prior to the live gate cut in the Figure. The 51% value represents the unbiased survival fraction after the live gate cut.
- More than twenty straw hits used in the fit (active).
- The "pitch cut" described elsewhere in the text ("reco pitch in Figure 3.15) is made after the twenty active hit cuts. One loses about another factor of two in acceptance. The active hit cut strongly selects for events in the pitch range and as a result the pitch cut does not appear to have a large effect in the Figure. The pitch cut is defined by $0.5 < \cos\theta = p_z/p < .707$, or $0.7 < p_T/p < 0.9$, and we use all these forms of expressing the cut in this CDR depending on context.
- The cut on momentum window 103.5 MeV/c < p < 104.7 MeV/c. Note Figure 3.14 shows a long tail at low energy that arises from energy loss in the stopping target and proton absorber. The window cuts off a significant portion of the tail, cutting ~ 40% of the events in order to minimize the DIO background. Lowering the momentum cut to increase acceptance rapidly increases the DIO background. Optimization of the proton absorber and stopping target are therefore high priorities for future work.

The best acceptance we can expect is about 10% (the product of these factors) and we are currently finding half that, about 5.25%. We expect most of that factor of two (5.25 to 10%) to be made up by a more sophisticated algorithm. Minimizing the material and consequent signal width arising from energy loss will require careful study of the design and may both increase the acceptance and lower the background by making the signal width smaller. Decreasing the signal width improves the sensitivity by significantly reducing the rapidly falling DIO background as well as most other backgrounds that fall linearly. These studies are underway.

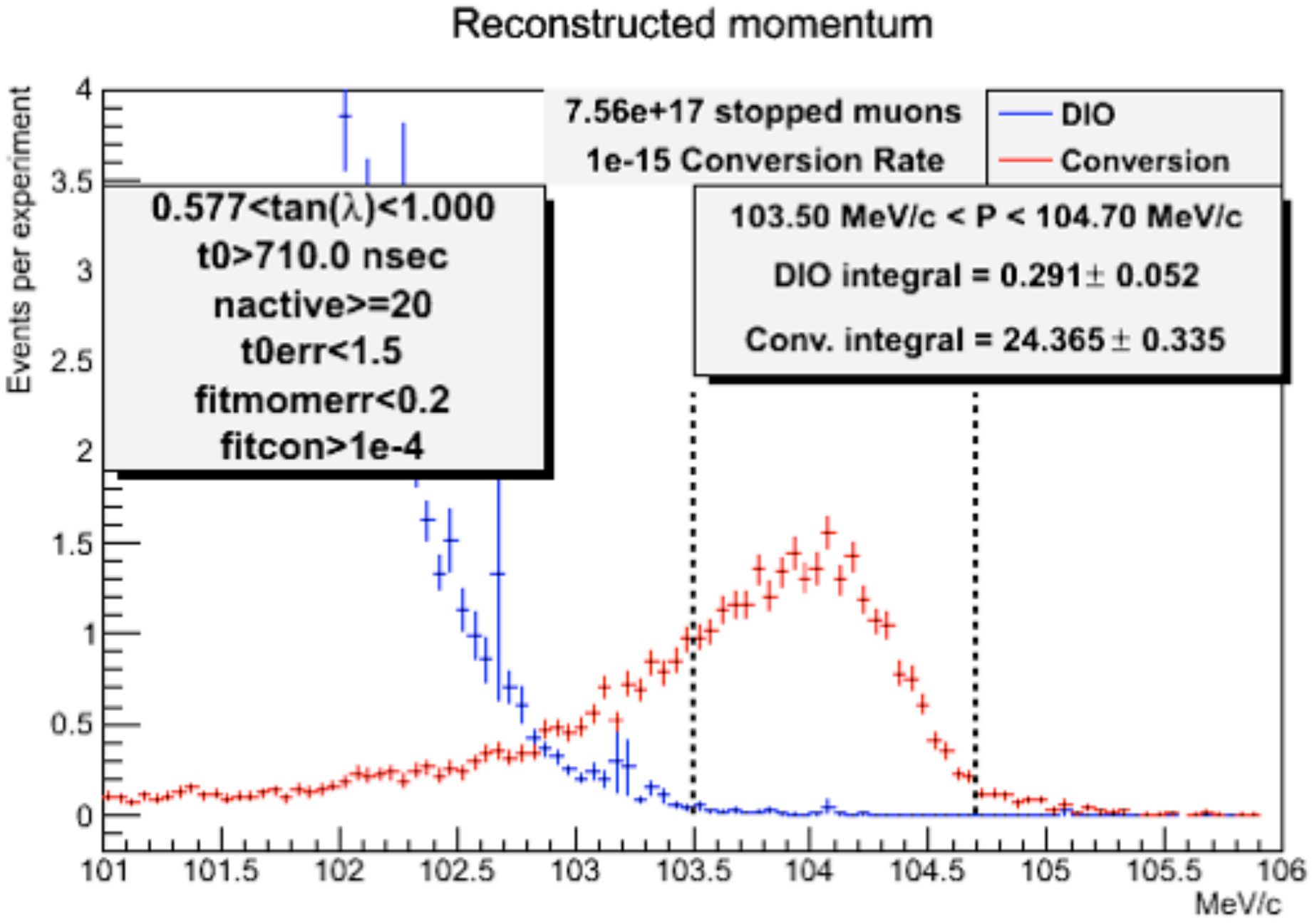

Figure 3.14. Reconstructed momentum of conversion electrons and DIO, scaled to the experimental number of stopped muons, assuming a conversion rate of 1 x$10^{-15}$.

### *3.5.3* Radiative Pion Capture

Radiative Pion Capture (RPC) can produce background through two processes:

1. $\pi^- N \rightarrow \gamma N^*$ with a subsequent conversion of the photon in the stopping target. The maximum photon energy is approximately equal to the pion rest energy at about 139.6 MeV, with peak photon energy near 110 MeV [26].
2. The internal conversion process $\pi^- N \rightarrow e^+ e^- N^*$.

The two effects are about equal for our stopping target mass and configuration. In either case, an asymmetric conversion of the photon can produce an electron with an energy that is consistent with a muon conversion. The calculation (before pattern recognition) was performed and cross-checked in two independent studies as described in Mu2e-doc-1087 and 1364. These calculations were done before the current reconstruction algorithm was written and the acceptance used was about twice as high as that of the current algorithm. We have *not* reduced the background from the acceptance as calculated at that time in order to be conservative; thus for the purposes of the background calculation, we have assumed the future expected pattern recognition.

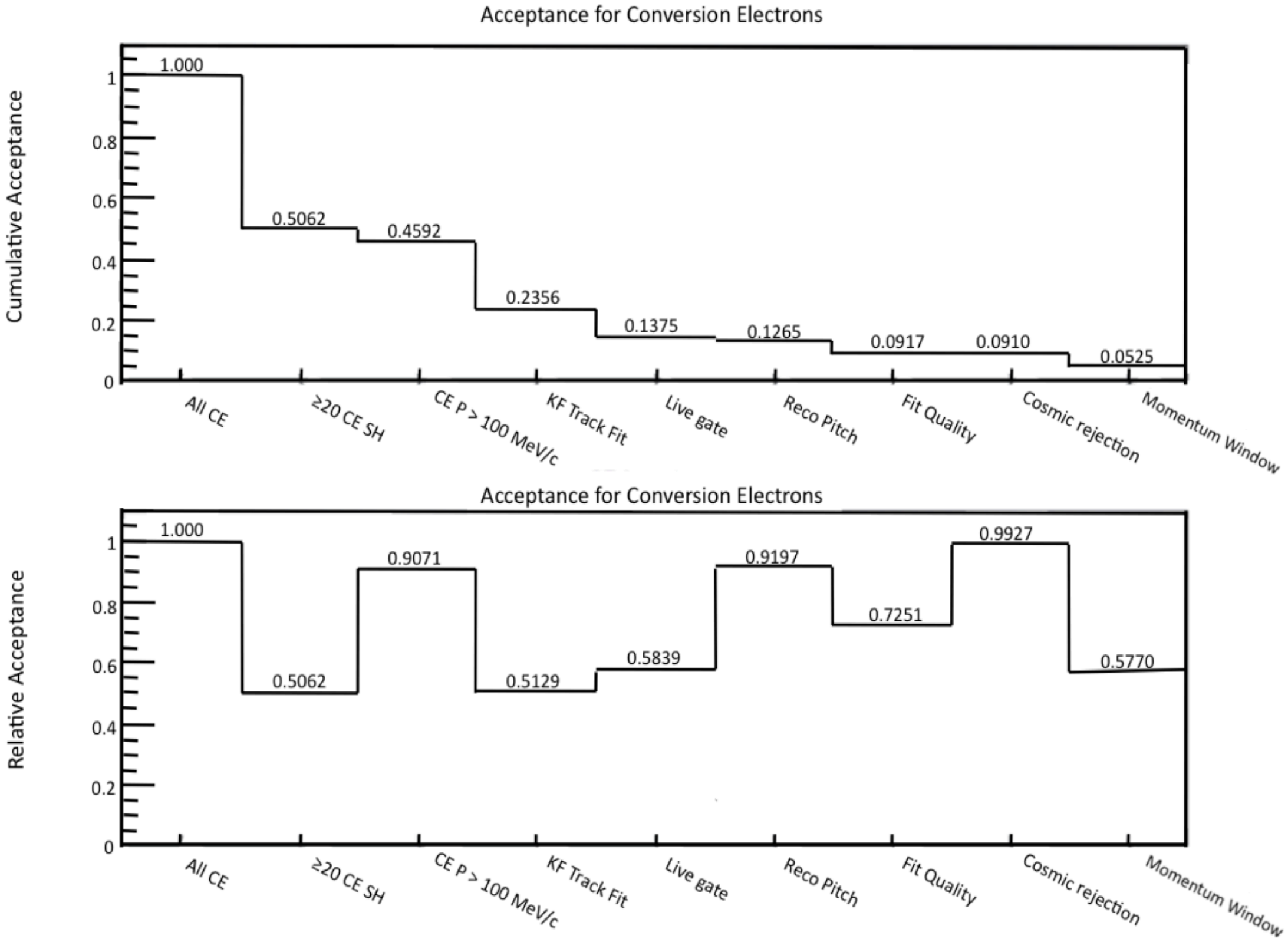

Figure 3.15. Cuts used in the simulation, applied in order (top panel) and stepwise (reduction from previous step.)

A small fraction of charged pions can traverse the Transport Solenoid and survive to reach the muon stopping target. The size of the resulting background is driven primarily by the arrival distribution of the pions at the stopping target. The lifetime of a captured pion is much shorter than the 864 nsec captured muon lifetime, so electrons that result from RPC will be detected in coincidence with the arrival time of the pion. Captured pions originating from the main proton pulse will produce electrons that are peaked in time long before the start of the delayed search window and are a negligible background source. However, protons that arrive late at the production target can produce pions that arrive late at the muon stopping target and produce electrons that fall into the delayed search window. The number of late-arriving protons and the arrival time of the resulting pions at the stopping target relative to the search window will ultimately determine the magnitude of the RPC background.

The fraction of stopped pions that radiate a photon is expected to be about the same 2.1% as is measured for Mg [26] and we calculate about $1.0 \times 10^{-6}$ pions per incident proton will stop in the Mu2e stopping target. Based on our simulation, we have chosen a measurement period (thereby waiting for RPCs from the beam pulse to end) and an extinction level of $10^{-10}$ (where extinction is defined as the ratio of protons between beam pulses to the protons contained in a beam pulse). Both internal and external conversions are modeled and outgoing electrons tracked and reconstructed in the full simulation. The background from radiative pion capture is then estimated to be $0.03 \pm 0.007$ events. The error is estimated by varying pion production models and the opening of the measurement period. There is an important issue: the size of the RPC background comes from two sources:

- pions arising from the late-edge of the main proton beam pulse. Although these are highly suppressed by the pion lifetime, one must set the opening of the measurement period so that this background is small and the width of the proton pulse is an important part of the simulation. We have chosen a reasonable model of the proton pulse width. The chosen opening of the measurement period yielded a negligible background.
- "unextinguished" pions produced by protons not in the beam pulse. As discussed in Chapter 5 on the Accelerator, these out-of-time protons can arise from sources such as beam gas scattering or RF jitter. We use methods such as momentum scraping to reduce the out-of-time background. We also use an external dipole whose field varies with time in such a way as to kick-out protons not in the pulse. The out-of-time level will be monitored as described in Chapter 5. For now we have modeled this source of pions as flat in time but more sophisticated models are under development.

Mu2e will not be forced to rely on estimates of this potentially serious background. The background from radiative pion capture can be determined from the data collected by Mu2e by taking advantage of its strong time dependence. We will measure the number of energetic electrons as a function of time after the arrival of the proton pulse and have designed the DAQ to handle the rates occurring earlier in the pulse in order to measure this background[27]. Furthermore, we can choose the earliest time (and the consequent rates in the DAQ) such that the number of background events from RPCs are at the current limit from SINDRUM-II, allowing us to create a "blinding box" based on both time and reconstructed momentum. Additionally, RPCs are expected to be one of the primary sources of high-energy electrons above the signal region (103.5–104.7 MeV/c), giving us an interpolation between "below" and "above" the signal region rather than relying on an extrapolation. Furthermore, we can intentionally spoil the extinction in a controlled way, raising the RPC background so it can be measured. Another way to measure the RPC background is to examine positrons above the signal region. The predominant source of such positrons is the RPC process. We thus have a number of handles, measurable within the data and part of this CDR design, to measure and control this background, which was the limit of the SINDUM-II PSI experiments. Going beyond these experiments demands the pulsed beam and therefore understanding and minimizing this background is crucial for the success of the experiment.

### *3.5.4* Antiprotons

Antiprotons are a source of background that did not occur in the SINDRUM experiments at lower proton energy. The 8 GeV kinetic energy proton beam can produce antiprotons in sufficient numbers to be a serious background. Further complicating the situation are:

- The fact that antiprotons do not decay. Only antiprotons with momenta below ~100 MeV/c can propagate through the Transport Solenoid, restricting the kinetic energy of the antiprotons to be less than ~5 MeV and speeds less than ~0.1c. The antiprotons spiral slowly through the Transport Solenoid; consequently the flux of antiprotons at the entrance to the Detector Solenoid would be nearly constant in time if allowed to arrive at the stopping target unimpeded. Note that the number of antiprotons produced by out-of-time protons is negligible: therefore the extinction system, which removes out-of-time protons, does not help with this background.
- The large acceptance of the solenoidal beam line for the negatively charged antiprotons; they pass easily through the sign-selecting collimator of the curved Transport Solenoid.
- The high probability to stop and annihilate on nuclei, releasing significant energy and a large number of secondary particles, which can include electrons near the conversion energy at 105 MeV.

These properties combine to make antiprotons a serious potential source of background. The proton kinetic energy is only slightly above threshold for antiproton production, so the production cross-section is small. At the same time, since the process has such a small cross section, it is difficult to measure. Consequently the cross sections are not well measured and flux predictions are relatively uncertain.

Most of the antiprotons that enter the Detector Solenoid have very low kinetic energies and almost all will stop in the muon stopping target and then be captured in an atomic orbit around the target nuclei. The antiprotons will quickly annihilate and produce a ~2 GeV shower of secondary particles including pions, kaons, neutrons, protons, electrons, positrons, and gamma rays. An electron near 105 MeV can be produced directly in the annihilation, through the decay of secondary particles, or through the interaction of secondaries in the stopping target or other nearby material, faking a conversion electron. For example, on average there are 0.9 $\pi^-$ and 1.2 $\pi^0$s produced in each annihilation on an Al nucleus. The $\pi^0$s decay immediately to photons that can produce 105 MeV electrons by converting to an electron-positron pair. The $\pi^-$ can stop in the stopping target and produce 105 MeV electrons in the radiative capture process, as described in Section 3.5.3.

A simulation combining G4Beamline and MARS was used to estimate the backgrounds from antiprotons; details are given in Ref. [34]. The existing production cross section data have been fitted, as described in Mu2e-doc-1776.

A large simulated sample of antiprotons was produced at the production target using the MARS simulation code, and are tracked to the entrance of the TS. The antiprotons are then tracked through the Production and Transport Solenoids to the Detector Solenoid and muon stopping target using a GEANT4 model that includes interactions with the walls, collimators, etc. Once the distribution of antiprotons stopped in the target is established, the spectrum of secondaries as predicted by another MARS simulation is tracked through the DS using the GEANT-based simulation.

A background conversion electron look-alike is then defined as an electron that passes the following cuts at a plane just upstream of the tracker:

- 103.5 MeV/c < p < 104.7 MeV/c (momentum cut)
- $0.7 < p_t / p < 0.9$ at the tracker (the pitch cut)
- Electron arrives during the live gate.
- The distance of closest approach of the helical trajectory to the solenoid axis is less than 100 mm, consistent with originating in the stopping target. This cut has little impact since most of the background electrons are in fact produced in the target.

The simulations tell us that if nothing were done to reduce the antiproton flux, then about $1\times10^9$ antiprotons would stop in the stopping target for $3.6\times10^{20}$ protons on the production target, producing a very large number of electrons in the signal window (~20,000).

Several means of eliminating antiproton background have been considered, including:

- Application of an electrostatic sweeper field in the first transport solenoid collimator. This is a relatively expensive and complicated option.
- Reduction of the incident proton beam energy to reduce the antiproton production cross-section. This is not practical in the Fermilab complex.
- Insertion of absorber material in the beam line.

The third approach is the preferred alternative. Because the kinetic energy of the transported antiprotons is low, it is possible to stop them efficiently in a low-Z window that can be thin enough to have little impact on the transported muon beam. We studied various locations for the absorber and determined the best location was the center of the Transport Solenoid (doc-db 2035). A wedge-shaped window ranging from 0.12 mm to 1 mm from the center to the top of the collimator reduces the number of antiprotons stopping in the stopping target to about 1000 for $3.6\text{x}10^{20}$ protons on the primary target, leading to about 0.02 background events. The presence of the absorber window reduces the rate of muon stops by an acceptable 9%.

It is also necessary to consider the background from the secondary particles produced by antiproton annihilation in the absorber window. A distribution of secondaries following antiproton capture on the absorber window was generated using MARS. The antiprotons were then transported to the entrance of the last Transport Solenoid collimator. The MARS simulation does not include a model of the magnetic field in the Detector Solenoid, therefore the distribution of secondaries at that point were taken as input to the GEANT simulation package which has a full magnetic field map and realistic geometry in the DS. It was found that the dominant source of background from these secondaries comes from pions stopping in the stopping target and undergoing radiative pion capture. The background is $0.08 \pm 0.04$ events where the error is dominated by the cross-section uncertainty.

Summarizing, our current design predicts the antiproton background from antiprotons annihilating in the window will be $0.08 \pm 0.04$ events, where the error is dominated by the cross-section uncertainty. The estimated background from the antiprotons that passed through the window and stopped in the stopping target would be $0.02 \pm 0.014$ electrons, where the error is again dominated by the cross-section uncertainty. The sum of antiproton-induced backgrounds is $0.10 \pm 0.05$ events. We have assumed the reconstruction efficiency is about twice what we obtained from simulations for DIOs and conversion electrons, since we anticipate large improvements in this efficiency as our track-fitting algorithms are improved, and it seems prudent to design the experiment with that safety margin. We are investigating strategies for measuring this background *in situ*, such as varying the window thickness in special runs in order to increase (and thereby measure) the antiproton background rate.

### *3.5.5* Radiative Muon Capture

The radiative capture of a muon can also produce electrons near the muon conversion energy. Radiative Muon Capture (RMC) is the process $\mu^- + \mathrm{Al} \rightarrow \gamma\,\nu + \mathrm{Mg}$. Background arises if the photon subsequently converts to an $e^+e^-$ pair in the stopping target material. Electrons can also be produced through internal conversion of the outgoing photon: $\mu^- + \mathrm{Al} \rightarrow e^+e^-\nu + \mathrm{Mg}$. As shown in Section 3.5.3, the rates for internal and external photon conversion are about equal for RPCs and the arguments are similar here; hence the external process is explicitly evaluated and then doubled to arrive at the total rate.

The shape and rate of the RMC process on medium mass nuclei is highly uncertain. The endpoint energy is given by a straightforward calculation:

$$k_{max} = m_\mu c^2 - \Delta M - R - B_\mu$$

where the terms are the muon mass, the nuclear mass difference, the recoil energy and the binding energy. The calculated endpoint value is 102.4 MeV for aluminum. Attempts to measure the spectrum near the endpoint have been limited by statistics. Bergbusch et al. [28] measured the fraction of radiative muon capture normalized to ordinary muon capture ($R_\gamma$) for photons > 57 MeV to be (1.43 ± 0.15) × $10^{-5}$. The spectrum, in the closure approximation discussed in Bergbusch et al., is given by

$$\frac{d\Lambda_\gamma}{dE_\gamma} = \frac{e^2}{\pi}\frac{k_{\max}^2}{m_\mu}(1-\alpha)\left(1-2x+2x^2\right)x(1-x)^2$$

where $x = E_\gamma / k_{max}$ and $\alpha = (N - Z)/A$ is the neutron excess. A fit of the data using the closure approximation results in a maximum photon energy of $k$max = 90.1 ± 1.8 MeV in aluminum. However, statistics in the measured RMC spectra are limited at high energy so sufficient data does not exist to determine the rate and spectra near the kinematical endpoint at 102.4 MeV; the observed spectra seem to cut off near 90 MeV. The calculations for higher energies vary by factors of two.

Radiative muon capture was the dominant background to the SINDRUM-II search for charge-changing muon conversion to a positron: $\mu^-(Z, A) \rightarrow e^+(Z\text{-}2, A)$ [29]. No events were observed past 95 MeV, so again there are no data to guide us near the kinematical endpoint.

A conservative set of assumptions was used to estimate the radiative muon capture background. The closure approximation with a kinematic endpoint of 102.4 MeV has been used along with the value of $R_\gamma$ above for the overall normalization. A Gaussian resolution function for the Mu2e spectrometer with a core σ =200 keV is assumed and integrated 3σ past the endpoint. A simple calculation allows us to estimate the background in the search window (103.5 – 104.7 MeV). Using this conservative set of assumptions the background from radiative muon capture is $< 2 \times 10^{-6}$, completely negligible.

The SINDRUM II data indicate there will be a measurable contribution from RMC in the 80 – 90 MeV energy range. Data from Bergbusch et al. [28], combined with the probability for an asymmetric photon conversion, predicts the rate of electrons from RMC is will decrease from $10^{-11}$ per conventional muon capture at 80 MeV to $10^{-13}$ at 90 MeV. This results in a rate of electrons that is similar to the rate of electrons from muon decay-in-orbit in this energy range.

Since there are no existing RMC data beyond ~90 MeV and the calculations near the endpoint are contradictory, RMC will have to be measured in Mu2e. Examining the behavior of the measured spectrum near the conversion energy can separate the contributions from DIO and RMC. In particular, studying the positron spectrum in this region can isolate the radiative muon capture contribution, since muon decay-in-orbit events do not produce positrons.

### *3.5.6* Pion Decay in Flight

A negative pion with momentum larger than 56.5 MeV/c can decay to an electron with an energy above 103.5 MeV through the decay channel $\pi^- \rightarrow e^- \nu$. The electron can be mistaken for a conversion electron if:

- It is produced by the decay of an out-of-time pion that arrives during the signal window.
- The fitted track extrapolates back to the stopping target (this cut is not used in the analysis but should be useful for this process).
- Its pitch, defined as the ratio of longitudinal momentum to total momentum, is in the range $0.7 < p_t / p < 0.9$ at the tracker.

An electron can fall into the accepted pitch range as a direct result of the pion decay or as a result of the electron scattering in the stopping target after pion decay. It is also possible for a pion in this momentum range to produce an electron with an energy near 105 MeV via the two-step $\pi \rightarrow \mu \rightarrow e$ decay chain, but the probability of this process producing a background electron has been determined to be negligible.

In order to estimate the background from pion decays in flight, a GEANT-based simulation was performed with $2 \times 10^8$ incident protons on the primary target. The pions are tracked down a fully modeled solenoidal beam line with all of the physics interactions turned on, but with pion decays turned off. Pions arriving at the Detector Solenoid are weighted according to their decay probability; the resultant phase space distributions of pions entering the Detector Solenoid are recorded in a data file. The pions in the data file are then used as input to a second GEANT-based simulation that tracks them from the beginning of the Detector Solenoid, applying the normal pion lifetime of 26.033 ns but forcing 100% of the pions to decay via the $\pi \to e\nu$ channel to reduce the computer CPU time required. The correction for the actual branching ratio for this decay mode ($1.23 \times 10^{-4}$) is applied later. A majority of these pions decay as they spiral downstream from the Detector Solenoid entrance to the upstream face of the tracker. The sum of the weights of events that produce an electron in the signal window (103.5 MeV to 104.7 MeV) and within the allowable pitch range (0.7 to 0.9) is 0.36. Scaling to $3.6 \times 10^{20}$ protons on target and an extinction of $10^{-10}$, we obtain

$$N_e \times BR(\pi \to e\upsilon) \times (N_{POT} / N_p) \times \chi \times \varepsilon_T = 0.0028 \text{ events}$$

where

- $N_e$ is the number of simulated events in the signal window = 0.36
- $N_{POT}$ is the total number of protons on target = $3.6 \times 10^{20}$
- $N_p$ is the number of simulated protons on target = $2 \times 10^8$
- $\chi$ is the extinction factor = $10^{-10}$
- $\varepsilon_T$ is the tracking efficiency after the pitch cut = 0.35 (consistent with 0.10 used elsewhere, here we have a larger number because some cuts, such as the pitch cut, have already been made on this data sample. Again this efficiency is higher than current efficiencies by about a factor of two to account for expected future improvements in track reconstruction efficiencies).

Based on this study, the background from pion decays in flight is negligible.

Several features distinguish pion decays in-flight from other sources of high-energy electrons. The number of events near the conversion energy increases rapidly at smaller pitch angles (small $p_t/p$). For helical pitch in the range $0.6 < p_t/p < 0.9$, the background increases by a factor of about six compared to the pitch range from 0.7 to 0.9 used to identify conversion electrons. This distinctive forward distribution, along with a distinctive energy distribution, will allow this background to be separated from other backgrounds, such as radiative pion capture and muon decays in-orbit where the electrons are produced more nearly isotropically at the stopping target. The large flux of incoming pions before the measurement period will provide information about the spectrum of electrons from in-flight pion decays. We can also study pion decay-in-flight by intentionally spoiling the extinction (for example by turning off the momentum scraping or AC-dipole extinction devices) to increase the pion rate by orders-of-magnitude.

### *3.5.7* Muon Decay in Flight

A negative muon with momentum larger than 76.7 MeV/c can decay to an electron with energy above 103.5 MeV through normal muon decay-in-flight. The electron can be mistaken for a conversion electron if:

- It is produced by a muon that arrives during the measurement period and decays in the vicinity of the stopping target.
- Its pitch, defined as the ratio of longitudinal momentum to total momentum, is in the range $0.5 < p_z / p < 0.71$ at the tracker, or in terms of the transverse momentum, $0.7 < p_t / p < 0.9$.

An electron can fall into the accepted pitch range as a direct result of a muon decay or as a result of scattering of the electron in the stopping target following a muon decay.

A GEANT-based simulation with a fully modeled solenoidal beam line was used to estimate the background from muon decays in flight. $6 \times 10^9$ protons incident on the production target were used to generate an ensemble of muons as they enter the central collimator of the Transport Solenoid. In order to achieve the necessary statistics in the simulation with a reasonable amount of computer CPU time, each muon was launched and tracked 10 times from the entrance of the central collimator in the TS, each time with a different set of initial pseudorandom number seeds. This method is valid because the probability of decay for one of these high-momentum muons is small (less than 10% on average). To further improve the statistics, all muons were forced to decay to electrons with the maximum possible rest frame energy, 52.8 MeV; these were the most likely to produce an electron near 105 MeV when boosted to the lab frame and led to another factor of ten improvement in computing efficiency. The resulting number of background events due to out-of-time muons, assuming an extinction of $10^{-10}$, is estimated at $0.003 \pm 0.001$, which is negligible.

There can also be background due to in-flight muons produced during the main proton bunch arriving during the live gate, which would require the muon to take a very long time to traverse the solenoidal beam line. To determine the number of background events in one simulation step from inflight muons decays, due to muons produced in the primary proton pulse, would involve simulating a huge number of muons and is not practical. Instead, the numbers of decay electrons per proton near 105 MeV were determined for two separate cuts, after a time cut t > 700 ns, and after a pitch cut 0.5 <$p_z/p$ < 0.71. These are relatively independent cuts and when multiplied together lead us to conclude that the estimated background from inflight muons produced by the primary proton pulse is negligible.

Background from muon decays in flight can be distinguished from other sources of high-energy electrons by the steeply falling energy distribution and a forward-peaked momentum distribution (small $p_t/p$). Spoiling the extinction will greatly increase the numbers of inflight muons during the measurement period for detailed study. The remarks of Section 3.5.6 on using the difference between the muon and pion lifetimes to separate their contributions apply here as well.

### *3.5.8* Beam Electrons

Electrons produced in the Production and Transport Solenoids are a potential source of background. These beam electrons can be produced in the production target, primarily through $\pi^0$ production followed by conversion of the decay photons. They can also be produced by decays or interactions of beam particles such as antiprotons anywhere upstream of the muon stopping target.

The collimators in the Transport Solenoid are designed to suppress the transport of particles with momenta above 100 MeV/c. Moreover, the magnetic field in the upstream section of the Detector Solenoid is graded, which shifts the pitch of electrons originating in the beamline to lie outside of the acceptable range for conversion electrons (The acceptable range for conversion electrons is 45° < θ < 60°, or equivalently 0.7 < $p_t/p$ < 0.9), provided the electron does not scatter in the stopping target. The magnetic field varies uniformly from 2 T at the Detector Solenoid entrance to 1 T near the upstream face of the tracker. As a result of this gradient, a beam electron that enters the Detector Solenoid cannot have a helical pitch larger than 45 degrees relative to the solenoid axis when it arrives at the tracker. The transverse component of the momentum in a uniform gradient is proportional to the square root of the magnetic field where

$$p_t = p_{t0}\sqrt{B/B_0}, \quad p_l = \sqrt{p^2 - p_{t0}^2 B/B_0}$$

and $B_0$ is the field at the entrance to the Detector Solenoid. In the worst case, $p_t = p$ for an electron at the entrance to the DS. When this worst-case electron arrives at the upstream face of the tracker, its pitch angle is reduced to a maximum angle given by

$$\sin(\theta) = \frac{p_t}{p} = \sqrt{\frac{B}{B_0}} = 0.707,$$

therefore the pitch angle (θ) for these electrons is smaller than 45°.

The muon stopping target is located in a magnetic field of about 1.5 T approximately midway between the Detector Solenoid entrance and the front face of the tracker. Conversion electrons emitted from the stopping target at angles $90^0 < \theta < 120^0$ wind up with pitch angles in the range $45^0 < \theta < 60^0$ at the tracker. By applying the standard pitch cut in the latter range, we effectively eliminate the beam electron background. The exception is if a beam electron scatters through a sufficiently large angle in the target or at the first crossing of the tracker. The probability for such a large scatter is quite small. A G4Beamline simulation (Mu2e-doc-2121) determined the beam electron background is $4.1 \times 10^{-3}$, negligibly small.

The distribution of background from beam electrons is characterized by the energy distribution of these events, which is falling very steeply with energy due to the beam acceptance, and by their small transverse momenta. This is different from the transverse momentum distribution of electrons from muon decay-in-orbit and radiative pion capture, which are more nearly isotropically distributed at the target. The energy distribution is also very different for electrons from radiative pion capture, which have an energy distribution that peaks at around 110 MeV and a maximum energy around 140 MeV. Beam electrons will also have a very different time distribution than electrons from muon decay-in-orbit. The distribution in time of the DIO background will follow the lifetime of the muon, while the out-of-time electrons will follow the time distribution of out-of-time protons, since high energy electrons move quite quickly down the beam line (< 75 ns). While the time distribution of out-of-time electrons cannot be measured on a pulse-by-pulse basis, the average time distribution will be measured using the data from the extinction measurement system monitoring the time distribution of out-of-time protons striking the target.

### *3.5.9* Long Transit Time Backgrounds

Long transit time backgrounds arise from particles produced in association with the primary proton pulse that take a long time to transit the beam line to the Detector Solenoid. If the transit time is long enough, the particle can arrive at the Detector Solenoid during the search window (live gate) that starts ~700 ns after the peak of the proton pulse arrives at the production target. Two sources of background from late arriving particles have already been described: the background induced by antiprotons (Section 3.5.4) and from late-arriving pions (Section 3.5.6). The antiproton background calculation is feasible because the number of antiprotons produced over the life of the experiment is relatively small and the problem can again be factorized into two independent pieces: the annihilation probability and the probability that annihilation products produced at the point of annihilation cause background. The late-arriving pion background is suppressed by delaying the measurement period for a sufficient duration after the arrival of the proton pulse until nearly all pions have decayed. The calculation is relatively straightforward, since large numbers of pions can be tracked down the beamline without decays and then weighted according to their decay probability. Backgrounds from other processes with long transit times are often more difficult to calculate, because large suppression factors cannot be calculated independently and then multiplied together to give an overall suppression factor. Backgrounds from late arrivals that can only be studied by simulating all processes starting at the production target and tracking particles through the muon beamline require generating up to $10^{18}$ protons on target, which is well beyond the available CPU capacity.

A number of potential sources of late-arriving backgrounds have been identified. The potentially largest of these backgrounds results from late-arriving muons with momenta above 76 MeV/c that can decay to electrons with energies in excess of 103 MeV. This channel was discussed previously in the section dealing with background from in-flight muon decays. Here, we examine in detail mechanisms that could cause a high-momentum muon associated with the proton pulse to be delayed sufficiently to arrive at the DS during the measurement period. Late-arriving muons can result from pion decays in areas with a relatively uniform field. If the muon is produced with a small longitudinal momentum in a uniform field region, it can take a very long time to progress down the beamline toward the muon stopping target. If the muon eventually decays in the Transport Solenoid to an energetic electron and the electron subsequently scatters in the stopping target or if the muon decays in the vicinity of the stopping target, the electron can be mistaken for a conversion electron.

Eliminating all uniform fields along the solenoidal beam path will greatly suppress late arriving backgrounds. It is for this reason that all straight sections in the Transport Solenoid have negative gradients and is one of the reasons for the negative gradient in the Production Solenoid as well. The interaction between the radial field component and the transverse component of the particle momentum leads to a net average longitudinal force on the particle that causes a steady increase in pitch (decrease in $p_t$ / p). The gradient in essence "pushes" the particles downstream. This significantly decreases the maximum time that a muon can spend in a straight section. For a 76 MeV/c muon the worst case delay is less than 50 nsec. The field gradient requirement can be relaxed in the toroidal sections of the Transport Solenoid. There the field varies as $1/r$, where $r$ is the distance from the toroid center of curvature. The radial gradient causes the spiraling particles to drift vertically. Particles with a small pitch progress slowly through the toroid and drift to the cryostat wall or to collimators where they are absorbed. It is not possible to avoid local regions of positive gradient due to the field matching requirements in the transition regions where the straight sections meet the toroidal sections of the Transport Solenoid.

Simulations indicate that these localize positive gradients are acceptable so long as the field components and variations meet requirements that are readily achievable, as described in the specifications for the Transport Solenoid.

The level of long transit time backgrounds has been estimated using the field specifications for the Mu2e solenoids. We first determined the maximum possible transit time in each of the straight and curved sections of the Transport Solenoid. All possible decay modes of $\mu \rightarrow e$, $\pi \rightarrow e$ and $\pi \rightarrow \mu \rightarrow e$ have been considered and scattering in collimators and targets were included. The presence of negative field gradients limits the backgrounds from late arriving charged particles to a negligible level (with the noted exception of background from antiprotons and radiative pion capture, treated elsewhere). An estimate of the background from neutral kaon decays was made, and found to be negligible. Examination of background from late arriving neutrons produced in the primary proton target and bouncing down the beam line to the Detector Solenoid is underway; initial estimates are that this is a negligible background source as well.

### *3.5.10* Cosmic Ray Induced Backgrounds

Background induced by cosmic rays has been observed in previous experiments and is a potentially limiting background [18]; cosmic rays must therefore be vetoed using detectors that cover a large portion of the solid angle around the Detector Solenoid. These detectors must perform with high efficiency despite a hostile environment that includes a large flux of neutrons emanating from the muon stopping target, muon beam stop, and the production target. Note that the cosmic ray background scales with live running time rather than muon intensity.

Cosmic rays can induce background through a number of mechanisms, including

- Muon decay in the Detector Solenoid.
- Muon interactions in the stopping target, proton absorber (Chapter 8), tracker, or other nearby material that produces electrons.
- Muons that enter the Detector Solenoid, scatter in the stopping target and are misidentified as electrons.

To understand the potential background from cosmic ray muons we used a GEANT4 based simulation [35]. The CRY code written by Lawrence Livermore National Laboratory [36] and a package written by the Daya Bay collaboration were both used to generate cosmic rays at the earth's surface and the results were compared for consistency. The two codes agreed to within 20%. The Daya Bay code was chosen to produce the incident cosmic ray muons and their energy, position and angular distributions were passed to the GEANT4 detector simulation package. The GEANT4 model included the detector hall with its earth and concrete overburden, a description of the solenoids and their magnetic fields, collimators, absorbers, the iron yoke surrounding the Detector Solenoid, the tracker, and the calorimeter. The track-finding and fitting code described earlier had not yet been implemented in the GEANT4 model, so track candidates were fit using an earlier version of that code using a fast simulation package (FastSim) and a Kalman filter package modified for Mu2e [37]. The results should not change significantly with the more advanced package; in fact, that version had about twice the track-finding efficiency since various sources of accidentals were not included. We keep the old numbers to be conservative.

We generated 1120 million cosmic-ray muon events. The simulation included all relevant muon decay and secondary particle production mechanisms. Charged particles that produce a minimum number of hits in the tracker were passed to FastSim. Normal track quality cuts were applied and no calorimeter requirement was imposed. A total of 35 candidate events with momenta between 80–120 MeV passed the track finding requirements. Of these, roughly 75% were from delta-ray production, with the balance from photon pair production, muon decay and Compton scattering. Nine were muons and three were positrons, which we assume can be eliminated by simple tracker cuts. Seven of the tracks either had additional tracks or were produced in the tracker and rejected, leaving 14 events that passed all cuts. One of the surviving events is shown in Figure 3.16. Assuming a signal window of 103.5 to 104.7 MeV/c and scaling this result to the total live time of the experiment, we find an upper limit of ~390 events at 90% CL (set by examining statistical and modeling errors) without a Cosmic Ray Veto (Mu2e-doc-944 and 1312, adjusted for later changes to the momentum window.)

The requirements for the Cosmic Ray Veto have been developed to limit the cosmic ray background to about 0.05 events. This requirement, combined with the simulation described above, set the efficiency requirements for the Cosmic Ray Veto. An overall inefficiency of ~0.01% is necessary to meet this requirement using the above 90% CL limit. Chapter 11 describes a Cosmic Ray Veto detector that meets this requirement.

Future simulations will include increased statistics and the real track-finding code and will map out the efficiency requirements as a function of location along the Detector Solenoid.

The cosmic ray background rate can be accurately measured when the beam is not being delivered. A direct measurement of the cosmic ray background will be performed as soon as the tracker and Detector Solenoid are in place and operating.

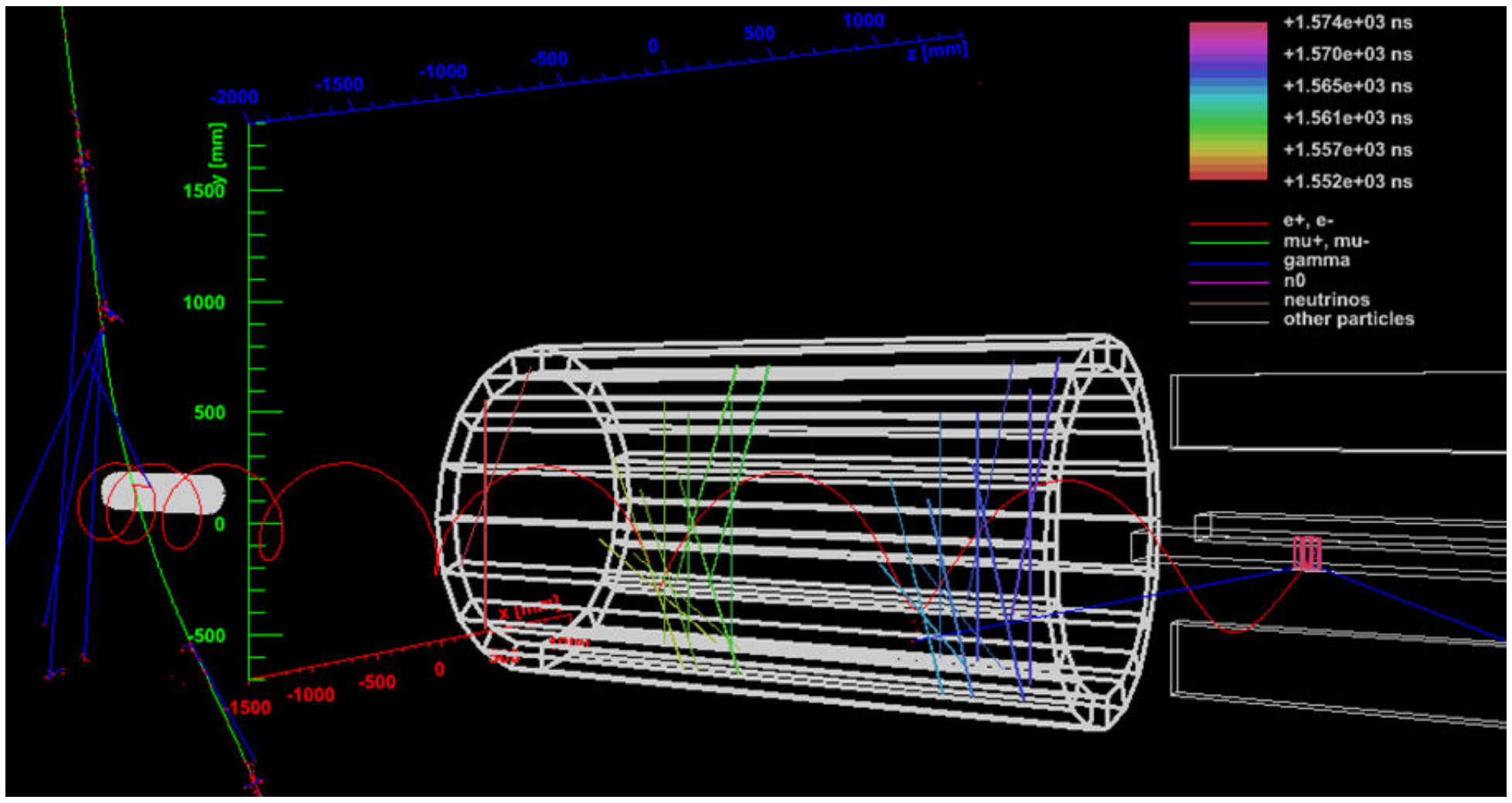

Figure 3.16. Monte Carlo cosmic ray muon induced background event. The muon enters the solenoid (not shown) at the top left, enters the target region (outlined in white), where it knocks off a 100 MeV electron that spirals into the tracker and enters the calorimeter. The frame of the tracker and outline of the calorimeter are shown in white. Colored lines within the tracker frame represent straws that register hits. The color of the straws indicates the time at which they were hit.

## 3.6 Sensitivity and Background Summary

The background estimates from this Chapter are summarized in Table 3.1. Mu2e's simulation capabilities are rapidly increasing in detail and sophistication and background estimates will continue to be evaluated as capabilities improve. In particular, the background from reconstruction errors in the tracker will be studied in detail and the results will refine future estimates. The current estimate is 0.45 background events so a considerable margin exists before the sensitivity of the experiment would be degraded.

### *3.6.1* Expected Sensitivity

The expected single-event sensitivity for a three-year run is $5.6 \times 10^{-17}$ as set out in Table 3.2. We assume a three-year run with an average beam power of 8 kW. This corresponds to two batches of $4 \times 10^{12}$ protons from the booster every 1.33 seconds. A year is assumed to consist of $2 \times 10^{7}$ seconds of running. An additional year is assumed for calibration runs, cosmic ray veto studies and special runs, for a total running time of 4 years.

## 3.7 Summary of Physics Requirements

The following physics requirements must be met in order to reject backgrounds to the required level and achieve the sensitivity set out in Table 3.2:

- Suppression of prompt backgrounds from beam electrons, muon decay in flight, pion decay in flight and radiative pion capture requires a pulsed beam where the ratio of beam between pulses to the beam contained in a pulse is less than $10^{-10}$. This ratio is defined as the beam *extinction*.
- To suppress backgrounds from muon decays in orbit, the width of the conversion electron peak, including energy loss and resolution effects, should be on the order of 1 MeV FWHM or better with no significant high energy tails.
- To suppress backgrounds from beam electrons the field in the upstream section of the Detector Solenoid must be graded. The graded field also increases the acceptance for conversion electrons.
- Suppression of backgrounds from cosmic rays requires a veto surrounding the detector. The cosmic ray veto should be nearly hermetic on the top and sides in the region of the collimator at the entrance to the Detector Solenoid, the muon stopping target, tracker, and calorimeter. The overall efficiency of the cosmic ray veto should be 0.9999 or better.
- Suppression of long transit time backgrounds places requirements on the magnetic field in the straight sections of the Transport Solenoid. The field gradient in the three TS straight sections must be continuously negative and relatively uniform.
- A thin window is required to absorb antiprotons.
- The capacity to take data outside of the search window time interval must exist.
- The capacity to collect calibration electrons from $\pi^+ \rightarrow e^+\nu$ is required [39].

| Background | Background Estimate | Error Estimate | Reference | Justification |
|---|---|---|---|---|
| Muon decay-in-orbit | 0.22 | ± 0.06 | 2085 | Acceptance and energy loss modeling, spectrum calculation; reconstruction algorithm |
| Cosmic Rays | 0.05 | ± 0.013 | CDR | Statistics of sample |
| Radiative Pion Capture | 0.03 | ± 0.007 | 2085 | Acceptance and energy loss modeling |
| Pion decay In-Flight | 0.003 | ± 0.0015 | 2085 | Cross-section, acceptance and modeling |
| Muon decay In-Flight | 0.01 | ± 0.003 | 2085 | Cross-section, acceptance and modeling |
| Antiproton Induced | 0.10 | ± 0.05 | 2121 | Cross-section, acceptance and modeling |
| Beam electrons | 0.0006 | ± 0.0003 | 2085 | Cross-section and acceptance (this is an upper limit) |
| Radiative muon capture | $< 2 \times 10^{-6}$ | – | 1230 | Calculation |
| **Total** | **0.41** | **± 0.08** | 2085 | Add in quadrature |

Table 3.1. Summary of background estimates and errors. Mu2e-doc-2085 is a more detailed summary with references.

| Parameter | Value |
|---|---|
| Running time @ $2 \times 10^7$ s/yr. | 3 years |
| Protons on target per year | $1.2 \times 10^{20}$ |
| $\mu^-$ stops in stopping target per proton on target | 0.0016 |
| $\mu^-$ capture probability | 0.609 |
| Fraction of muon captures in live time window | 0.51 |
| Electron Trigger, Selection, and Fitting Efficiency in Live Window | 0.10 |
| Single-event sensitivity with Current Algorithms | $5.6 \times 10^{-17}$ |
| Goal | $2.4 \times 10^{-17}$ |

Table 3.2. The expected sensitivities for a three year run. The numbers for the 'current algorithms' reflect results using the preliminary track recognition package, while the 'Goal' is the result when the anticipated level of efficiency for track recognition has been achieved. The preliminary package has met the interim goal of 50% of the eventual expected reconstruction efficiency.

## 3.8 References

[1] M. Raidal et al., Eur. Phys. J. **C57**, 13 (2008).
[2] J. Adam et al., Phys. Rev. Lett. 107 (2011) 171801. arXiv:1107.5547 [hep-ex]
[3] U. Bellgardt et al. (SINDRUM Collaboration), Nucl. Phys. **B299**, 1 (1988).
[4] C. Dohmen et al., Phys. Lett. **B317,** 631 (1993).
[5] J. Adam et al. (MEG Collaboration), Nucl. Phys. **B843**, 1 (2010).
[6] Y. Kuno et al., "An Experimental Search for Lepton Flavor Violating μ-e Conversion at Sensitivity of $10^{-16}$ with a Slow-Extracted Bunched Proton Beam", An Experimental Proposal on Nuclear and Particle Physics Experiments at the J-PARC 50 GeV Proton Synchrotron (2007).
[7] J. Appel et al., FERMILAB-FN-0904 (2008).
[8] M. Blanke et al., JHEP 0705, 13 (2007).
[9] L. Calibbi et al., Phys. Rev. **D74**, 116002 (2006).
[10] T. Suzuki et al., Phys. Rev. **C35**, 2212 (1987).
[11] A. Lagarrigue and C. Peyrou, Comptes Rendus Acad. Sci. Paris, 234, 1873(1952). See also J. Steinberger and H. Wolfe, Phys. Rev. 100, 1490 (1955).
[12] M. Conversi et al., Phys. Rev. **D122**, 687 (1961).
[13] R. Sard et al., Phys. Rev. **121**, 619 (1961).
[14] G. Conforto et al., Nuovo Cimento **26**, 261 (1962).
[15] J. Bartley et al., Phys. Lett. **13**, 258 (1964).
[16] D. Bryman et al., Phys. Rev. Lett. **28**, 1469 (1972).
[17] A. Badertscher et al., Phys. Rev. Lett. **39**, 1385 (1977).
[18] S. Ahmad et al., Phys Rev. **D38**, 2102 (1988).
[19] W. Bertl et al., Eur. Phys. J. **C47**, 337 (2006).
[20] R. Djilkibaev and V. M. Lobashev, Sov. J. Nucl. Phys. **49(2)**, 384 (1989).
[21] J. C. Gallardo in Beam Dynamics and Technology Issues for $\mu^+\mu^-$ colliders. Proceedings, 9th Advanced ICFA Beam Dynamics Workshop, Montauk, USA, October 15 – 20, 1995 (American Institute of Physics, Woodbury, N.Y., 1996).
[22] R. Palmer et al., Nucl. Phys. Proc. Suppl. **51A**, 61 (1996).
[23] Y. Kuno, COMET report to the J-PARC PAC, (2011), http://nuclpart.kek.jp/pac/1101/pdf/J-PARCPAC110115_KUNO.pdf
[24] A. Czarnecki, W.J. Marciano, and Xavier Garcia i Tormo, Phys. Rev. D84(2011) 013006, arXiv:1106.4756 [hep-ph]
[25] D. Brown, http://dnbmac3.lbl.gov/~brownd/SuperB/FastSim_workshop.pdf.
[26] J. A. Bistirlich et al., Phys. Rev. **C5**, 1867 (1972).
[27] R. Tschirhart, "Requirements of the Mu2e Trigger and data Acquisition System." Mu2e-doc-1150.
[28] P. C. Bergbusch et al., Phys. Rev. C **59**, 2853 (1999).
[29] J. Kaulard et al., Phys. Lett. **B422**, 334 (1998).
[30] A. Vaisenberg et al., JETP Lett. **29**, 661 (1979).

[31] B. Kopeliovich and F. Niedermayer, Phys. Lett. **B151**, 437 (1985).

[32] R. Armenteros et al., Phys. Rev. 119, 2068 (1960).

[33] G. J. Marmer et al., Phys. Rev. **179**, 1924 (1969).

[34] J. Miller and S. Striganov, “background from Pbars Stopping in the Stopping Target,” Mu2e doc-db 2121.

[35] V. Jorjadze, “Cosmic Ray Background Simulation for the Mu2e Experiment,” Mu2e-doc-1567.

[36] C. Hagmann, D. Lange, and D. Wright, “Cosmic-ray Shower Library,” Lawrence Livermore National Laboratory, UCRL-TM-229453.

[37] D. Brown and S. Siu, “Mu2e fastsim,” Mu2e-doc-915.

[38] R. M. Carey et al., “Mu2e Proposal,” Mu2e-doc-388.

[39] D. Brown et al., “Calibration Requirements of the Mu2e Experiment,” Mu2e-doc-1182.

[40] Y. Kuno and Y. Okada, “Muon Decay and Physics Beyond the Standard Model”, Rev. Mod. Phys. **53**, 802 (1991). arXiv:hep-ph/9909265v1". See eq. 111, which is commonly attributed to A. Van der Schaaf.

[41] David Measday, "The Physics of Muon Capture", Physics Reports **354**, 243-409 (2001).

# 4 Overview of the Mu2e Design

## 4.1 Introduction

The alternatives for a Mu2e facility that satisfy the mission objectives and are consistent with the documented requirements have been evaluated. An overview of the selected set of alternatives is presented in this chapter. A more detailed description of each significant project component, along with a detailed description of the requirements that drive the design and a discussion of the other evaluated alternatives, can be found in Chapters 5 – 12.

Mu2e proposes to measure the ratio of the rate of the neutrinoless, coherent conversion of muons into electrons in the field of a nucleus, relative to the rate of ordinary muon capture on the nucleus:

$$R_{\mu e} = \frac{\mu^- + A(Z,N) \rightarrow e^- + A(Z,N)}{\mu^- + A(Z,N) \rightarrow \nu_\mu + A(Z-1,N)}.$$

The signature of this process is a monoenergetic electron with an energy nearly equivalent to the muon rest mass (Section 3.2). The conversion process is an example of charged lepton flavor violation (CLFV), a process that has never been observed experimentally. The significant motivation behind the search for muon-to-electron conversion has been discussed in Chapter 3. The best experimental limit on muon-to-electron conversion, $R_{\mu e} < 6.1 \times 10^{-13}$ (90% CL), is from the SINDRUM II experiment [1]. Mu2e intends to probe four orders of magnitude beyond the SINDRUM II sensitivity, measuring $R_{\mu e}$ with a single-event sensitivity of $5.6 \times 10^{-17}$.

## 4.2 Proton Beam

The high flux of muons required for Mu2e result from the decay of pions generated by a high-intensity proton beam striking a production target. To reduce backgrounds, Mu2e requires bunches of protons that are separated by about twice the stopped muon lifetime (Section 3.4). The Fermilab accelerator complex is ideally suited to this task with the availability of the Recycler and Debuncher Rings at the conclusion of Tevatron Collider operations. The revolution period of 8.9 GeV/c protons in the Debuncher ring is 1685 ns, compared to the stopped muon lifetime in aluminum of 864 ns. The preferred alternative for delivering beam to the Mu2e detector is shown in Figure 4.1.

The Fermilab Booster accelerates batches of $4 \times 10^{12}$ protons to 8 GeV kinetic energy. Batches of protons from the Booster will be transported through the existing MI-8 line and injected into the Recycler Ring using an injector kicker being constructed as part of the NOvA Project. The Booster batches will be redistributed into 4 smaller bunches in the Recycler Ring by a 2.5 MHz RF system provided by g-2. The Recycler operations all take place while protons destined for NOvA are being ramped to 120 GeV in the Main Injector and the Recycler is otherwise empty. Each of the newly formed bunches will be kicked into the P1 line, one at a time, and transported to the Debuncher Ring through the existing P2, AP1 and AP3 lines. The proton bunches will be injected into the Debuncher Ring using a new injection kicker and captured by a new 2.4 MHz RF system consisting of RF modules that are identical to the RF modules being built for the Recycler Ring for g-2. The proton bunches will be slow extracted to Mu2e through a new external beamline. Third integer resonant extraction will be controlled using the RF knockout (RFKO) technique to heat the beam transversely upstream of the extraction septum. The RFKO system will provide feed-back for fine control of the spill rate. To excite and control the third-integer resonance, six harmonic sextupoles will be added to the base lattice. Three quadrupoles will be inserted into symmetric locations in the three straight sections for fine tune control throughout a spill.

The final run from the Debuncher Ring to the Mu2e detector requires construction of a new external beamline. Included in the beamline design is a fast AC dipole system that will deflect out-of-time beam into a collimation system ensuring that only in-time protons are transported to the production target. This *extinction system* is required to control prompt backgrounds (Section 3.2). Additional extinction is achieved in the Debuncher Ring. A momentum scraping system will remove particles that migrate outside of the RF buckets. The overall extinction, defined as the ratio of beam between pulses to the beam in each pulse must be smaller than $10^{-10}$ [2].

The primary production target is located in the evacuated warm bore of the Production Solenoid. The production target is a radiatively-cooled tungsten rod, 16 cm long with a 3 mm radius, compared to the beam profile of about 1 mm in radius. The target surface will be grooved to enhance emissivity. Tungsten is chosen for its high atomic number, necessary to optimize pion production, it's high melting point and it's high thermal conductivity that helps to achieve an acceptably low core temperature. The thin, pencil-like geometry is designed to minimize pion re-absorption.

Two Booster batches can be sequentially processed as described above during the part of the 1.33 second Main Injector cycle when the Recycler is not being used by NOvA. This corresponds to $8 \times 10^{12}$ protons per cycle for an average of $6 \times 10^{12}$ protons per second and a total of $1.2 \times 10^{20}$ protons per year ($2 \times 10^{7}$ sec.).

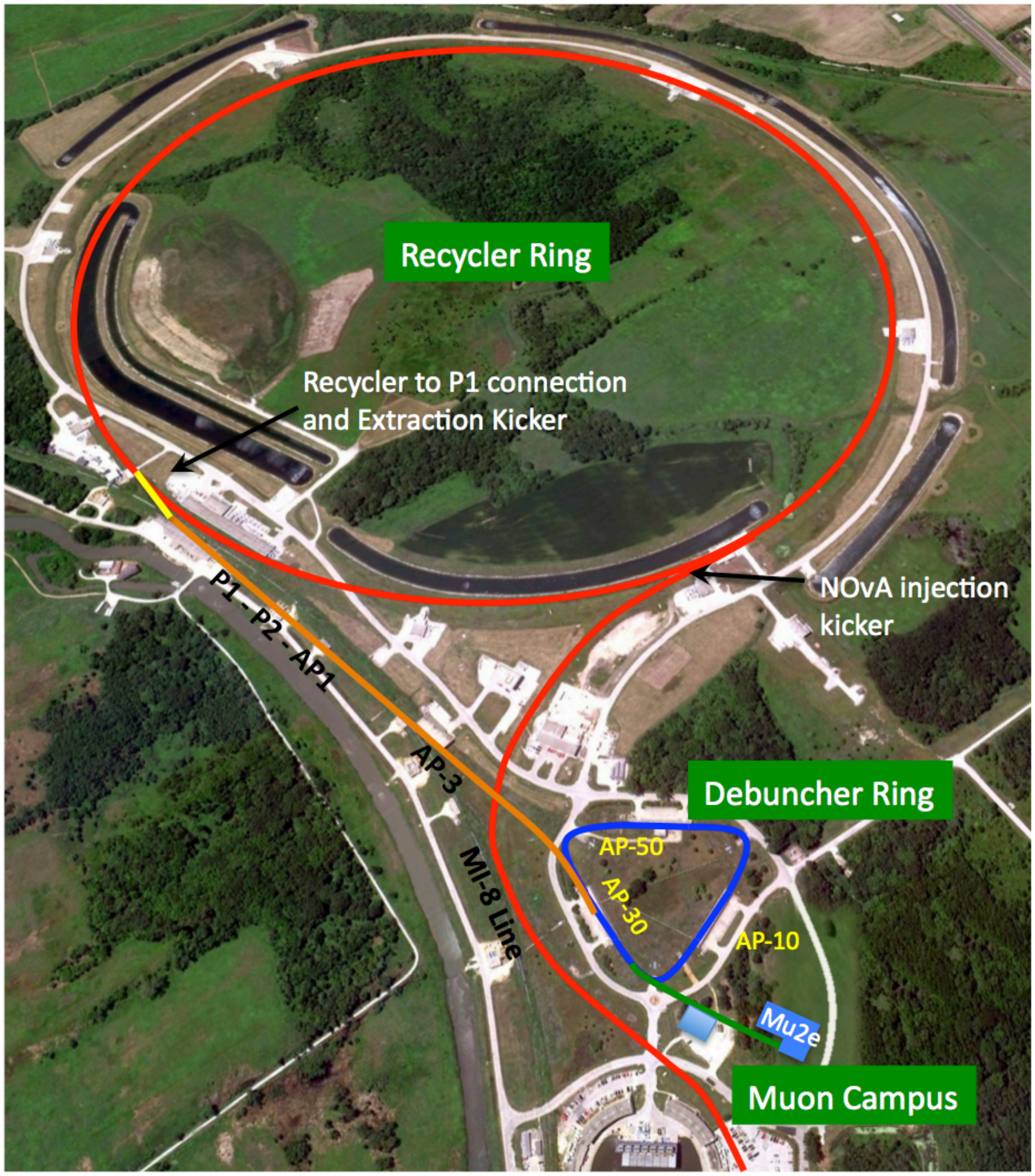

Figure 4.1. The path of protons from the Fermilab Booster to the Mu2e detector.

Nearly 8 kW of beam power could be delivered to the Mu2e production target. This is well beyond the beam power seen in the Antiproton facility during Tevatron collider operations. Sky shine that results from distributed losses around the Debuncher Ring is a concern because of the minimal shielding between the beamline enclosure and the 3 Antiproton service buildings (AP-10, AP-30 and AP-50, shown in Figure 4.1). The Mu2e beam power will be limited to the point where the exposure to the public is less that 1 mRem per year. This corresponds to continuous beam losses of 0.125% around the Debuncher Ring for 8 kW beam (2 Booster batches per Main Injector cycle) and 0.25% continuous losses for a beam power of 4 kW (1 Booster batch per Main Injector cycle). A system of Total Loss Monitors (TLMs) will be used to hold the losses to the appropriate level. Many months of beam commissioning will likely be required to optimize beam transport and minimize losses. However, there is a significant risk that Mu2e will not attain the level of beam losses required to operate at 8 kW. This risk can be mitigated with additional shielding at the AP Service buildings or with longer running time. The Mu2e Project has made no provisions for additional shielding. Point-source beam losses at the extraction septa and kickers will be many times larger than the distributed losses, but they will be locally shielded, mitigating their impact.

Fermilab personnel in the Accelerator Division will perform most of the work on the accelerator and beamline. This work will be executed in close coordination with the Accelerator Division, Facility Engineering Services, the g-2 Project and Fermilab Program Planning. The total scope of work required to deliver beam to Mu2e includes work included in the NOvA and g-2 projects as well as a number of common projects for the Muon Program. The Debuncher RF system, radiation shielding, the resonant extraction system, and all beamline components downstream of g-2, including the extinction system, are part of the Mu2e Project. Recycler upgrades, upgrades to the transfer lines between the Recycler and the Debuncher, the injection and abort kickers in the Debuncher Ring and the external beamline enclosure are all provided by other projects and are not included in the Mu2e project scope.

This proposed design satisfies all of the requirements developed by the Mu2e Collaboration [2]. A detailed description of the beam delivery scheme for Mu2e is described in Chapter 5.

## 4.3 Superconducting Solenoids

The dominant feature of the Mu2e apparatus is the system of three superconducting solenoids and their associated infrastructure (Figure 4.2). The solenoids are an integrated magnetic system that house the production target, the muon transport channel, the muon stopping target and the detector components that detect and analyze conversion electrons. The solenoids are divided into 3 functional systems; the *Production Solenoid*, the *Transport Solenoid* and the *Detector Solenoid*. The solenoid system also includes power supply and quench protection systems, cryogenic distribution, field mapping equipment and a significant installation and commissioning effort. A central cryo facility for the Muon Campus will be housed in the g-2 detector hall and will be funded off-project.

### *4.3.1* Production Solenoid

The Production Solenoid is shown in Figure 4.3. It is a relatively high field solenoid with an axial grading that varies from 4.6 Tesla to 2.5 Tesla. The Production Solenoid houses the production target. The purpose of the Production Solenoid is to trap charged pions from the production target and accelerate them towards the Transport Solenoid as they decay to muons.

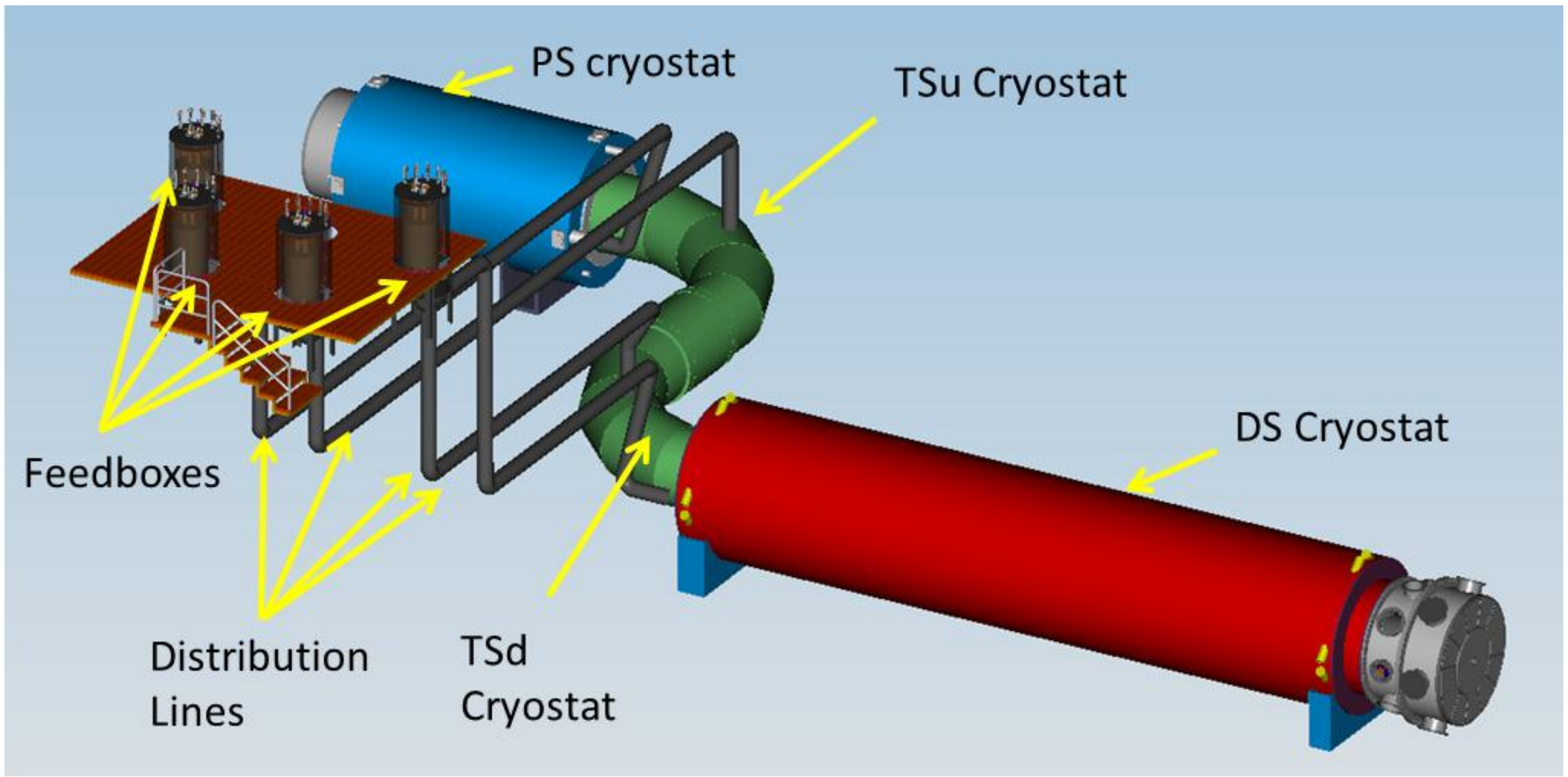

Figure 4.2. The Mu2e superconducting solenoid system, including the Production, Transport and Detector Solenoids and the cryogenic distribution system. Not shown are the power supply and quench protection systems.

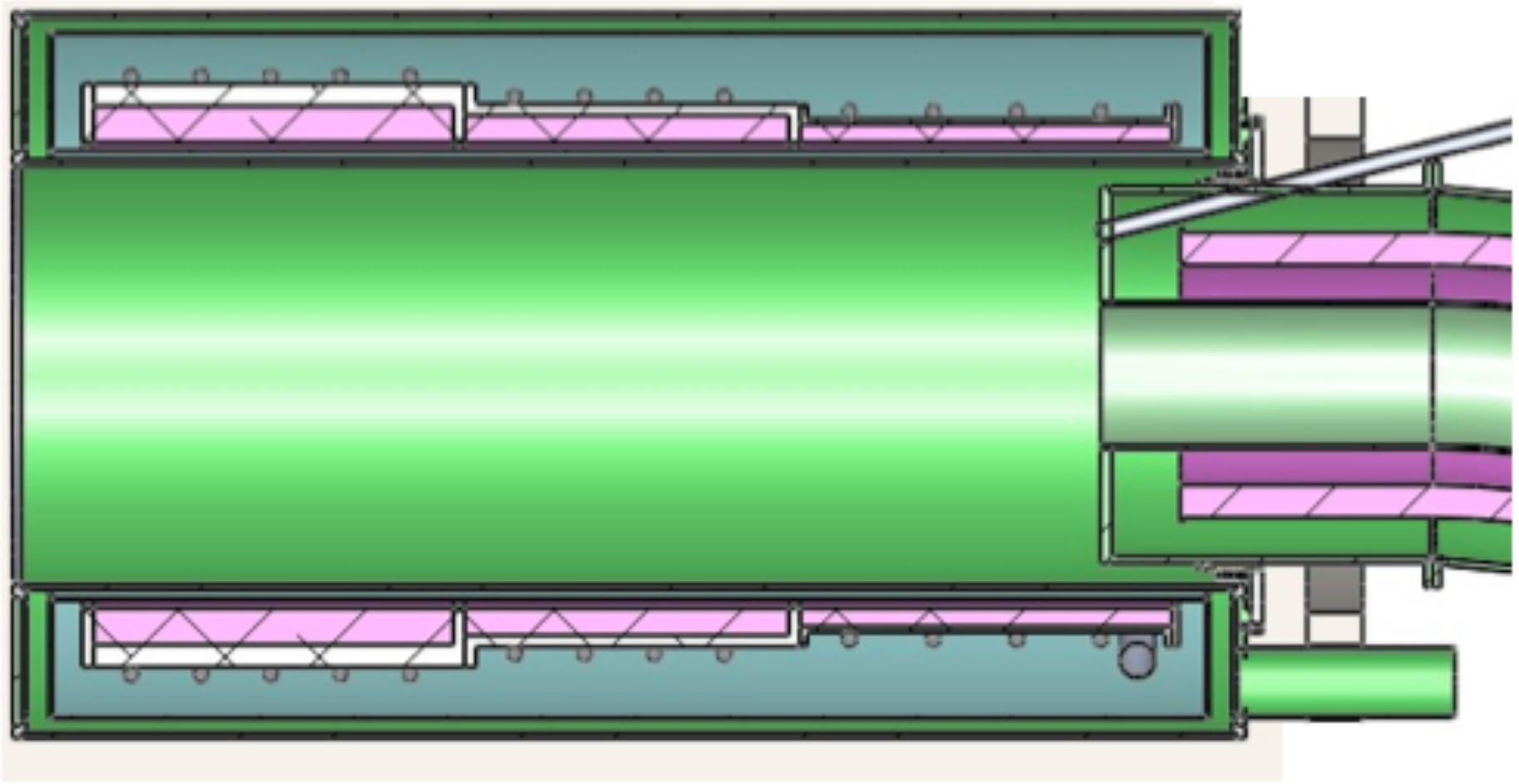

Figure 4.3. The Mu2e Production Solenoid. The beam tube for the incoming proton beam is shown in the upper right.

Protons enter the Production Solenoid through a small port on the low field side of the solenoid before intercepting the production target. Remnant protons that are not absorbed by the target and very forward-produced secondary particles exit at the high field end of the solenoid. Pions in the forward direction with angles greater than ~30°, relative to the solenoid axis, are reflected back by the higher field and move along with the backward produced particles in helical trajectories towards the Transport Solenoid.

The Production Solenoid must generate a uniform axially graded field ranging from 4.6 T to 2.5 T. This axial field change is accomplished using three solenoid coils with 3, 2 and 2 layers of aluminum stabilized NbTi superconducting cable, each coil with the same inner diameter. Aluminum stabilizer is used to reduce both the weight of the solenoid and the amount of nuclear heating around the conductor. Nuclear heating can result from the large flux of secondaries from the production target. Mu2e will use high-current, low-inductance cable that allows for efficient energy extraction during a quench, requires fewer layers to achieve the required field strength and minimizes thermal barriers between the conductor and cooling channels.

### *4.3.2* Heat and Radiation Shield

Lining the inside of the Production Solenoid warm bore is a heat and radiation shield designed to protect the solenoid's superconducting coils. The Heat and Radiation Shield is designed to limit the heat load in the cold mass to prevent quenching, limit radiation damage to superconductor insulation and epoxy and limit the damage to the superconductor's aluminum stabilizer. The shield is constructed from bronze. Because the proton beam is incident from one side of the Production Solenoid, the pattern of energy deposition in the heat shield is asymmetric with the largest depositions being near the target and collinear with the incoming proton beam direction. Even with the protection of the Heat and Radiation Shield, a significant number of atomic displacements will occur over time in the aluminum stabilizer surrounding the superconductor. The Residual Resistivity Ratio (RRR) of the aluminum, the ratio of the electrical resistance at room temperature of a conductor to that at 4.5 K, will decrease to the point where the stabilizer cannot adequately protect the superconductor in the event of a quench. The RRR can be completely recovered by warming the aluminum stabilizer to room temperature. Based on models of neutron production and energy deposition, it is anticipated that it will only be necessary to warm up once per year, coincident with annual accelerator shutdowns. The Heat and Radiation Shield is shown in Figure 4.4 and described in detail in Section 5.9.2.

### *4.3.3* Transport Solenoid

The S-shaped Transport Solenoid consists of a set of superconducting solenoids and toroids that form a magnetic channel that transmits low energy negatively charged muons traveling in helical trajectories from the Production Solenoid to the Detector Solenoid (Figure 4.5). Negatively charged particles with high energy, positively charged particles and line-of-sight neutral particles are nearly all eliminated by absorbers and collimators before reaching the Detector Solenoid. Selection of negatively charged muons is accomplished by taking advantage of the fact that a charged particle beam traversing a toroid will drift perpendicular to the toroid axis, with positives and negatives drifting in opposite directions. Most of the positively charged particles are absorbed in the central collimator.

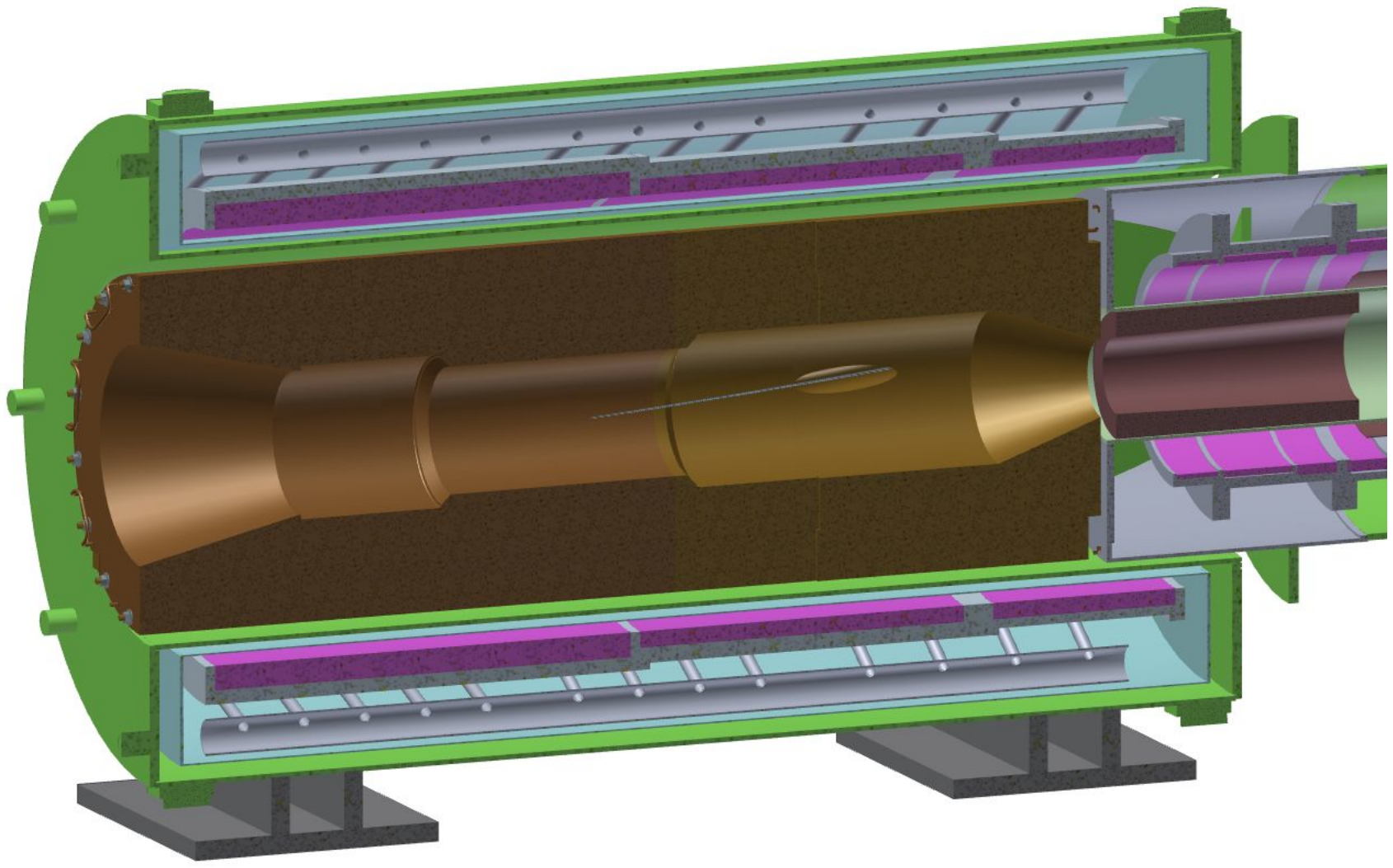

Figure 4.4. A section view through the Heat and Radiation Shield that lines the inside of the Production Solenoid.

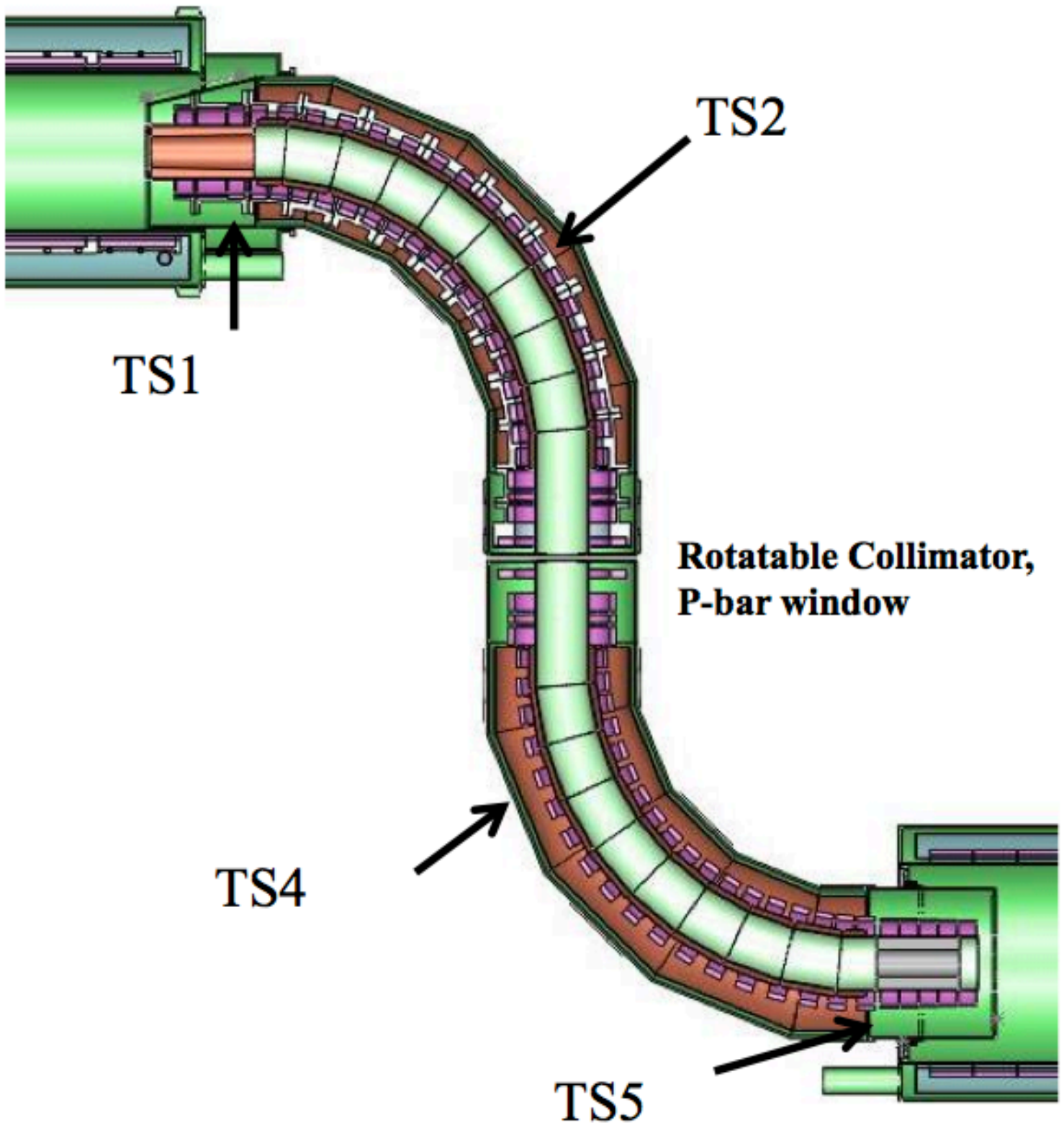

Figure 4.5. The Mu2e Transport Solenoid.

Late arriving particles are a serious potential background for Mu2e (Section 3.5.9). To minimize the transport of particles that spend a long time in the magnet system the magnetic field in the straight sections is designed to always have a negative gradient. This eliminates traps, where particles bounce between local maxima in the field until they eventually scatter out and travel to the Detector Solenoid where they arrive late compared to the beam pulse. The requirement on a negative gradient is relaxed in the curved sections of the TS because bouncing particles will eventually drift vertically out of the clear bore and be absorbed by surrounding material.

### *4.3.4* Detector Solenoid

The Detector Solenoid is a large, low field magnet that houses the muon stopping target and the components required to identify and analyze conversion electrons. The muon stopping target resides in a graded field that varies from 2 Tesla to 1 Tesla. The graded field captures conversion electrons that are emitted in the direction opposite the detector components causing them to reflect back towards the detector. The graded field also shifts the pitch of beam particles that enter the Detector Solenoid and travel to the tracker, playing an important role in background suppression. The actual detector components reside in a field region that is relatively uniform. The inner bore of the Detector Solenoids is evacuated to $10^{-4}$ Torr to limit backgrounds from muons that might stop on gas atoms.

The graded and uniform field sections of the Detector Solenoid are wound on separate mandrels but housed in a common cryostat. The conductor is aluminum stabilized NbTi. The gradient is achieved by introducing spacers to effectively change the winding density of the superconducting cable. The Detector Solenoid is shown in Figure 4.6.

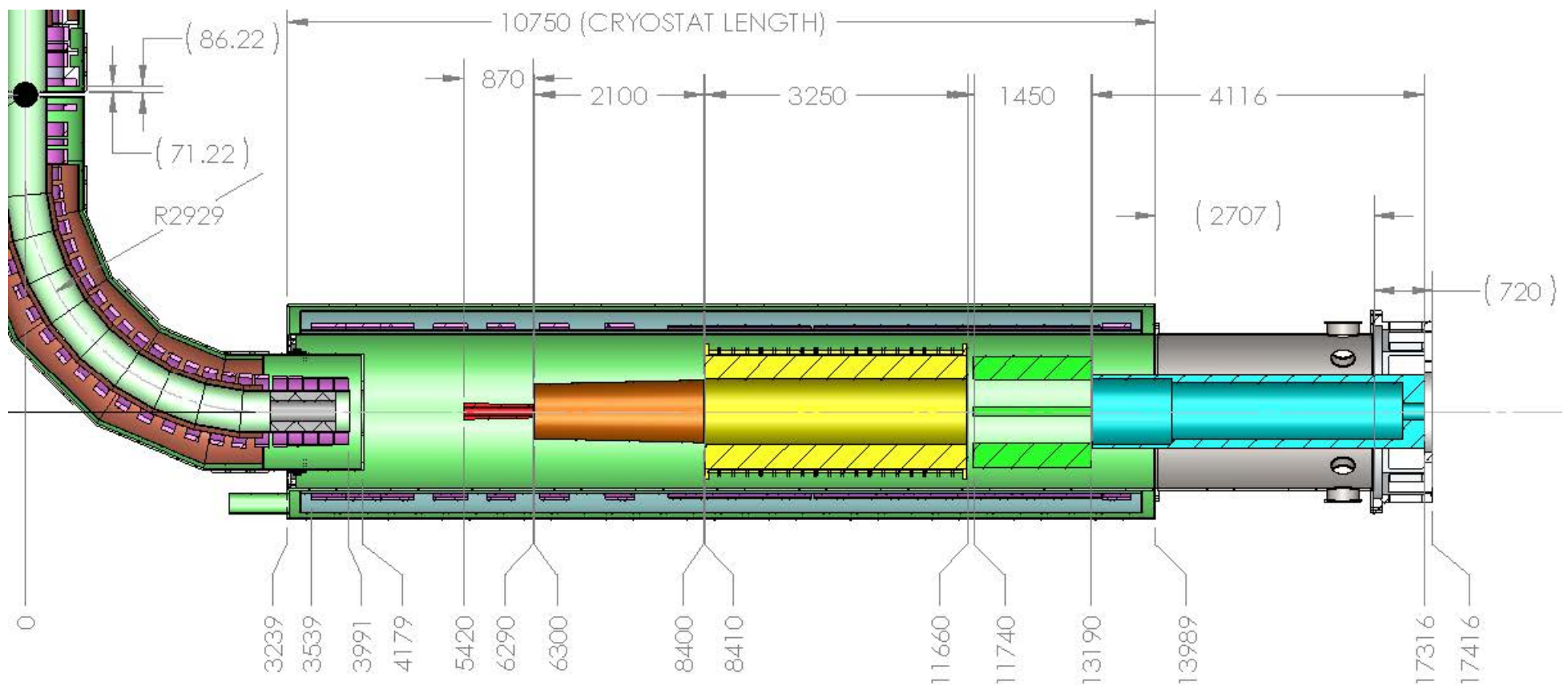

Figure 4.6. The Mu2e Detector Solenoid. The upstream section has a graded field to capture more conversion electrons. The downstream section has a uniform field in the region occupied by the detector elements.

### *4.3.5* Solenoid Acquisition Strategy

Extending the conceptual designs of the solenoid systems to final designs will require significant engineering and design resources that are not currently available to Mu2e. The solenoid acquisition strategy relies on the similarity of the Production and Detector Solenoids to magnets that have previously been designed and fabricated in industry. Neither solenoid is a carbon copy of a known, existing magnet as they each have their own unique design complexities. However, there are enough similarities in the size, geometry, mechanics, conductor and cooling techniques to provide confidence that vendors will be interested in the opportunity and will be able to deliver magnets that meet the Mu2e requirements. The procurement of the Production and Detector Solenoids will be similar to other successful magnet procurements at Fermilab.

The Transport Solenoid is unique and unlike the magnet systems commonly procured from industry. Resources exist at Fermilab to complete a final design on the Transport Solenoid. The various pieces would be procured from industry, including the conductor, the cryostats and possibly the wound coils. The Fermilab Technical Division would then act as the general contractor in assembling the pieces into the final system.

## 4.4 Secondary Muon Beam

Mu2e requires a significant number of negatively charged muons to be stopped in a thin target. In order to efficiently transport muons, minimize scattering off of residual gas molecules, minimize multiple scattering of conversion electrons and prevent electrical discharge from detector high voltage the Muon Beamline must be evacuated to the level of $10^{-4}$ Torr.

The muon stopping target must be massive enough to stop a significant fraction of the incident muon beam but not so massive that it corrupts the momentum measurement of conversion electrons that emerge. Lower energy muons allow for a thinner target to alleviate these concerns. The momentum distribution of muons at the Mu2e stopping target is shown in Figure 4.7. The number of muons that reach and stop in the stopping target depends on a number of factors. These include the target material and geometry, the proton beam energy, the magnetic field in the Production and Transport Solenoids, the clear bore of the solenoids and the design of the collimators.

In order to optimize the number of stopped muons a detailed simulation package with an accurate particle production model is required. The calculated values of particle fluxes in the secondary muon beam are based on GEANT4 simulations of proton interactions in a tungsten target. GEANT4 has a variety of hadron interaction codes and the cross sections and kinematic distributions can vary significantly between them. In order to reduce exposure to the uncertainty in the hadronic models of low energy hadron production, the results from GEANT4 have been normalized to data from the HARP experiment [3]. HARP measured the double differential cross-section for production of charged pions emitted at large production angles in proton-tantalum collisions at 8 GeV/c. The data from HARP does not cover the full kinematic range required for Mu2e. To cover the full range required for Mu2e the QGSP-BERT hadronic model [4] is used. QGSP-BERT is one of the physics lists available in GEANT4. QGSP-BERT and HARP are consistent in the region where they overlap. As a crosscheck, the production model is compared to the results from a Novosibirsk experiment [5] where measurements of pion production are reported in 10 GeV/c proton-tantalum interactions with more coverage in the backward direction than provided by HARP. This results in 0.0016 stopped $\mu^-$ per proton on target when all of the material in the muon beamline, including support structures and the antiproton window (see Section 3.5.4), are included. Errors on the double differential cross-section measurements by HARP are in the 10% range. The QGSP-BERT model and the difference between tungsten and tantalum introduce additional uncertainty. The overall uncertainty on the stopped muon rate is conservatively estimated to be at the 25% level.

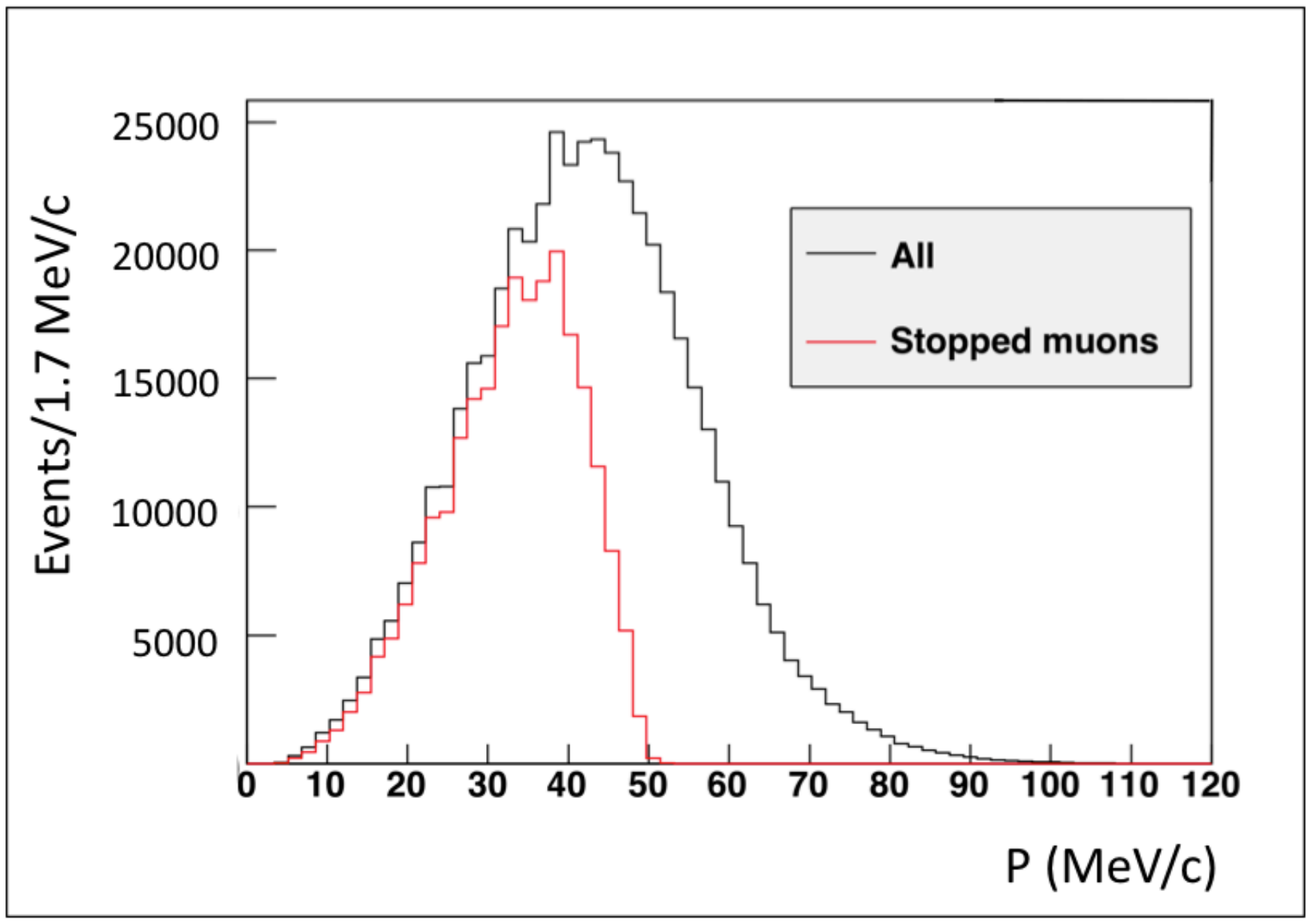

Figure 4.7. Particle momentum at the Mu2e stopping target. The black curve is the momentum of all muons that reach the stopping target and the distribution in red is the momentum spectrum of muons that stop in the target.

### *4.4.1* Muon Stopping Target

The muon stopping target should maximize the number of stopped muons while minimizing the amount of material traversed by conversion electrons that enter the acceptance of the downstream detector.

A distributed stopping target that consists of 17 aluminum disks, 0.2 mm thick, arranged coaxially along the Detector Solenoid axis has been designed. The disks are spaced 5.0 cm apart with radii that decrease from 8.3 cm upstream to 6.53 cm downstream. The tapered disks reduce the amount of material seen by conversion electrons produced in the target that are within the acceptance of the detector. The Mu2e stopping target is shown in Figure 4.8.

Energy loss and straggling in the stopping target are significant contributors to the momentum resolution function. It may be possible to optimize the geometry of the stopping target to stop more muons while minimizing the amount of material traversed by conversion electrons within the acceptance of the tracker. Conical geometries have been discussed in the past and will be investigated further before a final engineering design for the stopping target is completed.

Because of the diffuse nature of the muon beam a significant number of muons can strike the structure supporting the stopping target, producing DIO electrons at large radius where the acceptance for reconstruction in the detector is high. Because the endpoint energy of the DIO spectrum decreases for higher Z materials, and because the lifetime of the muons in muonic atoms decreases for higher Z materials, the supports must be constructed from high Z materials. Tungsten wires have been chosen for the target support.

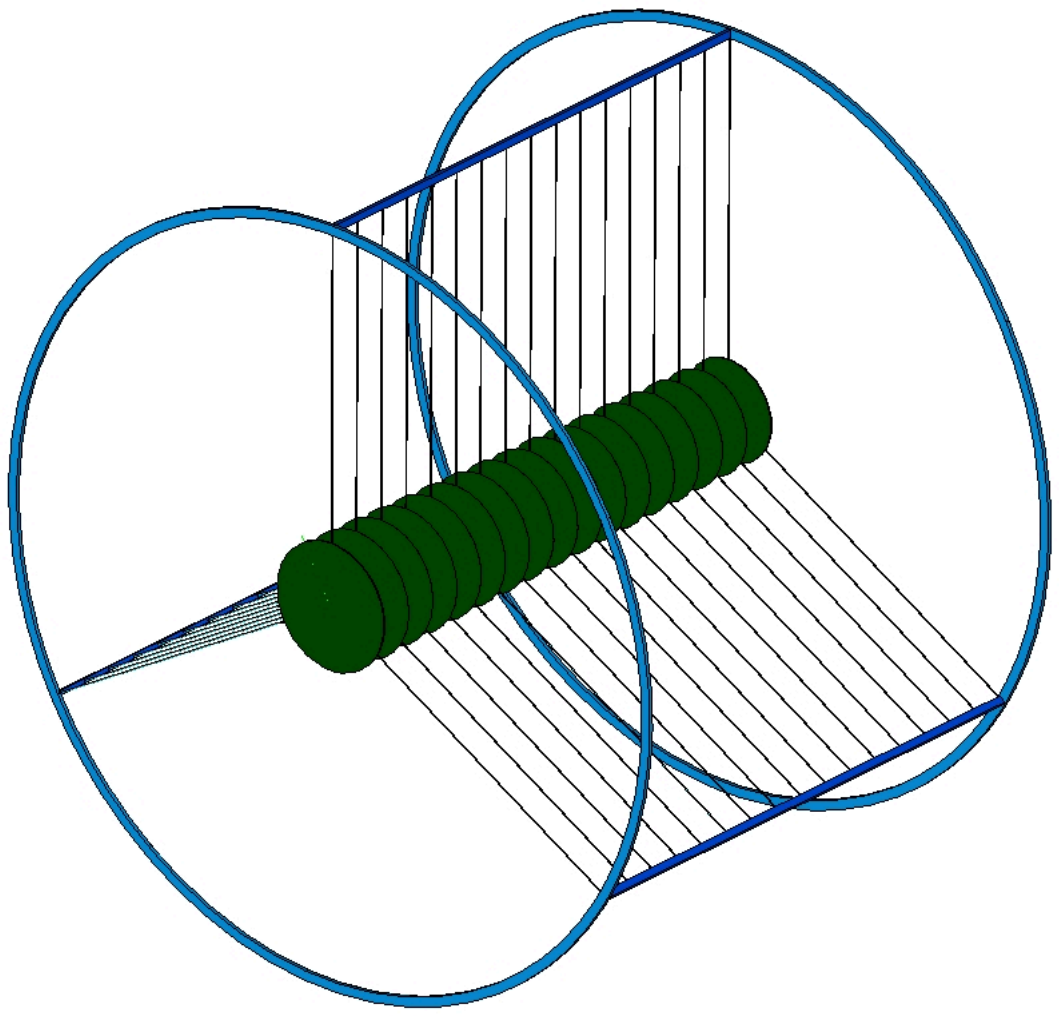

Figure 4.8. The muon stopping target and mechanical support.

### *4.4.2* Stopping Target Monitor

To measure $R_{\mu e}$ (Section 4.1), it is necessary to measure the number of muons that are captured in the stopping target. An effective and reliable Muon Stopping Target Monitor can be established by observing the prompt production of X-rays that signal the formation of muonic atoms in the target foils. The capture rate is derived from the rate of muons that stop and form muonic atoms. As an example, a spectrum of muonic aluminum X-rays measured with a germanium detector is shown in Figure 4.9.

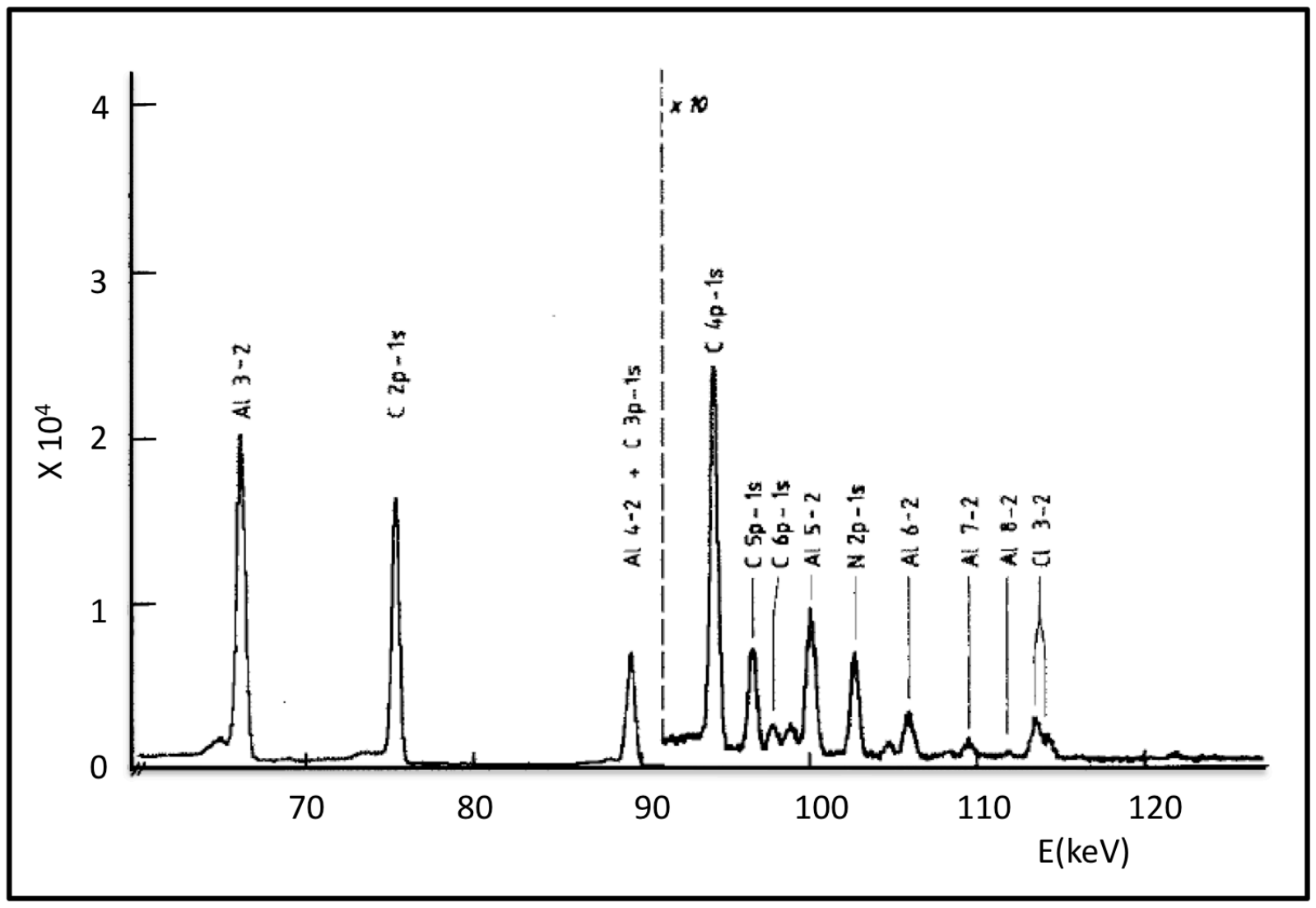

Figure 4.9. The X-ray spectrum from muonic aluminum [6].

A germanium crystal will be used to observe the energy spectrum of the muonic X-rays from the stopping target. The detector will be located near the axis of the Detector Solenoid, downstream of the endcap steel. The germanium detector will view the stopping target through a small diameter vacuum pipe, connected to the downstream end of the Detector Solenoid, that provides collimation for the X-rays. A window will separate the Detector Solenoid from the collimation pipe.

### *4.4.3* Proton Absorber

When muons are captured on a nucleus the process is accompanied by the emission of protons, neutrons and gamma rays. The produced secondaries are low in energy and cannot directly produce backgrounds, but by contributing rate to the detectors they have the potential to cause reconstruction errors that can result in backgrounds and reduce acceptance.

The largest potential contribution to the tracker rate during the measurement period is from protons that originate from muon capture. The protons are very slow moving and highly ionizing, so they can be easily identified in the tracking detector using pulse height information. The large pulse height associated with protons traversing the tracker will shadow hits from low energy electrons. This could lead to reconstruction errors that might create background. A thin polyethylene absorber between the stopping target and the tracking detector will significantly attenuate the proton rate. The proton absorber is a tapered cylindrical shell 0.5 mm thick with a radius slightly smaller than the inner radius of the tracker. However, conversion electrons also pass through the proton absorber and the associated energy loss and straggling can contribute to the momentum resolution function, so the geometry and material budget of the proton absorber must be precisely optimized before fabrication.

### *4.4.4* Muon Beam Stop

A significant fraction of the muons incident on the muon stopping target do not stop in the target but continue on through the evacuated center of the Detector Solenoid. The muon beam stop absorbs the energy of beam particles that reach the downstream end of the Detector Solenoid. The muon beam stop is made primarily of lead and plastic. Plastic is effective at absorbing neutrons and lead is used because it is effective in absorbing electrons and photons. Additionally, muons have a relatively short lifetime in lead, so the majority of electrons, protons, neutrons and gamma rays that result from muon decay-in-orbit or capture on the nucleus will appear before the beginning of the measurement period and have no impact. The Muon Beam Stop is located within the bore of the Detector Solenoid, downstream of the calorimeter where the magnetic field is falling to zero. The field gradient in the beam stop region helps to prevent particle splash-back from reaching the active detectors.

## 4.5 The Detector

The Mu2e detector is located inside the evacuated warm bore of the Detector Solenoid in a uniform 1 Tesla magnetic field and is designed to efficiently and accurately identify the helical trajectories of ~105 MeV electrons in the high, time varying rate environment of Mu2e. The detector consists of a tracker and a calorimeter that provide redundant energy/momentum, timing, and trajectory measurements. A cosmic ray veto, consisting of both active and passive elements, surrounds the Detector Solenoid.

### *4.5.1* Tracker

The Mu2e tracker is designed to accurately measure the trajectory of electrons in a uniform 1 Tesla magnetic field in order to determine their momenta. The limiting factor in accurately determining the trajectory of electrons is multiple scattering in the tracker. High rates in the detector may lead to errors in pattern recognition that can reduce the acceptance for signal events and possibly generate backgrounds if hits from lower energy particles combine to create trajectories that are consistent with conversion electrons. A low mass, highly segmented detector is required to minimize multiple scattering and handle the high rates.

The selected alternative for the Mu2e tracker is a low mass array of straw drift tubes aligned transverse to the axis of the Detector Solenoid, referred to as the T-tracker. The basic detector element is a 25 μm sense wire inside a 5 mm diameter tube made of 15 μm thick metalized Mylar®. The tracker will have ~22,000 straws distributed into 18 measurement stations across a ~3 m length. Planes are constructed from two layers of straws, as shown in Figure 4.10, to improve efficiency and help determine on which side of the sense wire a track passes (the classic "left-right" ambiguity). A 1 mm gap is maintained between straws to allow for manufacturing tolerance and expansion due to gas pressure. The straws are designed to withstand changes in differential pressure ranging from 0 to 1 atmosphere for operation in vacuum. The straws are supported at their ends by a ring at large radius, outside of the active detector region. The tracker is shown in Figure 4.11.

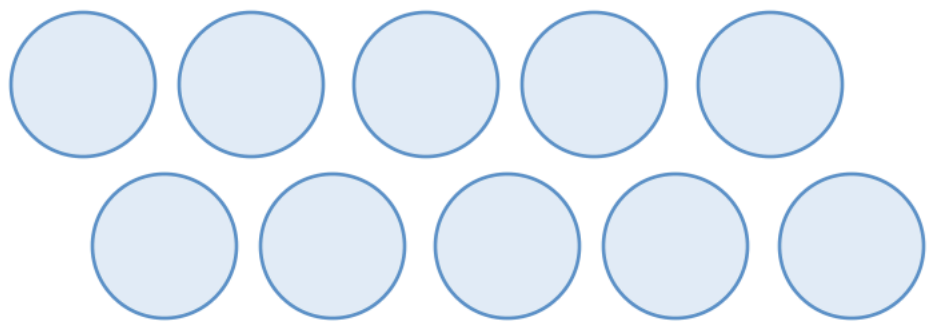

Figure 4.10. A section of a two-layer tracker straw plane. The two layers improve efficiency and help resolve the left-right ambiguity.

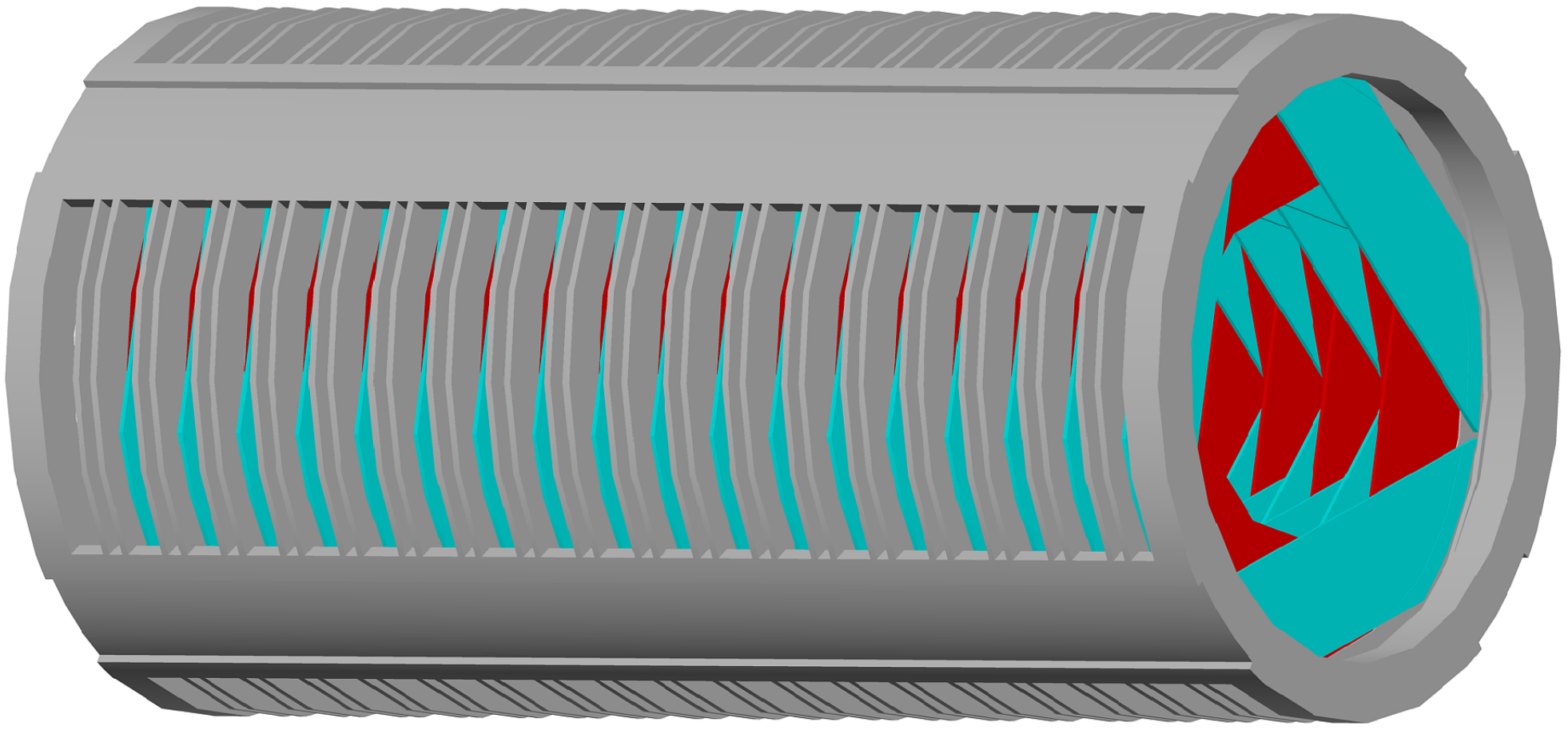

Figure 4.11. The Mu2e straw tube tracker. The straws are oriented transverse to the solenoid axis.

Each straw will be instrumented on both ends with preamps and TDCs that will be used to measure the drift time to determine the distance of approach of charged tracks relative to the drift wire. The arrival time of the signal at each end of the straw will be measured in order to determine the location of the track intercept along the length of the straw. Each straw will also be instrumented with an ADC for dE/dx capability to separate electrons from highly ionizing protons. To minimize penetrations into the vacuum, digitization will be done at the detector, with readout via optical fibers. A liquid cooling system will be required for the electronics to maintain an appropriate operating temperature in vacuum.

The tracker is designed to intercept only a small fraction of the significant flux of electrons from muon decays-in-orbit. The vast majority of electrons from muon decay in orbit are below 60 MeV in energy (Figure 3.4). Only electrons with energies greater than about 53 MeV, representing a small fraction of the rate (about 3%) will be observed in the tracker. Lower energy electrons will curl in the field of the Detector Solenoid and pass unobstructed through the hole in the center of the tracker. This is illustrated in Figure 4.12.

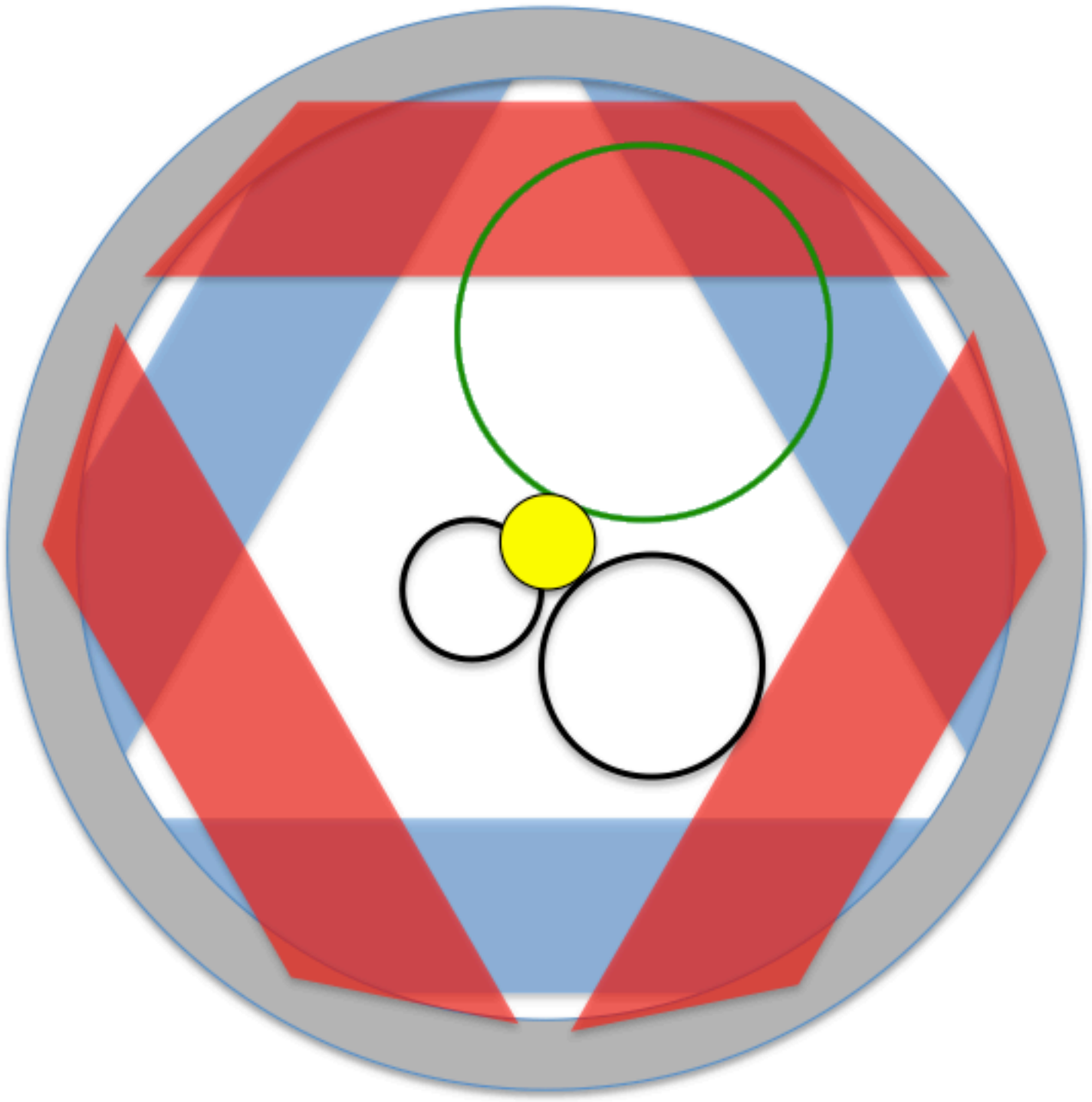

Figure 4.12. Cross sectional view of the Mu2e tracker with the trajectories of a 105 MeV conversion electron (top) and a 53 MeV Michel electron (lower right) superimposed. The disk in the center is the stopping target. Electrons with energies smaller than 53 MeV (lower left), representing most of the rate from muon decays-in-orbit, miss the tracker entirely.

Tracker resolution is an important component in determining the level of several critical backgrounds. The tracker is required to have a high-side resolution of $\sigma < 180$ keV [8]. The requirement on the low side tail is less stringent since it smears background away from the signal region while a high-side tail smears background into the signal region. Current simulations indicate that the high side resolution of the Mu2e tracker can be well represented by the sum of two Gaussians. The high-side resolution, which is the most important for distinguishing conversion electrons from backgrounds, has a core component sigma of 115 KeV/c, and a significant tail sigma of 176 KeV/c. The net resolution is significantly less than the estimated resolution due to energy loss in the upstream material. The Tracker is described in detail in Chapter 9.

### *4.5.2* Calorimeter

The Mu2e calorimeter supplies redundant energy, position, and timing information on tracks that have been reconstructed by the tracker. Calorimeters and tracking devices use different technologies and physical processes to measure the required information, so the sources of error from the two systems are quite different. These different but redundant measurements may help to eliminate backgrounds and provide a cross check to verify the validity of signal events. The calorimeter could also provide a fast trigger for high energy electron candidates, reducing the throughput requirements on the data acquisition system.

The calorimeter must operate in vacuum and in a magnetic field of 1 Tesla. There is a radiation load with the inner portions of the calorimeter, closest to the solenoid axis, receiving up to 200 Gy/year. It must handle a large, mostly low energy background of protons, neutrons and gamma rays produced by muon captures in the stopping target. It must deal with a large flux of electrons from muons decaying in atomic orbit in the muon stopping target. It must also handle a significant flux of particles during beam injection.

The proposed calorimeter consists of 1936 LYSO crystals located downstream of the tracker and arranged in four vanes as shown in Figure 4.13. Each crystal is $3 \times 3 \times 11$ cm$^3$ and will be equipped with two Avalanche Photodiodes (APDs) which work well in a magnetic field. Using two APDs per crystal increases the light yield, provides redundancy and allows for the correct energy to be determined if a charged particle passes through one of the APDs.

The performance of a LYSO array was evaluated in a tagged photon test beam at the MAINZ Microtron (MAMI). The measured resolution ($\sigma/E$) was $3.5 \pm 0.6\%$ (Figure 4.14). The measured position resolution was better than 4 mm. A detailed simulation of the matrix was also carried out and good agreement with the data was obtained. The calorimeter is described in detail in Chapter 10.

### *4.5.3* Cosmic Ray Veto

Cosmic-ray muons striking the muon stopping target and other materials in the detector region can produce delta rays that will occasionally be of the right energy and fall within the detector acceptance, producing conversion-like background events. Cosmic ray muons can also decay, producing electrons that could be mistaken for a conversion signal.

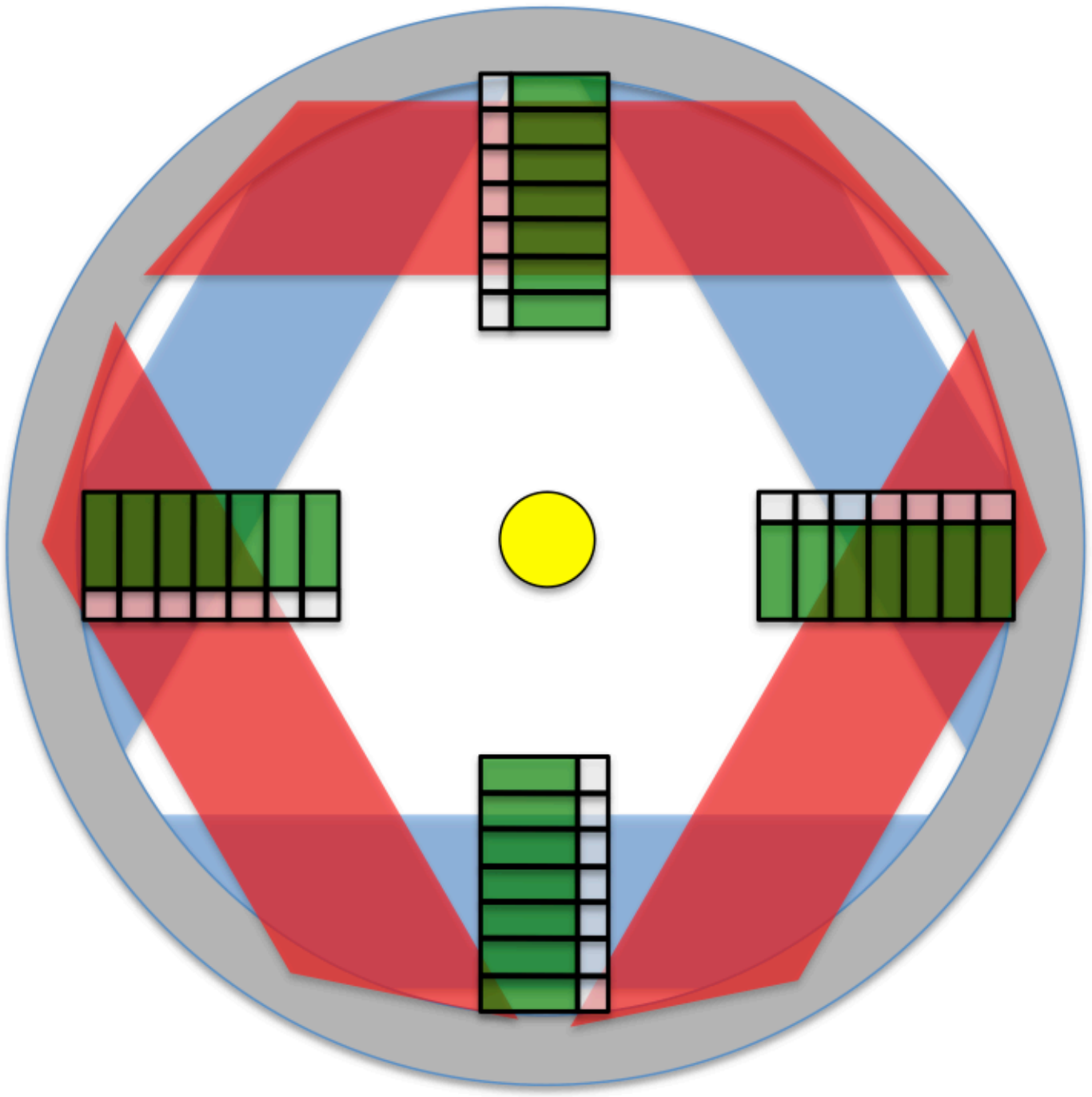

Figure 4.13. End view of the calorimeter. Electrons spiral in the counterclockwise direction.

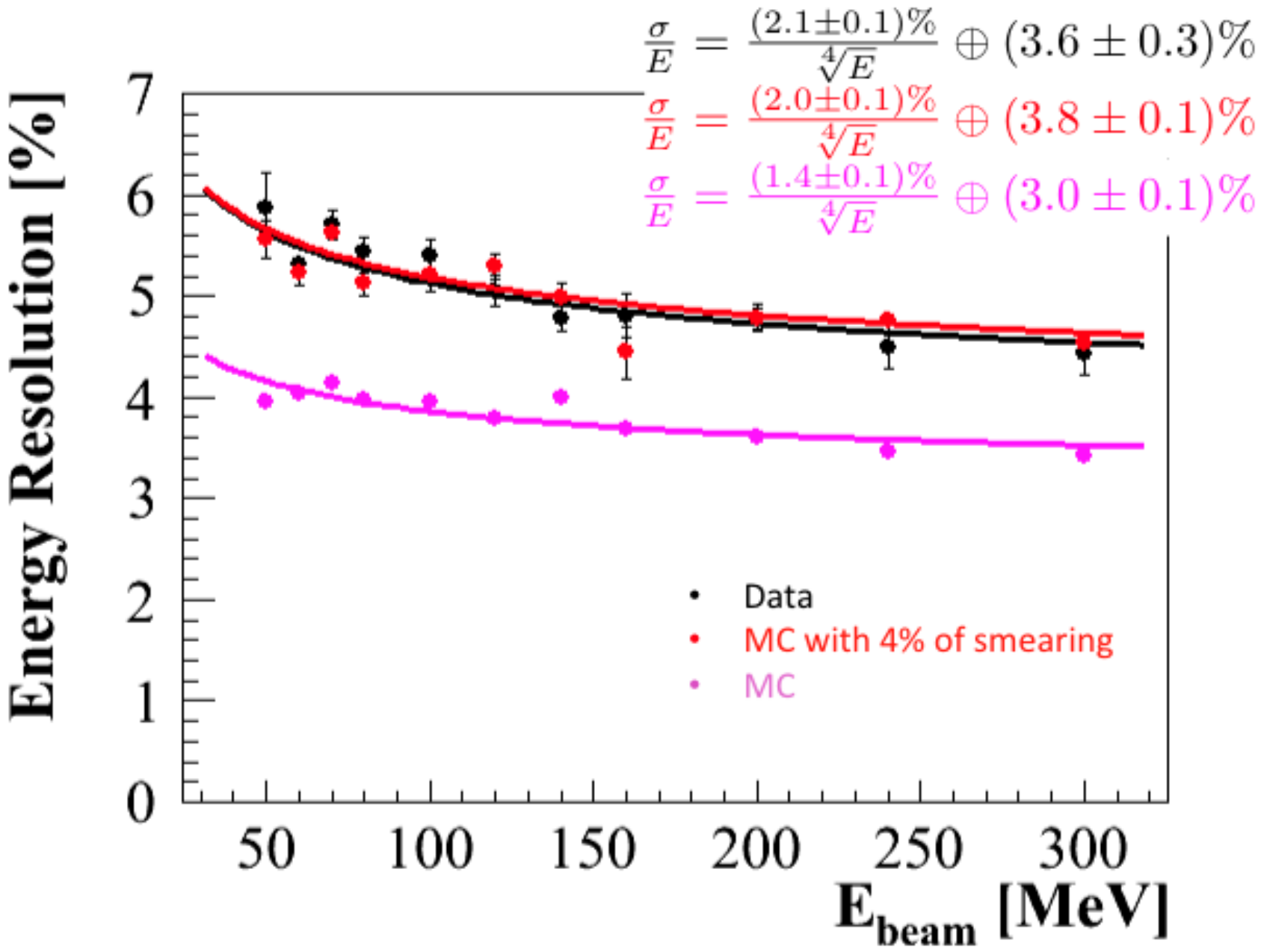

Figure 4.14. Test Beam results from the MAINZ Microtron (MAMI). The measured energy resolution of the overall LYSO crystal matrix (black points) is compared to simulations (red). To obtain reasonable agreement with the data, the energy response of each crystal was smeared by 4% in the simulation (blue).

The background from cosmic rays is directly proportional to the live time of the experiment, so the first layer of protection is the pulsed beam structure and the restricted time window when events are accepted. Passive shielding, including the overburden above and to the sides of the detector enclosure, and the neutron shield surrounding the Transport and Detector Solenoids, eliminates background sources other than penetrating muons, which cannot be suppressed, but rather must be identified. Approximately one conversion event per day from cosmic-ray muons is produced: to reduce that rate to 0.05 events during the entire running period the active shield must suppress the rate by a factor of 10,000.

The Cosmic Ray Veto for Mu2e will be constructed from three layers of extruded scintillator bars with embedded wavelength shifting fibers that are read out with Silicon photomultipliers (SiPMs). The CRV surrounds the Detector Solenoid on 3 sides (Figure 4.15) and extends up to the midpoint of the Transport Solenoid.

To reduce the background from cosmic rays to 0.05 events over the live time of the experiment the Cosmic Ray Veto must be essentially 100% hermetic on the top and sides. In the region of the muon stopping target the cosmic ray veto must be 99.99% efficient. To obtain this efficiency for a 3-layer detector, the single-counter photoelectron yield must be at least 10 for normally incident muons, when defining a veto as coincident hits in two-out-of-three layers. Prototypes indicate that this yield is obtainable.

The Cosmic Ray Veto must survive an intense neutron flux coming primarily from the muon stopping target. Most of the neutrons have kinetic energies below 10 MeV, with the most probable energy about 1 MeV. Polystyrene scintillator ($C_8H_8$) is sensitive to neutrons that elastically scatter on the hydrogen protons, although quenching (Birk's Rule) reduces the light output by an order of magnitude. Studies show that the rate in the counters comes primarily from gammas that are produced from neutron capture on hydrogen. Passive shielding outside the Transport and Detector Solenoids will moderate and capture most of the neutrons. The magnitude and pattern of energy deposition in multiple layers of scintillator is expected to be different for neutrons and muons, which can further help to eliminate false veto signals from neutrons. The Cosmic Ray Veto detector is described in detail Chapter 11.

## 4.6 The Mu2e Facility

The existing Antiproton facility can be used to provide the pulsed beam required for Mu2e in a relatively straightforward way. However, there are no nearby structures that could house Mu2e, so a new detector hall is required. The location of the Mu2e facility, just northwest of the Antiproton facility, is shown in Figure 4.1.

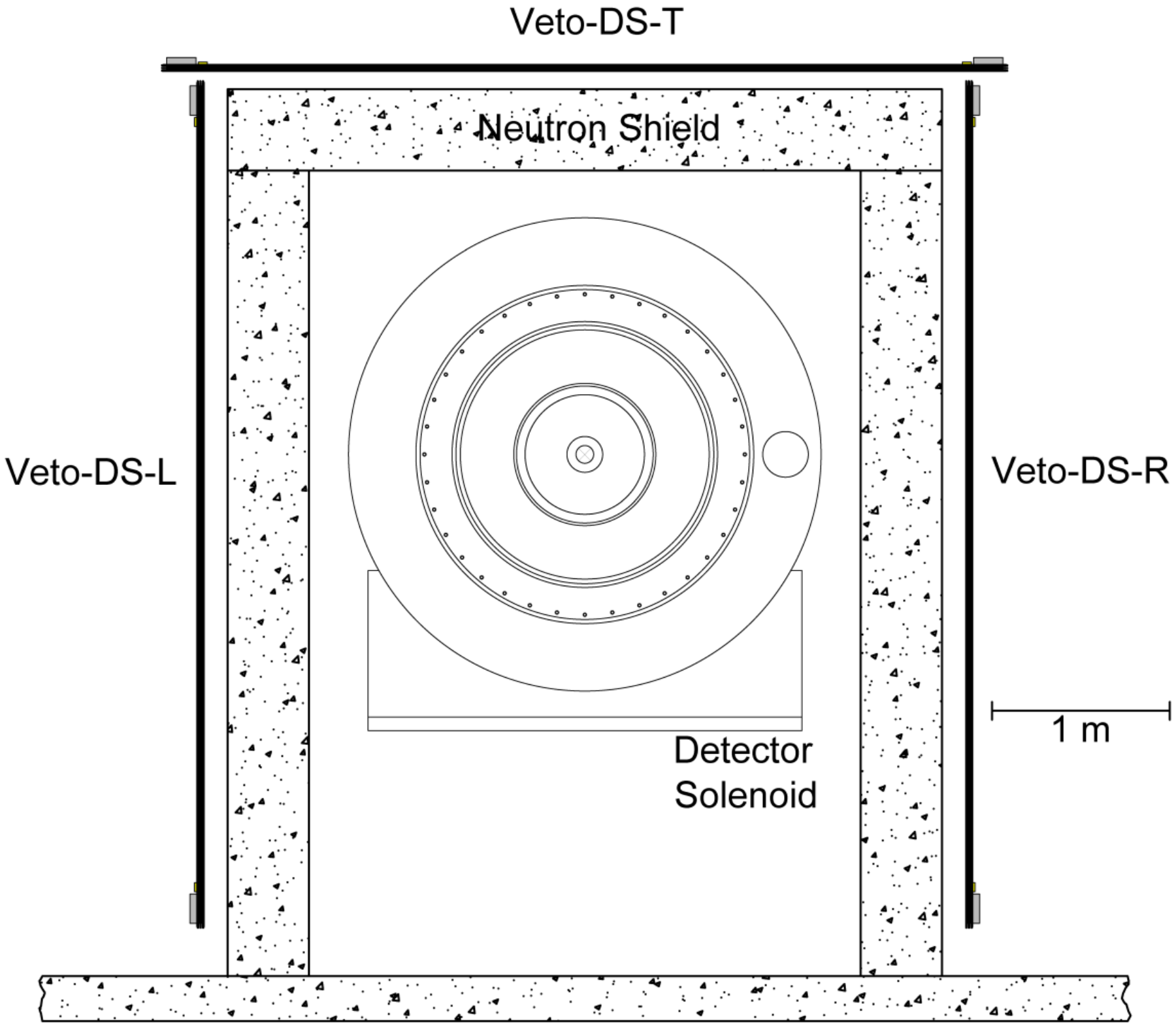

Figure 4.15. End view of the Cosmic Ray Veto.

The Mu2e facility will consist of an underground enclosure where the detector is located and a service building at grade level with infrastructure and services. A plan view of the Mu2e Detector Enclosure is shown in Figure 4.16. The primary proton beam incident on the Production Solenoid requires 16 feet equivalent of earth shielding and drives the depth of the enclosure, which is 25 feet below grade. The cast-in-place concrete enclosure houses the solenoids and provides two assembly spaces under crane coverage.

The at-grade service building consists of a high bay suitable for overhead bridge cranes and an adjacent low bay. The high bay provides space for unloading equipment from semi trailers and for staging and assembling the detector. The high bay will be equipped with two 30-ton capacity overhead bridge cranes. The adjacent low bay contains the electronics room that will house most of the off-detector electronics. Electronics in the low bay electronics room can be accessed when the beam is operating, while any electronics housed near the detector in the underground enclosure can only be accessed when the beam is off. The low bay also contains a mechanical room, an electrical supply room and janitorial facilities. Exit stairs and an elevator serve the underground enclosure. An elevation view of the facility is shown in Figure 4.17 and an architectural view of the facility is shown in Figure 4.18. The Mu2e facility is described in detail in Chapter 6.

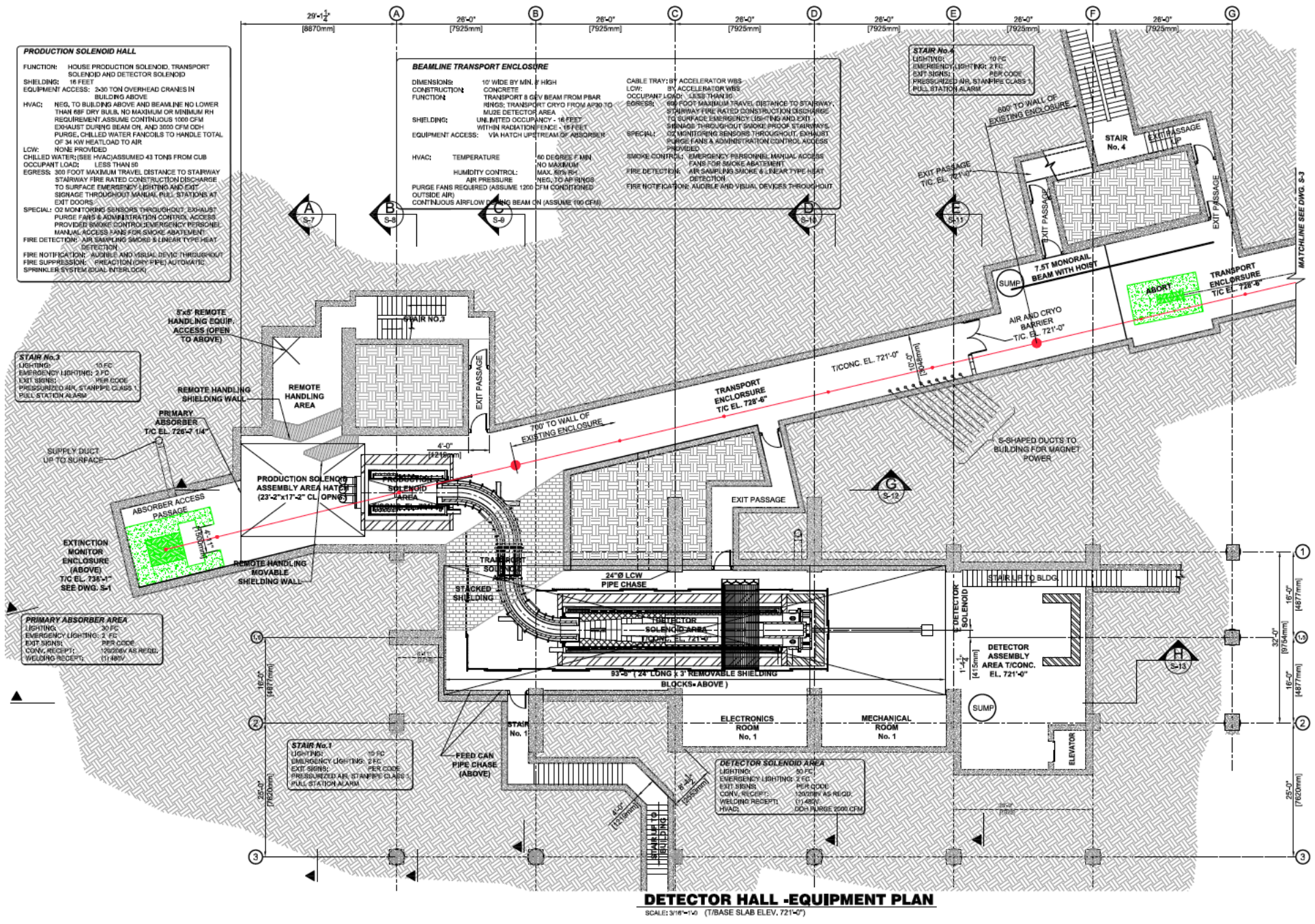

Figure 4.16. Plan view of the underground Mu2e detector enclosure.

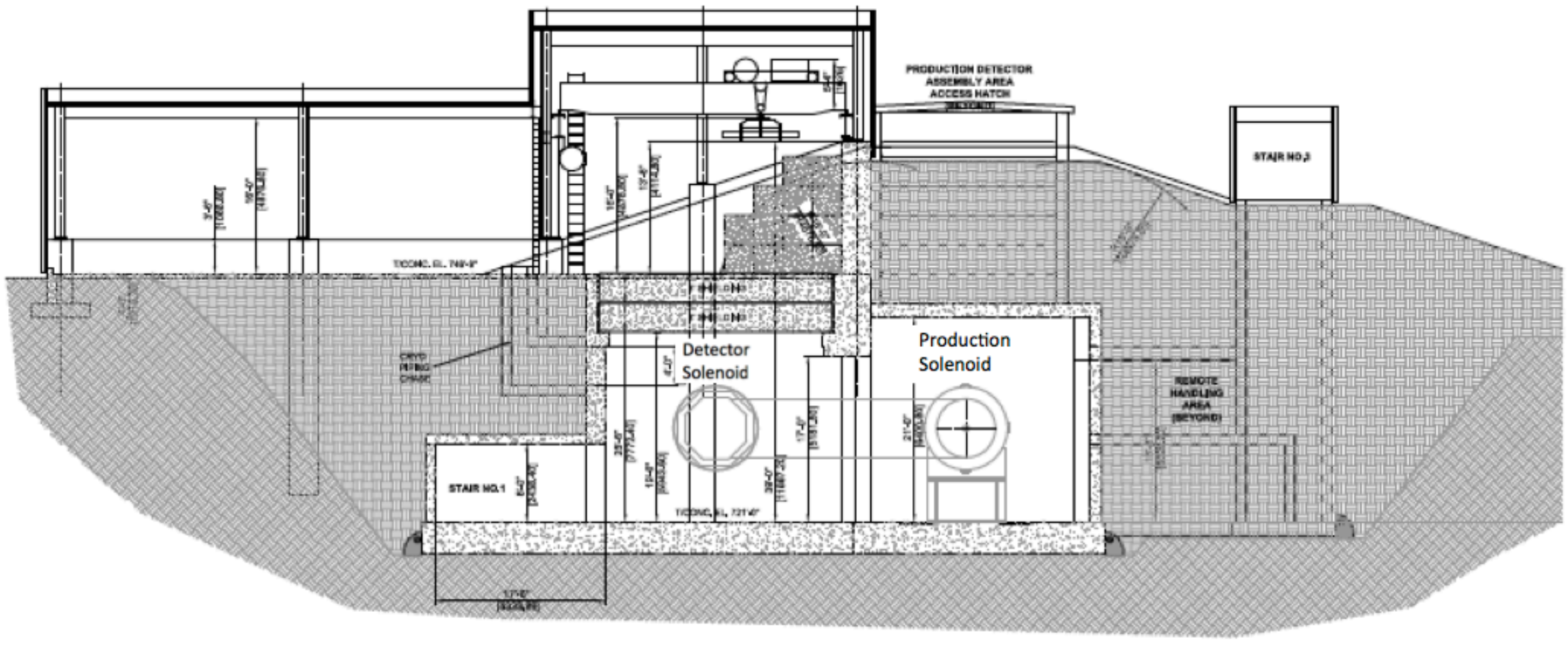

Figure 4.17. Elevation cut through of the Mu2e facility. The Production Solenoid, covered by a shielding berm, is shown of the right. The Detector Solenoid, located directly under the Mu2e Service Building, is shown on the left.

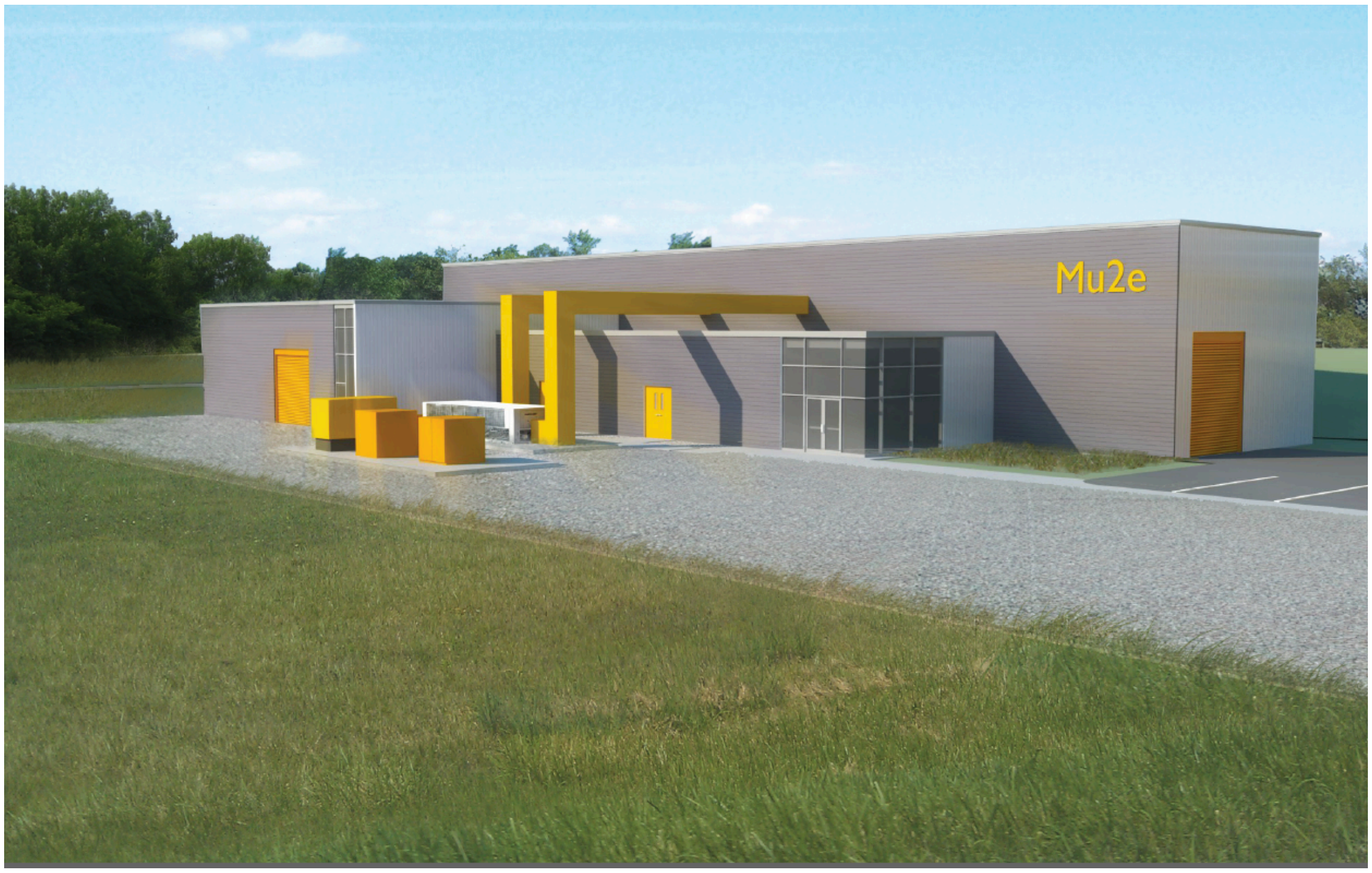

Figure 4.18. The Mu2e at-grade service building.

## 4.7 References

[1] C. Dohmen et al., Phys. Lett. **B317**, 631 (1993).

[2] J. Miller, "Beam extinction Requirement for Mu2e," Mu2e-doc-1175.

[3] R. Coleman et al., "Mu2e Proton Beam Requirements," Mu2e-doc-948.

[4] M. G. Catanesi et al., Phys. Rev. **C77**, 055207 (2008).

[5] A. Ribon et al., "Transition between hadronic models in GEANT4", 2009 IEEE Nuclear Science Symposium Conference Record, N13-89.

[6] D. Armutliiski et al., (in Russian), JINR-P1-91-191 (1991).

[7] F. J. Hartmann et al., Z. Phys. A **305**, 189 (1982).

[8] R. Bernstein, "Requirements Document for Mu2e Tracker," Mu2e-doc-732.

# 5 Accelerator Systems

## 5.1 Overview

The muon beam on the Mu2e stopping target is derived from the decay of pions produced by the interaction of an intense 8 GeV kinetic energy proton beam with a tungsten target. This chapter describes the upgrades to the existing Fermilab Accelerator facilities required for the delivery and targeting of the primary proton beam.

The baseline concept for the Mu2e accelerator upgrades has been changed significantly from what was originally envisioned in the Mu2e proposal [1]. In the course of the conceptual design process it became apparent that the costs of the accelerator upgrades required for the Mu2e experiment were much greater than expected. Consequently, a task force was formed to investigate alternatives to the original concept that would reduce the costs to an acceptable level. The findings and recommendations of the task force are contained in reference [2]. The accelerator systems upgrades presented here reflect the changes recommended by the task force.

The Mu2e proposal calls for a 25 kW proton beam on the Mu2e pion production target. This was to be accomplished by taking six proton batches from the Booster synchrotron using eight unused "ticks[a]" in a Main Injector NOνA cycle. These proton batches were to be momentum stacked three at a time in the Antiproton Source Accumulator Ring prior to being coalesced into four narrow 2.5 MHz bunches utilizing RF systems in the Accumulator. It is this basic scenario that was rejected as being too expensive.

The scaled-down concept presented here utilizes only two proton batches per NOνA cycle. The beam power on target is thus reduced by a factor of three to approximately 8 kW. The need for proton stacking has been eliminated, and bunch formation now takes place in the Recycler Ring. All of the functions originally proposed for the Accumulator Ring are performed elsewhere, thus obviating the need to use and upgrade the Accumulator Ring.

### *5.1.1* Beam Requirements

The Mu2e experiment requires a beam of narrow (in time) pulses separated by an interval that that is longer than the lifetime of a $\mu^-$ captured in the Aluminum stopping target (864 nsec). Moreover, the proton beam must be extinguished between these pulses such that the ratio of out of time beam to in time beam is less than $10^{-10}$. Table 5.1 summarizes the Mu2e proton beam requirements. These requirements are also specified in the Mu2e proton beam requirements document [3].

[a] A tick is $1/15$ sec = 67 msec = the length of a Booster cycle.

| Parameter | Design Value | Requirement | Unit |
|---|---|---|---|
| Booster synchrotron repetition rate | 15 | > 10.5 [b] | Hz |
| Booster synchrotron beam intensity | $4.0\times10^{12}$ | $4.0\times10^{12}$ | Protons/batch |
| Time between beam pulses | 1695 | > 864 | nsec |
| Out of time extinction factor | $10^{-10}$ | $\leq 10^{-10}$ | |
| Pulse full width | ±100 | ≤ ±130 | nsec |
| Pulse rms width | 40 | ≤ 50 | nsec |
| Duration of spill | 54 | > 20 | msec |
| Beam Power on Target | 8 | ----- | kW |
| Average proton intensity per pulse | 31 | < 50 | Mp/pulse |
| Pulse to Pulse intensity variation | 50 | < 50 | % |
| Minimum Target rms spot size[c] | 1 | 0.5 | mm |
| Maximum Target rms spot size[c] | 1 | 2.0 | mm |
| Target rms beam divergence | 0.5 | < 20 | mrad |

Table 5.1. Summary of the Mu2e Proton Beam Requirements

### *5.1.2* Operating Scenario

The proton beam will require considerable manipulation to produce the longitudinal structure required by the Mu2e experiment. These manipulations are performed in the Recycler and the Delivery[d] storage rings and in the beamline that connects the Delivery ring to the target. Figure 5.1 shows the layout of the Fermilab accelerator systems used to accomplish the Mu2e beam manipulations.

Protons designated for Mu2e are acquired from the Booster synchrotron by utilizing the unused portions of the Main Injector timeline during slip-stacking operations for NOνA (see Figure 5.2). Booster protons are extracted into the MI-8 beamline and injected into the Recycler Ring. The beam circulates in the Recycler Ring so that it can be bunched with a 2.5 MHz RF system to form the bunch characteristics required by the Mu2e experiment (see Section 5.5). The beam is then extracted from the Recycler, one bunch at a time, into a new beamline. This new beamline delivers the beam to the existing Antiproton Source beamlines. The beam is then transported to the Delivery ring from which it is resonantly extracted into an external beamline (see Section 5.6).

[b] The number given is the Mu2e requirement (2 batches) plus the NOνA requirement (12 batches) for each MI cycle.
[c] Assumes round beam
[d] The Antiproton Source Debuncher Ring has been renamed the Delivery Ring.

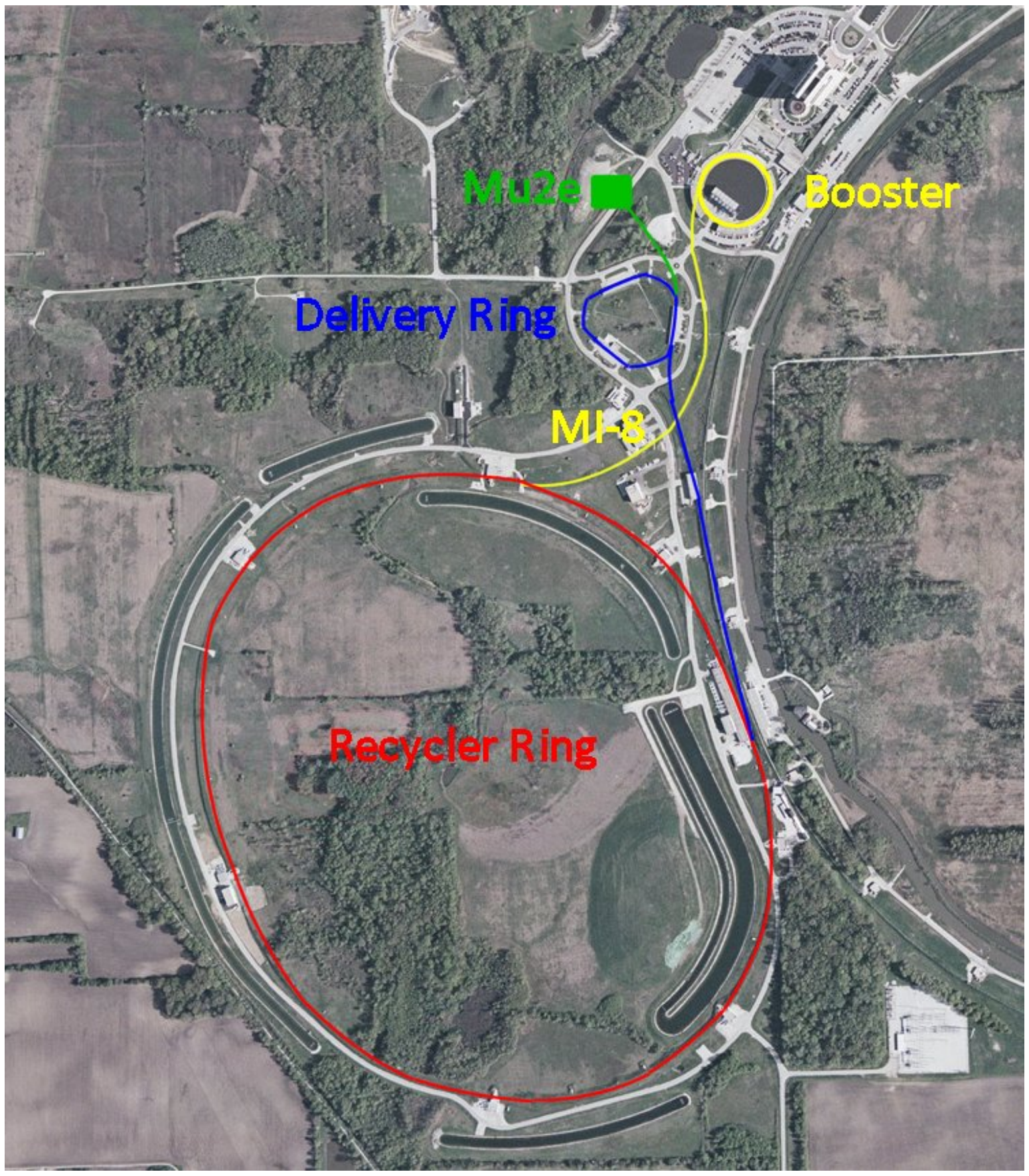

Figure 5.1. The components of the Fermilab accelerator complex used to deliver protons to the Mu2e experiment. The proton beam path from Booster to Recycler is shown in yellow. The beam path in the Recycler is in red. The beam path from the Recycler to the Delivery Ring is in blue, and the beam path from the Delivery Ring to the Mu2e target is in green.

The Recycler Ring 2.5 MHz bunch formation RF system will be built by the Muon g-2 project. The bunch narrowing requirements of the Muon g-2 experiment exceeds those of Mu2e (see Reference [4]). Thus, this system is adequate for the needs of the Mu2e experiment.

Extraction from the Recycler Ring and the beamline stub connecting the Recycler to existing beamlines are also being designed and built for the Muon g-2 experiment. Again, the Muon g-2 design requirements for Recycler Ring extraction exceed the requirements of the Mu2e experiment (see Reference [5]) so that this system adequately satisfies the requirements of the Mu2e experiment.

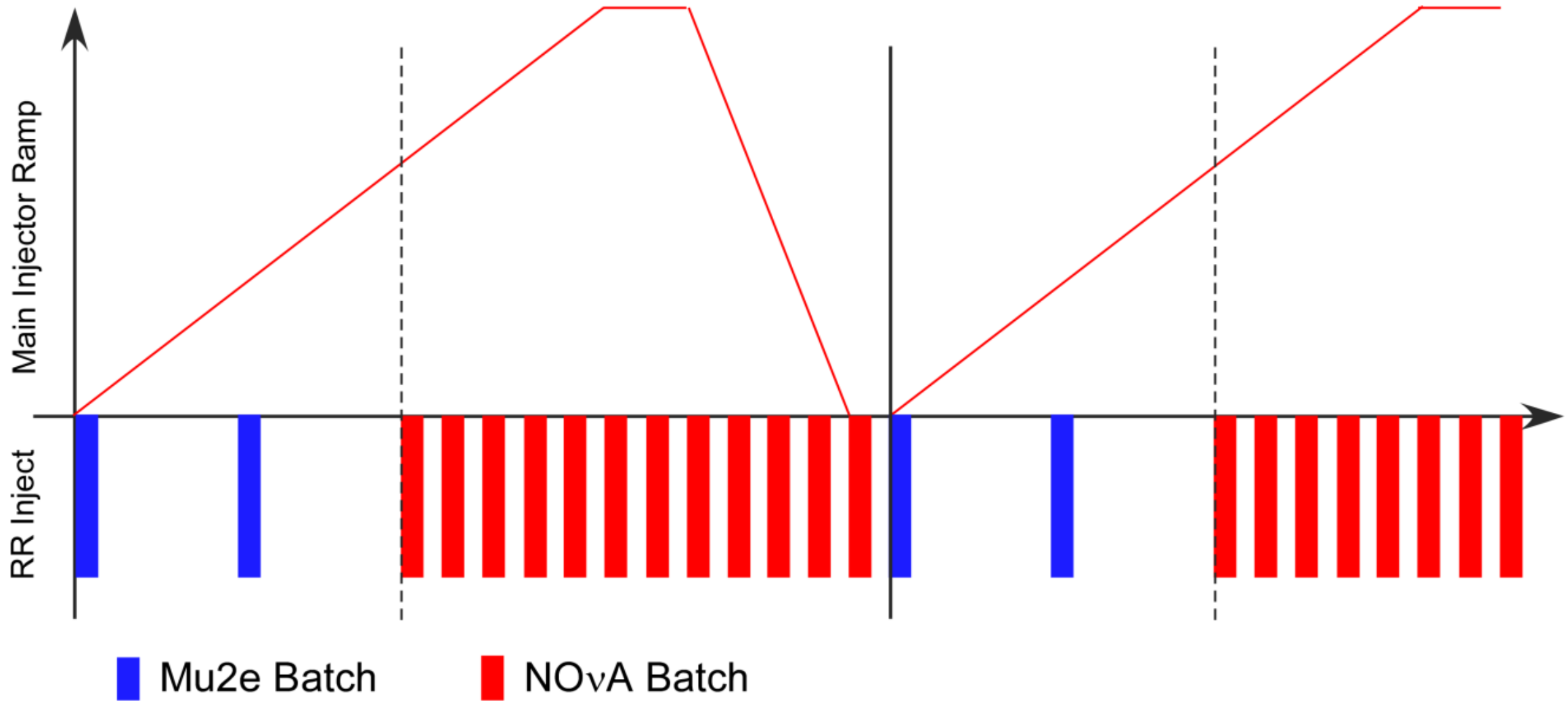

Figure 5.2. The accelerator timeline is shared between Mu2e and NOνA. The blue and red bars represent Mu2e and NOνA proton batch injections respectively. Mu2e Recycler Ring beam manipulations occur in the first eight 15 Hz ticks. NOνA proton batches are slip-stacked during the remaining twelve 15 Hz ticks. The total length of a cycle is 20 ticks = 1.333 sec.

The preparations required for the existing Antiproton Source beamlines and for the Delivery Ring for the Mu2e experiment are largely equivalent to the requirements of the Muon g-2 experiment. Thus, the Delivery Ring and proton transport preparations for both experiments will be accomplished as an Accelerator Improvement Project (AIP). The conceptual design for this AIP is given in Reference [6].

The Delivery Ring to Mu2e target external beamline is a new facility that transports the proton beam to the Mu2e pion production target (Section 5.7). The external beamline contains a beam extinction insert that removes out-of-time beam to the required level (Section 5.8). Upon arrival at the Mu2e pion production target, the beam interacts with a tungsten target inside the shielded super-conducting production solenoid (Section 5.9). The resulting pions decay, producing the muons that will ultimately constitute the muon beam for the experiment. A Heat and Radiation Shield (HRS) lines the inside of the production solenoid (Figure 5.3) to prevent quenches from the heat radiated from the target and to protect the solenoid super-conducting coils from radiation damage.

### *5.1.3* Macro Time Structure of the Proton Beam

The Mu2e experiment must share the Recycler Ring with the NOνA experiment, which uses the Recycler for proton slip-stacking. This sharing is accomplished by performing the required Mu2e beam manipulations in the Recycler prior to the injection of the first proton batch designated for NOνA. There are a total of twenty possible proton batch injections into the Recycler Ring from the Booster within each Main Injector cycle. These proton injections will occur at a maximum rate of 15 Hz (one batch every 67 msec)[e]. Of the twenty possible, NOνA requires twelve proton batches for slip-stacking. That leaves eight injection ticks (533 msec) for Mu2e to acquire its beam and complete the 2.5 MHz bunch formation process (see Figure 5.2).

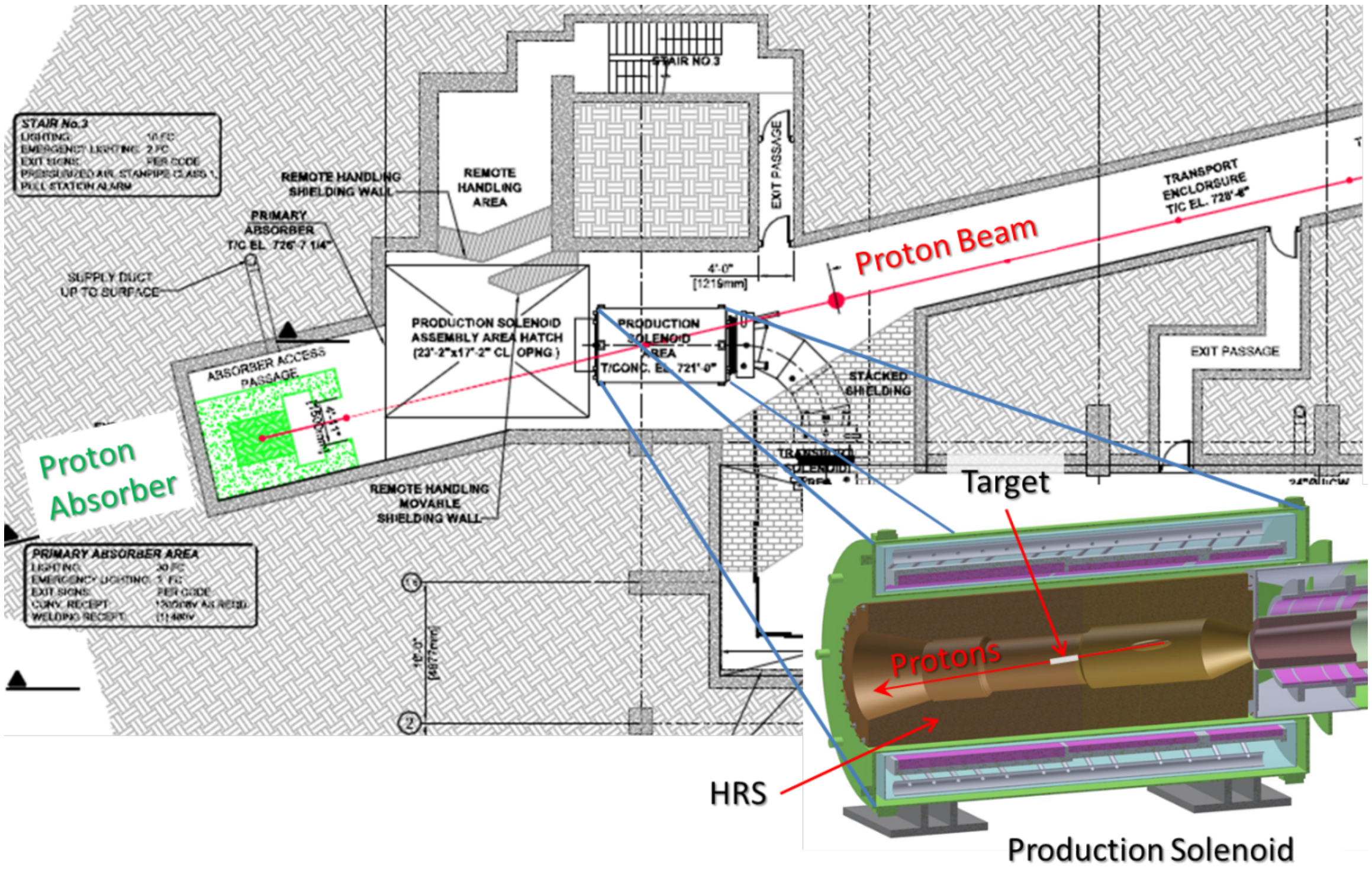

Figure 5.3. Plan view of the Mu2e pion production target station showing the proton beamline, production solenoid components, the Heat and Radiation Shield (HRS), and the proton absorber.

Figure 5.4 shows the utilization of the Mu2e portion of the Main Injector cycle. Two proton batches are injected into the Recycler – one at the beginning of the cycle and one four Booster cycles (ticks) later. After each injection, the beam circulates for 90 msec while the 2.5 MHz bunch formation RF sequence is performed. This RF manipulation coalesces the proton batch into four 2.5 MHz bunches occupying one seventh of the circumference of the Recycler Ring. These bunches are transferred, one bunch at a time, to the Delivery ring where the beam is slow-spilled to the experiment. Table 5.2 gives the parameters of the spill.

### *5.1.1* Accelerator Parameters

Table 5.3 gives a list of accelerator parameters pertinent to the Mu2e accelerator configuration. For a more comprehensive parameter list, see Reference [7]. Unless otherwise stated, these values are used in the calculations and simulations described in what follows.

[e] This statement assumes the successful implementation of the Proton Improvement Plan (PIP).

| Item | Value | Units |
|---|---|---|
| Number of spills per MI cycle | 8 | |
| Number of protons per micro-pulse | 31 | Mp |
| Maximum Delivery Ring Beam Intensity | 1.0 | Tp |
| Instantaneous spill rate | 18.5 | Tp/sec |
| Average spill rate | 6.0 | Tp/sec |
| Duty Factor | 32 | % |
| Duration of spill | 54 | msec |
| Spill On Time per MI cycle | 497 | msec |
| Spill Off Time per MI cycle | 836 | msec |
| Time Gap between 1st set of 4 and 2nd set of 4 spills | 36 | msec |
| Time Gap between spills | 5 | msec |
| Pulse-to-pulse intensity variation[f] | ±50 | % |

Table 5.2. Delivery Ring Spill Parameters

### *5.1.4* Project Management Overview

The accelerator systems upgrades required for the Mu2e experiment are managed and funded by three separate projects: the Mu2e project, the Muon g-2 project, and the Delivery Ring AIP. Table 5.6 shows the upgrades that are required for the success of the Mu2e project, but are managed by another project.

## 5.2 Beam Physics Issues

The beam intensities anticipated for Mu2e operation far exceed the intensities seen in the Antiproton Source during Collider running. Thus, intensity dependent effects must be given careful consideration. We discuss separately the impact of high intensity on the transverse and longitudinal degrees of freedom. Transverse effects predominantly manifest themselves in beam self-defocusing, which causes incoherent shifts in the betatron tunes of the circulating particles. In the longitudinal degree of freedom, we consider the synchrotron tune shift and space charge induced beam self-impedance. Longitudinal beam dynamics may cause collective beam instabilities in both longitudinal and transverse directions when certain intensity thresholds are exceeded.

The analysis reported in this section was completed before the cost reducing recommendations of the accelerator taskforce [2] were put in place. The maximum instantaneous intensity in the Delivery Ring for the operating scenarios assumed for the analysis below is $3\times10^{12}$ protons, whereas, the peak intensity of the present design is a factor of three lower ($1\times10^{12}$ protons).

[f] The pulse intensity is expected to be approximately uniform on short time scales (< 1 msec). The time scale of the variation in pulse intensity is expected to be of order a few msec.

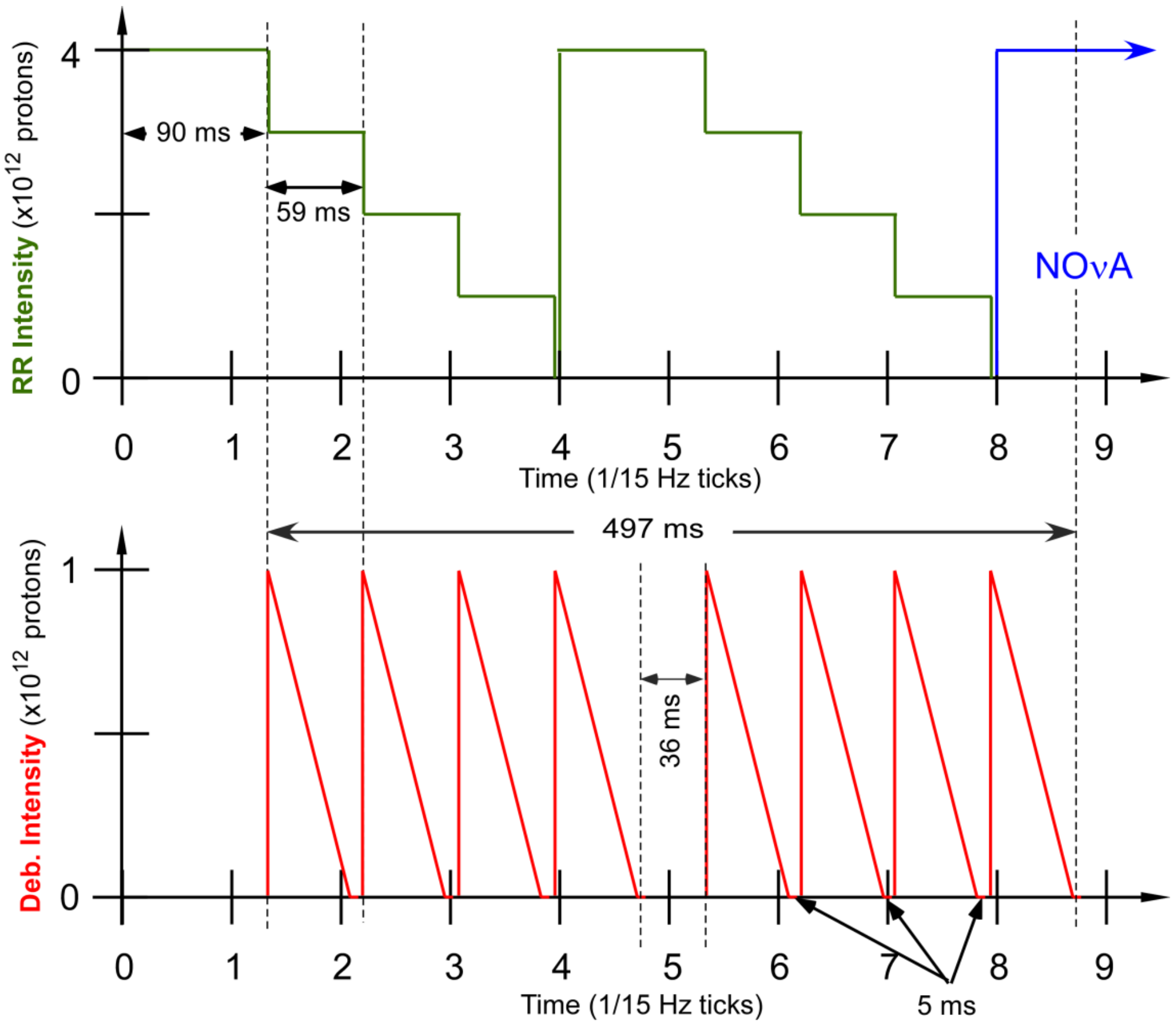

Figure 5.4. This figure shows the first eight Booster ticks of a Main Injector cycle. Proton batches are injected into the Recycler at the beginning of the cycle and again at the fourth tick. After each injection, the beam is bunched with 2.5 MHz RF and extracted one bunch at a time.

### *5.2.1* Space Charge

At Mu2e beam intensities the self-defocusing space charge field of the circulating beam is not small in comparison to the external focusing field of the lattice quadrupole magnets. Space charge defocusing shifts the betatron tune downward relative to the bare lattice tune. Furthermore, the amount of tune shift depends on the betatron amplitude of a circulating particle. Small amplitude particles are subject to the largest tune shifts while large amplitude particles undergo the smallest tune shifts. Thus, a beam containing particles with a wide distribution of betatron amplitudes will see a wide distribution of tune shifts.

| Parameter | Value | Units |
|---|---|---|
| **Booster** | | |
| Intensity per batch | $4\times10^{12}$ | protons |
| 53 MHz Bunches per batch | 81 | |
| Repetition rate | 15 | Hz |
| Average Repetition rate for Mu2e Beam | 4.5 | Hz |
| Transverse emittance | 15 $\pi$ | mm-mrad |
| Longitudinal emittance per 53 MHz bunch | 0.12 | eV-sec |
| **Recycler Ring** | | |
| Revolution Frequency | 89.824 | kHz |
| Beam Momentum (central orbit) | 8886.26 | MeV/c |
| Transverse emittance | 15 $\pi$ | mm-mrad |
| Longitudinal emittance per 53 MHz bunch[g] | 0.12 | eV-sec |
| Maximum Intensity (for Mu2e) | $1\times10^{12}$ | protons |
| **Delivery Ring** | | |
| Maximum Intensity | $1\times10^{12}$ | protons |
| Beam Momentum (central orbit) | 8886.26 | MeV/c |
| Revolution Frequency (central orbit) | 590018 | Hz |
| Orbit Length (central orbit) | 505.294 | m |
| $\nu_x$ | 9.653 | |
| $\nu_y$ | 9.735 | |
| Transverse emittance | 19 $\pi$ | mm-mrad |
| Bunch Length (rms) | 40 | nsec |
| Bunch Base width | 200 | nsec |
| Extracted Beam Power | 8 | kW |

Table 5.3. Accelerator Parameters for Mu2e operations.

The overall effect of space charge is the creation of a tune spread that extends from the bare lattice tune downward by an amount that depends on beam intensity and particle amplitude. Particles in the center of the beam will undergo the maximum tune shift because the charge density there is at its maximum. For particles in the high amplitude tails of the transverse distribution, the space charge field is weak and the resulting tune shift is small. Hence, the overall extent of the tune distribution is equal to the tune spread of the low amplitude particles at the core of the beam. This maximum tune shift can be estimated using the Laslett formula [8]:

$$\Delta\nu_x = \frac{3r_\rho N_{tot} L_R}{2\pi\gamma^2 \varepsilon_N L_b} \tag{5-1}$$

where $r_p$ is the classical proton radius, $N_{tot}$ is the total number of particles, $L_R$ is the orbit length, $L_b$ is the effective bunch length, and $\varepsilon_N$ is the normalized horizontal emittance.

[g] Prior to bunch formation

| **Item** | **Project** |
|---|---|
| MI-8 to Recycler transfer | NOνA |
| Recycler RF | Muon g-2 |
| Recycler Extraction | Muon g-2 |
| Transport to Delivery Ring Upgrades | Delivery Ring AIP |
| Delivery Ring Upgrades | Delivery Ring AIP |
| Upstream part of external beamline | Muon g-2 |

Table 5.4. Items funded and managed by projects other than the Mu2e project

Equation (5-1) assumes that the beam is round and that particle amplitudes are dominated by their betatron oscillations. The precise nature of the space charge tune shift must be obtained from the more accurate integration around the ring obtained by tracking simulations. Tracking simulations will also properly account for the effects of beam broadening in the high dispersion regions of the lattice.

Substituting the Delivery Ring parameter values from Table 5.3 into Equation (5-1) yields a Laslett tune shift for the Delivery Ring of $\Delta\nu_x = 0.007$. The Delivery Ring tune footprint from an ORBIT tracking simulation [9] is shown in Figure 5.5. The smaller tune shift shown in the tracking simulation is a consequence of the large energy spread of the beam after bunch formation. The beam spreads transversely in the arcs reducing the defocusing field felt by each particle. Since the arcs constitute a relatively large part of the Delivery Ring circumference, the effect is significant. The actual (simulated) space charge tune shift is about half of the Laslett value.

The increased tune footprint of the beam due to space charge constrains the choice of the operating point such that the entire tune footprint must lie to the right of the $2\nu_x + \nu_y = 2$ resonance line[h]. The tune footprint also must be in the vicinity, but to the left of, the $3\nu_x = 2$ line, which is the line used for resonant extraction. The greatest impact of space charge induced effects is on resonant extraction. This is discussed in detail in Section 5.6.

### *5.2.2* Coherent instabilities

Transverse stability in the Delivery Rings for Mu2e operating conditions is considered in References [10][i] and [11]. Reference [11] also treats longitudinal stability and accounts for space charge effects. The Delivery Ring is longitudinally and transversely stable for Mu2e beam conditions.

Betatron tune shifts due to space charge play a significant role in the treatment of the transverse stability. The space charge tune shift in the Delivery Ring significantly exceeds the synchrotron tune. In this case, for zero chromaticity, a bunch is stable up to the transverse mode coupling instability (TMCI) threshold [11], [12]. According to Reference [12], the TMCI threshold for a Gaussian bunch of rms length $\tau_b$ in a round chamber with conductivity $\sigma$ and radius $b$ occurs at[j]

$$K = \frac{N_b r_\rho \beta_x R_0 Q_{sc} \eta_{occ}}{4\pi^2 \gamma Q_s^2 b^3 \sqrt{\sigma \tau_b}} \approx 100 \qquad (5\text{-}2)$$

where the average Delivery Ring beta-function is: $\beta_x = 12$ m, and the conductivity is $\sigma = 1.3\times10^{16}$ sec$^{-1}$. The main contribution to the impedance comes from the fraction of the circumference occupied by the dipoles where the vertical aperture is $b = 2.6$ cm (the remaining 75% of the ring is a round chamber with a 6.4 cm radius). This yields $K \approx 10$. This is significantly smaller than the threshold value of Equation (5-2). Thus, under Mu2e operating conditions, the beam should be well below the TMCI threshold.

If the chromaticity is not zero, weak head-tail instability may be possible. The maximum growth rate of the weak head-tail instability is given in Reference [12]:

$$\Gamma\, T_0 = \frac{N_b r_\rho \beta_x R_0 Q_{sc} \eta_{occ}}{\pi \gamma b^3 \sqrt{\sigma \tau_b}} \left[\text{turn}^{-1}\right] \qquad (5\text{-}3)$$

---

[h] Note that this requirement would place the operating point to the right of the operating tunes shown in Figure 5.5.

[i] The analysis of reference [10] does not include the effects of space charge.

[j] Figure 4 of Reference [12] shows that for $K \lesssim 100$ the modes are uncoupled and therefore below the onset of TMCI.

This yields $\Gamma T_0 N_t = 1$ for the Delivery Ring. Thus the weak head-tail instability should not be an issue. In the event head-tail is an issue, the insertion of a small amount of chromaticity should be sufficient to damp this instability.

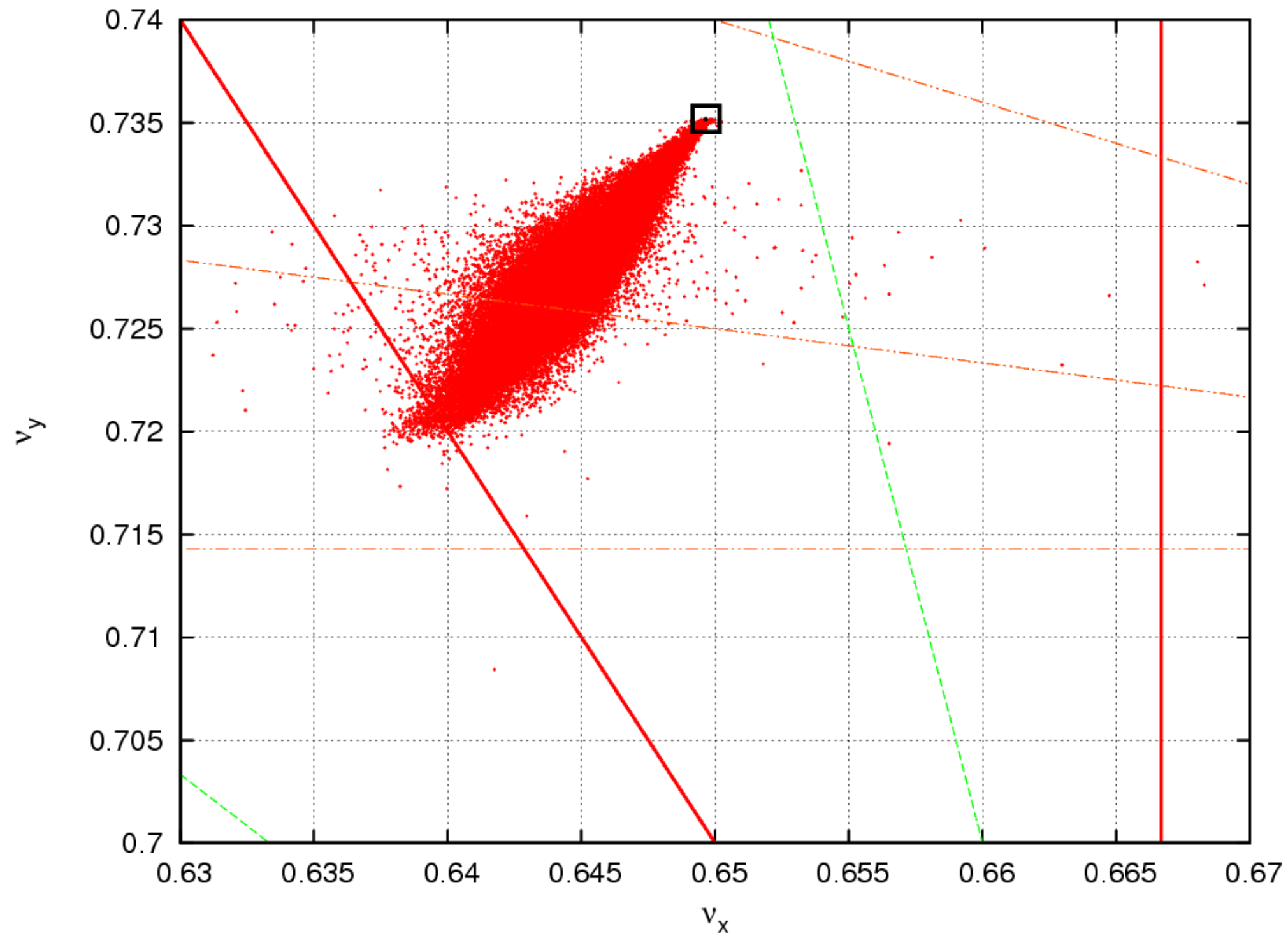

Figure 5.5. Debuncher tune footprint from an ORBIT simulation for a beam intensity of $3\times10^{12}$ protons[k]. The integer part of both tunes is 9. The black box in the upper right of the plot indicates the bare lattice tunes. The thick red lines are 3rd order resonance lines; the dashed green lines are 6th order resonance lines; and the orange dot-dashed lines are 7th order resonance lines. The resonance line used for third integer extraction is the $\nu_x = 2/3$ line at the right of the plot. The actual tune operating point will be chosen to keep the tune footprint well away from the 3rd order line ($2\nu_x + \nu_y = 2$) crossed by the high tune shift end of the footprint.

### *5.2.3* Other Intensity Dependent Effects

The bunched beam should not be affected by space charge or resistive wall longitudinal impedances. The space charge synchrotron tune shift is estimated to be as small as ~1% of the synchrotron tune, while the space charge resistive wall tune shift is even smaller.

The electron cloud instability should not be a big concern, since each bunch is short compared to the zero-current time for any visible cloud to be built [11].

[k] This intensity is a factor of three greater than the maximum Delivery Ring beam intensity in the present baseline operating scenario.

An analysis of intra-beam scattering shows that it is too slow to be seen, given the relatively short time the beam circulates in the Delivery Ring [14].

## 5.3 Recycler Extraction and Transport to the Delivery Ring

### *5.3.1* Introduction

The systems required for the extraction and transport of beam from the Recycler Ring to the presently existing beamlines will be entirely funded by the Muon g-2 project. A technical description of these systems is given in Reference [5]. Most of the planned upgrades to the existing beamlines are required by both the Mu2e and Muon g-2 experiments. Consequently, these upgrades have been incorporated into an AIP [6]. A brief overview of these systems is given here for completeness.

### *5.3.2* Recycler Extraction

The 2.5 MHz bunch formation RF system in the Recycler converts a single batch of 8 GeV protons into a short train of four bunches occupying a seventh of the circumference of the Recycler Ring. The proton bunches are separated by 400 nsec. Each of these bunches is extracted separately. The width of each bunch is approximately 200 nsec. Thus, the extraction kicker must have a rise time that is less than 200 nsec. A proton bunch is extracted every 59 msec (Table 5.2) during the Mu2e extraction part of the Main Injector cycle, therefore the extraction kicker must be capable of burst rates of about 17 Hz. These requirements are much less severe than the Muon g-2 requirements for the same kicker. Thus, the kicker that will be built for the g-2 project will be suitable for the Mu2e experiment.

### *5.3.3* Beam Transport to the Delivery Ring

Both Mu2e and g-2 need to transport beam from the Recycler Ring to the existing Antiproton Source beamlines. This requires a short beamline stub to be built at the MI-52 location to make this connection. This beamline stub is being designed and built for the g-2 project (Reference [5]). The layout of this new beamline is shown in Figure 5.6

Beam transport from the Recycler to the Delivery Ring (formerly Debuncher) closely resembles the 8 GeV "reverse proton" mode used for tune-up and studies during Collider operation. In the reverse proton mode, 8 GeV protons were transferred to the Main Injector, where they circulated for a short time before transfer into the P1 beamline. The proton beam passed through the P1 line, then continued through the P2, AP-1 and AP-3 lines. The AP-3 line connected to the Accumulator in the 30 straight section. The combined length of the beamlines connecting the Main Injector to the Accumulator is 974 m, as shown in Table 5.5.

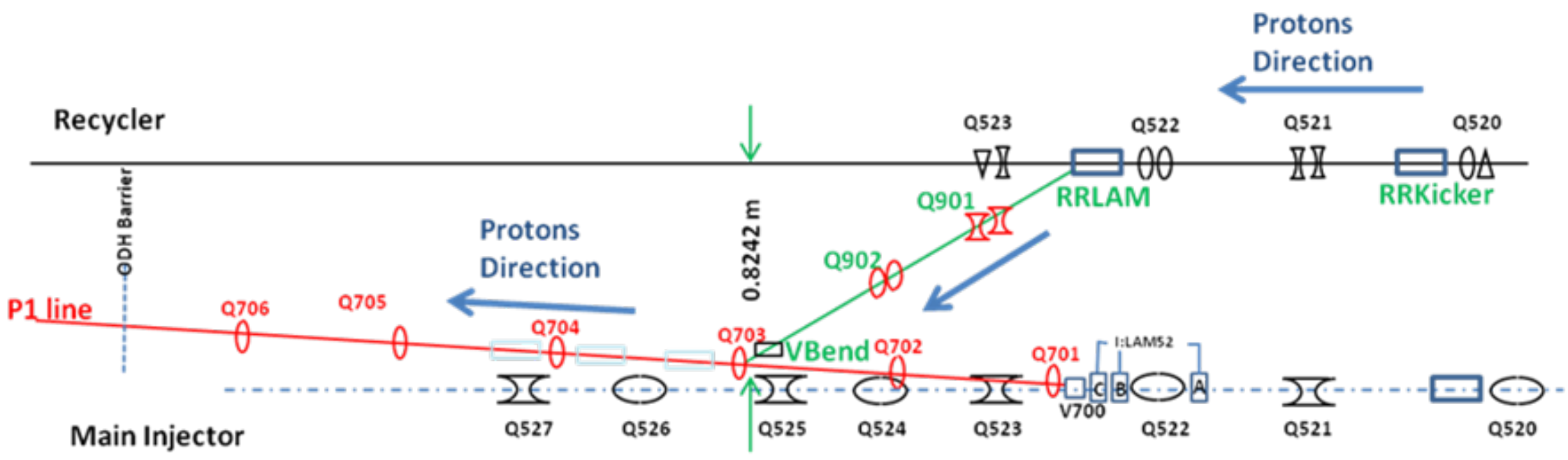

Figure 5.6. Schematic layout of the transfer line from the Recycler Ring to the P1 line.

Mu2e proton transport from the Recycler will primarily use the same beamlines, with modifications at the beginning and end of the combined beamline. The P1 line will be reconfigured to receive beam from the Recycler Ring instead of the Main Injector (see Section 5.3.1 above). The last 50 m of the line must be completely redesigned to accommodate injection into the Delivery Ring instead of the Accumulator. Because of the similarities with Collider reverse proton mode, there is very little technical risk involved in establishing 8 GeV protons from the Recycler to the Delivery Ring for Mu2e. However, the beamlines as they exist today lack sufficient aperture to transport the high intensity beam envisioned for Mu2e without significant beam loss. Although not needed for Mu2e operation, the old AP-3 line will also undergo a major design change for the Muon g-2 experiment, which requires a higher quadrupole density for their secondary beam.

| Beam Line | Length (m) |
|---|---|
| M.I. to P1 | 43 |
| P1 | 182 |
| P2 | 212 |
| AP-1 | 144 |
| AP-3 | 393 |
| M.I. to Accumulator Total | 974 |

Table 5.5. Existing beam line lengths

### *5.3.4* Beamline Changes from Collider Operation

During collider operation, the P1 line connected to the Main Injector at the MI-52 location. The P1 line supported operation with three different beam energies, 150 GeV for protons to the Tevatron, 120 GeV for Pbar production and SY120 operation, and 8 GeV for protons and antiprotons to and from the Antiproton Source. The junction between the P1 and P2 lines occurs at F0 in the Tevatron enclosure. The P2 line can run at two different beam energies, 120 GeV for antiproton production and SY120 operation and 8 GeV for protons and antiprotons to and from the Antiproton Source. The P2, P3 (for SY120 operation), and AP-1 lines join at the F17 location in the Tevatron enclosure. The AP-1 line also operates at 120 GeV and 8 GeV, but is not used for SY120 operation. The AP-3 line only runs at 8 GeV. It connects with the AP-1 line near the antiproton production target in the Pre-Vault beam enclosure and terminates at the Accumulator.

After the conversion from collider to NOνA and Mu2e operation, the Recycler will become part of the proton transport chain and will connect directly with the Booster. There will be a new beamline connection between the Recycler Ring and the P1 line. The P1 line will become a dual energy line, with no further need to deliver 150 GeV protons to the Tevatron. The F0 Lambertsons will no longer be needed to inject protons into the Tevatron and can be removed to improve the aperture. The C-magnets at the end of the P1 beamline can be replaced with larger aperture magnets, if desired, because the adjacent Tevatron beam pipe can be removed. The P2 line will remain a dual energy line supporting Mu2e and SY120 operation, so the junction between the P2, AP-1, and P3 beamlines will remain. AP-1 will only run at 8 GeV for Mu2e operation and AP-3 will remain an 8 GeV-only beamline. Elimination of AP-1 120 GeV operation for antiproton stacking provides an opportunity to improve the aperture with weaker magnets that previously were not practical for use as replacements. Figure 5.7 shows the horizontal and vertical beam size as a function of position for the combined P1−P2−AP-1−AP-3 beamline with the collider run lattice.

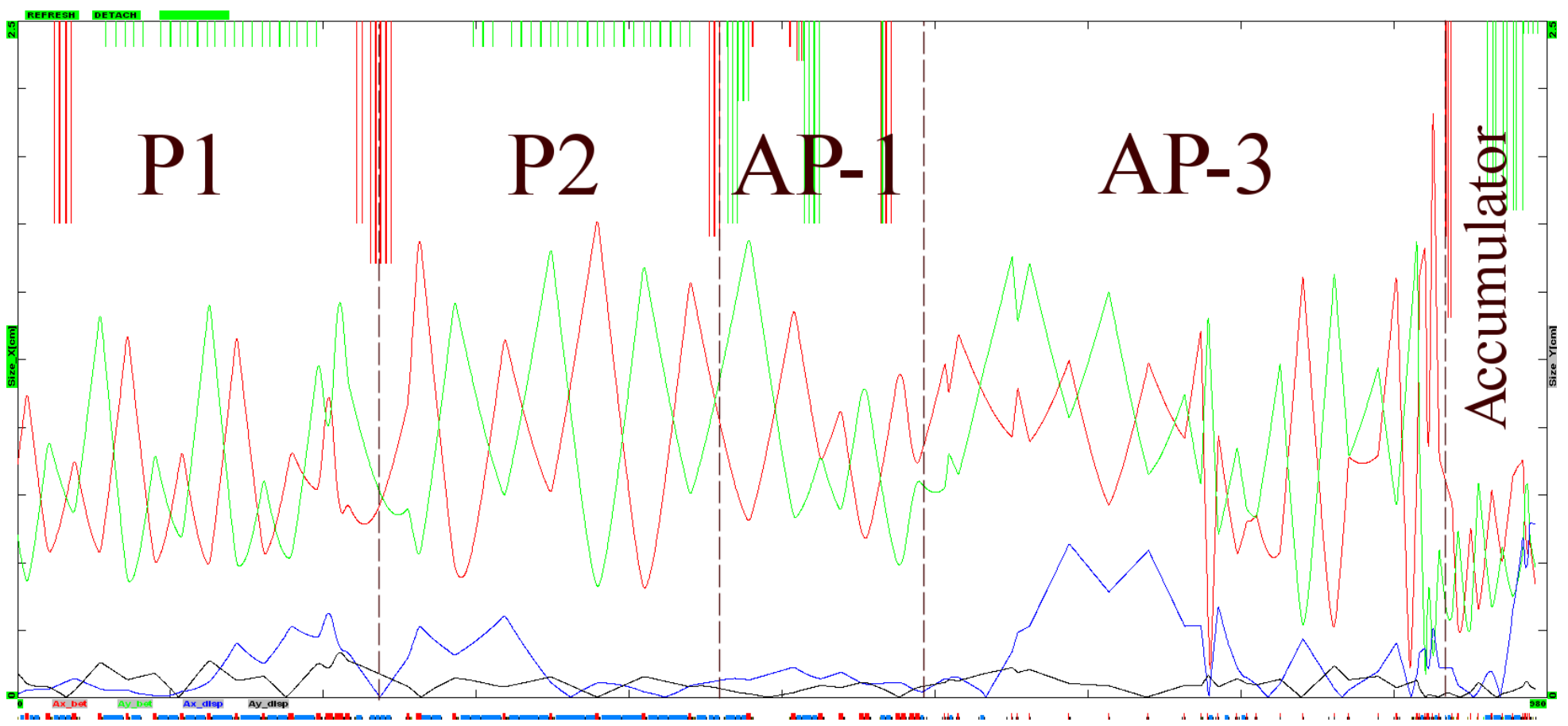

Figure 5.7. Existing Main Injector to Accumulator transverse beam size and aperture. The vertical scale is beam size in centimeters (full scale is 2.5 cm). The horizontal axis is position along the beamline in meters (full scale is 980 m). The red trace is the horizontal beam size, the green trace is vertical. The blue and black traces indicate the beam momentum/dispersion contribution to the horizontal and vertical beam sizes respectively. The red and green bars extending downward from the top of the figure represent the horizontal and vertical aperture limits respectively.

### *5.3.5* Naming convention of the reconfigured beamlines

The beamlines that made up the Antiproton Source (those that have an "AP" prefix) will be modified, reconfigured and renamed prior to Mu2e operation. The 8 GeV-only AP-1 line will be renamed M1. The AP-2 line will become two separate beam lines and will no longer be continuous. The upstream end of the line is needed for the g-2 experiment and will be renamed M2. It will provide a connection from the Pbar AP0 Target Station to the M3 line. The downstream section will become the abort line from the Delivery Ring. The old AP-3 line will be required to transport both 8 GeV beam for the Mu2e experiment and also a 3.1 GeV secondary beam for the g-2 experiment and will be renamed M3. Furthermore, the M3 line will connect to the Delivery Ring (formerly Debuncher) instead of the Accumulator. The extraction line connecting the Delivery Ring to the experiment will be called M4. Figure 5.8 compares the Pbar beamline configuration with that proposed for Mu2e and g-2 operation. In general, the AP-1, AP-2 and AP-3 lines will refer to the old Pbar beamline configuration and M1, M2, M3 and M4 will refer to the beamline configuration for Mu2e operation.

Most of the improvements to the beamlines and Delivery Ring that benefit Mu2e, g-2 and future experiments will be incorporated into an Accelerator Improvement Project (AIP). Table 5.6 summarizes which improvements are contained in the AIP, as well as those that will be managed as part of the Mu2e and g-2 projects. The Conceptual Design Plan for the AIP contains details about the beamline and Delivery Ring improvements being managed as part of that project and can be consulted for plan details.

### *5.3.6* Beamline Instrumentation

Much of the beam instrumentation needed for Mu2e operation already exists in the Antiproton Source beamlines. However, most of this equipment must be modified for use with the faster cycle times that will be seen in the Mu2e era. Additionally, there is an opportunity to take advantage of the shutdown of the Tevatron and acquire components for Beam Loss Monitors (BLMs) and Beam Position Monitor (BPM) systems.

***Beam Intensity Monitors (Toroids)***

Toroids are the essential beam diagnostics systems to monitor the beam current in the beam line. The plan is to continue using the commercial single-turn *Person* units for Mu2e operation and use as much existing signal processing equipment as possible. The present toroid installation locations will be reviewed and modified as needed to provide adequate coverage. Improvements on the analog signal-conditioning path, e.g. filters, chokes, and preamps, will be provided, as well as integrated gain calibration tools. The electronics will be modified where necessary, to adapt the toroids for the increased beam intensity expected for Mu2e.

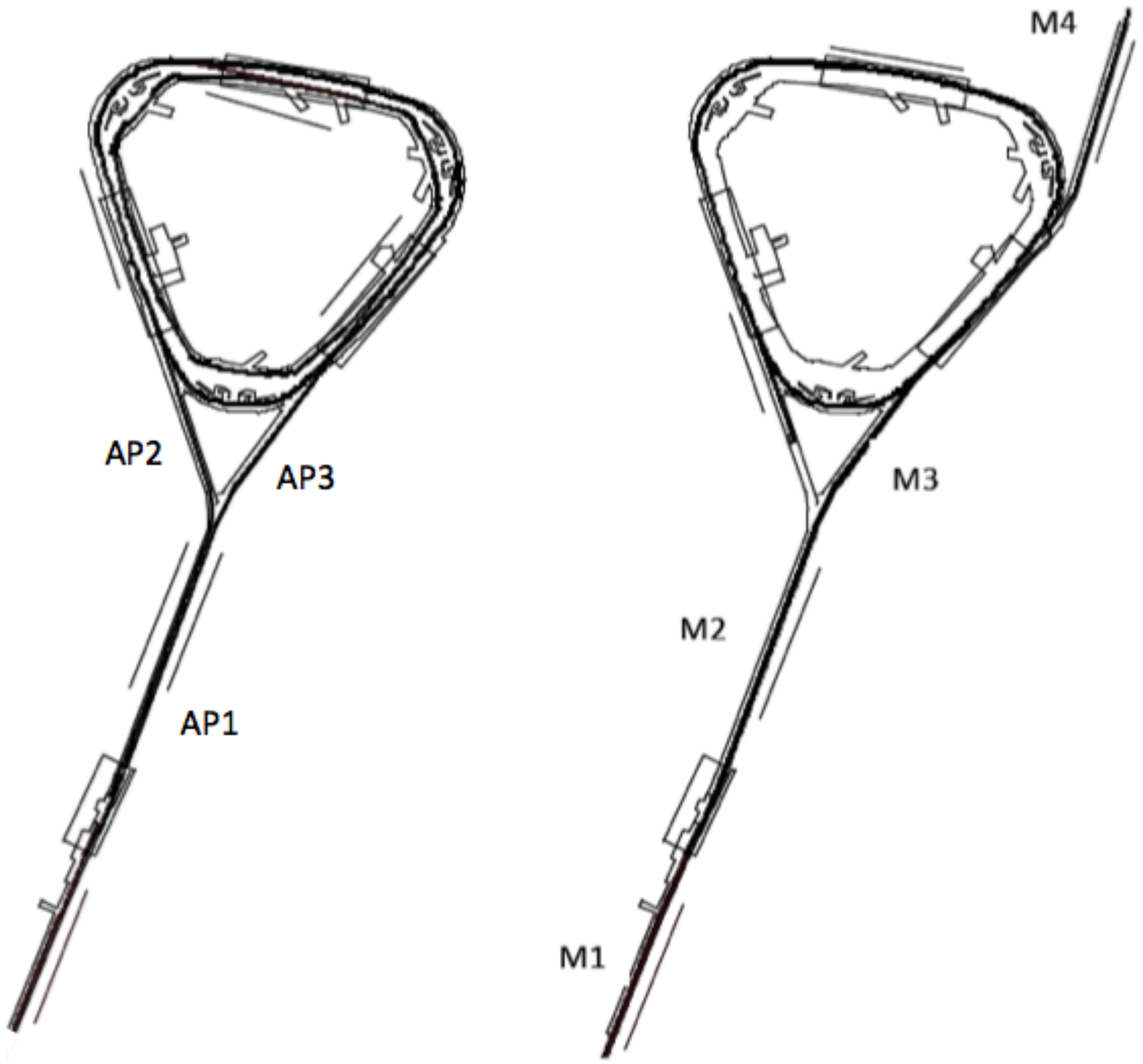

Figure 5.8. Layout of the Antiproton Source beamlines (left) and the reconfigured beamlines for Mu2e operation (right)

### *Beam Position Monitors (BPM)*

The existing BPM pick-ups were designed for a 53 MHz bunch structure and don't require modifications. This is also true for the read-out hardware; the analog signal conditioning "transition" module and the VME-based digital signal processing system (*Echotek* digitizer, CPU, timing board, etc.). However, modifications in the software are required to operate the BPM systems with the increased duty cycle. This kind of VME *Echotek* style BPM electronics in the P1, P2, AP-1, and AP-3 beamlines can be reused for Mu2e operations, only the AP-2 BPM electronics will need to be upgraded to the same standard. The AP-2 line will be used for the beam abort. BPM preamps in the AP-2 line can be eliminated due to the increased beam intensity expected. The BPM electronics for the A1 line, which are identical to those in P1, can be used for the AP-2 line.

| Description | Project | Comment |
|---|---|---|
| Recycler RF upgrade | g-2 | |
| Recycler extraction/beamline stub | g-2 | |
| P1,P2 and M1 aperture upgrade | AIP | M1 final focus quadrupoles are g-2 |
| Reconfigure AP-2 and AP-3 | g-2 | New lines are called M2 and M3 |
| Beam transport instrumentation | AIP | |
| Beam transport controls | Mu2e | |
| Beam transport infrastructure | AIP | |
| Delivery Ring injection | AIP | |
| Delivery Ring modification | AIP | |
| D.R. abort/proton removal | AIP | |
| Delivery Ring RF system | Mu2e | |
| Delivery Ring controls | Mu2e | |
| Delivery Ring instrumentation | AIP | DCCT and Tune measure are Mu2e |
| Resonant extraction from D.R. | Mu2e | |
| Fast extraction from D.R. | g-2 | |
| Delivery Ring infrastructure | AIP | |
| Extraction line to split | g-2 | Upstream M4 line |
| Extraction line from split to Mu2e | Mu2e | Downstream M4, including extinction |
| Extraction line from split to g-2 | g-2 | Beamline to MC-1 building |

Table 5.6. Beam line and Delivery Ring upgrades and associated project

### *Beam Loss Monitors (BLM)*

BLM's are already in place in the P1, P2, and AP-1 beamlines. The existing beamline and Tevatron ion chamber detectors will be reused for Mu2e operation. New ion chambers will be needed in the AP-3 line, A-to-D line and Abort Line. All read-out electronics will be upgraded using logarithmic amplifier techniques. An optional upgrade is being considered that would add snapshot capability to the BLMs.

### *Beam Profile Monitors*

All the transverse beam profile monitors located in the Mu2e beamlines work on the phenomenon of secondary emission. They are called either multiwires or SEMs (Secondary Emission Monitors). Depending on beamline requirements, various styles of detector assemblies have been built over time. Old detectors were constructed using 75 μm diameter tungsten wire and with 10 μm thickness Ti foils. The new detectors are all assembled with lighter materials i.e., Ti, Be, Cu or C to reduce beam losses. The physical location and some properties of the detector (e.g. wire spacing, materials) will be reviewed and modified if necessary. Some modifications and rearrangement of the electronics will be required because of the increase in beam intensity.

### *5.3.7* Beamline Controls

The existing infrastructure of CAMAC crates and timing links will be reused for Mu2e beam line operations, with a few updates to replace end of life hardware. The current CAMAC crates interface with either the Main Injector VME front ends or the Pbar VME front end. An inventory of existing CAMAC crates [15] show that we have on average about 25% of the available slots open, with more slots becoming available when equipment dedicated to collider operation is removed. It is anticipated that there will be ample CAMAC crate coverage for Mu2e operations, and that very few crates will need to be added or moved. A number of CAMAC modules are nearing end of life and may be upgraded to Hot Rack Monitor (HRM) cards in a VME platform. These installations will provide 16 bit A/D readbacks, DAQ, I/O and clock channels, but require the overhead of additional Ethernet connectivity.

The Mu2e service buildings will have numerous network nodes that will communicate via Ethernet to the control system. A map of the controls network is shown in Figure 5.9. Most service buildings have centrally located hardware that provides ample network bandwidth and connections. AP0, F23, and F27 are the only three buildings that do not have this functionality and have limited bandwidth. AP0 runs off 10Base5 Ethernet from AP10, while F23 and F27 run off 802.11b wireless from MI60. Both are 10 MBps shared networks. It is anticipated that the network in these three buildings may not be sufficient for Mu2e operations. The most economical solution for implementing a more robust network infrastructure in these buildings would be to run fiber messenger cable attached to the existing struts that support the cryogenic transfer line on the Tevatron berm. The cable would run from MI60 to AP0 and then to F23 and F27. New network switches will need to be installed in the buildings. Another option being considered runs rad-hardened fiber optic cable from the Pbar service buildings through the Pbar Rings and Transport enclosure back to AP0, and then on to F23 and F27.

## 5.4 The Delivery Ring

The Debuncher Ring will largely remain intact for Mu2e operation and will be renamed the Delivery Ring for its role in providing resonantly extracted short proton bunches to the experiment. A considerable amount of equipment will need to be removed from the Debuncher to improve the aperture and keep losses low. Most of the equipment targeted for removal was used for stochastically cooling the antiproton beam during collider operation and is not needed for Mu2e. The Accumulator Ring will not be needed for Mu2e and will become a source of magnets, power supplies and other components for use in the reconfigured beamlines. In particular, the M4 (extraction) line will be largely made up of former Accumulator components. Some larger aperture magnets will also be needed in the injection and extraction regions and will also come from the Accumulator.

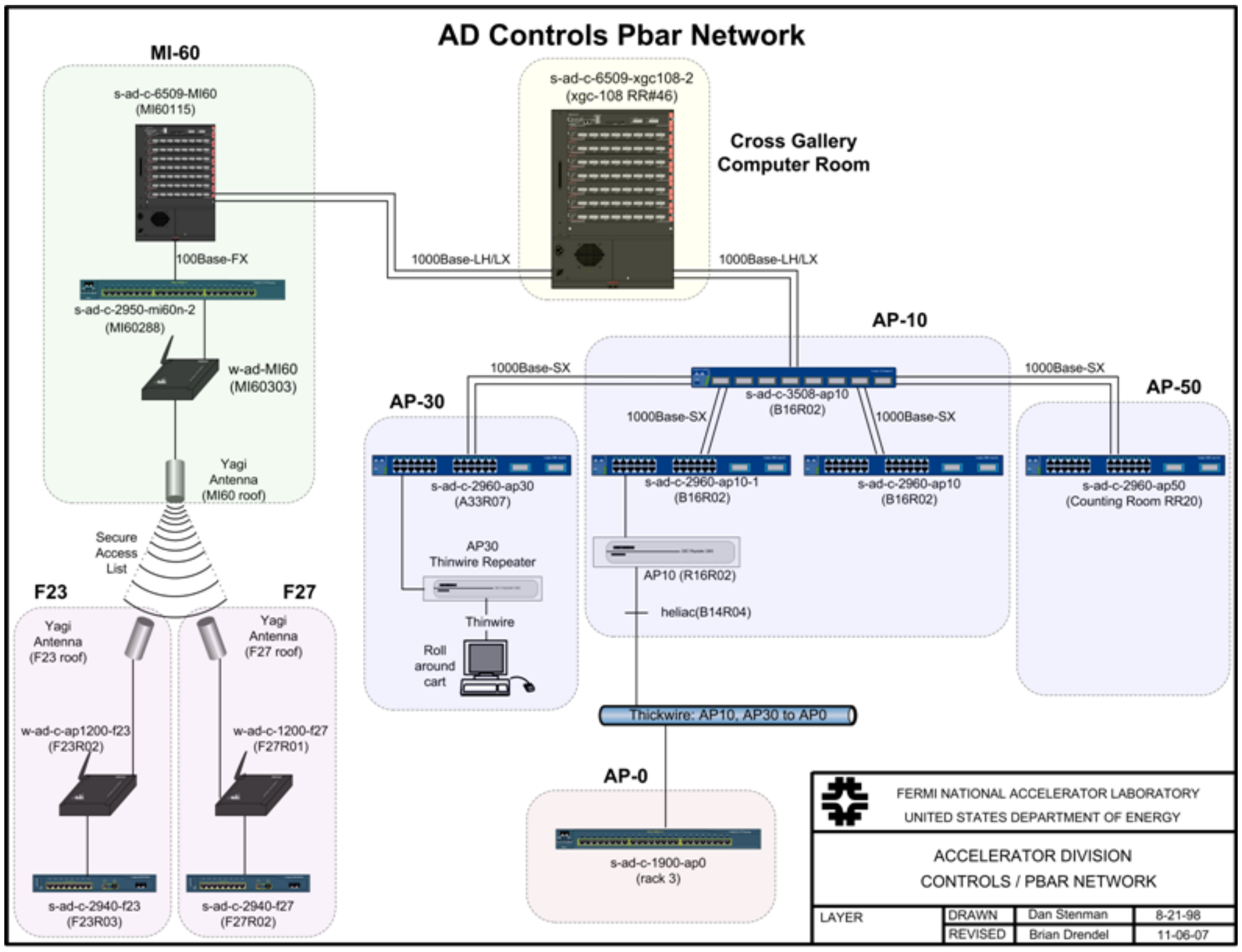

Figure 5.9. Accelerator Controls Network.

### *5.4.1* Injection

The old AP-3 line will be reconfigured to connect to the Delivery Ring instead of the Accumulator. This will require a change in the horizontal trajectory of the M3 line as it approaches the Delivery Ring. The M3 line will then pass above the 30 straight section, ending with a vertical translation into the ring. The plan is to use a Lambertson pair around the D3Q3 location in a similar fashion to what is planned for the extraction Lambertsons around D2Q5. Alternatively, a pulsed septum could be used instead of the Lambertsons if the aperture can be improved without additional expense. The injection process is completed with 3-module kicker system between the D30Q and D2Q2 magnets. The kicker magnets formerly used for extraction in the 10 straight section will be reused for injection.

### *5.4.2* Beam Abort System

A Beam Abort system will be required for the Delivery Ring to minimize uncontrolled beam loss and to “clean up” beam left after extraction. The abort system is required to minimize activation of tunnel components and minimize impact to ground and surface water. The resonant extraction process will not completely remove the entire beam, so what remains must be disposed of in a controlled way. The abort system will need to handle several percent of the total beam power to the experiment. The abort will also provide a mechanism to remove the beam when there is a loss of a beam permit due to power supply problems, beam loss, or other issues. A loss of beam permit will both inhibit further beam injections and trigger the abort process so that the beam already in the rings can be removed safely.

The old Debuncher injection point from AP-2 in the 50 straight section will be repurposed for the abort system. The existing kicker magnets will be reused, although a new power supply will be needed to operate at the frequency needed to support Mu2e. The septum magnet and power supply will also need to be upgraded for the same reason. The old AP-2 beamline will require the addition of a vertical bending magnet, to steer beam into the abort dump located in the middle of the Transport tunnel. A schematic drawing of the Delivery Ring abort system is shown in Figure 5.10.

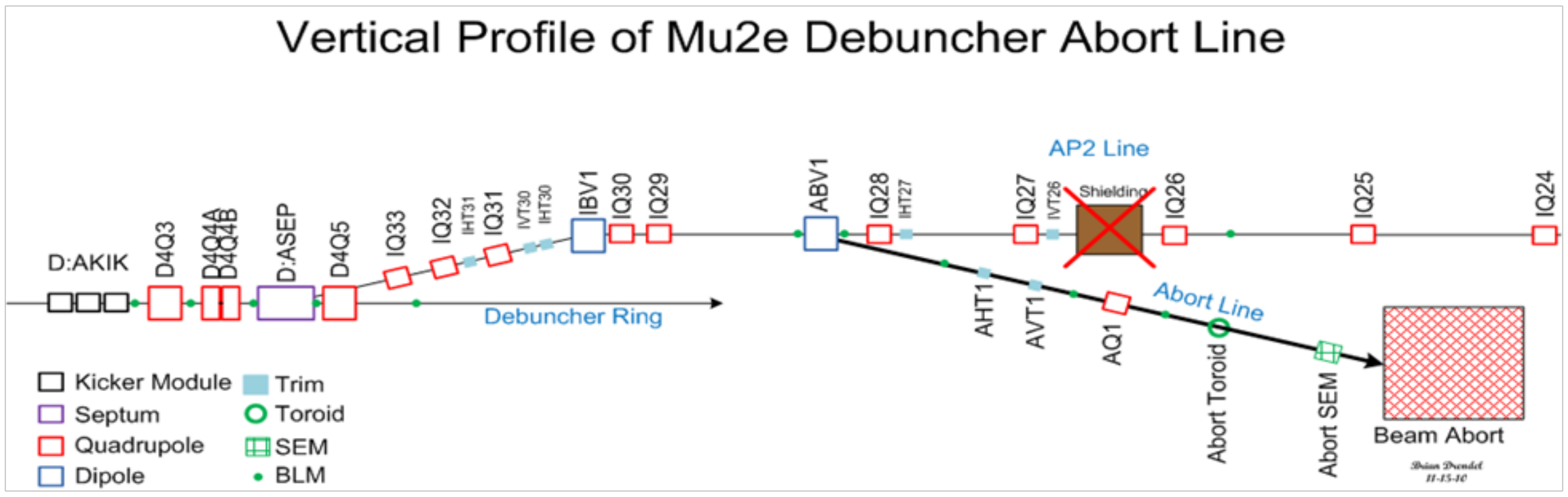

Figure 5.10. Delivery Ring abort layout

### *5.4.3* Kickers and Septa

The kickers and septa required for Mu2e operation will need to operate at a much higher frequency than that used for antiproton production, with peak rates as much as 30 times higher. Mu2e kickers typically will be required to produce the same magnetic fields and have lengths that are similar to those used for pbar operation. In an effort to make the new kicker systems more economical, existing magnets will be reused. Table 5.7 compares kicker parameters for existing pbar systems to the specifications for the Mu2e injection and abort kickers. The rise and fall time specifications for Mu2e are less strict than what was needed for antiproton production, due to the short bunch length needed for Mu2e. Although the Pbar kicker magnets are suitable for reuse, new power supplies will be needed to operate at the increased rate.

| Kicker | Integrated field (Kg-m) | Kick Angle (mrad) | Rise Time 95%/5% (nsec) | Fall Time 95%/5% (nsec) | Flat Top (nsec) |
|---|---|---|---|---|---|
| Debuncher Extraction | 1.34 | 4.6 | 150 | 140 | 1,700 |
| Debuncher Injection | 1.81 | 6.1 | 180 | 150 | 1,700 |
| Delivery Ring Injection | 1.34 | 4.6 | n/a | 400 | 400 |
| Delivery Ring Abort | 1.81 | 6.1 | 500 | n/a | 1,700 |

Table 5.7. Existing Pbar (top) and Mu2e (bottom) kicker strength and waveform specifications.

The septa and pulsed power supplies used during Pbar operation are not suitable for rapid cycling and can't be used for Mu2e. The septa have no internal cooling to handle the increased heat load from the planned high duty cycle and the power supplies are not able to charge quickly enough. The Booster-style septum magnets have the necessary size and field strength required for Accumulator extraction, and therefore are the preferred choice. The power supplies used in Booster to power the septum magnets also appear to be a good fit as they are designed to operate with the same cycle time and a higher duty cycle than those needed for Mu2e. The Booster septum magnets are the same length as their Pbar counterparts, so the abort septum magnet will fit in the same location as the old Debuncher injection septum. If a pulsed septum proves to be a better choice for Delivery Ring injection in the 30 straight section (instead of the Lambertson pair presently planned), there is ample space in the lattice to accommodate it.

### *5.4.4* Delivery Ring Lattice

The basic design of the Debuncher lattice will be used for the Delivery Ring and is described in [16]. The Ring will have a 3-fold symmetry with additional mirror symmetry in each of three periods and three zero dispersion straight sections. The original lattice design parameters were largely dictated by the requirements of the stochastic cooling and RF systems for antiproton production. The Debuncher was designed with a large transverse and momentum acceptance to receive secondaries from the pbar production target. Figure 5.11 shows the lattice functions of one period of the Debuncher.

#### *Original Rings Lattice Symmetry*

In the existing lattice used for antiproton production, the original periodicity and symmetry was broken in order to improve machine acceptance. Since all of the stochastic cooling arrays that represent major aperture limitations will be removed from the rings, the original design symmetry will be restored.

### *Operating Point*

The Debuncher operating point will be moved closer to the $\nu_x = {}^{29}\!/_{3}$ betatron resonance. While the horizontal tune is determined by the mode of extraction, the vertical tune should be chosen so to avoid crossing dangerous resonances during the extraction tune ramp. The exact choice of the base operating point in the Delivery Ring will be determined from beam studies when the Mu2e upgrades are complete.

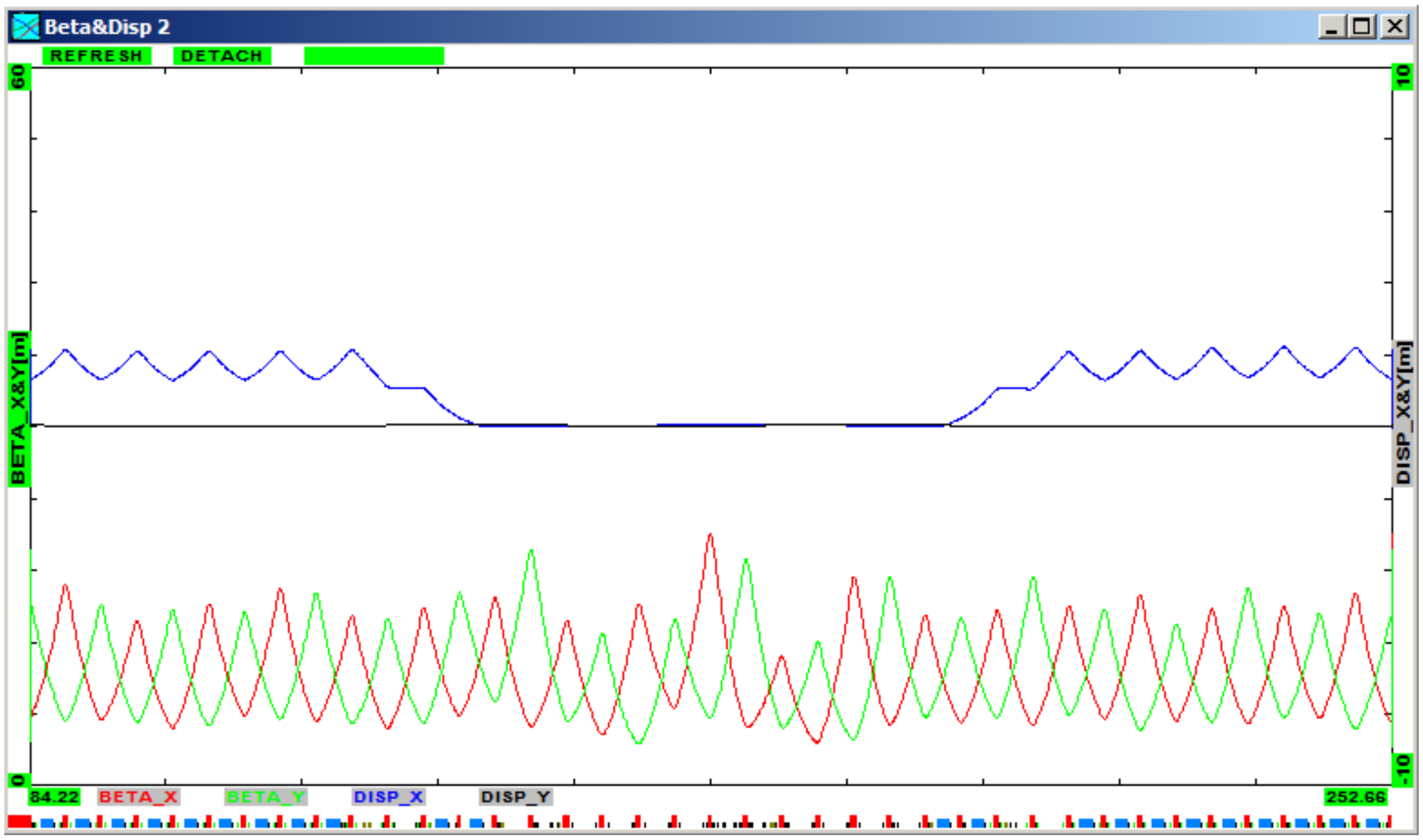

Figure 5.11. Present Debuncher lattice (one period of three). Blue and black traces show horizontal and vertical dispersion with zero in the middle, and red and green traces show horizontal and vertical beta functions respectively.

### *Delivery Ring Injection and Extraction Regions*

The Debuncher injection and extraction regions will both be located in the 30 straight section. In both cases, the tight quadrupole spacing in the Delivery Ring creates mechanical conflicts in the elevation change to ring level. There will need to be an alteration to the existing magnet layout to be able to clear the first quadrupole magnet upstream (downstream) of the injection (extraction) Lambertsons. The present FODO section length does not provide enough room for that in its present form. Either the extraction region optics will need to be redesigned to provide longer quad spacing or different quadrupole magnets will need to be installed to accommodate the extraction line.

### *5.4.5* Beam Instrumentation for the Delivery Ring

As is the case with the beam transport lines, most of the instrumentation needed for operation of the Delivery Ring already exists, but needs to be modified or upgraded to accommodate the faster cycle times. With the shutdown of the Tevatron, there is an opportunity to repurpose BLM and BPM hardware. To regulate and optimize the Delivery Ring resonant extraction process, a Spill Monitor and tune measurement scheme will be required.

***Beam Intensity Monitors (DCCT)***

The beam intensity in the Delivery Ring will be monitored using a DC current transformer (DCCT) technology, as outlined in Reference [17]. The existing tunnel device in the Debuncher can be reused, after applying some modifications. The Delivery Ring DCCT will not require any specific changes to the physical detector, but will need its analog conditioning and VME electronics modified for Mu2e operation. The Accumulator DCCT becomes a spare.

***Delivery Ring Tune Measurement***

The Delivery Ring is a resonant extraction machine that will require a tune measurement system with an accuracy of 0.001. The default tune measurement will use a kicker to excite the beam and calculate the tune from the resulting Beam Position Monitor (BPM) turn-by-turn data. This is a destructive measurement, but should allow a tune measurement of 0.001 at a 1 kHz update rate over the entire 58 msec resonant extraction cycle. Two other systems being considered are using a Schottky detector or a transverse damper. The Schottky would be less destructive, but has the challenge of achieving the desired accuracy at an adequately fast update rate. The transverse damper option requires a BPM pickup and a kicker system and will only be considered if the other two tune measurement systems are not able to meet their design goals. More information on each tune measurement system can be found in reference [18].

***Beam Position Monitors (BPM)***

The primary system to measure the beam orbit in the Delivery Ring will be the existing beam position monitors (BPM) distributed around the ring. The split-plate BPM pick-ups are suitable for Mu2e operation and will not require modifications.

The BPM read-out hardware is based on an analog differential receiver-filter module for analog signal conditioning, and a digital signal processing system, reusing the *Echotek* 8-channel 80MSPS digital down converter and other VME hardware from the Recycler BPMs. This system provides beam position and intensity measurements with a dynamic range of 55 dB and an orbit measurement resolution of ±10 microns. The position measurements can be performed on 2.5 MHz bunched beam, as well as on a 53 MHz bunched Booster batch. Data buffers are maintained for each of the acquisition events and support flash, closed orbit and turn-by-turn measurements. A calibration system provides automatic gain correction of the BPM signal path. The software will need to be modified to handle specific events and data acquisition for the Mu2e operation.

The upgrades to the Delivery Ring BPM read-out hardware are funded and managed in the Delivery Ring AIP.

***Beam Loss Monitors***

A plan detailing the implementation of the BLM's can be found in reference [19]. Although there are already BLM systems in place in the Debuncher, it will require significant upgrades for Mu2e operation. The existing photomultiplier tubes would not function well in the expected Mu2e radiation environment, so they will be replaced by ion chambers repurposed from the Tevatron. The electronics have to be re-designed to accommodate the fast cycle time planned for Mu2e. The system will provide a sample-and-hold acquisition technology on individual beam pulses. A high resolution plotting functionality is being considered, but would add significant cost to the design of the system.

The upgrades to the Delivery Ring BLM system are funded and managed in the Delivery Ring AIP.

### *5.4.6* Vacuum Systems

The existing vacuum systems in the rings and transport lines have performed very well during Pbar operation. Although the significant increase in beam power anticipated for Mu2e operation will increase the gas load, the existing vacuum systems should be more than adequate in their present form. Some of the vacuum equipment used in the rings for Pbar operation will no longer be adequate after the conversion to Mu2e operation. For instance, it will be important to identify and eliminate O-rings from vacuum connections, as they will not survive the increased radiation levels from the high beam power. Whenever possible, available vacuum equipment from the Antiproton Source should be reused.

The Debuncher Ring has good ion pump coverage that should generally be adequate for Mu2e operation. Stochastic cooling tanks, kickers and septa that will be removed during the conversion have built in ion pumps, so some of these pumps may need to be installed in the vacated spaces. The electrostatic extraction septum will need particularly good vacuum to minimize the risk of sparking. The septa should have ion pumps integrated into the design, but there should also be additional pumping capacity added to the surrounding area. The Accumulator has enough surplus ion pumps and vacuum pipe available to cover nearly all of the needs, so little expenditure will likely be needed. Additional vacuum valves from the Accumulator will be used to isolate the extraction septa and the Debuncher abort.

### *5.4.7* Infrastructure Improvements

Electrical power for the Antiproton Source is provided by Feeder 24, which used about 4.4 MW of power during pbar operation. In most cases, service buildings are expected to use approximately the same amount of power after the conversion for Mu2e. The exception is the AP-30 service building, where there will be a large increase in power load from the injection and extraction lines. A new transformer may be needed at AP-30 to provide the additional power. A power test was performed on the individual service building transformers to aid in predicting the power needs for Mu2e. Also, since the Accumulator will no longer be used, approximately 1.4 MW will be available for new loads.

Presently, Pbar magnets and power supplies get their cooling water from the Pbar LCW system. However, the removal of the heat load from the Accumulator should be enough to offset the extraction line and other new loads. It is also possible to design smaller closed-loop systems that exchange heat with the Chilled Water system, which has adequate capacity and is already distributed to the Pbar service buildings.

Tunnel components that have worked reliably for Pbar operation may be vulnerable with the elevated radiation levels anticipated for Mu2e. New systems will have radiation hardening as part of their design criteria, but existing systems carried over for Mu2e operation will need to be carefully examined for potential reliability issues. Two systems that have been identified as potential problems are LCW hoses/tubing and magnet shunts. To mitigate the problem with hosing and tubing, a more radiation resistant type of material will be used in higher radiation areas. The magnet shunts have electronics that have a greatly reduced service life when located in areas with elevated radiation. Most of the shunts will be relocated from the tunnel to the service buildings. Also, some shunts have not been used historically or operate at low currents, so they will be eliminated entirely.

### *5.4.8* Rings Controls

The existing infrastructure of CAMAC crates and timing links appear to be adequate for Mu2e storage ring operations, with a few upgrades to replace end of life components. Some CAMAC functionality may be replaced by Hot Rack Monitor (HRM) installations that run on a VME platform. Installing HRMs will provide 16 bit A/D readbacks, DAQ, I/O and clock channels, but have the overhead of requiring additional Ethernet connectivity. There should be no need to install additional CAMAC crates, as there is existing excess capacity. An inventory of existing CAMAC crates [20] shows that about 25% of the slots are unoccupied and could be used for additional CAMAC cards. After the removal of cards that are no longer needed after Collider Run II ends, there will be additional slots freed up. It is anticipated that there will be ample CAMAC crate coverage for Mu2e operation, and few crates will need to be added or moved.

There are six serial links distributed through and between the service buildings that will be used for Mu2e. In addition to the links used for communication with CAMAC, there are also links for distributing clock signals, a permit loop, and a link for remote ACNET consoles. The existing clock links distribute TCLK and MIBS, which will be adequate to synchronize most devices with beam for Mu2e operation. An RF marker system will also be needed for beam synchronization and is covered in the Rings RF section. The Permit loop will be a critical part of the machine protection scheme, providing a means of inhibiting incoming beam when there is a problem with the Mu2e beam delivery system. With the addition of a beam dump in the Delivery Ring, the beam permit system will need to be modified to accommodate the abort kicker at AP50.

The service buildings will have numerous nodes that communicate over Ethernet to the control system during Mu2e operation. A map of the controls network appears in Figure 5.9. All of the current Pbar service buildings have Gigabit fiber optic connections from the cross-gallery computer room to Cisco network switches centrally located in each service building. These will provide ample network bandwidth and connections after the reconfiguration for Mu2e. The central Ethernet switch for Pbar distribution is located in AP10. It is considered to be at the end of its life and will need to be upgraded.

Construction of the tunnel that houses the extraction beam line may interrupt the existing communications duct path that connects the Pbar service buildings to the control system and network. Controls Heliax cables from this controls duct may need be cut and spliced and network fiber optic cables will be replaced to restore control system operations. Additional controls fiber optic cables will need to be pulled to the Mu2e service building through existing duct banks, tunnels and service buildings. Fiber optic network cables that run through the Pbar Rings may need to be upgraded to radiation hardened versions due to the increased radiation levels expected during Mu2e operations.

## 5.5 RF Systems

### *5.5.1* Overview

The longitudinal bunch structure required for the Mu2e experiment will be largely accomplished by a 2.5 MHz RF re-bunching sequence in the Recycler Ring. This RF sequence will re-bunch the train of 53 MHz bunches that constitute a proton batch into four 2.5 MHz bunches that occupy one seventh the circumference of the Recycler Ring (see Reference [21]). Each of these bunches will be synchronously transferred, one bunch at-a-time, to the Delivery ring, where the beam is held in a 2.4 MHz RF bucket during resonant extraction. The final longitudinal phase space beam distributions in the Recycler prior to transfer to the Delivery Ring are shown in Figure 5.12. The proton bunches that are transferred to the Delivery Ring show artifacts of the original 53 MHz bunch structure. As can be seen in Figure 5.12, the entire bunch length, including most of the tails, is contained within the 200 nsec width requirement.

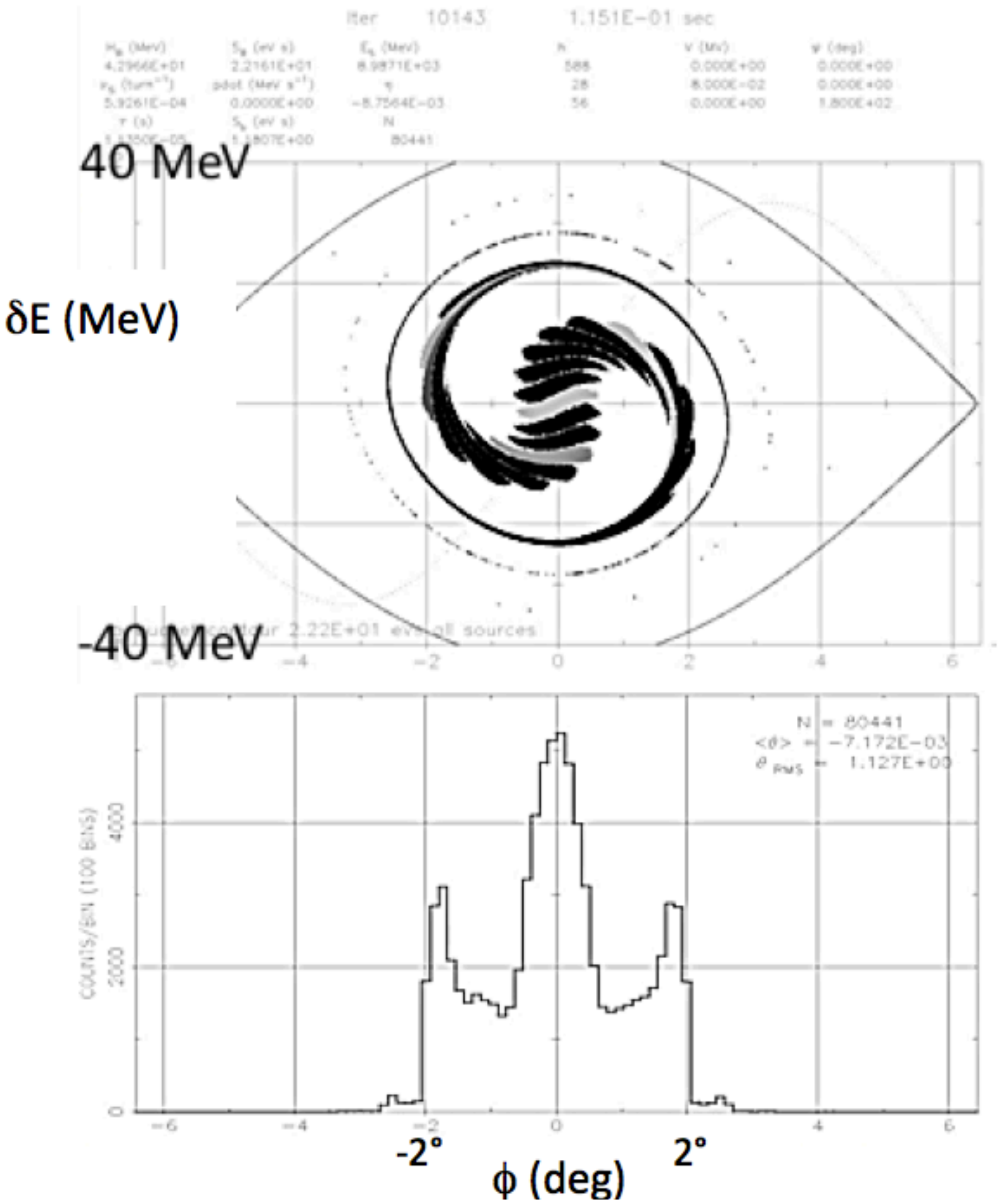

Figure 5.12. Longitudinal phase space distribution of protons in the Recycler Ring after the completion of the bunch formation sequence as calculated by an ESME [22] simulation. The top plot is a scatter-plot of the energy-phase coordinates of each proton. The bottom plot gives the projection of this distribution on the phase axis. Each degree of phase is approximately 31 nsec of pulse length. Artifacts of the original 53 MHz modulation of the beam are clearly visible.

Figure 5.13 shows the desired final longitudinal distribution of the beam to be delivered to the Mu2e pion production target. The required beam consists of a train of narrow (200 nsec FW) pulses separated by the revolution period of the Debuncher (1.695 μsec). The interval between pulses will ultimately be evacuated to the $10^{-10}$ level by the extinction system[l] (section 5.8). To minimize beam loss during extinction, the Debuncher RF system must synchronously capture the proton bunches from the Recycler and maintain a matched RF bucket throughout resonant extraction. A summary of RF system parameters relevant to Mu2e operation are given in Table 5.8 and Table 5.9.

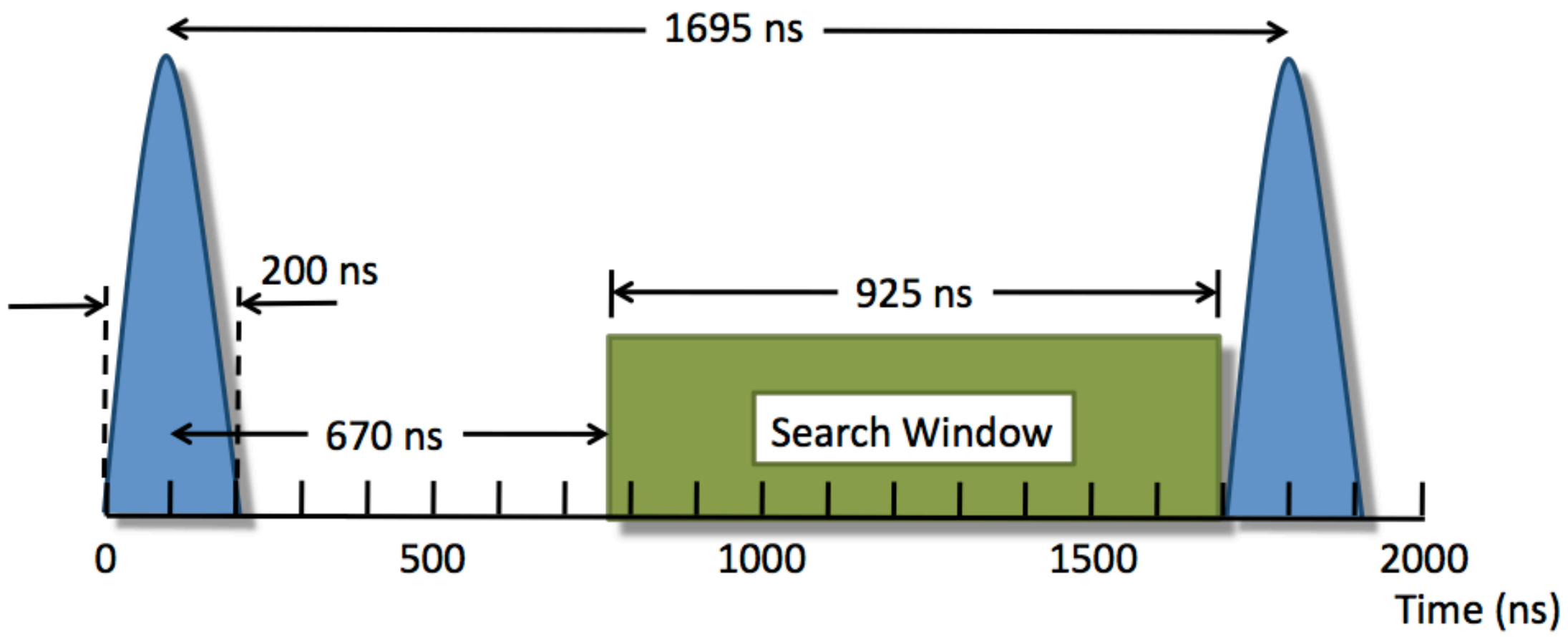

Figure 5.13. Longitudinal structure of the proton beam delivered to the Mu2e pion production target. The blue shaded structures are the beam pulses.

### *5.5.2* Delivery Ring 2.4 MHz RF System

***Overview***

The Recycler Ring is roughly seven times the circumference of the Delivery Ring. Consequently the Recycler RF harmonic number is seven times greater than that of the Delivery Ring. Moreover, the Recycler Ring is operated farther from transition than is the Delivery Ring[m]. These two circumstances mean that the Recycler RF bucket can be matched with a relatively small cavity voltage in the Delivery Ring. The flat-top voltage of the Recycler RF sequence is 80 kV. To match this, only 8 kV is required from the Delivery Ring RF cavity.

Since the proton bunch arrives in the Delivery Ring with a good deal of filamenting (see Figure 5.12), there will be significant modulation of the bunch length and energy spread as the bunch tumbles in the Delivery Ring RF bucket (Figure 5.14).

[l] $10^{-10}$ is the maximum allowed ratio of out-of-time beam to in-time beam.
[m] The Recycler Ring h = -0.00876, the Delivery Ring eta is 0.00607.

| Parameter | Value | Units |
|---|---|---|
| **Recycler Ring 2.5 MHz Bunch Formation RF System** | | |
| Harmonic Number | 28 | |
| Frequency | 2.515 | MHz |
| Peak Total Voltage | 80 | kV |
| Number of Cavities | 7 | |
| Duty Factor | 13.5 | % |
| Bunch Formation time | 90 | msec |
| **Debuncher 2.4 MHz RF System** | | |
| Harmonic Number | 4 | |
| Frequency | 2.360 | MHz |
| Peak Total Voltage | 10 | kV |
| Number of Cavities | 1 | |
| Duty Factor | CW | |

Table 5.8. Recycler and Delivery Ring RF Parameters

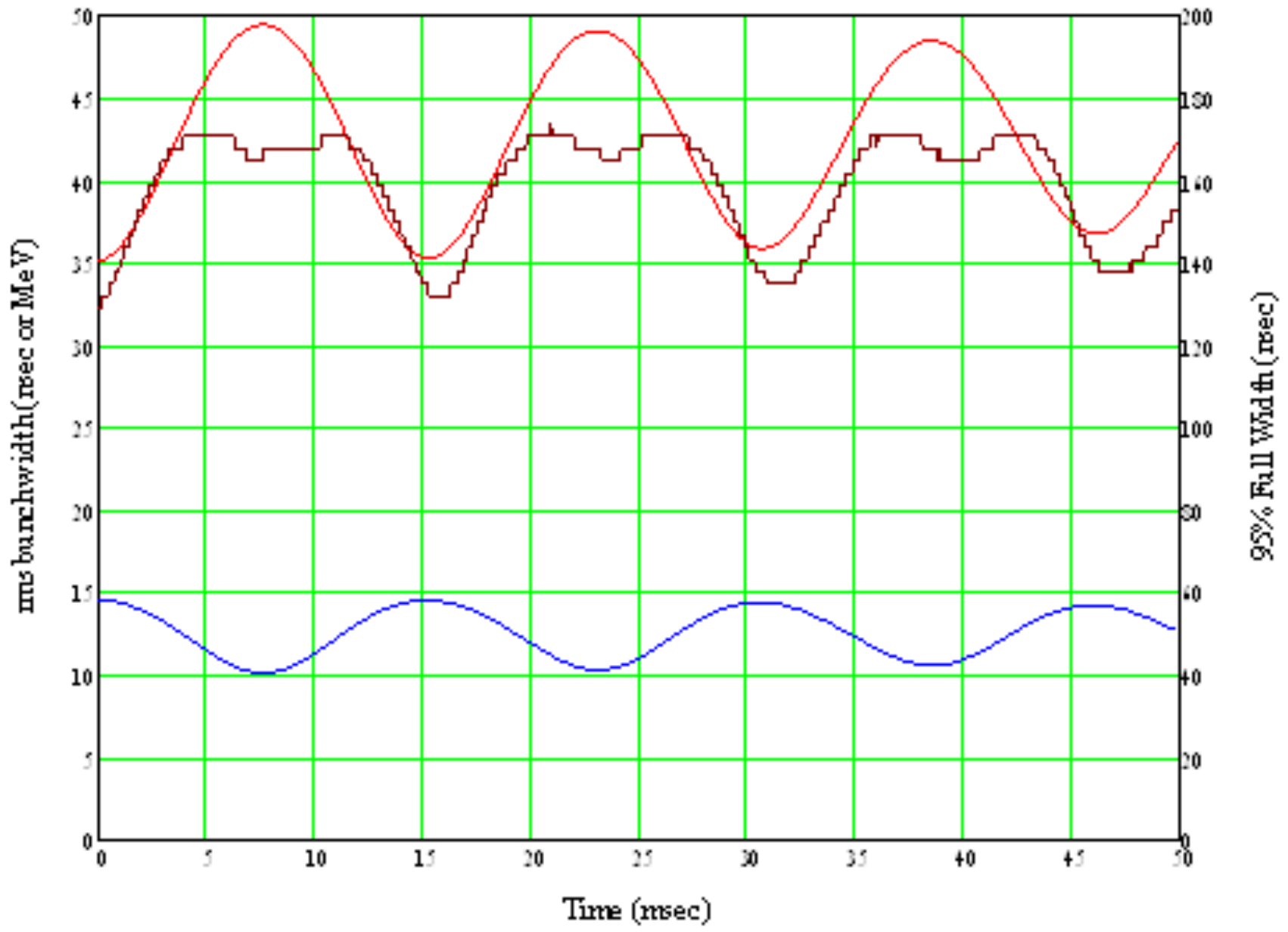

Figure 5.14. Proton bunch energy and time width as it circulates in the Delivery Ring RF bucket. The red trace is the rms time spread, the blue trace is the rms energy spread, and the brown trace is the 95% full width. The period of the modulation is one fourth of the synchrotron period ($T_{sync}$ = 29 msec).

This motion in the RF bucket causes substantial changes throughout the spill in the shape of the pulses delivered to the Mu2e target. Figure 5.15 shows the beam longitudinal phase at two positions in the synchrotron motion a quarter of a synchrotron period apart. The pulse time distributions are quite different at these two extreme positions in phase space.

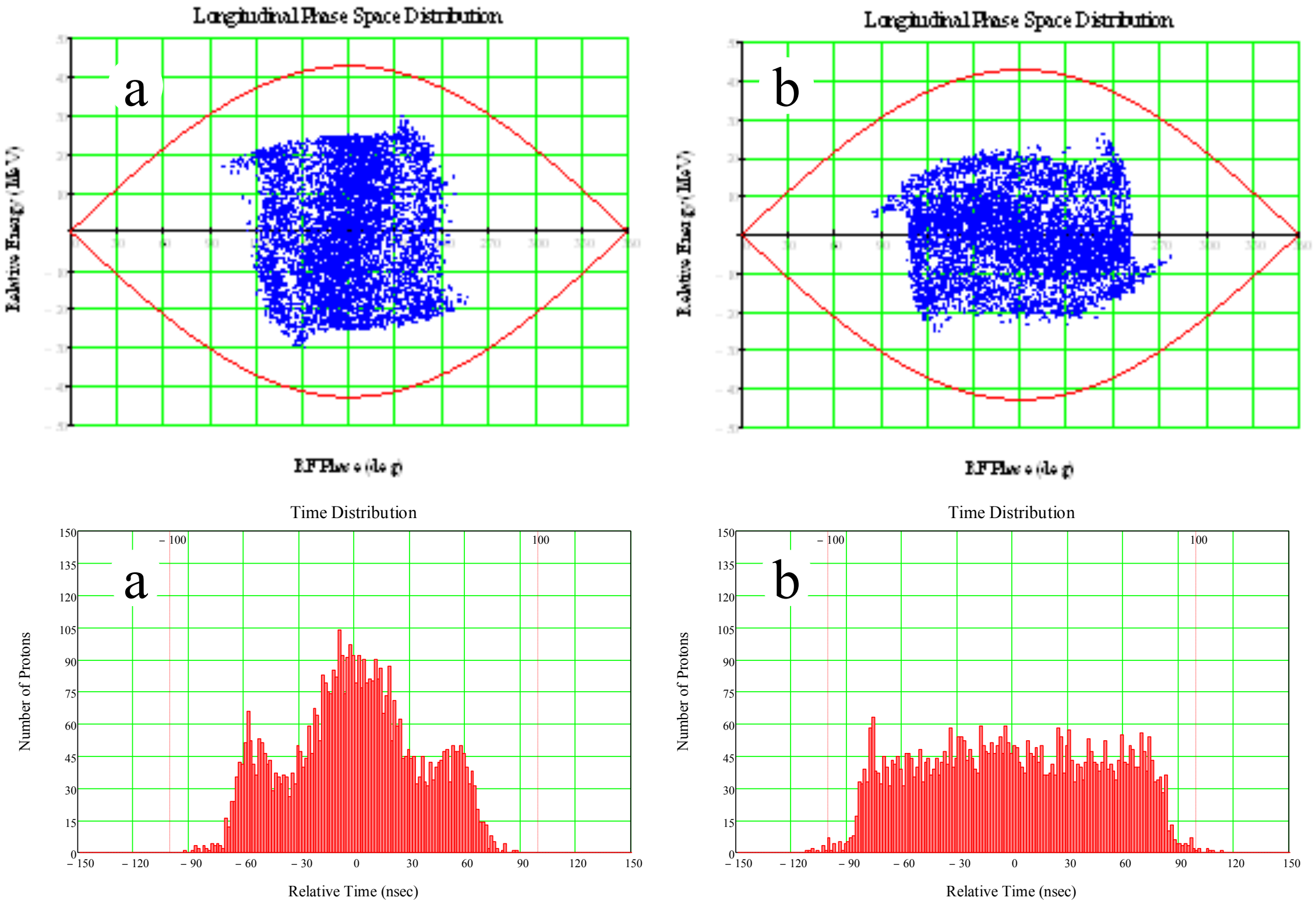

Figure 5.15. Results of a longitudinal tracking simulation the Delivery Ring. The extremes of the proton pulse time distributions are illustrated by two locations in the synchrotron motion of the beam a quarter of a synchrotron period apart. The (a) distributions occur at the minima of the time width plots of Figure 5.14. The (b) distributions correspond to the pulse width maxima. The top plots show energy versus RF phase for each proton tracked; the bottom plots show the projection of the phase space distribution on the RF phase axis, which is re-scaled in time. The vertical red dotted lines in the bottom plots show the ±100 nsec specification on the maximum beam width.

### *Synchronous single bunch transfer to the Debuncher (phase jump)*

Beam transfer to the Delivery Ring will involve a synchronous bucket-to-bucket transfer with a frequency hop due to the fact that the ratio of Recycler to Delivery Ring circumference is not an integer. The digitally synthesized LLRF will provide exact phase crossing alignment to facilitate the transfers. A similar system is presently in use for Main Injector to Debuncher beam alignment during antiproton stacking.

### *Delivery Ring 2.4 MHz RF Hardware*

Approximately 10 kV peak RF voltage will be needed on a continuous basis. A single ferrite-load cavity of the type being manufactured for the Recycler Ring 2.5 MHz RF system [21] is recommended. Transient beam loading will be pronounced as single bunches are circulating in the Delivery Ring. Power amplifier requirements are somewhat relaxed since the average beam current of a single bunch is only a quarter of a proton batch ($1\times10^{12}$ protons). The power amplifier for the Delivery Ring RF system will consist of a solid state amplifier that is similar to what is presently used for bunch coalescing in the Main Injector. Table 5.9 shows the physical parameters of the Delivery Ring 2.4 MHz RF system.

| Quantity | Value | Units |
|---|---|---|
| Beam Current ($I_p$) | $178.564\times10^{-3}$ | A |
| Number of cavities | 1 | |
| R/Q | 400 | Ω |
| Q | 125 | |
| Cavity Voltage | 10 | kV |
| Power Loss per Cavity | $1.0\times10^{3}$ | W |
| Total Apparent Power | $1.04858\times10^{3}\angle17.5089°$ | VA |
| Total Current | $209.716\times10^{-3}\angle17.5089°$ | A |
| Induced Mode Compensated | 3.786 dB = 35.3% | |
| Robinson Stable | 4 | |

Table 5.9. Parameters of the Delivery Ring 2.4 MHz RF system

### *5.5.3* Low level RF system

An FPGA based high speed LLRF system will generate the precise frequencies, phases, and amplitudes needed for beam bunching and coordinate the synchronous transfers between the various machines. The system will consist of the following modules:

- A supervisory processor to oversee the crate operations, communicate with ACNET and provide a local engineering interface.
- RF synthesizers for each cavity, one for 53 MHz and eleven for 2.5 MHz, a total of twelve, to provide independent phase and amplitude control.
- Input channels from each RF cavity for amplitude and phase feedback.
- Input signals from a resistive wall monitor for feed forward to compensate for transient beam loading.
- Local high speed processing to generate the active feedback and RF curves.

### *5.5.4* Studies RF System

Studies using the 2.4 MHz RF system are desirable to permit the use of the standard BPM systems and other RF dependent diagnostics. Since the Delivery Ring RF system is CW, extended use for beam orbit measurements and other uses is possible. The low lever RF system will be programmed to allow manual manipulation by studiers.

## 5.6 Resonant Extraction from the Delivery Ring

### *5.6.1* Introduction

The current design of beam preparation for the Mu2e experiment incorporates slow resonant extraction of protons from the Delivery Ring into the external beamline. The Delivery Ring will operate with beam intensities of $1\times10^{12}$ protons, approximately four orders of magnitude larger than its present value. The most challenging extraction requirements will be spill uniformity and the need for low losses in the presence of large space charge and momentum spread.

In the CD-0 proposal [1] only the third-integer resonance was considered for extracting the beam, because (a) the Delivery Ring's three-fold symmetry makes it more natural, (b) its operational horizontal tune, $\nu_x \approx 9.765$, is already within 0.1 of $\nu_x = 29/3$, and (c) third-integer extraction is simpler. Since then we have also considered the possibility of using a half integer resonance, at a horizontal tune of $\nu_x = 19/2$, with the result that the third-integer remains the preferred extraction resonance. This section reports considerations of the third-integer option only. Written below is a short summary; most items will be described further in following subsections.

***Machine models.***

Accumulator and Delivery Ring lattice files were written in MAD v.8 [24] syntax based on descriptions contained in Antiproton Source conceptual design reports. Control multipoles were placed in two of the three straight sections; "thin" extraction septum and Lambertson in the third. Special care was taken to preserve zero dispersion in these sections. These describe clean models, adequate at this stage for a conceptual study half-integer and third-integer extraction. However, moving forward, more realistic models should be used, including errors such as power supply ripple, misalignments, closed orbit control and error fields.

***Resonance species***

Currently, third-integer extraction is preferred to half-integer for slow extraction. Some attention was paid to half-integer extraction, but calculations and simulations indicated that, because the closed separatrix turns back on itself, it would be more difficult to reduce particle loss at the septum ("inefficiency") to 2%. Using an open half-integer separatrix might help but would require additional $19^{th}$ harmonic octupole circuits, in conjunction with the zeroth harmonic circuit needed for the closed separatrix. We did not study this more complicated (and more expensive) arrangement but would consider it if unexpected difficulties with the third-integer are encountered in the future.

***RFKO***

"RF knockout" (RFKO) technology will be added to the extraction protocol. With appropriate feedback and feed-forward, RFKO will help to control the spill structure, especially when emittance is small. The exact nature of the control systems is under investigation. The RFKO mechanism will be discussed below.

***Pre- and post- extraction.***

The time budget of the operating scenario (Section 5.1.2) allows several thousand turns for pre- and post-extraction operations. A few hundred to (less than) a thousand will be used to acclimate an unmatched injected bunch to the phase space of a resonance separatrix. After slow extraction and before the next bunch is injected, residual beam must be swept from the Delivery Ring and extraction circuits reset to their initial values.

***Losses***

Our objective is to lose no more than 5% of the protons at the extraction septum's wire, with a desired value of $< 2\%$. Losses at the septum will be more problematic for smaller emittances. We have yet to map the distribution of losses at the Lambertson or around the Delivery Ring; doing so will help to improve estimates of shielding requirements.

***Theory, quadrature, and simulation.***

Basic theories of half-integer and third-integer resonant extraction were reviewed and software written to analyze stepsize and inefficiency. Simulations using independent particles (CHEF [25]), based on our Delivery Ring lattice model, adequately validated theoretical predictions for these quantities. Programs for multi-particle simulations that include space charge – ORBIT [9] and SYNERGIA [26] – have been upgraded and are being used. Slow third-integer extraction in the presence of space charge has now been simulated successfully, but further upgrades in both quadrature and simulation software should be done as studies continue.

***Machine studies***

Delivery Ring tune scans were carried out in the vicinity of both half-integer and third-integer resonance lines. Half-integer extraction studies using the Main Injector were begun but not completed due to unexpected machine or control issues. We intend to use "RF knockout," a technique to be explained below, to help control resonant extraction. Recently, a Tevatron style damper/kicker was installed in the Delivery Ring as the device proposed to do this. Studies have just begun to examine its performance and suitability.

### *Hardware and costs*

Multipole requirements for both resonances have been written. Septum and Lambertson design bounds, based on stepsize, inefficiency, and aperture considerations, remain to be finalized. Though these devices are not yet designed, a first estimate of their costs has been carried out.

### *Instrumentation and control*

While we can control bussed multipoles to extract the beam, we have only begun to examine what instrumentation and control procedures will be needed to regulate the spill rate well enough to satisfy the experimental tolerance of ±50% variation per microbunch.

### *The Delivery Ring*

Our Delivery Ring lattice model was based predominantly on the description in the Tevatron 1 Design Report [27] (especially the SYNCH [28] file and output in Appendix D of the report), but with more realistic values for the quadrupole and dipole lengths. The ring has three-fold symmetry: three arcs and three straight sections. In addition, each arc and straight are mirror symmetric, giving the machine an overall dihedral symmetry. Machine optics were not quite symmetric during the Tevatron era in order to accommodate stochastic cooling and maximize the machine's acceptance.

The Delivery Ring's circumference has been measured to be 505.294 m, only 11 mm larger than its original design circumference. At 8 GeV, protons circulate at a frequency of 590.018 kHz. The lattice has a regular FODO structure with 60° cells in the arcs and simple "missing dipole" dispersion suppressor cells. In the long straight sections, six adjustable quadrupole circuits are available to adjust the machine tunes and match optics between sextants. Each sixth of the ring contains 19 quadrupoles - two of which are five inches longer than the others – 11 dipoles, and 23 chromaticity correcting sextupoles. Typical tunes are $(\nu_x, \nu_y) \approx (9.73, 9.77)$ and the maxima $(\beta_x, \beta_y)_{\max} \approx (17.8, 16.8)$. During the Tevatron era, the machine's large acceptance was quoted as 335 $\pi / \beta\gamma$ mm-mrad.

The Delivery Ring lattice is described more fully in Section 5.4.4. Importantly, dispersion in the Delivery Ring is fortuitously suited for resonant extraction of higher intensity beams. Large horizontal dispersion in the arcs and large momentum spread - $D_x \approx 2$ m with $\sigma_p / p \approx 0.004$ - will widen the beam horizontally, thereby reducing space charge induced tune spread. At the same time, dispersion is zero throughout the straight sections. Placing the septum and Lambertson in the straights assures that the separatrix's center *at the locations of these devices* will not depend on $\delta p / p$ to first order. However, its scale would still be affected if chromaticity were not also set to zero. Keeping chromaticity small will minimize chromatic "blurring" of the separatrix. We expect $(\beta_x, \beta_y) \approx$ ( 14-15, 4-6 ) m at the septum and at sextupole locations and ( 8-9, 10-11 ) m at the Lambertson.

The beam *current* in the Delivery Ring will have the same direction as before: viewed from above, antiprotons rotate clockwise; Mu2e's protons will rotate counter-clockwise. Under this assumption, the *negative x* direction, *not the positive*, of the Delivery Ring's local frames point toward the outside of the ring. (Thus, for example, dispersion in the arcs is negative).

To excite and control a third-integer resonance, six "harmonic" sextupoles will be added to the base lattice. Their locations are shown schematically in Figure 5.16. In addition, three quadrupoles will be inserted into symmetric locations in the three straight sections for fine tune control throughout a spill.

### *5.6.2* Quadrature and simulations

Estimating the orbits' stepsizes at the septum and the probability of hitting its wires are among the few important calculations that can be done in quadrature, by approximating the orbits of extracted particles as though they were exactly on the outgoing branches of the separatrix. This approximation is reasonable provided extraction is adiabatic. While crude, these estimates nonetheless will ultimately be useful for specifying the septum's geometry: i.e. placement and thickness of wires and the minimal extent of the field region. This section contains these calculations for the third-integer resonance. Equations are introduced briefly[n] to set up the calculations and to provide a basis for estimating required control parameters and hardware settings. They are reasonably valid only "near resonance" and only in certain physical situations, as described in an appendix of Reference [29]. Moreover, in this section attention will be confined to the horizontal plane.

The initial intensity of the beam upon injection into the Delivery Ring will be large enough that the effects of space charge cannot be neglected *a priori*. Theoretical and semi-analytic calculations do not take this into account. Doing so requires simulations, i.e. multi-particle tracking with interparticle forces included. Two programs, ORBIT and Synergia, are being employed to approach this problem. Their initial results are also presented below.

[n]These were derived and explained in Mu2e doc-556 (FERMILAB-FN-0842-APC-CD), "*Preliminaries toward studying resonant extraction ...*," [29].

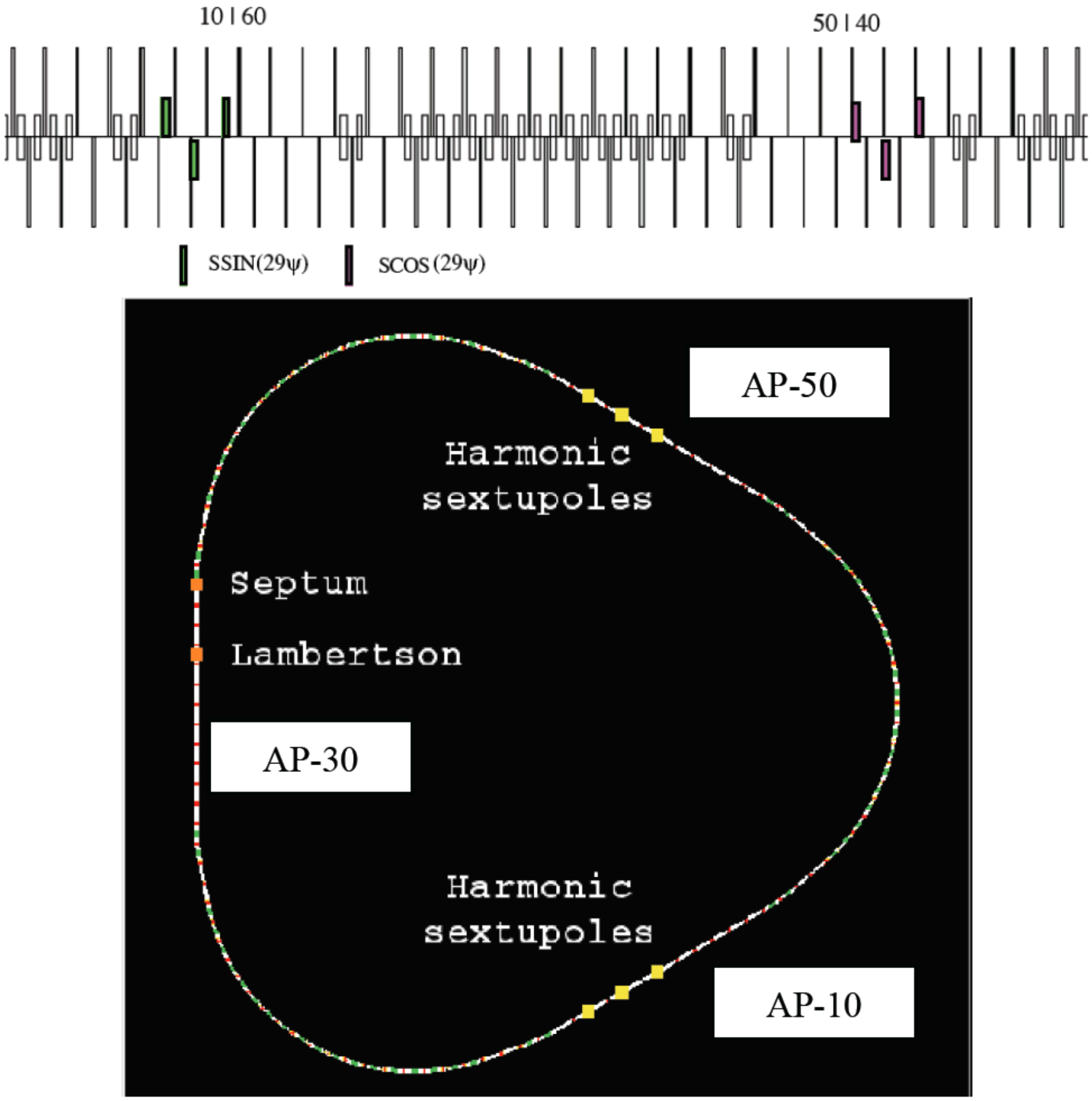

Figure 5.16. Locations of the harmonic sextupoles on two (approximately) orthogonal circuits in two of the Delivery Ring's straight sections. The septum and Lambertson will be placed in the third, though not necessarily placed as shown in the figure. The color scheme is: quadrupoles, red; dipoles, green; sextupoles, yellow; septum and Lambertson, orange.

***Separatrix***

Third-integer resonant extraction will be initiated by exciting harmonic sextupoles to drive the 29/3 horizontal resonance and create its separatrix. The proximity of the horizontal tune to the resonance and the strength of the sextupoles determine its initial size, which must initially accommodate the entire horizontal extent of the beam within its stable central region. Extraction proceeds as high amplitude particles encounter the separatrix and move outward along the separatrix branches. To maintain extraction, the separatrix is "squeezed" by using fast-ramping quadrupole magnets located in zero-dispersion regions of the Delivery Ring. These quadrupoles have the effect of decreasing $\Delta\nu \equiv \nu_x - \nu_r = \nu_x - 29/3$, thereby shrinking the size of the stable region of the separatrix. The quadrupole ramp continues until the bare lattice tune reaches the resonance value, $\nu_r$.

Figure 5.17. illustrates the idealized separatrix for the third-integer resonance, drawn in a complexified, normalized, horizontal phase space with coordinates as illustrated in Figure 5.18. Viz.,

$$a = \Re a + i\Im a = \sqrt{I}\, e^{i(\pi/2-\varphi)} \quad \text{and} \quad \sqrt{2\beta}\; e^{-i(\psi-\nu\theta)}\, a = x + i(\alpha x + \beta x') \;, \tag{5-4}$$

$\beta$, $\alpha$, and $\psi$ are the usual Courant-Snyder lattice functions, and $\theta = s/R$ represents azimuth. We have the freedom to set $\psi - \nu\theta = 0$ anywhere in the ring - in particular, at the location of the septum - so the extra phase is ignorable in this context.[o] The pair $(\varphi, I)$ are angle-action coordinates for the system. If $(x, x')$ are canonically conjugate, then so are $(\varphi, I)$ and $(a, a^*)$ as verified by their Poincaré invariants,

$$dx \wedge dx' = d\varphi \wedge dI = i\,da \wedge da^* = 2d(\Re a) \wedge d(\Im a) \;. \tag{5-5}$$

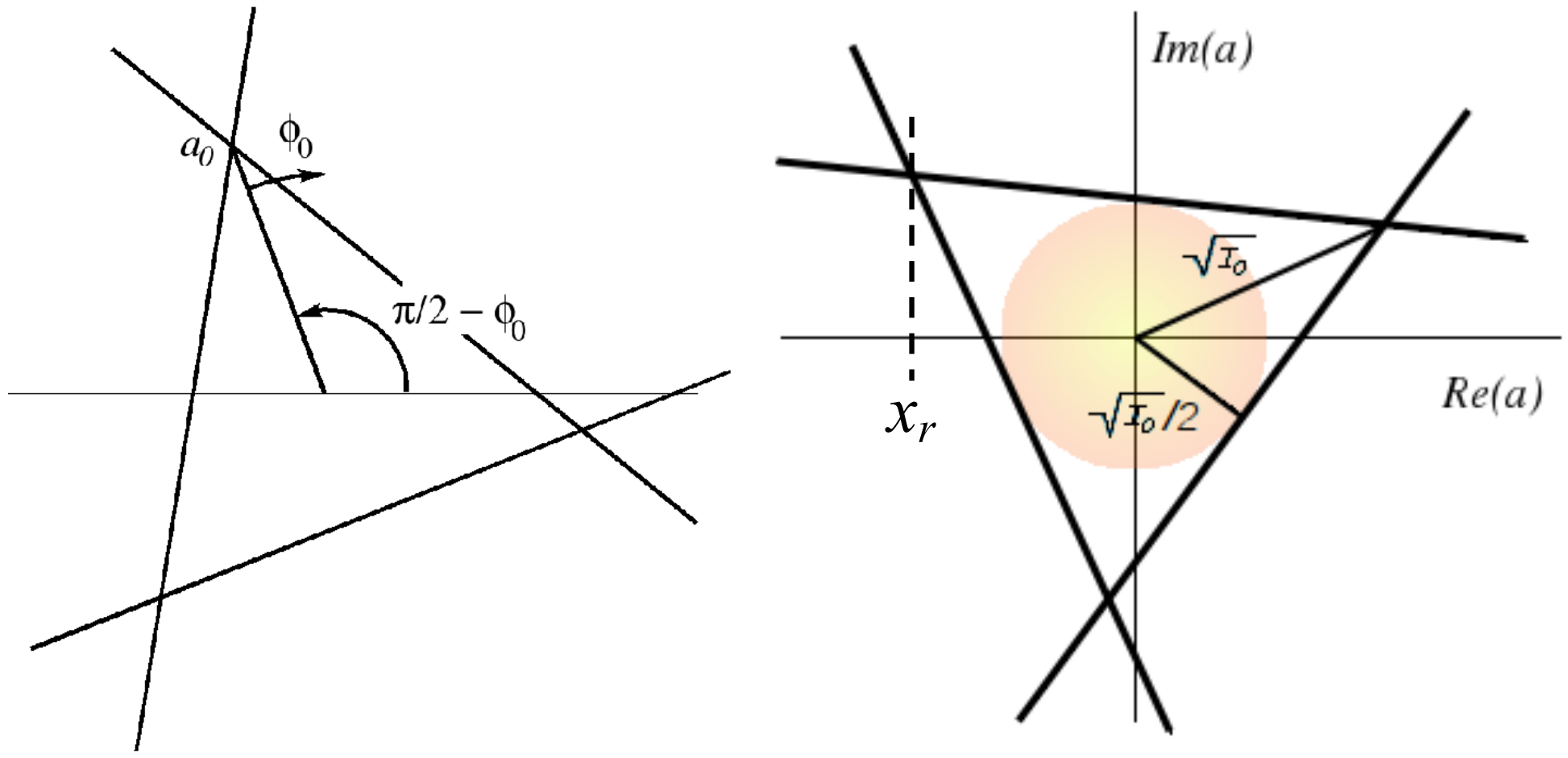

Figure 5.17. Left hand side: The angle connecting a “standard” triangle with the separatrix is $\pi/2 - \varphi_0$ not $\varphi_0$. Right hand side: The central stable region of the third-integer resonance is initially set to contain the injected bunch. Using these coordinates, “areas” are multiplied by 2 to obtain emittances. Alternatively, the length scale is multiplied by $\sqrt{2}$ .

[o]The extra phase is included to make $\varphi$ increase linearly with azimuth under linear dynamics.

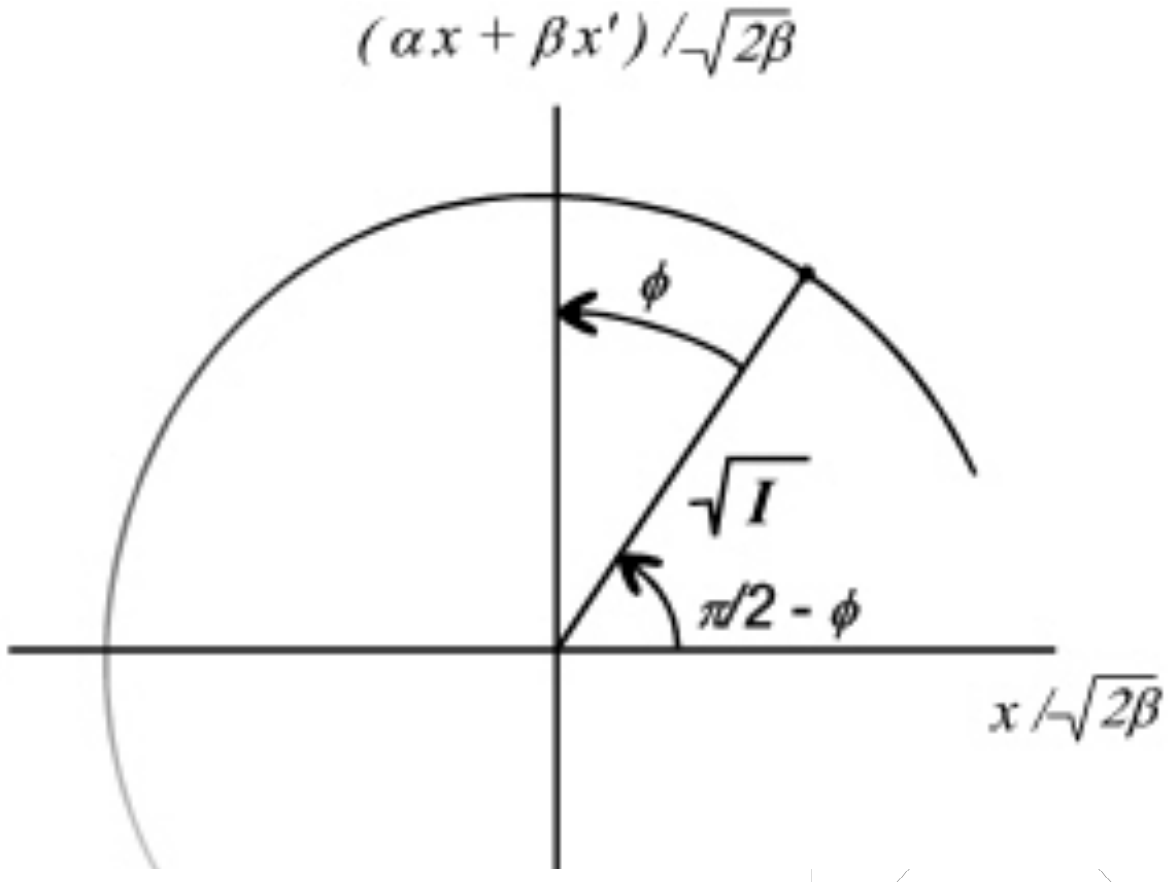

Figure 5.18. Defining the complex dynamical coordinate, .

The last expression on the right means that "area" in $(\Re a, \Im a)$ Cartesian space must be multiplied by two to be interpreted as "emittance." For example, a bunch contained within the (partly drawn) circle in Figure 5.18 has an emittance,

$$\varepsilon = 2\pi r^2 = 2\pi(\sqrt{I})^2 = 2\pi I$$

The orientation of the separatrix is at our discretion. It is parametrized by the phase of the resonant orbit, identified as $\pi/2 - \varphi_0 \pmod{2\pi/3}$ in Figure 5.17. At or shortly after injection, the size of the separatrix is just sufficient to enclose the proton bunch, as sketched on the right hand side of the figure. The beam's transverse distribution should accomodate itself to the distortion induced by the separatrix within less than a thousand turns., We intend to wait approximately 5 msec of the available time before beginning extraction in order to allow the bunch to fill the central stable region more uniformly. (See Figure 5.20 below.)

The Hamiltonian associated with the third-integer resonance model is [29],

$$\begin{aligned} H &= \Delta\nu\, a^* a - ig\, a^3 + ig^* a^{*3} + \ldots \\ &= \Delta\nu I - (g e^{-i3\phi} + g^* e^{i3\phi}) I^{3/2} + \ldots \quad . \end{aligned} \tag{5-6}$$

Here, $\Delta\nu \equiv \nu_x - 29/3 \approx 0$ is the difference between the linear (small amplitude) horizontal tune and the resonant tune and is presumed to be small; the "resonance coupling constant," $g$, is a linear functional of the sextupole field strength distribution.

$$g = \frac{i}{6\sqrt{2}} \frac{1}{4\pi} \sum \frac{B''l}{B\rho} \beta_x^{3/2}(\theta) e^{-i3(\psi_x(\theta) - \Delta\nu\theta)} \qquad (5\text{-}7)$$

where the sum is carried out over the locations of the sextupoles. The phase of the complex parameter *g* determines the orientation of the third-integer separatrix, as will be given below.

In normalized phase space, the idealized third-integer separatrix comprises three straight lines forming an equilateral triangle whose interior is the central stable region of phase space. This is illustrated in Figure 5.17, which also identifies the orientation angle, $\varphi_0$, and the amplitude of the resonant orbit, $\sqrt{I_0} = |a_0|$. The vertices of the triangle are the resonant orbit. These equations describe how the control parameters determine separatrix geometry:

$$|a_0| = \sqrt{I_0} = |\Delta\nu / 3g|$$

$$0 < \Delta\nu \;\Rightarrow\; \varphi_0 = \Psi/3 \mod\left(\frac{2\pi}{3}\right)$$

$$\Delta\nu < 0 \;\Rightarrow\; \varphi_0 = \Psi/3 + \pi/3 \mod\left(\frac{2\pi}{3}\right)$$

$$\text{where} \quad g \equiv |g| e^{i\Psi} .$$

The emittance of the central stable region is the “area” of an equilateral triangle with “radius” $\sqrt{2I_0}$ .

$$\varepsilon_c = \frac{3\sqrt{3}}{4} |\sqrt{2I_0}|^2 = \frac{1}{2\sqrt{3}} |\Delta\nu / g|^2 \qquad (5\text{-}8)$$

For comparison, the emittance, $\varepsilon_b$, of a bunch in equilibrium *with the linear machine* and totally contained within the central stable region is the “area” of a circle whose radius is $\sqrt{I_0/2}$, i.e. $\varepsilon_b = \pi I_0 / 2$, (See Figure 5.17)

$$\frac{\varepsilon_b}{\varepsilon_c} = \frac{\pi}{3\sqrt{3}} \approx 0.605. \qquad (5\text{-}9)$$

The orientation, $\varphi_0$, of the separatrix is also important. It determines the angle, in phase space, of extracted particles entering the septum area. The separatrix triangle must be oriented such that, at the extraction septum, the *x*-projection of the beam motion along an outbound branch of the separatrix lies within field-region of the septum (i.e. at values of $x$ that are less than the $x$ position of the septum wire-plane (denoted $x_w$ below). Conversely, one could say that, for efficient extraction, the septum wire plane must be placed somewhere between the $x$-coordinate of a vertex of the separatrix triangle and the maximum reasonable horizontal extent of the resonant motion (called $x_m$) in the third-integer resonance model[p].

Typically, the sextupole magnets are grouped into two circuits, each controlled with its own "knob." Let $k_c$ and $k_s$ represent the two "knobs" controlling the two (not necessarily orthogonal) harmonic sextupole circuits so that

$$g = k_c g_c + k_s g_s \tag{5-10}$$

where $g_c$ and $g_s$ are complex while $k_c$ and $k_s$ are real. (Subscripts "$c$" and "$s$" stand for "cosine" and "sine.") (see [29], p.25),

$$\begin{aligned} k_c &= \Im(g_s^* g) / \Im(g_s^* g_c) \\ k_s &= \Im(g_c^* g) / \Im(g_c^* g_s) \ . \end{aligned}$$

The six phasors that appear in the resonance sum, $g$, of Equation 5-7 are shown in Figure 5.19. Labels identify regions of the Delivery Ring's straight sections. By reversing the polarity of the "-2" locations relative to the "-1" and "-3", we see that sextupoles in the "20-x" and "50-x" locations form almost perfectly orthogonal circuits, though orthogonality is not essential. With them we can adjust both amplitude and phase of $g$ as desired. The phasors are followed in the figure throughout a squeeze, resulting in the dark traces. It is important to note their stability: i.e. individual phasors change only slightly, resulting in an even smaller variation in $g$. Because of this, it should not be necessary to ramp the sextupoles to keep $g$ sufficiently constant throughout extraction. However, it may still be desirable to rotate the separatrix slightly during extraction in order to improve efficiency. This would require changing the relative strengths of the two sextupole circuits (see below).

***Pre-extraction***

Upon injection into the Delivery Ring, the bunch distribution is not matched to the invariant orbits distorted by the separatrix in horizontal phase space. Accordingly, it will immediately begin to filament within the central stable region, as shown in Figure 5.20 and if not centered correctly on the closed orbit, to tumble as well. Extraction should be delayed until the microstructure has been reasonably smoothed out. Even without space charge, simulations suggest this should take less than 500-1000 turns. The randomizing effects of space charge forces will work to decrease this further. This fits easily within the 5 - 10% of the beam time budgeted for pre- and post-extraction operations.

[p] $x_m$ is defined as the maximum outward extent of the projection of motion on the x-axis. In actual (i.e. non-model) motion, $x_m$ is the $x$-coordinate of the horizontal aperture. The resonance model breaks down before this point.

At the onset of extraction proper, when the separatrix has been squeezed sufficiently to intersect the beam, there will be a tendency for an initial brief burst of protons at a higher rate than desired. This must be controlled via measurement and feed-forward of the tune control circuit's ramp.

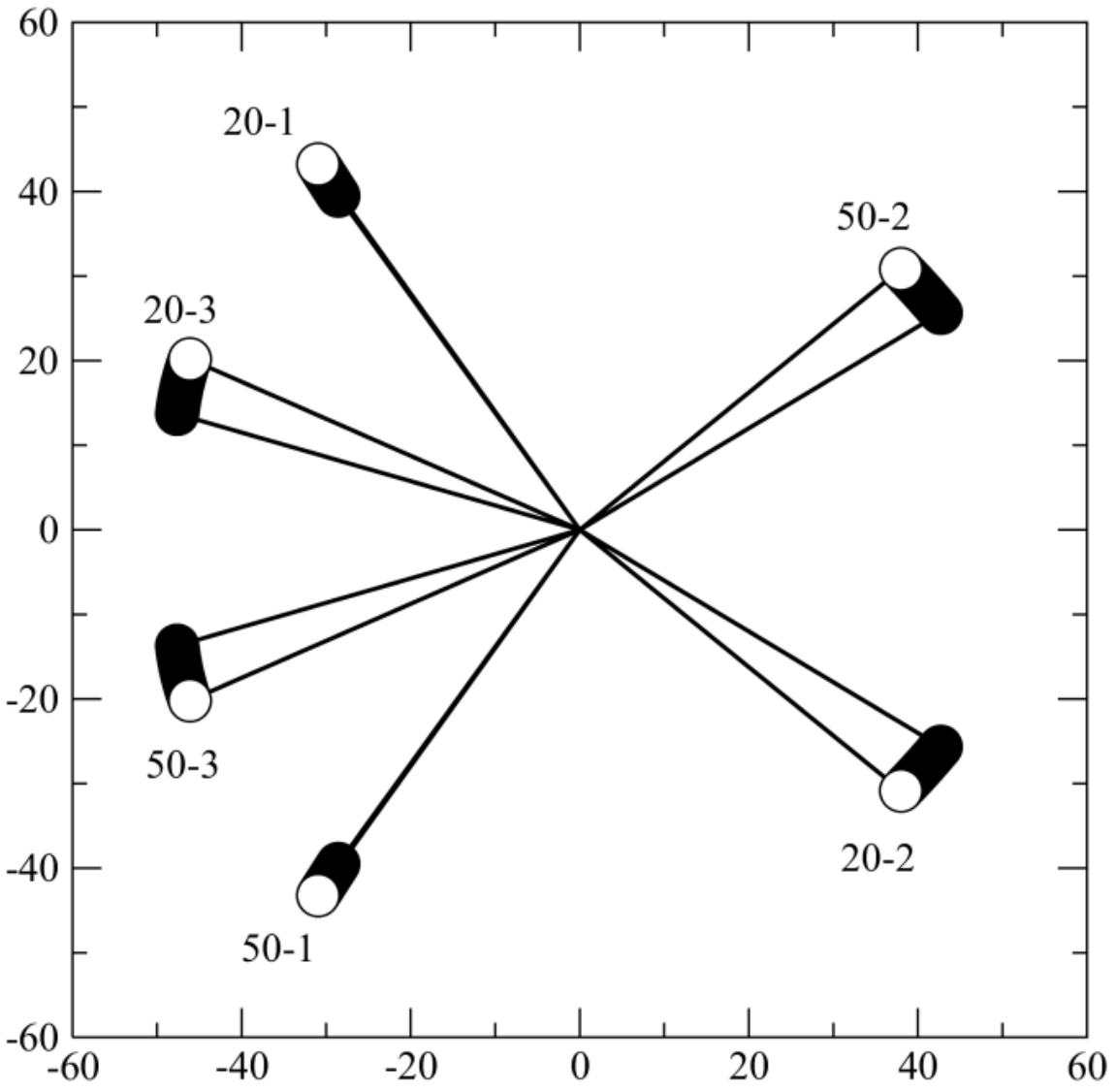

Figure 5.19. Phasor stability of the $\nu_x = 29/3$ resonance during the squeeze. Labels indicate locations in the Delivery Ring.

***Step size and (in)efficiency***

The horizontal coordinate of a proton on an outgoing branch of the separatrix is given by the expression,

$$x = \sqrt{2\beta}\, \Re\left\{ a_0 (1 + r e^{sgn(\Delta\nu) i\pi/6}) \right\} \tag{5-11}$$

where the complex $a_0$ was defined earlier and $r$ is a non-negative real number. The particle is on the vertex when $r$ = 0. The "stepsize" is the difference in $x$ – i.e. the change in the horizontal position – after three turns around the ring. For the third integer resonance, it can be expressed analytically as follows:

$$\Delta x = \sqrt{2\beta}\,\Re\left\{ a_0 e^{sgn(\Delta\nu) i\pi/6} \right\} \Delta r, \text{ where}$$

$$\Delta r = \frac{r(r+\sqrt{3})}{[\sqrt{3}/(\exp(6\pi\sqrt{3}\Delta\nu)-1)]-r}$$

The results are displayed in Figure 5.21 for $\varphi_o$ –120°. If the septum's wire is placed at 1.6 cm, and $\varepsilon_b$ = 30 $\pi$ mm-mrad, a proton just inside the wire (i.e. still in the septum's field-free region) will reach about 4.4 cm after three turns, requiring a minimum field region of 2.8 cm. Moving the wire closer to the central orbit reduces this number but, as will be described in the next section, increases losses at the wire. The final septum design will have to balance these two tendencies against each other.

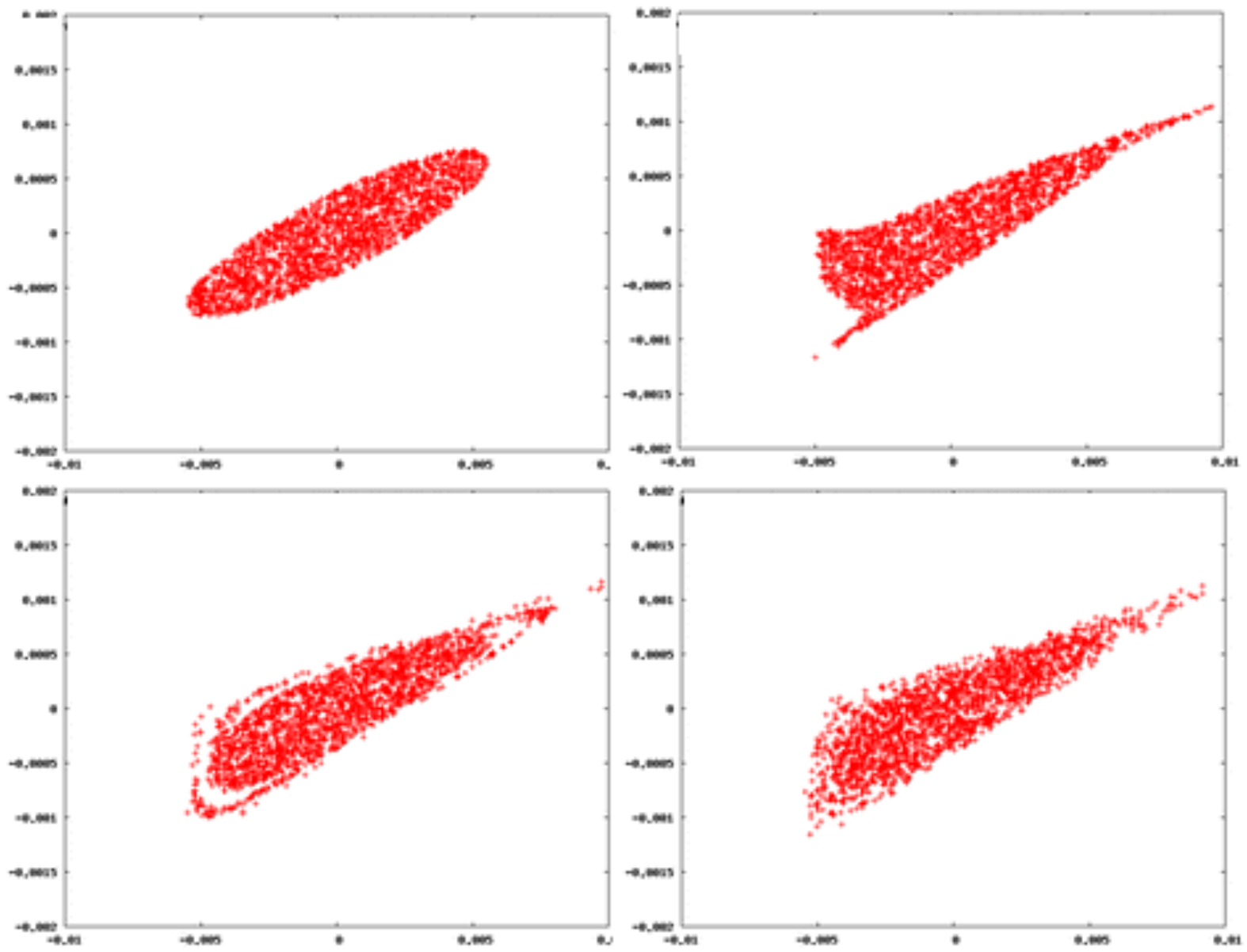

Figure 5.20. Time evolution in transverse horizontal phase space ($x$, $x'$) of the beam immediately after injection into the Delivery Ring. Upon injection, an unmatched bunch will filament until it fills the stable region within the separatrix. This should take less than 1000 turns.

A useful, rough estimate for the fraction of particles hitting a septum's wire, rather than falling into the field region, was first written in 1978 by Edwards [30]:

$$P = \frac{1}{N}\frac{w}{(dx/dn)_w} = \frac{1}{2\pi N}\frac{w}{(dx/d\theta)_w} \tag{5-12}$$

where $N$ is the order of the resonance (i.e. 2 for the half integer, 3 for the third), $w$ is the wire's width; $(dx/dn)_w$ is the "per turn" rate at which $x$ changes, evaluated at the wire position, $|x_w|$, for an orbit *on the separatrix.* Here, we shall simply use this expression without discussing the approximations used to obtain it or their limitations.

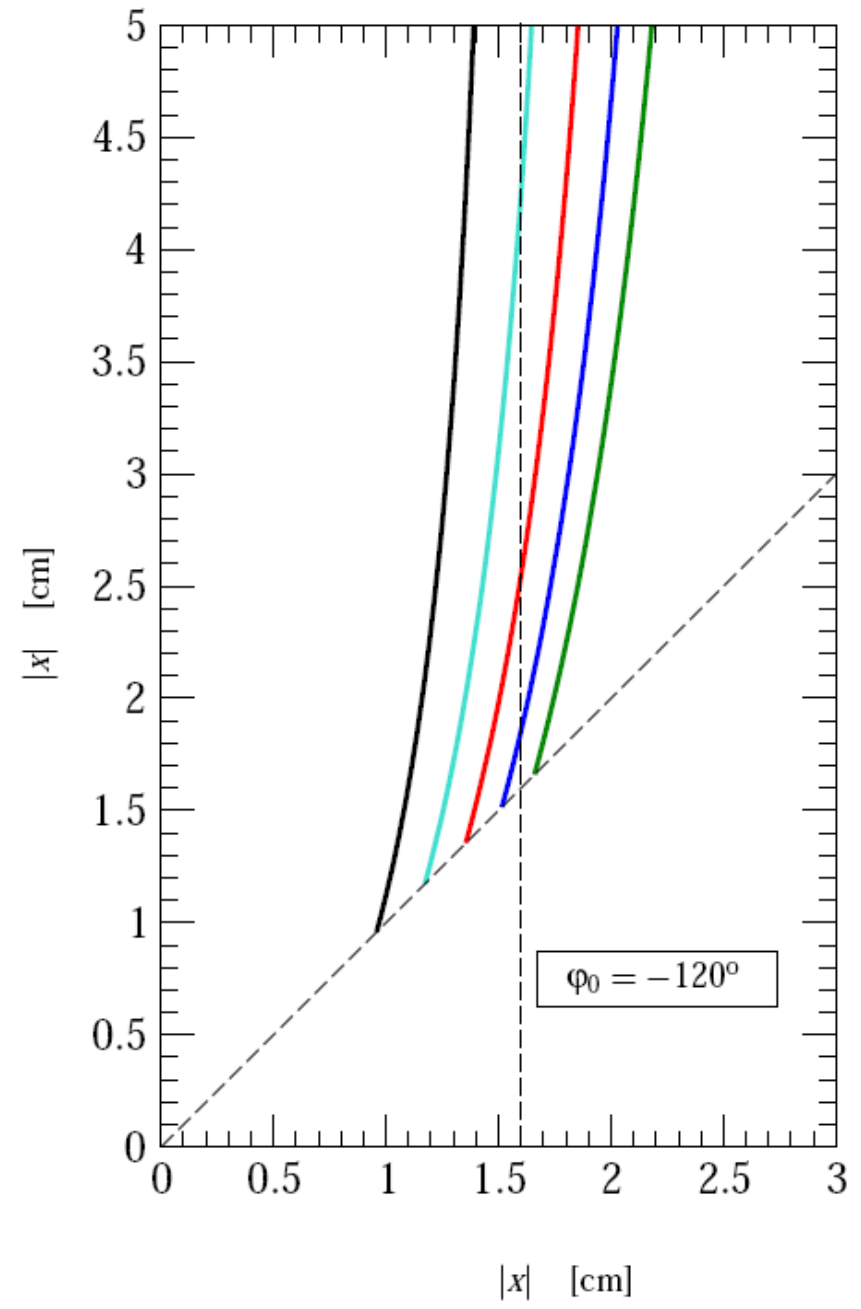

Figure 5.21. "Reach" of orbit on third-integer resonance separatrix for $\varphi_o = -120°$. The horizontal axis indicates the proton's initial position and the vertical axis its final position after three turns. Different colored curves indicate different emittances. The dashed vertical line shows that if we place the septum's wire at 1.6 cm, and $\varepsilon_b = 30\ \pi$ mm-mrad, a proton just inside the wire (i.e. still in the septum's field-free region) will reach about 4.4 cm after three turns, requiring a minimum field region of 2.8 cm.

For the third-integer resonance, the general expression used to estimate inefficiency, becomes [29]:

$$P = \frac{1}{6\pi}\frac{1}{\sqrt{2\beta}}\frac{w}{\Re(3g^* a^{*2} - i\Delta\nu a)_w} ,$$

where $a$ is placed on the separatrix and, again, the subscript $w$ means "evaluated at the wire." The results are plotted in Figure 5.22 for

(a) $\Delta\nu \in \{-0.01, -0.02\}$,
(b) $\varphi_o \in \{-180°, -160°, -140°, -120°\}$, and
(c) $\varepsilon_b / \pi \in \{20, 30, 40, 50\}$.

The color coding in Figure 5.22 identifies $\varepsilon_b$. The group of curves at the extreme left correspond to $\varphi_o = -180°$. They achieve the smallest inefficiencies because the separatrix's orientation in phase space means the particle moves most slowly along $x$ - permitting it to move farther along the separatrix before reaching the wire, but at the cost of rapidly increasing $x'$. Such particles would almost certainly be lost elsewhere in the machine before reaching the wire.

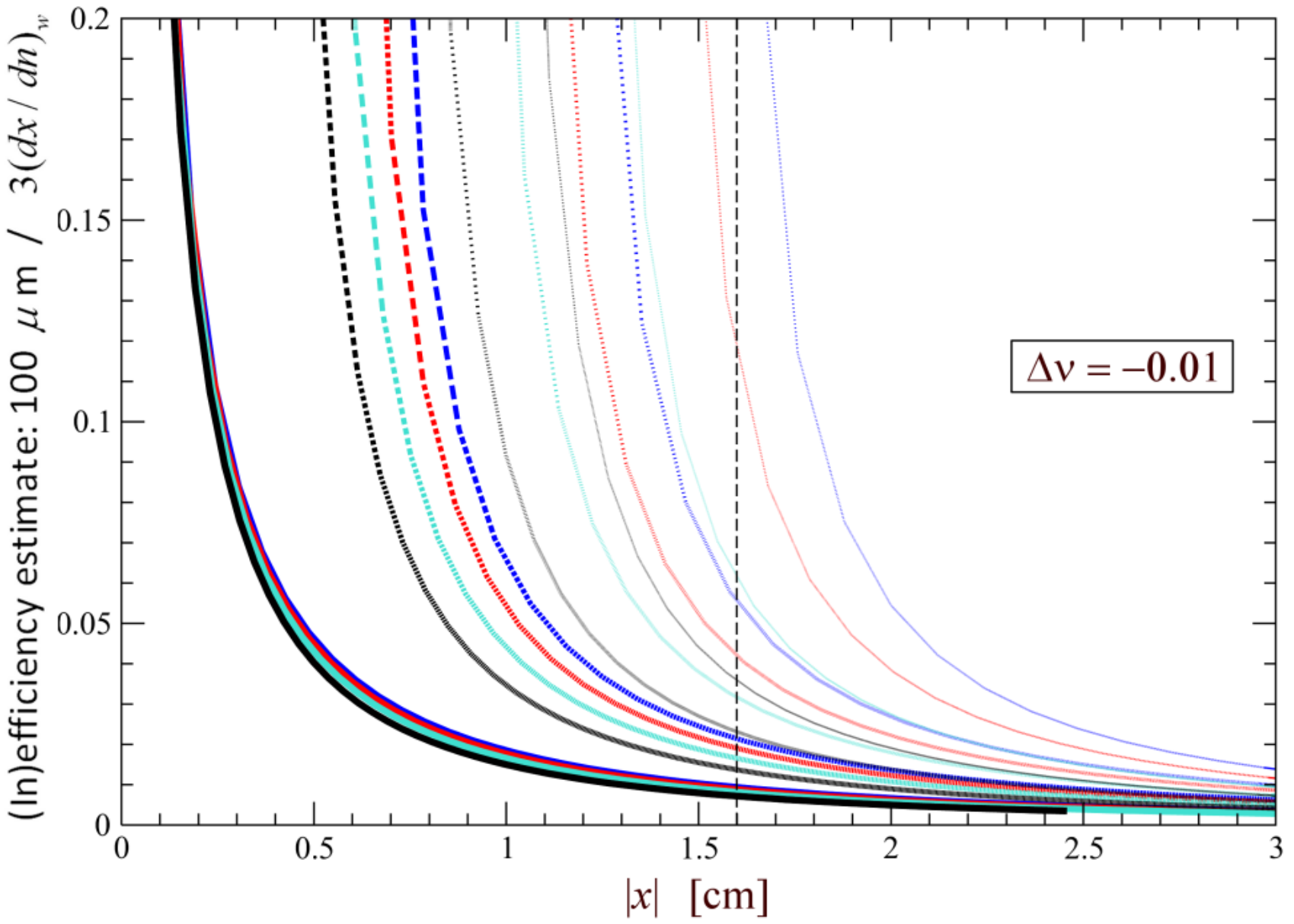

Figure 5.22. Inefficiency for third-integer resonance: $-180° \leq \varphi_0 \leq -120°$. The group of curves at the extreme left correspond to $\varphi_o = -180°$. The color coding identifies $\varepsilon_b$ (black = $20\pi$, blue = $50\pi$).

There is justficiation for healthy scepticism regarding such lowest order calculations. We considered it worthwhile to check them against tracking simulations involving random distributions of protons, but not including space charge effects. Results are shown in Figure 5.23 and Figure 5.24 below.

First, Figure 5.23 shows two representative comparisons of theoretical calculations of the stepsize, or "reach," with the results from tracking simulations. (Note: the axes record $|x|$, not $x$. Recall that $x$ is negative.) In these figures, the initial $\Delta\nu_x = -0.01$, and the sextupole circuit was set up to accomodate an initial invariant emittance, $\varepsilon_b = 10\pi$ mm-mrad. The separatrix's orientation was set to $\varphi_0 = -120^{\circ}$ on the left and $-130^{\circ}$ on the right. To mitigate statistical effects, and to capture the stepsize for these settings, simulations were carried out by populating protons in a thin layer just inside the stable region of the separatrix and then approaching the resonant tune until orbits of all the particles have diverged. As the particles leak past the stable region, their initial and final positions after three turns, i.e. $(x, x + \Delta x)$, are iteratively recorded for display. The black curves indicate the theoretical calculations; the scatterplot of "data" from simulations are in blue. The vertical dashed line marks the location of a hypothetical septum wire at 1.6 cm. Agreement is adequate for *practical* settings of the parameters, though in all cases studied, the theoretical estimates exceed those from simulations.

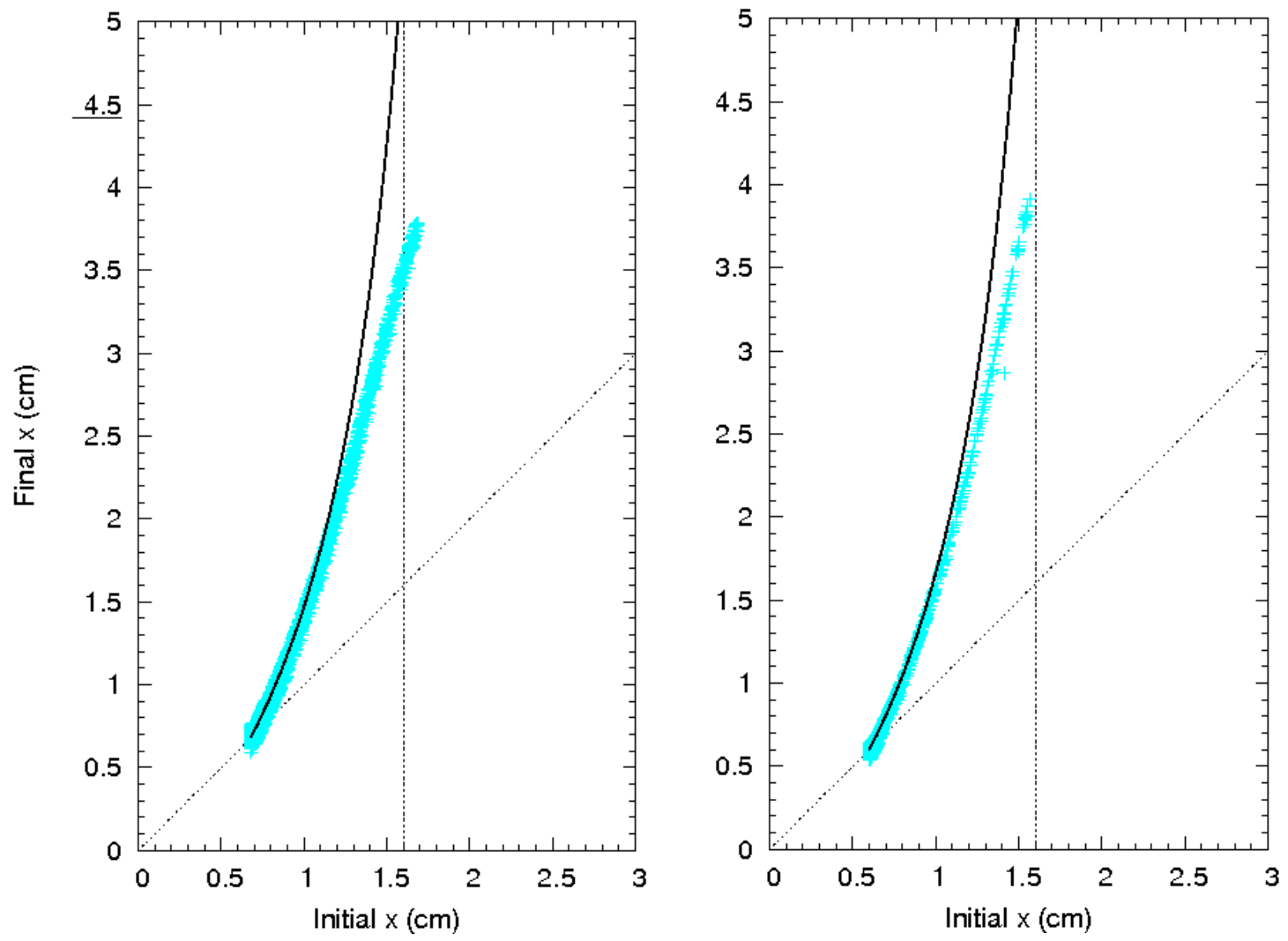

Figure 5.23. Comparisons of step size calculations between theory and simulations.

Figure 5.24 shows two representative comparisons of inefficiency calculations between theory and simulations. The simulations were set up as before, by populating protons near the edge of the separatrix to reduce statistical effects. In these two plots, the orientation was set to $\varphi_0 = -140^{\circ}$; the initial invariant emittance was $\varepsilon_b = 10\pi$ and $20\pi$ mm-mrad on the left and right, respectively. The agreement – better than expected, in fact – provides confidence that low order theory can be used for later estimates of hardware requirements.

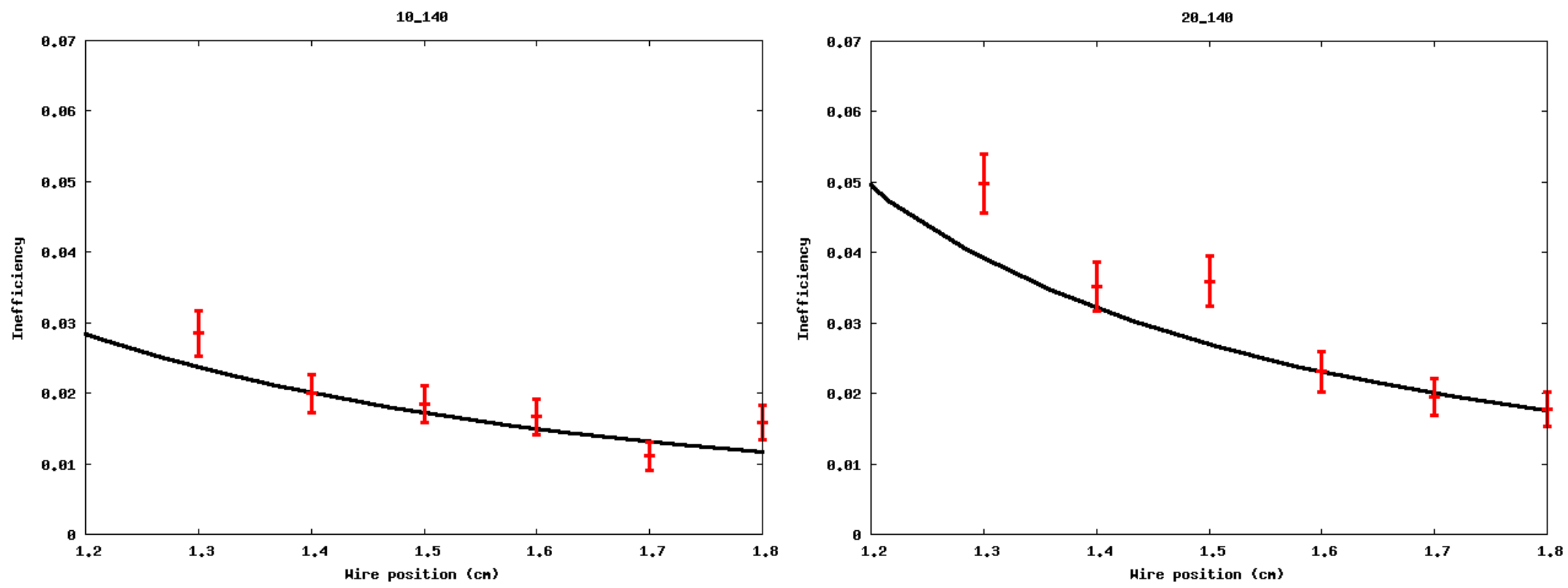

Figure 5.24. Comparison of step size calculations between theory and simulations.

As can be seen above, considerations of step size and inefficiency work against each other. This point is reinforced by the graphic shown in Figure 5.25 below. The red and blue curves indicate initial and final positions of an orbit on the separatrix as a function of the dimensionless parameter *r* appearing in Equation 5-11. The step size, $\Delta x$ is then the difference between the two. (Parameters for this calculation were $\Delta\nu = 0.02$ and $\varepsilon = 20$ mm-mrad, and "inefficiency" has been simplified to the frequently used expression $w/\Delta x$.) Designing the septum will require balancing step size and inefficiency against the other.

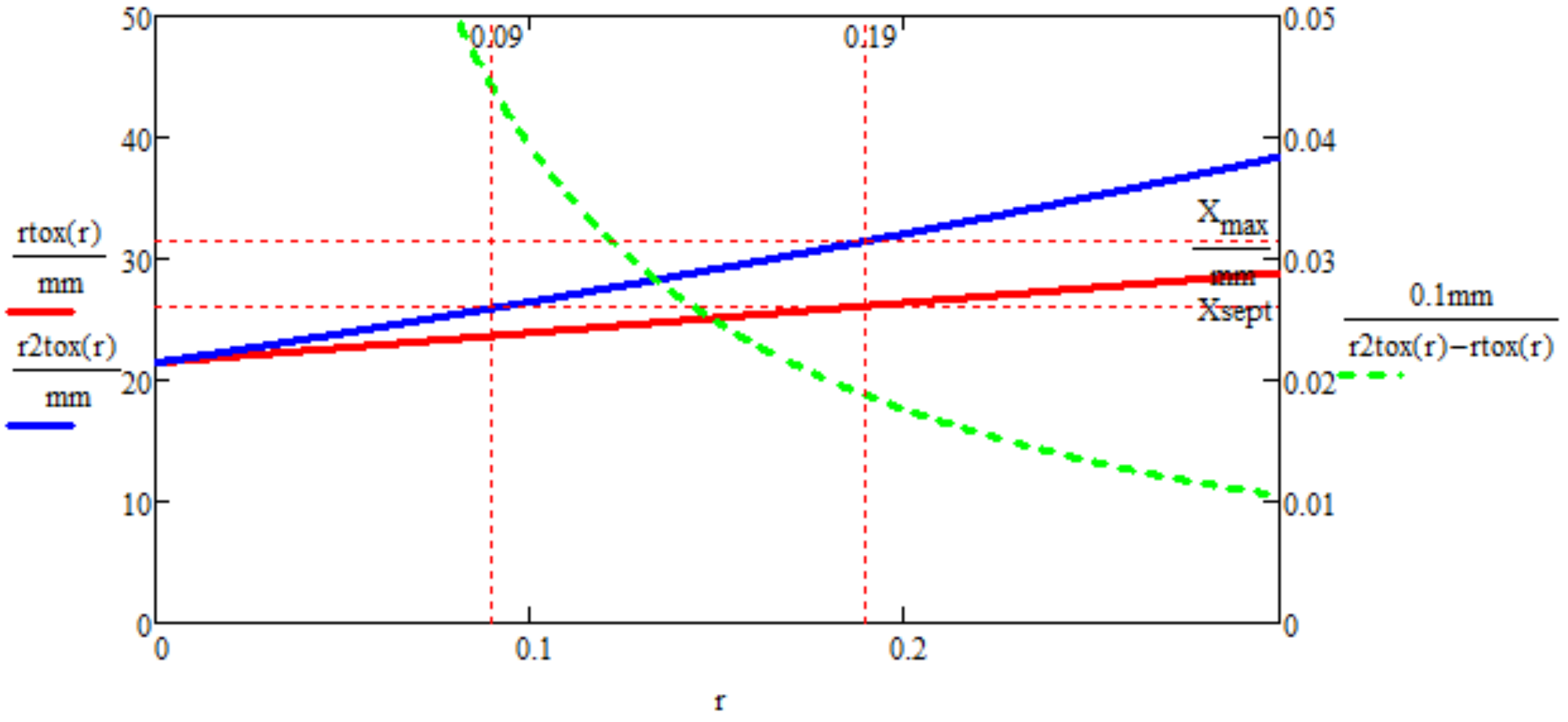

Figure 5.25. Alternative view of stepsize and inefficiency: the horizontal axis, r, is the parameter appearing in Equation5-11. Red and blue curves – scaled on left axis - indicate initial and final values of |x|, the horizontal projection of an orbit on probability of hitting the septum wire.

### *RF knockout (RFKO)*

Analytic calculations like those of the previous subsection assume a model in which extraction proceeds by squeezing the separatrix using trim quadrupoles, and possibly rotating it using the two sextupole circuits. However, limiting our extraction strategy in this way would encounter control difficulties, especially towards the end of the spill when the effects of residual tune spread and power supply ripple would be felt strongly. We propose to add another element to extraction which has already been successful in other applications.

One way of representing the tune spread issue is a "Steinbach diagram," as seen in Figure 5.26(a), showing the distribution of particle tunes versus horizontal action at the onset of the resonance. The starting machine tune is 9.650. Red lines show the 2/3 resonance unstable (i.e. extraction) area boundaries due to the sextupole field. Figure 5.26(b) shows this distribution after machine tune been ramped to the exact resonance. A substantial part of the beam remains, far from the resonance. After the tune ramp stopped at this point, extraction continues and the tune spread shrinks, which helps the extraction rate, but this rate is still very low. Extraction can be assisted with continuing the tune ramp and exercising multiple resonance crossing, but it is hard to control the spill rate uniformity in this case.

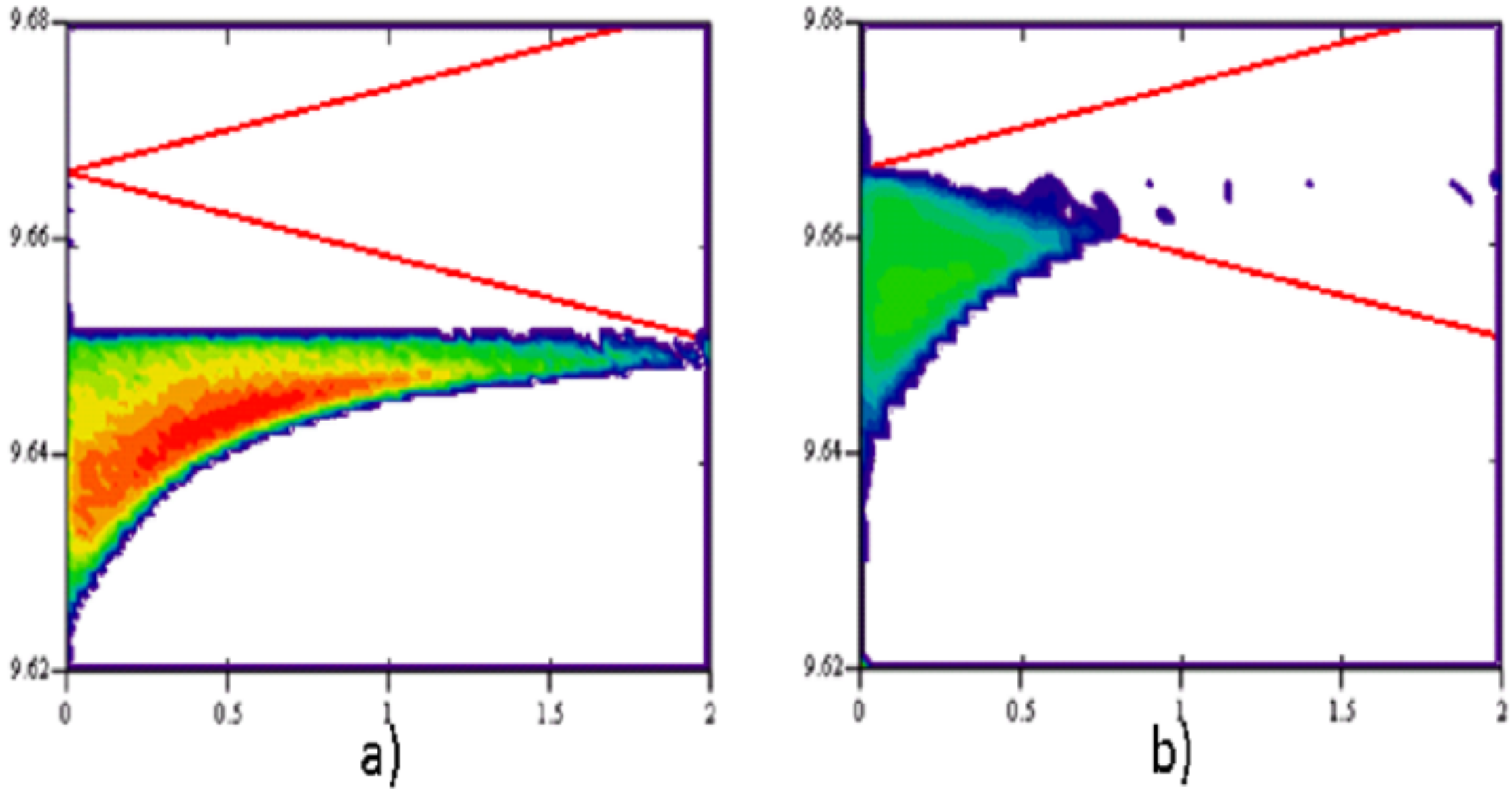

Figure 5.26. Tune distributions vs. horizontal action at: a) the onset of the resonance and, b) at exact resonance. Red lines show the 2/3 resonance unstable (i.e. extraction) area boundaries due to the sextupole field.

A technique known as RF knockout (RFKO is a way to assist extraction in this situation. If one succeeds to heat the beam transversely fast enough, the tune distribution would move to the right and up. This would make it closer to the extraction area on one hand, and with a proper mixing, would also reduce the space charge tune spread. This technique already has been used for slow extraction purposes in medical applications [31], although the primary goal of that was to turn beam extraction on and off. We intend to use RFKO as a feed-back tool for fine control of the spill rate. RFKO allows us to continue extraction with presence of the strong space charge while keeping the machine tune close to the resonance. In this case particles are extracted on the resonance, therefore the step size is maximized. Sufficient transverse electric field may be provided by a regular recycled Tevatron damper, in which case the cost of including RFKO would not be excessive.

To heat particles with a distribution of betatron frequencies, the transverse RF force must have a frequency modulation bandwidth covering the beam tune spread around a single betatron sideband. Two modes of modulation were examined: “colored noise” - i.e. a random signal within a given bandwidth - and frequency sweeping with phase randomization between sweeps. Colored noise would appear to be the better approach, however we did not observe a significant improvement in performance over normal sweeping with phase randomization. Thus, the latter was used in our simulations. Frequency spectra associated with these two approaches are shown in Figure 5.27.

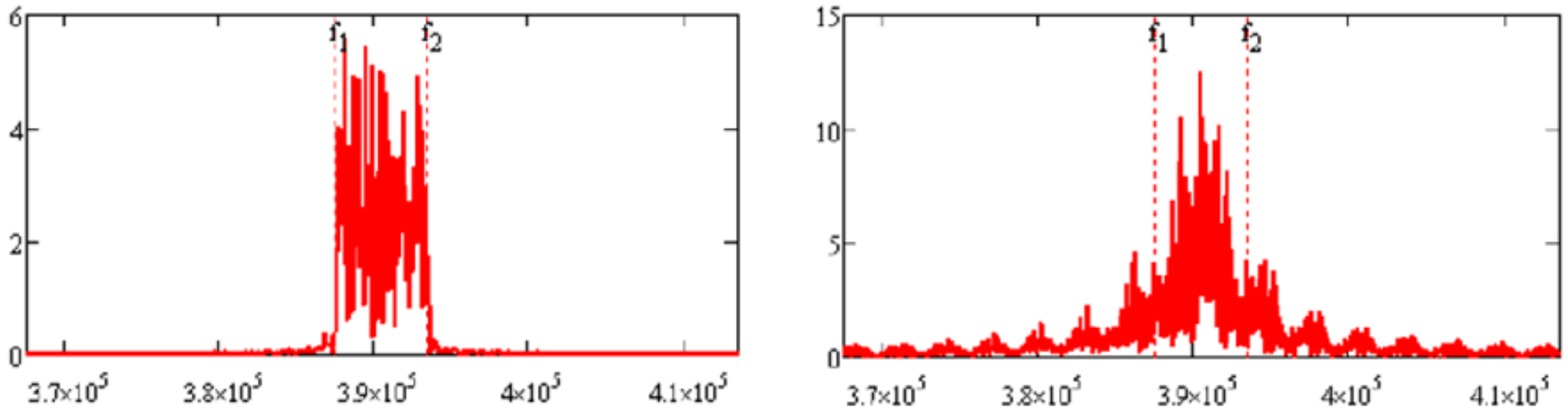

Figure 5.27. Spectra from two methods of modulating RFKO frequency: colored noise (left) and frequency sweeping (right).

***Simulations, including space charge***

Simulations of third-integer extraction have been done which demonstrate a possibility of successfully extracting beam from the Delivery Ring when "RF knockout" (RFKO) is included as part of the procedure. The sextupole field was formed by 2 orthogonal groups of 3 sextupoles, located in two straight sections; an extraction septum and Lambertson magnet were placed in the third. (See Figure 5.16). The septum wire width was assumed to be 100 μm. A circuit for ramping the horizontal tune comprised three trim quadrupoles inserted into the middle of each straight section.

In the Delivery Ring during extraction, space charge forces of the high intensity beam – nominally $10^{12}$ protons (1 Tp) per bunch – will induce betatron tune spread and eventually affect the resonant extraction processes, such as spill rate and beam losses. Separate simulations were carried out for $\nu_x \approx 29/3$ using the ORBIT and SYNERGIA software packages to include space charge effects. ORBIT uses a so-called "2.5D-mode," where the particle density in longitudinal bins is calculated according to the actual longitudinal distribution, and the transverse distribution is assumed to be the same along the bunch. SYNERGIA can (now) operate in a similar mode but also has the capability of performing fully 3D calculations of space charge forces.

An RF cavity is placed in one of the model's straight section, and operated at $h = 4$, $f = 2.36$ MHz, $V = 32$ kV[q]. With 40 ns RMS bunch length and 20 π mm-mrad normalized initial emittance, we have $\delta p/p = 0.002$.

[q] Note: This is not the present design 2.4 MHz RF voltage.

We originally tried approaching the resonance from above in order to expose particles in the core first. However, this method increased losses on the septum, because in this case particles are extracted away from exact resonance and the step size is therefore reduced. Typically in this case septum losses start with about 5% in the beginning of the ramp and reduce to 1-2% by the end of the ramp. Approaching the resonance from below is more promising in terms of losses, although in this case it is more difficult to control extraction of the core particles in a uniform way. The fact that space charge reduces the protons' tunes - especially in the core of the beam - means approaching the resonance from below initially places the core of the beam away from the resonance, assuring that particles will be depopulated in the order expected. Simulations exploring many initial conditions confirmed that starting with $\nu_x < \nu_{res}$ is preferable.

Figure 5.28 shows an extreme example of phase space distributions without (a) and with (b) space charge effects. These simulated data were obtained using the now obsolete high intensity scenario of the Mu2e proposal, with $3\times10^{12}$ protons per bunch. Each color in (a) represents proton states at a different point during extraction. As the central stable region is squeezed, protons leak out along the the separatrix's outgoing branches. Notice that - because both dispersion and chromaticity are zero - at each instant the separatrix is well defined, with cleanly dilineated branches, the separatrices are concentric, and their union covers the region of phase space traversed by the outgoing beam. In contrast, the distribution in (b) is plotted *at only one instant* during extraction. The "fuzziness" of the outgoing proton orbits is attributed to a "fuzziness" induced in the separatrix by the tune spread induced by space charge. In the presence of space charge extraction effectively occurs along different separatrices simultaneously.

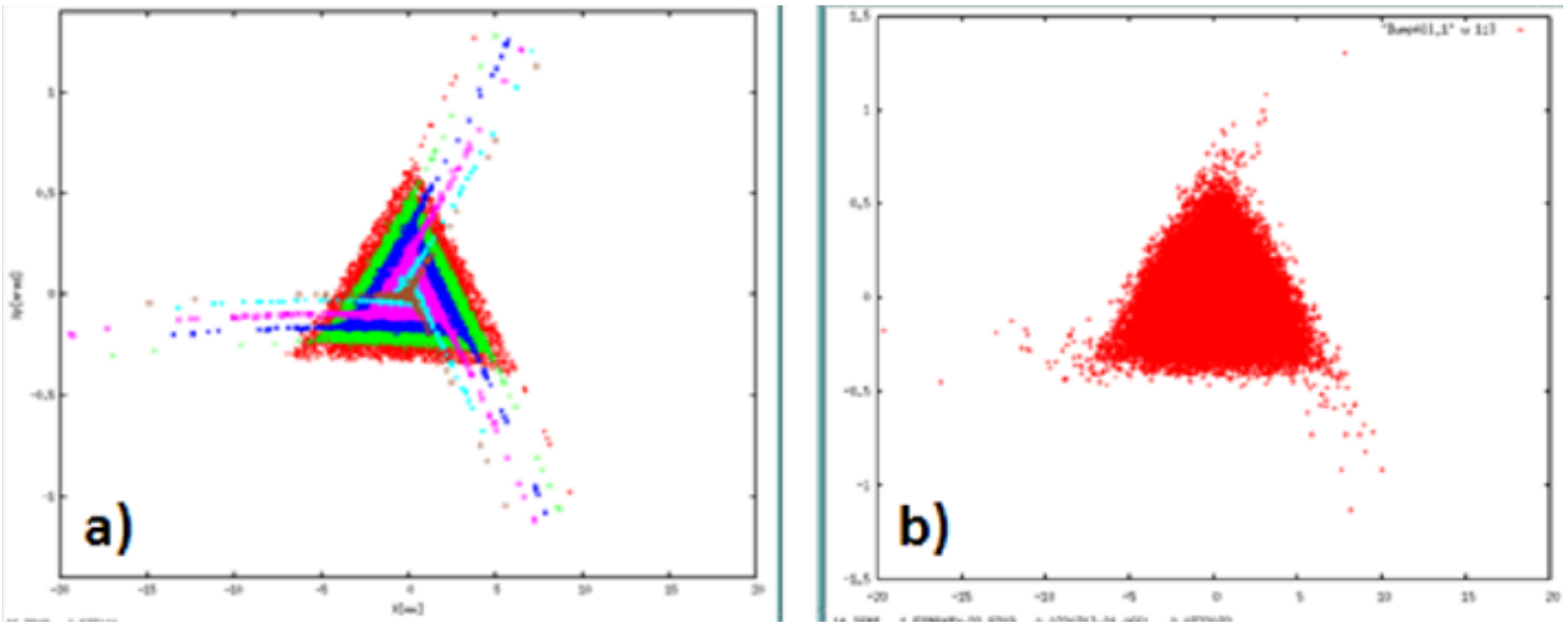

Figure 5.28. Normalized phase space of the beam at for: a) without space charge at different times during extraction; b) with space charge at the beginning of extraction. (These calculations were done with an increased initial intensity of 3 Teraprotons per bunch.)

Losses at the extraction point are an essential concern. From Table 5.3, the total beam power will be 8 kW, so losses at the level of 2% would produce a localized release of 160 W in the beam enclosure. Septum losses are estimated using the fraction of particles that hit the septum wires. One of the advantages of the third integer resonance is that step size grows rapidly with the betatron amplitude and can be made large far from the separatix. The septum's position and the sextupole field should be chosen to maximize step size within the limits of machine acceptance. In general, two ramps are available to control the spill: the sextupole field ramp and the tune ramp though, the sextupole field could be kept constant at an optimum setting. The spill has to start soon after injection. In the simulation presented below, the sextupole field ramps rapidly from zero to the nominal value and then stays constant.

A simulation including a 3D space charge solver, which is necessarily slow, examined the early part of extraction. Sextupoles were linearly ramped for the first 100 turns, then kept constant at their final settings. The separatrix was squeezed by ramping trim quadrupoles at a constant rate. From Table 5.2, 5 msec is reserved for non-extraction activities – this suggests that the spill rate is 31 Mp per turn, and it corresponds to about 32,260 turns for one spill with 1 Tp. The simulation was performed for the first 1000 turns after the sextupole-field rampings. 500,000 macro-particles were injected simulating 1 Tp. Lattice functions ( $\beta_x$, $\beta_y$ ) = ( 13.96, 5.34 ) m and ( $\alpha_x$, $\alpha_y$ ) = ( -1.77, 0.58 ) at the septum, and ( $\beta_x$, $\beta_y$ ) = ( 8.66, 10.00 ) m and ( $\alpha_x$, $\alpha_y$ ) = ( -1.27, 1.51 ) at the Lambertson magnet. In the simulation, the septum wire is -16 mm from the centerand has thickness 100 μm.

Figure 5.29 shows phase space plots at septum (left) and Lambertson (right), with (bottom) and without (top) space charge effects. While the circulating beam is enclosed by the separatrix lines, particles are streamed along branches by squeezing the separatrix, kicked at the septum, and then extracted at the Lambertson. After 1000 turns, $\approx 8.3\times10^9$ protons were extracted without space charge effects, and $\approx 4.0\times10^9$ with space charge effects. These differences are observed since fewer particles are populated near vertices of the separatrix with space charge effects. The observed loss rates for protons hitting the septum wire were 0.015 and 0.021 without and with space charge effects, respectively. In a simulation with the RF was turned off and δp/p was artificially set to zero, particle losses occurred only at the septum wire. When RF was turned on, i.e., with non-zero $\delta p/p \approx 0.003$, the observed loss rate for the first 1000 turns of particles hitting the beam pipe somewhere around the ring was 0.002.

The fractional number of extracted particles is plotted in Figure 5.30 as a function of turn number. There will be a tendency for an initial brief burst of protons at a higher rate than desired. This must be controlled via measurement and feed-forward of the tune control circuit's ramp. The spill rate with space charge effects is also less than that without space charge effects. This could be controlled by increasing the initial quadrupole ramp rate combined with using RFKO to move protons from the core closer to the separatrix.

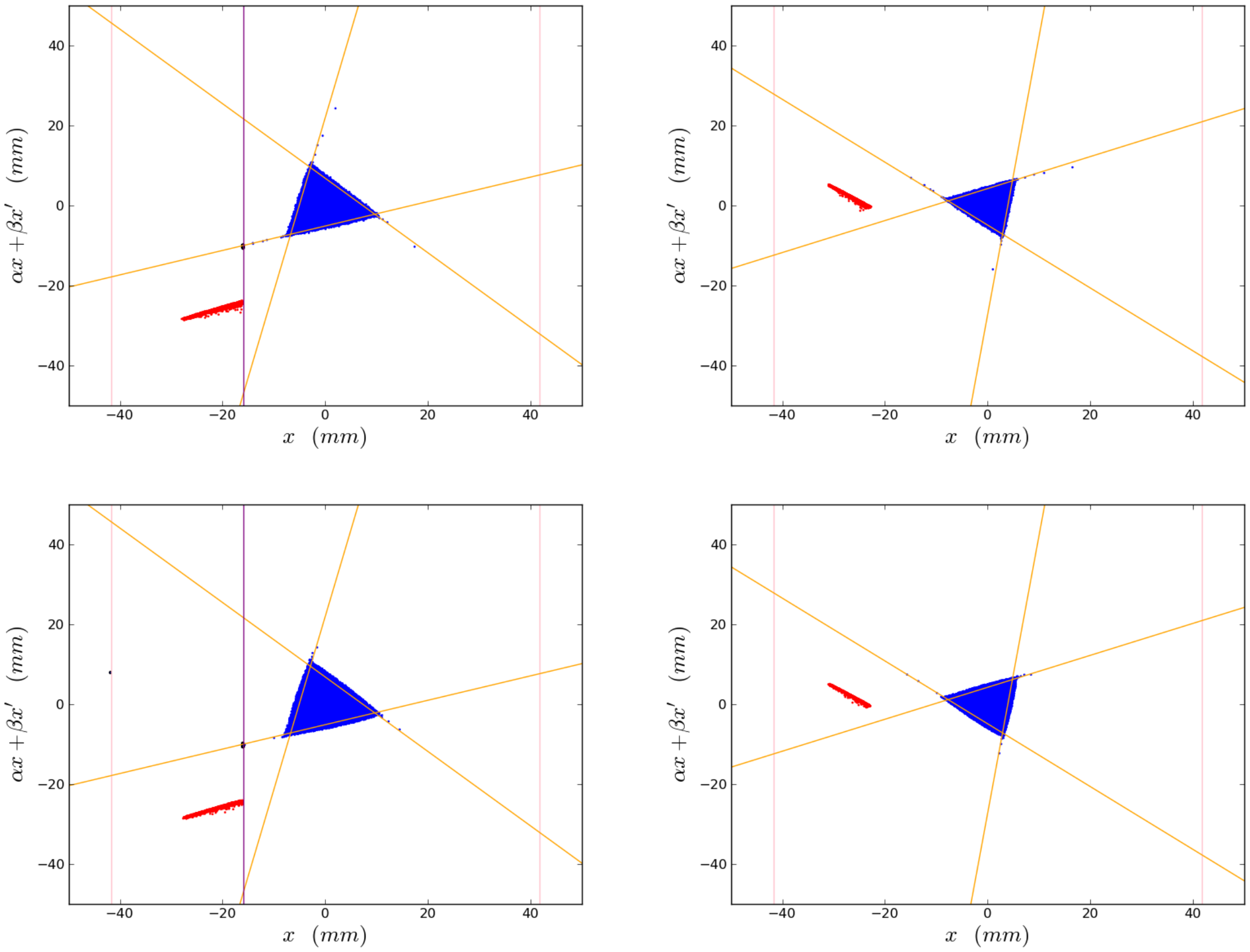

Figure 5.29. Phase space of the third order resonance extraction: circulating beam is shown as blue; kicked beam, red. The two figures on the left show the distribution at the septum; the septum's wire, shown as a vertical line, was placed at -16 mm. The distribution at the Lambertson is shown on the right. Space charge forces have been turned off in the two top figures.

The naïve strategy of ramping the tune at a uniform rate is obviously inadequate, as exemplified by the early spill rate of ≈ 8.3 Mp/turn in the previous exercise, about 20% of what is desired. One way of addressing that is to tailor the tune profile to accommodate the lower density of particles near the separatrix (for $\Delta\nu < 0$) during the early stage of extraction. Another is to use an RFKO device to "tickle" protons in the core to larger amplitudes, moving them toward the separatrix at an earlier stage. In another simulation, using 2.5D ORBIT, RFKO power was used as a feedback knob to control the spill uniformity. Although sophisticated and intelligent techniques exist to manage feedbacks, only a simple filter was used in simulations. The power setting was updated once every 100 turns. If the extraction rate is higher than a nominal target at the moment of update, power is reduced by a constant down-factor. Power was multiplied by an up-factor when the rate was lower than nominal and not growing. Normally it takes about 1 msec for the feedback to take effect, therefore the up-factor is chosen to be above and close to 1.0. When the rate started to grow, it grew quickly, therefore the down-factor was made small.

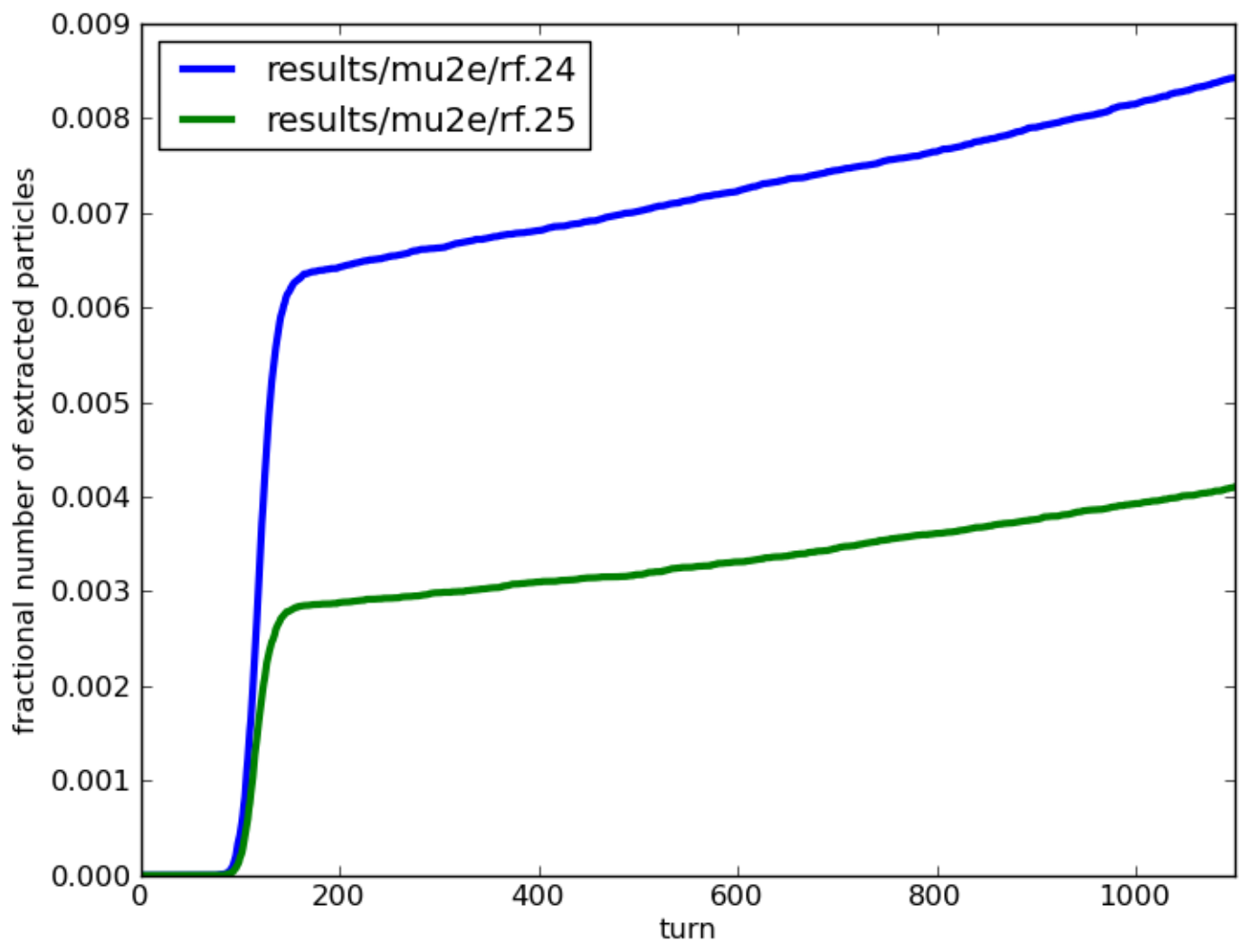

Figure 5.30. The fraction of extracted particles, with (green) and without (blue) space charge forces.

The result of these simulations is shown in Figure 5.31. RFKO alone is not strong enough to control the spill rate; too much power would be required. Careful adjustment of the tune ramp is also needed. The green trace in the larger graphic on the right shows the turn-by-turn structure of the spill, with its linear scale shown on the right axis, while the red shows RFKO power on a log scale (left axis). The dashed blue line shows the constant target extraction rate used in the feedback mechanism described above. Maximum RFKO power was limited by available hardware specifications.

As seen, RFKO power grows slowly, drops quickly, and remains under control. The effectiveness of the beam heating with RFKO is limited, so the tune ramp curve must be chosen carefully to facilitate as close to a constant rate of extraction as possible without it. On a macroscopic scale, a linearly decreasing spill was achieved. However, even in the best simulations, turn-by-turn variations in the spill rate were substantial and hard to avoid. An optimistic estimate places the experimental tolerance at ±50%; a more conservative one would require ±20%. Simulations carried out so far, including the one shown here, did not satisfy either criterion. Research has begun on how to control the spill using beam monitoring in the extraction line to regulate the RFKO device.

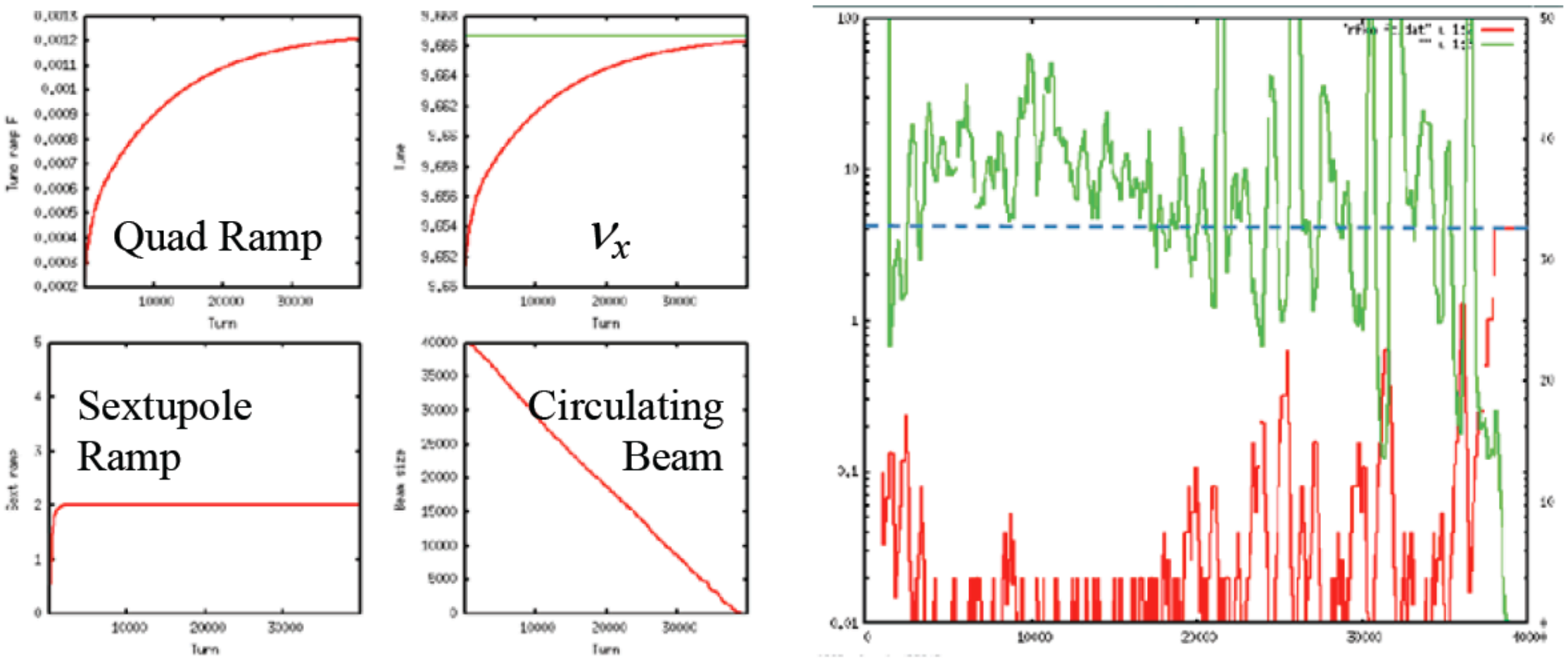

Figure 5.31. Simulations of third-integer extraction, including space charge. On the left are plots of the quadrupole circuit ramp, tune, sextupole ramp, and beam intensity during the spill. On the right, the green trace shows the turn-by-turn structure of the spill, while the red shows RFKO power on a log scale.

### *5.6.3* Estimated hardware attributes

This section estimates the field requirements of quadrupoles and sextupoles to be used for third-integer resonance control. To set the scale for hardware specifications, we have assigned values to parameters so as to maximize field strength requirements.

- $|\Delta\nu| \approx 0.015$ the difference between horizontal tune and the resonant tune at the onset of extraction;
- $\beta\gamma\varepsilon_b \approx 10\pi$ mm-mrad, invariant horizontal emittance at injection into the Delivery Ring;
- $\beta_x \approx 10$ m, horizontal lattice function;
- $B\rho \approx 30$ T-m.

We shall refer to these as the "scaling parameters." The hardware attributes listed below were used for estimating costs of construction. Most will be given to one significant figure, emphasizing their role as first order estimates.

Using a third-integer resonance for slow extraction utilizes three real control parameters. In the Hamiltonian of Equation 5-6 they appear as $\Delta\nu$ and the amplitude and phase of $g$.

- $\Delta\nu$ is the difference between the horizontal tune and the resonant tune: in our case, $\nu_{res} = 29/3$. It is assumed to be controlled by a zeroth harmonic quadrupole (tune control) circuit.

- g is the (complex) strength of the harmonic sextupole circuit, as defined in Equation 5-7.

The emittance of the central stable region in Figure 5.17 depends quadratically on the ratio $|\Delta\nu / g|$, as was written in Equation 5-8. It is impractical to suggest reducing it to zero by increasing the value of $|g|$. $\Delta\nu$ must approach zero in order to squeeze all the particles onto the separatrix for extraction.

***Quadrupoles***

Third-integer resonant extraction will require controlling the horizontal tune: i.e. initially placing it near a resonance and then carefully moving it into the resonance. The first operation will be accomplished using the normal bussed quads already in the Delivery Ring. The second operation can be facilitated by adjusting a small set of trim quads introduced specifically for this purpose. Specifications for those control elements can be estimated from the usual expression for tune change.

$$\Delta\nu = \frac{1}{4\pi}\sum_{\text{trim quads}} \beta_x \frac{B'l}{B\rho} \approx \frac{N}{4\pi}\beta_x \frac{B'l}{B\rho} \Rightarrow N\,|B'l| \approx 4\pi\,|\Delta\nu|\,B\rho / \beta_x$$

An estimate for $|B'l|$ is established by substituting values from the scaling parameters to get,

$$N|B'l| \approx 0.6 \text{ Tesla}.$$

In writing hardware requirements, we added an additional safety factor of approximately 30% to increase this to ≈ 0.8 Tesla. If we symmetrically insert a 33 cm trim quadrupole into each straight section in order to make this tune adjustment, each would have a maximum field gradient of approximately 0.8 T/m.

***Sextupoles***

Equations 5-7, 5-8 and Equation 5-9 provide a scale for the required sextupole strengths:

$$|g| = \frac{|\Delta\nu|}{3\sqrt{2}} \cdot (\varepsilon_b/\pi)^{-1/2} \approx \frac{1}{24\pi\sqrt{2}}\left(\frac{N}{2}\sqrt{2}\right)\beta_x^{3/2}\frac{|B''l|}{B\rho}$$

$$\Rightarrow N\,|B''l| \approx 8\pi\sqrt{2}\,|\Delta\nu|\frac{B\rho}{\beta_x^{3/2}(\varepsilon_b/\pi)^{1/2}}$$

The factor $N/2$ appears because, in order to estimate the maximum required field, we pessimistically assume the separatrix to be orientated such only one of the two sextupole circuits is used to drive the resonance; that is, half of the installed sextupoles are not powered. Using the scaling parameters,

$$|g| \approx 2.4\ m^{-1/2}\ \text{and}\ N|B''l| \approx 690\ T/m.$$

Adding the ≈ 30% safety factor increases this to ≈ 900 T/m. If three equal strength sextupoles are used, so that N = 6, then max $| B''l | \approx$ 150 T/m for each magnet.

***RFKO device***

The RFKO device must produce an oscillating electric field designed to gently increase the horizontal transverse emittance of the proton bunch, moving protons to the edge of the separatrix. Providing finer spill control will require a broadband device so as either to chirp the frequency or to produce a signal with a flat spectrum over a finite bandwidth (or both). The device's principal characteristics are conceptually estimated as follows:

| | |
|---|---|
| central frequency | ≈ 393 kHz ( = 2/3 × 590 kHz ) |
| frequency sweep | ± 6 kHz |
| power | ≈ 1.6 kW |
| maximum field | ≈ 8.6 kV/m |
| length | ≈ 1.4 m |
| field gap | ≈ 6.4 cm |

These values will result in a gentle horizontal kick of ~ 1-2 μrad.

***Septum and Lambertson***

While magnets will create a separatrix and the RFKO oscillator will "tickle" the beam toward its edges, septa and Lambertsons are the devices that will drive extracted protons into M4, the extraction transport line. The septum's electrostatic field should provide a small horizontal kick and the Lambertson's magnetostatic field will then provide a much larger kick. First order estimates of hardware characteristics to accomplish this at 8 GeV kinetic energy are as follows:

| | |
|---|---|
| septum: | length ≈ 1.5 m ( ×2) |
| | voltage ≈ 100 kV |
| | gap ≈ 1.4 cm ( E ≈ 71 kV/cm ) |
| | wire width ≈ 50-100 μm |
| Lambertson: | length ≈ 1.5 m ( ×2 ) |
| | magnetic field ≈ 1 Tesla |
| | thickness between central and field regions ≈ 2 mm. |

A septum wire's width is typically 50 μm, but construction errors and mechanical and electrical forces could double its "effective" width. The electrostatic septum is expected to provide 1.5 - 2 cm separation at the Lambertson, enough to center it in the field region. Recent considerations of geometric constraints in the Delivery Ring have suggested that two septa and Lambertsons must be used in order to provide sufficient integrated field while fitting into spaces available in the Delivery Ring. (It is considered impractical to increase a Lambertson's magnetic field beyond 1 Tesla.)

***Spill monitoring***

A number of methods of measuring the spill for monitoring and feedback purposes have been considered. We expect a resistive wall monitor (RWM) could provide a spill intensity measurement if the noise level on the spill signal is not too large. The pre-extinction monitor, described elsewhere, may also provide a signal for fast spill monitoring and control. Another alternative under consideration would be to use an optical transition radiation (OTR) monitor.

***Resistive wall monitor for RFKO Feedback***

A resistive wall monitor is a simple device that can be used to measure the beam intensity by measuring its image charge [33]. This device does not provide profile information. The resistive wall monitor was chosen as part of the baseline because there is extensive experience with it at Fermilab. Slow extraction requires measurement of the spill at the extraction channel where expected intensities in the order of $2.8\times10^7$ protons per micro-bunch every 1.69 μsec. A prototype Resistive Wall Monitor could provide micropulse-by-micropulse measurements to ~10 %. A prototype was constructed using CMD5005 magnetic core material (μ = 100,000) and a launcher with a ceramic gap. Pulses of amplitude and frequency were measured similar to the expected RFKO 3rd integer extraction. The measured micropulse across the ceramic was amplified 20 dB. Figure 5.32 shows the actual pulse measurements [34].

The wall current monitor was then simulated using SPICE, an electrical circuit simulator. The first step was to generate a train of Gaussian bunches with $2.8\times10^7$ protons per bunch, $\sigma = 40\times10^{-9}$ sec and period of $\frac{1}{f_{rotation}}$ equal to $1.694 \times 10^{-6}$ sec.

The bunch amplitude obtained was similar to the WCM mockup. Figure 5.33 shows 12 bunches at the output of the wall current monitor model. 20 dB gain was applied to the modeled bunches, and the output signal drove a 590 kHz bandpass filter of 150 kHz bandwidth producing an oscillation at the output of the filter. This ringing will continue for 34.852 k turns in the case of 58.9 msec spill. The output of the 590 kHz filter was further amplified by 32 dB producing an output in the order of – 4.0 dBm (400 mV peak-to-peak) as seen in Figure 5.34. This signal will then be digitized, down-converted to baseband, and filtered with a moving average filter, and used in the feedback process in the RFKO system.

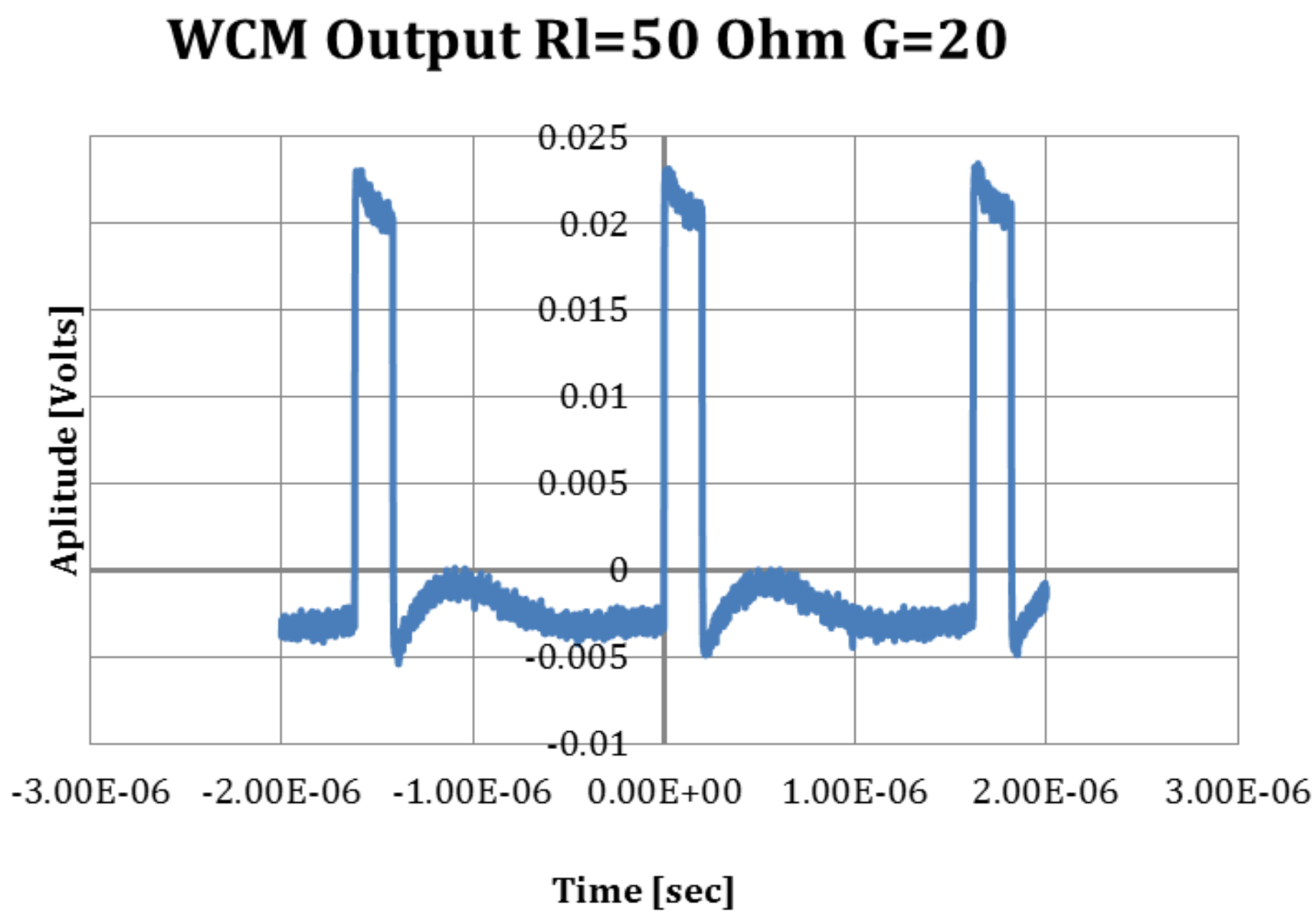

Figure 5.32. Pulses measured across ceramic gap of the prototype resistive wall monitor.

This excellent performance does not take into account consideration of noise, the actual level of which is difficult to predict. It is possible that it could mandate a reduction of the filter bandwidth or even seek an alternate solution. However, at this point, the results of prototyping and modeling are very encouraging.

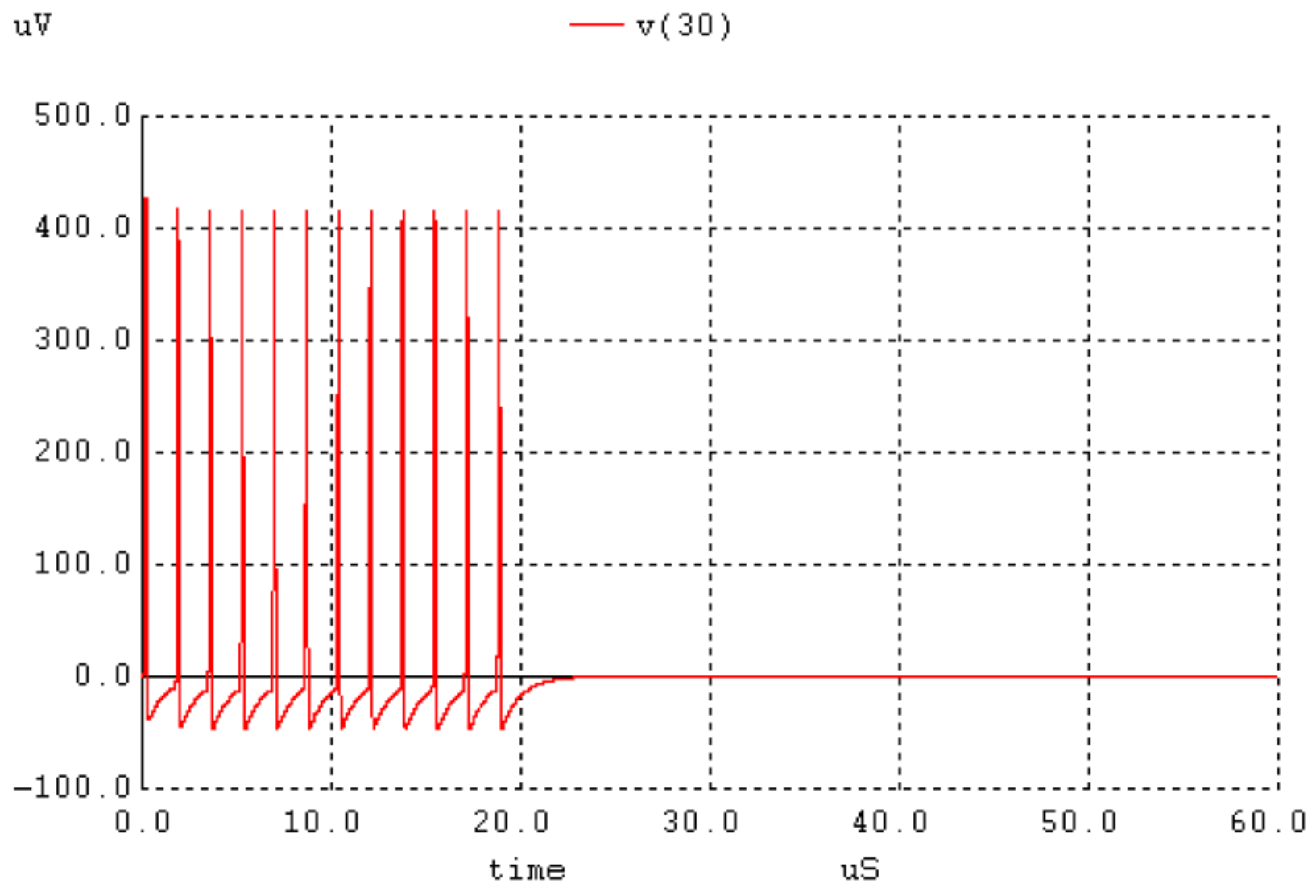

Figure 5.33. A train of 12 Gaussian pulses at the output of the resistive wall monitor model.

### *5.6.4* Machine studies

Calculations and simulations cannot take the place of operational studies using hardware, software and diagnostics as *similar as possible* to those to be employed by the Mu2e experiment. During Run II, collider-mode configuration of the Fermilab accelerator complex was not suitable for such studies. Nonetheless, we did have the opportunity to do two preliminary studies aimed at studying the Delivery Ring and testing the Mu2e extraction concept: (a) a study of half-integer extraction has begun using the Main Injector, and (b) tune scans, systematically searching for indigenous parasitic resonances in the Delivery Ring. Both of these are incomplete. The former has yet to produce useful data; preliminary measurements from the latter are summarized below.

After the end of Run-II operations on September 30, 2011 the former Fermilab Antiproton Source became an idle machine. The ability to stack antiprotons was disabled shortly after this with termination of the CRYO facility operations. Nevertheless there is still an ability to run 8 GeV protons into both rings of the Antiproton Source from the Main Injector, and these beams can be used for the parasitic studies. We are taking this opportunity to carry out studies, some of which address certain design goals of Mu2e slow extraction.

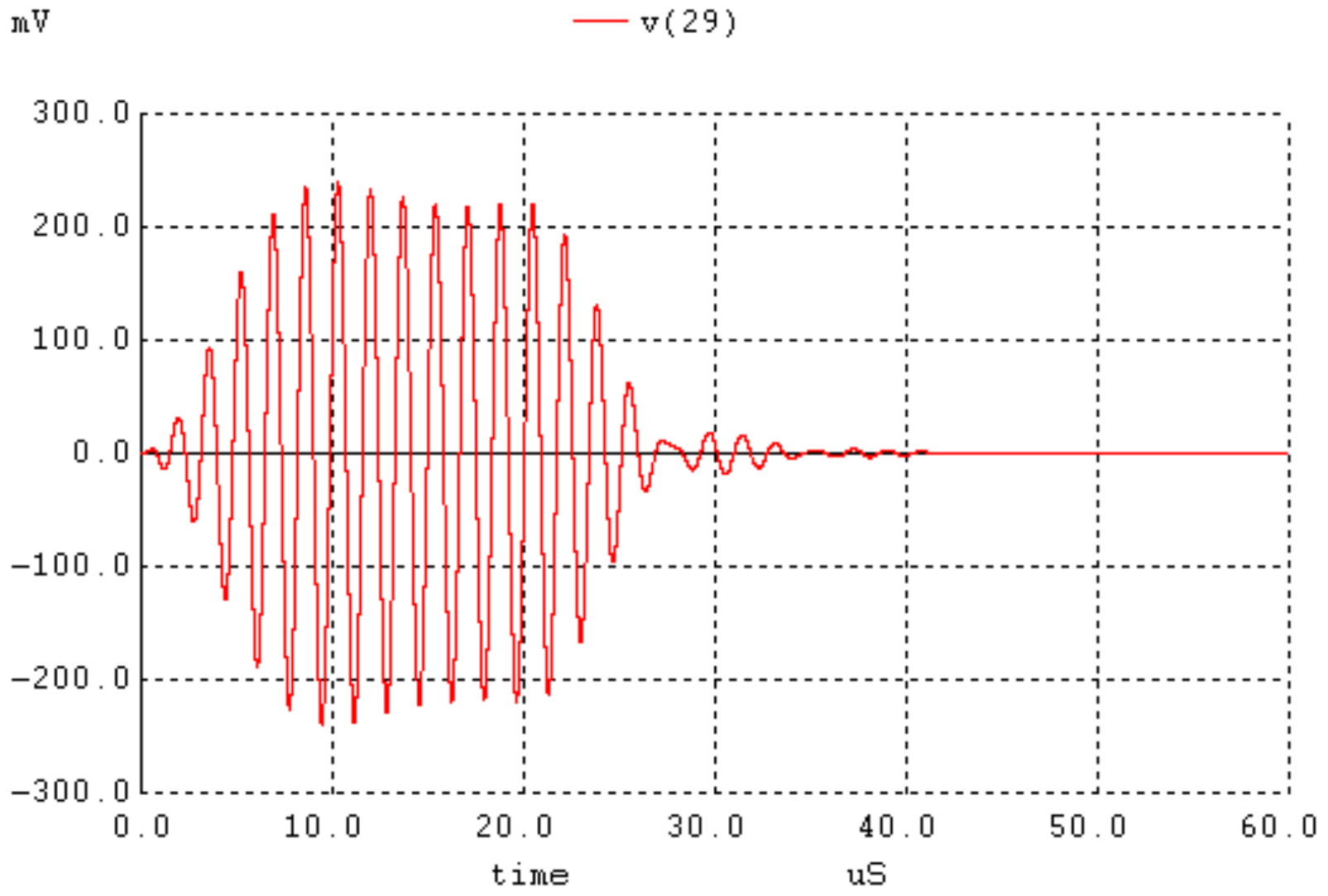

Figure 5.34. The filter response to the pinging of the filter by the extracted bunches. Gain of 52 dB was applied from extraction to output. This signal can then be digitized and down-converted for further processing by the RFKO control system.

***Tune scans***

Several tune scans were carried out in July of 2010, one of which is shown in Figure 5.35. Figure 5.35 displays a scan in the vicinity of $\nu_x = 29/3$. Each point represents the intensity of stable circulating beam – after approximately 2.2 seconds of machine time into the stacking cycle - at a fixed tune. The error in tune is approximately 0.002, both horizontally and vertically. (A tune measurement itself takes about ten minutes.) The pulse to pulse variation is normal and not part of the study. The significant features are the dips which occur near resonance lines: $3\nu_x = 2$, $\nu_x + 2\nu_y = 2$ and $2\nu_x + \nu_y = 2$ (mod 3) in Figure 5.35.

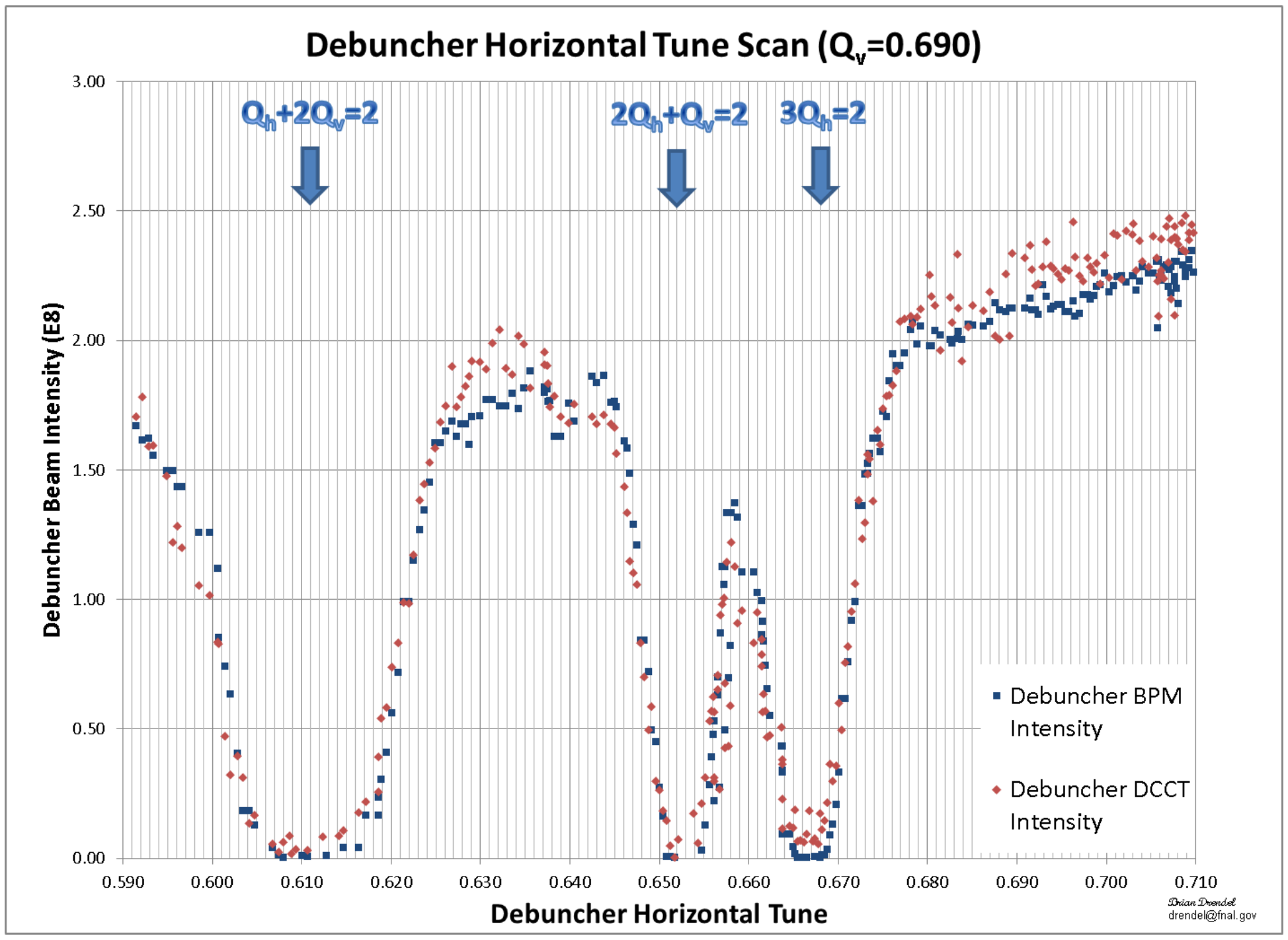

Figure 5.35. Tune scan in the Delivery Ring covering the 29/3 resonance. The vertical scale indicates the Delivery Ring circulating beam intensity; the horizontal axis is the fractional part of the horizontal tune.

These resonances are the result of pre-existing sextupole excitation in the Delivery Ring. The chromaticity was small during these studies, which means that chromaticity correcting sextupole circuits were contributing to these resonances. This study can be very complicated; our goal is to identify large problems as early as possible with a reasonable effort. The purpose of the scan is to move the beam along those paths in the tune diagram that are more likely to be real operational ramp paths and look for beam losses at particular points. Because the losses are beam size dependent, we combine intensity measurements with beam profile measurements using the IPM. We should be looking at resonances of 5$^{th}$ order and higher, as we certainly want to avoid anything below that.

These first studies started during Tevatron operation and were carried out during stacking cycles using antiproton beam. Scans took long because tune measurements were made using the stochastic cooling pick-ups. We now have improved the diagnostics and can use substantially higher beam intensity.

***Proton beam issues***

Although proton beam was made available in the Tevatron era for beam studies in the Delivery Ring, these studies were not frequent and available time was very limited by the collider operation schedule. In general, limited beam quality was never an issue. Now it becomes important: tune scans assume lengthy measurement with constant beam characteristics, and RF knock-out needs fine control on beam parameters. Ideally maintaining all three emittances close to those of the Booster would suffice; however there are certain challenges in doing so with the lack of diagnostics that storage rings normally have. The biggest problem was the presence of very large high-Q impedance of the RF rotator cavities in the ring; therefore the effort was made to remove those cavities. Another issue is a short beam lifetime of approximately 8 hours due to low vacuum, but we should be able to live with this through our current round of studies.

***Delivery Ring Preparation***

During anti-proton operation the level of instrumentation support was deliberately reduced in order to maximize the machines' acceptance. In particular, three Schottky detectors that we now need for the fast tune measurements had been removed in 2006. At the end of October, 2011, general maintenance was carried out during a one-week shutdown in the Pbar Source, and all Schottky detectors were reinstalled. At this time an old style Tevatron damper kicker was installed at the D3Q10 location as an RFKO element. Old Micro-Channel Plates (MCP) were replaced in the Ionization Profile Monitor (IPM), which also required opening the vacuum. (The decision to remove the rotator cavities from the Delivery Ring came later, by the end of January 2012, and required another one-week shutdown.)

***RFKO studies***

Although we do not yet have an opportunity to perform direct tests of resonant extraction, we can study the efficiency of beam heating with RF Knock-Out excitation. This will provide an important benchmark of our calculations and simulations and increase (or decrease) our confidence in the slow extraction concepts that are presented in this document.

The waveform is created by an AFG-3252 generator that has the capability to program arbitrary sequences, including the "colored" noise, in a wide frequency range. The high level is made with two 0.5 kW Amplifier Research solid state power amplifiers. Power to the RFKO kicker is gated by an RF switch that can be triggered by two T-Clock timers to synchronize it to arbitrary accelerator clocks. The beam excitation is made at one of the lowest sidebands. The frequency modulation depth, frequency and the modulation type are varied during the studies as well as beam parameters such as tunes and tune widths (using chromaticity). Beam width growth is recorded using ion profile monitors (IPMs) that sample profiles at the rate of 10 Hz for up to 5 seconds. IPM start time is synchronized with RF power using the same T-Clock events. These studies have just begun.

## 5.7 External Beamline

### *5.7.1* Considerations

The Mu2e beamline must cleanly separate and transport resonantly extracted beam to the Mu2e production target while minimizing the transport of out-of-time particles. This is accomplished through a number of specialized optical insertions and a series of collimators. To contain the overall length requires customizing insertions to fit seamlessly within the civil constraints of the local geography (see Figure 5.36). Another criterion has been added to what is now termed, the muon complex: a g-2 derivative line that utilizes the AP30 extraction magnetic components, mainly septa and Lambertsons, through an achromatic vertical bend section which separates and delivers beam into the external beamline. Physical separation of the two lines can only occur efficiently by utilizing the string of strong left horizontal bends that determines the direction of the line in the civil plan.

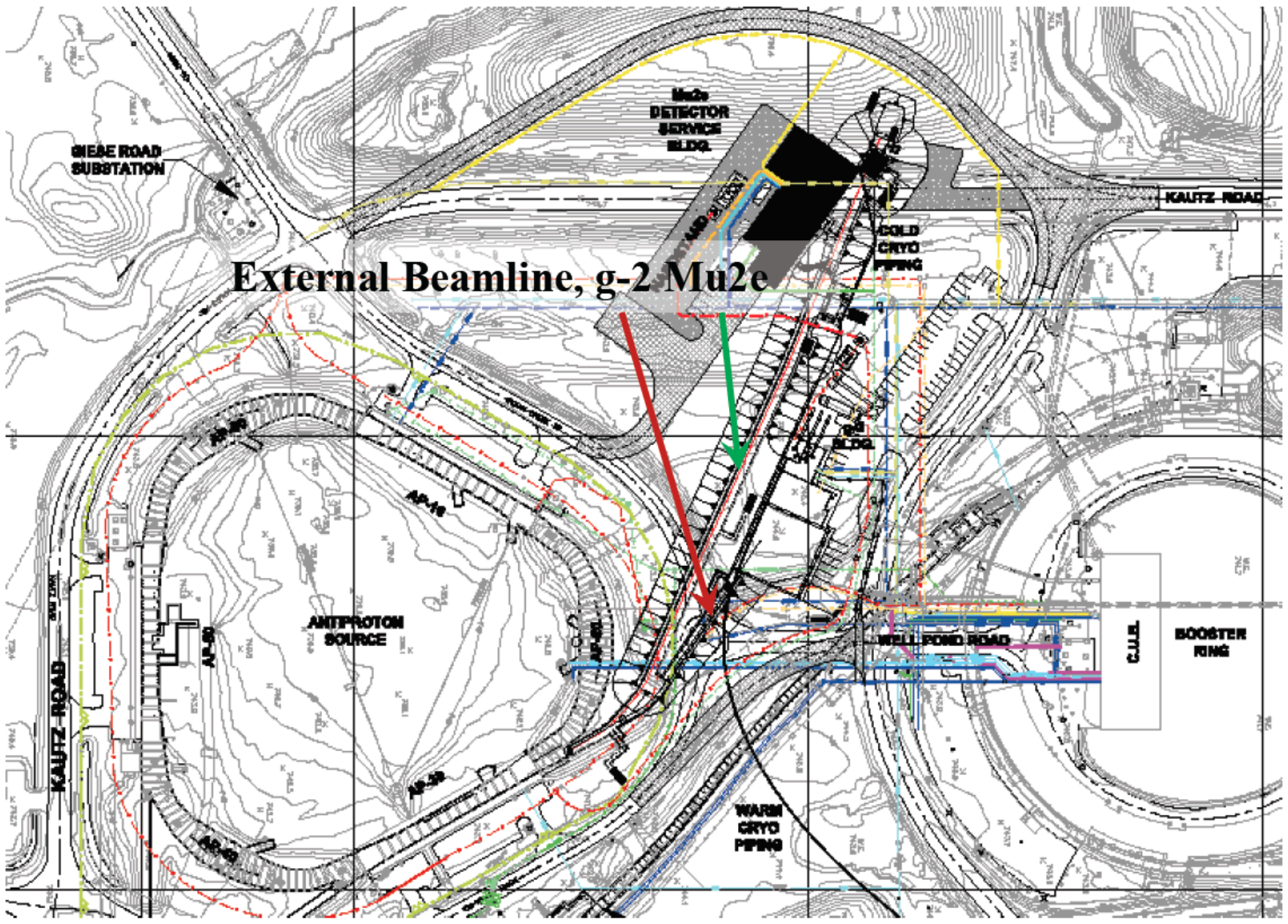

Figure 5.36. Layout of the Antiproton Source showing the Mu2e external beamline and experimental hall.

To appreciate the complexity of this beamline, the criteria that must be addressed are listed below. The first two sections are common to both Mu2e and g-2 and the third is designed to perform the separation between the two in addition to other required functions (momentum collimation).

***Civil Layout***

- Horizontal resonant extraction from the AP30 straight of the Delivery Ring.
- Extraction is followed by a 4′ vertical elevation change (up) to take advantage of existing civil construction (left over from a previous beamline).
- A 36° horizontal bend string is required to fix the direction of the beam line from the AP30 straight towards the optimal geographic location for the experiment. This section is also used to cleanly derive a separate beamline for g-2 by changing the bend strengths.
- A reduced vertical decline of 2′ in the beam line elevation to the elevation of the experimental target.
- A diagnostic beam absorber for commissioning, tuning, and beam studies to prevent spraying beam and unnecessary activation of the experiment.
- A 2.4° downward slope at the entrance to the solenoid to hit the target correctly.
- A shield wall to physically separate the beam line enclosure and restrict access from the experimental hall. The shield wall encloses a diagnostic/abort absorber.

***Beam Properties***

- Momentum collimation is needed to restrict the momentum spread of the extracted beam to ±1%. (Although there will be momentum collimation in the Delivery Ring, septum wires will also produce beam halo, for example.)
- Extinction Section, consists of pairs of 90° collimators (the second collimator intercepts forward angle scattered beam from the first – angle turns into a position offset):
  - Horizontal series of collimators, minimum: one 90° pair
  - Vertical collimation, one 90° pair.

***Beam Optics***

- Vertical achromat is required to suppress dispersion from the AP30 extraction to the upstream, beam line elevation.
- A horizontal achromat is required for momentum collimation and the 36° horizontal directional bend string. This bend string has been formatted into two separate tunable achromats to switch beam to the g-2 beamline by changing to 24° of total bend.
- A second horizontal achromat is required to bump the beamline around the diagnostic/abort beam absorber and restore the correct targeting trajectory to the experiment.

- A third vertical achromat controls the declination slope into the solenoid for proper targeting trajectory.
- A 250 m high-beta function is required to make the technical specification of the AC dipole feasible.
- A periodic, repetitive section with 90° of phase advance between alternating horizontal and vertical peak beam sizes is required for locating collimators.

Civil and geographical constraints (avoidance of wetlands, for example) dictate a 30° - 40° bend after extraction from AP30 to satisfy all of the constraints on the beamline length and the location of experimental hall. Only ~200 m are available for a beam line. Accommodating the many required insertions and beam manipulations is difficult and requires combining multiple functions in designing the insertions. Therefore, custom insertions and beam line sections must be designed with multi-function purpose. Further, robust optics design, particularly in this stringent environment, requires modular design with some independent variability ‘built-in”. For example, steering the beam correctly onto the target and through the strong fields of the Mu2e Production Solenoid requires achromatic variability of the beam in both position and angle. This inherently implies two vertical achromats: one for position and one for angle and the two cannot be decoupled since they share the same dipoles and optics. In this case, a special double-layered achromat was designed and is one example of the custom optics required to make this line work within the civil and experimental constraints.

### *5.7.2* Beam Line Specialty Sections

As stated above, the beam line is best described in terms of its modular functionality. Correspondingly, the following descriptions detail the important subsections, and discuss the rationale and justify the design approach for each subsection.

#### *Extraction from the Delivery Ring*

Addition of g-2 extraction into the AP30 straight g-2 has been carefully designed to not impact the present optimized extraction optics from the Delivery Ring for Mu2e. Presently g-2 extraction inserts and/or replaces elements in the Delivery Ring upstream of QS302 (the last quadrupole presently impacted is QS303). In principle, this leaves free the section from QS302 – QS207. Mu2e extraction begins with septa positioned upstream and downstream of QS203, which is only approximately one and a half FODO cells downstream from the last element in g-2 extraction.

Extraction from the Delivery Ring is accomplished via resonant horizontal extraction across a septum. A 1.5 m septum only delivers ~1 mrad of kick, so at least two septa are required. Positioning the septa about a horizontally-focusing quadrupole (QS203) - upstream of a horizontally defocusing quadrupole - exploits not only the maximum $\beta_x$ but the downstream defocusing quadrupole enhances the septum kick and thus maximizes the beam separation at the entrance to a Lambertson. The Lambertson must also be split and if positioned upstream and downstream of the next focusing quadrupole (QS205) in the Delivery Ring, then this quadrupole acts like a combined-function magnet and adds to the Lambertson kick. The combined effect of all three vertical bends allows beam to clear the next magnet – the last horizontally defocusing quadrupole in the AP30 straight (QS206).

As an alternative to splitting the extraction septum and Lambertson into two modules each, we are investigating the feasibility of moving the upstream and/or downstream quadrupole magnets near each device to make enough space for single, longer, septum and Lambertson modules. Such moves would spoil the symmetry of the Delivery Ring lattice, but may make it easier to achieve extraction without the use of large-aperture quadrupoles. Given the requirements for resonant extraction and precise phase advances, this approach may be too disruptive to implement.

The initial bend upwards is so strong (in order to clear the Delivery Ring components), it must be leveled before the exit beamline elevation can be achieved, or it is not possible to implement a vertical achromat, which requires significant phase advance generated by quadrupoles. Leveling the beamline reference trajectory at an intermediate elevation allows a straight to be inserted with sufficient space for a series of quadrupoles to generate the phase advance required to cancel vertical dispersion after the next set of vertical bends. After beam exits the Lambertsons, a "small" quadrupole can be centered on the extracted beamline just upstream of (QS206). Beampipe in either the Delivery Ring or the extracted line can now clear this small quadrupole or the next Delivery Ring quadrupole, QS206, respectively, with a beam center-to-center distance of 0.484 m. An EDWA dipole can then installed after QS206 in the subsequent AP30 straight with a bend equal and opposite to the combined bends of the Lambertsons and focusing quadrupole. This reverse bend provides for the long "intermediate" insertion straight at an elevation of 0.78 m (again extracted beam center to Delivery Ring center), and thus for installation of an independent extraction beamline without conflicts with the Delivery Ring line below. Now an achromat can be formed using 3 quadrupoles followed by two dipoles with reverse bends (up/down) that elevate the extracted beam to the final elevation of the external beamline: with a total change from the Delivery Ring to external beamline of 1.22 m (4′).

Figure 5.37 displays the achromatic optics of Delivery Ring extraction from the center of the first quadrupole upstream of the Lambertsons to the end of the achromat. Optical functions are assumed to be those that correspond to a perfect Delivery Ring as originally designed. Since extraction is horizontal, the vertical optical functions are likely to be close, but the horizontal phase space will not be elliptical and downstream matching sections must shape the beam as required for downstream specialized insertions; i.e. high beta, collimation, and final focus. The effect of the extraction septa does not generate significant horizontal dispersion.

Matching quadrupoles will be required after the vertical achromat. Resonant extraction will not produce elliptical phase space occupation in the horizontal (also unlikely in the vertical), so the optics at this time are left unspecified, but the nature of the quadrupoles will be to provide for arbitrary matching of optics to the downstream modules.

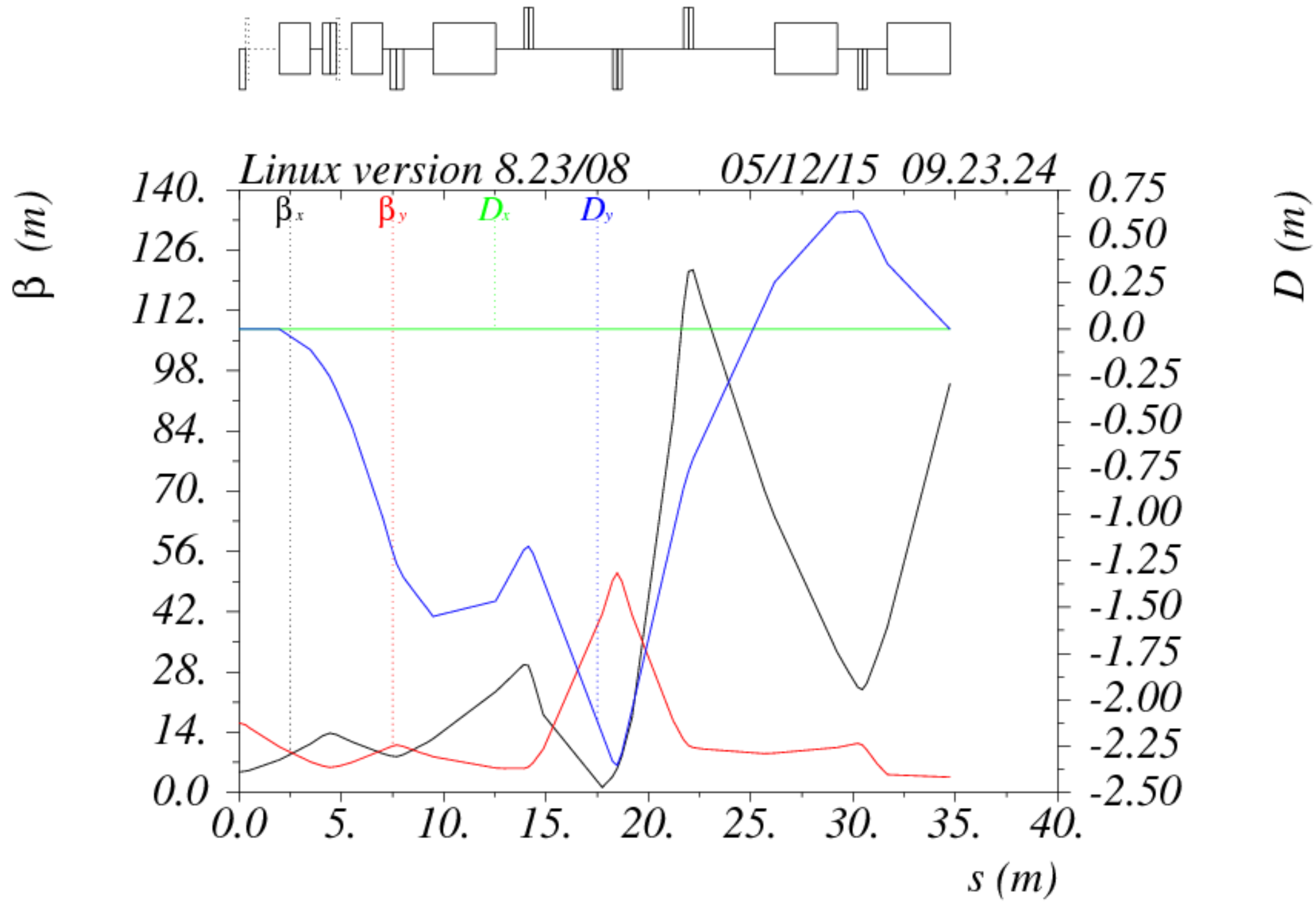

Figure 5.37. The extraction optics showing the two Lambertsons followed by an opposite-sign vertical bend, quadrupoles to form the achromat and a final bend up and then level again to the elevation of the beamline (all EDWA dipoles).

### *Horizontal Bend String/Momentum collimation*

One of the first opportunities for a dual-purpose section exploits the required ~36° of horizontal westerly bend to meet the constraints on the direction of the beam line and the location of the Mu2e experimental facility. Since momentum collimation is required for g-2 and is likely to be important to Mu2e as well (due to scattering on septum wires and other losses at extraction) it is incorporated most efficiently into this strong bend section. Further, the split to a dedicated g-2 beamline is and must be designed into this section and represents a significant impact. Such a split requires a total westerly bend of 24° to optimally locate the g-2 experimental ring as depicted in Figure 5.38.

The present approach employs 6-bends per line as shown in Figure 5.39 with the capability to steer both 8 GeV and 3 GeV beam into their respective Mu2e and g-2 beamlines. A common momentum collimator is inserted upstream of the split. The first two bends are comprised of a 6-4-120 powered in series with a SDEW to deliver 9.94° of left bend to both lines. The next bend, a 6-4-120, is variable, delivering 6.56° and 3.56° of bend, respectively to Mu2e and g-2. A large-aperture quadrupole (LQE) with a star chamber delivers an additional angular kick of +1.5° and -1.5°, respectively to Mu2e and g-2 beam. This off-center quadrupole kick is required to efficiently separate the two trajectories in ~4 m to clear the next quadrupole in the Mu2e line. In order to construct an achromat and to allow the bends to be tunable independently for Mu2e and g-2, a second set of 6-4-120 and SDEW dipoles follows the split in the Mu2e line, positioned 180° in phase advance from the first set. These two sets are powered together and dispersion is canceled independent of the other bends. The intervening 6-4-120 + quadrupole bend which totals 8.06° must now be paired with an equal downstream bend 180° away and this is a 6-3-120 (because it must be stronger than a 6-4-120). The same type of pattern is repeated in the g-2 line. The total bends for the Mu2e and g-2 lines are now 36° and 24°, respectively. For more precise momentum collimation, a section is reserved for a second collimator, again 180° from the first collimator. The optics of the line is given in Figure 5.39. Outside of the horizontal bend/momentum collimation section, there must be no residual dispersion. Therefore, this section must also fulfill conditions for a linear achromat. Phase advance and dipole placement for dispersion cancellation dictate this section to be restricted to the optics as designed. Matching to the optics of this section must occur on either side.

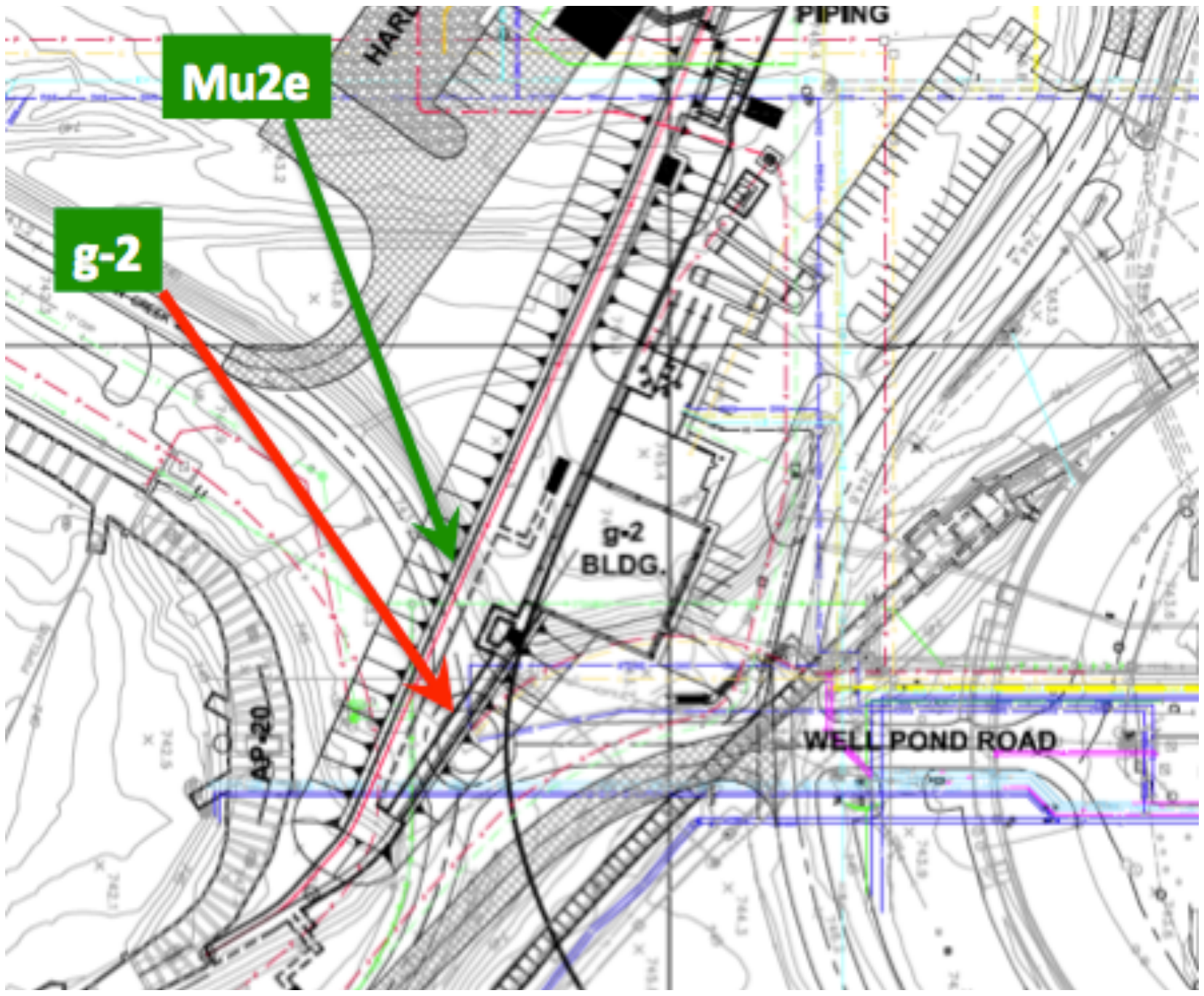

Figure 5.38. Total horizontal bends to the Mu2e and g-2 experiments.

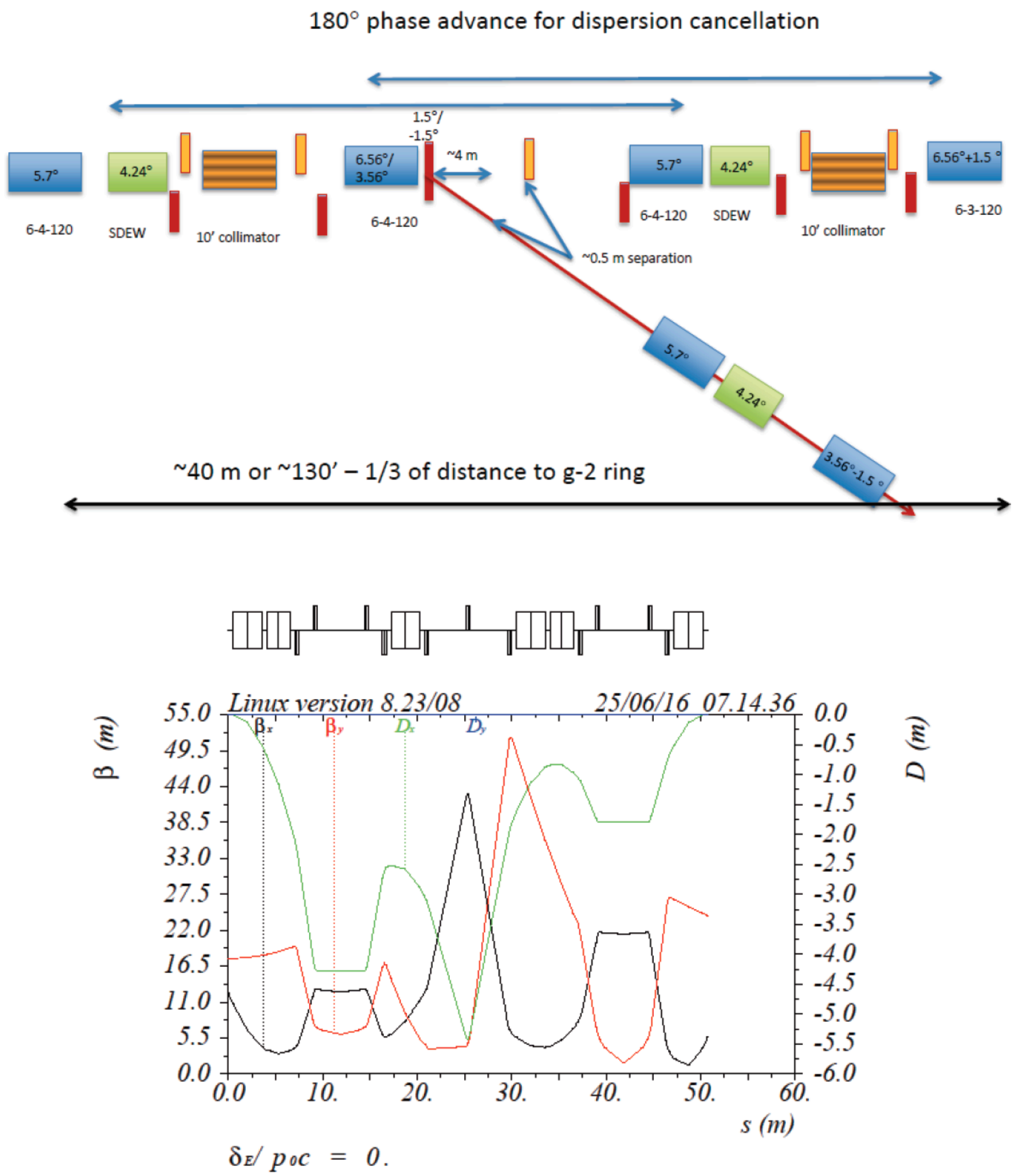

Figure 5.39. The optics of the horizontal bend and momentum collimation insert.

The collimators are a priori assumed to be 10 feet long; a potential reduction in length is possible pending detailed MARS simulations. Care must be taken that particles intercepted by the collimators are not simply scattered into the beam increasing the background. A 10 ft collimator accomplishes this. Accordingly, this length was assumed for all collimation. With such fine momentum resolution, the collimator can be set with millimeter accuracy and requires no exceptional calibration or other measures.

***High Beta/AC dipole Insertion***

The optics required to make the AC dipole (see Section 5.8.1) technically (and realistically) feasible pose the most challenging optical design problem for the line, resembling to some degree a collider interaction region. In this case, however, the horizontal plane must have a very large beam envelope (high beta) and a small beam size in the vertical (low beta – the vertical beam size is only about ±0.5 cm). The high horizontal beta enhances the extinction kick of the AC dipole and the small vertical beam size allows for a smaller dipole spacing, thus limiting the required dipole excitation to acceptable power levels. The high-beta insertion dominates, and largely determines, the physical length of the beam line. This effect can be seen in the unavoidable transition from the beta functions generally characteristic of the line to the high beta value (see Figure 5.40). Already strong focusing is employed to effect this rapid transition; further focusing makes the high beta increasingly achromatic (chromaticity $\propto \beta kl$, with $kl$ the normalized quadrupole strength times its length). This is to say, further focusing would cause an unacceptable distortion of the beamline optics for off-momentum particles. The focusing is already strong, achieving a rapid falloff in horizontal beta in 30 m. Because of the alternating high/low beta in the horizontal/vertical plane, there is an unavoidable seesaw effect; as the horizontal beta function drops, the vertical invariably rockets to a level comparable to the horizontal high beta before both can be focused to the main transport sizes. The overall length required by high betas is governed by the decline of betatron functions in a straight, from 250 m down to ~20 m on average requires 70 m ($\beta_{max} \approx s^2/\beta$).

***Extinction Collimation Sections***

A series of interleaved horizontal and vertical collimators have been staged after the high-beta region to perform the efficient extinction required. Interleaving minimizes the overall length of the collimation section (Figure 5.41). Scattering from the collimator edges in both planes will likely induce background particles downstream in either plane. Collimators are staged in 90° pairs to remove such particles. Positioning the beam accurately in the collimator apertures is important; therefore, appropriate diagnostics are critical in this section. As in the momentum collimation section, the collimator length has been set at 10 feet. Also indicated in Figure 5.41 is the dispersion wave generated by the AC dipole that displaces out-of-time particles onto the collimators. The number of collimators required and the extinction efficiency is discussed in a later section (see Section 0), but meets the criteria specified by the experimenters.

Centering the beam in the collimators and profiling the beam after the action of the collimators is very important to establish proper extinction and coordination with AC dipole operation. A multiwire is required downstream of each collimator to observe centering of the beam as the collimator is closed. An upstream multiwire would facilitate a more rapid beam line tune-up, but at least one multiwire per collimator is required. For the current instrumentation a loss monitor will be used in conjunction with the downstream multiwire to tune the beam interaction with the collimators.

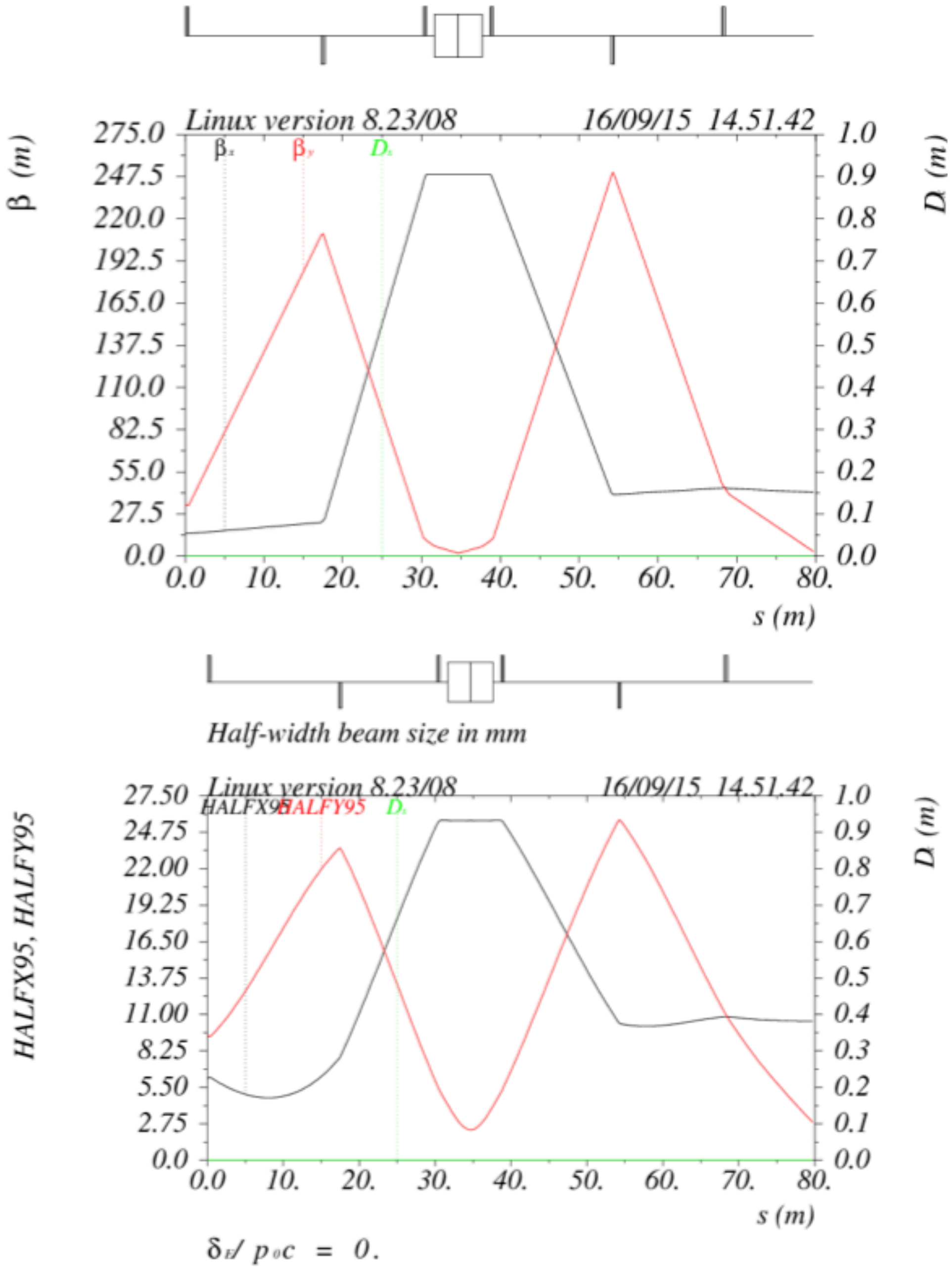

Figure 5.40. The high-beta optics (top) and beam sizes (bottom).

***Shield Wall and Diagnostic Absorber Section***

The next section is a prime example of designing broad functionality into a single insertion for compactness. Several criteria are combined in the following: a) a shield wall is constructed between the beamline enclosure and the enclosure housing the final focus and the experimental hall to separate the two areas, b) a diagnostic beam absorber is then embedded in the shield wall to permit commissioning, tune-up, and beam studies without spraying beam into the target station area, and c) a horizontal bypass around the absorber through the shield wall is designed into the beamline. Separation of the primary beamline enclosure from the experiment allows the beamline to be classified as ODH 0 facilitating maintenance, installations, and personnel access. A 4-bend dipole string is implemented in the bypass fulfilling the conditions for a geometric achromat, as shown in Figure 5.42.

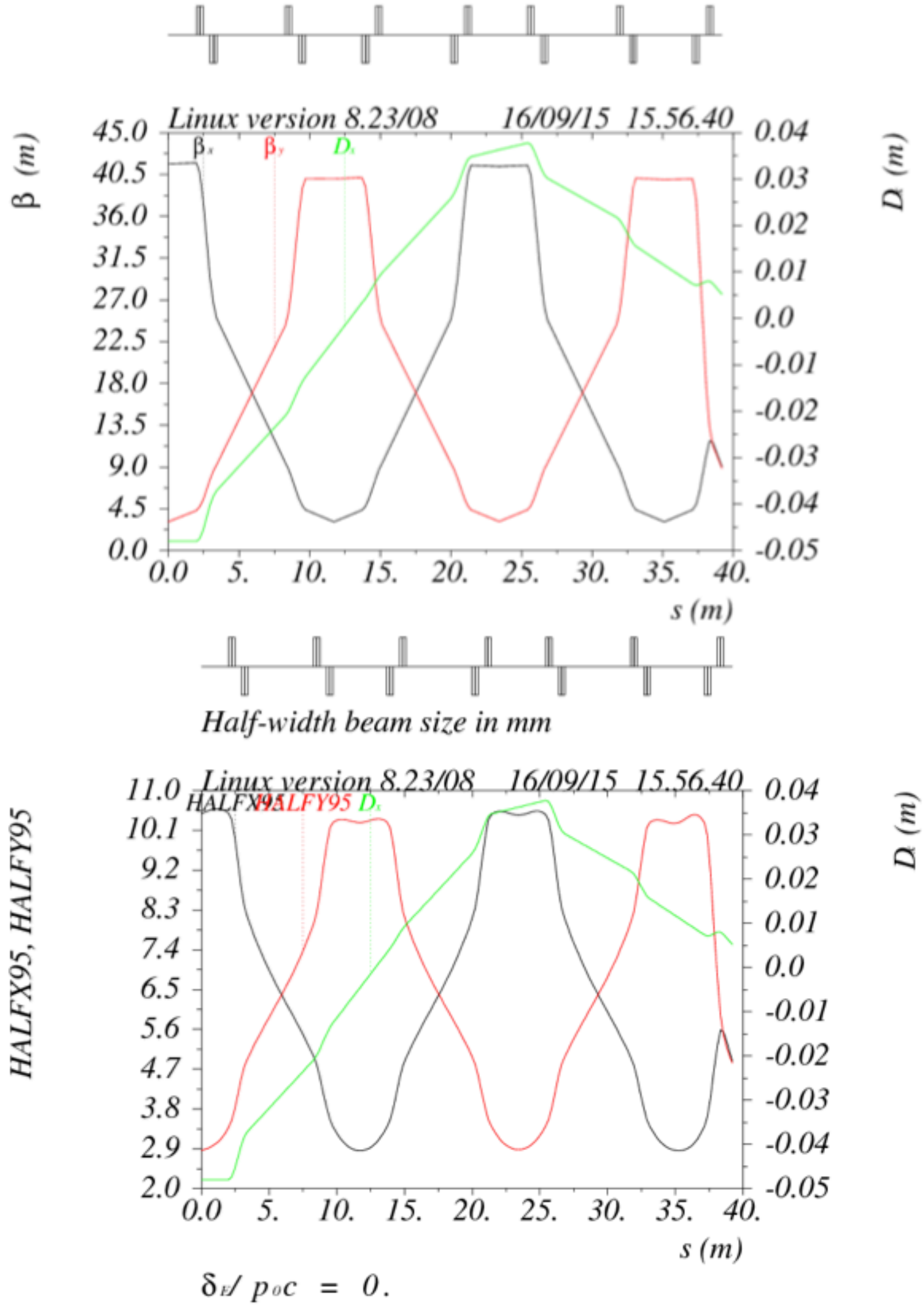

Figure 5.41. The extinction collimation optics (top) and beam sizes (bottom).

Advantage is then taken of the space required for shield wall to provide a magnet-free straight in the horizontal bypass. This straight is 6 m long and supports insertion of a diagnostic absorber. Without magnets, only a beampipe passes through the shielding and around the diagnostic absorber. The location of this insertion, and the fact that only beampipe is required, allows a shield wall to be built between the experimental hall and the majority of the beam line. Thus, the beam line can be an ODH 0 zone making maintenance and access much less restrictive. Without the shield wall, the beam line would be classified identical to the experiment: ODH 1.

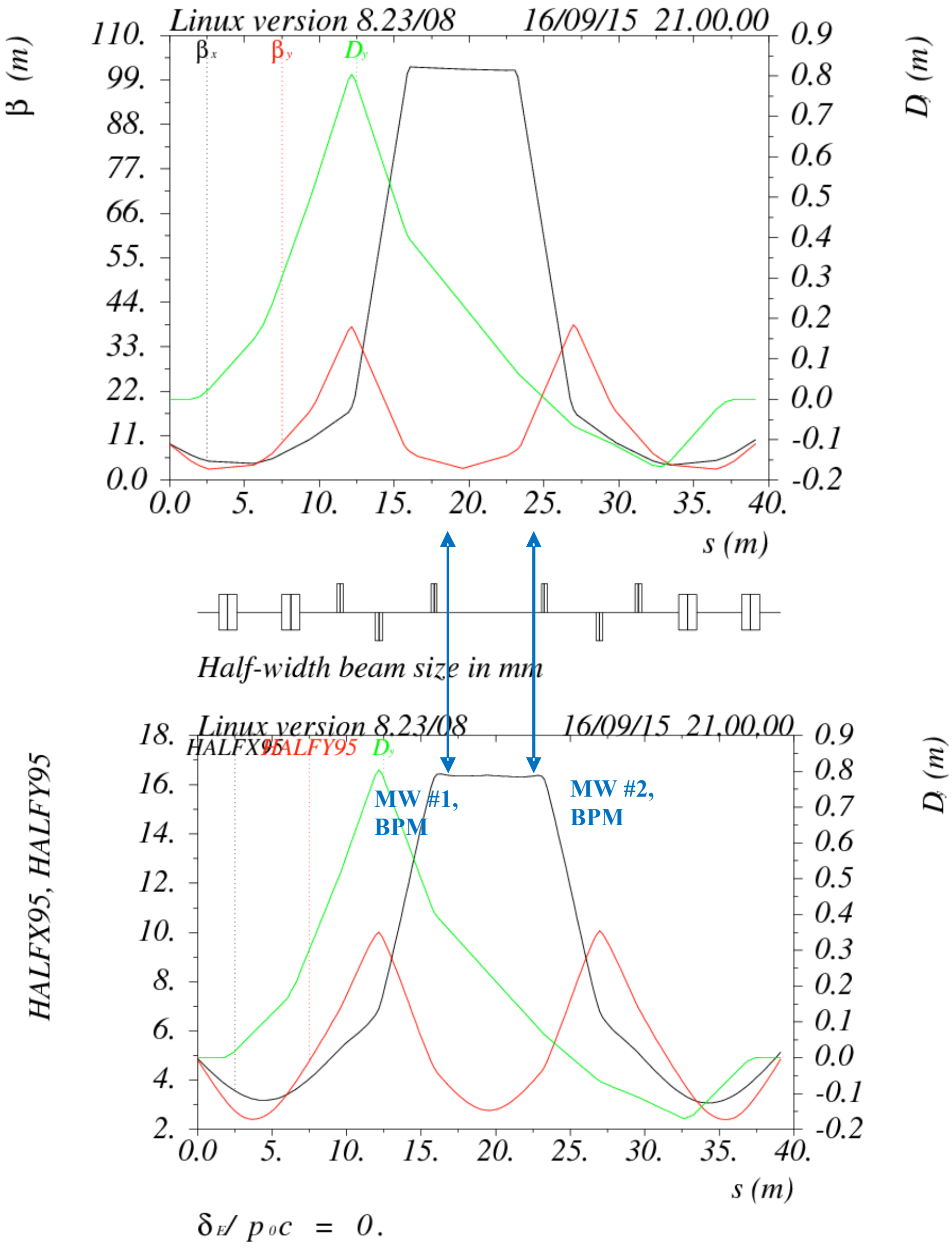

Figure 5.42. The horizontal bypass and absorber location optics (top) and beam sizes (bottom).

### *Final Focus*

The final focus is a typical "collision" type optics region using a quadruplet with point-to-point imaging with a demagnification factor of ~2.5, as shown in Figure 5.43. The telescope focuses beam to a round, achromatic waist with a 2 m low-beta function, which is consistent with the 1 mm rms required for the production target. Although point-to-point has almost double the chromaticity of parallel-to-point in a final-focus telescope, the chromaticity is still low so that chromatic and geometrical aberrations do not affect the quality of the beam spot size at the target. The target design radius is 3 mm, which corresponds to 3-5σ depending on the extracted emittance. The optics and corresponding beam size are plotted in Figure 5.43.

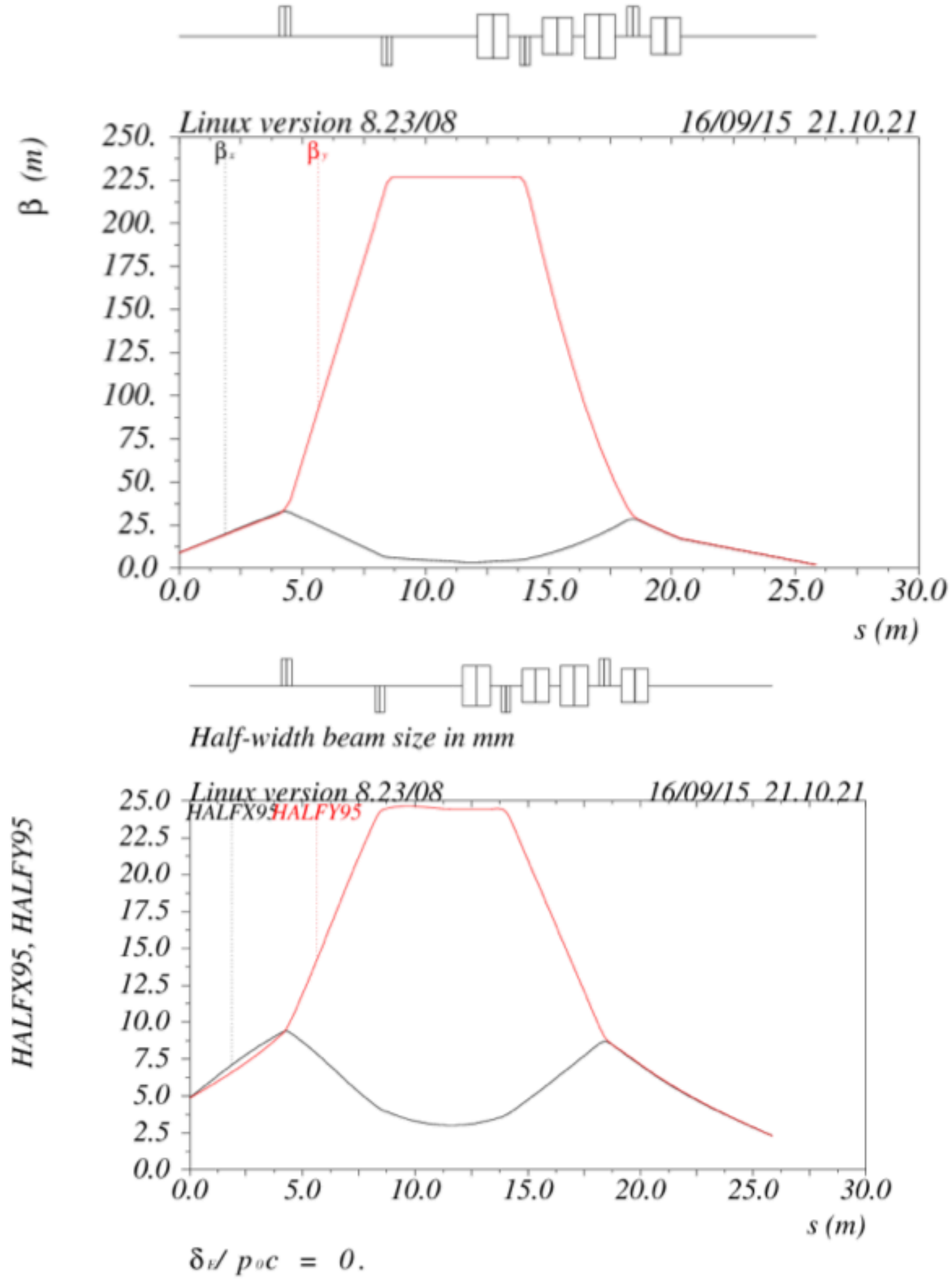

Figure 5.43. The final focus telescope optical functions (top) and beam sizes (bottom).

As was discussed in a previous section, a vertical declination angle is required at the entrance to the production solenoid. A compact final focus and reasonable quadrupole apertures require the 2.4° decline to be started further upstream in order to cancel vertical dispersion during nominal operating conditions and not impose additional achromatic constraints on the final focus. The entire final focus section is therefore mounted on a 2.4° slope generated by a pair of 1.2° dipoles in the diagnostic absorber section. The following summarizes the present demands on the final focus. Beam steering and target scans are addressed next.

- Small 1 mm transverse σ on target (2 m beta function)
- A 2.4° vertical decline at the entrance to solenoid – must be achromatic, no vertical dispersion at target.
- Independent position/angle controls in BOTH vertical and horizontal
- ±2 cm in horizontal/vertical for target scan.
- ±1° in horizontal/vertical
- Conventional design approaches; i.e. kicking upstream and through more than one of the final-focus quadrupoles, result huge offsets in the beam trajectory through these final focus quadrupoles.

The steering magnets in the final focus must be capable of adjustment in order to scan the target during commissioning. This is complicated by the fact that the beam deflects in the magnetic field of the Production Solenoid. The layout of the steering magnets designed to confine the orbit excursion satisfy all aperture requirements in all final focus quadrupoles. The difficulty of the steering magnet placement that simultaneously avoids large apertures in the quadrupoles and steering dipoles is presented below. Steering magnets must be interleaved not only between each plane but also with final focus quadrupoles for compactness. The steering magnet length has been assigned to be 4′, which is typical of an 8-GeV corrector.

Although a strong vertical steering capability of ±1° is in place, the vertical dispersion cannot be canceled away from the nominal operating point. The steering magnets cannot be positioned in an achromat configuration without extreme changes in aperture and length. Given these constraints, the present final focus design accomplishes its purpose very well.

The result of optimizing the position of the steering dipoles from the standpoint of minimal quadrupole apertures and gaps in downstream dipoles is presented in Figure 5.44 through Figure 5.47.

### *5.7.3* Beamline Optics Performance

The ±1% momentum performance is documented in Figure 5.48 through Figure 5.50. The line optics is exceptionally stable as a function of momentum.

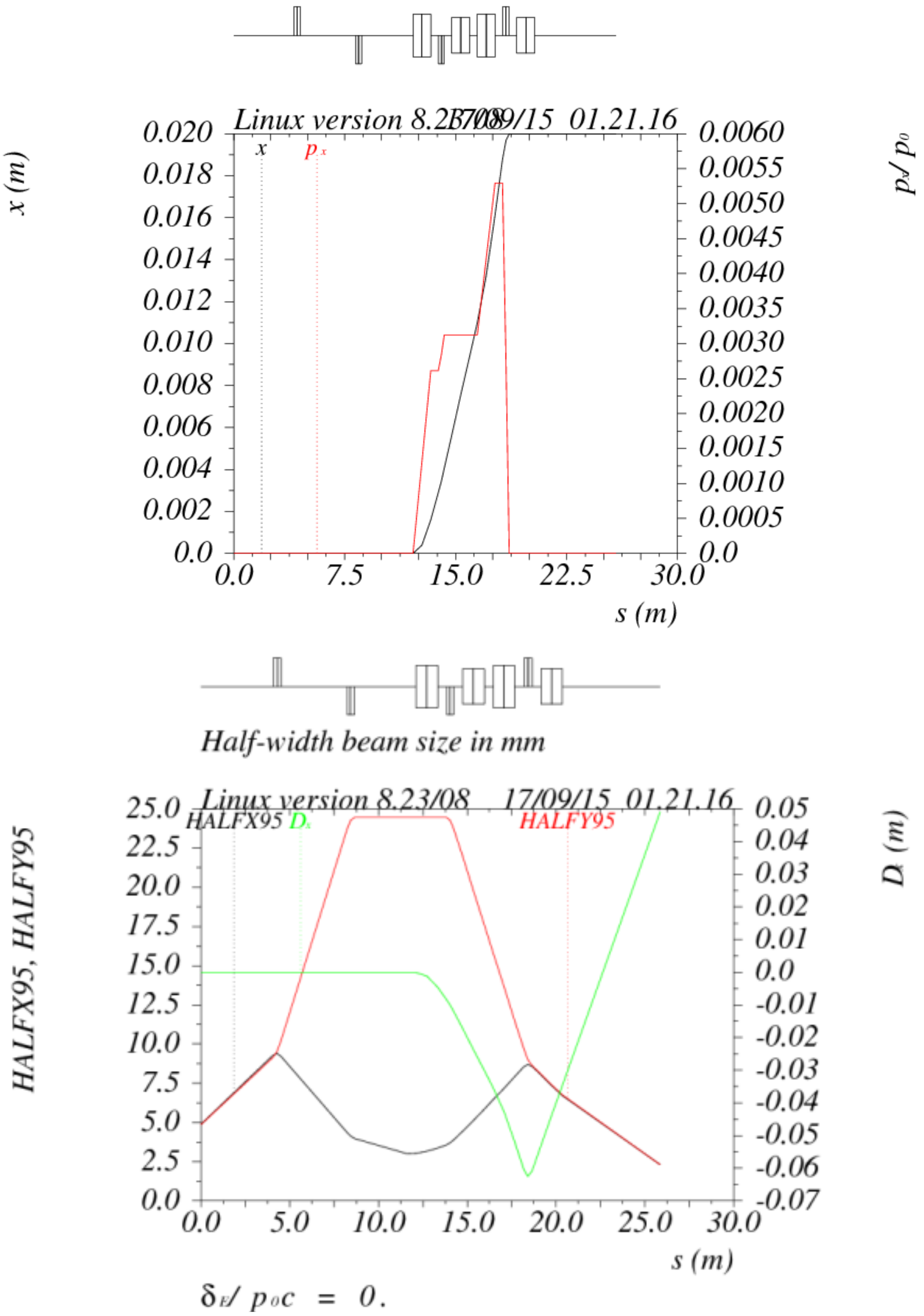

Figure 5.44. Independent position control in horizontal plane with ±2 cm capability at the target (top). The corresponding optical functions are shown in the bottom plot.

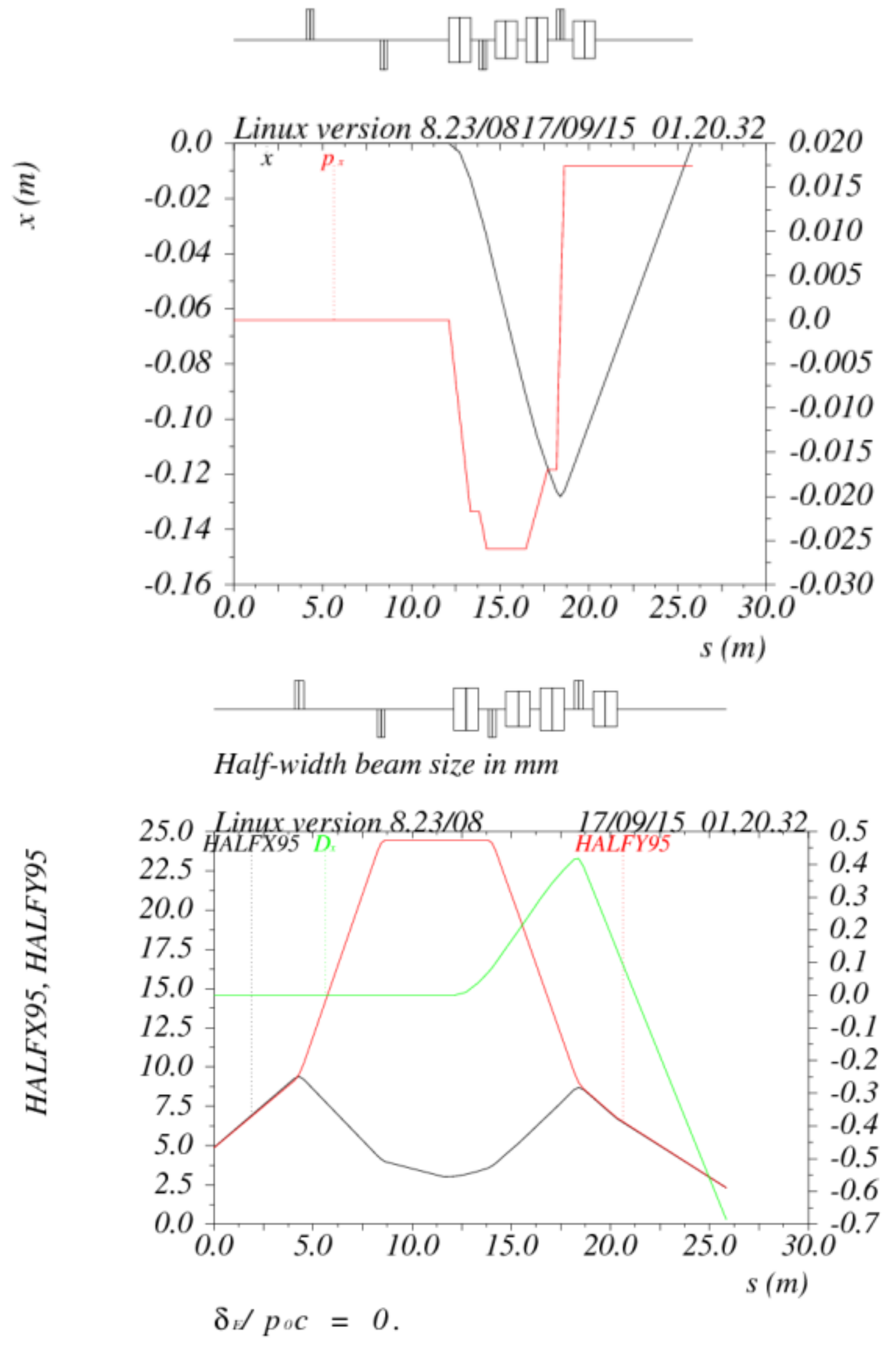

Figure 5.45. Independent angle control in horizontal plane with ±1° cm capability at the target (top). Corresponding optical functions (bottom).

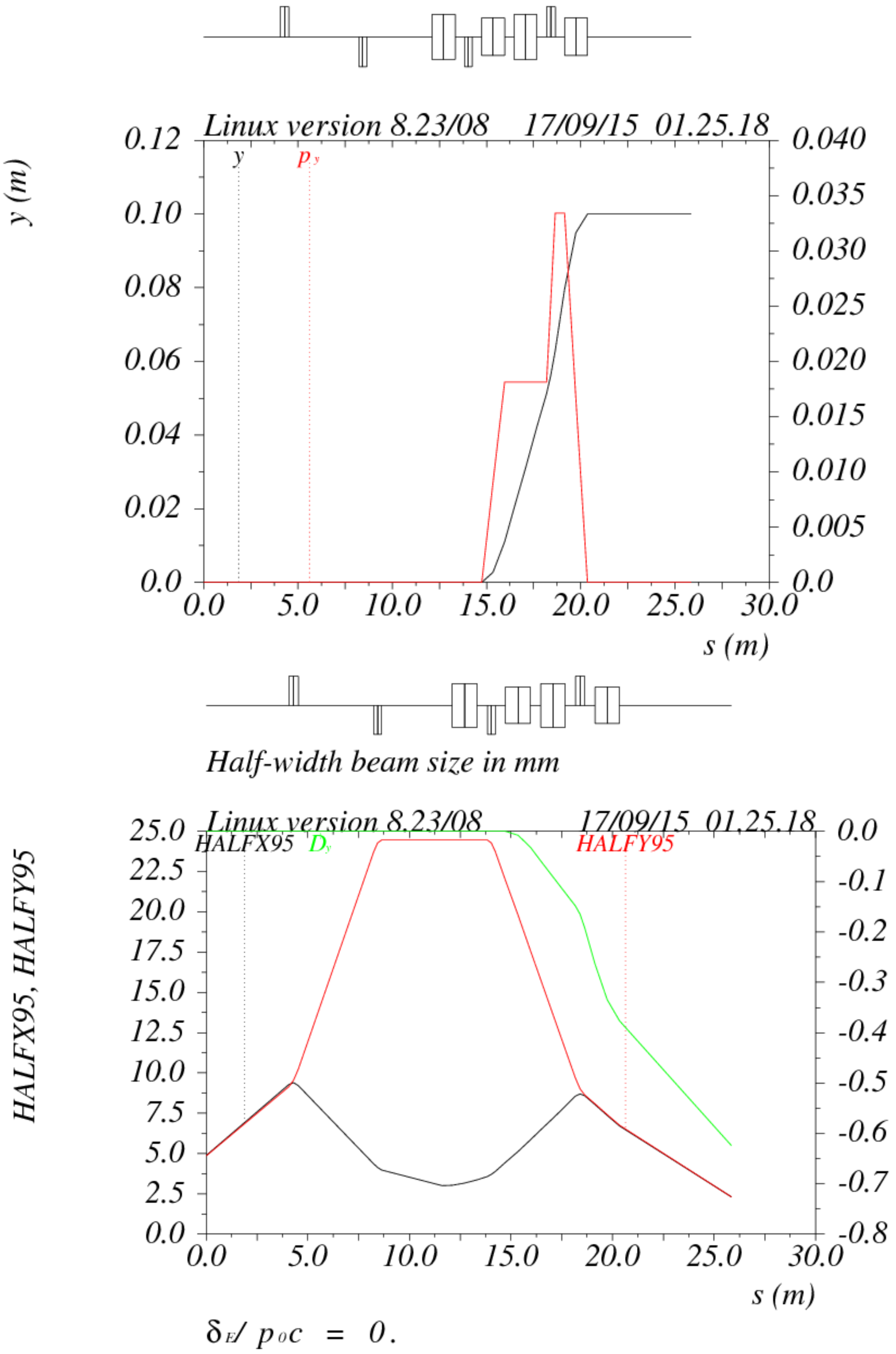

Figure 5.46. Independent position control in vertical plane with ±10 cm capability at the target (top). Corresponding optical functions (bottom).

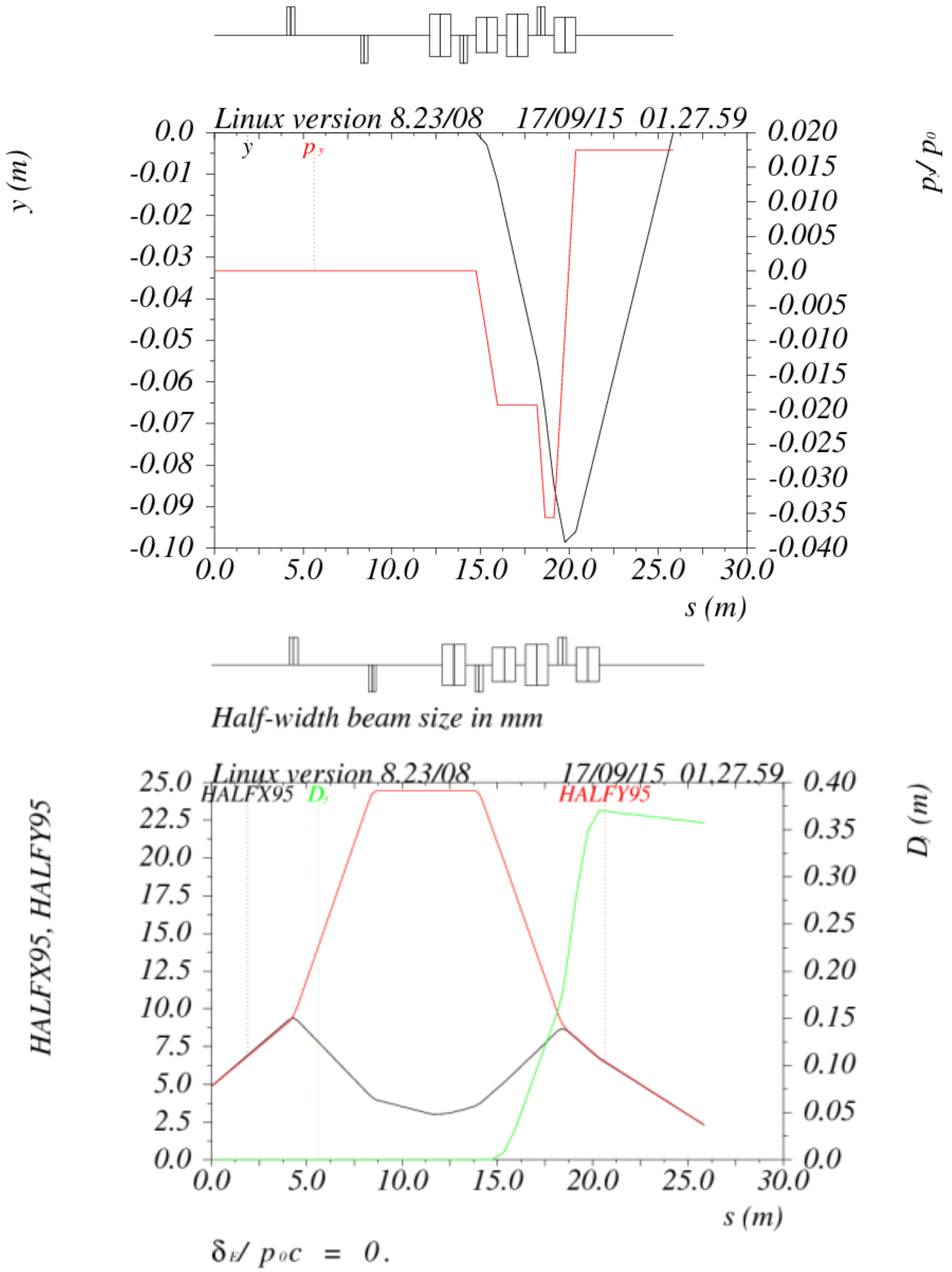

Figure 5.47. Independent angle control in vertical plane with ±1° capability at the target (top). Corresponding optical functions (bottom). Note that the dispersion from this angular kick is not insignificant - ±4 mm for a 1% $\delta p/p$.

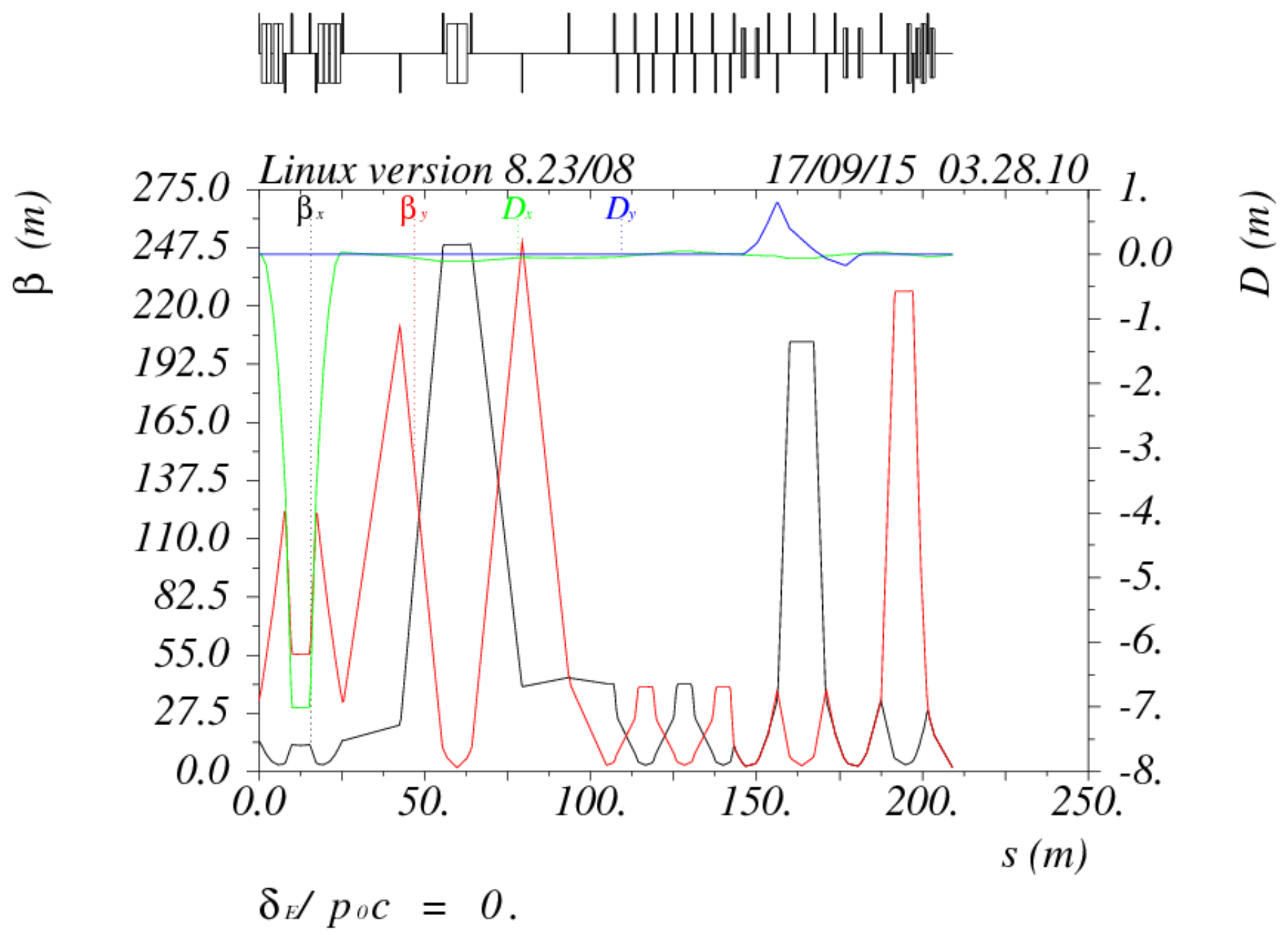

Figure 5.48. Optical functions for the entire beamline initiating just upstream of the momentum collimation section. The AC dipole is centered in the horizontal high-beta peak at ~55m location.

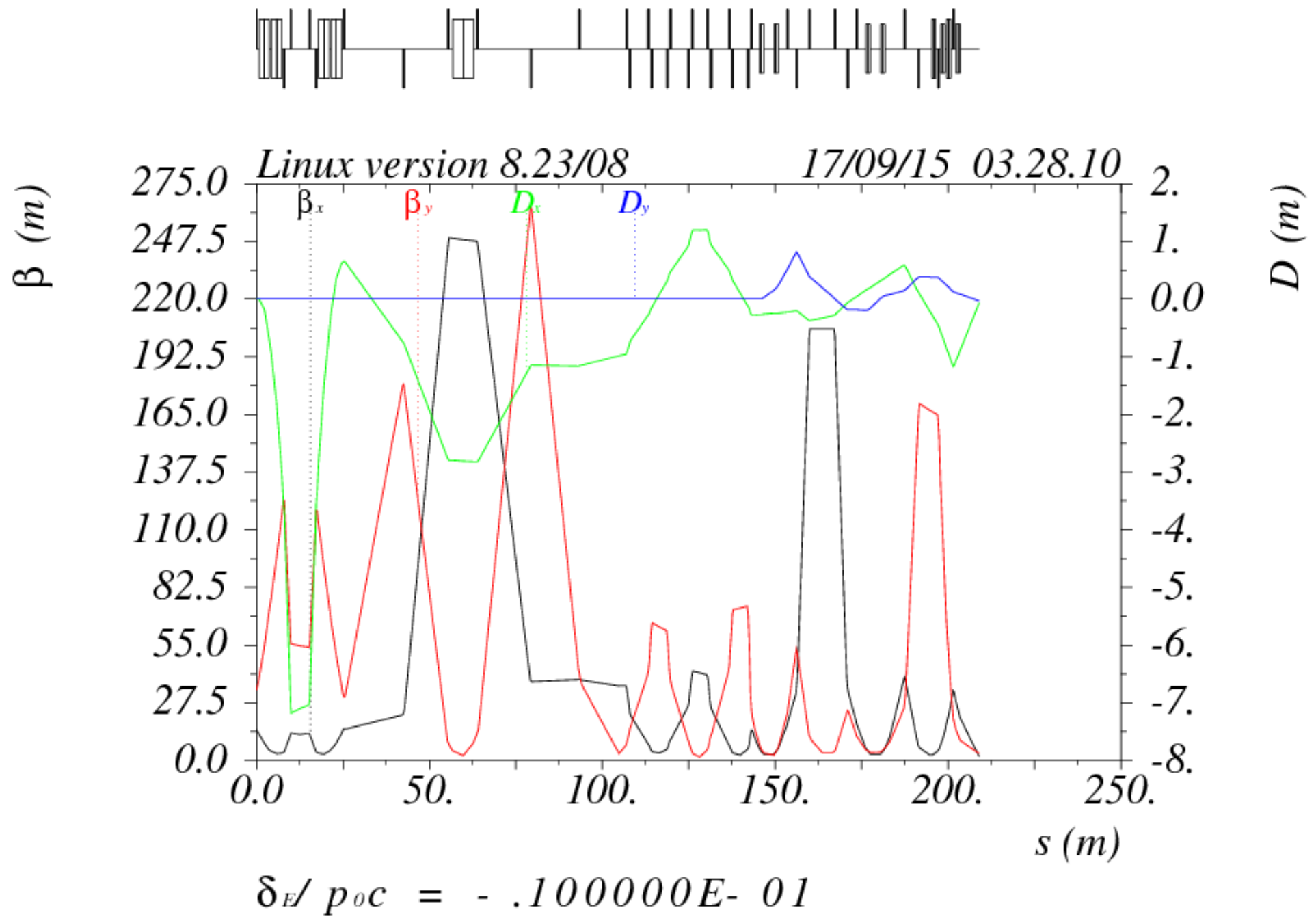

Figure 5.49. Optical functions for the entire beamline initiating just upstream of the momentum collimation section for a $\delta p/p$ of -1%.

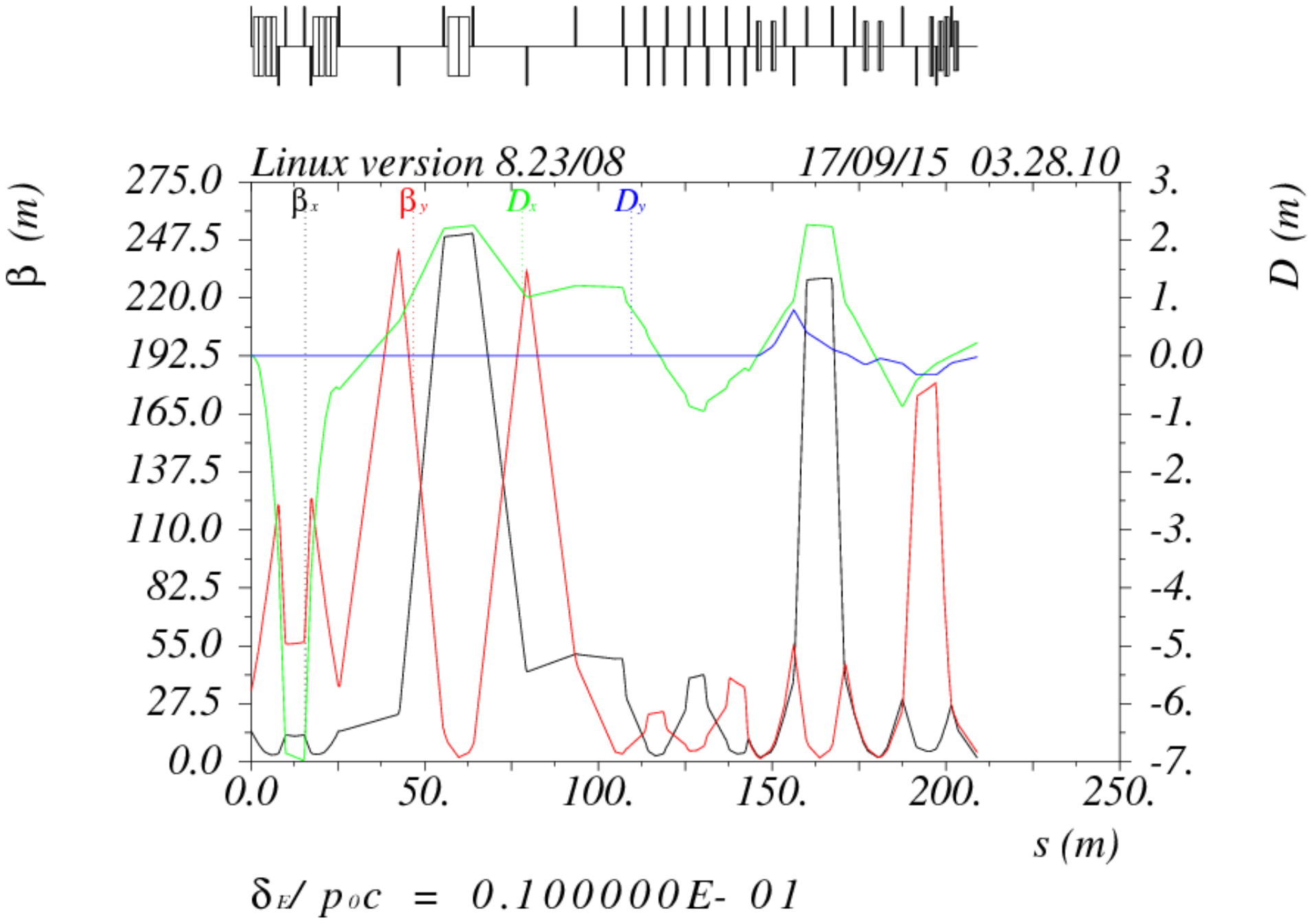

Figure 5.50. Optical functions for the entire beamline initiating just upstream of the momentum collimation section for a $\delta p/p$ of +1%.

## 5.8 Mu2e Beam Extinction and Extinction Monitoring

The Mu2e experiment proposes to use a proton beam pulsed at approximately 0.6 MHz. The use of a pulsed beam is motivated by the fact that a significant background is produced by secondary beam particles (primarily pions) that reach the detection region during a time interval starting shortly after pulses hit the production target and extending for about 700 nsec. By detecting conversion electrons only for times later than 700 nsec, this background is significantly reduced to an acceptable level.

The same type of background could be produced by protons hitting the production target during or somewhat before the detection interval; this puts a limit on the number of particles hitting the production target between pulses. We define the beam extinction as the ratio of the number of protons striking the production target between spills to the number striking it during the spills. It has been established that an extinction of approximately $10^{-10}$ is required to reduce these backgrounds to an acceptable level [35]. The extinction requirement varies with the exact time that the proton strikes the target between the pulses, and $10^{-10}$ is a representative number assuming that the out-of-time particles are distributed uniformly between pulses.

The extinction part of the Mu2e Project comprises both the generation and monitoring of the requisite level of extinction, both of which will be very challenging.

### *5.8.1* Beam Extinction

The required beam extinction will be achieved through the combination of two separate mechanisms:

- The technique for generating the required bunch structure, as described in Section 5.5.1, will naturally lead to a high level of extinction, which can be enhanced using momentum collimation in the Delivery Ring.
- A system of AC dipoles and collimators will be arranged in the proton transport such that only in-time beam is transported to the target.

***Beam Extinction in the Delivery Ring***

Recall that the beam is formed into four bunches in the Recycler and then transferred a single bunch at a time to the Delivery Ring. This means that there will be a very small number of particles outside of the nominal bunch at the time the beam is transferred, and out-of-time beam will result from beam migrating out of the nominal bucket. In general, this can occur in one of three ways:

- Space charge effects
- Intra-beam scattering
- Beam gas interactions
- Beam scattering from the electrostatic extraction septum

In all cases, the mechanism results in a change of energy of the protons that causes them to migrate out of the bucket along the separatrix, as shown in Figure 5.51. This migration can be ameliorated by momentum collimation in the high dispersion regions of the Delivery Ring. Simulations are still ongoing, but preliminary results show that an extinction level of at least $10^{-5}$ should be achievable in this way.

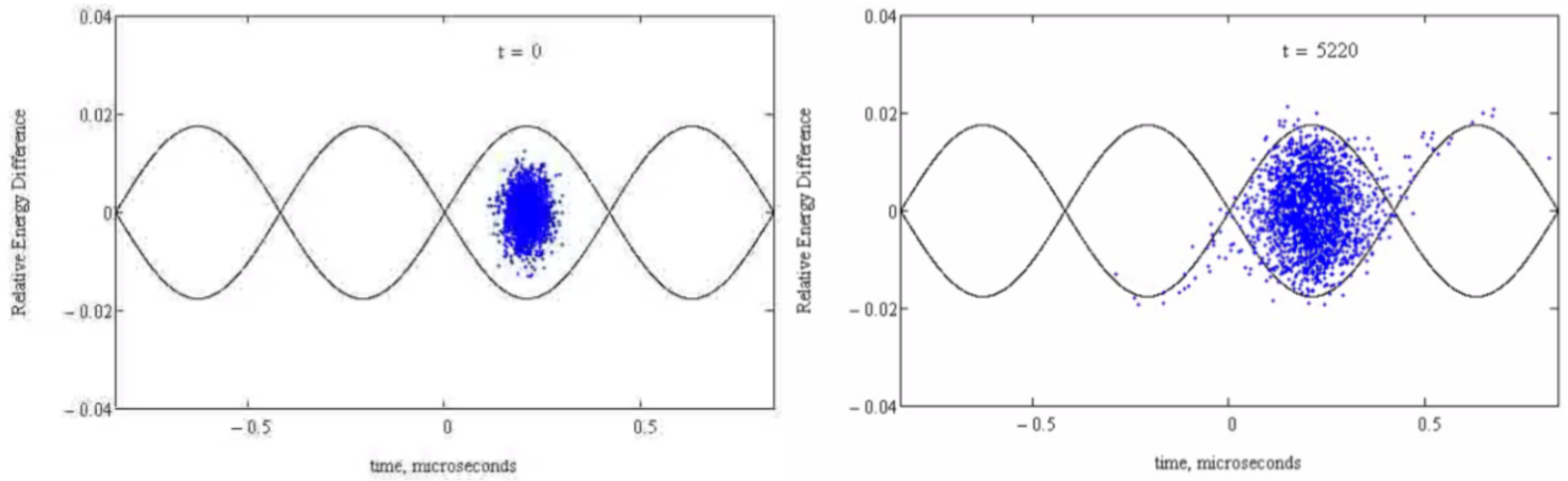

Figure 5.51. This shows the evolution of out of time beam, in this case generated by an unrealistically large amount of RF phase noise. Note that beam migrates along the separatrices.

***Beam Line Extinction***

Beam line extinction is based on deflecting magnets and a collimation system, such that only in time protons are transported to the production target. Ideally, one would like a kicker that would cleanly kick the out-of-time beam into an absorbing collimator, or equivalently kick in-time beam into the transport channel; however, such a kicker operating at the 600 kHz bunch rate is beyond the state of the art at the moment. Therefore, the system will utilize a pair of resonant dipoles to achieve the desired deflection.

These magnets still represent a technical challenge, so the design effort has focused on the optimization of the magnet specifications, which are tightly coupled to the optical parameters of the beam line [36]. To summarize, the minimum stored energy of the bending magnet scales roughly as

$$U \alpha \frac{1}{\sqrt{\beta_x L}}$$

where $\beta_x$ is the betatron function in the bend plane, and $L$ is the length of the magnet, assuming a waist in the non-bend plane. Thus, we are driven to long, fairly weak magnets and large betatron function in the bend plane. It was determined that a length of 6 m and a betatron function of 250 m were the largest practical values that could be achieved without unacceptable complication and increased length of the beam line [37], and these are used as the basis for the magnet design.

The baseline magnet configuration consists of two AC dipoles with the parameters shown in Table 5.10. Generally, the function of the lower frequency magnet is to sweep the out of time beam out of the transmission channel, while the higher frequency magnet serves to limit the slewing in the transmission window. The fact that the low frequency magnet runs at half the bunch frequency means that it crosses through 0, moving in opposite directions as the beam passes through the transmission channel on successive pulses. Therefore, the high frequency magnet must be an odd harmonic to maintain the proper relative sign.

| Magnet | Frequency (kHz) | Length (cm) | Aperture | | Peak B Field (Gauss) |
|---|---|---|---|---|---|
| | | | bend plane (cm) | non-bend (cm) | |
| A | 300 | 300 | 7.8 | 1.2 | 120 |
| B | 3800 | 300 | 7.3 | 1.2 | 15 |

Table 5.10. Specifications for each of the AC dipoles in the extinction system.

Details of the resulting waveform are shown in Figure 5.52. The boundaries indicated on the figure are the amplitude at which the centroid of the beam hits the collimator (~50% transmission) and the amplitude at which all beam will hit the collimator (full extinction). Figure 5.53 shows the transmission efficiency for this collimation system. The longitudinal distribution comes from a simulation of the bunch formation and development in the Recycler and Delivery Rings.

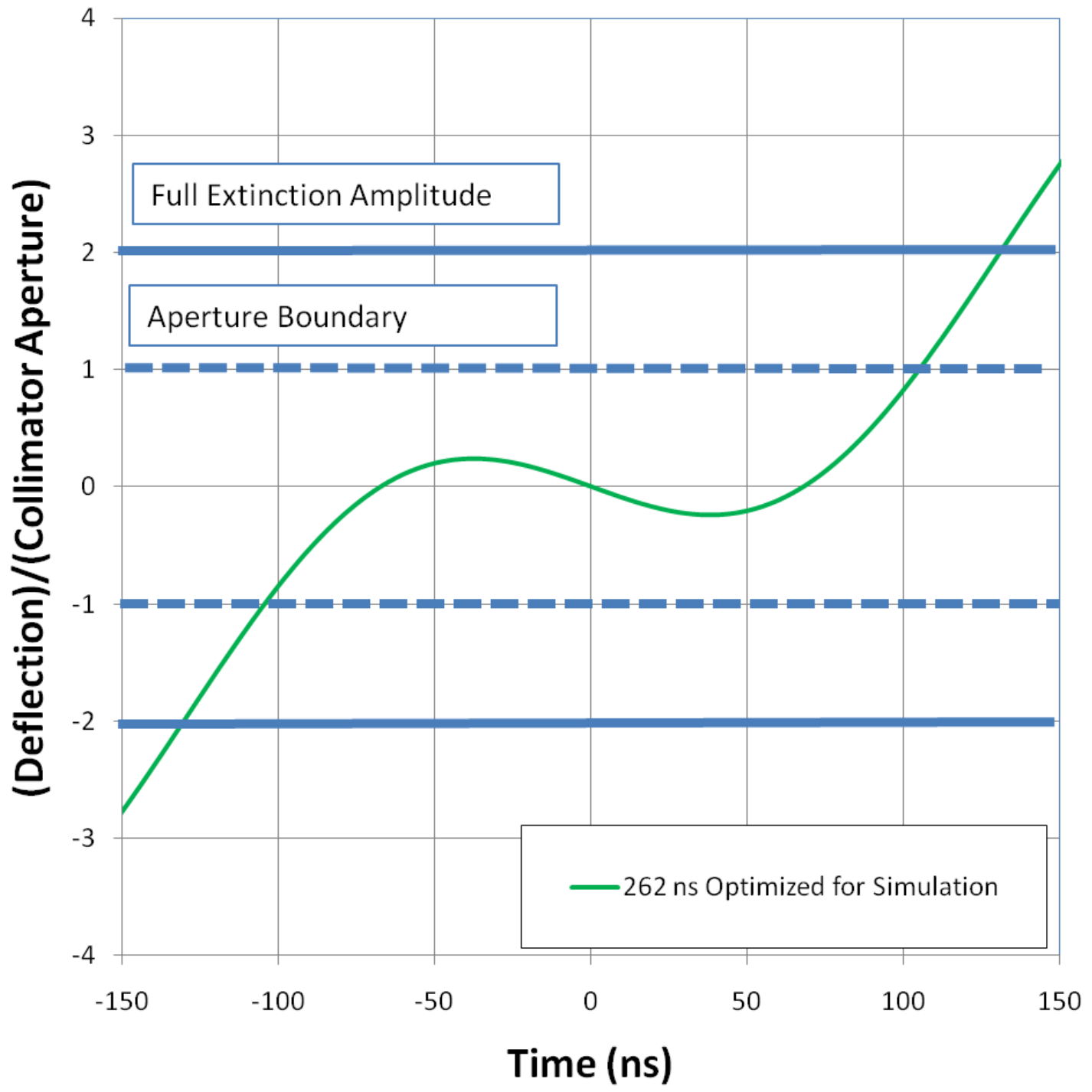

Figure 5.52. Details of the combined AC dipole waveform in the transmission window.

As discussed in Section 5.7, the optics requirements of the extinction AC dipole place a significant constrained on the beam line design. A simulation of the beam line and collimation found that with 210M particles incident on the first collimator, 27 hit the target, or $< 1 \times 10^{-7}$ [38].

### *5.8.2* Extinction Monitoring

Details of the extinction monitoring requirement are described elsewhere [39]. The most precise measurement of beam extinction would be some sort of detector with sensitivity to single out-of-time particles. However, because of the much larger number of particles that are in-time, such a detector would either have to have a very large dynamic range or somehow be "blinded" during the in-time window to avoid saturation effects. No workable solution has been found that adequately achieves either of these requirements [40].

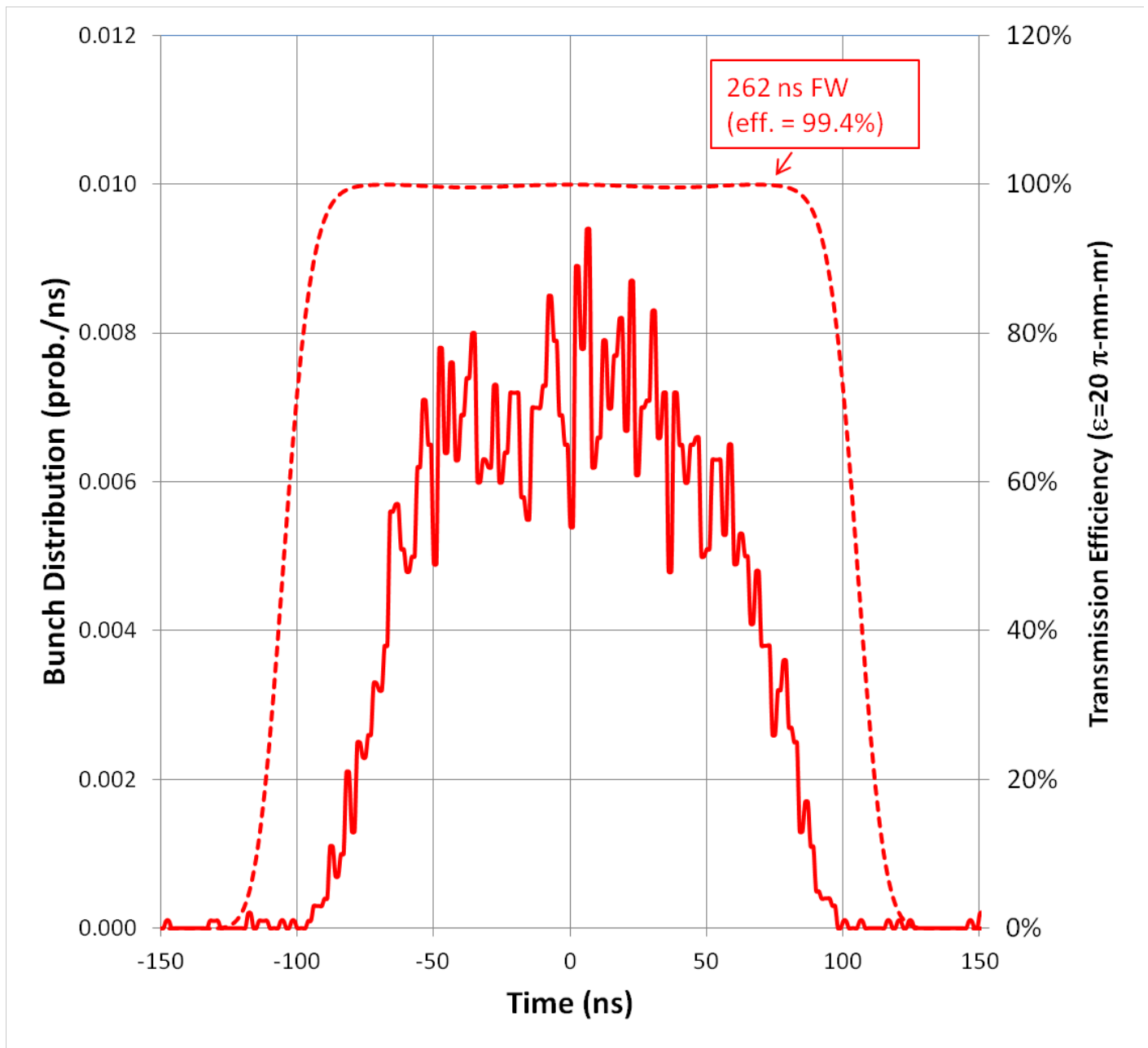

Figure 5.53. The beam transmission window, with the simulated bunch distribution of the extracted beam. The simulated distribution is the same as that shown in Figure 5-41. The overall transmission is 99.6%, assuming a normalized 95% transverse emittance of 20 π-mm-mrad,

The Mu2e baseline therefore focuses on a statistical approach. By measuring scattered particles, we can design detectors with good time resolution, but a very small effective acceptance, such that there are only, at most, a few hits for in time particles. Over many bunches, a statistical picture can be built of the out of time population with a precision limited only by the accidental rate.

There are two time scales to be considered in terms of extinction measurement. The relevant time scale for the precision measurement is the entire time over which data are taken; however, we choose roughly an hour as the time scale for precision measurement to avoid long running periods with unexpected anomalies. It is important to monitor for potential failures of the system on a much shorter time scale. The operation of the AC dipole system can be best monitored via an indirect means that measures the phase and magnitude of the magnetic field in the magnets. If these do not change, then it is difficult to imagine a scenario in which the system could malfunction in a way that would not cause large losses at the extinction collimator, which could be easily detected.

There could be much more subtle problems with the extinction in the Delivery Ring, so it is very important that this be monitored on a much shorter time scale. Fortunately, the precision needed is only about $10^{-5}$ or so.

It is likely that any out of time beam would develop over the time scale of the slow spill, so it is important that enough timing information exists for both extinction measurements to synchronize them to this time.

***Fast/Low Precision Extinction measurement***

To measure the level of extinction achieved in the Delivery Ring, a monitor will be placed upstream of the AC dipole. The baseline plane is to insert a thin foil into the beam, similar to the "Texas Multiwires" used in NuMI. High angle scatters from this foil would be monitored by a charged particle telescope, consisting of quartz Cerenkov radiators read out, by PMTs or Silicon photomultipliers (SiPMs), as illustrated in Figure 5.54.

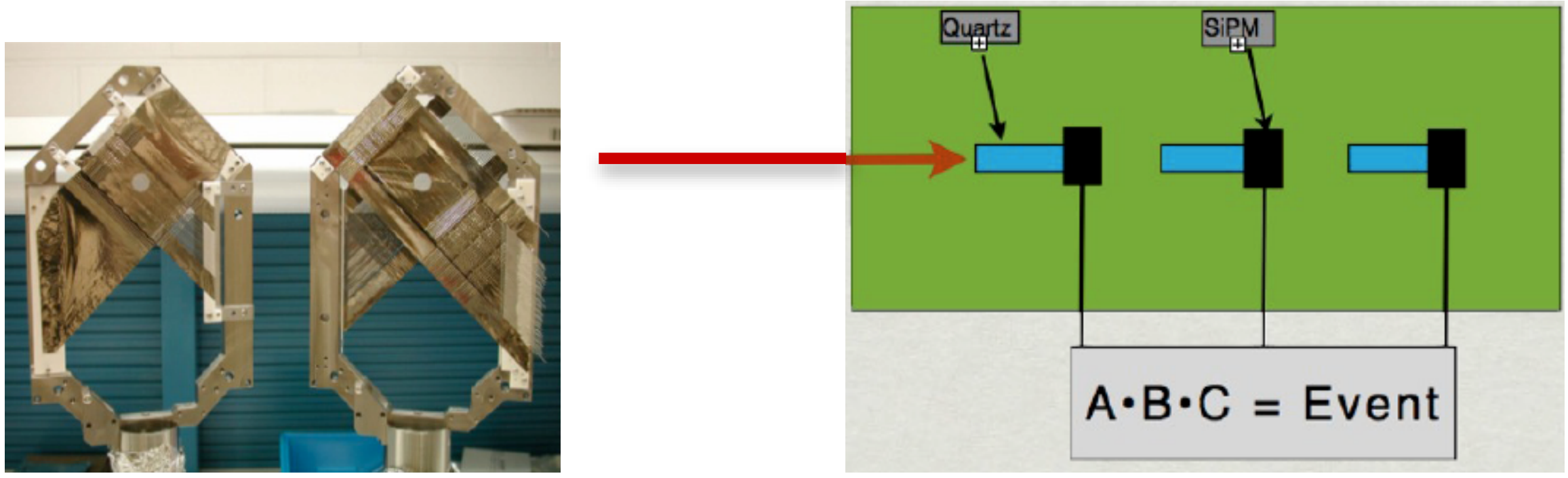

Figure 5.54: Conceptual schematic of fast extinction monitor.

***Long Time Scale/High Precision Extinction Measurement***

The primary tool for precision extinction monitoring will be a target monitor designed to detect relatively high energy (~4 GeV) particles from the target with a time resolution that is able to separate particles produced "in time" with the proton bunches from those produced between bunches. The extinction is then obtained by taking the ratio of "out of time" to "in time" signal events integrated over many bunches. This device must be able to measure extinction at the $10^{-10}$ level in about an hour. Assuming $7.2x10^{9}$ bunches per hour, this requirement can be met by designing a filter that provides at least 14 signal particles per bunch provided that the background rate in the detector is negligible. A background rate of 1 event per hour would require at least 30 signal events per bunch to achieve the same sensitivity. To accomplish this, a "filter structure" will be situated above the beam dump [41]. A schematic view of the filter is shown in Figure 5.55. A permanent magnet channel will be combined with entry and exit collimation regions in order to isolate a sample of charged particles with a well-defined direction. A detector consisting of 6 planes of 4x4 cm pixel detectors will be used to count signal particles exiting the filter by reconstructing straight tracks that are closely aligned with axis of the exit collimator. It will do so with a time resolution of 25 ns. The momentum spectrum and rate distribution of signal tracks exiting the filter are shown in Figure 5.56. This device is optimized to measure extinction at the $10^{-10}$ level in about an hour. In order to monitor the performance of the in-ring extinction on a faster time scale, a second monitor will be placed upstream of the AC dipole and extinction collimation system. It will utilize one of the technologies described in the previous sections.

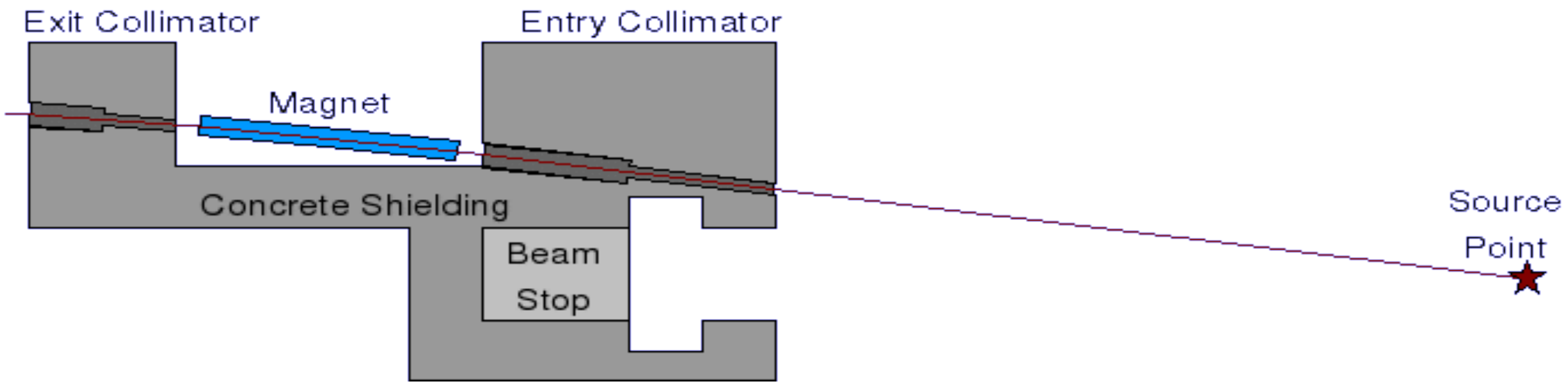

Figure 5.55. Conceptual design of target monitor “filter”.

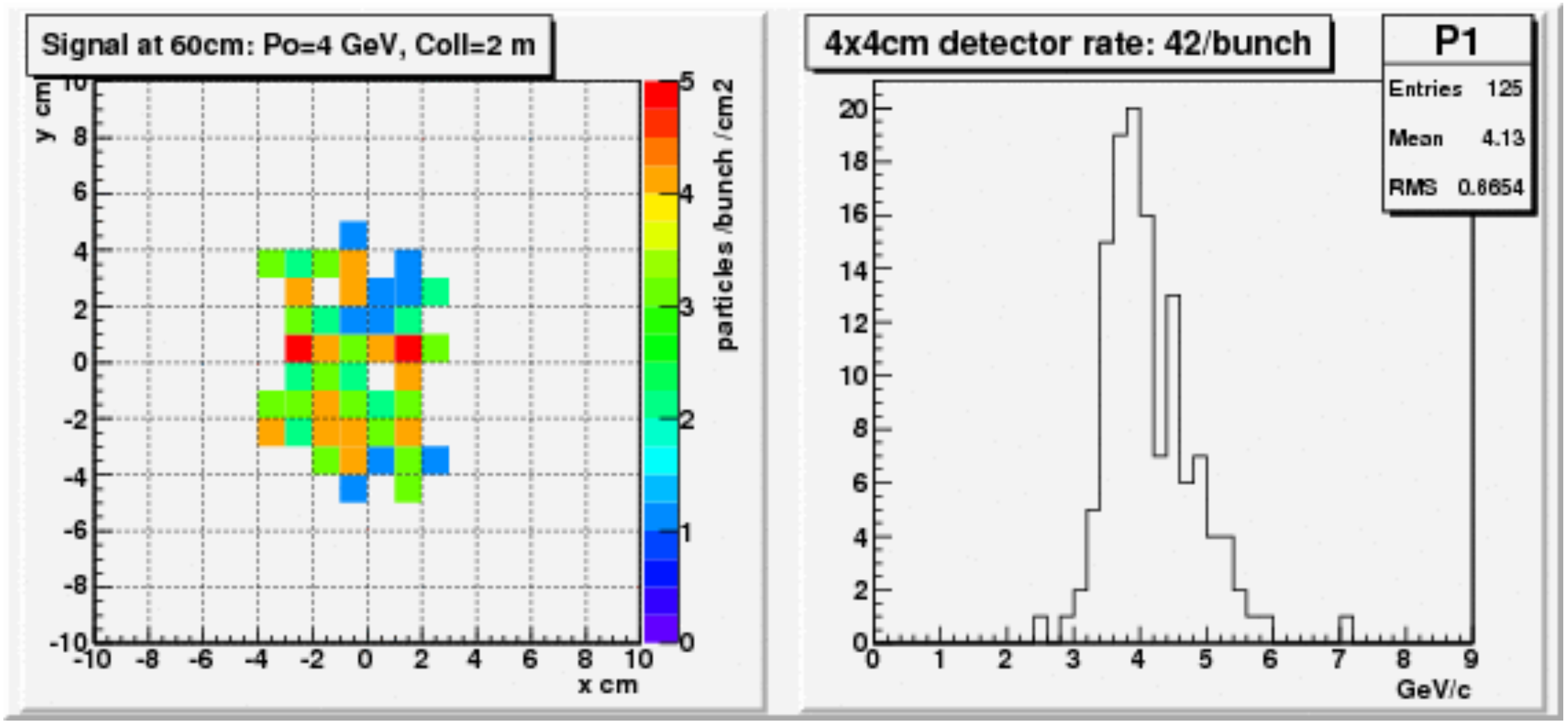

Figure 5.56. Momentum spectrum and rate distribution for particles passing through the filter structure within 0.015 radians of the nominal exit trajectory.

## 5.9 The Mu2e Pion Production Target Station

The Mu2e Production Target Station consists of three devices: the pion production target, the production solenoid heat and radiation shield and the proton beam absorber.

Figure 5.57 shows the Target Station layout. The target lies near the center of the cavity of the heat and radiation shield (HRS), and in turn, the HRS resides, under vacuum, in the inner bore of the Mu2e Production Solenoid cryostat. The proton beam enters the beam entry port in the heat shield, strikes the target, and produces a forward-going shower of particles, and a more diffuse spray of lower-energy secondary particles. Finally, the proton beam absorber intercepts the spent beam.

Water cooling is required only for the heat and radiation shield (HRS). The HRS and the PS vacuum closure bulkhead hold the water cooling system in place. The tungsten production target can be changed using a remote handling system. The proton beam absorber is unmovable and part of the concrete structures in the building.

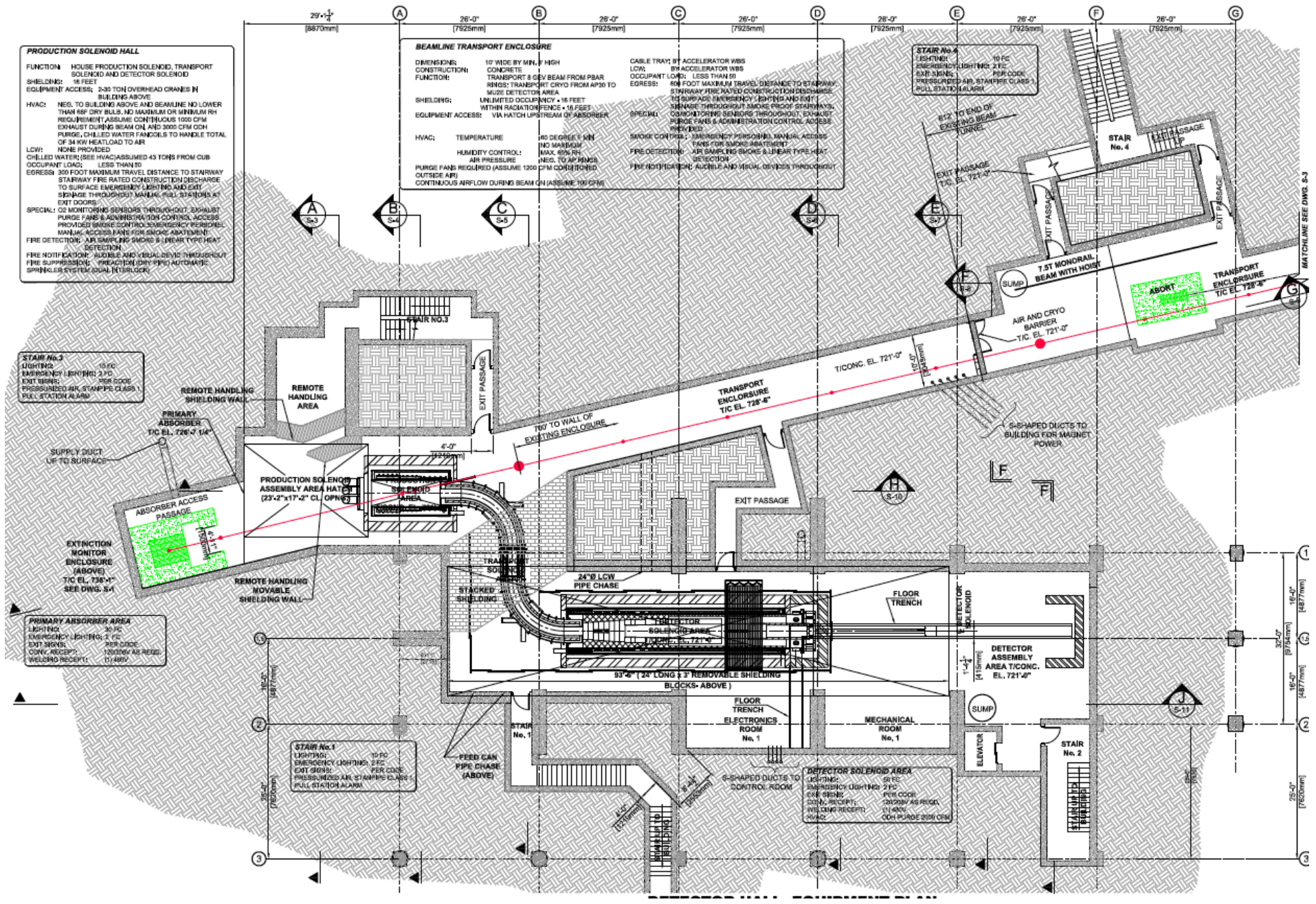

Figure 5.57. Layout of the Mu2e experiment and the proton beam line (in red).

### *5.9.1* The Mu2e Production Target

The Mu2e experiment requires about $4\times10^{20}$ protons over 3 years directed onto a pion production target. The proton interactions will produce pions that subsequently decay to muons. The primary beam will consist of 8 kW of 8 GeV (kinetic energy) protons distributed into micro pulses of ~$3.7 \times 10^{7}$ protons delivered every 1.7 μsec. The proton beam dimensions will be about 1 mm in radius (rms) at the target and about 200 nsec in duration (see Figure 5.13). Reference [1] contains a detailed description of the proton beam. The target requirements are given in Reference [43].

The production target must function in a harsh and complex environment presenting a number of technical challenges, including operation in vacuum. However, the basic requirements for the production target are quite straightforward. The technical details for a target system that satisfies these requirements are considerably more complex.

The production target must be designed to maximize the number of stopped muons at the stopping target in the Detector Solenoid. Maximizing the number of stopped muons depends on many factors that must be optimized simultaneously. These include the target material, density, shape, size, position, and orientation of the target, as well as the target support structures. Other factors (external design constraints) not directly related to the production target also have an impact, including the details of the proton beam, the magnetic fields of the solenoids, the clear bore of the Production Solenoid, the collimators in the Transport Solenoid and the details of the stopping target design.

The target material must have a high atomic number and density to ensure a high rate of beam-target interactions. The pion production cross section of the target material must be large enough to allow Mu2e to produce the required number of stopped muons. Operating in a solenoid with a graded magnetic field induces charged pions to follow spiral trajectories, which can return to the target region, possibly suffering re-absorption on some part of the target system.

The target system geometry is constrained by several factors. The effect of the overall alignment, thermal distortion of target and supports, and survey uncertainty on the muon stopping yield should be negligible. Given the choice of beam and target size, alignment tolerance must be less than about 0.5 mm to avoid losing more than a few percent of the targeting efficiency. The impact of pion re-absorption on stopped muons suggests a target that is compact, low mass, and presents a small geometric profile to the trajectories of pions in the Production Solenoid. The target, its supports, any associated infrastructure (e.g. cooling), and pion re-absorption must be included in calculations of stopped muon yield [44] and in heating calculations.

The main design challenge for the production target is direct heating from the energy deposited by the incident proton beam. The target must have high thermal conductivity to help achieve an acceptably low core temperature. The production target must be a material with a melting point well above the anticipated operating temperature. The target must maintain its mechanical integrity at high temperature. The production target is a high-risk device, which requires a thorough risk-failure analysis that clearly defines operating margins and develops procedures for managing target failure conditions.

The Mu2e pion production target is a radiation cooled tungsten rod, the size and shape of a pencil. The rod is 16 cm long with 0.3 cm radius.

The total power deposition in the rod for a steady state beam is 700 W. The power distribution is rotationally symmetric about its length; however, the power deposition rises for the first two cm along the length, reaches a maximum and decreases almost linearly to the end of the rod. The distribution also peaks radially at about 0.12 cm from the core. An ANSYS finite element model is shown in Figure 5.58 below [45]. Steady-state temperature and Von-Mises stress are shown in Figure 5.59.

The tungsten radiation cooled target is supported by tantalum spokes, as shown in Figure 5.60 and Figure 5.61. The target supports have a negligible effect on the stopped muon yield [45].

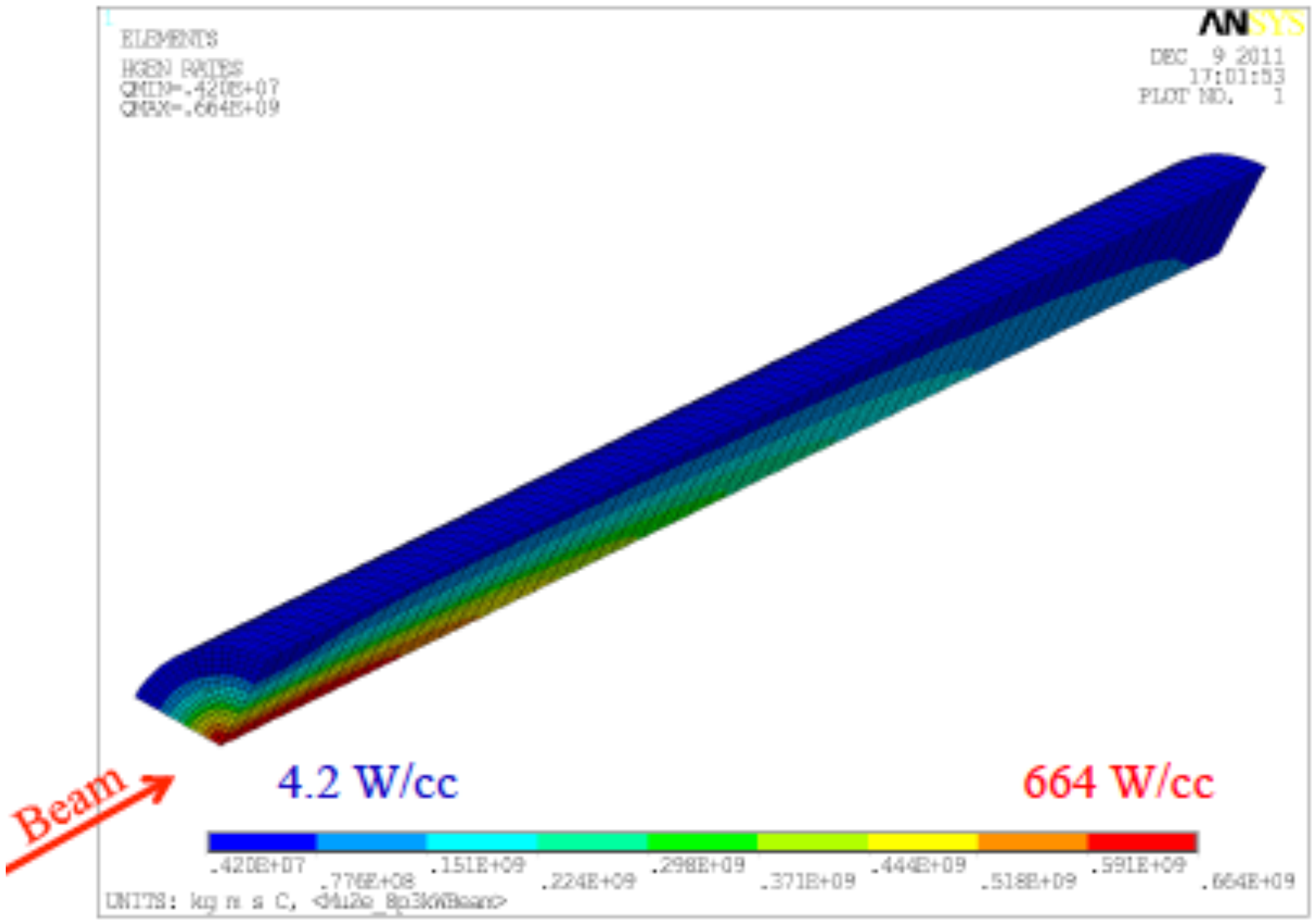

Figure 5.58: Thermal analysis showing the Watts per cc in the tungsten target from an 8 kW beam centered on the target.

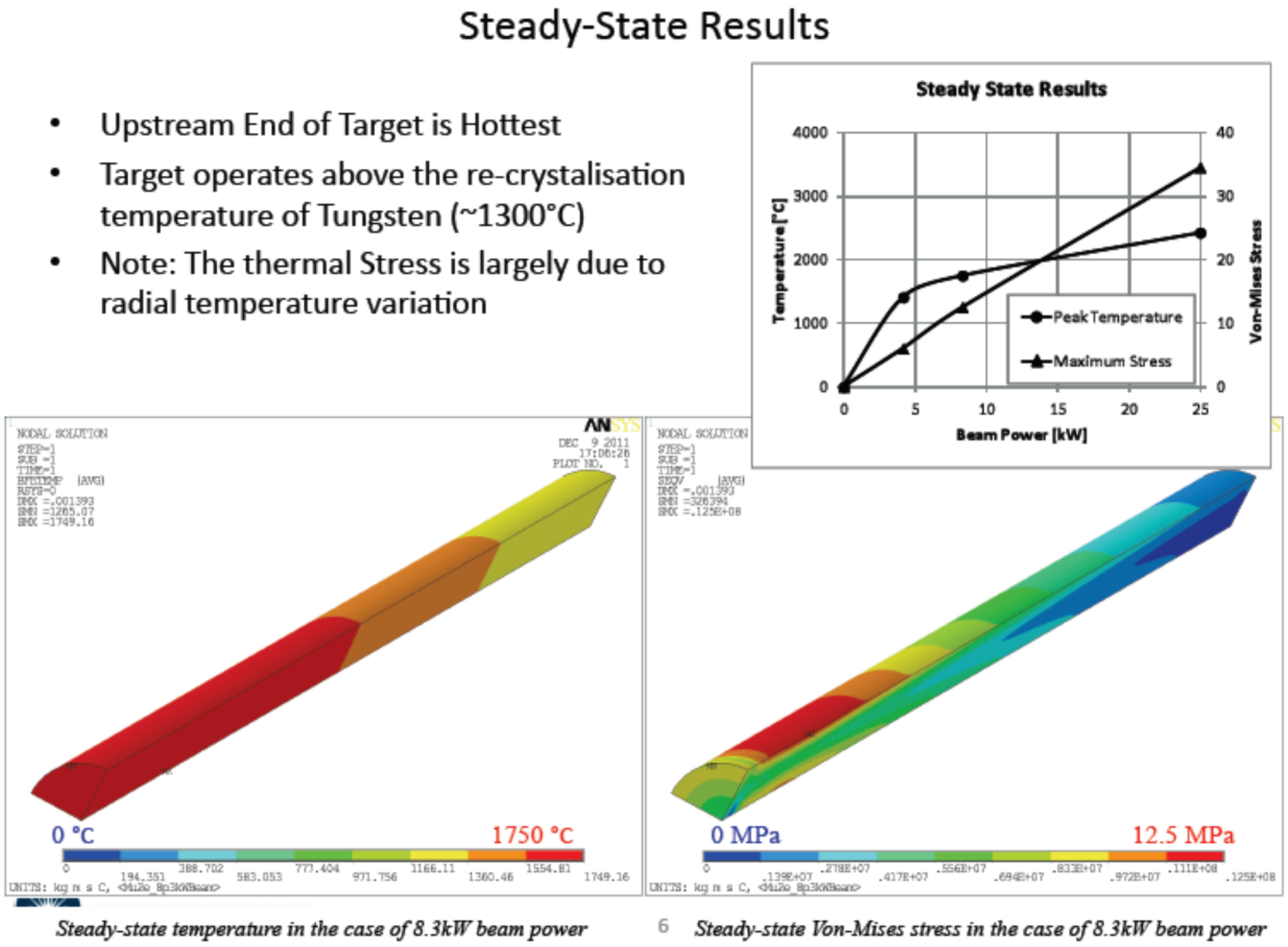

Figure 5.59: Steady-state temperatures and Von-Mises stress.

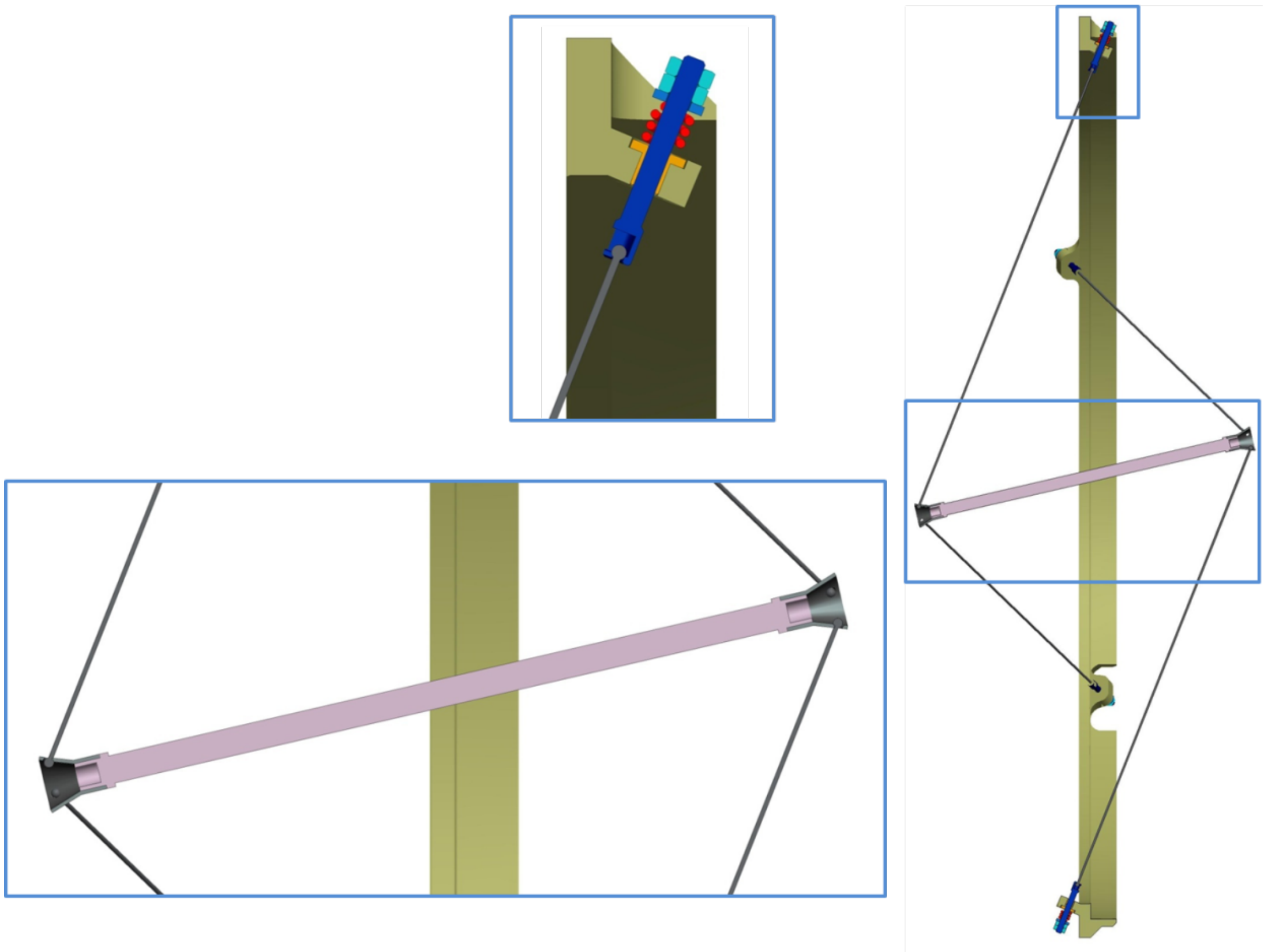

Figure 5.60: Radiation Cooled Tungsten Target supported by Tantalum spokes.

The production target region is a high-radiation zone that must be accessible for replacement and diagnostic purposes. A remote handling system has been developed to satisfy the technical need for access while ensuring personnel safety [46]. Figure 5.62 shows the major components of the remote handling system. The target support must interface with the heat and radiation shield, the hot cell & manipulator arms, movable shield wall and target insertion/extraction tool.

The target remote handling system provides a means to remove the downstream solenoid window, replace and dispose of the target. It also provides a means to replace the downstream windows in the solenoid end cap. The design assumes the use of a hot cell, manipulators, and a target insertion & extraction tool.

The facility design provides a surface hatch downstream of the production solenoid. During experiment operation, this hatch will be filled with concrete shielding blocks. In the event of a target change, a rented portable crane will be used to remove the hatch, unstack the shielding, and place the target insertion and extraction tool in the target hall. Technicians working within the hot cell would use the manipulators and the target insertion and extraction tool to access and remove the target. The spent target will be placed in a shielded cask, and removed by the crane. A new target would then be installed. At the conclusion of the target exchange, the target insertion and extraction tool would be removed by the crane. The shielding blocks would be replaced, and the surface hatch reinstalled. Remote handling equipment will be designed and built during construction. However, the rental of the crane (of order $100K per target exchange) would be an operations cost.

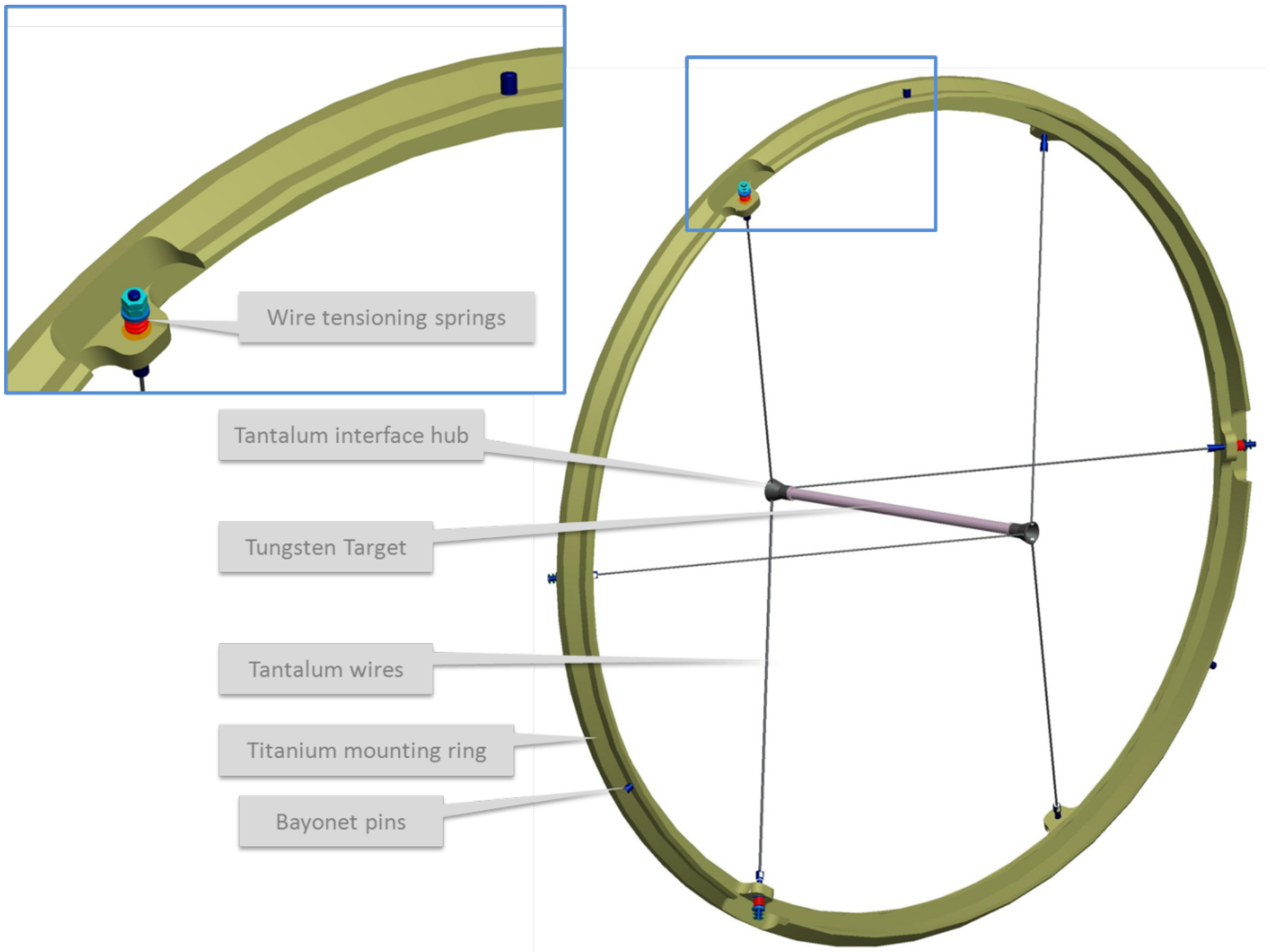

Figure 5.61: The engineering design (RAL) is shown here. The target is supported by tantalum spokes attached to the support ring (green).

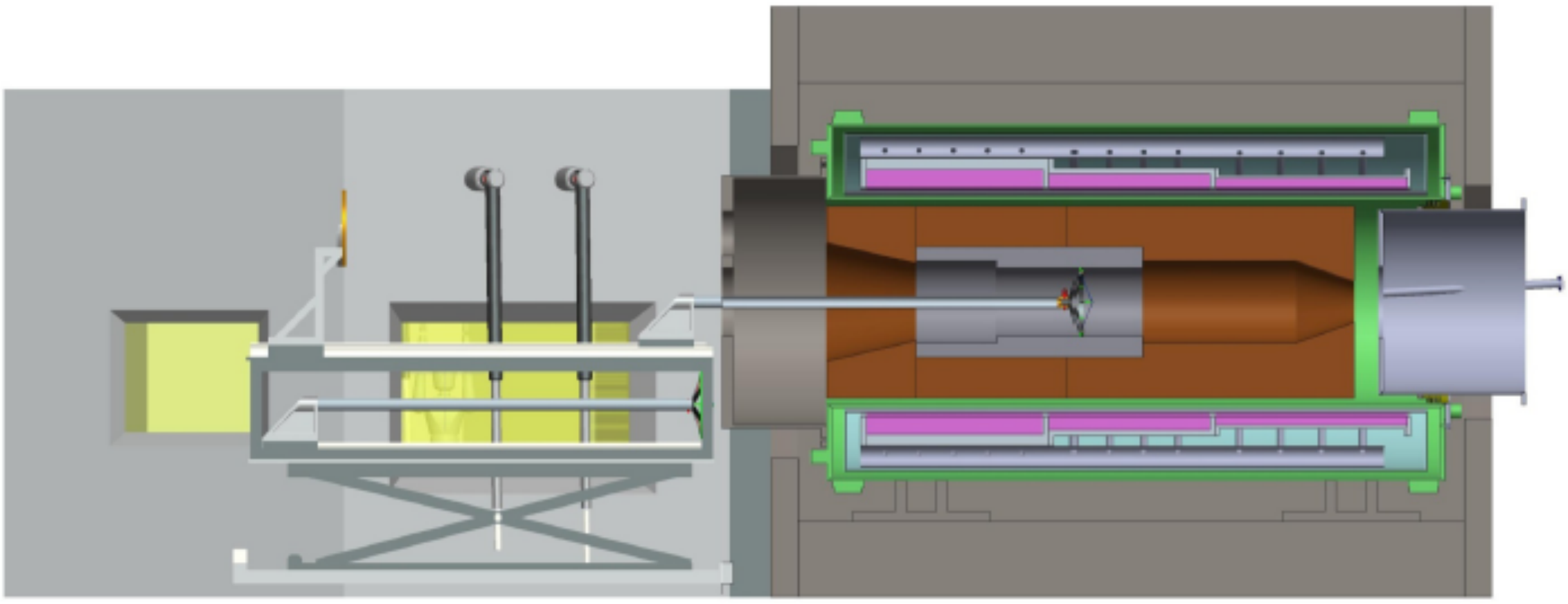

Figure 5.62: Cut-away view of the PS with the target "cartridge" inserted inside the heat and radiation shield, the target insertion & extraction tool, and the hot cell & manipulator arms.

### *5.9.2* The Mu2e Heat and Radiation Shield

The Heat and Radiation Shield (HRS) serves to protect the superconducting coils of the Production Solenoid (PS) from the intense radiation generated by the primary proton beam striking the production target within the warm bore of the PS. This shield also protects the coils in the far upstream end of the Transport Solenoid (TS). The HRS must have sufficient inner aperture to allow for efficient capture of pions and muons to maximize the muon stopping rate in the stopping target.

The heat shield is intended to prevent radiation damage to the magnet coil materials and prevent quenches of the superconducting PS. A detailed explanation of the HRS requirements is given in [47]. The HRS has four primary performance parameters:

1. Limit the continuous power delivered to the cold mass.
2. Limit the instantaneous local heat load allowed anywhere within the superconducting coils.
3. Limit the maximum local radiation dose to the superconductor epoxy over the lifetime of the experiment.
4. Limit the damage to the superconductor's aluminum stabilizer and copper matrix.

With regard to the fourth parameter above, at liquid helium temperature damage to the atomic lattice of a superconducting cable and its quench stabilizing matrix made from normal conductor takes the form of an accumulation of atomic displacements; i.e., tiny lattice defects. After exposing a metal sample to a given neutron flux spectrum, the damage can be characterized by the average number of displacements per atom (DPA). The DPA is directly related to electron transport in metals. The Residual Resistivity Ratio (RRR), defined as the ratio of the electrical resistance at room temperature of a conductor to that at 4.5 K, will decrease with exposure to radiation. However, warming the irradiated conductor to room temperature leads to recovery of the RRR [48] [49], but the degree of recovery depends on the metal. The PS utilizes superconducting cable embedded in an aluminum matrix for quench protection. Aluminum is one example of a material that shows complete recovery at 300 K. The annealing time is on the time scale of minutes at 300 K. The specification of the heat shield is that the RRR of the Al stabilizer in the coil package should not be reduced below 100 in a one year operating period. Warming-up to anneal once per year is consistent with the annual Fermilab Accelerator shutdowns that typically last for several weeks.

An acceptable shield design should establish the following limits for nominal operating conditions: the maximum allowable total heat load deposited in the PS coils must be less than 100 W. The most radiation sensitive material sets the lower limit of radiation tolerance; in particular, the epoxy used to bond the insulation to the superconducting cable can tolerate a maximum of 7 MGy before it experiences a 10% change in its shear modulus. This leads to a 350 kGy/yr limit that allows a conservative 20 years of operation. This limit of 350 kGy/yr is the equivalent of 30 μW/gm. The performance of the HRS is summarized in Table 5.11.

| | Peak DPA/yr* [E-5] | Peak Power Density [μW/g] | Rads/yr [MGy/yr] | Years Before 7 MGy | Watts |
|---|---|---|---|---|---|
| Specification | 4 to 6 | 30 | 0.35 | 20 | 100 |
| Performance [48] | 3.2 | 17 | 0.33 | 21 | 20 |

Table 5.11. Performance of HRS compared to specifications on the magnet coils.
* This is the DPA damage per year at which RRR degrades to 100. After this RRR reduction we must warm-up and anneal

The Heat and Radiation Shield, shown in Figure 5.63 and Figure 5.64, is constructed of several cylindrical bronze shells. The shield has an outer radius of 70 cm over most of the volume, but the radius is expanded to 73.75 cm in critical regions where the energy deposition is the largest. The PS inner cryostat wall, at a radius of 75 cm, will support the final HRS design. The length of the HRS is about 4 m. The pion production target is located in the HRS cavity. The inner shield wall is limited to no less than 25 cm [50] since smaller radii negatively affect the stopped muon yield. In addition, any acceptable shield design must avoid any line-of-sight cracks between components that point from the target to the inner cryostat wall and thus the magnet coils. The red block in Figure 5.63 at the proton beam entrance marks the location of the HRS protection collimator. The protection collimator is designed to ensure that no mis-steered proton beam directly hits the HRS or the solenoid coils.

The HRS must be dimensionally and otherwise mechanically stable. Current simulations suggest the shield will experience an average heat load of 3.3 kW. The HRS is inside the solenoid. Radiation cooling (passive) leads to unacceptably high operating temperatures in the HRS; therefore, the shield must be actively cooled. The cooling is accomplished by using water channels at the outer radius of the HRS. These water channels are alternated with support rod channels that connect the ring sections.

The materials used to construct the shield must not impact the magnetic field specifications of the solenoid system, therefore, non-magnetic materials must be used (magnetic permeability less than 1.05). All conducting materials must be designed to reduce eddy current forces that can arise during a quench. We have chosen bronze C63200 which satisfies these requirements. It can also be manufactured in large forged pieces. A thermal analysis has been completed and the results are shown in Figure 5.65 below [52]. The performance of the shield limits the heat and radiation to the superconducting to an acceptable level.

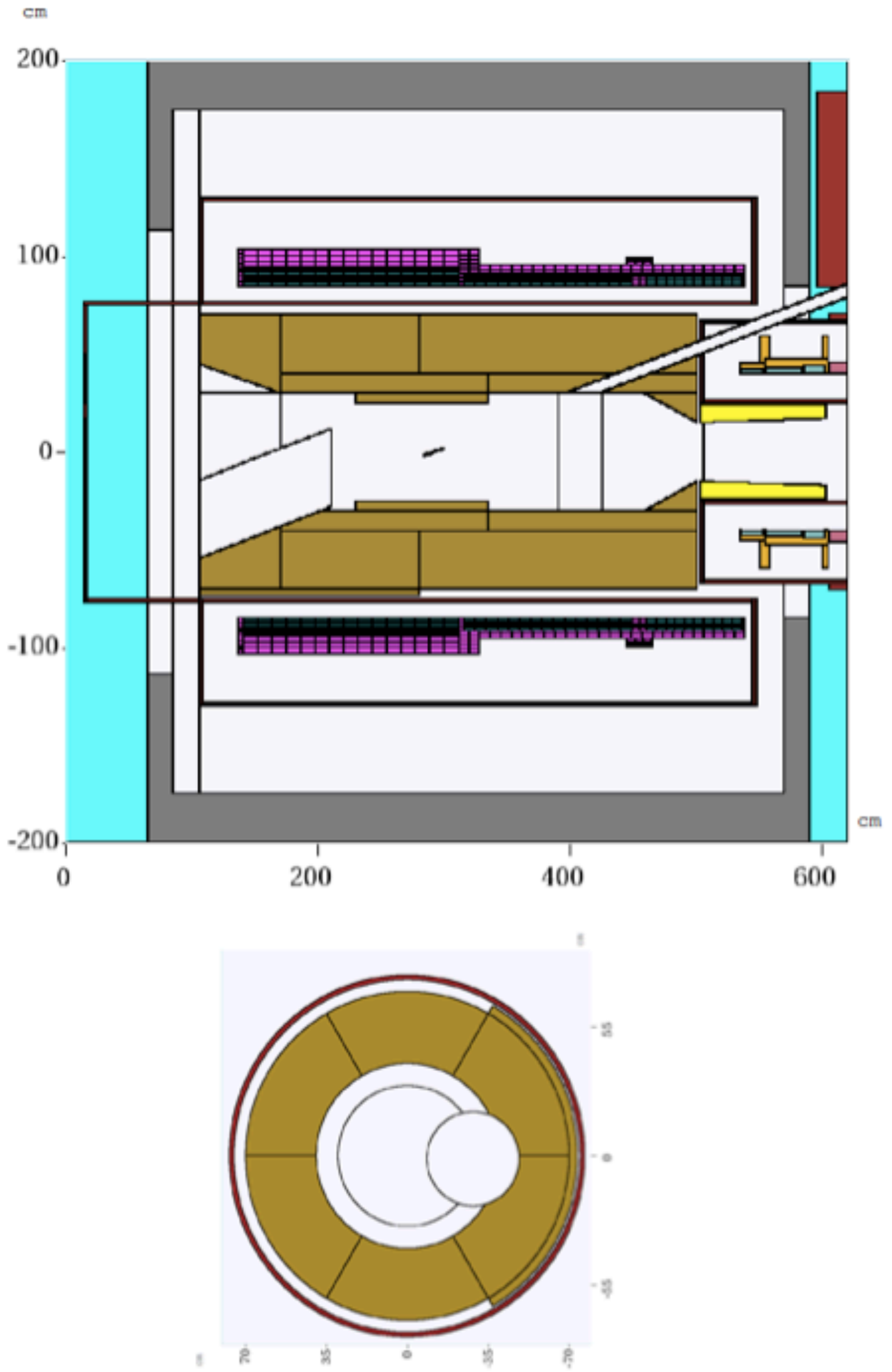

Figure 5.63: PS Heat and Radiation Shield geometry simulated in MARS15. Note the target lies on the magnet axis in the plane shown. The target is tilted 14° away from the axis. The HRS is shown in brown, which represents bronze material.

### *5.9.3* The Mu2e Proton Beam Absorber

The Mu2e beam absorber [53] stops the proton beam and secondary particles that make their way through and beyond the target in the forward direction. The beam power from the accelerator complex is 8 kW, and while 0.7 kW will be deposited into the target itself, and 3.3 kW more will be absorbed by the production solenoid heat shield, a significant amount of power is contained in the beam absorber. The beam absorber must be shielded so that its prompt and residual radioactivity does not significantly contribute to the radiation dose rate at the downstream end of the production solenoid enclosure.

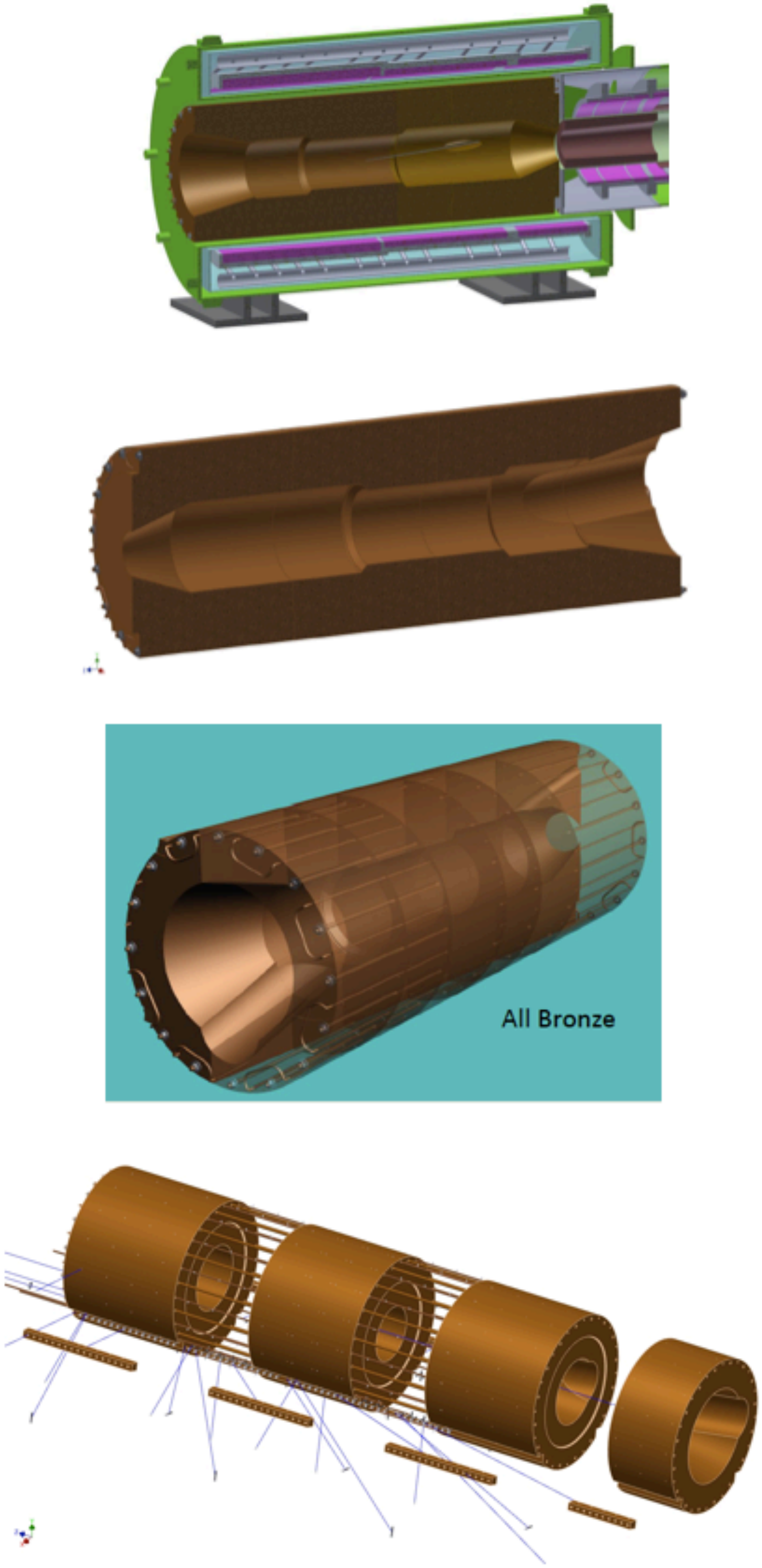

Figure 5.64. The Heat and Radiation Shield shown in 4 views.

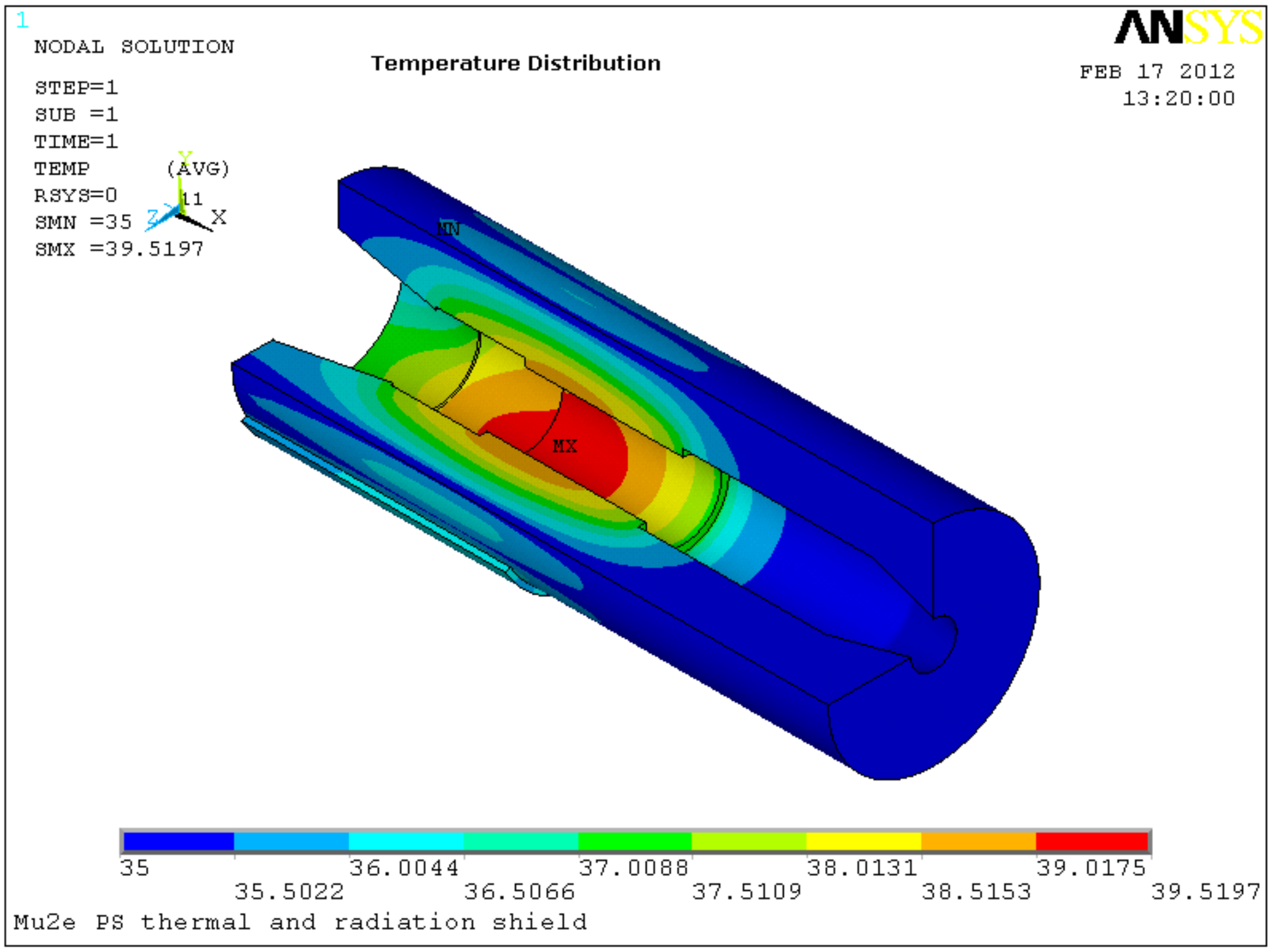

Figure 5.65: Temperature profile of the Heat and Radiation Shield.

The beam absorber must be able to accept the entire beam power in the event that the target is missed, or during pre-targeting beam tests. The beam absorber must be placed outside of and well beyond the Production Solenoid to allow access to the crane hatch and room for remote target exchange equipment. The beam absorber must be compatible with the extinction model located above and behind the beam absorber [54].

The performance of the beam absorber was evaluated for a beam power of 25 kW, following the initial proposal for Mu2e. Using the Revised Fermilab Concentration Model [55], the impact of the beam absorber on the surrounding environment was evaluated. The calculations demonstrated that average concentration of radionuclides in the sump pump discharge will be 33 pCi/ml due to tritium and 3 pCi/ml due to sodium-22. This is 30.5% of the total surface water limit if pumping is performed once a month (conservative scenario). Build-up of tritium and sodium-22 in ground water at $4\times10^{20}$ protons per year will be as low as 0.06% of the total allowed limit over 3 years of operation. Estimates show that airflow should be below 500 cfm in the configuration without a pipe connecting the target region to the beam dump (average flux over the whole hall volume is 1.65 $cm^{-2}s^{-1}$) [56].

The absorber must be able to accept the total number of protons required by the experiment, $4\times10^{20}$ over several years, plus an acceptable overhead to account for commissioning and tuning (100%), without replacement over the life of the experiment. The transverse dimensions of the absorber must be consistent with the beam properties as described in the Mu2e Proton Beam Absorber Requirements Document [53], accounting for distance from the target and divergence of the beam, including scattering in the target. The transverse proton beam size at the absorber face has a sigma of 1.3 cm in both planes.

The proposed absorber, shown in Figure 5.66, consists of an Al core (with Fe as an alternate) with the dimensions 1.5 × 1.5 × 2 m and concrete shielding with the dimensions 3.5×3.5×5 m, so that the core is surrounded by 1 m of concrete from the sides and the bottom. It has a 1.5×1.5 m opening toward the beam and also a 2.5 × 2.5 × 1 m albedo trap to protect the downstream end of the Production Solenoid from the secondary particles generated by the spent proton beam in the Al core.

Simulations [56], [58] show that peak power density in *accidental* mode (primary beam misses the target) will be 0.6 W/g for the Al core (Figure 5.67). In the *normal operation* mode (primary beam hits the target) where the spent beam hits the absorber, the respective values will be 7 mW/g for an Al core. Residual dose on contact with the concrete shielding of the beam absorber after 30 days of irradiation and 1 day of cooling will be at the level of few mSv/hr. Dynamic heat load in the dump is presented in Table 5.12.

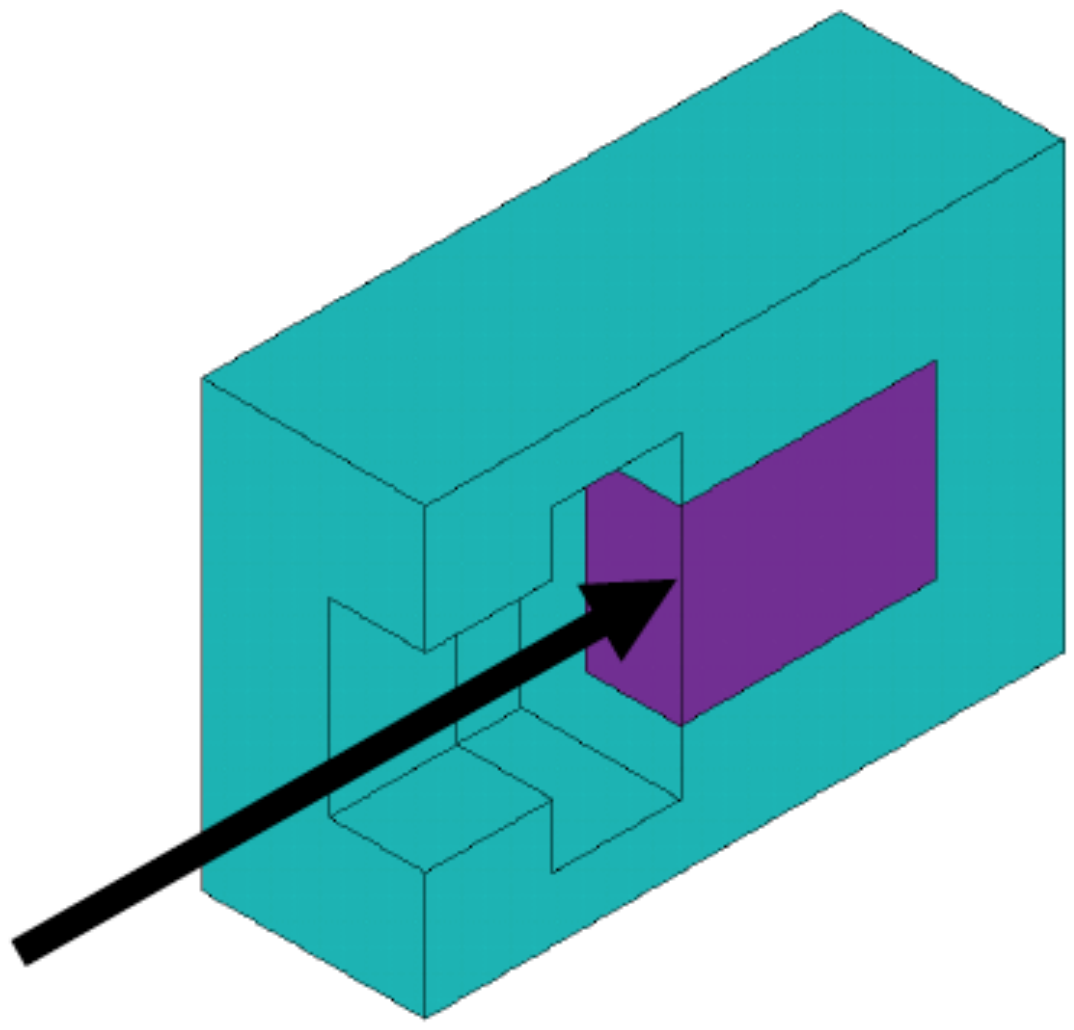

Figure 5.66: A cut-away view of the proton beam absorber. The purple rectangular volume is the aluminum absorber and the surrounding material and entrance are concrete.

| Dump/mode | Q(kW) |
|---|---|
| Al, accident | 5.8 |
| Fe, accident | 6.7 |
| Al, operation | 1.4 |
| Fe, operation | 1.7 |

Table 5.12: Power deposition in the proton beam absorber: The "accident" condition refers to the proton beam missing the target and the "operation" condition refers to the proton beam striking the target.

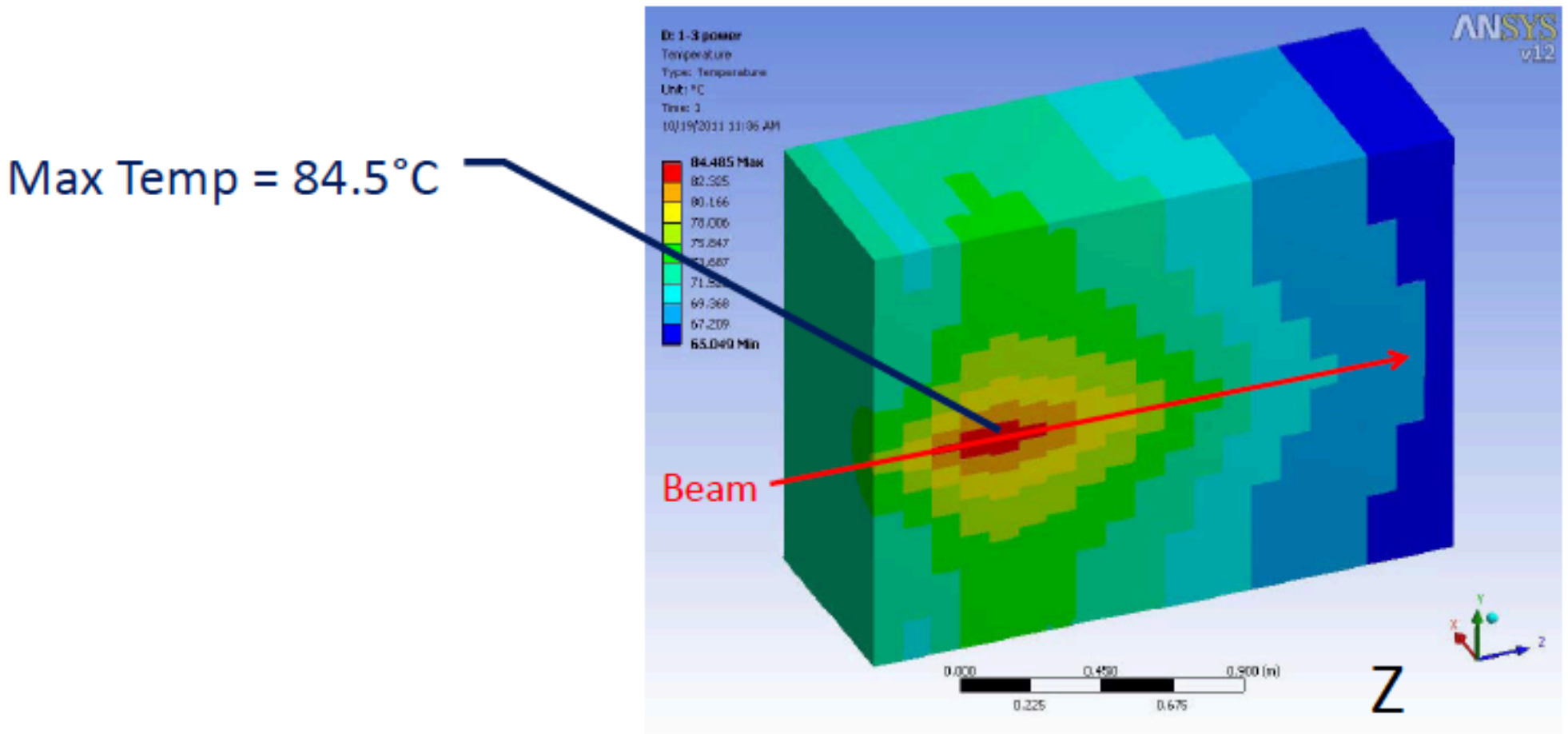

Figure 5.67: Temperature distribution for an air-cooled proton beam absorber for the case where the beam misses the target and the full beam power is deposited in the absorber.

The results of MARS15 [56] and ANSYS calculations [57] show that for passive cooling during operation mode, the peak power densities lead to temperature conditions below melting. The melting point for aluminum is 660°C. Though passive cooling would be preferable, there are a few drawbacks to passive schemes. For natural convection, one would want a reservoir air space above the absorber. This may interfere with shielding and/or the extinction monitor. Conduction to the concrete requires one to make assumptions about the quality of the aluminum/concrete thermal contact and thermal continuity of the concrete. Conduction to the earth through additional means may be promising and merits consideration. For now our baseline is the conservative solution of active cooling with air [58]. The aluminum core will be constructed using 20 plates of 100 mm thickness along the beam. To facilitate air cooling the plates will have machined gaps in some areas of plate-to-plate interfaces to provide channels for air flow. Air flows in the transverse direction to the beam. The air flow system consists of a Vortex blower with inlet filter, duct system, manifolds (integral with concrete shielding) and a vent to the target hall with outlet filter.

# 5.10 Radiation Safety Plan

## *5.10.1* Introduction

The Mu2e experiment requires the delivery of an 8 kW proton beam from the Fermilab Booster to the Recycler Ring, transport through the P1, P2, M1 and M3 lines followed by injection into the Delivery Ring. From the Delivery Ring, beam is to be resonantly extracted to a new external beam line and transported to a new experiment hall that is to be constructed near the Delivery Ring enclosure. The purpose of this section is to discuss the adequacy of the existing Accumulator/Debuncher facility shielding and to describe the modifications and controls that will be required to operate the facility for the Mu2e experiment within the requirements of the Fermilab Radiological Controls Manual (FRCM) [64].

The naming convention of various machines requires some explanation. Some beam line and accelerator names are changing, in part, because their repurposing from antiproton production to proton delivery for the Mu2e experiment requires reconfiguration. In some cases, beam lines are being removed and other lines are being installed. Reference is made to both the old and new configurations because radiation shielding measurements for the old configuration are directly applicable to the new configuration. Table 5.13 can be used as a cross reference for the old/new naming conventions.

| Old Name | New Name | New Purpose |
|---|---|---|
| Recycler Ring | Recycler Ring | 8 GeV beam RF manipulations |
| P1 line | P1 line | Transport from RR to Main Ring F0 |
| P2 line | P2 line | Transport from F0 to F17 |
| AP1 line | M1 line | Transport from F17 to M3 |
| AP3 line | M3 line | Transport from M3 to Delivery Ring |
| Debuncher Ring | Delivery Ring | Prepare beam for slow resonant extraction |
| - | M4 line | Beam line from Delivery Ring extraction to Production Solenoid |
| Accumulator Ring | - | |

Table 5.13. Naming conventions for antiproton source machines (old) and Mu2e/g-2 (new) machines.

## *5.10.2* Scope

Batches of 8 GeV protons will be produced in the Fermilab Booster and transported to the Recycler Ring where they will be re-bunched into four 2.5 MHz bunches. One of the four bunches will be kicked out of the Recycler and transported through the P1 line, P2 Line, M1 line, M3 line and then injected into the Delivery Ring. The bunch will be transferred from the Delivery Ring by slow resonant extraction through the new M4 beam line to the Production Solenoid. The remaining three bunches will be similarly transported to the Production Solenoid. 267 ms later, the next Booster batch will be delivered to the Recycler Ring where the process is repeated in the same sequence. A pair of Booster batches will be produced every 1.33 seconds or once every Main Injector acceleration cycle.

The primary focus of this Radiation Safety Plan is to assess the radiation safety concerns, primarily radiation shielding, for the downstream M1 line, the M3 line, the Delivery Ring, the new M4 extraction line, the production target and solenoid, and the beam absorber. The MI8 line, Recycler Ring, P1 line, P2 line, and upstream M1 line that are part of the beam transfer to the Delivery Ring are also considered.

A fundamental issue associated with repurposing the Debuncher Ring arises from the radiation skyshine phenomenon. A shielding assessment conducted for the Antiproton Source in 2000 for Tevatron Collider Run II [65] provides the basis for evaluation of the skyshine. The radiation shield thickness between the Accumulator/Debuncher Service Buildings and underlying tunnel is a total of 10 feet. The partial shield composition due to the concrete tunnel ceiling and the service-building floor is a total of 1.5 feet of concrete. The remaining shield space consists of river washed gravel that has a packing efficiency of perhaps 80%. A calculation [66] has shown that a total beam loss of 250 Watts due to normal operations distributed among the three service buildings would deliver an effective dose of 10 mrem per 4000 hours of operation at a distance of about 500 meters. If permitted to exist, such an annual dose triggers a DOE reporting level and/or would restrict public access to large areas of the laboratory that presently do not have restricted access. Since the anticipated normal beam loss in the Accumulator/Debuncher Rings for the Mu2e experiment may be considerably larger, an improved radiation-shielding scheme is required for the Accumulator/Debuncher Service Buildings.

The present shielding for the Accumulator/Debuncher arcs and beam transport lines, including sections of AP1 and AP3, while adequate for the Anti-proton production, will require some mitigation when repurposed for the Mu2e experiment.

### *5.10.3* Radiation Safety Plan Elements

The Radiation Safety Plan consists of 11 elements as follows:

- Radiation Skyshine
- Total Loss Monitoring system
- Beam collimation systems

- Supplemental shielding requirements
- Labyrinths and Penetrations Evaluation
- Radiological Posting and Fencing Requirements
- Entry Controls
- Residual Radioactivity Control
- Air Activation
- Surface Water Activation
- Ground Water Activation

Radiation Skyshine has become the dominant concern and is discussed first since the solution for the skyshine problem is the most challenging. A proposed Total Loss Monitoring system (TLMs) is a key element of the Radiation Safety Plan and is discussed next because the protection it provides sets the scale for the level of protection required by other elements of the Plan.

***Skyshine***

Radiation skyshine can occur when a shower of energetic neutrons exits a thinly shielded zone. The emitted neutrons can travel large distances since interactions in the atmosphere occur rarely due to the density of the atmosphere. It is the usual practice to design a radiation shield consistent with the power of the associated beam such that secondary particles produced by beam losses would be absorbed. It is a relatively simple matter to determine the thickness of the required radiation shield. For example, the radiation dose rate due to the total loss of a 25 kW, 8 GeV beam on an iron magnet five feet from a beam enclosure ceiling can be reduced to less than 1 mrem/hr by a radiation shield of nominal density that is about 21 feet thick.

Figure 5.68 shows a measurement from the 2000 pbar shielding assessment of a deliberate beam loss of about 12.8 Watts on an iron magnet about 5 feet from the tunnel ceiling in the Accumulator/Debuncher Ring. The total thickness of the shielding in this location is 10 feet, but the performance of the shield indicates that the shield is effectively 7 feet thick. The reduced performance is attributed to the use of river-washed gravel in the intervening shield with a packing efficiency of perhaps 80%.

The geometry shown in the Figure 5.68 is typical of the Accumulator/Debuncher storage rings. A calculation of the radiation skyshine emitted from such a shield with an incident beam power of 25 kW has been completed [66]. One result of this calculation, shown in Figure 5.69, is that significant radiation dose rates are possible at considerable distance from the Accumulator/Debuncher Rings. The potential for measurable radiation exposure to non-radiation workers as well as members of the public can be realized unless some protective measures are implemented.

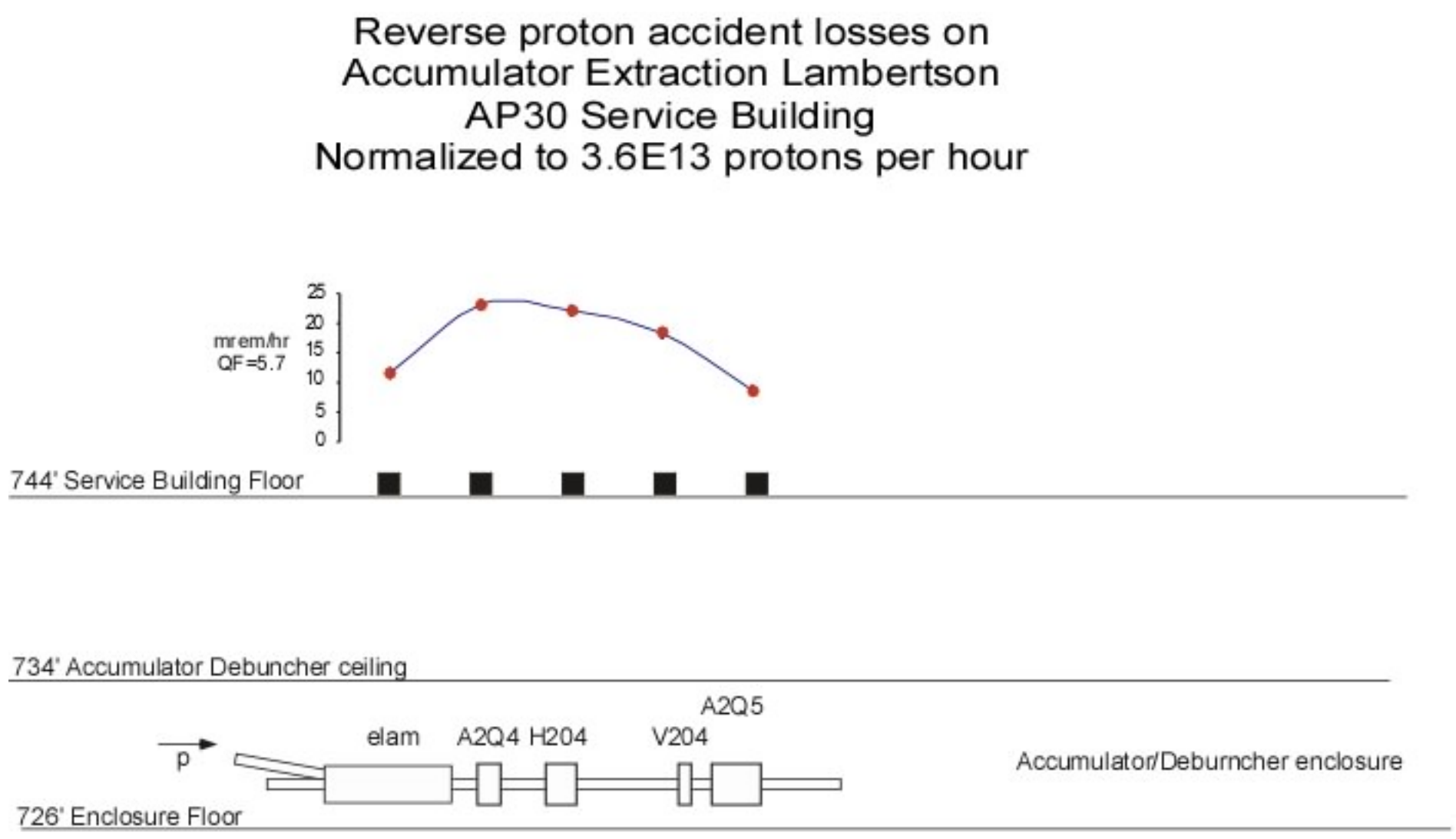

Figure 5.68. A typical measurement from the 2000 pbar shielding assessment. This particular measurement is of a deliberate beam loss of about 12.8 Watts on an iron magnet about 5 feet from the AP30 tunnel ceiling in the Accumulator/Debuncher Ring. The shield thickness is 10 feet at this location.

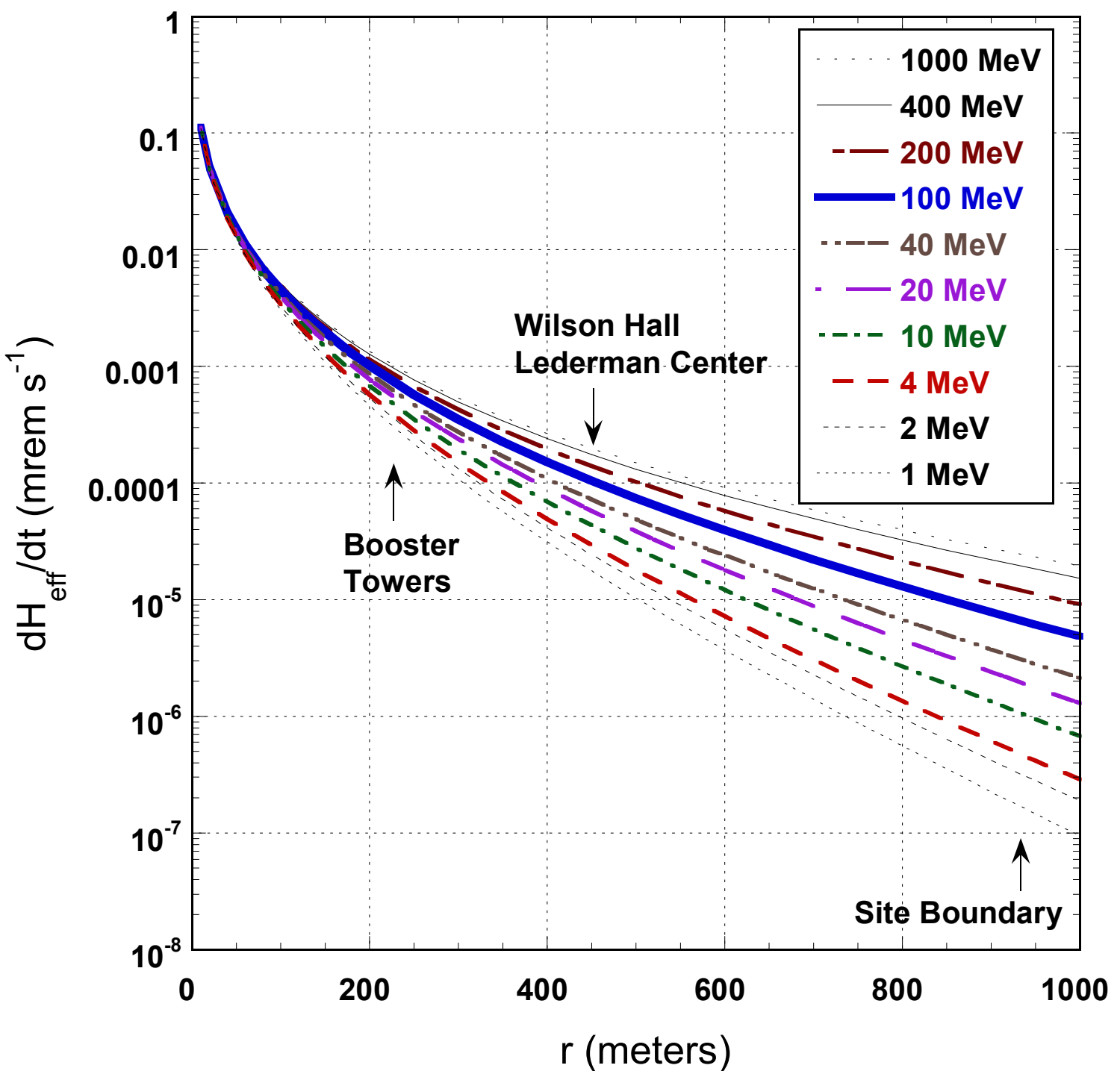

Figure 5.69. Radiation effective dose rate as a function of distance due to a continuous 25 kW beam loss at the Accumulator Injection Lambertson Magnet.

The mitigation factor required to protect against a 500 Watt beam loss distributed among the three Accumulator/Debuncher service buildings was calculated; the result is shown in Figure 5.70. Based on a 500 Watt beam loss persisting over an operational year of 4000 hrs, the shielding effectiveness must be improved by a factor of 30 to 50 in order to limit the effective dose rate to less than 1 mrem per year at a distance of about 500 meters from the Accumulator/Debuncher rings. Alternatively, limiting the lost beam to a maximum of 10 Watts would give essentially the same reduction factor of 50. It is this approach, the control and limitation of beam loss at the Accumulator/Debuncher service buildings, which is to be used in the radiation safety plan for Mu2e. The contributions of radiation effective dose from other radiation sources such as the M3 line berm and the Accumulator/Debuncher service building exit stairways must be included in this analysis. The sum of effective dose rates from all such sources must be considered to ensure that the annual effective dose guidelines are met.

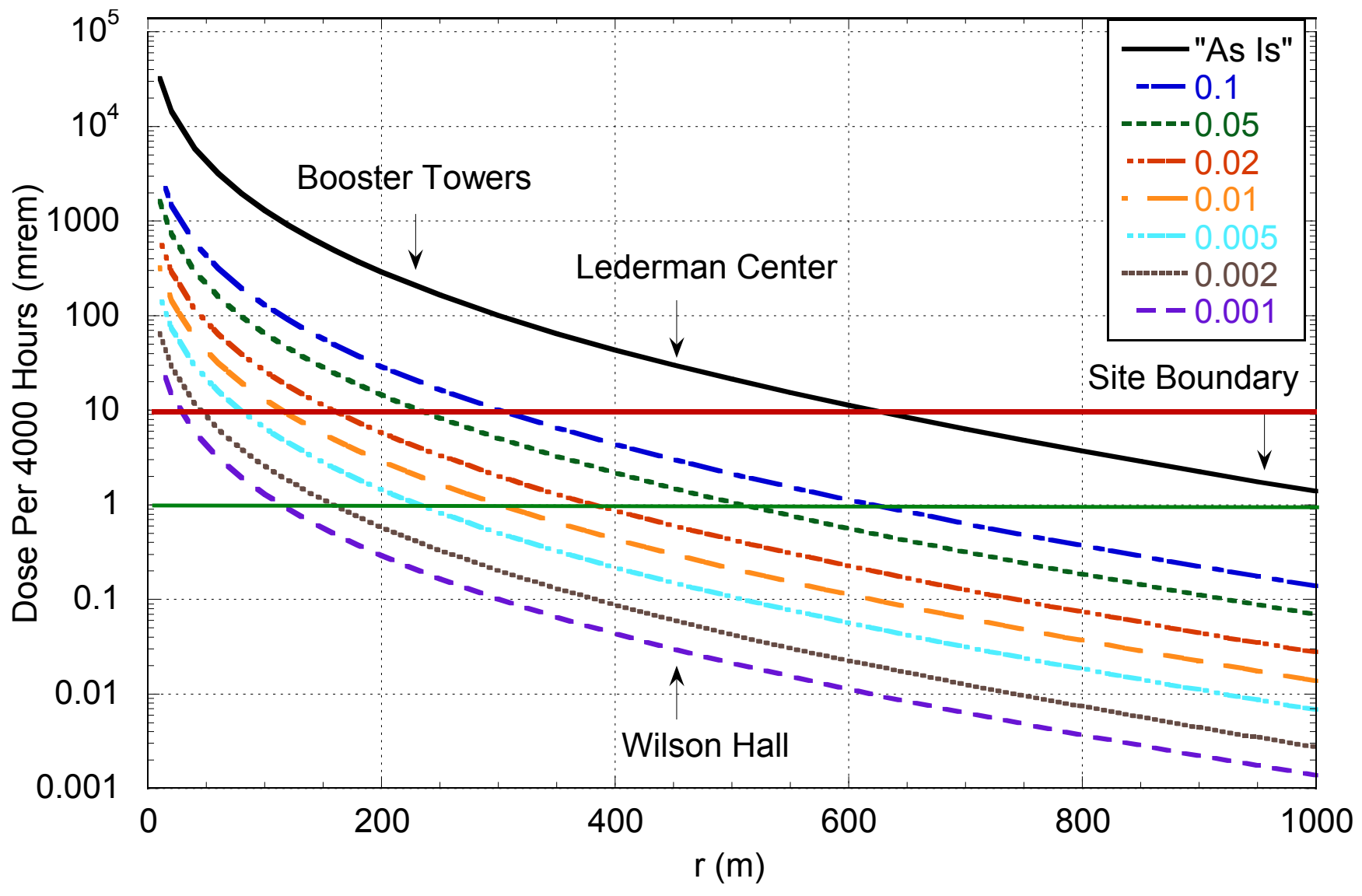

Figure 5.70. Calculated reduction factors to meet equivalent dose limits as a function of distance. The present shielding must be enhanced by a factor of 30-50 to reduce the dose at 500 m due to a 500 Watt beam loss to less than 1 mRem/year.

### *Total Loss Monitoring System*

A Total Loss Monitoring system (TLMs) is an ion-chamber-based Radiation Safety System that has been conceived to effectively limit the duration and intensity of uncontrolled or unintended beam loss. By limiting the duration and intensity of beam loss in the various beam enclosures, the need for additional passive shielding is reduced or eliminated. In addition, the transmission of radiation through various labyrinths, ducts, and penetrations can also be limited by to levels permitted by the FRCM. The TLM consists of an HJ5-50 heliax cable containing argon gas with a high voltage applied to the outer shield. Electric charge resulting from ionization of the argon gas due to interaction with beam shower is collected and measured with an electrometer. The amount of charge collected has been found to be directly proportional to lost beam. The relationship of beam loss and radiation dose rates outside of the passive shielding at the Antiproton Source Facility was extensively studied and documented in a shielding assessment conducted in 2000. By combining the results of these known relationships, a comprehensive system to limit radiation dose rates outside of the beam enclosures for the Mu2e experiment has been developed. Extensive testing for TLM systems has been ongoing since June 2011. The Total Loss Monitoring system can be used to limit beam loss for a wide variety of applications including at facilities with insufficient passive shielding, control of radiation skyshine, control of beam loss at labyrinths and penetrations, and the control of residual radiation levels for worker protection.

The TLM response has been measured for a number of beam loss geometries at an energy of 8 GeV. The charge collection per $10^{10}$ protons varies with the mass of objects into which the beam is lost. TLM response from beam loss in more massive objects tends to be lower than is the case for less massive objects. The TLM response for the accident condition shown in Figure 5.68 was measured using three different lengths of TLM. The result, shown in Figure 5.71, has been determined to be about 6 nC of charge per $10^{10}$ protons lost.

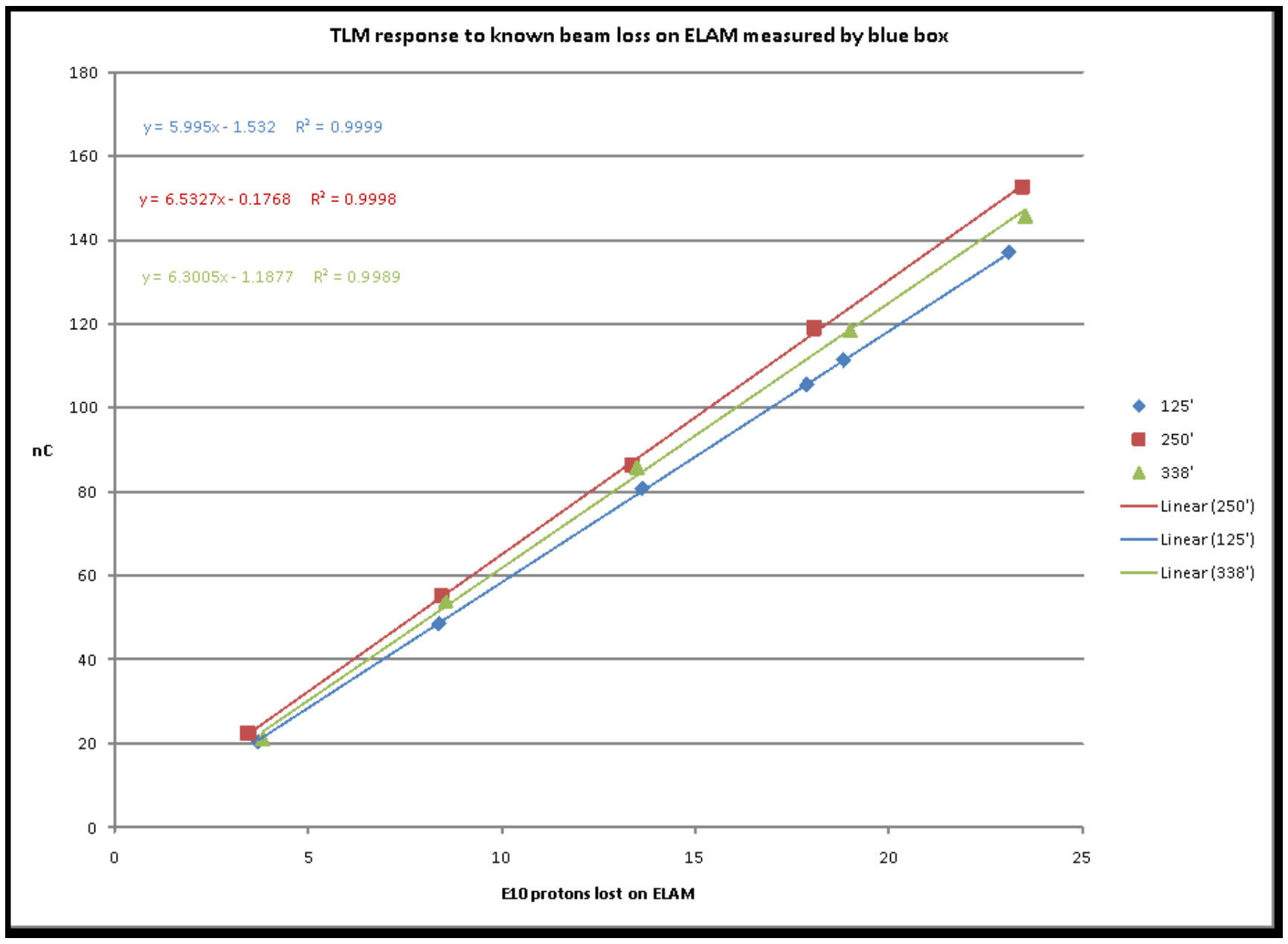

Figure 5.71. 125’, 250’, and 338’ TLM response to deliberate beam loss on a magnet (ELAM).

A similar determination has been made from another accident condition studied at the A2B7 magnet as shown in Figure 5.72; in this second case, the charge collected from a 250’ TLM and a 338’ TLM is about 3 nC per $10^{10}$ protons as is shown in Figure 5.73.

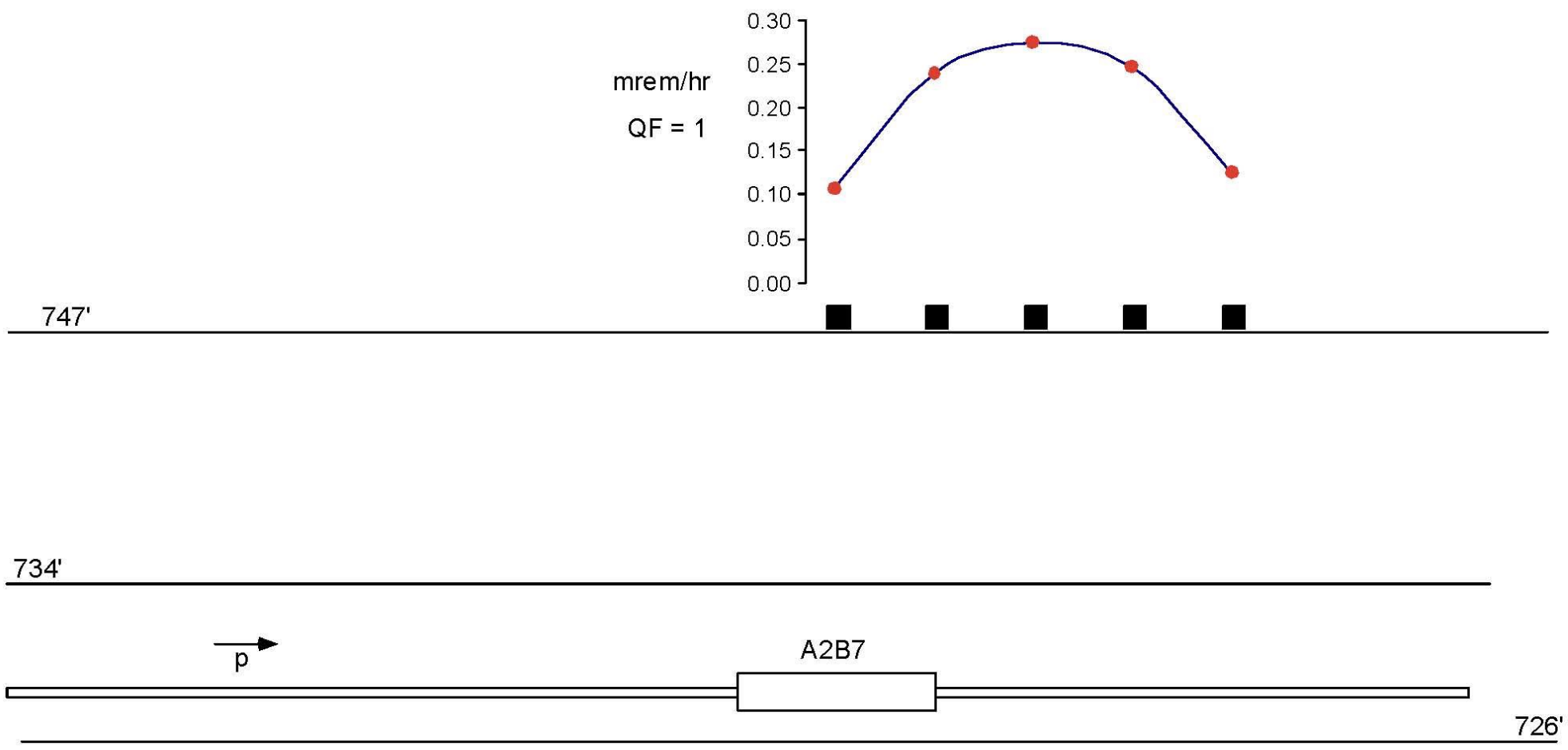

Figure 5.72. Measured radiation dose through a 13 foot shield due to a controlled 8 GeV beam loss at A2B7.

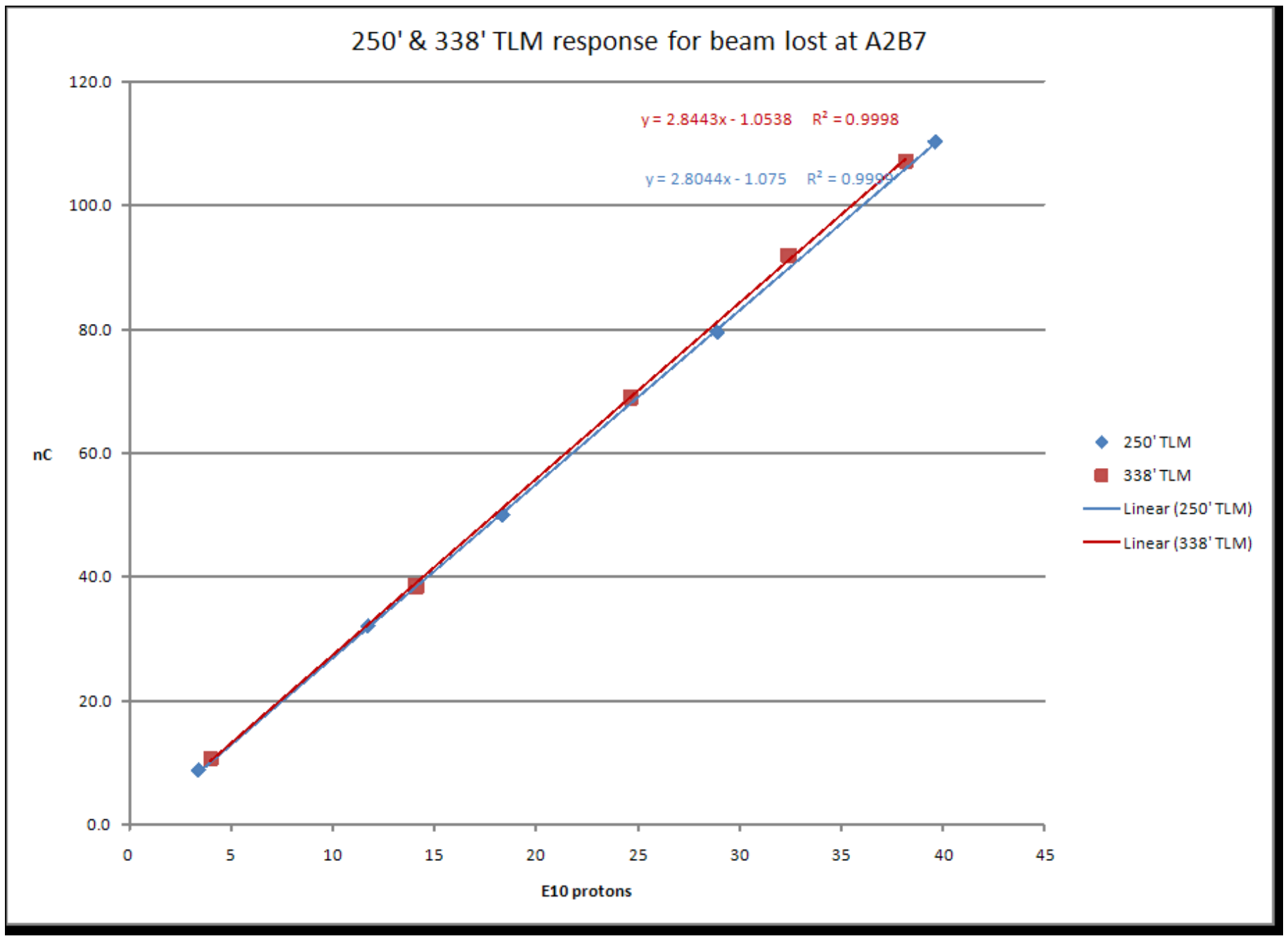

Figure 5.73. Measured 250’ and 338’ TLM response to a controlled 8 GeV beam loss at A2B7.

In order to use TLMs to control beam losses for the Mu2e experiment, it is important to understand their response as a function of beam intensity. The TLM response to 8 GeV beam loss as a function of low, medium, and high intensity and as a function of applied high voltage has been studied at the controlled beam loss location shown in Figure 5.72. It was found that as the single pulse beam intensity increases, the TLM response decreases unless the applied high voltage to the TLM is increased appropriately. The TLM response as a function of 3 beam intensities, roughly spanning 2 decades, is shown as a function of applied high voltage in Figure 5.74. At an applied high voltage of 500 volts, the response drops by about 20% for the high intensity beam loss case relative to the low and medium cases. With an applied high voltage of 1000 to 2000 volts, the difference in response is 10% or less. It will be shown below that the per pulse beam loss at the high intensity, shown in Figure 5.74, is well beyond tolerable beam loss levels for any of the beam lines under consideration for Mu2e. The TLM electrometer system will be required to disable the proton beam in most cases when either the low or medium level intensity beam loss is detected.

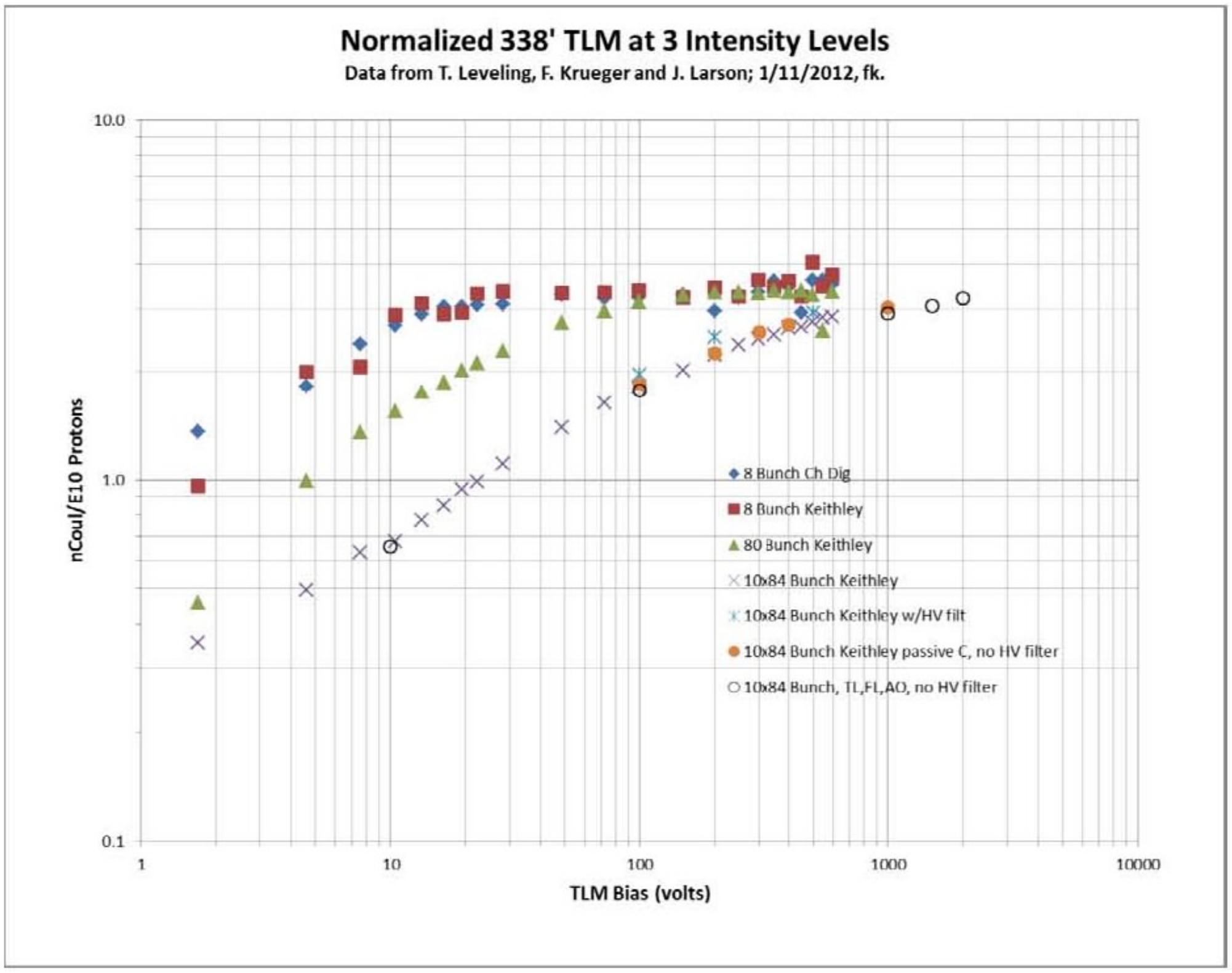

Figure 5.74. Measured TLM response for three beam loss intensity conditions as a function of applied TLM voltage. The response is approximately 3 nC/$10^{10}$ protons lost at A2B7.

Another important observation has been noted regarding TLM response as a function of loss location. Figure 5.75 and Figure 5.76 show coincident TLM response and chipmunk response for a series of beam pulses injected into the Accumulator Ring. The chipmunks indicated in the 0 to 25 foot positions are the same array as depicted in Figure 5.68. The chipmunks in the 100 to 125 foot location are the same array as depicted in Figure 5.72. These two figures demonstrate the following points:

a. In Figure 5.76, the response of chipmunks in the AP3 service building is greater in most cases than on the berm at A2B7, even though the beam is extinguished at A2B7 (except for the ELAM OFF) case. This clearly illustrates that individual TLMs must be placed and designated specifically considering the amount of shielding available. A TLM trip level established for the shielding berm would not be sufficient for a TLM trip level necessary for protection in the region of the service building.

b. The TLM response as a function of beam intensity is lowest for the A2B7 loss case; this is the most massive device in the region covered by the study. The response is approximately 3 nC/$10^{10}$ protons.

c. In Figure 5.75, a series of blue points for the 250' TLM shows TLM response for clean injection (all beam lost at A2B7) and for various trim magnet current changes intended to mis-steer the beam. The TLM response increases above the level for clean injection, but never drops below it. Inspection of Figure 5.76 for these corresponding points shows the chipmunk response at the AP30 service building is also increasing, attributed to beam scraping on ELAM. The largest TLM response for this series of measurements is about 5 nC/$10^{10}$ protons.

d. In Figure 5.75, two points labeled "before reverse proton tune-up" and "losses between ELAM and A2B7" show rather extreme TLM response resulting from grossly mis-steered beam. The TLM response is approximately 10 nC/E10 protons.

e. It is generally understood in the shielding of components in beam line enclosures that massive objects tend to concentrate losses and consequently, require thicker passive shielding. This is rationalized by the understanding that for a thin, less massive object such as beam pipe, the shower development spans a longer distance along the inside tunnel wall surface, effectively decreasing the spatial flux density. The resulting peak dose rate outside of a radiation shield is thus diluted by this geometry factor. Alternately, when a less massive object such as a beam pipe is buried in the shield, the spatial flux density is concentrated and the shielding requirement is increased; this is because the low density shield at the loss point less effectively attenuates the initial shower. These points are readily illustrated by examination of the shield scaling criteria commonly used for assessment of radiation shielding (Table 5.14).

Collectively, these observations indicate that setting a TLM trip level for the limiting case on the most massive object for the region (in this case, 3 nC/$10^{10}$ protons) guarantees that protection for the worst case beam loss condition is established.

As described previously, it has been determined that beam loss distributed among the three service buildings must be limited to a total of 10 Watts ($4.69 \times 10^{11}$ protons per minute at 8 GeV) in order to observe the annual radiation dose limit at publicly occupied locations such as Wilson Hall. The Debuncher circumference is 505 meters in length. The length of each service building where the shielding is thinnest is about 51 meters. Each service building would have its own TLM to cover the 51 meter length. By limiting the beam power loss at each service building to 3.3 Watts, it is assured that the 10 Watt beam loss total beneath the service building limit would be observed. Alternately, if the TLMs can be monitored from a central location, the sum of the three can be limited to 10 Watts.

| **Category** | **Dose Equivalent Rate** | **magnet in enclosure** | **pipe in enclosure** | **buried pipe** |
|---|---|---|---|---|
| No. Interlocked Detectors | | a | b | c |
| 1 | <1 | 22 | 20 | 24 |
| 2 | 1<D<5 | 19.9 | 17.9 | 21.9 |
| 3 | 5<D<100 | 16.5 | 15.5 | 18 |
| 4 | 100<D<500 | 15 | 13 | 16.5 |
| 5 | 500<D<1000 | 14 | 12 | 15.5 |

Table 5.14. Shielding requirements for a $2\times10^{13}$ proton beam loss occurring every 57 seconds for an hour for a 1 TeV beam.

The ELAM accident condition depicted in Figure 5.68 serves as a basis for limiting beam power losses at the service buildings. However, the final configuration of the Delivery Ring must be specified to determine the device type for which the protection basis is to be established. The TLM response to the ELAM loss condition is about 6 nC/$10^{10}$ protons, as can be determined from Figure 5.71. A 3.3 Watt beam loss corresponds to a total beam loss of $1.55\times10^{11}$ protons per minute that should yield about 93 nC/min of charge collected at the TLM.

TLMs can only determine the sum of charge collected for the regions in which they are installed. Consequently, a charge of 93 nC/min could be collected from a single point loss such as at ELAM as well as from a uniformly distributed loss throughout the entire straight section. To first order, the skyshine contribution for a uniformly distributed loss or a single point loss should be equivalent. However, a single point loss must be considered the limiting case for personnel protection locally in the service building. Returning to Figure 5.68, a normal loss of $1.55 \times 10^{11}$ protons per minute ($9.3\times10^{12}$ protons per hour) at ELAM would deliver a normal beam loss of about 6.5 mrem/hr. This radiation dose rate is acceptably low for normal beam operation in the service buildings if the buildings are locked radiation areas. The buildings would normally be accessible during 8 kW beam operation with a TLM based Radiation Safety System.

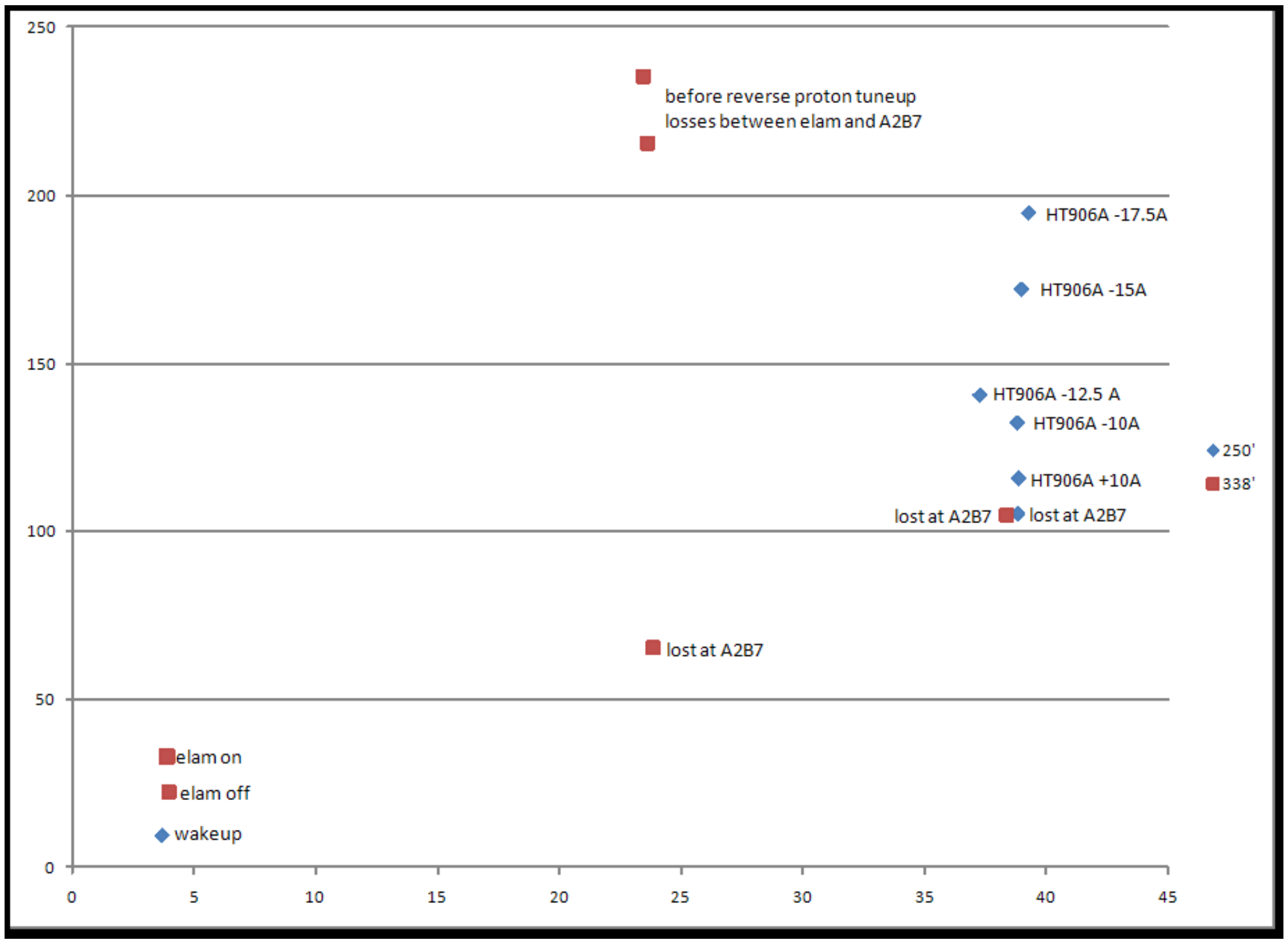

Figure 5.75. TLM response for various beam losses on 250' and 338' TLMs for varying beam loss conditions in the vicinity of ELAM and A2B7. The Accumulator bend bus is de-energized so that no beam is transported beyond A2B7.

A TLM based Radiation Safety System can similarly be used to protect the outdoor areas of the Delivery Ring, M3 line, and the AP0 service building. The first step is to determine what radiation dose rate is acceptable for outdoor radiation areas. For example, from the FRCM, a normal dose rate of 0.05 mrem/hr requires no extraordinary posting or fencing. Using the shielding scaling criteria, a beam loss of $1.96 \times 10^{12}$ protons per minute would result in a dose rate of 1 mrem/hr. A single point beam loss of $9.78 \times 10^{10}$ protons per minute (a 2 Watt beam power loss) would result in a dose rate of 0.05 mrem/hr. In this particular case, the shield scaling criteria can be compared against a measurement. Figure 5.72 shows that a beam loss of $3.6 \times 10^{13}$ protons delivers a peak dose of about 0.27 mrem. By simple scaling of the measurement, it follows that a single point beam loss of $9.78 \times 10^{10}$ protons per minute on the A2B7 magnet would give a radiation dose rate of 0.044 mrem/hr.

Another reasonable approach for setting a TLM trip level is to limit residual activation in the beam enclosure to ensure worker radiation exposure is kept at reasonable levels. 1 Watt per meter is typically used as a design basis for accelerators. The length of the Debuncher arcs, including the stairway sections of the service buildings, is a total of 352 meters in length. Three individual TLMs could be used to control beam loss in the arcs, each about 118 meters in length. This implies that the TLM trip level could be set at the equivalent of 118 Watts of 8 GeV protons or a total of $5.53 \times 10^{12}$ protons per minute.

TLMs cannot discriminate between a distributed beam loss and a single point beam loss, so it is possible that a single point beam loss of $5.53 \times 10^{12}$ protons per minute could occur. Since the 1 Watt per meter average beam loss is a normal condition, the resulting dose rate on the surface of the shielding berm due to a single point normal beam loss must be considered. Using techniques described above, the resulting dose rate from such a beam loss would be 2.5 mrem/h. Since the normal dose rate would be less than 5 mrem/hr under normal conditions, Controlled Area posting and a minimal occupancy requirement would be sufficient. It would be improbable, though acceptable, that the entire normal beam loss would occur randomly at some single point in the arcs. Consequently, TLMs set into each of the arcs with a trip level of 1700 nC/minute would provide sufficient protection against normal radiation beam loss. Further consideration of accident conditions for the Debuncher arcs is unwarranted.

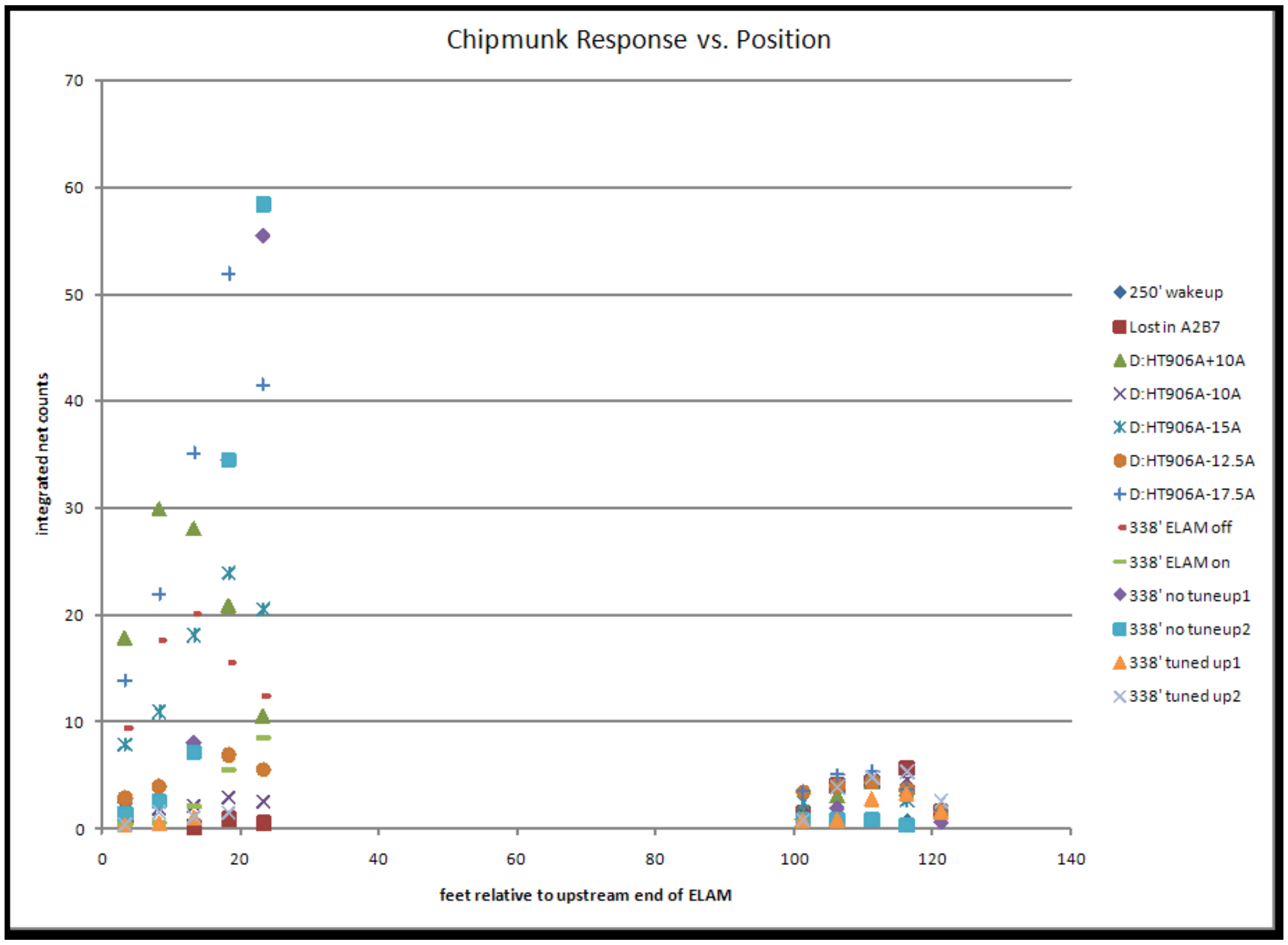

Figure 5.76. Simultaneous chipmunk responses for various beam injection conditions in the vicinity of ELAM and A2B7. Locations are in feet relative to the upstream end of ELAM. The Accumulator bend bus is de-energized so that no beam is transported beyond A2B7.

The remainder of the beam lines, including the section of the M1 line which lies beneath the AP0 service building, the M3 (Transport) line beneath the AP0 service building, and the M3 (Transport Enclosure) beam line can all be similarly protected using TLMs. A list of TLMs required to provide protection for the complex at 8 kW beam power operation is shown in Table 5.15.

| Location | Length (meters) | Basis (normal condition) | Peak Surface Dose Rate | TLM trip level nC/min |
|---|---|---|---|---|
| AP1 line at AP0 | 11.6 | 1 W/m | 3.75 | 2,200 |
| M3 (Transport) Beam line upstream | 138 | 5 mrem/h | 5 | 7,500 |
| M3 (Transport) Beam line downstream | 138 | 5 mrem/h | 5 | 7,500 |
| AP30 Service Building | 51 | 3.3 Watts | 6.5 | 93 |
| AP10 Service Building | 51 | 3.3 Watts | 6.5 | 93 |
| AP50 Service Building | 51 | 3.3 Watts | 6.5 | 93 |
| Delivery Ring 20 Arc | 118 | 1 W/m | 2.5 | 1,700 |
| Delivery Ring 60 arc | 118 | 1 W/m | 2.5 | 1,700 |
| Delivery Ring 40 arc | 118 | 1 W/m | 2.5 | 1,700 |
| M4 extraction beam line upstream | 138 | 1 W/m | 0.5 | 3,675 |
| M4 extraction beam line downstream | 138 | 1 W/m | 0.5 | 3,675 |
| g-2 Abort Line | | | | |

Table 5.15. TLM applications for 8 kW beam operation in the Accumulator/Debuncher. The choice of application is based upon the most limiting case among radiation skyshine, single point beam loss locations, and residual radiation levels in the beam enclosure.

The M4 extraction line from the Debuncher to the experiment hall is new construction and will be designed and built with 16 feet of passive shielding. TLMs could be used to limit the average beam loss to 1 Watt/meter. Further consideration of normal and accident conditions would be unnecessary.

The implementation of TLM based Radiation Safety System requires the development of an electrometer. The effort is to develop and electrometer is presently ongoing within the Accelerator Division. This effort is independent of the Mu2e Project. The present Radiation Safety System interface to the Chipmunk ion chamber would be used with only minor modifications to the currently available selectable trip levels.

***Beam collimation systems***

A set of transverse and momentum beam collimation systems can be used to capture beam losses at controlled locations that might otherwise occur at locations such as RF cavities and other limiting apertures, for example, momentum apertures. It is estimated that beam not captured in the momentum phase space is less than 0.1% of the total beam power or about $3.75\times10^{11}$ protons per minute. This level of beam loss is well below that described above the for 1 Watt/meter case in the Debuncher arcs. Therefore, momentum collimation systems should not be necessary for radiation protection purposes in the Debuncher Ring, although they could be added at some later date if necessary.

Transverse collimation systems could be useful, for example, to limit beam loss in RF cavities in the straight sections beneath the AP Ring service buildings. At this time, it is thought that control of beam size and available apertures in the straight sections is sufficiently good such that transverse collimation systems will not be required. Transverse collimation systems could be added in the event they are found to be necessary to limit radiation losses in the straight sections. An example of a collimator application is shown in Figure 5.77.

***Supplemental shielding requirements***

The primary safety system for beam operations in the Accumulator/Debuncher Rings will be the TLM system described above. The system will prevent beam operation when any beam power loss limits are exceeded. However the TLMs system will not guarantee that the beam can be operated; it can only limit operation in the event of excessive beam loss. Since the existing passive shielding is only sufficient for very low beam power losses, supplemental shielding will need to be provided inside the tunnel enclosure anywhere that significant beam losses might occur. Beam transfer junctions, kicker systems, and the electrostatic septa to be used at AP30 for extraction to the M4 beam line hall will require supplemental shielding.

Local shielding in the tunnel such as is currently used at the Debuncher injection septum (see Figure 5.78) will be required to limit normal radiation losses in the AP10, AP30, and AP50 service buildings. Each foot of steel shielding provides a factor of 10 reduction in the overall shielding thickness. Therefore, a three foot shield would provide a factor of 1000 reduction. The floor loading resulting from installation of such local shielding has been determined to be well within the design criteria [67].

Losses at the Debuncher extraction region due to beam interaction with electrostatic septa are expected to be about 2% to 5% under normal conditions. These losses will result in significant residual activity levels in the beam enclosure as well as very high prompt radiation levels in the AP30 service building. A shielding module or modules will be designed and built to both limit prompt radiation levels in the AP30 service building and to minimize residual radiation levels in the beam enclosure during beam enclosure access. Such a shielding module could be designed with features that permit remote disconnection of vacuum C-clamps and other bolted closures and plugged connections.

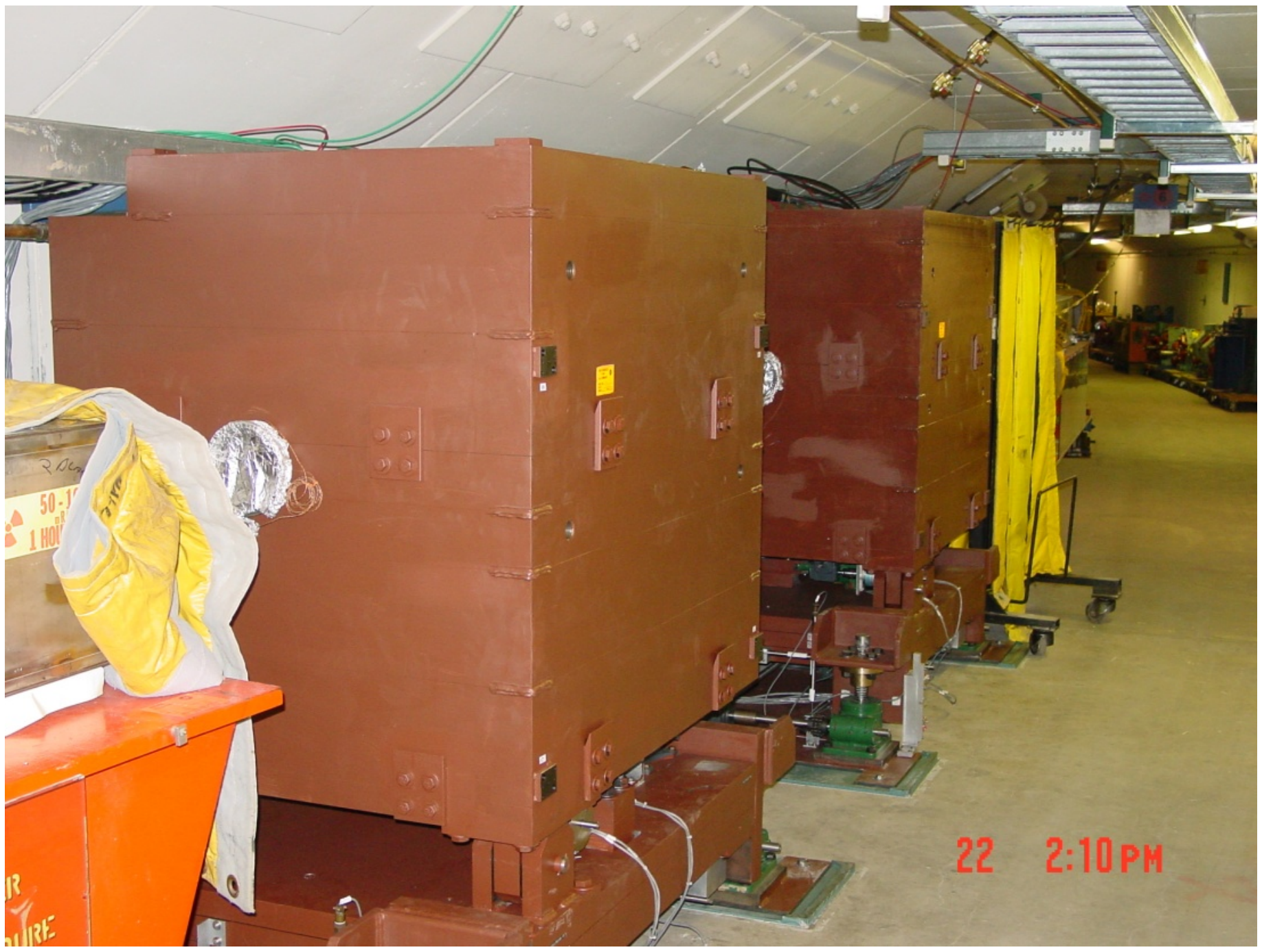

Figure 5.77. Two Beam collimation systems used in the Fermilab Booster Accelerator.

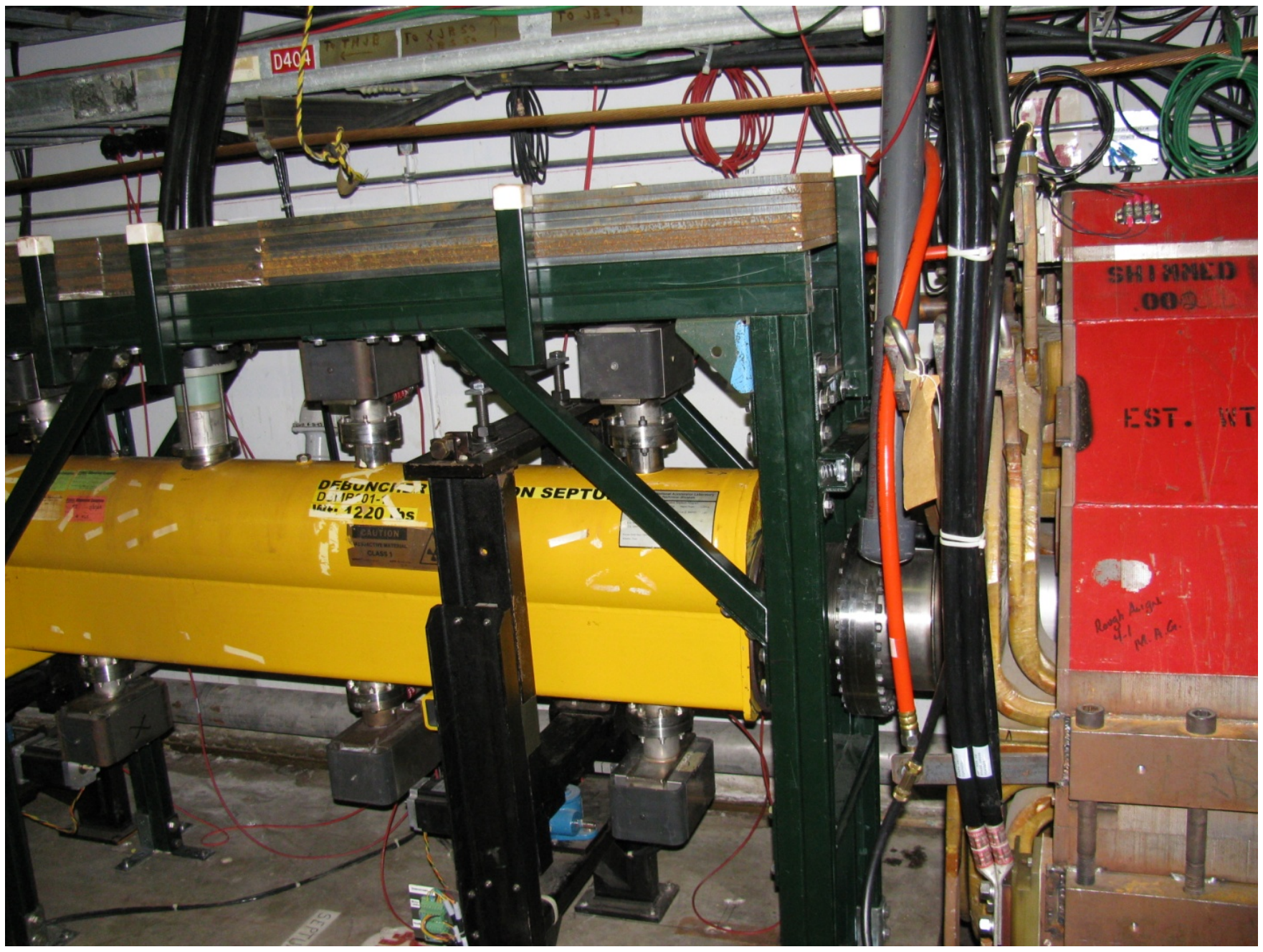

Figure 5.78. Four inch supplemental shield above the Debuncher Injection Septum

Table 5.16 contains a list of components that are likely to require supplemental steel shielding, an estimate of anticipated beam power loss in the Delivery Ring beam enclosure along with a preliminary required shield thickness. These components may all be required to be located in the AP30 straight section. Since the total beam loss in the straight section must be limited to 3.3 Watts, the beam power loss due to these known loss points must be limited sufficiently to prevent excessive skyshine losses. The shielding thicknesses given in Table 5.16 can be adjusted as the understanding of anticipated beam loss improves.

| Location | Component | % Beam Loss | Lost Beam power | Iron shield thickness | Effective Beam power loss (Watts) |
|---|---|---|---|---|---|
| Debuncher | Injection kicker | 0.5% | 40 Watts | 2 feet | 0.4 |
| Debuncher | Injection Septum | 0.5% | 40 Watts | 2 feet | 0.4 |
| Debuncher | Electrostatic Septum | Up to 5% | 400 Watts | 3 feet | 0.4 |
| Debuncher | Extraction line quad | 1% | 80 Watts | 2 feet | 0.8 |
| Debuncher | Extraction Lambertson | 1% | 80 Watts | 2 feet | 0.8 |

Table 5.16. Supplemental In Tunnel Steel Shielding in the AP30 straight section to limit beam power loss to 2.8 Watts. The total beam loss at AP30 must be limited to 3.3 Watts in order to observe skyshine limits.

The final configuration of supplemental shielding for the Mu2e project will be driven by the necessity to keep the TLM system from interrupting beam for normal operations. Considerations for supplemental shielding include control of radiation skyshine, limit worker exposure during tunnel access, and limit the production of surface water, ground water, and airborne radioactivity emissions.

In the present configuration of the M3 beam line, it will not be possible to add supplemental shielding, especially beneath the AP30 service building where the total shielding thickness is 10 feet. This beam line center is about 18 inches below the ceiling; consequently, there is no available space to add shielding. A shielding study was made for the existing AP3 line at AP30 that has a similar geometry to determine whether it should be possible to operate an 8 kW beam line with sufficiently low losses. The measurements were made after simple reverse proton tune up of the AP3 line followed by tuning up injection closure. After completing this tune up, the Accumulator bend bus was turned off to ensure that no circulating beam contributed to the chipmunk response for the measurement. An array of detectors was placed on the floor of the AP30 service building in the locations shown in Table 5.17. Several reverse proton injection of about $3.5\times10^{12}$ protons per pulse were made with the nominal tuning. This single pulse intensity is about 3.5 times the beam intensity intended for 8 kW operation of the M3 line. Consequently, this test of normal injection losses should be a severe one, especially because the line has not been optimized for the Mu2e experiment. Results for the measurement of normal injection, normalized for 8 kW operation, are shown in Figure 5.79. Normal injection losses appear to be reasonable until at location 13, about 3 feet past the beginning of the down bend. Once past the down bend, radiation dose rates quickly rise to unacceptably high levels. There are several points to make about the results of this normal beam loss injection study:

a. It should be possible to design the M3 line near the tunnel ceiling for clean beam injection to keep normal radiation levels acceptably low. The present aperture and optics should be used as a minimum standard for this section of the M3 beam line.

b. The losses in the down bend region for the existing line are unacceptably high. The optics and aperture of this section of the line need to be improved in order to maintain acceptably low beam losses, especially to minimize radiation sky shine.

c. The peak radiation dose rates for the existing line are well in excess of 1 rem/hr. However, it is possible to add steel shielding to the region to reduce injection losses. Some combination of improved aperture and steel shielding should be possible to reduce normal losses to acceptable levels.

A second study was done to understand the effectiveness of the service building radiation shielding for the region of the beam line near the ceiling where it is not possible to add steel shielding. For this study, the same conditions used for the normal injection loss study except that the vertical down bend power supply (EBV1) was turned off. The results of this accident condition study are shown in Figure 5.79. The resulting radiation dose rates are high but would be readily preventable with the proposed TLM system trip levels proposed in Table 5.15.

The production solenoid located in the target hall may require supplemental shielding. The initial conception of the production solenoid included an iron return yoke that would have doubled as a radiation shield in the target hall to reduce residual radiation levels around the production solenoid. The shield could serve to limit radiation worker exposure during maintenance activities in the target hall. It remains to be determined whether such a shield will be required. The need for such a shield can be determined after consideration of the following unknowns:

a. Residual radiation dose rates in the target hall around the production solenoid as a function irradiation and cooling time. This work is in progress.

b. Specific tasks to be performed including number of workers, duration of tasks, spatial location requirements for work

Residual radioactivity in the target hall is discussed further below.

| Location | Position | channel # |
|---|---|---|
| 1 | 4' DS of EQ8 | 2109 |
| 2 | plus 10' 3" | 2108 |
| 3 | plus 8' 1" | 2107 |
| 4 | minus 44" US of EQ7 | 2106 |
| 5 | plus 14" DS of HT906B and 14" beam left of the beam line | 2105 |
| 6 | plus 102" DS of cryo room wall | 2104 |
| 7 | plus 39" DS of VT906B | 2103 |
| 8 | at EQ6 | 2102 |
| 9 | plus 4' 2" DS of EBV1 | 2101 |
| 10 | plus 71" of location 9 | 2110 |
| 11 | plus 61" of location 10 | 2111 |
| 12 | minus 14" of EQ2 and 73" DS of location 11 | 2100 |
| 13 | plus 35" of EBV1 | 2099 |
| 14 | plus 51" of EQ1 | G:A3RD01 |
| 15 | minus 28" of cmag | G:A3RD00 |
| 16 | plus 51" of cmag | 2098 |
| 17 | nominal location - plus 20" of ELAM | 2097 |
| 18 | nominal location - 69" US of North wall | 2096 |

Table 5.17. Chipmunk locations for AP30 service building injection line shielding study.

### *Interlocked Detectors*

Historically, each service building of the Accumulator/Debuncher Rings has required the use of about 14 interlocked detectors to limit the intensity and duration of accidental beam loss. Interlocked detectors connected to the Radiation Safety System should no longer be required for beam operation in the Debuncher since TLMs will provide that function.

### *Labyrinth and Penetration Evaluation*

The shielding evaluation for the Accumulator/Debuncher Rings and related beam transport lines have been considered in conjunction with discussion on TLMs applications. Penetrations through the passive radiation shielding including stairways, various ducts, and cable penetrations are considered in this section. TLMs described above also play a role in limiting the radiation dose rate for these penetrations through the radiation shield.

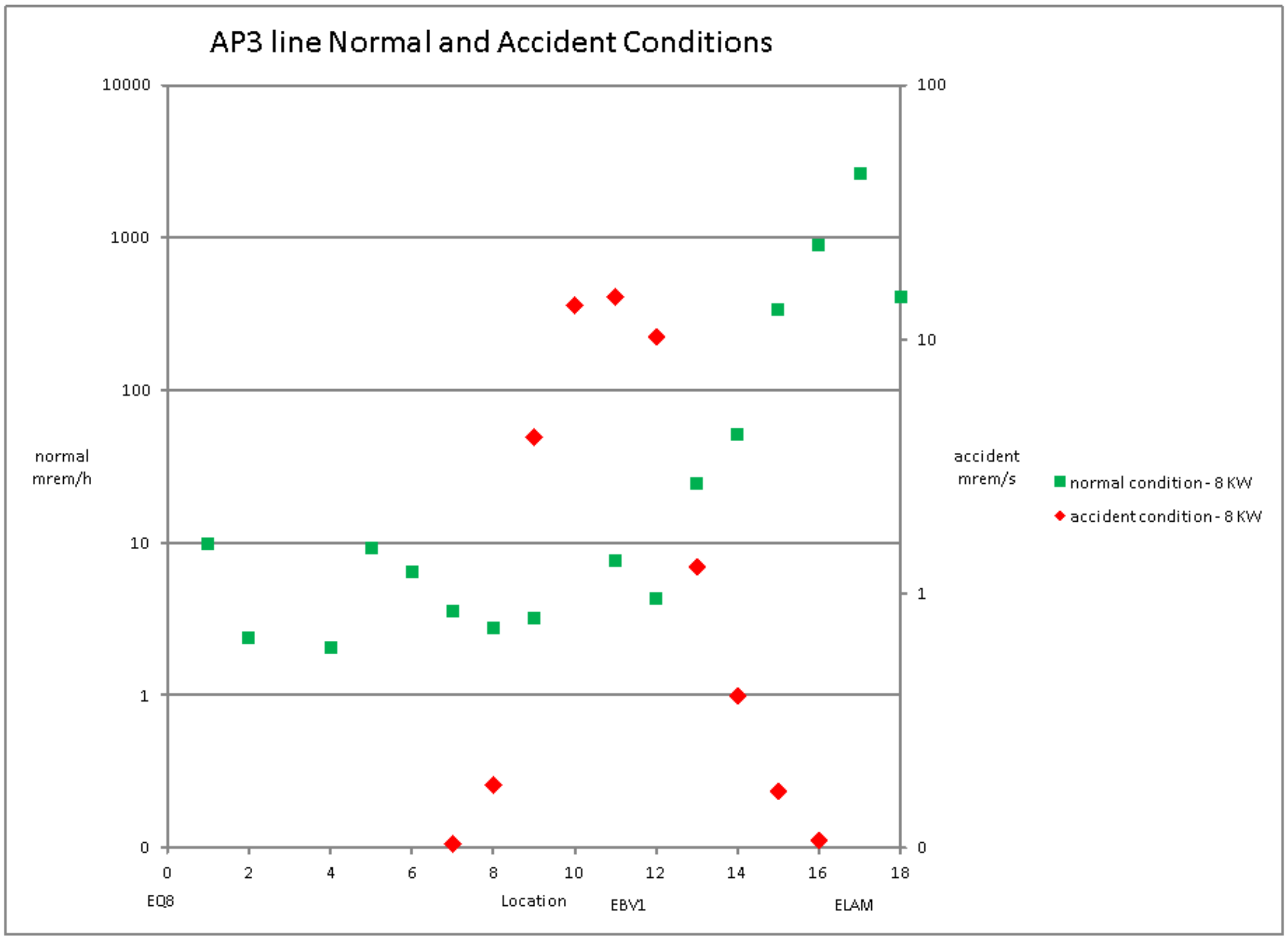

Figure 5.79. Radiation dose rate measurement for normal and accident conditions in the existing AP3 line beneath the AP30 Service Building. Green points are for optimized injection. Red points show the result with the vertical down bend magnets (EBV1) de-energized.

An Excel spreadsheet developed by the ES&H Section [70] was used to calculate the radiation dose rates at the exit of labyrinths and penetrations based upon user input parameters including the source term, aspect ratio, and length of each of the legs of the labyrinth or penetration. The evaluated penetrations are listed in Table 5.15 along with the resulting dose rate calculated for the 2000 Pbar shielding assessment. The third column of the table shows the resulting dose rate by scaling to the 8 kW beam power required for Mu2e. The fourth column shows the maximum number of protons lost per hour as limited by the TLMs trip levels established in Table 5.15. The fifth column shows the maximum possible dose rate (single point beam loss) at the exit of facility penetrations based upon the TLM trip levels established in Table 5.15. As indicated in Table 5.17, the resulting radiation dose rates at the exits of these penetrations are within limits prescribed by the FRCM. No addition remediation will be required for the existing facility including the three elevator shafts at the type 1 stairways.

| Penetration Name | Calculated exit dose rate from 2000 pbar shielding assessment | Scaled to 8 kW, 8 GeV proton beam loss | Max protons lost/hour limited by TLMs | Penetration dose rate limited by TLMs |
|---|---|---|---|---|
| **Determined for 3.6E13 8 GeV primary protons per hour** | | | | |
| ACC/DEB airshaft | 7.54E-02 | 47 | 3.32E+14 | 1 |
| ACC/DEB stairway type 2 | 1.85E-03 | 1 | 3.32E+14 | 0 |
| Transport to AP0 penetrations | 9.62E-02 | 60 | 1.115E+14 | 0 |
| Stub Room Penetrations | 2.00E-01 | 125 | 9.281E+12 | 0 |
| AP0 water pipe penetrations | 8.21E-01 | 513 | 1.115E+14 | 3 |
| Transport air duct vent to AP0 | 4.01E-03 | 3 | 1.115E+14 | 0 |
| Transport to F27 Penetrations | 6.32E-14 | 0 | 1.115E+14 | 0 |
| ACC/DEB elevator shafts | 5.09E-01 | 318 | 3.32E+14 | 5 |
| Transport stairway | 4.47E-02 | 28 | 1.115E+14 | 0 |
| ACC/DEB stairway type 1 | 1.41E-05 | 0 | 3.32E+14 | 0 |
| AP50 Pit Vent | 7.63E-07 | 0 | 9.281E+12 | 0 |
| AP50 Pit Labyrinth | 1.78E-02 | 11 | 9.281E+12 | 0 |
| **Determined for 1.8E16 120 GeV primary protons per hour** | | | | |
| PreVault stairway | 1.58E-02 | 0 | 3.26E+13 | 0 |
| Sweeping Magnet Penetrations | 1.23E+00 | 0 | 3.26E+13 | 0 |
| PreVault to F23 Penetrations | 5.12E-04 | 0 | 3.26E+13 | 0 |

Table 5.18. 2000 Pbar shielding assessment penetration dose calculations scaled to proposed TLM trip levels. Radiation dose rates at penetration exits would require no addition mediation if TLMs are used as described above.

For a cross check of the adequacy of the calculations of attenuation through the stairways shown in Table 5.18, one can review results of a set of measurements [73] made in January 2011 to measure radiation attenuation through the rings service building stairways. The dose rate outside the stairway exit door was measured to be about 4.5 rem/hr for a 25 kW beam loss on ELAM. The service building area TLMs will limit total proton beam loss to about $9.3\times10^{12}$ protons per hour. Scaling the result of the stairway measurement, the resulting dose rate just outside the labyrinth door at the proposed TLM trip limit would be about 0.6 mrem/hr; this result is somewhat higher than the dose rate calculated in the spreadsheet; however, the measurement result is consistent with radiation area posting levels proposed for the service buildings.

***Fence and radiological posting requirements***

The implementation of the TLM system along with in-tunnel, supplemental shielding will limit radiation dose rates in accessible outdoor areas. The Controlled Area posting would be required and the areas would need to be established as for minimal occupancy. No fences or other physical controls would be required by the FRCM. The new M4 beam line enclosure directed to the production solenoid should not require fencing or radiological posting since it will be designed (16’ shielding in conjunction with TLM system use) for the intended beam power for the Mu2e experiment.

***Entry controls - requirements for service building and shielding berm beam-on access***

The calculated radiation dose rates in the Accumulator/Debuncher service buildings including the AP0, AP10, AP30 and AP50 service buildings will be high enough to warrant the Radiation Area posting. Access to the buildings by trained Radiation Workers can be permitted since the anticipated radiation dose rates are within the limits for such activities prescribed by the FRCM.

The radiation dose rates in outdoor areas including the Debuncher berm and the Transport enclosure berm may be measureable during beam operations. The radiation levels there would be limited there by the TLM system to those requiring the Controlled Area posting. Other controls such as fences would not be required. The areas would be established as minimal occupancy areas as described by the FRCM. No other postings or controls would be required for these outdoor areas.

***Residual radioactivity control***

The use of supplemental shielding and the TLM system, both described above, will serve to limit residual radiation levels due to beam operations in the enclosures while workers are present. The further control of worker exposure for the Mu2e project can be accomplished with well-established work procedures and practices currently in place within the Accelerator Division.

Residual radioactivity in the target hall due to activation of the Production Solenoid beam absorber and the production solenoid/target will be significant. Some MARS calculations have been performed to understand the levels of residual radioactivity as a function of irradiation and cooling times. Examples of this work are shown in Figure 5.80.

The residual dose rates indicated in Figure 5.80 are taken from the surface of components within the target hall. Some additional work is required to determine the radiation dose rate contours in the spaces of the target hall where occupancy is possible or required. As mentioned previously in the section on supplemental shielding, a list of work activities for the target hall is necessary to understand the collective radiation dose that will accrue in support of the Mu2e experiment.

A concrete radiation shield is shown surrounding the production solenoid in Figure 5.80 left. If there are tasks to be completed on the Production Solenoid, the concrete shield would be an additional radiation source that would add to worker exposure rather than limit it. It is not clear at this time whether the concrete shield should be used. If the concrete shield is not installed, the radiation dose rates at the enclosure side walls would certainly increase. However, the radiation dose rate to workers in close proximity to the Production Solenoid would be reduced. Some additional study is required to understand how to control worker exposure to residual radioactivity in the target hall.

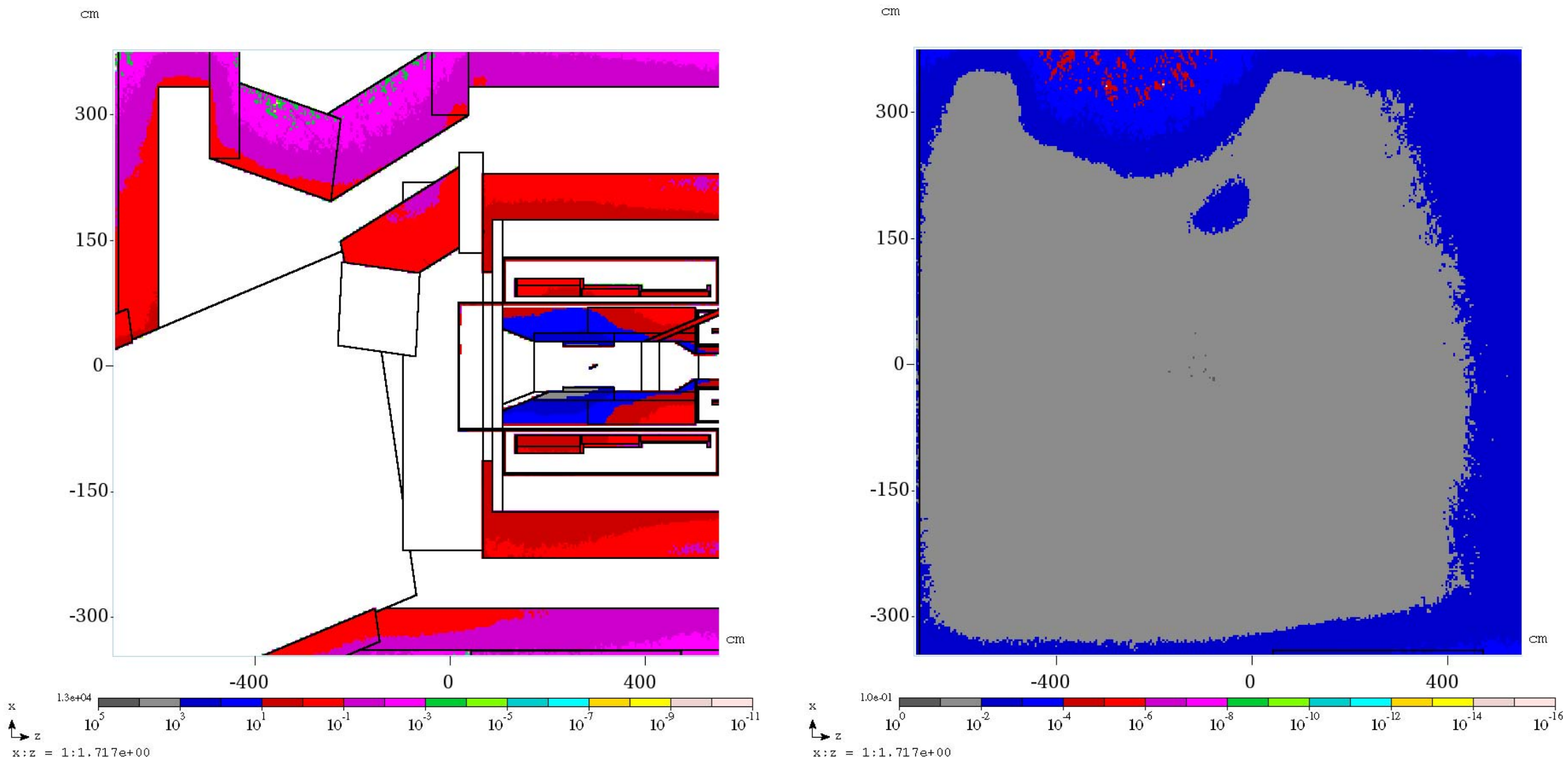

Figure 5.80. Plan view of target hall at beam elevation (left) and at floor elevation showing residual radiation dose rates (mSv/h) after 30 days irradiation and 7 days cooling time. Residual rates vary significantly as a function of elevation in the target hall. Radiation dose rates at the floor elevation beneath the production solenoid are much lower than at beam elevation. The beam absorber is not shown. Concrete shielding around the production solenoid may be undesirable.

### *Ground water activation*

The major sources of ground water activation due to beam operations at the Accumulator/Debuncher facility for the Mu2e experiment includes losses at the following locations:

- Debuncher beam absorber
- Diagnostic absorber
- Main beam absorber (downstream of the Production Solenoid)
- Debuncher Injection
- Debuncher extraction

Detailed calculations for ground water activation for Mu2e operation of the Accumulator/Debuncher Rings have been completed [74]. No ground water issues have been identified.

***Surface water activation***

The major sources of surface water activation due to beam operations at the Accumulator/Debuncher facility for the Mu2e experiment are the same sources as those listed for ground water activation. Detailed calculations for surface water activation for Mu2e operation of the Accumulator/Debuncher Rings have been completed [74]. No surface water issues have been identified.

***Airborne radioactivity***

The major source of airborne radioactivity due to beam operations at the Accumulator/Debuncher facility for the Mu2e experiment is from primary/secondary beam passing through the air volume between the Production Target Solenoid and the Main Beam Absorber. Other sources to be considered are due to beam losses at locations listed in the ground water section. Local shielding at the extraction region and other beam transfer locations may reduce airborne radioactivity levels significantly. Ventilation controls might be required depending upon the amount of activity produced. However, the need to limit beam losses related to the skyshine problem should eliminate any concern for airborne radioactivity other than the major source at the Production Solenoid. Detailed calculations for airborne radioactivity for Mu2e operation have been completed [74]. Engineered ventilation controls will be used to limit the impact of air emissions for the Mu2e project.

The Production Solenoid Enclosure air system will be separated from the M4 beam line enclosure by a gate/barrier located near the M4 beam line absorber. The air supply in the Production Solenoid Enclosure will come from a supply duct located adjacent to the Production Solenoid Beam Absorber. This air will pass through the Production Solenoid Enclosure and will be exhausted at an exhaust trunk located near the gate/barrier. The normal exhaust flow rate at the normal fan speed will be about 50 cfm. In the event of an ODH alarm in the Production Solenoid Enclosure, a high fan speed will exhaust air at 10,000 cfm from the same exhaust stack.

A separate ventilation exhaust stack, located on the M4 beam line side of the gate/barrier, will exhaust air from the Delivery Ring and M4 beam line. The nominal flow rate for this system will be designed and set to minimize emission of airborne radioactivity from the Delivery Ring and M4 beam line.

Finally, a third exhaust fan will be located in the Transport Enclosure (M3 beam line enclosure) as a second exhaust for the Transport beam line and the Delivery Ring.

### *5.10.4* Radiation Safety Plan Summary

Operation of the Mu2e experiment in a repurposed Antiproton Source Facility at a beam power of 8 kW appears feasible. The radiation skyshine issue is mitigated with the application of TLMs as described above in conjunction with local, in-tunnel steel shielding. Special attention will be required to the design of the M3 line to ensure that relatively loss free injection is possible. Supplemental steel shielding will be necessary to mitigate losses, especially at beam transfer points located beneath the service buildings where radiation shielding is thinnest. Entry controls like those used for the AP0 service building during the years of Pbar Source operation, including Radiation Area postings, Radiation Work Permits, and locked doors will be sufficient and will allow worker access to service buildings during beam operations. Controlled Area postings will be required for the Transport Enclosure shielding berm and the Debuncher shielding berm. Electrometers for the TLM system are being developed independently of the Mu2e project and will connect directly to the existing Radiation Safety System.

## 5.11 Operations Preparation

### *5.11.1* Machine Protection

The high beam power in the Mu2e transport lines and storage rings will require a sophisticated machine protection system to avoid damage of accelerator components due to errant beam pulses. The backbone of the machine protection scheme will be a beam permit system that will be based on the existing CAMAC serial loop permit system used in the Antiproton Source, with a number of upgrades to make it compatible with the operating scenarios required for Mu2e operations. Increased beam power levels to certain portions of the Mu2e complex, as compared to current Antiproton Source operation, will translate to an increased number of systems being monitored by the beam permit system. Systems that will be monitored need to be carefully defined for each beam line and the Delivery ring, and likely will include power supplies status and output, beam orbits, loss monitors, beam intensities, vacuum components, and safety system devices. Some systems, such as the TLM that is being considered in the radiation safety portion of this document, can also be used to augment the machine protection system.

The increased number of devices monitored by the beam permit will result in an increase in the number of required CAMAC 200 modules. This will only have a minimal cost due to the ample stock of spare CAMAC 200 modules that will be available from the Tevatron after HEP Collider operations are retired. These modules will only require an EPROM or PAL change in order to be used for Mu2e operations.

In cases where a changing analog signal must be sampled at a specific time, such as the flat top value of a ramping power supply, a CAMAC 204 module and associated chassis will be used to provide the permit to the CAMAC 200 module. Currently CAMAC 204 modules are used to sample dipole and quadrupole stacking ramps at flattop in the P1, P2, and AP-1 lines, with different ramps covering different beam modes. In the Mu2e era, the same CAMAC 204 modules can easily be used to monitor the Mu2e ramps as long as the same power supplies and MADC channels are used. If Switchyard 120 GeV operations are still in place, we will need to maintain this configuration in the P1 and P2 lines. The cost associated with implementing new CAMAC 204 modules can be significant since it includes the 204 chassis, a fan in box, a VME crate and VME processor and HRM data acquisition system. Since very few devices in the Mu2e rings will be ramped, it is anticipated that we will need very little, if any, additional CAMAC 204 module coverage.

***Beam Permits***

There will be two sets of beam permits required for Mu2e operation: The M1/M3/Delivery Ring Beam Permit and P1/P2 Beam Permit. The M1/M3/ Delivery Ring beam permit will use the existing Pbar permit loop that has a path the covers all service buildings that have equipment associated with the M1, M3 and Delivery Ring. This permit will provide a single input back to the Beam Switch Sum Box (BSSB) in the Main Control Room, which in turn will provide a single beam switch and beam permit indication at each of the Main Control Room beam switch boxes. When the Delivery Ring beam permit conditions are lost, the Delivery Ring permit drops, sending Delivery Ring beam to a beam abort located in the AP-2 beam line. Due to the addition of a Delivery Ring beam abort, the Delivery Ring permit will need CAMAC 201 and 479 cards located in a rack adjacent to the Delivery Ring abort kicker at AP-50. Extending the abort cabling to the Mu2e experimental hall will also allow us to send Delivery Ring beam to the Delivery Ring abort if Mu2e experimental hall loses its beam permit.

Once the P1 beam line is no longer used to transport beam to the Tevatron, the existing P1 and P2 beam line permits will be combined into a single P1/P2 beam permit for Mu2e operations, with a single input back to the Beam Switch Sum Box (BSSB) in the Main Control Room. When the P1/P2 beam permit conditions are lost, new Mu2e beam is inhibited until the permit conditions are restored.

***Operating Scenarios***

The beam permit system must be able to handle all possible operating scenarios. Normal Mu2e operations will have beam traverse the entire Mu2e accelerator chain, being sent down the extraction line to the Mu2e experimental hall. However, there are two other possible modes of operation corresponding to running beam to either the extraction line beam dump or the Delivery Ring beam dump. To send beam to the extraction line dump, there will be a switched magnet in the extraction line that will determine if the beam is sent to the Mu2e experiment or the dump. To send beam to the Delivery Ring beam dump, simple permit masking configurations will be created.

***Software***

There are a number of software issues to be addressed with the implementation of the machine protection scheme. An ACNET interface will need to be written for the beam abort, similar to the current Pbar interface page P67. Controls experts have written a generic template page for aborts, which will minimize programming time. The beam line auto-tune application will need to be upgraded for Mu2e operations to include the correct beam line path, beam energies, and cycle times. It has become the norm at Fermilab for beam line tuners to be written as Open Access Clients (OACs) which run in the background, so additional programming time may be required to migrate this application to an OAC platform.

# 5.12 R&D Plan for the Mu2e Accelerator Subproject

Many studies will be needed to bring the whole design to its maturity. In many cases prototype tests and machine studies are necessary to resolve important unknowns before final design decisions are made. In this chapter we summarize our current plan for these studies. Although their schedule will depend largely on the availability of resources and accelerator beam time, it is in the best interest of the project to be able to carry them out as early as possible.

### *5.12.1* Debuncher and Slow extraction

Further studies should be performed in the Debuncher to better understand its environment for slow extraction. One of these studies is to continue tune-space scans to determine potential impact of any existing high order resonances during slow extraction ramps near the operating point for Debuncher Mu2e operation. Two different techniques have been tested using pbar stacking cycles. We plan to continue this study as soon as machine time becomes available.

A second study that needs to be done is measuring the ripple in the Debuncher main quadrupole bus power supplies. This measurement can give us an idea what level of tune ripple to expect during slow extraction. This study does not even require the presence of beam in the Debuncher and can be executed as soon as “quiet time” becomes available.

A third extraction study is the testing of RFKO beam blowup efficiency. This study requires substantial resources to be available as it assumes the installation of the RFKO device in the Debuncher ring and the creation of the control infrastructure (power supplies, regulation, controls) required to operate it. These studies are presently underway.

In addition to these studies in the Debuncher, we would like to complete the half-integer resonance tests in the Main Injector using the existing slow extraction system. The novelty of the half-integer extraction scenario proposed for potential use in Mu2e is that it is using the zero-harmonic ramp exclusively of the harmonic quad ramp. Although the parameters of the Main Injector slow extraction are far from those of Mu2e, the existence of the slow spill operation in the Main Injector provides a unique chance to carry out proof-of-principle tests.

### *5.12.2* Extinction

Achieving an out-of-time beam extinction level of $10^{-10}$ will be an extremely challenging task that involves using two unprecedented techniques: a 300 kHz/5 MHz AC dipole system to remove out-of-time beam, and the operation of the extinction monitoring device at the required level of sensitivity. While the latter can only be demonstrated when beam of the desired quality becomes available, the feasibility of the former must be demonstrated on the test stand. In the current design, the two frequency components will be provided by the same type of magnet: a three meter magnet composed of half-meter segments. A prototype of one of these segments has been constructed and successfully tested. The full specified field of 160 G at 300 kHz has been achieved. High frequency tests will be conducted when a suitable power supply is located.

Because the high resolution extinction monitor relies on high momentum proton scatters that can be well modeled, no R&D is planned; however tests are planned for the low resolution particle telescope that will measure the extinction of the ring. A simple charged particle telescope will be installed in the MiniBooNE beam line. It will consist of three small quartz Cerenkov detectors that will be read out by Silicon photomultipliers (SiPM's). This telescope will be used to measure scattering from the prototype thin foil profile monitor which is still located in the beam line, and use this to reconstruct a statistical picture of the out of time beam. Once the measurement is successful, it can be used to test models of out of time beam formation in the Booster.

### *5.12.3* Radiation protection

Currently the radiation safety systems in Fermilab use "chipmunk" detectors for radiation level monitoring. This technology is well proven but limited in detection range and extremely expensive when needed to cover large areas. A new robust and money-saving technique using Total Loss Monitors (TLM) is proposed for the safety system. A TLM is a long ionization chamber built using heliax cable and filled with a noble gas.

This technique would enable us to cover large areas with very few detectors. Although very promising, the technique is not safety certified and needs to be tested. The full suite of required test has not yet been specified by the Fermilab ES&H office, nevertheless we have started some tests with 2 long detectors in the Pbar tunnel enclosure, using a real beam environment with controlled losses and radiation level monitoring in the service building.

### *5.12.4* Instrumentation

Slow spill rate regulation requires fast (a few hundred turns) and reliable spill monitoring. However this measurement will be difficult, as it needs to resolve intensities of protons per bunch, or peak current of about 25uA. Several techniques have been examined as candidates, but only one has the potential to provide the required performance at a reasonable cost and complexity; a modified Resistive Wall Monitor with resonant readout. Preliminary calculations and quick table tests have demonstrated potential viability of this technique. We need to conduct more detailed tests with more realistic prototype parameters before CD-2.

### *5.12.5* Infrastructure

We also have two sets of measurements that are needed to evaluate general system characteristics. One includes careful study of the electric loads on main parts of the Pbar source, and this one has been done already. Both resistive and reactive loads have been measured across service buildings, starting from main power supplies and going down to individual racks. This information will be used for calculating the needs of electric power, cabling, transformers, etc.

An "as found" survey of ring magnet elements is planned, and will be completed before the civil construction begins and after construction ends. This is especially necessary in the extraction section of the Debuncher, because the tunnel levels may be potentially affected by the construction of the external beam line.

## 5.13 References

[1] Mu2e Collaboration, Proposal to Search for $\mu^- N \rightarrow e^- N$ with a Single Event Sensitivity Below $10^{-16}$, Mu2e-doc-388, October 10, 2008.

[2] E. Prebys, et. al., Findings and Recommendations of the Mu2e Task Force, Mu2e-doc-1911, November 2011.

[3] M. Syphers, Mu2e Proton Beam Requirements, Mu2e-doc-1105-v1. September 2010.

[4] Reference to 2.5 MHz Recycler RF system

[5] Meiqin Xiao, Transport from the Recycler Ring to the Antiproton Source Beamlines, Beams-doc-4085, March 2012.
[6] Delivery Ring AIP CDR. In preparation.
[7] J. Dey, et al., Mu2e Accelerator Requirements and Derived Parameters, Mu2e-doc-1374, March 2011.
[8] L.J. Laslett, BNL Report 7534 (1963) p. 324.
[9] J. Galambos, *et al.*. ORBIT: a Ring Injection Code with Space Charge. *Proceedings of the Particle Accelerator Conference*, July 1999.
[10] K.Y. Ng, Stability Issues of the Mu2e Proton Beam, *Proceedings of the Particle Accelerator Conference (PAC09)*. Mu2e-doc-564, December 2008.
[11] A. Burov, Beam Stability for Mu2e Project, Mu2e-doc-1165, November 2010.
[12] A. Burov, Head-tail Modes for Strong Space Charge, *Phys. Rev. ST Accel. Beams 12, 044202,* December 2008.
[13] A. Chao, "*Physics of Collective Beam Instabilities in High Energy Accelerators*", World Scientific, 1993.
[14] V. Nagaslaev, Estimation for IBS effects in the Accumulator, Mu2e-doc-1138, October 2010.
[15] B. Drendel, Pbar CAMAC Crate Inventory, Mu2e-doc-1161, November, 2010.
[16] Design Report Tevatron 1 Project. Technical report, Fermilab, September 1983 and September 1984.
[17] B. Drendel, A. Ibrahim, Possible Toroid and DCCT Upgrades for Mu2e Storage Rings and Beam Lines, Mu2e-doc-1180, November 2010.
[18] B. Drendel, N. Eddy, Mu2e Debuncher Tune Measurement System, Mu2e-doc-1616, June 2011.
[19] B. Drendel, M. Olson, Possible BLM Upgrades for Mu2e Storage Rings and Beam Lines, Mu2e-doc-1179, November 2010.
[20] B. Drendel, Pbar Controls for Mu2e, Mu2e-doc-1161, November 2010.
[21] J. Dey, I. Kourbanis, Recycler RF technical report. In preparation.
[22] J. MacLachlan, J-F Ostiguy, User's Guide to ESME esmF95 (esme2009.3), http://www-ap.fnal.gov/ESME/#label4, September 2009.
[23] J. Dey, Mu2e RF Systems, Mu2e-doc-1372, March 2011
[24] H. Grote, F.C. Iselin, The MAD Program (Methodical Accelerator Design), CERN/SL/90-13.
[25] L. Michelotti, F. Ostiguy, CHEF: an Interactive Program for Accelerator Optics, FERMILAB-CONF-05-090-AD, May 2005.
[26] J.F. Amundson, et al., Synergia: an accelerator modeling tool with 3-D space charge. *J.Comput.Phys.*, 211:229-248, 2006.
[27] Design Report Tevatron 1 Project. Technical report, Fermilab, September, 1983 and September, 1984.
[28] A. Garren, et al., A User's Guide to SYNCH, FERMILAB-FN-0420, June 1985.

[29] L. Michelotti, J. Johnstone. Preliminaries toward studying resonant extraction from the Debuncher. Technical report, Fermilab, 2009. FERMILAB-FN-0842-APC-CD.

[30] D.A. Edwards. Comparison of Half Integer and Third Integer Extraction for the Energy Doubler. Technical report, FERMILAB-TM-842 (UPC 034), Fermilab, 1978.

[31] T. Furukawa *et al.* Global spill control in RF-knockout slow-extraction. *Nuclear Instruments and Methods in Physics Research*, A522:196-204, 2004.

[32] L. Michelotti, *Intermediate Classical Dynamics with Applications to Beam Physics*. John Wiley & Sons, Inc., New York, 1995.

[33] B. Fellenz and J. Crisp, An Improved Resistive Wall Monitor, *Proceedings of the Beam Instrumentation Workshop* '98 (BIW98), Stanford Linear Accelerator Center, May 1998.

[34] P. Prieto, FNAL/AD, Private communication.

[35] R. Bernstein, Estimate of Radiative Pi Capture Background, Mu2e-doc-1087, January 2011.

[36] E. Prebys, Extinction Magnet Specifications for the Mu2e Experiment, Mu2e-doc- 709, April 2010.

[37] C. Johnstone, FNAL/AD, private communication.

[38] A. Drozhdin and I. Rakno, Study of Extinction Efficiency in the Mu2e Experiment, Mu2e-doc-1206, November 2010.

[39] W. Molzon et al., Extinction Monitor Requirements, Mu2-doc-894-v4, May 2010.

[40] R. Bernstein, Conceptual Discussion of Intensity and Extinction Monitors, Mu2e-doc-1375, March 2011.

[41] P. Kasper, Extinction Measurement, Mu2e-doc-1426, March 2011.

[42] R. Kutschke, Some First Thoughts, Mu2e-doc-383, October 2008.

[43] J.L. Popp, Production Target Requirements, Mu2e-doc-887-v4, November 2010.

[44] V. Khalatian, Different Production Target Models Comparisons, Mu2e-doc-1483, April 2011.

[45] C. Densham et al., Radiative Cooled Target Concept, Mu2e-doc-2054, February 2012.

[46] C.J. Densham, et al., STFC-RAL Conceptual Design Study of Mu2e Experiment Pion Production Target Components and Systems – Second Intermediate Design Study Report, Mu2e-doc-1746, July 2011.

[47] G. Ambrosio, R. Coleman, V. Kashikhin, M. Lamm, N. Mokhov, J.L. Popp, V. Pronskikh, Requirements for the Mu2e Production Solenoid Heat and Radiation Shield, Mu2e-doc-1092, February 2012.

[48] M. Guinan, J. Kinney, R. Van Konynenburg, Defect Production and Recovery in FCC Metals Irradiated at 4.2 K, *Journal of Nuclear Materials* 133&134, p 357, 1985.

[49] J. Horak, T. Blewitt, Isochronal Recovery of Fast Neutron Irradiated Metals, *Journal of Nuclear Materials* 49, p 161, May 1973.

[50] V. Khalatian, Comparison of Different Heat Shield Radii, Mu2e-doc-1487, April 2011.

[51] L. Bartoszek, The All-Bronze Heat and Radiation Shield, Mu2e-doc-1907, October 2011.

[52] Z. Tang, Thermal Analysis of Mu2e PS Radiation Shield, Mu2e-dco-2059, February 2012.

[53] R. Coleman, J. Popp, V. Pronskikh, M. Syphers, Mu2e Proton Beam Absorber Requirements, Mu2e-doc-948, November 2010.

[54] P. Kasper, Extinction Monitor Status, Mu2e-doc-2024, January 2012.

[55] J.D. Cossairt, A.J. Elwyn, P. Kesich, A. Malensek, N. Mokhov, and A. Wehmann, The Concentration Model Revisited, Fermilab EP Note 17, June 24, 1999.

[56] N. Mokhov and V. Pronskikh, MARS15 simulations for the requirements on the proton beam absorber, Mu2e-doc-1072, September 2010.

[57] R. Wands, Proton Dump Thermal Analysis, Mu2e-doc-1201, November 2010.

[58] C. Baffe, Mu2e Proton Absorber- Preliminary Thermal Analysis of Low Power Scenario, Mu2e-doc-1895, October 2011.

[59] J. Popp and V. Khalatian, Comparison of Different Proton Target Types, Mu2e-doc-1475-v1.

[60] C.J. Densham, et al., Conceptual Design Study of Mu2e Experiment Pion Production Target Components and Systems – First Intermediate Design Study Report, Mu2e-doc-1577, April 2011.

[61] C.J. Densham, et al., STFC-RAL Conceptual Design Study of Mu2e Experiment Pion Production Target Components and Systems – Second Intermediate Design Study Report, Mu2e-doc-1746, July 2011.

[62] L. Bartoszek, Status of the Heat and Radiation Shield Design, Mu2e-doc-1789, July 2011.

[63] P. Kasper, Extinction Monitor Status, Mu2e-doc-2024-1, January 2012.

[64] Fermilab Radiological Controls Manual, http://esh.fnal.gov/xms/FRCM

[65] Antiproton Source Department Shielding Assessment, June 2000, http://www-bdnew.fnal.gov/pbar/documents/Antiproton%20Source%202000%20Shielding%20Assessment/Antiproton%20Source%202000%20Shielding%20Assessment.htm.

[66] D. Cossairt, Assessment of Neutron Skyshine Near Unmodified Accumulator/Debuncher Storage Rings Under Mu2e Operational Conditions, Mu2e-doc-1307-v2, January 26, 2011.

[67] Tom Lackowski, FNAL/FESS, private communication, November 2010.

[68] Generic Shielding Criteria for Compliance with Chapter 6 of the Fermilab Radiation Guide, a letter with attachment from D. Cossairt to J. Peoples dated December 11, 1990, unpublished.

[69] J. Anderson Jr., G. Federwitz, Radiation Protection Utilizing Electronic Berms, Mu2e-doc-1480-v1, submitted to docDB April 2011.

[70] Labyrinth and Penetration Spreadsheet, RP NOTE 118.

[71] A. Sondgeroth, A Collection of Useful Radiation Shielding Measurements for the Mu2e Project, Mu2e-doc-1232-v1, December 8, 2010

[72] A.F. Leveling, Shielding Materials Study, Mu2e-doc-1479, February 2011.

[73] A.F. Leveling, Radiation Attenuation through Stairway #2 at the AP30 Service Building, Mu2e-doc-1478, April 2011.

[74] K. Vaziri, Groundwater and air activation issues for Mu2e areas, Mu2e-doc-1553, May 2011

# 6 Conventional Construction

## 6.1 Introduction

The existing Fermilab Antiproton Source (FAS) and proposed elements of the Muon Campus Program play an essential role in the beam delivery strategy for the Mu2e Experiment. No enclosure currently exists in the vicinity of the FAS that would be suitable to house the Mu2e Experiment; therefore a new Mu2e facility is required.

The Conventional Construction sub-project includes the management, planning, design, and construction of new structures, buildings and utilities, as well as modifications to existing structures needed to house and support the assembly and operation of the Mu2e experiment. The Conventional Construction builds upon and extends the existing infrastructure and the proposed facilities provided in the Muon Campus projects.

The Conventional Construction scope includes the elements of work normally included in conventional construction such as earthwork, utilities, structural concrete, structural steel, architectural cladding, finishes, roofing, plumbing, process piping, HVAC, fire protection, lighting and electrical. Also included in this scope of work is the extension of existing utilities to the project site, excavation associated with the below grade cast-in-place concrete enclosures, creation of a shielding berm and site restoration.

Design activities have been packaged into two (2) distinct functional packages:

1. Detector Service Building and Detector Enclosure
   The Detector Enclosure will house the Mu2e detector including the Production Solenoid, Transport Solenoid and Detector Solenoid and will provide space for assembly and support functions. The Detector Enclosure adjoins to the existing Muon Campus Beamline Enclosure and will house the physics apparatus that comprise the proton beamline.

2. Antiproton Upgrades
   In order to accommodate the increased beam power required for the Mu2e Experiment, portions of the existing FAS infrastructure will be upgraded as part of the Conventional Construction scope of work. The Mu2e Accelerator sub-project has the responsibility to assess and provide the criteria for the conventional facilities portion of the required upgrade. Based on the increased beam power, the Antiproton Rings Service Buildings (ARSB) will be posted with entry controls to restrict access during beam operations. In addition, the increased beam power will require the installation of additional electrical power and cooling to specific areas within the ARSB. This includes the beamline power supplies located in the ARSB.

The preliminary and final designs for the Mu2e Conventional Construction subproject will be procured from one (1) or more Architectural/Engineering (A/E) firms currently under master contract with Fermilab. The subproject manager for Conventional Construction will manage the effort of the A/E firms. Subject matter experts within the Fermilab Engineering Services Section (FESS) will provide support and guidance to ensure that the Fermilab/Mu2e design, quality and configuration requirements are maintained throughout the design and construction process.

The subproject manager for Conventional Construction will also serve as the Construction Manager during the construction phase. The Construction Management Office (CMO) consists of a Construction Manager (CM), Construction Coordinator(s) (CC), and Procurement Administrator (PA). While the CM is ultimately responsible for all coordination and correspondence with the Subcontractor, the CM may delegate certain daily responsibilities to the CC and PA. Line management is directly linked from the CM to the Mu2e Project Manager and the FESS Section Head.

The scope of the Conventional Construction subproject will be realized through several construction packages. While the final configuration of the construction packages will be determined during the final design phase, it is anticipated that there will be between one (1) and three (3) construction packages for a project of this magnitude. This is intended to provide a logical and constructible sequence to reduce the construction period to a minimum. Further design iterations will be required to determine how to best optimize the construction packaging based on programmatic and funding limitations. The design methodology and construction means and methods for the Mu2e Conventional Construction work are expected to be similar to that which has been employed on the Fermilab site for decades.

## 6.2 Requirements

### *6.2.1* Technical Requirements

The technical requirements for the Mu2e Conventional Construction scope of work were developed from stakeholder input, organization processes and enterprise assets. The main sources of stakeholder inputs are the other Mu2e subproject managers. Regularly scheduled meetings are held with the Mu2e Collaboration, Mu2e

Technical Board, Mu2e Working Group, Mu2e Accelerator subproject leaders, Mu2e Solenoid subproject leaders and detector groups. The requirements employed in the design and construction of the Mu2e Conventional Construction are contained in the Mu2e document database [1] which provides a central, trackable repository for project data.

***Organizational Processes***

Organizational Processes provide institutional requirements for the design, construction and operations of all projects built and operated at Fermilab. For the Mu2e Conventional Construction these requirements are derived from the Policies and Procedures of the Directorate, Facilities Engineering Services Section (FESS), and FESS/Engineering. All applicable DOE orders and standards are included in these requirements. A selection of applicable standards is listed below:

- DOE Order 151.1C – Comprehensive Emergency Management System
- DOE Order 413.3A – Program and Project Management for the Acquisition of Capital Assets, Change 1 issued 7/28/06.
- DOE Order 414.1C – Quality Assurance
- DOE Order 420.1B – Facility Safety
- DOE Order 430.1B – Real Property Asset Management (2/8/08)
- DOE Order 430.2B – Departmental Energy, Renewable Energy and Transportation Management
- DOE Order 450.1A – Environmental Protection Program (6/4/08)
- DOE STD-1066-99 – Fire Protection Design Criteria
- DOE STD-1073-2003 – Configuration Management
- DOE Guide 420.1-2 – Guide for the Mitigation of Natural Phenomena Hazards for DOE Nuclear Facilities and Non-Nuclear Facilities
- 10 CFR 835 – Radiological Protection Program
- 10 CFR 851 – Worker Safety and Health Program
- 10 CRF 851.23 – Safety and Health Standards
- Internal Fermilab permits and work notifications as described in FESHM.

***Enterprise Standards***

Enterprise standards from regulatory agencies, code bodies and trade organizations also provide input for the design and construction of the Mu2e facility. The Fermilab Engineering Standards Manual provides a comprehensive listing of applicable and adopted building codes and design standards. The applicable standards are listed below:

- Codes, Standards, and Guidelines
- International Building Code (IBC) – 2009 Edition

- International Energy Conservation Code – 2009 Edition
- International Fire Code – 2009 Edition
- International Mechanical Code – 2009 Edition
- Minimum Design Loads for Buildings and Other Structures – ASCE 7-05
- Building Code Requirements for Structural Concrete – ACI 318-05
- Specification for Structural Steel Buildings – AISC 360-05
- Building Code Requirements for Structural Concrete and Commentary – ACI 318-08
- Building Code Requirements for Masonry – ACI 530-05
- Illinois Plumbing Code – 2004
- Illinois Department of Public Health Codes
- Illinois IEPA
- NFPA 101 Life Safety Code – 2009 Edition
- NFPA 13 – Standard for the Installation of Sprinkler Systems – 2010 Edition
- NFPA 24 – Standard for the Installation of Private Fire Service Mains and Their Appurtenances – 2010 Edition
- NFPA 30 – Flammable and Combustible Liquids Code – 2008 Edition
- NFPA 55 – Compressed Gases and Cryogenic Fluids Code – 2010 Edition
- NFPA 70 – National Electrical Code – 2008 Edition
- NFPA 70E – Standard for Electrical Safety in the Workplace – 2009 Edition
- NFPA 72 – National Fire Alarm Code – 2010 Edition
- NFPA 80 – Fire Doors and Fire Windows – 2010 Edition
- NFPA 90A – Standard for the Installation of Air-Conditioning and Ventilating Systems – 2009 Edition
- NFPA 90B – Standard for the Installation of Warm Air Heating and Air Conditioning Systems – 2009 Edition
- NFPA 92A – Standard for Smoke-Control Systems utilizing Barriers and Pressure Differences – 2009 Edition
- NFPA 92B – Standard for Smoke Management Systems in Malls, Atria, and Large Spaces – 2009 Edition
- NFPA 110 – Emergency and Standby Power Systems – 2010 Edition
- NFPA 115 – Standard for Laser Fire Protection – 2008 Edition
- NFPA 780 – Standard for the Installation of Lightning Protection Systems (and UL 96A) – 2008 Edition
- National Institute of Standards and Technology
- ASHRAE Standard 90.1-2004 Energy Standard for Buildings Except Low-Rise Residential Buildings
- ANSI/HFES 100-2007 – Human Factors Engineering of Computer Workstations
- ANSI 17.1 Safety Code for Elevators and Escalators

- ANSI/ASHRAE Standard 62.1-2004 Ventilation for Acceptable Indoor Air Quality
- ANSI/AIHA Z9.5-2003 Standards for Laboratory Ventilation
- ASME Boiler and Pressure Vessel Code (2007)
- ANSI/ASME B31.3 – Process Piping (2002)
- ANSI 31.9 – Building Services Piping (1996)
- Occupational Safety and Health Administration (OSHA)
- Underwriters Laboratory
- ICC/ANSI A117.1 – 2003 Standard for Accessible and Usable Buildings and Facilities Illinois Accessibility Code
- ADA Accessibility Guidelines for Buildings and Facilities (ADAAG) – 2004
- Illinois Accessibility Code

## 6.3 Recommended Design

The location for the Mu2e facility was selected primarily to facilitate the efficient transport of protons from the Debuncher Ring of the FAS to the Mu2e experiment. Figure 6.1 is an aerial photo of the southwestern portion of the Fermilab site with the location of key landmarks indicated along with the location of the envisioned Mu2e Facility. The connection between the Mu2e external beamline and the FAS is made at the upstream end of the Muon Campus Beam Enclosure. The availability of an existing beamline stub and the length of the external beamline, driven by the physics requirements of the delivered proton beam, determine the location of the Mu2e Experiment. A segment of Kautz Road will be relocated to the west of the existing location and west of the proposed location of the Mu2e facility.

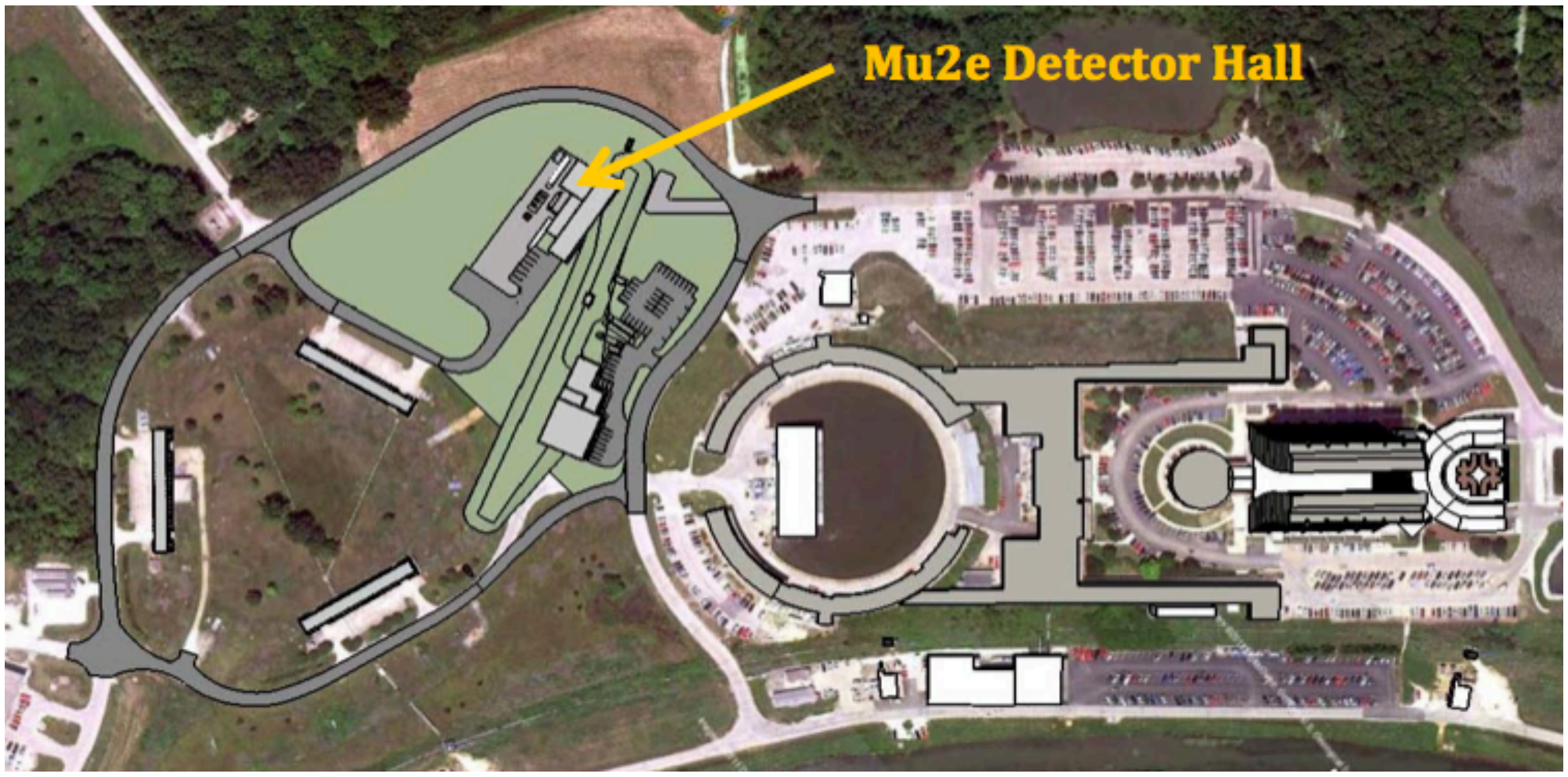

Figure 6.1 Aerial Photo showing the project location

The selected location is at a confluence of existing utilities serving the Main Injector, MiniBooNE, SciBooNE, the Antiproton Ring and the MC-1 Building. These utilities include Industrial Cooling Water (ICW), Domestic Water Service (DWS), Natural Gas (Gas), Low Conductivity Water (LCW) and electrical power. Figure 6.2 is a site utility plan that indicates the location of existing and proposed utilities, surface buildings and the below grade enclosures. The selected location provides access to the majority of the utilities needed to support the assembly and operation of the Mu2e project.

The Industrial Cooling Water (ICW), Domestic Water Supply (DWS), Chilled Water Supply/Chilled Water Return (CHWS/CHWR), Natural Gas (Gas) and Sanitary Sewer (SS) services for the Mu2e project will be extended from existing services along Kautz Road or Well Pond Road. Primary Electric Power will be extended as a "looped" feed from a manhole near the MC-1 Building. Additionally, data and communication services will be extended from the existing FAS. These services will be routed via new or existing utility corridors. Details of the anticipated utility work are listed below:

- ICW will be used primarily for fire protection in the sprinkler system and hydrants. A small, 30 GPM flow of ICW will be utilized to cool a vacuum pump that will service the physics apparatus. The ICW return will be routed via ditches to the existing Swan Lake system, upstream of the designated outfall to waters of the state.
- Adequate cooling capacity for the Low Conductivity Water (LCW) system required for the external beamline components currently exists at the Central Utility Building. To fully utilize this capacity, the system will be re-programed to take advantage of the flow and heat rejection capacity of the abandoned Accumulator Magnets in the existing FAS. The new LCW service will be routed through the underground enclosure to the Mu2e Experiment.

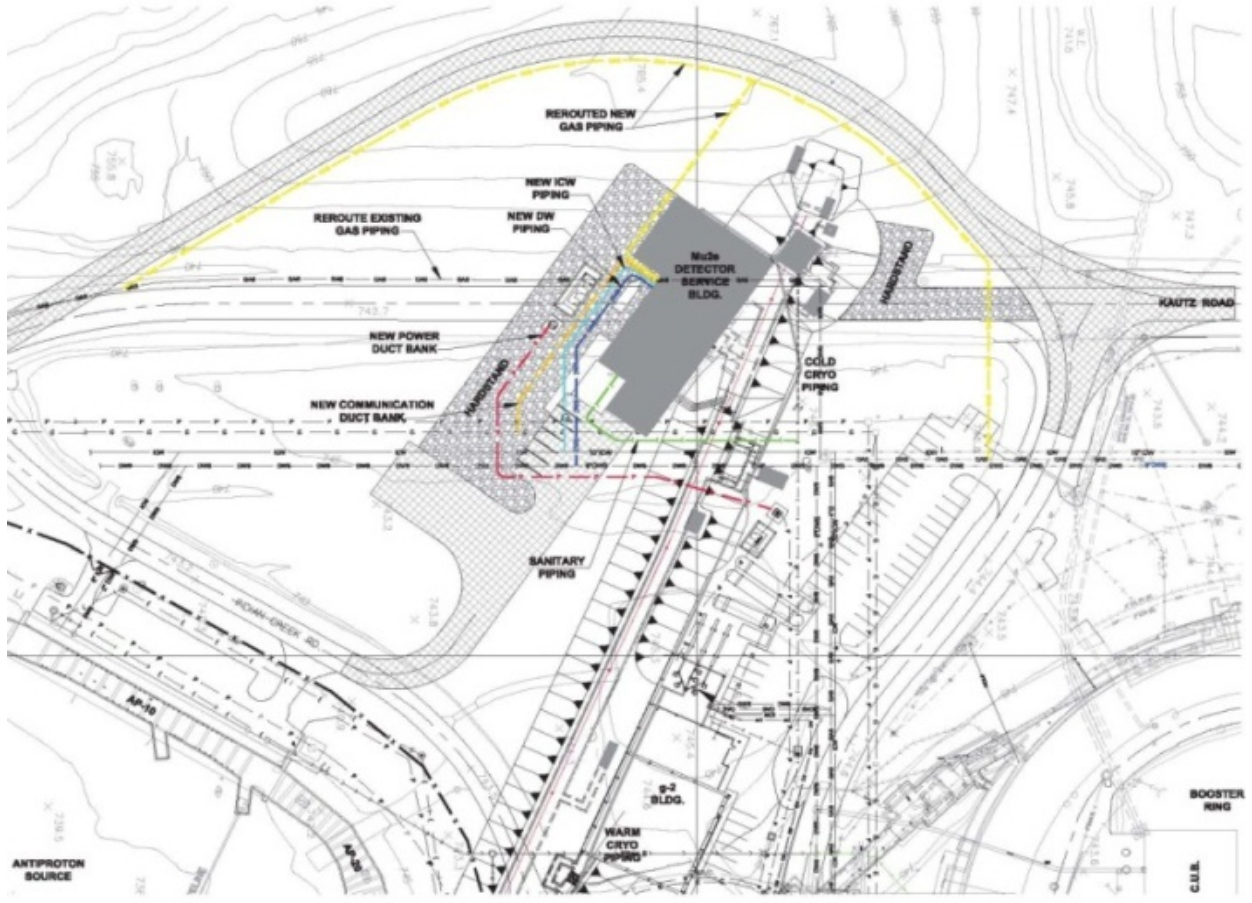

Figure 6.2. Utility Site Plan.

- An adequate supply of drinking water is available through an existing 6" DWS supply line that runs along Kautz Road. This existing line will be extended into the Mu2e facility for domestic uses.

- Natural Gas from an existing underground line will be extended to the Mu2e facility for HVAC heating.
- The connection to the Sanitary Sewer service will be made at the existing lift station near MC-1 via an underground gravity pipe.
- An adequate supply of Chilled Water (CHW) is available from the Central Utility Building (CUB). Connection to the CHW Supply and CHW Return lines will be made near MC-1 and extended via underground lines to the Mu2e facility. CHW will be used for cooling of the radioactive water systems.
- Radioactive Water (RAW) systems will be utilized for the programmatic equipment. The RAW system, based on existing Fermilab system designs of similar size, will be isolated from surface water and will reject the heat to the chilled water system originating at the CUB. While the design, procurement and installation of the RAW systems are included in the Accelerator Subproject, the Conventional Construction subproject will provide the CHW piping to a location within the Mu2e facility that will allow for final connections to the RAW system.
- The FAS is currently powered from the 13.8 KV feeder, Feeder 24, originating at the Master Substation (MSS) through the air switch at F-3. A “looped” feed runs around the Antiproton Ring from F-3. Feeder 52 from the Kautz Road Substation is connected at the F-3 air switch and provides backup power for lighting and water sumps. The electric power for the Mu2e Experiment will be provided via an underground duct bank and feeder loop extended from the MC-1 building.

A new access road will provide vehicular access to the Mu2e Experiment from existing Fermilab roads. This new road will be constructed in a similar manner to existing Fermilab roads and will be suitable for all weather access. The access road will intersect existing Indian Creek Road adjacent to the AP-10 parking lot. Paved parking will be provided for six (6) vehicles at the facility along with a gravel hardstand that will provide a staging area during detector assembly. A paved approach to the at-grade loading dock with suitable truck maneuvering space is provided. Figure 6.3 shows the overall area site plan indicating proposed roads and buildings.

A combination of earth, concrete and steel shielding will be provided for below grade beamline enclosures in order to provide the equivalent of sixteen (16) feet of earth shielding for all primary beam transport and targeting enclosures. An earth berm with maintainable side slopes will be used.

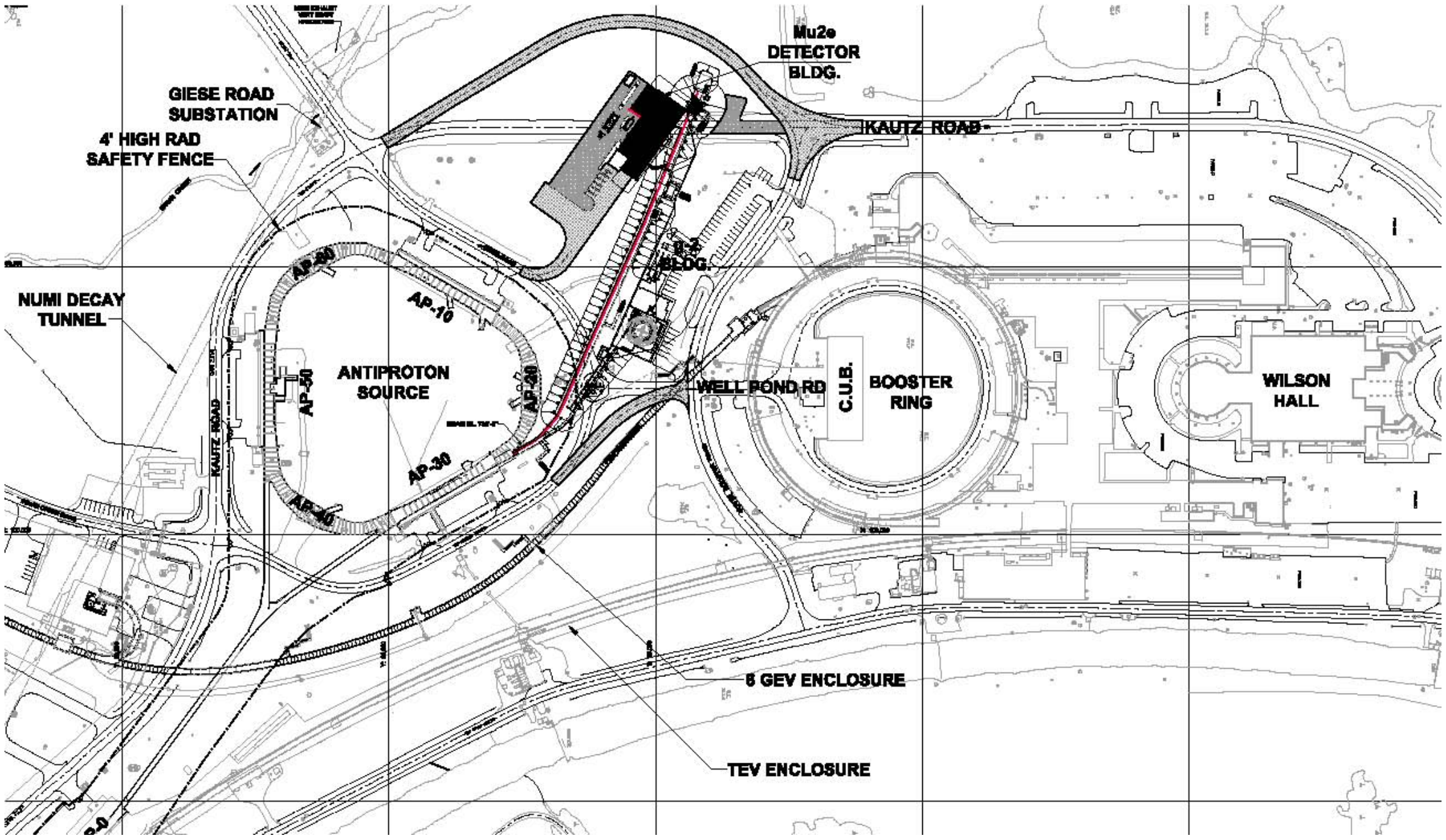

Figure 6.3.The overall area site plan indicating the proposed Mu2e facility and surrounding roads.

### *6.3.1* Mu2e Detector Service Building and Detector Enclosure

***Mu2e Detector Enclosure***

The Mu2e Detector Enclosure will consist of a below-grade, cast-in-place concrete enclosure located below the Mu2e Detector Service Building. The Detector Enclosure, shown in Figure 6.4 consists of the spaces required to house, support and accommodate assembly and installation of the proton beamline, Mu2e solenoids and Detector components as well as the detector support equipment.

The finished floor of the Mu2e Detector Enclosure will be located at an elevation of 721'-0", 25.5 feet below the floor of the Mu2e Detector Service Building and approximately 25 feet below grade level.

The Detector Enclosure adjoins to Muon Campus Beamline Enclosure to house the programmatic beamline components that will be required to transport the proton beam from the existing Debuncher Ring to the Mu2e production target internal to the Production Solenoid. The connection to the Muon Campus Beamline will be made with a ten-foot wide by eighteen-foot high concrete enclosure approximately 100 feet long, running from the northwest end of the Muon Campus Beamline Enclosure to the Mu2e Detector Enclosure. The enclosure is designed to support 16 feet of earth and concrete shielding to grade. This amount of shielding will allow for "unlimited occupancy" rating of all above ground areas accessible to the general public. A labyrinth style corridor between the enclosure and detector enclosure will provide the Detector Enclosure a code required second means of egress.

The Production Solenoid and related experimental equipment, shown in Figure 6.5, are contained within a cast-in-place concrete enclosure that has been designed to be isolated from the downstream detector components by way of modular precast concrete shielding installed after the detector installation is complete.

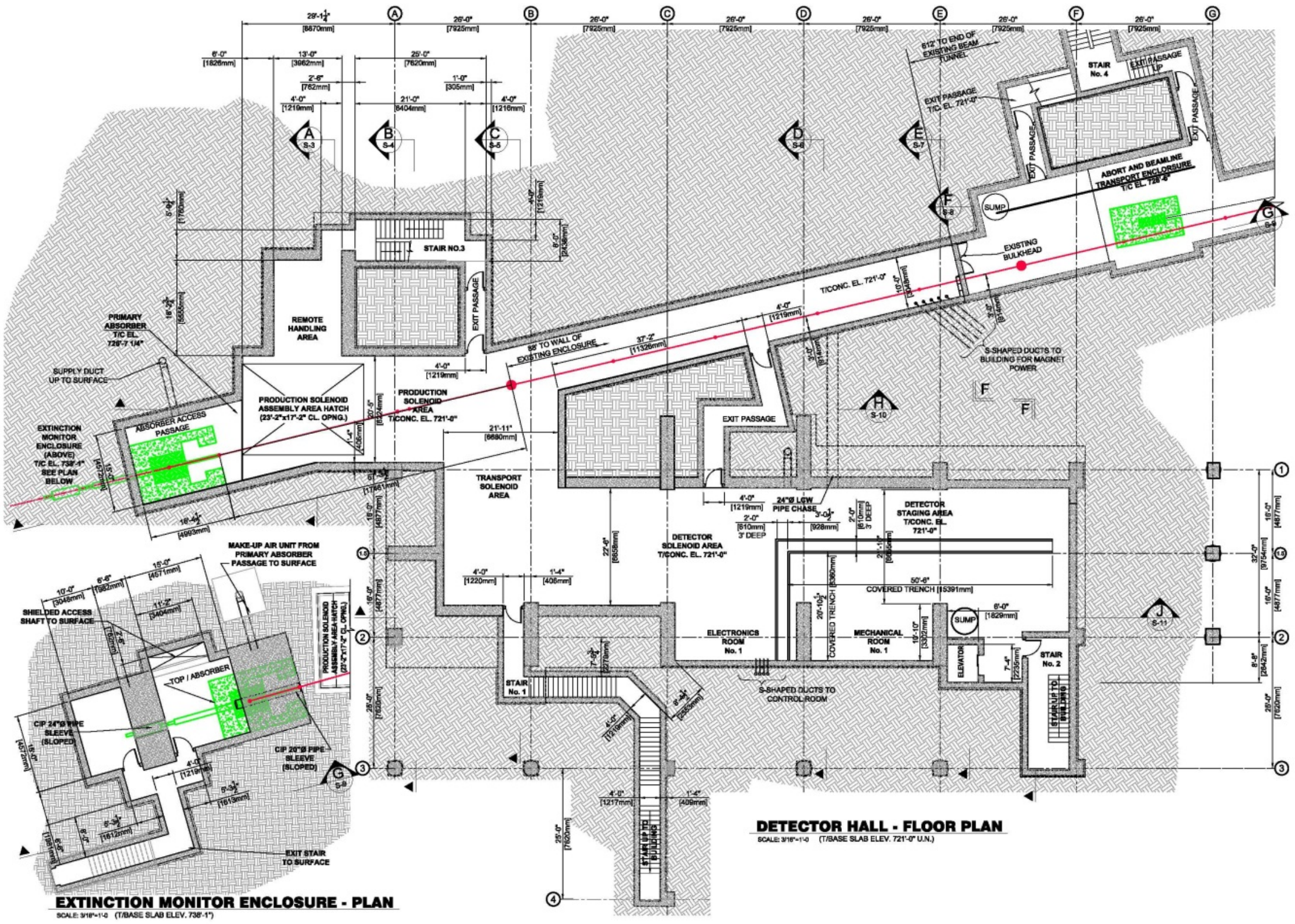

Figure 6.4. Detector Hall Underground Plan.

After entering the Production Solenoid the proton beam interacts with the production target. The fraction of the beam that does not interact in the target exits the Production Solenoid and terminates in the downstream proton Beam Absorber. This area, downstream of the Production Solenoid, consists of two cast-in-place concrete enclosures that will house the Beam Absorber, Extinction Monitor as well as space for the remote handling equipment required to replace components inside the Production Solenoid.

Located downstream of the Production Solenoid is a shielded hatch sized to accommodate the installation for the Production Solenoid as well as providing access to accommodate the remote handling operations. An exterior crane will facilitate lifting and lowering of equipment through the hatch. A small detector to monitor the beam extinction is positioned above the Beam Absorber.

The Transport Solenoid and the Detector Solenoid are housed under precast concrete shielding at the Detector Service Building floor level. The above grade building's two 30 ton bridge cranes will be used for vertical transportation of material and equipment required for the installation and maintenance of the Detector and Transport Solenoids. These solenoids will be shielded during operations by modular concrete shielding blocks.

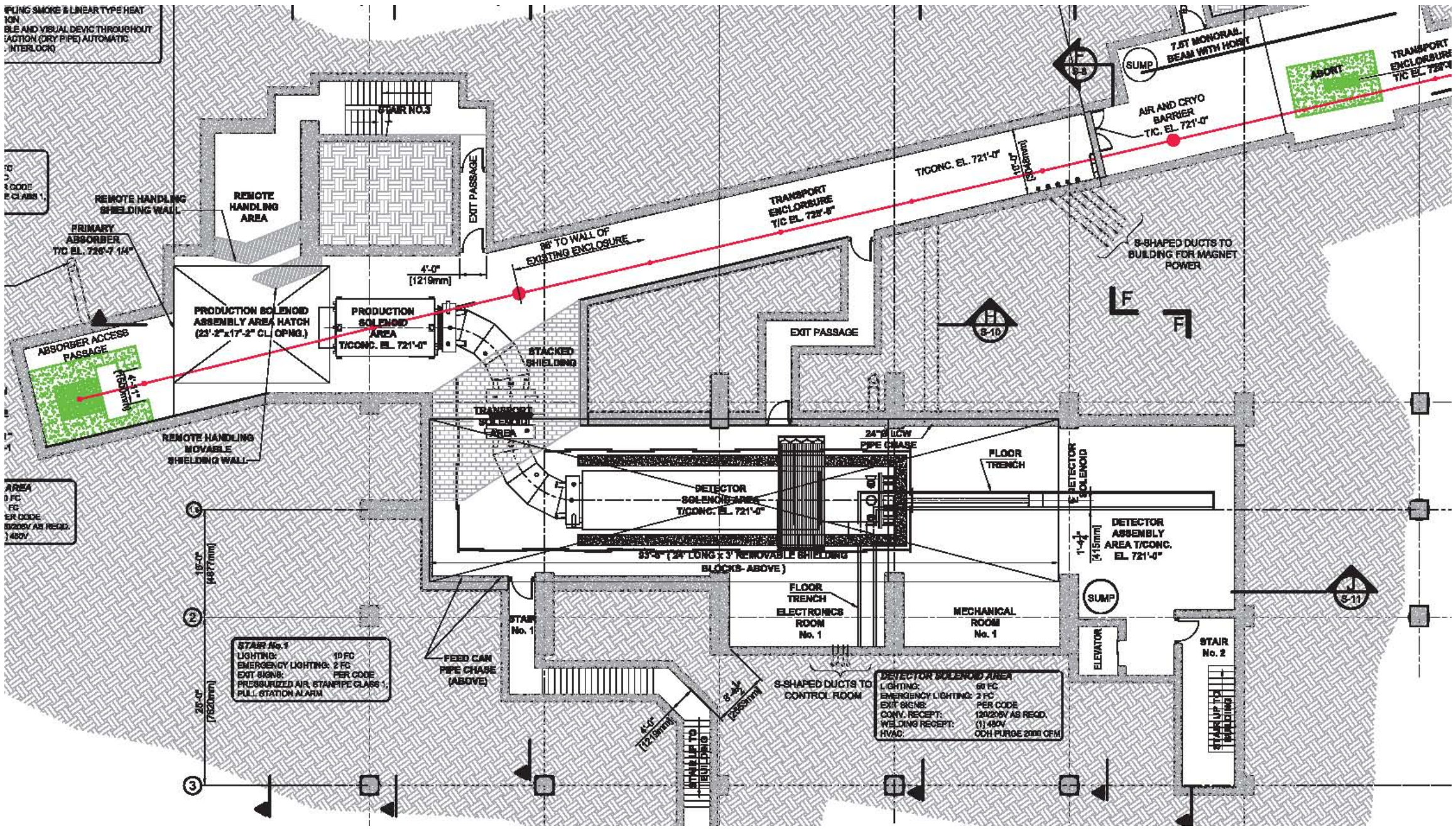

Figure 6.5. Underground plan showing the Mu2e apparatus.

The Detector Solenoid area shown in Figure 6.5 has been sized to accommodate the Detector Solenoid and the surrounding Cosmic Ray Veto as well as an associated staging area. Steel track plates anchored to the concrete base slab will accommodate the transport and support of the solenoid and detector elements. The staging area will allow the interior detector components to be commissioned outside of the Detector Solenoid prior to installation and removed later for repair or adjustment, if necessary. Space for electronics, vacuum pumps and other infrastructure necessary to support detector operations will also be available in this area.

The construction of the below grade structure will utilize traditional "open cut and cover" method. This method has been used successfully at Fermilab for the construction of the majority of shielded enclosures on-site. Listed below are specific features of the Detector Enclosure:

- The below grade enclosures will have code compliance exit corridors and stairways provided to conform to the required maximum distance to an exit. These exit stairs will be configured to maintain the shielding requirements.
- An earthen shielding berm with maintainable side slopes will be used.
- The interior surfaces of the walls and ceiling will be painted.
- The exterior concrete wall surfaces will be moisture proofed to provide a safe and dry semi-conditioned space for personnel and equipment.
- The below grade structures will be flanked with underdrain piping that will negate the hydraulic pressure on the walls and roof of the enclosure. The underdrains will be routed to sump basins with duplex sump pumps that will discharge water onto grade and away from the structure.
- The walls and ceiling of the enclosure will be fitted with embedded channel inserts to allow for the support of cable trays, piping, electrical conduits and fire detection equipment.
- Convenience outlets, 120/208VAC, will be provided at least every sixty (60) feet along the walls.
- In addition to required emergency and exit lighting, light fixtures will be provided to supply a minimum of 20 foot-candles. A percentage of these lights will be on UPS circuits to provide emergency lighting during power failures.
- The enclosure will be ventilated with neutral, dehumidified air.
- Fire detection will be via air sampling and line type sensors. Fire protection will be provided for the detector and solenoids.

Underground airflow has been developed to provide a required oxygen deficiency hazard (ODH) purge, minimum temperatures and maximum relative humidity below the dew point.

***Grade Level Structures***

An above-grade Detector Service Building (DSB) will provide space for the various support services required for the Mu2e Experiment. A 3-D depiction of the grade level structure is shown in Figure 6.6. The Mu2e Detector Service Building will be constructed as a braced frame, steel construction with prefinished metal siding and a built up roof system. The construction type and style will match that of adjacent facilities on the Fermilab site.

Figure 6.7 provides a plan view of the Mu2e DSB. The above grade Mu2e DSB will include a high bay and an adjacent low bay.

The Mu2e DSB high bay will provide space for unloading, staging and assembling the detector components. The high bay will be equipped with two 30-ton capacity overhead bridge cranes, similar in style and construction to cranes installed with the Main Injector project.

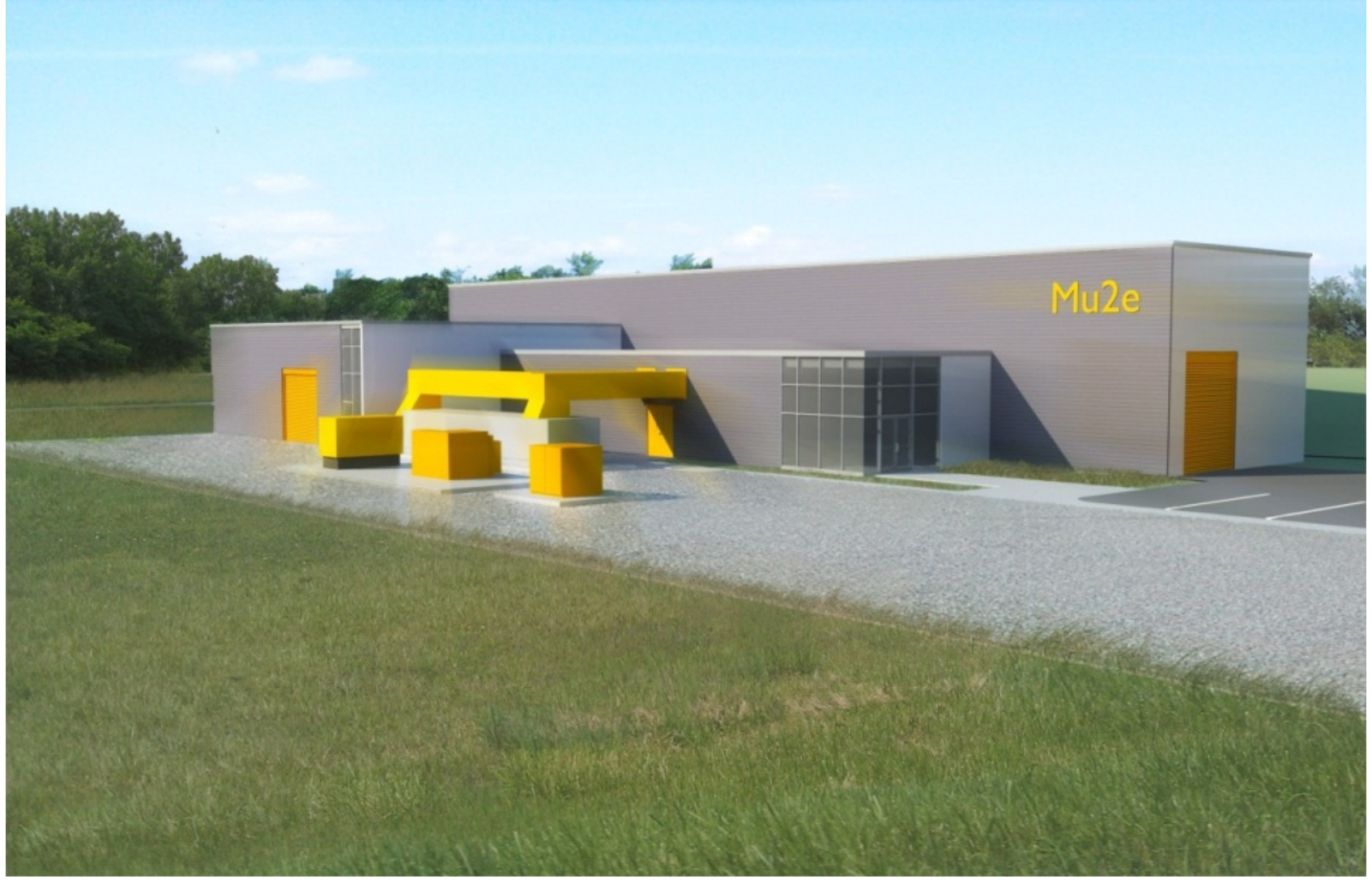

Figure 6.6. A 3-D depiction of the grade level structure.

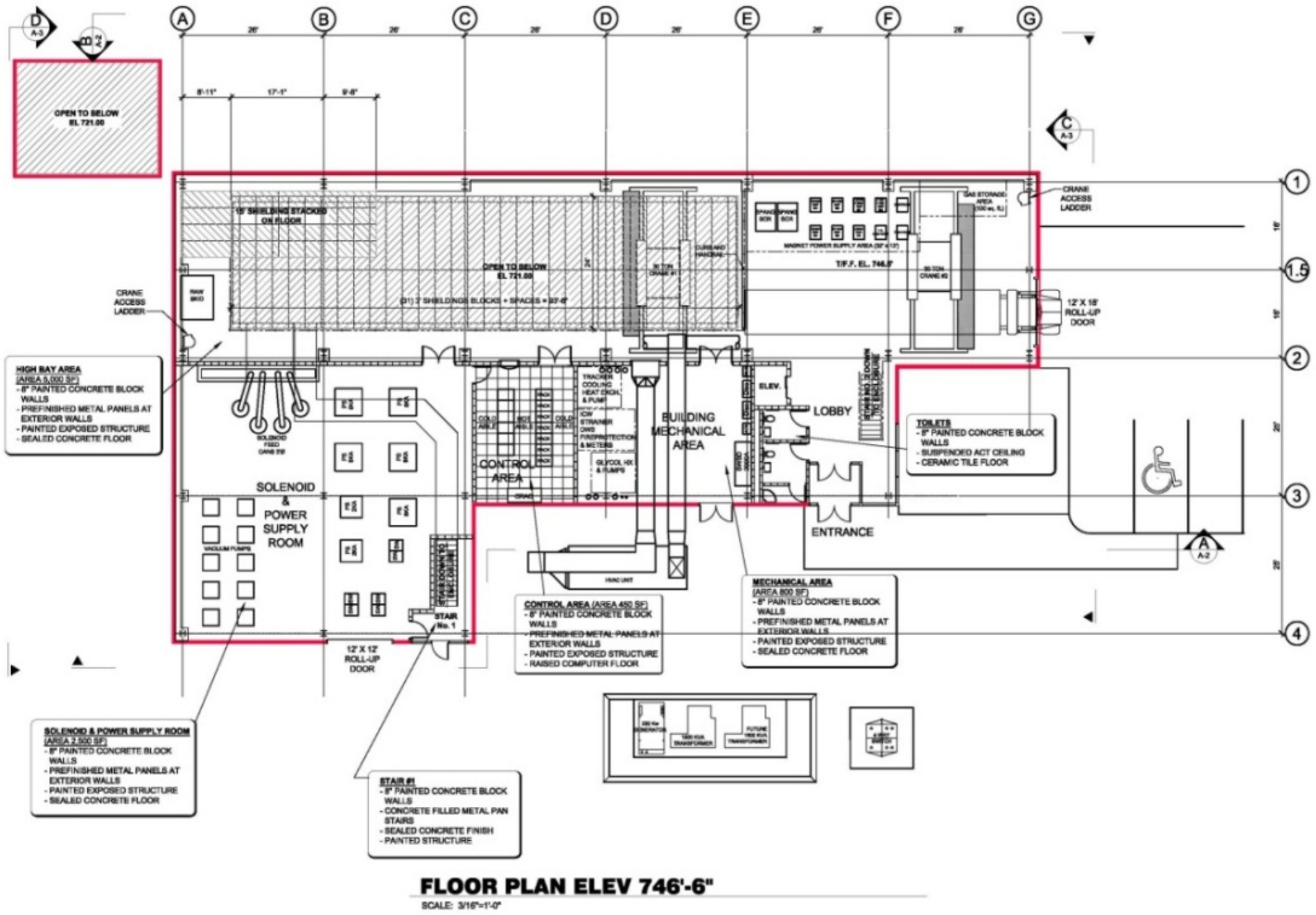

Figure 6.7. Detector Service Building Grade Level Plan.

A portion of the high bay, shown in the upper right hand corner of Figure 6.6 above, will be used as a loading dock as well as housing beamline power supplies and a gas storage area. The area above the below grade Mu2e Detector Enclosure and below grade staging are intended to remain open during assembly and installation process. Once the Mu2e detector installation is complete, the opening will be filled with modular precast concrete shielding blocks to a depth of three (3) feet to provide the required shielding during operations. The configuration provided will allow for up to six (6) feet of shielding if needed in the future. The high bay contains space for temporary staging of the shield blocks should removal be required for maintenance and/or repair of the detector components.

The Mu2e DSB low bay area will contain the spaces required to support the assembly, installation and operation of the Mu2e detector including a Solenoid Power Supply Room, a Mechanical Room for mechanical and electrical equipment, an Electronics Room, toilet rooms, janitor's closet and exit stairs from the below grade Mu2e Detector Enclosure. An elevator is provided for vertical transport of personnel and equipment between the surface and below grade portions of the building.

The Solenoid Power Supply Room is sized based on the criteria provided by WBS 4, Solenoid subproject and will house the electrical equipment to power the below grade solenoids. Electrical power for the power supplies will be brought to this area and terminated at a disconnect switch for distribution. A shaft like penetration provides connection to the below grade enclosure to the Cryo piping with strip lines.

The Mu2e DSB low bay will contain an Electronics Room that will house the electronic racks for the solenoid controls and data acquisition system including the online processing farm, data transfer controllers and networking equipment. This equipment, primarily computers, requires strict environmental controls for proper operation. This space will be conditioned with dedicated equipment to provide suitable environmental control. This space is intended to house equipment only and will not be occupied during normal operations. During normal operations the Mu2e Experiment will be run from a remote control room located within Wilson Hall.

The Mechanical Area will house the equipment that supports the assembly and operation of the detector and operation of the facility. This includes the cooling equipment for the experiment, ICW strainer, fire protection riser, heat exchangers, pumps and the huge number of meters and monitoring equipment required as part of the "smart lab" initiative. The Mechanical Area will also contain the electrical switchgear, transfer switch and panelboards to distribute the incoming electrical service. The electrical switchgear will serve conventional facilities equipment, the programmatic equipment for the Mu2e Experiment and the 480 V HVAC systems. New electrical panels serving the lights, outlets and general house power will be included in the electrical power distribution system.

The Heating Ventilation and Air Conditioning (HVAC) systems for the Mu2e DSB will conform to ASHRAE 90.1, ASHRAE 62, applicable NFPA requirements and applicable sections of the Fermilab Engineering Standards Manual. The Heating, Ventilation and Air Conditioning Design Parameters for the Mu2e DSB are listed below:

- The temperature shall be maintained between 68 to 78 degrees Fahrenheit.
- The relative humidity shall be maintained below 50%. There is no minimum requirement.
- All plumbing work will be installed in accordance with Illinois Plumbing Code and Standard Specifications for Water & Sewer Main Construction in Illinois.
- The Mu2e DSB will be conditioned with a packaged HVAC system.
- A duplex sump pump system will be installed to collect subsurface water from around the Mu2e DSB and discharge to grade.

The Mu2e DSB egress, construction type, emergency lighting, exit signage and smoke control ventilation will be designed in accordance with the International Building Code (IBC) and National Fire Protection Association (NFPA) standards. Fire alarm and suppression systems for the Mu2e facility will be designed in accordance with the applicable sections of the Fermilab Engineering Standards Manual. Automatic sprinkler systems will comply with the standards for an Ordinary Hazard Group 1 classification, in accordance with latest edition of the NFPA Codes and Standards. The most commonly used NFPA standards relative to automatic sprinkler systems are: 13, 20, 25, 70, 72, 90A, and 101. A life safety consultant has developed a Preliminary Fire Protection / Life Safety Recommendations Report that have been incorporated into the conceptual design.

Fire alarm systems will be designed with a minimum standby power (battery) capacity capable of maintaining the entire system in a non-alarm condition for 24 hours, in addition to 15 minutes in full load alarm condition. The most commonly used NFPA standards relative to fire alarm systems are: 70, 72, and 90A.

The Mu2e DSB will be equipped with an addressable evacuation fire alarm system consisting of:

- Manual fire alarm stations at the building exits.
- Sprinkler system water flow and valve supervisory devices.
- A combination fire alarm horn and strobe located throughout the building.
- Addressable emergency voice and visual fire alarm system.
- Connection to the site wide FIRUS monitoring system
- Smoke and heat detection as required.

***Electrical Distribution***

A new 1500 kVA transformer will provide the electrical power for the Mu2e Experiment. The transformer has been sized to accommodate the anticipated electrical power for both the conventional facilities and the Mu2e programmatic equipment. A new concrete encased power duct bank will be installed to connect the Mu2e facility to Fermilab's existing 13.8 kV electrical power grid. This connection, to Feeder 24, will originate at a manhole near the MC-1 Service Building. Figure 6.8 below depicts the single line diagram for the Mu2e Experiment.

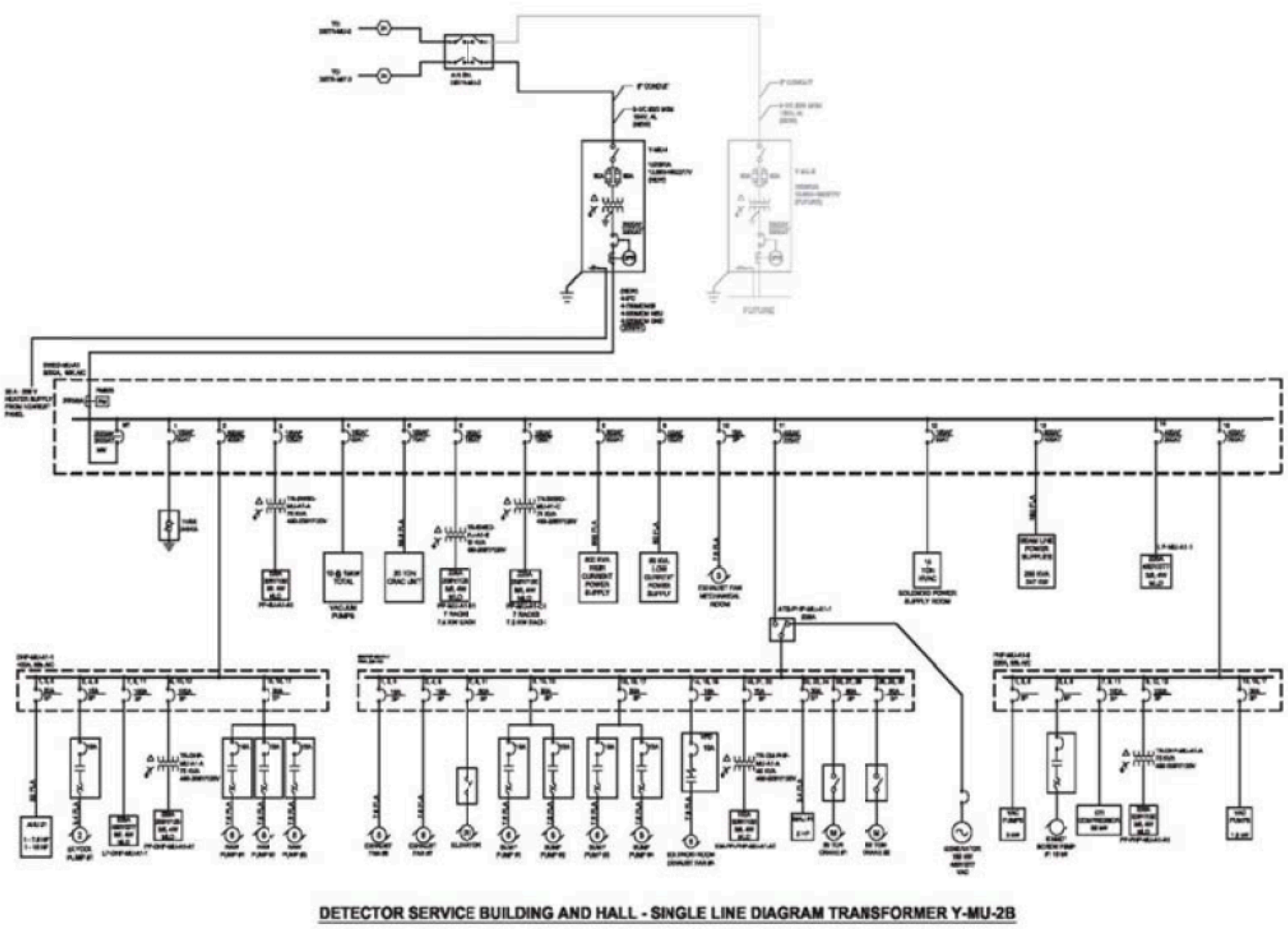

Figure 6.8. Detector Service Building Electrical Single Line Diagram.

The new Mu2e electrical substation will be located adjacent to the Mu2e Detector Service Building. The substation will be designed to accommodate the new transformer and associated 4-way air switch. In addition, the substation will be designed to accommodate a second future transformer. The new electrical substation will be connected to the Mu2e Detector Service Building via a new concrete encased duct bank and will be routed to new electrical service switchgear or disconnects inside the Mu2e Detector Support Building. Future loads and reasonable redundancy have been considered in the design of the electrical system.

### *6.3.2* Antiproton Upgrades

The Conventional Construction upgrades to the FAS includes the two (2) components listed below:

1. Electrical upgrade to provide an additional 1500 KVA transformer and the associated connections.
2. Mechanical upgrades including increased fan coil units in the existing AP-30 and AP-50 Service Buildings.

***Service Building***

The existing FAS was constructed in the early 1980's and was designed to handle a beam current of antiprotons that is significantly smaller than the 8 kW of proton beam power required for Mu2e (see Section 5.10). The requirements for the FAS Service Buildings are described in Section 5.10.3.4. The Service Buildings will have Rad Posting and controlled entry.

***Electrical Upgrade***

In order to support the new Mu2e beamline components that will be installed in the existing FAS, a new electrical substation with a 13.8 kV four-bay air switch and 1500 kVA pad mounted, oil filled transformer will be installed adjacent to the existing AP-30 Service Building. The four-bay air switch will provide additional configuration control as well as a means of serving the new transformer. New incoming service disconnects will be installed inside the existing AP-30 Service Building.

***Mechanical Upgrade***

The increased heat load produced by the beamline equipment housed in the existing AP-30 and AP-50 Service Buildings will exceed to capacity of the existing cooling equipment. The Conventional Construction subproject will install additional cooling capacity to provide cooling for this equipment.

# 6.4 Considered Alternatives to the Proposed Design

## *6.4.1* Alternatives Designs and Locations

As part of the conceptual design process, two (2) Project Definition Reports (PDR) were developed for the Mu2e facility. The first PDR was produced in September 2008 and the second in May 2009. A brief description of each alternative is described below. A complete copy of these reports can be found in [2].

***September 2008 Project Definition Report***

The September 2008 Project Definition Report positioned the facility located in an area west of Kautz Road and south of Giese Road on the Fermilab site. The proton beam is extracted from the existing Antiproton Source near AP-60, directed downward and beneath Indian Creek and transported to a new below grade enclosure housing the Mu2e Experiment. The location of the proposed facility is shown in Figure 6.9.

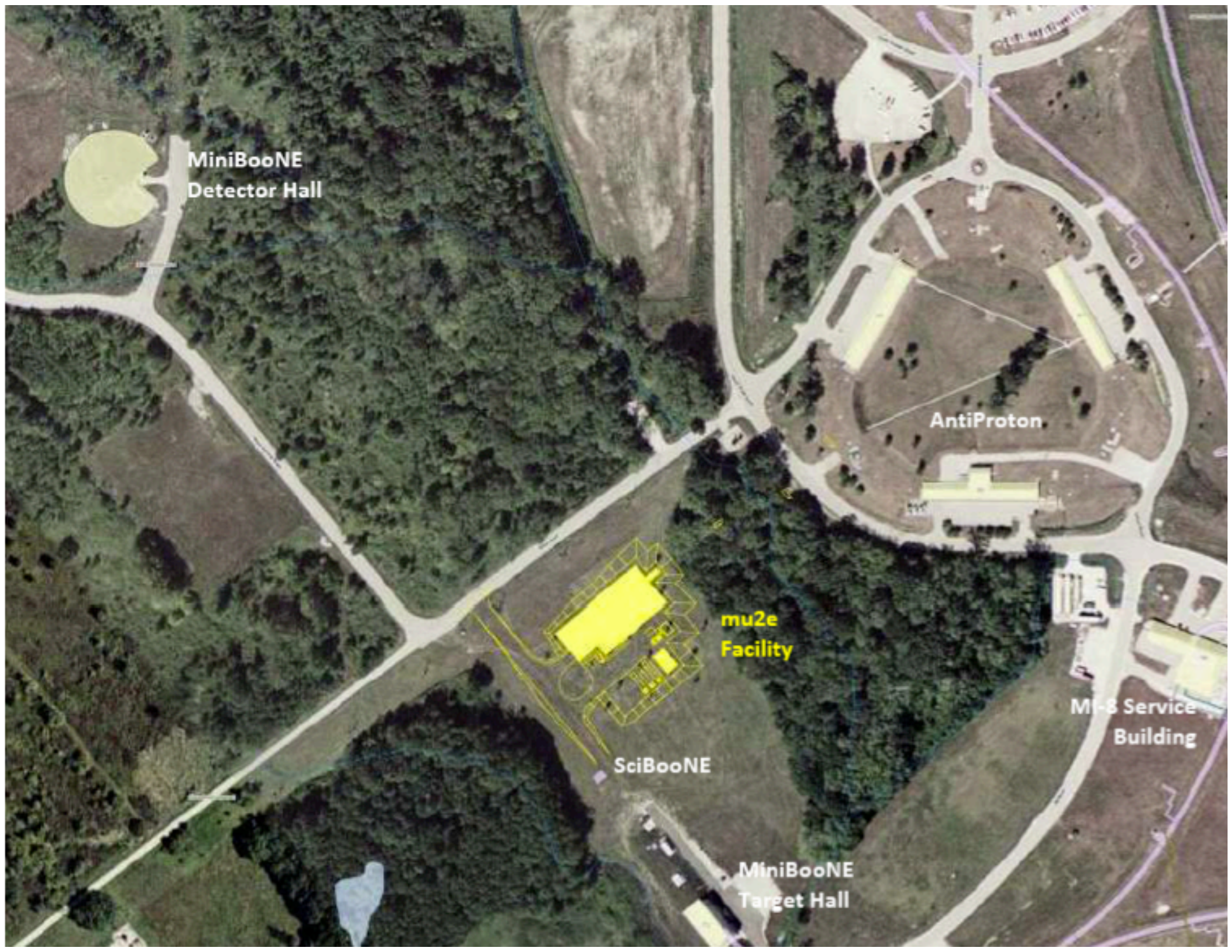

Figure 6.9. Alternative location for the Mu2e facility from the September 2008 Project Definition Report.

The Production Solenoid, Transport Solenoid and Detector Solenoid were contained within a cast-in-place below grade concrete enclosure. The below grade portion of the Mu2e Detector Support Building was separated from the above grade portion of the building by 21 feet of earth equivalent shielding. In order to accommodate installation and maintenance of the Mu2e detector components, the shielding over the Production Solenoid and the Detector Solenoid was removable precast concrete shielding blocks. This provided complete crane access to the entire detector. The high bay provided space for unloading the detector components from semi-trailers. The adjacent low bay contained the Control Room, Mechanical Room and Electrical and Power Supply Room.

The plan described in the 2008 PDR was abandoned for several reasons. It was recognized that the beam extraction from the AP-60 sector of the Antiproton Source was problematic. From a conventional construction standpoint, the cost and environmental risk of extending the proton beamline under Indian Creek was deemed excessive. Additionally, the cost associated with full access to the detector through removable shielding across its entire length was very high due the cost of the shielding blocks themselves as well as the cost of the re-enforced side wall structure required to withstand the load from lateral earth pressure when the shielding blocks were removed.

***May 2009 Project Definition Report***

A second Project Definition Report, completed in May 2009, addressed many of the issues that had been identified in the 2008 PDR. A more suitable location for extracting the beam was identified at the north end of the existing AP-30 enclosure using an existing beam pipe, used in the past for studies with injected protons from the Booster Ring.

It was recognized that the connection to the existing FAS could be made at this location with a minimum of demolition and component reconfiguration. The Mu2e facility was intended to be located in an area east of existing Kautz Road and north of Giese Road. Figure 6.10 includes a site photo depicting the new construction required for the Mu2e beamline and facility indicated in yellow.

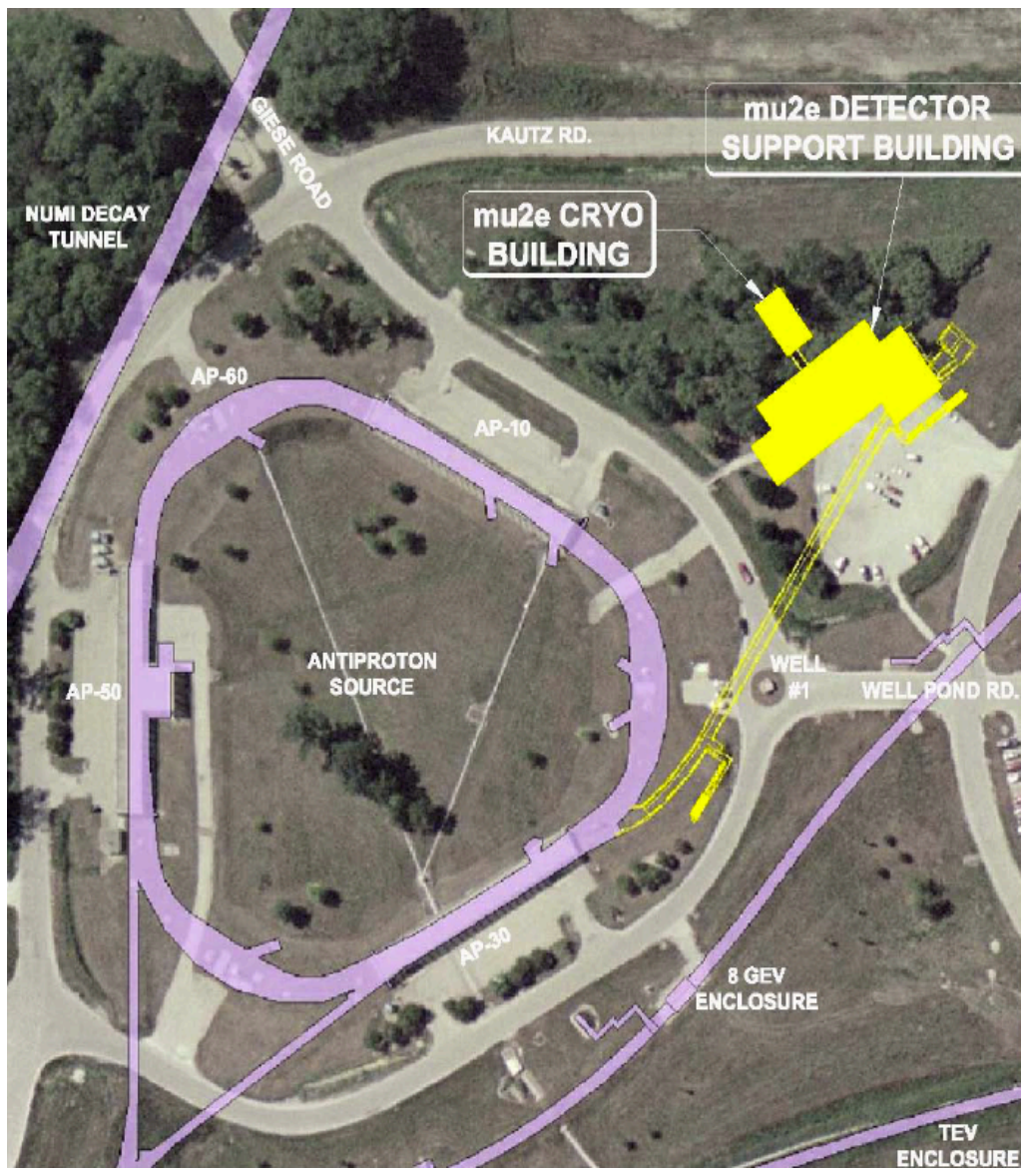

Figure 6.10. Alternative location for the Mu2e facility from the May 2009 Project Definition Report.

Because it would no longer be necessary to for the beamline enclosure to pass beneath Indian Creek, the new beamline would have been shorter and at a shallower depth, proving more cost effective with fewer construction complexities. To further reduce cost, the design of the Mu2e Detector Support Building was modified to require fewer removable shielding blocks. In order to accommodate installation and maintenance of the Mu2e detector components, some removable shielding over the Production Solenoid and the Detector Solenoid was preserved. The below grade portion of the Mu2e DSB is separated from the above grade portion of the building by 21 feet of earth equivalent shielding above the beamline components and the Production Solenoid for primary beam shielding. A plan view of the above grade portion of the Mu2e DSB is shown in figure 6.11 below. The Detector Solenoid is separated by 15 feet of concrete shield blocks from the above grade portion of the building. The high bay is equipped with a 35-ton capacity overhead bridge crane. The high bay has a limited staging area for the precast concrete shield blocks.

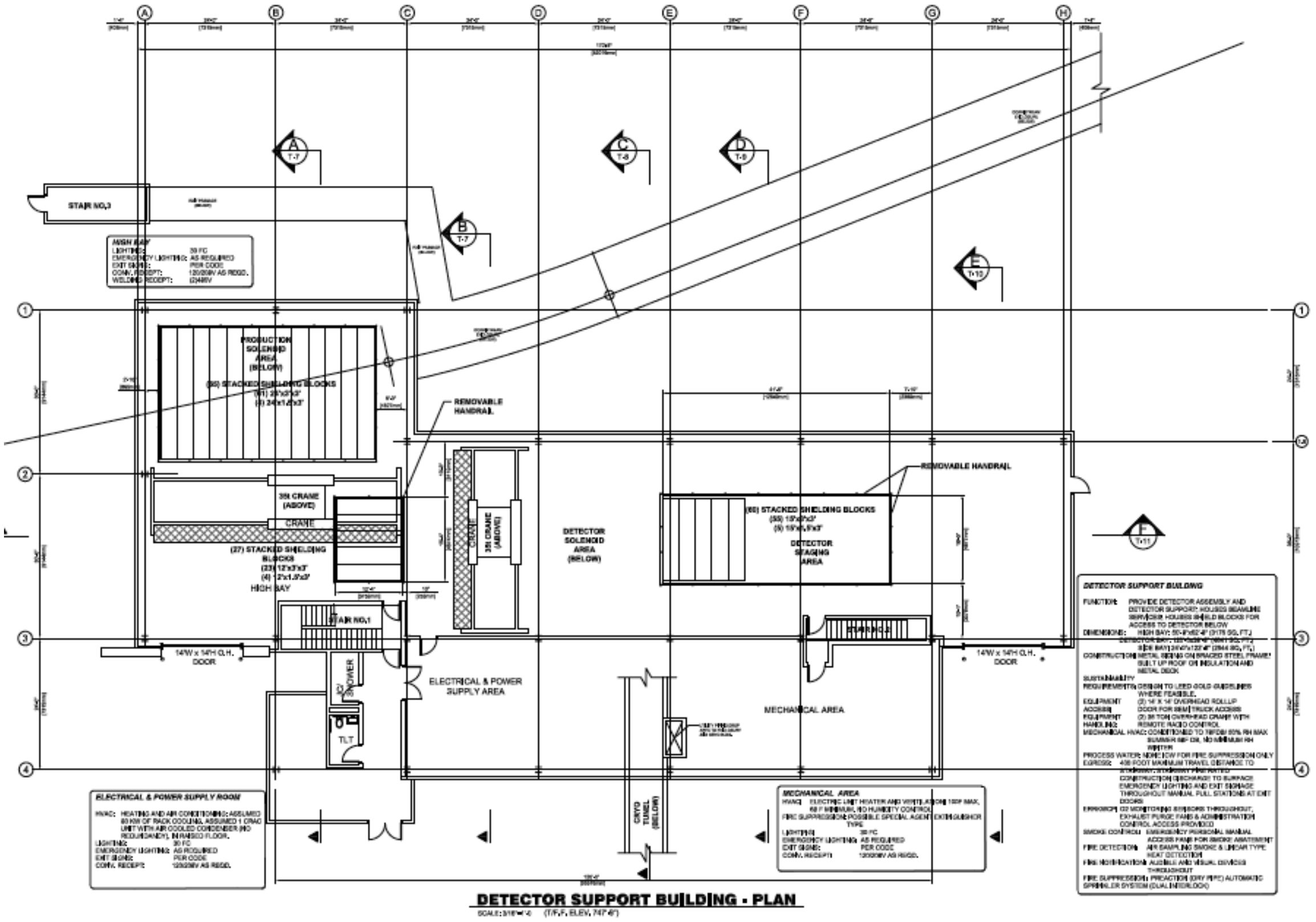

Figure 6.11. Plan view of the alternative design for the Detector Service Building as shown in the September 2009 PDR.

The adjacent low bay contains the Mechanical Room and Electrical and Power Supply Room. The low bay also contains support facilities including a bathroom, janitor's closet and exit stairs serving the below grade portion of the building.

The current design for Mu2e project is similar to the alternative documented in the May 2009 PDR. The length of the beamline has increase based on physics requirements to accommodate beamline components. This has resulted in a shift of the facility to the north.

***Previous Design Iteration***

The current design reflects the design refinements and simplifications that have resulted from continued discussions with the other Mu2e subproject leaders and related stakeholder input. Figure 6.12 shows the design of the Detector Service Building prior to the Value Management effort. The design shown below accommodated a 25 KW proton beam on target that required a larger structure, deeper excavation with more shielding. The Value Management studies described in Chapter 6.8, Value Management, describes additional design changes that optimized the layout to the current configuration.

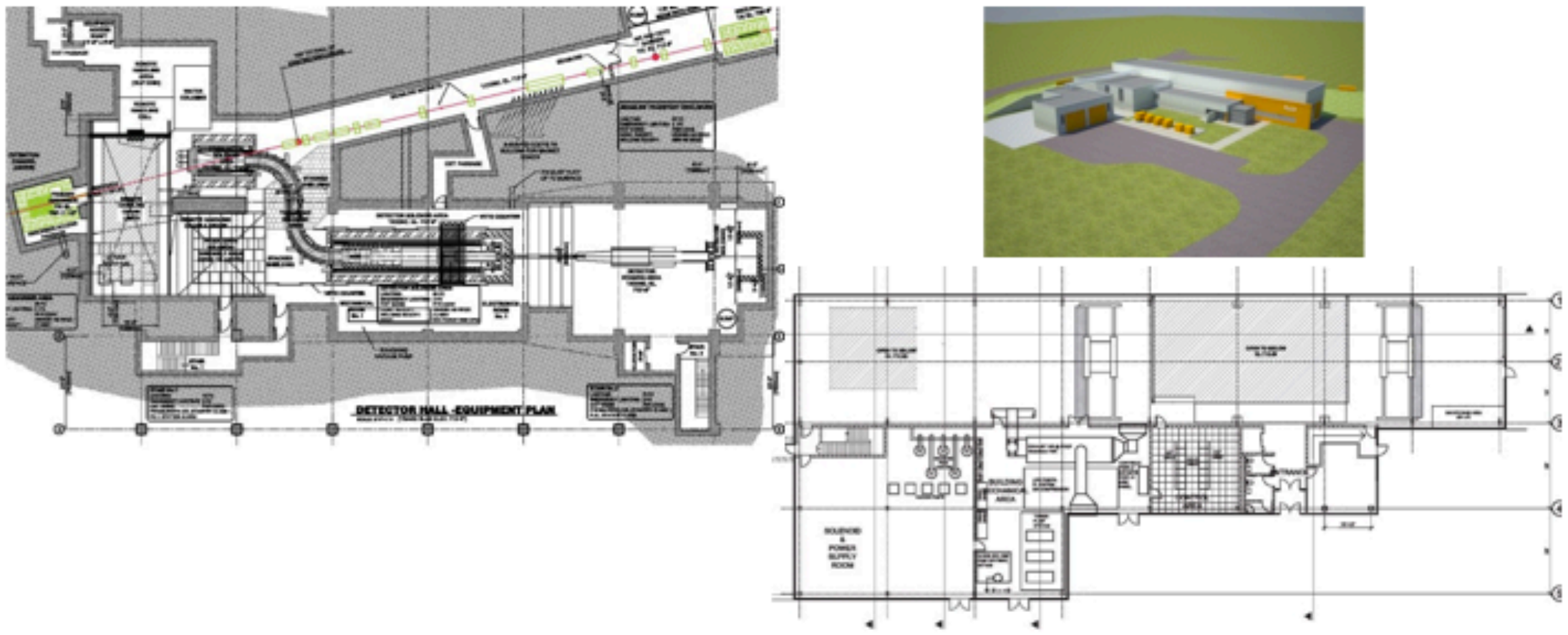

Figure 6.12. The design of the Mu2e facility prior the Value Management effort.

### *6.4.2* Systems Alternatives

The following system alternatives have been evaluated for the Conventional Construction subproject portion of the Mu2e Experiment. In subsequent phases of the project, these ideas will be explored and documented in greater detail through the value management process.

- Dedicated Chiller
  The preferred alternative is to obtain chilled water from the existing Central Utility Building (CUB) through an existing system. The existing 1400-ton capacity chiller at CUB currently serves the Main Injector and FAS and is currently at 80% capacity. While this is adequate to serve the needs of the Mu2e project, other operational impacts could reduce this capacity. An alternate is the use of a dedicated air-cooled chiller to provide the required cooling for the Mu2e Experiment. This alternate will have a higher initial cost as well as higher maintenance and replacement costs.

- LCW from CUB
  An alternative to extending the existing Low Conductivity Water (LCW) system from the FAS to the Mu2e facility is to increase the LCW capacity at CUB in order to accommodate the Mu2e requirements. Preliminary calculations indicate that the existing CUB system has limited capacity in both heat exchanger and flow (pipe size). An alternate to the current design of a dedicated stand-alone LCW plant in the Mu2e DSB is to upgrade the existing CUB LCW plant adding new or larger heat exchanger and increase the pipe sizes to that required for Mu2e. The cost and schedule impact of this option require additional investigation.

- Capacitors on Electrical Equipment
  Both the existing PBar Rings and the planned power supplies are estimated to have power factors below eighty percent (80%). Capacitors located at or near the Mu2e facility could reduce the requirements for the transformers and/or feeders. The cost, schedule and technical impact of this option require additional investigation.

- DSB High Bay Cooling
  The preferred HVAC design includes cooling of the high bay portion of the Mu2e DSB. An alternative would be to provide heating and ventilation only for this space resulting in an upfront and operation cost savings. The requirements for this space will be reviewed prior to CD-2 and the preferred HVAC design could change as a result.

- HVAC High Bay Cooling Unit Location
  The current design locates the HVAC unit for the DSB high bay at grade south of the Mu2e DSB. A preliminary investigation indicates that a location on the roof of the DSB low bay would be more efficient. The location issue will be investigated to better understand the roof top location impact on maintenance access and operation.

- Exterior Crane
  The current design assumes that a mobile crane will be used to access the removable hatch that serves the underground portion of the area downstream of the Production Solenoid. Under consideration is the use of an exterior crane supported on fixed runway to serve this area.

## 6.5 Risks

Approximately 35 risks have been identified for the Conventional Construction subsystem. Most of the risks are those typically seen with construction projects of this size and complexity including the risk of cost overruns, unforeseen subsurface conditions, stakeholder requirement changes, and delays in the CD process.

Based on the ranking criteria of the Mu2e project, none of the identified risks have an overall ranking of “high”. A complete listing of the risks for the Conventional Construction can be found in reference [3].

The top ranked risks include significant “cost overrun” or “cost underrun” due to market conditions, or material costs and the inadequate capacity in the ICW and electrical power 15 KV feeder systems.

Cost overrun is an accepted risk that will be carefully considered and monitored during subsequent design phases with a focus on the regional and local labor markets. Construction proposals received at Fermilab are trended to help understand potential costs impacts on projects. During the Final Design phase the project will identify possible scope reduction or enhancements and likely issue the request for proposal with several “deduct” options in order to align the costs with the project goals.

Costs have been associated with the identified risk threats as well as opportunities and are considered in the determination of the cost estimates lower and upper bounds.

## 6.6 ES&H

As with all Fermilab projects, environmental, safety and health will be integrated into all aspects of the Conventional Construction subproject. The primary set of building construction codes used to review the design of the Mu2e project will be the International Building Code (IBC) – 2009 Edition, including all referenced Codes within the IBC, and the 2009 version of National Fire Protection Association “Life Safety Code” (NFPA 101). All other applicable NFPA documents will be used to evaluate specific design features of Mu2e for compliance. The Mu2e facility will be classified by IBC as a Business Occupancy and by NFPA 101 as General Industrial Occupancy.

### *6.6.1* Fire Protection Assessment

The purpose of the Fire Protection Assessment (FPA) is to comprehensively and qualitatively assess the risk from fire within the proposed Mu2e facility and to ensure that all DOE and Fermilab fire safety regulations are met. The FPA documents the overall design features of the facility and evaluates them against the applicable codes, standards, orders, and directives. DOE fire protection criteria are outlined in DOE Order 420.1B and DOE Standard 1066. The FPA includes identification of the risks from fire and related hazards (direct flame, hot gases, smoke migration, fire-fighting water damage, etc.) in relation to the designed fire safety features to assure that the facility can be safely controlled and stabilized during and after a fire. The FPA will be updated as the design of the Mu2e facility evolves.

### *6.6.2* Hazard Analysis Approach

A principal component of an effective ES&H program is to ensure that all hazards have been properly identified and controlled through design and procedure. To ensure that these issues are understood at the conceptual phase, a Preliminary Hazards Analysis [4] has been conducted to identify potential hazards that could be encountered during the project's construction and operational phases.

Conventional construction hazards pose the potential for serious injury, death, and damage to equipment and schedule delays. Due to the preponderance of operational controls rather than design controls, the post mitigation risk will be relatively high even though the probability of occurrence will be significantly reduced. Fermilab has a mature construction safety program with many recent experiences that provide input for future projects. Lessons learned from these experiences combined with experience from other construction projects in the DOE complex will help manage the risk at the Mu2e facility. Typical construction hazards anticipated at the Mu2e construction site include:

- Site Clearing
- Excavation
- Work at elevations (steel, roofing)
- Utility interfaces, (electrical, steam, chilled water)
- Material Handling
- Misc. finishing work
- Weather related conditions
- Transition to Operations.

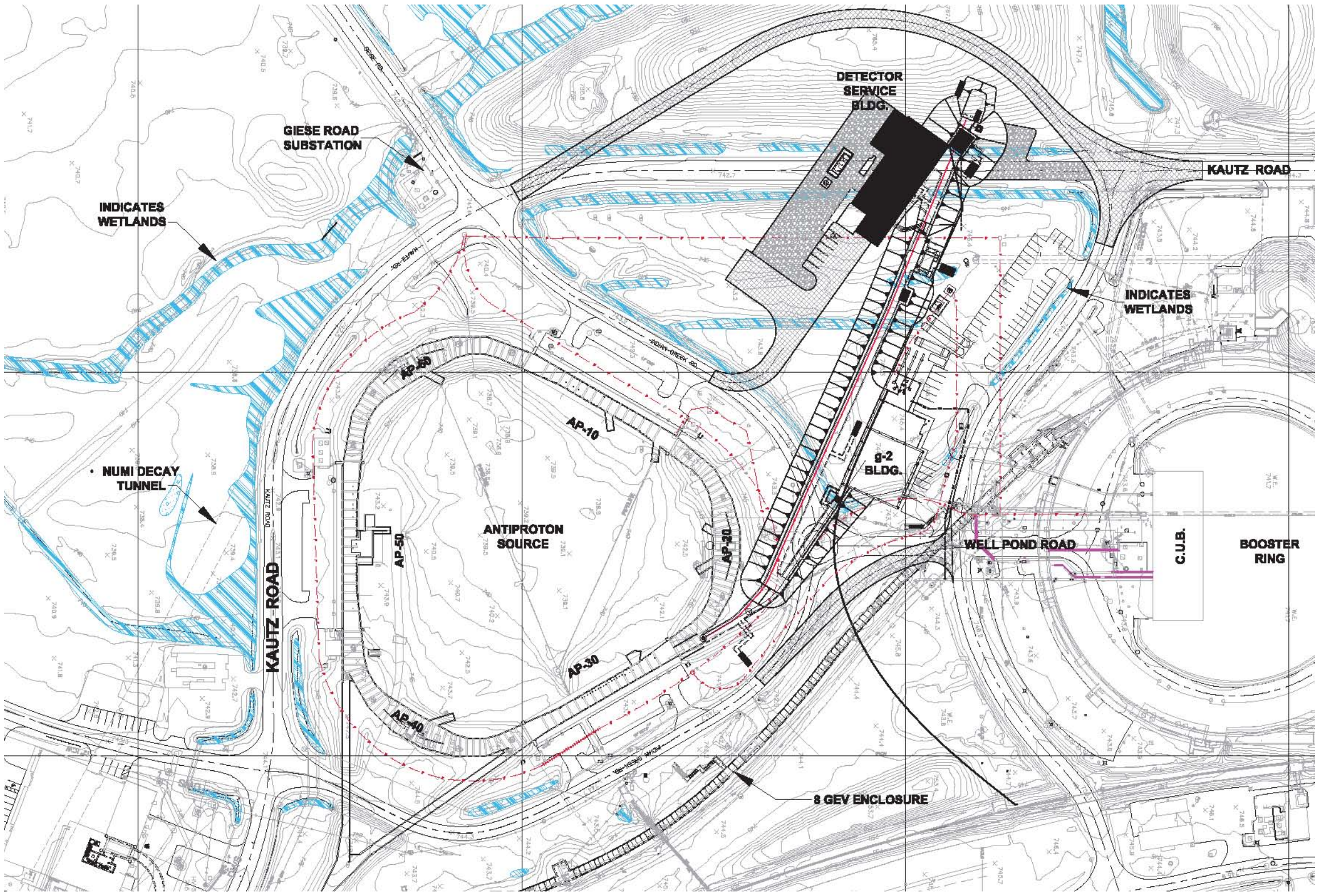

Figure 6.13. Surveyed Wetlands Map

Records from the Illinois Department of Natural Resources (IDNR) were searched to determine the presence of species over the past 3 years that have been identified by the State as threatened or endangered. The Illinois Natural Heritage Database indicated four protected resources potentially near the project area. The United States Fish and Wildlife Service (USFWS) technical assistance website was searched for federally listed threatened and endangered species. The USFWS lists two species in Kane County that are protected by the Endangered Species Act. None of the identified species are expected to be impacted by the Mu2e facility.

### *6.6.4* Sustainability

The project scope incorporates sustainable design principals into all phases of planning, design and construction. Sustainability is broadly defined as the design and implementation of projects to simultaneously minimize their adverse environmental impacts, maximize occupants' health and well-being, and improve bottom line, life cycle and economic performance. The United States Green Building Council (USGBC) has developed the Leadership in Energy and Environmental Design (LEED) standard to provide guidance for builders who wish to incorporate sustainable elements into their projects. LEED for new construction (and remodeling)

Site clearing and all excavation work will require planning and to work around utilities (e.g. electrical, steam, water) and other potential legacy systems so as to prevent unwelcome intrusions.

One of the keys to controlling construction risks is to ensure that safety requirements effectively flow down to subcontractors and that sufficient supervision of subcontractors is in place. The subcontractors performing construction work at the Mu2e facility will be pre-qualified and each will develop a hazard analysis and a safety plan for the work to be done. These plans will comply with Fermilab and applicable OSHA requirements. Fermilab will appoint Construction Coordinator(s), who will oversee the subcontractor's QA and ES&H programs in accordance with FESHM 7010. They will fulfill an auditory function to ensure that all work is carried out in accordance with the subcontractor's ES&H and Quality Assurance plans. Per FESHM Chapter 7010, the ES&H Section will appoint a construction safety expert, who is available to provide ES&H support for the Construction Coordinator. This individual will provide ES&H oversight of construction activities as well. The subcontractor will be required to have someone competent in appropriate safety procedures, with appropriate authority on site at all times while work is ongoing. The qualifications of this competent person will be commensurate with the hazards of the work activity in progress. All subcontractors will undergo on-site training per NFPA 70E (for electrical workers) and Fermilab CSO training for all workers. All subcontractors will perform daily work planning. All Construction Progress Meetings will have construction safety as a standing agenda item. Subcontractors will perform and document toolbox safety meetings. Daily safety inspection reports will be prepared by the contractor and Fermilab safety inspectors. Safety performance will be assessed regularly and corrective action or incentive reduction assessed.

Any Fermilab employees or users seeking access to the construction site must have the appropriate safety training. The subcontractor will control access to the work areas and will set the minimum requirements for entry. The subcontractor will establish a training program to meet these requirements.

The Mu2e design team will consider the relevant natural phenomena hazards. Those hazards of interest to this project include seismic events, flooding, high winds and snow loading. All natural phenomena are normally considered to be 'Low' hazard.

### *6.6.3* Environmental

During the construction phase of the Mu2e facility the disposition of spoils from excavation, dust, noise, impacts on ground and surface water, chemical use, frequent transport of components, spills and disposal of waste are issues that must be addressed. A full National Environmental Policy Act (NEPA) review of the Mu2e facility will be performed prior to CD-2 to ensure that there are no significant impacts to the site, the surrounding waterways or wildlife. It is anticipated that there will be concurrence that the project will conform to a categorical exclusion.

Planning Resources Inc. has performed wetlands delineation around the proposed Mu2e site (Figure 6.13) [5]. The wetland delineation was conducted in accordance with the Interim Regional Supplement to the Corps of Engineers Wetland Delineation Manual: Midwest Region (COE 2008) and the wetland delineation and reporting guidance provided by the Chicago District Corps of Engineers on 13 April 2010. The wetland delineation identified one high quality wetland and two wetlands associated with roadside storm water drainages, two sections of non-vegetated drainage way, one artificial pond/ditch and Indian Creek near the Mu2e site. The Corps of Engineers has reviewed the results of the wetlands study and visited the project site and ascertained that the wetlands are non-jurisdictional.

As part of the wetland delineation a Floristic Quality Assessment (FQA) was conducted of the area surrounding the proposed Mu2e facility to determine the relative vegetative quality of each wetland. Plant species are given numerical values and species that are very specific to natural community rate higher values than those tolerant of a wider range of habitat conditions and disturbances. Taking into account the total number of species present a Floristic Quality Index (FQI) is calculated. These values may be used to compare the relative quality of a given area. If the Floristic Quality Index of an area registers in the middle 30s or higher, there is sufficient native character to be important in terms of a regional natural area perspective. None of the surveyed areas had an FQI higher than 29.4.

is a set of specific and quantifiable measures, each of which confers a credit towards certification of a project as a “LEED-certified” building.

Following DOE Order 430.2B, new construction and major renovation projects over $5M are required to achieve LEED-NC Gold certification whenever appropriate. Projects that exceed the $5M threshold that are not appropriate for LEED certification, such as experimental and/or industrial buildings, must apply the principles of sustainability wherever appropriate and cost-effective and must describe the applied measures in project documents. The Mu2e Project cannot feasibly meet the Leadership in Energy and Environmental Design (LEED)-Gold certification due to the function and operation of the facility. Specifically, the facility will not be occupied on a regular basis. In lieu of LEED-Gold certification, the Project plans to utilize guiding principles and ASHRAE recommendations to meet sustainability goals.

## 6.7 Quality Assurance and Quality Control

All aspects of this project will be periodically reviewed with regard to Quality Assurance issues from Conceptual Design through Close-out. Design and construction documents are reviewed for appropriateness of the proposed systems, impacts on existing systems and operations, specific technical requirements to be incorporated into the design and compliance with best and required practices of the authority having jurisdiction. This review process will be completed in accordance with the applicable portions of the Fermilab Director’s Policy Manual, Section 10. The following elements will be included in the design and construction effort:

- An identification of staff assigned to this project with a clear definition of responsibility levels as well as delineated lines of communication for exchange of information.
- Configuration management of design criteria and criteria changes and a record of all standards and codes used in the development of the criteria.
- Periodic review of design process, including drawings and specifications, to ensure compliance with accepted design criteria.
- Identification of underground utilities and facility interface points prior to the commencement of any construction in affected areas.
- Conformance to procedures regarding updates to the project plan and compliance with the approved construction schedule;
- Conformance to procedures regarding the review and approval of shop drawings, sampling results and other required submittals.

- Conformance to procedures for site inspection by Fermilab personnel to record construction progress and adherence to the approved contract documents.
- Verification of project completion, satisfactory system start-up and final project acceptance.

Comments that result from a Comment and Compliance Review will be entered into the electronic comment database. This will clearly document the names, organizations and dates of commenter’s for specific projects and allow for a formal tracking of comments. All comments entered into the electronic database will elicit a response.

## 6.8 Value Management

A series of value management efforts have already resulted in a reduced footprint of the underground portion of the Mu2e Project. Reduction of the cosmic shielding from twelve (12) feet to three (3) feet above the detector, combined with elimination of the steel and shielding below the Production and Detector Solenoids allowed the base slab to be raised eighteen inches.

The reduction of beam power on the Mu2e target has resulted in a reduction of the proton beam shielding requirements from twenty one (21) feet to sixteen (16) feet. This facilitated a change in the beam optics at the diagnostic dump, midway down the Muon Campus External beamline, from a vertical bend downward to a horizontal bend. This change in optics allowed the base slab to be raised an additional five and a half (5.5) feet, for a total reduction in depth of seven (7) feet, reducing the required excavation and concrete volumes. The reduction in beam power also reduced the need for additional shielding over the FAS Service Buildings and the elimination of fencing around the Debuncher Ring.

The reduced beam power and the associated elimination of the need for the PBar Accumulator Ring resulted in the re-programing of the CUB supplied LCW for use in the external Beamline and power supplies.

It is anticipated that separate value management exercises will be conducted for various aspects of the Conventional Construction subproject as the Project moves forward. A function-oriented, systematic team approach will be applied to analyze and improve the value of the conventional facilities. The focus will be on reducing cost while achieving the same level of quality and performance. Several value management studies are currently being considered and are listed below:

- Optimization of technical system requirements to reduce life cycle costs.
- A cut and fill study to balance the amount of excavated material with the volume of overburden.
- A study to evaluate using existing spoils on the proposed Mu2e site for landscaping rather than moving it to another location.
- A study to add capacitors to reduce the power factor employed in the design of transformers and feeders.
- Reuse of Tevatron 1500 KVA transformers currently in the Tevatron Compressor Buildings.
- Elimination of the raised computer floor in the Electronic Room.
- Reuse of existing concrete shield blocks.
- Dedicated Mu2e chiller to provide chilled water.
- Upgrading the existing LCW system in CUB.
- Mu2e DSB HVAC modifications.
- Mu2e DSB high bay cooling unit location.
- Exterior crane to service the removable hatch.

## 6.9 References

[1] T. Lackowski, “Mu2e Conventional Facilities Requirements,” Mu2e-doc-1088.
[2] S. Dixon and T. Lackowski, “Project Definition Report,” Mu2e-doc-357.
[3] M. Dinnon, “Risk Register,” Mu2e-doc-1463.
[4] R. Ray, “Preliminary Hazard Analysis Document,” Mu2e-doc-675.
[5] Planning Resources Inc., “Wetland Report,” Mu2e-doc-1286.

# 7 Solenoids

## 7.1 Introduction

The solenoids perform several critical functions for the Mu2e experiment. Magnetic fields generated from these magnets are used to efficiently collect and transport muons from the production target to the muon stopping target while minimizing the transmission of other particles. Electrons are transported from the stopping target to detector elements where a uniform and precisely measured magnetic field is used to determine the momentum of electrons. The magnetic field values range from a peak of 4.6 T at the upstream end to 1 T at the downstream end. In between is a complex field configuration consisting of graded fields, toroids and a uniform field region, each designed to satisfy a very specific set of criteria.

Mu2e proposes to create this complex field configuration through the use of three magnetically coupled solenoid systems: the Production Solenoid (PS), the Transport Solenoid (TS) and the Detector Solenoid (DS). Requirements for the magnets, as well as a design concept that meets these requirements are described below. The Mu2e Solenoid system also includes all ancillary systems such as magnet power converters, a cryogenic plant, cryogenic distribution and quench protection instrumentation and electronics. The Solenoid system is shown in Figure 7.1.

## 7.2 Requirement

### *7.2.1* General Requirements

The Mu2e solenoids and their supporting subsystems are designed to meet a complex set of requirements. The requirements are defined so that the deliverables will meet the physics goals of the experiment. The requirements are explained in detail in several reference documents and summarized below [1] - [8].

Because of the high magnetic field and large stored energy, the solenoids will be made from superconducting NbTi coils, indirectly cooled with liquid helium and stabilized with either high conductivity aluminum or copper. It must be possible to cool down and energize each solenoid independent of the state of the adjacent magnets. Individual magnets will have different schedules for installation and commissioning, requiring independent operation. It will also be necessary to warm-up individual magnets to repair detector components housed inside or to anneal the conductor. Furthermore, it may be required that the magnets be operated in special field configurations for detector calibration.

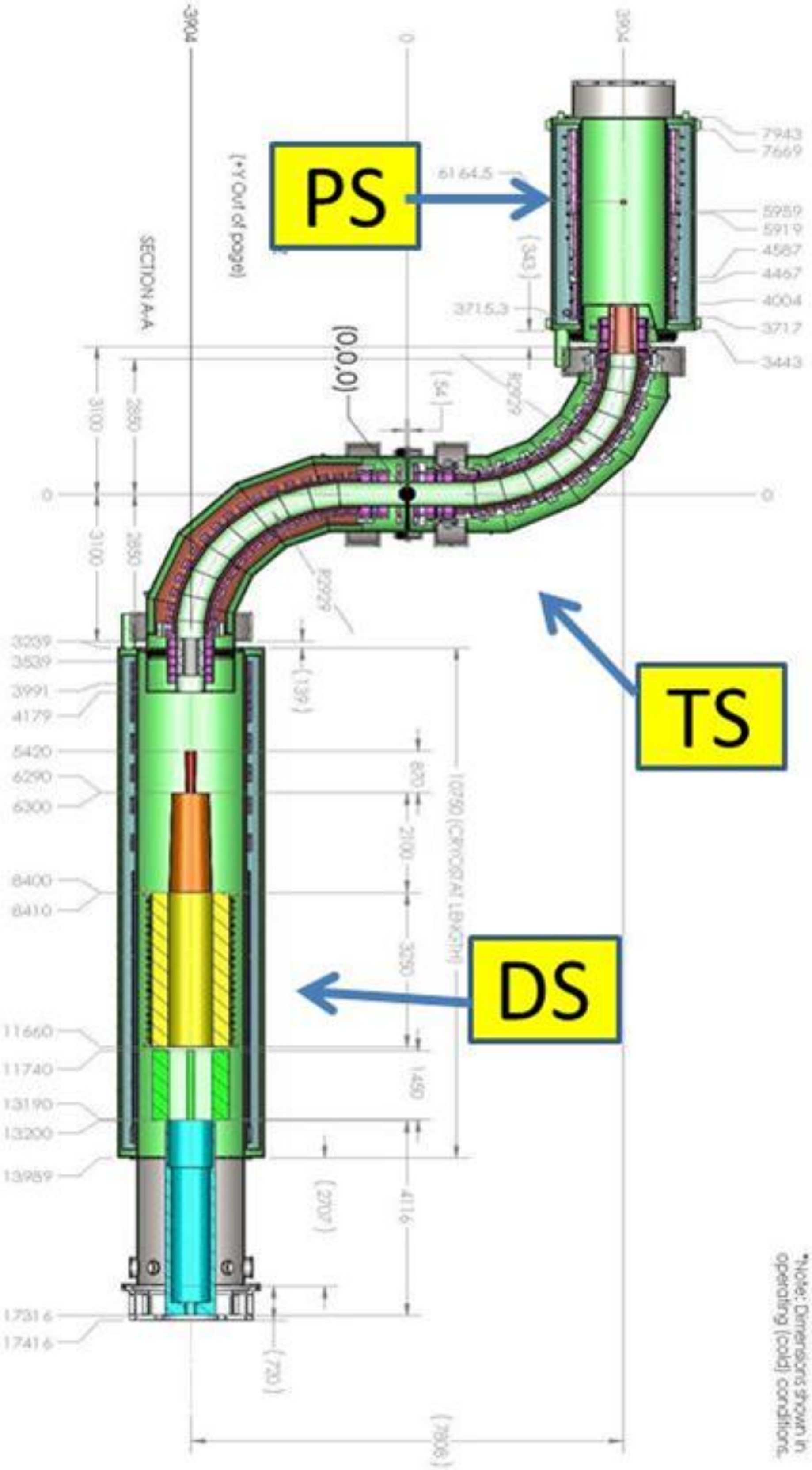

Figure 7.1. The Mu2e Solenoid System.

Significant axial forces will be present between these magnetically coupled systems and these forces will change if the fields are changed. The magnets must be designed to withstand this range of forces when they are being operated in their standard configuration as well as in the various configurations described above. The mechanical support for each of the magnets will be independent and will not depend on adjacent magnets. This simplifies integration issues but complicates the mechanical support system. The bore of the magnets share a common beam vacuum but the magnet vacuums will be bridged with bellowed connections.

The solenoid coils must be designed for repeated full field quenches and thermal cycles, without degradation in performance, over the lifetime of the experiment. The expected duration of the experiment is 4 years at full luminosity; however the magnets should be designed for the possibility of an extended physics run at the maximum design luminosity. The primary consequence of extended running will be the need for repeated thermal cycles to anneal the damage from irradiation. Quenches may occur during the initial campaign to full field as well as during normal operation conditions. The lifetime requirements are summarized in Table 7.1.

| | |
|---|---|
| Design lifetime (min-max) | 5 – 20 years |
| Number of full thermal cycles after commissioning | Up to 50 |
| Number of full quenches after commissioning | > 50 |

Table 7.1: Lifetime requirements for Solenoids.

### *7.2.2* Magnetic Field Requirements

Each of the solenoids performs a different set of functions and each has a unique set of field requirements. The requirements for each solenoid are described in the sections that follow and are summarized in Table 7.2.

***PS Uniform axial gradient (PS2)***

The Production Solenoid is a relatively high field solenoid with an axial grading that varies from 4.6 Tesla to 2.5 Tesla. The purpose of the Production Solenoid is to trap charged pions from the production target and direct them towards the Transport Solenoid as they decay to muons. The nominal peak field of 4.6 Tesla provides the high end of the field gradient while still allowing for sufficient operating margins for temperature and current density with NbTi superconductor. There is a ± 5% requirement on the deviation from a uniform gradient along the axis (*dB/B*). At radii less than 0.3 m there can be no local field minimums where particles might get trapped.

| Region | Length | $s_{min}$ | $s_{max}$ | $B_{initial}$ | $B_{final}$ | $R_{max}$ | Uniformity Requirement |
|---|---|---|---|---|---|---|---|
| | (m) | (m) | (m) | (T) | (T) | (m) | |
| PS1 | 1.5 | -10.58 | -9.08 | > 4.5 T at s=9.4m | n/a | ~0.5 | No local minimum anywhere |
| PS2 | 2.5 | -9.08 | -6.58 | n/a | 2.50 | 0.25 | On axis, dB/B < 0.05 about a uniform negative axial gradient from the peak field in PS1 to the TS $B_{initial}$ field. No local minima off axis. |
| TS1 | 1.0 | -6.58 | -5.58 | 2.50 | 2.40 | 0.15 | \|dB/B\| < 0.05 about a uniform negative axial gradient. dB/ds < -0.02T/m everywhere. |
| TS2* | 4.6 | -5.58 | -0.98 | n/a | n/a | 0.15 | Ripple \|dB\| < 0.02 T |
| TS3 | 1.95 | -0.98 | 0.98 | 2.40 | 2.10 | 0.15 | \|dB/B\| < 0.05 about a uniform negative axial gradient. dB/ds < -0.02 T/m everywhere. |
| TS4* | 4.6 | 0.98 | 5.58 | n/a | n/a | 0.15 | Ripple \|dB\| < 0.02 T |
| TS5 | 1.0 | 5.58 | 6.58 | 2.10 | 2.00 | 0.15 | \|dB/B\| < 0.05 about a uniform negative axial gradient. dB/ds < -0.02 T/m everywhere. |
| DS1 Gradient | 3.0 | 6.58 | 9.58 | 2.00 | 1.18 | 0.3-0.7 cone | dBs/ds = -0.25 ± 0.05 T/m, \|dB/B\| < 0.05 |
| DS2 Transition | 1.2 | 9.58 | 10.78 | 1.18 | 1.00 | 0.7 | Magnitude of gradient decreasing |
| DS3 Uniform | 3.6 | 10.78 | 14.39 | 1.00 | 1.00 | 0.7 | \|dB/B\| < 0.01 |
| DS 4 Uniform | 1.5 | 14.39 | 15.85 | 1.00 | 1.00 | 0.7 | \|dB/B\| < 0.05 |

Table 7.2: Summary of Mu2e Solenoid field specifications. The "s" coordinate is the path length along the central axis, referenced from the geometric center of TS3; "r" refers to the perpendicular distance from the solenoid axis. With the exception of the PS1-PS2 interface, the required magnetic field values at the beginning and end of each straight section are shown. *Curved sections TS2 and TS4 are defined by |dBs/dr| > 0.275 T/m for r = 0.

***Solenoid straight sections in the transport solenoid (TS1, TS3, TS5)***

Particles produced with a small pitch in a uniform field region can take a very long time to progress down the beamline toward the muon stopping target. This can result in background (Section 3.5.8). To suppress background from late arriving particles the 3 straight sections in the Transport Solenoid have negative axial gradients. The gradients in TS1 and TS5 are required to be more uniform than the gradient in TS3. TS3 has a complicated interface to make it accessible to experimenters to service the rotating collimator and antiproton window housed inside. However, the field gradient must be negative at all locations in the straight sections for radii smaller than 0.15 m. The radius is set by the geometry of the beam collimators. This requirement is intended to eliminate traps, where particles bounce between local maxima in the field until they eventually scatter out and travel to the Detector Solenoid where they arrive late and potentially cause background.

***Toroid sections (TS2 and TS4)***

In the toroidal sections of the Transport Solenoid, the field varies as ~1/r, where r is the distance from the toroid center of curvature. In a toroid region, spiraling particles drift up or down depending on the sign of their charge, with a displacement that is proportional to their momentum and inversely proportional to their pitch. Particles with small pitch progress slowly through the toroid and drift to the wall where they are absorbed. This allows for a relaxed gradient specification in the toroid sections, defined by dBs/dr > 0.275 T/m, where s is the coordinate along the beam path. There is an additional requirement on the field ripple, dB, within a 0.15 m radius transverse to the central axis of the magnet system. Large field ripples can trap particles.

***Detector Solenoid Gradient region (DS1, DS2)***

The muon stopping target resides in a graded field provided by the Detector Solenoid that varies from 2 Tesla to 1 Tesla. On the Transport Solenoid side of the muon stopping target, the graded field captures conversion electrons that are emitted in the direction opposite the detector components causing them to reflect back towards the detector. On the downstream side of the stopping target, the graded field focuses electrons towards the tracker and calorimeter. The graded field also plays an important role in background suppression by shifting the pitch of beam particles that enter the Detector Solenoid out of the allowed range for conversion electrons before they reach the tracker. The muon stopping target is located approximately in the middle of DS1. The uniformity requirement for the graded field is $dB/ds = 0.25$ T/m, where s is the direction along the solenoid axis. DS2 is the transition region between the graded and uniform fields. It should be as short as possible without introducing a local minimum in the axial field.

***Uniform field section (DS3 and DS4 Uniform).***

To accurately determine the conversion electron momentum and energy, the magnetic field in the region of the tracker is required to be uniform to within ± 1% inside a radius of 0.7 m. Furthermore, the absolute field in the vicinity of the tracker must be measured to within ± 0.01%. In the vicinity of the calorimeter, the field uniformity requirement is relaxed to ±5% and the field measurement requirement is relaxed to ± 0.02%.

### *7.2.3* Alignment Requirements

Magnetic elements must be properly aligned to one another as well as external interfaces such as beam collimators, the proton beam line and internal detector elements. This is required to assure optimal muon transmission, suppression of backgrounds, minimization of forces amongst magnetically coupled systems and minimization of radiation damage due to improperly located collimators. Alignment requirements, which vary from amongst magnetic elements, are the subject of several ongoing analyses [8], [27]. Generally speaking, alignment tolerances between magnetic elements (PS, TSu, TSd, DS) in their cold and electrically powered nominal positions are ~10 mm. Alignment tolerances between coils within a cryostat are ~1 mm. Alignment between magnets and tracker elements are ~0.1 mm.

### *7.2.4* Radiation Requirements

Radiation damage to the solenoids must be carefully considered during the design process. The coil insulation, superconductor and superconductor stabilizer are the biggest concerns. The largest radiation dose is estimated to be 0.3 MGy/year to the Production Solenoid at full beam intensity [9]. Materials with poor radiation properties must be avoided. Irradiation of the conductor and insulation is not expected to be a major concern over a 20-year life cycle. There is a significant concern about damage to the superconductor stabilizer in the Production Solenoid, causing a significant reduction in electrical and thermal conductivity at low temperature [11]. For aluminum, the stabilizer room temperature resistivity ratio (RRR) must not fall below ~100 over the lifetime of the experiment. For copper the requirement is for the RRR to stay above 30.

The experimental hall in the vicinity of the PS will be highly activated and accessible only to highly trained personnel under strictly controlled conditions. Therefore, every effort must be made to reduce the need for access. Cryogenic valves, power and instrumentation connections should be located outside of this area.

### *7.2.5* Electrical Requirements

The following electrical requirements have been developed for the solenoid system.

- The magnets will be designed with sufficient superconductor margin to allow operation without quenching at full field during the delivery of peak beam intensities. The target operating $J_c$ margin is 30% and the required $T_c$ margin is 1.5 K.
- The Superconductor will be "standard" copper stabilized NbTi strand. In order to achieve the $J_c$ and $T_c$ margins for the Production Solenoid, the design $J_c$ (4.2 K, 5T) value should be greater than 3000 A/mm$^2$ for conductor in the peak field region of the PS. The magnet will be operated DC.
- For the Production and Detector Solenoids, the NbTi strand will be woven into a "Rutherford cable". The cable will be further stabilized with low resistivity aluminum. For the Transport Solenoid, NbTi can be cabled and soldered into a copper channel. This technology has been used successfully in MRI magnets with similar currents and fields [10].
- The solenoids will be divided into several independent power circuits. Each circuit will have an external energy extraction resistor. The value of the resistor will be chosen so that the peak voltages will be limited during a quench to less than 300 V to ground and then 600 V across the magnet terminals. For DS and PS depending on the intra-coil connection scheme, there is the possibility of ~100 V between adjacent coil layers. Coil-to-ground and layer-to-layer insulations must be sized to meet these requirements. Turn-to-turn voltages are not expected to exceed 10 V. However, cable insulations must be designed conservatively as these potentially damaging turn-to turn electrical breakdowns will be difficult to detect during fabrication.
- The peak coil temperature must not exceed 130 K as the result of a quench. Stabilizer will be sized to meet this requirement.
- To insure that there is no net transfer of magnetic stored energy between magnets during a quench, the stand-alone time constant for energy extraction for each power circuit must be set to approximately 30 seconds.

### *7.2.6* Cryogenic Requirements

The solenoids will be divided into 4 cryogenic units. All solenoid coils will be potted with epoxy and indirectly (conduction) cooled by liquid helium. The coils can be cooled using either a "force flow" or a "thermal siphoning" system. 80 K thermal intercepts in the cryostat will be cooled by liquid nitrogen.

Refrigerators (not included in the solenoid project scope) located in a separate cryo building will supply liquid helium for the entire solenoid system. Based on the estimated heat loads, the required liquid helium can be supplied by one Tevatron "satellite" refrigerator [12]. This means that during normal operations, the 4.5 K heat load cannot exceed 600 Watts. A second refrigerator will double the available capacity for use during initial cool-down and quench recovery. The time required to cool an individual magnet from room temperature to liquid helium temperature should be no more than a week. Cryogens will flow to/from each cryostat via a single cryo-link chimney. This chimney must be routed from the magnet cryostat through the magnet concrete shielding and cosmic ray telescope (DS), up to the cryogenic feed box located at ground level. Care must be taken in locating these gaps in the shielding to avoid "line of sight" paths for neutron and cosmic ray backgrounds. Chimneys must be routed to minimize interference with utilities and crane coverage.

## 7.3 Proposed Design

For the purpose of this report the solenoid system can be divided into the following areas: Production Solenoid, Detector Solenoid, Transport Solenoid, cryogenic system, power systems/quench protection, magnetic field mapping and monitoring and solenoid installation and commissioning. The solenoids are shown in Figure 7.1.

### *7.3.1* Production Solenoid (PS)

The Production Solenoid is a wide aperture superconducting solenoid with an axially graded field. Its primary role is to maximize the stopped muon yield by efficiently capturing pions and focusing secondary muons towards the Transport Solenoid. The PS also provides a clear bore for beam line elements including the primary production target and radiation shield (not shown). The shield and target will be mechanically supported from the PS cryostat. The salient features of the PS are shown in Table 7.3 and pictorially in Figure 7.2.

The Production Solenoid must generate a uniform axially graded field ranging from 4.6 T to 2.5 T. This axial field change is accomplished using three solenoid coils with 3, 2 and 2 layers of aluminum stabilized NbTi superconducting cable, each coil with the same inner diameter. The axial length of the PS coils range from 0.75 to 1.8 meters. The coil lengths are a design parameter used to achieve the required gradient uniformity and field matching at the Transport Solenoid.

While the maximum required on-axis field is 4.6 T, it is highly desirable to be able to adjust the peak field by ~10% and still be able to meet the field uniformity requirements. An adjustment upward will allow for a ~10 % increase in the stopped muon yield at the stopping target; a lower field value would still allow the experiment to operate, albeit with a ~10% decrease in stopped muons. This range is reflected in the PS parameter table where applicable.

In this section the various Production Solenoid design features (conductor, coil, and cryostat) are presented. This is followed by a summary of the studies that have been performed to show how the PS design meets the project's requirements. Details of these studies have been documented in various design notes [9][13][14][15].

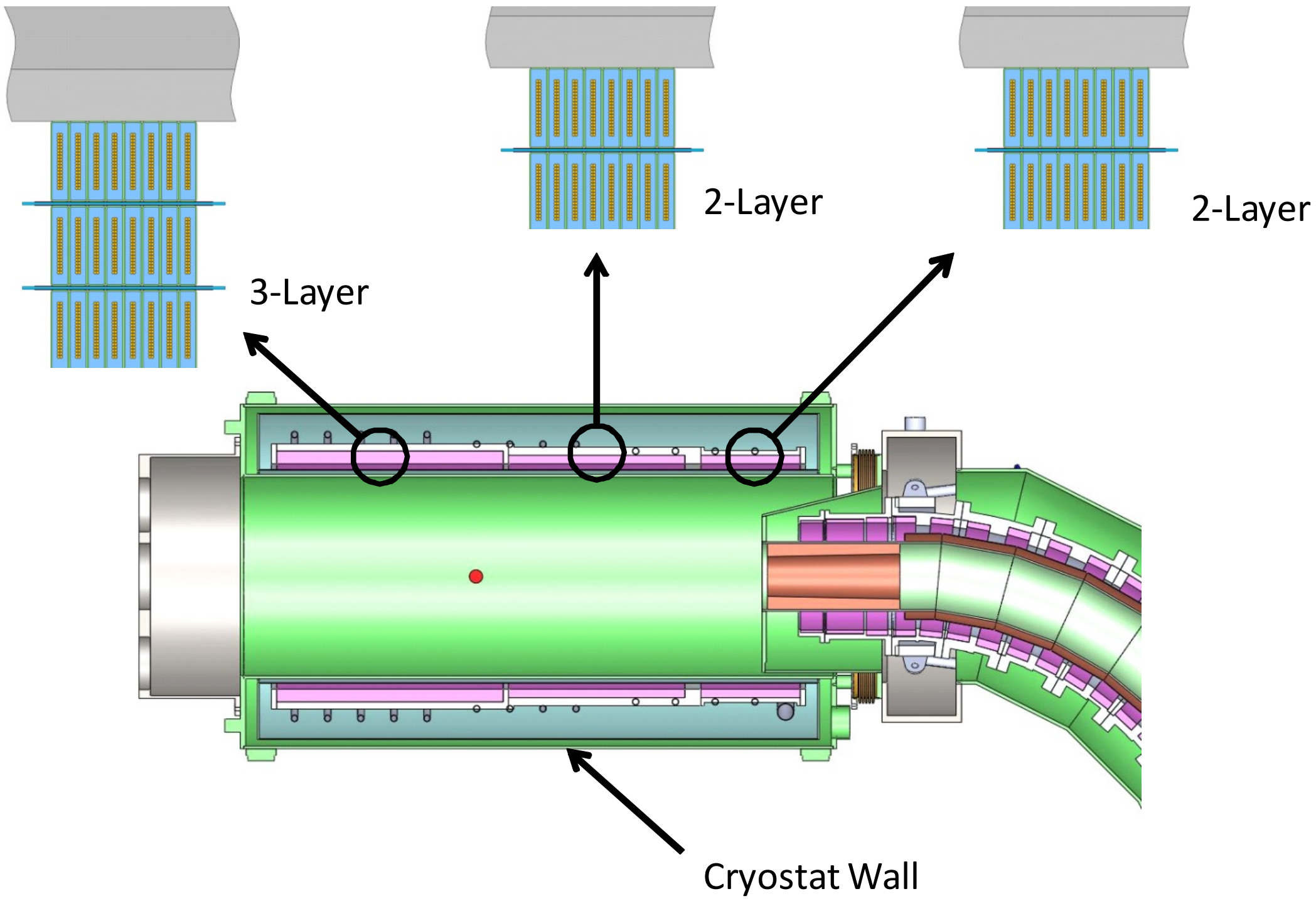

Figure 7.2. Cross Section of the 3-coil design of the axially graded Production Solenoid.

***Conductor Design***

Figure 7.3 shows a cross section of the Production Solenoid conductor. The conductor consists of copper-clad NbTi superconducting strand formed in a Rutherford cable and stabilized with structurally enhanced aluminum. The nickel doped aluminum alloy was chosen to be a compromise between a high RRR and good mechanical strength. The insulation type and thickness were chosen to meet the required voltage standoff while minimizing the thermal barriers that could impede efficient conduction cooling.

The PS employs a composite cable insulation made of polyamide and pre-preg glass tapes. This type of insulation, originally developed for the TRISTAN/TOPAZ solenoid, was also used in the ATLAS Central Solenoid [18]. The cable is insulated with two layers of composite tape consisting of 25 μm of a semi-dry (BT) epoxy on one side of 25 μm Kapton tape and 75 μm of pre-preg E-glass on the other side.

| Parameter | Units | Value |
|---|---|---|
| Cable bare width | mm | 30.00 |
| Cable bare thickness | mm | 5.50 |
| Overall Stabilizer/non-stabilizer ratio in bare cable | | 5.56 |
| Radial insulation thickness | mm | 0.25 |
| Axial insulation thickness | mm | 0.25 |
| Coil inner diameter | m | 1.70 |
| Coil length | m | 4.01 |
| Operating temperature | K | 4.6 |
| Operating current | kA | 9.20±0.95 |
| Peak axial field at the operating current | T | 4.56±0.46 |
| Peak coil field at the operating current | T | 4.97±0.51 |
| Quench current at the operating temperature | kA | 15.81 |
| Current sharing temperature at the operating current | K | 7.04-6.50 |
| Minimum temperature margin | K | 1.50 |
| Maximum allowable temperature | K | 5.54-5.00 |
| Fraction of SSL at the operating temperature | | 0.52-0.64 |
| Fraction of SSL at the maximum allowable temperature | | 0.63-0.69 |
| Stored energy | MJ | 55.15-79.74 |
| Self-inductance | H | 1.58 |
| Net axial force with TS powered on | MN | 1.28-1.36 |
| Cold-mass weight | tonnes | 10.93 |

Table 7.3: Parameters for the Production Solenoid. The range, where specified, denotes variations of the trim current.

To facilitate the heat extraction from the coil and increase the structural integrity, all gaps between turns and layers are to be filled with epoxy resin during vacuum impregnation. Since the composite cable insulation is impermeable for epoxy, the sheets of dry E-glass insulation are introduced between the coil layers and between coil and support structure to provide paths for epoxy penetration, as shown in Figure 7.3. It will ensure a good thermal and structural contact between the cables, thermal bridges, and the support structure. The purpose and design of the thermal bridges, as well as the ground insulation between the thermal bridges and the coil are discussed later in the text.

The ground insulation between the cables and the support structure with the total thickness of 2 mm consists of dry E-glass and 2x25 μm layers of Kapton. The extra thickness of E-glass between the coil and support structure allows for the machining the outer coil surface after the impregnation for precise fitting the surfaces of the support shells and the coils.

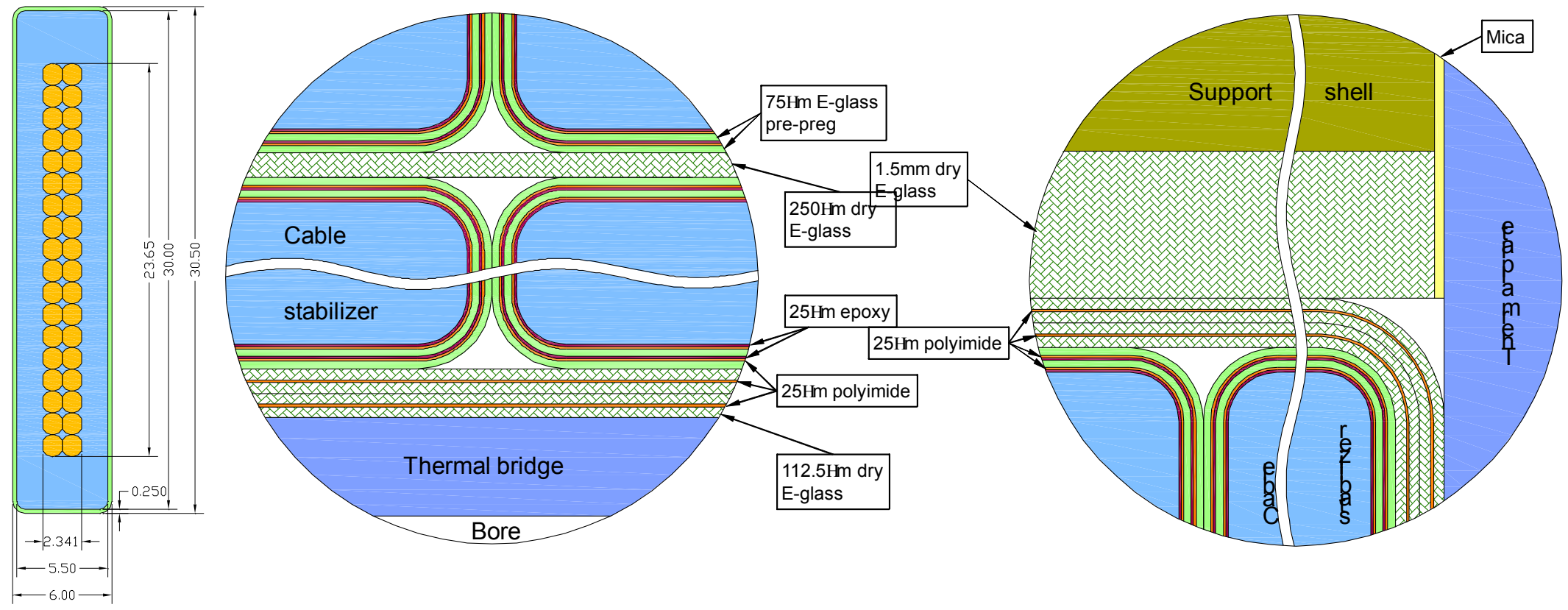

Figure 7.3. Production Solenoid conductor cross section (left) and insulating scheme (middle-right).

### *Coil Design*

As shown in Figure 7.2, the Production Solenoid consists of three coils that all use the same conductor. The coil parameters are shown in Table 7.4. Each coil has an aluminum outer support structure to manage hoop stress. The cylinder thickness ranges from 83 mm for the 3-layer coil to 40 mm for the 2-layer coils. The material of choice for the coil outer support shells is Al 5083-O. This particular grade of aluminum was selected because of the material strength enhancement at cryogenic temperatures, and because it does not exhibit a reduction of strength after welding.

| Coil No. | ID (mm) | OD (mm) | Length mm | Peak Current Density (A/mm$^2$) |
|---|---|---|---|---|
| 3-Layer | 1700 | 1892 | 1710 | 52.86 |
| 2-Layer | 1700 | 1828 | 1332 | 52.86 |
| 2-Layer | 1700 | 1828 | 750 | 47.92 |

Table 7.4: Production Solenoid Coil Parameters.

The final coil design will be developed in conjunction with the Production Solenoid manufacturer. For the current baseline design each coil will be wound on a collapsible mandrel. The coil layers are separated from one another by 0.25 mm thick sheets of dry E-glass insulation that provide paths for epoxy during the impregnation. Layers of aluminum made of 1 mm thick strips are installed on the inner coil surfaces either prior to or after the impregnation. These aluminum layers form thermal bridges by connecting to 4 mm sheets of aluminum placed between the coil ends and the end flanges. This is shown pictorially in Figure 7.4. Once wound, the outer support structure will be placed around the coil and will form the outer surface of the fixture used to pot the coils. The ground insulation between the coil and the support structure includes 2 mm of composite fiberglass/polyimide insulation.

The three-coil structure will have aluminum end flanges between the coil modules and at both ends of the cold mass. The coil modules will be bolted together in line to form the 4-meter long cold mass assembly. The coil conductors will be bused in series and powered by a power converter through a pair of HTS leads rated for 10 kA. In addition, the 3-layer coil and the middle 2-layer coil will be connected to a bipolar trim power converter rated at +/- 1 kA, either through a dedicated pair of leads or through one main lead and one additional lead. This is shown schematically in Figure 7.5.

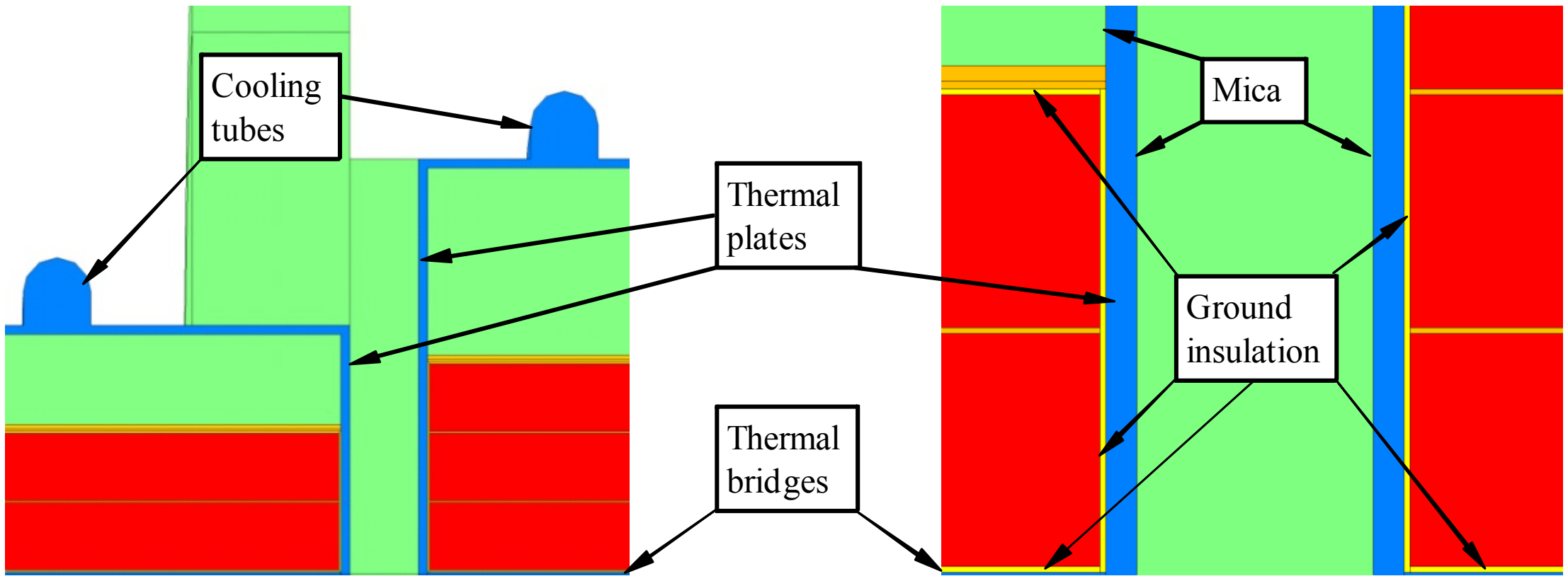

Figure 7.4. Production Solenoid coils with insulation and cooling features.

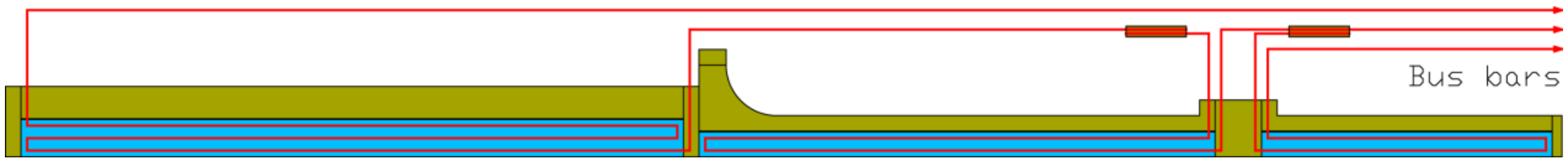

Figure 7.5. Electrical connection scheme for the Production Solenoid coils.

### *Cryostat and Mechanical Support Design*

The superconductor will be indirectly cooled via heat exchange from the coils to helium - filled aluminum tubes welded to the aluminum outer support structure (see Figure 7.6). A "thermal siphoning" cooling scheme will be used to cool the Production Solenoid.

Also shown in Figure 7.6 are the axial and radial support structures. The PS axial anchors run the entire length of the cryostat and are anchored to the cold mass at the center of the vacuum vessel. Eight axial rods made from Inconel 718, four running from each end of the vacuum vessel and attaching to the center of the cold mass, constitute the axial anchor system. The center attachment point allows for radial shrinkage of the cold mass during cool-down. A series of Belleville springs at the end of each rod accommodate approximately 3 mm of axial contraction of the rods themselves during cool-down. In this configuration, each set of four rods resist axial loads in each of the two directions. In both cases, the set that is not in use slides, unloading the Belleville springs rather than allowing them to go into compression.

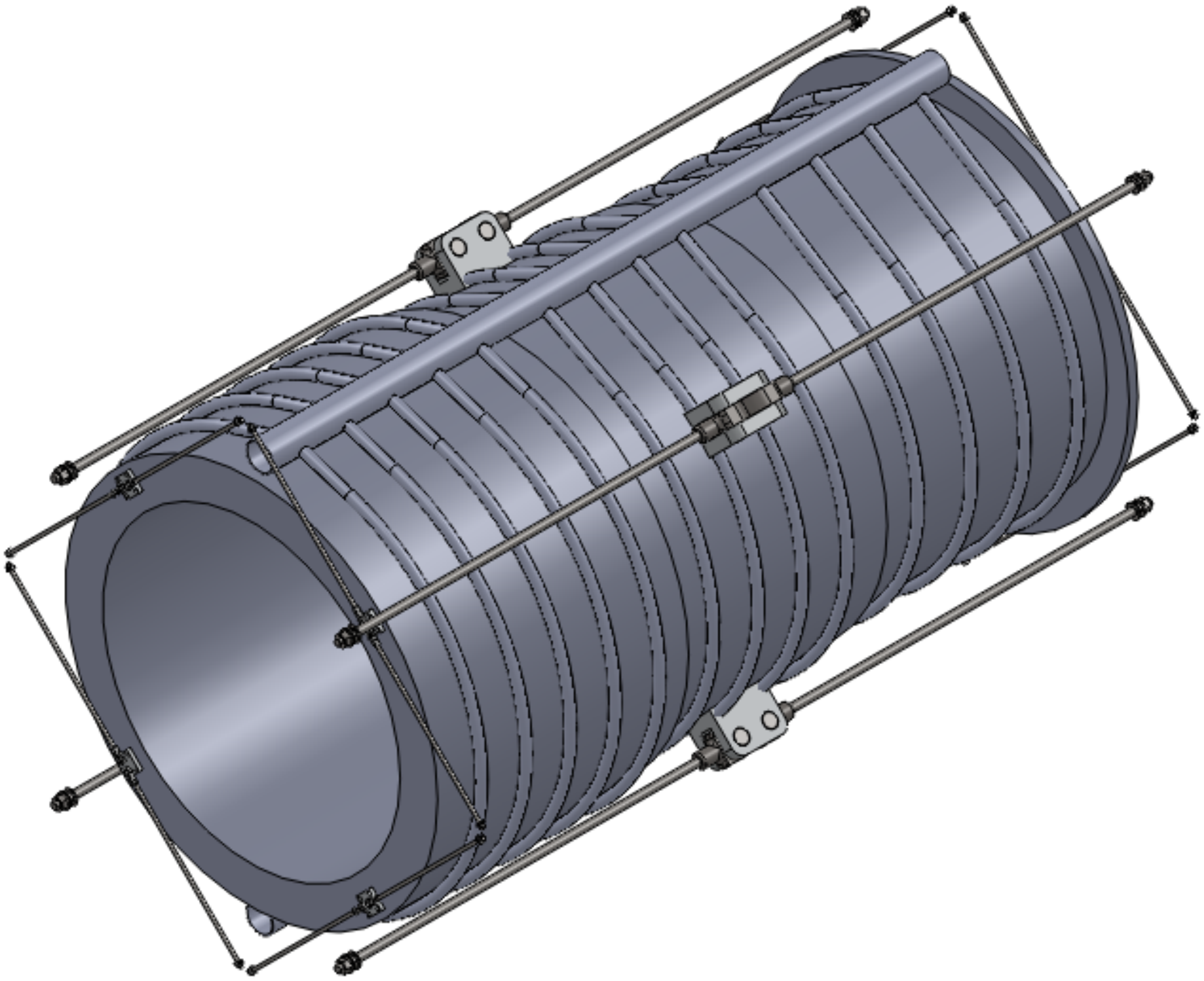

Figure 7.6. Coil cold mass with cooling tubes and support structure.

The radial supports have been designed to resist the weight of the cold mass and any off-center load due to magnetic field. As with the axial supports, Inconel 718 was chosen for a series of 16 tension rods arranged in pairs at each end of the cold mass.

Several candidate materials were studied for the anchor rods. Inconel 718 was chosen because it is structurally strong and represents a relatively low heat load to the cryogenic system. Inconel 718 has the added benefit that it is readily available and is well suited for welding. The allowable stress for Inconel 718 in the annealed and precipitation hardened condition is 425 MPa or 62,000 psi. The lengths of the support rods are summarized in Table 7.5.

The coils will be housed in a stainless steel cryostat. The thickness of the stainless steel will be chosen to support the weight of the expected 60 tonne internal heat shield. Cryostat parameters are shown in Table 7.5 and are shown pictorially in Figure 7.7.

| **Cryostat Component** | **Value (mm)** | ***Material*** |
|---|---|---|
| Vacuum Vessel Outer Shell (OD / Wall thickness) | 2600 / 20 | *Stainless Steel* |
| Vacuum Vessel Inner Shell (ID / Wall thickness) | 1500 / 20 | *Stainless Steel* |
| Vacuum Vessel End Wall (Thickness) | 30 | *Stainless Steel* |
| Thermal Shield Outer Shell (OD / Wall thickness) | 2460 / 6 | *Aluminum* |
| Thermal Shield Inner Shell (ID / Wall thickness) | 1600 / 6 | *Aluminum* |
| Thermal Shield End Wall (Thickness) | 6 | *Aluminum* |
| Coil Axial Support Rods, 8X (Diameter / Length) | 38.1 / 2100 | *Inconel 718* |
| Coil and Thermal Shield Radial Support Rods, 32X (Diameter / Length) | 15.88 / 875 | *Inconel 718* |

Table 7.5: Production Solenoid cryostat dimensions and materials.

***Magnetic Analysis***

In addition to providing the necessary field distribution, the design includes the possibility of changing the peak axial field and gradient with minor impact on the transition field at the TS interface. That is accomplished by changing the current in two coil sections connected to the trim power supply. The peak axial field in the Production Solenoid is 5.02 T and the peak field at the conductor is 5.48 T in the first (upstream) coil when operating at the maximum trim current.

Figure 7.8 shows the magnetic model created within COMSOL Multiphysics code with the flux density diagram at the maximum operating current. Note that the picture also shows the straight section of the Transport Solenoid that was included in the model for the field matching purposes and the Heat and Radiation shield made from C63200 bronze with the relative magnetic permeability of 1.04.

The field in the PS is a combination of the field from the PS coils and a contribution from the adjacent Transport Solenoid. (See Figure 7.9) In the PS2 region, the gradient is negative and the 1.45 m long region has a gradient variation within ± 5% from the value of -1.0 T/m (as shown in Figure 7.10). The magnet is designed to operate at 52 - 64 % of the short sample limit along the load line at the liquid helium temperature of 4.6 K and at 63-69 % of the short sample limit at the maximum allowable temperatures of 5.0 - 5.5 K.

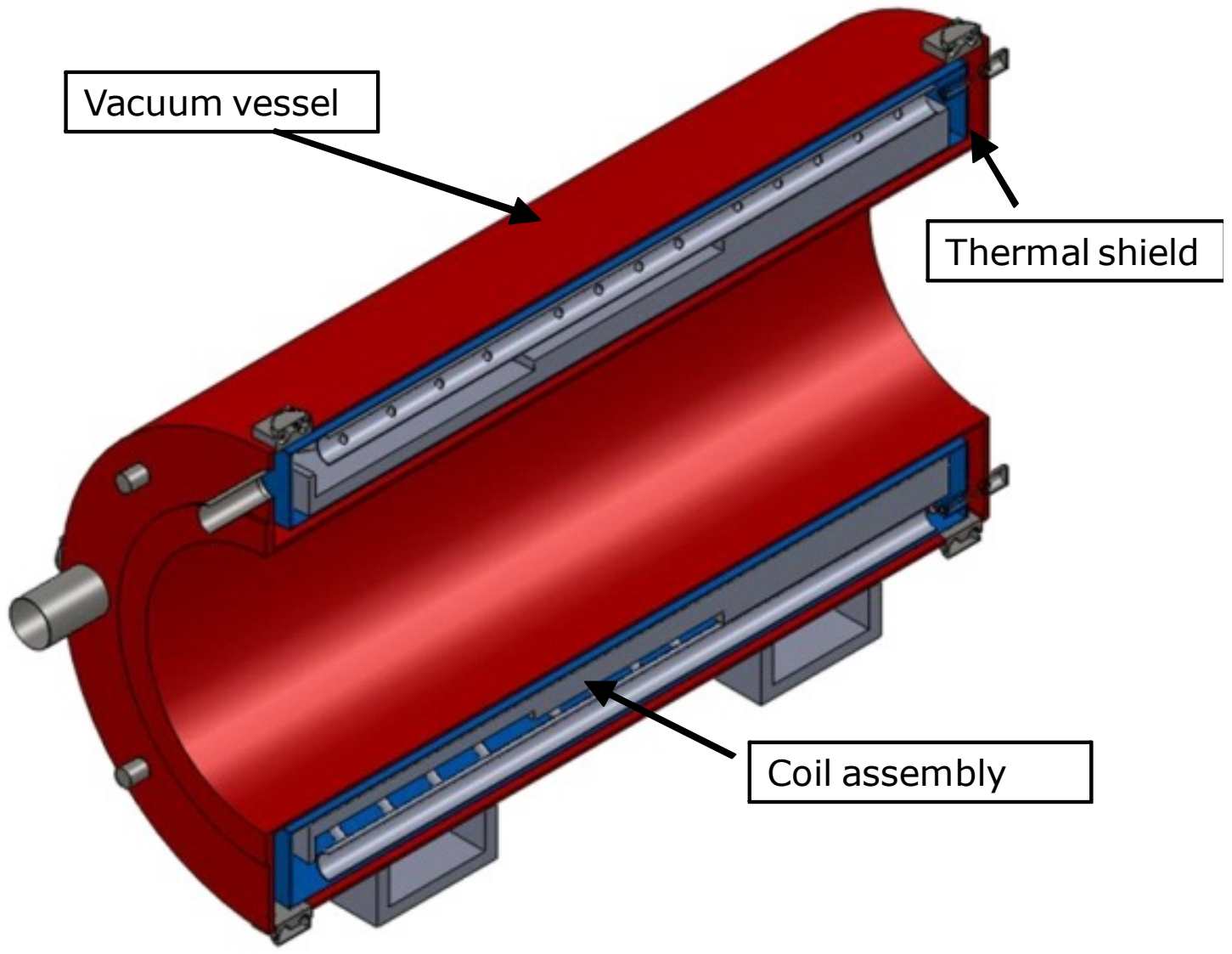

Figure 7.7. Production Solenoid Cryostat.

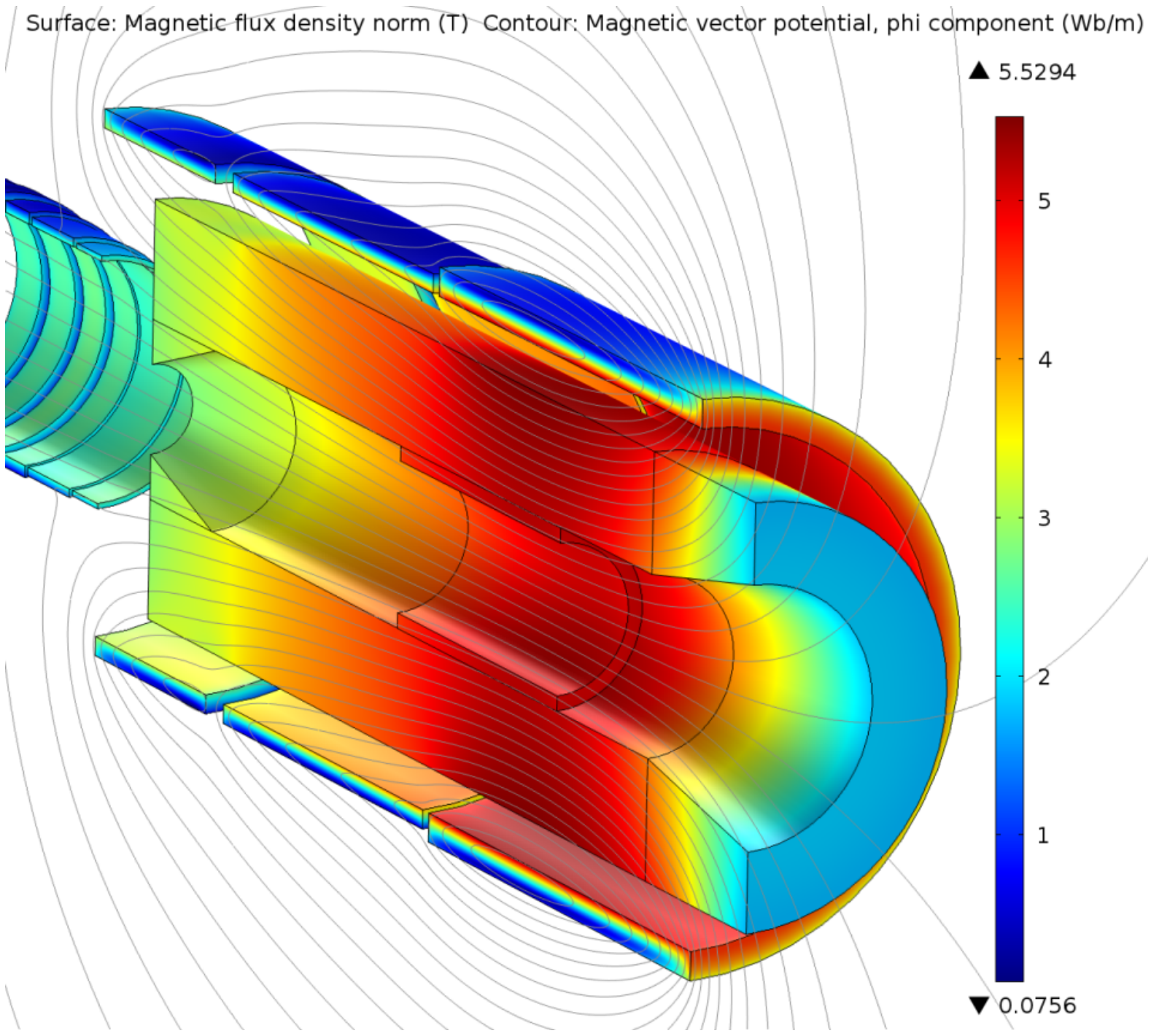

Figure 7.8 FEM magnet model with the flux density diagram

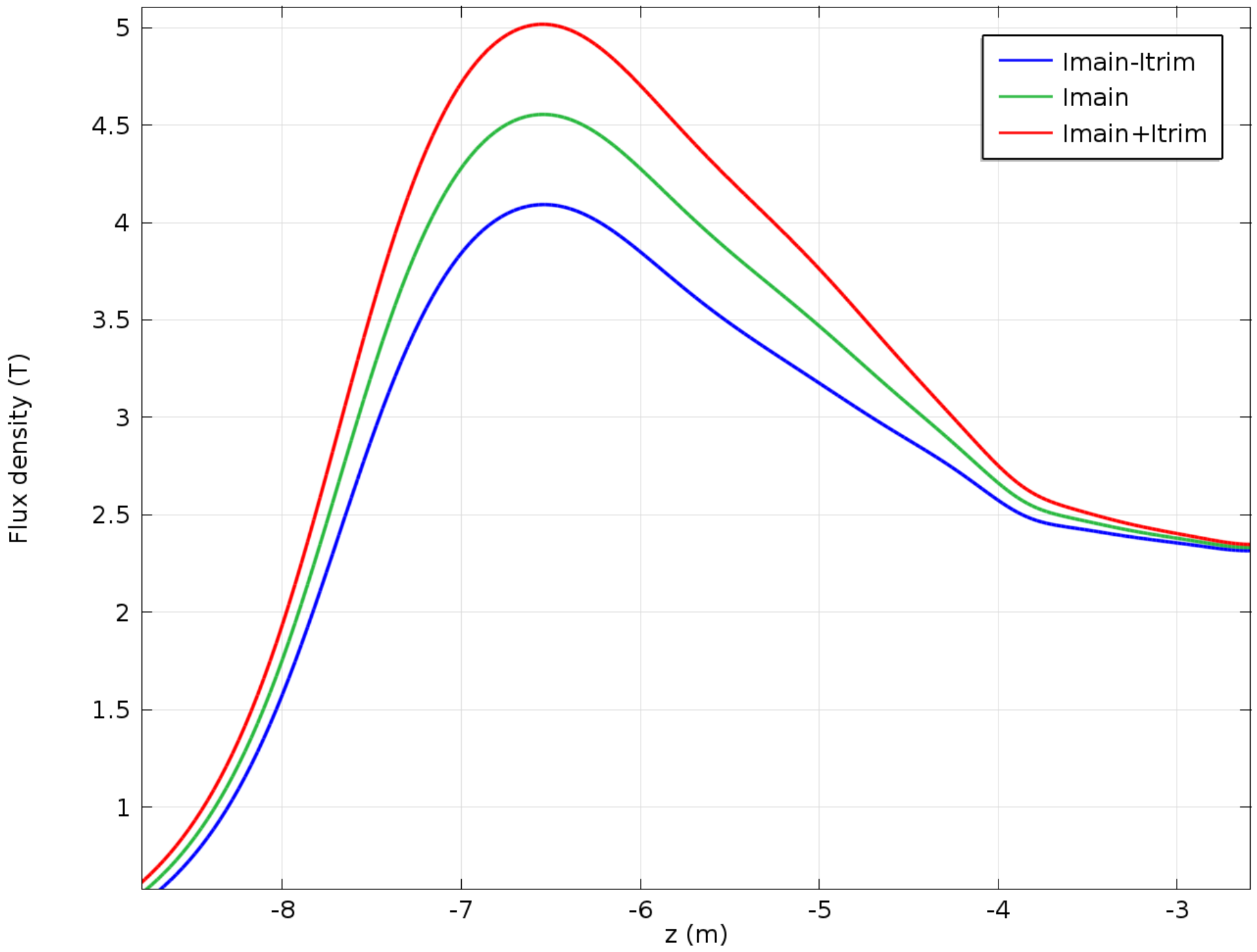

Figure 7.9. Axial field profiles on the Production Solenoid axis. Different lines correspond to the maximum, zero and minimum trim currents.

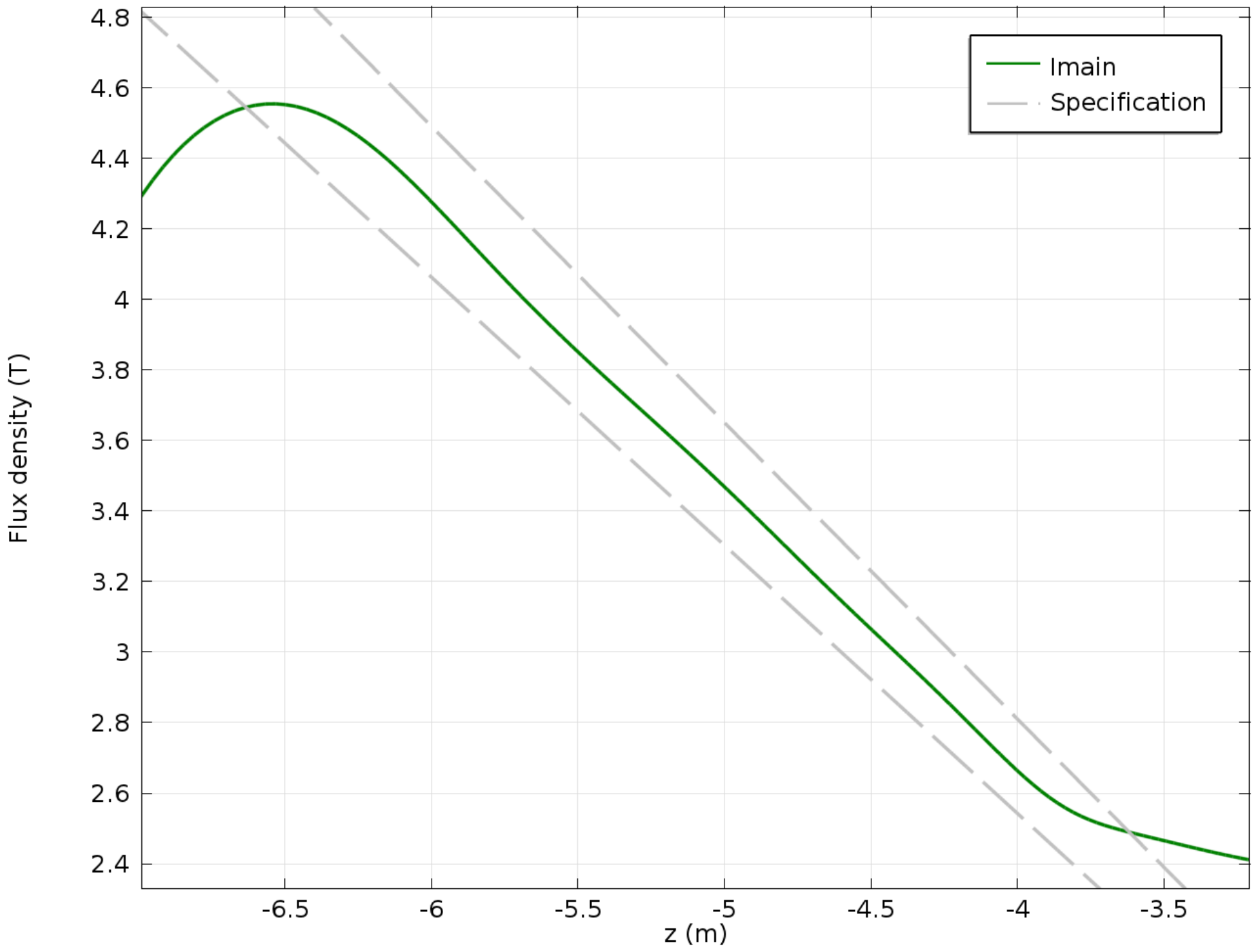

Figure 7.10. The design magnetic field Bz(r=0) in the PS2 region compared to field tolerances (dashed lines) at zero trim current.

***Mechanical Analysis***

Several studies were performed to understand the performance of the Production Solenoid conductor, coil and support structure during cool-down from 300 K to 4.6 K and during coil excitation. A finite element model was developed using COMSOL. The measured mechanical properties for the conductor, insulator and support materials were used as inputs.

The mechanical strength of the conductor is an important contributor to the overall mechanical integrity of the magnet. Lacking data for the actual PS cable, the material strengths are derived from the ATLAS Central Solenoid cable that has the same width but is a factor of 1.3 thinner. This is a conservative assumption since the PS cable has a factor of ~3 lower stabilizer-to-strand ratio than the ATLAS cable.

Table 7.6 lists the measured ATLAS cable parameters from reference [16]. Since there is no data for the ultimate cable strength at 4.2 K, the maximum allowable stress for the cable is based on 2/3 of the yield strength. It is expected that this criterion will be more conservative than using 1/2 of the ultimate strength, a scaling that is often used when the ultimate cable strength is known. As shown, the maximum allowable stress for the cable is 98 MPa.

| Material/Property | Temp, K | Al. Stabilizer | Overall Cable |
|---|---|---|---|
| Yield Strength, MPa | | 81 | 128 |
| Ultimate Strength, MPa | 300 | 86 | 184 |
| Maximum allowable stress, MPa | | 43 | 85 |
| Yield strength, MPa | | 110 | 147 |
| Ultimate Strength, MPa | 4.2 | 294 | - |
| Maximum allowable stress, MPa | | 73 | 98 |

Table 7.6: Measured strength of aluminum-stabilized ATLAS cable.

Table 7.7 shows the expected mechanical forces on the Production Solenoid cold mass during normal operation and quench [9]. When the Transport Solenoid is also powered, there will be a net force of ~1.4 MN pulling the PS towards the TS. Transverse forces are possible due to misalignment with the TS coils, however these forces self-center the two magnets with respect to each other. Note that because of the curved nature of the TS, there is a slight horizontal force when the Transport Solenoid is energized. These forces are well within the structural limits of the Inconel Rods in Table 7.7.

| Total force on PS coil | At nominal position, kN | Sensitivity to the coil displacement, kN/cm |
|---|---|---|
| $F_x$ | -15.7 | -9.0 |
| $F_y$ | 0.0 | |
| $F_z$ | 1361.9 | 18.0 |

Table 7.7: Summary of the expected mechanical forces on the Production Solenoid cold mass during the operation at the maximum trim current and force sensitivity to the displacements from the nominal position.

A structural analysis of the PS coils was performed using the material properties defined in the coil and conductor section. The peak equivalent stresses are 21 MPa in the coil and 51 MPa in the support structure after cool-down. Energizing the coil with the maximum operating current brings the stresses to 73 MPa and 96 MPa for the coil and support structure, respectively, lower than the maximum allowable stresses at the nominal operating temperature.

Another important structural parameter is the shear/de-bonding stress in the coil insulation. Since the main purpose of the insulation is to create the electrical barrier between turns and the support structure, it is desirable to minimize the shear/de-bonding stresses under all conditions since these can significantly degrade the effectiveness of the insulation over time.

Since the coils have lower thermal contraction coefficients than the support shells in the radial and azimuthal directions, the interfaces between the coils and the support shells are always under normal compression. The shear stresses at these interfaces are within 5 MPa.

In order to avoid creating tensions and shear stresses between coils and end flanges that can be the sources of quenches, there are layers of mica paper between the thermal plates and the end flanges.

***Thermal Analysis***

Secondary particles generated from the 8 kW proton beam interacting with the production target will deposit energy in the Production Solenoid coils. A bronze heat shield lining the warm bore of the PS will intercept most of the energy. However, an expected 20 W of continuous beam power will still be deposited in the cold mass. The radiation heat is extracted from the coil through the system of thermal bridges and plates. Because of the irradiation-induced degradation, one cannot take advantage of thin layers of high-purity Al as it is often done in conduction-cooled magnets; the thermal bridges and plates need to be sufficiently thick to conduct the heat even after irradiation.

The thermal bridges, made from 1 mm sheets of the same aluminum used for the cable stabilizer, are installed on the inner coil surface and extend throughout each coil length. The ends of the thermal bridges are connected to the thermal plates by either welding or soldering, as shown in Figure 7.4. The composite ground insulation, with a total thickness of 500 μm, separates the insulated cables from the metal parts on all sides. There are two 25 μm layers of Kapton between the layers of fiberglass in the ground insulation. The 125 μm thick dry E-glass sheets placed between the coil layers are to be filled with epoxy during the coil impregnation. The outer coil surface is wrapped with a sufficient thickness of fiberglass prior to impregnation that will be machined down to an average of 2 mm after the coils are impregnated.

The thermal plates, made from 4 mm sheets of the same aluminum alloy as the cable stabilizer, protrude outside of the coil and are welded to the outer surface of the support shells, where the cooling tubes are installed. The ends of the thermal plates are stress-relieved by providing a clearance at the corners of the support shells to accommodate the differential contraction between the coils and shells due to cool-down and Lorentz forces. Layers of mica paper are introduced between the thermal plates, flanges and support shells to avoid accumulation of shear stresses at these interfaces. There is also a layer of high-purity Al installed between the support shells and cooling tubes, helping to equalize the thermal loads between the tubes.

A 3D finite element model created within the COMSOL Multiphysics code was finalized to the level of individual cable layers and included all the cooling/insulation features described earlier. The simulation was made for the worst expected case when the RRR of all Al elements (excluding the support structure made of Al 5083-O) was degraded to 100. The law of mixtures was used to define the equivalent thermal conductivities of the insulated cables, the interlayer insulation, and the ground insulation. All other elements had the actual thermal properties of the corresponding materials.

In addition, the relevant static heat loads were applied to all external surfaces to model the thermal radiation/gas conduction, to the middle support ring to model the heat load through the axial supports and to the end flanges to model the heat load through the transverse supports. It was assumed that the cooling tubes are kept at constant temperature $T_0$ by the cryogenic system. One half of the cold mass was modeled since the production target lies in the horizontal plane and the heat deposition map is symmetric with respect to that plane.

The resulting temperature in the cold mass is shown in Figure 7.11 for $T_0$ = 4.6 K. The maximum temperature is in the middle of the inner surface of the thickest coil; that location coincides with the peak field location, and, therefore, directly affects the thermal margin. In order to determine the thermal parameter space, the radiation power was scaled up to a factor of 16. Figure 7.12 shows the peak coil temperature as a function of radiation power factor.

Under the static heat load, the peak temperature increment between the coil and the cooling tubes is ~30 mK, indicating that most of the static heat load is intercepted by the thermal bridges and plates before it enters the coil. Under the nominal dynamic heat load, the temperature increment increases up to ~236 mK. Nevertheless, the peak coil temperature is below the maximum allowable temperature by ~164 mK that provides an additional thermal margin to offset the uncertainly in calculating the power depositions, fabrication tolerances and material properties. As shown, the radiation power can be increased by a factor of ~2 before the peak coil temperature starts to violate the maximum allowable value based on the 1.5 K thermal margin.

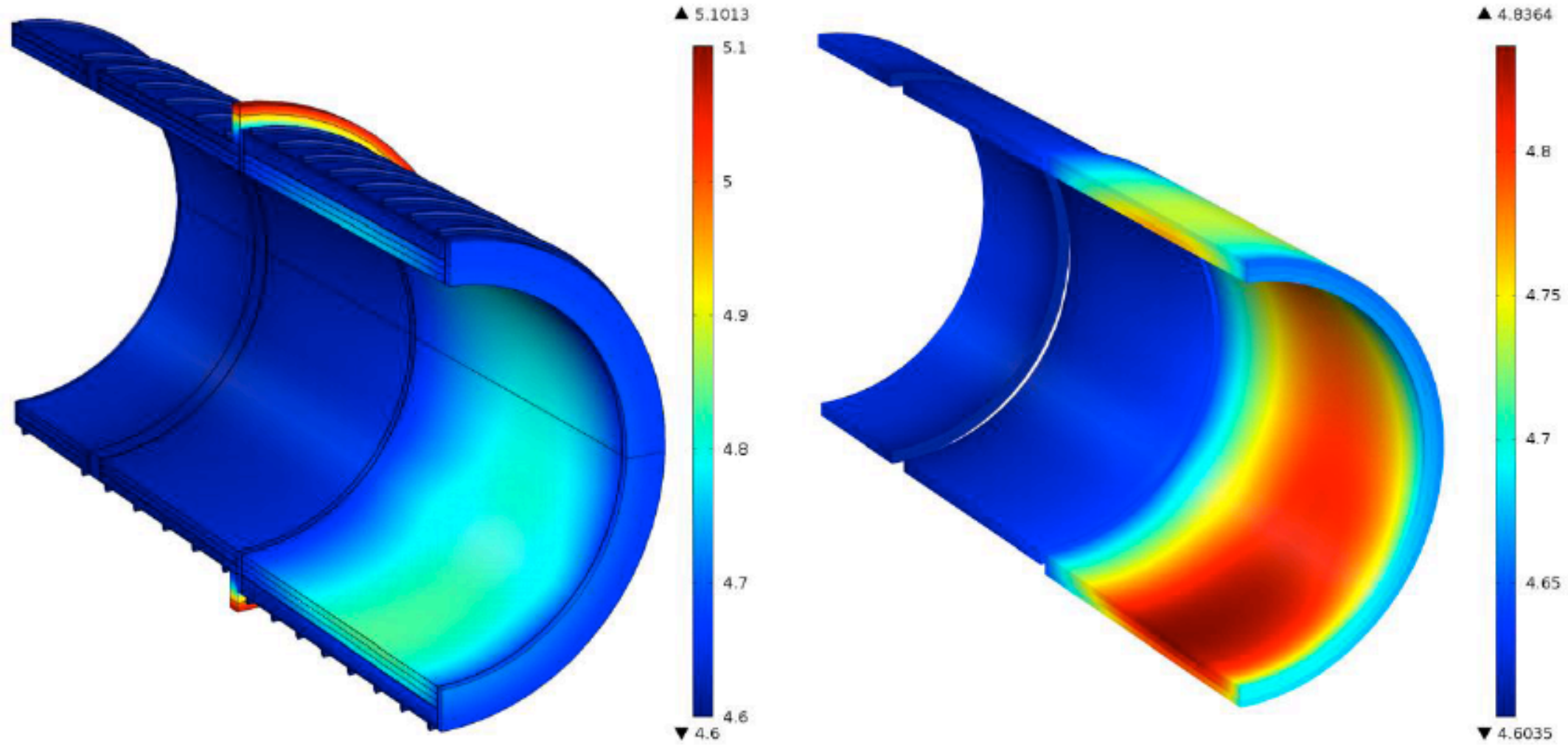

Figure 7.11. Temperature distribution in the PS coldmass (left) and in the coil (right) at $T_0$ = 4.6 K.

***Quench Analysis***

The quench protection system is designed to extract energy to dump resistors outside of the cryostat. The quench protection system continuously monitors the voltage across the magnet leads. If the resistive voltage component exceeds the quench detection threshold of 0.5 V for more than 1 second, the power supply is disconnected and the current starts flowing through the dump resistor. The resistance of the dump is driven by the maximum allowable voltage of 600 V which, for the given operating current, corresponds to a resistance of ~ 0.06 Ω.

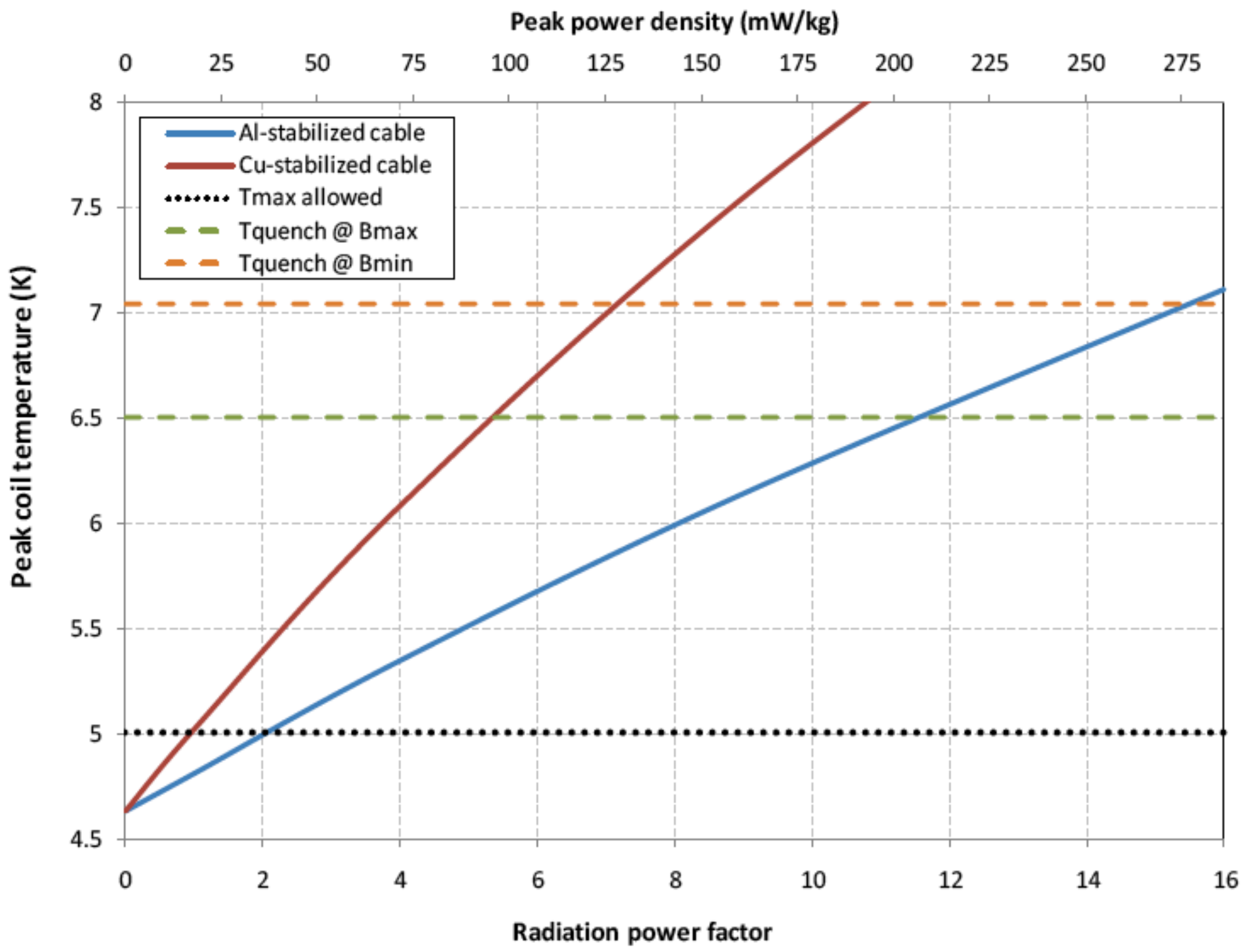

Figure 7.12. Peak coil temperature as a function radiation power factor that is the scale factor applied to the nominal radiation power. A curve for the Cu-stabilized cable is shown for relative comparison.

A 3-D quench propagation model was built using QLASA [17] for the PS coil geometry described in Table 7.3. It was assumed that quenches start at the peak field point in the middle of the upstream coil section when it is energized with the maximum operating current. The results are shown in Figure 7.13 for aluminum and copper stabilizer with various room temperature resistivity ratio (RRR) values. For the nominal values of RRR for aluminum (1000) and Copper (100), the peak temperature is ~80 K. As stated in Section 7.2.4, the RRR will be degraded during operation due to neutron radiation from the production target. Reducing the RRR in the PS coils to the minimum values achieved during one year of operation of 200 (Al) and 73 (Cu), raises the peak temperature to ~108 K.

The radiation deposition in the PS coils will not be uniform. To simulate non-uniform degradation of electrical conductivity in the aluminum stabilizer, the RRR of a short (10 m) cable segment was reduced to the minimum one year value, leaving the RRR of the remaining coil at the average one year values of 400 and 88 for Al and Cu, respectively. This represents the worst-case scenario compared to uniform degradation of the whole coil since the quench detection time is increased while the local heating power in the degraded segment remains the same. As shown, the peak temperature rises to ~132 K, which is close to the peak temperature requirement [2].

The quench analysis is conservative because it does not include quench-back due to heating of the coil support structure by the eddy currents during the current discharge. This effect would create a more uniform heat distribution in the coil and lower the peak temperature.

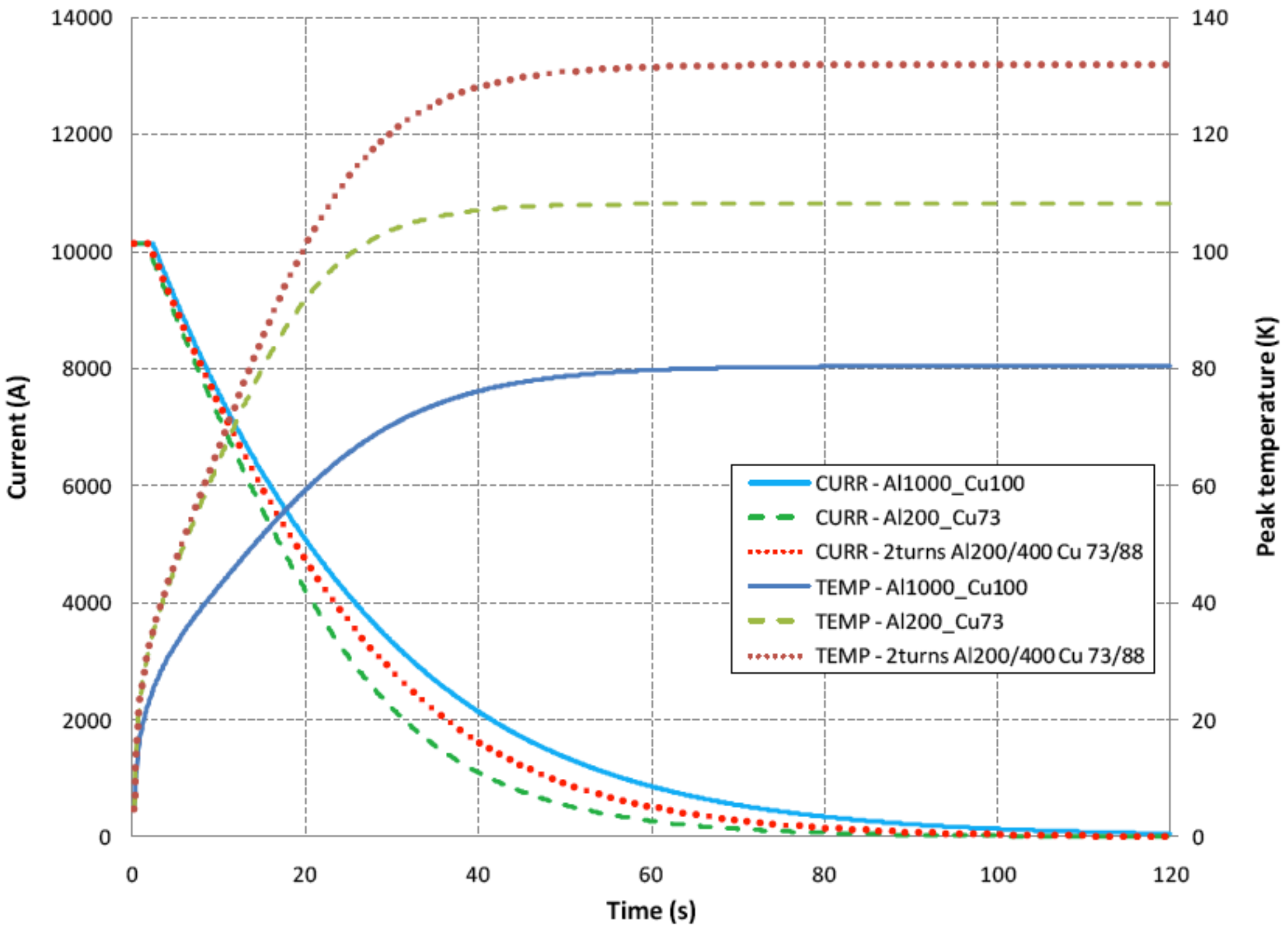

Figure 7.13. Production Solenoid quench current decay and peak coil temperature for various coil RRR values: Initial (solid), RRR uniformly degraded to the minimum value (dashed), RRR degraded in 10-m section of coil (dotted).

### *7.3.2* Transport Solenoid

The Transport Solenoid consists of a series of wide aperture superconducting solenoid rings arranged into two cryostats. Each cryostat has a chimney for superconducting leads, helium supply and return lines and instrument ports. Internal mechanical supports in each cryostat transmit forces to external mechanical supports that connect to the experiment enclosure structure. The Transport Solenoid does not have an iron return yoke.

As shown in Figure 7.14, the Transport Solenoid is segmented into the following set of components:

- TS1 - Straight section that interfaces with the Production Solenoid.
- TS2 - Toroid section downstream of TS1.
- TS3 - Straight section downstream of TS2 (TS3u coils are in the TSu cryostat, TS3d coils are in the TSd cryostat).
- TS4 - Toroid section downstream of TS3.
- TS5 - Straight section downstream of TS4 that interfaces with the Detector Solenoid.

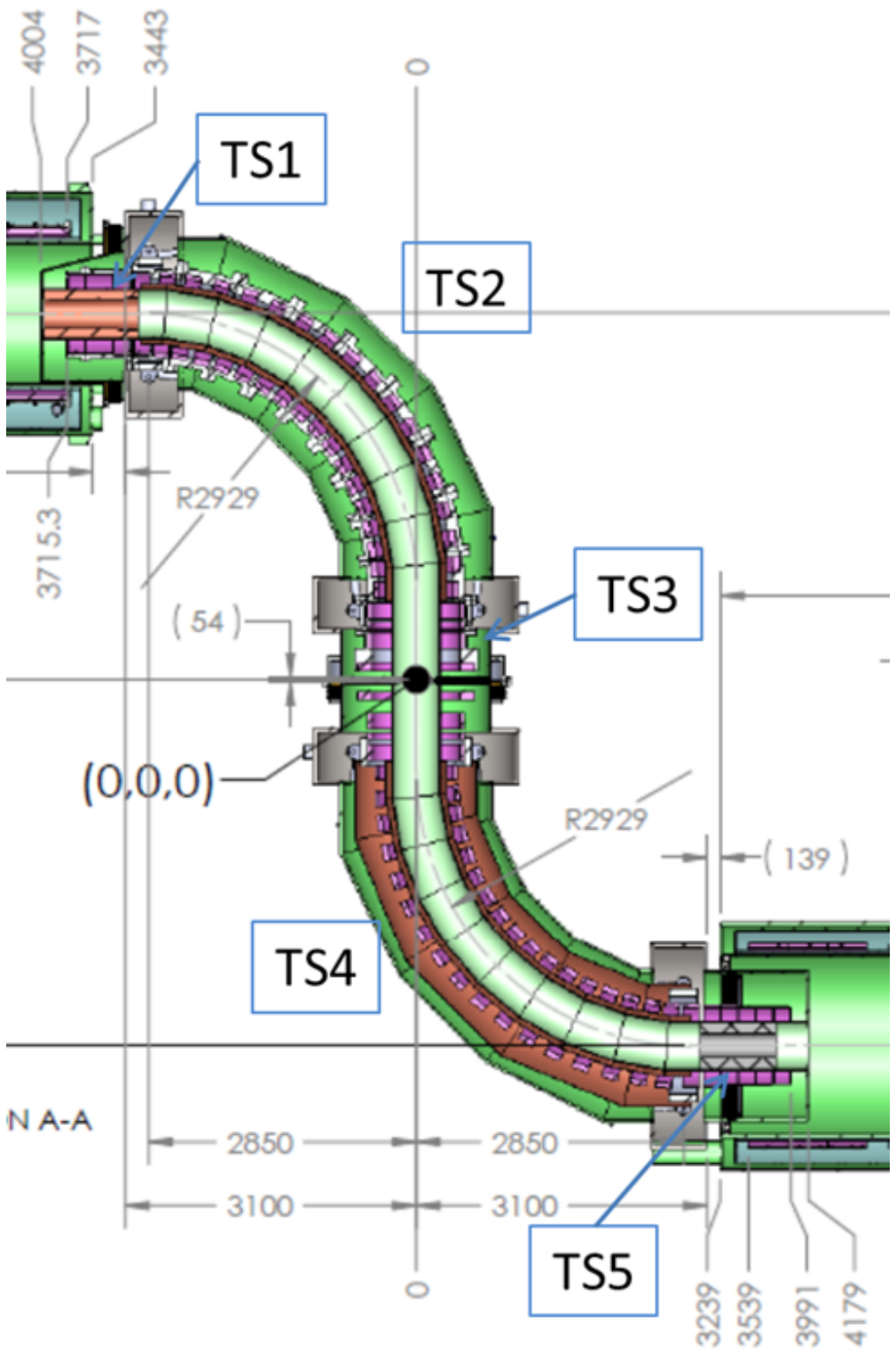

Figure 7.14. The Transport Solenoid with the significant components identified.

The Transport Solenoid performs the following functions:

- Pions and muons are created in the production target in the Production Solenoid. The Transport Solenoid maximizes the muon yield by efficiently focusing these secondary pions and subsequent secondary muons towards the stopping target located in the Detector Solenoid. High energy negatively charged particles, positively charged particles and line-of-sight neutral particles will nearly all be eliminated by the two 90° bends combined with a series of absorbers and collimators.
- The TS1 field must be matched to the field of the Production Solenoid at the interface for optimum beam transmission.
- There must be a negative axial gradient at all locations in the straight sections (TS1, TS3 and TS5) for radii smaller than 0.15 m to prevent particles from becoming trapped or otherwise losing longitudinal momentum.

- Through the first toroid section (TS2) the beam will disperse vertically, allowing a collimator in TS3 to perform a sign and momentum selection.
- The second toroid section (TS4) will nearly undo the vertical dispersion, placing the muon beam centroid near the TS5 axis.
- The TS5 field must be matched to the field of the Detector Solenoid at the interface for optimum transmission of the muon beam to the stopping target.
- The Transport Solenoid acts as a beamline interface for the antiproton window and various collimators, including the rotating collimator in TS3.

The Transport Solenoid consists of two independent cryostats and power units. The TS1, TS2 and TS3u coils are assigned to the TSu cryostat. The TS3d, TS4 and TS5 coils share the TSd cryostat. Each cryostat will have its own superconducting link, feed box, power converter and extraction circuit.

All TS coils use the same conductor design and similar cooling schemes. The TSu unit and the TSd unit are nearly identical, so only the conceptual design of TSu will be presented.

***TSu Design Concept***

TSu is shown in Figure 7.15 and includes the following design features:

- A single cryostat to avoid gaps between coils and reduce complexity and cost.
- The coils are powered in series to minimize the number of leads and the complexity of the power and quench protection systems.
- The quench protection strategy is based on extracting most of the energy to external dump resistors.
- Coils are preassembled and tested inside modules (mostly with two coils per module) in order to reduce complexity during cold mass assembly.
- The mechanical support system consists of 15 supports: four supports along the toroid main radius, four axial supports close to each end, and three gravity supports.

There is a gap between the TS3u and the TS3d coils as a result of the cryostat interfaces and the mechanical hardware necessary to actuate the rotating collimator and to insert the anti-proton window. In an early study of the TS magnetic design [20] it was shown that if this gap is larger than 150 mm a significant increase in complexity is needed to meet the field requirements. Subsequent studies have shown that the present coil layout can accommodate a larger gap if the inner radius of the TS3 coils is larger than the inner radius of the other TS coils. To allow for a 220 mm gap, the inner radius of the TS3 coils have been increased to 465 mm, compared to an inner radius of 405 mm for the remaining TS coils. Further details about the TSu design can be found in reference [19].

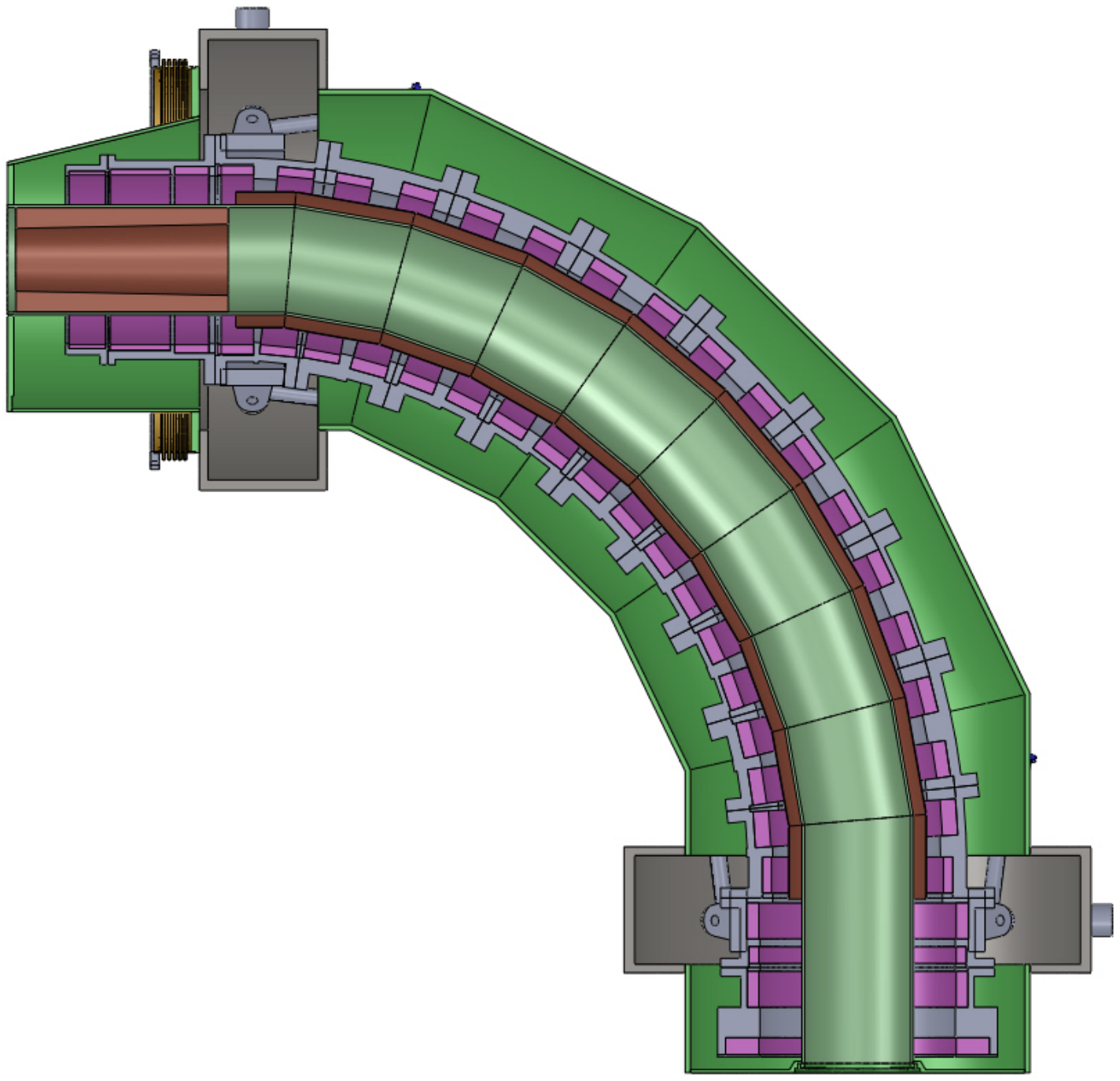

Figure 7.15. TSu cross section along the horizontal plane.

***TSd Design Concept***

The TSd coil and cryostat designs are very similar to TSu. TS3d coils are connected to a toroidal section (TS4) and to a straight section (TS5) in a single cryostat and powered in series.

***Conductor Design***

The conductor used for the TS is an aluminum stabilized NbTi Rutherford cable. This kind of conductor is typically used for detector systems in particle accelerators and colliders. The strand diameter and Rutherford cable thickness proposed have been used for the conductor of the BELLE detector solenoid at KEK [24]. A cross section of the cable is shown in Figure 7.16.

The cable parameters are shown in Table 7.8. The insulation is made of fiberglass braided (or tape with some overlap) with a thickness of 0.125 mm per side. Epoxy impregnation completes the insulation.

The TSu is powered by a dedicated power supply with an operating current of 1730 A. With this conductor and insulation configuration the critical engineering current density is 47 A/mm$^2$ with a peak field of 3.4 T. The operating current fraction on the load line at 5.1 K is 58%. The temperature margin at 5.1 K and 3.4 T is 1.82 K.

| Conductor Parameter | Unit | Value | Comment |
|---|---|---|---|
| Cable critical current (at 5T, 4.2K) | A | 5570 | For reference |
| NbTi critical current density (at 5T, 4.2K) | A/mm$^2$ | 2700 | After coextr. |
| Cable critical current (at 3.5T, 4.2K) | A | 7140 | |
| Number of NbTi strands | | 12 | |
| Strand diameter | mm | 0.67 | |
| Strand copper/SC ratio | | ~1/1 | |
| Copper RRR | | > 150 | |
| Filament size | um | < 30 | |
| Strand twist | mm | 50 | Typical |
| Rutherford cable width | mm | 4.23 | |
| Rutherford cable thickness | mm | 1.15 | |
| Cable width (bare) | mm | 9.9 | |
| Cable thickness (bare) | mm | 3.15 | |
| Overall Al/Cu/SC ratio | | 13/1/1 | |
| Aluminum RRR | | 500 | |
| Aluminum 0.2% yield strength at 300 K | MPa | 30 | |
| Aluminum 0.2% yield strength at 4.2 K | MPa | 40 | |
| Shear strength btw aluminum - strands | MPa | > 20 | |

Table 7.8 TS Conductor Parameters

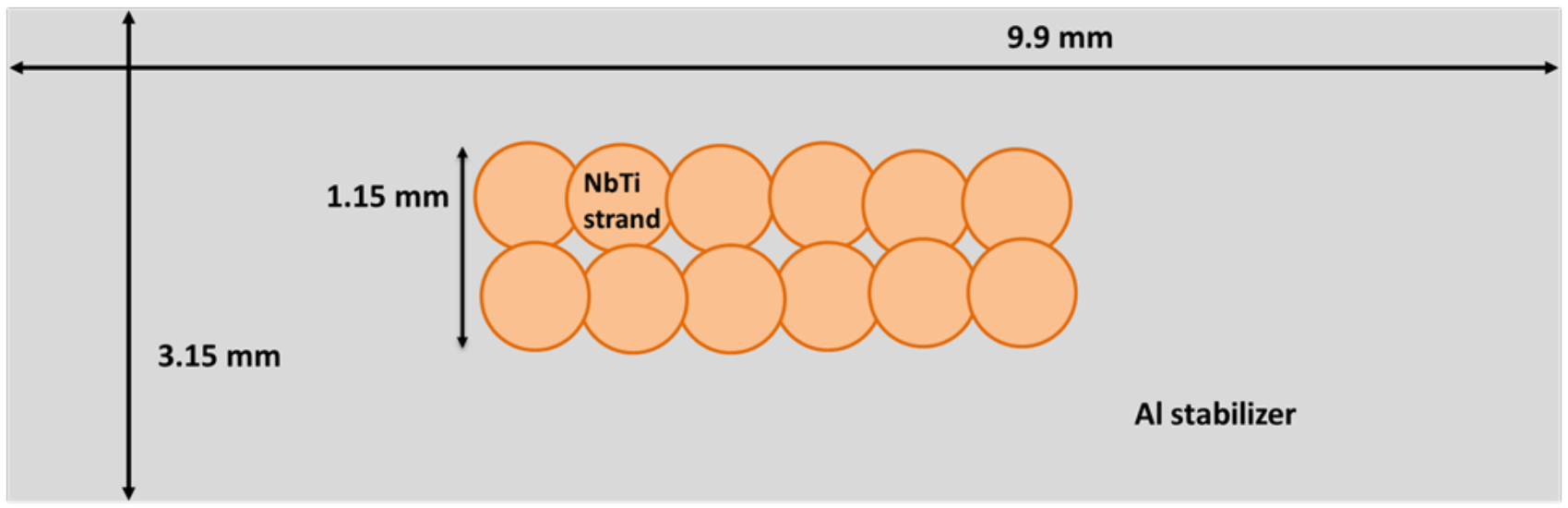

Figure 7.16 Cross Section of TS Cable.

### *TSu Coil Design*

The TS coils will be wound on collapsible mandrels and then inserted into aluminum shells (modules). The modules are assembled into a single cold mass and power unit. TS1 is a straight solenoid made of 3 coils (Figure 7.17) with different outer diameters and separated by flanges. Table 7.9 lists the main coil parameters for TS1.

TS2 is a quarter of a toroid made of 18 coils (Figure 7.18). Table 7.10 lists the main coil parameters for TS2.

TS3u is a straight solenoid made of four coils (Figure 7.19). Table 7.11 lists the main coil parameters for TS3u.

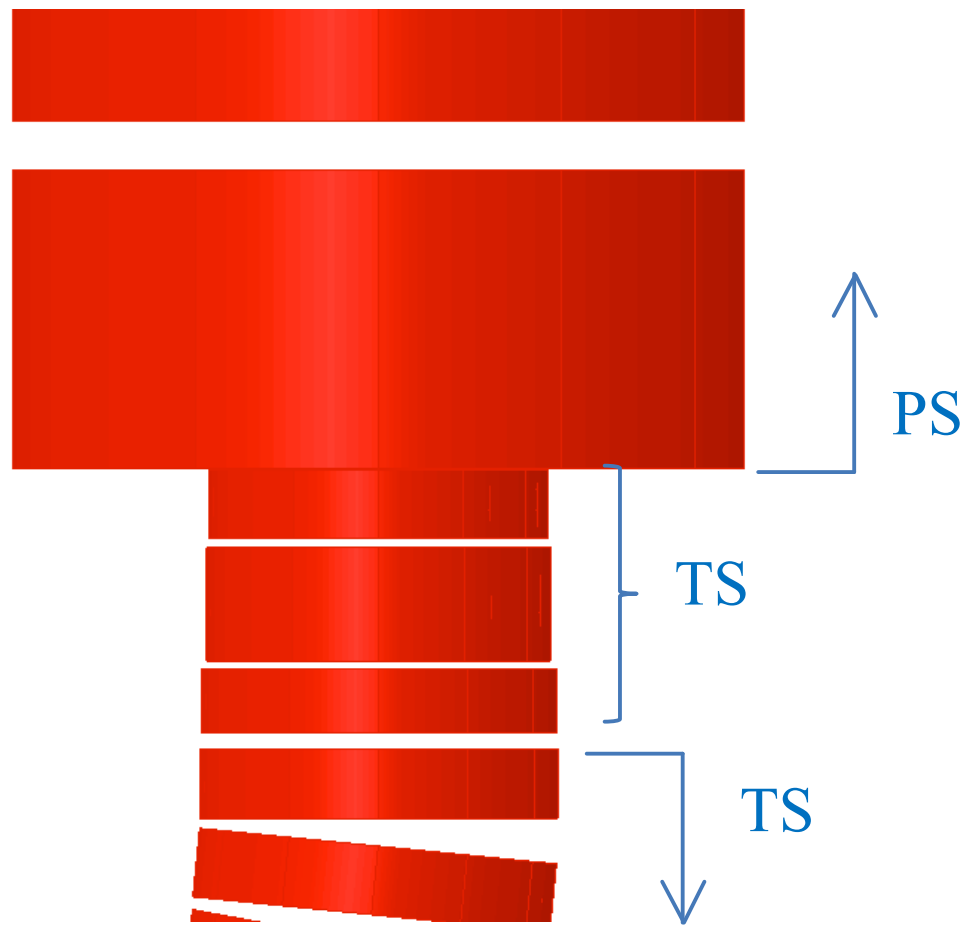

Figure 7.17 TS1 Coils.

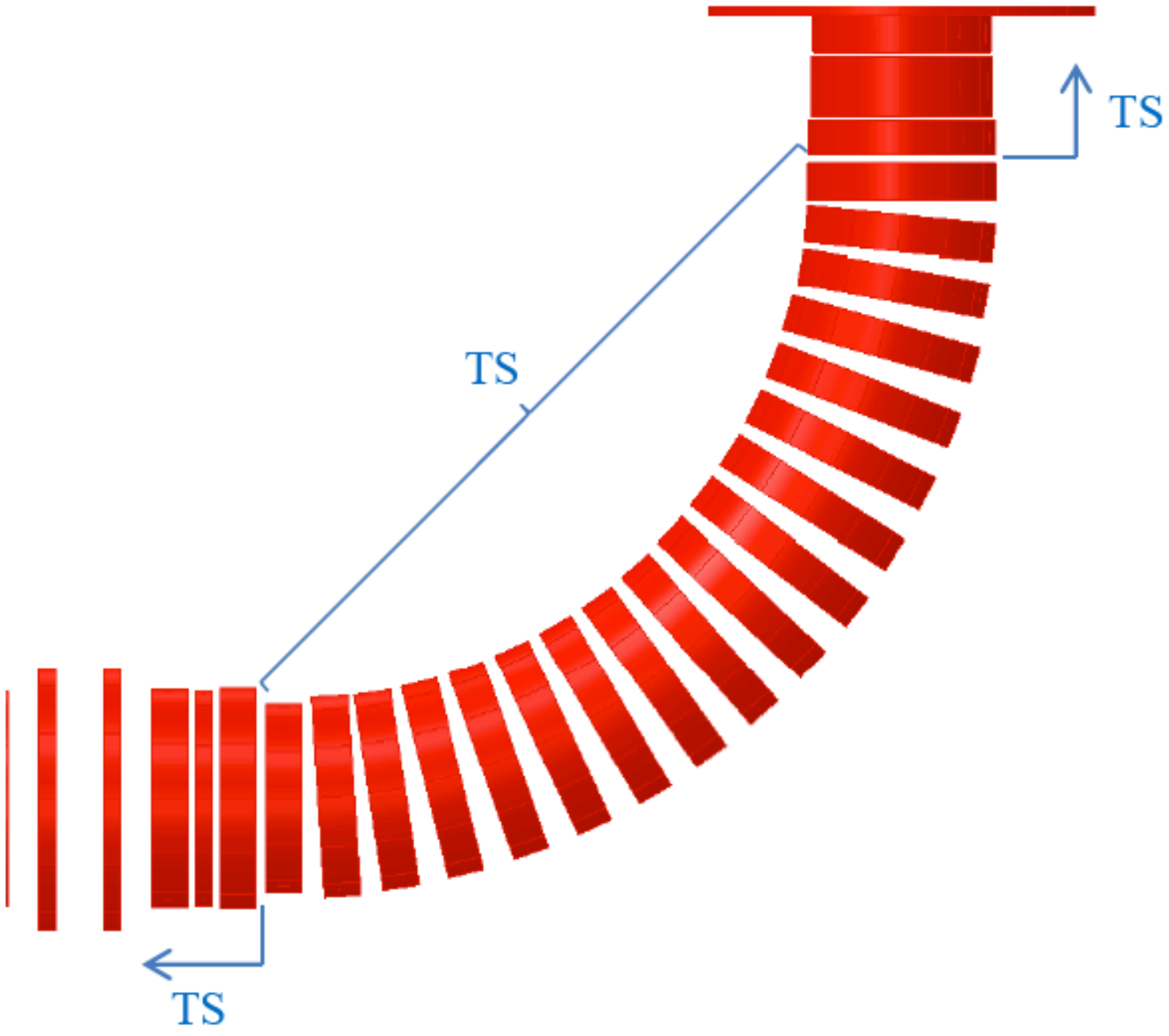

Figure 7.18. TS2 coils. The upstream end is at the top.

| COIL No. | Inner radius (mm) | Outer radius (mm) | Length (mm) | Layers | Turns/layer |
|---|---|---|---|---|---|
| 1 | 405 | 423.0 | 172.6 | 5 | 17 |
| 2 | 405 | 430.3 | 284.2 | 7 | 28 |
| 3 | 405 | 444.9 | 162.4 | 11 | 16 |

Table 7.9. TS1 coil parameters

| COIL No. | Inner radius (mm) | Outer radius (mm) | Length (mm) | Layers | Turns/layer |
|---|---|---|---|---|---|
| 4 | 405 | 448.6 | 172.6 | 12 | 17 |
| 5 | 405 | 448.6 | 172.6 | 12 | 17 |
| 6 | 405 | 448.6 | 172.6 | 12 | 17 |
| 7 | 405 | 463.2 | 172.6 | 16 | 17 |
| 8 | 405 | 463.2 | 172.6 | 16 | 17 |
| 9 | 405 | 463.2 | 172.6 | 16 | 17 |
| 10 | 405 | 463.2 | 172.6 | 16 | 17 |
| 11 | 405 | 466.8 | 172.6 | 17 | 17 |
| 12 | 405 | 466.8 | 172.6 | 17 | 17 |
| 13 | 405 | 466.8 | 172.6 | 17 | 17 |
| 14 | 405 | 466.8 | 172.6 | 17 | 17 |
| 15 | 405 | 470.5 | 172.6 | 18 | 17 |
| 16 | 405 | 470.5 | 172.6 | 18 | 17 |
| 17 | 405 | 470.5 | 172.6 | 18 | 17 |
| 18 | 405 | 470.5 | 172.6 | 18 | 17 |
| 19 | 405 | 470.5 | 172.6 | 18 | 17 |
| 20 | 405 | 477.8 | 172.6 | 20 | 17 |
| 21 | 405 | 448.6 | 172.6 | 12 | 17 |

Table 7.10: TS2 Coil parameters.

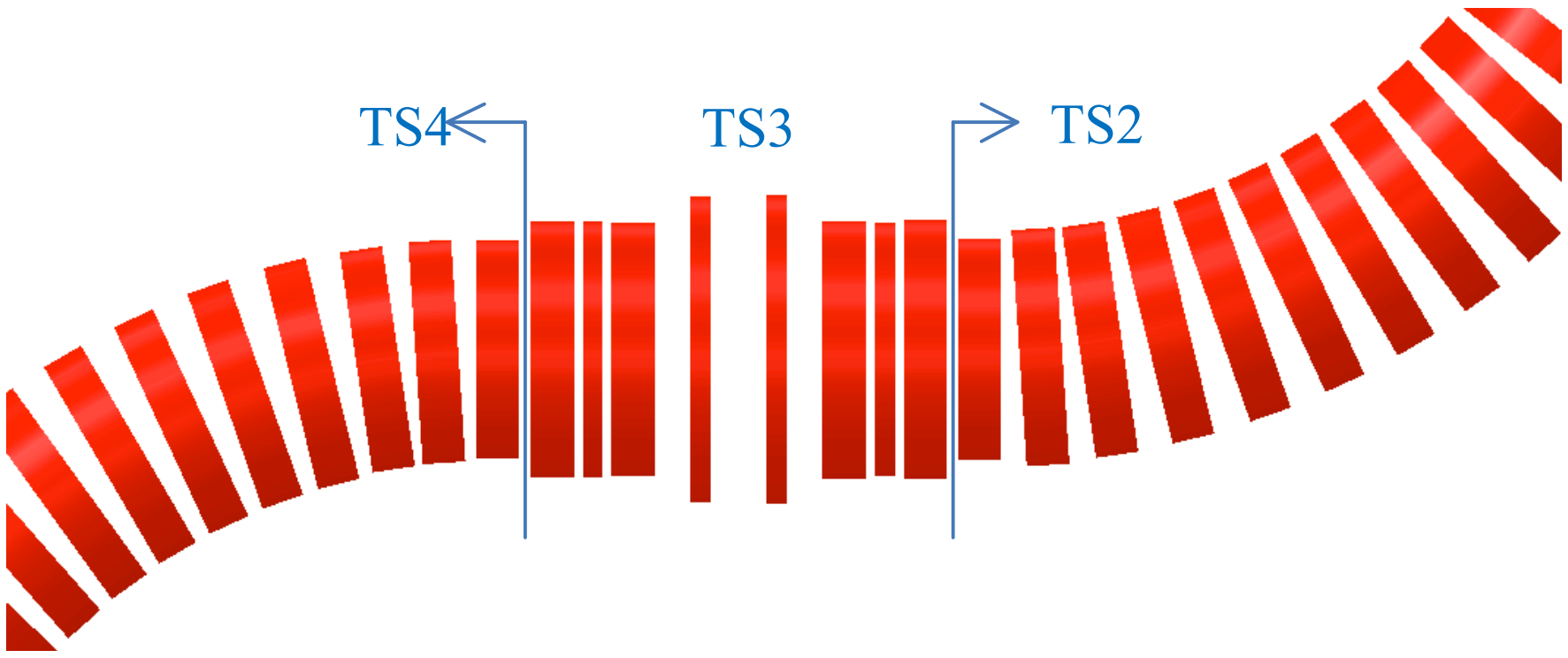

Figure 7.19. TS3 Coils. TS3u coils are adjacent to TS2; TS3d coils are adjacent to TS4

| Coil | Inner radius (mm) | Outer radius (mm) | Length (mm) | Layers | Turns/layer |
|---|---|---|---|---|---|
| **22** | 465 | 523.2 | 172.6 | 16 | 17 |
| **23** | 465 | 512.2 | 81.2 | 13 | 8 |
| **24** | 465 | 519.5 | 172.6 | 15 | 17 |
| **25** | 465 | 621.7 | 81.2 | 43 | 8 |

Table 7.11 TS3u coil parameters

Each module in the toroid and in TS3 can house two coils, which are inserted from each end. Each module will be warmed up, allowing sufficient clearance for coil insertion followed by a shrink fit. A typical TS2 module without coils is shown in Figure 7.20. The modules can be fabricated by using a 5-axis industrial milling machine and a CNC lathe. The cross section of all TSu coils and modules is shown in Figure 7.21. The cooling system is attached to each module's outer surface. Heat will be conducted from the coil to the cooling system by conduction through the module (made of Al 5083-O) and by strips of pure aluminum connected to the coil inner surface.

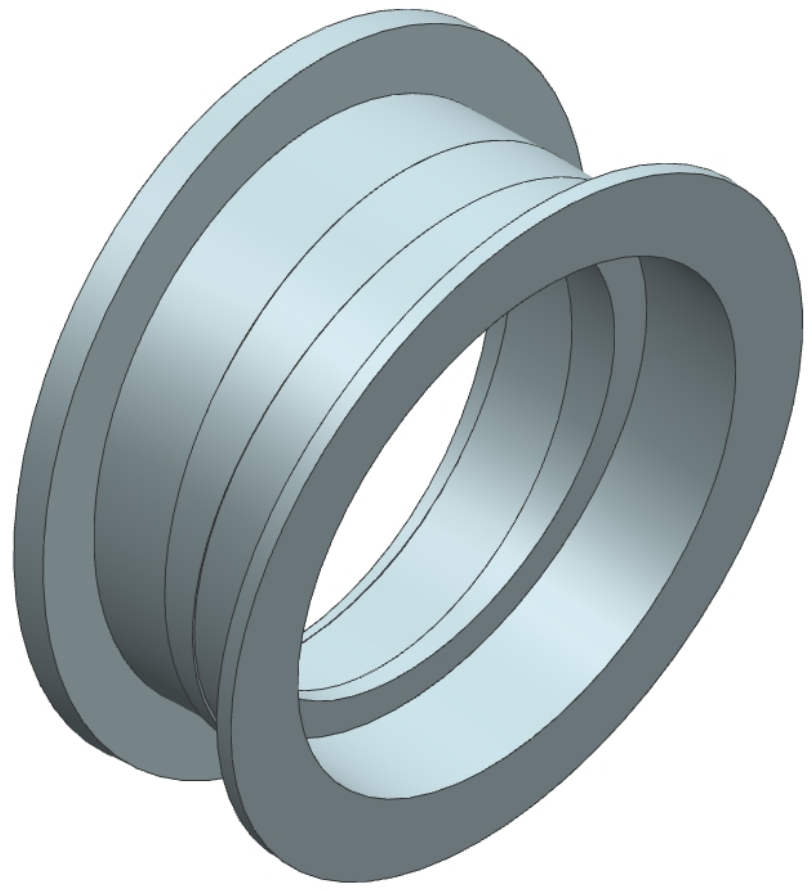

Figure 7.20. Typical TS2 Coil Module. One coil will be inserted from each side.

***TSu Mechanical Design and Cryostat***

The TSu cryostat contains the TS1, TS2 and TS3u coils. It consists of the components and systems listed below.

- Structural supports for the magnetic coils and the vacuum vessel.
- A 4.5 K cooling circuit.
- An 80 K thermal shield.
- A vacuum vessel with a warm bore.
- Interface to the PS cryostat.
- Interface to the proton beam line.

- Interface to the TSd cryostat and to the antiproton window.
- Interface to a cryogenic transfer line.
- Support for a Collimator in the warm bore at the PS interface.
- Support for a Rotatable Collimator in the warm bore at the TSd interface, and interface with rotating mechanism.

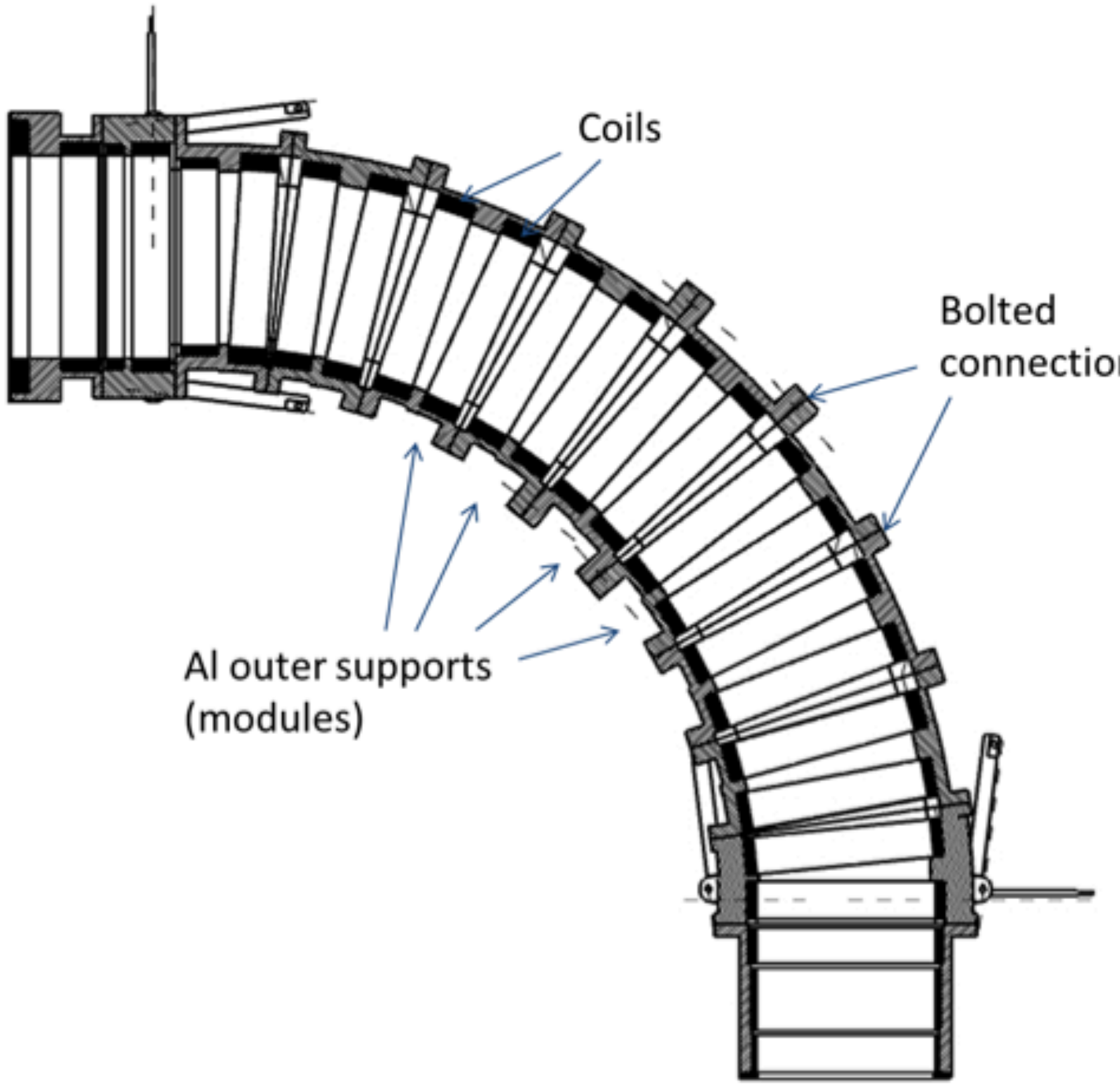

Figure 7.21. Cross Section of TSu showing modules and coils.

TS1 is a straight section with a length of 704 mm with a free end flange that interfaces with the Production Solenoid. The other end has a flange that bolts to a mating flange on TS2, a toroid with a global centerline bend radius of 2.929 m. TS3u is a straight solenoid 750 mm long with a free end flange that interfaces with the TS3d. The other end of TS3u has a flange that bolts to a mating flange on the TS2. TS3u is made of two modules housing two coils each. TS2 is made of 9 modules housing two coils each. TS1 is made of a single module housing three coils.

The mechanical support system for TSu consists of four radial supports (in the direction of the toroid main radius), eight axial supports and 3 gravity supports, as shown in Figure 7.22. The radial supports react only against tensional loads. The axial supports operate under both tension and compression. The orientation and configuration of the axial supports are designed to allow a clear path for the proton beam tube that passes nearby as it intersects the PS. A picture of the TSu support system is shown in Figure 7.23. The dimensions of the supports are listed in Table 7.12. The radial and gravity supports are made from rods (the table shows outer diameter and length), the axial supports are made from pipes (the table shows outer diameter, wall thickness and length). All supports are made of Inconel 718.

The TSu cryostat is shown in Figure 7.23 and Figure 7.24. Table 7.13 lists the dimensions and materials for the various cryostat components.

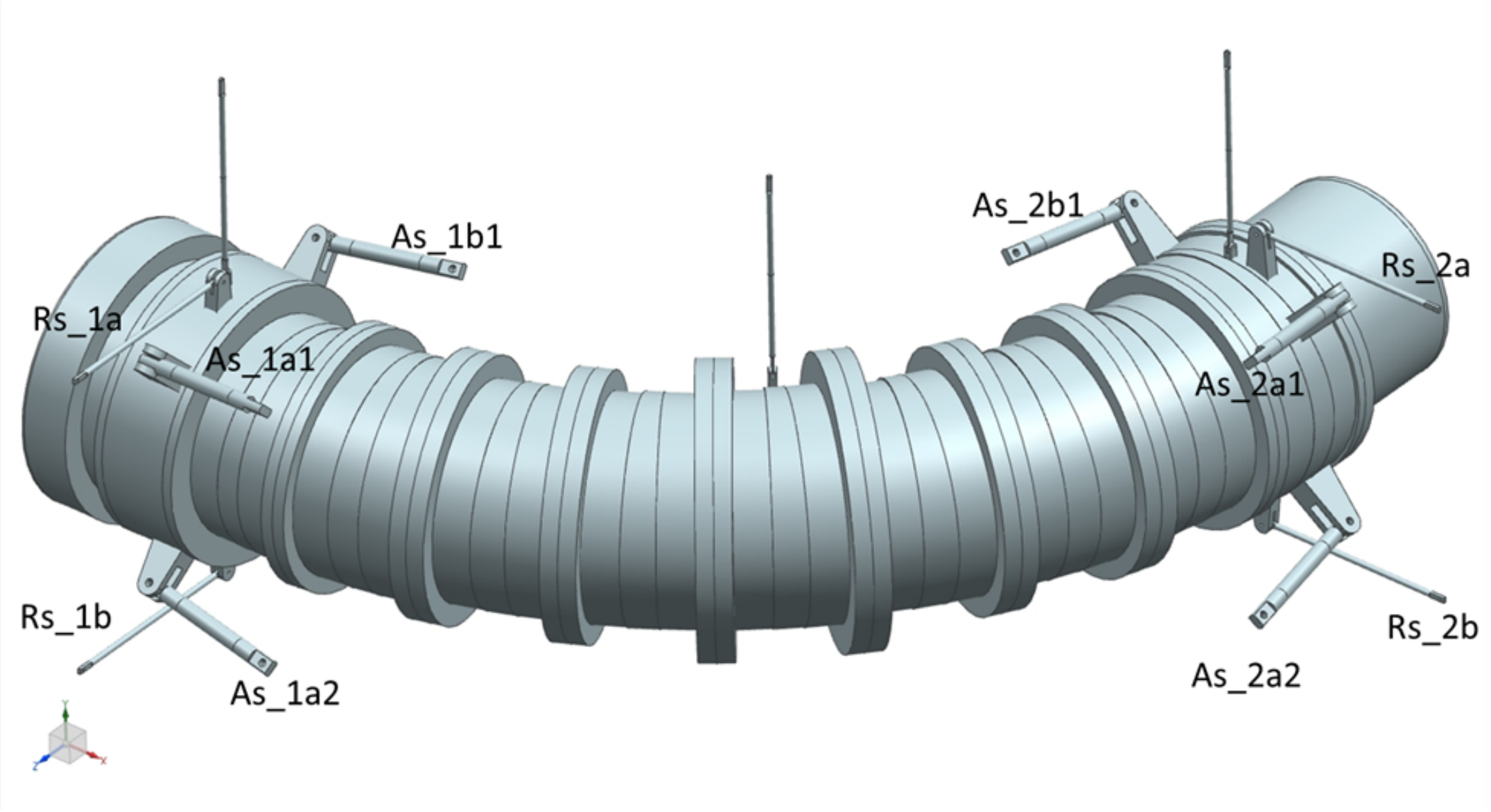

Figure 7.22. TSu support structure.

| Name | Function and Location | Dimensions |
|---|---|---|
| Rs_1a | Radial at TSd end | 28.6 mm OD, 1 m length |
| Rs_1b | Radial at TSd end | 28.6 mm OD, 1 m length |
| Rs_2a | Radial at PS end | 28.6 mm OD, 1 m length |
| Rs_2b | Radial at PS end | 28.6 mm OD, 1 m length |
| As_1a1 | Axial at TSd end | 73 mm OD, 3 mm thick, 0.75 m |
| As_1a2 | Axial at TSd end | 73 mm OD, 3 mm thick, 0.75 m |
| As_1b1 | Axial at TSd end | 73 mm OD, 3 mm thick, 0.75 m |
| As_1b2 | Axial at TSd end | 73 mm OD, 3 mm thick, 0.75 m |
| As_2a1 | Axial at PS end | 73 mm OD, 3 mm thick, 0.75 m |
| As_2a2 | Axial at PS end | 73 mm OD, 3 mm thick, 0.75 m |
| As_2b1 | Axial at PS end | 73 mm OD, 3 mm thick, 0.75 m |
| As_2b2 | Axial at PS end | 73 mm OD, 3 mm thick, 0.75 m |
| Gr_1 | Gravity at TSd end | 25 mm OD, 1 m length |
| Gr_2 | Gravity in the center | 25 mm OD, 1 m length |
| Gr_3 | Gravity at PS end | 25 mm OD, 1 m length |

Table 7.12. Locations and dimensions of TSu supports.

| Cryostat Component | Dimension (mm) | Material |
|---|---|---|
| Vacuum Vessel Outer Shell (OD / wall thickness) | 1350 / 25 | Stainless Steel |
| Vacuum Vessel Inner Shell (ID / wall thickness) | 500 / 12.7 | Stainless Steel |
| Vacuum Vessel End Wall (upstream / downstream thickness) | 30 / 15 | Stainless Steel |
| Thermal Shield Outer Shell (OD / wall thickness) | 1100 / ~2 | Aluminum |
| Thermal Shield Inner Shell (ID / wall thickness) | 650 / ~2 | Aluminum |
| Thermal Shield End Wall (thickness) | 3 | Aluminum |

Table 7.13 TSu Cryostat Parameters

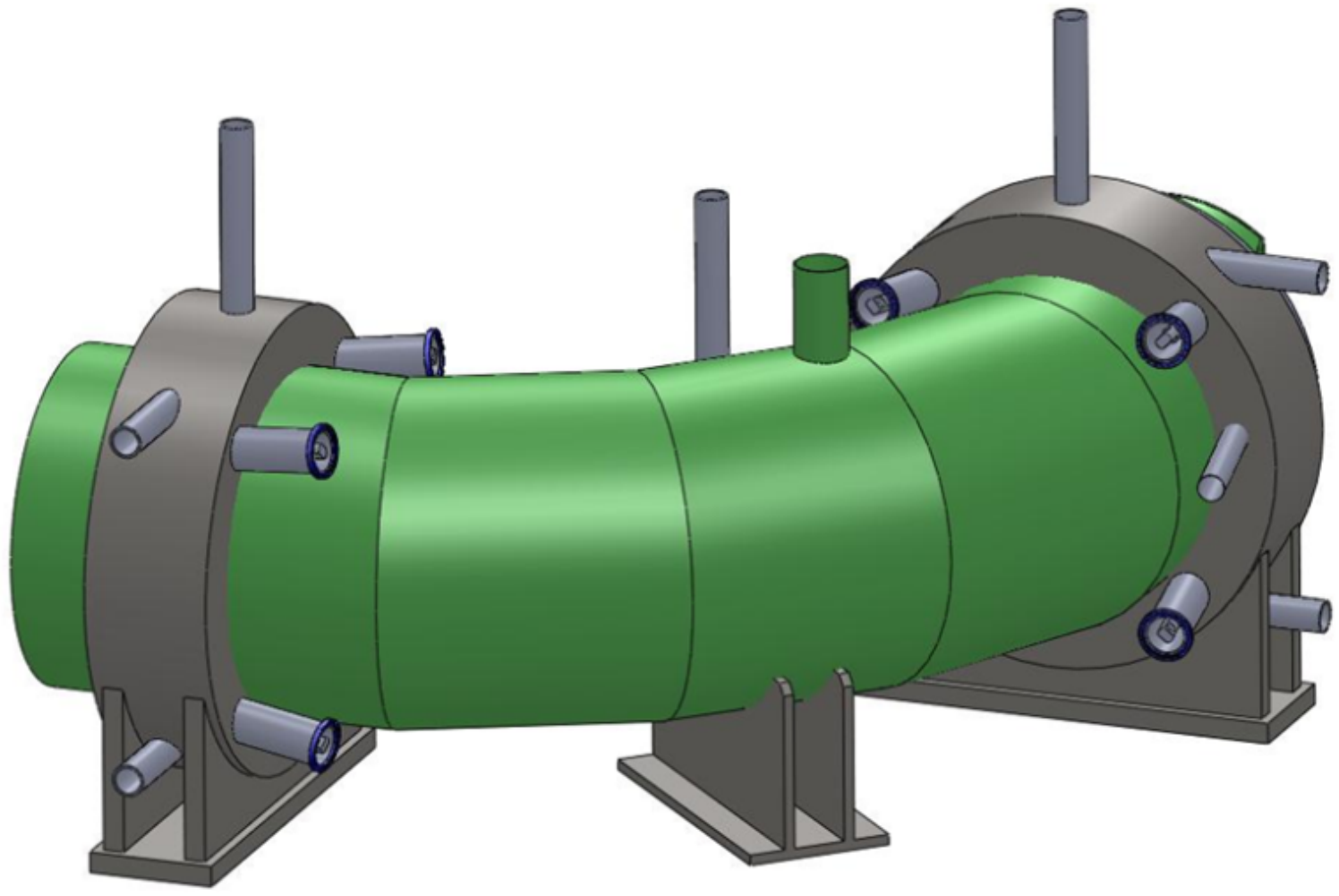

Figure 7.23. View of TSu cryostat. Note proton beam line on the right.

The interface between TS3u and the PS cryostats will be flanged connections with bellows. The incoming proton beam line passes through the TSu cryostat where there is the large section housing the axial supports at the ends of the PS. The cryostat around the TS1 coils has a cutout to avoid interference with the proton beam.

The interface between TS3u and TS3d cryostats will be flanged connections with bellows housing the frame of the antiproton window (also used to separate upstream and downstream vacuum) between mating flanges. The bellows will allow for up to 20 mm of axial offset (See Figure 7.25).

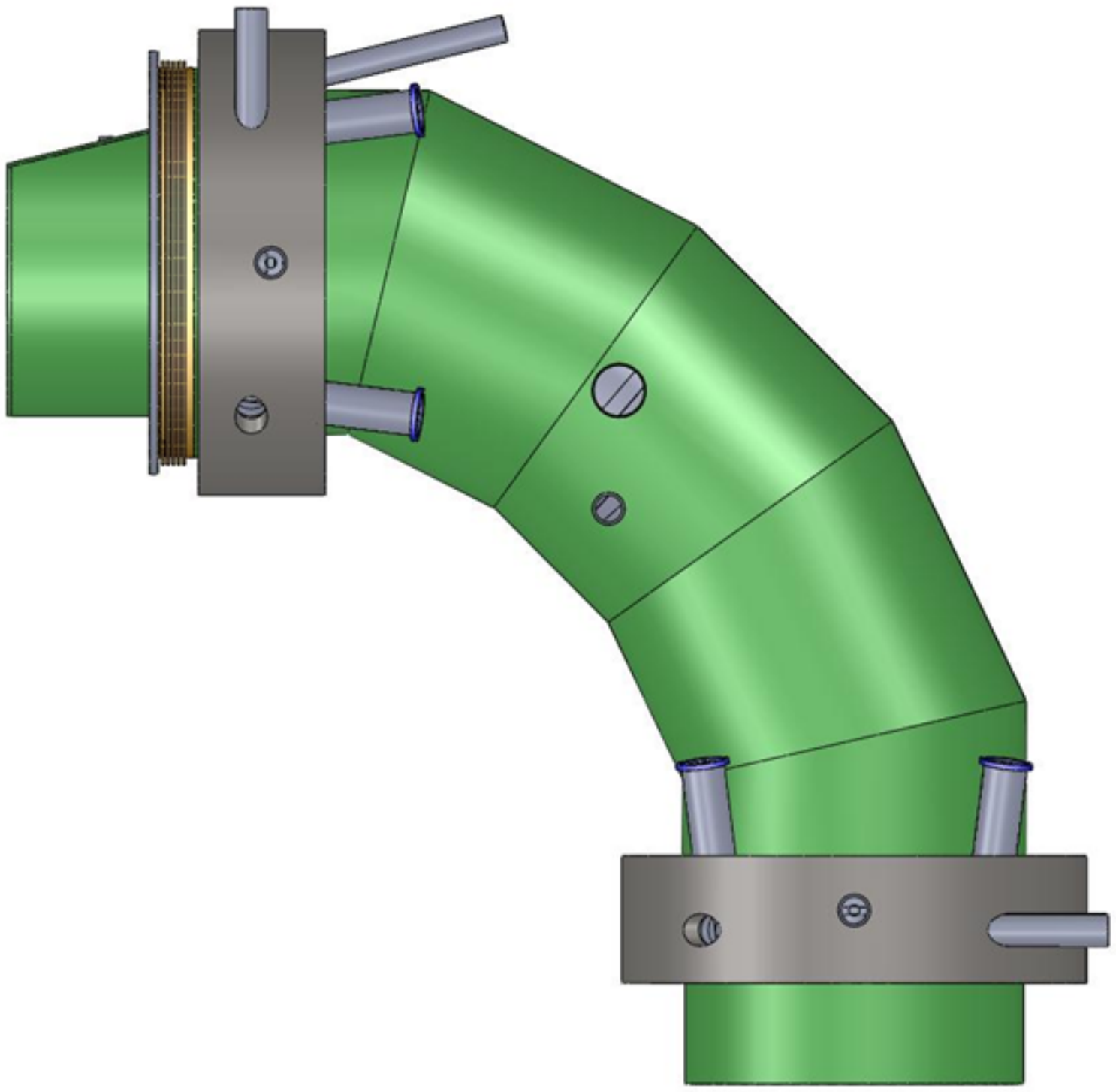

Figure 7.24. Top view of the TSu cryostat. Note the proton beam tube near the top of the figure.

***TS Magnetic Analysis***

As shown in the requirements section (7.2.2) the straight sections and the toroid sections have unique field requirements. The straight sections require a field gradient that is negative everywhere inside a radius of 0.15 m. In the toroid sections the field ripple must be small and the radial gradient must satisfy $|dBs / dr| > 0.275\ T/m$.

An OPERA model of all Mu2e Solenoids was generated using the design coil geometries. There is no iron yoke around any Mu2e magnet and the fringe field of the PS and DS magnets has a significant impact on the trajectory of the particles in the Transport Solenoid causing some horizontal drift. This drift is corrected by a small rotation around the vertical axis of the TS2 and TS4 coils. A pictorial description of the OPERA model is shown in Figure 7.26. The results of the magnetic analysis are shown in Figure 7.27 to Figure 7.30.

In each straight section (TS1, TS3 and TS5) the generated fields and gradients are shown and compared to requirements (black lines). For the toroid sections (TS2 and TS4), the field ripples are compared to requirements. In all cases the field specifications are met.

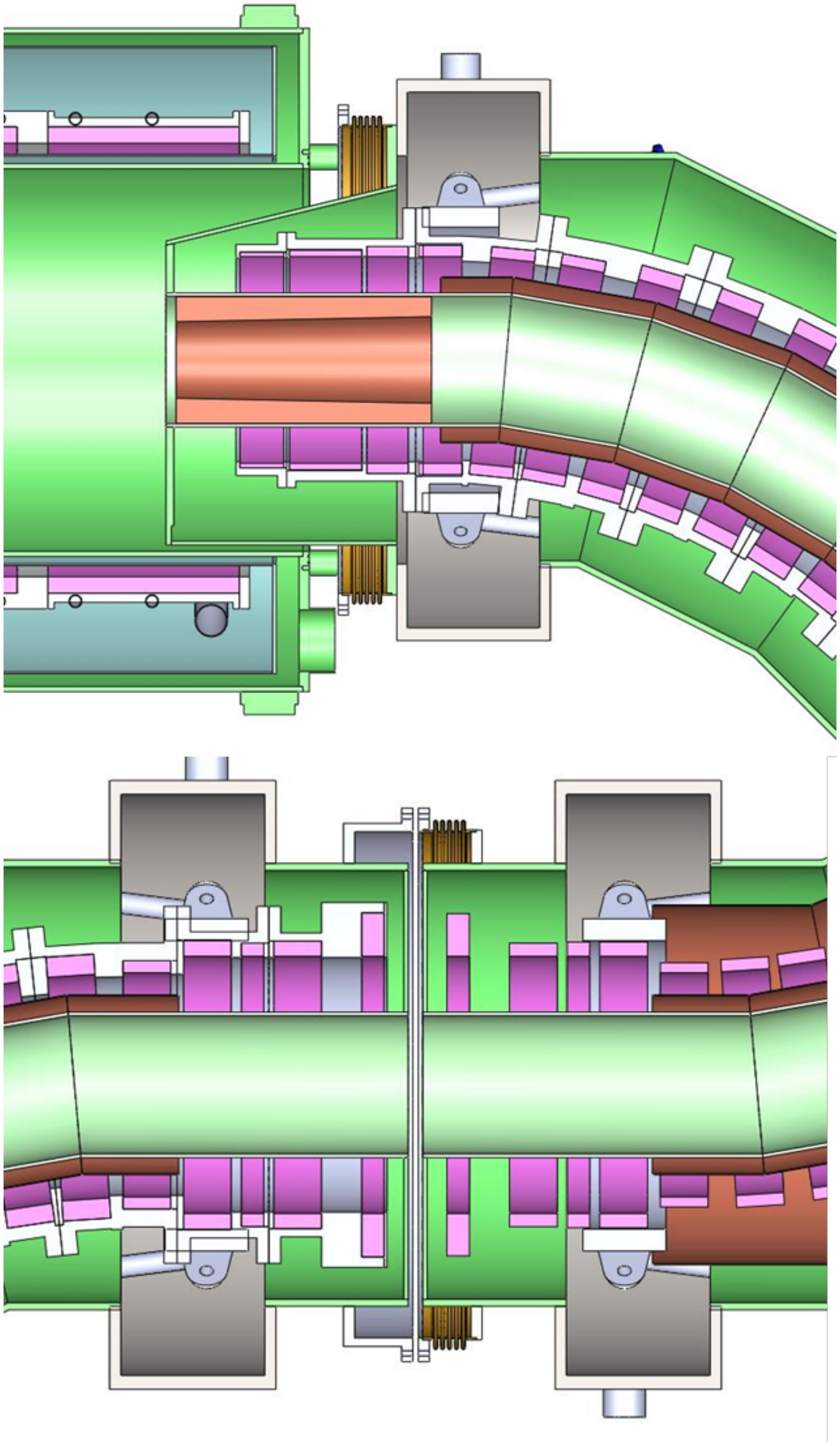

Figure 7.25. TSu Cryostat Interfaces. Top: TSu-PS interface; Bottom; TSu-TSd interface.

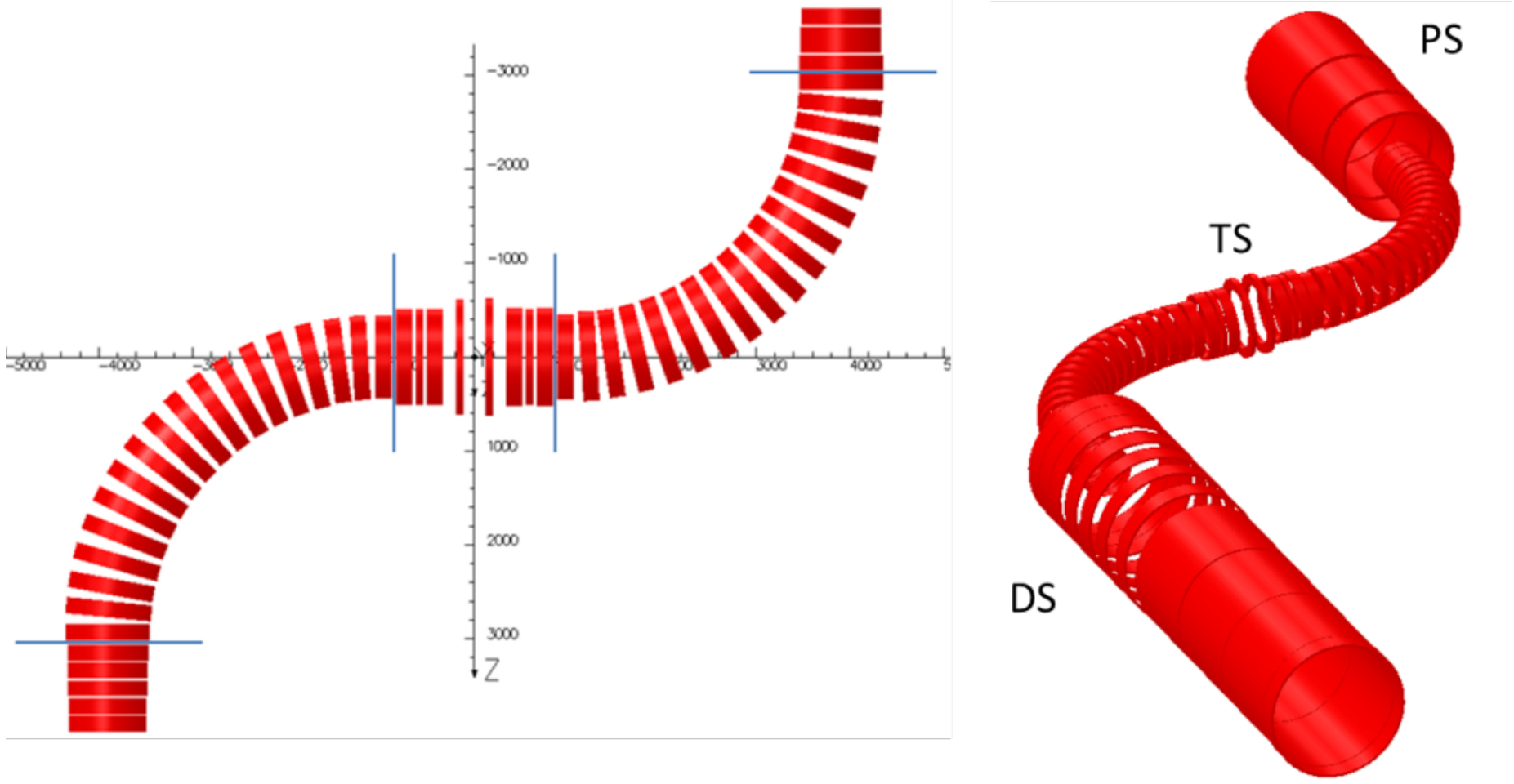

Figure 7.26. Left: OPERA model of the Transport Solenoid. Right All Mu2e Coils.

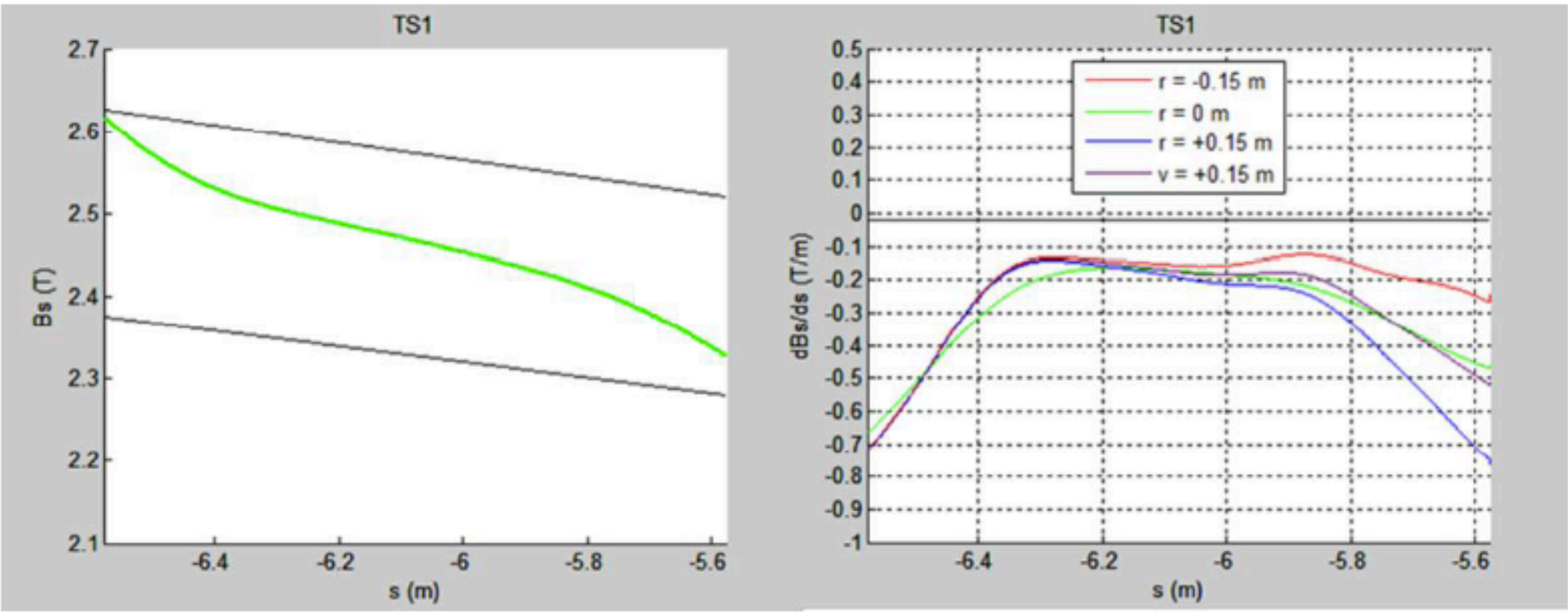

Figure 7.27. Axial field distribution at the center of TS1 (left). Axial gradient along TS1 (right).

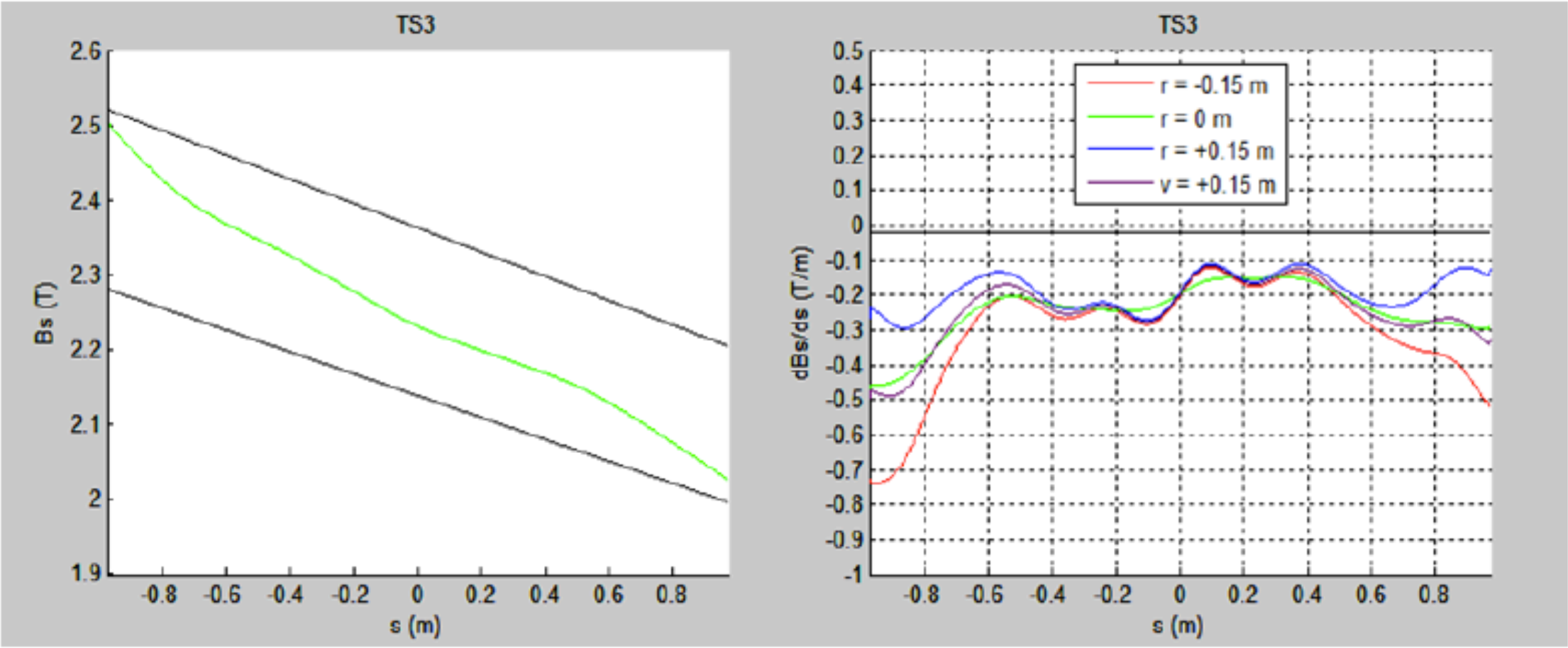

Figure 7.28. Axial field distribution at the center of TS3 (left). Axial gradient along TS3 (right).

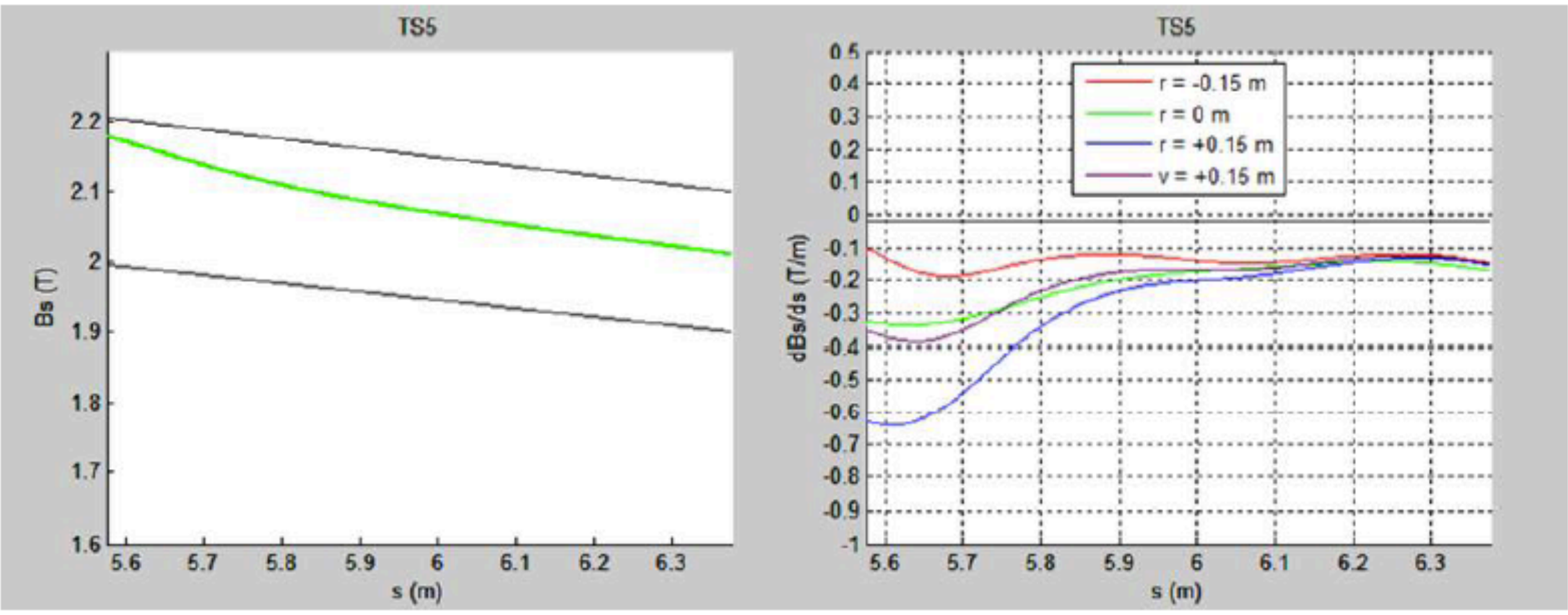

Figure 7.29. Axial field distribution at the center of TS5 (left). Axial gradient along TS5 (right).

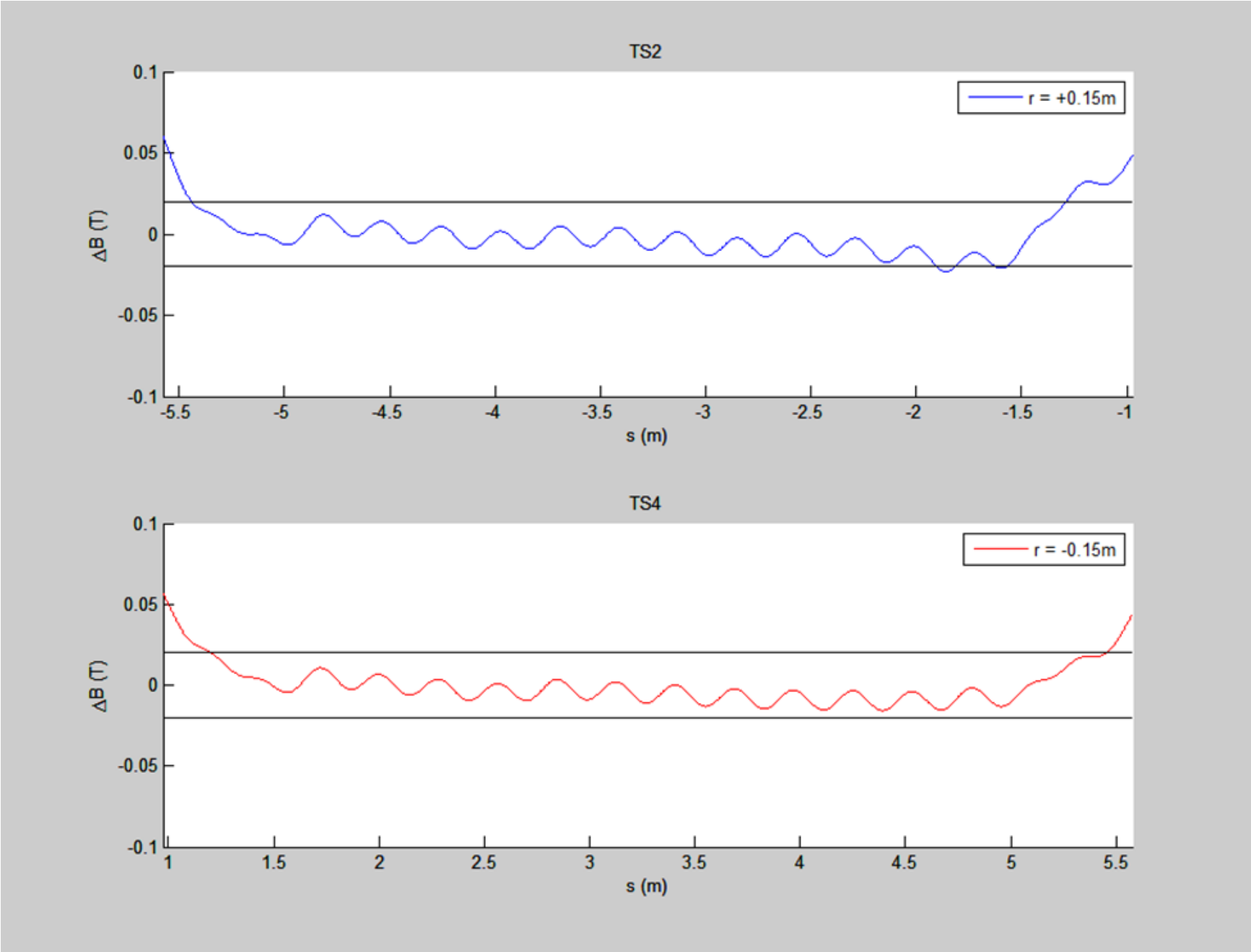

Figure 7.30. Ripple in the TS2 and TS4 curved sections.

### *TS Mechanical Analysis*

Several load cases were simulated in order to study the TS coil displacements, the stresses in the coils, the structure and support rods during normal operation and various failure scenarios. Large variation of the forces can be generated when the magnetic systems adjacent to TSu are powered off. These conditions will be avoided during commissioning and normal operation. Nonetheless, they may occur in the case of a quench or the failure of adjacent systems. The TSu structure and supports have been designed for normal operation with sufficient margin to withstand these failure modes.

After cooldown (Figure 7.31) and during excitation (Figure 7.32) the von Mises stress in the coils is less than 25 MPa (excluding stress concentrations points due to the mesh). Table 7.14 shows the stress in the supports during normal operation (all magnets at operating current) and in case of failure scenarios (PS off or TSd off, with all other magnets on). The radial supports see the highest tensile load during normal operation. The axial supports could see the highest tensile stresses or some compression if adjacent magnets fail. The supports are made of Inconel 718 (allowable stress when cold: 531 MPa), and are designed to withstand the full load of these failure modes.

Table 7.15 shows the von Mises stress (MPa) in the support structure after cooldown, during normal operation, and for failure modes. In all conditions the stresses are below the allowable stress for Al-5083 (107 MPa at 4 K).

| **Name** | **Function and Location** | **All On** | **PS Off** | **TSd Off** |
|---|---|---|---|---|
| Rs_1a | Radial at TSd end | **293** | 132 | 182 |
| Rs_1b | Radial at TSd end | **324** | 154 | 213 |
| Rs_2a | Radial at PS end | **319** | 189 | 145 |
| Rs_2b | Radial at PS end | **354** | 210 | 180 |
| As_1a1 | Axial at TSd end | 356 | **472** | -134 |
| As_1a2 | Axial at TSd end | 339 | **461** | -152 |
| As_1b1 | Axial at TSd end | 430 | **434** | -8 |
| As_1b2 | Axial at TSd end | 413 | **421** | -24 |
| As_2a1 | Axial at PS end | 346 | -95 | **446** |
| As_2a2 | Axial at PS end | 322 | -107 | **422** |
| As_2b1 | Axial at PS end | 374 | -6 | **399** |
| As_2b2 | Axial at PS end | 351 | -19 | **376** |

Table 7.14. Peak stress (MPa) in TSu supports during normal operation (All On) and for the failure modes discussed in the text. Numbers in bold show the highest load condition.

### *TSu Quench Protection and Analysis*

The quench protection strategy in the Transport Solenoid is based on extracting most of the energy to external dump resistors rather than relying on the quench to propagate through a series of small coils, which is inefficient and could damage the magnets. If the resistive voltage component exceeds the quench detection threshold of 0.5 V for more than 1 second, the current is shunted through a 0.34 Ohm dump resistor and the power supply is switched off. Symmetric grounding will be used to keep the maximum coil-to-ground voltage well below the requirement with 600 V at the leads when the dump resistance is inserted into the circuit. Symmetric grounding may be disconnected in case of a single-point coil-to-ground failure. Therefore, the ground insulation should be 2 mm thick and will be made of several layers (minimum 6 layers) of fiberglass cloth or G10. The layer-to-layer insulation should be 0.25 mm thick and will be made from 0.125 mm thick fiberglass tape with a 45% overlap.

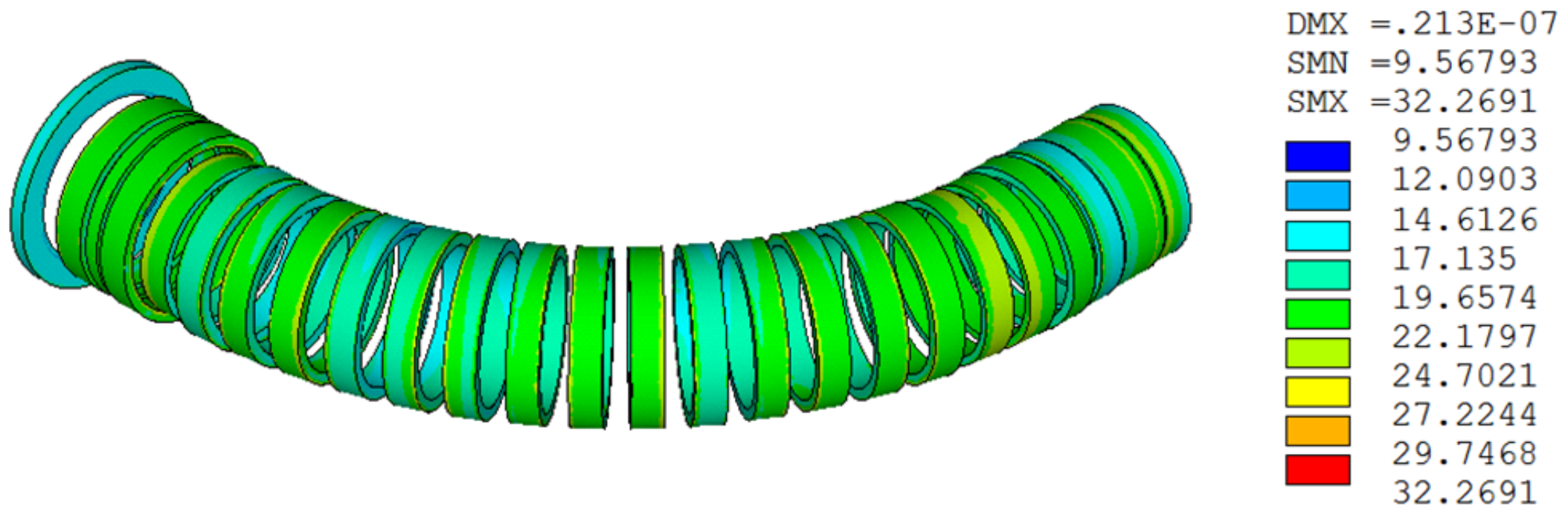

Figure 7.31. Equivalent stress (MPa) in the TS coils after cooldown.

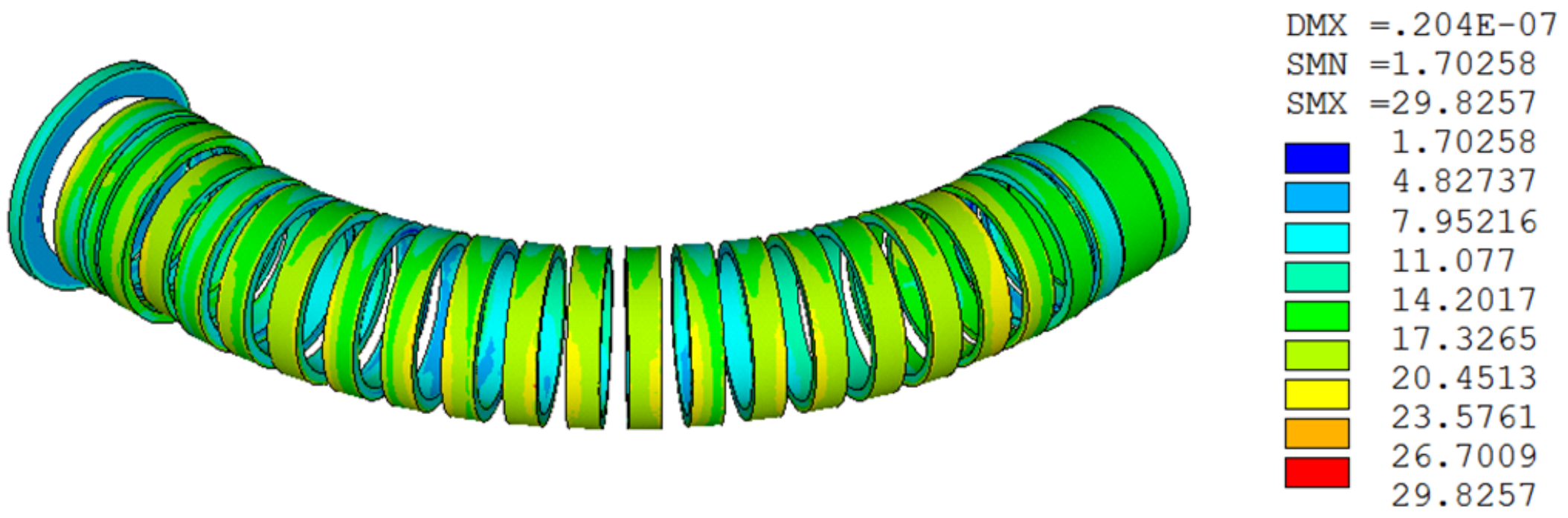

Figure 7.32. Equivalent stress (MPa) in the TS coils during excitation.

| Case | Stress |
|---|---|
| After cooldown | 40 MPa |
| All magnets at nominal current | 49 MPa |
| PS off | 57 MPa |
| TSd off | 44 MPa |

Table 7.15. Peak stress in the TSu structure during normal operation and for failure modes.

The main parameters of the TSu quench protection system are shown in Table 7.16. The TSu stored energy at nominal operating current (1730 A) is 7.1 MJ (with the adjacent magnets powered off). In this analysis, 10.4 MJ of stored energy was used to account for coupling with the adjacent magnets and to provide some margin. The inductance was scaled accordingly.

The hot spot temperature was computed using the numerical code QLASA [17] in the adiabatic condition, assuming that the quench propagates only in the quenching coil. The "High Field" column shows the results when the quench starts in the peak field area of the coil with the highest field (3.4 T). The "Low Field" column shows the results for the case when the quench starts in a low field (1.0 T) area of a coil with a low peak field (2.2 T). The hot spot temperature is well below 120 K (the maximum acceptable temperature) in both cases, even with these conservative assumptions.

| | Units | High Field | Low Field |
|---|---|---|---|
| Copper RRR | | 100 | |
| Aluminum RRR | | 200 | |
| Operating current | A | 1730 | |
| B max in quenching coil | T | **3.4** | **2.2** |
| B where quench starts | T | **3.4** | **1.0** |
| Energy (with coupling & margin) | MJ | 10.4 | |
| Inductance (with coupling & margin) | H | 6.95 | |
| Dump resistance | Ohm | 0.34 | |
| L/R | s | 20.4 | |
| Quench detection threshold | V | 0.5 | |
| Time above threshold before QPS activation | s | 1 | |
| Voltage at leads at dump insertion | V | 590 | |
| Hot spot temperature (QLASA) | K | **70** | **69** |
| $J_{eng}$ | A/mm$^2$ | 47 | |
| $I_{op}$ / $I_c$ (5.1 K) on load line | | 58% | |
| Temperature of heat generation (3.4 T, $I_{op}$) | K | 6.92 | |

Table 7.16. TSu quench protection parameters. The *High Field* column shows the results for a quench that starts in the peak field area of the coil with the highest field. The *Low Field* column shows the results for a quench that starts in a low field (1.0T) area of a coil with a low peak field.

### *7.3.3* Detector Solenoid

The main functions of the Detector Solenoid (DS) are to provide a graded field in the region of the stopping target and to provide a precision magnetic field in a volume large enough to house the tracker downstream of the stopping target. The inner diameter of the magnet cryostat is 1.9 m and its length is 10.75 m. The inner cryostat wall supports the stopping target, tracker, calorimeter and other equipment installed in the Detector Solenoid. This warm bore volume is under vacuum during operation.

It is sealed on one side by the muon beam stop, while it is open on the other side where it interfaces with the Transport Solenoid. The last section of the Transport Solenoid protrudes into the DS cryostat.

The Detector Solenoid is designed to satisfy the field and operational requirements defined in the DS requirements document [3]. The overall structure of the solenoid is shown in Figure 7.33. It consists of two sections: a “gradient section”, which is about 4 m long, and a “spectrometer section” of about 6 m. The magnetic field at the entrance of the gradient section is 2 T and decreases linearly to 1 T at the entry of the spectrometer section, where it is uniform over 5 m.

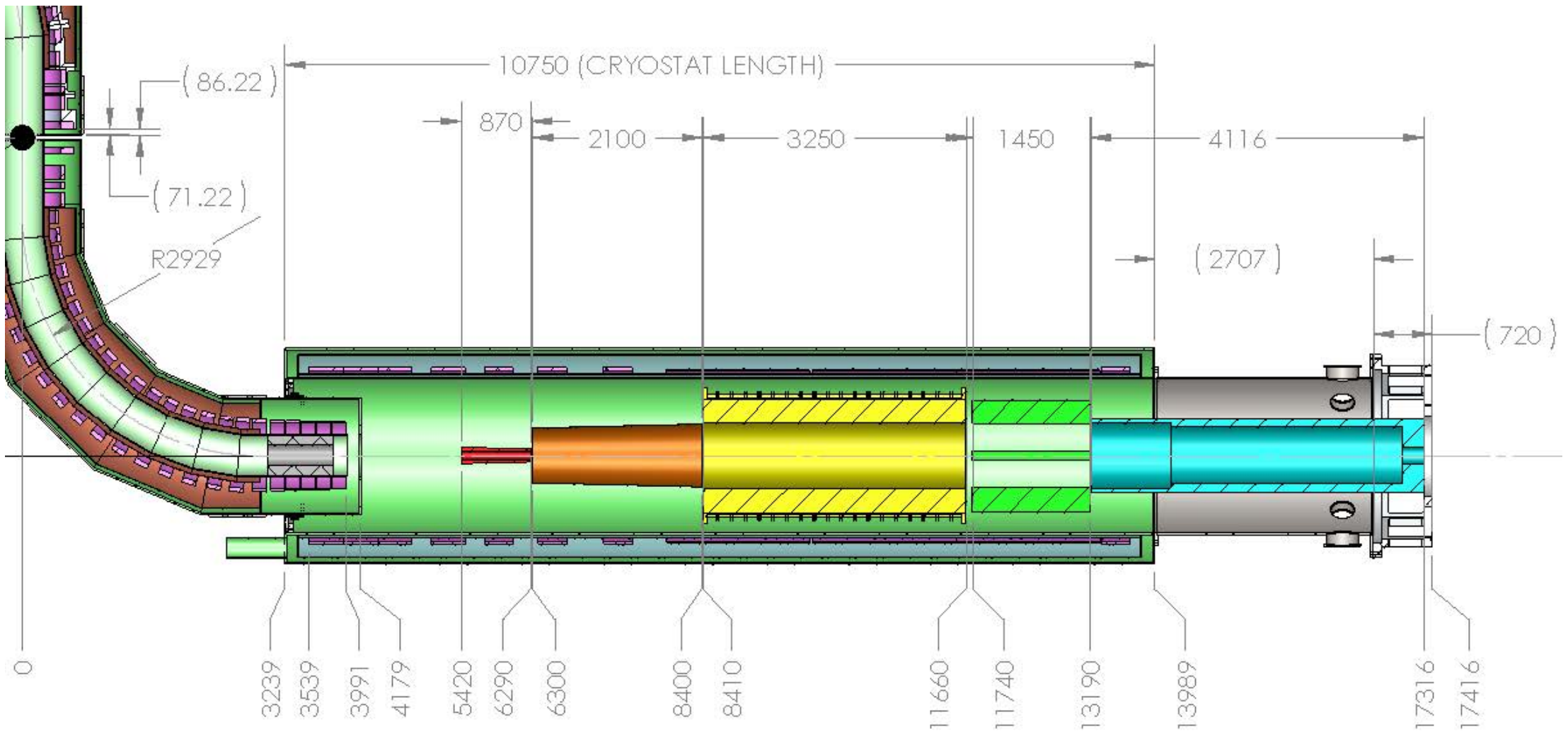

Figure 7.33. Overall structure of the Detector Solenoid coils and cryostat.

The Detector Solenoid coil design is based on a high purity aluminum sheath surrounding a NbTi Rutherford cable. This type of conductor has been used successfully in many similar superconducting detector solenoids. Aluminum has very small resistivity and a large thermal conductivity at low temperatures providing excellent stability. Furthermore, aluminum stabilized conductors can be extruded in lengths of several kilometers. Precise rectangular conductor shapes can be obtained, allowing for high accuracy in the coil winding.

Two types of conductor are required; both 20 mm high. The “narrow” (5.25 mm wide) conductor will be used in the DS gradient section, while the “wide” (7 mm wide) conductor is used in the spectrometer section. The dimensions are optimized to give the required field when identical current is transported in both conductors. The conductors contain Rutherford-type NbTi cables with 12 and 8 strands, respectively. The strands have a diameter of 1.3 mm, a SC/Cu ratio of 1, and a critical current of 2750 A/mm$^2$ (4.2 K, 5 T). As a result, the conductors have significant stability and safety margins in case of a quench.

In the baseline design, the gradient section is wound in two layers using the “narrow” conductor (20 mm × 5.25 mm), which is necessary to obtain a field of 2 T. The field gradient is obtained by introducing several sets of spacers between coil turns. The field uniformity in the spectrometer section is achieved with a “wide” conductor (20 mm × 7 mm), wound in a single layer coil.

It is envisaged that the DS coil will be wound in standardized modules on accurately machined collapsible mandrels. After curing, the winding mandrels are extracted and the outer aluminum support cylinders are placed over each module and the assembly epoxy bonded. The preassembled modules are then electrically connected and bolted together with spacers in a single cold mass before installation in the cryostat. The Detector Solenoid (cold mass and cryostat) weights about 39 tonnes. Other parameters of the magnet are summarized in Table 7.17.

| Parameter | Units | Value |
|---|---|---|
| **Coil** | | |
| Inner radius | mm | 1050 |
| Thickness | mm | 43 |
| Length | mm | 10,150 |
| Mass (cold mass) | tonnes | 8.8 |
| **Cryostat** | | |
| Inner diameter | mm | 1900 |
| Outer diameter | mm | 2656 |
| Length | mm | 10,750 |
| Mass | tonnes | 30 |

Table 7.17. Summary of the Detector Solenoid parameters.

***Coil Layout***

The DS coil layout, shown in Figure 7.34 consists of 11 modules, seven in the gradient section and four in the spectrometer section. Two module types used in the gradient section differ only in number of turns (that is active length) and are wound in two-layers. The modules are separated by pre-machined spacers to give the correct field profile. The spectrometer section contains three identical modules, wound as single layer coil. Finally, to provide a sharp field fall-off, a short two-layer module, identical to that used in the gradient section, is mounted on the far end of the spectrometer section. The parameters of the Detector Solenoid coils are listed in Table 7.18.

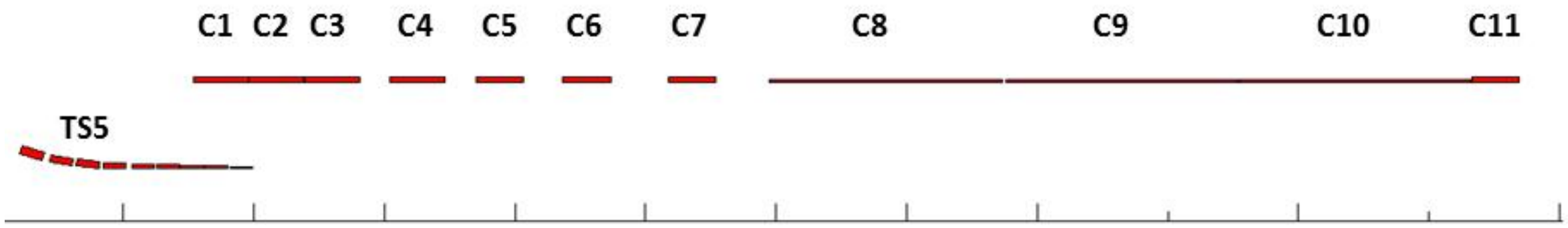

Figure 7.34. Layout of the Detector Solenoid coils. Coils segments are labeled C1-C11. See Table 7.18. TS5 coils are shown on the left.

| Segment Number | $Z_{min}$ (m) | Length (mm) | Inner radius (m) | Current (A) | Turns | Conductor (mm/mm) | Current density (A/mm$^2$) |
|---|---|---|---|---|---|---|---|
| 1 | 3.539 | 419.75 | 1.050 | 6114 | 2x73 | 20 × 5.25 | 58.23 |
| 2 | 3.964 | 419.75 | 1.050 | 6114 | 2x73 | 20 × 5.25 | 58.23 |
| 3 | 4.389 | 419.75 | 1.050 | 6114 | 2x73 | 20 × 5.25 | 58.23 |
| 4 | 5.042 | 419.75 | 1.050 | 6114 | 2x73 | 20 × 5.25 | 58.23 |
| 5 | 5.699 | 362.25 | 1.050 | 6114 | 2x63 | 20 × 5.25 | 58.23 |
| 6 | 6.369 | 362.25 | 1.050 | 6114 | 2x63 | 20 × 5.25 | 58.23 |
| 7 | 7.176 | 362.25 | 1.050 | 6114 | 2x63 | 20 × 5.25 | 58.23 |
| 8 | 7.949 | 1777.5 | 1.050 | 6114 | 1x237 | 20 × 7.00 | 43.67 |
| 9 | 9.761 | 1777.5 | 1.050 | 6114 | 1x237 | 20 × 7.00 | 43.67 |
| 10 | 11.544 | 1777.5 | 1.050 | 6114 | 1x237 | 20 × 7.00 | 43.67 |
| 11 | 13.326 | 362.25 | 1.050 | 6114 | 2x63 | 20 × 5.25 | 58.23 |

Table 7.18: Parameters of the Detector Solenoid coil segments.

The two-layer modules are bonded with epoxy to the inner surface of a 5083-0 aluminum support cylinder. The thicknesses of the support cylinders range from 1-3 cm and are chosen so that they can resist the full stress generated by the magnetic pressure. The cylinder also provides a load path to the cold mass support system. This aluminum grade is non-heat-treatable and can be reliably welded without loss of strength. It is also one of only three materials for which the ASME Code, Section VIII, Div. 1, allows higher working stresses at cryogenic temperatures. The layout of the coil modules, spacers and support cylinder is shown in Figure 7.35.

***Cryostat Design***

The Detector Solenoid cold mass is held within the cryostat by radial and axial support systems, as shown in Figure 7.36. The radial support system consists of 8 pairs of tangentially opposed Inconel 718 rods, 4 pairs at each end. The rods are 66 cm long, and 12.7 mm in diameter. At the ends of each rod there is a spherical bearing to accommodate motion from thermal contraction. The axial support system consists of eight 1.25" Schedule 10 Inconel 718 rods, located on the downstream end only. The axial forces are in all cases directed upstream where the DS interfaces with the TS and the axial support system is always in tension. The axial supports use spherical bearings at the rod ends to accommodate the radial thermal contraction of the cold mass. The warm ends of the axial supports connect directly to the cryostat outer shell.

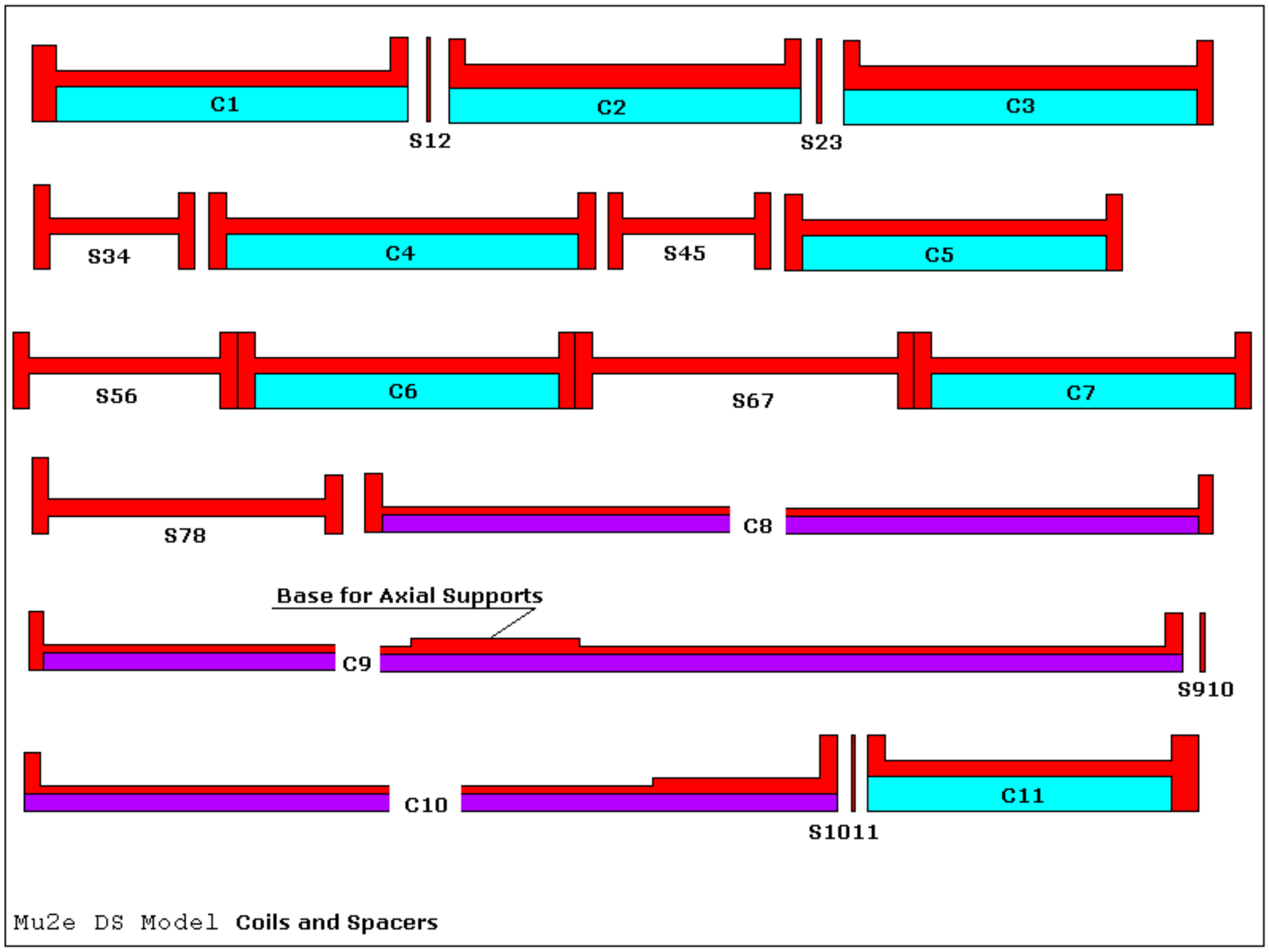

Figure 7.35. Layout of the Detector Solenoid coil modules, spacers and support cylinder.

The cryostat consists of concentric 2 cm thick stainless steel cylindrical shells connected by annular end rings. The shells are sized according to ASME Section VIII, Div. 1 rules [23] for cylindrical shells under external pressure. Given that the bore of the solenoid may be evacuated while the magnet is warm (at room temperature and pressure), both the inner and outer shells have a design external pressure of 1 atmosphere. The cryostat sits on two saddles, positioned very close to the ends of the vessel, also shown in Figure 7.36.

The relative permeability of the candidate 300-series (austenitic) stainless steels can range up to 1.02 at 200 Oersteds. A 2-D finite element analysis shows that the effect of this permeability on the solenoid field is about 0.01% [23].

The cryostat provides the load path for cold mass reactions (weight and magnetic force) through its support system. The axial supports bear directly against the cryostat outer shell, and transmit the forces to the saddle support. This arrangement essentially produces no stresses on the cryostat. The warm ends of the radial supports attach to the cryostat through towers, which transmit the load through the outer shell to the saddles.

The bore of the cryostat must accommodate approximately 10 tonnes of detectors, shielding and other equipment. This load rests on rails attached to the inside of the inner vacuum vessel [25].

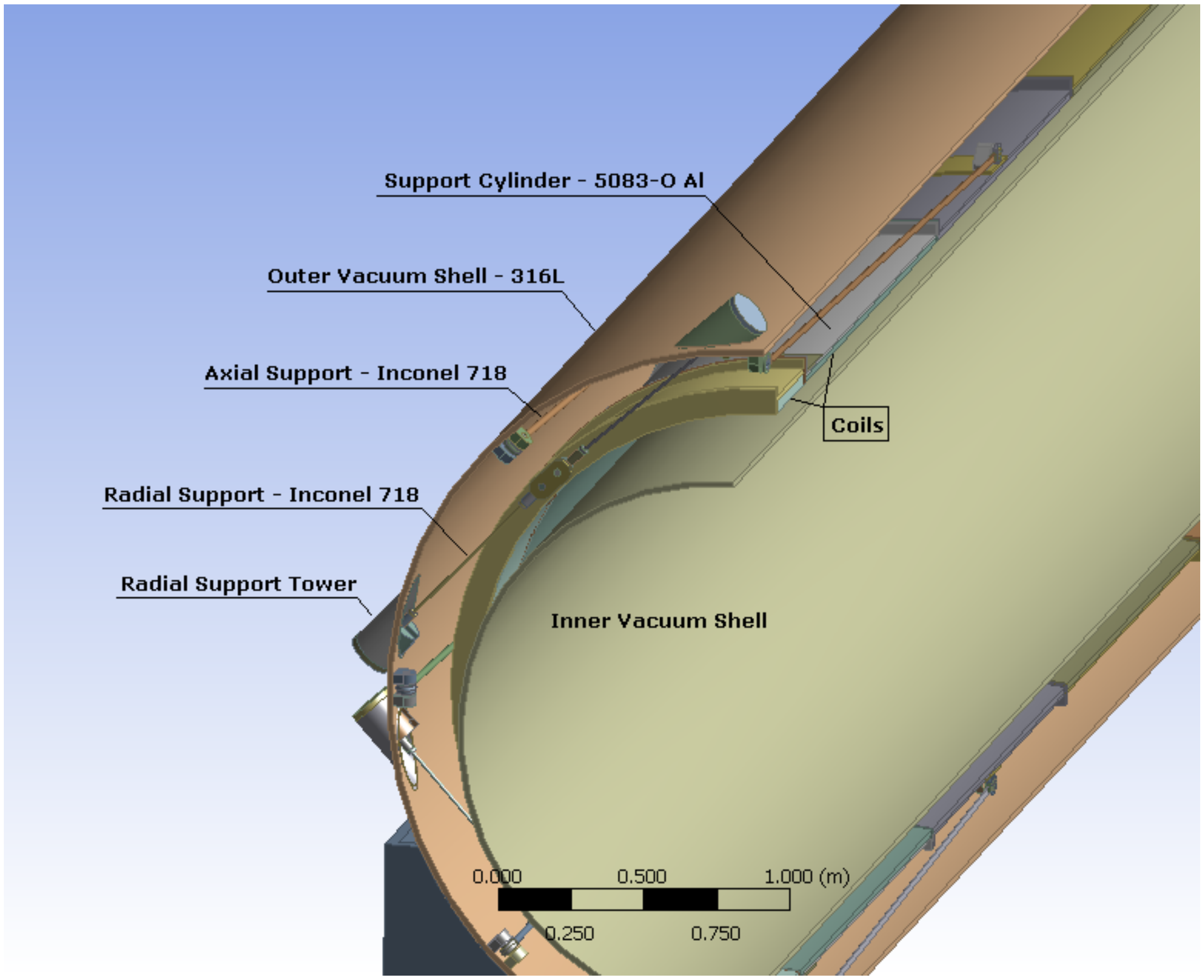

Figure 7.36. Radial and axial supports for the cold mass at the downstream end of the DS magnet. Also shown are the support saddles for the cryostat.

### *Magnetic Analysis*

For field characterization, the Detector Solenoid can be divided into four axial sections: DS1-DS4. Note that the coils upstream of DS2 have a large impact on the field quality of the Transport Solenoid. At the same time, there is a contribution from the Transport Solenoid that must be accounted for in the DS magnet design. Using a 3-D finite element model it was possible to determine the distribution of the conductors and spacers in both sections of the DS coil to satisfy the requirements given in reference [1] in the presence of field contributions from the TS. This is shown pictorially in Figure 7.37.

The peak field in the Detector Solenoid coils is 2.2 T and occurs in the first module of the gradient section. At the operating current of 6.1 kA, the magnet is at 45% of conductor quench current, with a temperature margin of 2.5 K with respect to the assumed conductor temperature of 5.0 K. The field in other segments of the gradient section is lower with a consequently larger temperature margin.

Figure 7.38 and Figure 7.39 compare the calculated field with the field requirements in the spectrometer and gradient sections. As shown, the field on axis meets the field tolerance specifications. In the spectrometer section, the field at the outer radius may need to be adjusted slightly at the axial extremes.

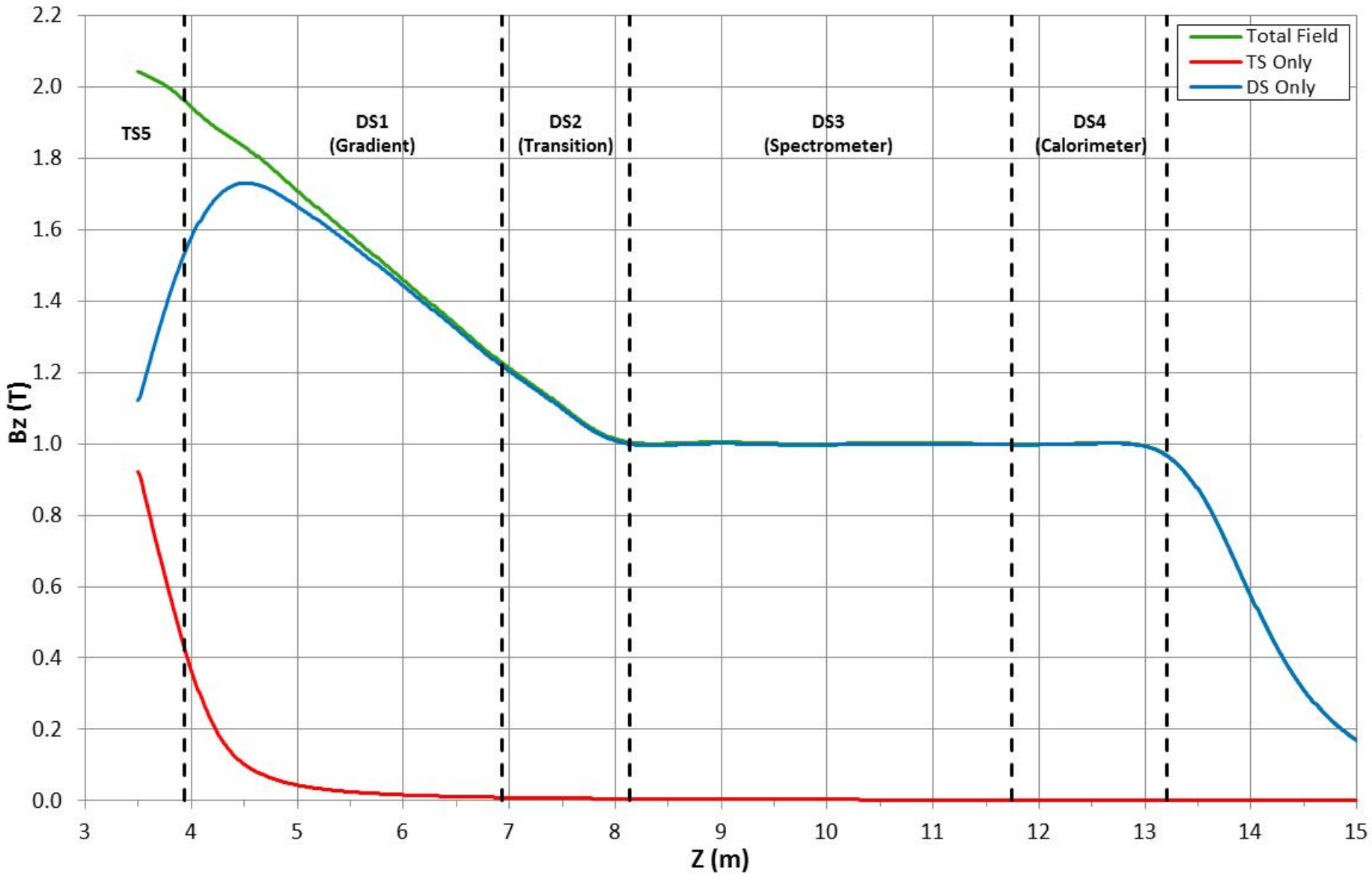

Figure 7.37. Magnetic field on the central axis of the Detector Solenoid. As shown, there is an important contribution to the gradient section form adjacent TS coils.

### *Mechanical Analysis*

The primary direction of Lorentz forces in a solenoid is radially outward, tending to dilate the coils. Flux also bends around the magnet ends, producing additional forces that tend to compress the solenoid axially. If these radial or axial forces are imbalanced, due to non-uniform field and current density or a non-symmetric positioning of the magnet relative to external fields, net forces result which must be reacted by the cold mass support system.

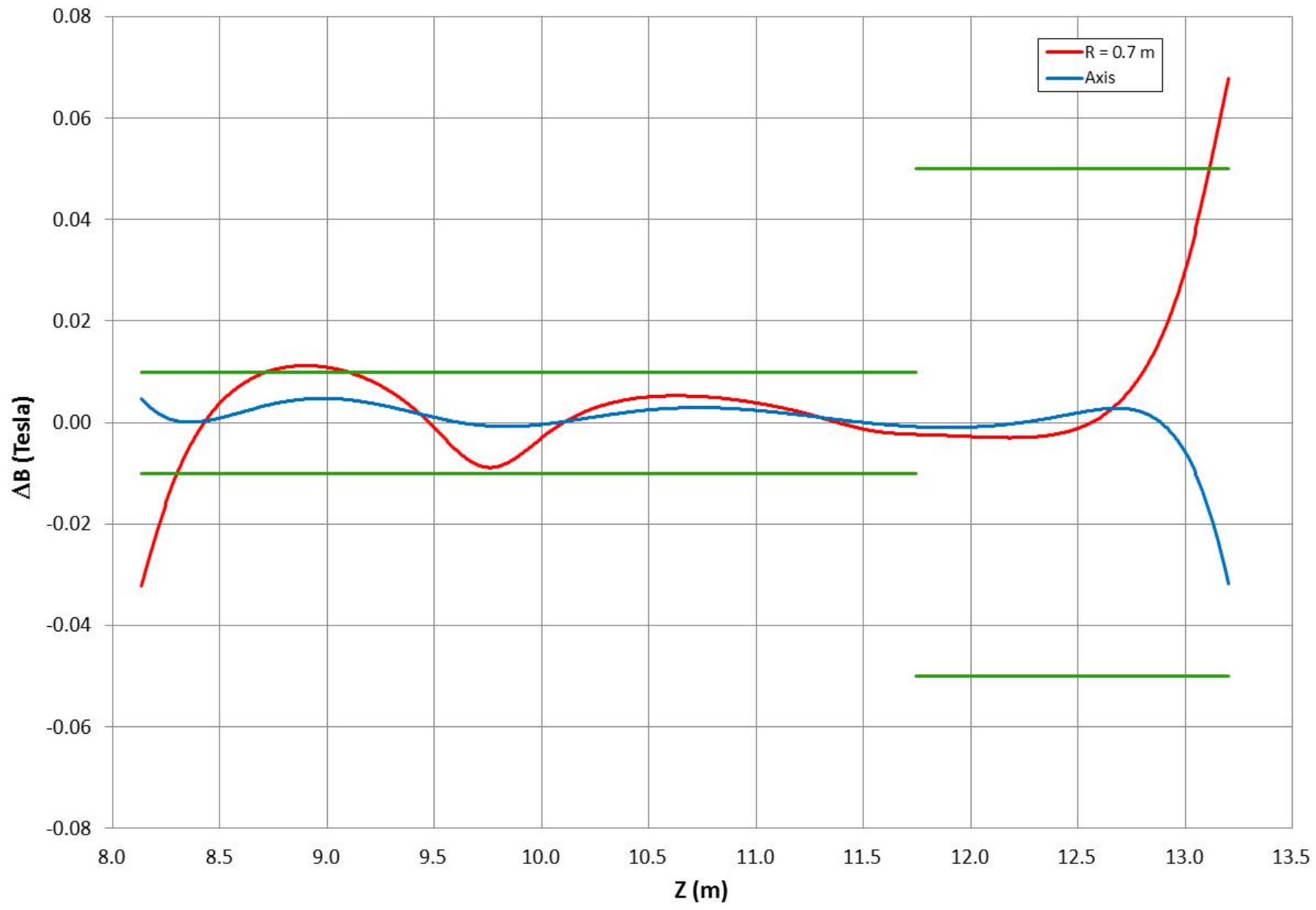

Figure 7.38 Comparison of the magnetic field with the field requirements in the DS spectrometer section (DS3-4 Uniform). Field requirements from Table 7.2 are shown in green. ΔB is relative to a uniform field of 1.0 T on axis (blue) and at a radial distance of 0.7 m (red).

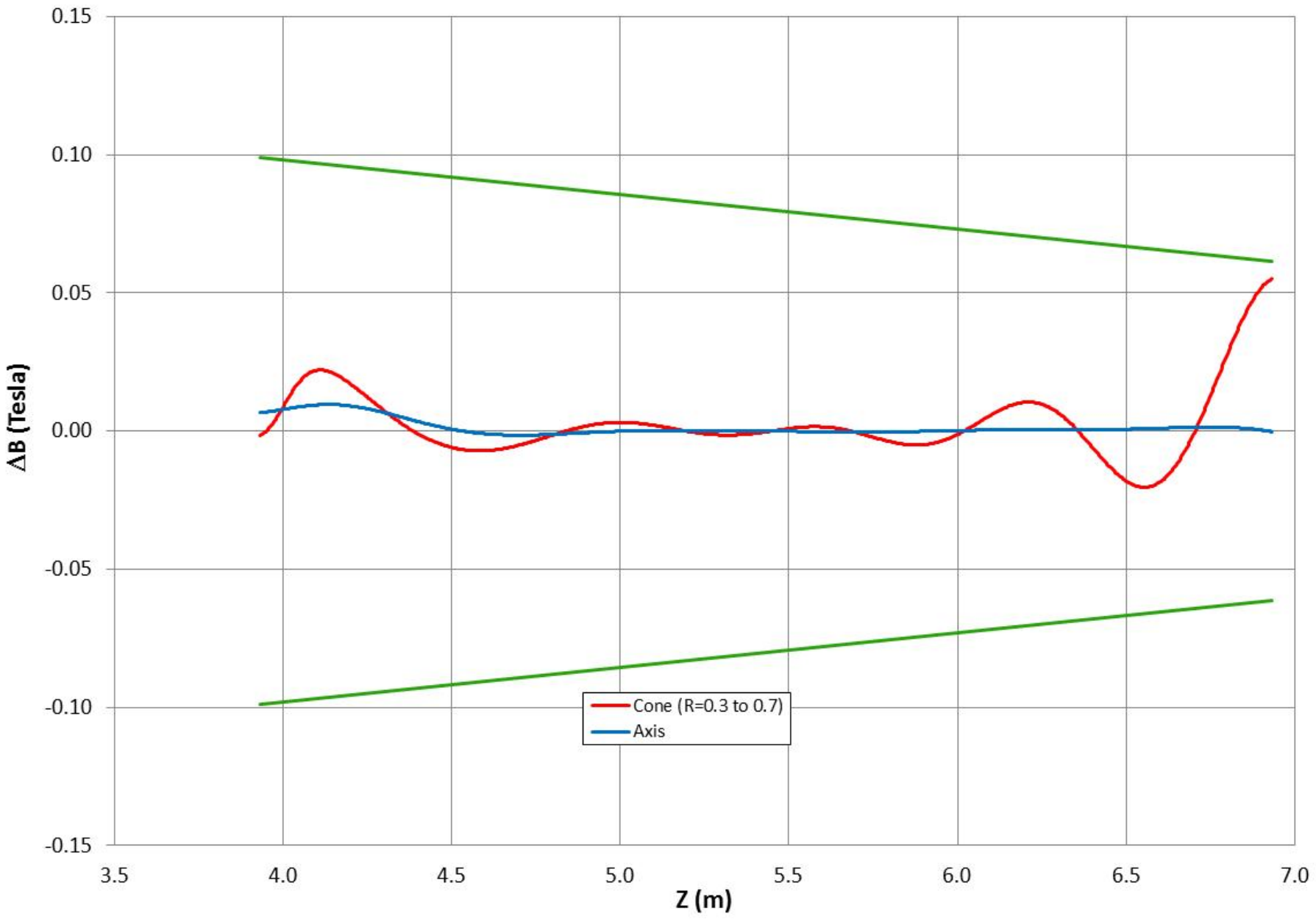

Figure 7.39. Comparison of the magnetic field with the field requirements in the DS gradient section (DS1 Gradient). Field requirements from Table 7.2 are shown in green. ΔB is relative to uniform gradient of -0.25 T/m and a field value of 1. 5 T at the stopping target on axis (blue); on a radial cone from 0.3m to 0.7 m starting at the upstream end of DS1 section (red).

The asymmetry of the Detector Solenoid inherently forces the DS toward the TS. However, this axial force will vary depending on how accurately the solenoid is positioned relative to the TS. A radial centering force is associated with misalignment of the Detector Solenoid with respect to the axis of the Transport Solenoid. The radial centering forces reacted by the radial support system are not divided equally between the upstream and downstream ends of the magnet. The upstream end reacts approximately 80% of the total radial centering force.

Axial and radial centering forces were calculated with a 3-D finite element model that included the Transport Solenoid and the Detector Solenoid. The assumption was made that the misalignments would not exceed ± 2 cm. It should be noted that the radial misalignments studied were uniform along the length of the magnet; rotation of the solenoid was not considered.

A summary of the forces used for the design of the cold mass support is given in Table 7.19. The axial force is at a maximum when the TS is operating, and equals 100 tonnes. The radial centering force with the TS off can reach 1 tonne for an installation 2 cm from the nominally centered position.

| **Description** | **Calculated Force (Tons / kN)** |
|---|---|
| Axial Force – 0 cm offset | 98 / 960 |
| Axial force – 2 cm offset | 100 / 980 |
| Dead Weight | 8.8 / 86 |
| Radial Centering Force | 1 / 9.8 |

Table 7.19: Summary of forces used for the design of the DS cold mass support system.

The equivalent stress in the coil and its support cylinder at various loading stages is shown in Figure 7.40 and Figure 7.41. The maximum equivalent stress in the conductor is 35 MPa, while it is 42 MPa in the support cylinder. Both maxima occur in the first segment of the gradient section. The stresses in the remainder of the magnet are much lower. The maximum allowable working stress of 5083-0 aluminum at 77 K is 107 MPa, so the support cylinder provides structural integrity of the coil even in the case of full load.

***Quench Analysis***

The Detector Solenoid must be protected in case of a quench, whether occurring in the body or in the leads of the magnet, and from any malfunctions of the cooling and powering systems. The Detector Solenoid will be powered with a dedicated power convertor. The circuit will contain protection elements for the solenoids (circuit breakers and extraction resistors) as well as for ancillary equipment (current leads, bus bars, grounding circuits) as shown in section 7.3.5. The normal powering and discharge of the circuit will be performed with ramp rates that pose negligible risk of quench. In case of minor faults, a slow discharge of the circuit will be launched without provoking a magnet quench. Only in case of a major fault shall the fast discharge be engaged and the magnet energy extracted quickly in a controlled way. This mode of protection relies on extraction resistors, dimensioned such that the peak voltage to ground is less than 500 V.

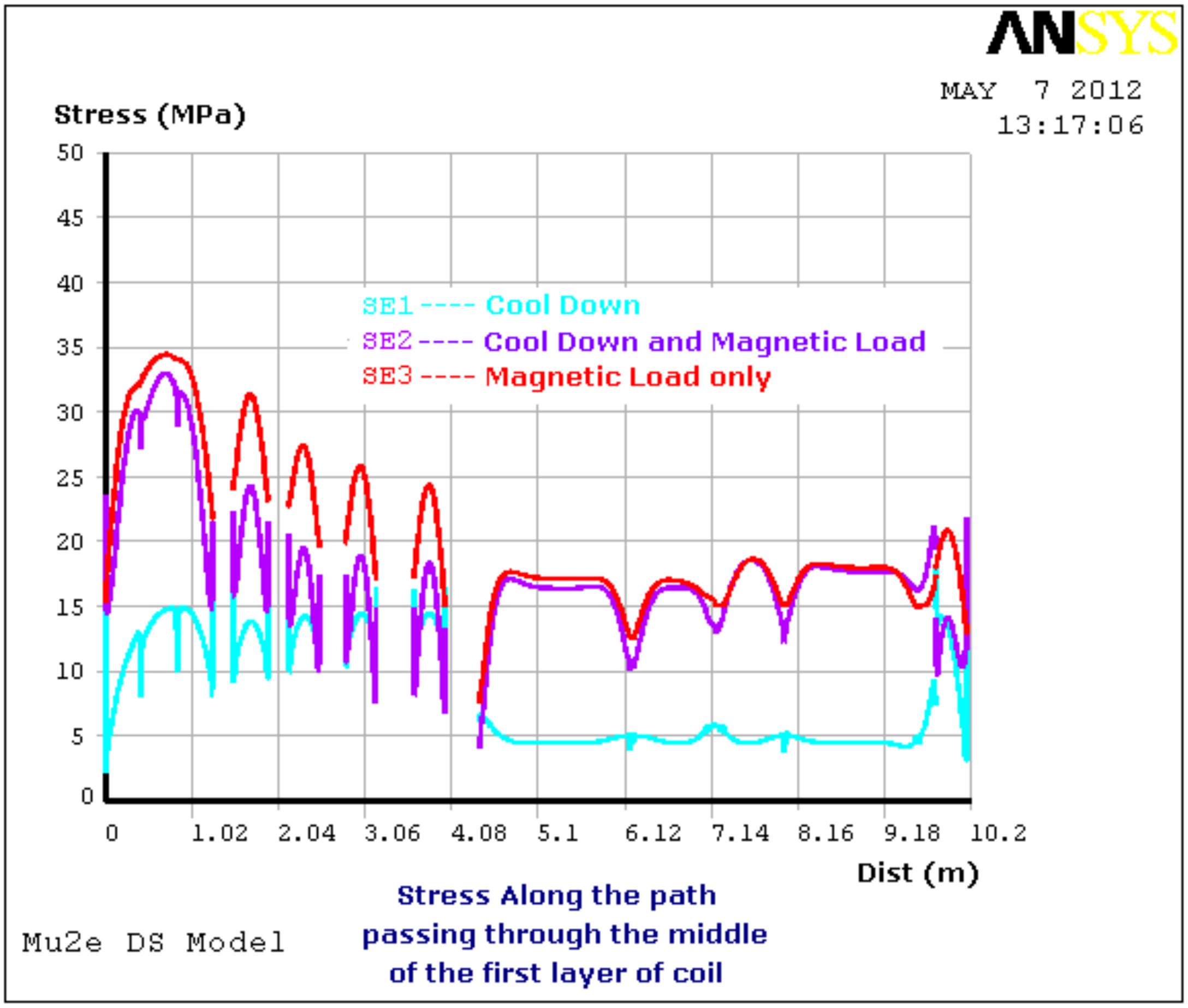

Figure 7.40. Equivalent stress inside the first DS coil layer for the three load cases; cooldown, powering, and combined cooldown and powering.

An analysis of the quench propagation using the QLASA code [17] in the adiabatic regime indicates that for a wide range of parameters (starting quench positions and different RRR of the Al-stabilizer) the peak temperature in the coil remains below 85 K. As the Detector Solenoid design relies on Al-stabilized conductor and an Al-support cylinder, it is expected that the protection properties of the magnet will be strongly enhanced by quench-back. A quench simulation model based on ANSYS was developed to study this case [22]. The model includes all material and dimensional features of the conductors, and the field and eddy current distribution in the coil and its support cylinder. This model shows that when quench-back is suppressed, the maximum temperature in the coil is 80 K, very close to the adiabatic result. In case of quench-back, initiated by discharging the magnet through a dump resistor, all sections of the magnet are resistive within a few seconds after the quench begins. The maximum temperature in the coil in that case is 30 K. These results indicate that the DS magnet can be safely protected whatever the origin of a quench.

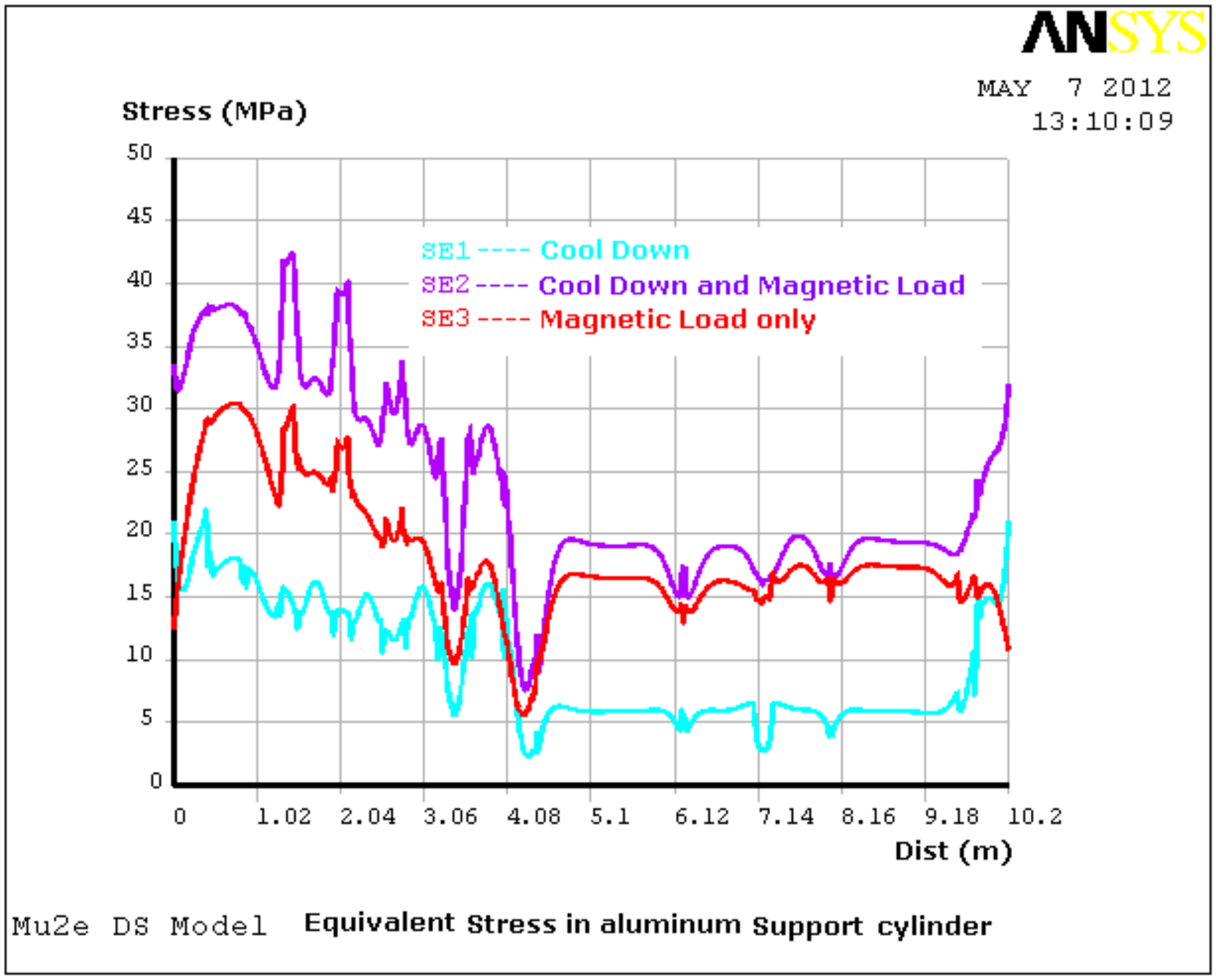

Figure 7.41. Equivalent stress in the DS support cylinder for the three load cases: cooldown (SE1), powering (SE2), and the combined cooldown + powering (SE3)

### *Thermal Analysis*

The Detector Solenoid cold mass is indirectly cooled by liquid helium flowing in tubes welded to the outer surface of the support cylinders. The liquid helium will be supplied by either a forced-flow or thermosiphon system. At present, the analysis has been completed for the case of forced-flow cooling with a mass flow rate of 50 g/s and a pressure of 0.23 MPa. Under these conditions and given the expected heat load of 20 W from mechanical supports and radiation, the temperature of the liquid helium leaving the magnet will be 0.1 K above the entry temperature. Under these conditions the maximum coil temperature rise is 0.49 K and occurs in the region of the axial supports. Due to the superior performance associated with thermosiphon cooling, further work will be focused on this option.

### *7.3.4* Cryogenic Distribution

#### *Introduction*

The superconducting solenoids require a cryogenic distribution system and supporting cryoplant for liquid helium and liquid nitrogen. The requirements for this system are described in Section 7.2. The scheme is to divide the solenoids into 4 semi-autonomous cryostats. Cryostats can be cooled down or warmed up independent of the state of the other cryostats. Each cryostat will require 4.5 K liquid helium as well as 80 - 90 K liquid nitrogen for the cryostat thermal shields. This system is shown in block diagram form in Figure 7.42 and described in detail in the sections below.

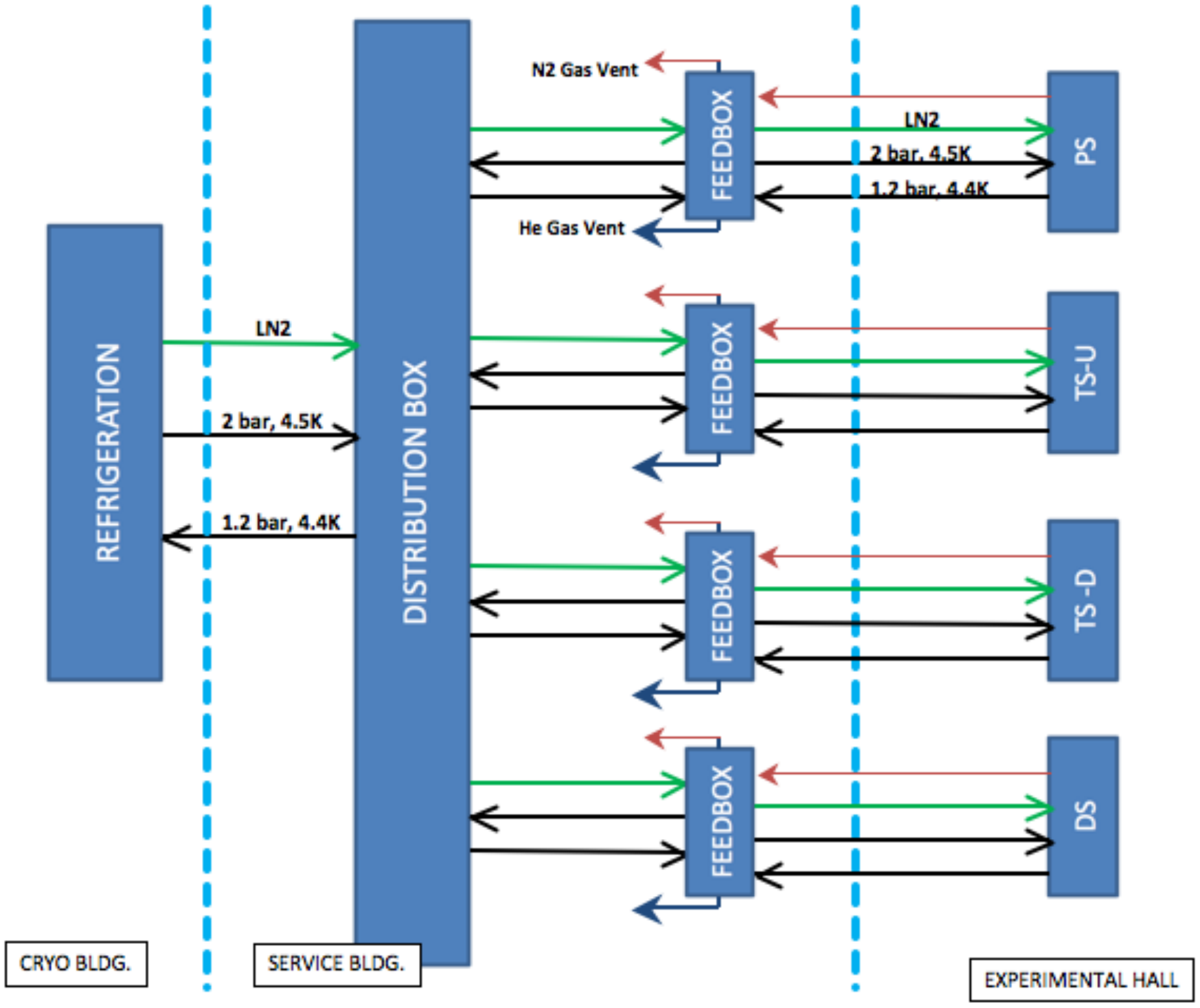

Figure 7.42. Block Diagram for the Mu2e Cryogenic System.

#### *Cryoplant description*

While the cryoplant is beyond the scope of this project, it is an important interface to the cryogenic distribution system. Thus, a brief description of the system is provided here. Refrigerators and refrigerator components from the Tevatron will be recycled and refurbished to provide liquid helium for the solenoids. As shown below, one satellite refrigerator appears to be sufficient for steady state operation with a second satellite for redundancy (hot spare) as well as added capacity for cool-down and quench recovery. The Refrigeration vs. Liquefaction curve for a satellite refrigerator is shown in Figure 7.43. As shown, a steady state operation of 450 Watts with 1.5 g/s liquefaction can be comfortably achieved with one refrigerator.

A separate building will be dedicated to the refrigerator and support equipment for Mu2e and other Muon Campus experiments. Three modified satellite refrigerators will be installed to support operations of the cryogenic components. To sustain reliable operation, these refrigerators will be modified to have redundant sets of dry and wet expanders. New valve boxes to interface refrigerators, expanders and the distribution system must be built.

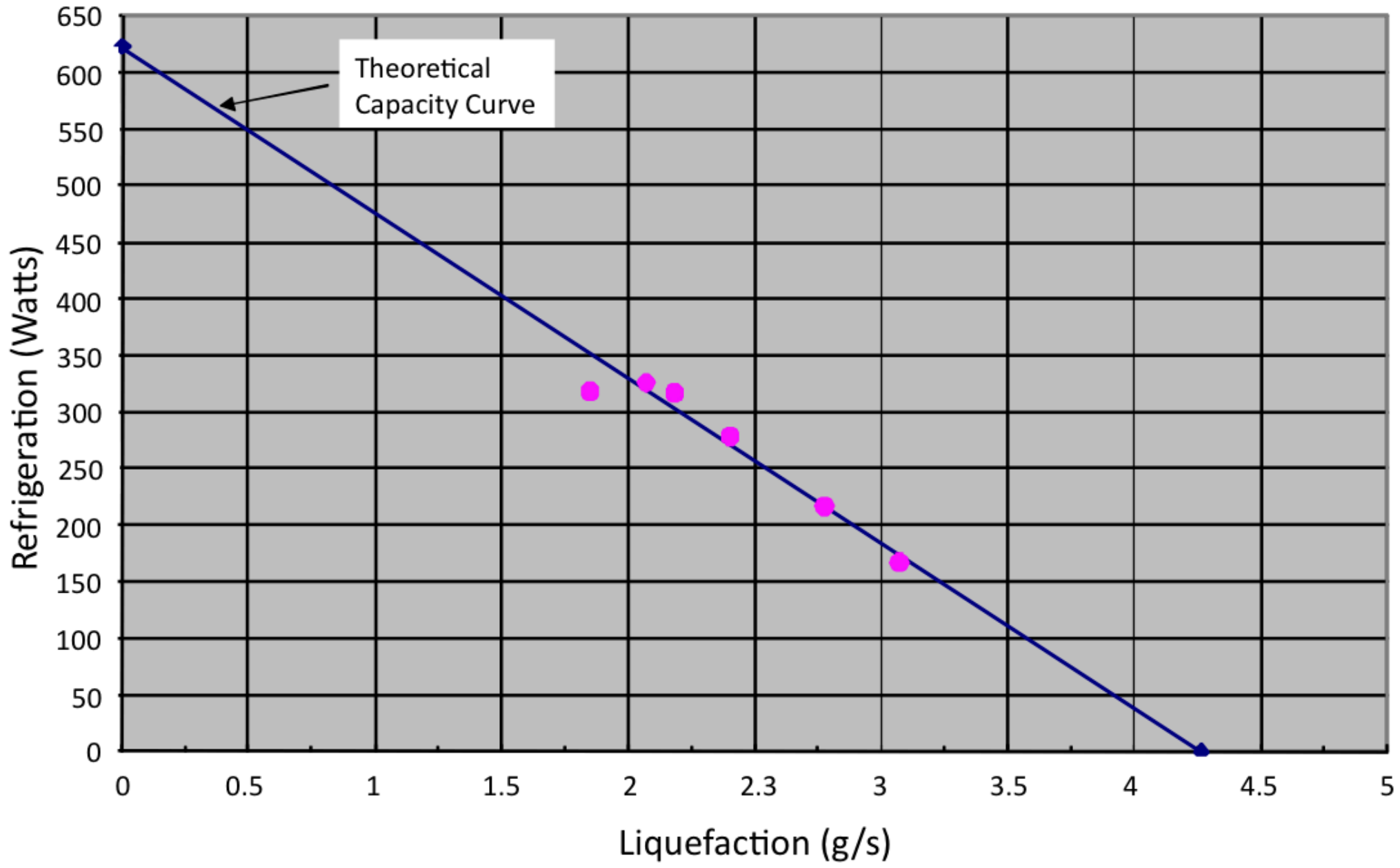

Figure 7.43. Satellite Refrigerator Stand Alone Capacity.

***Cryogenic Distribution System***

The cryogenic distribution box will be located between the feedboxes and the refrigerator building, shown schematically in Figure 7.42. The distribution box will contain electronic cryovalves for controlling the flow of liquid helium to the individual cryostats. In this way, magnets can be individually warmed up or cooled down, as required. The distribution box will require liquid nitrogen to cool the 80 K helium lines for the cryostat thermal shields. Figure 7.44 is a model showing the cryogenic distribution system. The feedboxes will be located in a room in the above grade detector service building. From the feedboxes, cryogenic distribution lines will run to each cryostat. A chase will be designed for this above-grade to below-grade transition to minimize the line of sight for radiation from the detector. The horizontal runs will be located near the ceiling of the below-grade detector hall. The locations of the cryostat penetrations will depend on the individual cryostat, but will be chosen to avoid interference with mechanical supports or required radiation shielding.

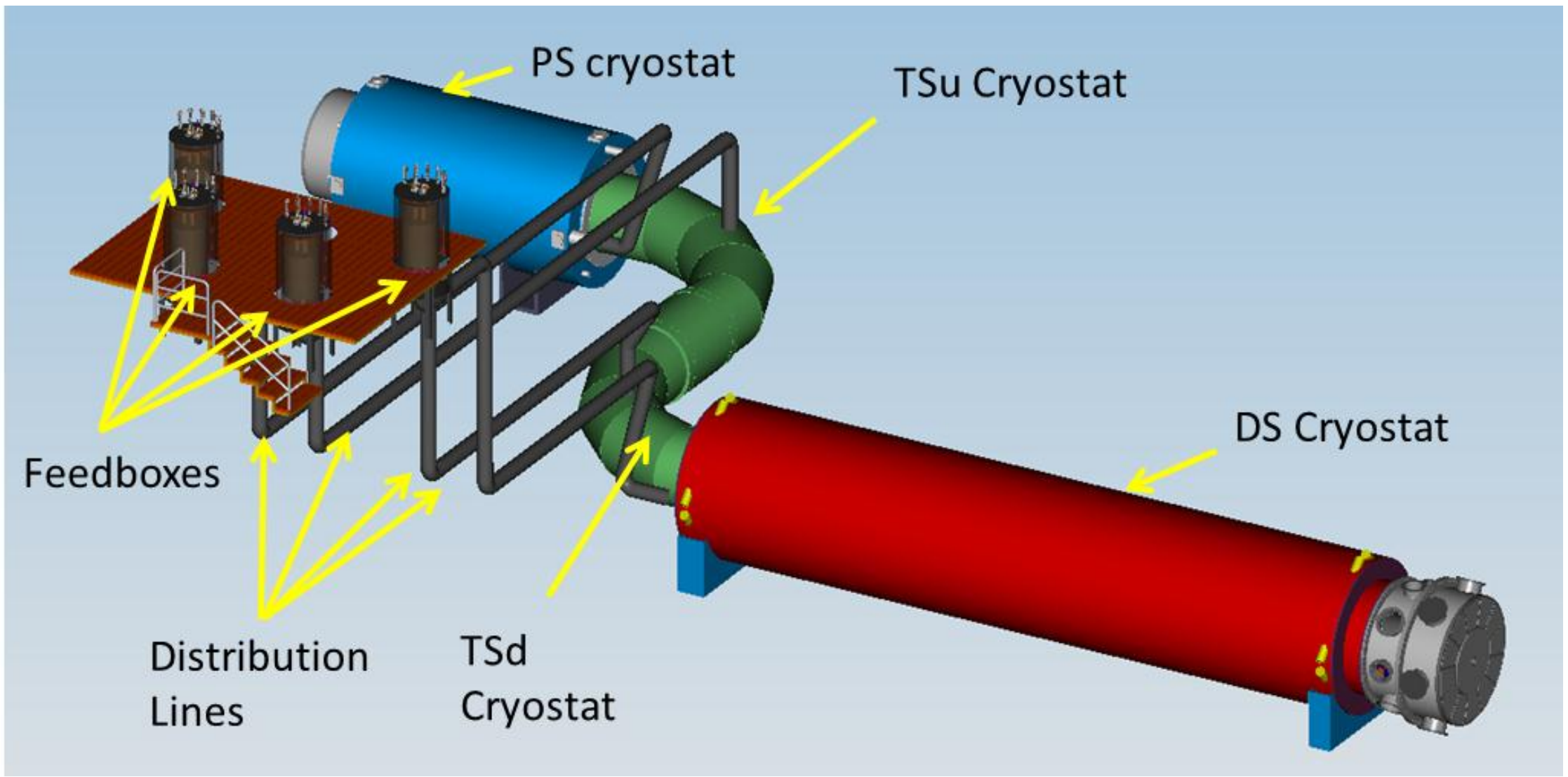

Figure 7.44. Layout of Cryogenic Distribution System.

***Feedbox Distribution***

The cryo distribution feedbox is modeled after previous designs of similar systems. Aside from the local cryogenic distribution, it serves as the cryo-to-room-temperature interface for magnet power supplies and instrumentation for thermal and quench systems. A feedbox schematic is shown in Figure 7.45. Note that this feedbox is configured for thermal siphoning conduction cooling. With a simple modification the box can be reconfigured for forced flow application.

It is envisioned that recycled HTS leads from the Tevatron will be used, at least for the high current Detector and Production Solenoids. It has been demonstrated that these leads are capable of 10 kA DC operation [26]. Additional studies are needed to validate their application with the large Mu2e solenoids with significant time constants and stored energy. Liquid nitrogen will be required to cool the HTS section of the leads.

***Cryogenic distribution lines***

The cross section for the cryogenic distribution line is shown in Figure 7.46. The line contains two 0.625" lines for liquid helium. These lines, while supplying helium to the cryostat, will conductively cool the magnet supply and return electrical bus.

Additional lines are shown for the 4 K return and the 80 K for the cryostat shield. The distribution lines will be vacuum jacketed to minimize the heat load. Each line will be approximately 20 meters in length.

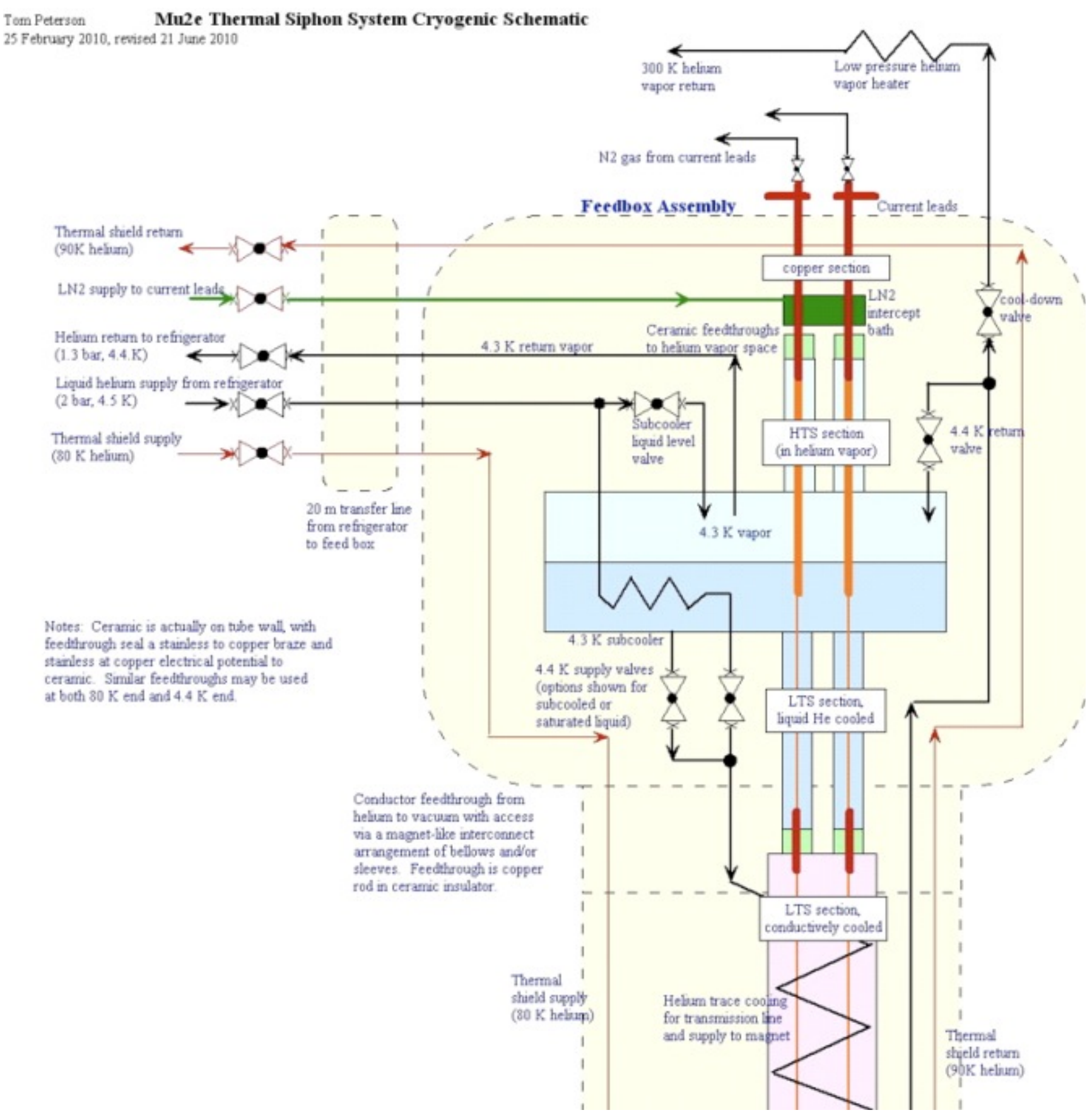

Figure 7.45 Schematic of Feedbox with thermal siphoning.

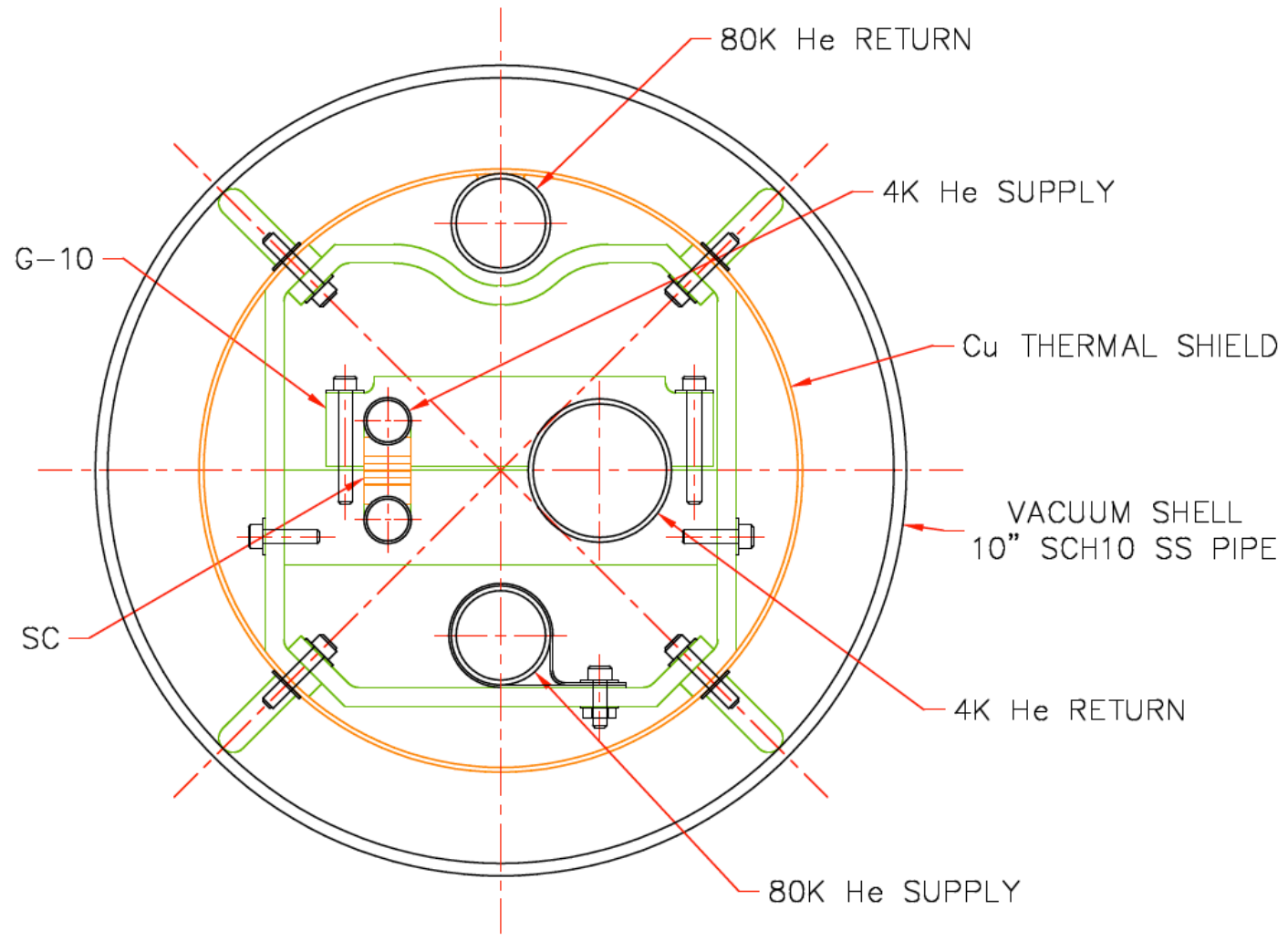

Figure 7.46 Cross Section of cryogenic distribution line.

***Cryogenic analysis of Magnets***

Studies have been performed on each of the magnet subsystems (PS, TS and DS) to evaluate the basic cooling schemes (thermal siphoning vs. forces flow) as well as the geometry, sizing of pipes, etc. Estimates of the static and dynamic heat loads and cool-down times were obtained from these studies. Details of these studies are reported in the reference design reports but summarized here.

As an example, Figure 7.47 shows a schematic of a thermal siphoning piping scheme for the Production Solenoid. The piping on the upper right is a continuation of the piping shown on the bottom right of Figure 7.45. Note that the liquid supply control valves shown are not physically located on the cryostat. They must be placed in a low radiation area so they can be serviced in the event of a failure.

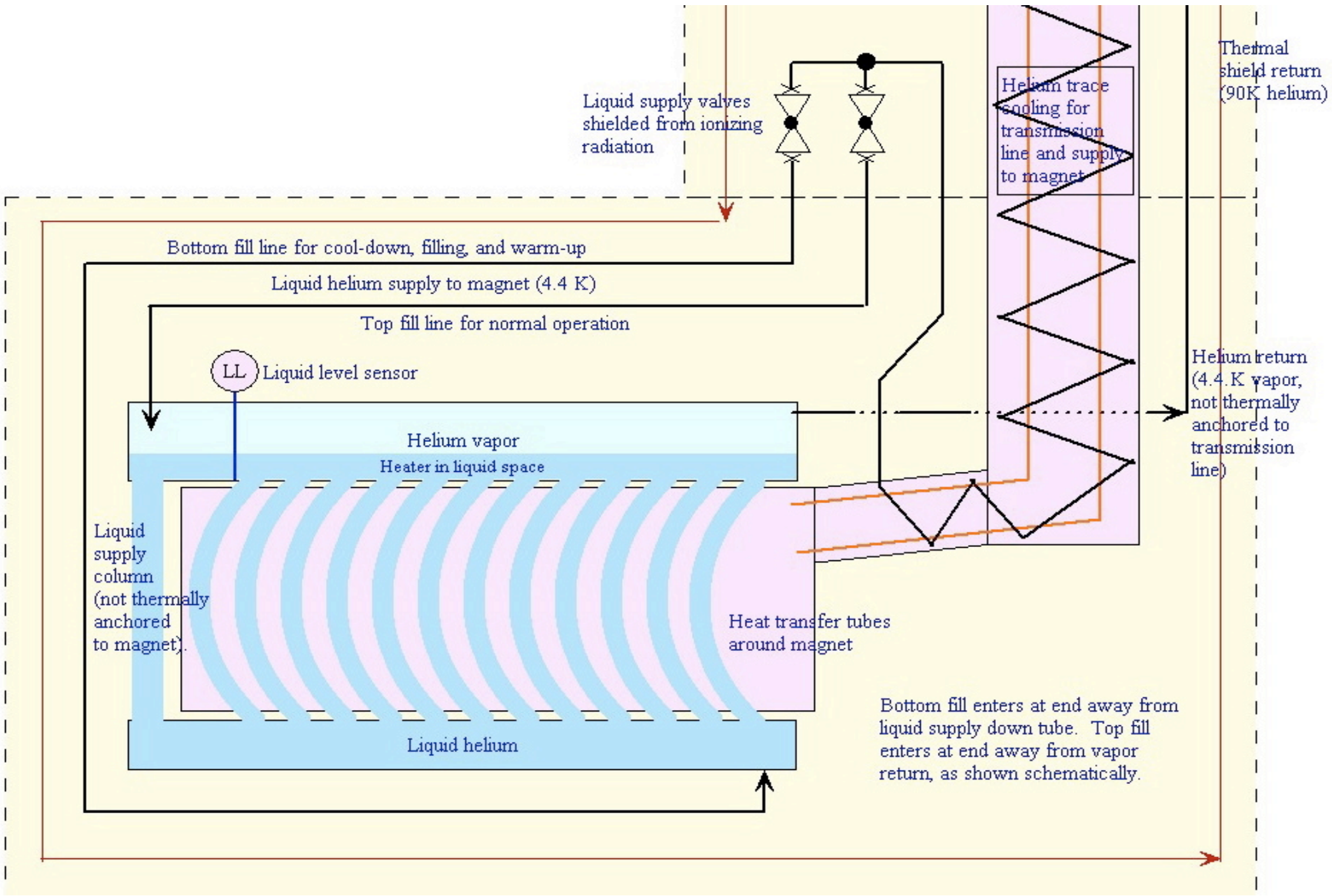

Figure 7.47 Thermalsiphon magnet cooling scheme for the Production Solenoid.

Table 7.20 and Table 7.21 include estimates of the heat loads and liquid helium requirements for the solenoid system. In this study, it is assumed that the PS will utilize thermal siphoning while the TS and DS will use a forced flow system. As shown, the total heat load to 4.5 K is 436 W. As shown in Figure 7.43, this is well within the capacity of one satellite refrigerator running in steady state. It assumes a heat load of 65 W in the PS, which is dominated by dynamic (beam) heating. This value is strongly correlated to the heat shield design and is subject to uncertainties in the particle production model used in simulations of the proton beam interacting in the production target. The heat load to 80 K is dominated by the cryostat design and cryostat surface area. As shown, the heat load is 1.6 kW, which translates into a required liquid nitrogen flow of ~550 liters/day.

| **Best Estimates (no contingency)** | **Production Solenoid** | **TSu** | **TSd** | **Detector Solenoid** | **Total** |
|---|---|---|---|---|---|
| **Nominal Temperature** | 80 K | | | | |
| 80 K magnet heat (W) | 130.7 | 252.0 | 252.0 | 500.0 | 1134.7 |
| 80 K feedbox and link heat (W) | 140.0 | 140.0 | 140.0 | 140.0 | 560.0 |
| Total 80 K heat (W) | 270.7 | 392.0 | 392.0 | 640.0 | 1694.7 |
| N2 usage for shield (liquid liters/day) | 149.93 | 217.11 | 217.11 | 354.46 | 938.6 |
| | | | | | |
| Number of 10,000 A HTS leads | 2 | 0 | 0 | 2 | 4 |
| Number of 2000 A HTS leads | 0 | 2 | 2 | 0 | 4 |
| N2 lead flow per magnet (g/s) | 2.2 | 0.2 | 0.2 | 2.2 | 4.8 |
| N2 usage for leads (liquid liters/day) | 237.6 | 21.6 | 21.6 | 237.6 | 518.4 |

Table 7.20: Heat load estimate at 80 K.

| **Best Estimates (no contingency)** | **PS** | **TSu** | **TSd** | **DS** | **Total** |
|---|---|---|---|---|---|
| **Nominal Temperature** | | | | | |
| 4.5 K magnet heat (W) | 64.9 | 44.0 | 42.0 | 22.5 | 173.4 |
| 4.5 K feedbox and link heat (W) | 14.0 | 14.0 | 14.0 | 14.0 | 56.0 |
| Thermal Siphon | | | | | |
| Total heat load (W) | 78.90 | | | | 78.9 |
| Total helium flow (g/s) | 4.20 | | | | 4.20 |
| 2.3 bar to 2.0 bar forced flow | | | | | |
| Helium inlet temperature (K) | | 4.5 | 4.5 | 4.5 | 13.5 |
| Total heat added (W) | | 58.0 | 56.0 | 36.5 | 150.5 |
| Selected flow rate (g/s) | | 50.0 | 50.0 | 50.0 | 150 |
| Exit temperature (K) | | 4.68 | 4.67 | 4.61 | |
| Circulating pump real work (W) | | 25.0 | 25.0 | 25.0 | 75.0 |
| Circulating pump system static heat (W) | | 15.0 | 15.0 | 15.0 | 45.0 |
| Total refrigerator cooling load (W) | | 98.0 | 96.0 | 76.5 | 270.5 |

Table 7.21: Heat load at 4.5K. The total heat load is 349.4 W.

### *7.3.5* Magnet Powering and Quench Protection

#### *Introduction and Scope*

The Mu2e solenoid system, shown schematically in Figure 7.48, is powered in four independent sectors:

- The Production Solenoid.
- Two sectors in the Transport Solenoid (TSu and TSd).
- The Detector Solenoid.

As can be seen in Table 7.22 there is a large variation of electrical parameters between the four sectors, with currents ranging from 1730 A for the TS, to 10.2 kA for the PS running at the highest field (5T). As a result of the different requirements, the possibilities for standardization across the four sectors are limited.

The Solenoids must be protected in case of a quench, whether occurring in the body or in the leads of a magnet, and from any malfunctions of the cooling and powering system. Furthermore, due to strong coupling between the PS, TS and DS appropriate measures have to be taken to protect each component in case a malfunction or quench occurs in any part of the solenoid system.

| **Parameter** | **Unit** | **PS** | **TSu** | **TSd** | **DS** |
|---|---|---|---|---|---|
| Max. allowable conductor temperature | K | 5.0 | 5.4 | 5.5 | 6.1 |
| Maximum current | A | 10150 | 1730 | 1730 | 5976 |
| Peak field in coil | T | 5.48 | 3.4 | 3. | 2.2 |
| Current fraction along load line at 4.5 K | % | 63 | 58 | 50 | 45 |
| Inductance | H | 1.58 | 7 | 7 | 1.61 |
| Stored energy | MJ | 79.7 | 10.4 | 9 | 28.7 |
| Cold mass | tonnes | 10.9 | 13 | 13 | 8 |
| E/m | kJ/kg | 7.31 | 0.8 | 0.7 | 3.6 |

Table 7.22. Main electrical parameters of the Mu2e magnets.

***Power Supply System***

As shown in Figure 7.48, each of the four power sectors has a dedicated power source that is sized to its nominal current requirement. This allows the current in each sector to be set arbitrarily to any level between the minimum and maximum value. Each circuit contains protection elements for the power sources (e.g. freewheeling diodes), the magnets (circuit breakers and extraction resistors) and ancillary equipment (current leads, bus bars and grounding circuit). All of these elements will be sized for the required operating conditions, which includes significant coupling between sectors, and will include a high level of redundancy.

The main elements and parameters of the power supply system are shown in Figure 7.49 and listed below.

1. Power converters:
   a. Two-quadrant, 6 kA/+10 V/-4V and 1 kA/+10 V/-4V, thyristor based modules are envisaged. With appropriate module configuration, all nominal currents can be provided.
   b. All power converters will have internal DC current transformers (DCCT) for current regulation. Additional driving signals for regulation will be provided by a DCCT directly linked to the magnet leads.
   c. The power converters will be water cooled.
   d. Internal freewheeling thyristors, mounted on air-cooled heat sinks, will provide protection in case of a water fault or other internal converter failure. The thyristors will be rated to carry nominal currents for a duration that is equivalent to about one time constant of the circuit.

2. Dump resistors:
   a. Air-cooled resistors in parallel with the magnets are foreseen for all circuits. The discharge time constants will be closely matched for all circuits.
   b. Peak voltages across the resistors will be less than 600 V. All circuit elements will be tested to a standoff voltage of about 1.5 kV.
   c. The maximum temperature of the resistors following a fast dump will be less than 200° C. The recovery after the fast dump will a few hours, shorter than the expected recovery of the cryogenic system. In the case of slow dump, the resistors are expected to recover in less than an hour.

3. Circuit breakers:
   a. Each leg of the two buses leading to a power converter will have an independent mechanical breaker. The main dump resistor will be permanently connected to the magnets. The circuit for slow dump will be activated with a separate breaker.
   b. The breakers will be air-cooled and normally in an open state.
   c. An opening time of about 0.1 s is required for all circuit breakers.

4. Busbars:
   a. The busbars will be made of copper and will be air-cooled. Their total resistance will be significantly less than the resistance of the circuit in case of slow dump, but sufficient for converter regulation.

5. Grounding circuit:
   a. The grounding circuit divides symmetrically the voltage across the magnet terminals through a 1 kΩ resistance. The leakage current to ground is continuously monitored and will trigger a magnet discharge if a preset limit is reached.

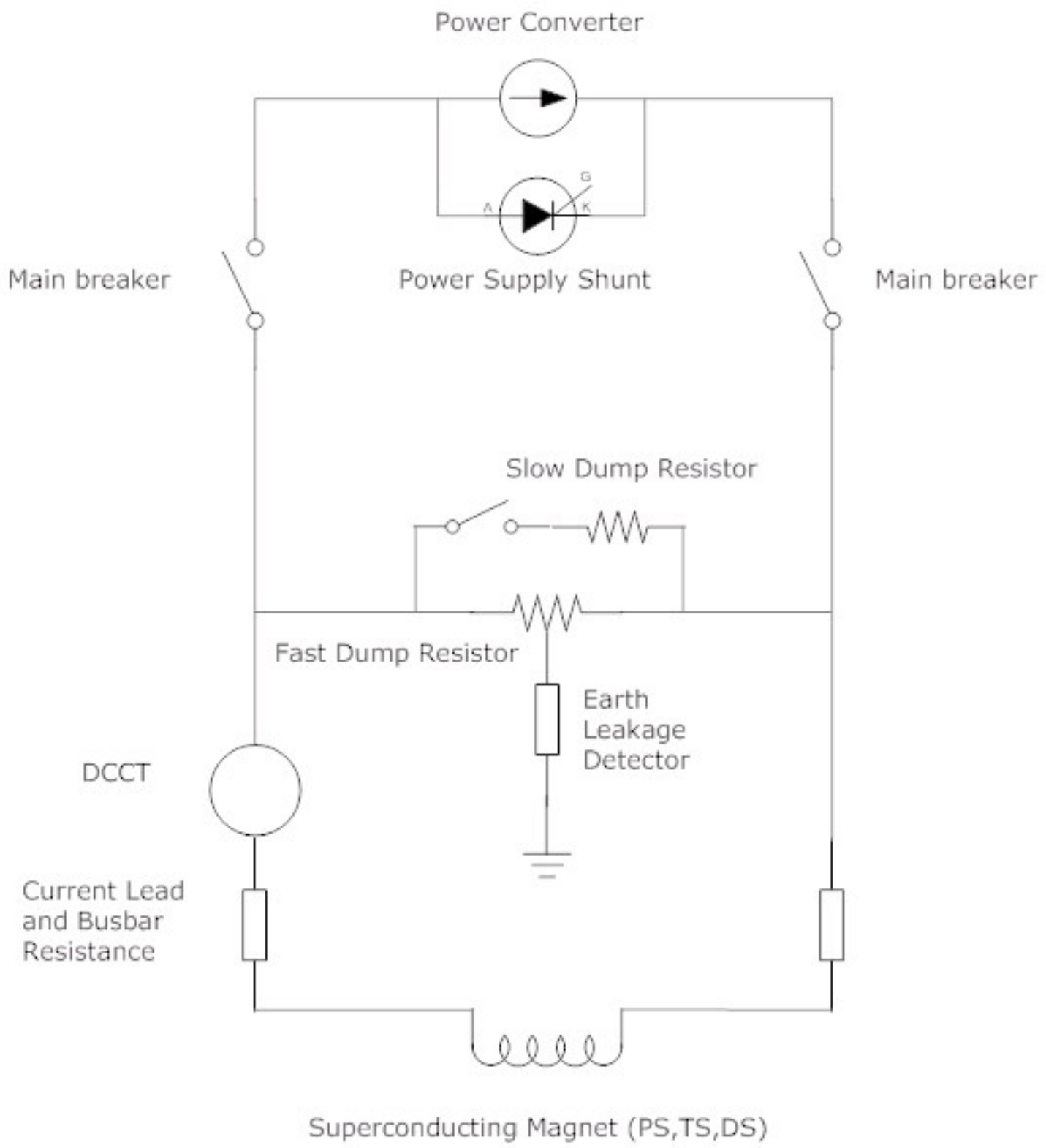

Figure 7.48. Schematic of the electrical circuit for a Mu2e magnet for the case where each magnet is powered individually.

***Quench Protection System***

As can be seen from Table 7.22 and in reference [7], the set of magnet requirements lead to a large variation of electrical parameters between the four independent power sectors, including a range of stored energy that varies from 9 MJ (TSd) to 80 MJ (PS). As shown in Section 7.3.4, there are also long superconducting busbars that connect the magnets to the cold-to-room-temperature power transitions. A quench could result in a fast deposition of this energy into the cold mass of the solenoids or busbars. Such a transition, if uncontrolled, could damage the magnets. In order to avoid such situations, a system that effectively protects the magnet at all times must be devised.

Both the Power Supply System and the Quench Protection System are designed to protect the solenoids including the superconducting busbars. They are both integrated with the Magnet Control System (MCS), which combines the supervision of all systems necessary for safe operation of the magnets. The elements of the protection systems are schematically represented in Figure 7.49.

The function of the Power Supply System, described in [6], is to safely operate the magnets during both nominal and transitory conditions and in the case of faults, to safely discharge the magnets and allow for their quick recovery. This system also includes the dump resistors used for discharging the magnets in case of a minor fault (slow dump) or emergency discharge (fast dump). The latter occurs if a quench is detected in any of the superconducting magnets, or in any part of the cold electrical circuit (cold busbars or current leads). It is the job of the quench protection system to detect and signal such a resistive transition.

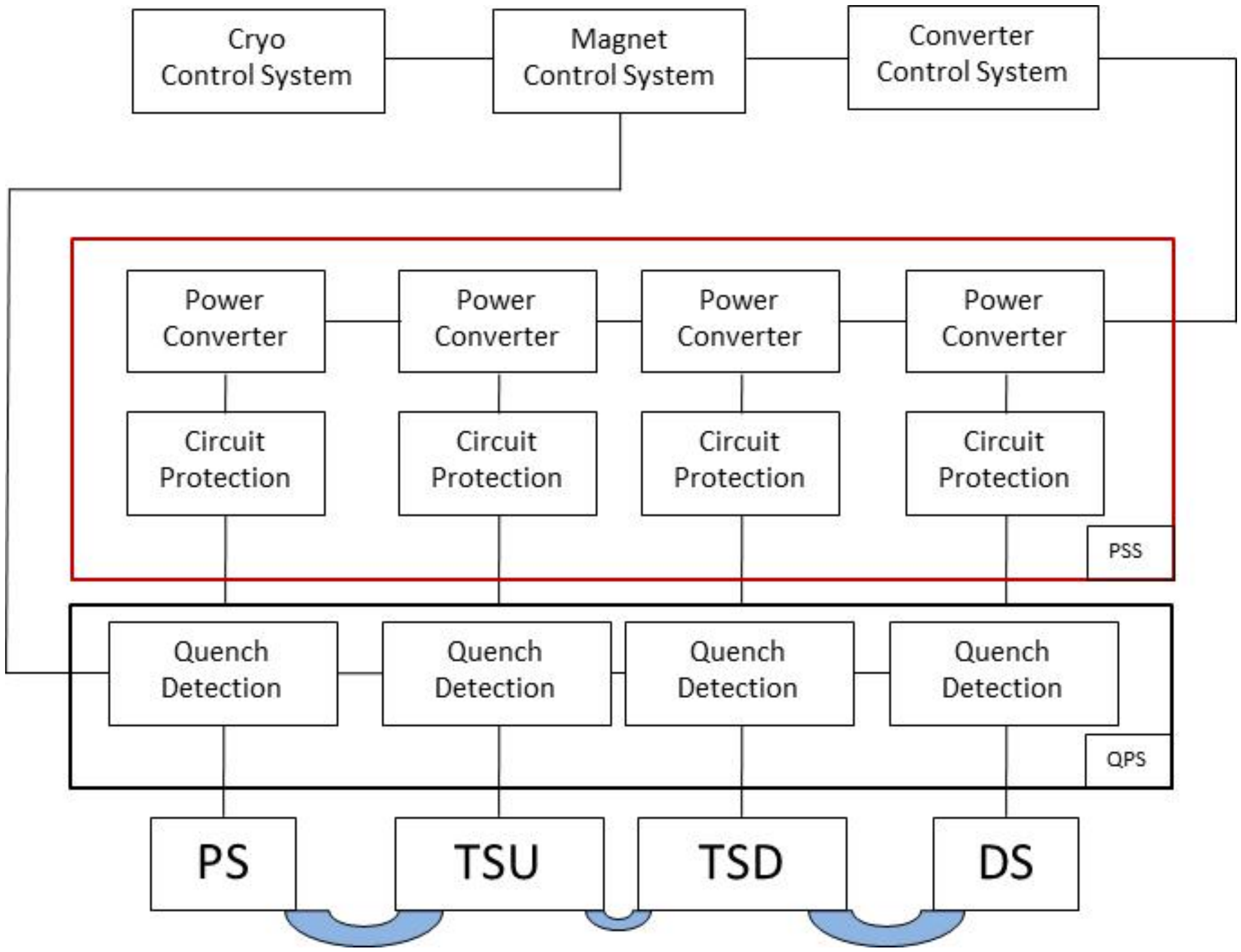

Figure 7.49. Schematic of the elements of the Power Supply System (PSS) and other systems relevant for the protection and powering of the mu2e magnets. The thick lines at the bottom of the figure represent the magnetic coupling of the solenoids.

In case of a quench, the quench protection system could also activate additional protection measures that are not part of the electrical circuit. For example, quench heater power supplies could be activated when the order is given to the power supply system to discharge the magnets. Quench heaters are a well-known technique for protecting superconducting magnets, relying on resistive layers imbedded in the coil windings. A short and intense current pulse through the resistive layers generates a rapid temperature rise and provokes a quick spread of resistive volume in the superconducting coil. The number and location of quench heaters in the Mu2e magnets is still under study.

Although several redundant levels of quench detection and protection elements are envisaged for the Mu2e magnets, it is always possible that some or all of them may fail to act. The magnets themselves must be designed to maintain the temperature of the coils below 130 K in the case of a full energy discharge. This is best accomplished if quench-back is enhanced by the design of the superconductor and its support structure. Also, magnet protection can be improved through passive thermal bridges between turns, which enhance the spread of the dissipated energy throughout the coil.

### *7.3.6* Magnetic Field Mapping

Requirements for the Mu2e solenoids [1] are derived from a set of unique magnetic field specifications that must be met in order for Mu2e to satisfy its physics goals. Efficient transport of negatively charged muons to the stopping target combined with effective suppression of positively charged particles, line-of-sight neutral particles, and high energy negatively charged particles is required. Throughout the solenoid system the specifications on absolute field strengths and local gradients are at the 1 to 5 percent level. These levels are consistent with known manufacturing, assembly and alignment tolerances as well as uncertainties in the final geometry due to thermal changes and large mechanical forces from magnetic interactions and vacuum systems. Furthermore final state electron tracking and momentum determination relies upon a precise field map of the DS spectrometer region, at the level of 0.01 percent.

To ensure that Mu2e physics goals can be met, it is necessary to validate the fields produced by the solenoids; the specific requirements are elaborated in [8]. Clearly for such a large and complex system of magnets, achieving the desired field profile across the entire transport channel in the final magnet configuration will require great attention to quality control during the fabrication of individual components. Measurements of coil conductor geometry and electrical (insulation) properties will be required at certain steps, to recognize any errors or problems and corrective action will need to be taken before repairs become difficult and expensive.

Once the large PS and DS solenoids are built, there is a possibility to perform a magnetic field map at the vendor location with the coils at room temperature; such a map could be performed with a relatively simple apparatus, and would provide some confidence prior to shipping the magnets to Fermilab. However the maximum current in the room temperature coils is limited by stabilizer cross sectional area and cooling conditions. Prior to installation in the cryostat (with air flow cooling), the coils may tolerate 1% of the nominal operating current. This, of course, does not ensure the magnet performance when cold; however, it is not expected that vendors will have the necessary cryogenic, power, and quench protection system infrastructure needed to confirm cold test performance prior to shipping to Fermilab. Therefore, we assume that all cryogenic performance tests and full-field magnetic measurements will be conducted at Fermilab; this entails risk to magnet cost and schedule associated with possible need to return a magnet to the vendor for repair.

The situation with the TS toroidal solenoid is slightly different: we envision that (pairs of) individual coil modules will be tested under cryogenic conditions at Fermilab before they are put together into their final TSu and TSd arrangements. Thus, coil quench performance and basic magnetic field profiles will be confirmed at an early stage. However, there are constraints that severely limit what direct field measurements can be made within the fully assembled Transport Solenoids. First, there will be no infrastructure for cold power testing of the completed TS magnets except at the Mu2e experimental hall. Second, current plans for installation and commissioning make it unlikely there will be an opportunity to measure magnetic fields in either of the TSu or TSd sections. TSu will be installed first (after PS), and could be tested in a stand-alone fashion (with PS not powered, due to force imbalance); however once TSd is installed, it would be time consuming and impractical to re-position the magnets to install and remove a field mapping system. Thus, a different solution is envisioned for validating the final field configuration throughout the transport channel which involves monitoring of coil positions (and adjustments) under thermal and magnetic force conditions, with fixed arrays of in-situ Hall probe sensors in each straight section, and finally a test of low energy electron transport through the channel.

The field mapping effort is intertwined with the installation and commissioning of the Mu2e magnets, beam line, and detector systems. Mapping of the PS and DS fields is important not only for qualifying the magnetic profile, but also to determine the magnetic axis for proper positioning of each solenoid. Therefore a reasonably quick coarse-grid map of the stand-alone DS and PS solenoids is envisioned to be performed immediately following (or during) the cold power and quench performance testing of these solenoids. These field maps would be followed by magnet position adjustments, and a thermal cycle to room temperature and back to 4.5 K. A second round of very thorough field maps would then be performed, once the entire magnetic system is installed and commissioned. This “final” configuration is likely to be approached in an iterative fashion, since the TS coil positions will be monitored at each interface and adjustments made to achieve the desired result. For final configuration magnetic measurements several steps are required: first, the DS and PS volumes would be mapped; second the regions within the TS1 and TS5 collimator holes at the interfaces with PS and DS would be mapped. The in-situ arrays in these regions will provide complementary information and monitor field conditions during operation (although sensor calibration changes, from beam irradiation, are expected to introduce uncertainty over time). Note that field maps of the PS will be required at three possible operating points, with the peak field at 4.6 T (nominal), 4.2 T, and 5.0 T. The DS spectrometer region will be mapped at the nominal 1 T field, and also at the physics calibration field of 0.7 T.

***Design of Field Mapping System***

The Mu2e field mapping system is conceptually similar to systems previously developed for mapping the large detector solenoids for the CMS and ATLAS experiments at CERN [31][32]. The field mapping system, shown schematically in Figure 7.50 will utilize arrays of Hall and NMR probes that are rigidly mounted and accurately oriented along arms of a non-magnetic mechanical motion carriage device capable of positioning the probe array precisely in azimuthal and axial coordinates. The motion carriage and probe arrays will be supported on, and will translate along, precision non-magnetic rails that are securely mounted to the magnet cryostat or inner vacuum bore. The carriage, probes and rails will all be referenced to one another by survey, and will have precision encoders for position and angle determination. For actuating the transporter motion, both pneumatic and non-magnetic cable drive (with external motor) options can be considered.

Since PS and DS have very similar geometry and volumes to be mapped, a single (adjustable) mapping system will be used for both. Since the detector elements will be installed on a precisely aligned (~0.125 mm) non-magnetic rail system (responsibility of the Muon Beamline group [33]), it is natural for the magnetic field mapper to utilize these rails for positioning along the axial direction. In the PS, the Heat & Radiation Shield (HRS) will also be installed on support rails (once field mapping is completed), and these can serve to guide the mapping system motion in the PS. For the PS and DS mapping, Hall probes would be mounted at fixed radii along "propeller" arms that will be stepped through a series of angles at each axial Z position. To map the PS-TS1 and DS-TS5 interfaces (inside the collimator holes), the transport system will be reconfigured with a different array of Hall probes positioned on a long (counterweighted) extension that will extend through the collimator. The transporter design needs flexibility to adjust for differences in rail width and magnetic axis height; these differences are relatively small. Relying upon a single field mapper saves on component and labor cost, but it entails some risk to schedule because using this device will be on the critical path to proceeding with final steps in the solenoid and detector assembly. It will need to be relocated, reconfigured, set up, surveyed, and operated several times during the mapping program.

The most demanding Mu2e measurement requirements are in the DS spectrometer uniform field region, where the field components must be determined to 1 G in a 1 T field (0.01% precision); the PS 2.5 to 5 T field must be measured at ~1% precision. Therefore the field mapping system will use modern, commercially available 3-axis Hall probes that are calibrated to 0.1% linearity over the required range (0-5 T) of magnetic fields. NMR probes will accompany the Hall probe arrays to provide the required 30 ppm absolute field reference measurement in the DS spectrometer region; to get a sufficiently uniform field for the NMR system to lock on to a signal in the DS gradient section, "gradient compensation coils" will be used.

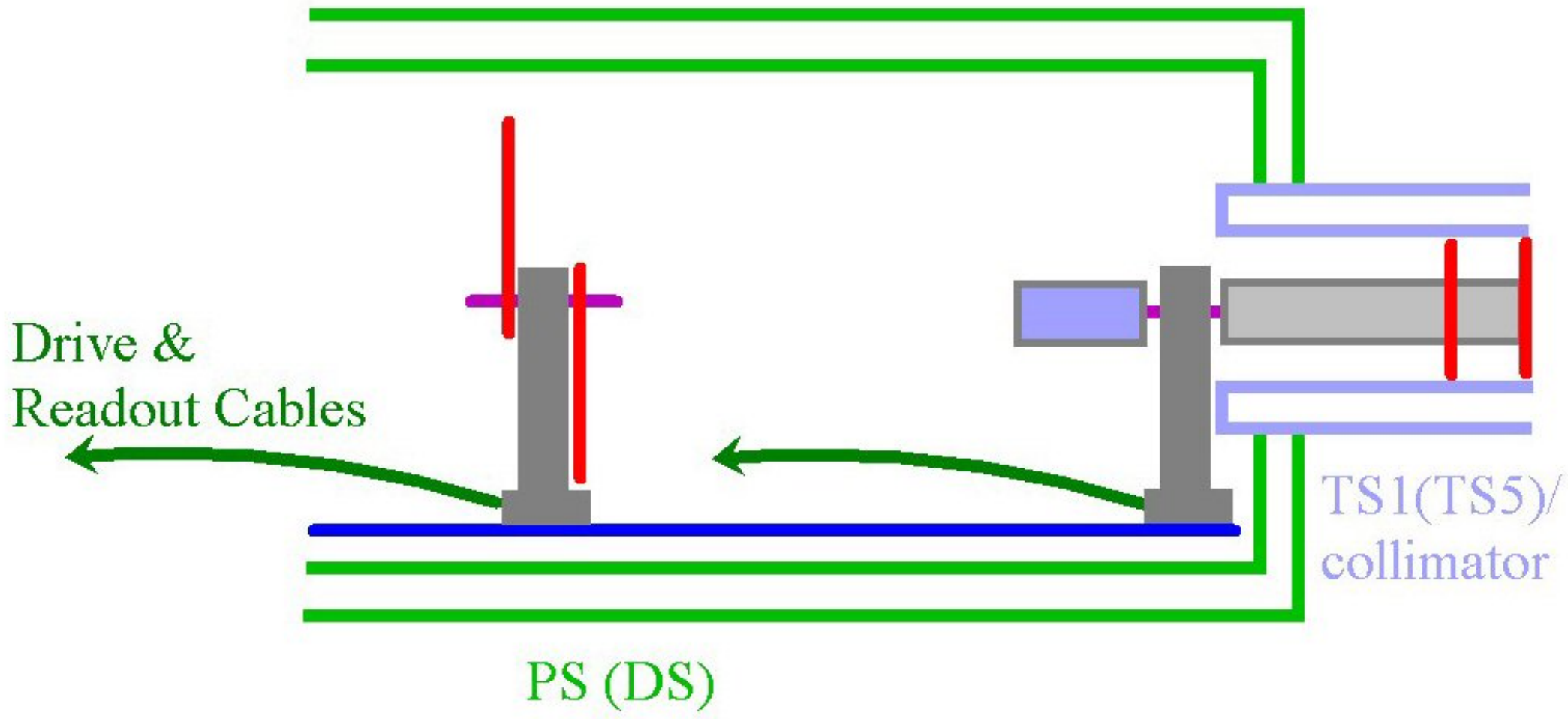

Figure 7.50. Schematic side view of a mapping system that translates along rails (dark blue) within the large aperture solenoid. Two arrangements of Hall probe arrays allow the same system to map the large solenoid volumes as well as the TS collimator interface regions.

***Fixed Array Design***

Fixed arrays of magnetic probes will be permanently deployed in the solenoid system to monitor long-term stability. The field at interface regions at the ends of each Transport Solenoid cryostat may be sensitive to small changes in geometry. Fixed arrays in these regions will determine how much magnetic or vacuum forces affect the field profiles, and if they agree with the expected fields. In the TS3 straight section, in-situ Hall sensors are the only practical way to directly verify that the field profile has the required continuous negative gradient. To make the (strength and gradient) measurements, linear arrays of several precisely spaced 1-D Hall probes will be installed to measure the axial field strength at the beam tube radius on either side of each interface; a platinum RTD will accompany each Hall probe to measure and compensate for the sensor temperature.

The arrays will be mounted on the outer surfaces of the collimator support shells. This will allow them to be pre-assembled as a robust mechanical structure, and mounted in a precisely measured geometry that will be well defined with respect to the cryostat. Furthermore, this scheme will simplify routing and protection of wiring, and will provide shielding of the sensors by the collimators to minimize radiation damage. The same scheme, shown in Figure 7.51, can be applied at each of the collimators, TS1, TS3u, TS3d, and TS5. Four arrays (up, down, left, right) of 4 or 5 probes will be mounted on the shells at each of the 4 collimators to give sufficiently redundant information to determine and understand the field profiles.

Accurate electron momentum reconstruction requires a precise knowledge of the magnetic field in the tracker region of the Detector Solenoid. A fixed array of probes will be deployed in the tracker region to monitor the stability of the field over time. Two (or more) NMR probes will be rigidly mounted to each end of the tracker frame along with a set of 3-axis Hall probes. This arrangement will allow for monitoring the field stability over time, and will also be sensitive to changes in the orientation of the tracker. The precise positions of these Hall probes should be determined by survey in order to accurately relate them to the mapped field coordinate system. The Muon Beamline group has responsibility for coordinating survey and alignment specifications and efforts [27].

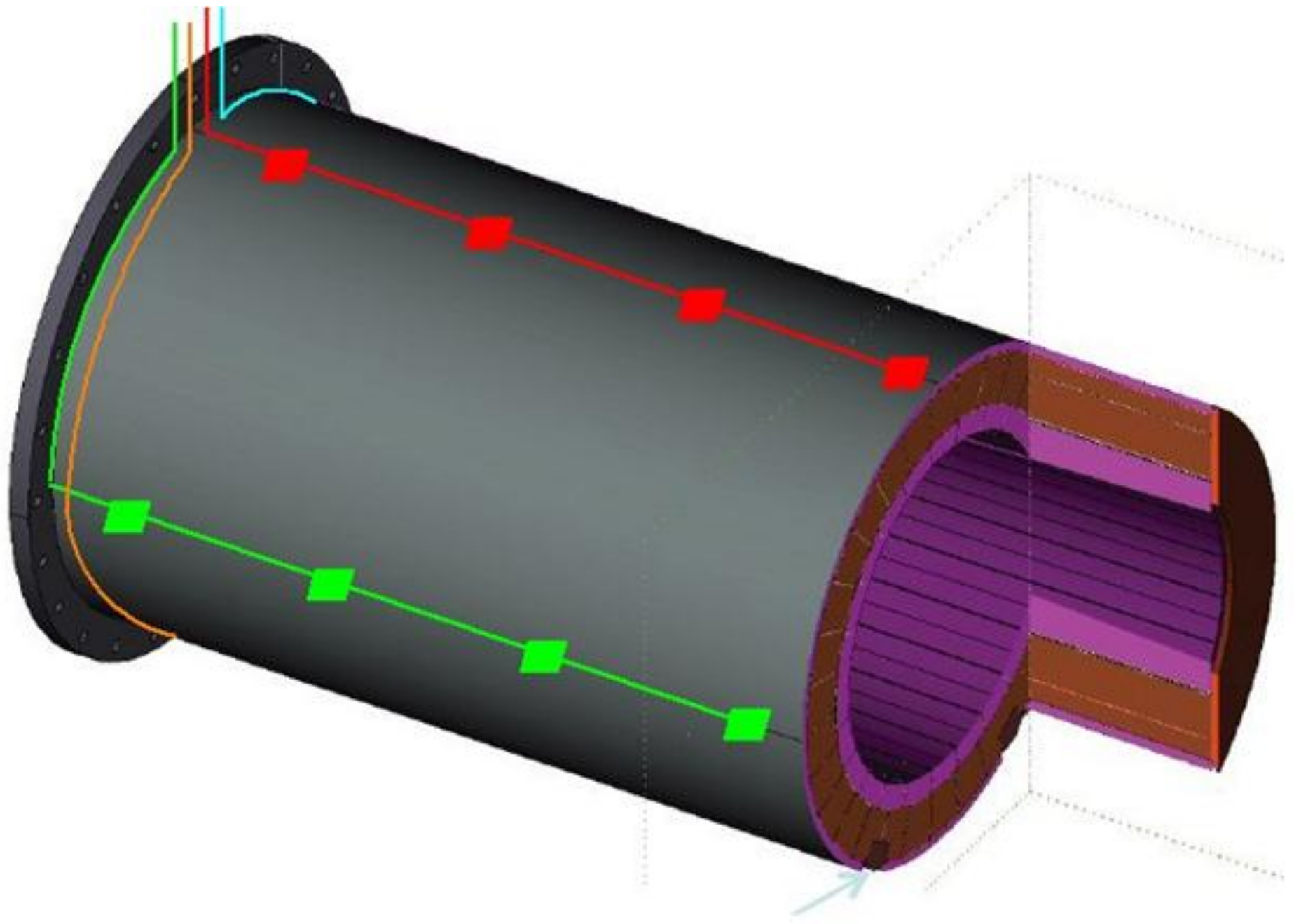

Figure 7.51. Concept for in-situ Hall probe arrays mounted on the outside of collimator support shells.

***Coil Displacement Sensors***

Displacement sensors will be used to monitor the coil positions with respect to cryostat walls at each of the TS interfaces (TS1, TS3u, TS3d, and TS5). This will allow comparison with expectation under conditions of cool down and when energizing the fields. If adjustment of the coil positions and angles is necessary, these sensors will provide feedback on resulting coil positions. The preferred approach is to utilize commercial fiber optic displacement sensors, which have very good resolution up to the required range of expected TS coil displacements and have been adapted to MRI applications – they therefore work in a magnetic field and are available with vacuum feed-throughs. To fully determine coil positions and angle displacements, four axial and four transverse sensors are required at each of the four interfaces, for a total of 32 sensors. A readout system is also required.

***Electron Source and Detectors***

The concept for testing the final magnetic configuration is to study the transport of low energy electrons from the PS production target region all the way to the DS stopping target region. To do this, once the HRS is installed in the PS, an electron source would be moved into the location where the production target will be installed later. Possible electron sources include a Beta source (~1 MeV average energy), or an electron gun (up to ~25 keV). It will be necessary to measure where electrons end up from different transverse starting positions, so measurements at the center position and translations (up, down, left, right) of the beta source would be performed using a ceramic/piezo motion stage. In the case of the electron gun, a fixed array of 5 guns would explore the same dependence on position. For the electron gun, the entire transport channel would need to be evacuated; this may not be necessary (but would be beneficial) for the beta source option. Downstream, detectors would be deployed to determine where the electrons end up. One option is to tile the upstream face of the TS1, TS3u, and TS5 collimators (outside the hole) with (phosphor screen plus) scintillators, as shown in Figure 7.52. HPD photo detectors, which have been shown to work in fields up to 5 T, can be used to detect the scintillation light. Four quadrants would indicate if, and where, electrons from each source position hit the collimator. These detectors could simply be left in place after the test. At the stopping target location, another array of more finely segmented scintillation-based detectors would be placed, and removed after completion of the test. Thus, with the full magnetic configuration and evacuated channel, the transport of low energy electrons (which will closely follow the magnetic field lines) will be studied as a function of the electron starting position, to confirm expectations and validate the final configuration.

### *7.3.7* System Integration, Installation and Commissioning

***Mu2e Building Layout***

A detailed and fully integrated installation procedure will be developed for the Mu2e solenoids and supporting equipment as part of the preliminary and final design process. The solenoids will be constructed elsewhere and installed in the Mu2e detector enclosure at the appropriate time. The detector hall will contain two independent 30 ton cranes for handling the components. These two cranes will have the capability to be connected and controlled as one 60 ton crane for handling large components like the Detector Solenoid. The layout of the underground detector enclosure is shown in Figure 7.53.

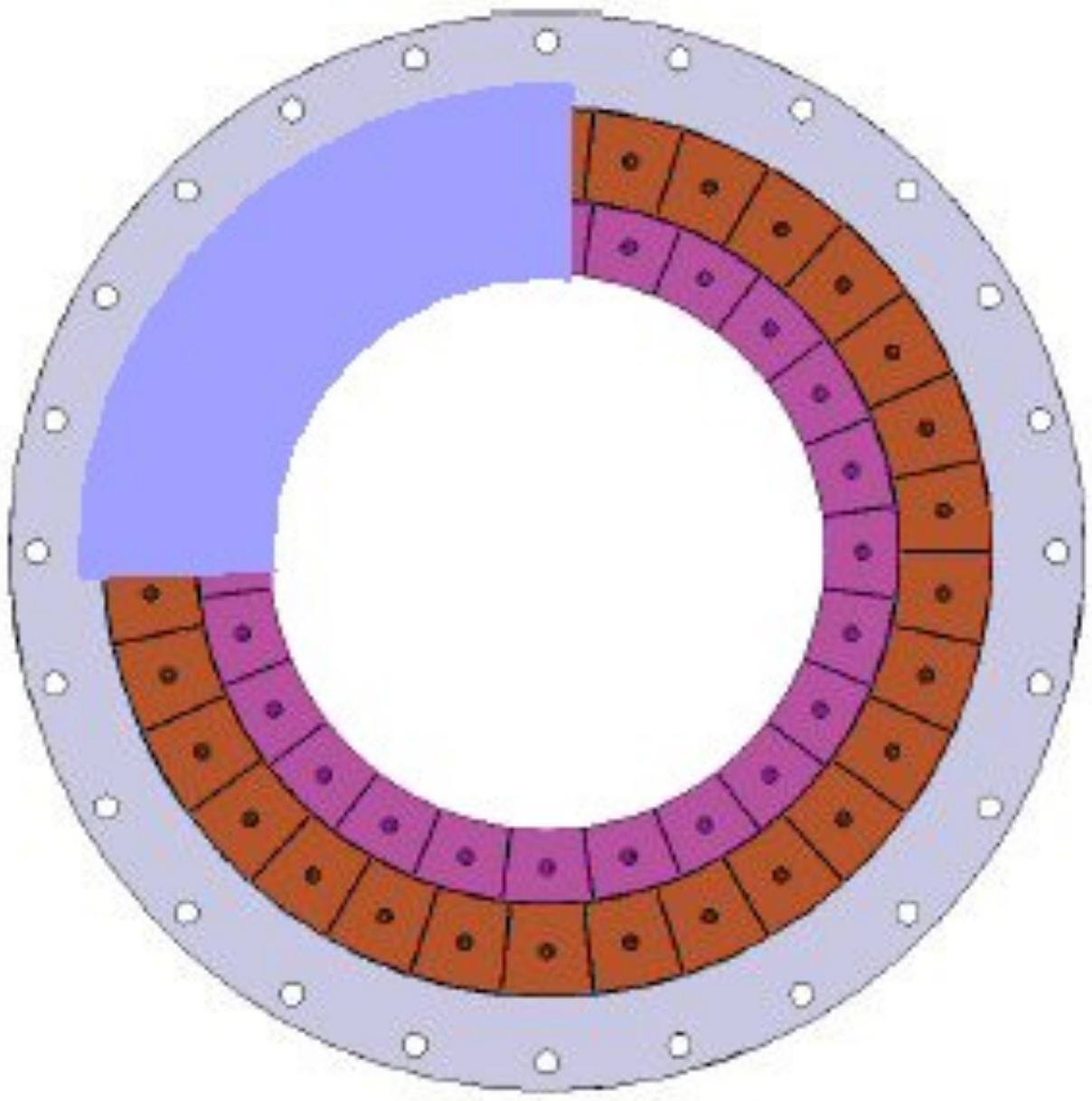

Figure 7.52. Concept for one (of four) quadrant scintillation tile at upstream side of a collimator.

The building is open over the Detector Solenoid during installation and will be filled in with shielding blocks during operations. This ensures there is direct crane coverage for installing the Detector Solenoid. There is no crane overage over the Production or Transport Solenoids so each component must be moved into place using a rail system. The rail system can be seen in Figure 7.53. This system is similar to other installations at Fermilab, including CDF and D0. The rails are imbedded in the concrete floor during construction. Each rail has holes for mounting a hydraulic cylinder for pushing or pulling the heavy components into place.

***Magnet and System Installation***

Once the magnet assemblies are accepted by Fermilab and delivered to the Mu2e building, installation can begin. Installing the Detector Solenoid assembly is straight forward as it can be lowered directly into place. Installation of the Production Solenoid assembly is a little more difficult. The Production Solenoid must be lowered down a hatch outside of the building and then rolled into place by way of a rail system built into the floor of the enclosure. Installation of the TS magnets will also utilize a rail system.

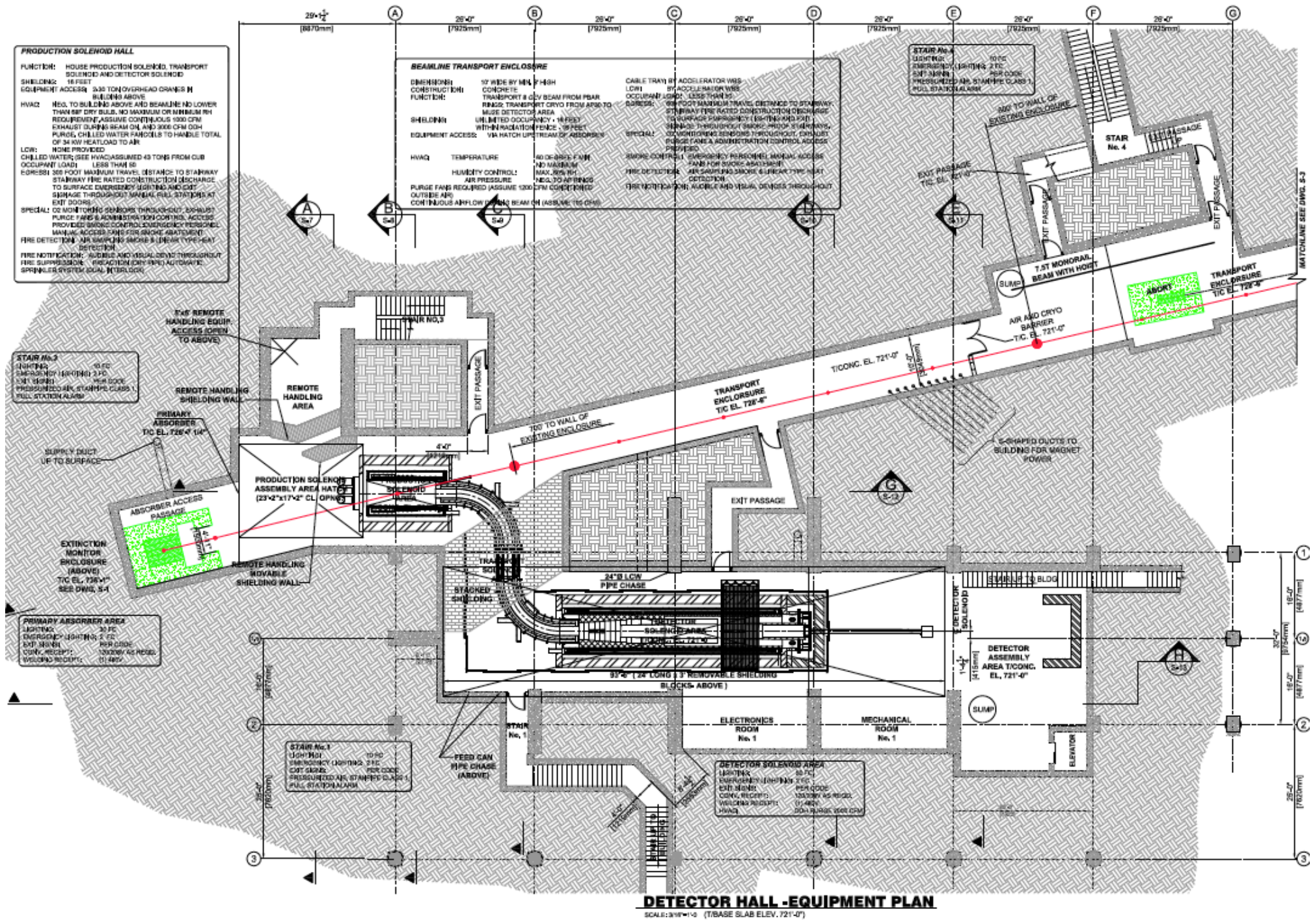

Figure 7.53. Layout of the Mu2e detector enclosure layout.

The installation plan is broken down into magnets and all other solenoid systems. The magnet sequence is important but the other systems can be worked on in parallel. A detailed installation plan that optimizes the assembly and installation tasks will be developed during the preliminary and final design process. The general installation plan is as follows.

Magnet installation sequence:

1) Install DS and PS assemblies (order does not matter here).
2) Install TSu magnet.
3) Install TSd magnet.
4) Make up magnet interconnects.
5) Connect magnets to cryogenic and vacuum system.

Other system installations:

1) Feed box installation.
2) Transfer line installation.
3) Magnet power system installation.
4) Cryogenic instrumentation and controls.
5) Insulating vacuum system installation.

Once all systems are installed and preliminary testing is complete, the magnet system commissioning can begin.

***Magnet System Commissioning***

The solenoids will be fully assembled as a magnet system and cold tested in the Mu2e detector enclosure. The Production Solenoid heat shield and detector components will not be present during initial magnet commissioning. Clear magnet bores are required for mapping the magnetic field of each solenoid. Once the magnets are cold tested and the field maps measured, installation of the heat shield and detector components can begin.

## 7.4 Considered Alternatives to the Proposed Design

During the process of developing the conceptual design, several alternate designs have been considered. These designs were not selected as the preferred alternatives for several reasons. Primary considerations were cost, design feasibility and ease of fabrication. A few alternative designs (more precisely alternative materials) are still under consideration for the Transport Solenoid. The choice of cryogenic cooling schemes for Transport and Detector solenoids (thermal siphoning vs. forced flow) is still under consideration.

### *7.4.1* Production Solenoid Alternative Designs

A number of alternative design solutions were considered in the course of the PS magnet design studies. These were developed to various levels of detail, but always to a level that was sufficient to discriminate between alternatives.

***4-layer Coil***

A coil design with 4-3-2 layers has been studied (shown in Figure 7.54) [9]. The extra layer of cable allows increasing the Al stabilizer thickness by 0.5 mm and to operate at a lower current while fulfilling the design requirements. As a result, the peak quench temperature is reduced. However, adding an extra layer of cable increases the magnet cost, and, at the same time, reduces the thermal margin because of the larger coil volume impacted by the radiation power and a larger number of insulation layers in the radial direction that impedes the heat transfer. Thus in order to achieve the same thermal margin as the 3-layer design, at least one thermal bridge has to be embedded between the cable layers, potentially impacting electrical reliability.

***Continuous Coil***

A continuous coil concept, shown in Figure 7.55 was investigated where the complete Production Solenoid coil is wound from a single piece of cable. This assumes that continuous conductor of the required length can be obtained. The overall coil length, cable dimensions, number of layers and main operating parameters are the same as the baseline concept. The difference is in the absence of axial gaps separating the individual coil sections. The result is a slightly smoother field profile. However, it was recognized that this approach has a number of challenges absent in the baseline design. The primary disadvantage is that the mechanical support across such a large span is more difficult because the forces are different in regions that have a different number of windings. The 3-coil baseline design has full-radius flanges between coils that allow for radial and axial supports that are sized for the expected force on each coil.

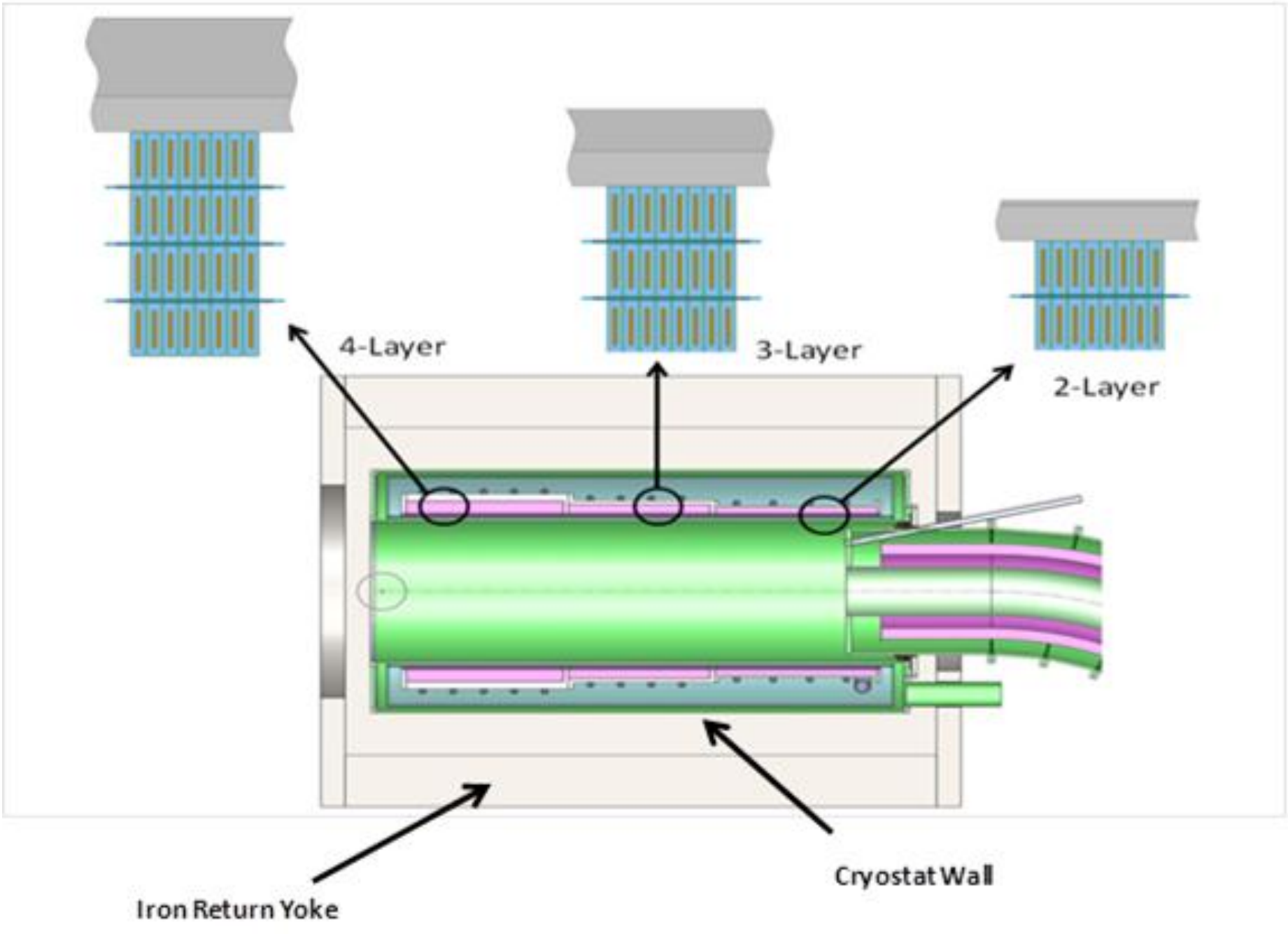

Figure 7.54. Alternate PS “4-3-2” layers with iron yoke.

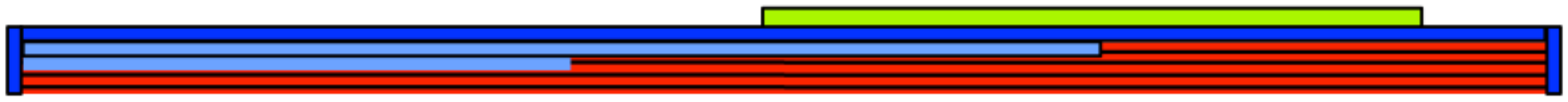

Figure 7.55. Possible support structure for the continuous coil design. The coils are in red. The structures depicted in blue and green provide mechanical support.

### *Copper Stabilized Cable*

Copper stabilized cable was considered for the MECO Production Solenoid [29]. The advantages of this approach might be a potentially larger number of vendors who are capable of producing the cable as well as the potential for cost savings. However, copper-stabilized cable has a number of important disadvantages. The total dynamic heat load in copper-stabilized cable is a factor of three larger than in aluminum-stabilized cable due to larger energy loss from secondary particles that interact in the stabilizer [30]. Because of the relatively low margin between the required cooling temperature and the helium boiling temperature, the increased heat load from copper stabilizer is likely to require either a sub-cooled helium supply or the pool-boiling cooling scheme considered by MECO. Both options would have an impact on the magnet cost that could partially or completely eliminate the cost advantage of using the copper stabilizer. Another disadvantage of using copper is an increase in the weight of the cold mass of the Production Solenoid by approximately a factor of two, requiring a proportional increase of the radial support strength (and the corresponding static heat load).

Perhaps the most important impact of using copper-stabilized cable is in the incomplete recovery of the resistivity after irradiation and subsequent annealing. According to [10], copper only recovers to ~ 90% of the initial resistivity after room temperature annealing leading to a degradation of the quench performance after prolonged irradiation. Aluminum does not exhibit such behavior, recovering to 100% of the initial resistivity after a thermal cycle.

***G10 Spiders as the Cold Mass Support***

An alternative means of supporting the cold mass in the radial direction is through the use of "spiders". Spiders are rings made of low thermal conductivity material (usually G10 or G11) that are anchored to the cold mass and the cryostat wall at several points. A possible spider concept is illustrated in Figure 7.56.

One disadvantage of a spider support is a potential interference with the thermal siphon cooling tubes. Another disadvantage is the potential for large thermal stresses to develop in the spider due either to a mismatch in the thermal expansion coefficients of the spiders and the cold mass or potential magnet shifts during transitions from warm to cold.

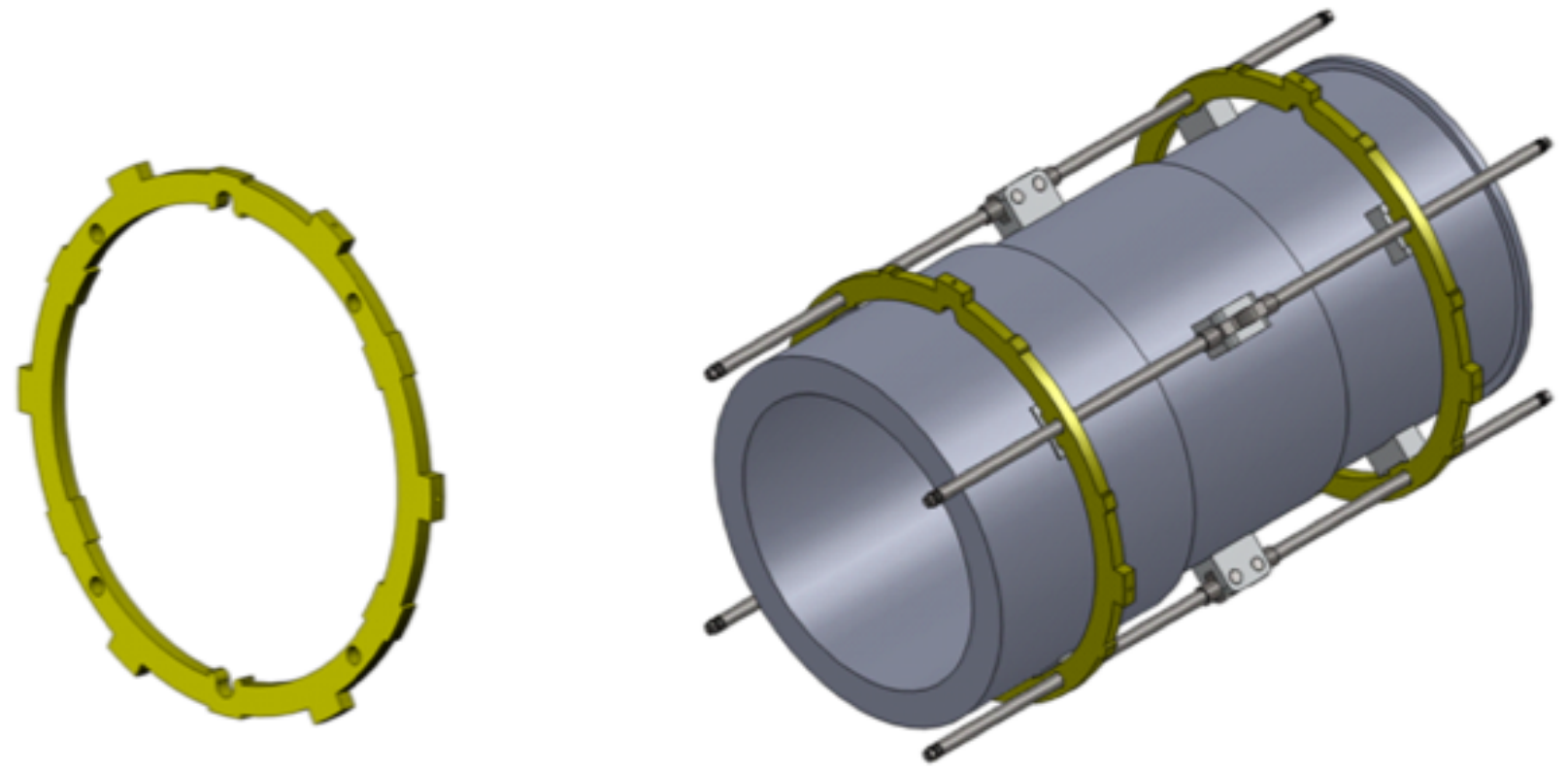

Figure 7.56. Alternative cold mass support design: G-10 spider ring (left) and spider attachments to the cold mass (right).

### 7.4.2 TS Design Alternatives

Several alternative designs have been considered for the Transport Solenoid. Aside from a number of studies to optimize the number of individual coils required in the Transport Solenoid to meet the field specifications, most of the alternative studies centered on the complicated central straight section (TS3), and on the design of the modules.

***TS3 in a separate cryostat.***

In this design, all TS3 coils are in a dedicated cryostat (see Figure 7.57). This cryostat is connected with bellows to the adjacent cryostats (TS1-2 upstream, and TS3-4 downstream). It can be quickly disconnected from the adjacent cryostats and can be lowered (or moved sideways) in order to access its warm bore for maintenance and replacement of the rotatable collimators and the anti-proton window.

The drawbacks of this design are the extra costs due to the separate cryostat, its support system, the dedicated feedbox, cryogenic link and service lines. More details about this alternative design can be found in [35].

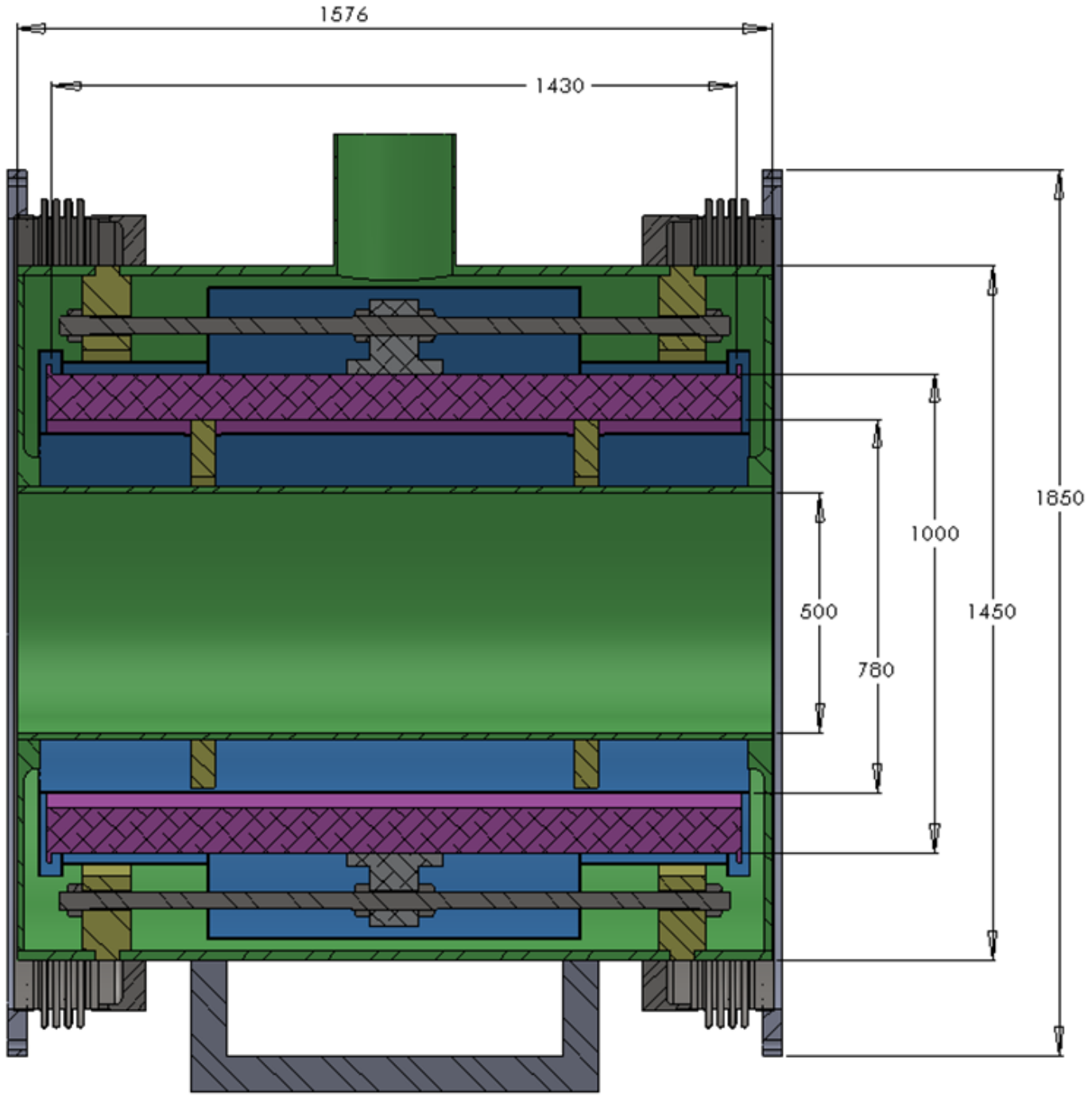

Figure 7.57 Alternative design with TS3 coils in dedicated cryostat.

***Copper stabilized conductor***

Copper stabilized conductors (for instance: Wire-in-channel, or Rutherford-in-channel) have some advantages with respect to aluminum stabilized conductors: they are less expensive and can be provided by a larger number of vendors. But they also have drawbacks. The RRR degradation of copper due to radiation does not anneal completely at room temperature, whereas aluminum does. Additionally, aluminum stabilized conductors can easily provide more stability than copper.

An example of a TS design using a copper stabilized conductor is presented in [35]. Since the TS coils will be coupled with the larger adjacent magnet systems, a large stability margin is recommended.

***TS modules made of stainless steel used as bobbins***

This alternative design employs a stainless steel structure to support the TS coils, which are wound and impregnated in the modules. The modules can house three or four coils each (see Figure 7.58). The use of stainless steel makes it easy to fabricate the TS modules, but makes it hard to provide axial pre-stress to the coils. The axial thermal contraction of the coils is larger than the thermal contraction of stainless steel, therefore axial pre-stress must be provided by using bladders or slices of material with a very low thermal contraction coefficient (for instance INVAR). More details about this design are presented in [35].

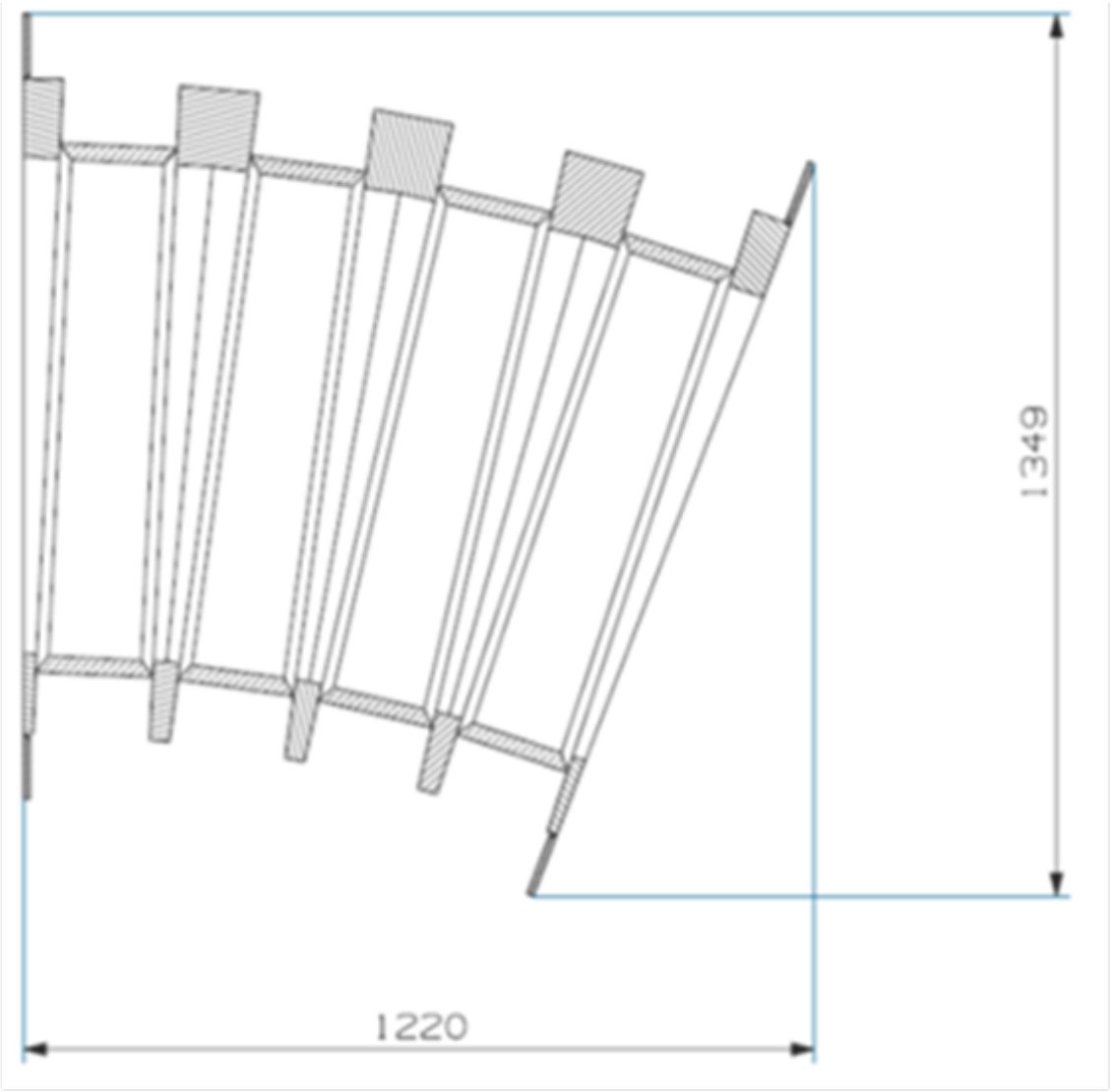

Figure 7.58 Cross-section of 4-coil stainless-steel module/bobbin for TS.

***TS modules made of aluminum used as bobbins***

In this design the aluminum-stabilized cable is wound on bobbins made of aluminum alloy (Al-5083) that can house three or four coils. These modules are very similar to the modules presented in the previous alternative design, with the exception of the material and the fabrication method. This concept solves the issue of the axial thermal contraction. On the other hand there is a risk of separation between the coil ID and the bobbin OD during cooldown and powering. This risk can be reduced by using a bandage to pre-stress the coil after winding and/or impregnation. In order to completely eliminate this risk, a stainless steel liner should be inserted in the ID of the modules. By using both features (bandage and liner) this design [35] meets all requirements, although the module fabrication is quite complicated and expensive.

### *7.4.3* Detector Solenoid Design Alternatives

***Two Layer spectrometer section***

In the process of developing the Detector Solenoid conceptual design, several alternatives were considered. The two-layer design for both the gradient and spectrometer section, powered in series was found to satisfy the field requirements with satisfactory margins. This two-layer design avoids the problem of having an uncompensated return bus, whose field is on the order of 25-40 G. However, the present field uniformity requirement of |dB/B|< 1.0% or 100 G (see Table 7.2) means that this field non uniformity is not an important consideration.

For this two layer design, a very wide cable is needed in the spectrometer section to power the gradient and spectrometer sections from a single supply, (see Figure 7.59). The wide spectrometer cable is almost three times wider than the gradient section conductor. This added aluminum is not needed for quench protection or stability, but is used to achieve the required engineering density. Thus, the weight of the magnet becomes very large, on the order of 50 tonnes. Furthermore, the cost of the magnet conductor increases with the additional conductor-quality aluminum volume. The baseline "single-layer" spectrometer is a much more efficient use of conductor, while still meeting the Mu2e magnetic field requirements.

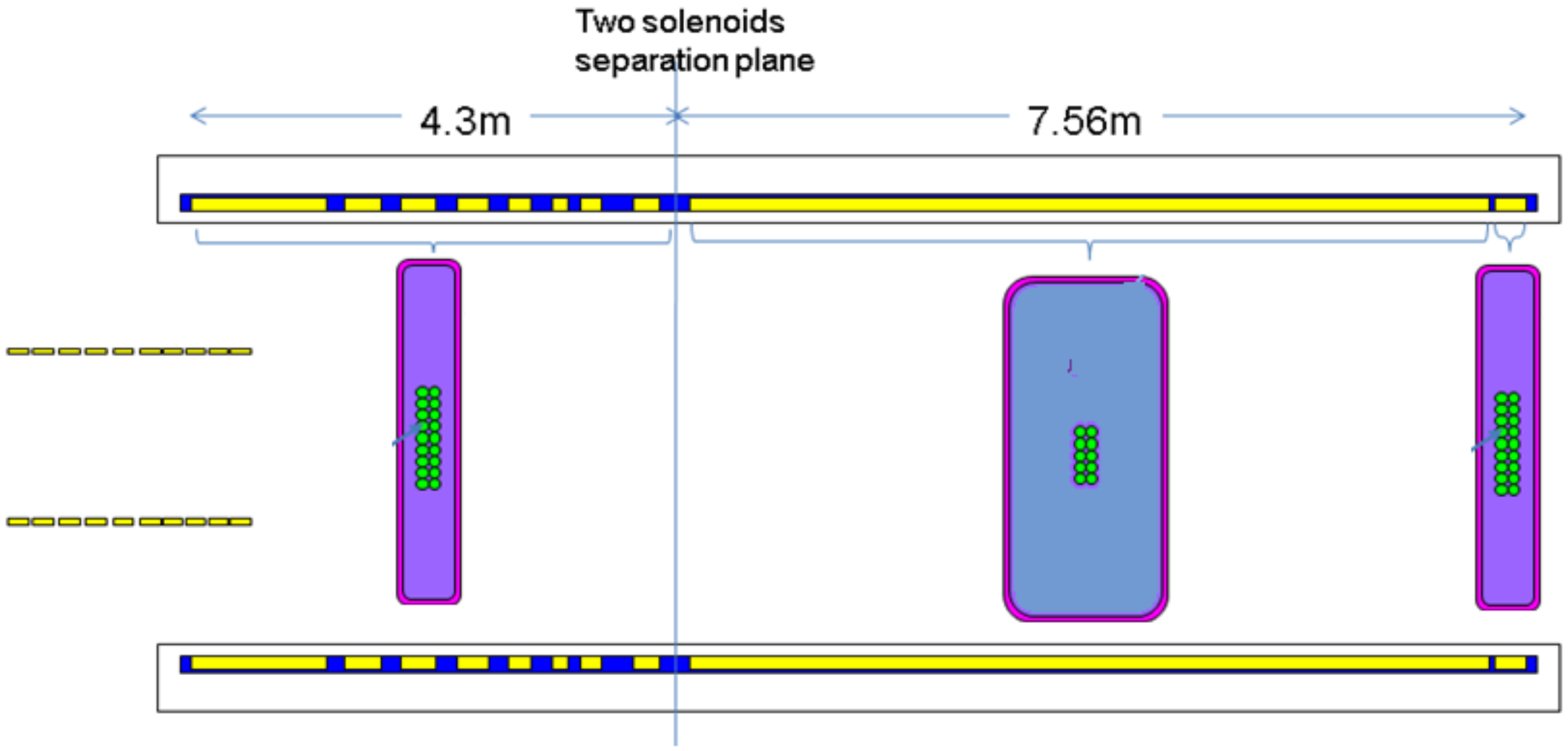

Figure 7.59. An alternate design for the Detector Solenoid coil featuring two- layer gradient and spectrometer section.

Another possibility was to power the DS spectrometer and gradient on separate power supplies. In this concept, the DS could be fabricated with a single conductor geometry, possibly resulting in savings in conductor fabrication cost. The disadvantage of this approach is the additional superconducting leads for the independently powered coil segments. Furthermore, the gradient section would have to be powered at a much high current, on the order of 15 kA. The added complication of high current power leads and high current-rated DC breakers offset the cost benefits of this approach relative to the baseline design.

***Copper stabilizer***

The primary reason to choose Aluminum stabilizer over copper for the Detector Solenoid is weight. Copper stabilizer would increase the weight of the coil itself to almost 60 tonnes. Adding in the weight of the cryostat would render the Detector Solenoid too heavy to lift with the proposed 60 Ton crane in the experimental hall.

### *7.4.4* Iron Return Yokes for PS and DS

An earlier baseline design for Mu2e called for iron return yokes for the PS and DS (shown for PS in Figure 7.54). This iron contributed modestly to the PS and DS fields (1-5%), and through pole pieces helped to shape the transition field between PS/DS and the neighboring TS coils. They also performed secondary roles as a radiation shield and reduced the amount of stray fields in the experimental hall. The disadvantage was the cost, driven by the very large volume, and the requirement that the iron must be mechanically supported against the significant forces from the Mu2e magnetic elements. The iron-less baseline is able to provide the same magnetic field. Care must be taken to guard magnet-sensitive objects in the experiment hall.

### *7.4.5* Cryogenic Alternatives

***Thermal siphoning vs. forced flow***

The baseline cooling scheme for the Production Solenoid is thermal siphoning; the baseline cooling scheme for the Transport and Detector Solenoids is a forced flow system. The cryogenic feedboxes will be designed to accommodate both types of cooling (with some modifications). It is unlikely that forced flow cooling would be adopted for the Production Solenoid. Thermal siphoning provides liquid helium at a lower inlet temperature to the magnet and the PS requires as much temperature margin as possible because of heating from secondaries from the production target. It is possible that thermal siphoning could be adopted for the Detector Solenoid after further analysis and depending on vendor preference.

### *7.4.6* Power

***Recycle existing power converters from Tevatron***

With the decommissioning of the Tevatron, there will likely be many 5000 A/30 V power converters available. Some cost savings may be realized by using existing boxes. The significant downside is that these power converters were all purchased in the 1980's or earlier, so they have been in service for 25 years or more. While it is difficult to predict the lifetime of a power converter, the primary failure mode is deterioration of the transformer's electrical insulation. Since the cost of a power converter is relatively small compared to the cost and difficulty of repairing one the Mu2e solenoids, it seems prudent to invest in new devices. Furthermore, the existing power converters do not have the "two quadrant" capability required for Mu2e (see Section 7.3.5).

## 7.5 R&D Plan

The purpose of the solenoid R&D program is to validate critical design choices. These items can be divided into the following categories: magnet, cryo and magnetic measurement and control.

### *7.5.1* Magnet-related R&D

***Aluminum Stabilized Conductor***

The Production Solenoid, Detector Solenoid and possibly the Transport Solenoid will use aluminum stabilized superconductor. The superconductor can be a single NbTi strand or several strands combined into a cable. The aluminum, which is coextruded over the NbTi cable, provides both electrical stability and mechanical strength to the superconductor. Since the electrical, thermal and mechanical properties of the cable are integral to the solenoid performance, it is important to demonstrate that a cable can be made within specifications and at a reasonable cost.

Work has already begun on high current superconducting cable, applicable to the Production Solenoid, as part of the US-Japan HEP collaboration. 200 meters of cable, as shown in Figure 7.60, has been fabricated in Japanese industry. At the same time, there is a newly formed collaboration in Europe to develop a wide range of aluminum stabilized conductors for future applications.

The Mu2e plan is to work with both the Japanese and European collaborations to validate the technology for the Production, Detector and Transport Solenoids. The result of these collaborations will be one or two cables of the order of 100-200 meters in length that are suitable for the PS, DS and if needed, the TS. Tests will be performed on these cables to verify that they meet the specifications outlined above. Additional mechanical studies will be performed to measure the composite elastic modulus and thermal contraction coefficients. From these studies a final design and vendor for the conductor will be chosen for each conductor application.

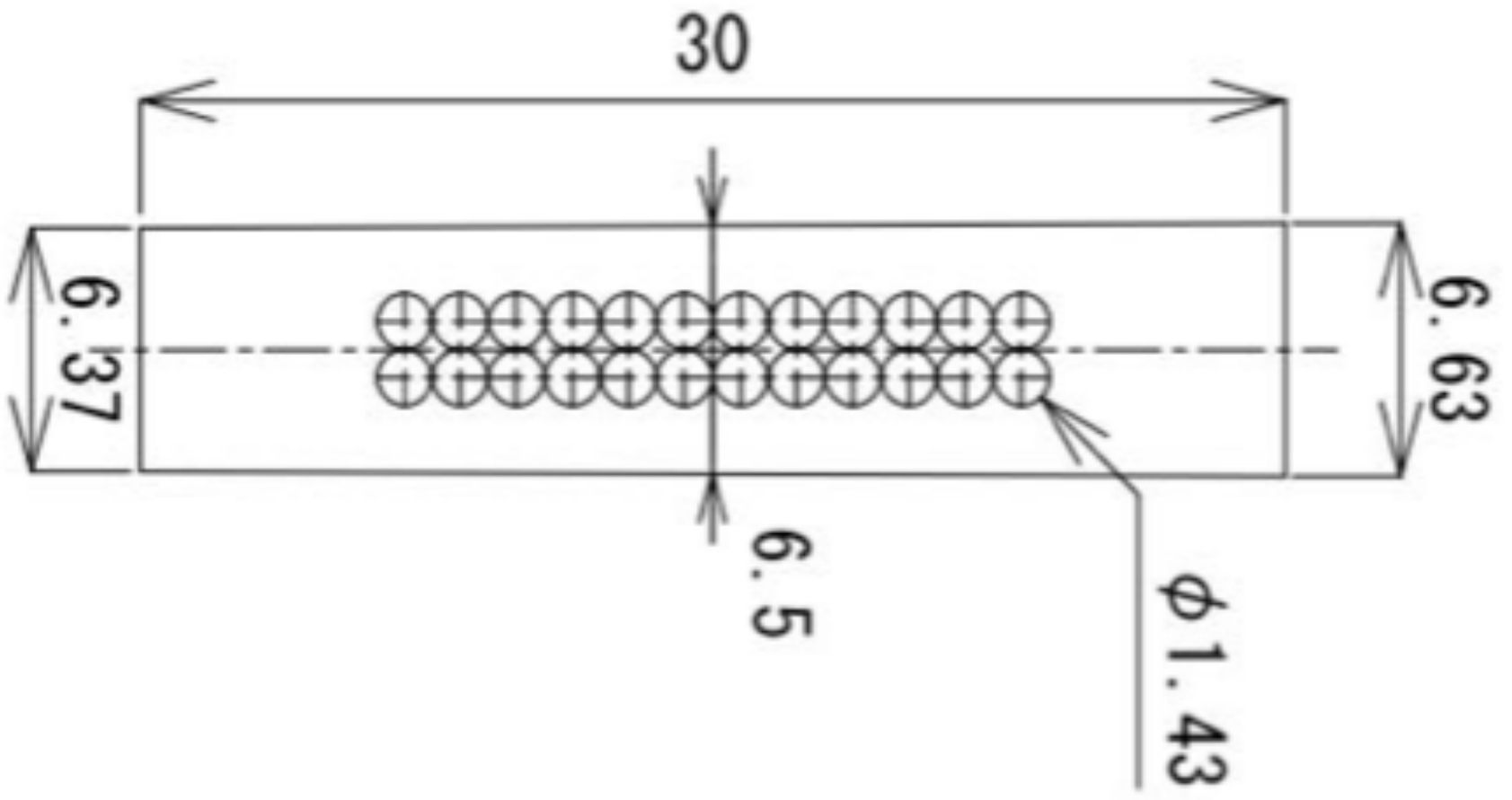

Figure 7.60. Prototype conductor for the Production Solenoid fabricated in Japanese industry.

***TS Coil Prototype***

The design concept for the Transport Solenoid toroidal sections calls for the use of solenoid rings implemented in a curved geometry. As shown in section 7.3, non-linear layout of these coil results in complicated stresses during cool down and excitation.

A two-coil prototype is proposed, using the mandrel and coil wind for the two largest coils in TS2. A schematic of the prototype coil bobbin is shown in Figure 7.61. The purpose of this coil is to validate the TS mechanical, electrical and thermal design as well as to gain experience in the fabrication of this unique coil structure. The coil will contain as many features of the final design as possible, including cooling tubes to simulate the proposed force flow conductive cooling scheme.

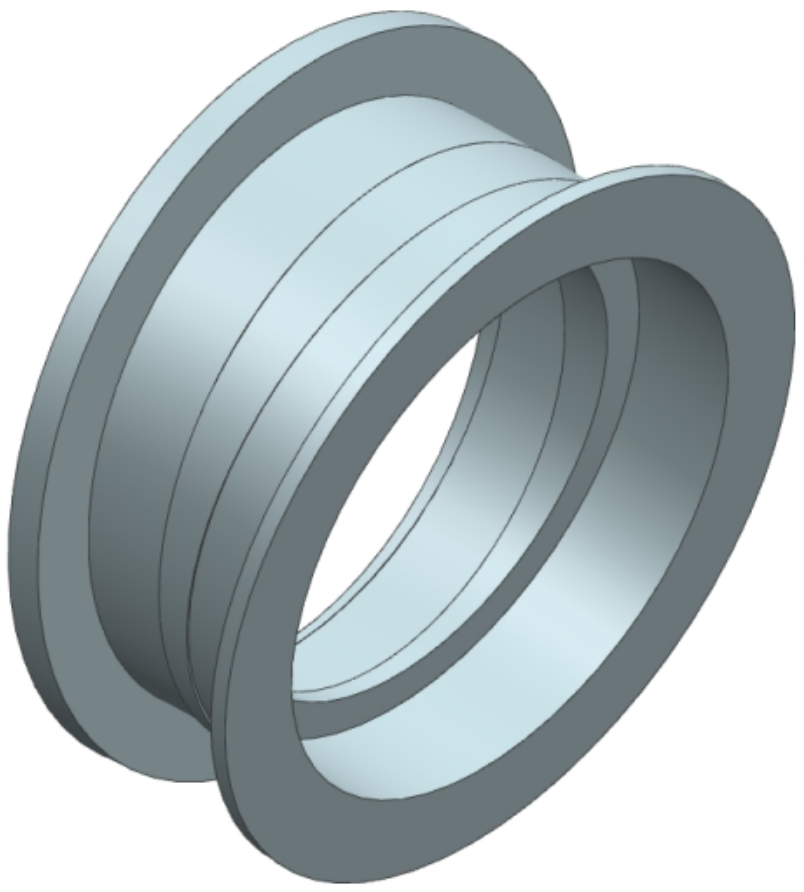

Figure 7.61. Schematic of Transport Solenoid test coil bobbin.

The completed two-coil prototype will be tested in the Fermilab Magnet Test Facility. Strain gauges will be mounted on the coil and mechanical structure to study the force during cooldown and excitation. Tests will include current excitation and magnetic field measurements. Afterwards, the coil will be sectioned to allow for visual inspection of the potting after a cool down and excitation.

***PS Coil Prototype***

Using the conductor that was developed as part of the US-Japan agreement, a prototype coil similar in cross section to the PS coils will be built in industry. A schematic of the coil is shown in Figure 7.62. As shown, this coil will incorporate several of the features used in the baseline PS coil as discussed in Section 7.3.1. A film heater will be inserted in between coil turns. This film, when connected to an external power source, can be used to study the quench and thermal properties of the coil. The plan is to build this coil in FY2012. The coil will then be shipped to Fermilab for testing. Depending on the results of these tests and the availability of PS conductor, a second coil is envisioned.

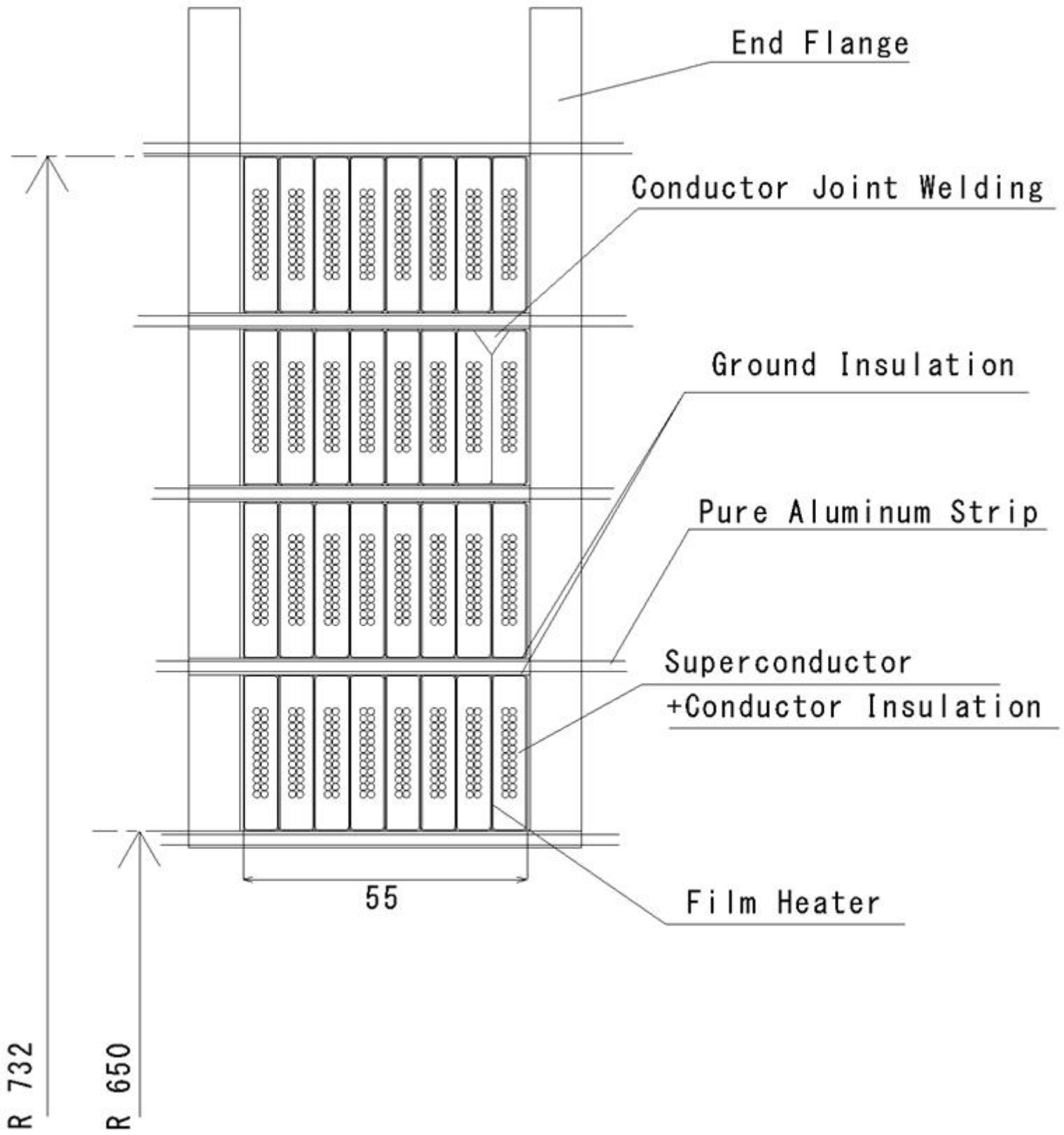

Figure 7.62. Production Solenoid prototype coil cross section.

***Splicing Superconducting Cables***

The electrical operation of the solenoids will require several electrical joints. A summary of the joints required for the solenoids is shown in Table 7.23. Connections will occur between adjacent coils within a solenoid; between coil and superconducting bus; between bus and cryogenic-to-room-temperature power leads; even inside of a coil winding a splice may be needed in the case of conductor breakage. Because of the large amounts of stored energy and the difficulty of repairing a coil once in operation, it is essential that each splice be very reliable.

| Solenoid | Intra-Coil | Magnet lead to SC Bus | SC Bus | SC bus to Feedbox | Feedbox SC to HTS Power lead |
|---|---|---|---|---|---|
| PS | 6 | 2 | 6 | 2 | 2 |
| TSu | 24 | 2 | 6 | 2 | 2 |
| TSd | 26 | 2 | 6 | 2 | 2 |
| DS | 10 | 2 | 6 | 2 | 2 |
| **Total** | | | | **106** | |

Table 7.23 Summary of Mu2e Conductor Joints

Electrical joints in superconductors encapsulated in aluminum require special consideration. The joint must be mechanically strong and be of low resistance (~1 nΩ at cryogenic temperatures) without damaging the NbTi superconductor due to overheating. Experience from the two large aluminum stabilized solenoids for CMS and ATLAS show that reliable joints can be made using TIG welding [39],[40]. In the case of CMS, joints were made between coil layers and between each of the 5 longitudinal segments. Conductor segments were routed to the outer support cylinder. The segments were oriented in a “praying hands” configuration, which allows for long joint lengths to minimize resistance. For the ATLAS central solenoid, the splices were made internal to the coils. This was accomplished by butting the conductor units together in a continuous wind, then placing a weld between an adjacent turn to the joint. This has the effect of shorting one turn (acceptable within the ATLAS field quality specs) but results in a long, low resistance joint. In both the CMS and ATLAS solenoids, joints on the order of 1 nΩ were achieved.

Splices between copper and aluminum stabilized bus (for example, at the power lead to the superconducting bus connection) can only be made if the aluminum stabilizer is removed either mechanically or by chemical etching. Care must be taken not to damage the copper coated NbTi strand. As shown in Figure 7.63 and in [41], tests were performed at Fermilab to study the chemical etching process. A combination of acid and base treatments were performed. The result is a mechanically strong splice with no visible damage to the NbTi conductor copper stabilizer. While the results are encouraging, cold resistance measurements have not yet been performed to verify the electrical integrity of the splice.

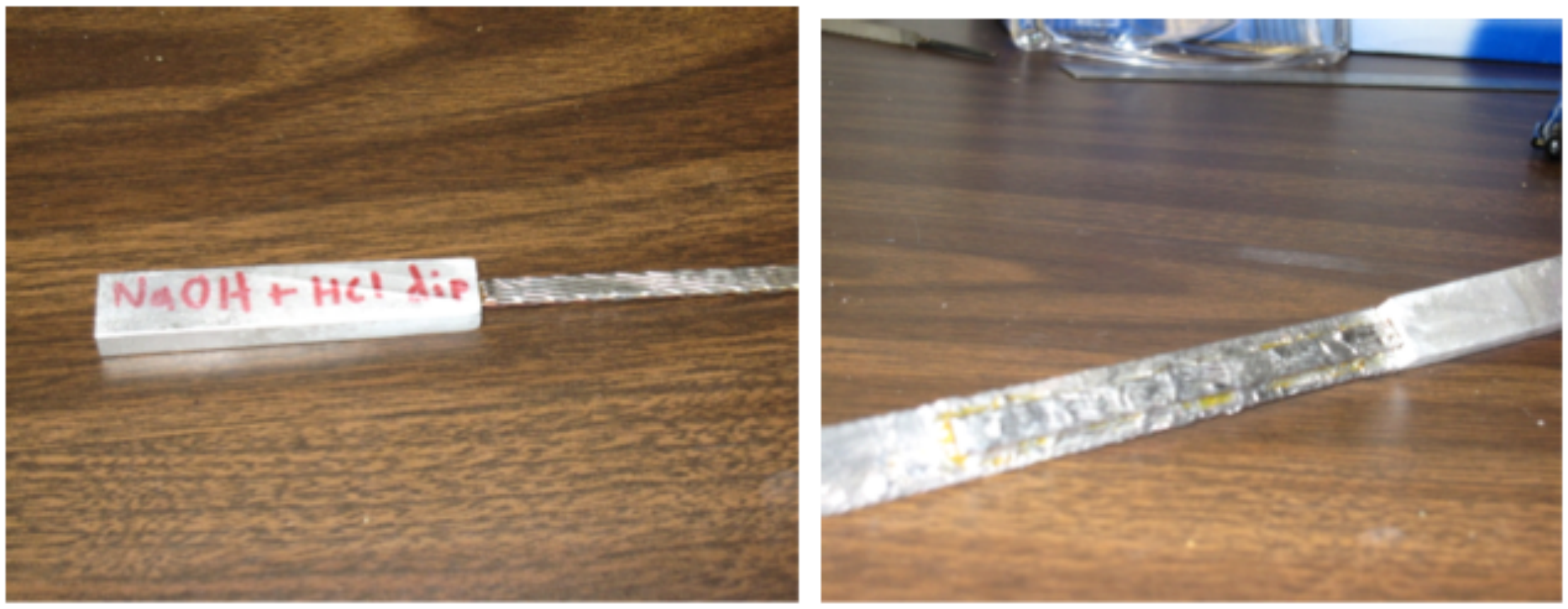

Figure 7.63 a) Chemical Etching of Aluminum Stabilized NbTi Conductor. b) Successful Solder Joint.

Using as a guide the experience on recently built detector solenoids, a test program is planned to develop in-house expertise to perform these joints. Elements of this program include:

1. Aluminum-to-aluminum joints on prototype conductor samples. Joints will be tested for resistance as well as superconductor performance.
2. Additional studies on "etched aluminum" NbTi-NbTi soldered joints will be performed, including investigation of direct Aluminum-NbTi connections. Low temperature, high current tests will be performed.
3. Tests of PS and TS prototype magnets will include both intra-coil (aluminum-aluminum) and Superconducting bus to lead (aluminum-NbTi) joints. These joints will be monitored during high current testing for resistance/power dissipation.

Results of these studies will be incorporated into the technical specifications that will eventually be passed on to vendors. For the PS and DS, these splices will be performed by the vendor and will not be easily tested until the magnet is already in a cryostat. Thus, we will require vendors to perform splice joint inspections during fabrication. Splice inspection as well as an approved weld test regimen will be incorporated into the QA/QC plan for the PS and DS vendor. For the TS, all splices will be performed at Fermilab. All TS production coils will be tested in 4-coil units prior to installation into cryostats.

### *7.5.2* Cryo-related R&D

***Cryo feed-throughs***

Magnet and superconducting power bus will require a transition from the bath-cooled lead cryostat to the indirectly-cooled cryo link. This transition will occur in the cryogenic feedbox, through the stainless steel wall of the HTS power lead Dewar, as shown in Figure 7.45. Prototype feed-throughs will be developed during the project design phase. These feed-throughs will be used during the test of the previously mentioned PS and TS prototype magnets.

***HTS Power Lead Testing***

The baseline plan is to re-cycle High Temperature Superconducting (HTS) power leads from the Tevatron. These leads were designed for 6 kA operation. However, during bench testing, one set of leads were shown to comfortably operate at close to 10 kA, making them a possible candidate for Mu2e. An additional constraint for Mu2e, in addition to the high current, is that the leads will be connected to a large inductive load. Each HTS lead needs to be retested, to verify that they can operate safely at 6 kA or higher. During a power-lead quench, the leads must not overheat or develop excessively high voltage to ground. The plan is to evaluate each set of HTS leads in the Fermilab Magnet Test Facility using an existing test apparatus.

### *7.5.3* Magnetic Measurement and Power Supply Control R&D

***Field Mapping Prototype***

As described in Section 7.3.6, each Mu2e solenoid section has a unique set of field measurement requirements. Field mappers will be developed to perform point scans over the entire magnetic volume. While each mapper will be unique, they will all utilize 3-D Hall and NMR probes that must be remotely actuated. Prior to the fabrication of the production mappers, sample 3-D Hall and NMR probes will be acquired and evaluated for this application. The Fermilab Zip Tracker may be employed to evaluate these devices for precision use over a large magnetic volume.

***Power Supply Control***

Power supply control for the solenoids is significantly complicated by the large magnetic coupling between adjacent elements. While the final list of operating scenarios is still being studied, it is likely that sub-elements of the magnetic system may be powered individually or in groups for special beam operation or diagnostics. A prototype power controller and data acquisition system will be built to understand these issues. Prior to use on the actual Mu2e solenoid system the controls will be tested using computer simulations of coupled magnetic systems and on coupled resistive loads.

## 7.6 ES&H

While there are several potential hazards associated with large, superconducting magnet systems, the hazards are well understood and have documented and proven mitigation strategies. The primary areas of ES&H concerns are listed below. A more detailed analysis for the solenoids, along with the rest of the Project, is covered in the Mu2e Preliminary Hazard Analysis Document [36].

***Radiation***

An 8 kW proton beam will interact with a target in the bore of the Production Solenoid. A sizable fraction of the generated secondary beam will be intercepted by the PS heat shield and surrounding structures. The PS and its surroundings will become radioactive and access will be restricted in accordance with well-established Fermilab procedures.

***Cryogenics***

Cryogenic hazards associated with the Mu2e solenoids will include the potential for oxygen deficient atmospheres due to catastrophic failure of the cryogenic systems, thermal (cold burn) hazards from cryogenic components, and pressure hazards. Initiators could include the failure/rupture of cryogenic systems from overpressure, failure of insulating vacuum jackets, mechanical damage/failure, deficient maintenance, or improper procedures.

An Oxygen Deficiency Hazard (ODH) analyses will be conducted in areas at risk for releasing significant amounts of inert gases. ODH mitigation could include deployment of oxygen deficiency sensors that alarm locally and read back into the Control Room at times when oxygen levels fall below preset levels (OSHA 29CFR 1910.146 considers atmospheres containing less than 19.5% oxygen by volume to be oxygen deficient). Emergency response procedures will be developed for responding to these alarm systems.

Dewar vessel filling stations and piping systems will be evaluated for ODH risk and handled as described above. Use of these systems will be limited to trained personnel. Personal Protective Equipment (gloves, goggles, face shields) will be supplied at the station and their use required. A written operating procedure will be posted at the fill station.

All piping and storage systems will be designed and installed to comply with applicable ASME and ANSI standards. The Fermilab Cryogenic Safety Subcommittee is the Authority with Jurisdiction for reviewing and approving all cryogenic systems and operational procedures.

With any cryogenic system there is a potential for thermal burns. The facility's distribution systems will be adequately marked and protected from inadvertent contact. All staff, visitors, and users of the Mu2e facility, as with all accelerator complexes, will undergo general hazard awareness training that will cover the potential hazards of cryogenic material.

***Stored Magnetic Energy***

The Mu2e solenoid system will have over 100 MJ of stored energy that must be controlled and, if necessary, safely dissipated under electrical fault conditions. Operating currents will be as high as 10 kA. Sophisticated quench protection systems have been designed to dissipate the stored energy in the solenoids in a controlled and safe fashion (see Section 7.3.5). Under quench conditions the power will be dissipated in approximately 30 seconds.

***Coupled Magnetic forces***

Forces between solenoid elements are in excess of 100 tonnes. The solenoids are designed to resist these forces, but access to the solenoids will be restricted during times when they are powered.

***Stray Magnetic Fields***

One concern associated with large magnets is the strength and extent of the fringe fields and how they may affect persons and equipment in their vicinity. Fringe fields in excess of 5 gauss are of particular concern because they could affect medical electronic devices (pacemakers), while fields over 600 gauss could impact ferromagnetic implants (artificial joints) and other material (tools).

The Production Solenoid, Detector Solenoid and the Transport Solenoids have no return yokes. There are stray magnetic fields that are as large as 10 kG (1 Tesla) falling off to about 1 kG at a distance of 100 cm [38]. Access will be restricted when the solenoids are powered and warnings will be posted for people with pacemakers.

***Pressure Safety***

In order to operate at cryogenic temperatures, the coils must be thermally insulated from the ambient surroundings. This requires a very large cryostat with an enclosed vacuum volume that encapsulates the coils. Vacuum vessels pose a potential hazard to equipment and personnel from collapse, rupture or implosion. This danger is greatest near thin windows that are often required by experimenters. The vacuum system associated with the solenoid cryostats pose little danger because of the substantial material used for the cryostats that form the vacuum barrier.

***Very heavy objects***

The solenoids are very heavy. For example, the Detector Solenoid weights ~40 tonnes. There are significant mechanical hazards associated with the transport of heavy objects from one location to another. The solenoids will be moved by crane and by guide rails in the floor of the detector enclosure.

The potential consequences of mechanical hazards include serious injury or death to equipment operators and bystanders, damage to equipment, and interruption of the program. A dropped or shifted load, equipment failure, improper procedures or insufficient training/qualification of operators could initiate these hazards.

A hazard risk assessment will be conducted to evaluate the specific hazards to personnel for each of these activities and determine the means to mitigate the hazards. Any support structure that must carry over 10 tonnes will be reviewed for adherence to standard Fermilab engineering practices. Special lifting fixtures and transports will be designed for use with the solenoids. All lifting fixtures will be engineered, fabricated, and tested in accordance with ANSI/ASME Standard B30.20 (*Below-the-Hook Lifting Devices*). All applicable Fermilab design standards governing lifting devices will also be met. The maximum allowable lifting load will be legibly marked on each fixture. Sufficient space will be available inside the enclosures for personnel to remain clear of all lifting operations. Crane training, crane interlocks, inspections, and periodic maintenance will follow the procedures listed in the Fermilab ES&H Manual (FESHM). Personnel requiring authorization to use material handling devices must complete laboratory specified training and pass a qualification practical exam conducted by a skilled operator.

***Electrical Hazards***

Significant power and current are required to operate the Mu2e solenoids. Electrical hazards include the potential for serious injury, death, and equipment damage. Exposed conductors, defective or substandard equipment can cause electrical shock and arc flash lack of adequate training, or improper procedures.

The electrical systems used to power the Mu2e solenoids and the hazards associated with them are similar to those in other experimental areas at Fermilab. Power distribution systems for the solenoids will be designed in strict compliance with applicable code. All systems will be grounded. Particular care will be provided for cable distribution to ensure code compliance with cable tray loading and content requirements. All electrical equipment and cables and cable trays will be protected against mechanical hazards. Established Fermilab ES&H procedures and Lock out/Tag out procedures will be employed to assure safety.

## 7.7 Risk

There are several potential risks associated with the acquisition of a large, superconducting magnet system. These risks are understood and have been documented and proven mitigation strategies have been developed. The primary identifiable risks are listed below with an estimate of their severity. A more detailed analysis for the solenoids along with the rest of the project is covered in the Mu2e Risk Analysis Documentation [37].

***Schedule delay (high)***

Unanticipated technical difficulties or overly aggressive schedule estimates can lead to schedule delays. This is the highest risk to the solenoid system. The solenoids are on the project critical path, so any delay to the solenoid schedule will likely result in a delay in the entire project. This can be mitigated through careful analysis of vendor's capabilities prior to awarding contracts, and monitoring vendor progress throughout the project. Built into the project plan is the ability to install and commission individual magnet components in the order that they arrive from the vendor.

***Interface problems (moderate)***

There is a moderate risk that solenoids built by different vendors might not fit together mechanically or magnetically. Detailed and carefully reviewed interface specifications, QA and QC requirements built into the vendor contracts are important elements to mitigating this risk.

***Too few or no bidders on larger solenoid or solenoid component acquisitions (moderate)***

Fabrication of complicated one-of-a-kind items requires non-standard tooling and design capabilities. This poses a risk that the number of interested vendors might be limited. To mitigate this risk, the Project has initiated communications with a number of qualified vendors to gauge interest. Additionally, conceptual designs for the solenoids have been developed using, whenever practically possible, established fabrication methods commonly employed in existing large aperture solenoids.

***Underestimate force issues on magnetic coupling of adjacent magnets (moderate)***

There are large magnetic forces that are a strong function of the fabrication details. Building adequate safety margin into the design specifications mitigates this risk. A second mitigation strategy is to perform a thorough engineering analysis of the final design of the entire system before authorizing fabrication.

***Properties of aluminum stabilizer under high radiation (moderate)***

Experiments have shown that the electrical conductivity of aluminum degrades under high levels of radiation. This could have a deleterious effect on the magnets ability to respond to a quench. The Production Solenoid, because of its large stored energy and anticipated radiation exposure, is particularly susceptible to damage. The PS uses aluminum for its stabilizer. It has been shown that warming the aluminum to room temperature repairs the damage. Partial warm-up to temperatures below room temperature results in partial repair [10]. The risk can be mitigated by scheduled room temperature or partial warm-ups, based on measured or calculated radiation deposition. To understand the required frequency of warm-ups several studies must be performed. These include studies of the effect of radiation on the conductivity of the chosen aluminum alloy, a study of the radiation deposition for the final heat shield design for the Production Solenoid, and a thorough study of the quench protection strategy for the anticipated local increases in aluminum resistivity.

***Planned cryogenic capability vs. calculated heat load (moderate)***

Cryogens will be delivered to the solenoids by reconfiguring existing Fermilab satellite refrigerators. Preliminary heat load estimates and expected refrigerator capabilities indicate that one refrigerator will be sufficient for operations. If there are additional heat loads, one refrigerator may not be enough. This risk is mitigated by performing detailed studies of a refurbished refrigerator and by installing a hot spare that can be used in parallel with the primary refrigerator to provide additional cooling capacity, if needed.

***Tevatron HTS leads cannot be used (moderate)***

The baseline plan is to use the HTS leads designed and fabricated for the Tevatron. One set of leads was used successfully in powering Tevatron circuits with currents on the order of 5 kA. The other leads have been in storage for the past 10 years. For Mu2e, the currents will go as high as 10 kA. Bench tests of one pair of HTS leads demonstrate that they can operate at this current. Others have only been tested to the nominal Tevatron operating current. This risk is mitigated by re-testing the Tevatron leads under the expected Mu2e conditions at an early stage in the Project.

***Fabrication errors in TS lead to low muon transmission(moderate)***

The “S shaped” TS is used to select muons of the proper momentum and sign, and reduce backgrounds. The field requirements are complex, and the field is difficult to measure and verify once in place. Misplaced coils might lead to unexpected fields, resulting in poor muon transmission. The risk is mitigated by performing detailed simulations, considering all realistic scenarios for misalignment. Mechanical adjustments to the coil position will be designed into the system. In situ magnetic measurement probes will be placed in the system to monitor the field.

## 7.8 Quality Assurance

Quality Assurance (QA) will play a very important role in all phases of the Mu2e solenoid program: Design, Procurement, Production and Inspection/Testing. Several of the deliverables will be “one-of-a-kind” objects critical to the experimental program and will be extremely difficult to service once the experiment is in operation. Therefore it is imperative that an effective QA program be implemented to guarantee that the solenoids meet their detailed specifications and operate reliably for the duration of the experimental program.

The Mu2e solenoid project, for both in-house and vendor-provided components, will be executed in a manner that is consistent with the Fermilab Integrated Quality Assurance (IQA) Program as well as the Fermilab Technical Division TD-2010 Quality Management Program. Details of the program will be described in the Mu2e QA plan [33]. Highlights of the QA implementation for the Mu2e solenoids are outlined below.

***Document Control and Records Management***

Control documents will be created and maintained at a level commensurate with the level of work performed. At the highest level, documents must go through version control, and change approval by Project Managers. These documents include engineering specifications, engineering drawings, travelers and operating procedures. The Fermilab Engineering Data Management System (EDMS) and the Mu2e document database will be used as the repository for these documents. Records of less formal documentation, such as presentations from in-house design meetings, vendor contacts and in-house reviews will also be stored in the document database.

It is important to keep records to provide evidence of the work performed. As much as possible, the records should be kept in an electronic format so all interested parties can access them. Records include documentation from all phases of the project: design, procurement, production and inspection/testing.

***Functional and Interface Specifications***

Engineering requirements will be conveyed to designers, procurement and production staff through functional and interface specification documents. Documents will specify the component performance requirements, mechanical tolerances and in some cases the required material properties. These documents will be formally controlled documents, subject to review and approval by project management.

***Testing***

Because of the high degree of reliability required of the solenoid components, a program of testing will be implemented throughout the production process. Specific to the solenoid fabrication, the following tests will be performed:

- Electrical measurements will be performed to assure that the coils have the proper electrical insulation to withstand the anticipated high voltage from coil to ground as well as between coil turns. Tests will be performed both at room temperature and in a cryogenic helium environment.
- Vendors will perform magnetic measurements after the coils have been fabricated and installed in cryostats but prior to shipping to Fermilab. Measurements will be compared to calculated values for acceptable field quality as defined by the functional specifications.
- Mechanical dimensions will be measured during fabrication to ensure that the magnet coils have the proper dimensions. This is critical due to the tight mechanical tolerances between components. Additional material tests will be performed to validate structural integrity at cryogenic temperatures.

***Calibration and Monitoring***

Instrumentation will be developed and/or commercially acquired for evaluating the electrical, magnetic and mechanical properties of the solenoid components. This instrumentation will be periodically calibrated. Calibration records will be kept along with the instrumentation data in an approved database. A subset of the instrumentation will be used post-production, as part of final vendor tests prior to shipment, acceptance tests upon arrival at Fermilab, and during installation and commissioning.

***Review Process***

Progress on design, procurement and production will be periodically evaluated through a series of reviews. Reviews serve to verify that the project is ready to proceed to the next phase. Reviews will range from formal Mu2e project reviews to informal in-house reviews. Review presentations, review evaluations and review responses will be reported to the Mu2e project management. They will also be recorded in the Mu2e document database.

***Vendor Responsibility***

Key provisions of the Mu2e QA plan will be written into contracts with vendors. The level of QA implementation will be sized according to the project risk associated with the procurement. For procurements greater than $250k (the PS, DS, TS, power

supplies, cryo feed boxes, etc.) vendors will be required to submit their own written QA plan.

## 7.9 Value Management

The Solenoids are the major cost driver for the Mu2e project. Therefore, it is essential to provide the most economical solutions while not compromising the project's success. Because the solenoids are also the schedule driver for Mu2e, value management must be applied to the design at an early stage. Value management is an ongoing enterprise that is continuously applied to the design process. The current conceptual design includes a number of features that have resulted from value management considerations. Value management principles will continue to be applied as the design advances.

The following are some of the areas where value management has already been applied to the design and some areas where further value management principles may be applied.

***Production Solenoid***

The length of the Production Solenoid was shortened when a new model for pion production, based on data from the HARP experiment, was implemented in the Mu2e Monte Carlo. The new production model indicated that an adequate number of stopped muons could be obtained with a shorter Production Solenoid. Reducing the length of the Production Solenoid not only reduces the cost of the solenoid itself, but of the heat shield inside of the PS as well. It also simplifies the cooling system. A final set of studies will be performed to optimize the length before fabrication.

Additionally, the nominal peak field of the solenoid was lowered from 5.0 T to 4.6 T. This is accomplished by reducing the maximum number of coil layers from four to three. While this reduces the muon yield by about 10%, it saves money by simplifying the coil fabrication and reducing the required amount of superconductor.

***Transport Solenoid***

A decision was made to use the same conductor for all sections of the Transport Solenoid. Conductor can be purchased in bulk at a lower unit price. All TS magnets will operate at the same current, saving on the spare inventory of power leads and power converters.

***Detector Solenoid***

A potential cost driver for the Detector Solenoid is the required field uniformity in the downstream section occupied by the tracker. Based on a set of initial studies of momentum resolution in the tracker the uniformity requirement in the spectrometer section of the DS was relaxed from 0.1% to 1%. In the calorimeter section of DS the uniformity requirement was relaxed to 5%, allowing the DS length to be reduced by about 1 meter. Relaxing these requirements has also made it possible to move from a two-layer coil to a single layer coil, resulting in significant cost savings.

***Cryogenic Distribution***

Liquid helium will be distributed from a Fermilab-supplied refrigerator system to four identical cryogenic feedboxes. This will reduce design costs as well as the number of required spares. The tradeoff between this approach and building feedboxes to meet the specific cryogenic requirements of each sub element (PS, TSu, TSd, and DS) will be evaluated.

Studies will be performed to confirm that the existing Tevatron HTS power leads can be used for Mu2e's application. The alternative is to purchase new HTS leads specifically designed for the Mu2e feedboxes.

***Power Distribution***

To power the solenoids, 6 kA and 2 kA power converters will be purchased. The 6 kA boxes can be used in parallel for the Production and Detector Solenoids. The Transport Solenoid is configured for all sub elements to operate at 2 kA. This will reduce the number of required spares. The cost benefit of both air-cooled and water-cooled bus work and extraction circuits will be evaluated.

***Magnetic Measurement***

The magnetic field inside the various regions of the solenoids will be measured and monitored using a series of 3-D Hall probes (Section 7.3.6). Studies will be performed to determine the optimum number of sensors required to meet the needs of the experiment. A major cost driver for this system will be the accuracy required of the Hall probes and the position resolution required for the Hall probe arrays. These requirements will be evaluated together with the tracking group and a deployment strategy developed that balances the cost benefits vs. program risks.

## 7.10 References

[1] R. Coleman et al., "Mu2e Magnetic Field Specifications," Mu2e-doc-1266.
[2] M. Lamm, "Production Solenoid Requirements," Mu2e-doc-945.
[3] M. Lamm, "Detector Solenoid Requirements," Mu2e-doc-946.
[4] M. Lamm, " Transport Solenoid Requirements," Mu2e-doc-947.
[5] T. Peterson, "Mu2e Cryogenic Distribution Requirements," Mu2e-doc-1244.
[6] R. Ostojic, "Mu2e Quench Power Supply Systems Requirements," Mu2e-doc-1237.

[7] R. Ostojic, "Mu2e Quench Protection System Requirements," Mu2e-doc-1238.
[8] D. Kawall et al., "Magnetic Field Measurement Requirements," Mu2e-doc-1275.
[9] V. Kashikhin, "Design Studies of the Mu2e Production Solenoid," Mu2e-doc-1110.
[10] C. Berriaud et al., IEEE Transactions on Applied Superconductivity, **20** No. 3, 1408 (2010).
[11] Bruce S. Brown, Journal of Nuclear Materials **97,** 1 (1981).
[12] A. Klebaner and J. Theilacker, "Mu2e Cryogenic Cost Estimate," Mu2e-doc-276.
[13] T. Nichol, "Production Solenoid Cryostat and Suspension System," Mu2e-doc-1249.
[14] N. Mokhov and V. Pronskikh, "Mu2e Production Solenoid MARS15 Studies," Mu2e-doc-1060.
[15] T. Page, "Magnet iron, Installation and Ancillary Equipment," Mu2e-doc-1255.
[16] A. Yamamoto et al., IEEE Transactions on Applied Superconductivity, **Vol. 9**, No. 2, 852 (1999).
[17] L.Rossi, M.Sorbi, INFN/TC-04/13 (2004).
[18] ATLAS Central Solenoid Technical Design Report, ATLAS Magnet Project TDR Vol. 4, April 1997.
[19] G. Ambrosio et al., "TS123u Design Study", Mu2e-doc-2075.
[20] G. Ambrosio, "TS3 pre-CDR Design," Mu2e-doc-959.
[21] R. Yamada, CDF-SDS-12 (1982).
[22] M. Wake, "Quench Calculations for the DS Magnet", in preparation.
[23] B. Wands, "Mechanical Support of the DS Cryostat and Cold Mass Support System," Mu2e-doc-1274.
[24] I. L. Horvath et al., "Aluminum Stabilized Superconductor for the BELLE Detector at KEK-B", Proceedings of the Fifteenth International Conference on Magnet Technology, p. 1044-1047.
[25] R. Bossert, "Detector Support and Installation System", in preparation.
[26] S. Feher et al., "HTS Power Leads for the BTeV Interaction Region", PAC'05, Knoxville, 1147 (2005).
[27] S.Feher, "WB 5.11 Muon Beamline System Integration, Test and Analysis", Mu2e-doc-1168
[28] G. Ambrosio, "TS1-2 Design Study," Mu2e-doc-1239.
[29] "MECO Superconducting Solenoid System Conceptual Design Report", MIT PSFC, June 6 (2002).

[30] V. Pronskikh and N. Mokhov, “Optimization of a Mu2e Production Solenoid Absorber using MARS15,” Mu2e-doc-1133.
[31] M. Aleksa et al., Journal of Physics Conference Series **110** 092018 (2008).
[32] V. I. Klyukhin et al., IEEE Trans. Appl. Supercond.**18**, No. 2, 395 (2007).
[33] R. Ray, “Quality Assurance Program for the Mu2e Project,” Mu2e-doc-677.
[34] N. Andreev et al., “WBS 5.10, Detector Support and Installation System,” Mu2e-doc-1383.
[35] G. Ambrosio, “TS3 pre-CDR Design,” Mu2e-doc-1134.
[36] R. Ray, “Preliminary Hazard Analysis Document,” Mu2e-doc-675.
[37] M. Dinnon et al., “Risk Register,” Mu2e-doc-1463.
[38] M. Lopes, “Mu2e Stray Magnetic Field Maps”, Mu2e-doc-1381.
[39] A. Yamamoto et al. “Progress in ATLAS CENTRAL Solenoid”, IEEE Trans. Appl. Supercond. **10**, No. 1, 353 ( 2000).
[40] P. Fabbricatore et al., “Electrical Characterization of S/C Conductor for the CMS Solenoid”, IEEE Trans. Appl. Supercond. **15**, No. 2, 1275 (2005).
[41] N. Andreev et al., “Superconducting Solenoid Magnet R&D for Muon Conversion Experiments”, Mu2e-doc-2237.

# 8 Muon Beamline

## 8.1 Introduction

The fundamental goal of the Muon Beamline is to deliver a stopped muon rate of approximately a few times $10^{10}$ per second to the muon stopping target, located in the Detector Solenoid (DS), and to reduce the background in the tracker, calorimeter and cosmic ray veto detectors to a level sufficient to achieve the desired experimental sensitivity (see Chapter 3).

It is important to identify all possible background sources and equip the Muon Beamline vacuum space with elements that can produce a muon beam with the requisite cleanliness while guiding negatively charged muons to the stopping target. The stopping target and the surrounding absorbers must be designed to maximize the capture of muons and transmission of conversion electrons to the detector while minimizing harmful particles that can result in background.

The Muon Beamline is essentially a vacuum space, which serves as a free path for negatively charged muons within the desired momentum range. The muons spiral to the detector area along the streamline of a B-field (created by superconducting solenoid magnets) parallel to the direction of the muons. The S-shaped muon channel is surrounded with coils that form a toroidal B-field in the two curved sections and solenoidal fields in the three straight sections. The upstream toroidal field, with strategically placed collimators, filters the particle flux producing a momentum- and charge-selected muon beam, with good reduction in contamination from positively charged and neutral particles. A thin Kapton window resides in the central straight section of the Muon Beamline. The window stops antiprotons from reaching the stopping target and creating background. It also serves to separate the upstream and downstream vacuum volumes to prevent radioactive ions or atoms from the production target from contaminating the detector solenoid volume. The muon stopping target consists of thin aluminum discs placed in the detector region of the muon channel in a graded magnetic field. The muons have high efficiency for stopping in the muon stopping target. Muons not stopped in the target mostly bypass the detectors and are transported to the muon beam-stop. Protons and neutrons originating from the muon capture process in the stopping target, final collimator and muon beam stop are attenuated by absorbers to minimize detector background rates.

The Muon Beamline Level 2 system has been divided into 10 Level 3 sub-projects:

- WBS 5.2 Vacuum system
- WBS 5.3 Collimators
- WBS 5.4 Beamline Shielding
- WBS 5.5 Stopping Target
- WBS 5.6 Stopping Target Monitor
- WBS 5.7 Proton Absorber
- WBS 5.8 Muon Beam Stop
- WBS 5.9 Neutron Absorber
- WBS 5.10 Detector Support and Installation
- WBS 5.11 System Integration, Tests and Analysis

In Section 8.2 the requirements for the Muon Beamline are presented. In Section 8.3 the proposed design and the alternatives to the proposed design are discussed. In Section 8.4 Risk Management for the Muon Beamline is described. Section 8.5 discusses Quality Assurance, Value Management is described in Section 8.6, ES&H is discussed in Section 8.7 and the Muon Beamline R&D plan is described in Section 8.8.

## 8.2 Requirements

In this section the overall requirements for the Muon Beamline are described. A detailed description of the requirements for each Level 3 deliverable can be found in the various Requirements & Specifications documents [1] - [10].

### *8.2.1* Vacuum System

A detailed description of the requirements and specifications for the Mu2e vacuum system can be found in [1].

The Muon Beamline must be under vacuum. There are a number of factors, listed below, that must be considered to determine the requirement on vacuum system pressure.

- The muon beam must be efficiently transported from the production area to the stopping target with minimum loss of the muons in the momentum range below 50 MeV/c.
- The background rate from beam particles scattering off residual gas molecules shall be less than the background generated by electrons from muon decays in orbit.
- The vacuum level must be below the Townsend limit to prevent electrical discharge from detector high voltage.

- Multiple scattering and energy loss must be minimized for conversion electrons to limit the overall contribution to the energy resolution of the Tracker to less than 10 keV, which is negligible compared to the required Tracker resolution.

It has been shown that by keeping the vacuum ≤ $10^{-4}$ Torr for the downstream vacuum volume the physics requirements can be met [1]. Consequently we require $10^{-4}$ Torr or lower pressure at the downstream part of the vacuum volume for the Muon Beamline. For the upstream part of the vacuum volume ≤ $10^{-1}$ Torr pressure is sufficient.

The detector vacuum volume must be isolated from the upstream part of the beamline to prevent migration of radioactive molecules generated at the production target area to the detector area.

The vacuum system includes a closure at the upstream end of the Production Solenoid. The closure shall contain a vacuum window to allow beam protons that are not absorbed by the production target to exit the PS. The Proton Beam Exit window will also serve as the Extinction Monitor window. Another vacuum window will be incorporated into Production Target Extraction port, centered on the solenoid axis, for target insertion/removal via remote handling. An additional port for a radiation-hard camera will allow safe visual monitoring of the production target. The vacuum closure must also provide a vacuum pump duct. The duct must have a conductance that is suitable to provide the required evacuation when the vacuum pumps are located some distance away in regions that allows adequate shielding from radiation and magnetic fields.

The Detector Solenoid vacuum enclosure must support all signal, power, high-voltage, detector gas and cooling lines from the interior of the DS to the outside. Furthermore, the enclosure must accommodate the muon beam stop so that it may lie in the region where the magnetic field falls to zero; this prevents many charged particles from returning upstream towards the detectors. The DS vacuum enclosure requires a thin window at its center for the muon stopping target monitor system.

The vacuum system must work in concert with the detector high voltage (HV) distribution systems to either isolate the HV from the vacuum or ensure that the HV is turned off before sparks occur in the event of a sudden vacuum loss.

The pump down time is to be of order several hours so as not to significantly impact experimental operations.

### *8.2.2* Collimators

The Muon Beamline must be equipped with collimators to define the muon beam and a thin window to absorb antiprotons. The collimators are required to perform a number of functions.

- Select the charge and momentum range of particles. The muon beam line suppression of electrons (the ratio of electrons that enter the TS to the number that traverse the muon beam line and enter to the DS) with energies above 100 MeV needs to be suppressed by a factor of 106 or better. The probability for particles with the wrong charge to pass through the central collimator (COL3) must be less than 2% for every proton that hits the production target.
- Antiprotons must be absorbed in a thin window located between TSu and TSd. The annihilation products must be absorbed in the collimation system before reaching the Detector Solenoid. The antiproton window must also prevent radioactive molecules from migrating from the production target region to the Detector Solenoid. The window must be thick enough to prevent antiproton transmission and molecules migrating through the window. On the other hand, the window needs to be thin enough to attenuate the muon yield by no more than 15%.
- The detectors should be shielded from neutrons and low momentum muons that do not hit the stopping target.
- The heat load in the Transport Solenoid coils from particle debris originating from the production target must be minimized. The peak power density deposited into the TS coils must be less than 5 μW/g at the highest proton beam intensity.

A detailed description of the requirements and specifications for the Collimators can be found in [2].

### *8.2.3* Muon Beamline Shielding

The S-shaped Transport Solenoid filters unwanted charged and neutral particles from the beam. Shielding might be required both inside and outside the Transport Solenoid to further reduce the yield of neutral particles wandering downstream (like thermal neutrons) and to shield the superconducting coils. An intense neutral particle beam hits the first curved section of the Transport Solenoid. It is necessary to reduce the thermal load in the superconducting coils to less than 5 μW/g at any point within the coil volume. A detailed description of the requirements and specifications for the Muon Beamline Shielding can be found in [3].

### *8.2.4* Muon Stopping Target

The stopping target material, shape and location must be chosen to stop as many muons as needed (~$10^{18}$) while minimizing the energy loss and maximizing the acceptance for conversion electrons. The stopping target requirements are strongly coupled to the choice of the experimental layout (B-field requirements, Tracker and Calorimeter design and placement, neutron and proton absorbers design and placement etc.). The detailed description of the requirements and specification that takes into account the current Mu2e design is described in [4].

### *8.2.5* Muon Stopping Target Monitor

A Muon Stopping Target Monitor is required to measure the muon stopping rate during the live time of the experiment by measuring the characteristic X-ray spectrum from the muon atomic cascade process. The capture rate is derived from the stopping rate. The requirement is to achieve a measurement accuracy of 10%. The actual choice of the measurement technique strongly depends on the experimental layout (target material choice, dimension and location, possible locations for the monitor etc.). The detailed description of the requirements and specifications that takes into account the current Mu2e design is described in [5].

### *8.2.6* Proton Absorber

Protons that result from muon captures on target nuclei in the muon stopping target can negatively impact the performance of the tracker and the calorimeter and possibly lead to backgrounds. The proton absorber must meet the following three requirements:

- Reduce the rate of stopping target protons that reach the tracker to a level where reconstruction of electron tracks is reliable and robust. For the current tracker design this rate is 60 kHz.
- Minimize the energy loss and multiple scattering of conversion electron candidates at 105 MeV that pass through the proton absorber below 300 keV.
- Ensure by the shape and location of the proton absorber that muons do not stop on the proton absorber.

The first two requirements work contrary to one another. A thicker absorber will stop more protons but will further degrade the momentum resolution. A detailed optimization of the stopping target, absorbers and tracker is required to finalize the proton absorber design. Further detailed requirements and specifications for the Proton Absorber can be found in [6].

### *8.2.7* Muon Beam Stop

The Muon Beamline must be equipped with a Muon Beam Stop to absorb the energy of secondary beam particles (primarily muons and electrons) that reach the end of the Detector Solenoid to reduce the background to the detectors from the muon decays and captures in the beam stop. This is especially important during the detector live-time, which begins about 700 ns after the proton micro bunch hits the target. Near the downstream end of the Detector Solenoid the uniform field transitions to a graded field that drops off along the beam direction; this is a critical feature of the muon beam stop. The field gradient reflects most low energy charged particles produced in the beam stop away from the detectors. A detailed description of the requirements and specifications for the Muon Beam Stop can be found in [7].

### *8.2.8* Neutron Absorbers

Cosmic rays can cause backgrounds and must be vetoed (Section 3.5.9). A Cosmic Ray Veto (CRV) has been designed for Mu2e that surrounds the Detector Solenoid. The CRV must operate in an environment rich with neutrons from the production target, the final collimator in the Transport Solenoid, the muon stopping target and the muon beam stop. About 1.2 neutrons are produced for each muon that undergoes nuclear capture in the muon stopping target. The neutrons range from thermal to a few tens of MeV. Neutrons from the production target and muon beam transport must be shielded to below the rate of neutrons from muon capture in the stopping target. The preferred alternative design for the CRV is a detector sensitive to these neutrons (scintillator), though the pattern of energy deposition in the multiple layers of active detector will be distinct from cosmic ray muons in most cases. Sufficient absorbing material must be introduced to reduce the overall rate in the CRV and to limit the number of neutrons mistaken for muons.

Neutrons that result from muon captures on target nuclei in the muon stopping target can negatively impact the performance of the tracker and the calorimeter and possibly lead to backgrounds. Further simulations are needed to show whether it is required to place Neutron absorbers in the Detector Solenoid bore together with the muon beam stop to reduce the accidental hits to $6 \times 10^6$ hits/sec/m$^2$ in the tracker.

The Stopping Target Monitor is sensitive to radiation damage caused by neutrons emerging from the muon stopping target, the production target and the beam stop. It is important to reduce this neutron rate to be able to operate the detector for at least a month continuously without annealing the detector crystals. We anticipate having two detectors so that one is in operation when the other is being annealed.

A detailed description of the requirements and specifications for the Neutron absorbers can be found in [8].

### *8.2.9* Detector Support and Installation System

The Detector Support Structure is required to transport and align components within the Detector Solenoid warm bore. The muon stopping target, proton absorber, tracker, calorimeter, and muon beam stop must all be moved into position and aligned to the standard Mu2e coordinate system. The components vary significantly in mass (from less than 160 g to over 4000 kg) as well as required alignment accuracy. These components must be supported by the inside wall of the Detector Solenoid cryostat and be moved accurately and safely inside the bore. Physics requirements dictate the overall size, location and placement accuracy of the individual components within the DS bore. In addition, the support structure must not impede particle trajectories or lead to an enhancement of detector rates or physics background as a result of interacting particles. A detailed description of the requirements and specifications for the Detector Support Structure can be found in [9].

## 8.3 Proposed Design

### *8.3.1* Vacuum System

The Beamline Vacuum system consists of two distinct vacuum volumes, an upstream volume labeled PS + TSu and a downstream volume labeled TSd + DS. A sealed vacuum window located between Transport Solenoid sections physically separates the two volumes. The vacuum window also serves to stop antiprotons that can lead to a dangerous physics background. The vacuum volume is defined by the inner cryostat walls of the various solenoids, the Production Solenoid enclosure on the upstream end, and the Detector Solenoid enclosure on the downstream end. Connections between solenoids are made with flexible bellows.

The two vacuum volumes are connected by a bypass valve that only opens during pump down or vacuum loss to minimize the pressure differential across the vacuum window.

Not all components of the vacuum system will be ASME code stamped, but they will be designed and manufactured according to the 2010 ASME Boiler and Pressure Vessel Code Section VIII. All components will also be designed according to the standards of the Fermilab Engineering and ES&H Manuals, particularly Vacuum Vessel Safety chapter 5033.

The Production Solenoid Vacuum Enclosure, shown in Figure 8.1, is a 316L stainless steel weldment bolted directly to the upstream end of the PS cryostat. It closes the PS + TSu vacuum volume, provides a place to connect vacuum pumps and provides openings and mounting locations for several PS vacuum windows and feed-throughs.

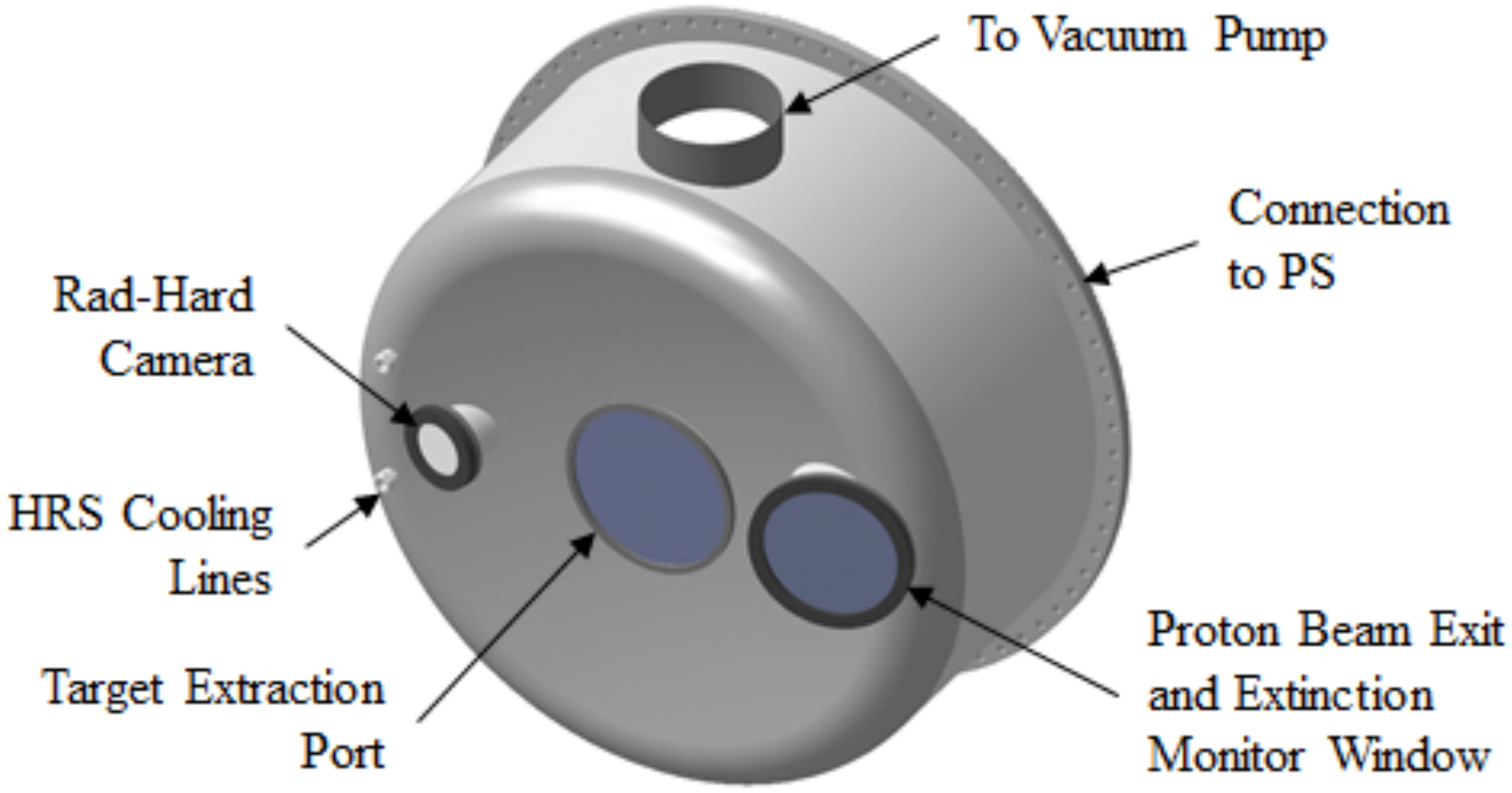

Figure 8.1. Model of the vacuum enclosure on the upstream end of the Production Solenoid (PS) showing the vacuum pump piping, windows, and feedthroughs.

Radiation levels in the Production Solenoid are expected to be high enough to require metal vacuum seals at all locations. One 406.4 mm port with circular cross-section is provided to connect to vacuum and roughing pumps located some distance away. Two 34.93 mm diameter feed-throughs are provided for water cooling lines for the Production Solenoid Heat and Radiation Shield (HRS).

The proton beam exit and extinction monitor window is 400 mm diameter × 2.84 mm thick Grade 5 titanium. The window for a radiation hard camera is 200 mm diameter × 25.4 mm thick fused quartz. The target extraction window is 500 mm diameter × 3.58 mm thick Grade 5 titanium incorporated into the design of the Production Target extraction port. After activation, remote handling equipment will access the production target through this port, should a replacement be required.

The Detector Solenoid Vacuum Enclosure, shown in Figure 8.2, is a two-piece 316L stainless steel weldment bolted directly to the downstream end of the DS cryostat to close the TSd + DS vacuum volume. The Vacuum Pump Spool Piece (VPSP) bolts directly to the downstream end of the DS cryostat and remains fixed, becoming an extension of the cryostat and the internal rail system. Four 406.4 mm diameter flanged ports are provided for direct connection of vacuum pumps and four 304.8 mm diameter ports are available to connect to roughing pumps located some distance away.

The removable Instrumentation Feed-Through Bulkhead (IFB) provides an opening and mounting location for a Stopping Target Monitor (STM) vacuum window. The STM window is 100 mm diameter × 1.09 mm thick 316L stainless steel. The IFB also provides an axial connection for the Muon Beam Stop (MBS) and a feed-through area for the many cables, gas lines, and cooling lines for the detectors mounted inside the Detector Solenoid. Metal seals will be used for the large diameter connections, but radiation levels in this region are expected to be low enough to allow O-ring seals on feed-throughs.

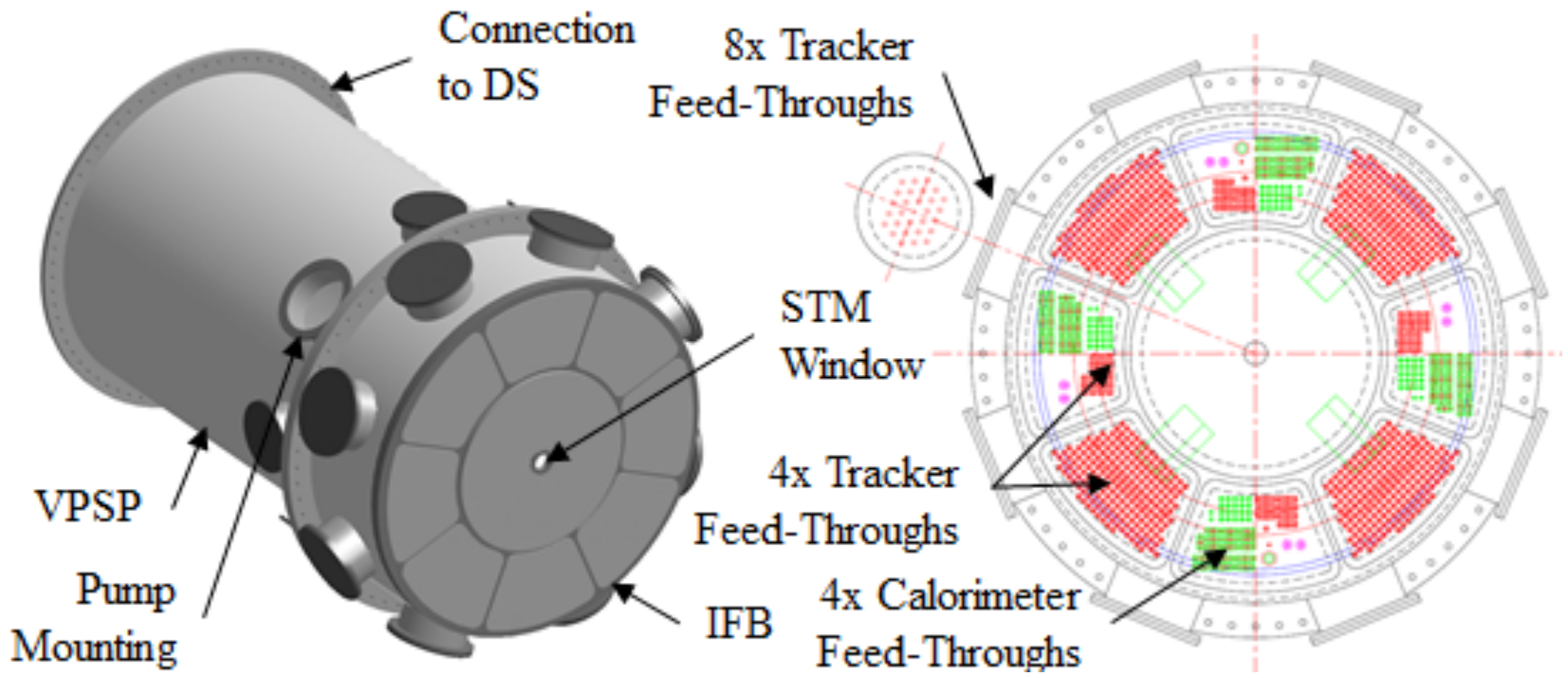

Figure 8.2. Model of the vacuum enclosure on the downstream end of the Detector Solenoid (DS) showing the vacuum pump mounting ports and window. Layout of the downstream end of the Instrumentation Feed-Through Bulkhead showing all Tracker, Calorimeter, and Solenoid feed-throughs.

The Transport Solenoid interconnects, to the Production Solenoid at one end and the Detector Solenoid at the other end, are accomplished by flanged, bolted connections as shown in Figure 8.3. Bellows are built into these connections to allow movement of the TS relative to the PS and DS. Metal seals will be used in these large diameter connections.

The interconnects between the Transport Solenoid sections and the antiproton stopping window module are accomplished by flanged, bolted connections as shown in Figure 8.4. Metal seals will be used in these large diameter connections.

The antiproton stopping window module is designed to be removable for service or replacement, and a bellows is built into this connection design to facilitate this.

Because the space between TSu and TSd is limited, appropriate fixturing and handling equipment will be provided to protect the window, seals, and sealing surfaces during this process.

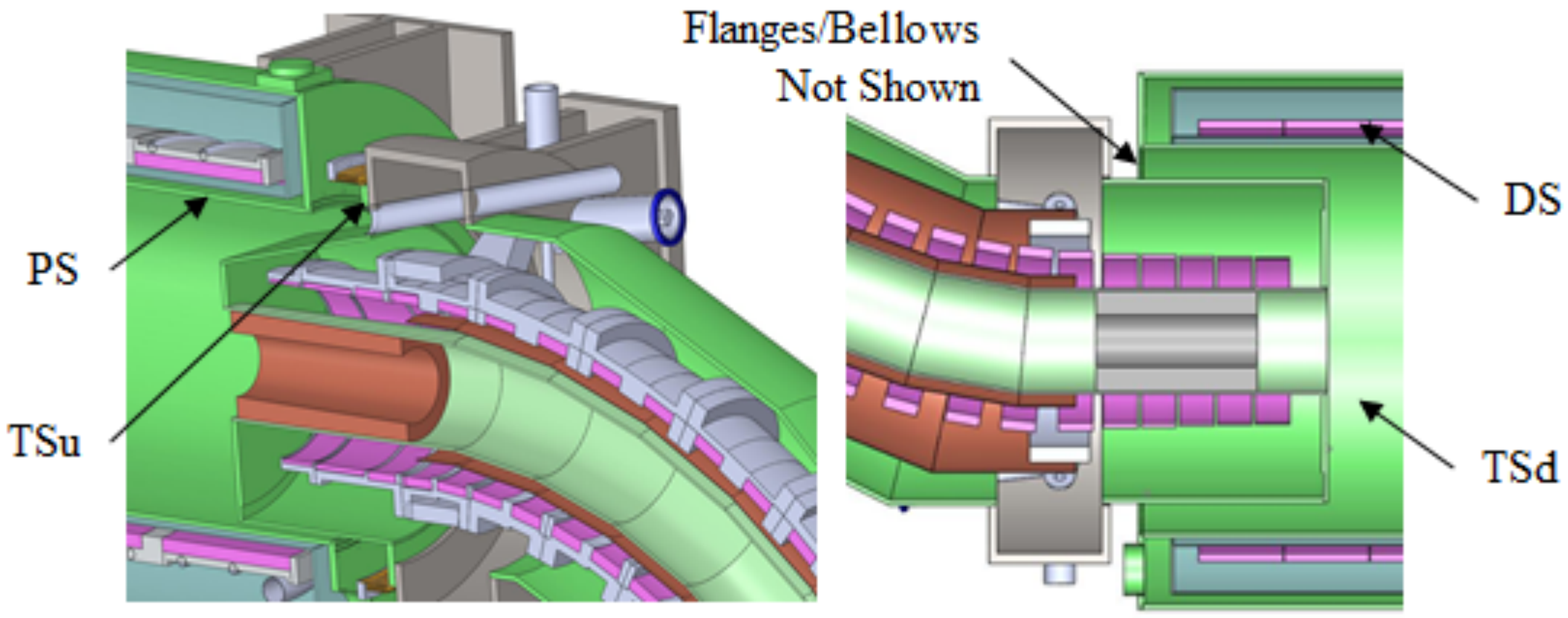

Figure 8.3. Models of the connection of the first Transport Solenoid section (TSu) to the Production Solenoid (PS), and the connection of the last Transport Solenoid section (TSd) to the Detector Solenoid (DS).

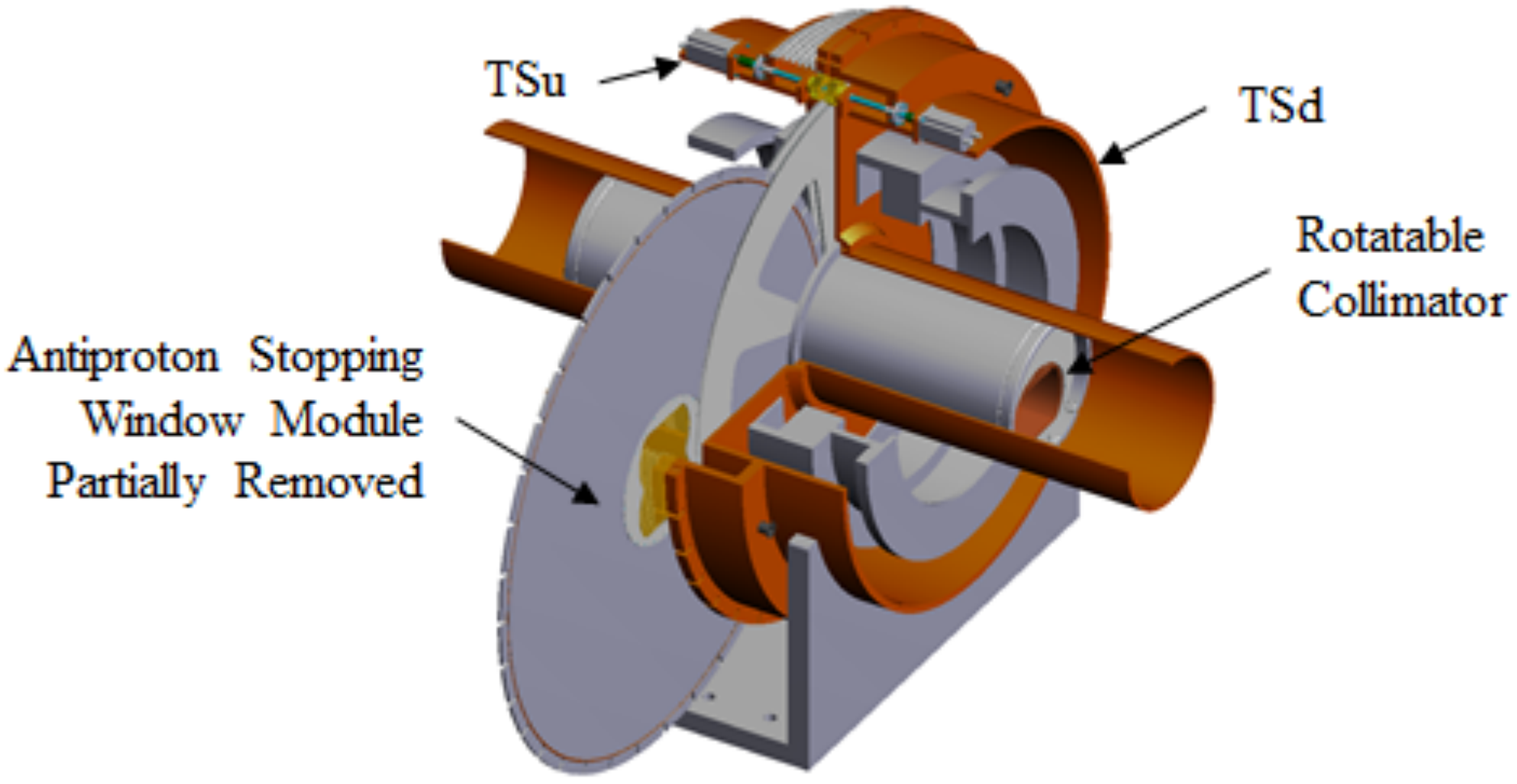

Figure 8.4. Models of the connection of the Transport Solenoid sections (TSu and TSd) to the Antiproton Stopping Window Module.

The flanges welded onto TSu and TSd must accommodate the collimator rotating mechanism, as well as the bypass piping necessary to maintain the 1 torr or less pressure differential across the antiproton stopping window. A design concept for the bypass piping is shown in Figure 8.5. A control system will be required to safely operate the bypass system and protect the antiproton stopping window from pressure damage.

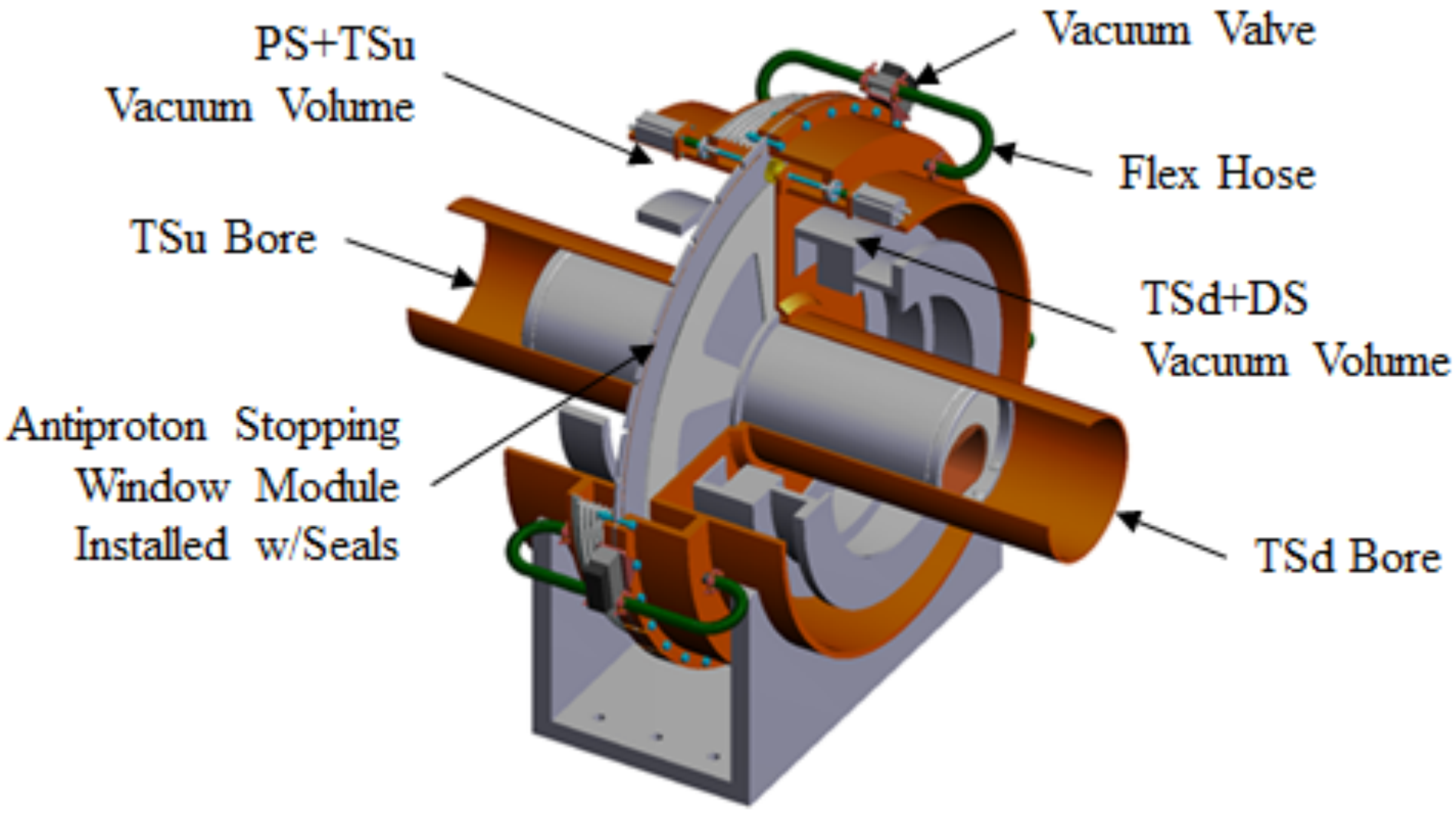

Figure 8.5. Model showing the proposed method of protecting the Antiproton Stopping Window from pressure differential using a system of bypass pipes and control valves.

Identification of the appropriate gas loads, establishment of the specific vacuum equipment, specification of appropriate materials, and definition of interfaces are all factors in selecting the external vacuum system components. To satisfy the requirements the nominal pressure in the detector solenoid must be adjustable from $10^{-4}$ to $\sim 10^{-5}$ Torr. Gas loads are the most difficult factor to predict in a system like the Mu2e Beamline. Therefore, the design allows for a flexible pumping capacity, incorporating appropriate sensors, pumps, ducts, valves and gas abatement with function, assembly, installation, decommissioning, and ES&H requirements in mind.

The two vacuum volumes share the same dry screw roughing pump. Each vacuum volume has a magnetic levitated hybrid viscous drag-turbo molecular pump to speed the transition to base pressure. Gate valves are located across the intake and exhaust of each high vacuum pump. Pressure sensors are mounted between the gate valves and pumps, near each pump intake.

Materials under vacuum in the PS + TSu volume include copper, brass, tungsten, and stainless steel and the TSd + DS volume contains Kapton, stainless steel, copper, lead, aluminium, lead tungstate crystal, and polyethylene. Nearly every surface of the PS + TSu volume is metal; however, the heat and radiation shield has some obscured surfaces and may turn out to be a minor source of virtual leaks. The challenge to pumping down the PS + TSu volume will simply be the pump-volume distance. Vacuum equipment for the PS + TSu volume must be located far from the intense radiation and strong magnetic field near the Production Solenoid leading to longer ducts and reduced effective pumping speed. In the TSd + DS volume, material out-gassing will present a significant load for the first ten hours at base pressure, then the dominant load will be steady-state leaks from the tracker.

A partial description of the gas loads in the TSd + DS volume has been compiled. The pump-vacuum vessel distance for PS + TSu is driven by the need to protect the equipment from the radiation and magnetic fields present. The DS vacuum volume has two cryo pumps: one for normal operation and one spare that also operates as a backup pump during regeneration. While gas loads in the PS + TSu volume are dominated by transient outgassing from all metal surfaces, the TSd + DS volume is filled with materials that will outgas significantly. The gas loads in the TSd + DS volume will also include leaks from the large number of feed-throughs required in the DS Vacuum Enclosure as well as the tracker gas distribution system and straw end fittings. If the rates in the detectors from scattering on gas atoms require adjustment (e.g., reducing the vacuum level below $10^{-4}$ Torr), then another pump may be necessary. This risk is mitigated by the inclusion of a third high vacuum pump for the DS.

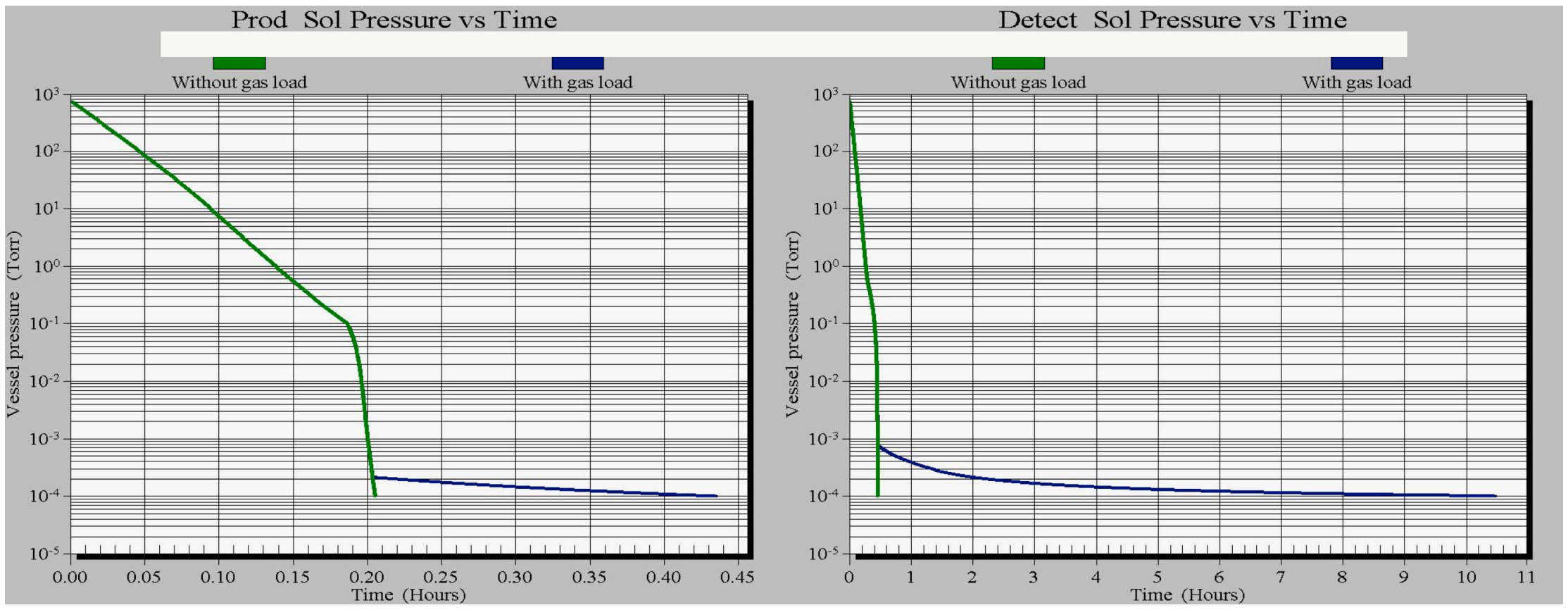

Figure 8.6. Pump down for Mu2e. The current calculated outgas rate only includes all surfaces are assumed prepared to minimize outgassing and considered un-obscured, the maximum tracker leak rate, and displacement volumes we have already designed for Mu2e.

The pump down performance is shown in Figure 8.6. A Kinney SDV 800 roughing pump services both volumes. Each volume has its own Shimadzu Magnetically Levitated Hybrid Molecular Drag/Turbo-Molecular pump model TMP-1003LM and three CTI – Cryogenics (Helix Technologies) model 400 high vacuum cryo pumps. The effective pumping speed of a single cryo-pump attached to the TSd + DS volume is determined from the conductance of the ports attached to the vacuum spool piece.

- Cryo-Torr CTI-400, 6000 l/s
- Orifice, 14,144 l/s
- VAT 16” pneumatic gate valve, 47,638 l/s
- 16” circular duct, 37,803 l/s.

Combining the above information results in a required effective pumping speed of 3509 l/s. At the DS base pressure, the selected cryo pump model can move 0.35 Torr l/s of air [14]. The current estimate of gas loads for this volume is 0.12 Torr litre/sec, so as the gas load estimate becomes more refined it may be possible to eliminate at least one cryo pump or substitute smaller capacity, less expensive pumps.

The effective pumping speed of a single turbo pump servicing the PS + TSu volume is determined as above. Although the PS+TSu vacuum only requires a modest pressure, 0.1 Torr, the turbo-molecular drag hybrid selected can attain pressures just above $10^{-4}$ Torr quickly. Hence, the estimates for the PS vacuum performance uses this as the upstream base pressure. We make the conservative estimate that the PS + TSu pump will be approximately 6 m away from the PS vacuum ports, behind shielding. The estimated distance has at least a 50% uncertainty. The conductance is estimated using the following input:

- TMP1003LM, 1100 l/s
- Orifice, 14,144 l/s
- VAT 16” pneumatic gate valve, 47,638 l/s
- 16” circular duct, 1453 l/s.

Combining this information gives an effective pumping speed of 592 l/s for a single turbo-pump. At base pressure, the pump moves 0.06 Torr litres/sec in the PS + TSu volume. The gas loads for this volume are estimated to be less than 0.08 Torr l/s [14]. The gas load and pumping rate are comparable in size; as we shall see in Figure 8.6, this poses no challenge to turbo pump operation. However, the required duct length could be much longer and the conductance correspondingly lower. Hence, the PS vacuum closure can accommodate addition pumps and ducts if necessary.

The pump down performance for both vacuum volumes was calculated using VacTran, a commercial vacuum simulation program. Figure 8.6 shows the pump down performance of each volume with a Kinney SDV 800 pump. PS + TSu required 5.5 minutes to reach approximately 10 Torr and TSd + DS required about 10 minutes to reach the same pressure. At 10 Torr the TMP1003 turbo pump is turned on and the SDV 800 is switched off. In the DS vacuum system, the TMP1003 runs for approximately 20-30 minutes, as seen in Figure 8.6, to reach the crossover region for the cryo pumps. Once the cryo pumps are switched on, the turbo pump is deactivated. At that point, the PS + TSu volume reaches the desired $10^{-4}$ Torr level in about 12 minutes while the TSd + DS volume requires about 10.5 hours.

During all phases of the pump down (as well as the process of coming up to atmospheric pressure), it is critical that the pressure differential across the antiproton Stopping/Vacuum Window located between TSu and TSd never rise to a value greater than 50 Torr. The window is at significant risk if a leak develops that causes a huge pressure rise and large pressure differences across the window over a short time period. Another significant risk is present during the time of initial pump down. Uncontrolled pump-down speed could exert viscous drag forces on fragile components in the vacuum volumes, particularly in the Detector Solenoid.

The strategy during initial pump down calls for two pressure control valves that are actuated by a control algorithm residing in a Programmable Logic Controller (PLC) used to control and monitor all of the Beamline Vacuum System instrumentation. The algorithm will monitor the measured vacuum pressure in the Production and Detector Solenoids. Since the conductance paths to the two areas are not identical, it is critical that the pressure control valves be stepped from closed to open in a manner that allows for pump down of the entire mass to be as uniform as feasible. The algorithm will adjust each pressure control valve to maintain a constant profile. During the initial pump down stage, valves bypassing the antiproton/vacuum window volume will be open to reduce the risk of a large differential pressure.

The differential pressure (dP) across the antiproton/vacuum window must be maintained below approximately 50 Torr. This dP is approximately 12.5% of the yield point of the window. When the pressure differential across the window is less than 1 Torr, as measured by pressure transducers in the Production and Detector Solenoids, the pressure control valves can be fully opened and the bypass valves across the window may be closed.

***Considered Alternatives to the Proposed Design of the Vacuum System***

The PS Enclosure may be designed using a thick flat plate instead of an elliptical reverse-domed head. The flat plate design would simplify the vacuum window design and mounting scheme, but would retain higher activation levels from the proton beam.

If verified by gas load studies, the Vacuum Pump Spool Piece may be rotated 45° to allow mounting four 304.8 mm diameter vacuum pumps and gate valves instead of 406.4 mm diameter pumps and valves, resulting in a significant cost savings. At the present time, calculations show that 406.4 mm diameter equipment is required to handle the large amount of material outgassing within the Detector Solenoid.

A faceted Instrumentation Feed-Through Bulkhead (IFB) design made from aluminum plate may be considered. This design would save considerable weight and may minimize the support structure necessary to hold up the Detector Solenoid Vacuum Enclosure to avoid stressing the DS cryostat inner shell. The disadvantage of this faceted IFB is its manufacturability, requiring 100 mm thick aluminum plate to be mitered and welded while maintaining vacuum integrity and ASME code compliance.

### *8.3.2* Collimators

There are two types of collimator assembly configurations. The first type is a relatively simple collimator (Figure 8.7) that will be used for COL1 and COL5, the collimators in the first and last straight sections of the upstream (TSu) and downstream (TSd) Transport Solenoid. COL1 and COL5 are essentially copper (COL1 has an additional 10 mm graphite liner – inner tube) tubes with a flange and insertion rollers. They are about one meter long with a 0.48 m outer diameter. The COL1 inner bore diameter is 0.3 m and the COL 5 inner bore diameter is 0.256 m. COL1 may require active cooling and monitoring to control heating that results from the radiation load due to its proximity to the production target. A radial gap of 1-2 mm will be provided between the ID of the cryostat bore and the OD of the collimators to allow for insertion during assembly. COL1 and COL5 (Figure 8.7) are permanently installed. These collimators are not required to be accessible or repairable during the lifetime of the experiment. The flanges will be bolted to the cryostat to prevent axial motion, particularly during a quench when eddy currents may be induced in the collimators that lead to axial forces. The location of the various collimators is shown in Figure 8.8.

The detector area is required to be shielded from neutrons. In additional to its role in collimating the muon beam, the neutron rate is an important consideration in the design of COL5. COL5 could be made from polyethylene or a combination of polyethylene and a high Z material for better neutron absorption. The high Z material will also help to reduce background from beam electrons.

The central straight sections of TSu and TSd contain two collimators, COL3u and COL3d, which rotate (see Figure 8.9 and Figure 8.10). The antiproton stopping window module is located in between TSu and TSd cryostats and must have collimators on both sides, necessitating two rotating collimators rather than just one.

The collimator downstream of the antiproton window is necessary to absorb annihilation products from the interaction of antiprotons in the window. The upstream collimator absorbs higher momentum antiprotons that might otherwise penetrate the window and make their way to the Detector Solenoid.

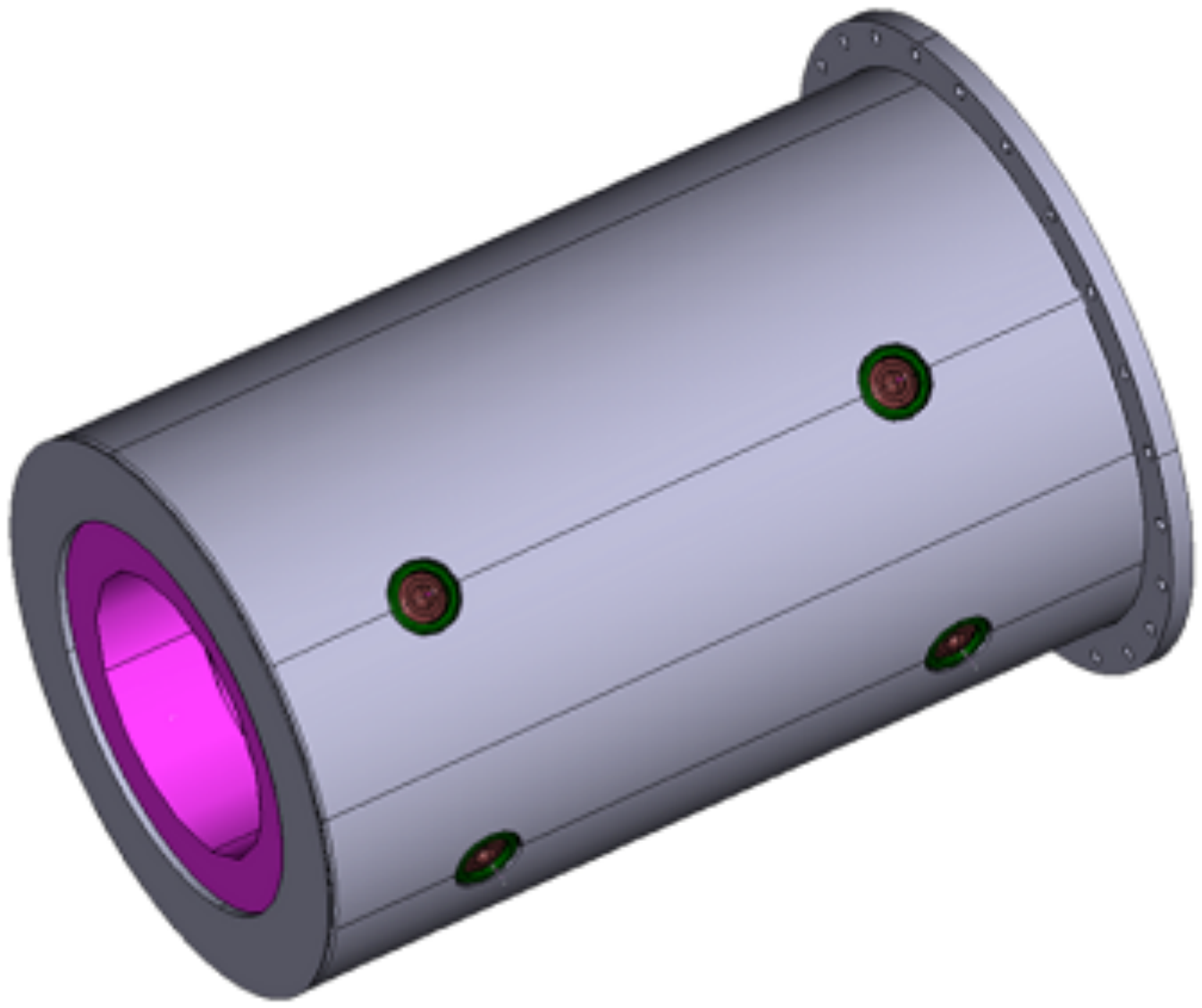

Figure 8.7. 3-D model view of the collimator (COL1 & COL5).

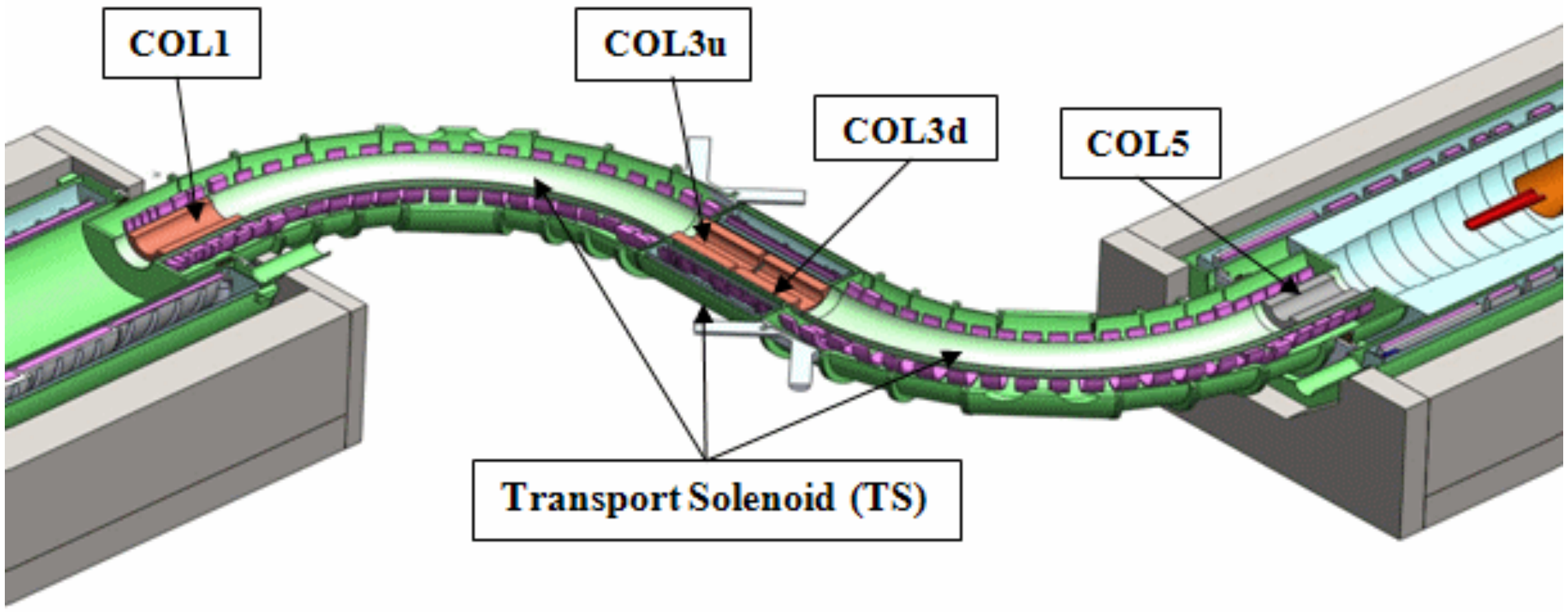

Figure 8.8. Overview of the collimators positioned within the warm bore of the Mu2e Transport Solenoid.

COL3u and COL3d each consist of two major components:

- A stainless steel container with a flange for the connection with the TSu or TSd cryostat flanges. This container will have axial insertion rollers similar to COL1 and COL5 for insertion in the cryostat.

- A copper collimator body with an asymmetric clear bore shown in Figure 8.9. This asymmetric bore provides charge and momentum selection. The collimator will allow passage of low momentum negative particles (or positive particles if the collimators are rotated 180 degrees) while strongly suppressing positives (or negatives). The collimator bodies have bearings to allow 180-degree rotation inside the stainless steel container for detector calibration using μ+ decays. The rotation mechanism consists of a large gear that is attached to the collimator body (shown in figure 8.9).

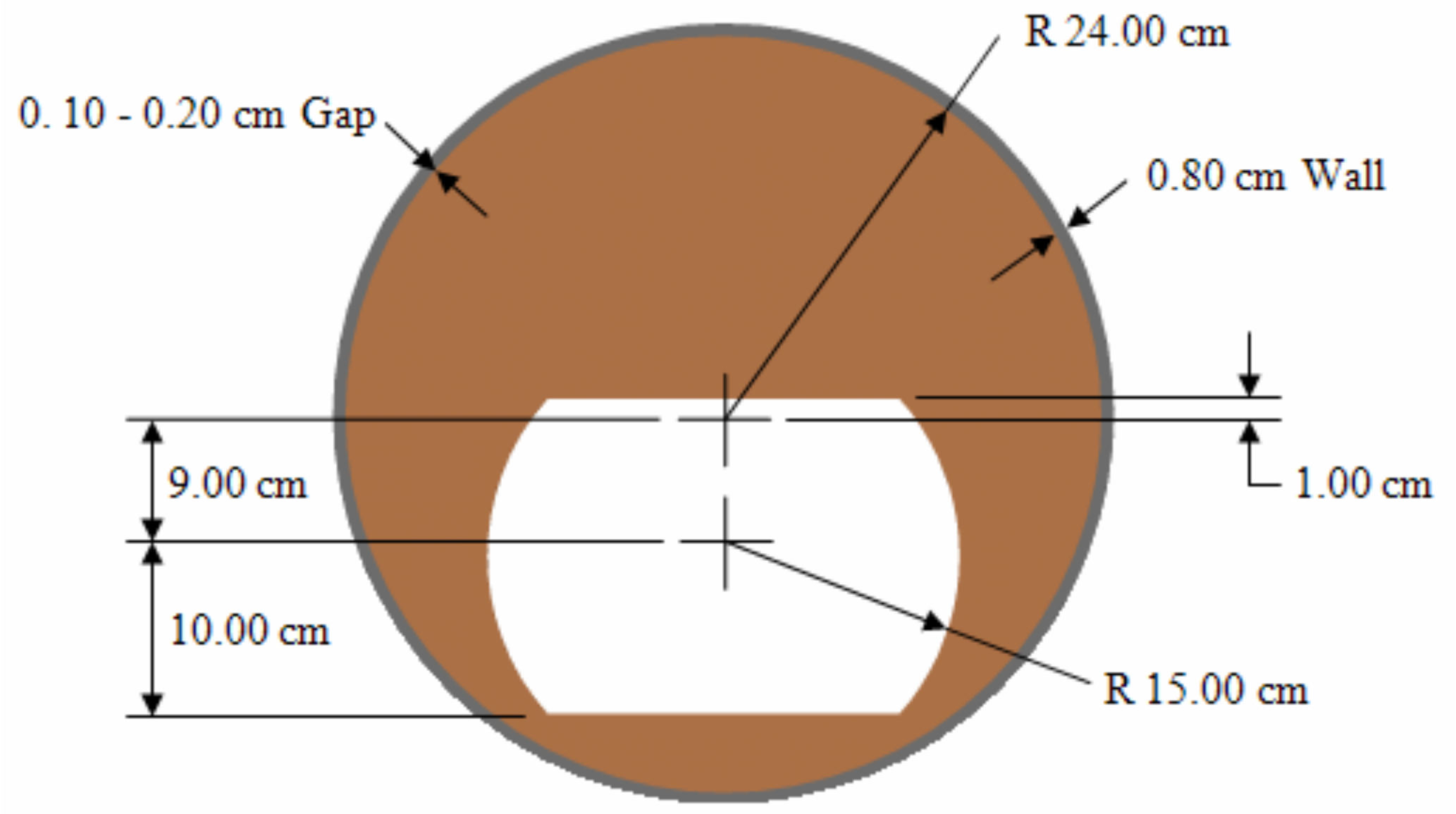

Figure 8.9. Cross-sectional view of the central collimators COL3u and COL3d.

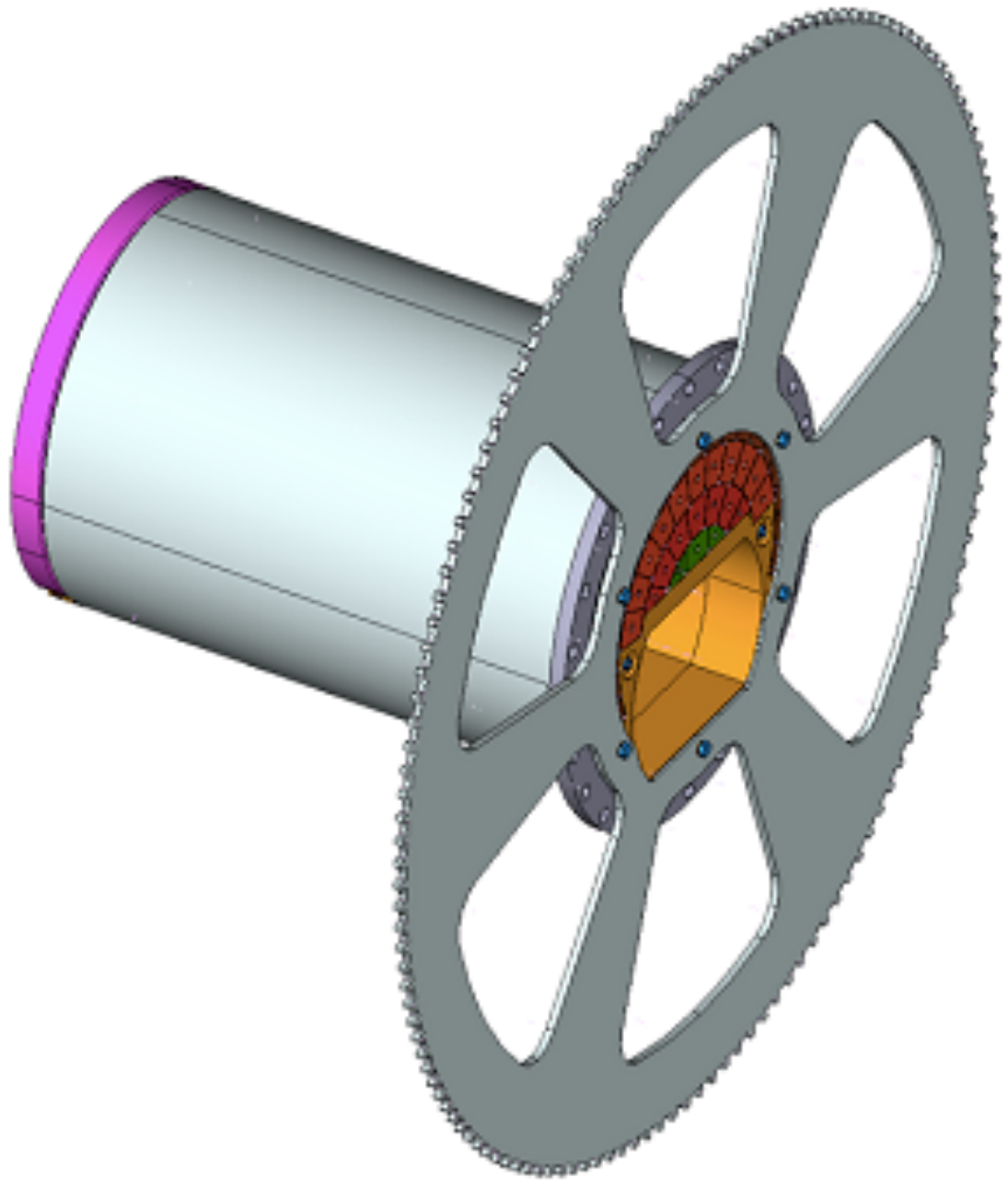

Figure 8.10. Schematic view of the COL3u collimator.

A section view through the TSu and TSd straight sections, including the rotatable collimators, is shown in Figure 8.11. The collimators (COL3u and COL3d) will be supported by the inner tube of the TSu and TSd cryostats and installed during TS assembly. The insertion clearance between the OD of each collimator and the inner TS cryostat wall is 1-2 mm radially. Tolerances for collimator positioning are ±1mm inside the cryostat bore. Two rows of bearings will be used to support the collimator body inside the container and provide easy azimuthal rotation inside the container tube. Rotation will be provided by an electrical motor, that will electronically control the orientation of the collimators. There is a 50 mm gap between TSu and TSd cryostats. This gap will provide sufficient room for the two large gears and for the thin antiproton stopping window module that is located in the center of the gap. There will be two drive shafts. Each drive shaft will be rotated by a servomotor. The other end of the drive shaft has a smaller gear that is in connection with the larger gear. Each collimator can be rotated separately.

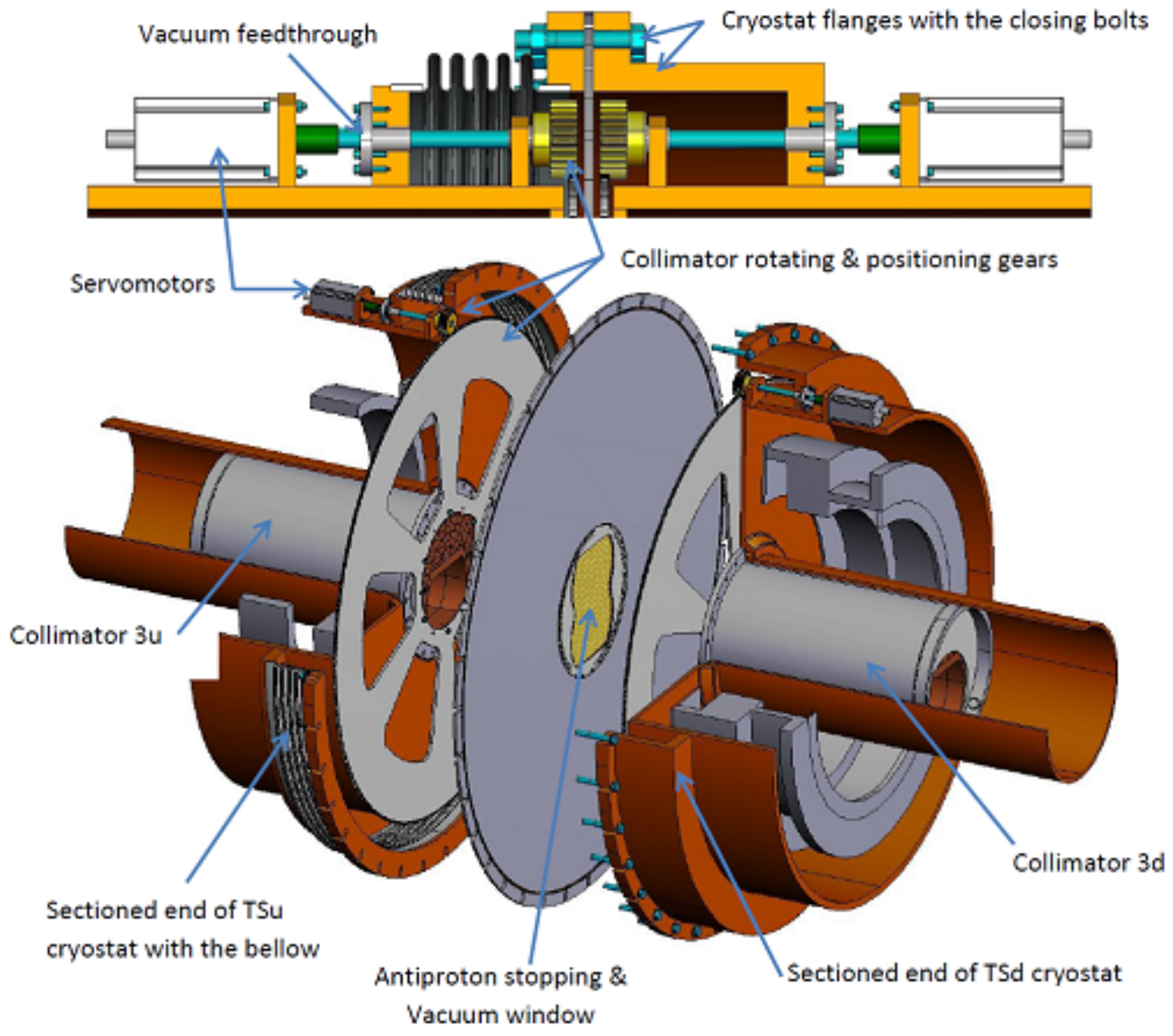

Figure 8.11. Conceptual section view through the TSu and TSd straight sections, antiproton stopping window module, and rotatable collimators.

The antiproton stopping window module will be located in the center of the TS solenoid bore (see Figure 8.11). The window support structure must provide adequate vacuum seals to isolate the upstream and downstream vacuum volumes and also isolates the vacuum volumes from atmospheric pressure. This will be accomplished by large diameter seals between the window plate and the cryostat flanges. To be able to install/remove the antiproton stopping window module requires breaking the vacuum seal and creating a ~5 mm gap. This can be achieved by installing a bellows on one side of the joint. The antiproton stopping window must be made from low-Z material (for example, from Kapton) with a thickness of about 0.67 mm.

The effectiveness of the collimators to select muons of the appropriate charge sign and reduce wrong sign muons is shown in Figure 8.12 and Figure 8.13. In Figure 8.12 the upper curve is the spectrum of negatively charged muons as they exit the Production Solenoid and are incident on the collimation system. The lower curve is the spectrum of negatively charged muons that emerge from the collimation system and enter the Detector Solenoid. Low energy muons with a high probability to stop in the muon stopping target are preferentially selected. In Figure 8.13 the same plot appears for positively charged muons, whose rate is significantly attenuated by the time they reach the Detector Solenoid.

***Considered Alternatives to the Proposed Design***

Alternatives exist for the composition of COL5, which could be made from polyethylene or a mixture of copper and polyethylene. Additional simulations, optimization of the muon beam, and further studies of backgrounds are required before these alternatives can be fully evaluated.

An alternative design for COL3 has been considered which incorporates a movable central block as shown in Figure 8.14. This design allows fine-tuning of the collimator aperture and might be an advantage during the experiment run time.

Both collimator alternatives described above are mechanically and technologically feasible. More information can be found in the Requirements and Specifications document for the collimators [2].

### *8.3.3* Muon Beamline Shielding

The heating of the Transport Solenoid coils and other structures is due to primary radiation from the production target and secondary production on the collimators. The Internal Muon Beamline Shield shall limit the instantaneous local heating (as described in [3]) in the Transport Solenoid coils. Simulation shows that the heating is sufficiently low that no internal shielding will be required.

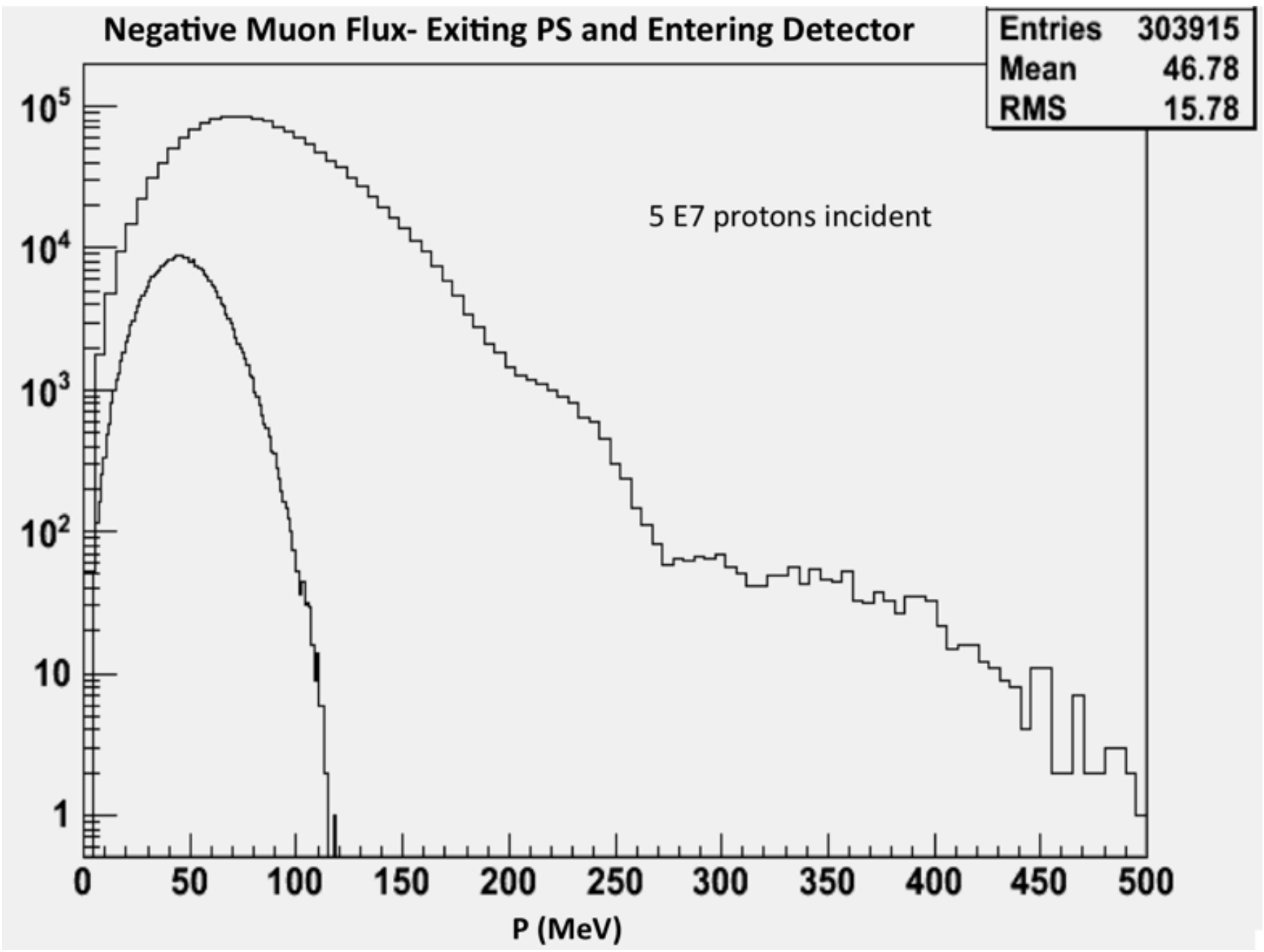

Figure 8.12. The momentum spectrum of negative muons exiting the Production Solenoid and incident on the collimation system (upper curve) and the momentum spectrum of the negatively charged muons that emerge from the collimation system and enter the Detector Solenoid (lower curve).

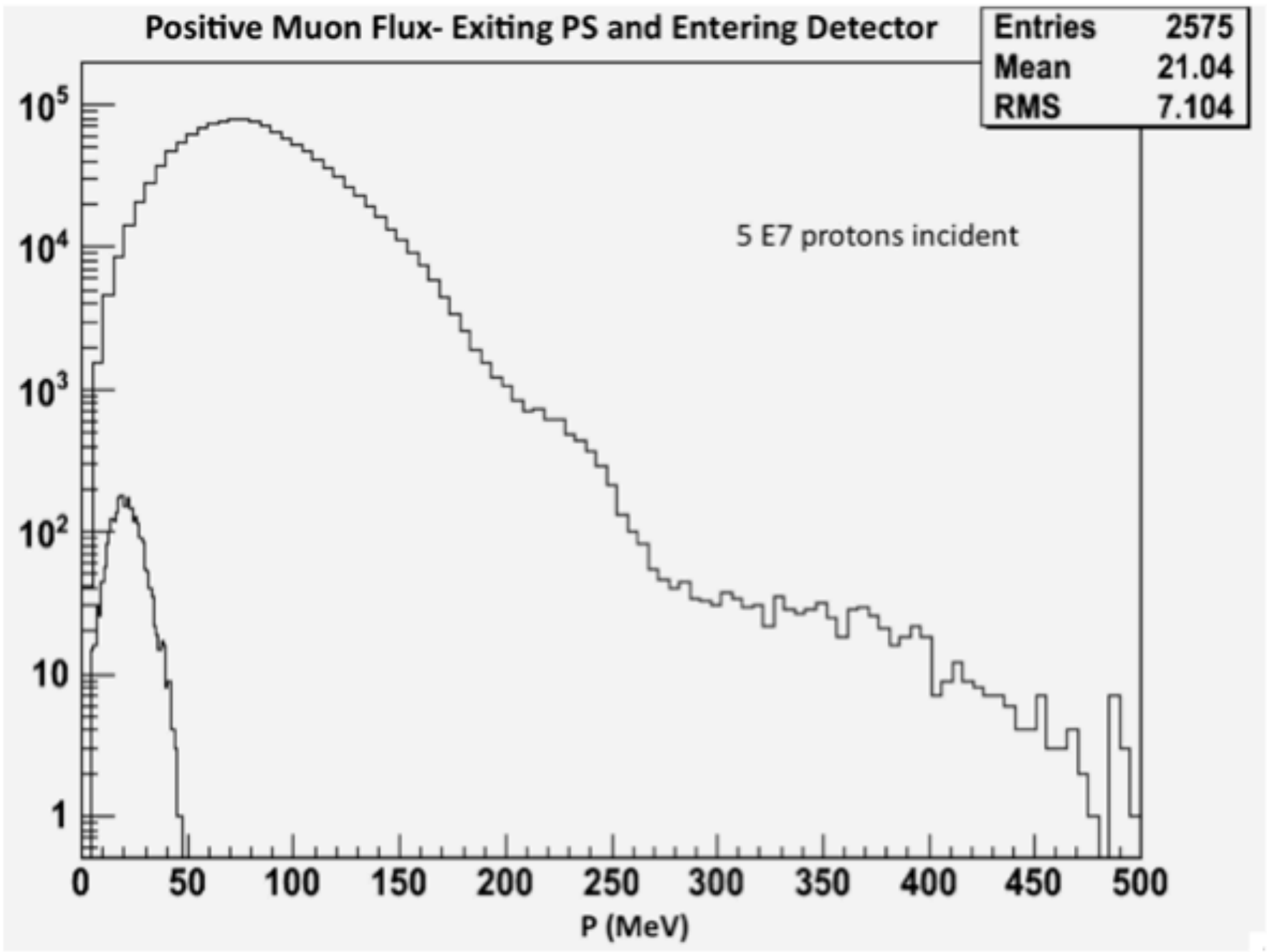

Figure 8.13. The momentum spectrum of positive muons exiting the Production Solenoid and incident on the collimation system (upper curve) and the momentum spectrum of the positively charged muons that emerge from the collimation system and enter the Detector Solenoid (lower curve).

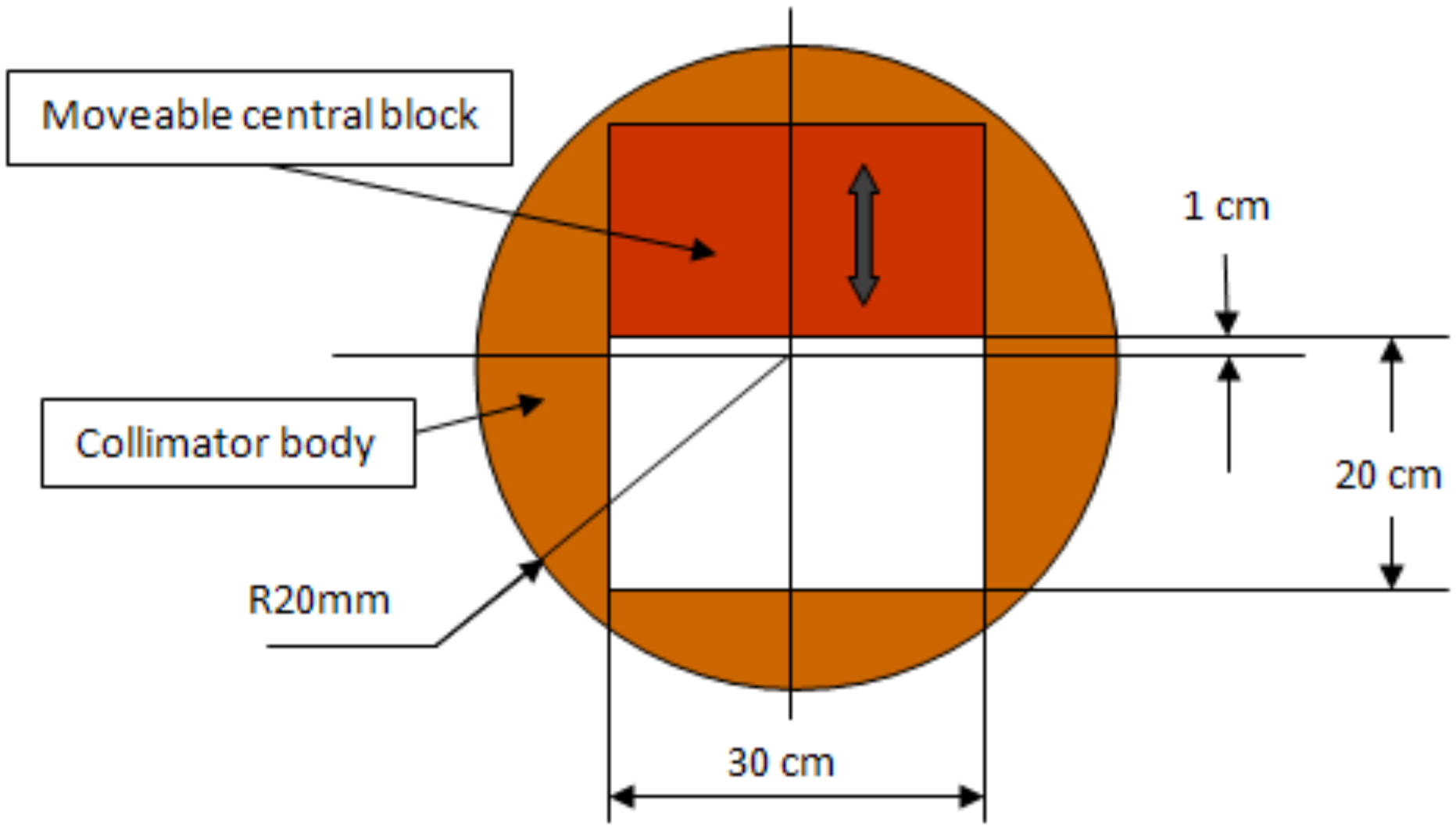

Figure 8.14. Conceptual cross-section of the asymmetric collimator.

The External Muon Beamline Shielding Elements surround the Transport Solenoid and are supported off the building floor as close as possible to the TS cryostat (see Figure 8.15). The primary purpose of the External Muon Beamline Shield is to reduce the flux of neutrons that reach the Detector Solenoid region. The concern is that neutrons could fire the Cosmic Ray Veto or scatter into the sensitive detectors in the Detector Solenoid.

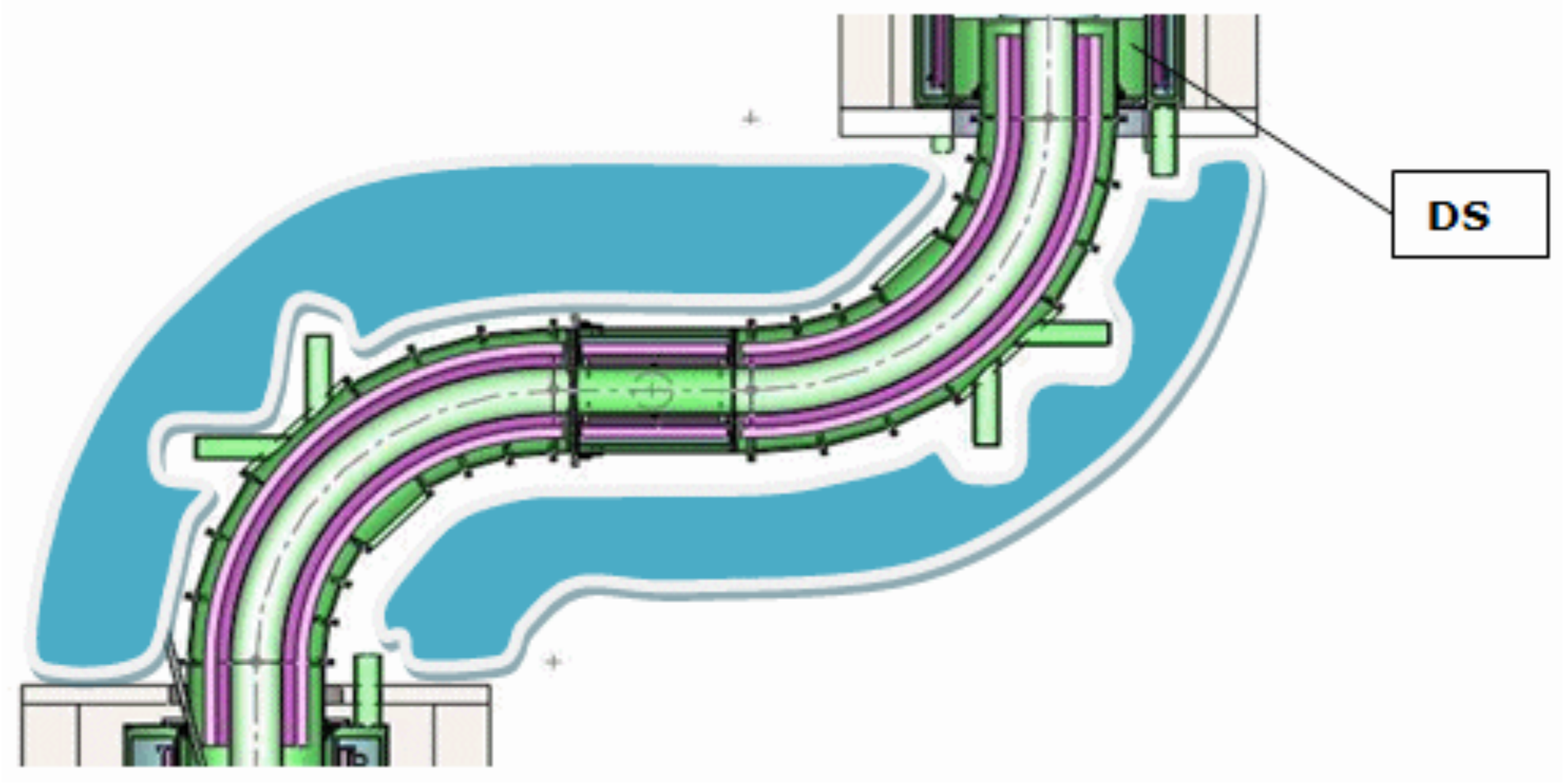

Figure 8.15. Planned area for external Muon Beamline Shielding (blue area).

External Muon Beamline Shielding could be constructed from concrete or polyethylene bricks assembled inside support frames. The geometry of this external shielding will be defined after a neutron background simulation is completed. The position and mass of each of the external shielding components will be defined after a preliminary engineering shielding design is completed. The structural support and installation fixtures for the External Muon Beamline Shielding will be designed to accommodate the gravity loads of the Shield. Parts of the External Shielding may be manufactured in sections, which will be individually moved into the area around the Transport Solenoid.

***Considered Alternatives to the Proposed Design***

Since further evolution of the Mu2e Experimental Setup supported by beam studies might reveal a need to introduce Internal Muon Beamline shielding, two alternative solutions, shown in Figure 8.16 and Figure 8.17, have been developed. The first alternative design of the Internal Muon Beamline Shielding is integrated with the design of the Transport Solenoid cryostat vacuum tube (see Figure 8.16). The segmented vacuum tube will have similarly segmented shielding blocks. Shielding blocks will be attached to the inner surface of the vacuum tube for each segment before connecting them to the whole assembly. The Internal Muon Beamline Shielding will be supported by the TS cryostat vacuum tube.

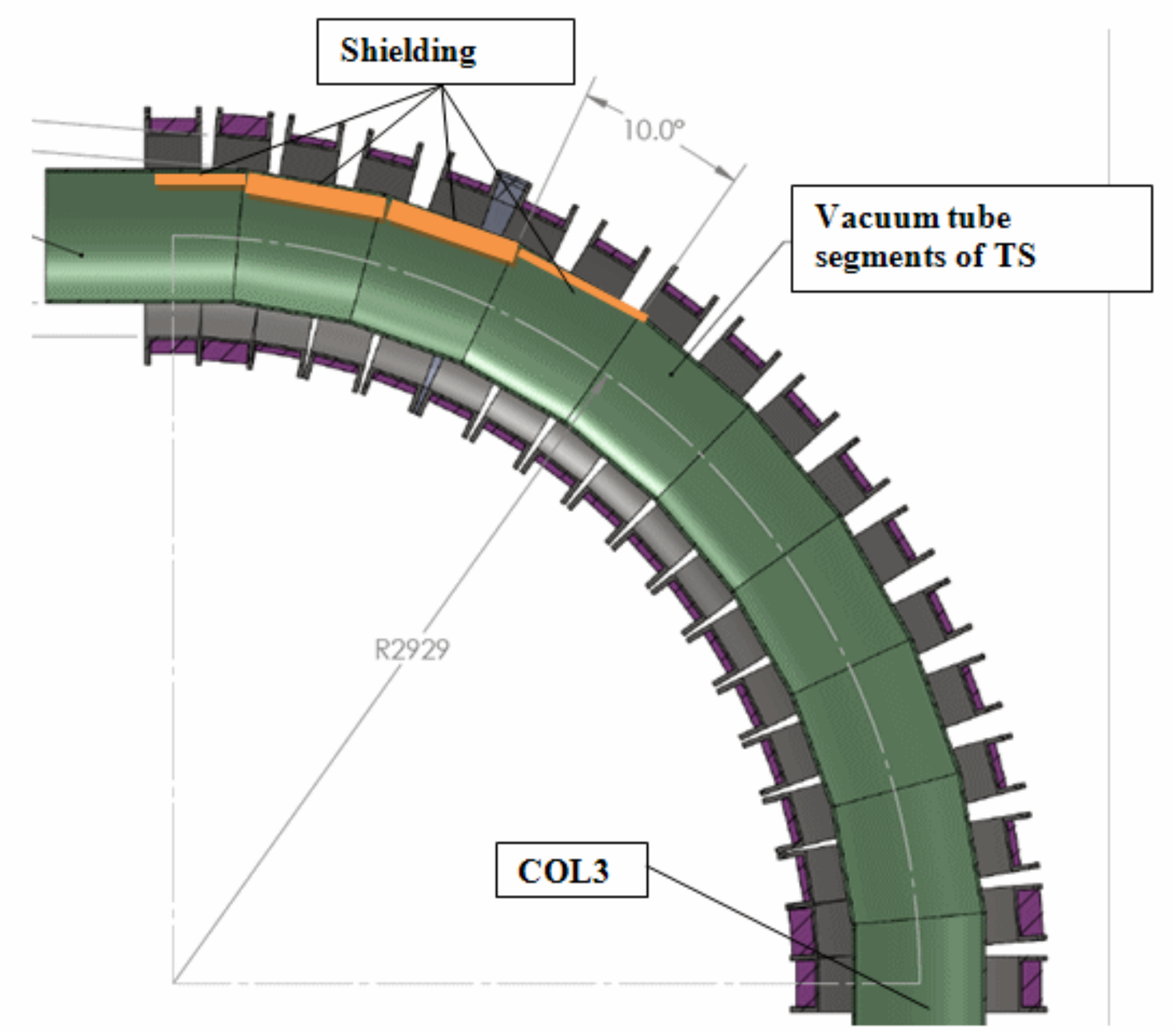

Figure 8.16. Conceptual Design for the Internal Muon Beamline Shielding.

Parts of the Beam Shielding may need to be manufactured in sections, which will be individually installed in the bore of the Transport Solenoid. The beam shielding does not require active cooling or monitoring, thus there are no cooling or electrical interface concerns.

For the second alternative design the Transport Solenoid cryostat vacuum tube is made from two halves and the shield tiles and collimators are installed inside each half tube before final mid-plane welding of the bore assembly. This design might be necessary if a continuous smooth inner surface is required for the cart used to access the Transport Solenoid to map the magnetic field (See Section 7.3.6). The primary disadvantage of this alternative is the high cost of constructing the cryostat in two halves as well as a possibility of tube shape deformation during welding. Installation of this completed tube assembly inside the TS solenoid is challenging as well. The proposed method for this installation is shown in Figure 8.18. Rollers or bearing pads will be used to support the tube assembly.

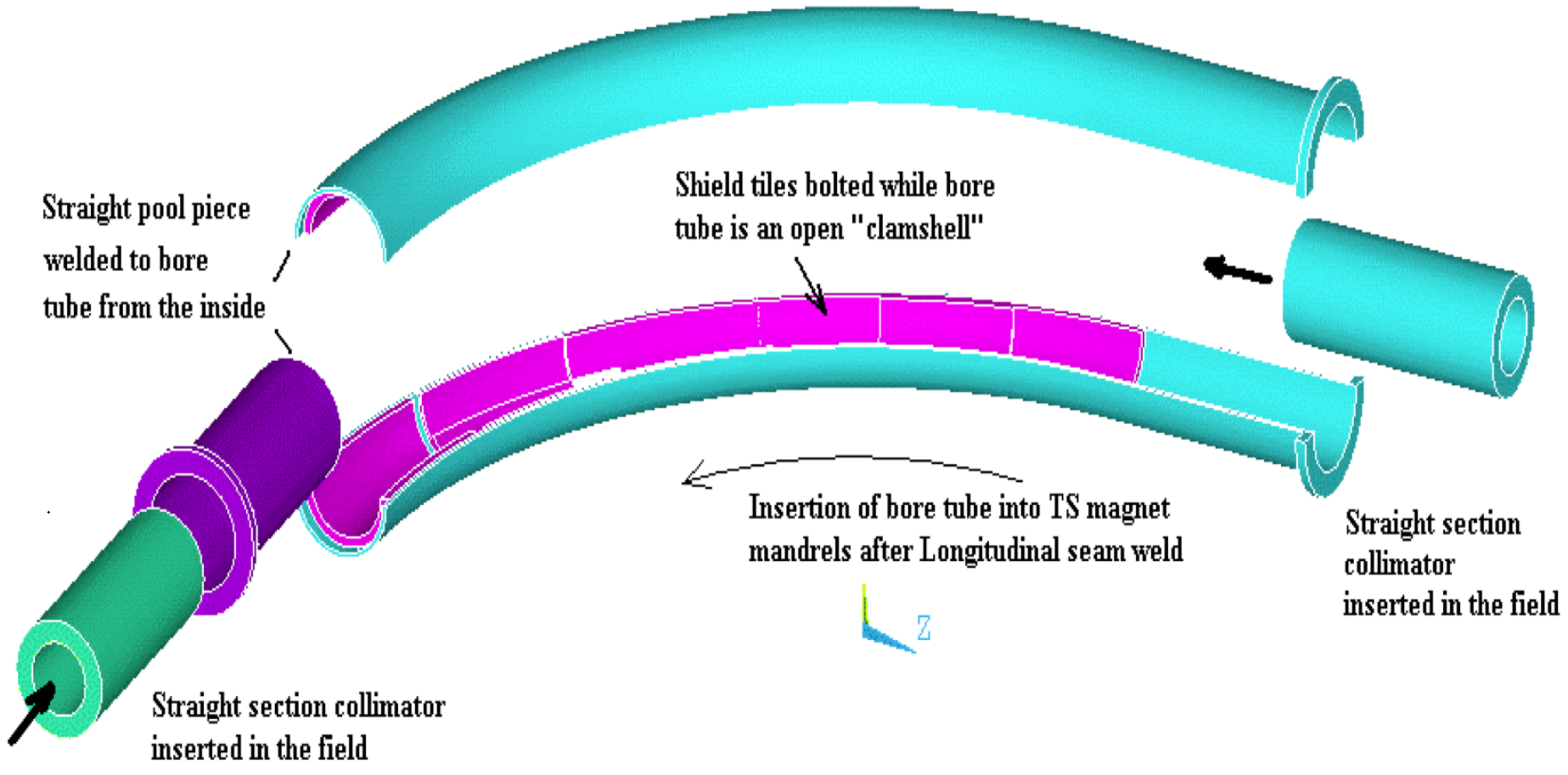

Figure 8.17. Design of the muon beam line shielding from MECO.

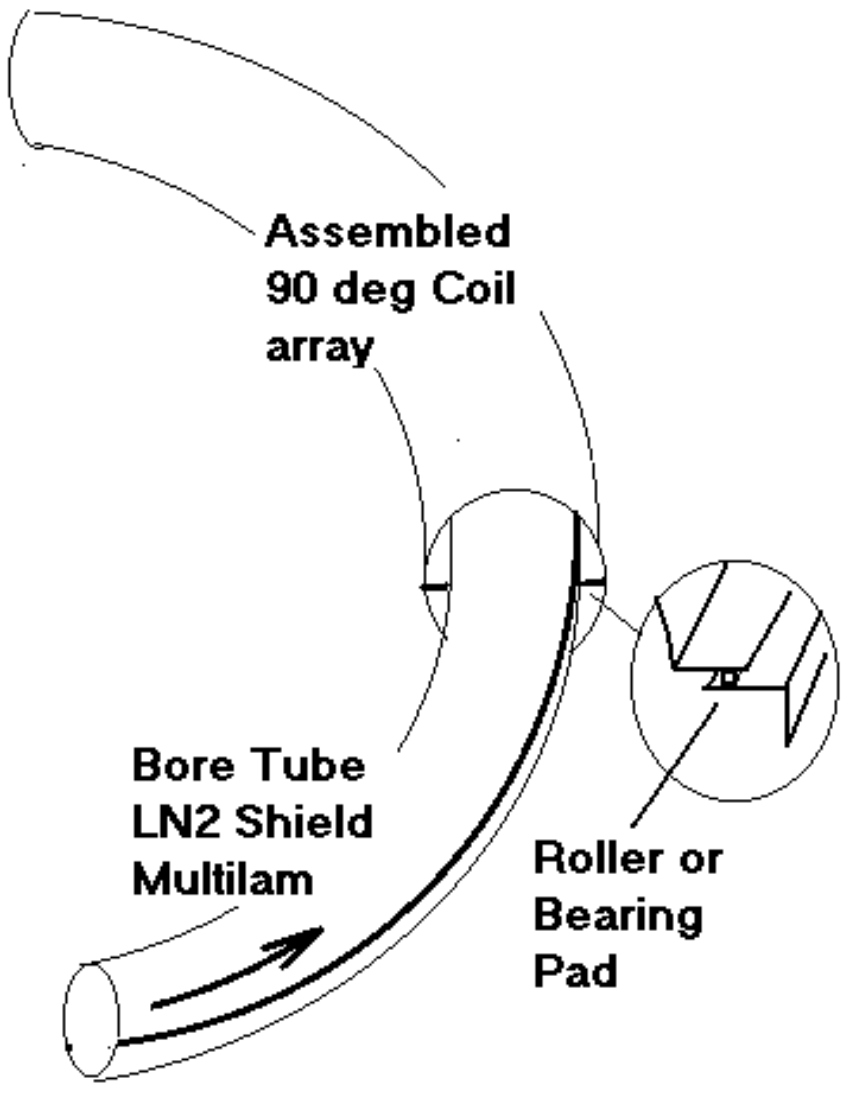

Figure 8.18. Bore tube insertion method from MECO.

### *8.3.4* Muon Stopping Target

The muon stopping target consists of 17 circular aluminum foils that are arranged coaxially. They are equally spaced 50 mm apart and have a thickness of 0.2 mm. The radii range from 83 mm to 65 mm and are tapered with decreasing radii in the direction of decreasing magnetic field. The position of the target in the Detector Solenoid is such that the first foil is at 1.57 T and the last at 1.30 T.

There are several physics requirements [4] that limit the choice of target material as well as the geometry. The selected material must have a conversion energy that is higher than the maximum photon energy from muon radiative capture ($\mu^- + (A, Z) \rightarrow (A, Z\text{-}1) + X + \gamma$), which can induce background. To avoid prompt backgrounds from the beam, data taking begins about 700 ns after the peak of the proton beam pulse. The lifetime of the muon in the target material (which decreases with increasing Z) must be long enough that a significant portion of the muons remain after 700 ns, but short enough that most decay before the next arriving proton pulse at about 1700 ns. However, the expected conversion rate increases with increasing Z, so that it is advantageous to choose a material with high Z. To reach the required sensitivity, at least 40% of the muons must stop in the target. Finally, the target geometry must be chosen to minimize energy loss from potential conversion electrons, minimize background contamination from sources passing through the target (beam electrons, cosmic rays, etc.), maximize the interception with the muon beam, and minimize the rate of DIO electrons that can reach the tracker. A schematic of the proposed design is shown in Figure 8.19.

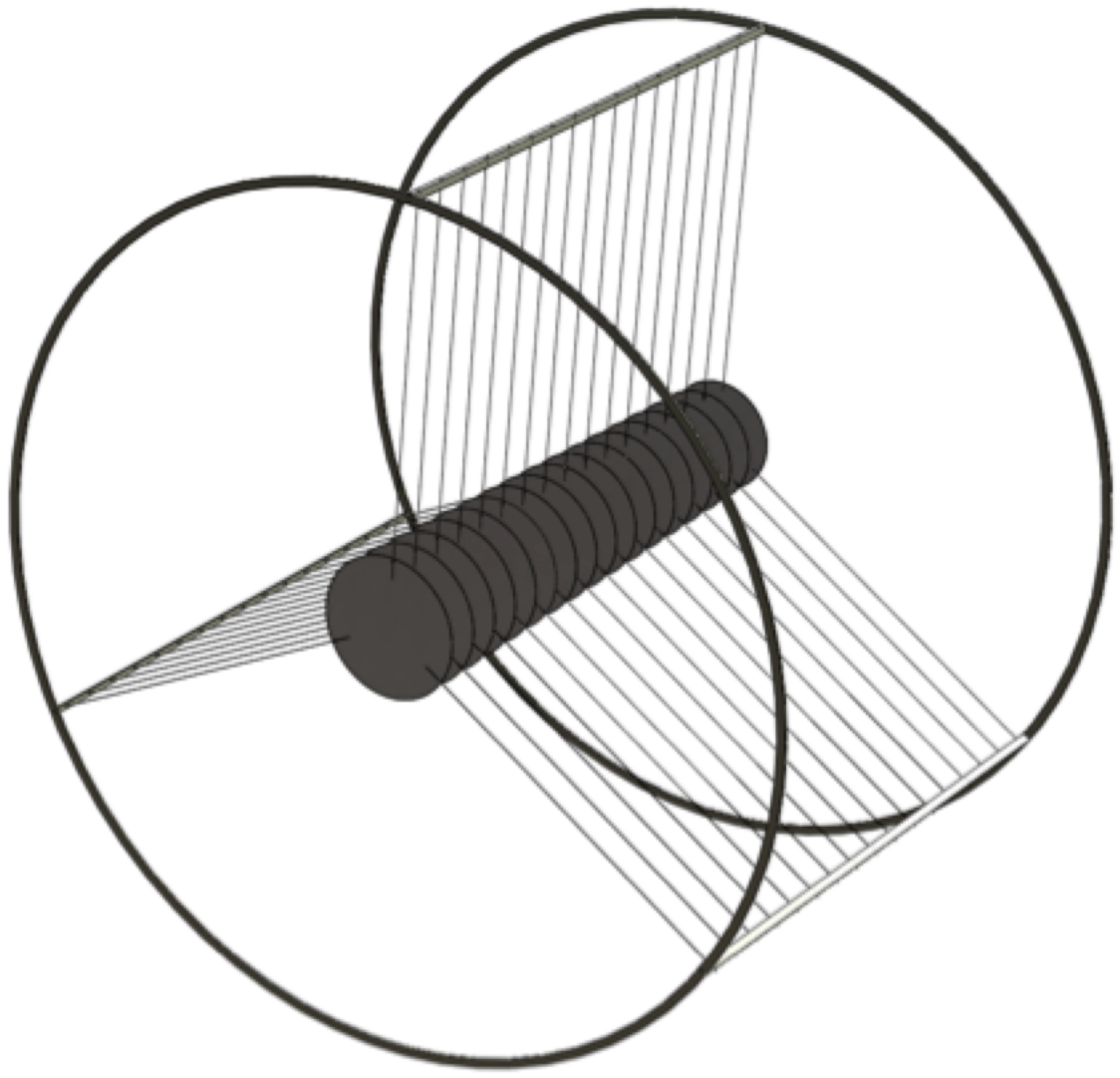

Figure 8.19. Schematic of the stopping target and support.

***Stopping Target Material***

The stopped muon lifetime and the muon conversion rate as a function of Z are shown in Figure 8.20 and Figure 8.21. Selenium and antimony have the highest conversion rates, but they also have short muon lifetimes. The requirement that the muon radiative capture endpoint energy of the $\gamma$ be below the conversion energy is equivalent to the condition that the rest energy of the nucleus (A, Z) be below the rest energies of the possible combinations of (A, Z-1) + X. For the target materials that also satisfy the other requirements, the differences are 2.5 MeV (aluminum), 600 keV to several MeV (titanium, depending on isotope). However, in the case of titanium it is possible for oxygen to penetrate the foil and contaminate the material. Since oxygen has a lower Z than aluminum the endpoint of the DIO spectrum is higher, so the oxygen contamination can lead to a severe background. Taking these various factors into account, aluminum is chosen as the target material.

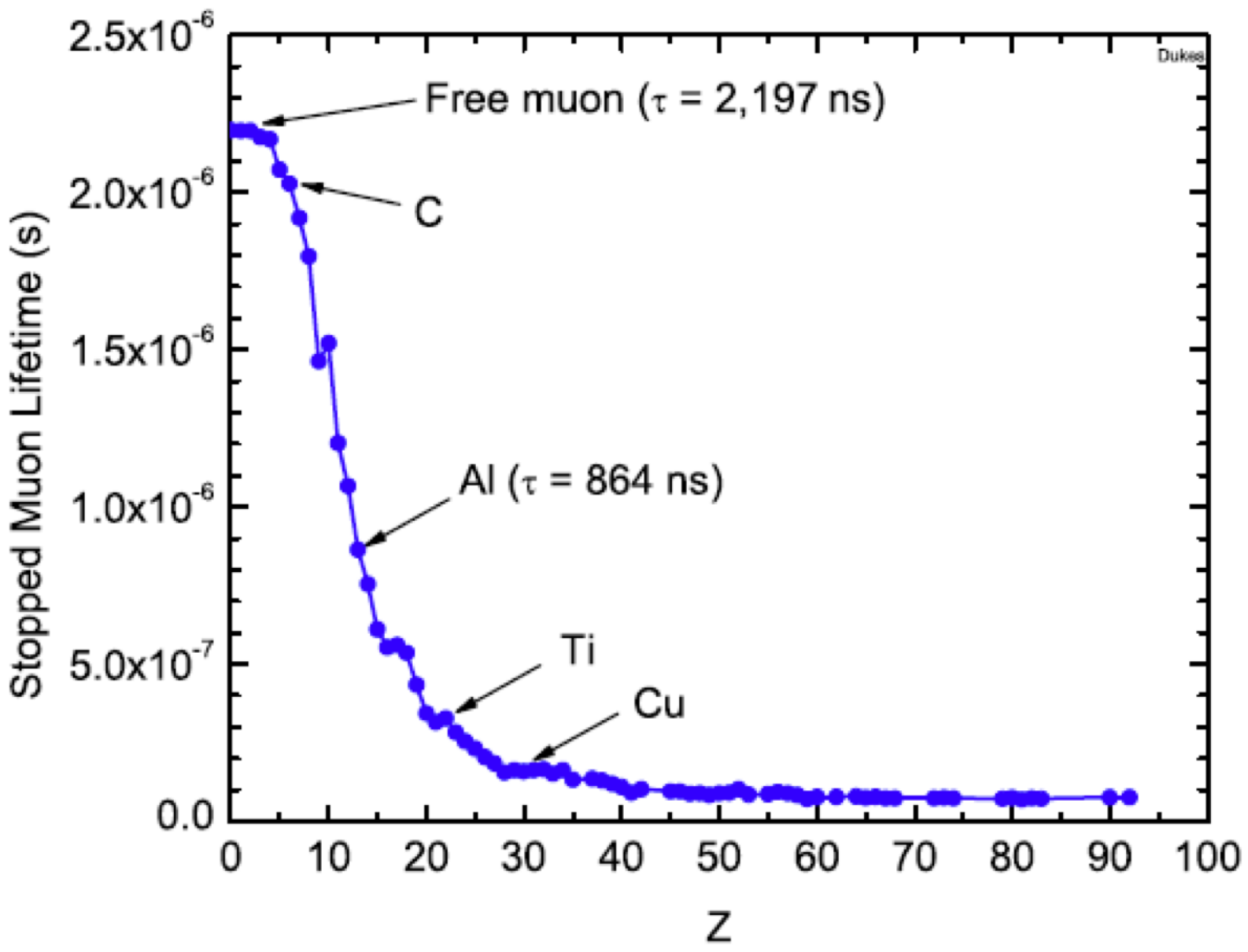

Figure 8.20. The stopped muon lifetime vs. Z.

***Position***

The graded magnetic field reflects electrons that are emitted in a direction away from the tracker. The position of the target in the field is chosen to be such that the first foil is at 1.57 T and the last at 1.30 T.

***Geometry***

The geometry of the target must be chosen so that the maximum number of muons are captured while minimizing the energy loss of the conversion electrons. Figure 8.22 shows the conversion electron energy spectrum for a configuration of 8 foils (each 400 microns thick), 33 foils (each 100 microns thick), and 17 foils (each 200 microns thick). The mass of the each target configuration is kept the same. An optimal geometry is found to be 17 foils.

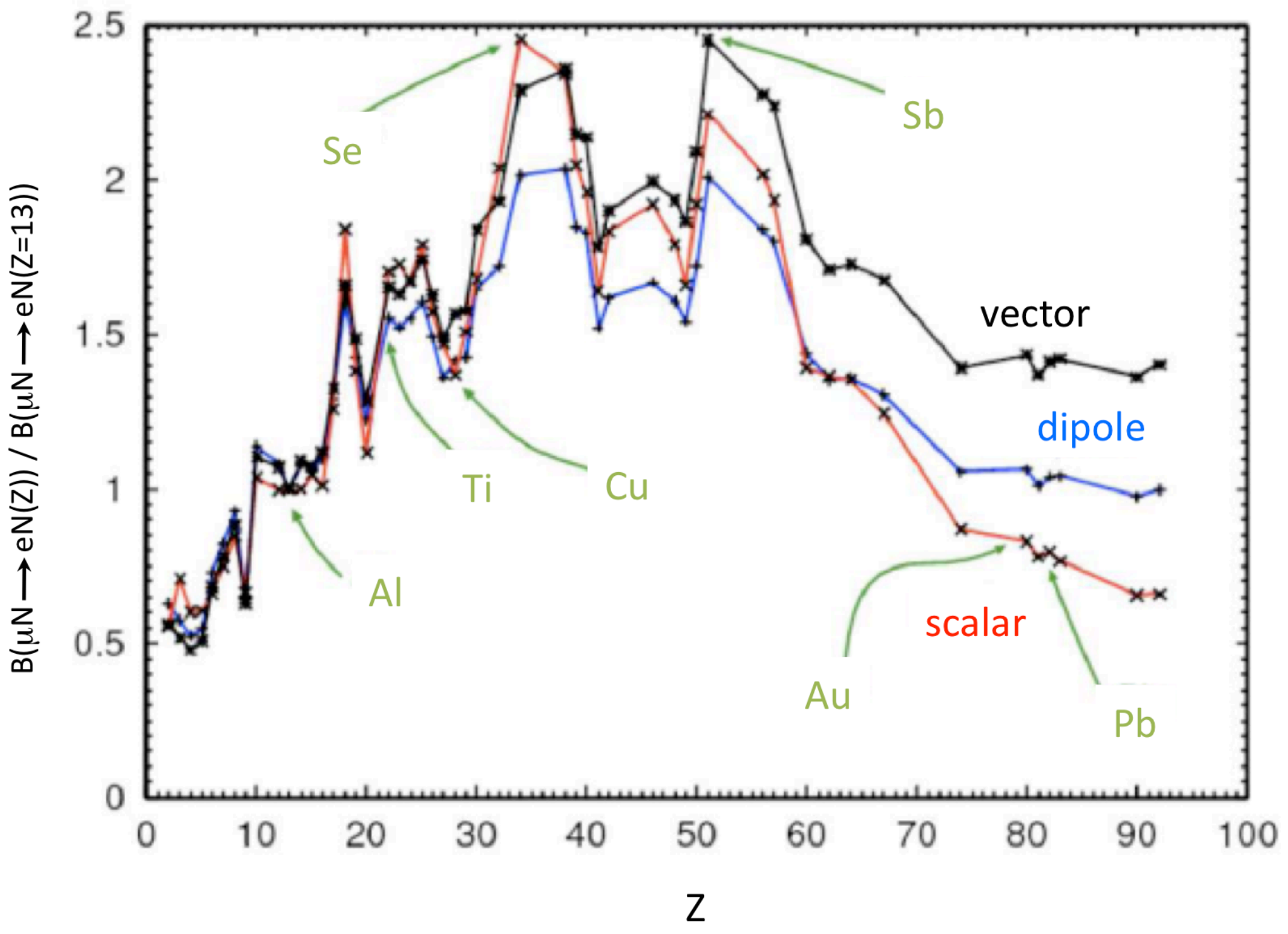

Figure 8.21. The muon conversion rate vs. Z, normalized to the rate in aluminum.

As is stated in the requirements, the target needs to stop at least 40% of the muons exiting the TS to achieve the desired sensitivity. A simulation of the momentum distribution for muons, overlaid with the momentum distribution of stopped muons, that encounter the target is shown in Figure 8.23. According to simulation, approximately 49% of all muons exiting the TS are stopped in the target with the geometry of 17 foils.

The effects of target misalignment (conservatively estimated to be ± 2 mm) would be to increase or decrease the number of DIO electrons striking the tracker. The reconstruction cannot place the origin of the electron to smaller than the spacing between the foils, so that misalignment does not have a significant effect. Misalignment of 2 mm has minimal impact on effort to track muons back from the tracker to the stopping target because the tracking uncertainty is as large or larger.

***Support Structure***

The support structure holds the stopping target in place in the center of the Detector Solenoid. Because of the diffuse nature of the muon beam as it enters the Detector Solenoid, a significant number of muons will strike the support structure and produce DIO electrons. Therefore, the support structure must be made of a high Z material because the endpoint of the DIO spectrum and the muon lifetime decreases as Z increases. The chosen material is tungsten.

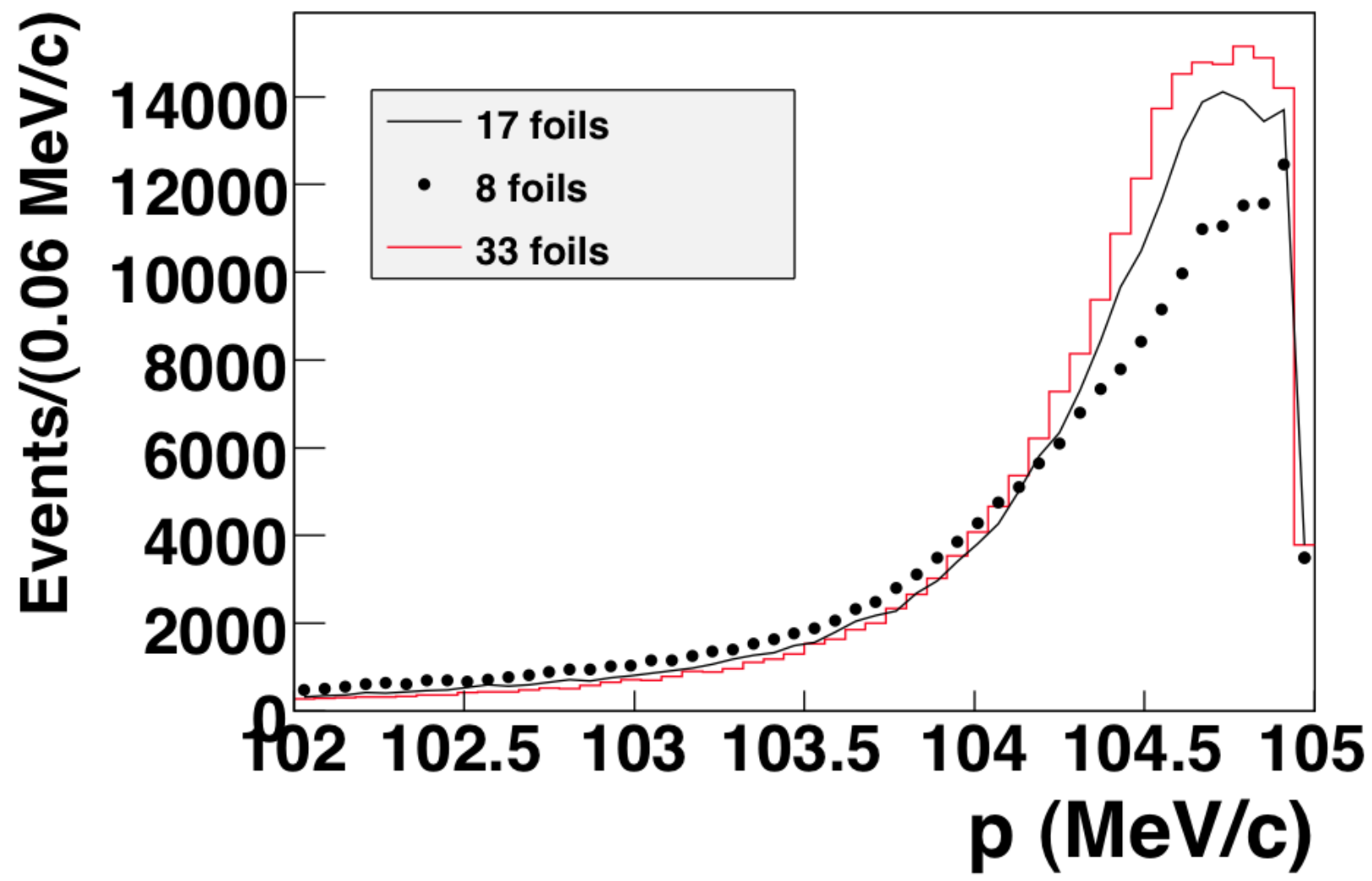

Figure 8.22. The conversion electron momentum spectrum of a target configuration of 17 foils (black line), 8 foils (black dots), and 33 foils (red line). For each target configuration, there are 500k muons that are required to convert to an electron when stopped in the target.

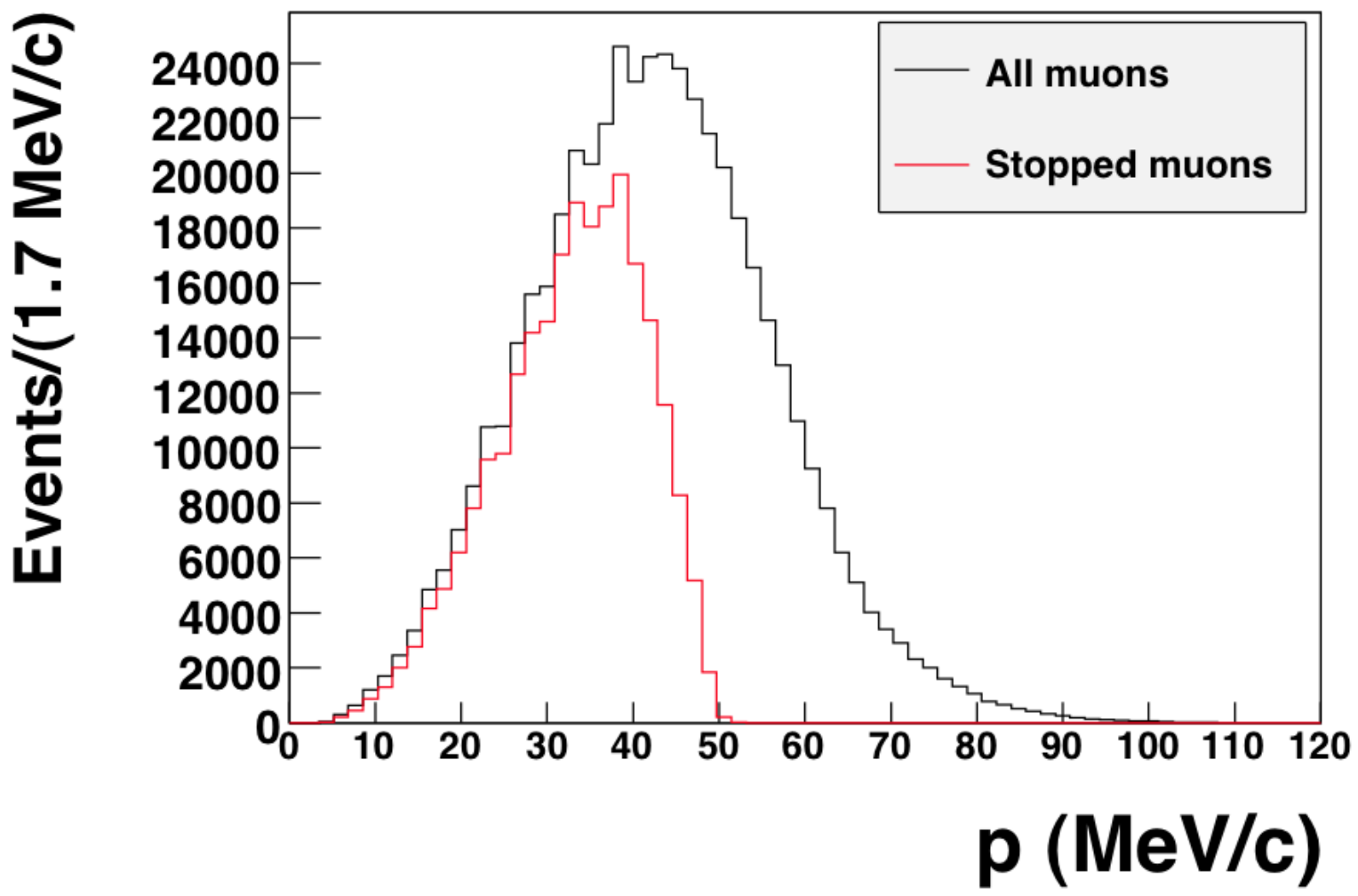

Figure 8.23. The incident muon distribution overlaid with the stopped muon distribution for the geometry of 17 foils. There are approximately 500k incident muons.

***Thermal Properties***

Beam electrons and muons deposit about 400 mW of heat in the muon stopping target. The heat must be dissipated through a combination of radiation and conductive heat transfer through the support structure. Taking into account the surface properties of aluminum, Ref [3] indicates that radiation alone is sufficient to dissipate the heat and keep the target temperature near 60° C.

***Considered Alternatives to the Proposed Design***

Some studies of a helical stopping target configuration have been done, with the goal of increasing the muon stopping rate but reducing the amount of material traverse by conversion electrons. The geometry was found to be helpful for these issues, but the drawback is that it is not optimal for positive particles. One of the main calibrations for the experiment uses positive pions that are stopped in the target. The positrons that result from the two-body decay provide an energy calibration. There are also plans to study alternative geometries where each target disk is replaced with a conical shaped element. From geometric considerations, incoming muons would see more material, increasing the stopping rate, while conversion electrons in the acceptance of the tracker would see less target material, improving the momentum resolution function. If additional studies indicate that these geometries are more favorable than the current design, it will be adopted.

### *8.3.5* Stopping Target Monitor

Muons come to rest in the stopping target and almost immediately emit X-rays, which signal the formation of muonic atoms. The estimated muon stopping rate is few times $10^{10}$ Hz. From aluminum, these X-rays are in the range of 66 keV to 446 keV. Oxygen, which can be an impurity in the target foils, has a characteristic x-ray at 134 keV (2p → 1s). In order to study these X-rays, a germanium detector (shown in Figure 8.24) will be utilized, which can detect photons with an energy of a few 100 keV to ~ 1 MeV with sufficient photopeak efficiency and energy resolution.

There are several requirements [5] that determine how the germanium detector should be placed. Sufficient collimation should be provided so that the detector only views the target. Additionally, the rate of the X-rays at the crystal must be compatible with the detector's capability. The detector should be placed far from the target because of the high X-ray rate. The material between the target and the detector (collimators, windows) should emit muonic X-rays that do not fall too close to the X-rays from aluminum. This material should also not absorb X-rays from the target. Finally, the detector must lie outside the enclosed Detector Solenoid so it can be serviced periodically to repair the damage from incident neutrons.

A significant background to the muonic X-rays is bremsstrahlung photons coming from electrons that intercept the target. This background should be kept to a minimum to reduce the flux of photons arriving at the germanium detector. Most electrons arrive about 100 ns before the muons, therefore the germanium detector must recover in this short time period from the effects of the electron flush.

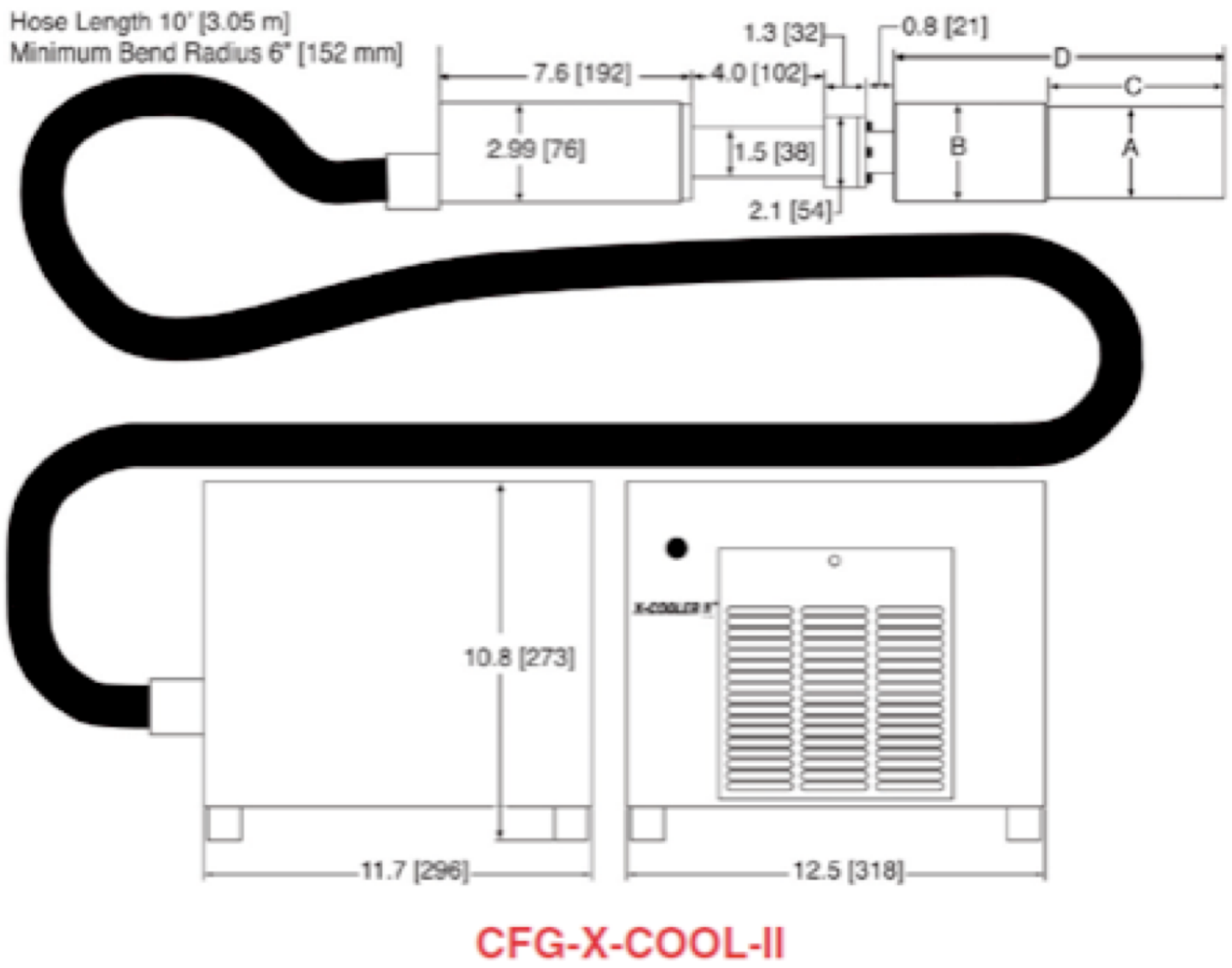

Figure 8.24. Configuration of ORTEC GMX HPGE detector fitted with X-cooler system.

***Surrounding Material***

X-ray collimation will be provided by a pipe inserted into the steel and concrete shielding downstream of the detector Solenoid. A window will separate the vacuum of the Detector Solenoid from the entrance of the collimation pipe. From these materials, the muonic X-rays have an energy that is well separated from the characteristic X-ray spectrum from aluminum.

***Radiation Damage***

Neutrons, coming from sources such as muon capture, will damage the germanium detector. Once the efficiency loss from radiation damage reaches unacceptable levels the detector has to be removed and annealed. According to the ORTEC catalog, a neutron fluence of $10^9$ will degrade the detector efficiency to less than 70%. It is estimated that annealing will be required every couple of months. The plan is to have two germanium detectors so that there is no down time associated with the annealing process.

***Cooling***

Germanium detectors operate at liquid nitrogen temperature. A commercially available mechanical cooler, such as the X-cooler system from ORTEC Inc., will be used.

### *Considered Alternatives to the Proposed Design*

In the event that radiation levels from neutrons and Bremstrahlung photons are too high to operate a germanium detector, an alternative mode of operation is possible. The germanium detector could be used to benchmark the rate of muon captures to rates in detector elements at reduced proton beam intensities. The germanium detector can be removed during production running at full intensity to avoid damage and properly scaled detector rates can be used to estimate the number of captured muons. The germanium detector can be remounted periodically as a cross check.

## *8.3.6* Proton Absorber

The proton absorber (shown in Figure 8.25) made of polyethylene, is a tapered cylindrical shell 0.5 mm thick with a radius slightly smaller than the inner radius of the tracker. The proton absorber is 210 cm in length and extends from the end of the stopping target to the beginning of the tracker. The physics requirements [6], that constrain the design of the proton absorber mandate that it should reduce the rate of protons at the tracker to allow proper reconstruction of electron tracks while minimizing the energy loss and straggling of conversion electrons that intercept it. Additional rate in the tracker can result in reconstruction errors that add a tail to the momentum resolution function and creates background. Energy loss and straggling in the proton absorber can also add tails to the resolution function. The proton absorber must be designed to balance these two effects. The proton absorber should not intercept the muon beam.

### *Material*

To minimize the impact on the momentum resolution function for conversion electrons the proton absorber should be constructed from a low Z material. This is satisfied by the choice of polyethylene. HDPE is the initial choice pending simulations, because it represents an acceptable balance between ease of construction and density.

### *Support Structure*

Since the support structure of the proton absorber lies well outside the extent of the muon beam, it is not constrained by the same requirements as the support structure for the stopping target. However, the wires that connect the proton absorber to the target frame are also made of tungsten. Figure 8.25 shows the space frame that supports the target and proton absorber. It will be constructed from stainless steel and will be mounted to the rail system internal to the Detector Solenoid. Figure 8.26 shows a detail of the connection between the space frame and the rail system bearing block.

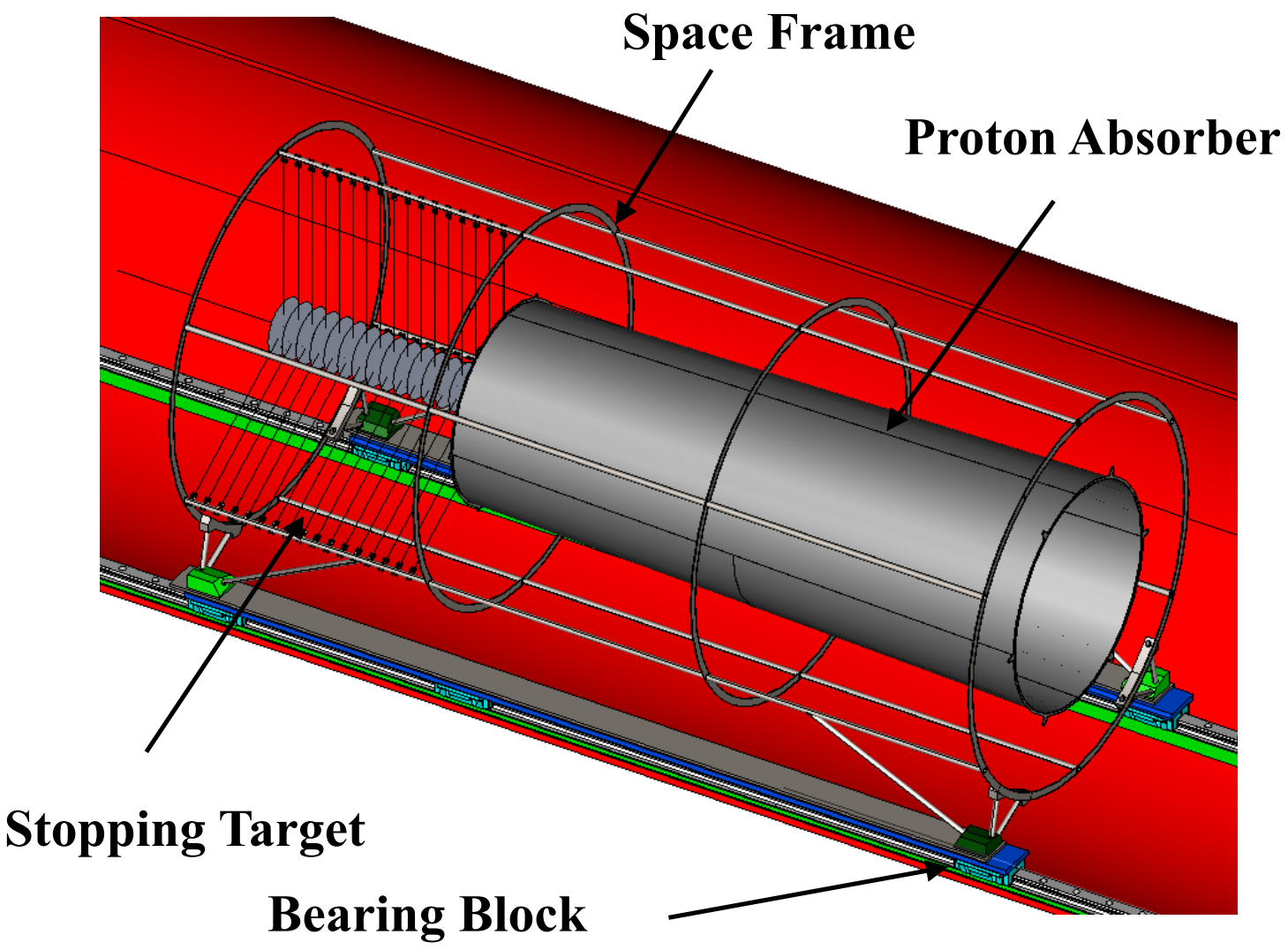

Figure 8.25. The proton absorber, stopping target and stainless steel space frame that supports the target and proton absorber.

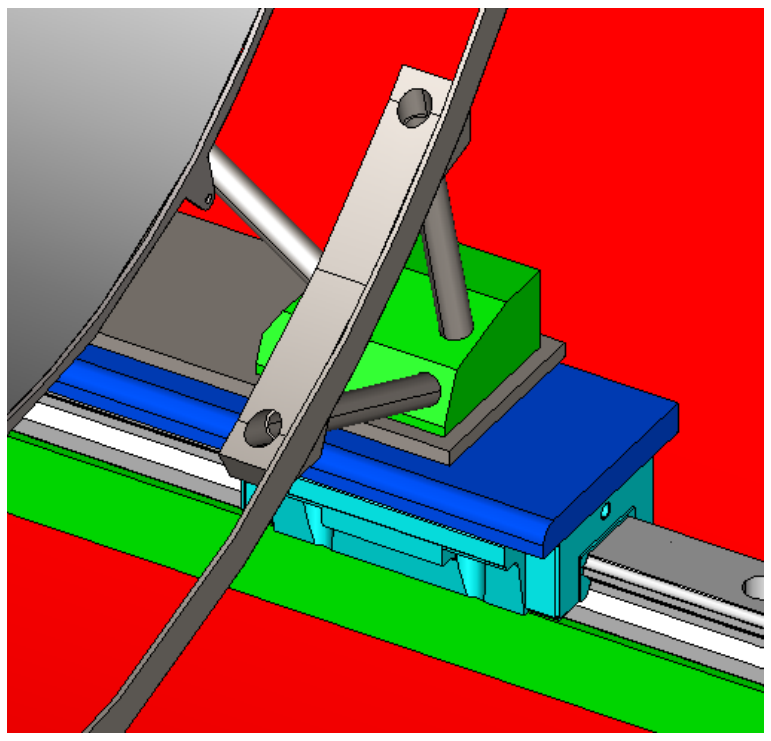

Figure 8.26. Detail of the connection between the space frame and the internal rail system.

***Considered Alternatives to the Proposed Design***

Thin polyethylene sheet is rather difficult to support. A prototyping effort is under way to evaluate this concern. Alternate low Z materials with better mechanical properties exist (other hydrocarbon polymers or Styrofoam) and will be studied if necessary.

The nominal shape of the proton absorber is a simple hollow cone. Other shapes, such as the “blade” configuration shown in Figure 8.27, are being considered. The blade alternative would allow conversion electrons to spiral through unaffected while still intercepting most protons. Simulations will determine which configuration is preferable.

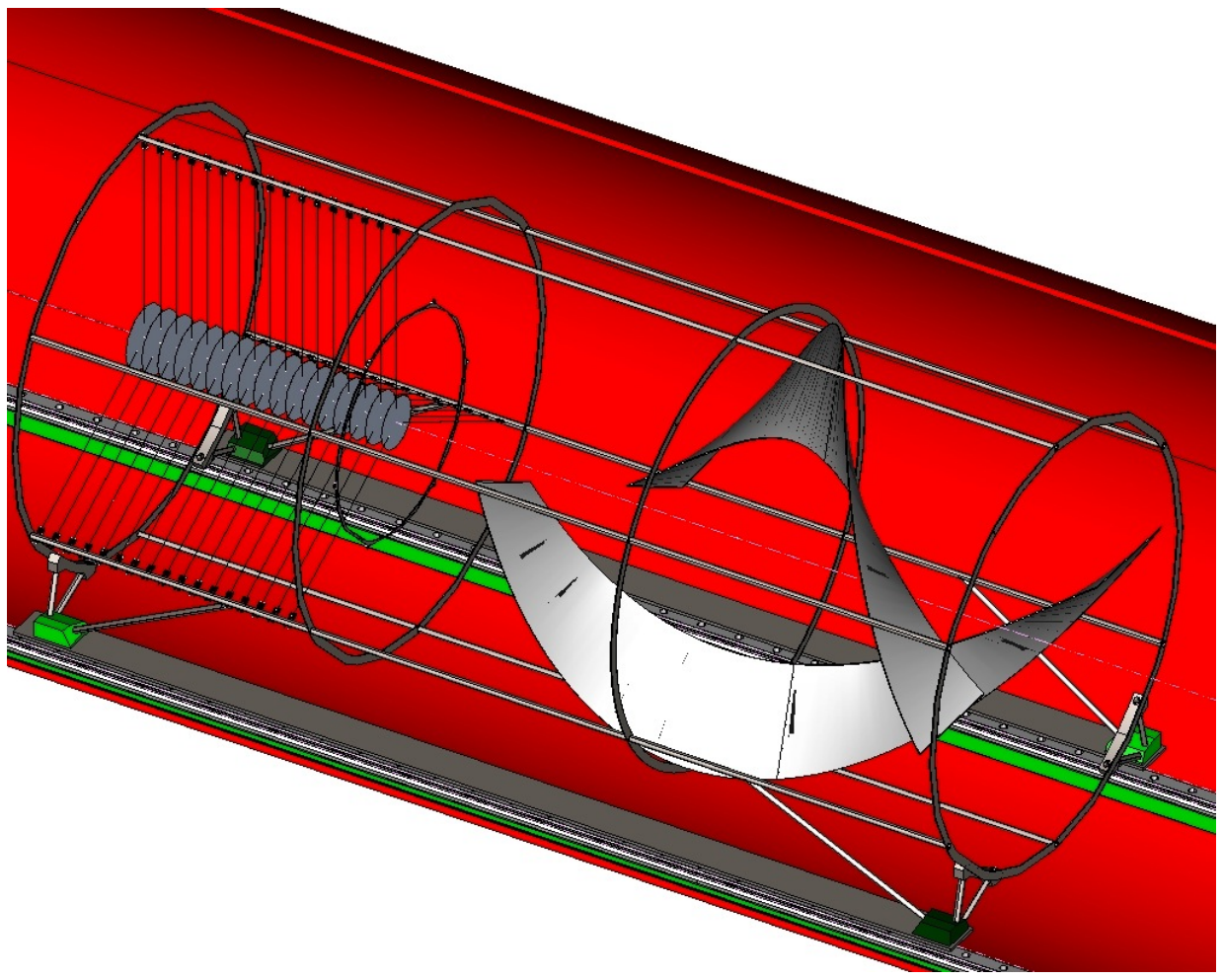

Figure 8.27. Alternate spiral "blade" configuration of the proton absorber.

### *8.3.7* Muon Beam Stop

The purpose of the Muon Beam Stop (MBS) is to absorb the energy of beam particles that reach the downstream end of the Detector Solenoid to minimize the rate of accidental particles in the active detectors from muon decays and captures in the Beam Stop. The Muon Beam Stop is located within the warm bore of the Detector Solenoid, downstream of the calorimeter. It is supported and aligned by the rail system internal to the Detector Solenoid and moves into place on ball bearing blocks. A pictorial view of the muon beam stop is shown in Figure 8.28.

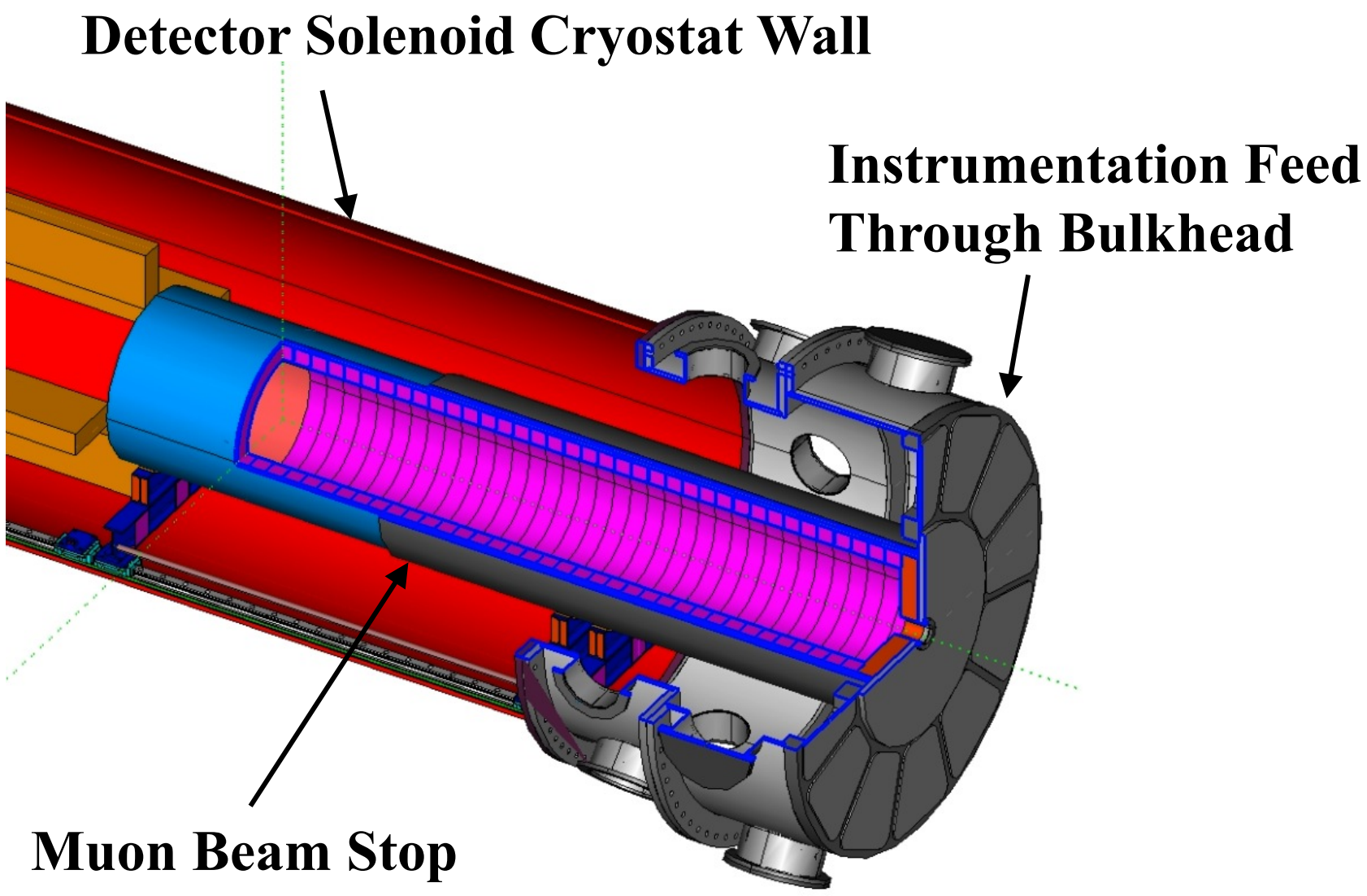

Figure 8.28. Muon Beam Stop in position.

The Muon Beam Stop consists of several cylinders, composed of different materials. Components consisting of stainless steel, lead and high density polyethylene (HDPE) will be assembled and bonded together. The HDPE may be doped with either boron or lithium. Figure 8.29 shows the beam stop with the individual component names and materials labeled.

The end plug (labeled CLV2 in Figure 8.29) contains a 10 cm diameter hole through the center to provide a line-of-sight for the Muon Stopping Rate Monitor located on the muon beam axis downstream of the MBS. The precise sizes, volumes and masses of the Muon Beam Stop components are described in [7]. Volumes and diameters of each component are derived from simulations.

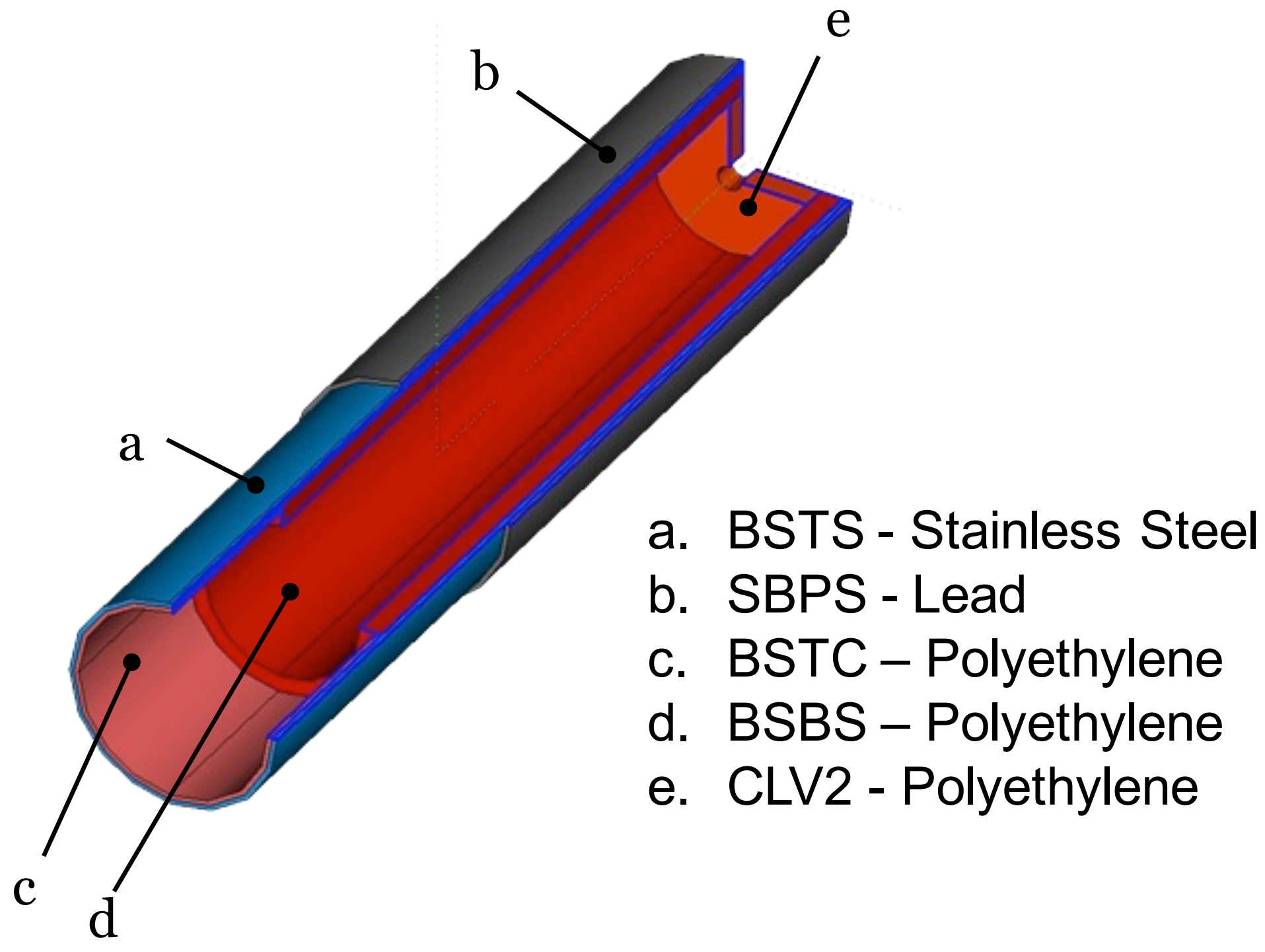

Figure 8.29. Muon Beam Stop components.

***Manufacturing and Assembly***

The Muon Beam Stop will be manufactured and assembled in outside facilities and transported as an assembly to the experiment hall. The stainless steel outer shell will consist of a single piece that is rolled and welded. The polyethylene pieces will be made of a series of rings, separately machined and bonded together with no line-of-sight cracks, except for CLV2, which will be made from a single sheet. The thickness of the rings will be limited to 100 mm, the manufacturing limits on the raw material. A drawing of an individual ring is shown in Figure 8.30.

The Muon Beam Stop will be made in several steps. First, the stainless steel tube will be manufactured and shipped to Fermilab for inspection. Measurements of the inside diameter of the tube will be used to determine the precise outside diameter of the HDPE parts that will fit inside it. The stainless tube will then be sent to a lead vendor, where the lead tube will be rolled and welded around the existing stainless tube. Finally, the stainless/lead tube will be shipped to Fermilab for assembly with the HDPE parts.

***Support and Alignment***

The Muon Beam Stop will be transported into position, supported and aligned by a linear rail system and bearings, manufactured to fit inside the Detector Solenoid warm bore and to support the full mass of the beam stop. The Beam Stop will rest on and be aligned to the linear rail system using a support structure that contacts the linear bearings at four positions as shown in Figure 8.31.

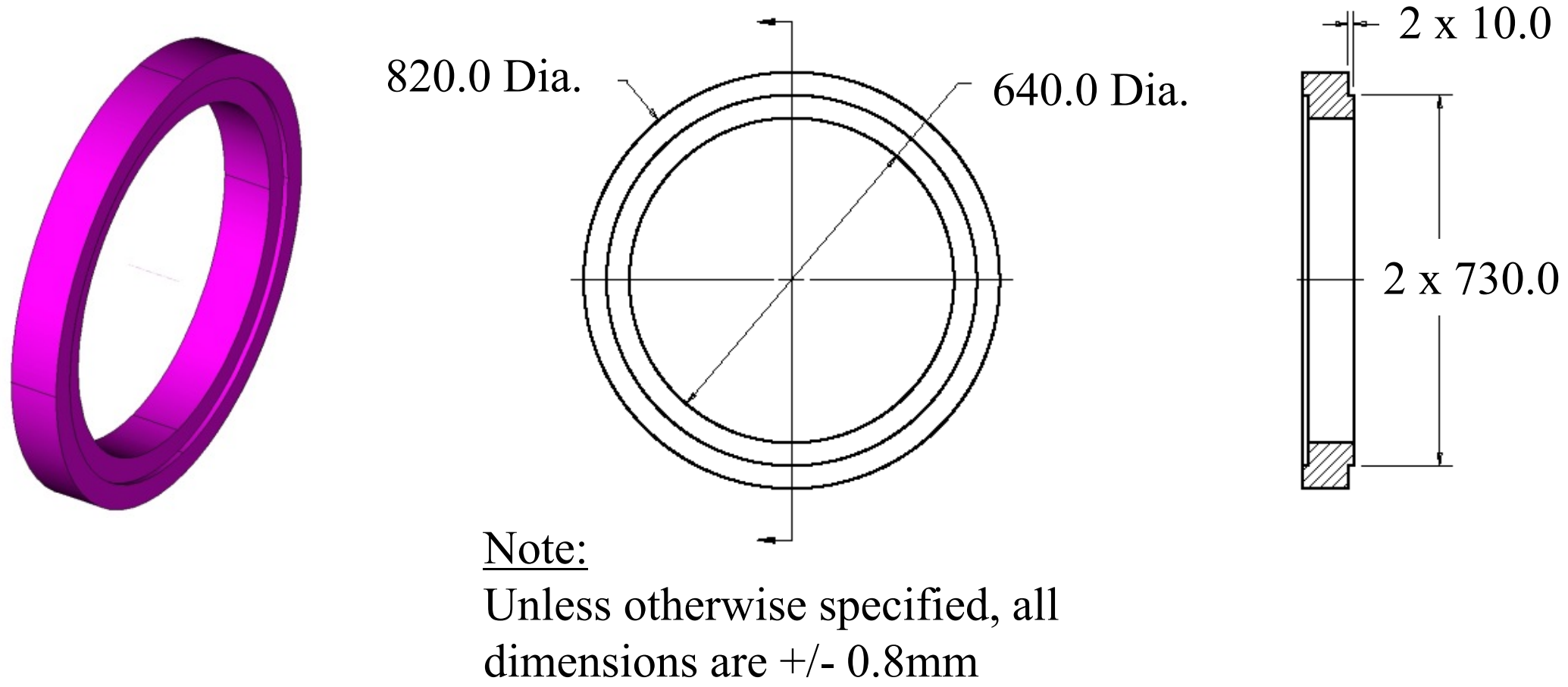

Figure 8.30. Muon Beam Stop in final position with internal support structure.

The Muon Beam Stop will be rolled into position while connected axially to the Tracker, Calorimeter and Instrumentation Feed Through Bulkhead (IFB). This is necessary because the cables and cooling tubes from the Tracker and Calorimeter are terminated at and permanently attached to the IFB. Although connected in the axial direction, these components will be aligned separately in x and y. The support positions and installation procedure for the MBS are given in [9]. The MBS will be aligned with respect to the center of the Detector Solenoid magnetic field as described in [10].

***Considered Alternatives to the Proposed Design***

The Muon Beam Stop component sizes and materials have been chosen based on a series of design studies, simulations and manufacturing considerations. Further study may call for some modification in the size and shape of the rings. Proposals to alter the length and/or diameter of the MBS are under consideration [13].

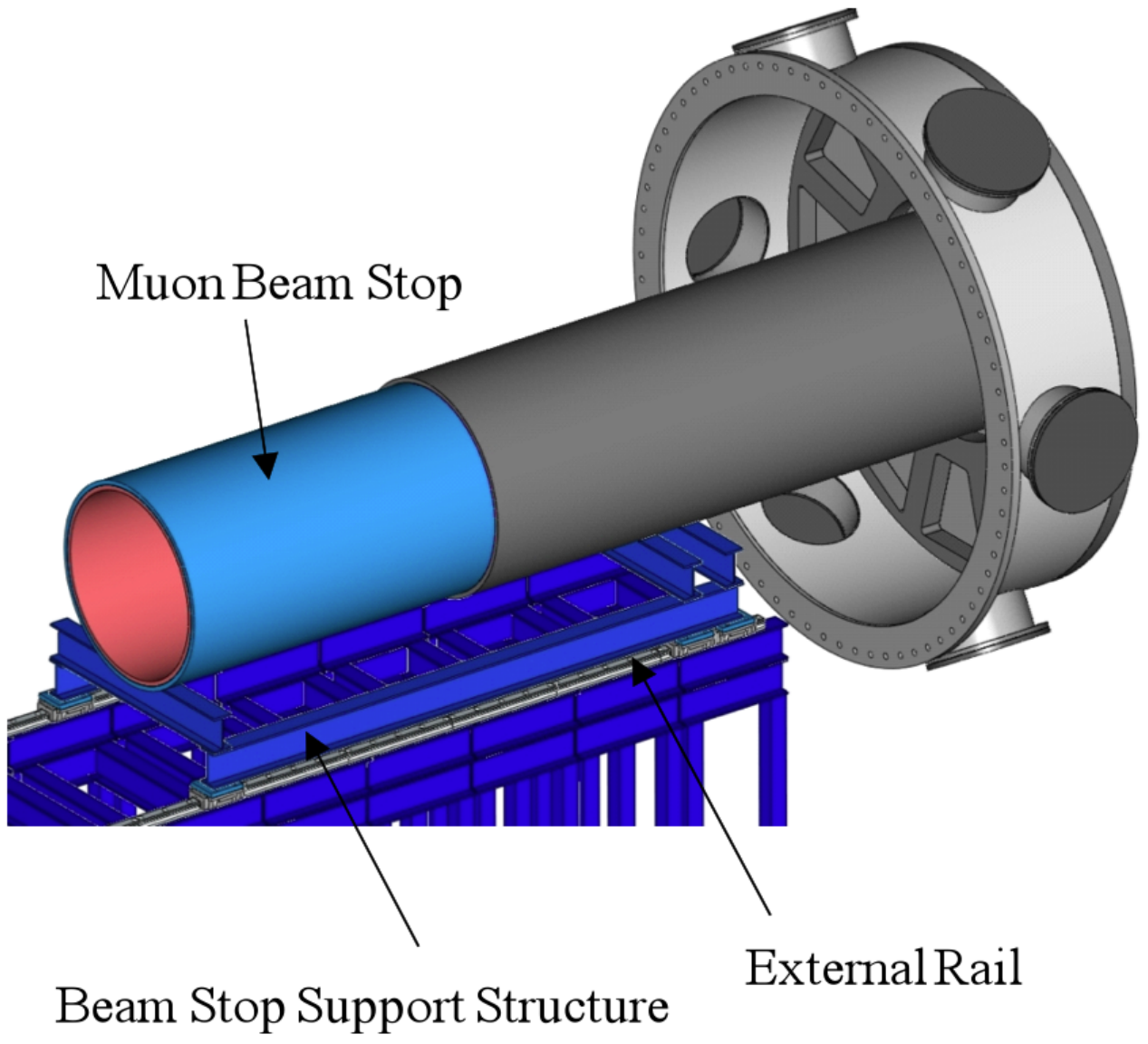

Figure 8.31. Muon Beam Stop in final position with internal support structure.

Although the baseline material of the polyethylene parts is HDPE, several types of doping are being considered. Boron doping between 5% and 30%, lithium doping of 7.5% natural lithium (92.6% lithium-7 and 7.4% lithium-6), or enriched lithium, with higher percentages of lithium-6, may be used for all or part of the polyethylene parts.

In the baseline design, the MBS is completely supported by the rail system. Partial support from the cryostat wall, by attachment to the IFB on the downstream end, is being considered. This would allow the rear bearing blocks to bear less weight and possibly improve the mechanical stability of the MBS, but would result in a more complicated support system. This alternative may be necessary if the length of the MBS is increased.

### *8.3.8* Neutron Absorbers

Absorbers must be placed around the Detector Solenoid to limit the number of neutrons reaching the Cosmic Ray Veto. These absorbers surround the DS vacuum enclosure and will be supported independently from the Detector Solenoid.

The Absorbers are constructed primarily of concrete blocks. The blocks include magnetic steel reinforcement bars and brackets. Analysis will need to determine whether the incorporation of the steel parts so close to the DS solenoid field will be acceptable. An illustration of a “typical” concrete block is shown in Figure 8.32. A specification for the composition of the concrete and steel to be used is given in [8].

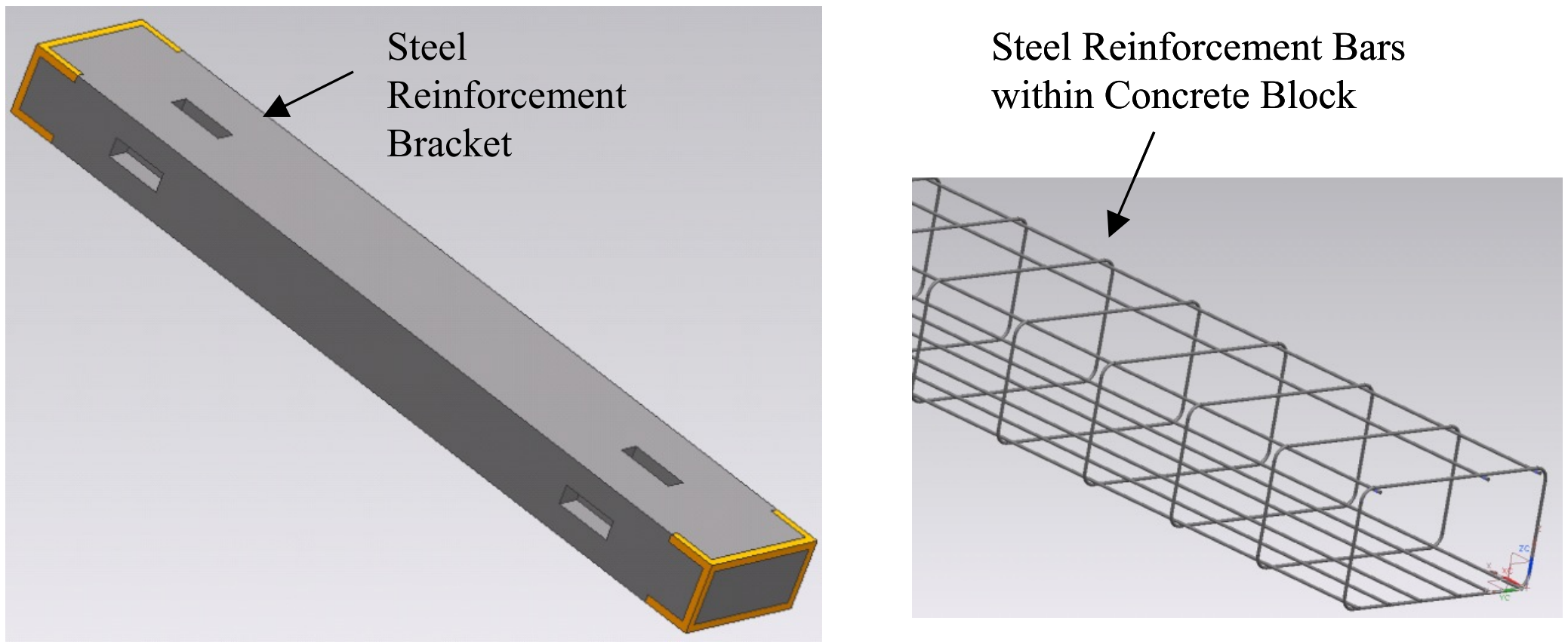

Figure 8.32. Neutron Absorbers and Shielding.

The neutron absorbers are made of two different sections. They are shown assembled in Figure 8.33. The central section encompasses the entire DS and a section of the VPSP. This section occupies the axial space that was formerly taken by the cryostat iron. The end cap shielding encloses the IFB and the termination of the connectors for the cables and pipes from the Tracker and Calorimeter.

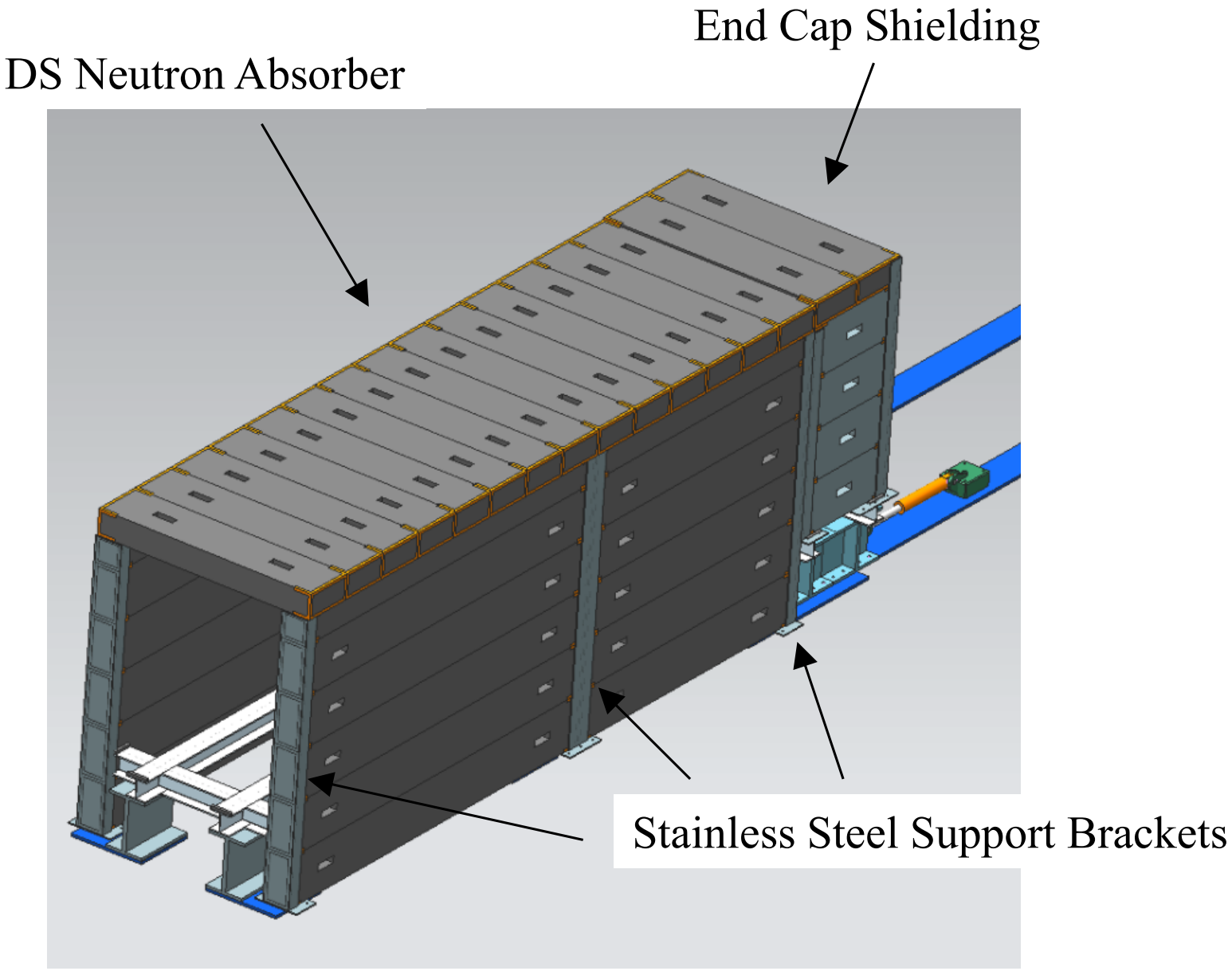

Figure 8.33. Overall view of Neutron Absorbers and Shielding.

***Manufacturing and Assembly***

Due to manufacturing constraints and to facilitate installation, the absorbers and shielding are made of many pieces. Figure 8.34 and Figure 8.35 show in more detail how the central section and the end cap shielding are constructed. These components will be assembled from many concrete blocks of different sizes. The stacks of concrete blocks are supported by stainless steel brackets which provide structural support and make assembly easier. An opening in the bottom of the end cap shielding allows an exit area for cable and cooling tubes from the Tracker and Calorimeter.

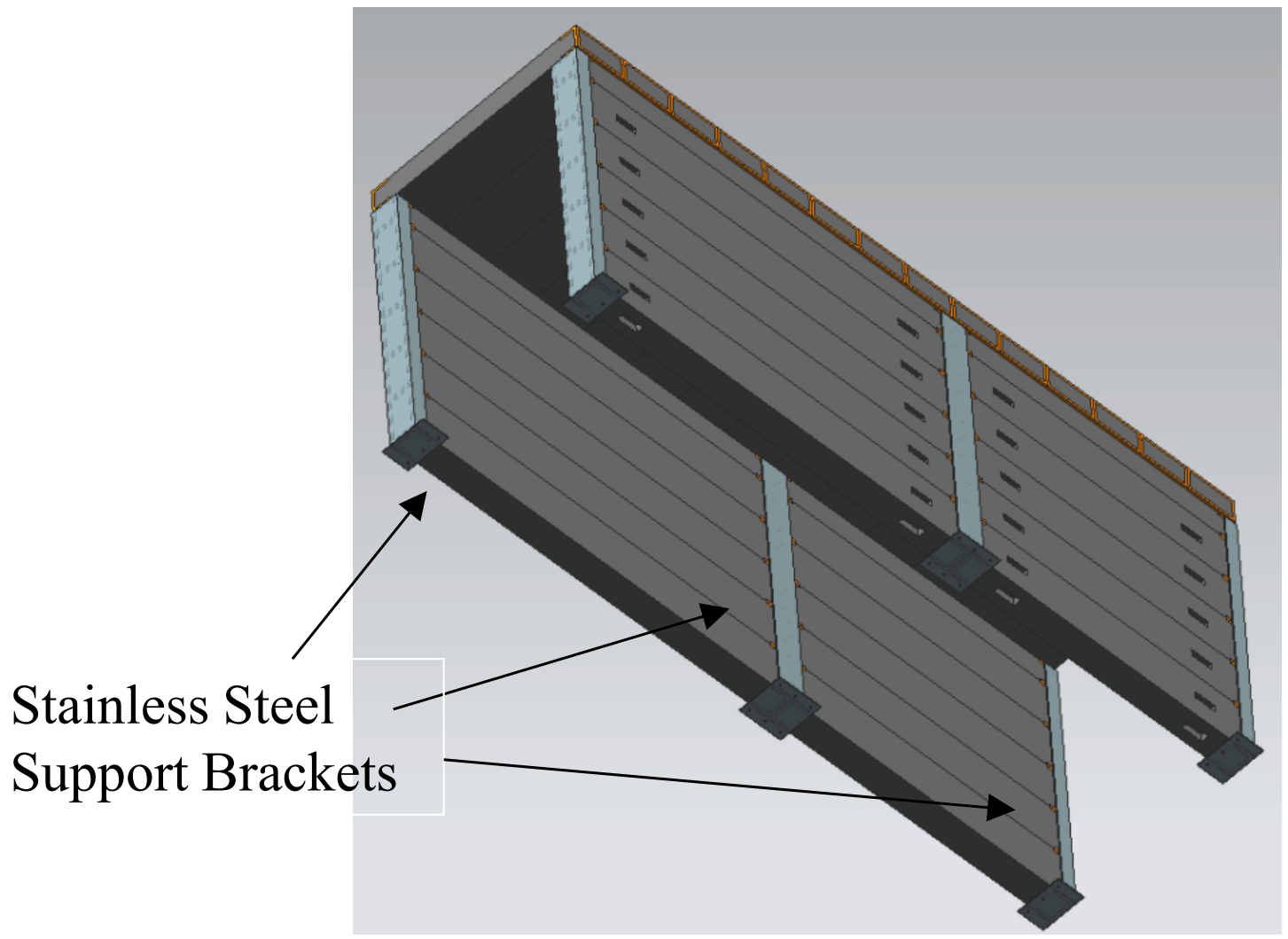

Figure 8.34. Bottom view of Central Section of the Neutron Absorber.

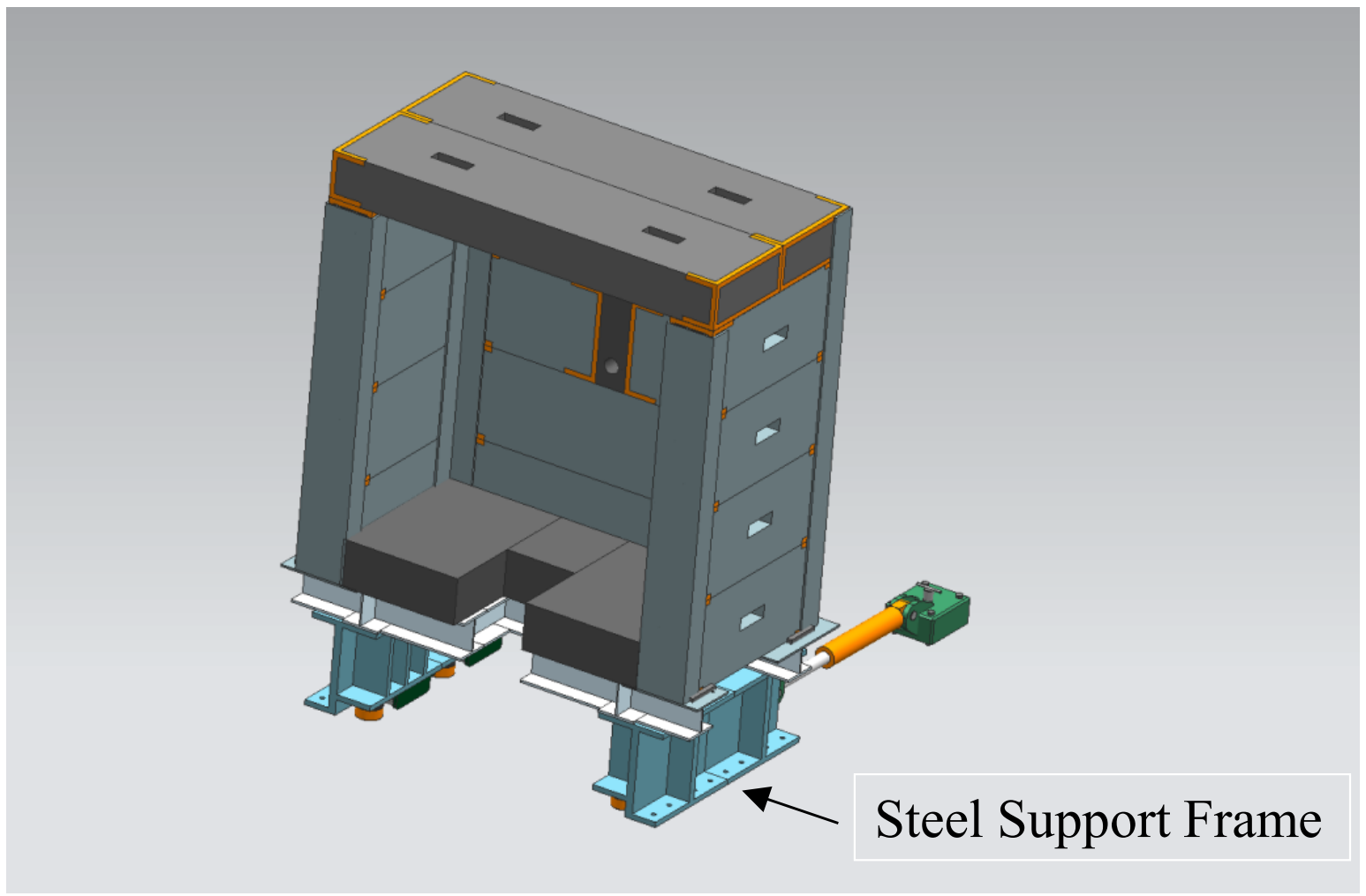

Figure 8.35. End Cap Shielding.

***Support and Alignment***

The concrete blocks that compose the central absorber and end cap shielding will be generally self-supporting. Both sections will be built in the detector building using the available overhead crane. The central area will rest upon the floor of the detector building, while the end cap shielding will be supported by a steel frame as shown in Figure 8.35. As noted above, some stainless steel brackets will be necessary to facilitate support. Structural analysis to determine the size and frequency of supports is taking place.

The absorbers will be aligned with respect to the center of the Detector Solenoid magnetic field using fiducials mounted to the exterior of the DS cryostat [10]. The concrete end cap shielding will be rolled into position on a track containing Hillman rollers and supported by pads that will be shimmed to the correct height. It can then be rolled back to allow access if necessary for maintenance.

***Considered Alternatives to the Proposed Design***

Although the concrete blocks contain magnetic steel reinforcement bars and brackets, austenitic stainless steel (304 or 316) could be substituted if the magnetic steel is found to be unacceptable. Additional cost would be incurred for the manufacture of blocks with stainless steel. Ongoing studies will determine whether stainless steel is necessary for this application.

It is possible to design the end cap shielding to rest on the floor of the detector hall, as does the central section. This may be slightly cheaper to build, and will not require a frame or hydraulic equipment to allow axial movement. However, this alternative would require that the structure be disassembled with the overhead crane to allow access to the DS internal area. This would in turn require disassembly of the ceiling of the detector hall.

Water is a good absorber of neutrons and could be used if an additional external absorber is desired. A water tank can be built with walls of HDPE. Such a tank could surround the DS vacuum space between the outside wall of the cryostat and the concrete enclosure. A conceptual design of the tank is shown in Figure 8.36.

If an absorber is desired within the vacuum space of the DS cryostat, HDPE tanks could be built for this application. If used internally, water is not desirable because, although the possibility of a leak is very small, any risk of a water leak in this area is unacceptable. Internal tanks could be filled with a powder containing boron or lithium to eliminate this possibility.

### *8.3.9* Detector Support Structure

The detector support and installation system will be used to transport and align components within the Detector Solenoid. The muon stopping target, proton absorber, tracker, calorimeter, and muon beam stop will be moved accurately and safely into position and aligned with respect to the standard Mu2e coordinate system [9]. The components vary significantly in mass (from less than 3 kg to over 4000 kg) as well as in their alignment accuracy requirements. These components will be supported by the inside wall of the Detector Solenoid cryostat.

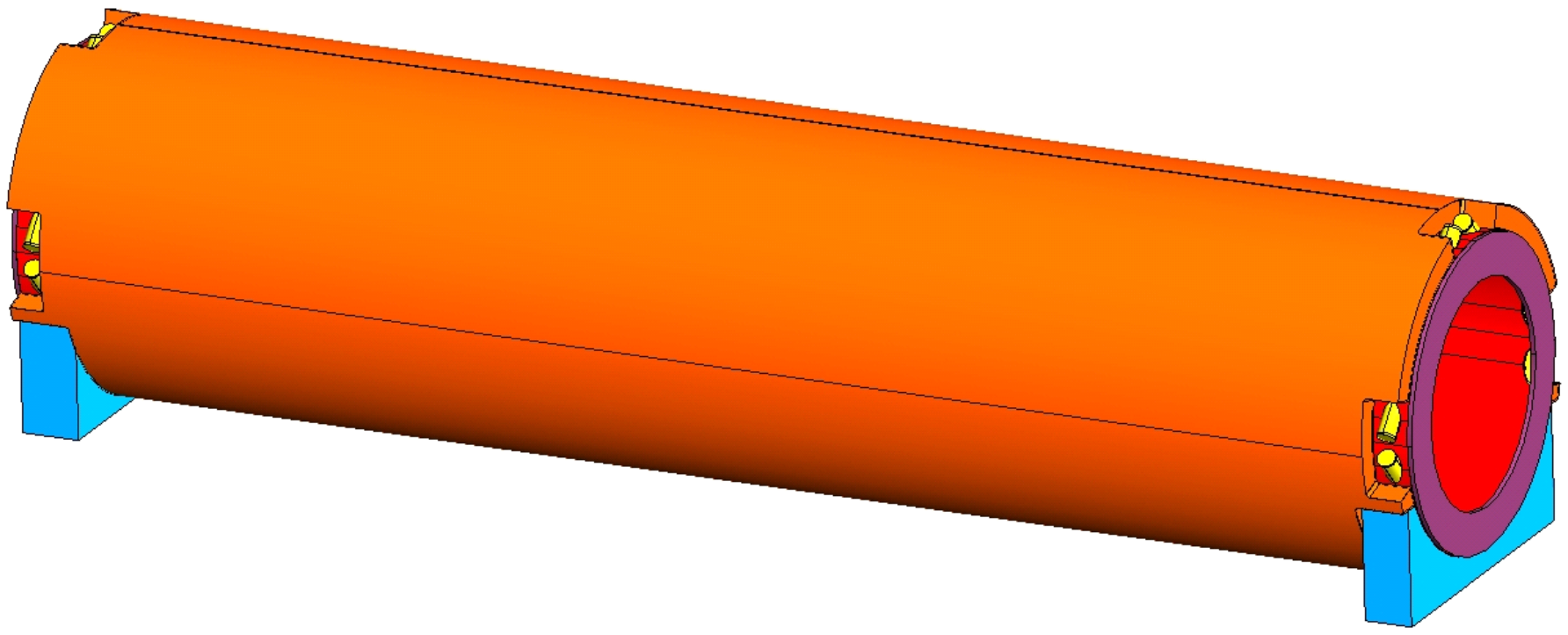

Figure 8.36. Water tank (shown in orange) surrounding the DS cryostat.

An overall view of the detector support and installation system is shown in Figure 8.37. The components are supported by two rails and transported by linear ball bearing blocks. Two separate rail systems will be implemented, the “internal” and “external” systems. Once transported, the alignment of all components will be maintained by the internal rail system. In Figure 8.37, the muon stopping target and proton absorber are shown in their final positions, while the remaining components are still located on the external rail system.

The internal system is shown in cross section in Figure 8.38. It will be attached to stainless supports that are welded onto the inside wall of the DS cryostat, also shown in Figure 8.38. It will support the weight of each component, allowing all alignment criteria to be achieved. The alignment criteria for each component are given in [9]. In the proposed design, the rails and blocks are made exclusively of non-magnetic components. Areas of the neutron absorber are cut away to provide clearance for the rails in a way that minimizes any line-of-sight cracks.

The external rail system (Figure 8.37) is located outside of the Detector Solenoid and is used to transport components into position inside the DS warm bore. It consists of a series of stands, each of which can be installed or removed as needed. The external stands are made of structural steel, each with sections of rails mounted to the top surface that can be connected and disconnected accurately. The rails will be identical to those used for the internal system, and the last stand, closest to the cryostat, will be attached to the internal system during installation.

Figure 8.39 and Figure 8.40 show a single external stand with the rail system attached, and the connection between the final external stand and the internal system, respectively.

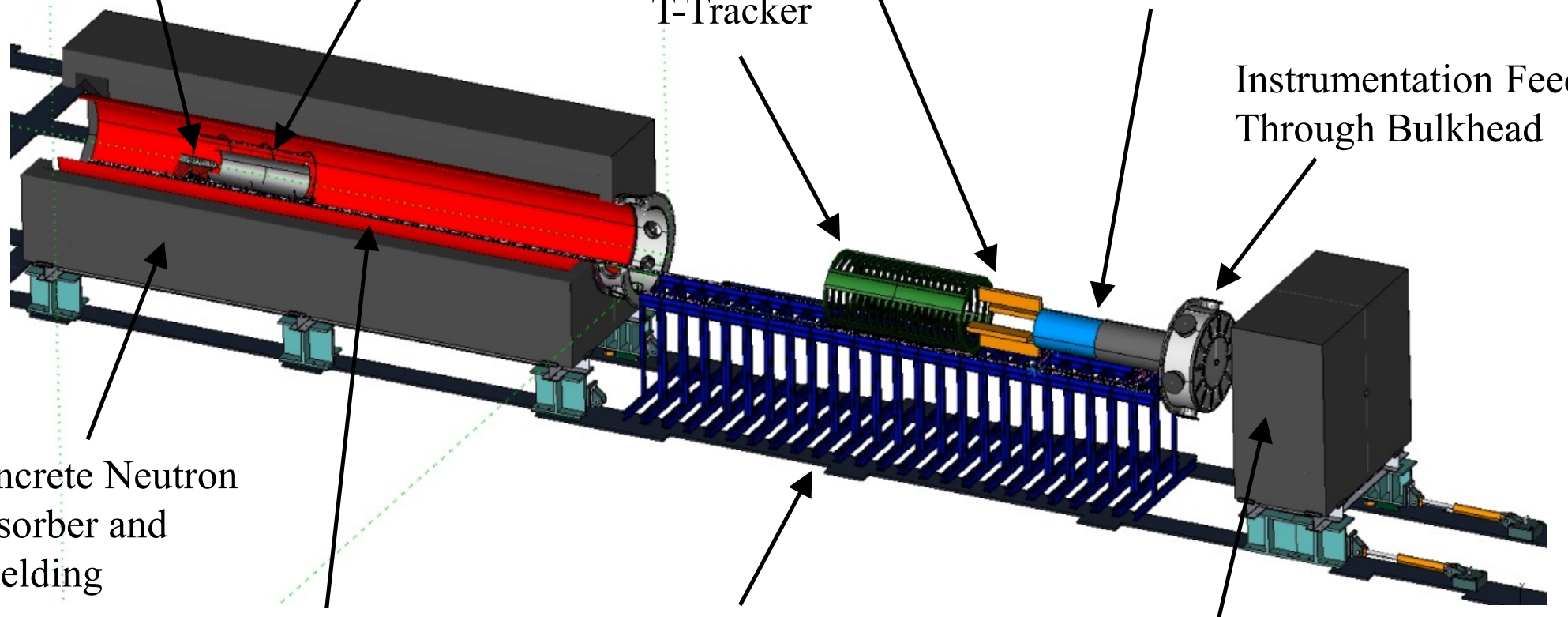

Figure 8.37. Conceptual Design of the Rail System.

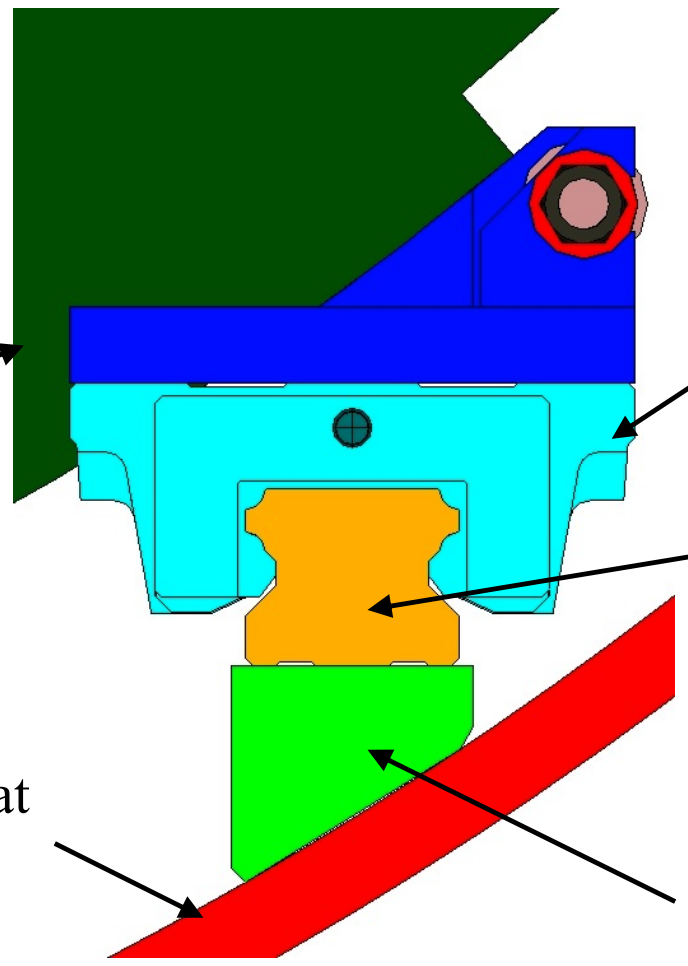

Figure 8.38. Cross Section of the Internal Rail System.

Each component will be aligned to the Detector Solenoid warm bore and attached to the rail system separately. Each will include support elements that provide the necessary structural support. Four leveling feet will allow for alignment of each component. The leveling feet will contact the linear bearing blocks. The specific position of each bearing block has been determined from a structural analysis described in [9]. Descriptions of the support structures are included in the conceptual design sections for each component.

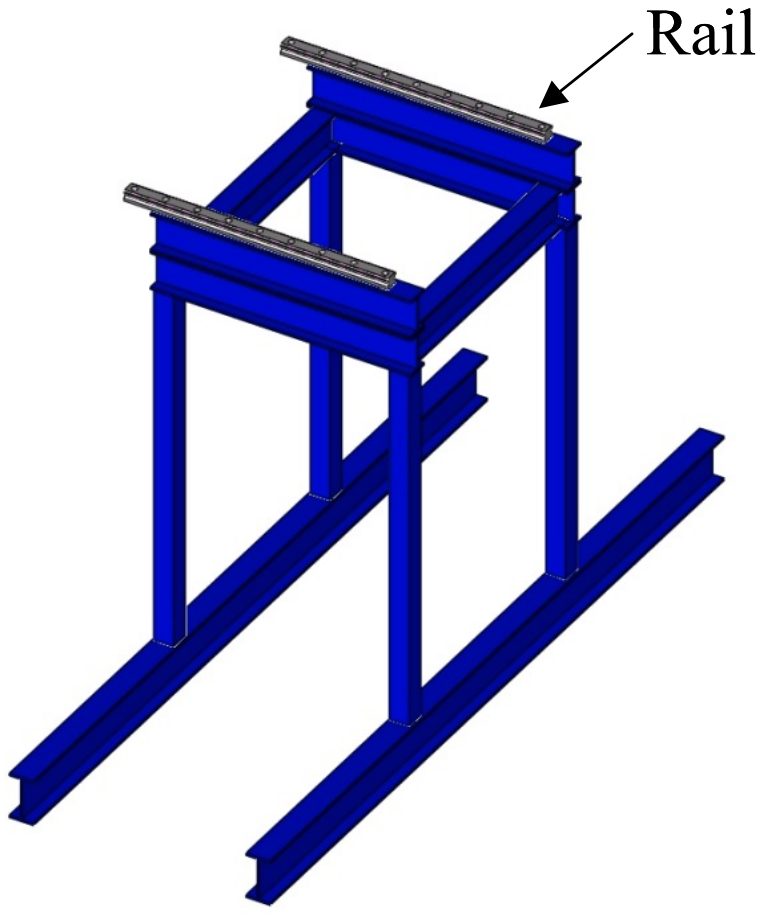

Figure 8.39. Individual external rail system stand.

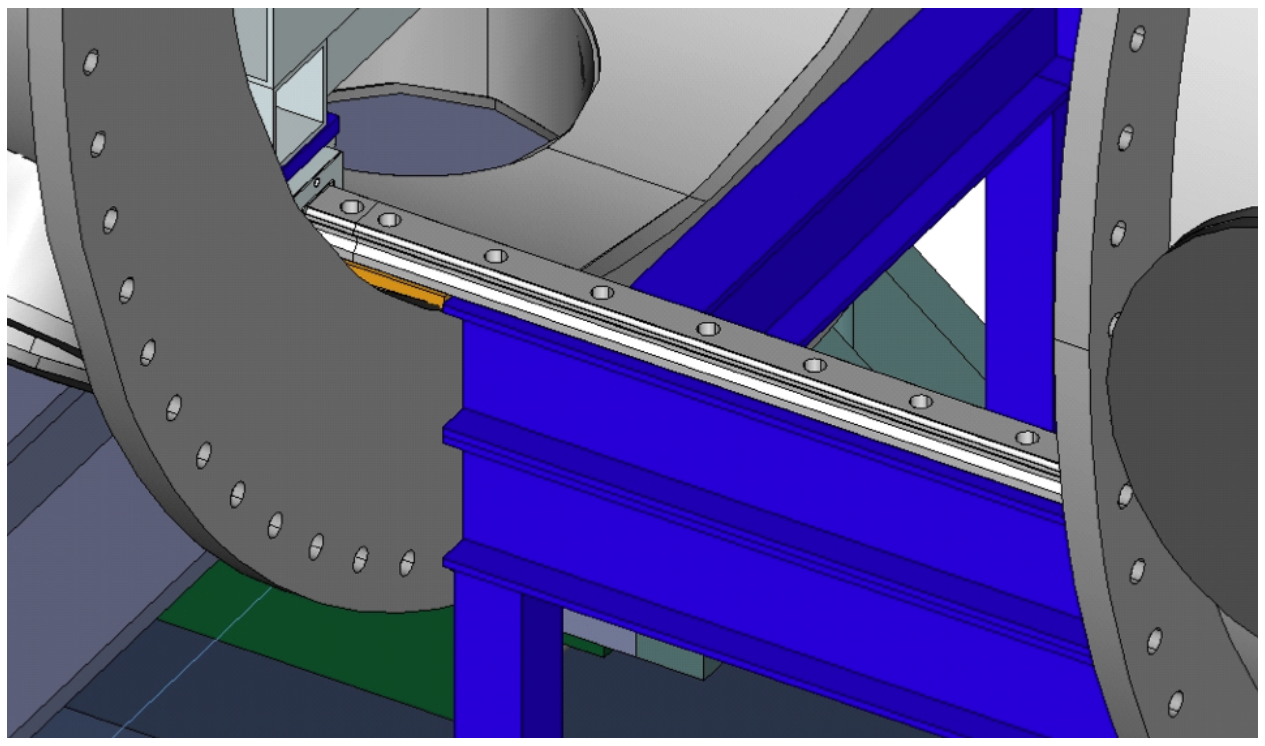

Figure 8.40. External to Internal rail system attachment.

Cabling and cooling tubes from the tracker and calorimeter will be installed and attached to the Instrumentation Feed-Through Bulkhead (IFB) before moving the entire assembly, including the detector components, into position [9]. The cables and tubes will pass over the Muon Beam Stop before being terminated in the IFB. This mandates that the tracker, the calorimeter, the beam stop and the IFB be rigidly attached to one another and moved into place as a single unit. After being individually aligned, these components will be connected axially by attaching their respective bearing blocks, as shown in Figure 8.41. Figure 8.37 shows these components in their positions before being rolled into the Detector Solenoid bore.

***Manufacturing and Assembly***

Several manufacturers have been identified who make rail systems that will potentially fit the requirements of the project with respect to accuracy, load and magnetic properties. The proposed design has been developed in conjunction with THK Co., LTD.

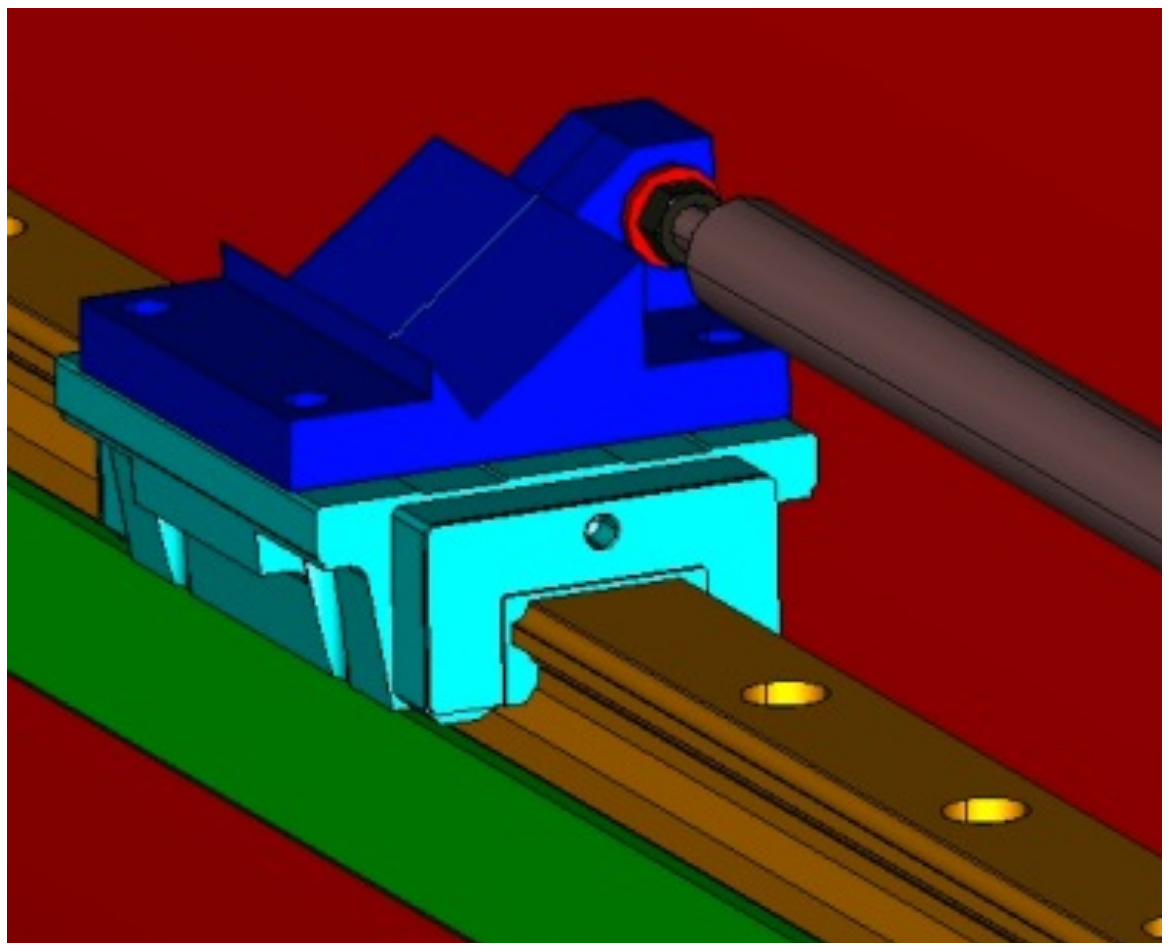

Figure 8.41. Axial Component Attachment System.

The supports for the internal rail system will be welded to the inside wall of the Detector Solenoid by the cryostat manufacturer before the DS coil is wound. The rail system and bearing blocks will be manufactured and tested outside Fermilab by the rail system vendor. Individual parts will be inspected upon their arrival. The parts will then be mounted by Fermilab personnel to the existing supports, aligned and shimmed into position with respect to the center of the DS magnetic field. The components of the internal and external rail systems will be identical in cross section.

***Detector Solenoid Component Installation Procedure***

The components to be installed by the system described in this section are the muon stopping target, proton absorber, tracker, calorimeter, and the muon beam stop. The Instrumentation Feed-Through Bulkhead and end cap shielding will be moved on a different rail system, discussed elsewhere, although the IFB is required to be connected axially to the other components. The assembly procedure for installation of components is described in [9].

***Considered Alternatives to the Proposed Design***

A rail system using Teflon guides instead of nitride linear ball bearing blocks has been considered. Such a system would be equally non-magnetic, but not as accurate, and involve a much higher coefficient of friction that requires more complicated methods of moving the heavy objects.

A similar rail system comprised of magnetic components that would have to be removed after installation has been evaluated. Systems such as these are more readily available at lower cost, have an equally low coefficient of friction and are just as accurate. However, the requirement to remove the rail system prior to operation would make it necessary to place the components on a separate set of fixed supports before removing the rails. This would make the procedure for installation and alignment, as well as removal and replacement of components, significantly more complex.

Bolting the rails to the walls of the cryostat has been evaluated as an alternative to welding. This would eliminate any possibility of distortion to the inner wall from welding as well as eliminate any risk of increased magnetic permeability after welding. However, the difficulty of placing the holes accurately, as well as structural issues related to the wall thickness of the tube, make this solution less desirable. It is still a possibility if tests reveal that the amount of magnetic permeability created by welding is unacceptable.

Instead of extending the rail system to the upstream end of the cryostat, the rails could terminate at the downstream end of proton absorber. In this case, the muon stopping target and proton absorber would be mounted to fixed supports directly attached to the DS cryostat wall. This alternative has the advantage of having a slightly lower cost, due to the shorter length required for the rails and the elimination of a few bearing blocks. However, it has several disadvantages. The magnet measurement system would not be able to extend to the upstream end of the DS. Also, the movement of the stopping target and proton absorber into the bore would become more difficult. They would either need to be transported by hand or cantilevered from the rail system before being set on the fixed supports. In either case, it would be more difficult to align the components with multiple "iterative" adjustments, because the alignment adjustments would have to be done in place, rather than outside the bore area. Replacement or realignment, if necessary, would also be more difficult.

### *8.3.10* System Integration, Tests and Analysis

The Muon Beamline is a WBS Level 2 subproject that consists of many different deliverables. The Muon Beamline subproject is divided into ten Level 3 deliverables. The integration task has been developed to manage the relationship of the deliverables with one another and with systems in other Level 2 subprojects, recognizing that proper integration requires effort, resources and manpower.

Alignment is another task that spans deliverables and subprojects so it is necessary to have an integrated, global approach across the entire Muon Beamline subproject. System tests and global failure mode analysis are also generalized tasks that require integration across the entire subproject. A detailed description of the System Integration, Tests and Analysis is described in [10].

## 8.4 Risks

The Muon Beamline deliverables, design and construction (L3) sub-projects of the Muon Beamline sub-project are well within the experience and expertise of the technical staff and physicists who are participating. Every effort has been made to specify these (L3) sub-projects in a manner that reduces the risk to an acceptably low level.

The risks identified in the Muon Beamline are summarized in the Mu2e Risk Registry [11]. The greatest identified risk is to the vacuum system if gas loads due to detector material, especially the tracker components that can be subject to strong outgassing and gas permeation through the thin wall of the individual straws, are higher than anticipated.

This risk, before mitigation, has a significant probability to occur. This is considered primarily as a technical risk. If the gas load exceeds $10^{-3}$ Torr the High Voltage interlock system prevents power from being delivered to the tracker. This could also impact the Project cost and schedule. Mitigation of this risk includes building in extra contingency in the pumping capacity of the vacuum system, strict specifications for gas loads, a program of outgassing tests for materials inside the vacuum and determination of the permeation properties of the straw tube material.

A second significant risk is that a vacuum window is damaged or breaks when the system is evacuated. Physics concerns often demand that vacuum windows be as thin as possible, making it more likely that a break might occur. Should a vacuum window break it could result in harm to nearby individuals, damage to equipment and the spread of radioactive materials. This risk will be mitigated by understanding the physics and technical requirements of each window and designing them to be no thinner than necessary. Operational and administrative controls will also be in place to limit access and activity in the vicinity of vacuum windows when the system is evacuated.

## 8.5 Quality Assurance

The Muon Beamline sub-project has developed a quality assurance plan [12]. In this section, as an example, the application of the quality assurance plan to the design, procurement and installation of the Muon Beam Stop is described.

- The Muon Beam Stop (MBS) consists of several parts made of three different materials: stainless steel, lead and high-density polyethylene (HDPE). The stainless steel and lead tubes will be designed per ASTM specifications (ASTM A240 and ASTM B749, respectively) and all design procedures will conform to the Fermilab Engineering Policy Manual.
- Engineering drawings for each individual component, sub-assemblies and assemblies will be completed. Tolerances are specified on each drawing in accordance with the Requirements and Specifications (R&S) document. The Fermilab Technical Division Design and Drafting Department will make the drawings, and formally check them according to applicable geometric tolerance standards.
- The drawings will be released to the Technical Division Quality and Materials (Q&M) Department. They will be sent to several outside manufacturers who have been approved by Q&M personnel. After quotations are received, they will be reviewed by personnel from the Fermilab Engineering and Q&M department, and a manufacturer will be chosen.
- The stainless steel tube will be manufactured first. After completion, they will be inspected by the Technical Division Inspection Department, and a formal inspection report will be electronically generated.
- If the tube passes inspection, it will be shipped to the vendor to have the lead tube rolled and welded around it.
- The HDPE parts that fit inside the stainless tube will not be manufactured until the stainless tube has been inspected. The actual manufactured dimensions from the stainless tube will be used to apply the specific tolerances and dimensions to the HDPE. The HDPE rings may need to be made of several pieces and fit together at assembly, due to size limitations from manufacturers.
- The stainless steel/lead tube assembly and the HDPE parts will be manufactured, inspected and approved or rejected by the same process as has been previously described for the stainless steel tube.
- Once all the parts have been received and approved, they will be assembled in an assigned area in the Fermilab Technical Division to ensure that all parts fit and that any required shimming has been established and documented. Cleanliness issues during assembly will be followed according to section 6.2 of the Muon Beam Stop R&S document [7].
- A prototype assembly of the rail system to be used in the Detector Solenoid bore will also be assembled in the Technical Division. The assembled Muon Beam Stop will be mounted to this system to ensure that all alignment issues are completely understood.

- A controlled document, or “production report”, will be written during the pre-assembly and alignment, and placed into the Fermilab Engineering Data Management System. Reports of these steps will be included in the general review process of the Muon Beamline sub-projects.
- Finally, the Muon Beam Stop Assembly will be shipped to the Mu2e experimental hall to be installed. During pre-assembly, decisions will be made regarding whether the Beam Stop will need to be partially or fully disassembled for shipment, or whether it can be shipped to the experimental hall as a complete assembly.
- Installation in the experimental hall will be done according to the steps outlined in sections 4.3 and 6 of the Detector Support and Installation System R&S document [9]. Alignment will be done in accordance with Appendix A of the same document.

## 8.6 Value Management

Value Management is defined as an organized effort directed at analyzing the functions of systems, equipment, facilities, services, and supplies for the purpose of achieving the essential functions at the lowest life-cycle cost consistent with required performance, quality, reliability and safety. Application of Value Management principles to the Muon Beamline sub-project includes physics and engineering studies and project and subsystem reviews to address technical, cost and schedule issues. Value Engineering has been applied in preparing the conceptual design for the Muon Beamline and will continue to be applied throughout the life of the Project.

The Mu2e Project has implemented an internal design review system where each subproject is closely examined to obtain optimal value for the system, given the technical requirements and schedule constraints imposed on it. These reviews are documented in the project’s document database. Documentation and updates are available to the project management staff, subsystem managers and other project personnel.

Value Management is applied for each WBS element at Level 3 or lower (except for WBS 5.1, Project Management). The following items provide a snapshot of Value Management considerations.

***Vacuum System***

A careful estimate of the vacuum load and selection of materials with known levels of out-gassing will help to properly specify the vacuum system components needed to meet the requirements with an appropriate level of performance margin. Less performance margin is necessary if the loads are well understood. Selection of materials with low out-gassing rates would reduce the load on the system and might also result in cost savings.

The conductance through the various system components (ducts, ports, etc.) will be studied in detail as part of the optimization of the vacuum pumps.

***Collimators***

Various drive mechanisms could be used to rotate the collimators in the TSu and TSd straight sections. The available options will be evaluated for cost and reliability.

***Muon Beamline Shielding***

The External Muon Beamline shielding material around the Transport Solenoid must be optimized to effectively shield the detectors while allowing access to the Antiproton Stopping Window Module region for repairs. The Internal Muon Beamline shielding assembly procedure must be optimized.

***Stopping Target***

Stopping target tolerances could drive the design and production costs. A careful simulation is required to determine appropriate tolerances that don't compromise performance.

***Stopping Target Monitor***

Utilizing existing detector solutions for the stopping target monitor should minimize the cost.

***Proton Absorber***

Styrofoam will be analyzed as an alterative approach to Polyethylene for the proton absorber. This would simplify the mechanical support since Styrofoam is largely self-supporting across the required span, while the required thickness of Polyethylene may not be.

***Muon Beam Stop***

Simulations of the Muon Beam Stop will be performed to optimize the material dimensions and locations.

***Neutron Absorbers***

The installation of the neutron absorbers is a complicated procedure due to the number of pieces, their size and shape. The installation procedure still requires optimization to reduce the necessary time and effort.

An attempt will be made to use cheaper, magnetic reinforcement bars in the concrete blocks that surround the solenoid rather than stainless steel. Studies will determine whether this is possible.

***Detector Support Structure***

The Detector Support Structure will be optimized through close coordination and integration with the components that it must transport and support. Prototypes will be built prior to procurement of final parts to verify performance.

## 8.7 ES&H

There are several potential hazards associated with the Muon Beamline component fabrication, installation and operation. These hazards are well understood, being similar in nature to other hazards frequently encountered by Fermilab Technical Division personnel. Proven mitigation strategies have been developed and documented. The primary areas of ES&H concern are listed below. A more detailed analysis for the Muon Beamline components is covered, along with the rest of the project, in the Mu2e Preliminary Hazard Analysis Document [15].

***Radiation***

The high intensity and high energy initial proton beam that hits the production target located in the PS bore is the primary source of radiation that will directly and indirectly (through secondary particles) irradiate the Muon Beamline components. Although the highest radiation dose will be accumulated in the PS side, all other Muon Beamline components will be exposed to a sufficiently high dose that proper radiation safety procedures will be required throughout the life of the project.

***Pressure Safety***

The Muon Beamline is essentially a large vacuum space formed by the bore of the solenoid system. Vacuum vessels pose a potential hazard to equipment and personnel from collapse, rupture or implosion. This danger is greatest near the thin windows that are required by the Mu2e experiment. Proper hazard and failure mode analysis will be performed to avoid the risk of generating vacuum failure.

***Very heavy objects***

Many of the Muon Beamline components (e.g., solenoid enclosures) are very heavy. There are significant mechanical hazards associated with the transport of heavy objects from one location to another. The Muon Beamline components will be moved by crane and by guide rails within the detector enclosure.

The potential consequences of mechanical hazards include serious injury or death to equipment operators and bystanders, damage to equipment, and interruption of the program. These hazards could be initiated by a dropped or shifted load, equipment failure, improper procedures or insufficient training/qualification of operators.

A hazard risk assessment will be conducted to evaluate the specific hazards to personnel for each of these activities and determine the means to mitigate the hazards. Any support structure that must carry over 10 tons will be reviewed for adherence to standard Fermilab engineering practices. Special lifting fixtures and transports will be designed for use with the solenoids. All lifting fixtures will be engineered, fabricated, and tested in accordance with ANSI/ASME Standard B30.20 (*Below-the-Hook Lifting Devices*). All applicable Fermilab design standards governing lifting devices will also be met. The maximum allowable lifting load will be legibly marked on each fixture. Sufficient space will be available inside the enclosures for personnel to remain clear of all lifting operations. Crane training, crane interlocks, inspections, and periodic maintenance will follow the procedures listed in the Fermilab ES&H Manual (FESHM). Personnel requiring authorization to use material handling devices must complete laboratory specified training and pass a qualification practical exam conducted by a skilled operator.

***Electrical Hazards***

High voltage and current are required to operate the Mu2e detectors. Since the Muon Beamline is responsible for providing the electrical feed-through ports, it is also exposed to electrical hazards. Electrical hazards include the potential for serious injury, death, and equipment damage. Electrical shock due to high voltage can be caused by exposed connectors, defective or substandard equipment, lack of adequate training, or improper procedures.

The electrical systems used to power the Mu2e solenoids and the hazards associated with them are similar to those in other experimental areas at Fermilab. Power distribution systems for the detectors will be designed in strict compliance with applicable codes. All systems will be grounded. Particular care will be provided for cable distribution to ensure code compliance with cable tray loading and content requirements. All electrical equipment, cables and cable trays will be protected against mechanical hazards according to established Fermilab ES&H procedures.

## 8.8 R&D Plan

***WBS 5.2 Vacuum system***

An R&D Test Vessel is proposed to test and verify all critical elements of the Muon Beamline Vacuum System. The general layout and components for the R&D Test Vessel are shown in Figure 8.42. The pumps, valves, and compressors shown are the actual ones specified for the experiment and will be used when Mu2e is assembled for production.

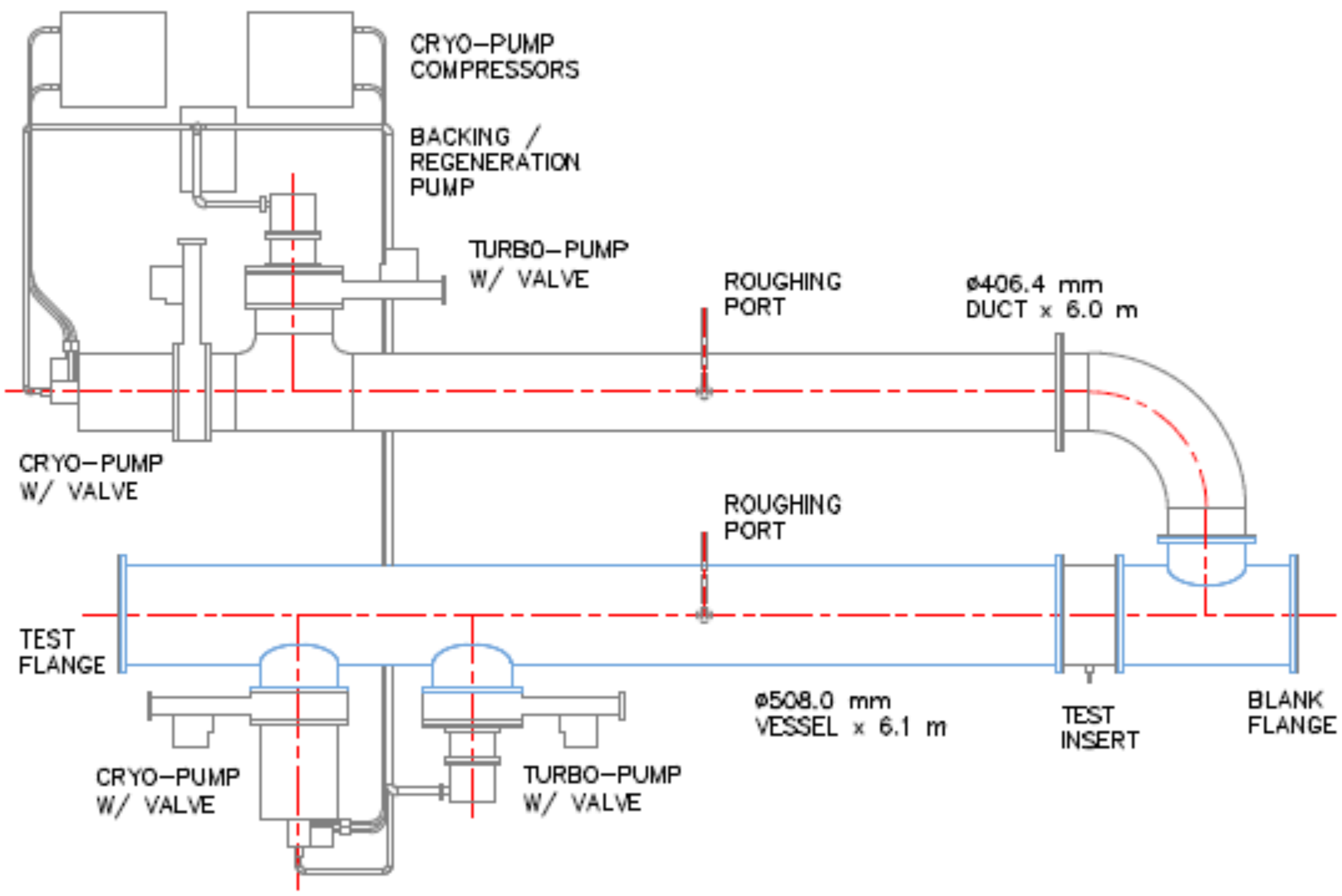

Figure 8.42. Configuration 1 of the Beamline Vacuum System R&D Test Vessel

The R&D Test Vessel is designed with flexibility to accomplish a large number of tests. The Configuration 1 apparatus shown can be used to verify the applicability and performance of the selected vacuum pumps. If the Vessel side pump valves are closed, the Duct side pumps will simulate and verify the conductance and pump-down performance of the PS + TSu vacuum system. In this configuration, the Test Insert is merely a through pipe, and the Test Flange is merely a blank flange.

If the Duct side pump valves are closed, the Vessel side pumps will simulate and verify the conductance and pump-down performance of the TSd + DS vacuum system. The Vessel can be loaded with a volume of outgassing material proportional to that found in the DS. In this configuration, the Test Flange is merely a blank flange and the Test Insert a through pipe, but a port on the Test Insert will allow the influx of a test gas. A proportional volume of tracker gas may be introduced to simulate straw leakage. Voltage breakdown measurements may be performed for the selected detector gases as a function of gas pressure by inserting a special High Voltage spark gap into the vacuum chamber. Air, water vapor, helium, or any combination of gases may be introduced to simulate any leakage or out-gassing effect.

The Configuration 2 apparatus, shown in Figure 8.43, can be used to verify the performance of the two most critical vacuum windows specified for Mu2e. In this configuration, the blank Test Flange is replaced by a 500 mm diameter Titanium window assembly, and the through Test Insert is replaced by one with a Kapton antiproton window assembly and a series of bypass valves and piping installed. The Test Insert may be fitted with optical viewing ports to allow visual monitoring of the antiproton window.

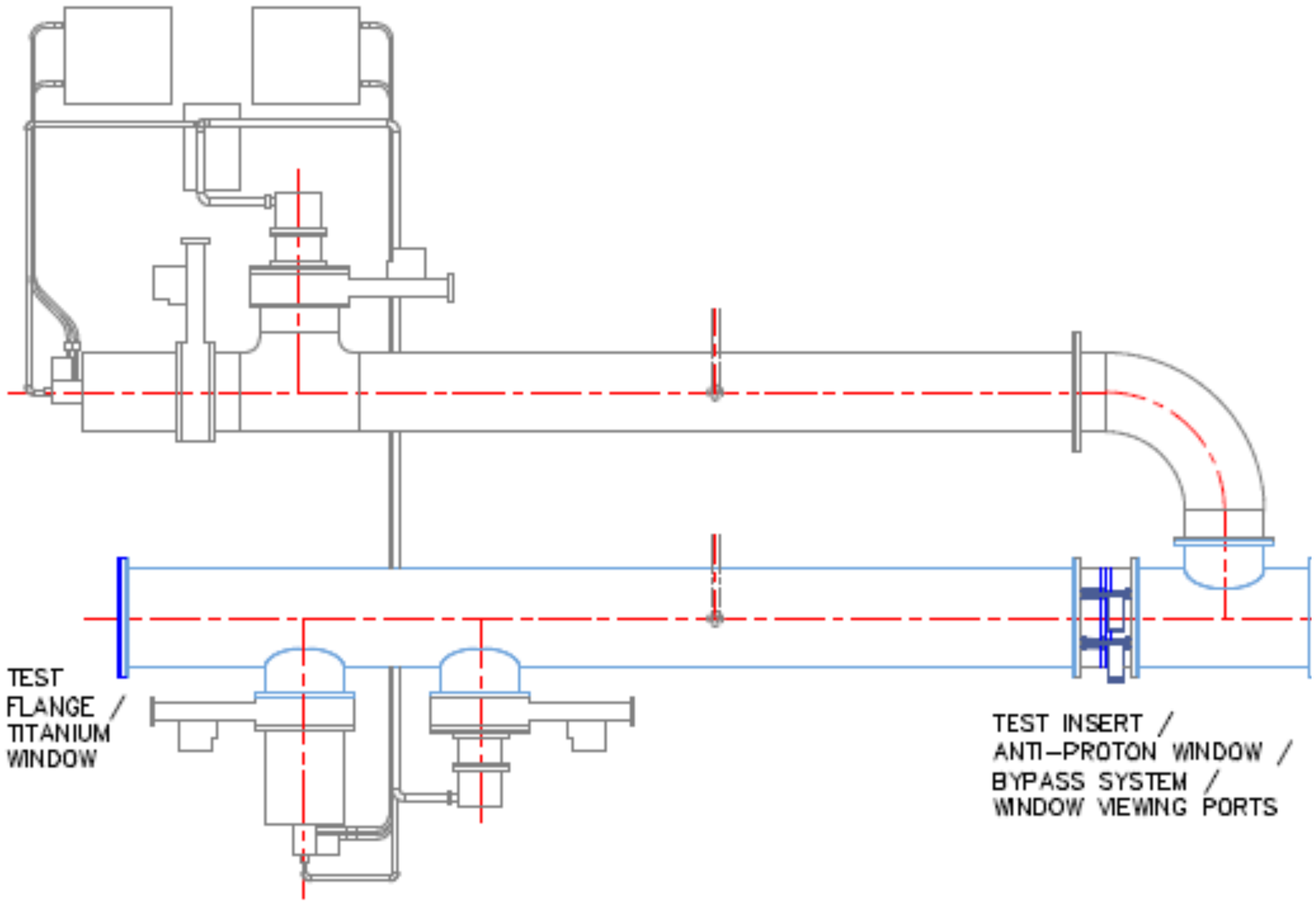

Figure 8.43. Configuration 2 of the Beamline Vacuum System R&D Test Vessel

In this configuration both the Duct side and Vessel side pump valves are open and all four pumps are in operation. One test will verify the performance of the bypass system and its associated control system in protecting the antiproton window from high differential pressures. The antiproton window may also be purposely subjected to high differential pressure to observe if failure occurs.

In this configuration, the largest PS Enclosure Titanium vacuum window can also be tested to ensure it survives the normal operating condition of 1 atm. differential pressure. Suitable ES&H restrictions will be maintained to ensure that no personnel are allowed near the Test Flange in case of Titanium window failure. In addition, a controlled failure of the Titanium window may be initiated to verify that the antiproton window survives the pressure wave produced in this kind of failure, and that the control system effectively shuts down the vacuum pumps. Mesh filters will be installed to protect the equipment from the potential of Titanium debris during a catastrophic window failure.

***WBS 5.3 Collimators***

Build and test a prototype drive mechanism for Col 3 to study the functionality of the collimator rotating mechanism, particularly the operational behavior of the electric motor under cryostat vacuum conditions.

Development of the antiproton stopping window design including:

- Window material selection and design.
- Window mounting and sealing design.
- Operational and failure pressure tests.

This work will be coordinated with the Vacuum System R&D efforts under WBS 5.2.

***WBS 5.4 Muon Beamline Shielding***

External Shielding Design:

- Study of the external shielding materials and alternate designs.
- Prototype development for an easily assembled and movable shielding system.

***WBS 5.5 Stopping Target and WBS 5.6 Proton Absorber***

Build prototype stopping target and proton absorber with their support structures. Verify that positioning and alignment tolerances can be met.

***WBS 5.7 Stopping Target Monitor***

The stopped muon rate in Mu2e is obtained by monitoring the rate of muonic xray production in the stopping target. At the same time, a major source of background comes from the muon capture process, where protons, neutrons and gamma rays are emitted in quantity. To verify the feasibility of this concept of monitoring the muon cascade and the rate of secondary production in the capture process, we need to perform a series of tests using a low energy muon beam. To verify the monitoring process, we propose to utilize the same Germanium detector (part of STM) as is planned to be used in the Mu2e experiment. The fluxes of secondary protons and neutrons will be monitored with silicon detectors and neutron detectors, respectively, while high energy photons will be monitored with a sodium iodide detector. It will be difficult to find a large solenoid system to use for the tests so the tests very likely will not be performed under the influence of the magnetic field. We might also be limited in our ability to obtain the right bunch structure for the muon beam. Nevertheless the results we obtain will be essential to properly predict how the ST and the STM will work under the Mu2e experimental conditions. An additional simulation effort – to predict STM behavior under Mu2e experiment conditions based on the obtained test data – will be also included in this plan.

Two distinct periods of tests are envisioned:

1. Tests at TRIUMF
2. Tests at PSI

***Tests at TRIUMF***

Osaka University plans to obtain beam time at TRIUMF and we will join them to perform the following tests:

- Measuring the muon beam flux
- Measuring the X-ray spectra from different target material
- Measuring high energy photon spectra

***Tests at PSI***

The PSI (Paul Sherrer Institut, Villigen Switzerland) test can be divided into three phases:

- Test Preparation
- Data taking at PSI
- Offline Data Analysis

***Test Preparation***

In this phase we will procure the ST and STM detector including the readout system that allows us to collect and record the data. We also need to work out the interface issues with PSI to make sure that the proper infrastructure is present for our equipment.

***Data Taking at PSI***

In this phase we will travel to PSI and perform the following measurements:

- Measure muon flux via muonic X-rays and beam counters
- Measure proton flux
- Measure neutron flux
- Measure gamma flux and delayed gamma flux
- Measure high energy (above 100 MeV) photon flux.
- Repeat for stopped pions.

To perform these measurements we need to spend 3 weeks at the PSI facility.

***Offline Data Analysis Effort***

Most of the effort will be done back at the home institutions where the infrastructure is well established so we don't expect any additional M&S expenses.

The offline analysis will include:

- Processing the raw data
- Extracting capture signal rates and background rates
- Estimating capture rates at Mu2e experimental setup

***WBS 5.8 Muon Beam Stop***

- Simulations will be completed to determine the type and level of doping of HDPE parts.
- Simulations will be completed to determine the level of magnetic permeability that can be accepted from the stainless support ring. If slightly higher levels of permeability can be accepted, 304 rather than 316 stainless can be used, decreasing cost.
- Sample HDPE parts will be fabricated to ensure tolerances can be held.
- A larger ring (short mockup of the stainless ring) will be built and assembled with the HDPE rings to understand fit and shimming of HDPE parts within the stainless ring.
- Calculations will be completed to ensure the beam stop will not deflect excessively under its own weight and that the support structure for the beam stop is structurally adequate.

***WBS 5.9 Neutron Absorbers***

- The current design utilizes concrete blocks with reinforcement bars of magnetic steel. Studies will determine whether the location of the magnetic bars will be acceptable with regard the DS field quality and/or will have structural issues due to eddy currents. If field quality is an issue, stainless steel bars are available, at a higher cost.

***WBS 5.10 Detector Support and Installation System***

- Structural calculations of the entire system, internal and external, will be completed and documented to ensure the structural adequacy of the system.
- Simulations will be completed to determine the level of magnetic permeability that can be accepted from the rail system components. If slightly higher levels of permeability can be accepted, some alternate materials may be used, decreasing costs.

- Analysis will be done to ensure that temperature variations within the DS bore do not cause binding or other problems associated with the rail system, to a degree that causes the alignment specifications to be violated.
- Critical support elements will be purchased and tested to prove the concept of both the internal and external rail system. These will include a section of the rail system at least 3 meters long, made out of standard magnetic material, as well as at least 4 bearing blocks. The “mockup” rails can be re-used as the external system for the actual installation, recouping the money spent for these parts.
- This system will be set up at a location in the Industrial Area at Fermilab, and mockups of critical components will be placed on them with their support structures, aligned, and realigned, using the proposed alignment system, to test wherever possible through “dry run”, the entire assembly procedure, and to verify that the specified tolerances on placement can be achieved.
- Two small sections of rail will be fabricated, attached and tested to ensure that this connection can be made adequately.
- In the actual DS bore, the rail system will be supported by a set of stainless steel supports to be welded to the inner wall of the DS cryostat. As part of the R&D plan, a section of stainless steel, the same thickness as the DS inner wall, will be welded to a mockup of the supports by the same process proposed for the actual supports. Measurements of distortion and magnetic permeability will be taken before and after welding to ensure that the welding process does not cause a violation of the DS specifications.

## 8.9 References

[1] J. Brandt, B. Norris and J. Popp, “Requirements and Specifications Document, WBS 5.2, Beamline Vacuum System,” Mu2e-doc-1481.

[2] N. Andreev, “Requirements and Specifications Document, WBS 5.3, Collimators,” Mu2e-doc-1044.

[3] N. Andreev et al., “Requirements and Specifications Document, WBS 5.4, Beamline Shielding,” Mu2e-doc-1506.

[4] T. Ito, “Requirements and Specifications Document, WBS 5.5, Stopping Target,” Mu2e-doc-1437.

[5] T. Ito, “Requirements and Specifications Document, WBS 5.6, Stopping Target Monitor,” Mu2e-doc-1438.

[6] T. Ito, “Requirements and Specifications Document, WBS 5.7, Proton Absorber,” Mu2e-doc-1439.

[7] R. Bossert, “Requirements and Specifications Document, WBS 5.8, Muon Beam Stop,” Mu2e-doc-1351.

[8] R. Bossert, “Requirements and Specifications Document, WBS 5.9, Neutron Absorber,” Mu2e-doc-1371.

[9] R. Bossert, “Requirements and Specifications Document, WBS 5.10, Detector Support and Installation System,” Mu2e-doc-1383.

[10] S. Feher, “Requirements and Specifications Document, WBS 5.11, System Integration, Tests and Analysis,” Mu2e-doc-1168.

[11] M. Dinnon, “Risk Registry,” Mu2e-doc-1463.

[12] S. Feher, “Quality Assurance Plan for the Muon Beamline sub-project of the Mu2e Experiment,” Mu2e-doc-1471.

[13] D. Hedin, “Muon Beam Stop Update”, presented at Mu2e collaboration meeting, Jan 28, 2012, Mu2e-docdb-2003.

[14] J. Popp and M. McKeown, “Vacuum Studies II”, Mu2e-doc-1267.

[15] R. Ray, “Preliminary Hazard Analysis Document,” Mu2e-doc-675.

# 9 Tracker

## 9.1 Introduction

The Mu2e tracker provides the primary momentum measurement for conversion electrons. The tracker must accurately and efficiently identify and measure 105 MeV/c electrons while rejecting backgrounds and it must provide this functionality in a relatively unique environment. The tracker resides in the warm bore of a superconducting solenoid providing a uniform magnetic field of 1 Tesla; the bore is evacuated to $10^{-4}$ Torr. A key feature of Mu2e is the use of a pulsed beam that allows for elimination of prompt backgrounds by looking only at tracks that arrive several hundred nanoseconds after the proton pulse (see Figure 3.8). The tracker must survive a large flux of particles during the early burst of "beam flash" particles that result from the proton pulse striking the production target, but it does not need to take data during this time. The Mu2e signal window is defined as $670 < t < 1595$ (Figure 3.8), where $t = 0$ is the arrival of the peak of the beam pulse at the stopping target. However, in order to study backgrounds such as radiative pion capture, the tracker must be fully efficient during the interval from $500 < t < 1700$ nsec.

The dominant interactions for stopped muons are radiative muon capture – $\mu^{-}N \to \gamma\,\nu\,N'$ – and decay in orbit: $\mu^{-} \to e^{-}\bar{\nu}_e\nu_\mu$. The former frequently leads to ejected protons from nuclear breakup; these are a source of "noise" hits but do not mimic the signal. Decay in orbit, or DIO, produces electrons that are distinguishable from the signal only by their momentum. The differential energy spectrum of DIO electrons, shown in Figure 9.1, falls rapidly near the endpoint, approximately[1] proportional to $(E_{\text{endpoint}} - E_e)^5$. The most prominent feature is the well-known Michel Peak. Nuclear recoil slightly distorts the Michel peak and gives rise to a small recoil tail that extends out to the conversion energy. The Mu2e tracker is optimized to distinguish conversion electrons from DIO electrons. The key to this is momentum resolution, with particular emphasis on DIO electrons that smear up in energy and appear to be signal electrons. A precision, low mass tracking detector in a magnetic field is the most practical way of achieving the required precision. A tracking system has been designed to meet this requirement while being blind to most of the rate from DIO electrons.

## 9.2 Requirements

Requirements for the tracker have been documented elsewhere [1] and are only summarized here.

[1] The full Shanker form, not this simplified power-law approximation, is used in Mu2e simulations.

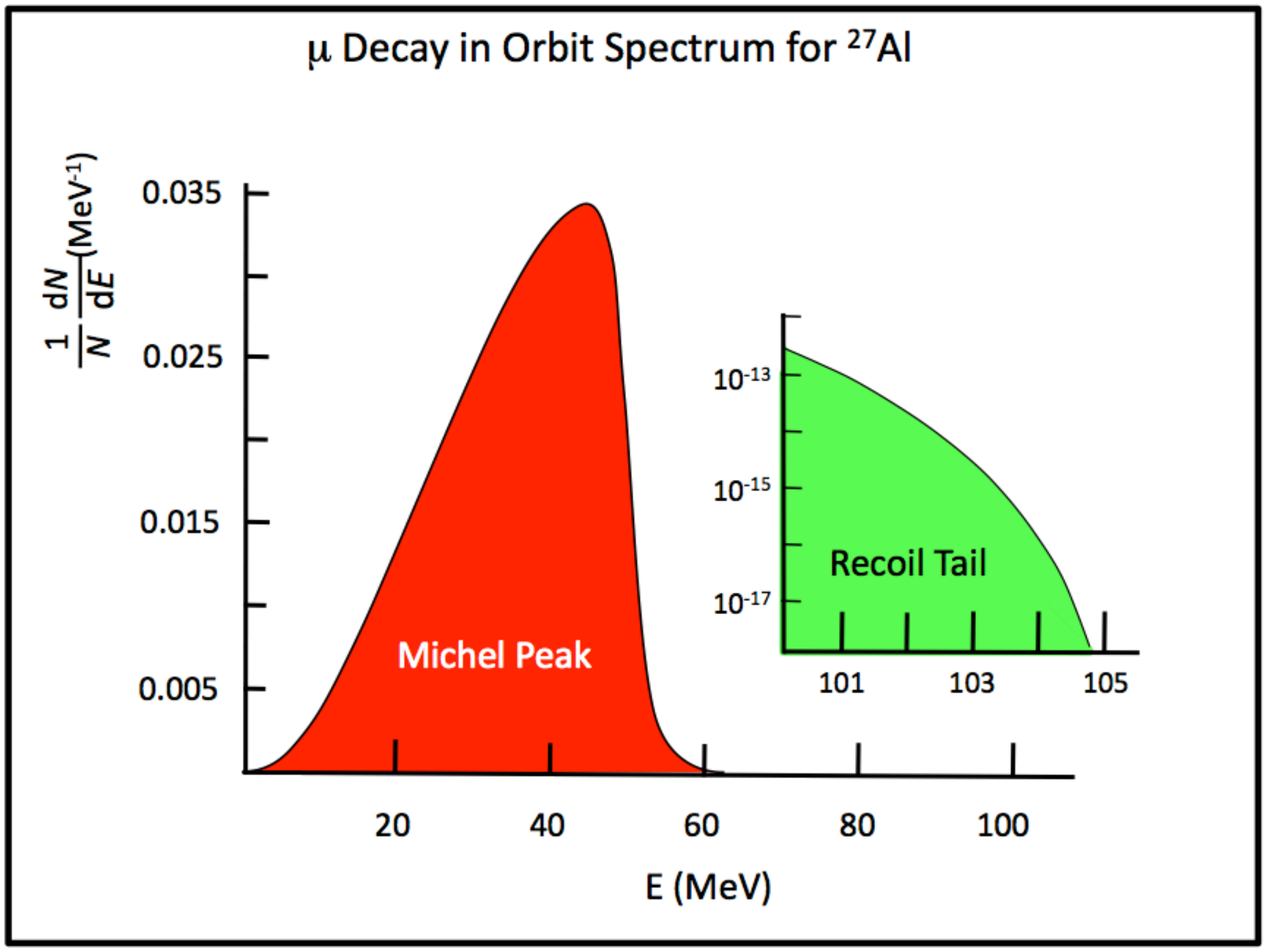

Figure 9.1. Electron energy spectrum from muon decay in orbit. Recoil against the nucleus results in a small recoil tail that extends out to the conversion energy (inset).

The Detector Solenoid [2] will provide a uniform 1 Tesla field in the region occupied by the tracker. To have good acceptance for signal electrons without being overwhelmed by DIO electrons (including electrons scattered into the active region), the active area of the tracker should extend from about $40 < r < 71$ cm (where radius r is measured from center of the muon beam)[2]. The final optimization of these dimensions depends on the size and geometry of the muon stopping target, the size of the muon beam, and the magnetic field properties, but the goal is to maximize the acceptance to conversion electrons while minimizing the number of low energy electrons that intersect the tracker.

The momentum resolution requirement is based on background rejection: the signal is sharply peaked, whereas backgrounds are broad (cosmic rays, radiative pion capture) or steeply falling (DIO electrons). For a Gaussian error distribution the requirement is that $\sigma < 180$ keV/c. This is simply a convenient reference point; the actual resolution is not Gaussian and may be asymmetric. Furthermore, scattering and straggling in material upstream of the tracker are significant contributors to the final resolution.

[2] Reference 1 specifies 71 cm. Here we use 70 cm based on a cut on polar angle. 105 MeV/c tracks reaching 71 cm have a helix that is too tight to be effectively fit.

To obtain the required precision, multiple scattering through the measurement region must be minimized, as must d*E*/d*x* straggling throughout the entire path for signal electrons. Therefore the detector elements must be surrounded by vacuum.

Since neither signal nor DIO electrons reach $r > 70$ cm, mass at these larger radii does no harm. Therefore mechanical support, readout electronics, etc. are to be placed at $r > 70$ cm.

Because of the need to break and later re-establish vacuum, access to the detector is expected to require several days of downtime. Therefore the tracker will be designed for an overall mean time to failure (MTTF) of >1 year. A few dead channels do not constitute a tracker failure if they can be kept isolated. The ability to isolate local failures is an important part of the design. In particular, the capability to remotely disconnect high voltage to each straw will be implemented so that a few broken or otherwise defective wires will not necessitate an access.

The device must operate at the peak expected rate of ~20 kHz/cm$^2$ at the inner radii at the beginning of the live window; an average rate of ~15 kHz/cm$^2$ during the live window; and it must tolerate (but need not take useful data during) a 3 MHz/cm$^2$ "beam flash" prior to the live window.

Roughly half of the hits in the detector are from slow protons (≲100 MeV/c momentum or ~5 MeV kinetic energy) ejected from the stopping target. The tracker must have d*E*/d*x* capability to distinguish such protons from electrons.

## 9.3 Recommended Design

The selected alternative for the Mu2e tracker is a low mass array of straw drift tubes aligned transverse to the axis of the Detector Solenoid, referred to as the T-tracker. The basic detector element is a 25 μm sense wire inside a 5 mm diameter tube made of 15 μm thick metalized Mylar®, referred to as a straw. This choice is based on several points.

- The straw can go from zero to 1 atmosphere pressure differential (for operating in a vacuum) without significant change in performance.
- Unlike other types of drift chambers, each sense wire is mechanically contained within a straw. Thus, failures remain isolated, improving reliability.
- The transverse design naturally places mechanical support, readout electronics, cooling, and gas distribution at large radii.

The detector will have ~20,000 straws distributed into 18 measurement stations across a ~3 m length. Each station will provide a ~200 μm measurement of track position.

Each straw will be instrumented on both sides with preamps and TDCs. Each straw will include one ADC for d*E*/d*x* capability. To minimize penetrations into the vacuum, digitization will be done at the detector, with readout via optical fibers. Electronics at the detector will *not* require an external trigger: all data will be transferred out of the vacuum to the DAQ system, and a trigger may be implemented as part of the DAQ.

### 9.3.1 Mechanical Construction

***Straws***

The T-tracker is made from 5 mm diameter straws. Each straw is made of two layers of ~6 μm (25 gauge) Mylar®, spiral wound, with a ~3 μm layer of adhesive between layers. The total thickness of the straw wall is 15 μm. The inner surface has 500 Å aluminum overlaid with 200 Å gold as the cathode layer. The outer surface has 500 Å of aluminum to act as additional electrostatic shielding and improve the leak rate. The straws vary in length from 334 mm to 1174 mm active length. The straws will be tensioned to 500 g and supported only at the ends.

The sense wire is 25 μm gold plated tungsten, centered in the straw. The wire will be tensioned to 80 g and supported only at the ends.

The drift gas is tentatively taken as 80:20 Argon:$CO_2$ with an operating voltage of ≤1500 V. However, the straw allows for operation with $CF_4$ for higher drift speed.

***Panels***

Groups of 100 straws are assembled into roughly trapezoidal panels as shown in Figure 9.2. Each panel has two layers of straws, as shown in Figure 9.3, to improve efficiency and help determine on which side of the sense wire a track passes (the classic “left-right” ambiguity). A 1 mm gap is maintained between straws to allow for manufacturing tolerance and expansion due to gas pressure. This necessitates that individual straws be self-supporting across their span.

A plane consists of 6 panels on two faces of a support ring, three panels per face, as shown in Figure 9.4 and Figure 9.5. Note the aluminum ring fattens at the OD to help make it stiff as well as to facilitate attachment to support beams, resulting in the apparent overlap with one of the panels in Figure 9.5.

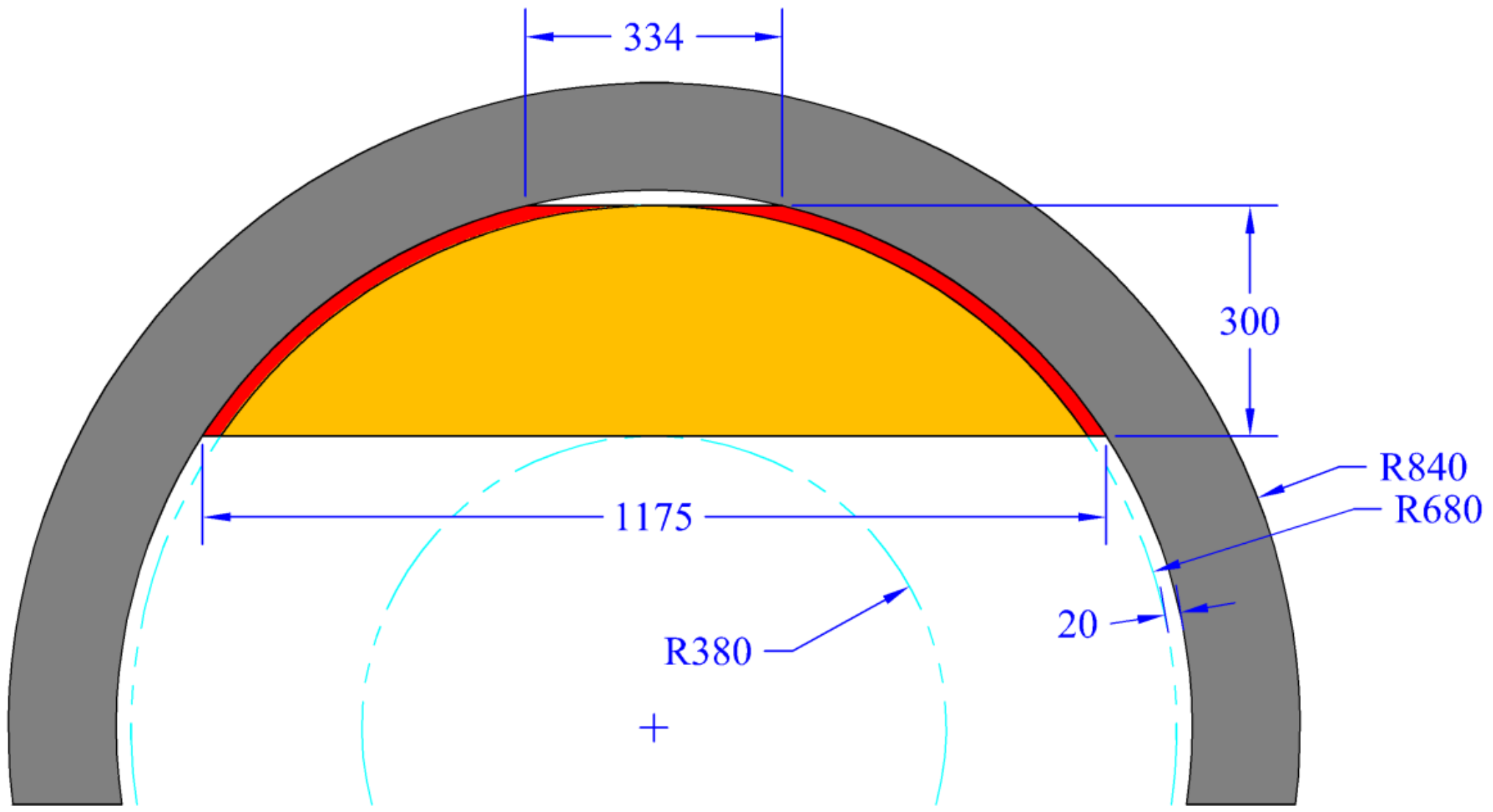

Figure 9.2. Outline of a tracker panel. Dimensions are in millimeters.

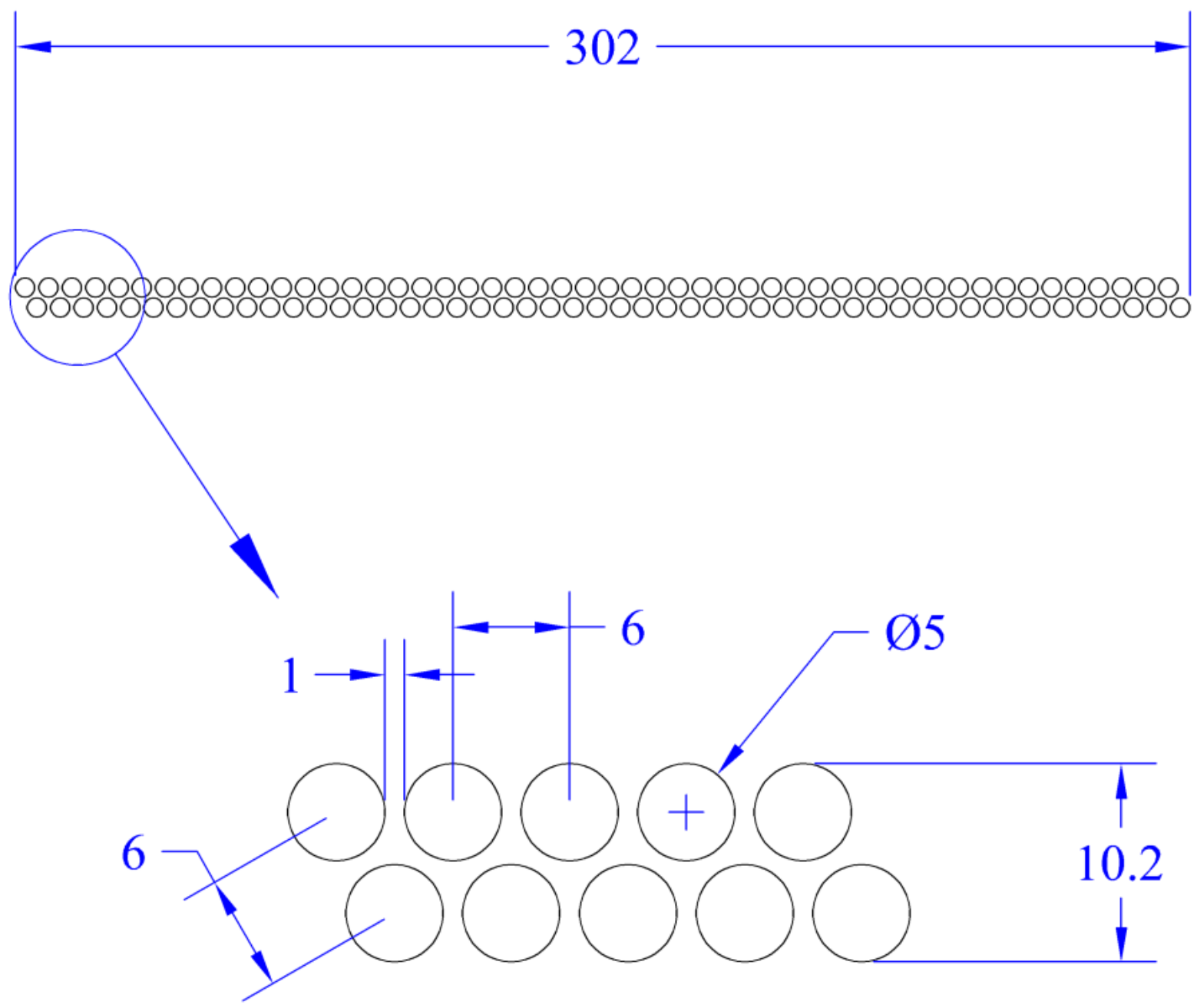

Figure 9.3. Edge view of a panel showing the arrangement of straws within a panel. Dimensions are in millimeters.

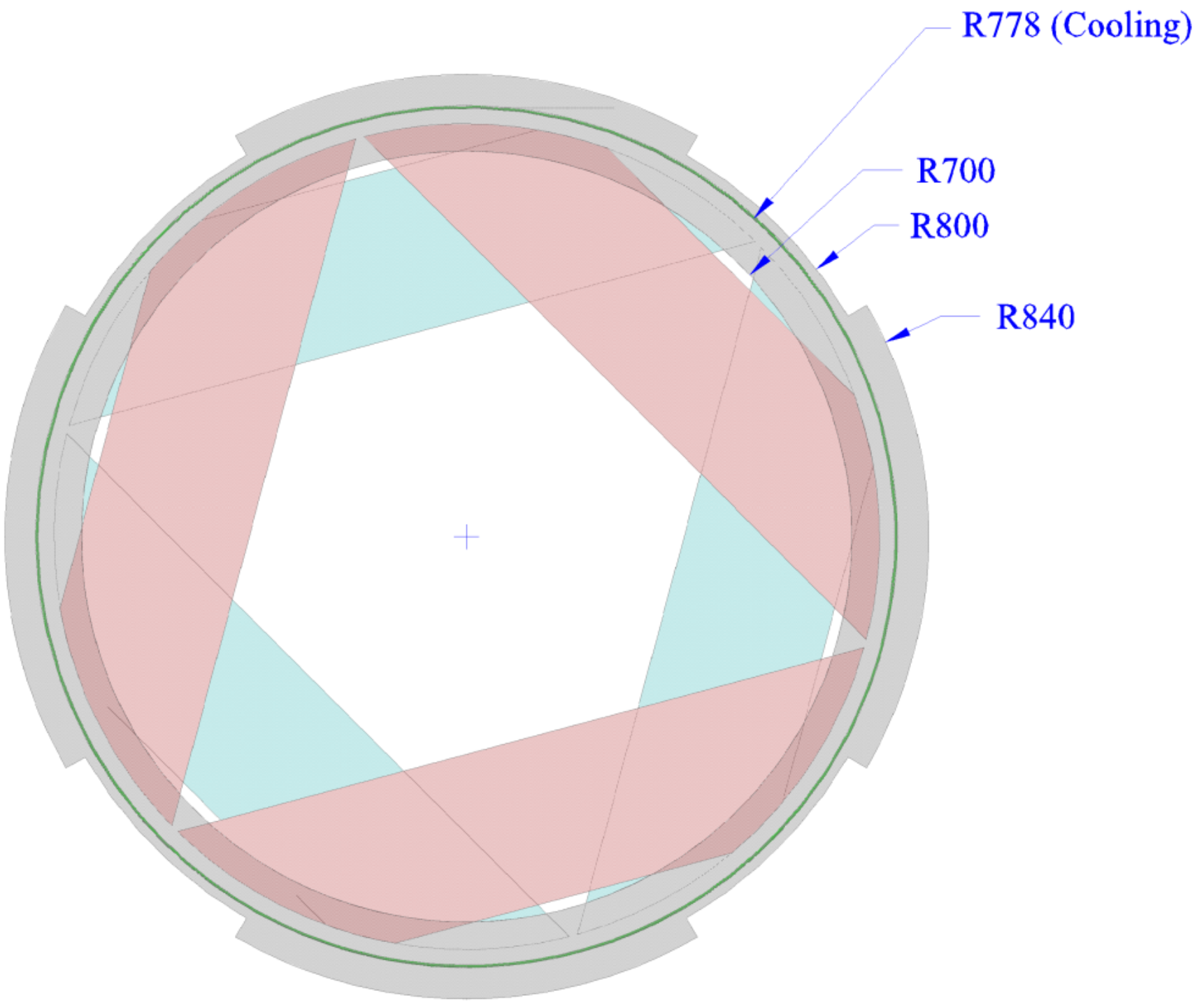

Figure 9.4. Front view of a tracker plane. The grey ring represents the support structure. Panels shown in red are on the front face; panels in light blue are on the back face. The green tube is for cooling. Dimensions are in millimeters.

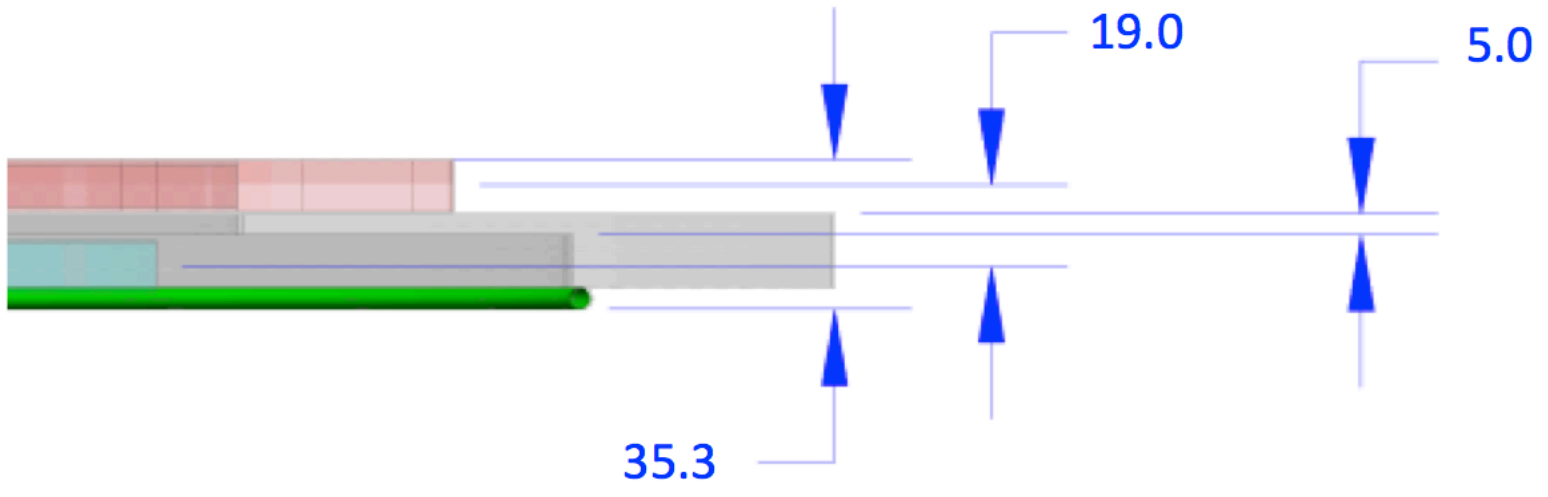

Figure 9.5. Edge view of a plane showing placement of the cooling ring and panels mounted on both sides. Dimensions are in millimeters.

***Station***

A pair of planes forms a station. The 2nd plane is rotated 30° relative to the first. The edge view shows spacing within a station. The intent is to pack all views into as small a $\Delta z$ as possible to improve stereo robustness.

Panel orientations within a station are shown in Figure 9.6 and Figure 9.7 and tabulated in Table 9.1. $\varphi$ is measured from the origin (detector center) to the middle of the panel, i.e. a panel with vertical wires will have $\varphi = 0$ or 180°. The local z coordinate, $z_L$, is measured from the station center to the center of the panel (midway between layers).

***Assembled Tracker***

The completed tracker, shown in Figure 9.8, consists of 18 stations with 170 mm gaps between stations. Horizontal beams are added to maintain longitudinal alignment of the rings. The thicker rings seen at the end stiffen the structure. The completed tracker has 21,600 straws as shown in Table 9.2.

| | | 15 | 45 | 75 | 105 | 135 | 165 | 195 | 225 | 255 | 285 | 315 | 345 |
|---|---|---|---|---|---|---|---|---|---|---|---|---|---|
| $Z_L$ | -29.5 | | | Y | | | | Y | | | | Y | |
| | -10.5 | Y | | | | Y | | | | Y | | | |
| | 10.5 | | Y | | | | Y | | | | Y | | |
| | 29.5 | | | | Y | | | | Y | | | | Y |

Table 9.1. Orientation of straws within a station. $Z_L$ refers to the local z coordinate, i.e. relative to the center of the station.

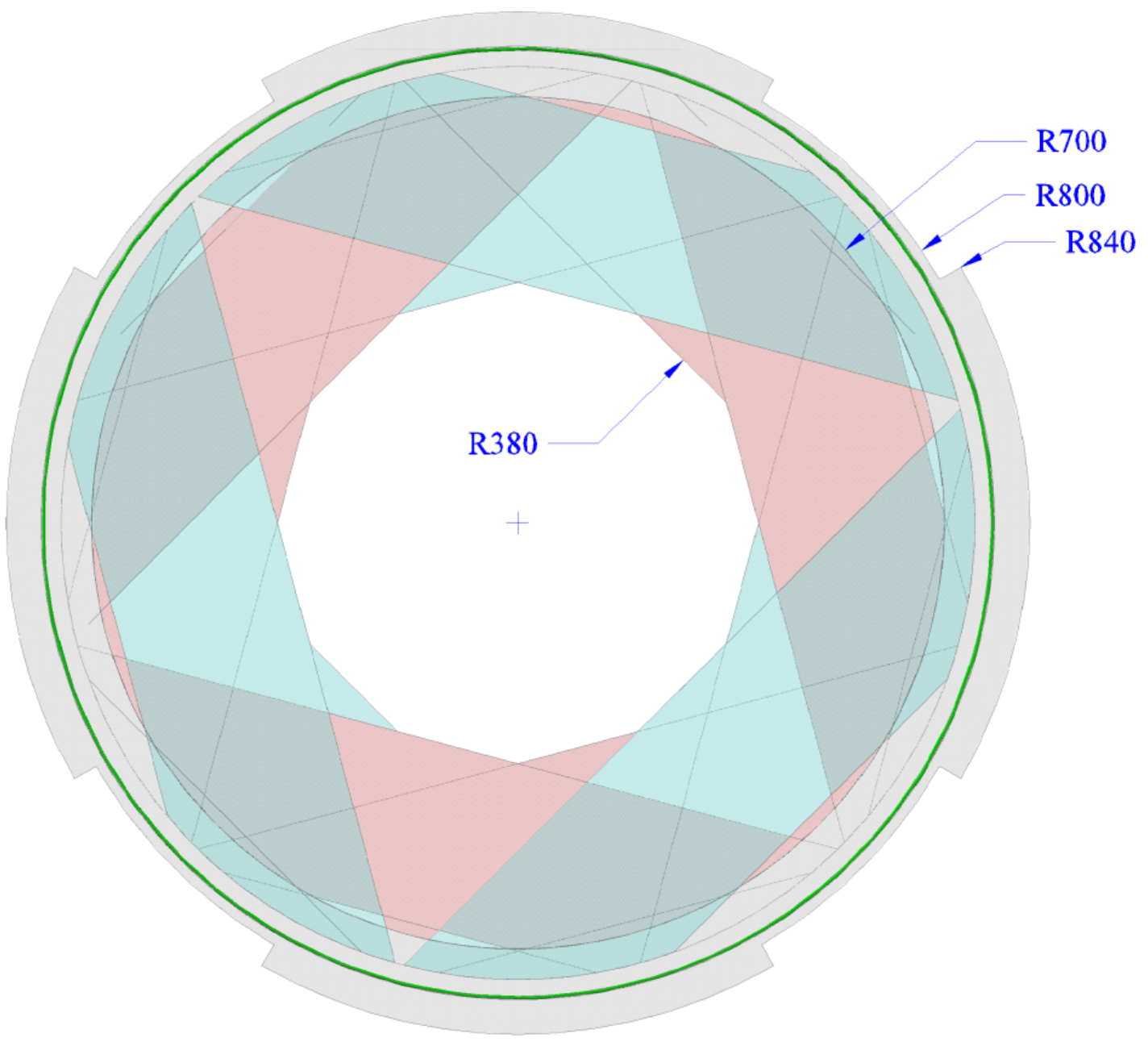

Figure 9.6. Front view of a tracker station. The grey ring represents the support structure. Panels shown in red are on the front face; panels in light blue are on the back face. The green tube is for cooling. Dimensions are in millimeters.

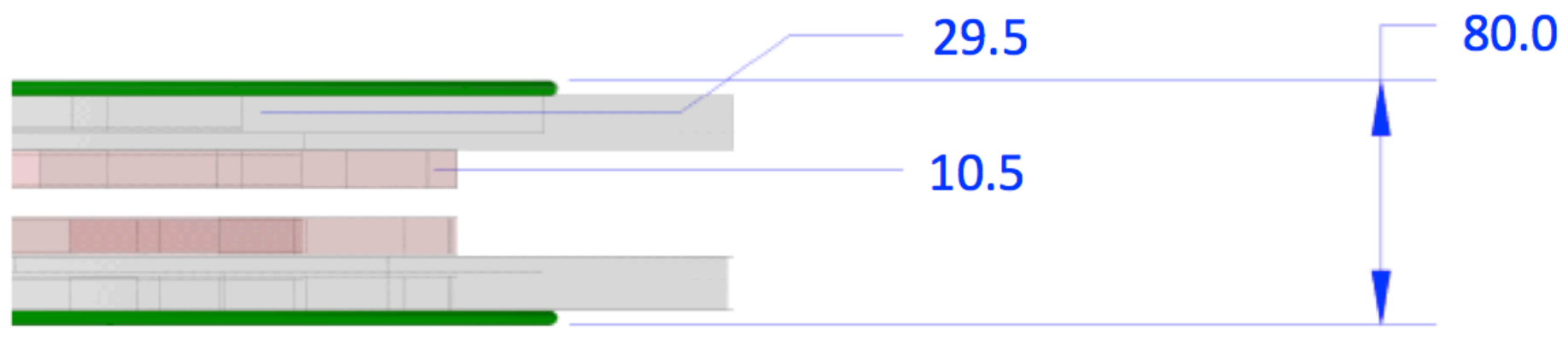

Figure 9.7. Edge view of a tracker station. Note that planes are assembled "back-to-back" to move the cooling rings to the space between stations. Dimensions are in millimeters.

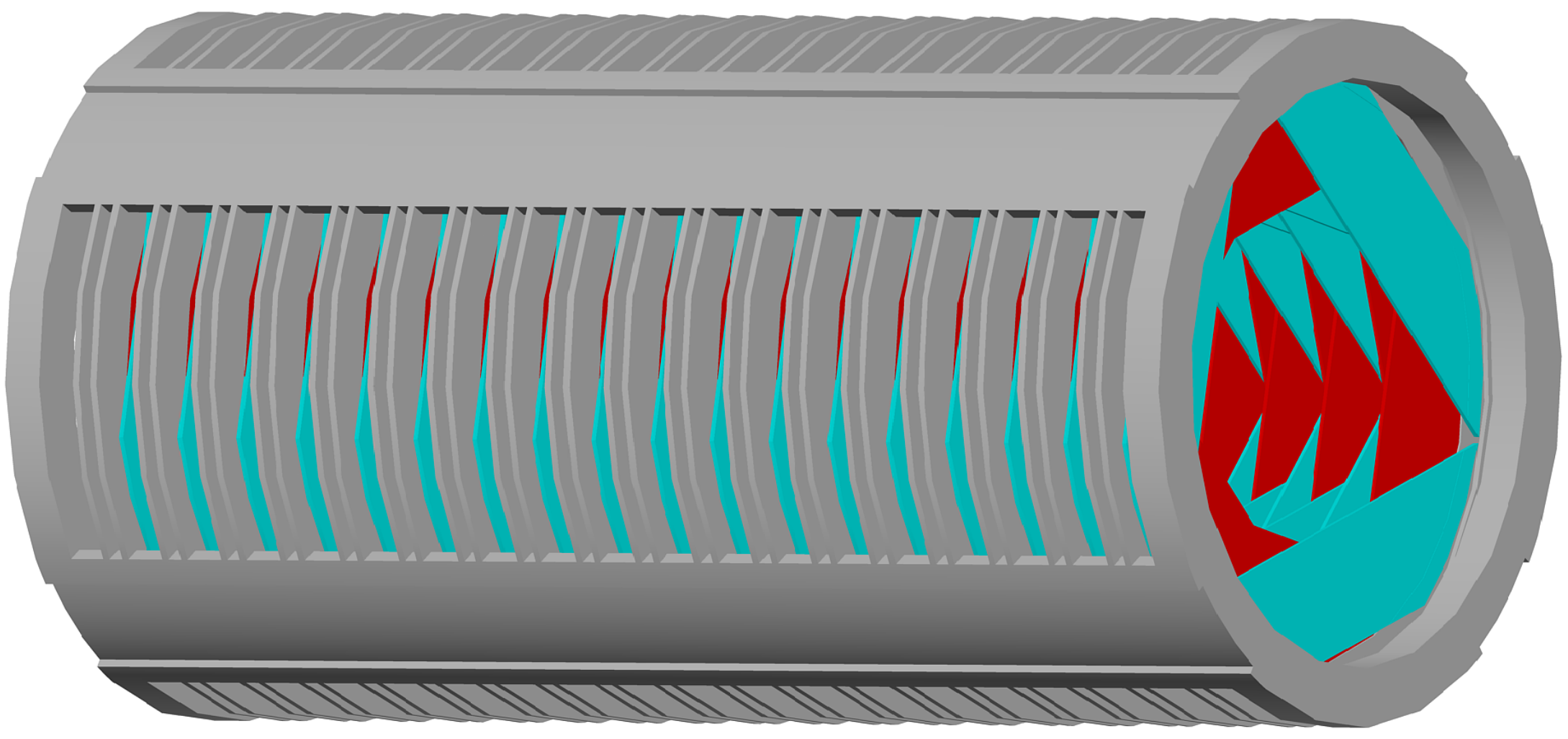

Figure 9.8. The assembled tracker.

| | |
|---|---|
| Stations | 18 |
| Planes per station | ×2 |
| Panels per plane | ×6 |
| Layers per panel | ×2 |
| Straws per layer | ×50 |
| **Total Straws** | **21,600** |

Table 9.2. Breakdown of the number of components in the Tracker. The straw total of 21,600 at the bottom of the second column is the product of the numbers above.

***Straw support***

Each straw has a 5 mm outer diameter brass tube inserted in each end. As shown in Figure 9.9, inside the brass tube is a longer plastic (or other insulating material) tube to deaden ~1 cm past the gas manifold, and to protect against breakdown.

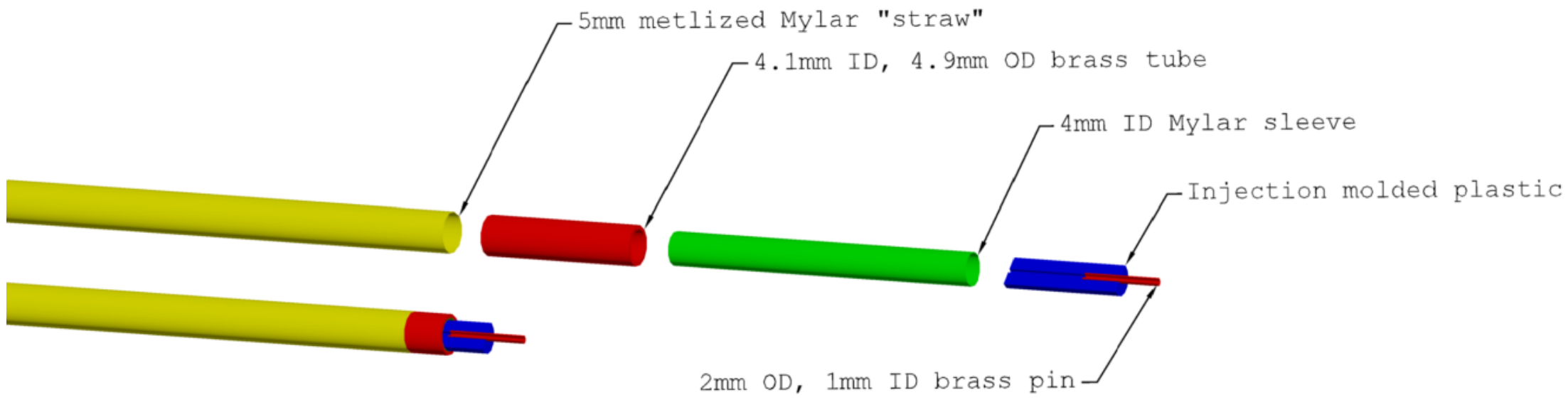

Figure 9.9. Straw termination. The green insulator slips inside a brass tube (red) to deaden the wire near the gas manifold.

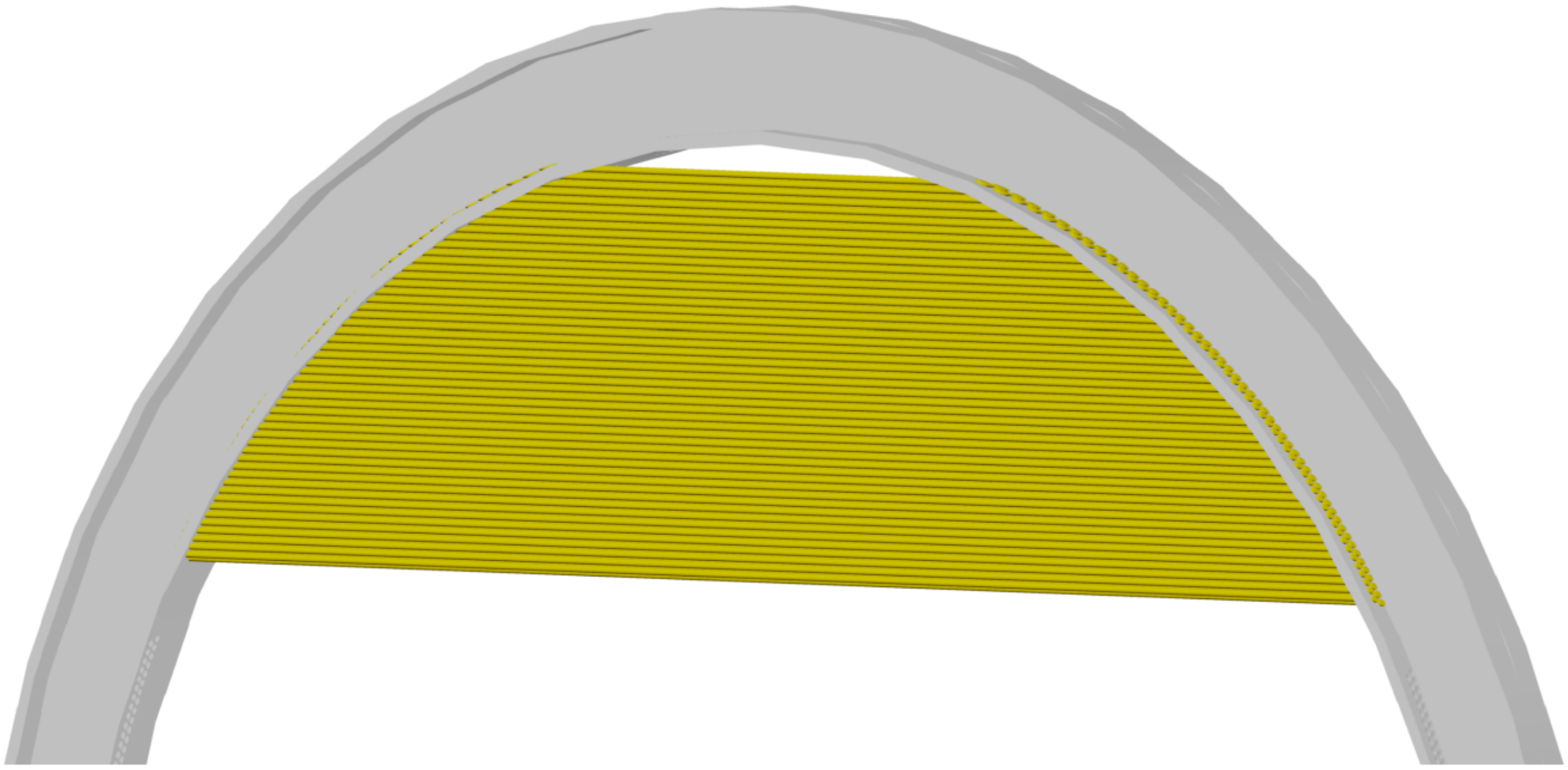

Figure 9.10. A panel of straw tubes (100 straws) attached to a gas manifold.

The gas manifold, besides feeding gas, positions the straw tube. A preliminary manifold design is shown in Figure 9.10 with one panel (100 straws) installed. Each panel requires 200 holes. Straws are epoxied to these holes, with the epoxy also forming the gas seal. The interface between a straw tube and the gas manifold is shown in Figure 9.11. The gas manifold will be made of aluminum and anodized to be electrically insulating. The straw is positioned via an external alignment jig (not shown). The location of the holes is not, therefore, critical; however, to avoid weakening the plate, we require ±0.25 mm placement and size accuracy.

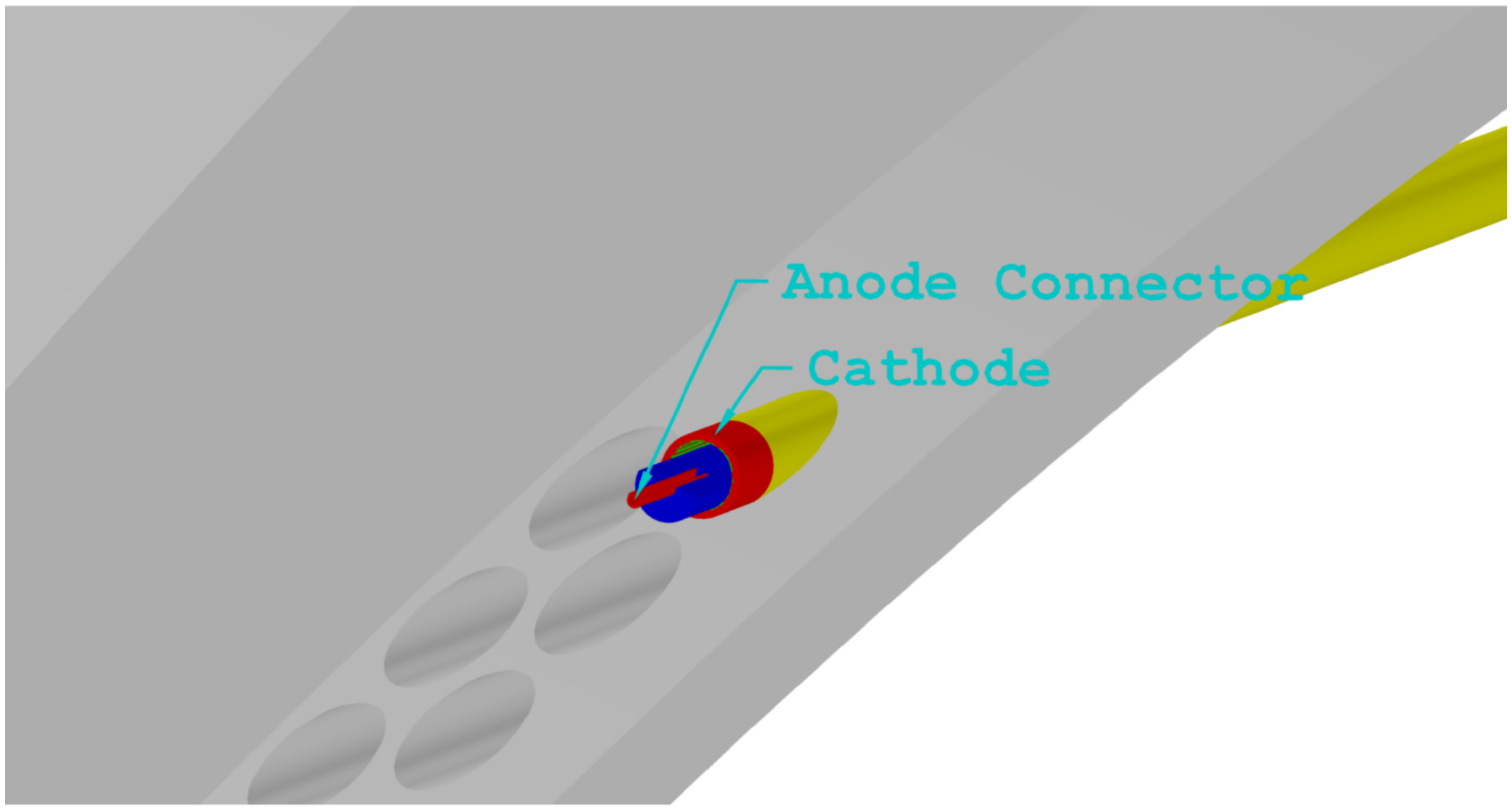

Figure 9.11. Interface of straw end with gas manifold, viewed from inside the manifold.

To avoid cumulative errors from tolerances on each part, sense wires will be aligned separately using another external alignment jig. The wire is then soldered to the pin. A groove in the injection molded endpiece is filled with epoxy to provide additional support. The wire placement jig can then be reused.

***Electronics and Readout***

To minimize penetrations through the cryostat, digitizers and zero-suppression logic are located on the detector. The electronics are described in this section. The digitizers are read out into the DAQ system which is described elsewhere [3].

Each straw requires:

- 2 preamp channels, 1 for each end.
- 2 TDC channels, 1 for each end.
- 1 ADC channel, measuring sum of both ends.
- 1 High voltage feed, with disconnect.

There are a total of 43,200 preamp and TDC channels, and 21,600 ADC channels. The two TDC channels allow for a measurement of hit position along the wire using propagation time, i.e. with "time division." Each panel requires one controller for a total of 216 controllers. Each controller includes communication via an optical cable (one or more fibers) to outside the cryostat, resulting in 216 optical cables penetrating the cryostat.

As shown in Figure 9.12, preamps are located at each straw end; this is required to get proper (~300Ω) termination of the straw. After amplification and some shaping, the analog signal is sent on a cable, or micro-strip transmission line on a board, to the digitization electronics. Bringing signals from the two sides to a single digitizer board avoids concerns of clock synchronization at the sub-nanosecond level, needed for time division. Centering the digitizers, along with the controller and communication electronics (as shown), helps distribute the heat load more uniformly, minimizing thermal gradients. The flow of signals through the readout chain is shown in Figure 9.12.

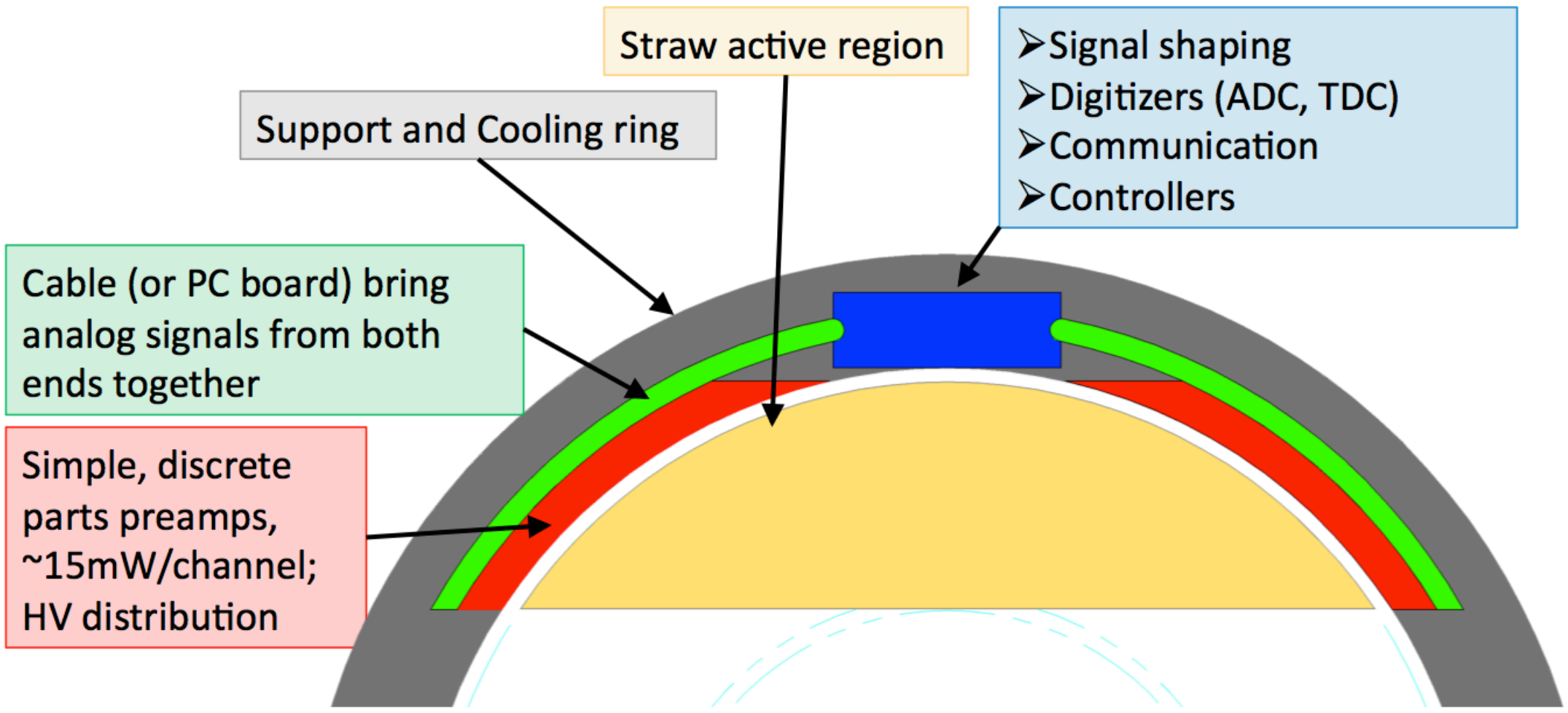

Figure 9.12. Layout of the electronics for a single straw panel.

***Preamp and High Voltage***

The preamp is simple, and the overall size of the circuit is dominated by high voltage components. It therefore makes sense to implement this with discrete electronics rather than a custom chip. This also permits shorter traces between the straw and the preamp input than is possible with a multi-channel preamp chip. The preamp output is differential analog.

High voltage is fed to each straw through a 100 kΩ current limiting resistor. A 1 nF blocking capacitor on each straw end couples to the preamps. The two layers of a panel are fed from opposite sides. The resulting 100 μsec *RC* time constant is chosen to be long compared to the 1.7 μsec microbunch spacing to average voltage sag over several microbunches. The voltage sag, including the burst from beam flash, is estimated to be ≲120 mV and to vary by ≲6 mV over the live window [4].

High voltage to each straw passes through a “fuse” that can be blown remotely to isolate broken wires without the need for an access. The fuse consists of a miniature beryllium-copper spring with one side soldered to the PC board using a low-temperature solder. To blow the fuse, that joint is heated by a resistor till the solder melts. A layer of Teflon over the connection prevents splattering of molten solder without preventing the spring from pulling back. Current to each heating resistor is controlled by an addressable switch with a unique address. To avoid accidentally blowing fuses, power for this circuit is separate from the preamp power and is normally locked out.

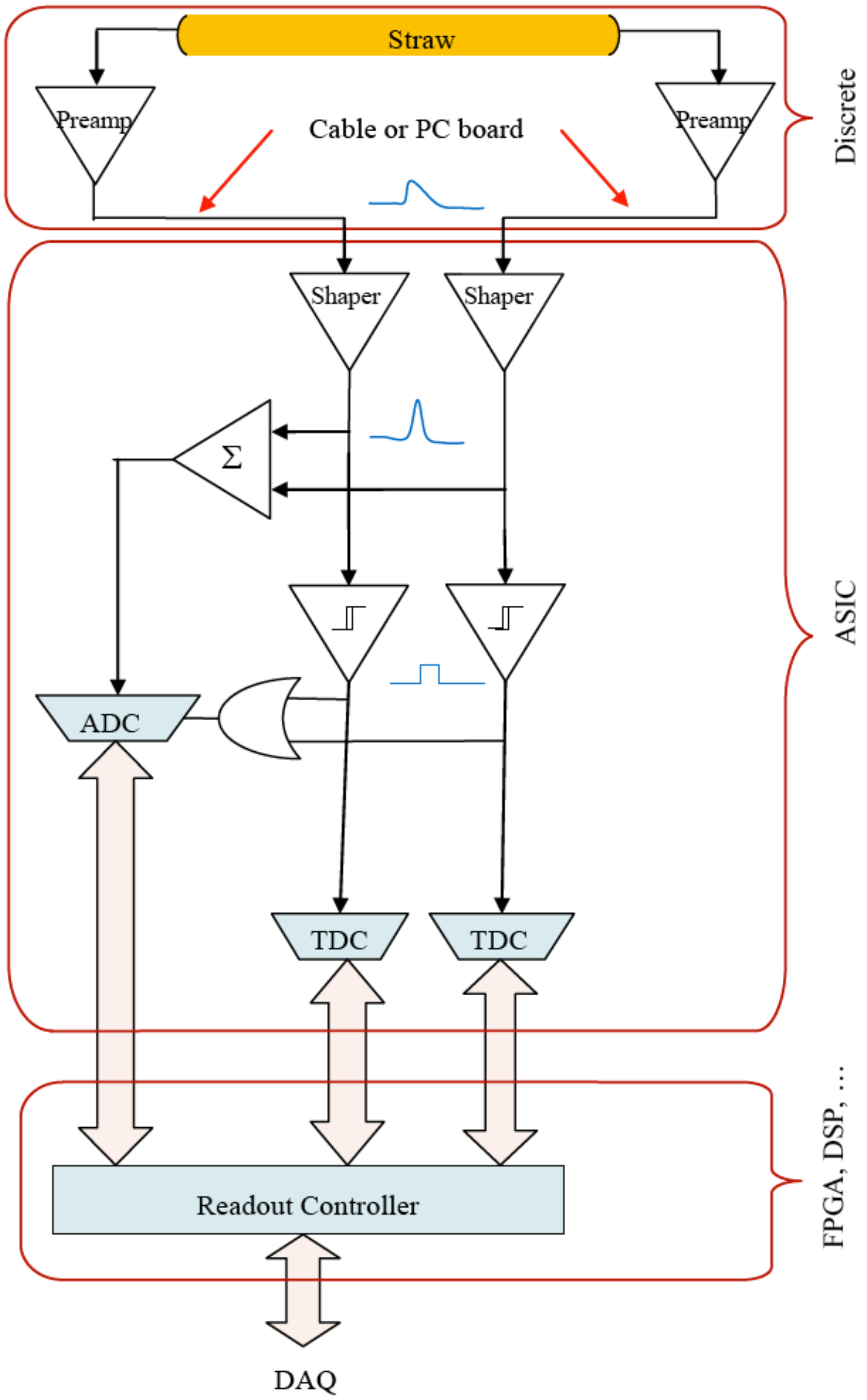

Figure 9.13. Flow of signals through the tracker readout chain.

***TDC***

The principal measurement of the tracker is drift time relative to a timing reference provided near the beginning of a beam microbunch. The TDC should not contribute significantly to the resolution, so it based on the most optimistic expectation for spatial resolution; 100 μm for the standard gas (Ar:$CO_2$, 80:20) with a drift velocity of 50 μm/nsec. Resolution with a "fast" gas (drift velocity ~100 μm/nsec) is generally worse, but again the optimistic assumption is that it remains at 100 μm. The corresponding drift time resolution is 1 nsec. The TDC will digitize drift time relative to a detector-wide timing reference to the nearest 0.5 nsec. This results in ≲1% degradation in resolution. The drift time range is 1300 nsec as shown in Table 9.3.

| | |
|---|---|
| 1700 nsec | Microbunch spacing |
| – 500 nsec | Minimum "beam flash" dead time |
| + 50 nsec | Maximum drift time |
| + 50 nsec | Pedestal ($t_0$) uncertainty |
| **1300 nsec** | **Full range** |

Table 9.3. Calculation of the drift time range. The full range is the time between microbunches when the detector is live, corrected for drift time and pedestal uncertainty.

To measure position along the wire, the difference in signal arrival time at the two ends is compared. Signal shape for the most part affects the two sides equally, so the difference in arrival time can benefit from higher precision than is useful for drift time measurement. An optimistic expectation for time division resolution is ~100 psec. A TDC that digitizes time difference to the nearest 50 psec would give ≲1% degradation in resolution. If the time-difference between the two ends is determined by subtracting the measurements from two independent TDCs the resolution will be degraded by a factor of $\sqrt{2}$ . In this case the TDCs are required to digitize time to the nearest 35 psec.[3]

The longest straw has an active length of about 117 cm for a maximum propagation time of about 4 nsec, so the time difference has a range of 8 nsec. Unlike the drift time measurement, there are no external sources of pedestal variation (cable lengths match), and pedestal variation within a chip is necessarily small. Therefore, 9 nsec is the full range necessary for this measurement.

[3] In the event the TDC does not use uniform bins, as is the case for some implementations in a FPGA, the corresponding requirements expressed as r.m.s. are <14 psec for a single TDC measuring time difference, or 10 psec for two independent TDCs.

### *ADC*

Each straw has one ADC, fed by the sum of the two sides. The intrinsic d$E$/d$x$ fluctuations are too large for pulse height information to be beneficial to fitting. The purpose of the ADC is to reject hits from protons, a significant source of "noise" hits.

The d$E$/d$x$ for protons from the muon stopping target is higher than for an electron by as much as ×50. However, it is not necessary to measure protons without saturation. "Too large," i.e. saturated, is sufficient to flag a pulse as being from a proton. The dynamic range of electrons is large due to intrinsic d$E$/d$x$ fluctuations and the variable path length of a track through a straw, but it is also not necessary to precisely measure charge at the low end. "Too small", i.e. no response, is sufficient to flag a pulse as *not* from a proton. The planned system is therefore modest: $\lesssim$100 MHz, 8-bit ADC (if the ADC has > 8-bits, record only the 8 most significant bits). To simplify merging of ADC and TDC data, the ADC and TDC clocks will be synchronized.

The ADC digitizes continuously; however not all data needs to be stored or transferred to the readout controller. The ADC must retain data for any time bin in which either TDC input (which are generated by the ADC chip, thus readily available) is high, and for the preceding time slice. The shaper, also in the ADC chip, should shape a typical pulse to a width that is on the order the maximum drift time, i.e. 50 nsec. Thus, the typical hit has 6 ADC readings, at 8-bits each. However, the number can be as low as two and as high as 170 (for 1700 nsec bunch spacing).

### *Readout Controller*

The readout controller pulls data from the TDC and ADC chips, formats it, and transmits it to the DAQ system. This controller also resides on the detector. The TDC and ADC designs are still in progress so it is too early to define the communication protocols between these chips and the controller. However, for the purpose of estimating data volume, an output format for the controller is proposed.

The data for each controller begins with a 113 byte "Controller Header" given in Table 9.4. The format constrains the hit count in any one straw to < 255. In reality, a lower limit would likely be set to avoid overwhelming the DAQ with oscillating channels.

The data hit format shown in Table 9.5 assumes two independent TDCs per straw and no merging within the controller. In this case, each TDC requires 0.035 nsec bins and a 1300 nsec range, requiring 16 bits per TDC for a total of 32 bits per hit. Each TDC hit has a variable number of 8-bit ADC values associated with it.

The typical number of ADC readings per hit is 6, so the typical hit requires 11 bytes. For an estimated 300 KHz rate [5], each controller must transfer nearly 500 Bytes per microbunch. The hit rate data is shown in Table 9.6.

With a microbunch spacing of 1700 nsec, the data transfer is 292 MByte/sec. With 216 controllers, the total rate from the T-Tracker is:

$$292 \times 216 = 63 \text{ gByte/sec.}$$

| Byte | Contents |
|---|---|
| 0-7 | 64-bit time stamp (100 nsec ticks) |
| 8-11 | 32 bit Checksum |
| 12 | Controller ID |
| 13 | Number of hits in straw 0 |
| … | |
| 112 | Number of hits in straw 99 |

Table 9.4. Output data controller header.

| Byte | Contents |
|---|---|
| 0 | n: Number of bytes in this hit, including this one |
| 1-4 | Drift time |
| 5 | First ADC reading |
| … | |
| n-1 | Last ADC reading |

Table 9.5. Format for hit data in the tracker.

| | |
|---|---|
| 100 | Straws |
| × 300 KHz | Average rate/channel |
| × 1200 ns | Live gate |
| × 11 | Bytes/hit |
| × 100 | Bytes/header |
| **496** | **Bytes/microbunch** |

Table 9.6. Hit rate data and transfer rate for the Readout Controller. The total rate is the product of the first 4 entries in the left column added to the fifth column entry.

### 9.3.2 Infrastructure

The tracker must bring power, signals, gas, and cooling through vacuum penetrations and past the beam stop and calorimeter. An overview before detector insertion into the detector solenoid is shown in Figure 9.14. Utilities use the horizontal support beams as "cable trays". A tentative layout is shown in 9.15.

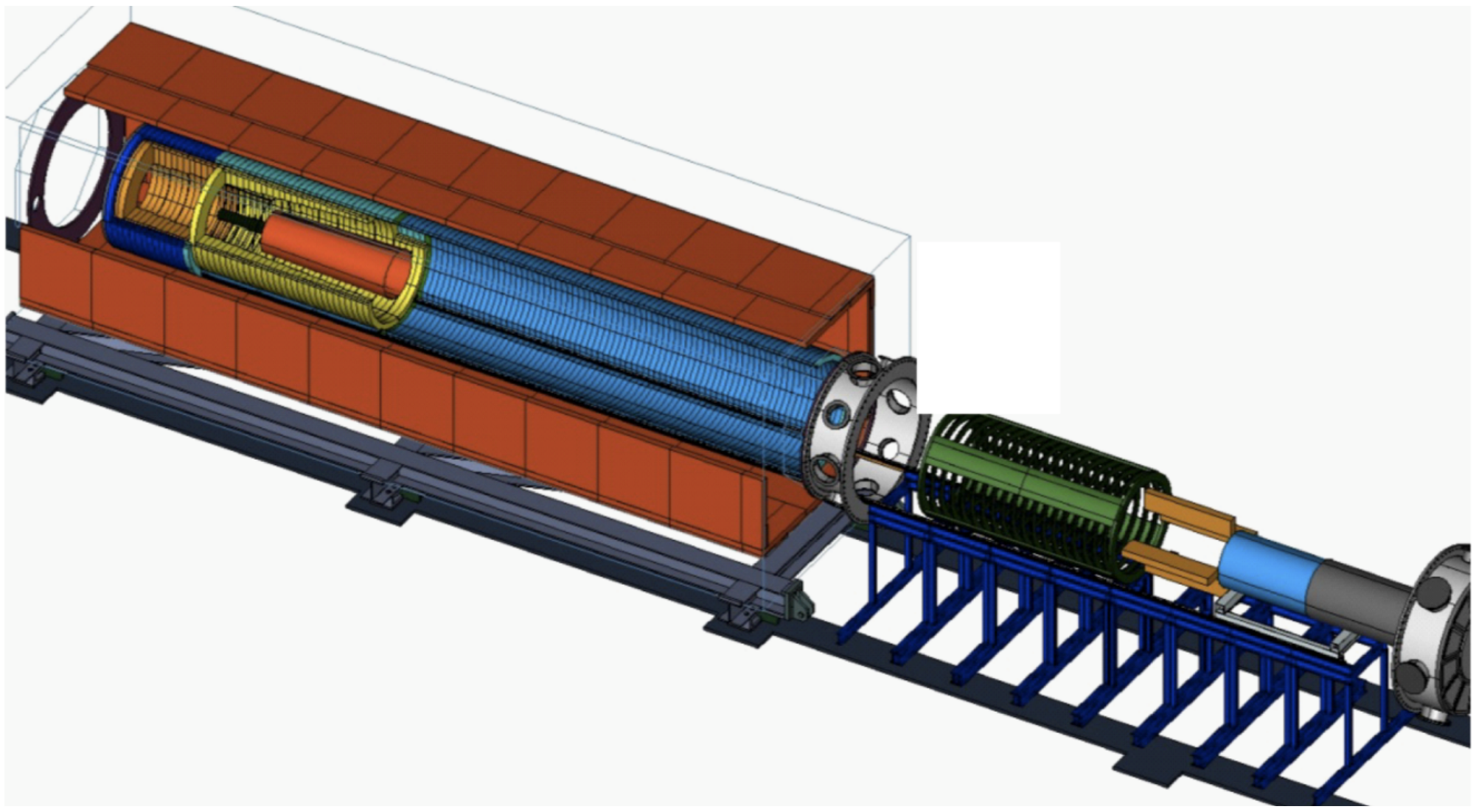

Figure 9.14. Detectors in position for insertion into the Detector Solenoid (cutaway)

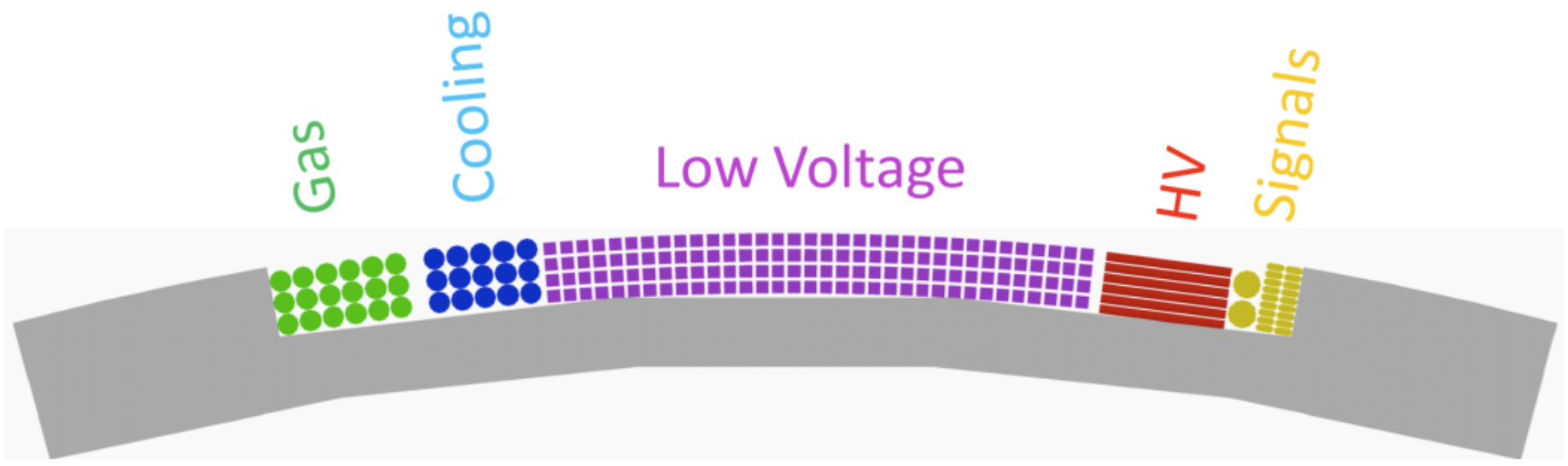

Figure 9.15. Layout of utilities within the horizontal support beams. There are four such beams around the detector as seen in Figure 9.8.

***Low Voltage***

The expected power consumption is given in Table 9.7. The controller is not yet designed. We assume it uses as much power as all other components combined; we believe this is a conservative assumption. The TDC and ADC, as currently envisioned, use both 3.3 V and 1.8 V; the ratio of currents drawn from each is not yet settled. To get an upper bound on current we assume all digital power is supplied at 1.8 V.

Each of the four beams can accommodate 136 × 3.5 mm square solid copper wire with 0.25 mm Nomex® insulation, as shown in Figure 9.16. This gives a total of 272 pair (supply and return) of power cables to carry the required 2121 A, or 7.8 A per wire. Resistance per pair is 27 mΩ for a 10 meter run (certainly sufficient to exit the vacuum) for a voltage drop of ~200 mV. Power loss on the cables will contribute an additional ~400 W heat load.

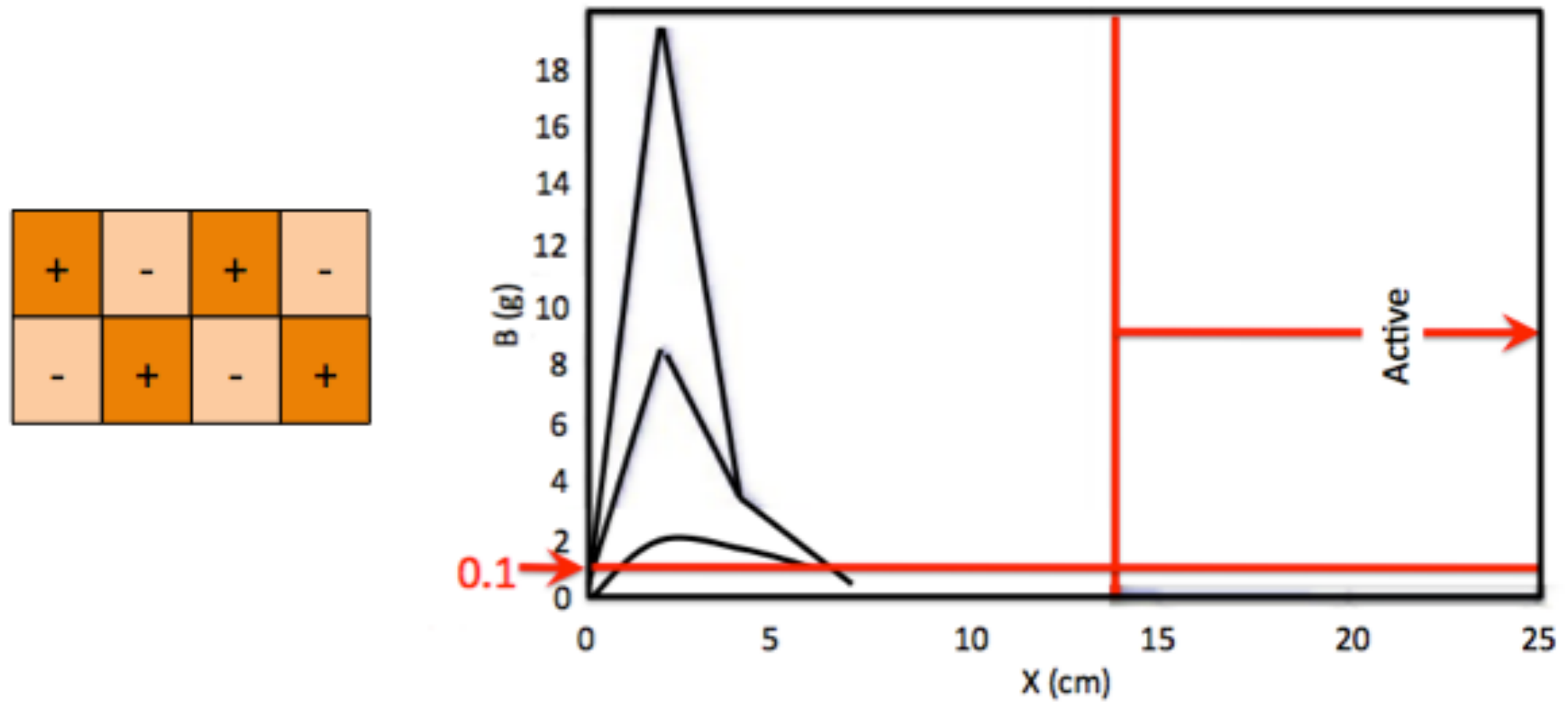

Figure 9.16. Quadrupole arrangement of power lines. B field perturbations in this configuration are < 0.01%.

In addition to voltage drop and power loss, a concern in the low voltage power distribution is perturbations to the magnetic field. This can be minimized by balancing supply and return currents in a "quadrupole" array as shown in Figure 9.16. Note the cable geometry is grossly pessimistic in that it assumes a smaller number of cables carrying the current.

| | mW per channel | mA per channel @ V | Channels per straw | Total Power (W) | Supply Voltage (V) | Total Current (A) |
|---|---|---|---|---|---|---|
| Preamp | 15 | 3 @ 5 | 2 | 600 | 5 | 120 |
| TDC | 25 | 14 @ 1.8 | 2 | 1000 | 1.8 | 556 |
| ADC | 25 | 14 @ 1.8 | 1 | 500 | 1.8 | 278 |
| Controller | | | | 2100 | 1.8 | 1167 |
| **Total** | | | | **4200** | | **2121** |

Table 9.7. Power consumption of tracker components

***Cooling***

Since the detector is in a vacuum, heat must be removed with an active cooling system. To maintain the most uniform temperature with the smallest volume in plumbing and smallest vacuum penetrations we intend to use a Suva® system. Entering the vacuum are one pair of lines, supply and return, as shown in Figure 9.17 [6]. Each ring has a cooling loop at the outer periphery, as seen in Figure 9.4, which taps into the cooling lines. The expected temperature gradient is shown in Figure 9.18 for 400 mW/straw heat load (80 kW total, compared to the expected 10 kW), distributed uniformly around the inner edge of the ring. Note ΔT<1.5 ºC even in this extreme case.

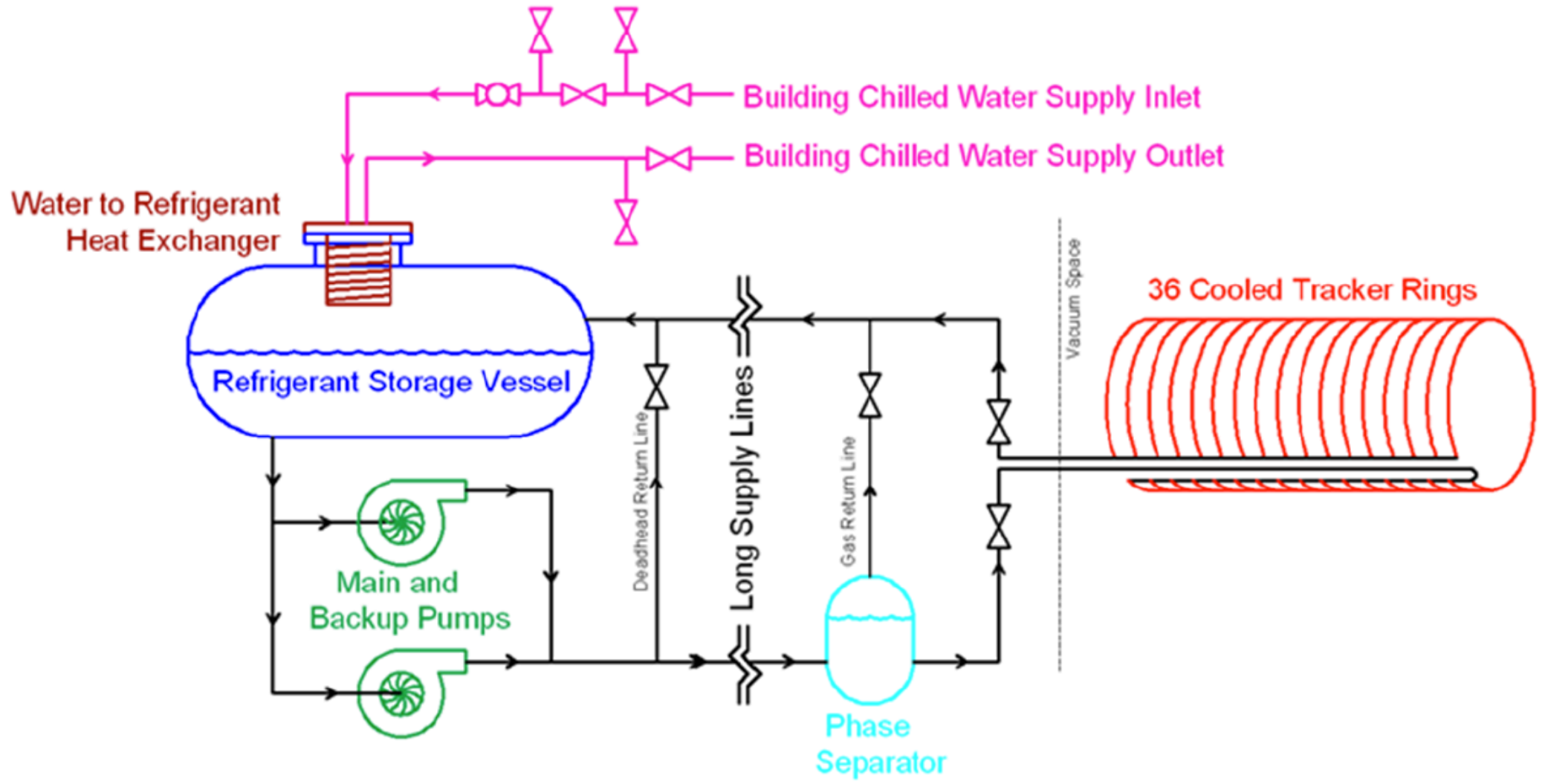

Figure 9.17. Conceptual design of the tracker cooling system.

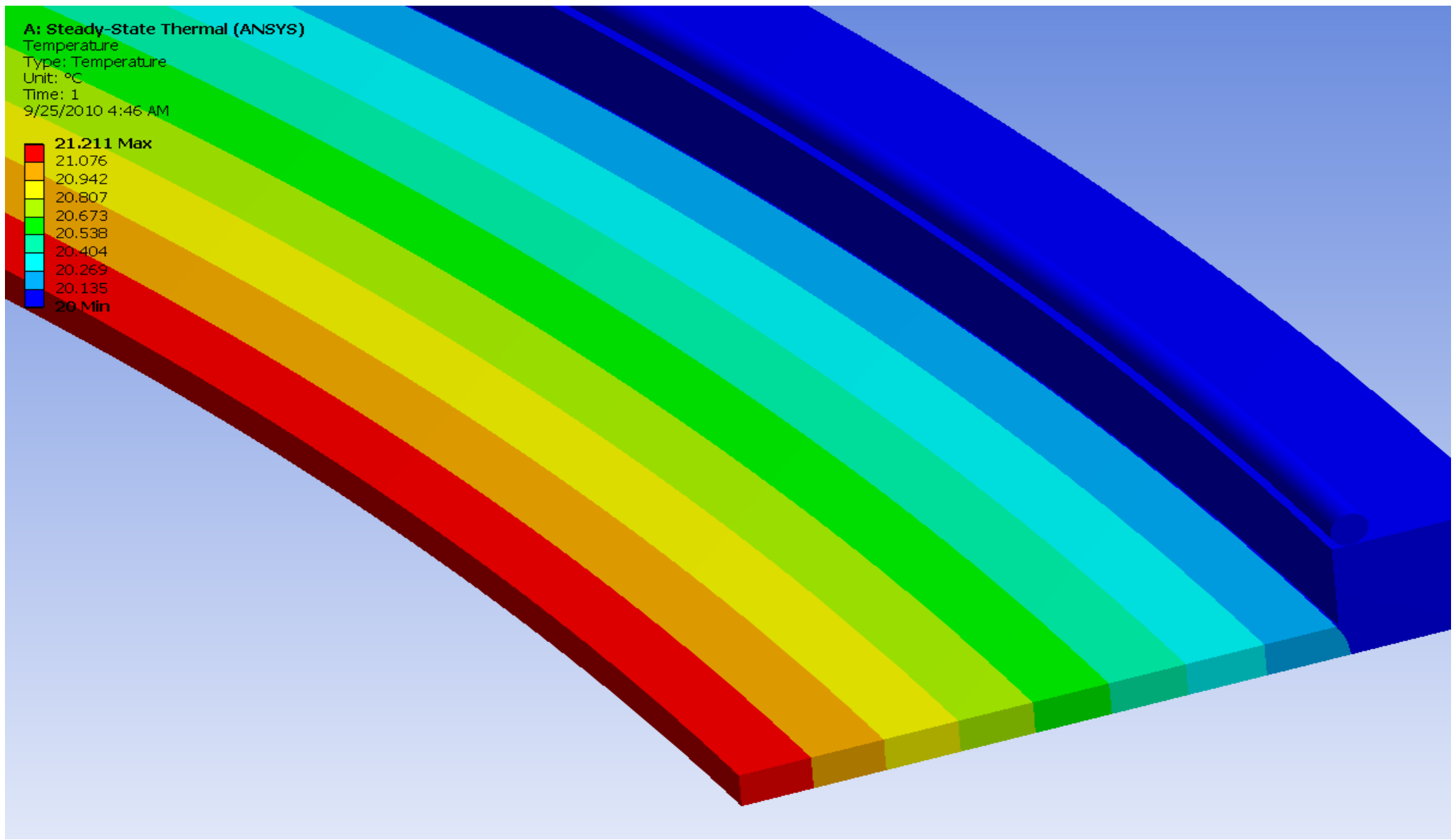

Figure 9.18. Temperature gradient across support ring. Heat applied at inner diameter and removed with Suva® around outer diameter.

## 9.4 Performance

A rigorous QA/QC program combined with a series of R&D prototypes and a final pre-production prototype will ensure that the Mu2e tracker operates with the required reliability in the Mu2e environment. The key performance criterion for the tracker is resolution: how well conversion electrons can be separated from all other tracks. Mu2e includes a Cosmic Ray Veto (Chapter 11) to reduce background due to cosmic rays and a calorimeter (Chapter 10) to confirm the trajectory of tracks, but the primary tool is the momentum resolution of the tracker.

The geometry of the tracker [7] as well as components such as the muon stopping target [8] and proton absorber [9] will continue to be optimized as simulation tools improve. The system as currently envisioned and described in this document, meets the resolution requirements.

### 9.4.1 FastSim

The primary simulation package for Mu2e is GEANT 4, which can simulate the geometry and detector response with high fidelity. However, for general detector optimization studies, such detail is not required. For those studies we use FastSim [10], a simulation package developed for SuperB detector optimization and physics reach studies. FastSim runs quickly and can be easily reconfigured to study alternative detector geometries, materials, or resolutions. It has realistic though simplified models of particle interactions with materials and detector element resolution, including non-Gaussian tails. It has a built in interface to a Kalman track fitter that can be used to measure the reconstructed momentum resolution, absent the effects of pattern recognition. Details about the Mu2e studies done with FastSim are described in a separate document [11].

### 9.4.2 Intrinsic Tracker Resolution

The tracker, the muon stopping target and the proton absorber all contribute to the momentum resolution. To study the intrinsic resolution of the tracker, we use FastSim to simulate the conversion of muons that stop in the stopping target and produce an electron that enters the tracker, ignoring the effect of the target and all other material upstream of the tracker. An event display of a single conversion is shown in Figure 9.19. The intrinsic tracker momentum resolution (Kalman fit result minus true) is shown in Figure 9.20, fit to a split double-Gaussian. The high-side resolution, which is the most important for distinguishing conversion electrons from backgrounds, has a core component sigma of 115 KeV/c, and a significant tail sigma of 176 KeV/c. The net resolution is significantly less than the estimated resolution due to energy loss in the upstream material.

***Tracker Layout Optimization***

FastSim has been used to study the effect of alternate layouts of the tracker on the resolution. For instance, we have investigated how the resolution changes with fewer or more stations, holding the volume of the tracker constant. We use the core sigma of the high-side resolution as the figure of merit for these studies. The results are shown in Figure 9.21. We find that the choice of 18 stations is near the asymptotic optimum. We also studied the impact of varying the spacing between stations and alternate configurations of panels within the stations. In all cases the tracker layout described above was found to be optimum.

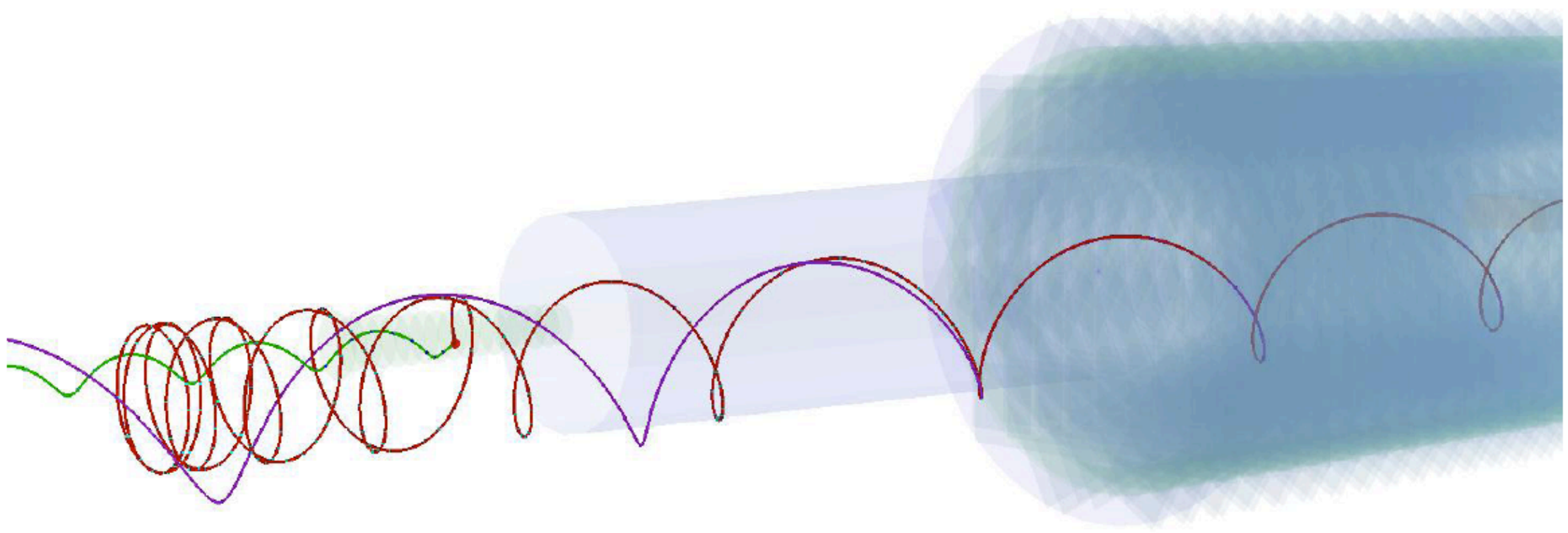

Figure 9.19. Fastsim event display. Muon (green) shown converting to an electron (red). The purple helix is the trajectory of the reconstructed Kalman filter track.

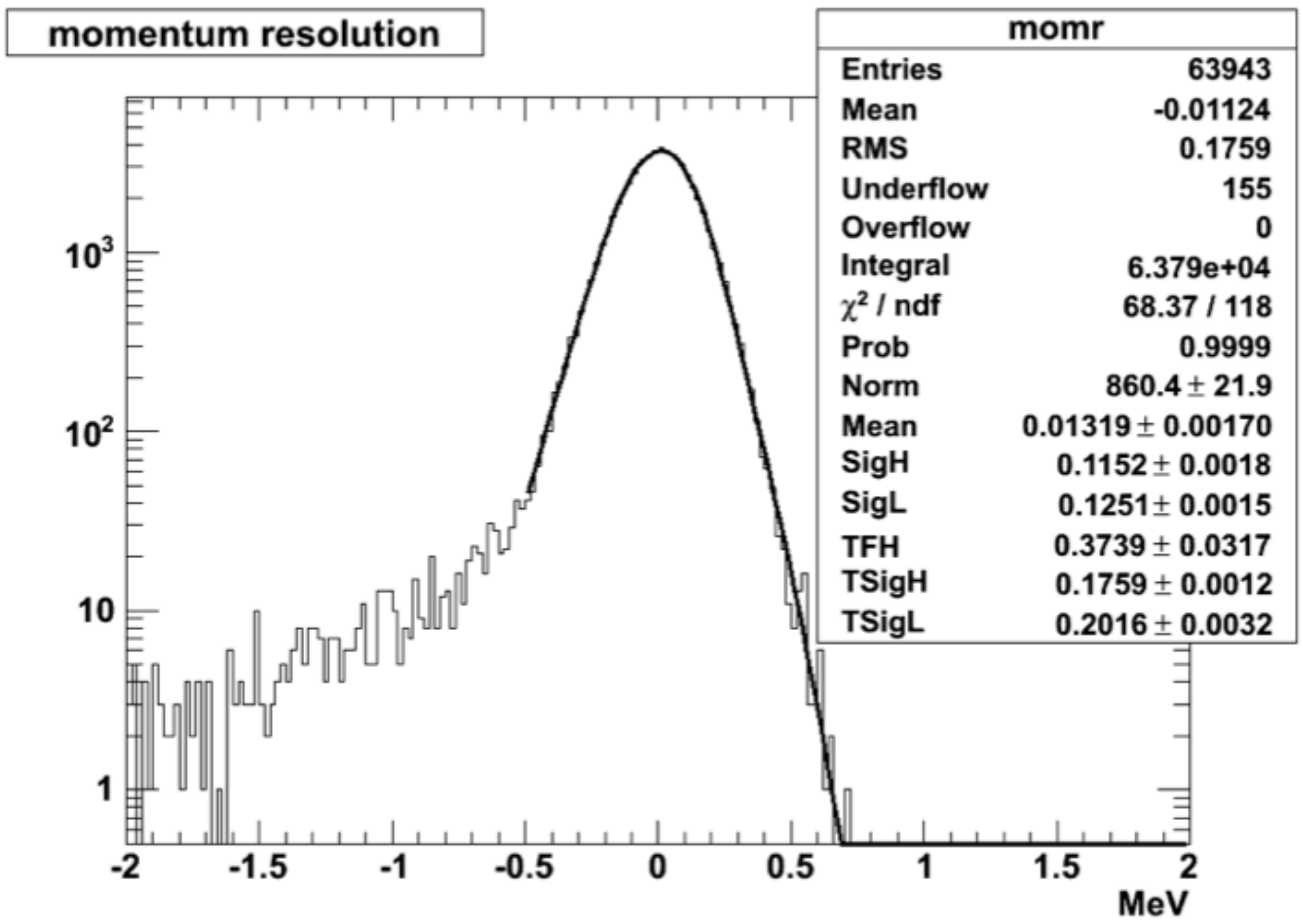

| | | |
|---|---|---|
| SigH | σ of core Gaussian, high side | 115 KeV/c |
| SigL | σ of core Gaussian, low side | 125 KeV/c |
| TFH | Fraction in the tail | 37% |
| TSigH | σ of tail Gaussian, high side | 176 KeV/c |
| TSigL | σ of tail Gaussian, low side | 202 KeV/c |

Figure 9.20. Intrinsic tracker momentum resolution for 105 MeV/c muons computed using Fastsim.

***Detailed Simulation***

The detailed simulation of the tracker uses the Geant4 [12] package. We use the detailed simulation to study pattern recognition and resolution in the presence of backgrounds, as described below. All of the relevant materials in and around the tracker are simulated. We use the QGSP_BERT_HP physics list to get accurate estimates of neutron interactions. The intrinsic transverse resolution of a single straw is taken to be 100 μm, independent of drift distance, which is conservative compared to results achieved in a prototype of the NA62 straw chamber [13]. We assume a resolution along the straw of between 6 and 9 cm, depending on position, based on measurements made with a straw prototype using time-division readout [14]. The tracker simulation assumes a modest d*E*/d*x* resolution, capable of distinguishing hits from highly-ionizing particles from conversion electron hits, but not performing true particle identification.

***Noise Hits***

Our studies show that tracker hits caused by beam particles and the daughters of stopped muons are several orders of magnitude more numerous than all other sources. Based on the experience of NA62, ATLAS and CDF, we believe the electronics gain and threshold can be set high enough to render the intrinsic electronics noise hit rate negligible, and so that is at present ignored.

Table 9.8 lists the estimated tracker hit rate induced by the dominant background sources. Hits produced directly and indirectly by these sources are counted in the rates. The effect of earlier bunches in the bunch train is included by folding the hit time distributions at the microbunch repetition period. At early times, beam electrons produced by the initial protons generate a high hit rate, but fall off quickly. Particles produced by stopped muons are the largest source of background hits during the experimental live time window, which starts nominally 600 ns after the proton pulse. Muon decay in orbit (DIO) produces low-momentum electrons. Protons, neutrons and γ particles are produced during nuclear breakup and de-excitation following muon capture. Neutrons generally produce hits indirectly through daughter electrons and photons, at a roughly constant rate over the microbunch due to the long neutron absorption time. Figure 9.22 shows the distribution of neutron hits in the tracker. The rates at low radii closest to the target are generally a factor of 10 higher than the average rate. The hit pattern is similar for all the background sources. More details on backgrounds can be found in several Mu2e documents [15].

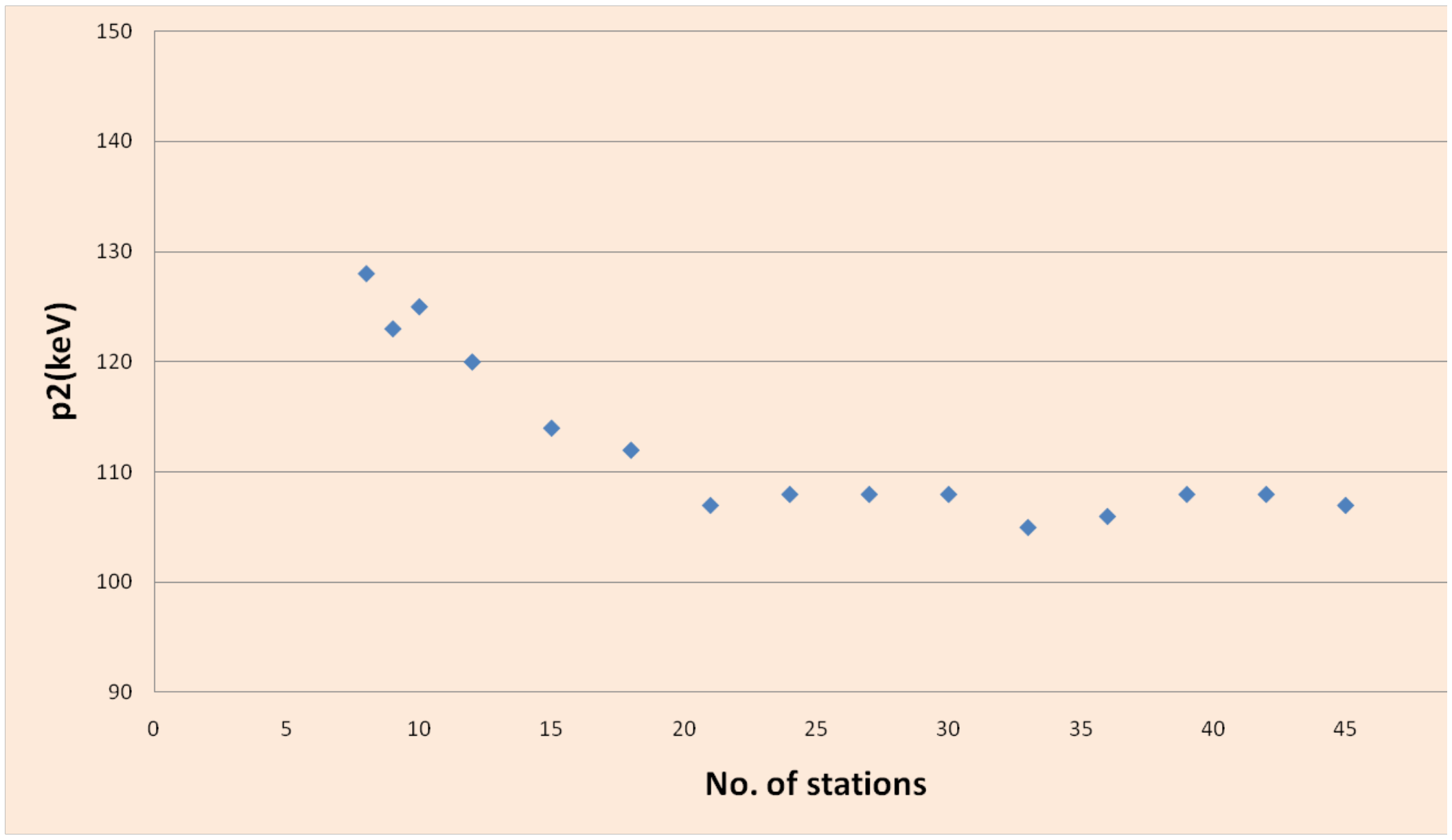

Figure 9.21. Intrinsic tracker momentum resolution vs. number of stations.

| Bkg source | Rates (KHz) (0 – 300 ns) | Rates (KHz) (300 – 600 ns ) | Rates (KHz) ( 600 – 1694 ns ) |
|---|---|---|---|
| DIO | 2.04 | | 2.45 |
| p | 11.8 | | 14.2 |
| n | 38.6 | | 42.9 |
| $\gamma$ | 30.3 | | 28.4 |
| Beam $e^{\pm}$ | 8445 | 61 | 14 |

Table 9.8. Average tracker hit rate per straw for different time intervals relative to the peak of the proton pulse.

### 9.4.3 Pattern Recognition

On average, background processes produce >2000 hits per microbunch in the live window. Distinguishing the roughly 40 conversion electron hits from these requires algorithms that exploit their distinct spatial and temporal properties. To demonstrate that the tracker design is adequate for finding conversion electrons and measuring their properties to sufficient accuracy, we have developed a prototype pattern recognition algorithm. To test the algorithm, a random sampling of all the background processes described above are overlaid on top a single conversion electron. Simulated track hits are produced from these using the detailed G4 simulation described above. The effects of hit overlaps in the straws are modeled assuming a 100 ns two-hit resolution in the electronics. We stress that the results below are extremely preliminary, and are intended only as a proof-of-principle, not a measurement of the true capability of the experiment. As listed at the end of this section, we know of many areas where significant improvements can be made with little cost or risk. These will all be studied as the Project moves forward.

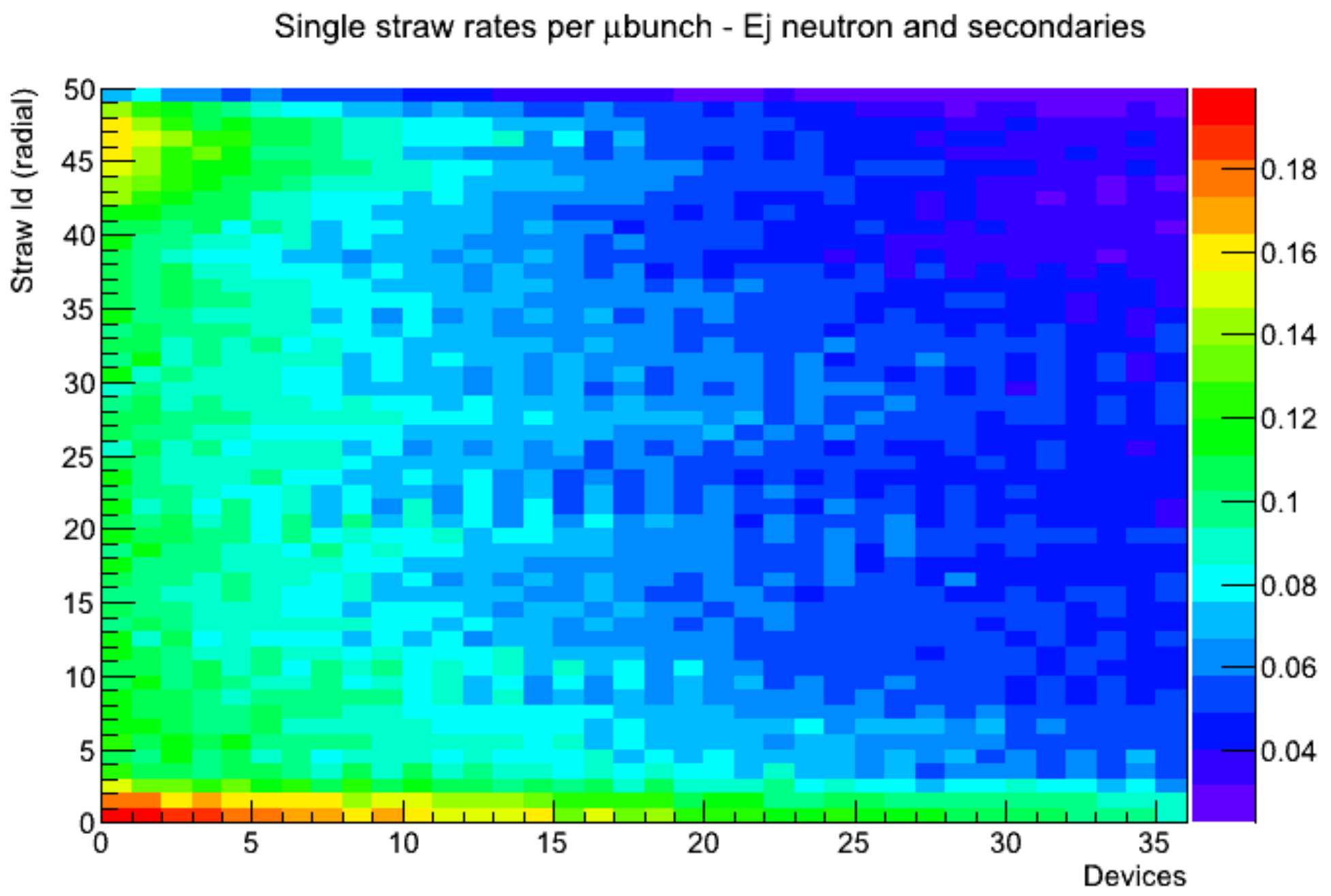

Figure 9.22. Neutron induced hit rate from muon induced Al nuclear breakup, shown per straw per micro-bunch vs. straw position. Horizontal axis is 2× station number, which rises linearly with z. Vertical axis is straw number, which rises linearly with radius. Mean hit density is indicated by colors

Pattern recognition begins with a pre-selection of hits. Because the inner straw hit rates are high and dominated by backgrounds, we require the hit position have a minimum transverse radius of 410 mm, effectively eliminating the inner 5 layers of straws. Hits from these straws are added back to tracks once they have been found and fit. Hits are also required to have a d*E*/d*x* measurement consistent with an electron. Pre-selection eliminates 95% of the hits from protons, and 20% of hits from other background sources.

The dominant source of background hits after pre-selection are low-momentum electrons which move in tight spirals through the detector. We identify these spirals using a dedicated algorithm that looks for clusters of hits consistent in azimuth and time. In principle, the radial position of the hits can also help identify electron spirals, but the straw geometry makes the radial resolution non-Gaussian and thus difficult to combine with other information. Removing identified tight spirals rejects 75% of the remaining backgrounds while keeping 85% of conversion electron hits.

We identify potential conversion electrons as peaks in the time distribution of the remaining hits. This is demonstrated in Figure 9.23, which shows the hit time spectra for three micro-bunches after pre-selection (Loose Selected) and spiral removal (Tight Selected). Conversion electron hits are found in a time peak with 80% efficiency.

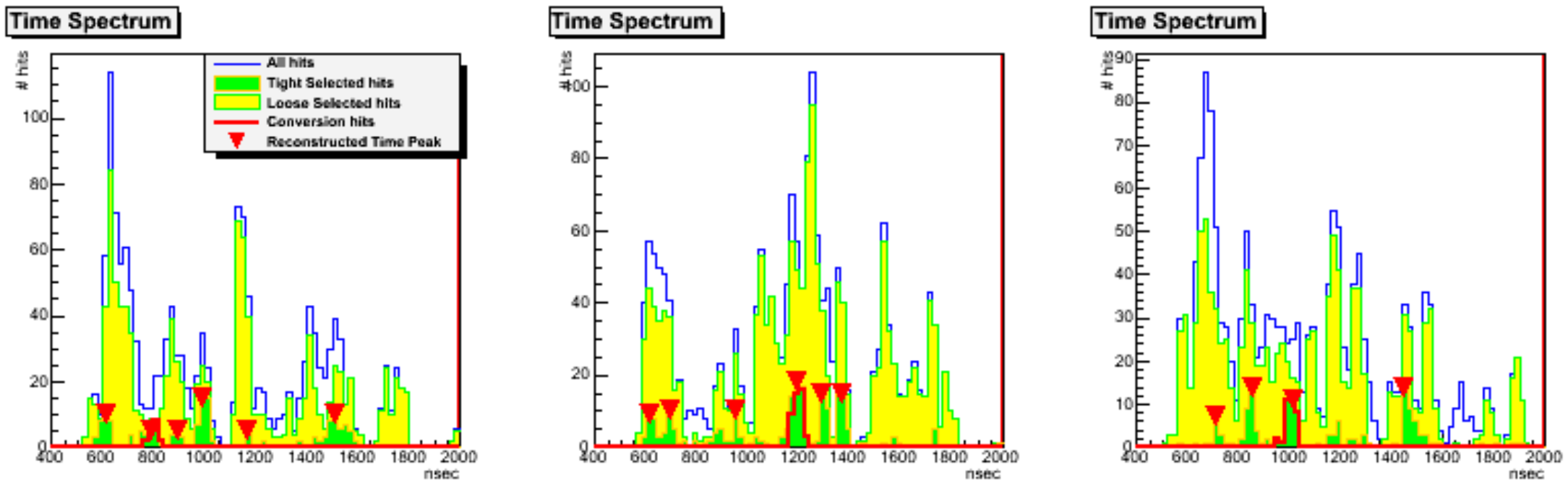

Figure 9.23. Tracker hit time spectra. Three representative microbunches show, after pre-selection and spiral removal.

Hits in time peaks are then fit to a helix using a series of successively more precise geometric fits. The first fit uses a "robust helix" algorithm adapted from optical pattern recognition [16], which describes the hits as space points given by the wire position and time division. The parameters obtained from that fit are used as input to an iterative least-squares helix fit, which describes the hits as lines in space, defined by the wires, with an error equal to the straw radius. The results of that fit are used as input to a Kalman filter fit, which describes the hits as drift circles around the wire, with 100 μm equivalent spatial error. The $t_0$ value used to define the drift radius is fit iteratively with the Kalman filter. Outlier hits are iteratively removed, based on their contribution to the fit**.** This multi-stage fit procedure is successful for roughly 60% of the conversion electrons. When run on a sample of 10,000 background-only events this procedure finds no tracks above 80 MeV/c momentum[4].

Figure 9.24 shows the conversion electron momentum resolution achieved by this algorithm, defined as the momentum returned by the Kalman filter fit minus the true momentum at the entrance of the tracker. The resolution is fit with a split double-Gaussian that allows for a separate core and tail sigma on either side of 0. The high-side core sigma (SigH), after some basic fit quality cuts, is roughly 140 keV. This is significantly less than the energy straggling caused by the target and the proton absorber, which is roughly 350 keV. A significant high-side resolution tail of roughly 600 keV sigma is present.

[4] The tracks found are from DIO electrons from overlaid background frames.

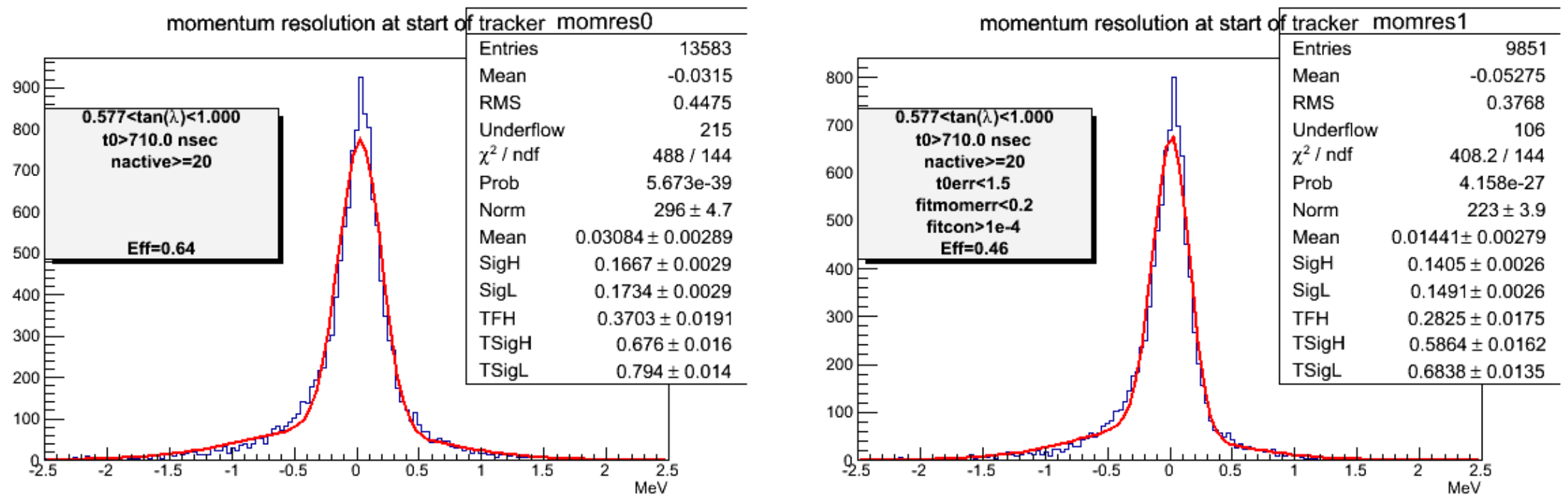

Figure 9.24. Conversion electron momentum resolution. Full background overlay and pattern recognition included, fit to a split double Gaussian. Left is with no cuts, right is with standard track fit quality cuts.

We have tested the robustness of the pattern recognition by varying the inputs. Doubling the total background rate results in a 40% relative drop in reconstruction efficiency, but essentially no change in the resolution. Similarly, degrading the time division resolution by a factor of 2 results in a 40% relative loss in efficiency, again with no appreciable change in resolution. By comparison, using the time division resolution we expect to obtain with final electronics results in a 15% relative improvement in the efficiency.

Several improvements to the pattern recognition algorithm are planned. Use of the full spatial information and more sophisticated clustering will improve the electron spiral background removal. Using multi-dimensional clustering will improve the conversion hit time peak finding. Alternate forms of the robust helix fit will be explored. Integration of $t_0$ as a fit parameter in the Kalman filter fit will reduce the number of iterations required. Using a global algorithm, such as simulated annealing to remove outlier hits, will improve the efficiency and purity of the hits used in the final fit.

# 9.5 Alternatives

## 9.5.1 I-Tracker

For a low mass tracking detector, a natural choice is an open cell device using an array of wires for both cathodes and anodes, the “hex cell” design being the classic example [17]. Mu2e has done extensive R&D on such a device [18] The “I-Tracker,” as it is known, is shown in Figure 9.25 would yield many more hits per track than the proposed straw tube tracker, resulting in more robust pattern recognition and fewer reconstruction errors. The increased number of hits may also make a shorter chamber and a correspondingly shorter Detector Solenoid possible.

The Mu2e environment, however, presents a formidable set of challenges.

- The device must operate in a vacuum. Tracks must traverse a containment vessel that can tolerate ~1 atmosphere pressure, which must be kept extremely light.
- Wires supports lie within the trajectory of electrons; therefore endplates and wire anchor points must also be made extremely light.
- To keep the mass of cathode wires low requires thin (~40 μm) aluminum wires, which are more prone to breakage than most drift chamber wires (tungsten, steel).
- Since the wires are not mechanically contained, a single broken wire would disable an unacceptably large segment of the detector, and require prompt repair.
- To control the mass contribution of the drift gas requires helium (plus a quenching gas). A helium-tight gas seal that can be opened to repair, or at least remove, a broken wire is difficult.

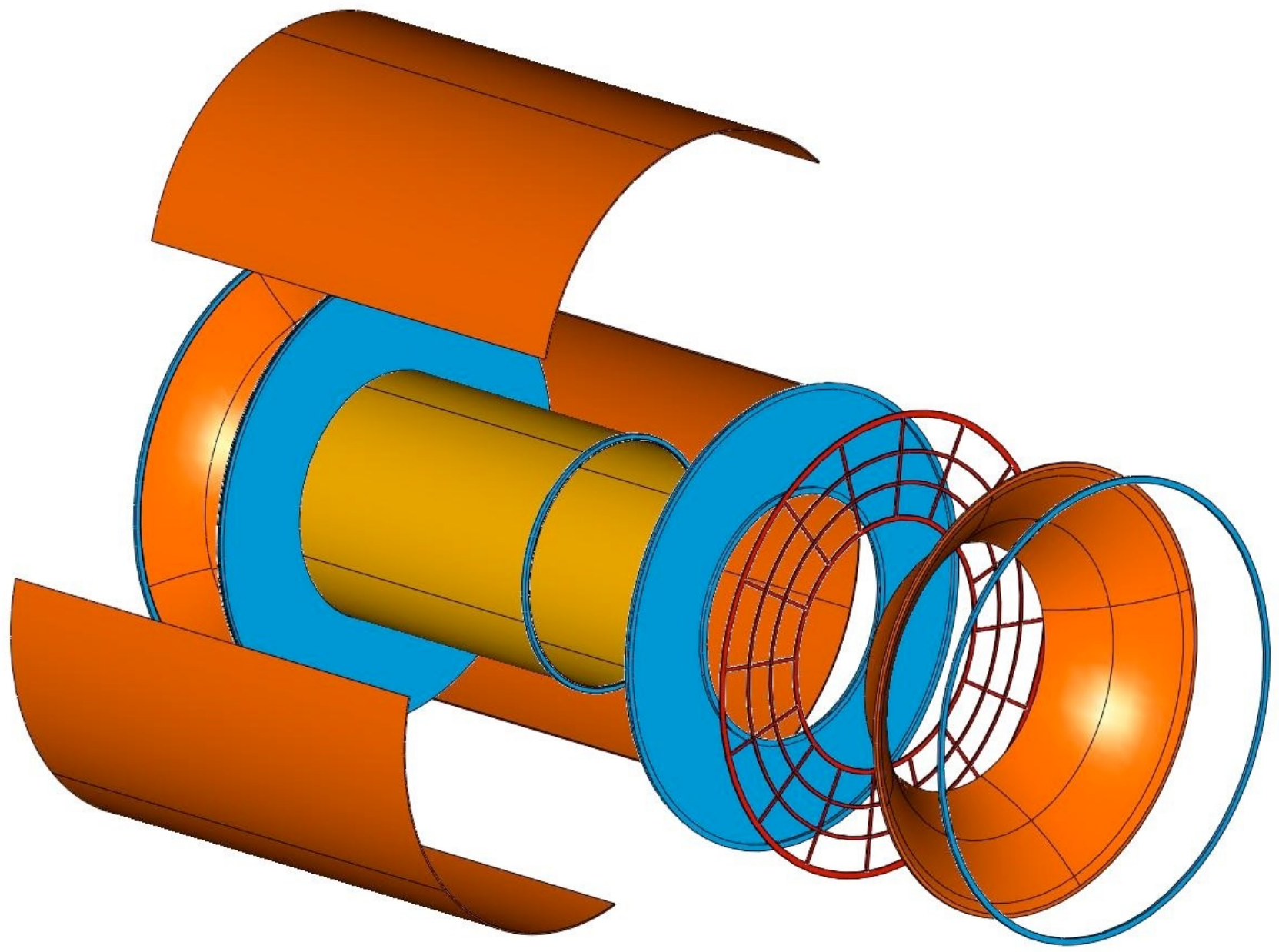

Figure 9.25. Exploded view of the I-Tracker, with separate tension carrying (blue) and pressure carrying (orange) end plates.

This option is still being explored because of its numerous attractive features. However, due to the many novel but untested aspects of this design, it has not been selected at this time.

### 9.5.2 Waveform Digitizers

The preferred readout system presented here is based on ADCs and TDCs residing on the detector. An alternative is ≲1 GS/sec waveform digitizers. Such a device would be good enough for a drift time measurement but not good enough for implementing time division. Examples exist of devices running up to 5 GS/sec for timing precision better than 200 ps, which would be sufficient for time division as well. However, 1 GS/sec – and even more, 5 GS/sec –waveform digitizers consume more power than can be tolerated within the vacuum: they require analog signals to be brought off the tracker to digitizers located outside the evacuated Detector Solenoid. On-chamber digitizers are preferred for several reasons.

- Space constraints limit the cable choice to "micro-coax" (~1 mm OD). Such cable has relatively high dispersion and would likely make time division impossible despite the availability of suitable digitizers.
- Such cable is also not very well shielded; to get reasonable signals out requires a differential drive, thus ~40,000 cables without time division, ~80,000 with time division. Getting this many high-speed signals through the vacuum seal is difficult and expensive.
- This large number of cables would block access to the detector, making repairs difficult.
- The weight of the cables is large and difficult to support without disrupting the detector alignment.

This is retained as an alternative in case the reliability required for on-chamber digitizers cannot be attained; however, this alternative has not been selected for the reasons given above.

### 9.5.3 Commercial (non-ASIC) Front End

An alternative to developing an ASIC is using commercial ADC chips and FPGA-based TDCs, both still residing on the detector. The advantages are: elimination of the ASIC development and testing cost; and, because the lead time is short, benefiting from advances in technology. Disadvantages are higher power (possibly five-fold) and more engineering effort on board layout.

We have greater confidence in the ASIC solution and take it as our baseline. However, both approaches are being explored.

## 9.6 ES&H

The Mu2e tracker is similar to other gas-based detectors that are commonly used at Fermilab. Potential hazards include power systems, compressed gas and the possible use of flammable gas. These hazards have all been identified and documented in the Mu2e Preliminary Hazard Analysis [19].

The detector requires both low voltage, high current (~5 V @ 2 kA total) and high voltage, low current (~1.5 kV @ 500 mA) power systems. During normal operation the tracker will be inaccessible, inside the enclosed and evacuated Detector Solenoid. Power will be distributed to the tracker through shielded cables and connectors that comply with Fermilab policies. Fermilab will review the installation prior to operation.

The tracker will require a supply of chamber gas to be supplied from compressed gas cylinders. The current plan is to use a non-flammable mixture of Argon and $CO_2$. A decision could be made to use a different gas mixture that may contain flammable components, for instance to achieve higher drift speed to accommodate high rates. Flammable and non-flammable gases that will be used for Mu2e will be kept in DOT cylinders in quantities limited to the minimum required for efficient operation. The cylinders will be stored in a dedicated location appropriate to the type of gas being used. Flammable gases will be segregated from oxidizers. The storage area will be equipped with a fire detection and suppression system. The installation, including all associated piping and valves, will be documented and reviewed by the Fermilab Mechanical Safety Subcommittee. If flammable gas is used, the system will be reviewed prior to operation by the flammable gas review committee to ensure strict adherence to Fermilab safety policies and procedures.

The cooling system requires SUVA® or similar refrigerant. Plumbing will be designed to tolerate a wide range of pressures, allowing for a cost effective and environmentally friendly solution based on product availability and EPA guidelines at the time of operation.

The T-Tracker itself does not have any radioactive sources; however, there will be sources used in monitoring chambers. Usage of radioactive sources will be reviewed to ensure adherence to Fermilab safety policies.

## 9.7 Risks

### 9.7.1 Performance Risk

Extensive simulations of the tracker have been performed, providing confidence that it will deliver the required resolution. However, work on pattern recognition is incomplete. There is a risk that it will prove difficult to identify tracks with the required efficiency. Work continues on simulations and pattern recognition that could lead to design modifications, if needed. This translates to a cost and schedule risk.

The tracker performance is sensitive to beam properties as well as upstream components like the proton absorber and stopping target. Work on beam simulation, and optimization of the target and proton absorber continue. Although these risks are not strictly part of the tracker, changes elsewhere (for instance, eliminating the proton absorber) could result in higher rates in the tracker. This risk is mitigated by working closely with other groups to ensure the combined system meets requirements.

Inability to hold alignment in the detector would lead to degraded resolution and unaccounted for background. The problem can be solved using in-situ, track-based alignment. This is a difficult and slow process for Mu2e, resulting in delays in publishing results. This risk is mitigated by continuous monitoring using precision electronic levels and Hall probes installed on each station.

### 9.7.2 Technical and Operational Risk

Contaminated gas is a serious risk for any drift chamber. This risk is mitigated in several ways. First, by using Ar:$CO_2$ as the drift gas: it is one of the least prone to harmful contaminants. Second, by performing detailed analysis on each batch of gas. Finally, monitoring chambers will be included in the system, illuminated with radioactive sources, to give early warning of problems.

Depending on results of ongoing studies, it may be desirable to use a gas with higher drift velocity to reduce overlapping hits. Such gas mixtures typically include $CF_4$ which, combined with small amounts of moisture, forms hydrofluoric acid, leading to etching of the cathode surface. Our gold over aluminum cathode is less prone to etching than the more common copper cathode. Never the less, using $CF_4$ would necessitate a drying unit immediately before the tracker (just outside the vacuum) and extensive baking of that part of gas system to remove moisture. It may also require adding a gold coat to the cathode, further increasing the cost.

Wire chambers always suffer the risk of a broken wire. The loss of a single wire is a minor problem, but the wire shorting other electrodes can leave a large region of the detector inoperable. The selected straw tracker mechanically contains each wire; however the resulting short to the cathode surface would upset the high voltage distribution. The ability to disconnect individual wires from high voltage without the need to access the detector has been included in the design to mitigate this risk.

The primary mechanical component of the straw is Mylar®. As with any plastic, Mylar® under stress creeps (gradually stretches), and is sensitive to humidity. However, both creep and hygroscopic expansion is less with Mylar® than the more widely used Kapton® straws. At worst it may be necessary to pre-tension (or re-tension) the straws for an extended period. This will impact the construction and assembly schedule.

## 9.8 Quality Assurance

Proper quality assurance is essential to construct a tracking detector that meets Mu2e's requirements for performance and reliable operation. Quality Assurance will be integrated into all phases of tracker work including design, procurement, fabrication and installation.

It is important that individual straw tubes be leak-tight, straight and have the proper wire tension. The straws will be leak tested before being installed. The straws will be connected to a clean gas system and over pressured. The leak rate will be measured over an appropriate time interval by measuring the pressure drop. After assembly of a plane, the entire plane leak rate will be tested again.

The straws must maintain their shape and be mounted straight to operate efficiently and to maintain an appropriate distance between the wire and the grounded Mylar® surface to avoid high-voltage breakdown. The straws will be visually inspected for roundness and straightness before assembly. Flawed straws that escape detection during visual inspection can be identified by non-uniform gas gain and resolution when exposed to a radioactive source. This will be done as part of the wire position measurement.

The appropriate tension must be applied and maintained in a straw tube for efficient, stable operation. Tension is applied through calibrated mechanical force but tension can be lost through slippage, creep, or other relaxation mechanisms. Both wire and straw tension can be measured after assembly via vibration resonance. Oscillation is induced by driving a current through the wire or straw in the presence of a magnetic field. The amplitude and frequency of vibration is measured by turning the drive current off and measuring the current induced by the conductor moving in a magnetic field.

All electronics components will be tested prior to installation on the tracker, including a suitable burn-in period. High voltage boards will be tested for leakage current. The threshold characteristics of each channel will be tested with a threshold scan. A noise scan will be performed for various threshold settings to identify channels with large noise fractions. A complete battery of tests will be developed once the specifications are fully understood.

## 9.9 Value Management

The tracker technology chosen for Mu2e is well established and has been implemented in other high energy and nuclear physics experiments [20]. Value management principles have been applied over time during the development of the technology and there is little left to be gained. In addition, the tracker is the most important detector element for identifying and analyzing conversion electrons yet it comprises only about 2% of the total project cost. Compromising tracker performance and reliability for a small cost savings would be unwise.

Two areas where it might be possible to apply value management are the use of time division and nature of the ADC.

Studies to date indicate time division is a significant assist to pattern recognition [21]. If further developments in pattern recognition indicate time division is of no benefit to pattern recognition, then time division could be dropped and the straws would only require single-sided readout, resulting in a small cost savings and reduction in complexity.

The value of pulse height measurement is unambiguous; however, the nature of the ADC needs to be studied. The selected option, analog shaping and ADC combined in a single ASIC, is conservative in the sense of optimizing resolution, power, and compactness. However, custom chip development costs are high and come with some risks. Alternatives exist depending on how these considerations are balanced against cost and reliability.

- An ASIC with a lower precision, or lower speed, ADC design.
- Use of commercial ADC chips instead of an ASIC. This would entail a separate shaping and discriminator circuit, either an ASIC or an implementation in discrete parts. In the case of an ASIC, there is the possibility of using an existing design, resulting both in cost savings and lower risk.
- Use of a “charge-to-time” converter, effectively encoding pulse height into pulse width, followed by a TDC. The TDC would be identical to what is needed for drift time measurement, resulting in simplification and cost savings. Shaping and the charge-to-time converter would need to be implemented separately, requiring an ASIC; however it may be possible to use an existing design [22].

## 9.10 R&D

Work is under way on straw termination, tensioning, and alignment. Results are in agreement with calculations [23]. The present effort will culminate in a full-panel prototype in June 2012. Between CD-1 and CD-2, this prototype will be used as a platform for testing electronics and overall performance, first in air (for ease of access) followed by tests in vacuum. The tests will use cosmic rays and no magnetic field. The vacuum vessel will be a medium size liquid helium cryostat available after CDF decommissioning.

Fermilab has a large number of magnets that can provide a 1 T field, however none of the available magnets can accommodate a T-Tracker panel. To validate operation of straws in a magnetic field, a mini-panel will be built, tentatively two layers of five straws each. The length of the straws is not yet determined; the one magnet known to be available (a test solenoid removed from CDF) would necessitate short (~30 cm) straws.

## 9.11 References

[1] R. Bernstein, "Requirements Document for Mu2e Tracker," Mu2e-doc-743.

[2] M. Lamm, "Detector Solenoid Requirements Document," Mu2e-doc-946.

[3] R. Tschirhart, "Trigger and DAQ Requirements," Mu2e-doc-1150.

[4] A. Mukherjee, "R&D for T-Tracker," Mu2e-doc-873.

[5] R. Tschirhart, "Data flow and volume estimates for consideration of a streaming DAQ architecture for the mu2e experiment," Mu2e-doc-916.

[6] E. Voirin "Mu2e Cooling system Conceptual Design and Cost Estimate," Mu2e-doc-2068.

[7] S. Siu, "T-tracker Studies," Mu2e-doc-983.

[8] K. Yarritu, "Stopping Target Studies," Mu2e-doc-1339.

[9] J. Alsterda et al., "Analysis of a Novel Proton Absorber Geometry for the Mu2e Experiment," Mu2e-doc-683.

[10] R. Andreassen et al., Journal of Physics: Conference Series **331**, 032038 (2011).

[11] D. Brown, "Tracker Resolution Studies using FastSim," Mu23-doc-1446.

[12] S. Agostinelli et al., NIM A **506**, 250 (2003).

[13] S. Movchan, NIM A **604, 307** (2009).

[14] V. Rusu, "Straw Properties," Mu2e-doc-1524.

[15] G. Onorato, "Background Rates in the T Tracker,"Mu2e-doc-1838, G. Onorato, "Beam Flash Simulation," Mu2e-doc-1865.

[16] I. Ladrón de Guevara et al., J Math Imaging Vis DOI 10.1007/s10851-010-0249-8.

[17] D. Boutigny et al., SLAC-R-457 (1995).

[18] F. Grancagnolo, "Status of the I-Tracker," Mu2e-doc-1158.

[19] R. Ray, "Preliminary Hazard Analysis Document," Mu2e-doc-675.

[20] G. Aad et al., Journal of Instrumentation **3** S08003 (2008).

[21] H. Nguyen and H. Wenzel, "T-Tracker Pattern Recognition Status," Mu2e-doc-1416.

[22] W.M. Bokhari et al., IEEE Nuclear Science Symposium, (1998)

[23] Seog Oh et al, "Properties of Mu2e straws from PPG", Mu2e-doc-2048, A. Mukherjee, "Straw Mechanical Properties", Mu2e-doc-1343.

# 10 Calorimeter

## 10.1 Introduction

The design of the Mu2e detector is driven by the need to reject backgrounds to the $\mu \rightarrow e$ conversion electron signal. A background of particular concern is false tracks arising from pattern recognition errors that result from high rates of hits in the tracker. These accidental hits could combine with or obscure hits from lower energy particles to create a trajectory consistent with a higher energy conversion electron. Even with modern computing resources, it is difficult to simulate this background to the level required to demonstrate that it is not a concern. Thus the primary purpose of the Mu2e calorimeter is to provide a redundant set of measurements that complement the information from the tracker, and enable rejection of background due to reconstruction errors. A second important function of the calorimeter is to provide an efficient trigger for the experiment.

## 10.2 Requirements

A set of requirements for the calorimeter has been developed by the Mu2e collaboration [1]. The primary functions of the calorimeter are to provide energy, position and timing information to confirm that events reconstructed by the tracker are well measured and are not the result of a spurious combination of hits, and to provide a potential trigger for the experiment. This leads to the following requirements:

- A position resolution of $\sigma_{r,z}$ = 1 cm or better to allow comparison of the position of the energy deposit to the extrapolated trajectory of a track reconstructed in the tracker.
- An energy resolution of $\sigma_E$ = 2% (FWHM/2.35, since the response is non-Gaussian) or better at 100 MeV is desirable to confirm the much more precise energy measurement from the tracker. The uncertainty in the energy scale should be small compared to the resolution;
- A timing resolution of ~1 ns to ensure that energy deposits in the calorimeter are in time with events reconstructed in the tracker, enabling rejection of backgrounds.
- The calorimeter should provide additional information that can be combined with information from the tracker to distinguish muons or pions near the conversion momentum from electrons with 99% C.L.
- Provision for a trigger, either in hardware, software, or firmware that can be used to identify events with significant energy deposits.
- The calorimeter must operate in the unique, high-rate Mu2e environment and must maintain its functionality for radiation exposures up to 50 Gy/year/cm$^2$, surviving a nominal run with a loss of light output of ≤ 10%;

- Have a temperature and gain variation such that the combined response of a calorimeter cell and its readout does not vary by more than ±0.5%, *i.e.,* is small compared to the required energy resolution.

The requirement on the calorimeter's position resolution is based on the error associated with extrapolating a track from the tracker to the calorimeter (Figure 10.1). There is no need for the calorimeter position resolution to be better than the extrapolation error, driven by multiple scattering in the tracker. Based on this study, a position resolution of 1 cm is sufficient.

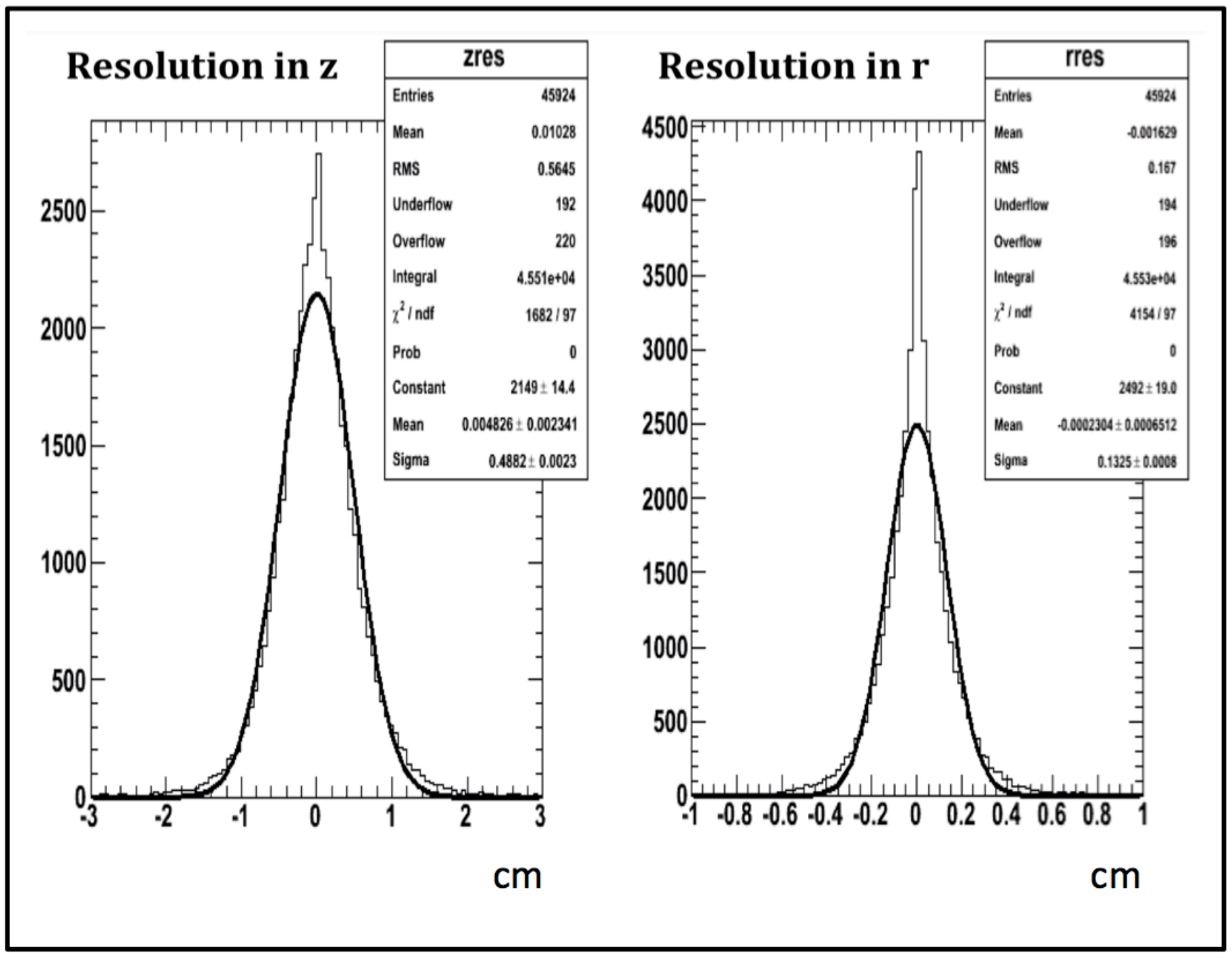

Figure 10.1. The error on the extrapolated position of tracks from the tracker to the calorimeter. The tracks were fitted with a Kalman filter and extrapolated to the calorimeter using the full covariance matrix. The *z* direction is along the solenoid axis and the *r* direction is transverse to the solenoid axis.

The energy resolution of a crystal calorimeter complements, but is not competitive with, that of the tracking detector. Even a coarse confirmation of track energy by the calorimeter will, however, help reject backgrounds from spurious combinations of hits from lower energy particles. The Mu2e simulation is not yet at the stage where this can be explicitly demonstrated, but 5% energy resolution has been achieved by other experiments operating in a similar energy regime [2]. With an energy resolution of $\mathcal{O}(2\%)$ we could use the energy information in combination with the tracker momentum to build a likelihood function for distinguishing signal from background events.

For real tracks, activity in the tracker and in the calorimeter will be correlated in time. The time resolution of the calorimeter should be comparable to the time resolution of extrapolated tracks from the tracker, estimated to be a few ns [1]. A calorimeter timing resolution of about 1 ns is consistent with the tracker and can easily be achieved.

In Section 10.5, after introducing the calorimeter layout, we provide a few detailed examples to demonstrate the impact of the calorimeter performance on event reconstruction, and describe the acceptance for the signal in various detector configurations. In Section 10.6 we summarize our estimates for the background rate on the calorimeter and its effect on calorimeter clustering reconstruction. An estimate of the radiation dose is also shown.

## 10.3 Proposed Design

In the 100 MeV energy regime, a total absorption calorimeter employing a homogeneous continuous medium is required to meet the resolution requirement. This could be either a liquid, such as xenon, or a scintillating crystal; we have chosen to investigate the latter. Two types of crystals have been considered for the Mu2e calorimeter: lutetium-yttrium oxyorthosilicate (LYSO) and a new version of lead tungstate ($PbWO_4$), called PWO-2. The design selected for the Mu2e calorimeter uses an array of LYSO crystals arranged in four vanes of 11 × 44 crystals that are approximately 1.3 m long. Electrons following helical orbits spiral into the side faces of the crystals, colored red in Figure 10.2. Photodetectors, electronics and services are all arranged on the opposite face. A lead tungstate-based alternative will also be briefly described, as will an alternative disk-based geometry for the crystals.

The 4-vane geometry has been optimized (see below) for the best acceptance at a given crystal volume (*i.e.* cost). The alternative disk geometry allows a further improvement in acceptance. Each vane is composed of a matrix of LYSO crystals, described below. The crystal dimensions are 3 × 3 × 11 $cm^3$; there are a total of 1936 crystals. Each crystal is read out by two large area APDs. Solid-state photo-detectors are required because the calorimeter resides in the 1 T magnetic field of the Detector Solenoid (DS). Front end electronics reside on the detector and digitizers for each channel are placed inside the DS. A flasher system provides light to each crystal for relative calibration and monitoring purposes. A source system provides absolute calibration and an energy scale. The crystals are supported by a lightweight carbon fiber support structure. Each of these components is discussed in the sections that follow.

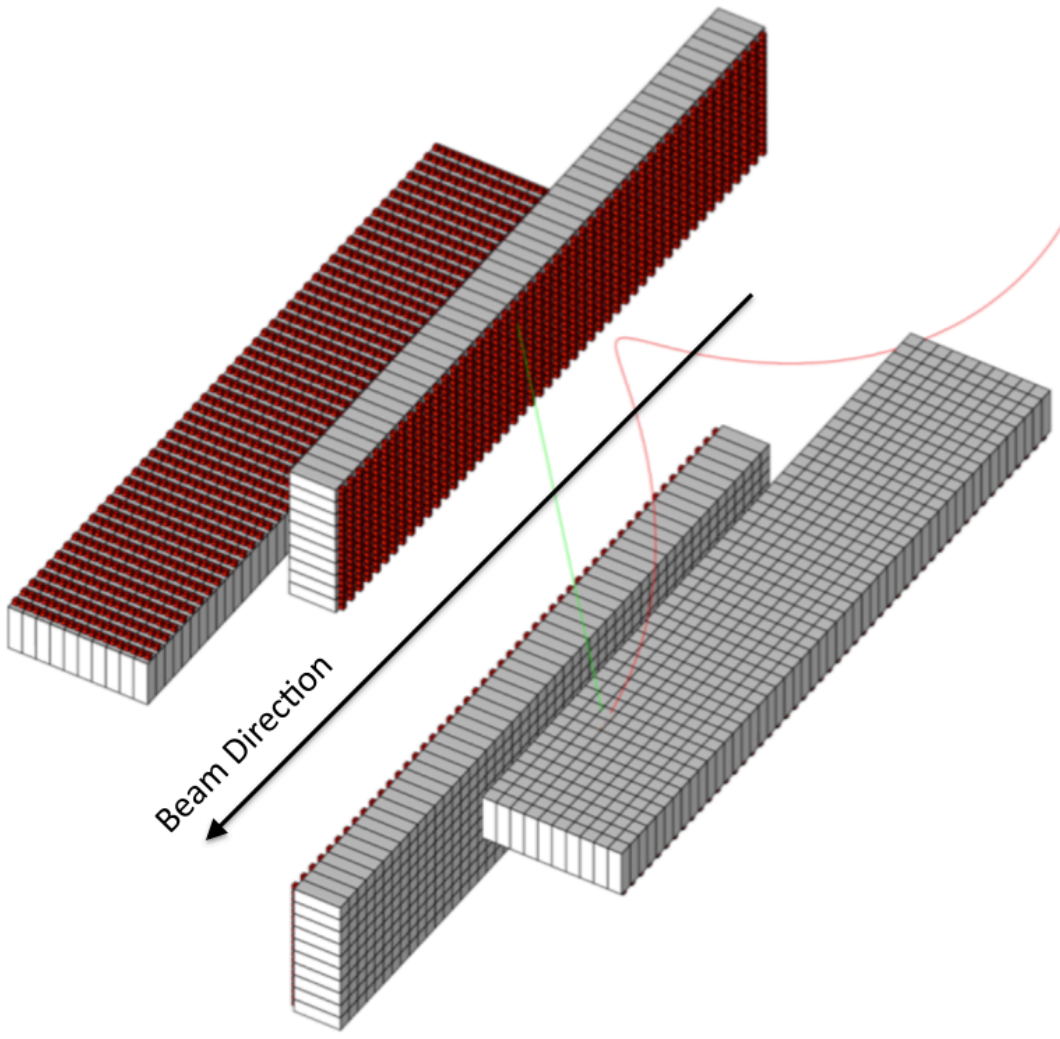

Figure 10.2. The Mu2e calorimeter, consisting of an array of LYSO crystals arranged in 4 vanes. Electrons spiral into the red faces.

### *10.3.1* Crystals

The basic calorimeter element is a lutetium-yttrium oxyorthosilicate (LYSO) crystal. LYSO is an excellent match to the problem at hand: it has a very high light output, a small Molière radius ($R_M$), a fast scintillation decay time, excellent radiation hardness, and a scintillation spectrum that is well-matched to readout by large area avalanche photodiodes (APDs) of the type employed in the CMS and PANDA experiments. LYSO is also the preferred option for the KLOE-2 upgrade and the forward endcap of the new Super*B* detector being designed for the Cabibbo Laboratory Super *B* factory.

LYSO is a material that combines excellent scintillation and physical characteristics. The properties of LYSO are summarized and compared to those of lead tungstate in Table 10.1. Electrons that hit the upstream face of a calorimeter vane in Mu2e will be poorly measured, so to maximize the acceptance of the calorimeter a high premium is placed on dense crystals with a short radiation length. The short radiation length and the fast emission time of LYSO are well matched to the Mu2e environment.

LYSO and $PbWO_4$ share certain desirable characteristics: they have fast scintillation decay times, have similar Molière radii ($R_M$), are not hygroscopic, and have reasonable mechanical properties. There are differences, however: LYSO has a slightly lower density and a slightly longer radiation length, but has a much higher light yield. The light yield of LYSO is a factor of ~200 better than the PWO-2 variant of $PbWO_4$ at room temperature. It is more radiation hard, and the scintillation light output is not rate-dependent, as it is for $PbWO_4$. The light emission spectrum of LYSO peaks at 402 nm, slightly lower that of $PbWO_4$, but both are compatible with APD readout. LYSO has a slower emission time ($\tau = 40$ ns), but the optimal integration time of 200 ns is compatible with the expected signal and background rate.

| **Crystal** | **LYSO** | **$PbWO_4$** |
|---|---|---|
| Density (g/cm$^3$) | 7.28 | 8.28 |
| Radiation length (cm) $X_0$ | 1.14 | 0.9 |
| Molière radius (cm) $R_m$ | 2.07 | 2.0 |
| Interaction length (cm) | 20.9 | 20.7 |
| *dE/dx* (MeV/cm) | 10.0 | 13.0 |
| Refractive Index at $\lambda_{max}$ | 1.82 | 2.20 |
| Peak luminescence (nm) | 402 | 420 |
| Decay time $\tau$ (ns) | 40 | 30, 10 |
| Light yield (compared to NaI(Tl)) (%) | 85 | 0.3, 0.1 |
| Light yield variation with temperature(% / °C) | -0.2 | -2.5 |
| Hygroscopicity | None | None |

Table 10.1. Crystal parameters for LYSO and $PbWO_4$.

The greatest advantage of LYSO crystals is the excellent light yield that allows one to achieve excellent energy resolution without having to operate the calorimeter at -25 °C, which is necessary if $PbWO_4$ is employed. This greatly simplifies the calorimeter design. Temperature stability requirements are also substantially less stringent.

The much larger LYSO signals provide greater flexibility in the choice of photosensors and front end electronics (FEE). Several alternatives have been considered:

- Use of a simple voltage amplifier in place of a charge sensitive amplifier and shaper. Tests show that even a single APD per crystal can provide an Equivalent Noise Charge (ENC) of 100 keV.
- Use of the front end electronics developed for PWO-2 together with a large area APD can reduce the ENC, due primarily to the APD leakage noise, to 30 - 40 keV.
- Retain the front end electronics developed for PWO-2 and reduce the APD area from 10 × 10 mm$^2$ to 5 × 5 mm$^2$, keeping the ENC at the level of ~150 keV.

For any of these options the overall noise for a group of 25 crystals will be below 1 MeV, allowing the energy resolution to be pushed close to the intrinsic photoelectron statistics limit of 1%.

A third advantage of LYSO is the excellent radiation hardness, which has been measured for both $\gamma$s and neutrons. Negligible deterioration of signals (10% loss in light yield) is observed with $\gamma$ exposures of 10,000 Gy (*i.e.* 15 years of Mu2e running). A factor of 5 smaller induced absorption than PWO-2 is seen after irradiation with a flux of $10^{13}$ $n$/cm$^2$. Therefore, with LYSO no stimulated recovery mechanism is required and there will be no reduction of running time due to this issue.

Figure 10.3 shows the response to a $^{22}$Na source of a LYSO crystal read out by a conventional PMT. The energy resolution is excellent. The same technique is used to measure the Longitudinal Response Uniformity (LRU) by scanning the crystal along its axis [3]. Control of the Ce concentration in the growing process has improved the longitudinal response uniformity in current production crystals to better than 2 to 3%.

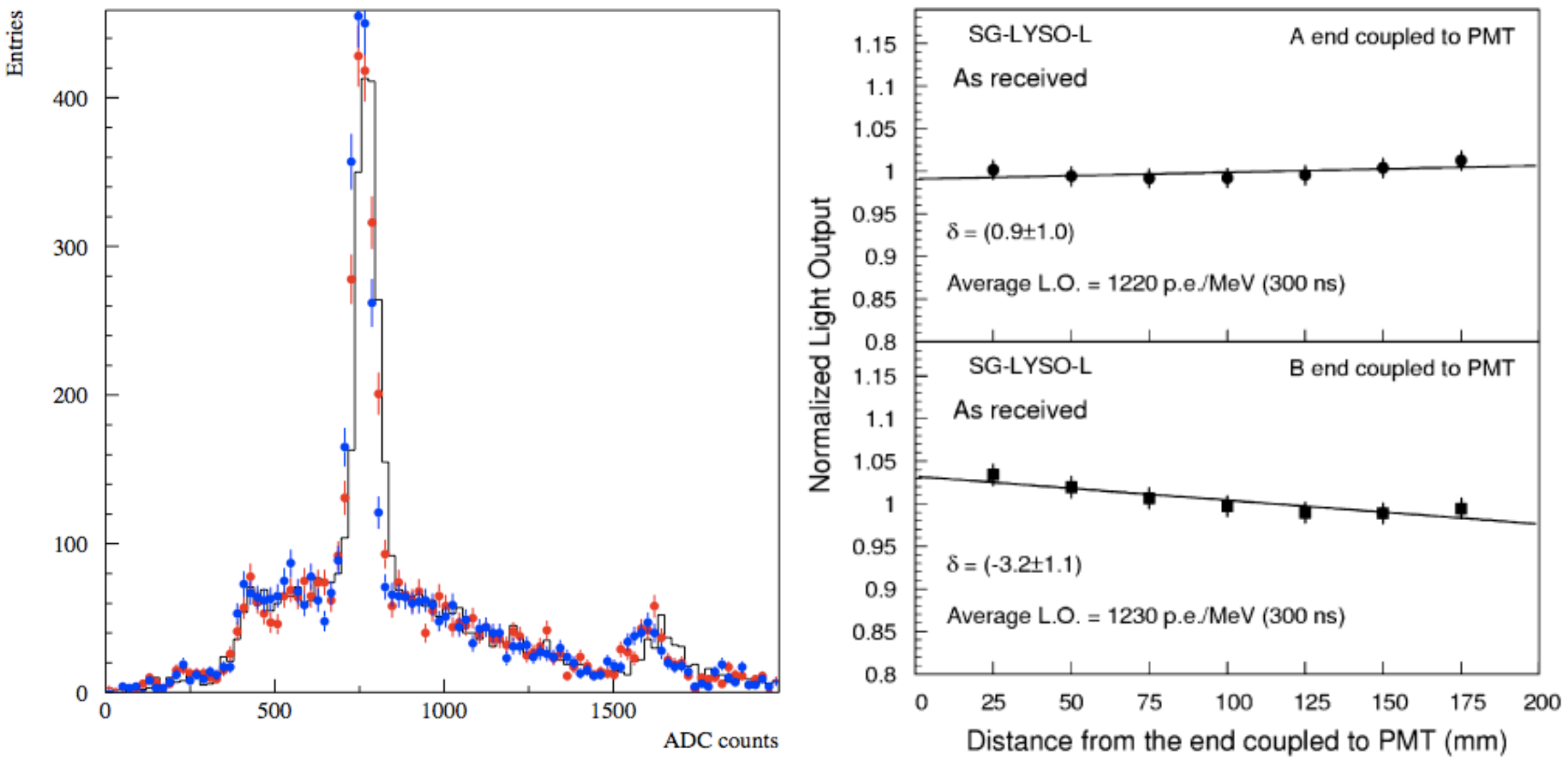

Figure 10.3. Charge response to a $^{22}$Na source for a LYSO crystal readout by a PMT (left). Longitudinal response uniformity measurement for a Saint-Gobain crystal (right).

The main disadvantage of LYSO is the cost. LYSO crystals are commercially available from Saint-Gobain, SICCAS (Shanghai Institute of Ceramics), SIPAT (Sichuan Institute of Piezoelectric and Acousto-optic Technology) and Zecotek. An active R&D program is underway at Caltech, in cooperation with SICCAS and SIPAT, aiming to reduce the commercial price of LYSO. This effort has produced full size crystals of excellent quality that have been employed in beam tests at Frascati, CERN and MAMI. The Frascati group has also carried out an extensive search of producers for similar crystals (Saint-Gobain, Zecotek, SICCAS, and SIPAT), testing large and small samples for each of the producers. They have found that the Chinese producers provide good quality material at reduced cost. Both SIPAT and SICCAS are able to produce the ~ 2000 crystals in a time span of 1 - 1.5 years. The LYSO crystal cost is of the order of a factor 2 more than PWO-2. There are, however, compensating cost reductions. It is not necessary to cool the crystals to -25°C, no provision for radiation damage recovery need to be provided, and the performance and running efficiency of the experiment will be improved. INFN is interested in the use of LYSO and will surely participate, especially given the synergy with the development of the Super*B* endcap calorimeter.

### *10.3.2* Photosensors

The presence of a 1 T magnetic field in the detector region precludes the use of conventional photomultipliers, requiring solid state photosensor devices that are insensitive to magnetic fields. Since PIN diodes show a large response to traversing charged particles ("*nuclear counter effect*"), their usage has been excluded: only large area avalanche photo diodes (APDs), and the newest type of Silicon Photomultipliers (SIPMs) have been considered. Large area SIPMs are still in the development phase at the moment and will be discussed only as an alternative to the basic design.

Large area APDs, with an active area of 5 × 5 $mm^2$ (Hamamatsu S8664-55) are used in quantity by the CMS experiment at the LHC, after a long R&D phase performed in conjunction with Hamamatsu Photonics [4] in the late '90s. APDs are reverse-biased diodes with an internal electric field used for avalanche multiplication of charge carriers. Typically a reverse type APD is composed of three parts, as shown in Figure 10.4:

1. A conversion layer (~2 μm thick) where the electron-hole pairs are generated.
2. A high electric field region (~6 μm), where the amplification of carriers occurs.
3. A drift-region of ~ 200 μm where the carriers drift towards the collection electrode.

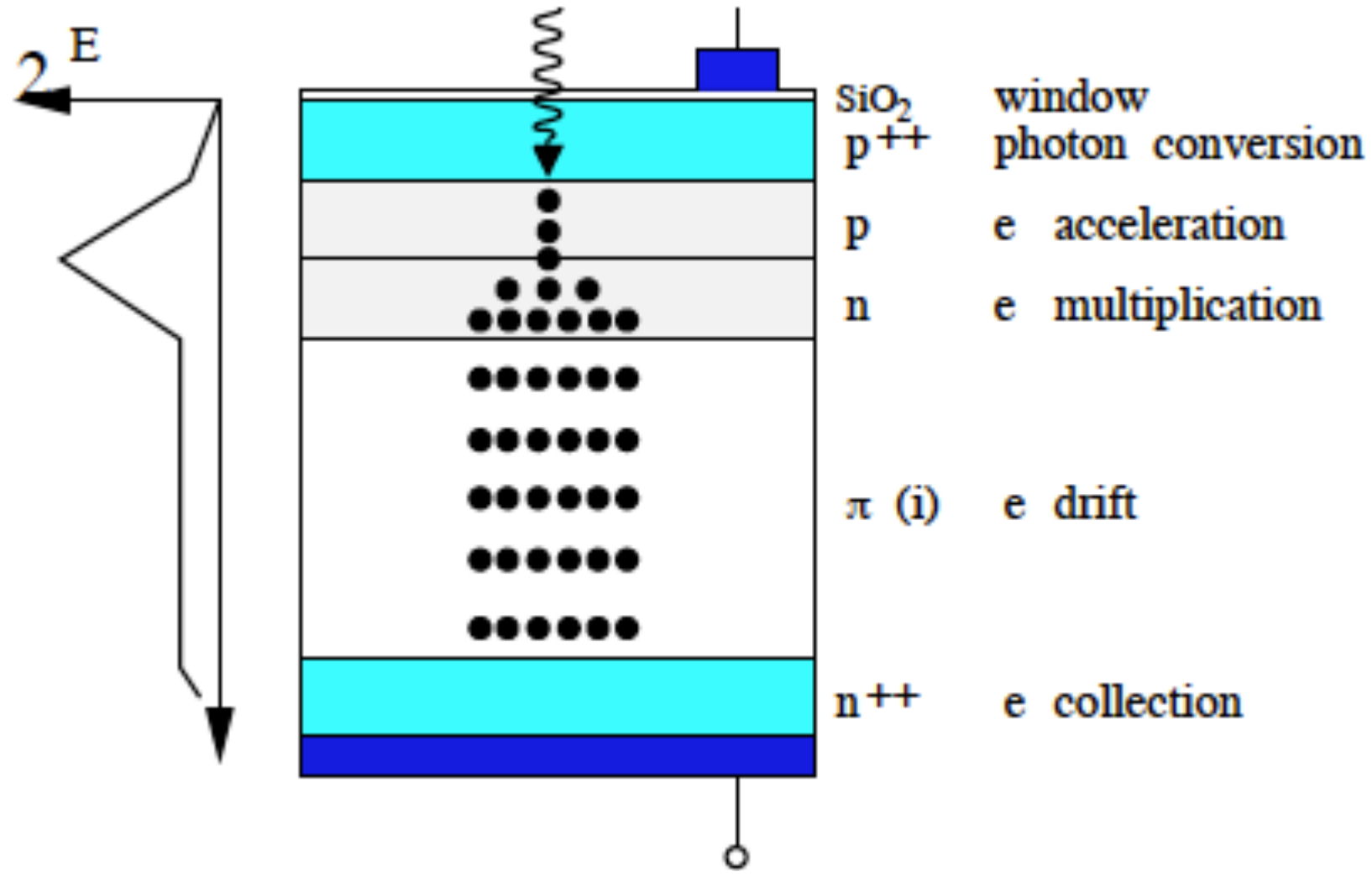

Figure 10.4. Schematic of the structure of a "Hamamatsu-like" APD.

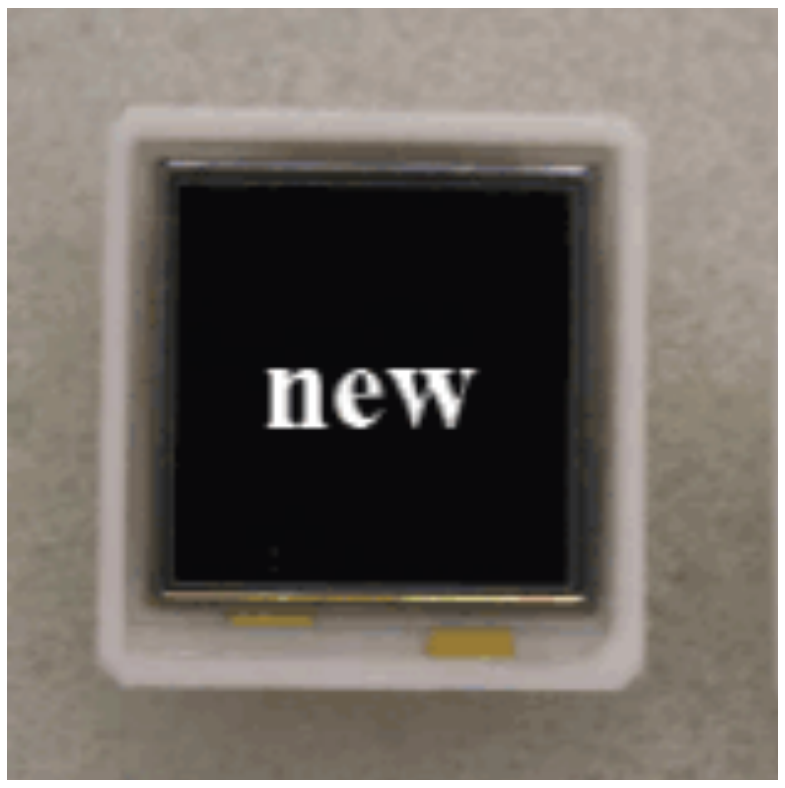

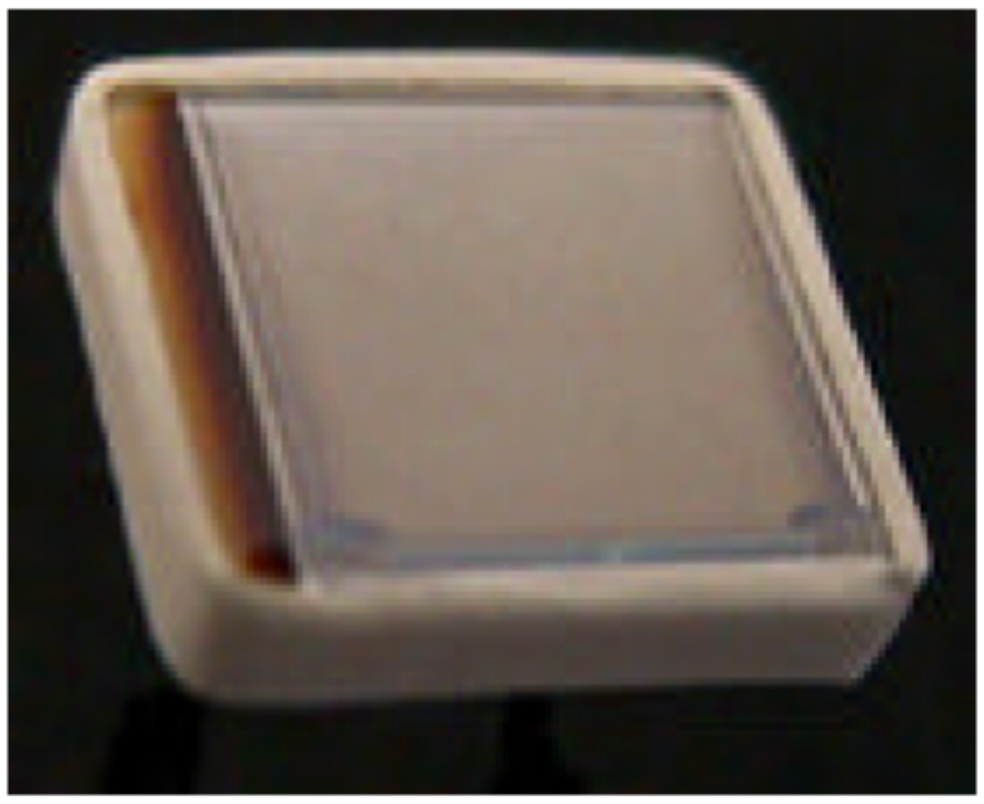

Figure 10.5. An RMD S1315 APD (left) and a Hamamatsu S8664-1010 APD (right).

Larger area APDs, developed by Hamamatsu for the PANDA calorimeter [5] are now available in different sizes (10 × 10 mm$^2$ and 7 × 15 mm$^2$) in both standard and low capacitance versions. Other producers are able to provide similar size devices, such as Radiation Monitoring Devices (RMD) in the US, which offers 8 × 8 mm$^2$ and 13 × 13 mm$^2$ (S0814 and S1315) devices.

Figure 10.5 shows two photosensor candidates for Mu2e: the RMD S1315 and the Hamamatsu S8664-55-1010. The relatively low gain of these APDs requires the use of a front-end amplification stage. Table 10.2 lists the properties of these devices. The large area of these devices provides high light collection efficiency from the 30 × 30 mm$^2$ area of the crystal: 19% for the S1315, 11% for the S8664-1010. The quantum efficiency of the two candidate APDs is shown in Figure 10.6 as a function of wavelength.

| **Properties** | **S8664-55** | **S8664-1010** | **S1315 (RMD)** |
|---|---|---|---|
| Active area (mm^2) | 5×5 | 10×10 | 13×13 |
| QE (~ 405 nm) | 0.65 | 0.65 | 0.65 |
| $I_d$ (nA) | 5 | 10 | Not measured |
| Capacitance $C_d$ (pF) | 80 | 270 | 120 |
| Gain | 50 @ 350 V | 50 @ 350 V | 100 @ 1700 V |
| Excess noise F | 2.0 @ gain =50 | 1.38 @ gain =50 | Not measured |

Table 10.2. Properties of the RMD S1315 and the Hamamatsu S8664 APDs.

These APDs match most of the requirements needed for a LYSO based calorimeter, with a high quantum efficiency that is well-matched to the emission spectra of the LYSO crystal (402 nm). They are also fast and radiation resistant. The gain and the dark current, however, have a strong temperature dependence. Measurements done by CMS and PANDA indicate a gain dependence of ~ 2%/°C. Good temperature and voltage stability are therefore required. Temperature stability of ± 0.2° C and voltage stability of ± 20 mV are adequate to achieve a 0.4 per mil gain stability. This is not difficult to achieve in practice. Two other relevant parameters that drive the APD choice are the excess noise factor (*F*) and the nuclear counter effect (*NCE*).

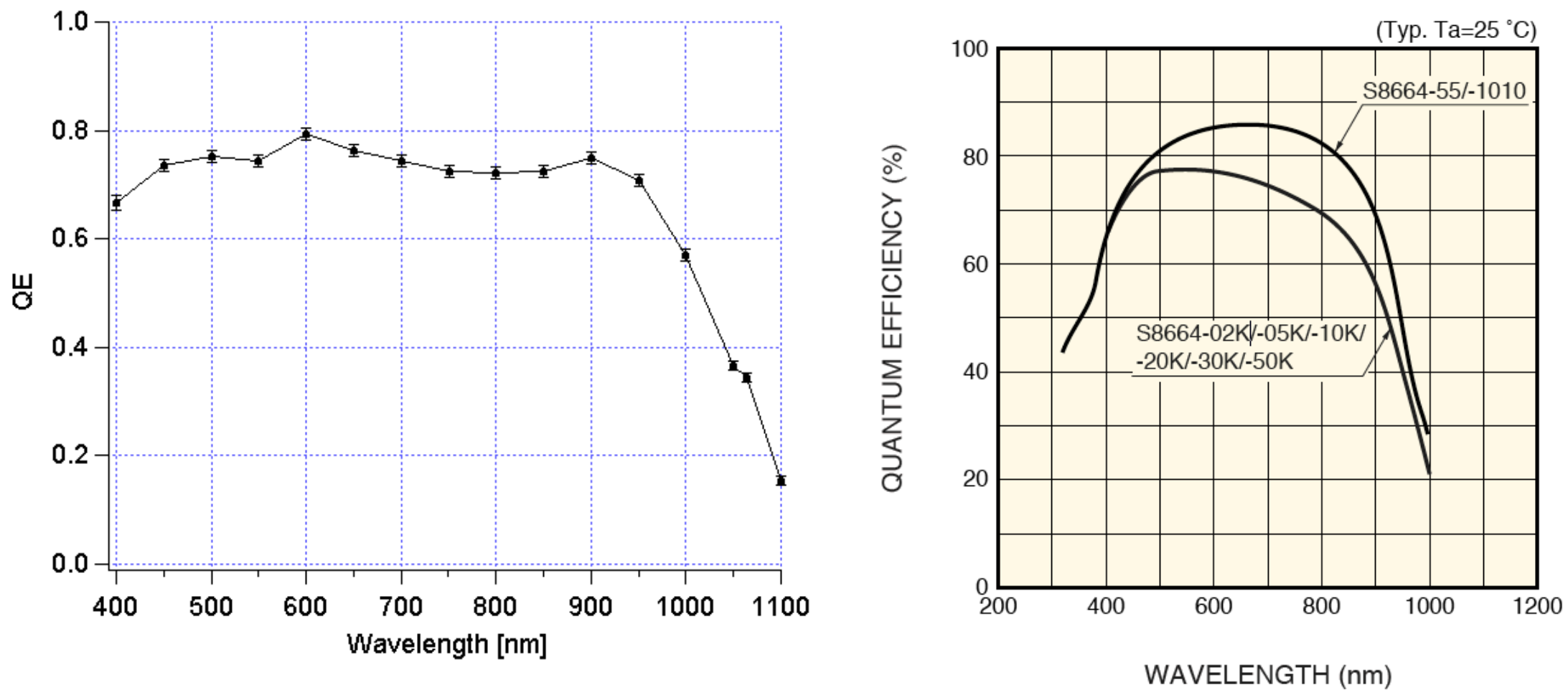

Figure 10.6. Quantum efficiency as a function of wavelength for the Radiation Monitoring Devices S1315 (left) and the Hamamatsu S8664-1010 (right).

For a light yield (*L*) of $N_{pe}$ / MeV, evaluated at the APD conversion layer, the total signal for a shower of energy *E* is *Q*=M*EL*, where *M* is the gain. The standard deviation of the signal is

$$\frac{\sigma}{E} = \frac{1}{\sqrt{EL}} \frac{\sqrt{M^2 + \sigma^2}}{M} = \frac{\sqrt{F}}{\sqrt{EL}} \qquad (10\text{-}1)$$

due to the combination of photoelectron statistics and fluctuations in the amplification process. The *F* factor is relevant for the achievable energy resolution and depends on both the device and on the applied voltage. In Figure 10.7 the response of a 20 × 20 × 150 mm3 LYSO crystal illuminated with a UV LED and read out by a Hamamatsu S8664-1010 APD is shown. Correcting for APD size ~ 2400 p.e./MeV are expected. A fit to the energy resolution following equation 10-1, after adding a constant term due to LED fluctuations is also shown in Figure 10.7. A test station is being designed to determine the Nuclear Counter Effect (NCE) in a more automatic way and compare the functionality of different APDs as a function of bias voltage and temperature.

The *NCE* is the charge produced by a photosensor when a charged particle directly hits its surface. When this happens, either from shower leakage or from external accidental background, the photosensor can generate unwanted signals that deteriorate the energy resolution. In the APD, only the carriers produced before the amplification layer experience full amplification; carriers produced in the avalanche region have an amplification that depends on their location at creation. A quantitative *NCE* measurement is given by $d_{\mathrm{eff}}$ (the effective thickness of the Si amplification layer). This is obtained by exposing the APD to a $^{90}$Sr source and by comparing the charge with that measured by a PIN diode of known thickness $d_{\mathrm{PIN}}$. The *NCE* is reduced by using an APD with a smaller $d_{\mathrm{eff}}$ and, as an overall effect in the resolution, by increasing the crystal light yield. The normal capacitance of the S8664-1010 APD [5] shows a reasonably small *NCE*. However, to minimize the effect of the Nuclear Counter Effect, each crystal will be fitted with two APDs. This will also improve the overall number of photoelectrons/MeV, and the reliability of the calorimeter.

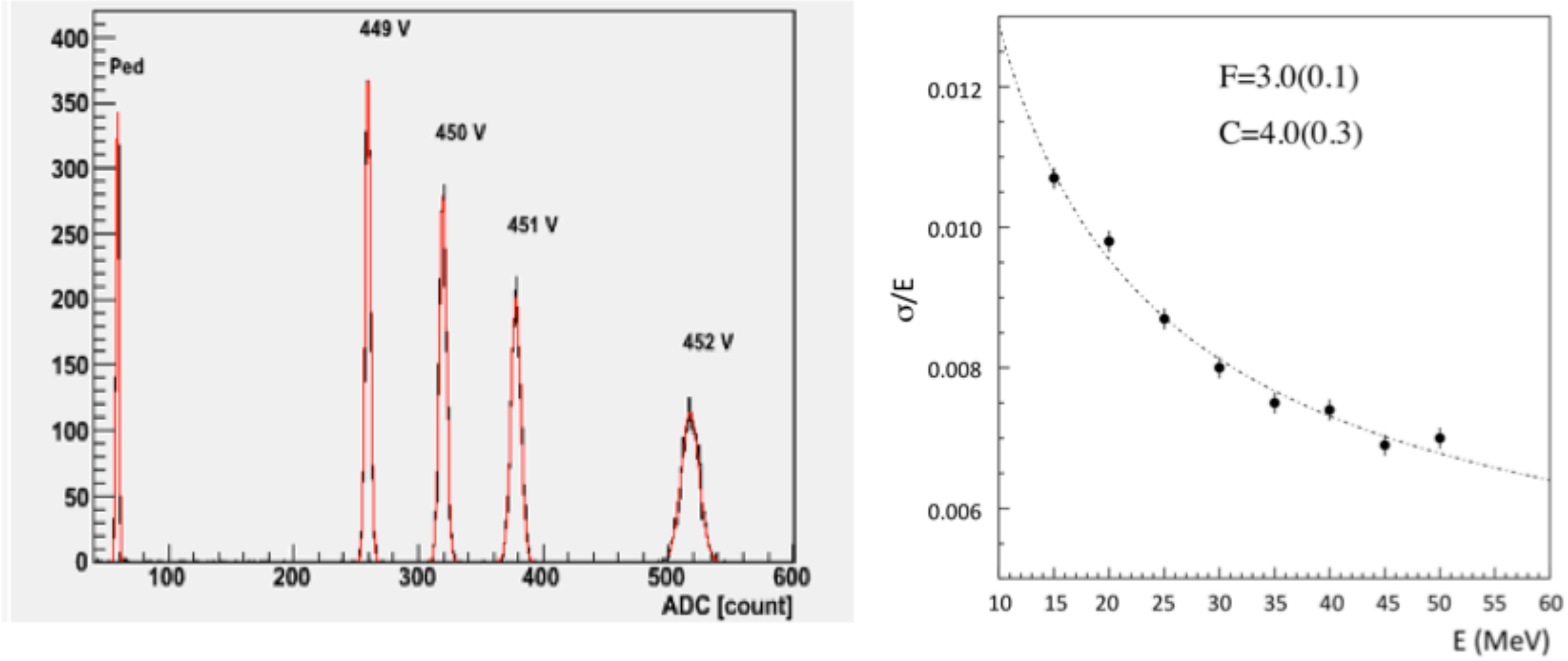

Figure 10.7. Response of a LYSO crystal read out by a Hamamatsu S8664-1010 APD to different amplitudes of a UV light source (left) and the energy resolution of the same crystal and APD (right). The fit parameters for the energy resolution are the excess noise factor F and the LED fluctuation, C, in %.

### *10.3.3* Electronics

The general scheme for the calorimeter readout electronics is shown in Figure 10.8. The front end electronics consists of a bias network for the photo-sensor followed by a preamplifier and shaping stage organized in FEE boards reading out 11 channels (one calorimeter column). Groups of 44 amplified signals are sent to a digitizer module, where they are sampled and processed before being optically transferred to the DAQ system.

For test purposes, a preliminary version of a voltage preamplifier with a gain of 20 has been developed. The resulting noise level is ~120 keV, as measured in the laboratory and at the test beam. Recently, progress has been made on the amplifier design. A novel low-noise low-consumption voltage preamplifier has been developed by the Rome II University (Tor Vergata), which shows characteristics matching our needs. One prototype has been tested to measure the response to a $^{22}$Na source of a LYSO crystal read out by a S8664-1010 APD. The optimal resolution [6] was achieved at an APD standard gain setting ($M$=50-100), with the amplifier at a working point of 3.2 volts, corresponding to a gain of ~ 3000. A noise of 30 keV has been measured, which corresponds to an equivalent noise charge, ENC, of around 4000 $e^-$, matching a similar test done with a commercial charge preamplifier. We expect the next FEE prototype to improve on this performance and reach a ENC of 1000 $e^-$ when working at a typical amplifier gain of 2000. The collected charge from the APD will be $Q = LMe$, *i.e.* (2400 p.e. / MeV) × 50 × (1.6 × $10^{-19}$ C) or 20 *fC* / MeV, corresponding to 2 *pC* at 100 MeV. The total charge after amplification will be of ~4 nC at 100 MeV. Shaping the signal to get a 200 ns maximum width, we will have a peak signal of ~5 mV/MeV.

The FEE electronics (Figure 10.9) will be placed very close (at a distance less than 60 cm) to the APD in order to minimize pickup noise. The maximum power dissipation will be of order 30 mW / channel. Two boards of 11 channels/each will collect the output signals for a calorimeter column, one board serving the left, the other one serving the right APD of the crystals. The FEE will be located on the top of the support structure in the back side of the vane. Cooling for the FEE electronics will be in common with the one for the Digitizer and will be connected to the same cooling system used for the tracker.

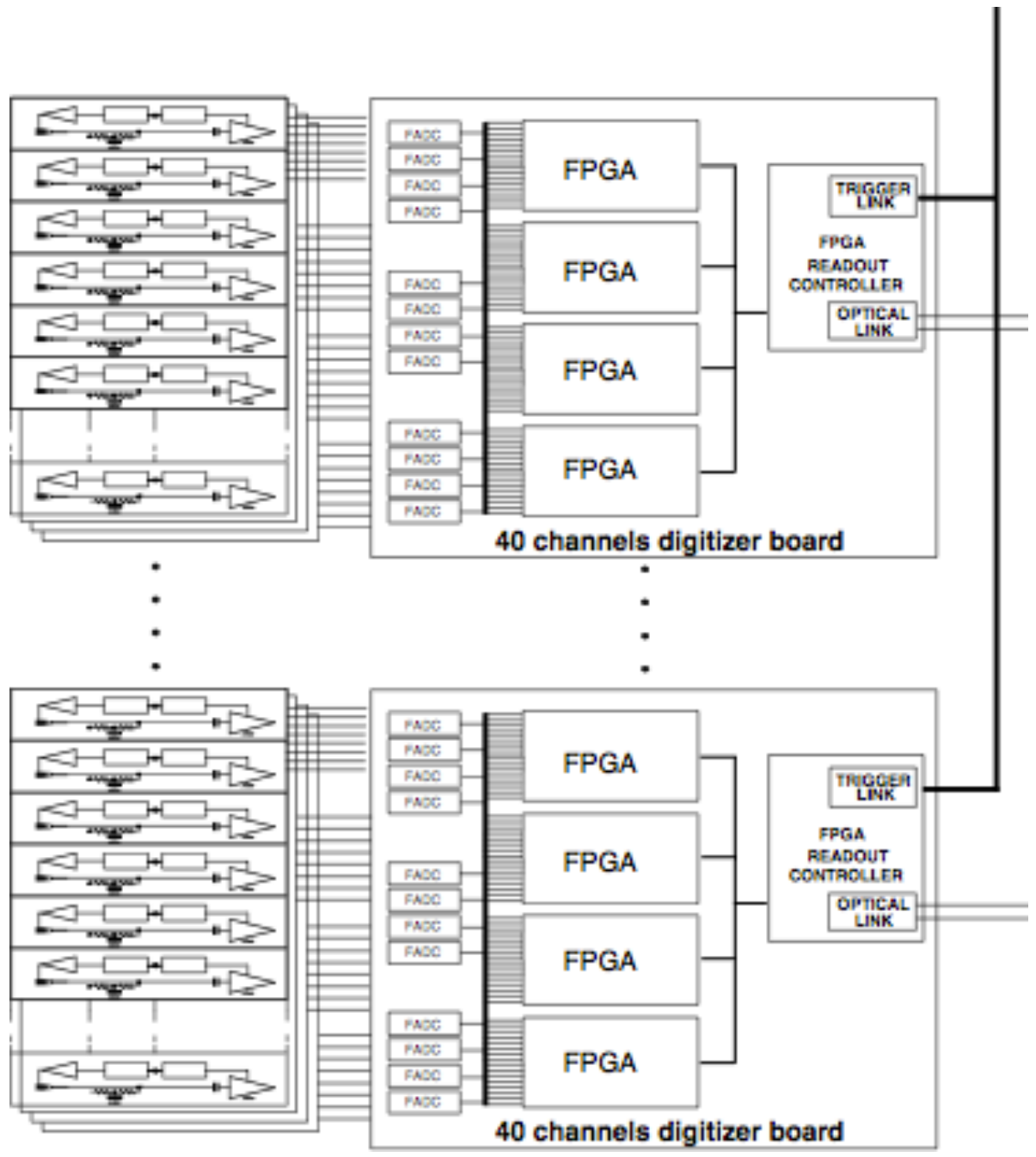

Figure 10.8. Overall schematic of the EMC readout electronics.

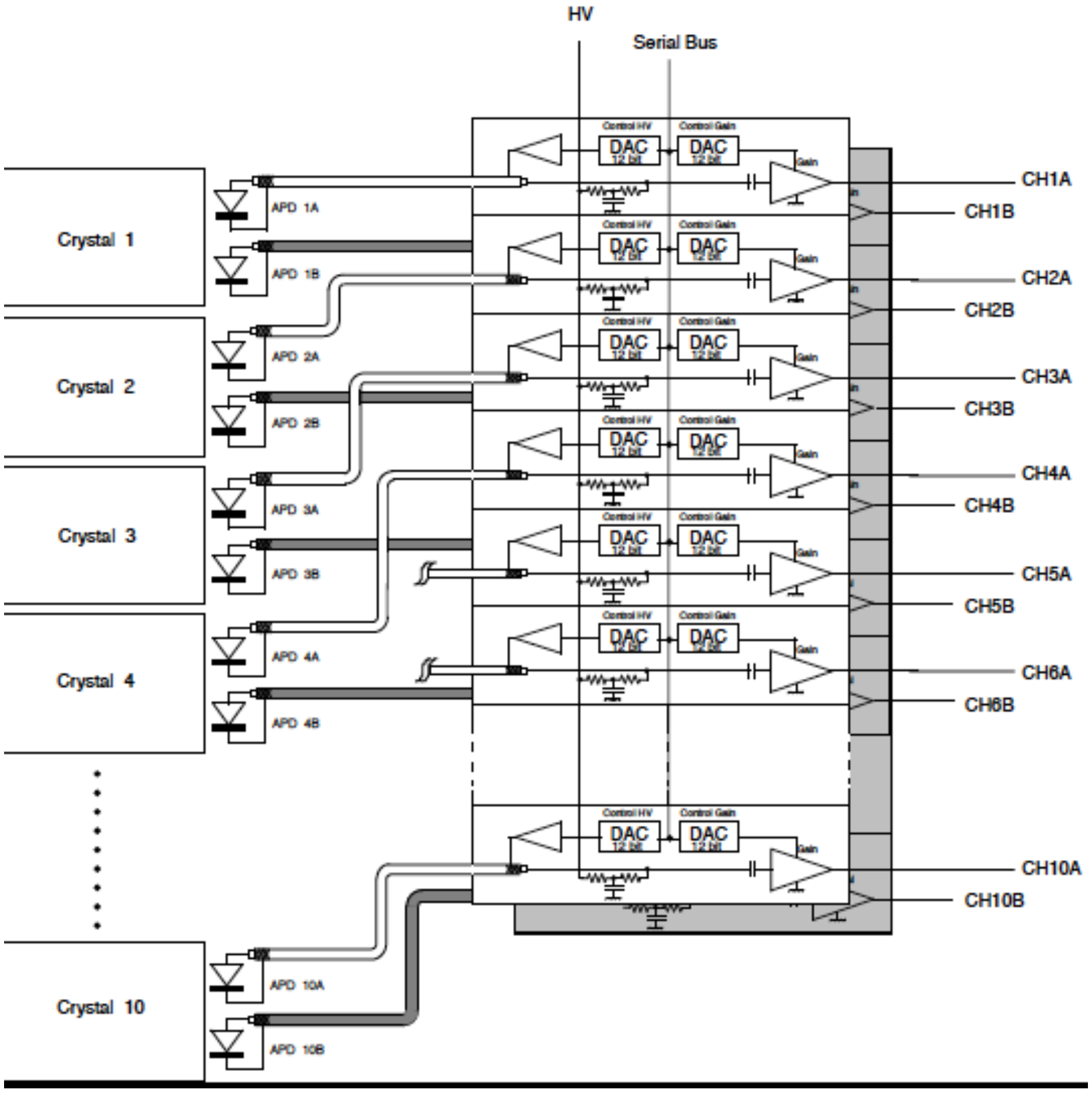

Figure 10.9. General schematic of APD cabling, the FEE stage and HV distribution.

The digitization electronics will be placed on circuit boards, DIGI, positioned as close as possible to the FEE electronics and integrated with the support structure for the vanes. Optical cables will transmit the digitized signals out of the Detector Solenoid to the data acquisition system. Each DIGI board will consist of 44 flash ADCs, FADC, which will sample the shaped signals from the FEE. An FADC with 11-bit resolution, is required to reach a LSB of 25 keV. A zero-suppression algorithm will be performed in the processing stage, removing data for all channels below threshold (<3σ, 100 keV) and preserving only samples related to the full signal width. A 200 msps FADC, for example the T.I. ADS58C48, is planned to execute the digital signal processing, DSP, via algorithms running in 4 low cost FPGA's (Spartan-6 LX25). Each DIGI board will also have an additional, more expensive, layer of FPGA's to drive the optical links and eventually form a trigger (Spartan 6 LXT45). The DSP should also provide mean charge and time by means of running averages. The total expected power dissipation per board is ~ 15 W, with 9 W for the FADC and 6W for the FPGA's. For a 200 ns signal shaping time, we will have about 50 samples per pulse (~80 Bytes), allowing for identification of pileup due to accidental activity in the offline analysis. Assuming an occupancy of about 10%, we expect an average of 17 kHz / channel of random hits. This corresponds to about 17 × $10^3$ × 84 (crystals) × 2 (L/R) × 80 Bytes/sec/vane (1.4 GBytes/sec/vane, 11 Gbits/vane) when streaming all events. Since there are 22 DIGI boards/vane, the average throughput / board will be ~ 0.6 Gbits. Groups of 4 DIGI boards will be organized in one optical ring so that, for each vane, there will be 6 rings for a total count of 24 optical fibers (2 MTP cables).

For the entire calorimeter, 88 DIGI boards are required. About 1300W of power dissipation is expected for the full system. The cooling system will be in common with the FEE.

To maintain the required gain stability, good voltage precision and stability are required of the bias voltage supplies. A typical bias voltage setting for the S8664-1010 APD will be of ~400 V with a $dM/M/dV$ of ~ 3%/V, for an APD gain M of ~ 50. A 50 mV precision/stability is required to reduce the gain spread below 0.3%. In Figure 10.10, dM/M/dV as a function of the gain is shown for several APDs from Radiation Monitoring Devices (RMD). A value of 2% / V is found for gains ~ 300, so a precision of 100 mV is sufficient in this case.

A 12-channel prototype HV supply board has been developed for the S8664-1010, based on a novel DC-to-DC converter chip that is able to supply from 0 to 500 V with a residual ripple below 20 mV and a setting precision of 30-40 mV. The performance of this chip in a magnetic field and a radiation environment remains to be investigated. Pending such studies and for cost optimization, we have chosen the option of regulating the voltage close to the calorimeter. The HV regulation circuit will be on the FEE boards. The master voltage for every 4 FEE boards (44 channels), will be provided by a single DC-to-DC converter positioned in a board outside the Detector Solenoid by means of a HV cable via a feedthrough. This granularity is reasonable for the chosen DC-to-DC converter, since we expect a maximum current of 100 μA /channel x 44 channels = 4.4 mA for each chip. A total of 11 × 2 = 22 HV cables/vane are needed, which corresponds to 88 cables for the entire calorimeter. The master voltage board will have 12 channels, so that 8 HV master boards, organized in a crate, will serve the calorimeter.

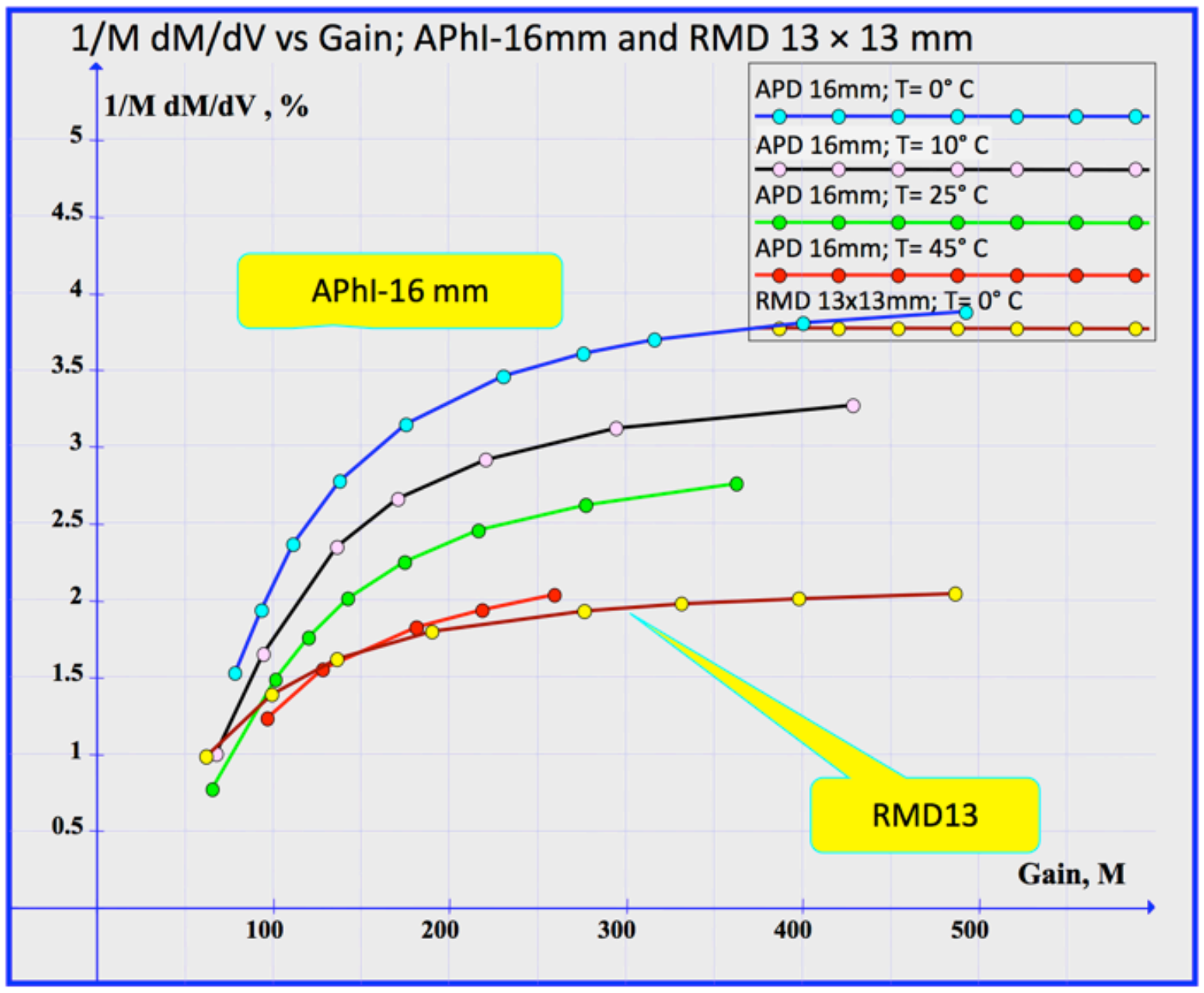

Figure 10.10. Fractional gain change with respect to $V_{bias}$ as a function of gain for RMD devices.

### *10.3.4* Mechanics

Figure 10.11 shows the arrangement of the vanes inside the barrel support structure. The preferred option is to suspend each vane inside the barrel at a 45° angle to reduce the mechanical stress on the stainless steel rings supporting the vanes. This also allows a more uniform weight distribution and uniform illumination by cosmic rays for debugging and calibration purposes. We have designed a self-standing structure, without supports in the inner region, to minimize passive material where spiraling background events will be concentrated. Two outer stainless steel rings hold each vane in position. Each vane is built up starting from a single unit composed by the crystal inserted in a 100 μm thick carbon fiber case, sealed at the extremities by two support caps. The back cap, made of 5 mm thick plastic material, provides the housing for 2 APDs and for the fiber connector carrying laser light for calibration. The back cap is also connected to the cooling system to set the APD working temperature. A tilted orientation of the vanes, which can increase the acceptance, and is compatible with this mechanical design, is under discussion.

The layout of the calorimeter support structure is a cylindrical barrel, 1450 mm long, with an outer radius of 810 mm, holding four parallelepiped vanes of dimension 330 × 1320 × 110 mm$^3$. Each vane consists of 484 crystals organized in 11 rows × 44 columns, for a total of ~360 kg/vane). The two outer support rings are 50 mm wide stainless steel hoops that are 30 mm thick and connected by four 30 × 130 × 1360 mm$^3$ stainless steel beams. The FEE boards will be located in the region of the outer support rings, on top of the vanes. In this outer region, special plates will also provide support for the digitizers and for the electronics cables, in order to place them outside the detector volume and avoid conversion electrons and physics background (Figure 10.12). This support structure will be installed and supported inside the Detector Solenoid by means of the common internal rail system discussed in Section 8.3.9.

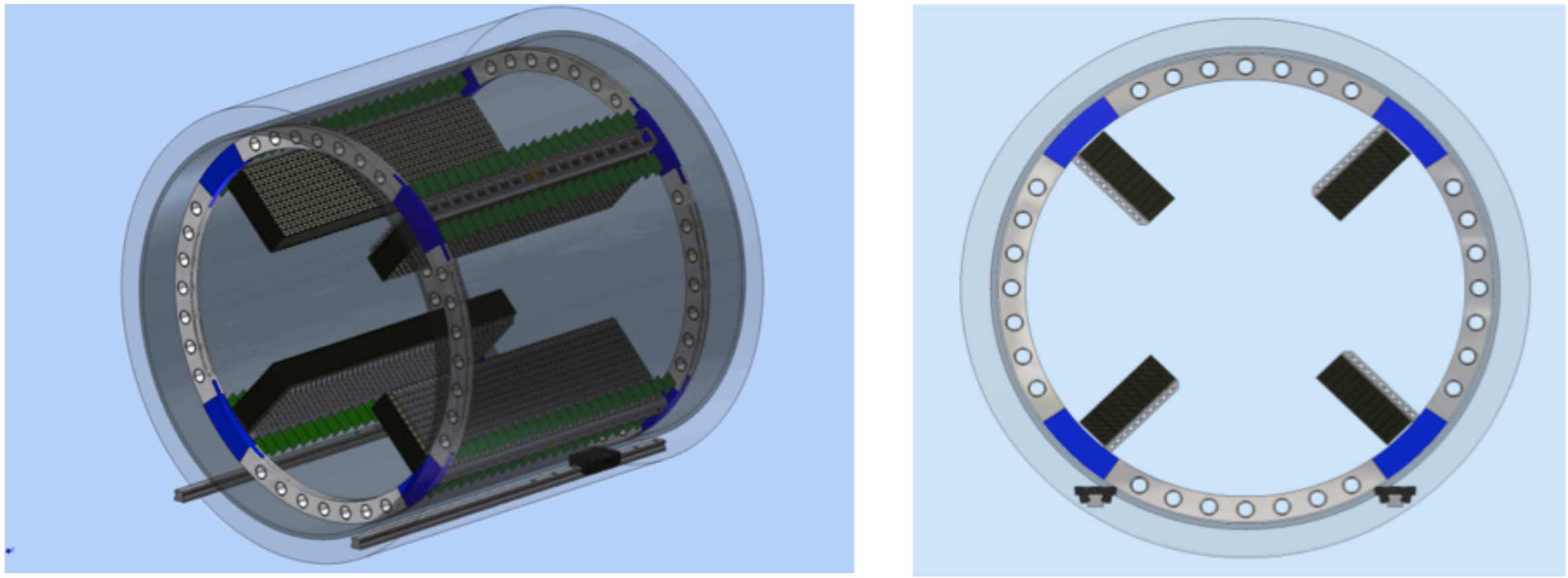

Figure 10.11. 3D view of the calorimeter barrel support structure for the vanes (left), and a cross section of the overall structure.

Each crystal is wrapped with a 0.1 mm thick layer of Tyvek and inserted in a 0.1 mm thick carbon fiber case (Figure 10.13). We are also considering the option of painting the inner region of each case with a diffusing material. Eleven crystal cells are stacked to form one of the 44 vertical columns of crystals composing the vane. The 11 crystal cells are held together by means of stainless steel profiles attached to the support caps of each carbon fiber case, as shown in Figure 10.14. Once the vertical columns are assembled, they are paired to each other by means of horizontal bars pulled in the holes present in the vertical profiles and connected to the support rings through appropriate plates, thus holding the vane in position. This layout ensures the rigidity of the back face of the vane. To provide support to the front face of the crystals, they will be connected to each other with a thin carbon fiber plate fastened at each cell edge. This supporting plate will also contain the circulating tubes of the source calibration system.

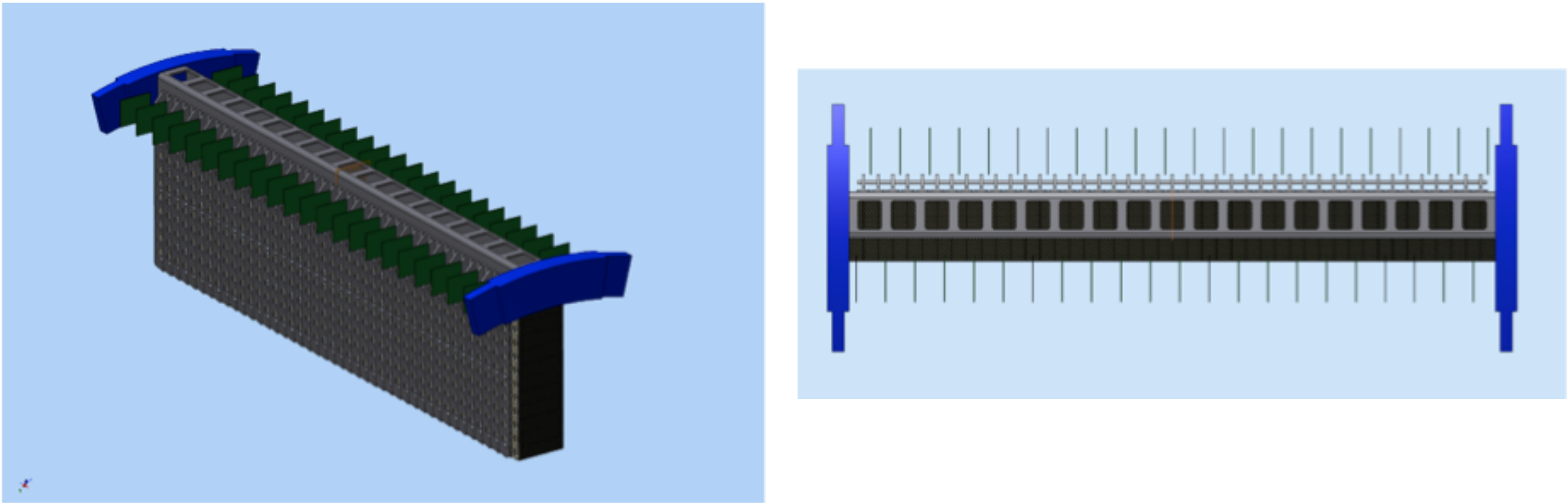

Figure 10.12. Single vane layout (left). Also FEE electronic boards (one per each column, 11 channel each) are shown. Top view (right) with a detail of the supporting rail.

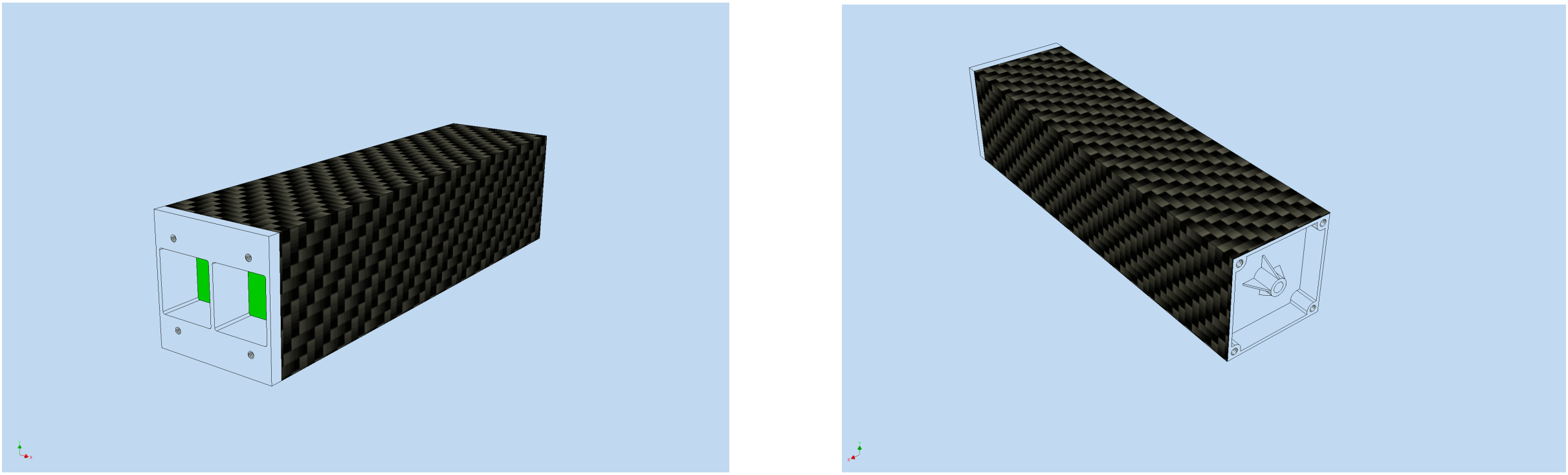

Figure 10.13. Exploded view of a crystal unit with a carbon fiber case and APD holder (left) and with a front face cover (right).

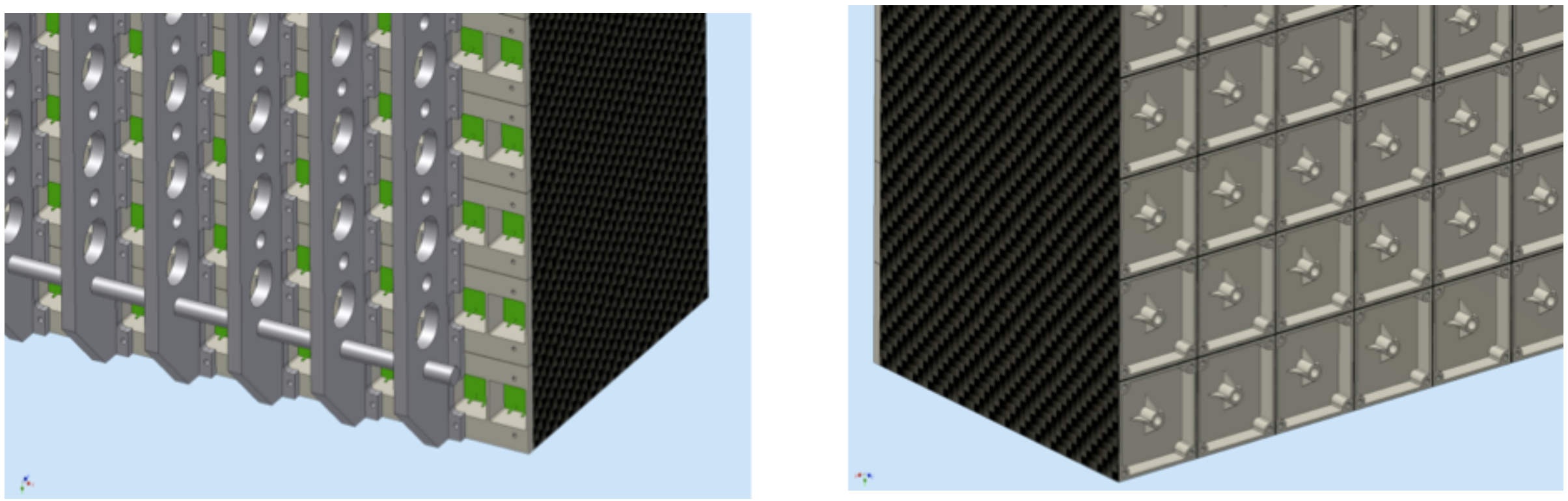

Figure 10.14. Details of the back (left) and front (right) face of the vane.

## 10.4 Expected Performance

The energy resolution of a calorimeter is usually parameterized by three terms added in quadrature as follows:

$$\frac{\sigma}{E} = \frac{a}{\sqrt{E/GeV}} \oplus \frac{b}{E/GeV} \oplus c.$$

The first term is the *stochastic term* related to fluctuations in the signal, in our case the Poisson distribution of the light yield L. Assuming a light yield of 2000 p.e./MeV, the *a* term is negligible, even when the excess noise factor *F* is included. In this case the Poisson fluctuations are negligible and in practice an empirically determined $a/E^{1/4}$ dependence replaces the Poisson term, accounting for intrinsic signal fluctuations that result from light attenuation and non-uniformity in the crystals.

The second term, *b*, is due to the electronic noise. For a given readout technology (photosensors and front end electronics) the noise remains essentially constant, so its relative contribution to the resolution drops linearly for increasing energy. The noise is not a limiting factor to the energy resolution of a LYSO calorimeter for energies around 100 MeV.

The third term, *c*, is due to leakage or calibration errors and describes the calorimeter's intrinsic resolution limit, since it is the only relevant term at very high energies or for calorimeters with excellent light output and good uniformity.

To understand the relevance of these contributions, a GEANT4 simulation of a LYSO cylinder, 15 cm long and 10 cm in radius (~5 $R_M$) (see Figure 10.15 left), has been performed. Electrons with energies between 100 to 500 MeV, incident on the center of the LYSO cylinder were simulated. The long tail at low energy in the spectrum, due to the longitudinal leakage, was taken into account by fitting with a standard log-normal function. For a reasonable cluster size of 4 $R_M$, equivalent to ~5 × 5 crystals in the baseline design, an intrinsic resolution limit of 1.2% results. In Figure 10.16 the dependence of the resolution as a function of energy and transverse dimension is summarized. To understand the impact of a non-uniformity in the longitudinal response of the crystal, the cylinder was sliced along the crystal axis, *z*. The energy was recorded for each slice and then smeared with an exponential attenuation along *z*. For a non-uniformity of 10% a contribution to the resolution of 1.8 % to 2.4% was observed for 100 MeV electrons with impact angles between 0 to 60 degrees.

A GEANT4 simulation also has been performed to study a large 11 × 11 matrix of crystals [7]. The crystals had transverse dimensions of 20 × 20 $mm^2$ and a length between 11 and 15 cm (Figure 10.15 - right). The matrix was evaluated in a magnetic field of 1 Tesla. A light yield of 200 p.e./MeV, distributed using Poisson statistics and an excess noise factor of 1.3 was simulated. A gain spread of 1% was applied and a noise term with a σ of 150 keV was added to each channel. 105 MeV electrons intersected the matrix center at an average angle of 55° with respect to the normal with a Gaussian spread of 1.5°. The energy deposited in the crystal array was summed to form a cluster of energy ($\Sigma E_{Clust}$). Only cells with energy above a 200 keV threshold were used.

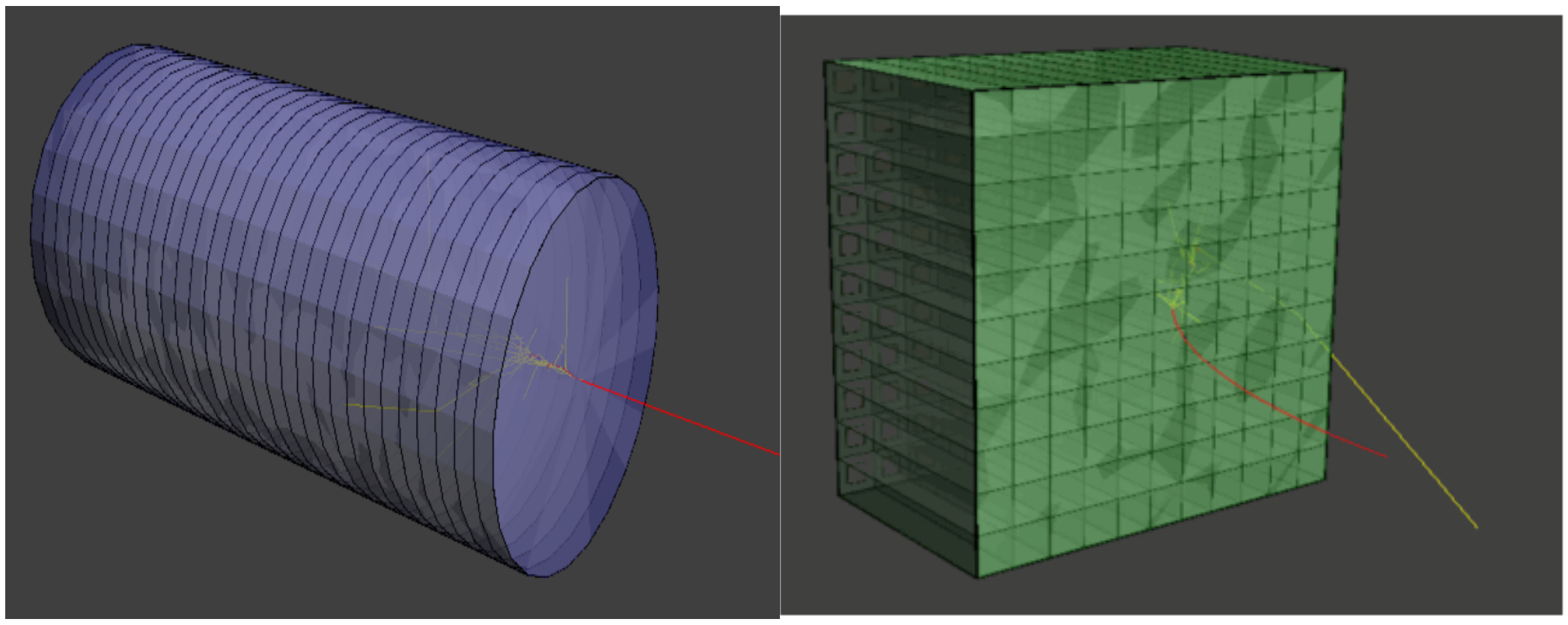

Figure 10.15. GEANT4 simulations of a large cylinder of LYSO (left), and a large LYSO/PWO-2 matrix (right) were performed. In each case, a 105 MeV electron impinging at 55 degrees is shown in red. Yellow tracks represent photons escaping from the shower.

In Figure 10.17 (top-left), the energy sum distribution for the simulated events is shown with a log-normal fit superimposed. An energy resolution of 1.7 MeV is obtained. The long tail at low energy is due to the escaping energy, *i.e.* lateral and longitudinal leakage plus "albedo" from back-scattered particles near the surface of the crystal, as shown in the correlation plot of Figure 10.17 (bottom-left). An attempt to correct for the mean energy loss is also shown (bottom-right). This results in a Gaussian response with a resolution of 0.6 MeV, thus demonstrating the promise of such a calorimeter.

### *10.4.1* Test Beam Results

To validate the reliability of the calorimeter simulation, a LYSO array was exposed to a test beam at the MAINZ Microtron (MAMI) in March 2011. For this beam test 9 new LYSO crystals were added to a crystal matrix built at Frascati National Laboratory (LNF) for the KLOE-2 experiment [8]. The new LYSO crystals, procured from the Shanghai Institute of Ceramics, Chinese Academy of Sciences (SICCAS), were 20 × 20 × 150 $mm^3$ and were surrounded by a leakage recovery matrix of $PWO_4$ crystals. Each LYSO crystal was read out by a single Hamamatsu S8664-1010 APD followed by a discrete voltage amplifier, while the $PWO_4$ crystals were read out by conventional Hamamatsu PMTs. The bias to the APDs was provided by a Mu2e prototype high voltage board while the PMT high voltage was supplied by a CAEN HV board. The total coverage for the matrix was ~ 2.5 $R_M$. Each channel was calibrated to approximately 2% by means of cosmic rays. The APDs were operated at ~ 50 V below the breakdown voltage at an average gain of ~300. The temperature stability of the APDs was maintained by adding two Peltier junction cells to a copper mask positioned on the calorimeter face and monitored by two thermo sensors.

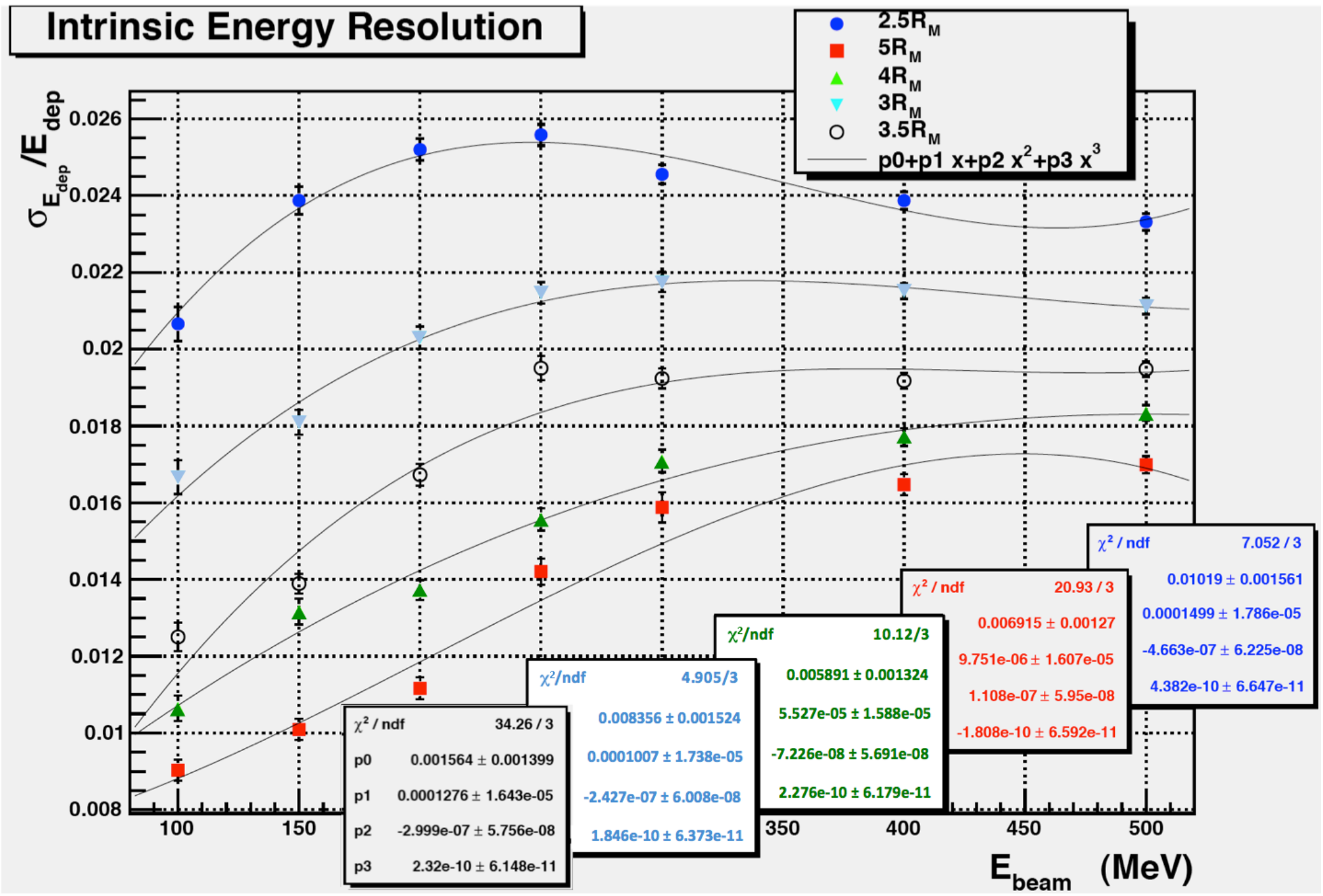

Figure 10.16. GEANT4 study of the intrinsic resolution limit for a 15 cm long LYSO crystal as a function of the crystal radius.

The data were taken with the APD temperature held at 24.5° C with an uncertainty of 0.5° C. The crystals were exposed to a tagged photon beam with energies ranging from 20 up to 400 MeV. A trigger was formed by a coincidence between the discriminated sum of the matrix and the reference tagging signal. The beam spot was ~8 mm in diameter. Data were taken at twelve different energies over a period of 2 days. Approximately 10,000 events were collected at each energy.

Figure 10.18 shows the dependence of the energy resolution as a function of beam energy for test beam data (black points) and for the simulation (magenta curve). A fit to the data results in a stochastic term of 2.4% with an $E^{1/4}$ dependence, a negligible electronic noise term and a constant term of 3.2% due to shower leakage. To obtain reasonable agreement with the data, the energy response of each crystal was smeared by 4% in the simulation using a Gaussian distribution. This was the fastest way to simulate the expected non-uniformity inside the crystals. Good agreement between data and Monte

Carlo is found, as shown in Figure 10.18 and Figure 10.19, for the raw energy distribution of the 9 inner LYSO crystals and their sum.

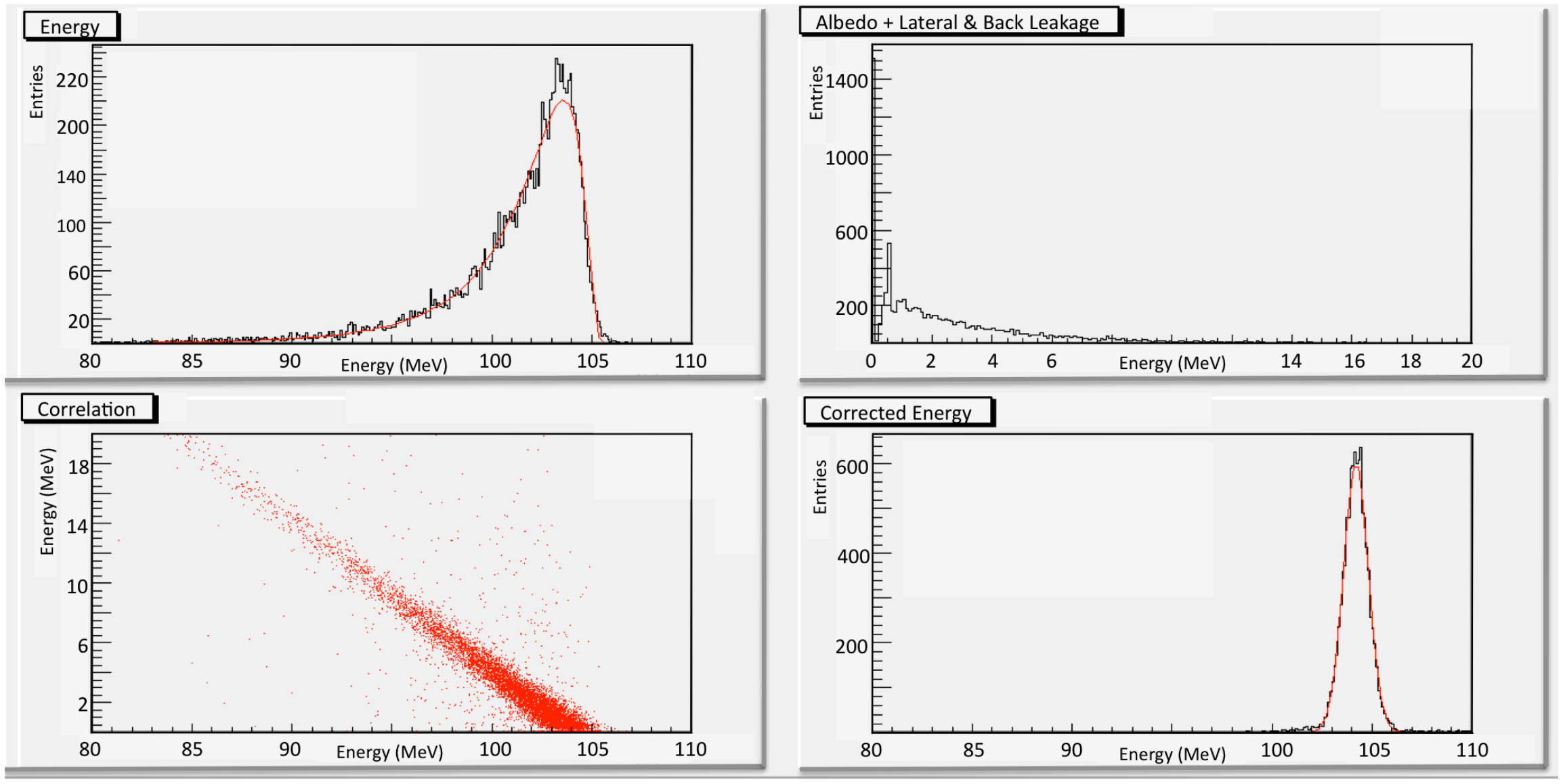

Figure 10.17. GEANT4 simulation of a LYSO Matrix for 105 MeV electrons: (top-left) energy sum distribution (MeV), (top-right) escaping energy (MeV), (bottom-left) correlation between deposited (horizontal) and escaping (vertical) energy, (bottom-right) energy sum corrected for mean energy loss.

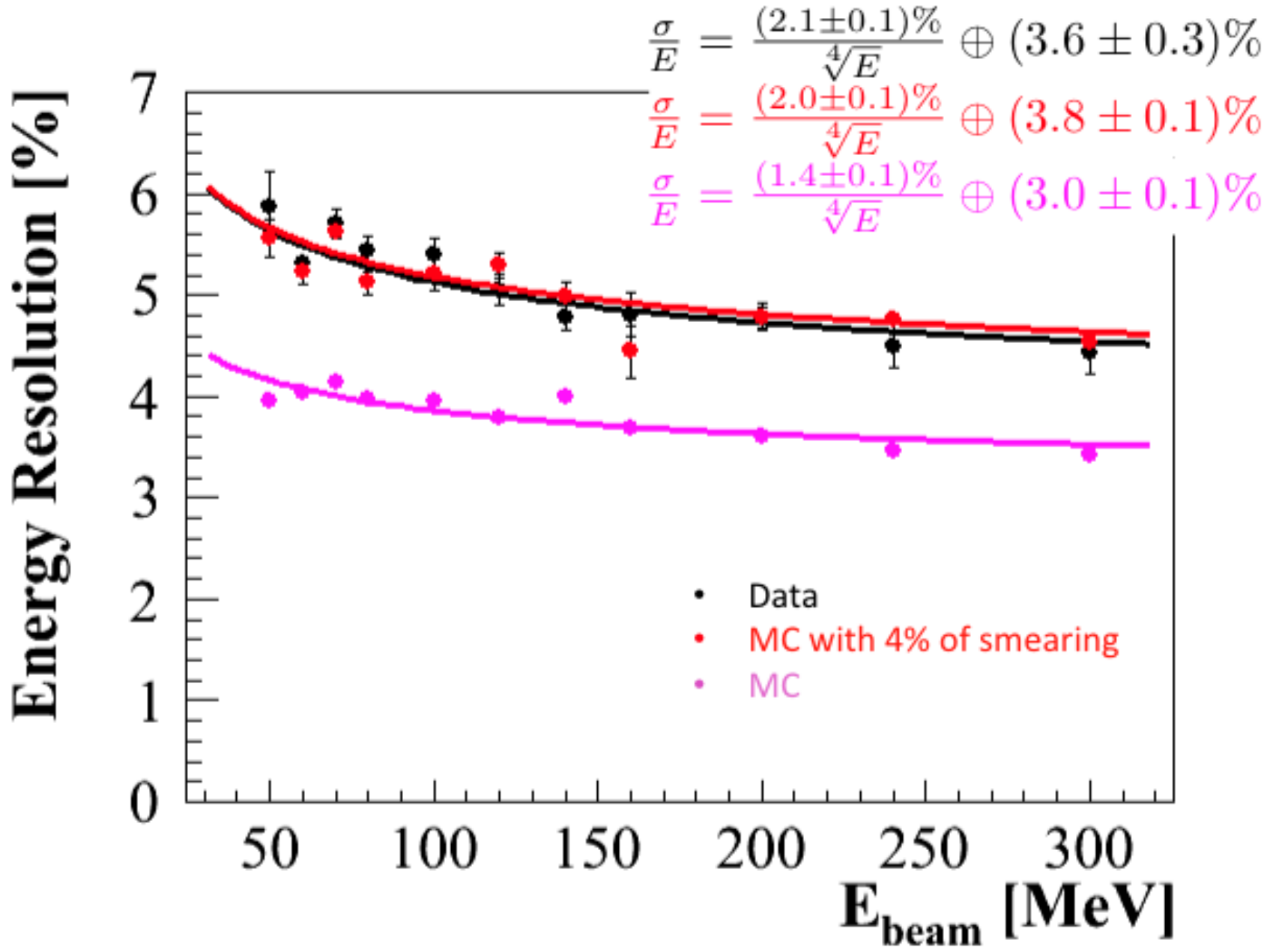

Figure 10.18. Test Beam results from MAMI. The measured energy resolution of the overall LYSO crystal matrix (black points) is compared to simulations (magenta curve). To obtain reasonable agreement with the data, the energy response of each crystal was smeared by 4% in the simulation (red curve).

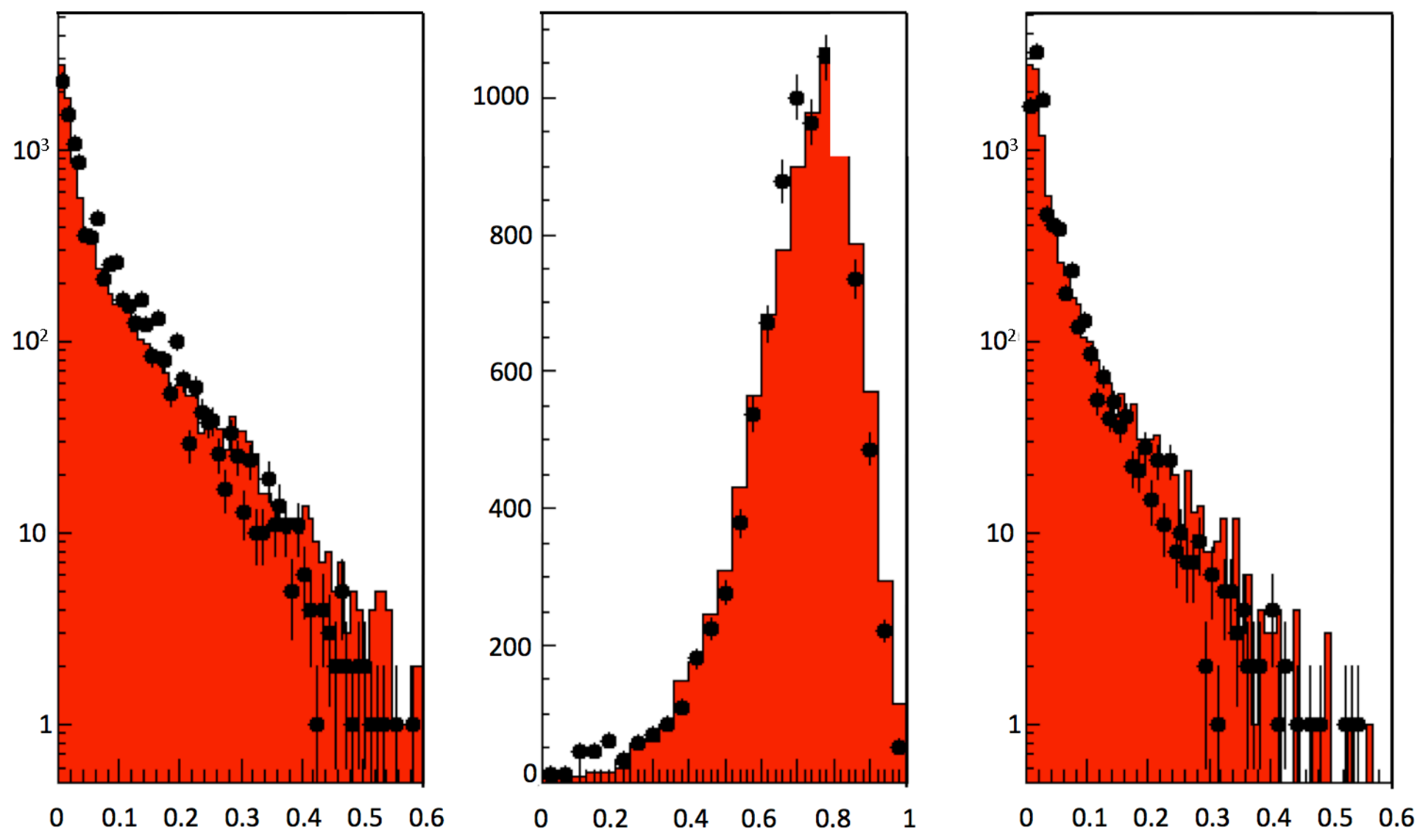

Figure 10.19. Data taken with a LYSO crystal array using a tagged photon beam at MAMI. The three plots correspond to the 3 crystals in the center row of the LYSO array. The ratio of the crystal response to the beam energy is plotted in each case. The photons are incident on the center crystal (center plot). The raw energy distributions of the LYSO matrix (black data points) are compared to a GEANT4 simulation of the array (red histogram).

The position resolution of the LYSO array was also studied at MAMI and good agreement between data and Monte Carlo was obtained. Position reconstruction was done using a simple energy weighted centroid method. Figure 10.20 shows the average reconstructed position for 100 MeV electrons hitting the central crystal in the array at normal incidence. The reconstructed position as a function of the impact point exhibits a typical S-response curve response. Although the overall resolution will be worse for larger transverse crystal dimensions, resolution below 1 cm can be achieved for the crystal sizes anticipated for Mu2e.

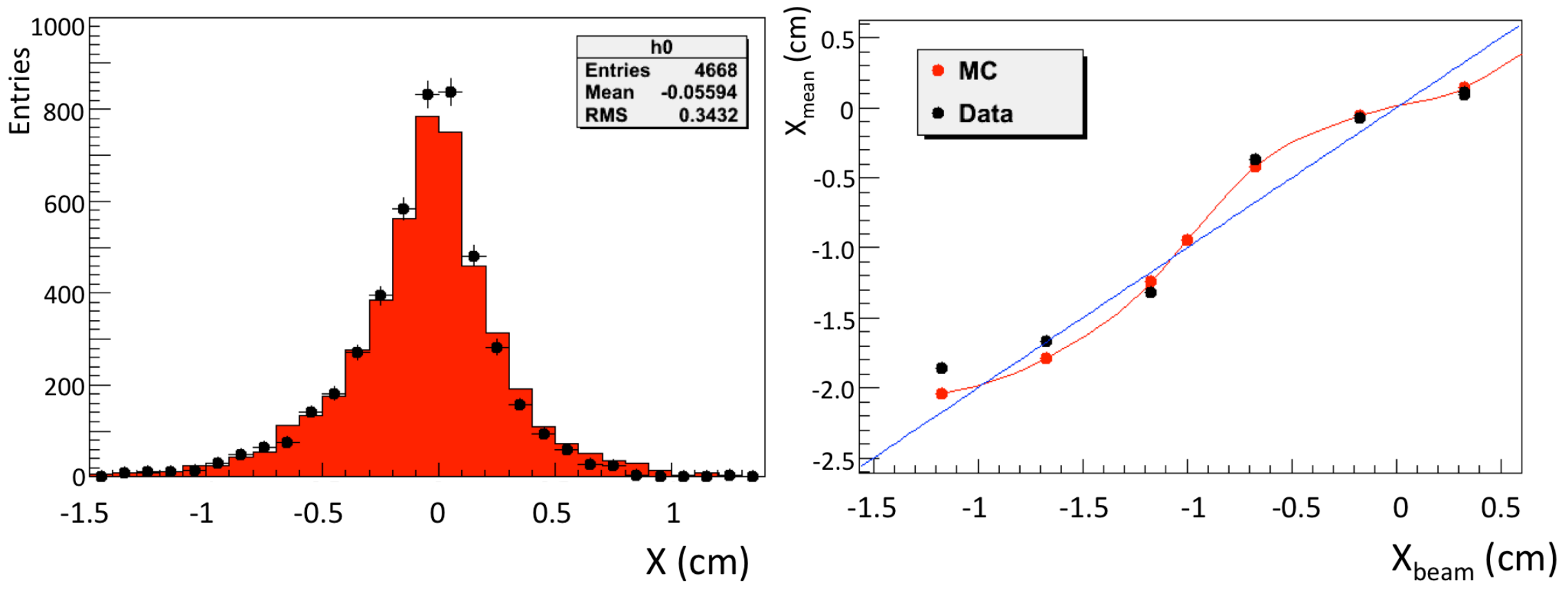

Figure 10.20. Position reconstruction for 100 MeV electrons at normal incidence with the beam hitting the central crystal. Data were taken at MAMI beam test.

Studies of the time resolution of the LYSO array at MAMI were limited by the large intrinsic timing jitter of the reference tagging signal (~800 ps). In a previous beam test at the Frascati Beam Test Facility [8], time resolution below 200 ps was obtained with an LYSO array exposed to 100 MeV electrons.

## 10.5 Calorimeter Optimization Studies

The baseline calorimeter geometry design (henceforth *vane* geometry) for Mu2e is similar to that originally proposed for MECO. The vane geometry presents minimal area for the interception of backgrounds to particles entering the Detector Solenoid and originating in the muon-stopping target or the beam dump. This section will discuss the optimization of the vane geometry. An alternative disk geometry that can increase the efficiency for signal tracks by as much as 10% will also be discussed.

In optimizing the vane design, the variables considered are the number of vanes, the placement and orientation of the vanes, their length and width and the crystal depth. The first question to be addressed is the optimal number of vanes. Versions with three, four, six and eight vanes have been considered. Figure 10.21 shows the cumulative efficiency for detection in the calorimeter of good signal electron tracks found in the T-tracker, as a function of the length of the vane, for differing numbers of vanes. A good hit is defined as a high quality T-tracker track that projects onto the front face of the vane. Hits on the lateral or bottom edges of the vane are not scored as good hits. We find that four vanes is the optimum number and that a vane length of ~130 cm is required to reach the highest achievable efficiency of 73% after fiducial cuts, at a nominal vane height of 36 cm. These cuts ensure that showers originating on the face of the vane are contained within the crystal volume, by removing tracks within 3 cm (~1.5 Molière radii) of the edge of the vanes.

Since the conversion electron tracks enter the vanes at an average angle of about 50°, the next question is whether tilting the plane of the vanes can improve the efficiency for good hits by increasing the fraction of front face hits relative to lateral or bottom edge hits. Figure 10.22 is a comparison of the nominal orientation in the four-vane case with a tilted vane configuration. The vanes are tilted about an axis parallel to the beam-line centered at 54 cm from the beam. Figure 10.23 demonstrates that tilting the vane by ~0.4 radians improves the reconstruction efficiency for good tracks after fiducial cuts to 78%, an improvement of 6.8%. The improvement is due to the reduction of the number of hits on the inner edge of the vane, with a corresponding increase of hits on the face. The downside of this configuration is that the effective thickness of the calorimeter in radiation lengths is reduced, since the tracks enter the vane face at closer to normal incidence. This effect will be addressed below. Reducing the length of the vanes from the nominal 132 cm to 126 cm, *i.e.,* removing two rows, reduces the untilted/tilted reconstruction efficiency to 71% and 77%.

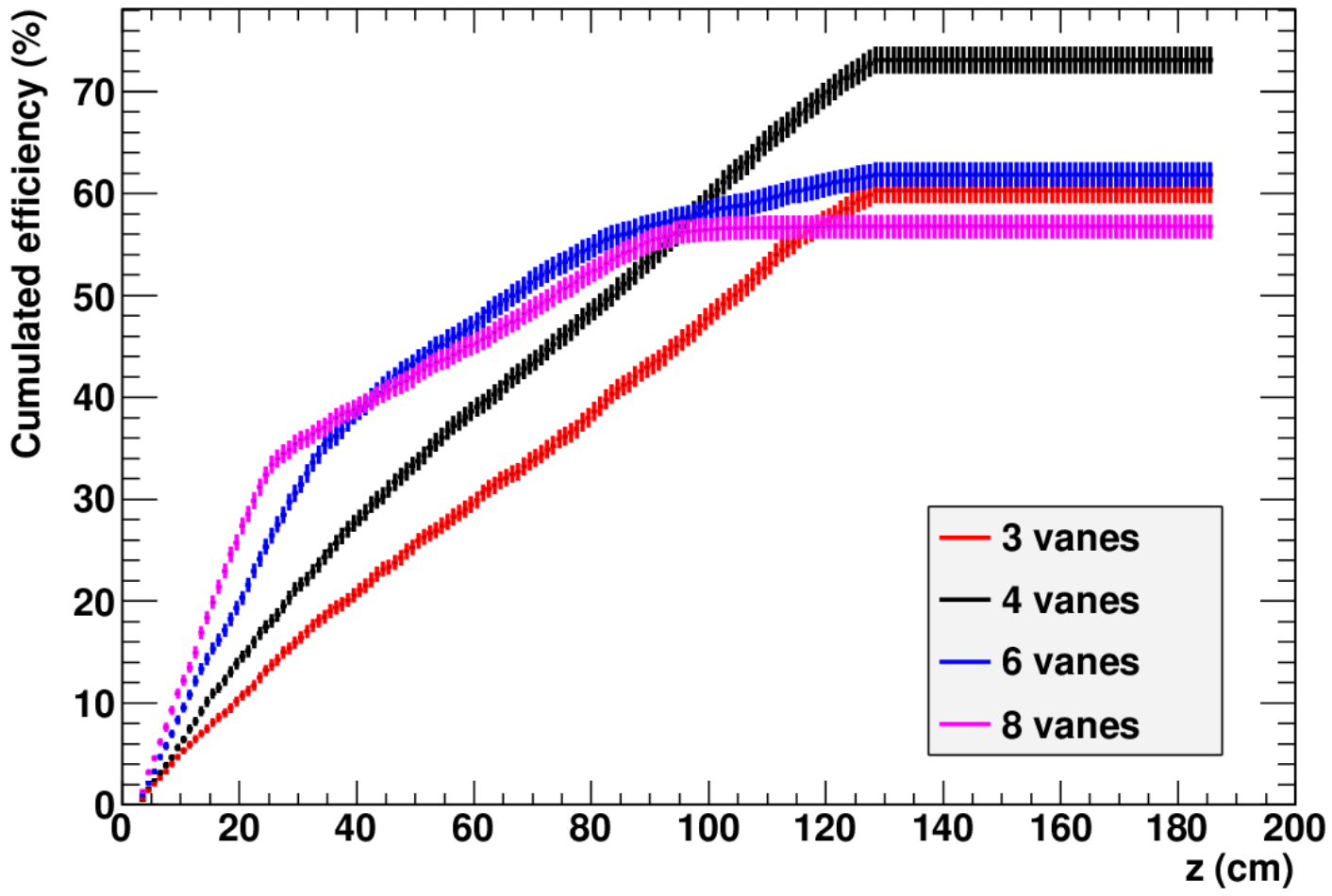

Figure 10.21. The cumulative efficiency for detection in the calorimeter of good signal electron tracks first found in the T-tracker, as a function of the length of the vane, for differing numbers of vanes.

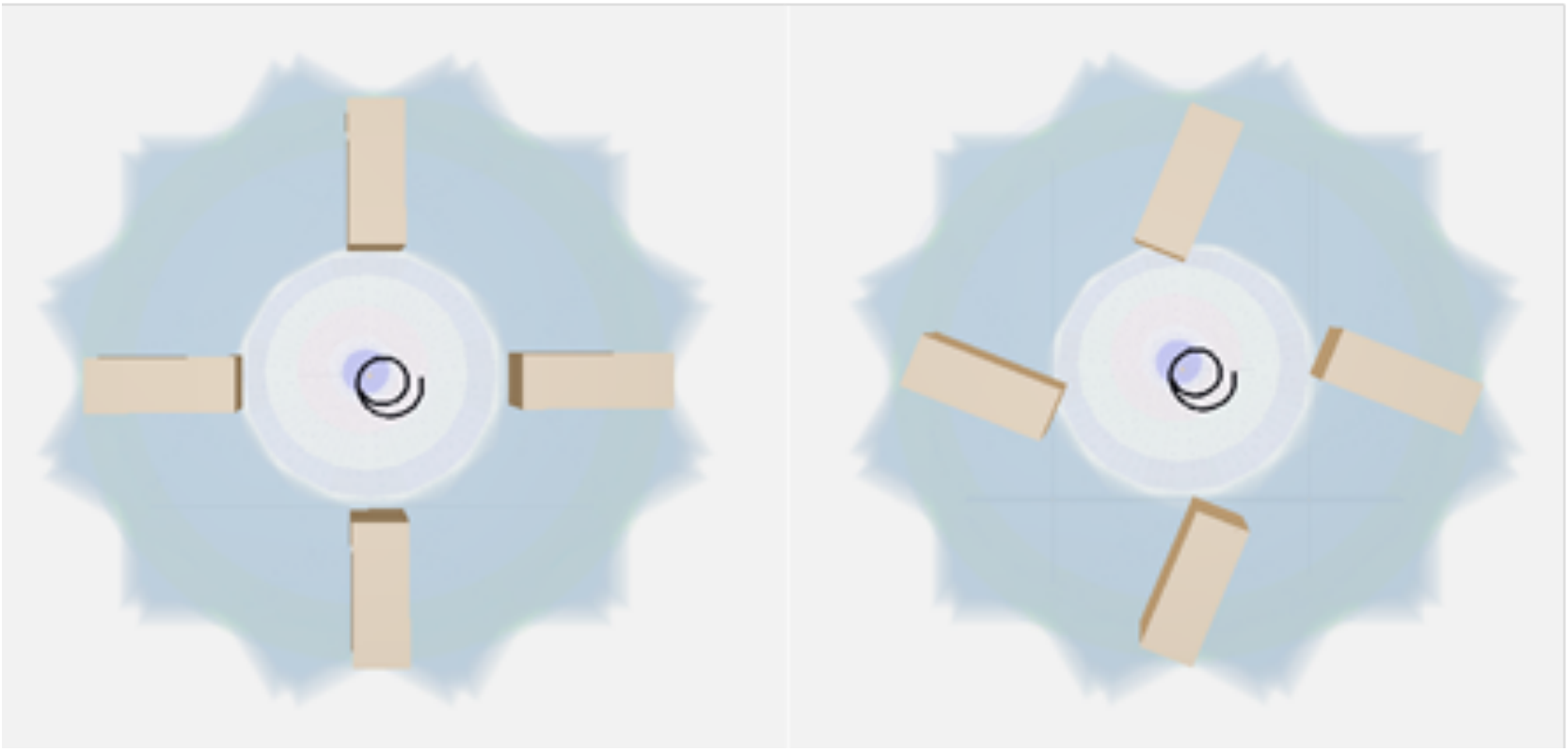

Figure 10.22. Nominal four vane configuration (left) and tilted vane configuration (right).

Another degree of freedom is the position of the axis about which the vane is rotated. Figure 10.24 shows the variation of reconstruction efficiency, after fiducial cuts, as a function of the radial position of the tilt axis for the nominal vane height of 36 cm (78% maximum at 54 cm) and a 30 cm vane with two fewer rows of crystals (maximum 75% at 50 cm).

Note that it is also possible to tilt the crystals within the plane of each vane in order to have the nominal direction of incidence closer to normal in both directions. This variant has not been studied.

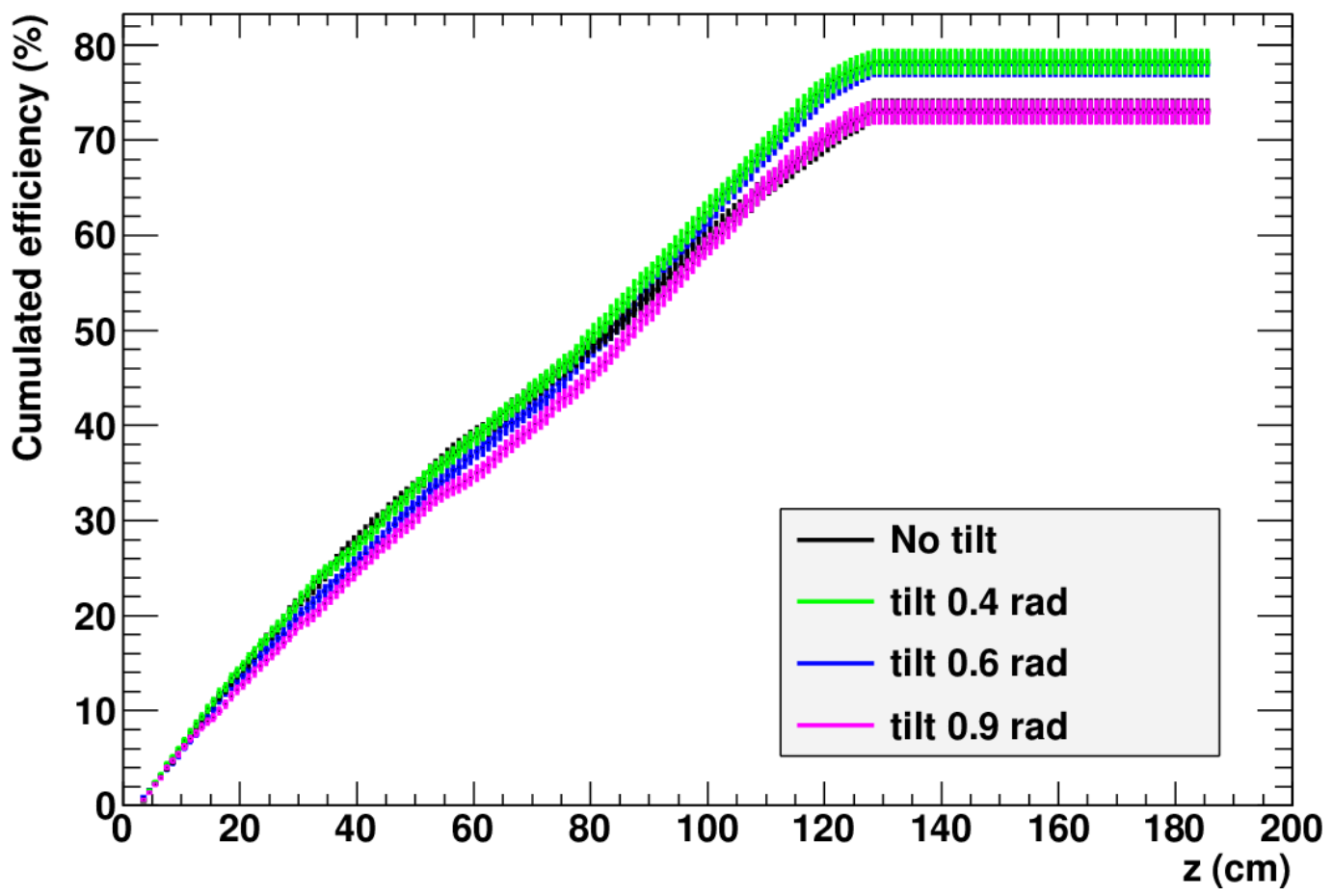

Figure 10.23. Cumulative efficiency, after fiducial cuts, for four orientations of the vanes.

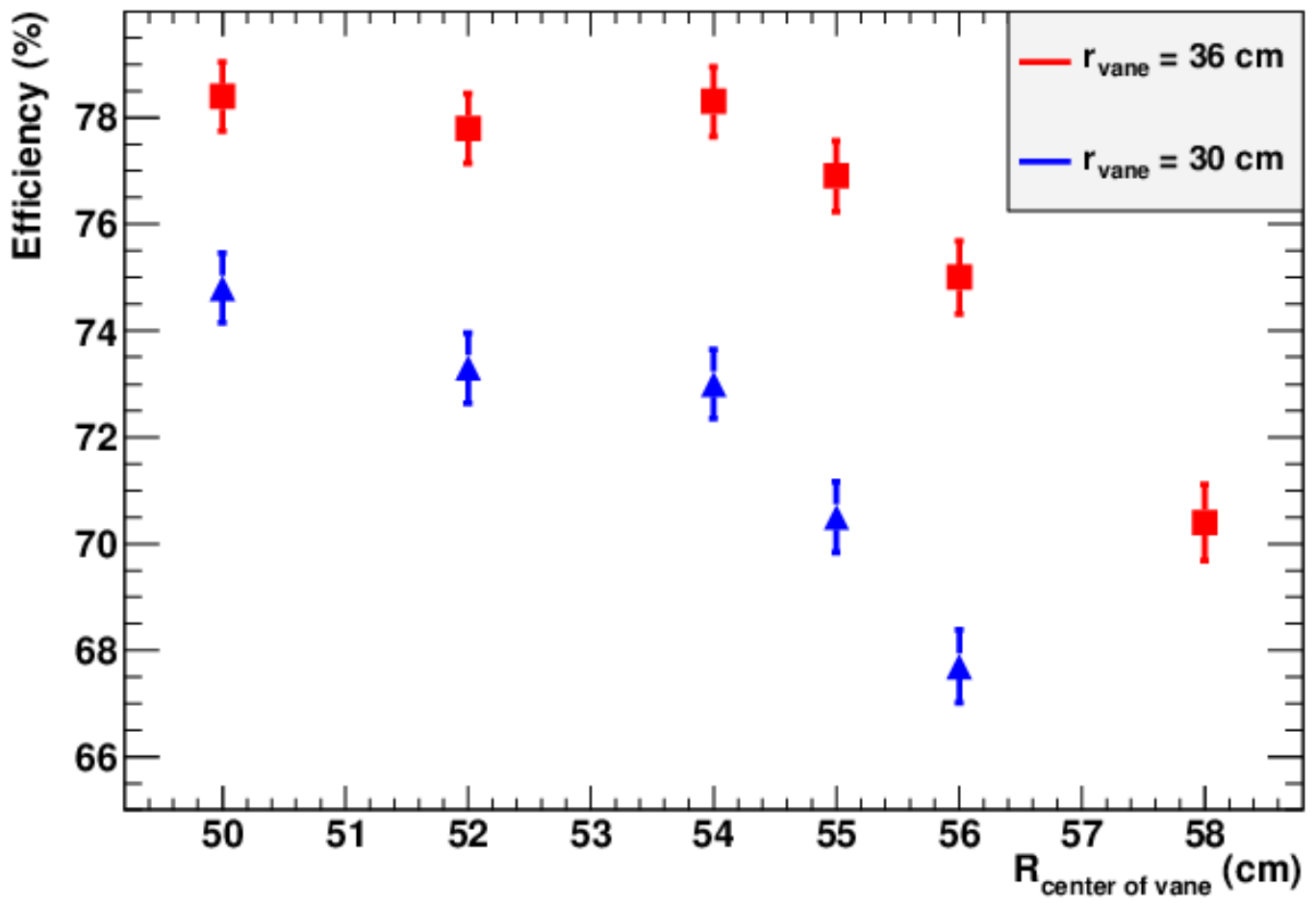

Figure 10.24. Reconstruction efficiency, after fiducial cuts, as a function of the radial position of the vane tilt axis.

By using the Mu2e software framework, a first version of a clustering algorithm has been developed. Each simulated hit contains information about the vane number, space position, time and energy deposits. A crystal with a deposited energy above zero is defined as a hit cell. We have not yet simulated the noise and photoelectron statistics, which previous tests show to be negligible. Moreover, we have not yet inserted the time evolution of the scintillation light in the simulation. In the case of more than one contributing particle per cell, we sum all hits that arrive within 100 ns of one another. We define as a cluster a group of topologically connected hits, by joining contiguous crystals around the cell with maximum energy deposition. Once the cluster is formed, we estimate the energy sum ($E$) and the arrival time of the incident particle (T). To find other clusters, we exclude cells that have already been used, and repeat the procedure around any additional energy maxima until all hit cells have been considered. On average, we find around 8 cells/cluster.

Using the first version of the clustering algorithm, we have obtained the best compromise between signal acceptance and cost in the baseline geometry of untilted vanes. In the following, we restrict our sample to conversion events that reconstruct in the tracker and deposit energy in the fiducial volume of the calorimeter. A hit in the fiducial volume is defined as a cluster in which the highest energy cell is located at least one row or one column away from an edge. In Figure 10.25 the reconstructed calorimeter energy divided by the reconstructed momentum from the tracker (*E/P*) is plotted.

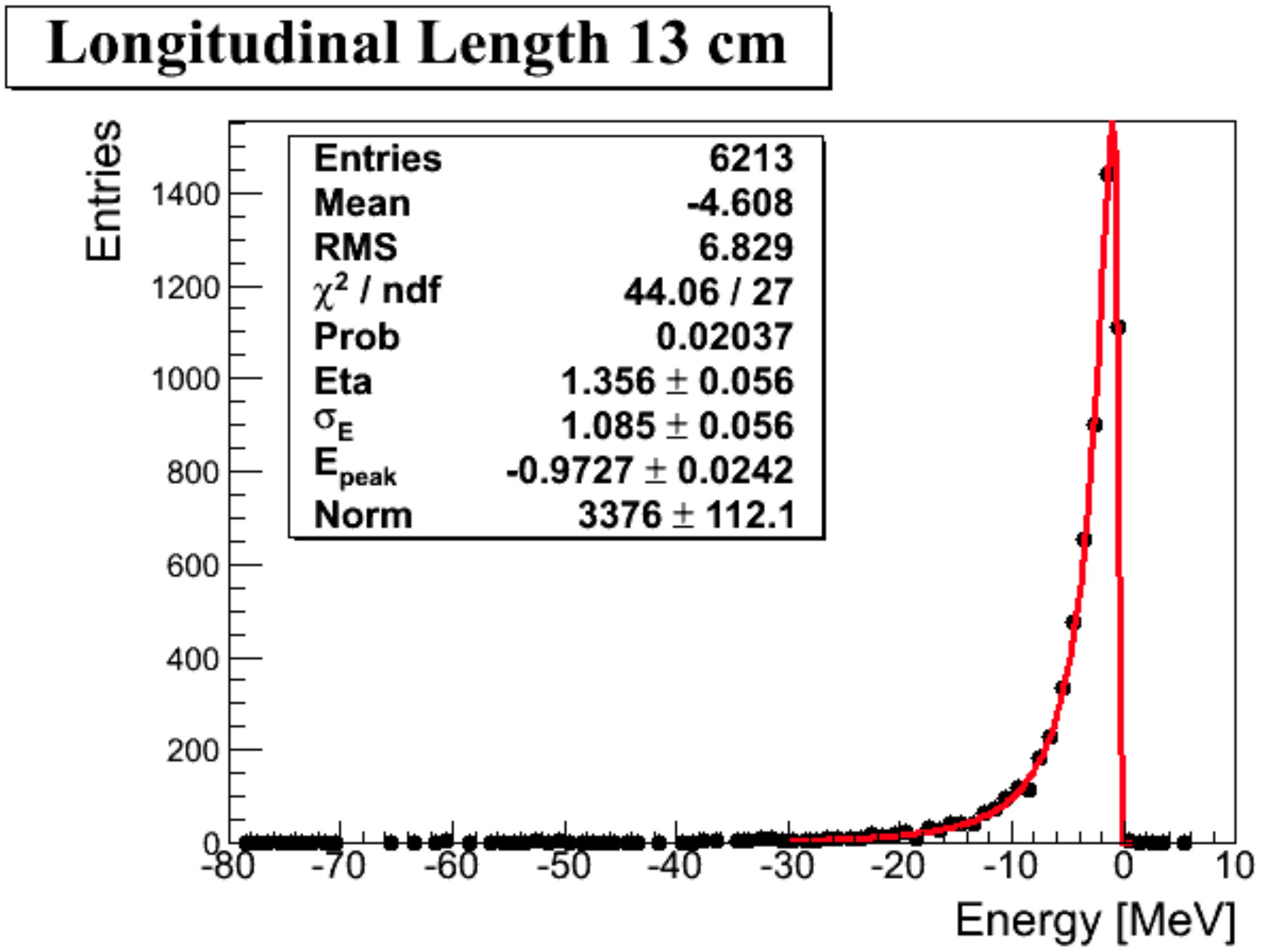

Figure 10.25. *E/P* distribution for conversion electrons that reconstruct in the tracker and deposit energy in the fiducial volume of the calorimeter.

Figure 10.26 shows the distributions of energy resolution and relative acceptance as a function of the crystal length. The energy resolution deteriorates below 10 cm, while the acceptance shows a slight increase with reduced crystal size. In the following, we use 11 cm long crystals as the baseline design. The number of rows and the inner calorimeter radius have then been varied and the result evaluated. Matrices of 10 × 44, 11 × 44 and 12 × 44 crystals/vane have been studies, with inner radii of 360 and 390 mm. The optimum is 11 × 44 crystals with an inner radius of 360 mm, which provides an acceptance of 75% in the fiducial volume, in good agreement with earlier studies.

The dependence of the energy resolution as a function of row (Figure 10.27 - left) and column (Figure 10.27 - right) has also been evaluated. The resolution is nearly constant along the vane, apart from the bottom row and the first column, justifying the a-priori definition of the fiducial volume.

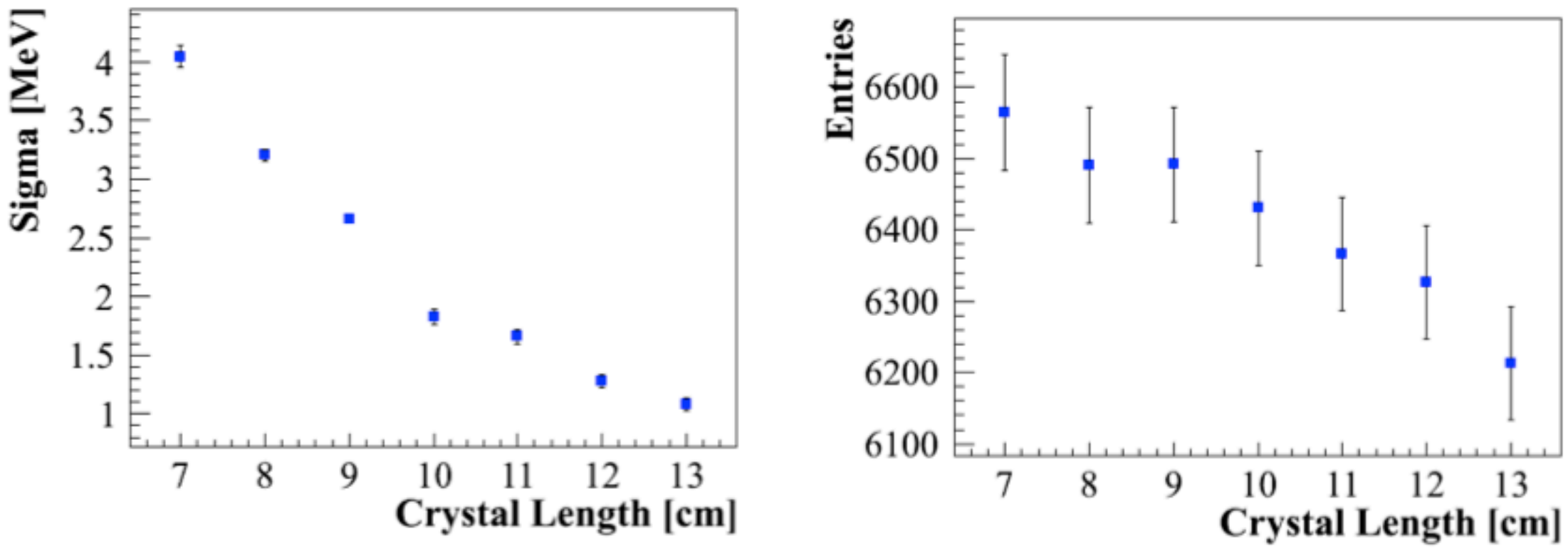

Figure 10.26. The energy resolution (left) and relative acceptance (right) as a function of crystal length.

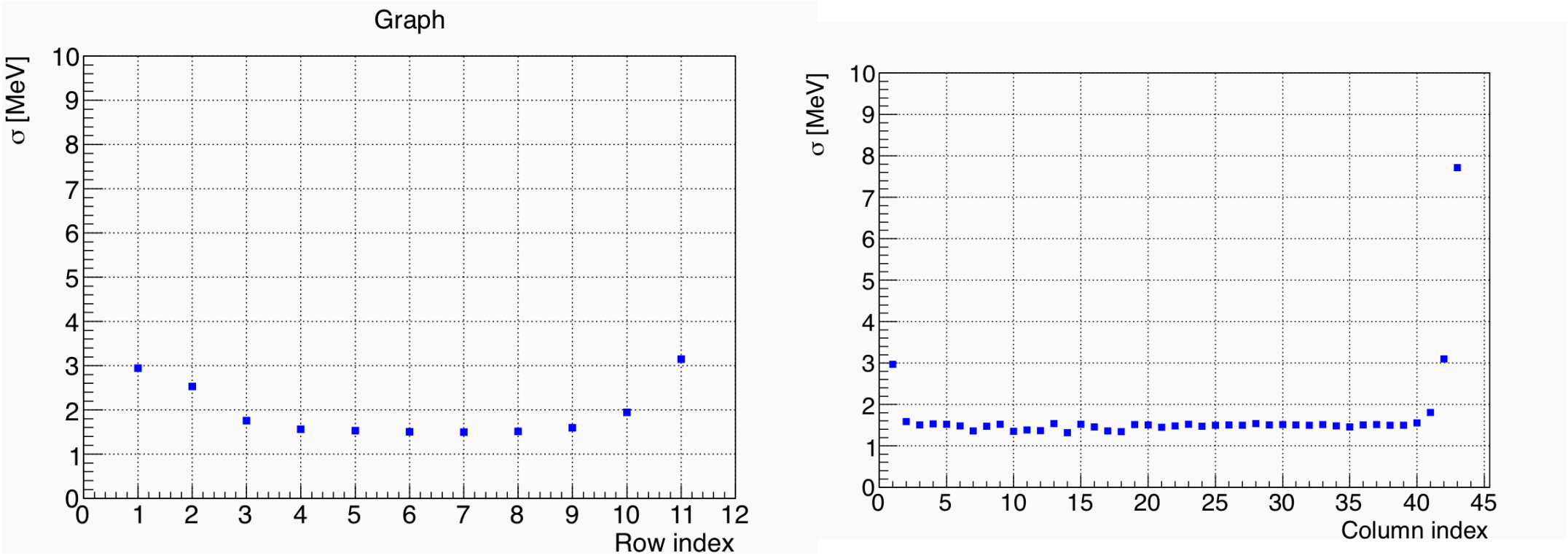

Figure 10.27. Dependence of the energy resolution on the row (left) and column (right) within a calorimeter vane.

### *10.5.1* A Calorimeter based trigger

The DAQ system requires the development of a trigger/filter algorithm to reduce the data throughput while restricting the maximum data storage to $\mathcal{O}$(1 PB). The calorimeter can provide a robust, fast and simple method to accomplish this by applying appropriate thresholds to the reconstructed clusters.

In order to evaluate the threshold response of the calorimeter, 100,000 signal events with DIO overlays were generated. It is assumed that DIOs deposit more energy in the calorimeter than any other background process. We have carried out this study for the nominal calorimeter resolution in the Mu2e simulation package and for 3 additional cases (2, 5 and 10 MeV resolution), where it is assumed that noise or instrumental effects have degraded the resolution. A full description of the analysis can be found in reference [9]. The main results are summarized here. In Figure 10.28, the signal efficiency and the rate of surviving DIO electrons are shown as a function of the applied energy threshold. A rejection factor of ~200 is obtained with a threshold of 64 MeV, which corresponds to a signal efficiency of 91% for nominal energy resolution. The energy threshold is applied to the deposited energy, so conversion electrons (or DIOs) that strike near the outside edge of a vane have lower efficiency than those in the nominal fiducial volume. The efficiency (DIO survival rate) is reduced (increased) when the resolution deteriorates. A factor of ~10 increase in DIO survival rate is observed for an energy resolution of 10 MeV.

The development of an additional trigger algorithm based on the tracker system could provide independent samples to determine the trigger efficiency. However, a calorimeter based trigger has many advantages. It is simple and can be implemented at the FPGA level to reduce the overall DAQ throughput, it does not suffer from spurious additional hits, and control samples can be collected by pre-scaling events with lower energy cuts.

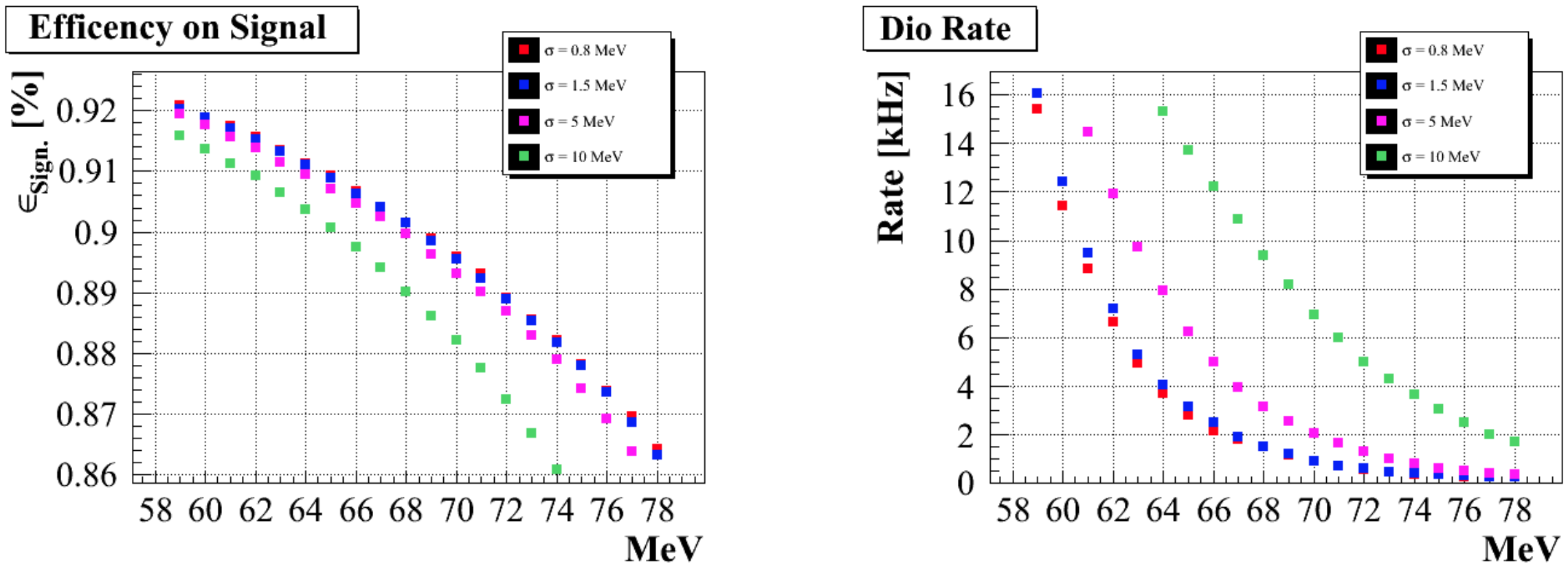

Figure 10.28. The signal efficiency (left) and the rate of surviving DIO electrons (right) as a function of the applied energy threshold in the calorimeter.

### *10.5.2* Muon Rejection

It is possible for muons passing close to the target to enter the tracker with the correct pitch angle and momentum and be misidentified as electrons. These events could be due to either cosmic ray muons that escape detection in the Cosmic Ray Veto or primary beam muons that do not interact in the target and diffuse at wide angles. A small fraction of beam muons (~$10^{-4}$) can reach momenta larger than 100 MeV. In reference [10], it was estimated that a small fraction (~$10^{-5}$) of such muons could diffuse to angles larger than 30°, corresponding to ~200 Hz of muons reaching the tracker and calorimeter.

Fortunately, most of these events are prompt and are eliminated by the beam extinction, but there is a small probability that a few could survive, resulting in background to the experiment. It is possible to combine information from the tracker and the calorimeter to form a powerful $\mu/e$ discriminant to reject any potential background from misidentified muons.

The calorimeter can provide a powerful "standalone" means of separating muons from electrons based on the total energy deposition. A dedicated simulation was performed by firing 105 MeV/c $\mu^-$ over the 5 × 5 crystal matrix described in Section.10.4. The resulting distribution of energy for the muons ranges from 44 MeV, corresponding to the kinetic energy lost by ionization, up to 65 MeV when nuclear remnants ($n$, $p$, $\gamma$) from muon capture contribute. In Figure 10.29 the evaluated $\mu/e$ rejection factors are shown as a function of the cut on deposited energy, for the case without (left) and with (right) application of the time window. A decrease of the rejection up to a factor five is observed when artificially deteriorating the calorimeter resolution (from 1 to 10 MeV). We obtain a $\mu/e$ rejection factor ranging from 80 to 400 with an energy cut of 100 MeV. An additional improvement between 3 to 5 is obtained by applying the time window cut.

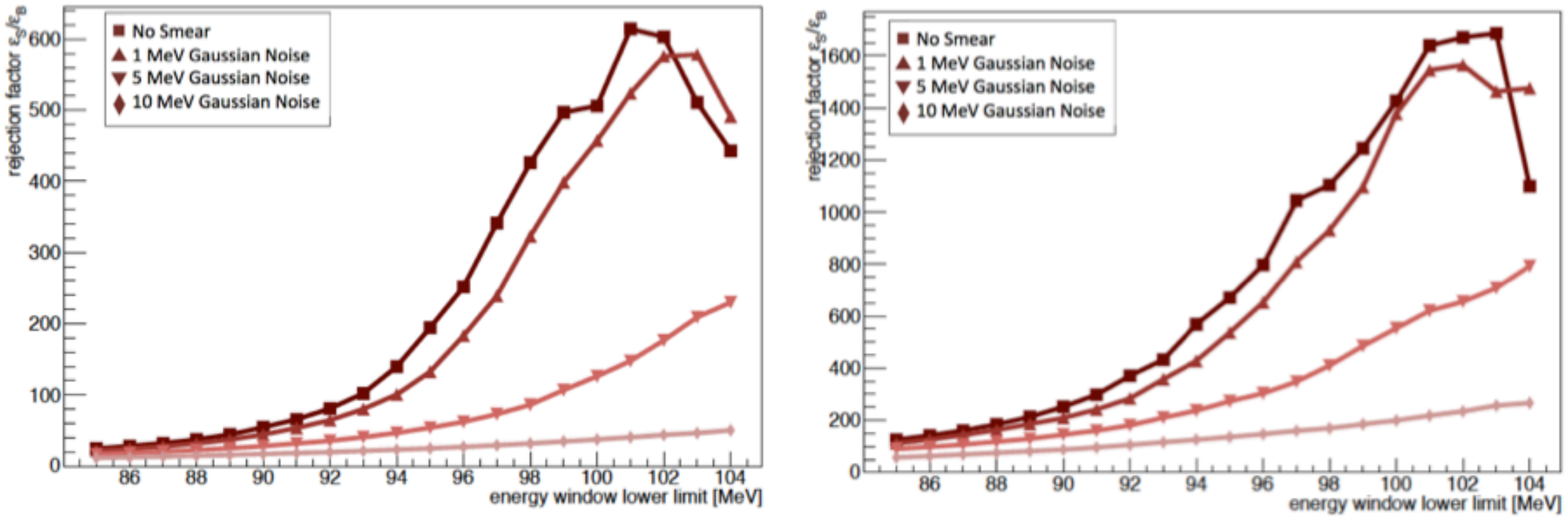

Figure 10.29. Dependence of the $\mu/e$ rejection factor in the calorimeter as a function of energy threshold. The plot on the left is made without a cut on the signal timing window and the plot on the right is for events within the timing window.

## 10.6 Calorimeter Rates and Radiation Environment.

One important requirement for the calorimeter is that it be able to survive the radiation environment expected in Mu2e. Dedicated studies of tracker and calorimeter rates have been reported in reference [12]. The main results are summarized here.

The Mu2e GEANT4 simulation framework is used to produce single particle background sources and transport them through the detector. Typical statistical sample sizes are $\mathcal{O}(10^7)$ events/background type, which is roughly what is expected for a micro-bunch. These studies were done for a 25 kW proton beam, based on an early configuration of Mu2e. The radiation dose results will be divided by 3 to account for the lower beam power that is now the Mu2e baseline (8 kW). The instantaneous rates remain unchanged, however, so the pile-up calculations require no modification.

The radiation background is divided in two main categories: (a) activity associated with the flash that occurs in the first ~200 ns after the arrival of the proton pulse, and (b) activity due to the muons stopping in the target or in the muon beam dump.

For the beam flash, the highest source of background is electrons, positrons and gamma rays that contribute to an average rate/crystal of 9, 2 and 0.2 MHz, respectively. These rates are strongly reduced to 30, 70 and 40 kHz by the time of the live gate. The pileup contribution due to the beam flash is small, and is highest on the calorimeter lateral edge pointing toward the target. We estimate single cluster pileup probabilities of 20% and 10% for positrons and gamma rays, respectively, with an average energy deposition of 300 keV and a large spread of values. The addition of thin lead sheets on the lateral edge of each vane will reduce this problem to a negligible level.

Muon captures in the stopping target will produce neutrons and gamma rays that reach the calorimeter with estimated rates of ~300 kHz and 85 kHz, respectively. This corresponds to pileup probabilities of 40% and 20%, respectively. The average energy deposition from neutrons and gamma rays in the calorimeter is estimated to be 0.5 MeV and 0.7 MeV, respectively, with an rms of 1.2 and 0.9 MeV.

From this simulation the dose/year along the calorimeter surface has been calculated, assuming a microbunch every 1700 ns for $10^7$ sec/year. For the beam flash, the maximum dose is ~100 Gy/year at the calorimeter edge pointing towards the target (Figure 10.30- top) while 33, 16 Gy/year are expected in the bottom edge and front face, respectively. The maximum dose from the muon stopping target is from DIOs and protons that accumulate (Figure 10.30- bottom) in the innermost row and contribute a total dose of 40 Gy/Year. An additional 15 Gy/year is expected in the lateral edge due to neutrons from muon capture. These calculations have been carried out with an old "transport" code for neutrons and in the presence of a neutron absorber around the target. A more refined calculation of neutron related background is currently underway.

## 10.7 Calibration and monitoring methods

In order to provide a precise, independent crystal-by-crystal calibration, a relatively low energy source is desirable, but with an energy sufficiently high to be well above electronic noise. The 6.13 MeV photon line from $^{16}O^*$ is an ideal match to this requirement. Such an approach has been successfully used for routine weekly calibrations in the *BABAR* experiment [13], and we have drawn upon this experience for Mu2e. This system, shown schematically in Figure 10.31, is referred to as the “source calibration (system)”.

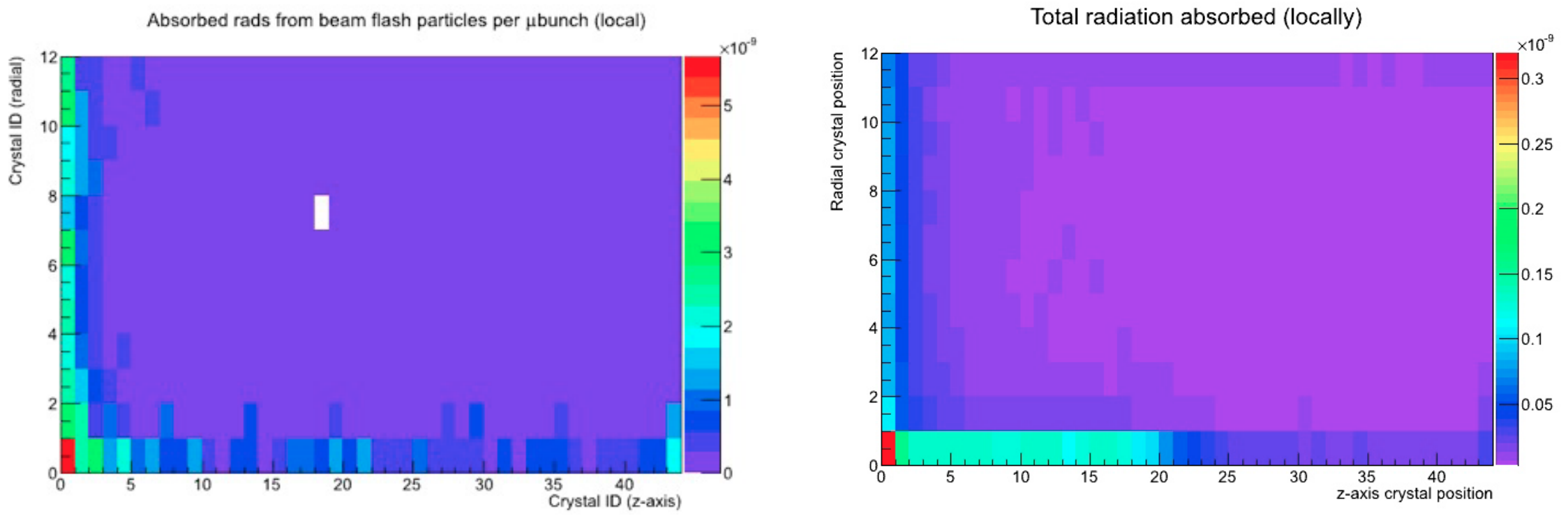

Figure 10.30. Radiation dose for a single microbunch distributed over a calorimeter vane due to the beam flash (left) and activity in the muon stopping target (right).

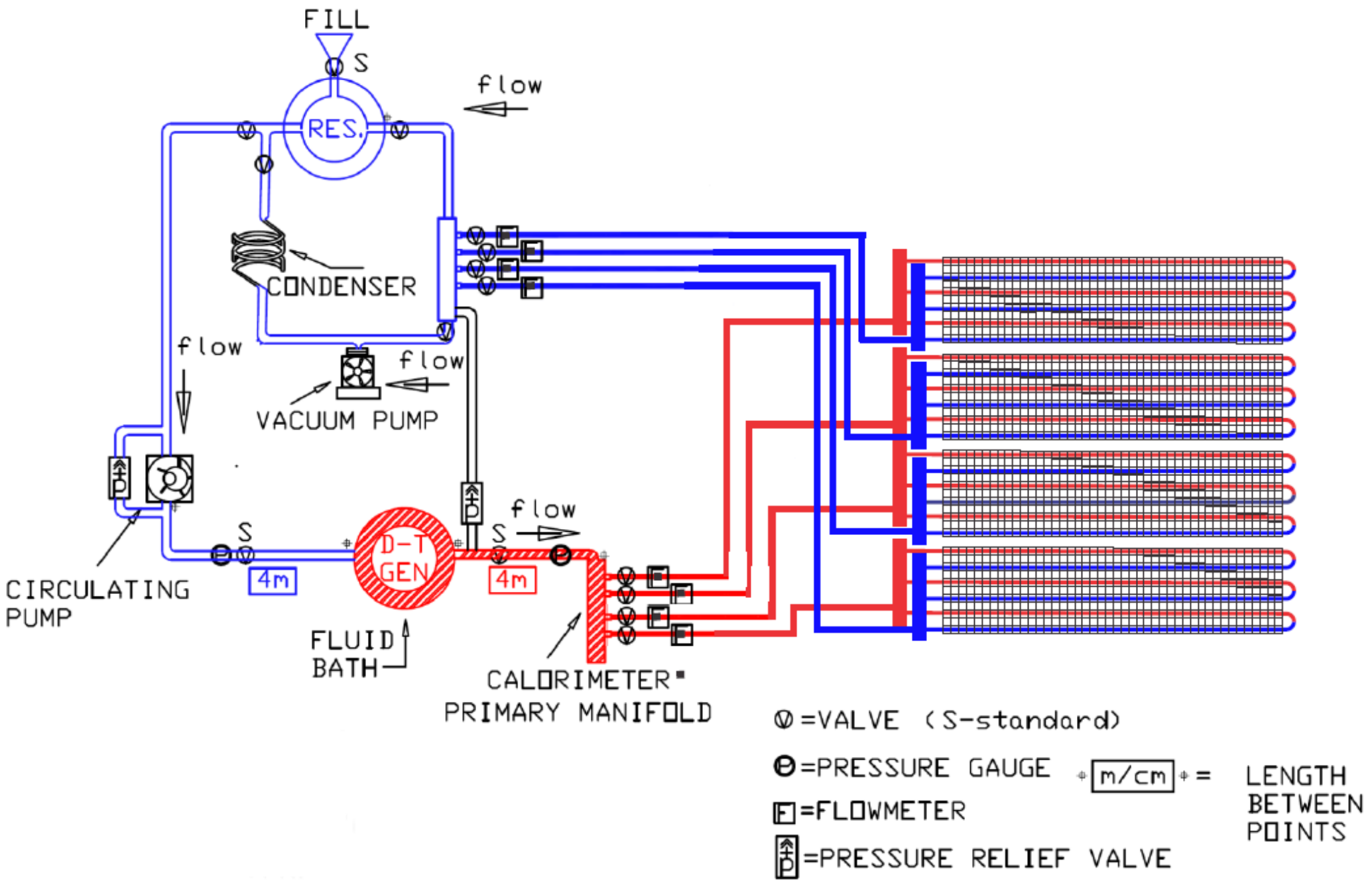

Figure 10.31. Schematic of the source calibration system.

The decay chain producing the calibration photon line is shown below:

$$^{19}F + n \rightarrow {}^{16}N + \alpha$$

$$^{16}N \rightarrow {}^{16}O^{*} + \beta \quad (\tau_{1/2} = 7\ s)$$

$$^{16}O^{*} \rightarrow {}^{16}O + \gamma\ (6.13\,MeV)$$

Fluorine is activated with a neutron source, producing the $^{16}N$ isotope. The $^{16}N$ beta decays with a half-life of seven seconds to an excited state of $^{16}O$, which in turn emits a 6.13 MeV photon as it drops to its ground state.

The source spectrum, as seen with a CsI(Tl) crystal (used in *BABAR*) with PIN diode readout is shown in Figure 10.32 There are three principal contributions to the overall peak, one at 6.13 MeV, another at 5.62 MeV, and the third at 5.11 MeV, the latter two representing escape peaks. It is important to note that all three peaks have well-defined energies and thus all are useful in the calibration, providing both an absolute calibration and a measurement of linearity.

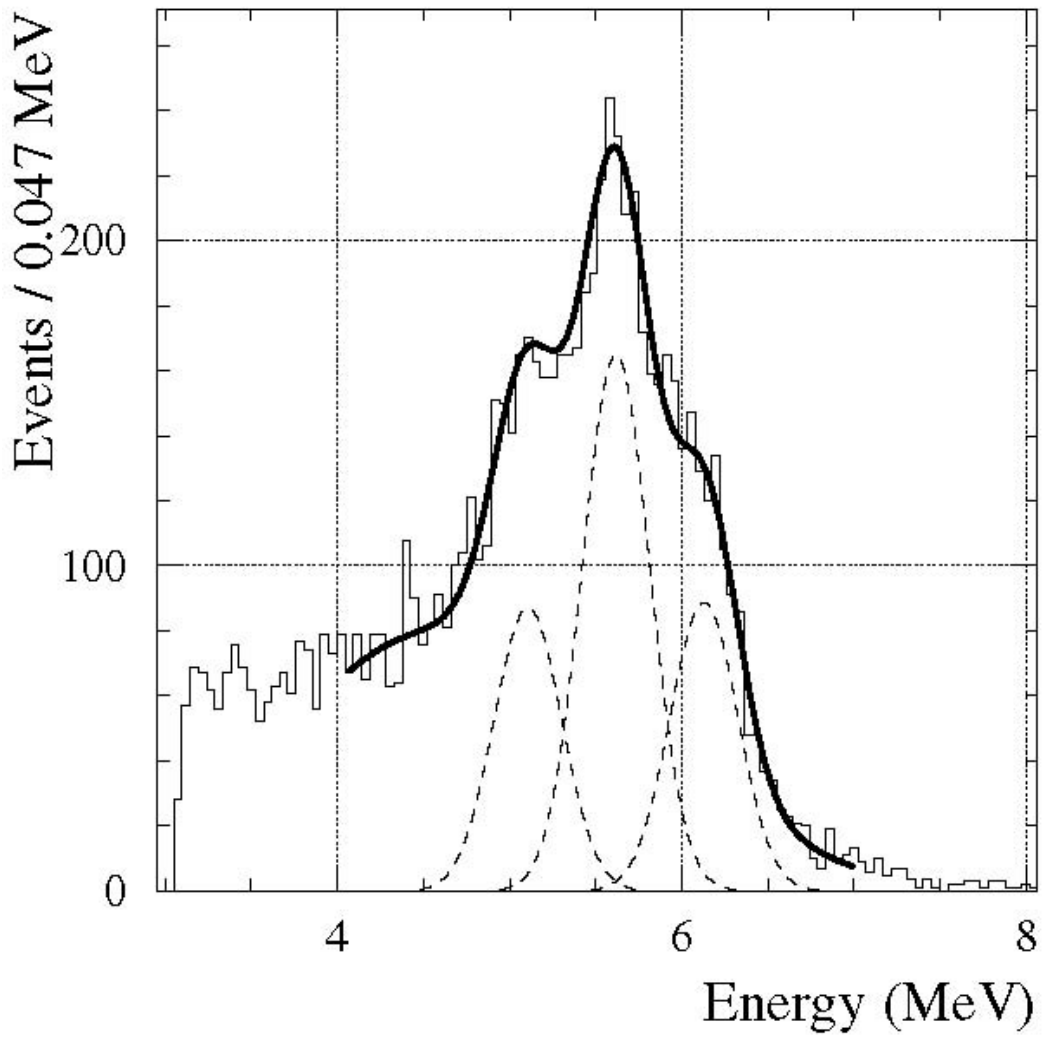

Figure 10.32. Energy spectrum in a *BABAR* CsI(Tl) crystal irradiated with 6.13 MeV photons from an $^{16}O^*$ source [13]. Readout is with a PIN diode. The solid curve is a fit to the data, including Gaussian contributions at 6.13 MeV, 5.62 MeV, and 5.11 MeV, indicated by the dashed curves.

The neutrons used to activate the fluorine are conveniently provided by a commercial deuterium-tritium (DT) generator. This device produces 14.2 MeV neutrons by colliding deuterons with a tritium target. Typical rates are several times $10^8$ neutrons/second. The DT generator is surrounded with a bath of fluorine-containing material.

The fluorine, in fluid form, is circulated from its activation point to the crystals. There are many possible fluorine compounds available from the refrigerant industry, for example “Fluorinert FC77” [14], which was used in *BABAR*.

The fluid is stored in a reservoir. When a calibration run is started, the DT generator is turned on, together with a circulating pump. Fluid is pumped from the reservoir through the DT bath to be irradiated, and thence to the calorimeter. The system is closed, with fluid returning from the calorimeter to the reservoir. For an idea of the capabilities, in *BABAR*, the fluid is pumped at 3.5 l/s, resulting in a rate of approximately 40 Hz in each of 6500 crystals, where the crystals are approximately 12 m from the DT generator. A statistical uncertainty of 0.35% is obtained in 30 minutes [13] (this was later reduced to under ten minutes [15]).

In *BABAR*, thin-walled (0.5 mm) aluminum tubing, 3/8 inch diameter, is used to transport the fluid across the crystals. The tubing was flattened to meet space constraints. One millimeter of Al is 1.2% of a radiation length. This material was in front of the *BABAR* crystals, as was an additional 2mm of Al for the structural support of the tubes. Simulations are underway to determine the effect of locating source calibration material in front of the calorimeter. The *BABAR* system was designed after the calorimeter design had been finalized. Mu2e has the advantage of designing the calibration system concurrently with the calorimeter; several options therefore exist to minimize the material in the system, including better integration with the calorimeter structure and/or using carbon fiber for structural support. Another possibility is to locate the source behind the calorimeter, instead of in front, thus reducing the material to zero.

We plan to monitor, in a more continuous way than with the source, the variations of the crystal optical transmittance and of the APD gains by means of a Laser system, following a scheme similar to the one used for the CMS calorimeter [16]. To achieve this goal, we need to illuminate each crystal with blue and red light by transporting it, with optical fibers, to the back of each LYSO crystal. A schematic diagram of the overall system is shown in Figure 10.33. Two high-precision pulsed lasers will funnel light in a integrating sphere of 2” diameter through a dichroic prism outside of the DS. From each output channel of the sphere, a group of six 1 mm diameter fused silica fibers (BigFiber) will bring the signals inside the DS and back to a reference crate where PIN diodes will monitor the change of the calibration signal. Inside the DS, each BigFiber will send light inside an integrating sphere that uniformly distributes light over a bundle of 150 smaller size fibers (200 μm diameter core). Four of such bundles will serve a vane. For each bundle there will be 121 fibers arriving to the back of the crystals and 4 fibers used as a reference; the rest will be spares.

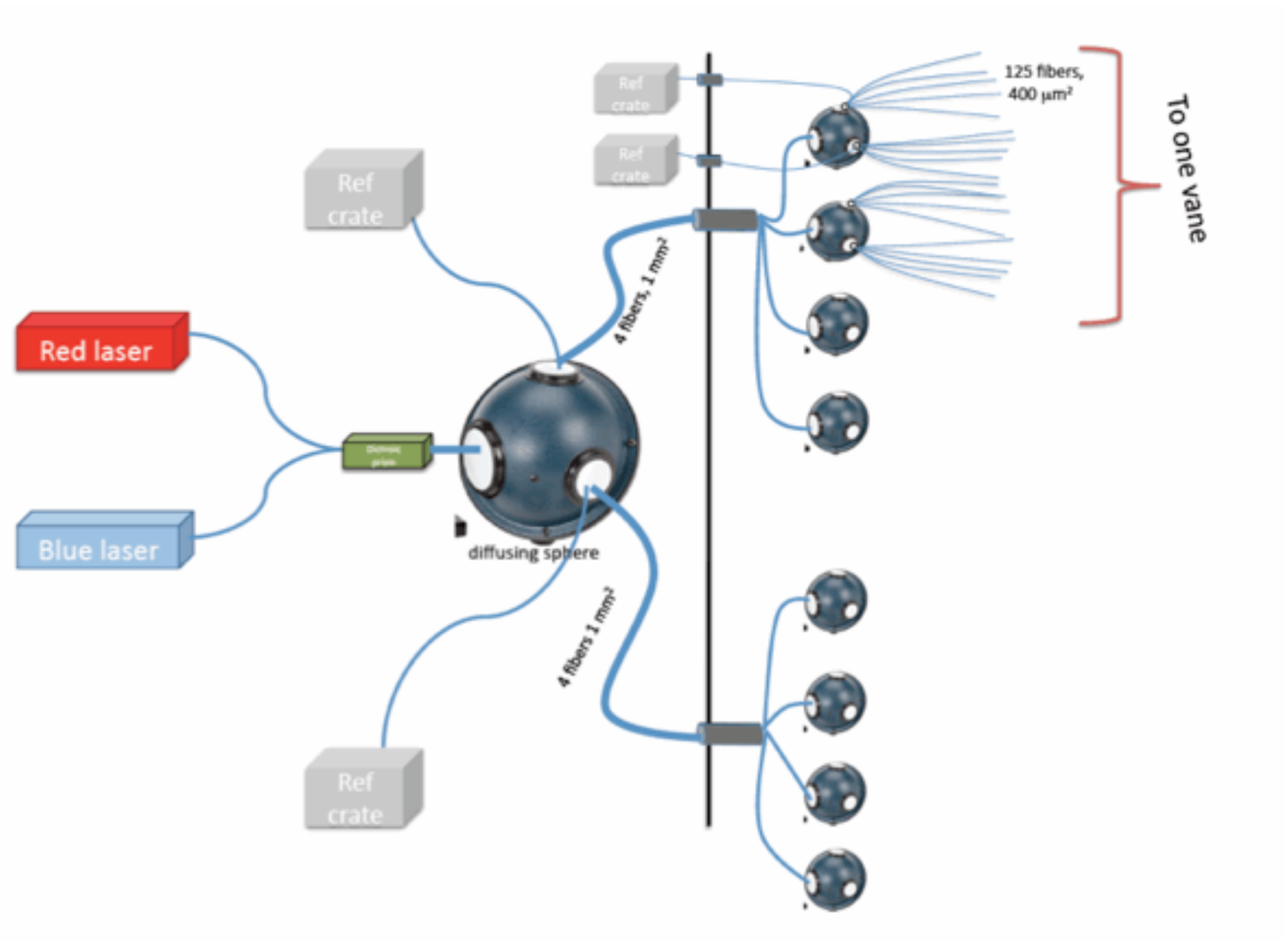

Figure 10.33. A schematic diagram of the calorimeter laser monitoring system that distributes laser light to each individual crystal.

## 10.8 Alternatives to the Proposed Design

During the process of developing the conceptual design, several alternative designs have been considered. The most attractive is the alternative disk geometry. The others, PWO-2 crystals and large area SiPMs are still under consideration.

### *10.8.1* Disk geometry

The baseline vane geometry for the calorimeter of Mu2e is actually derived from the MECO experiment. MECO also had a vane-based tracker geometry (L-tracker), which has been replaced in Mu2e by the planar T-tracker. It is then natural to explore calorimeter geometries for Mu2e to see whether improvements over the vane geometry are possible as well. The most promising direction appears to be a design based on two disks, spaced apart by one half wavelength of the conversion electron helical trajectory. This design provides improved reconstruction efficiency per unit volume of crystals and can reach absolute efficiency values higher than the vane geometry. Figure 10.34 shows the nominal placement of the two disk calorimeter placed downstream from the T-tracker.

We now proceed to optimize the two disk configuration. Figure 10.35 shows that the angle of incidence of conversion electron tracks on the disks is similar to the angle of incidence on the vanes. Thus appropriate clustering algorithms and crystal thickness optimization should be similar for the two options.

We assume that the inner radius of the disks is the same as that for the vanes: 36 cm. The outer radius of the two disks does not have to be identical. Figure 10.36 shows the reconstruction efficiency for conversion electrons as a function of the separation between front faces of the two disks, for an outer radius of the near disk of 70 cm, and three choices of outer radii for the far disk.

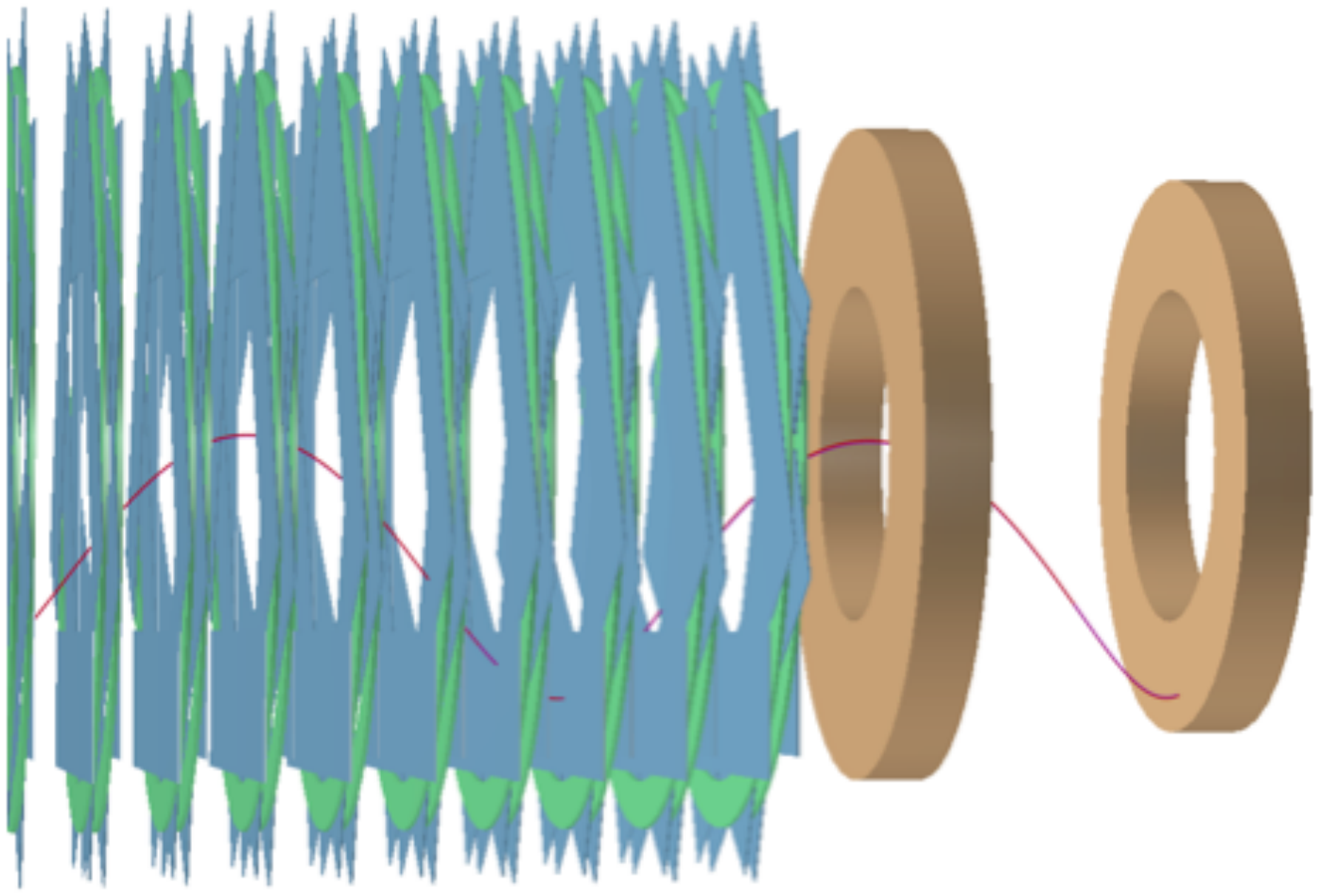

Figure 10.34. Two disk calorimeter configuration in position downstream of the tracker.

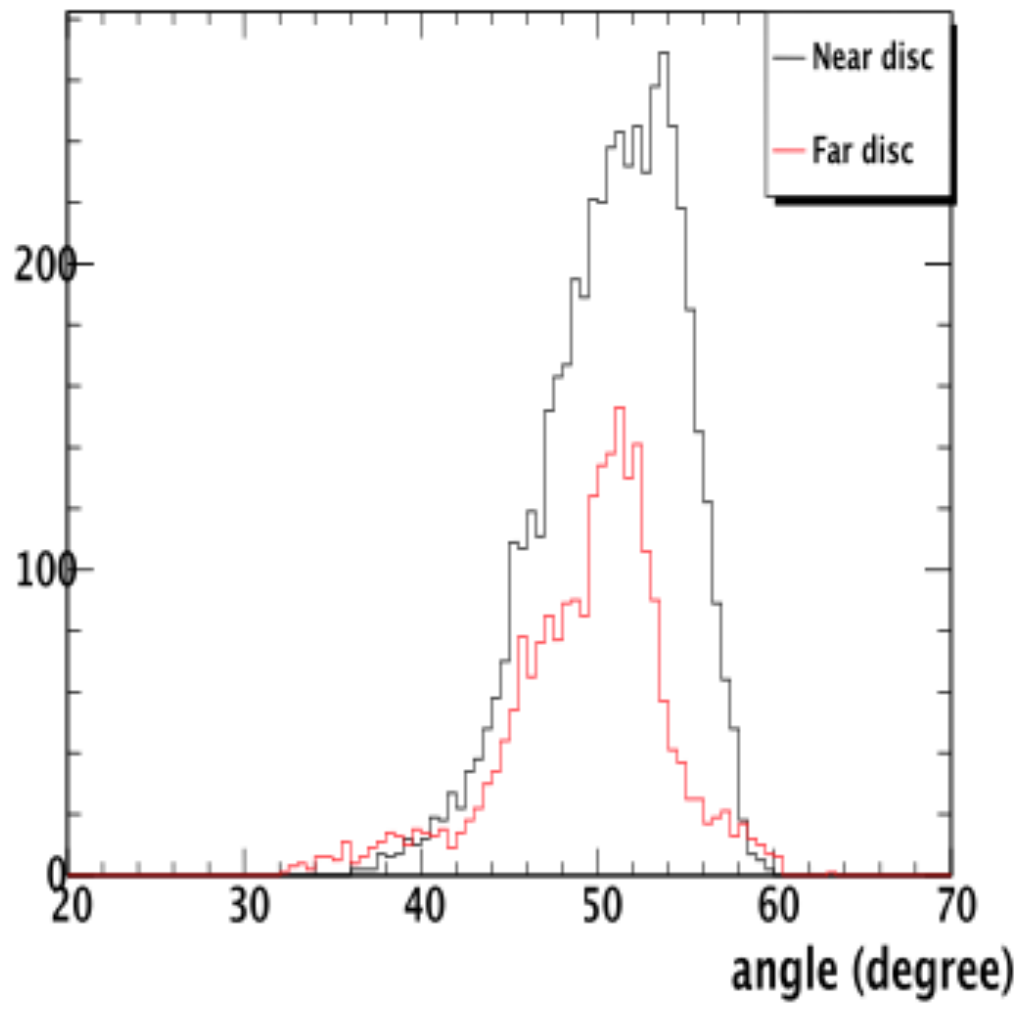

Figure 10.35. The angle of incidence of conversion electron tracks on calorimeter disks.

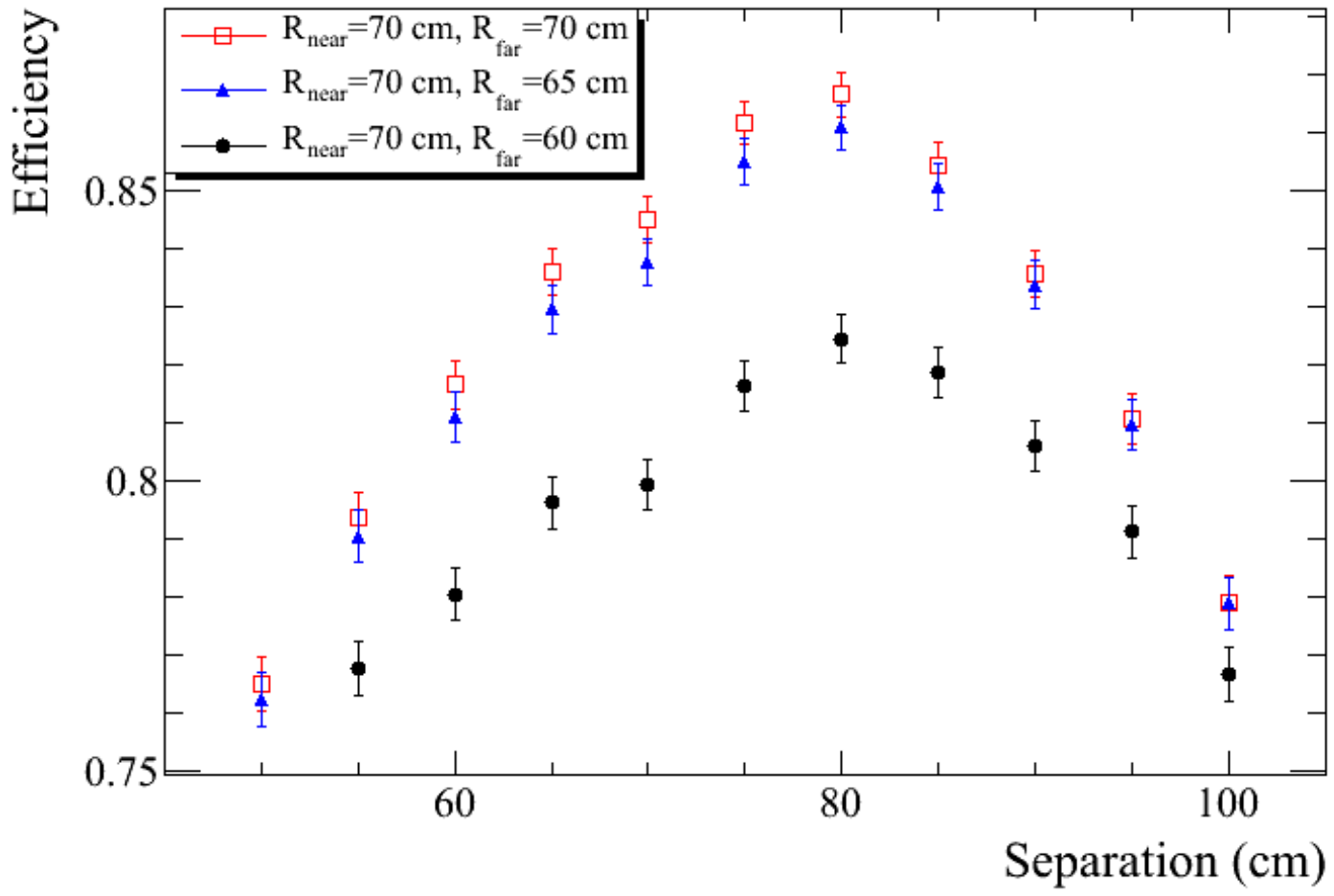

Figure 10.36. The reconstruction efficiency for conversion electrons as a function of the longitudinal separation between the front faces of the two disks, for an outer radius of the upstream disk of 70 cm, and three outer radii for the downstream disk.

An outer radius of 70 cm produces an efficiency of 84%, after fiducial cuts, at a disk separation of 80 cm. Reducing the outer radius to 65 cm lessens the efficiency to 83% but uses a smaller volume of crystals. This is a substantial efficiency gain over the maximum vane efficiency of 78% for four tilted vanes, *albeit* with an increased crystal volume. This is demonstrated in Figure 10.37, which compares the reconstruction efficiency for a variety of four vane and two disk configurations of differing volume. The disks are shown as blue circles, the vanes as red squares. Clearly, at any given crystal volume the reconstruction efficiency of a two-disk design surpasses that of four vanes. In addition, it can be seen that the efficiency of a four vane design peaks at 78%, while the two disk design can achieve an efficiency of as much as 10% higher.

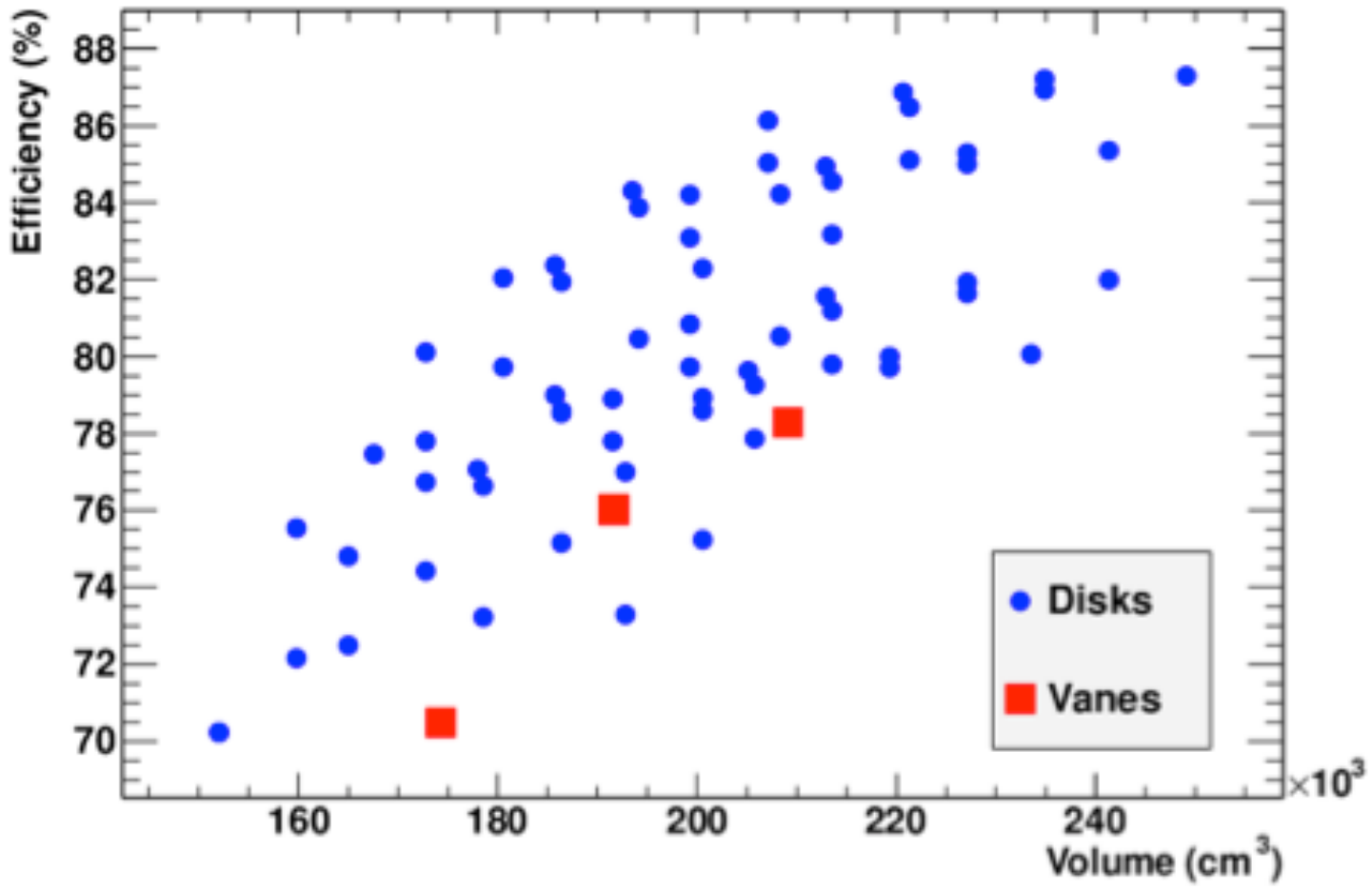

Figure 10.37. The reconstruction efficiency for a variety of four vane and two disk configurations of differing volume. The disks are shown as blue circles, the vanes as red squares.

Since at the inner radius, there is a substantial background from DIO electron tracks, we have investigated the change in efficiency that results from increasing the inner radius of the disks and vanes by 3 cm, from 36 cm to 39 cm. The result is shown in Figure 10.38. It is clear that again, for a given volume, the disk design offers higher efficiency.

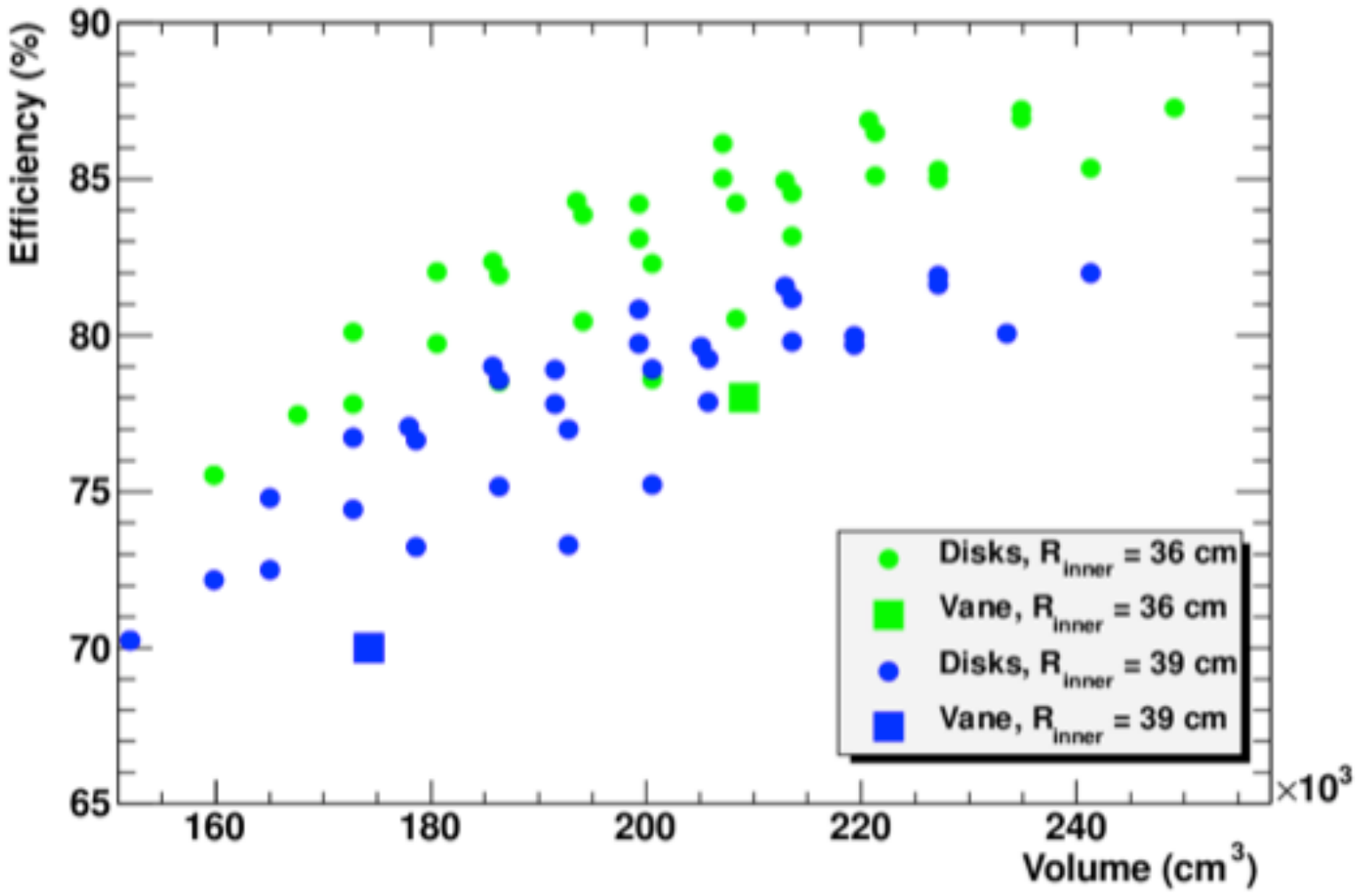

Figure 10.38. The reconstruction efficiency of several vane (squares) and disk (circle) configurations.

While the obvious tessellation of a rectangular vane is a square crystal face (we use 3 × 3 cm$^2$), the problem of tessellation of an annular disk is somewhat different. We have therefore considered the use of hexagonal faces, which provide a more natural tiling, as well as improved light collection efficiency. Figure 10.39 shows an experimental study of light yield performed by the *BABAR* experiment on CsI(Tl) crystals [17]. The two crystals, one square and one hexagonal, had equal face areas, length and tapers. It can be seen that, due to the improvement in the average reflection angle compared to total internal reflection, the hexagonal crystal produces a 22% higher output, a bonus of the disk configuration.

The crystal volume also depends on the length of the crystals. If the crystal is not long enough to fully contain the shower, fluctuations in the leakage energy will influence the resolution. The non-normal incidence angle of the conversion electron tracks in either the vane or disk design is a help in this respect. The energy resolution for the disk system is similar to that for the vane; it is also clear that a crystal length of 11 cm suffices, representing a 15% reduction in volume over the original 13 cm long crystal design in either case.

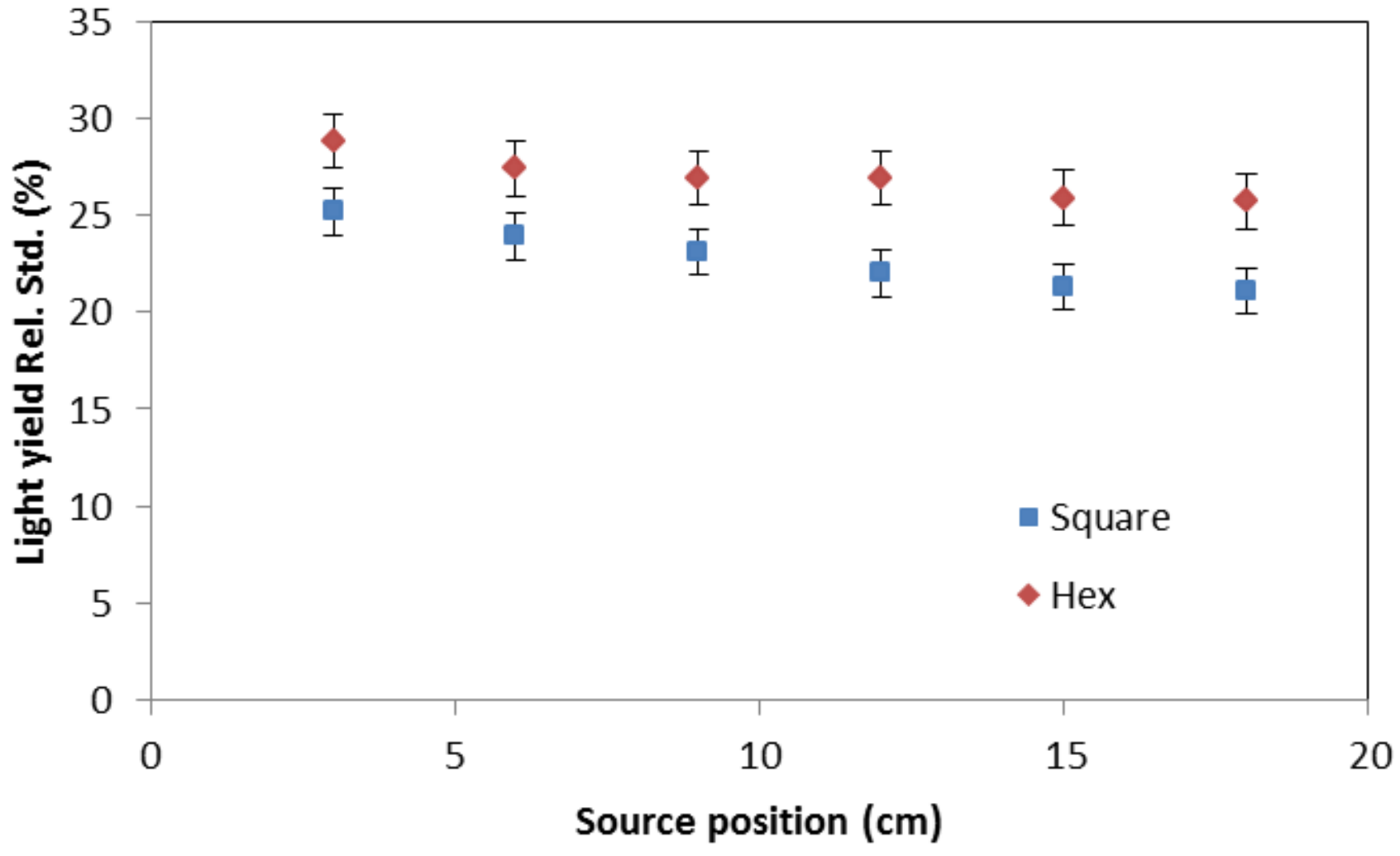

Figure 10.39. *BABAR* results on light yield for a square and a hexagonal CsI(Tl) crystal of equal length and taper.

In choosing a hexagonal crystal size, there are three obvious choices: equal area, equal dimension across flats or equal dimension across points of the hexagon. Each choice has characteristics to recommend it. For the following, we have chosen hexagonal crystals that are 3 cm across flats. The crystal arrangement for the near disk, with an inner radius of 36 cm and an outer radius of 70 cm is shown in Figure 10.40. It uses 1494 crystals. A far disk with an outer radius of 60 cm uses 954 crystals.

As of this writing, the main point that needs to be checked in the disk design is the tolerance of background, particularly for DIO events and for pileup with $n$ and $\gamma$ produced in the muon capture process. Since the disk design sees the beam face on, it will have greater occupancy than the vane design, in which it is the edge that sees most of the background. This is currently being studied using the G4 framework.

### *10.8.2* PWO-2 Crystals

Lead tungstate, $PbWO_4$, which has been used in the CMS, ALICE, PrimeX and PANDA experiments has been evaluated as an alternative to LYSO. The light yield from standard lead tungstate is marginal at Mu2e energies. The photon emission spectra have a maximum yield at 420 nm. The CMS experiment measured the light yield of lead tungstate with a standard bialkali photomultiplier tube with a quantum efficiency of ~20%, to be 8 - 12 p.e./MeV [18] for 23 cm long samples at room temperature. R&D carried out by the PANDA experiment over the past decade, in close collaboration with INP Minsk (Belarus) and the technical facilities at the Bogoroditsk Technical Chemical Plant (Russia), has led to the development of a new crystal, PWO-2, with a light yield that is double that of standard lead tungstate. The light yield was increased by improving the quality of the crystal structure and by modifying the lanthanum (La) concentration. PANDA reports that over a large sample of crystals the light yield has been improved to 17 - 22 p.e./MeV [19].

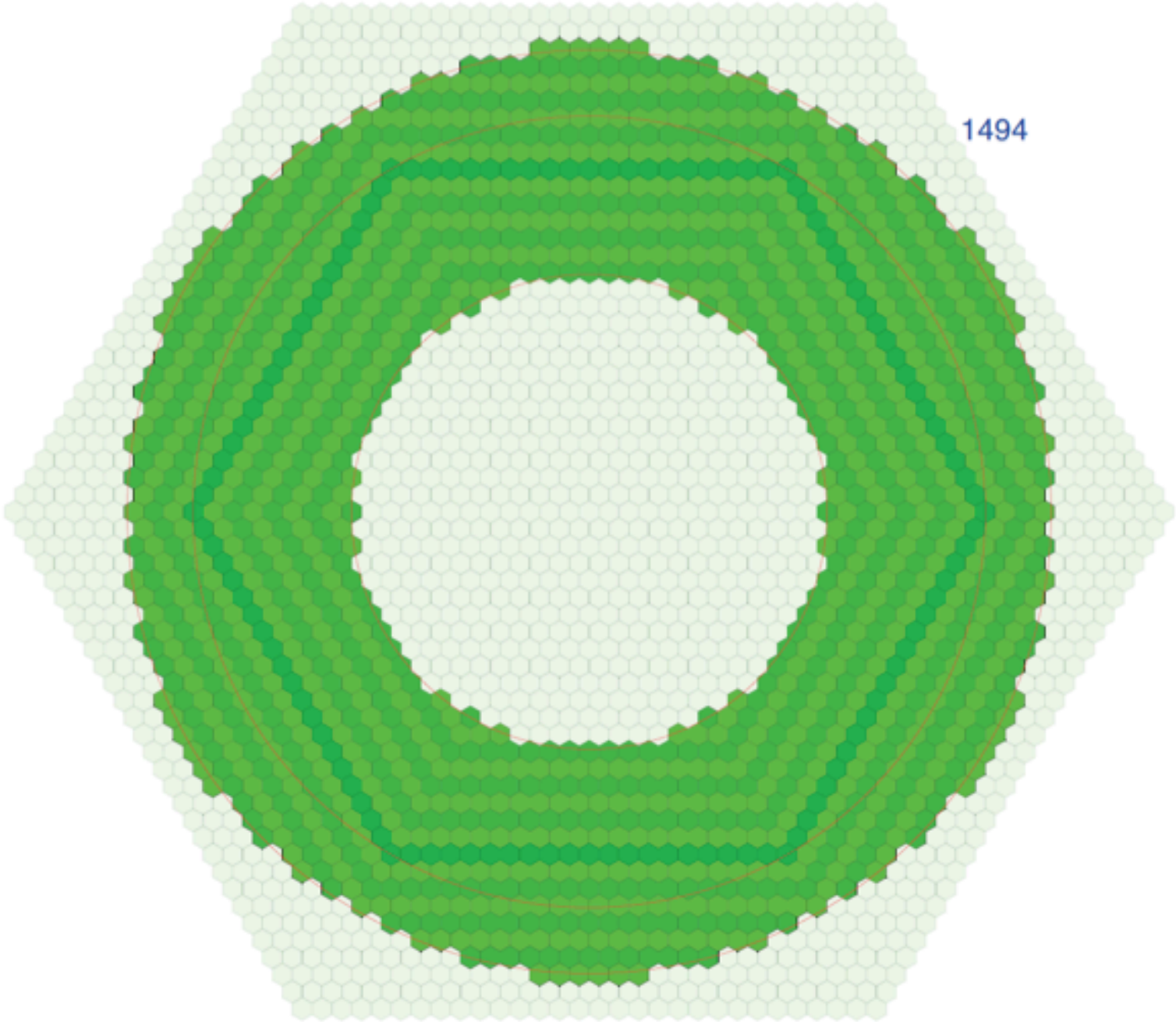

Figure 10.40. Tiling of the near disk annulus with hexagonal crystals that are 3 cm across flats.

While the light yield of LYSO crystals has only a slight dependence on temperature, PWO-2 crystals have a strong dependence. Variations of 0.2%/°C have been measured for LYSO; –2.5%/°C variations have been measured for $PbWO_4$. A LYSO crystal calorimeter can be operated at room temperature; for use in the 100 MeV regime, a $PbWO_4$ crystal calorimeter must be operated at reduced temperature.

In order to increase the functionality and the energy resolution at low energies, PANDA is designing a calorimeter that will operate at -25°C. The benefits are three-fold:

- The light yield doubles, but is still less than 1% that of LYSO.
- A cooling system provides a means of regulating the crystal temperature to 0.1° C to reduce variations in light output.
- The same cooling system can also be used to cool the APDs to increase their gain and reduce the noise.

The design for the alternative $PbWO_4$-based calorimeter would be similar to the design for PANDA; PWO-2 crystals, cooled to -25° C, with APDs readout, deployed in 4 vanes.

The limiting factor in resolution for a $PbWO_4$ calorimeter will be the electronic noise, which is expected to be of O (1 MeV)/channel. The NYU group from MECO [20] built a system with two 3.0 × 3.0 × 13 $cm^3$ $PbWO_4$ crystals read out by means of two 13 × 13 $mm^2$ RMD APDs and their own charge preamplifier, achieving an ENC of 0.7 MeV and a light yield of 38 p.e./MeV in a cosmic ray test. From these numbers, they estimated an energy resolution of 4.1 MeV at 100 MeV. Moreover, a more dedicated test with a 3 × 3 matrix of 20 × 20 × 200 $mm^3$ PWO-2 crystals has been performed by the PANDA collaboration. At 100 MeV, they measure an energy resolution of 5.5 MeV (3.9 MeV) with the crystals cooled to 10 (-25)° C when reading them out with conventional PMTs [5]. A slightly worse result, 7.5 MeV at -25 °C, is obtained when reading out each crystal with a 10 × 10 $mm^2$ APD [21].

The crystals must also survive and function normally in an expected radiation environment of 160 Gy (16 krad)/year/crystal., or 2–3 rad/h/crystal. Many measurements have been carried out on radiation damage of these crystals by the PANDA collaboration [5]. The ionizing radiation creates a variety of color centers in the crystal with absorption bands in a wide spectral region. Figure 10.41 shows the light output for cooled and room temperature crystals. A loss of ~30% in two months of running could be extrapolated from this plot for the Mu2e calorimeter.

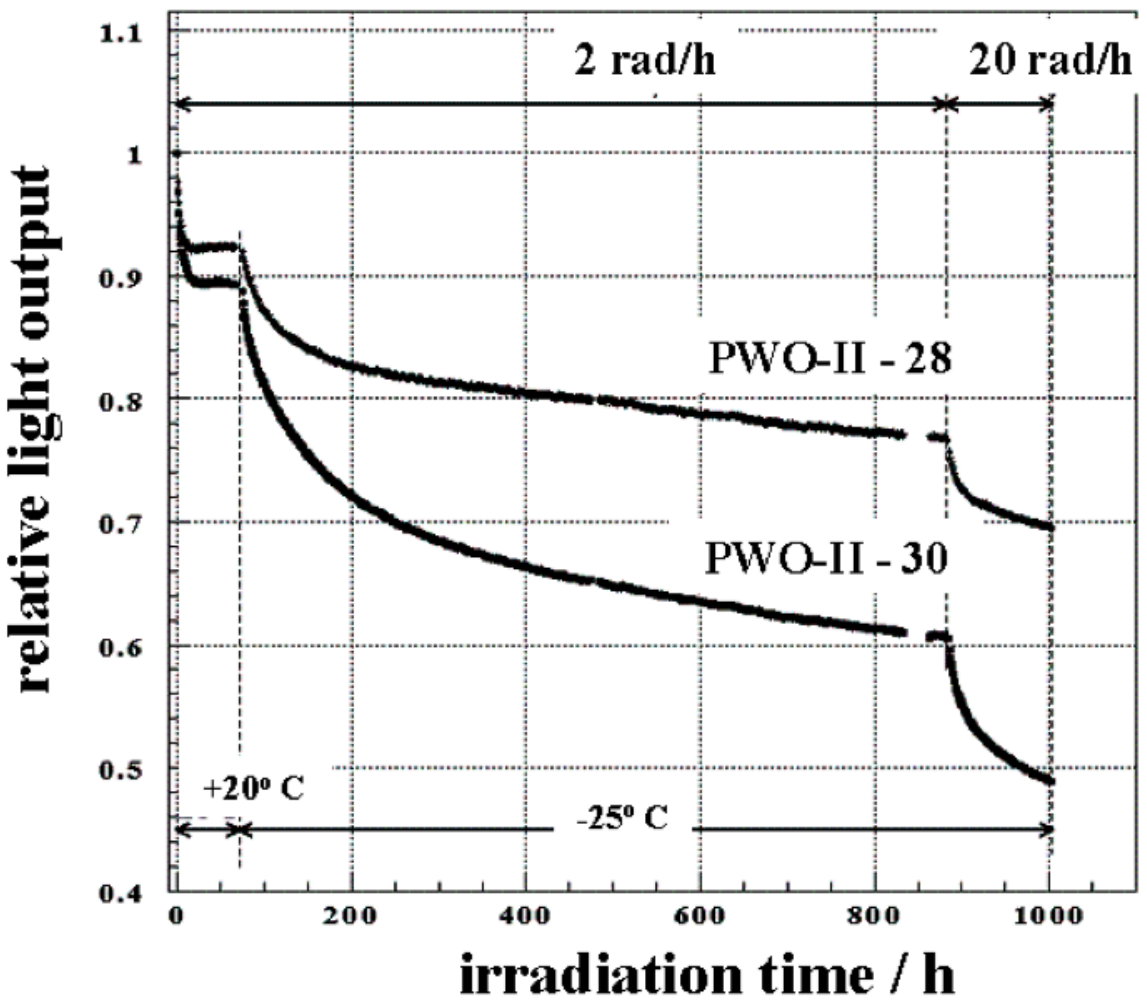

Figure 10.41. Relative change of light output for two samples of PWO-2 at room temperature and at -25°C, for two different doses.

When the irradiation stops, a spontaneous relaxation of the color centers takes place via thermo-activation, the recovery effect being slower in cooled than in room temperature crystals (from a few days to many weeks). The reasoning is that the thermo activation is a function of temperature as:

$$N_i = N_0 e^{-w_i(T)} - \sum (b_j I_j)t,$$

where $N_i$, $N_0$ are the current and initial population of color centers, $w_i$ the spontaneous relaxation probability, $b_j$ the interaction of a photon with the color centers, and $I_j$ the flux of a specific photon wavelength. To recover the light output for a cooled crystal, it is possible either to warm them up, which is not a simple operation in a large size detector, or to inject energy in the appropriate way. A practical method, being developed for the PANDA calorimeter, uses the principle of "*stimulated recovery*" [22] by delivering photons of selected wavelength to the crystals. In Figure 10.42 the recovery time of the light output as a function of LED wavelength is shown for samples irradiated with 30 Gy, and a typical LED intensity of $O$ ($10^{16}$ photons/second). Even infrared light can fully recover the losses in 1-2 days of "*stimulation*". This suggests that this procedure could be used to recover light output while the calorimeter is taking data.

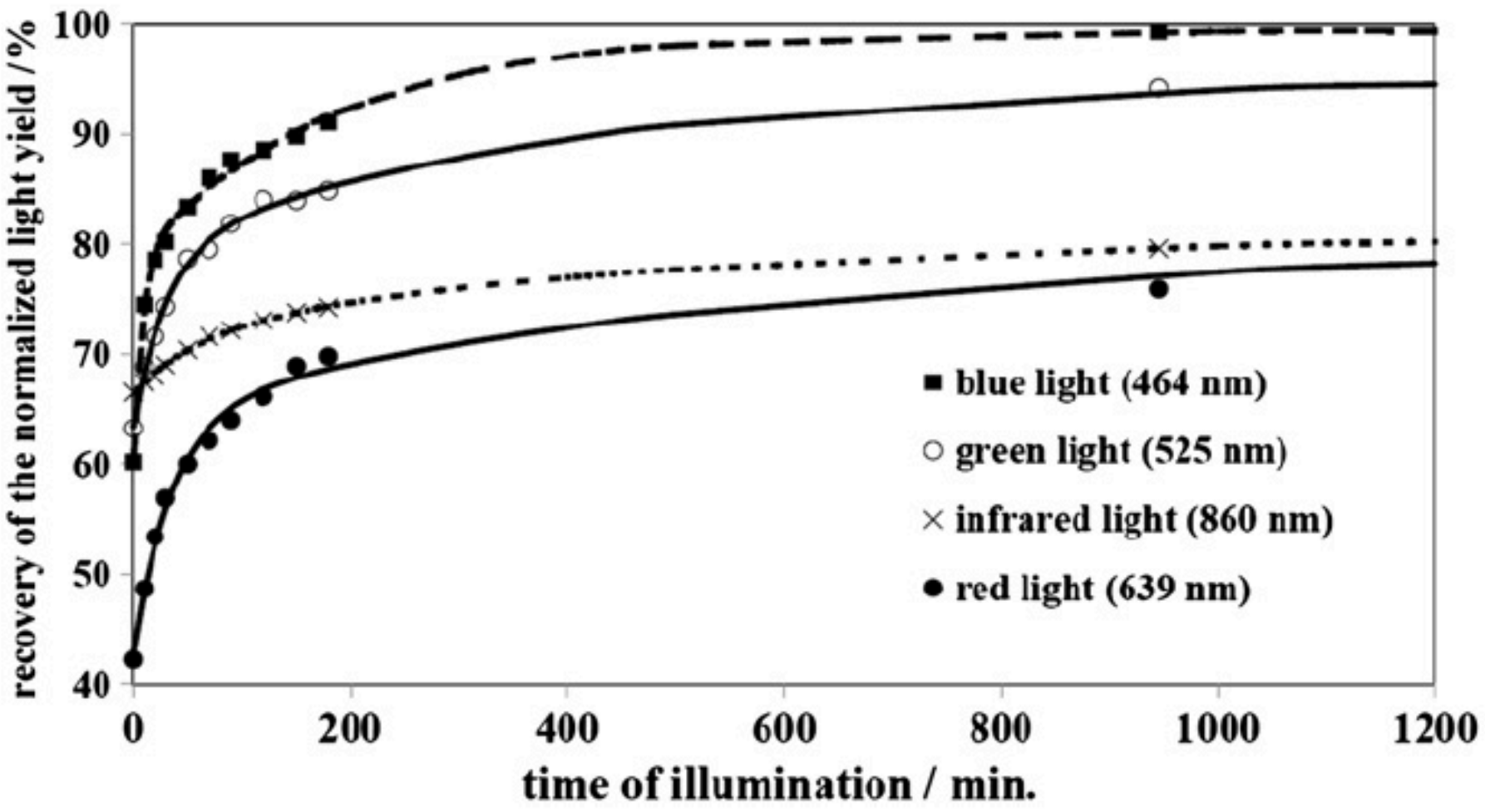

Figure 10.42. Stimulated recovery of PWO-2 crystal as a function of LED illumination time. The crystal had been exposed to 30 Gy prior to exposure to laser light. The recovery is shown for different wavelengths of light.

### *10.8.3* Large Area Silicon Photomultipliers (SIPMs)

The second option is to replace the APDs with the new generation large area silicon photomultipliers (SIPM). The SIPM, or Geiger-APD, is a digital device containing a matrix of APDs working in Geiger Mode. Large gains are reached for these devices, of $O(10^6)$, with functionality close to a conventional PMT. The dark noise of a SIPM is a general concern when running with many pixels. Non-linear response and saturation problems related to the digital nature of the device are additional concerns. A few large area SIPMs have been tested, including the Hamamatsu MPPC with a 6 × 6 mm$^2$ active area and an SMD device (4 × 4 mm$^2$) from IRST/FBK (Trento, Italy). The gain has been tested along with the time resolution and rate-dependence when coupled to a LYSO crystal excited with a UV LED. A beam test of a new matrix prototype equipped with Large Area SIPMs is being planned. The results will be compared with beam test measurements already made on an array equipped with APDs.

## 10.9 ES&H

The Mu2e calorimeter is similar to other crystal calorimeters that are commonly used at Fermilab. Potential hazards include power systems, mechanical hazards, lead exposure, and toxic fumes. These hazards have all been identified and documented in the Mu2e Preliminary Hazard Analysis [23].

The Mu2e calorimeter includes both low and high voltage power systems. During normal operation the calorimeter will be inaccessible, inside the enclosed and evacuated Detector Solenoid. Power will be distributed to the calorimeter through shielded cables and connectors that comply with Fermilab policies. Fermilab will review the installation prior to operation.

The size and weight of individual assembled vanes and the fully assembled detector require special precautions during handling. Explicit procedures for safely handling individual crystals as well as assembled arrays will be developed as part of a series of time-and-motion studies.

Lead sheets will likely be used for shielding the inner edge of the calorimeter from low energy photons. Standard handling precautions including PPE, ventilation, and disposal will be implemented for the use and handling of lead. Help and guidance will be provided by the Division ES&H group. The lead will be deployed inside of the evacuated Detector Solenoid during operation and will not be easily accessible once installed.

Small quantities of adhesive will be used in various applications including the attachment of photo detectors to crystals. Ventilation appropriate for these quantities will be installed in the assembly area and personnel working with adhesives will wear the appropriate personal protective equipment.

A laser could be used as a calibrated light source for the calorimeter. Lasers used for detector calibration will be completely enclosed in light-tight boxes that occupy permanent locations in the Mu2e detector hall. These locations will be identified during the design process and appropriate safety measures (*e.g.* control of exposed beams) will be incorporated into the design of the facility. All uses of lasers will be reviewed and controlled as required in national consensus standards and Fermilab regulations. Lasers will require written operating procedures.

The source system includes a deuterium-tritium (DT) generator to activate a fluorinated liquid that is circulated to the crystals. The half-life of the activated liquid is 7 seconds, hence it is not a substantial concern when the DT generator is not operating. The DT generator produces 14 MeV neutrons at a rate of $\sim 10^9$/sec. Operation of the source will be done remotely, in a no-access condition. The DT generator itself will be shielded according to Fermilab regulations. The shielding will be interlocked such that the DT generator cannot be operated if the shielding is not in place. A reservoir capable of holding the entire volume of fluid will be installed. Operation of the system is anticipated to occur approximately weekly. In the event of a fluid leak, the maximum exposure will be computed; the number for the similar *BABAR* system is a maximum integrated dose of less than 1 mrem. A detailed hazard analysis will be performed in collaboration with Fermilab ES&H.

## 10.10 Risks

There are several risks that could jeopardize the success of the calorimeter subproject. The risks as well as potential mitigation strategies are described below.

There is a risk that deliveries of crystals from vendors could be delayed. To mitigate this risk the crystal procurement should begin as early as possible. This requires the crystal specifications to be finalized as early as possible. There is a preference for multiple producers to help mitigate this risk. Close communication with crystal vendors will be required to obtain crystals that meet the specifications in a timely fashion.

The radiation hardness of lead tungstate crystals is a serious concern. If lead tungstate is ultimately selected the calorimeter will be shielded with lead on the side nearest the beam to minimize the dose from prompt-photon flashes and a stimulated recovery system using infrared light will be deployed.

Neutrons incident on the APDs could increase the dark current, deteriorating the calorimeter's performance. The radiation hardness of candidate APDs to neutrons will be evaluated and the potential benefit of cooling the APDs to improve the radiation hardness will be evaluated.

Accidental hits in the calorimeter due to a flux of neutrons or albedo from the beam stop could add (pileup) unwanted energy to a cluster, deteriorating the energy resolution of the calorimeter. Events of this type will be simulated to understand the optimal detector granularity and to determine if the pileup can be identified in the offline reconstruction of the shaped signals.

## 10.11 Quality Assurance

For a calorimeter of such a complexity, quality assurance (QA) is a fundamental part of the procurement, fabrication and assembly phases. Quality Assurance will be applied to all components and subsystems and will build on the relevant experience from the CMS and PANDA calorimeters. Expertise in the construction of the KLOE-2 calorimeter upgrade, as well as the *BABAR* and Super*B* calorimeters also exists within the Mu2e Collaboration.

Given the cost of the scintillation material a three-step QA procedure will be followed for all of the received crystals:

1. QA at the site of production:
   Crystals will be tested at the vendor site before they are shipped to Mu2e. A test station will be provided to the vendor along with a set of specifications that each crystal must pass before it can be shipped. The test station will consist of a light tight box and a stepping motor assembly for moving a radioactive source ($^{137}$Cs) along the crystal axis. The crystal response will be measured at several locations along its axis to determine the uniformity and light yield. The crystals will be optically connected to a 2 inch bialkali PMT readout by a PC controlled data acquisition system.

2. QA upon receipt by Mu2e:
   Mu2e will repeat the QA tests performed by the vendor for each delivered crystal using an identical test station. The biggest difference will be to measure the longitudinal uniformity both with PMT and APD in final readout configuration. Additional tests will also be performed, including a measurement of the crystal transmission properties by illuminating the front face with a tungsten lamp [3]. The dimensions of the crystals will also be measured. All measurement will be carried out in a temperature-controlled environment. Final crystal acceptance will be based on the tests made by Mu2e. At the end, we will have two test stations, one in Caltech and one in Italy or at FNAL in order to share the work on testing the large sample of crystals. We will also export much of our techniques to the final provider(s) for an agreed QA system.

3. Radiation Hardness testing:
   If lead tungstate crystals are used for the Mu2e calorimeter, it will be necessary to measure their radiation hardness and recovery time. Gamma irradiation will be performed, similar to what has been done by the LHC*b* experiment, using a high-intensity $^{137}$Cs source either in Italy at La Casaccia or at Caltech where a motorized source is already available. Neutron irradiation can be performed at JINR, Dubna. The recovery time will be measured for each crystal. If LYSO is selected we will control the radiation hardness by periodic sampling during production of the crystals.

QA of the photo sensors will include a measurement of the gain and its dependence on temperature and bias voltage. The APDs will be illuminated by blue and red Laser light and we will test with final preamps: the Gain and the Dark rate dependence on Temperature and bias voltage; the ENF as a function of bias voltage and the NCE. A light distribution system will allow to fire $\mathcal{O}(5)$ APDs at a time. We will add also a UV LED, a calibrated light attenuator to make linearity tests of the device and a Keithley amperemeter for current measurement. The system will be in a tight controlled temperature environment.

The preamplifiers and HV boards will be validated using standard bench test measurements of amplification and noise. A burn-in test of the HV board will be also employed.

A final test will be performed on crystals that have been fitted with APDs and readout electronics before they are inserted into a vane. The response of each individual assembly will be tested with a radioactive source. A system test will be performed on the assembled calorimeter using cosmic rays prior to installation in the Detector Solenoid.

## 10.12 Value Management

Value management for the calorimeter will consist of a careful examination and validation of detector requirements coupled with an alternative analysis of engineering and design choices with special attention to cost. The choice of crystal drives much of this work.

An R&D program in conjunction with industrial crystal growers to reduce the cost of LYSO crystals is envisioned. Mu2e, Super*B* and CMS are all interested in LYSO crystals and plan to collaborate on this R&D. If the cost can be reduced to be relatively competitive with lead tungstate then the various tradeoffs will have to be evaluated for the two applications. LYSO does not require a cooling system, though the APDs may still require cooling to reduce noise. The tradeoff between a combined cooling system and a dedicated system for the APDs would have to be evaluated in detail for cost savings. From the perspective of light output, only 1 APD per crystal would be required for a LYSO calorimeter, but two APDs per crystal are desirable for redundancy (and hence reliability) and to effectively deal with the nuclear counter effect from the large neutron flux. The increased light output allows for simpler front end electronics. Operationally a LYSO calorimeter would provide significant benefits because of the radiation hardness and the absence of rate-dependent light output issues.

Should low-cost, large-area SiPMs that meet Mu2e's specifications become available in the future they would be considered for use with the calorimeter. The inherent gain and lower noise of SiPMs might allow for a simpler design of the front end electronics.

## 10.13 R&D program

R&D to evaluate crystals for Mu2e started 2 years ago. The final phase of that R&D program is to test and characterize a reasonable sample size of crystals from different vendors. These crystals would also be used to test electronics and mechanical prototypes and would be used to develop QA stations that would be used to qualify the production crystals. The QA stations will be automated and designed to test ~5 crystals at a time, similar to the QA stations developed for the CMS crystals [3]. The Longitudinal Response Uniformity (LRU) will be measured with both with PMT's and APD's; the latter with the final readout in order to take into account the correct dependence of the quantum efficiency on the emitted wavelength. The construction of the QA station will be progressive, starting from a manual test station that evaluates a single crystal to an automated system that can process several at a time.

The *"roughness"* technique, developed for the Super*B* tapered crystals, will be evaluated as a means of improving the LRU. The procedure for exposing crystals to both gamma rays and neutrons and evaluating their radiation hardness will also be developed. Radiation hardness tests will be performed on samples provided by every vendor.

A parallel R&D effort will continue the characterization of photodetectors by acquiring and testing *O*(100) photo-sensors of the final size that can later be reused with prototype crystals for beam tests. A dedicated QA station is required that will evolve, similar to the crystal QA station, starting with a simple device that tests a single APD. The final QA station will be automated and will test multiple photosensors. Tests of the Nuclear Counter Effect (NCE) will be performed by exposing the APD to a $^{90}$Sr source and comparing APD and PIN diode signals. Neutron irradiation tests will be carried out at the JINR Institute in Dubna, while gamma irradiation tests will be carried out at Caltech and at LNF. Dependence of the radiation hardness on the APD temperature will be also evaluated.

R&D on electronics will focus on development of a low noise amplifier that is well matched to a suitable Wave Form Digitizer. Tests will be carried out on the noise level, signal shape and pileup discrimination.

The large area SiPM alternative will also continue to be studied for large area SIPMs. From simulation results, it is clear that pixel sizes below 25 μm are needed in order to obtain the required linearity. A dedicated R&D program has been developed by INFN and JINR, in collaboration with the Italian firm IRST/FBK to produce and test $\mathcal{O}(30)$ 8 × 8 $mm^2$ large area SIPM's and compare them with few samples of the best 6 × 6 $mm^2$ SIPM's from Hamamatsu, either in monolithic or array packaging. The best performing devices will be used to read out a prototype crystal matrix. This will allow the linearity and energy response to be characterized at a test beam.

R&D on the mechanical support for the crystals will lead to the development of a prototype vane. Two square matrices that each support a 5 × 5 array of crystals will be constructed and tested. The impact of various wrapping materials on the crystal light yield will be studied by comparing Tyvek with diffusing paint on the inner walls of the carbon fiber cells. Systems for mechanical support of the APDs and the light calibration system will also be tested. The modules will also be tested in a temperature controlled environment as well as in a vacuum, to simulate the Mu2e environment and to ensure that there are no unexpected problems.

Prototypes for the light and source calibration systems will also be developed and evaluated. A small part of the laser light distribution system will be assembled and the light losses along the distribution chain will be measured to optimize the required laser power, the diffusing sphere and the connectors. The source calibration system requires a minimum amount of material between the tube carrying the fluorinated liquid and the crystal. Prototypes will be developed to evaluate tubing and coupling to the crystal.

The two modules described above will be exposed to photon and electron test beams. This will provide a format for testing all of the prototype parts in a realistic setting.

## 10.14 References

[1] R. Bernstein, "Calorimeter Requirements," Mu2e-doc-864.
[2] R. W. Novotny et al., IEEE Trans. Nucl. Sci., 57 No. 3, 1441 (2010).
[3] J. Chen et al., NIM A 572 218 (2007).
[4] F. Cavallari CMS CR 1997/10.
[5] Panda TDR, http://www-panda.gsi.de/html/det/emc/tdr/panda_tdr_EMC.pdf
[6] Mu2e-Doc in writing "The electronic system of the mu2e calorimeter".
[7] A. Luca et al, "GEANT4 Simulation of a LSO Crystal Array," Mu2-doc-1080.
[8] M. Cordelli et al., Nucl. Instr. Meth. A617, 109 (2010).
[9] S. Miscetti et al," Calorimeter trigger study", Mu2e-DOC-1974-v (2012).
[10] S. Miscetti et al,"Simulation of the energy deposit of 105 MeV muons in a LYSO matrix", Mu2e-DOC-2071-v1 (2012).

[11] Mu2e-DOC in writing “Sensitivity reach with the mu2e calorimeter”

[12] G. Onorato, “Beam-flash simulation for the calorimeter,” Mu2e-DOC-1929-v1 (2011) and “ background rates for the calorimeter”, Mu2e-DOC-1885-v1(2011).

[13] B. Aubert et al., Nucl. Instr. Meth. A479, 1 (2002).

[14] Fluorinert is the trademark name for polychlorotrifluoroethylene), manufactured by 3M Corporation, St. Paul, MN, USA.

[15] M. Weaver, private communication (2011).

[16] M. Anfreville et al., Nucl. Inst. Meth. **A594**, 292 (2008).

[17] C. R. Wuest, “The BaBar Cesium Iodide Electromagnetic Calorimeter,” UCRL-JC-119320, (1994).

[18] A. Auffray et al., Nucl. Instr. Meth. **A523** 355 (2004).

[19] A. Borisevich et al., Nucl. Instr. Meth. **A537**, 101 (2005).

[20] R. Djikibaev et al., JINST **5** p01003 (2010).

[21] R. Novotny et al., IEEE Trans. Nucl. Sci. **NS-55**, 1295 (2008).

[22] V. Dormenev et al., Nucl. Instr. Meth. **A623**, 1082 (2010).

[23] R. Ray, “Mu2e Preliminary Hazard Analysis,” Mu2e-doc-675 (2010).

# 11 Cosmic Ray Veto

## 11.1 Introduction

Cosmic ray muons are a known source of potential background for experiments like Mu2e. A number of processes initiated by cosmic-ray muons can produce 105 MeV particles that appear to emanate from the stopping target (Section 3.5.10). These muons can produce 105 MeV electrons through secondary and delta-ray production in the material within the solenoids, as well as from muon decay-in-flight. These events, which will occur at a rate of about one per day, must be suppressed in order to achieve the sensitivity required by Mu2e. Backgrounds induced by cosmic rays can be defeated by both passive shielding, including the overburden above and to the sides of the detector enclosure and the 0.46-m-thick shielding concrete surrounding the detector solenoid, and an active veto detector whose purpose is to detect penetrating cosmic-ray muons.

Unlike the other backgrounds to the Mu2e conversion signal, the cosmic-ray background scales as the detector live time. The cosmic-ray background can be measured when the beam is not being delivered and at times outside the signal window (Figure 3.8) when the beam is being delivered. A direct measurement of the cosmic-ray background without beam will be done as soon as the tracker and the detector solenoid are in place and operating.

## 11.2 Requirements

The Mu2e collaboration has developed a complete set of requirements for the cosmic ray veto [1]. The outstanding performance requirement is that the cosmic ray veto limit the cosmic-ray background to no more than 0.05 events over the duration of the data taking period, and do so without reducing the overall detector live time by more than 1%. The derived requirements that must be met to achieve this performance requirement are listed below.

- The overall efficiency of the cosmic ray veto should be 0.9999 or better [2].
- The cosmic ray veto should be nearly fully hermetic on the top and sides in the region of the collimator at the entrance to the Detector Solenoid, the muon stopping target, tracker, and calorimeter.
- The time resolution of the cosmic ray veto should be less than 5 ns in order to reduce the random two-counter coincidence rate from background neutrons and photodetector noise.
- The tracker must be able to resolve upstream vs. downstream particle trajectories to eliminate background from positrons moving upstream.
- The tracker must be able to distinguish electrons from muons.

- Penetrations into the Detector Solenoid should not protrude through the top of the neutron shield and cosmic ray veto and should not be in the region of the stopping target, tracker, and calorimeter.
- The neutron fluence over the course of the run at the photodetector must be less than $1\times10^{10}$ n/cm$^2$.

The conceptual design developed by Mu2e satisfies these requirements as well as the other requirements detailed in the requirements document for the cosmic ray veto [1].

## 11.3 Proposed Design

### *11.3.1* Overview

The baseline design of the cosmic ray veto is described below. The layout of the cosmic ray veto is shown in Figure 11.1–Figure 11.3 and the system parameters are listed in Table 11.1. A more complete list of parameters is given in Ref. [3].

The cosmic ray veto is made of three layers of inexpensive extruded scintillation counters with embedded waveshifting fibers read out by silicon photomultipliers (SiPMs), sometimes called Geiger-mode avalanche photodiodes or GAPDs. The design is simple, robust and should operate with a high efficiency. It can be tuned to give the required veto efficiency, either by adding more layers or by modifying the individual counter efficiency, for example, by changing the waveshifting fiber diameter. There is considerable experience with this technology at Fermilab. Similar detectors have been fabricated for the MINOS, MINERvA, and NOvA experiments and quite a bit of tooling and infrastructure exists at Fermilab for the fabrication of the detector components. The quantity needed for Mu2e is well within the demonstrated capabilities of Fermilab; 2088 scintillator counters comprising 990 m$^2$ and 41 km of wavelength shifting fiber is needed for Mu2e compared to 100,000 counters over 28,000 m$^2$ and 730 km of fiber for MINOS.

The cosmic ray veto is positioned just outside of the concrete neutron shield surrounding the Detector Solenoid, and extends up to the midpoint of the Transport Solenoid. The coverage is nearly 100% on top and sides; the only gaps are at the Detector Solenoid endcap and the center of Transport Solenoid. Accessibility is good.

The scintillation counters are grouped into self-contained, 3-layer modules that are 12 counters wide for a total of 36 counters. Between the layers is an inert absorber layer of 4.8-mm-thick Al that prevents electrons produced from neutron capture gammas from traversing more than one counter and causing spurious coincidences. As shown in Figure 11.11, the counters within a module are staggered such that there are no projective cracks for muons incident at normal incidence. Muons incident at oblique angles along the cracks between counters produce an amount of light in the scintillator that is equivalent to that produced from normally incident muons.

| | |
|---|---|
| Scintillator layers | 3 |
| Scintillator counter size | 4.700 × 0.100 × 0.010 $m^3$ |
| Module size | 4.766 × 1.241 × 0.041 $m^3$ |
| Total number of modules | 58 |
| Total module active area | 330 $m^2$ |
| Counter (module) mass | 4.982 (328) kg |
| Counters per module | 36 |
| Total number of counters | 2088 |
| Total counter length | 9814 m |
| Total scintillator mass | 10,402 kg |
| Fiber diameter | 1.0 mm |
| Fibers per counter | 4 |
| Total number of fibers | 8352 |
| Total fiber length | 40,925 |
| Fibers per SiPM | 1 |
| Fiber ends read out | 2 |
| Readout channels per module | 288 |
| Front-end boards per module | 6 |
| Total number of channels (SiPMs) | 16,704 |
| Total number of front-end boards | 348 |
| Total number of readout controllers | 15 |

Table 11.1. Cosmic ray veto system parameters.

| Sector | Modules | Area ($m^2$) | Counters | Fibers | SiPMs | FEBs |
|---|---|---|---|---|---|---|
| DS-R | 18 | 103 | 648 | 2,592 | 5,184 | 108 |
| DS-L | 14 | 80 | 504 | 2,016 | 4,032 | 84 |
| DS-D | 1 | 6 | 36 | 144 | 288 | 6 |
| DS-T | 18 | 103 | 648 | 2,592 | 5,184 | 108 |
| TS-T | 1 | 6 | 36 | 144 | 288 | 6 |
| TS-R | 5 | 28 | 180 | 720 | 1,440 | 30 |
| TS-L | 1 | 6 | 36 | 144 | 288 | 6 |
| **Total** | **58** | **330** | **2,088** | **8,352** | **16,704** | **348** |
| Spares | 3 | 17 | 108 | 432 | 864 | 18 |
| **Total** | **61** | **347** | **2,196** | **8,784** | **17,568** | **366** |

Table 11.2. Cosmic ray veto sectors.

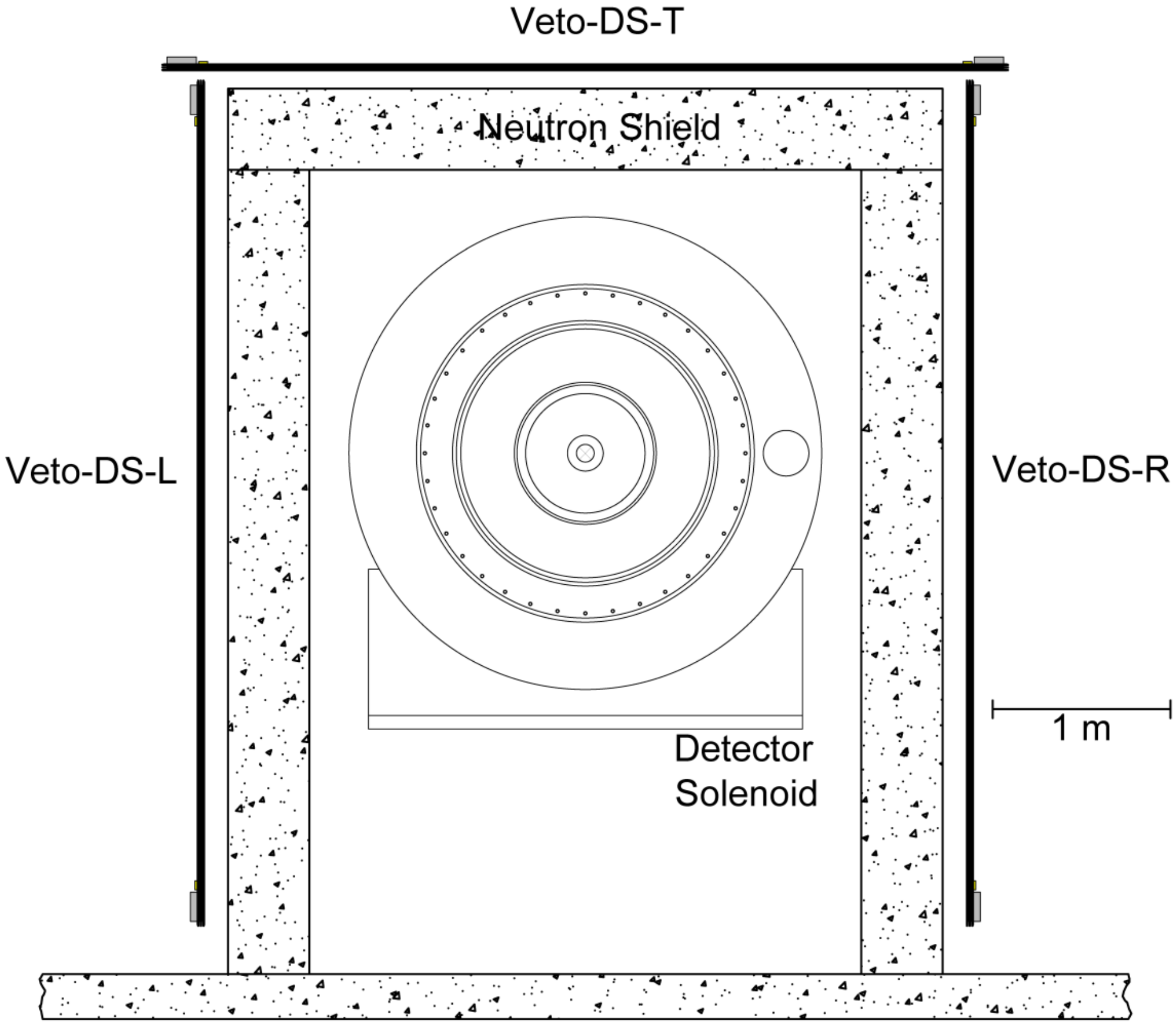

Figure 11.1. Section view of the cosmic ray veto looking downstream. The support structures for the modules and detector solenoid have been omitted.

Every counter has four embedded waveshifting fibers, each read with individual SiPMs on both ends. The fibers are glued in place by epoxy and the entire assembly is then cut and polished using a custom diamond-bit fly cutter, similar to those used by the MINOS and NOvA experiments [4].

The cosmic ray veto consists of a total of 58 modules. The modules are grouped into 7 sectors, listed in Table 11.2. They will be mounted just outside the neutron shield surrounding the Detector Solenoid using a commercial framing system such as Unistrut. Modules have an active area of 4.700 × 1.211 $m^2$. The full module size, including all inactive material, is 4.766 × 1.241 × 0.041 $m^3$. The readout electronics are designed to be easily accessible as it is anticipated that the greatest cause of inefficiency will be electronics failure. Components will be replaced promptly when they fail.

### *11.3.2* Scintillation Counters

The fundamental element of the cosmic ray veto is the scintillation counter, which is 4700 mm long, 100 mm wide, and 10 mm thick. Each counter contains 4 channels of 2.6 mm diameter into which the waveshifting fibers are inserted (Figure 11.4). The counters are composed of a polystyrene base doped with 1% PPO, and 0.03% POPOP.

To limit costs and to facilitate production of the fiber channels the counters will be extruded at the Fermilab-NICADD Extrusion Facility [5] using a procedure essentially the same as that used to produce the similar MINERνA counters. A 25 μm thick coating of titanium-dioxide ($TiO_2$) doped polystyrene, 15% by mass, is co-extruded on the surface of the counters providing a diffuse reflector for the scintillation photons while protecting the scintillator from physical damage [6]. A total of 2088 counters are needed for the veto system, with a total length of 9814 m and total scintillator mass of 10,402 kg.

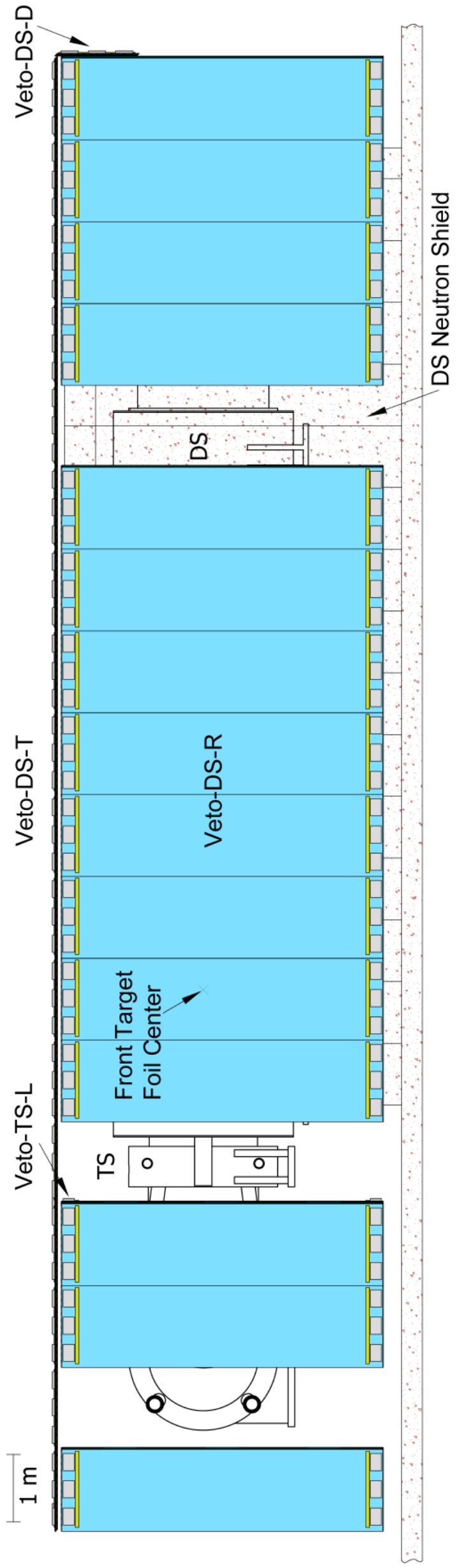

Figure 11.2. Elevation view of the cosmic ray veto. The support structure is not shown. Several modules have been omitted as well as the neutron shield in the TS region.

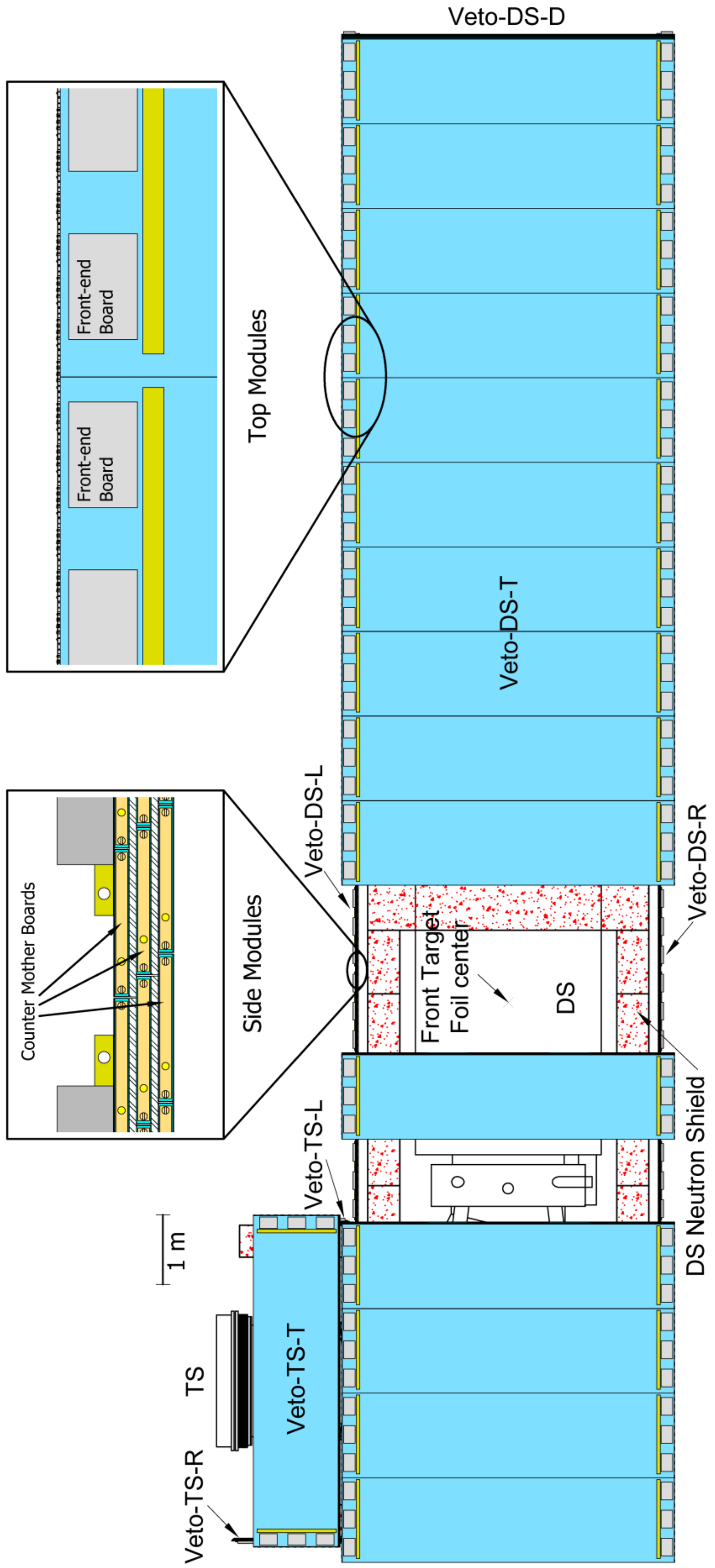

Figure 11.3. Plan view of the cosmic ray veto. The support structure is not shown. Several modules have been omitted.

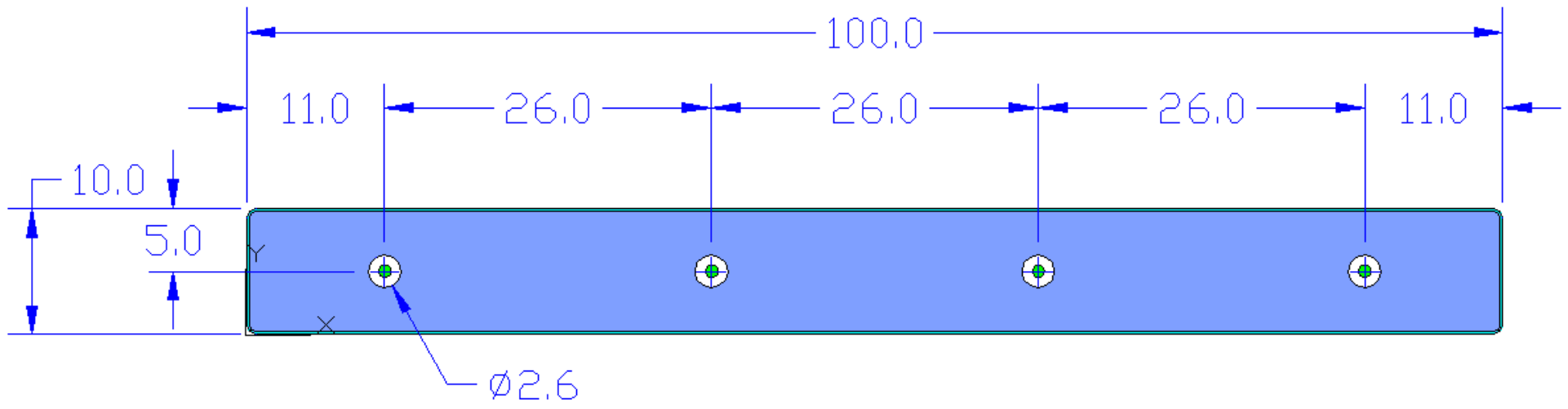

Figure 11.4. Profile of an extruded scintillation counter. All units are mm.

The waveshifting fiber absorbs the violet (400-450 nm) light produced in the scintillator extrusions and re-emits it in the green (500-600 nm), trapping the light by total internal reflection. Suitable fiber is available from Kuraray [7], which has made fiber for MINOS, MINERvA and NOvA. The selected fiber is double-clad, 1.0 mm in diameter with a polystyrene core covered with a thin (~3% of the fiber diameter) inner polymethylmethacrylate (PMMA or acrylic) cladding and an outer fluorinated-polymer shell (1-3% of the fiber diameter). The relative indices of refraction are 1.59, 1.49, and 1.42 for the core, inner, and outer cladding, respectively. The polystyrene is doped with ~175 ppm of Y11 (K27) waveshifting dye. Non-S type fiber with a nominal S-factor of 25 is preferred because of the longer attenuation length (S-type fibers have the polystyrene chains oriented longitudinally making them somewhat more flexible, at a cost of 10% less attenuation length). The fiber emission spectrum peaks at around 500 nm. The short-wavelength part of the spectrum is significantly attenuated as the light travels down the fiber, as shown in Figure 11.5.

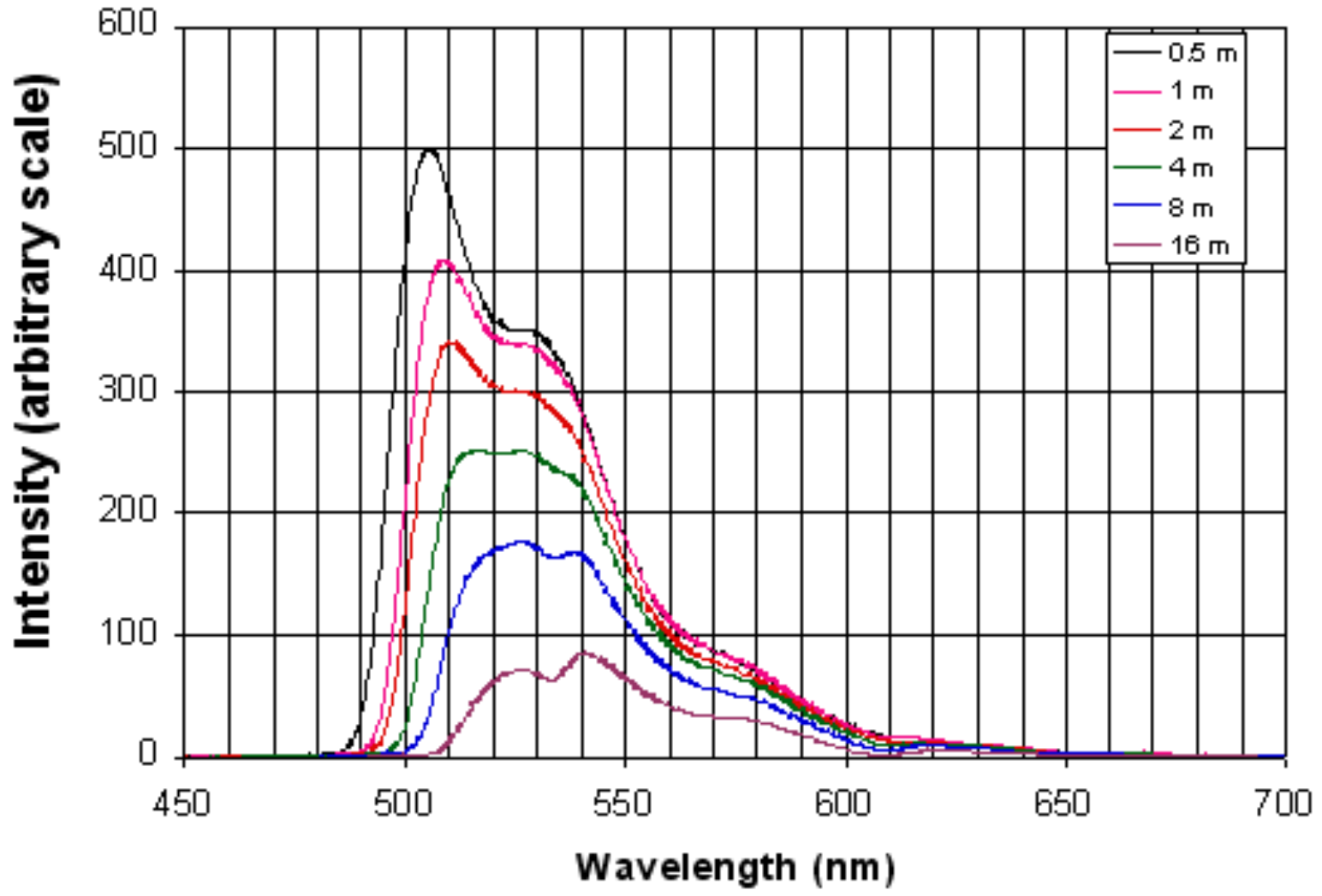

Figure 11.5. The spectrum of wavelength shifted light that survives transport through a fiber as a function of the fiber length. These measurements were made with a MINOS 1.2 mm diameter fiber for fiber lengths of 0.5 m, 1 m, 2 m, 4 m, 8m and 16 m.

Each end of the scintillator extrusions has a fiber guide bar glued to it with funnel-shaped holes that serve to guide the fibers to well-defined positions relative to the SiPMs (see Figure 11.6). The fibers are potted with an opaque epoxy and then cut and polished. The SiPMs are mounted onto the SiPM mounting blocks that are attached to the fiber guide bars, with registration holes to exactly position the SiPMs relative to the fibers. The SiPMs are mounted in cans with lips on which rubber O-rings press to keep them flush against the fiber guide bars. A flasher LED, which is used to monitor the SiPM performance, is inserted into the fiber guide bar. The SiPM mounting blocks are designed to be removed easily in the case of a SiPM failure.

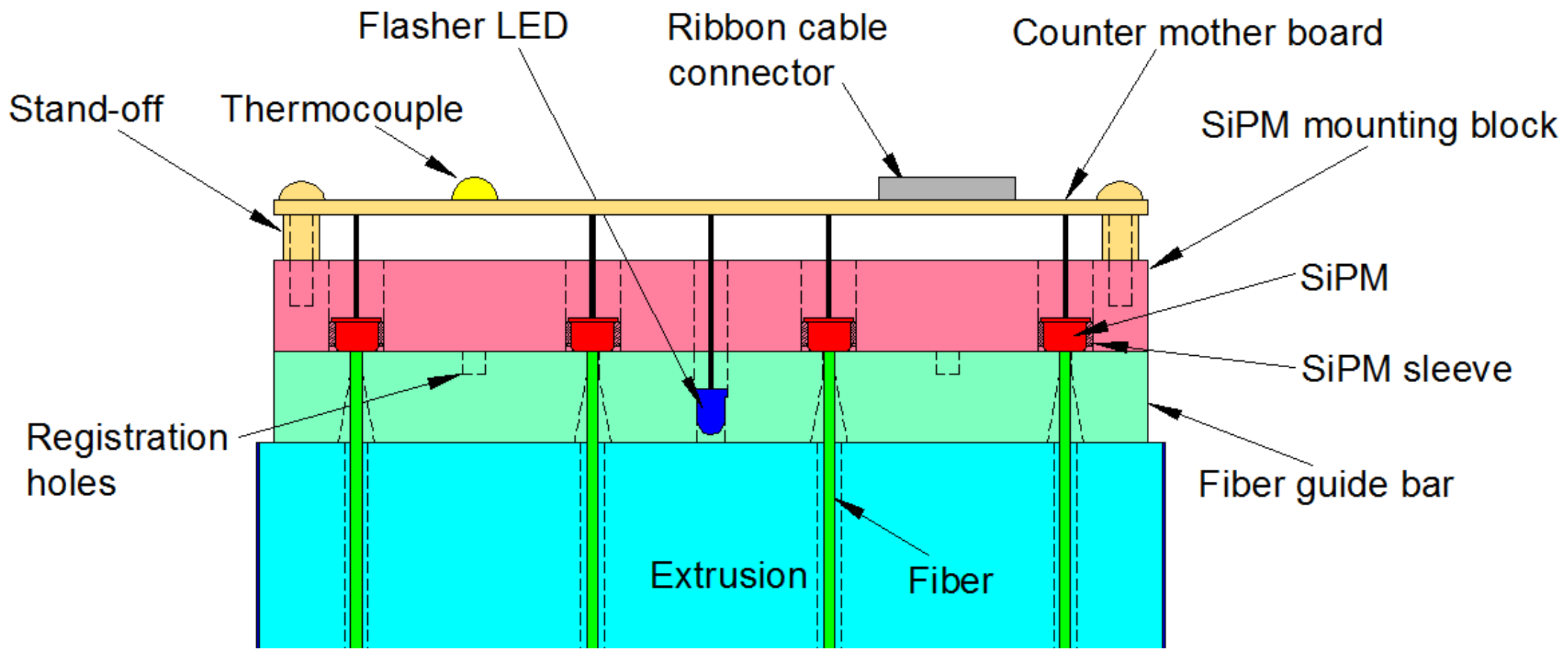

Figure 11.6. The end of a counter showing the mounting of the SiPMs and flasher LED.

The electrical connection to the SiPMs and the flasher LEDs is through the counter mother board, which is screwed to the SiPM mounting block. A temperature sensor is used to measure the local temperature that is needed for adjusting the SiPM bias. An HDMI cable provides the connection to the front-end boards.

Three layers of counters are grouped together to form a module, which is 4766 × 1241 × 41 $mm^3$ in size and weighs 328 kg (Figure 11.7). Modules are 12 counters wide and contain a total of 36 counters. The counters in each layer are glued to 4.8-mm-thick Al plates, which serve two purposes. First they provide mechanical support for the modules. Second, they prevent electrons from neutron capture gammas from traversing multiple counters to produce spurious coincidences. The counters within a module are offset by 20 mm in order to prevent projective cracks for muons that are close to normal incidence. The front and back of the top and bottom layers of counters are covered by a thin Al sheet onto which mounting bars are attached which allow the modules to be hung. Enclosures for the front-end boards are attached to the front Al cover. HDMI cables run from the counter mother boards to the front-end boards.

The scintillator extrusions will be fabricated at Fermilab and shipped out to the module fabrication factory in Virginia, where the counters and modules will be assembled and tested before being shipped to Fermilab to be installed in the detector.

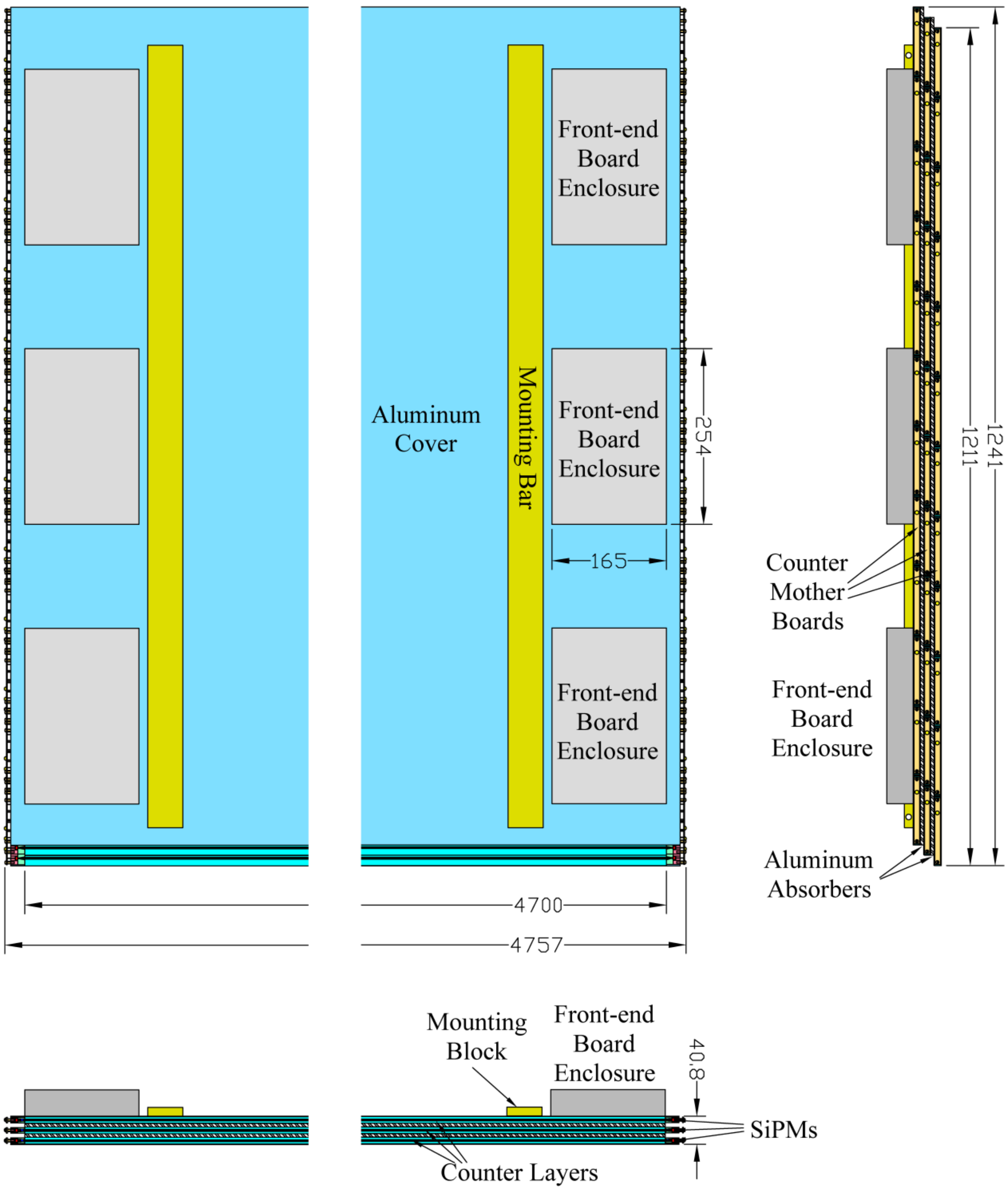

Figure 11.7. Front, top, and side views of a cosmic ray veto module. The cables from the counter mother boards to the Front-end boards are not shown. All units are in mm.

### *11.3.3* Photodetector

Readout of the cosmic ray veto is simplified by the fact that it resides on the outside of the Detector and Transport Solenoids and is easily accessible, the channel count is small, and the detector rates are low. The baseline photodetector is the so-called silicon photomultiplier (SiPM) [8] as it offers excellent reliability and ruggedness, immunity to the magnetic field and simple self-calibration. SiPMs are also very compact in size, which makes double-ended readout of the cosmic ray veto modules possible without unacceptably large dead regions in the coverage. This adds redundancy and makes the overall detector more reliable as each counter end will have a completely different readout chain. Hence, single device failures will not preclude attaining the desired detector efficiency requirement. Another advantage is that SiPMs can be customized in both size and packaging at low cost. Large-pixel (>50 μm) SiPMs are preferred because of their higher effective quantum efficiency (due to a better geometrical packing fraction). The larger dynamic range of small-pixel devices is not needed. Although the use of SiPMs is relative new in particle physics, there is considerable experience at Fermilab with these devices.

For the baseline design, the ASD-SiPM1C-M-40, a circular SiPM with a diameter of 1.2 mm in a TO-18 package from AdvanSiD [9] has been chosen. The parameters of this device are given in Table 11.3. The device is covered with a clear epoxy layer to protect the wire bond. These SiPMs have been evaluated in a test beam, which successfully demonstrated that the required efficiency can be achieved with single-ended readout. However, the availability of similar devices from other manufacturers (e.g. the CPTA 151-30 [10] or the HPK MPPC S10362-11-100U [11]) will allow us to choose the most economical detector to meet our needs. These detectors are designed to be interfaced to 1 mm WLS fibers and are well matched in sensitive area and wavelength and have higher effective quantum efficiency than most photomultipliers. The gain is sufficient to make the readout relatively simple. The excellent signal dispersion (the ability to distinguish one avalanche from two, three and so on) makes the calibration and monitoring of the photodetector straightforward. While the noise rate of single pixel avalanches is very high when compared to, for example, photomultipliers, the noise rate at the proposed threshold of 3 photoelectrons is much lower, and it is expected the rate will be dominated by signals from neutrons produced in the muon beamline. To overcome the sensitivity of the breakdown voltage on temperature (typically 2% per degree C) a digital temperature sensor, located on each counter, will be used as input to automatically adjust the operating bias on all photodetectors.

The large number of SiPMs in the cosmic ray veto demands a clear workable plan for photodetector quality assurance. Simply put, for every SiPM used in the detector, there must be an operating point where the electrical response to photons is big enough and the noise is small enough. Furthermore, it is a requirement of the design that the electronics can perform the necessary calibration and monitoring functions. The plan for calibration and monitoring is described in Section 11.8.

| | |
|---|---|
| Effective active area | 1.13 mm |
| Cell size | 40 × 40 μm$^2$ |
| Cell number | 660 |
| Spectral response range | 350 to 900 nm |
| Photon detection efficiency | 18% (@ 480 nm) |
| Breakdown voltage (BV) | 35±7 V |
| Working voltage range | BV+2 V to BV+7 V |
| Dark count | 2.5-4.0 × 10$^6$ |
| Gain | 1.6 × 10$^6$ |
| Breakdown voltage sensitivity | 76 mV/°C |

Table 11.3. Typical AdvanSiD ASD-SiPM1C-M parameters.

To fully characterize the SiPMs, a fairly sophisticated setup is required. A stable light source with the correct spectral characteristics is required. The temperature of the SiPMs must be controlled to about 1 °C. The bias will need to be adjustable with an accuracy of about 5 mV and a current measurement with an accuracy of about 1 nA is required. A high-speed digitizer (such as a digital oscilloscope) and a wide-bandwidth, low-noise amplifier are also required. These exist at the SiDet facility at Fermilab. This setup will allow a quantitative comparison of SiPMs from various vendors as well as a measurement of all relevant parameters for a sample of SiPMs used for production. This will also provide a sample of well characterized photodetectors for system integration tests such as prototype counters and eventually a full prototype module.

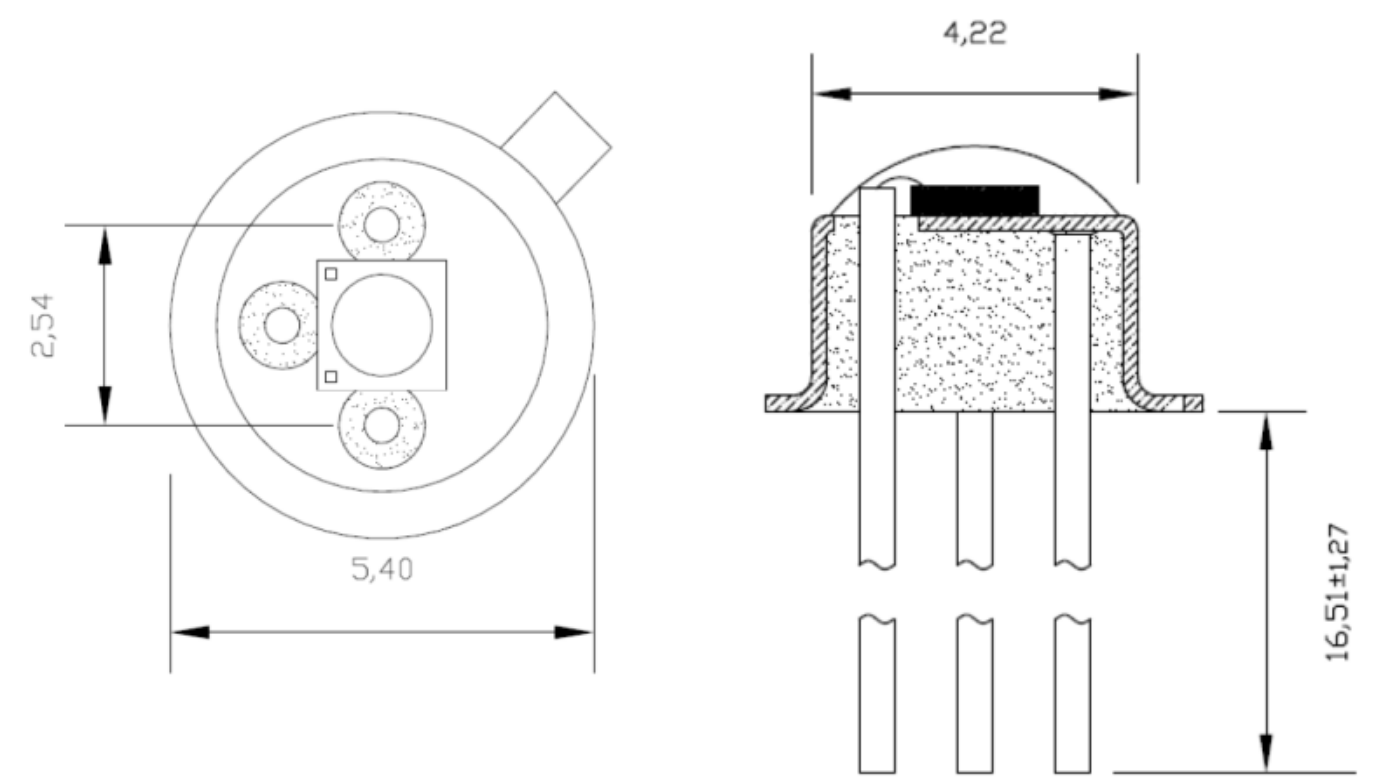

Figure 11.8. AdvanSiD ADD-SIPM1C-M-40 in TO-18 package.

### *11.3.4* Front-end Electronics

The front-end electronics consists of: (1) a mother board mounted directly on both ends of the counters onto which the SiPMs, flasher LEDs, and temperature sensors are connected; (2) a front-end board, which reads out and digitizes the data from the SiPM, both in time and charge, controls the flasher LED and provides bias to the SiPMs; and (3) a readout controller, which takes the data from the front-end boards and sends it to the data acquisition system and provides a means of communication with the front-end boards. The system has been designed to be simple, redundant, and inexpensive.

The front-end readout board uses design elements from the cosmic ray veto used for the Chicagoland Observatory Underground Particle Physics experiment (COUPP) [12] and the Fermilab/Northern Illinois University proton tomography project. This board will supply the necessary SiPM bias voltages, read out the SiPM signals, control the LED flasher, and read the thermosensor. One of the salient features of this design is the use of commercial off-the-shelf parts; no custom parts are employed. A block diagram of the board is shown in Figure 11.9. The SiPM signals are fed to commercial ultrasound processing chips, each of which has eight sets of low-noise preamplifiers, programmable gain stages, programmable anti-alias filters and 80 Msps, 12-bit ADCs. The digital portion of the card uses a DDR SDRAM for buffering data and FPGAs for converting the serial ADC data to parallel, applying thresholds to the digitized data for zero suppression, controlling the SDRAM and implementing the serial data links for communication to the readout controllers. A microcontroller is used for status and slow control. Power is supplied over the same category 5 cables used for the data link to the readout controllers, simplifying the cable plant. Commercial Power over Ethernet protocol will be used. One front-end board can accommodate 48 SiPMs. A total of 366 boards are required for the entire cosmic ray veto, including three spare modules.

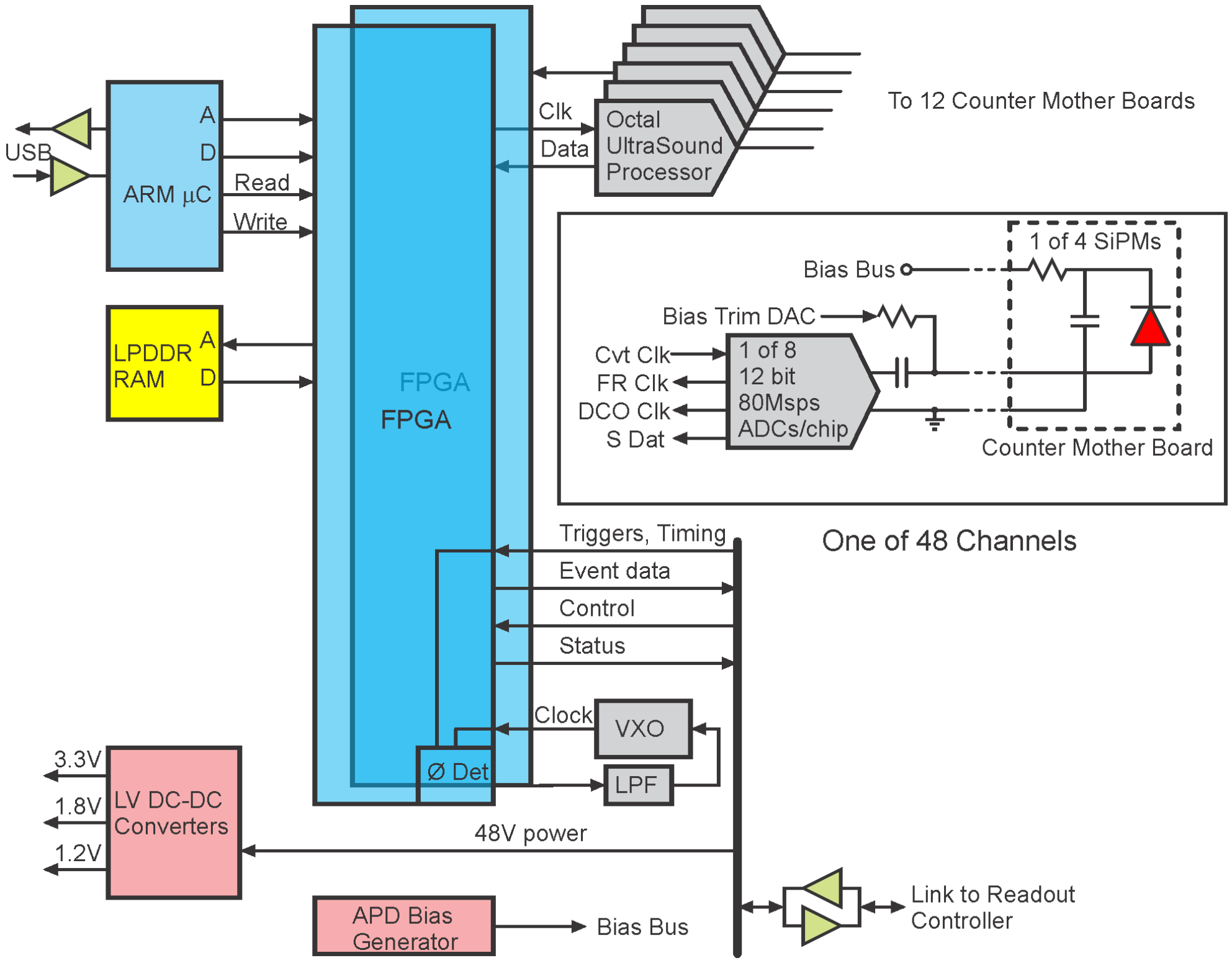

Figure 11.9. Block diagram of the front-end board.

Although the cosmic ray veto is an inherently digital device – it is only necessary to know that a counter has fired – for calibration purposes it is vital that the number of SiPM pixels that have signals be known in order to monitor the photoelectron yield. Moreover, the performance and level of integration available in the ultrasound processor chips is such that it is cheaper and requires less power to use ADCs to take a waveform trace of the SiPM signal and use a centroid finding algorithm implemented in the FPGAs to perform the timing function. The expectation is that with an 80 Msps sampling rate, 3 ns timing resolution will be achievable.

The outputs of the front-end boards go to a readout controller, which also serves as the link between the front-end boards and the slow and fast DAQ systems. Figure 11.10 shows a block diagram of a readout controller. Each readout controller will have 24 ports for front-end boards and two fiber-optic links. Fifteen readout controllers are needed for the entire cosmic ray veto. The topology of the fiber-optic links depends on the data rates and our desire for maximum redundancy. Initially, the plan is to daisy chain at least two readout controllers onto each link to the DAQ system.

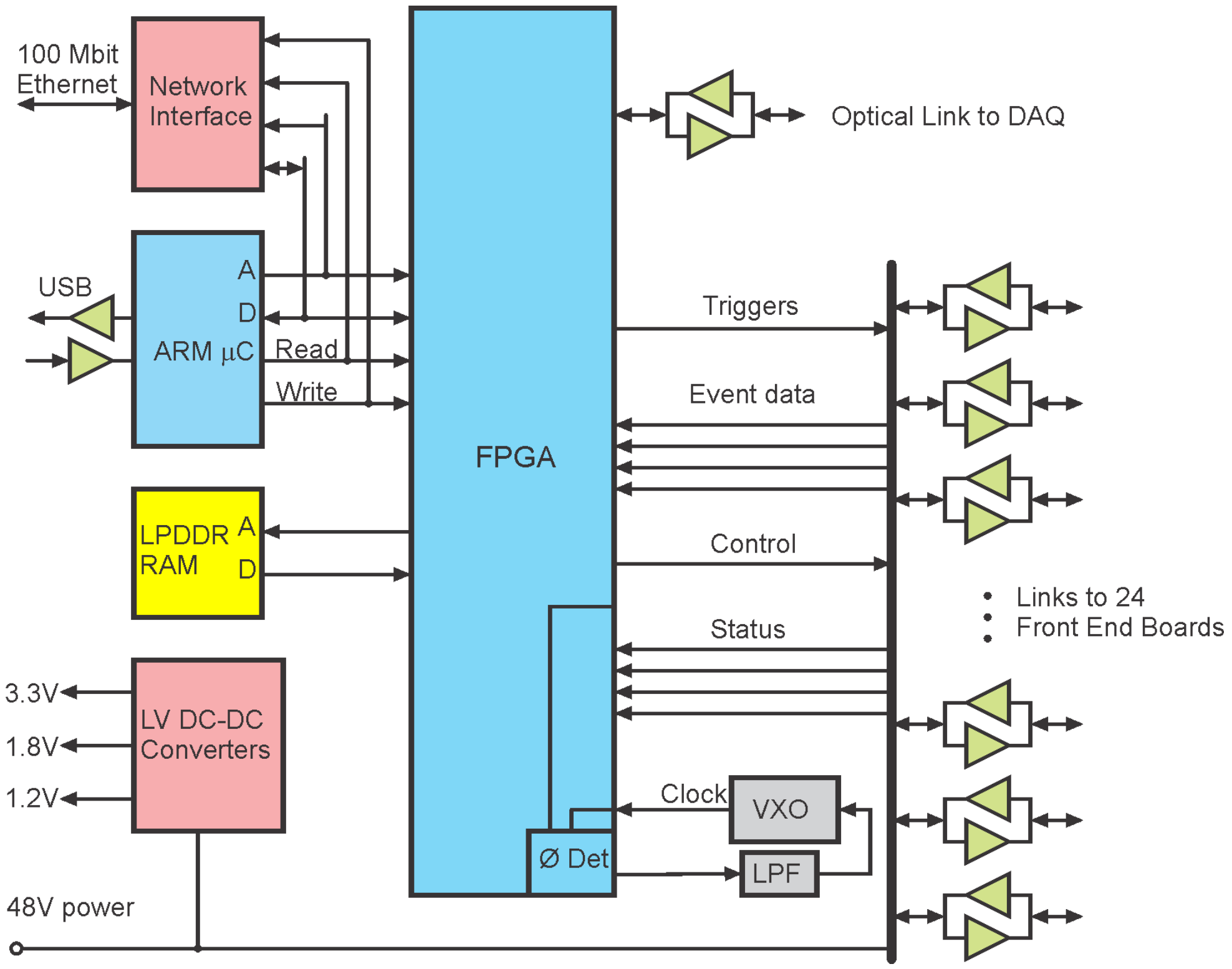

Figure 11.10. Block diagram of the readout controller.

The maximum rate in a counter is expected to be about 6 kHz, resulting primarily from background neutrons [13]. Assuming a 6-byte event header word and three 2-byte trailer words, the maximum front-end board readout rate will be about 0.8 MB/s, assuming that all SiPMs fire at the same rate. Simulations predict that the *total* hit rate in the cosmic ray veto will be about 4.5 MHz [13]. This produces an *overall* readout rate for the cosmic ray veto of 55 MB/s. The system will be designed to handle at least twice this rate to provide adequate margin for the uncertainty associated with the predicted neutron rate.

### *11.3.5* Trigger

The cosmic ray veto will be used to produce cosmic-ray muon triggers for calibration purposes. A simple coincidence between pairs of adjacent counters on different layers will be used to indicate the presence of a cosmic-ray muon. The spatial positions of the hit counters will allow a very crude track stub to be formed. Timing resolution of 5 ns on individual counter hits will greatly reduce the false coincidence rate from noise and backgrounds. A window of ~50 ns will be formed around the coincidence time. Events falling within that timing window will generate a signal that will be passed on to the data acquisition system where it can be used in firmware or software. Monte Carlo studies are planned to refine the veto criteria.

### *11.3.6* Calibration

The copious rate of cosmic-ray muons traversing the cosmic ray veto allows it to be calibrated during periods between beam pulses and when the beam is not operating. Beam-off running also allows the cosmic-ray background to the conversion signal to be directly measured, although the rate is quite low (about one event in the signal region per day). With relaxed cuts on the signal region roughly thirty times more off-spill than on-spill cosmic-ray produced conversion-like events, are expected or about one event in a nominal three-year run, assuming a $10^{-4}$ inefficiency.

The performance of the photodetectors and readout electronics must be carefully monitored and bad channels must be fixed promptly. The redundancy built into the design both allows for some failures and greatly facilitates calibration. In addition to the calibration provided by cosmic-ray muons, an active calibration system that provides rapid feedback will be implemented. LEDs on both ends of each counter will be pulsed and the SiPM response will be checked. Experience from the DØ fiber tracker has shown that LEDs can be very stable over a number of years [14].

## 11.4 Expected Performance

### *11.4.1* Prototype Studies

A prototype module similar to the baseline design was fabricated and tested using both cosmic rays and beam at the Fermilab Meson Test Beam facility [15]. The prototype used scintillator extrusions fabricated by Itasca Plastics [16]. The counters were co-extruded with $TiO_2$ and contained surface grooves for the fibers, rather than embedded channels, with the fibers glued in place. The counters had the same length, profile and chemical composition as the baseline design described above. Kuraray double-clad Y11 fibers with a 1.2 mm diameter and a reflective Al coating on the far end were read out with 1.2 mm diameter AdvanSiD silicon photomultipliers (SiPMs). The photoelectron yield for a minimum ionizing particle at normal incidence was found to be 13 at the far end of the counter.

It is difficult to extrapolate precisely this result to our baseline design, which has one additional fiber, fibers in the volume rather than the surface and double-ended readout. Note that we expect no decrease in light yield in using a smaller diameter fiber – 1.0 mm rather than 1.2 mm – because the SiPM was undersized in the prototype and hence light was lost. Conservatively, we expect the photoelectron yield to be at least the same as the prototype, if not larger. Prototypes of the baseline design will be fabricated and tested. If the photoelectron yield does not meet the design requirements then the yield will be tuned by increasing the fiber diameter, as well as the SiPM size. With the number of SiPMs needed for production, a custom diameter will be available at little additional cost.

### *11.4.2* Meeting the Efficiency Requirement

To determine the expected overall efficiency we have run a Monte Carlo simulation in which cosmic-ray muons were generated with the standard $\cos^2\theta$ zenith angular distribution and propagated through three layers of cosmic ray veto counters, corresponding to the configuration of the baseline design. The number of photoelectrons produced in each counter by a traversing muon was randomly determined using a Poisson distribution with a mean of 13 photoelectrons per cm at the far end of the counters, as measured with the prototype counter described above. The counter response was assumed to be uniform over the entire transverse profile. The simulation was used to: (1) determine the minimum offset distance between counters, (2) the maximum gap that can tolerated, and (3) the photoelectron yield needed to achieve the required efficiency. See Figure 11.11 for a definition of the terminology used.

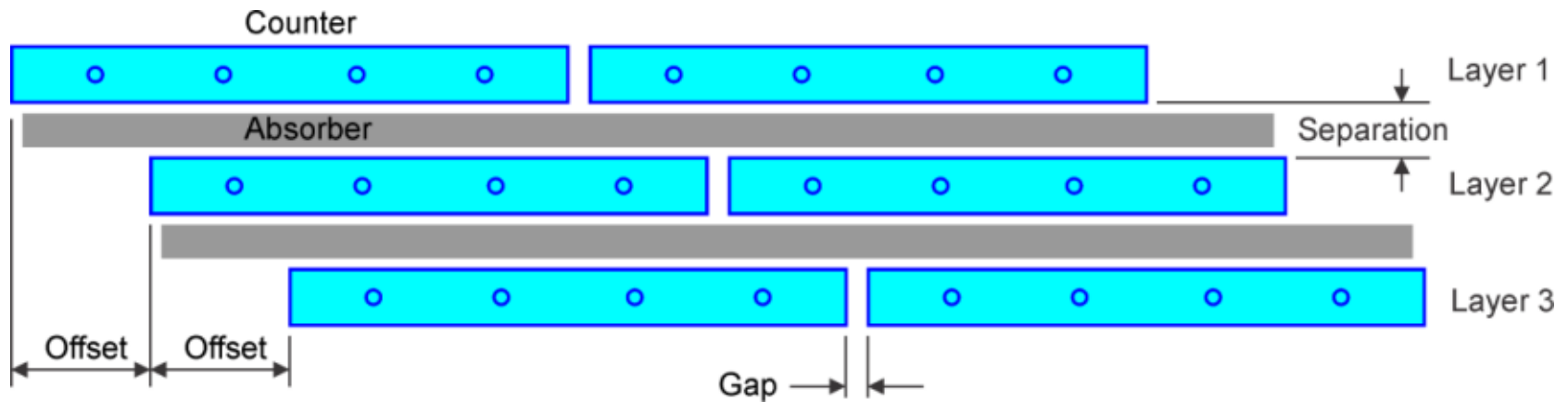

Figure 11.11. Cosmic ray veto nomenclature.

The effect of the offset distance between counters in different layers is shown in Figure 11.12 for a gap of 2 mm, the expectation from the fabrication process, and for photoelectron thresholds ranging from 1 to 6. An offset of 20 mm is sufficient for all gaps, after which there is no gain in efficiency. Using a somewhat smaller offset of 15 mm, the photoelectron yield required to produce an overall inefficiency of less than $0.5\times10^{-4}$ was determined. It depends on the photoelectron threshold and the gap separation between counters, as shown in Figure 11.13. With a threshold of 3 photoelectrons – which is needed to reduce the SiPM noise rate to an acceptable level – and a gap of 2 mm, at least 10 photoelectrons per cm are needed to achieve an inefficiency of $0.5 \times 10^{-4}$, which is the expectation based on prototype tests. Unpublished MINOS studies have shown that the scintillator and fiber exhibit a 3% light yield reduction every year. Over a 10-year period this would result in a 26% decline. Hence, an initial yield of 13 p.e. is necessary to maintain the required inefficiency for 10 years. Figure 11.12 shows that with 13 photoelectrons per cm and a gap of 2 mm an inefficiency of $1 \times 10^{-4}$ can be achieved with a threshold as high as 5 photoelectrons.

Note that the above simulations were done for counters normal to the zenith direction. For the side counters the more oblique angles of the cosmic-ray muons will produce more light and hence, be detected with a higher efficiency.

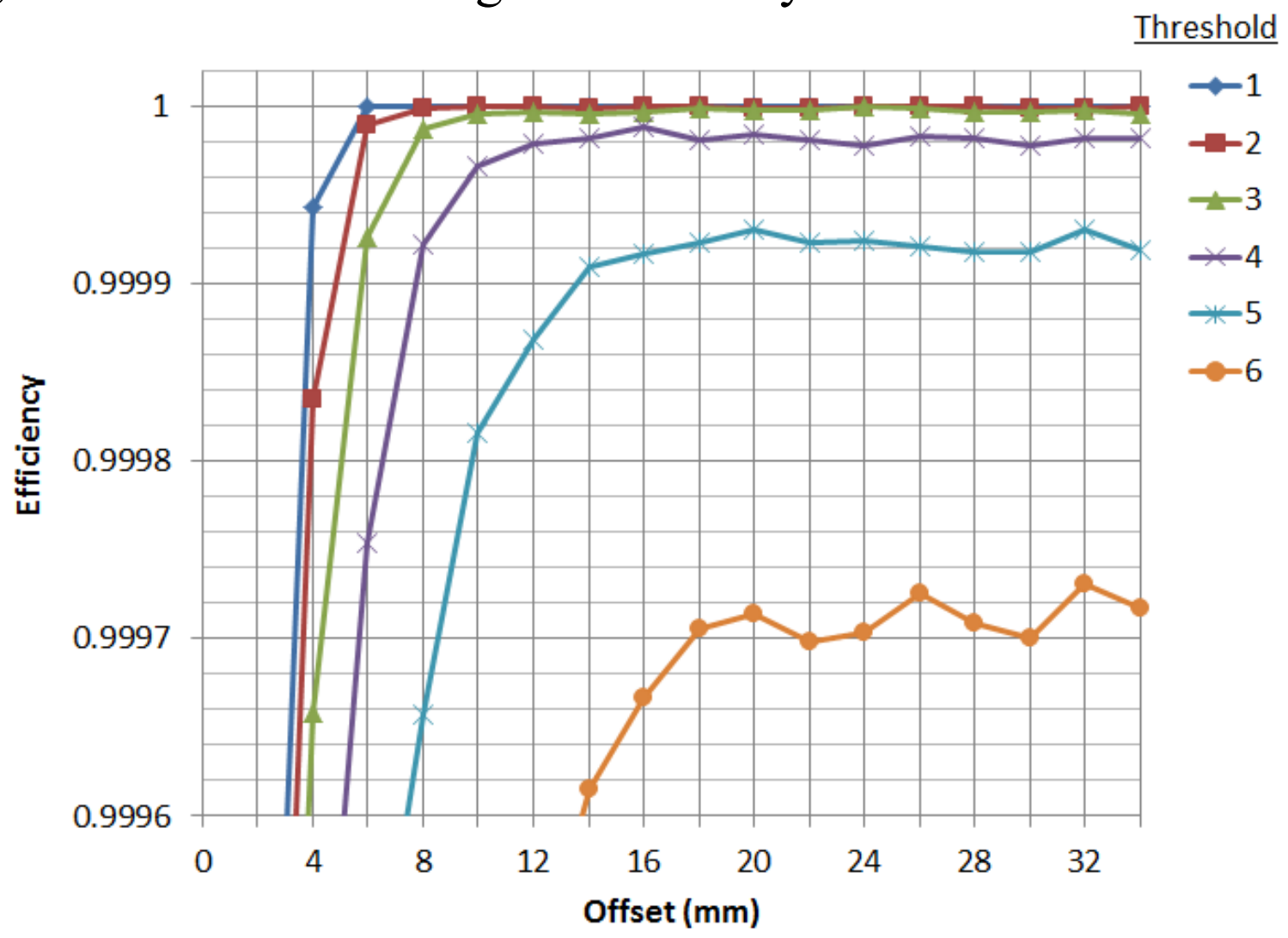

Figure 11.12. Efficiency vs. counter offset distance for six different photoelectron thresholds. A gap of 2 mm has been assumed and a mean of 13 p.e./cm was used in the simulation.

### *11.4.3* Neutron Rates

Although the cosmic-ray-muon rate at the detector is modest (~70 $m^{-2}$ $s^{-1}$ $sr^{-1}$ at the surface), the production target, collimators, muon stopping target, and muon beam dump are copious sources of background particles. In particular, neutrons are produced at a rate of about 60 billion per second from the stopping target alone. The cosmic ray veto must survive this neutron induced background rate with no untoward effects. The requirement is that the total veto rate from neutrons, and other sources, should not reduce the overall detector live time by more than 1%.

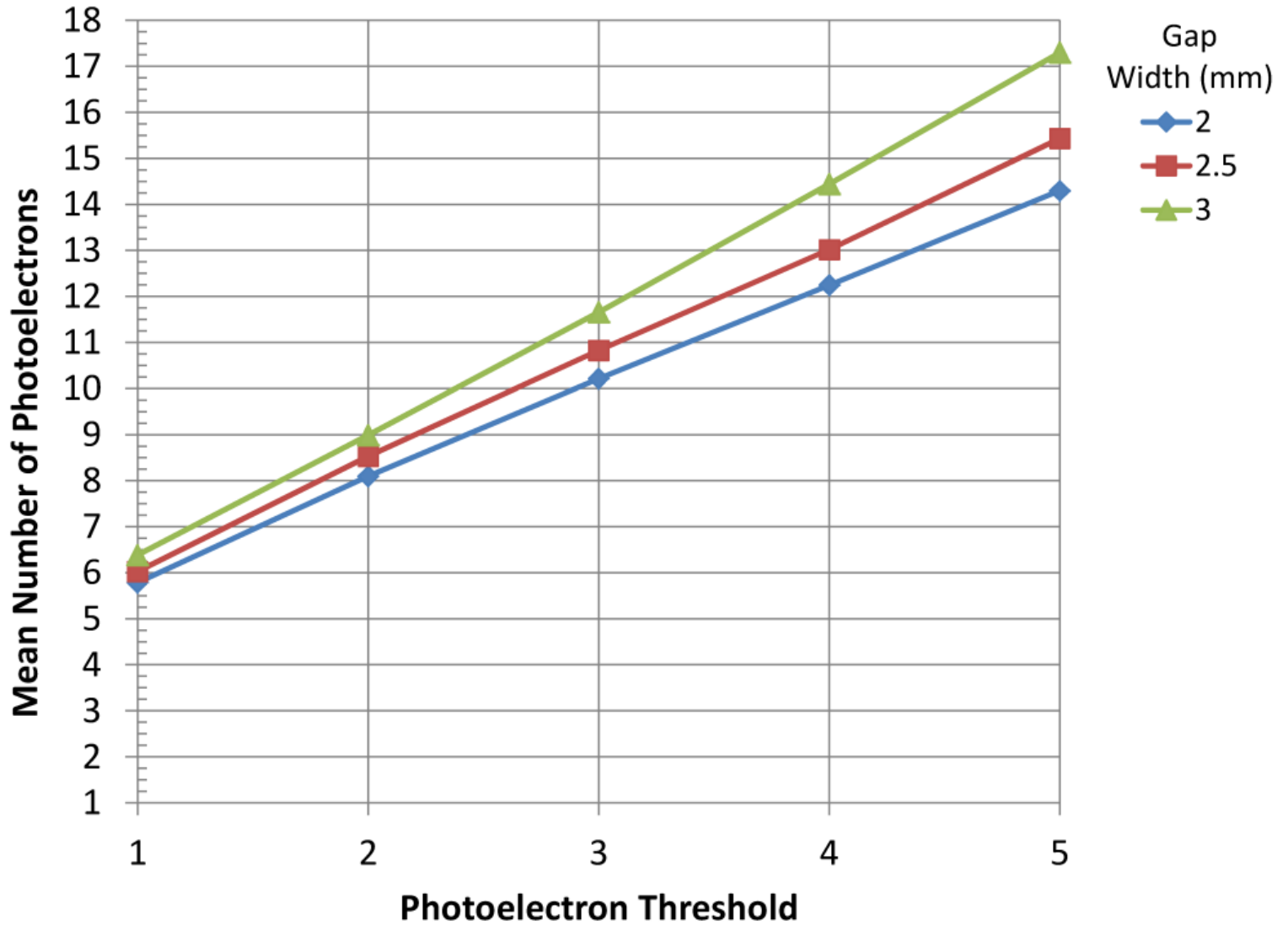

Figure 11.13. The required mean number of photoelectrons per centimeter of scintillator traversed by a cosmic-ray muon vs. the photoelectron threshold needed to achieve an overall cosmic-ray veto inefficiency of $0.5\times10^{-4}$, for three counter gap widths. The counter offset is 15 mm.

Every muon capture on Al produces, on average, 1.2 neutrons resulting in a total rate of 60 billion per second. Most (85%) of the neutrons have kinetic energies below 10 MeV, with the most probable energy about 1 MeV (see Figure 11.14) [17]. Polystyrene scintillator ($C_8H_8$) is sensitive to neutrons that elastically scatter on the hydrogen protons. The neutron cross section is large at low energies (~ 4 barns at 1 MeV), but falls off as $1/\sqrt{E_n}$ with increasing energy. Hence, scintillation counters typically have neutron efficiencies that peak at a few MeV and slowly fall off at higher energies. The scattered protons deposit their energy locally. The high ionization density quenches the light (Birks' Rule) reducing the output by about an order of magnitude compared to the light from a minimum-ionizing particle. Because of this suppression, typical scintillation counters have vanishingly low efficiencies for neutron kinetic energies below about 1 MeV.

The neutron shielding surrounding the Transport and Detector Solenoids is described in the chapter 8. A concrete neutron shield of 45.7 cm thickness (18") outside the transport and detector solenoids moderates and captures most of the neutrons. To mitigate the rate of 2.2 MeV gammas from neutron capture on hydrogen in the concrete, we may load the concrete with boron, which produces a 0.478 MeV neutron capture gamma, or with lithium, which although expensive, produces no capture gammas.

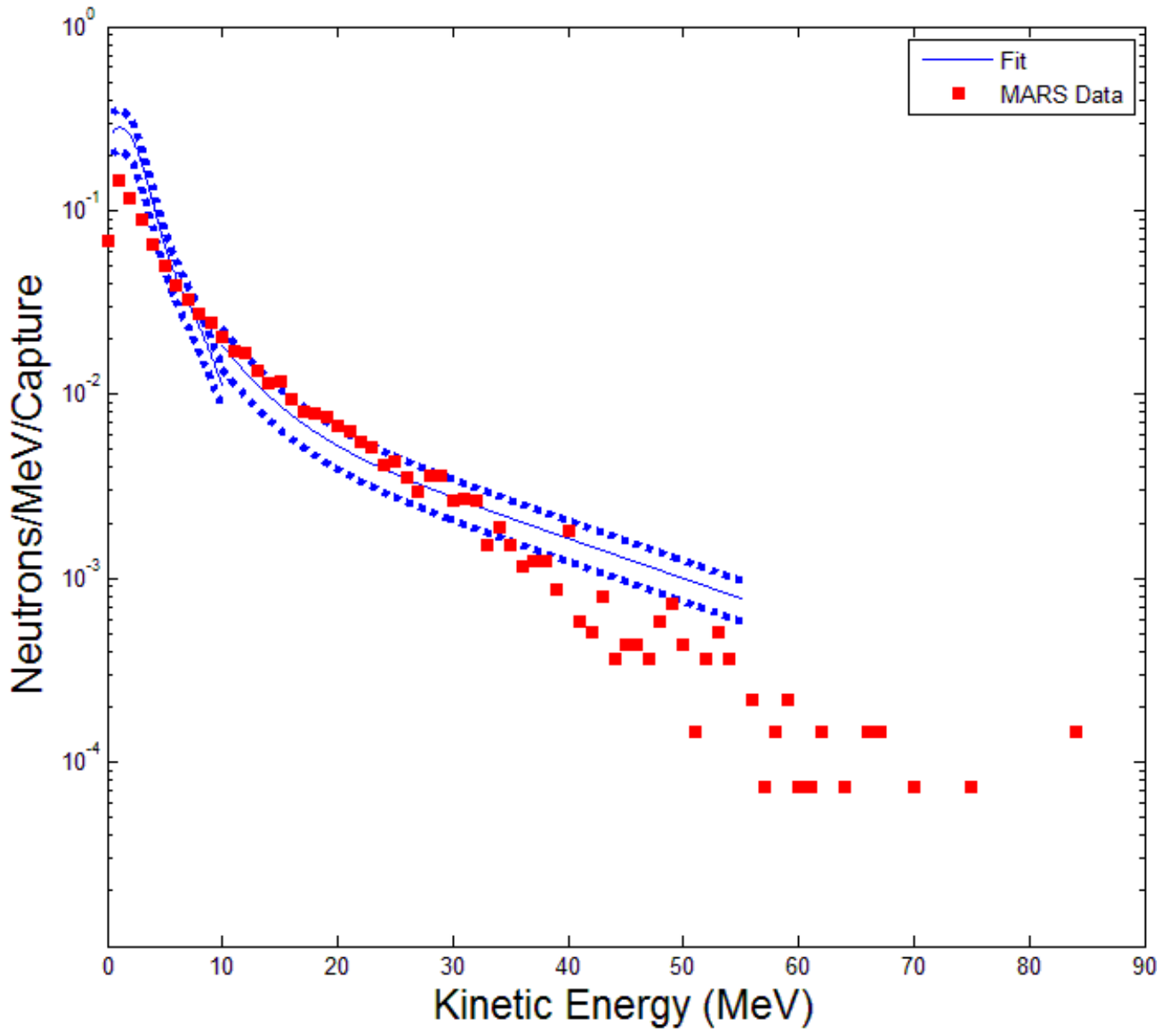

Figure 11.14. The kinetic energy spectrum of neutrons from the Al muon stopping target [17]. The blue points and associated fit come from an extrapolation of the measured Ca spectrum to Al; the red points are from a MARS calculation.

The rate of neutrons incident on the cosmic ray veto with 50 cm of Fe shielding has been simulated using Geant3 GCALOR [18], which was found to describe the experimental data well down to thermal energies [19]. We expect the rates due to concrete shielding to be similar based on the studies of Ref. [19] (Figure 11.15). The rates from the production target, muon stopping target, and the collimators in the Transport and Detector Solenoids were simulated. The rate from the muon beam dump was not estimated, but it can be easily shielded and should not be a significant source of neutrons in the cosmic ray veto. Since Geant3 does not simulate the products from negative muon capture the produced neutron kinetic energy spectrum was put in by hand (Figure 11.14). Birks' Rule was implemented in the code to correctly model the response of the scintillator. Events that deposited more than 50 keV in the scintillator were recorded. This corresponds to about 1/40 the energy from a minimum-ionizing particle at normal incidence. Light attenuation was not simulated. Neutrons from the muon stopping target dominate the rate in the cosmic ray veto. The maximum counter rate was a modest 6.2 kHz. The average counter rate was 1.7 kHz in the innermost layer, and slightly less in the outer layers. The simulation did not determine the pair-wise coincidence rate between individual counters in different layers. To estimate that rate, uncorrelated singles rates were assumed and the pair-wise coincidence rate was found to be 4.4 kHz. Even assuming a generous 50 ns veto window around each coincidence results in a negligible deadtime for the cosmic ray veto. The simulation is being updated using MARS [20]/MCNP [21], which does a better job of tracking thermal neutrons. The neutron efficiency of the prototype cosmic ray veto counters will be measured as well as the shape of resultant neutron-induced signal.

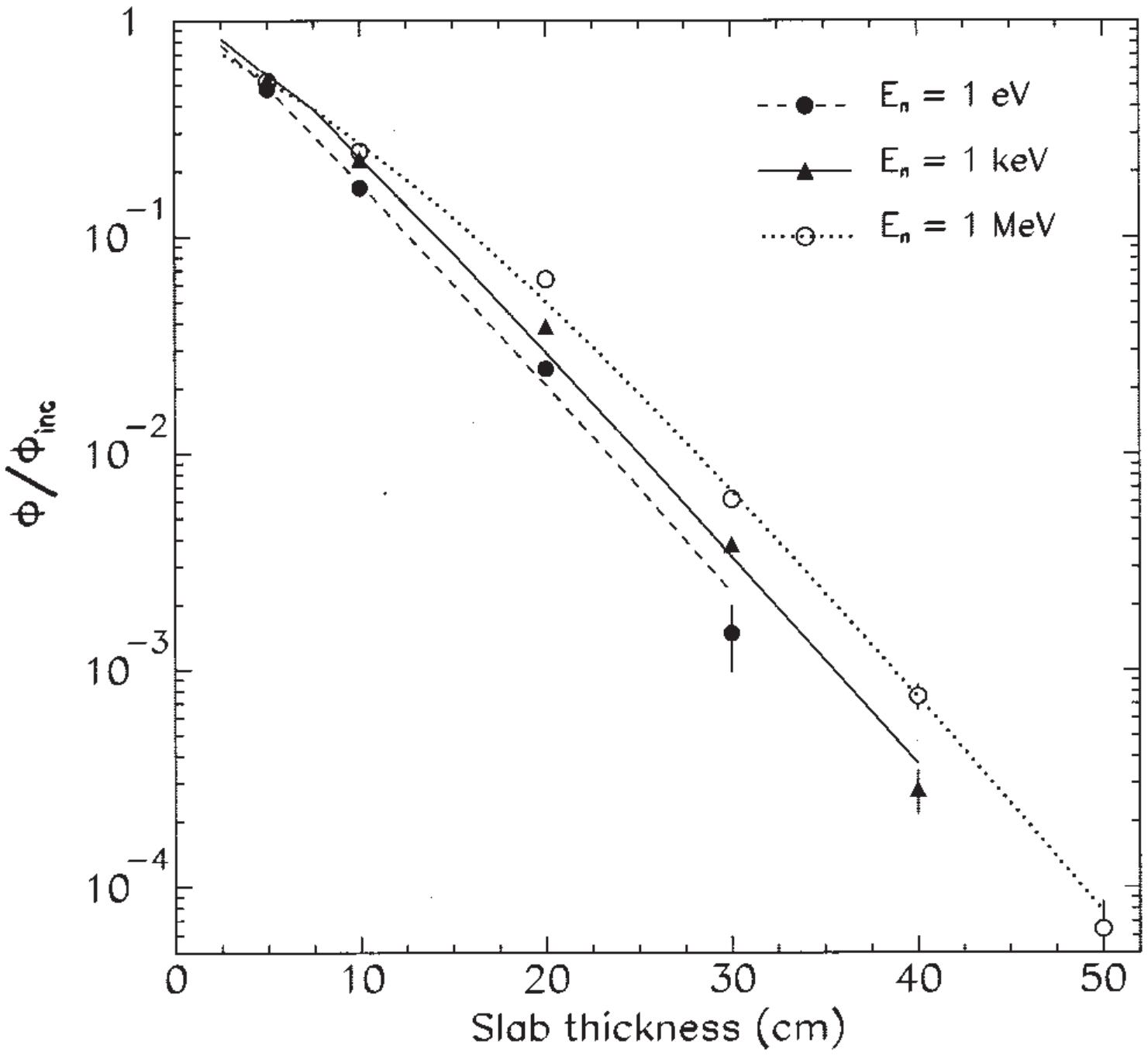

Figure 11.15. The neutron flux through a slab or ordinary concrete, from a study comparing a GEANT/MICAP simulation (points) to a prediction from MCNP-4B (lines) [19].

## 11.5 Considered Alternatives to the Proposed Design

Two alternatives to the baseline cosmic ray veto design have been thoroughly evaluated. These designs were not selected as the preferred alternatives for the conceptual design for several reasons. Primary considerations were cost, design feasibility and ease of fabrication.

Traditional photomultipliers are an alternative to SiPMs and gas-based detectors are an alternative to extruded scintillator. An extensive design exercise was made using multianode photomultiplier tubes in place of the SiPMs and it was found that their use would preclude doubled-ended readout due to space considerations. The lack of redundancy was the major factor in rejecting this option. The cost was found to be slightly higher.

Gas-based detectors have the advantage of being inexpensive, relatively neutron blind, and have better spatial segmentation. However, they are more difficult to operate and maintain at a high efficiency and have a higher channel count. Large area detectors have been fabricated for many experiments. We have considered the use of both resistive plate chambers and cathode strip chambers. We present below a conceptual design alternative using cathode strip chambers, which in general are more robust and have a more successful history than resistive plate chambers. It was found to be about one-and-a-half times more expensive due to the large labor costs associated with fabrication.

### *11.5.1* Photomultiplier Readout

Photomultipliers have several desirable features: they are robust, have low noise, and high gain. They do not have a high quantum efficiency at the wavelength shifting fiber wavelengths and would have to be shielded from the large fringe magnetic field. They are also bulky and expensive; to control costs a single photomultiplier would have to read out multiple fibers, and the fibers would have to be routed to them. Multianode photomultiplier tubes would be the preferred form the perspective of cost and space, although a drawback is the optical and electronic crosstalk between pixels that would produce a large rate of spurious coincidences from the neutron background.

Figure 11.16 shows the design of a cosmic ray veto module employing multianode photomultiplier tubes. To achieve the required photoelectron yield six 1.2 mm fibers would be needed for each counter. Each 16-pixel photomultiplier reads out 15 counters, leaving one pixel free for calibration purposes. The fibers extend beyond the scintillator by some 25 cm to a fiber coupler, an injection molded part that aligns the 6 fibers from each counter to a single pixel of the photomultiplier photocathode. The radius of curvature required of the fibers is sufficiently large to prevent any light loss. The fibers are fixed in place by epoxy that is injected into the fiber coupler mold. The entire assembly is then cut and polished using a custom diamond-bit fly cutter. The fibers are routed to minimize the noise-induced pair-wise coincidence rate from different layers due to cross talk in adjacent pixels. The other end of the fibers would be polished and sputtered with reflective Al. The counters, fibers, and photomultipliers all lay inside a light-tight box. A light-tight feed-through connects the photomultiplier to the front-end board, which is outside the enclosure, and is easily accessible so it can be quickly replaced in case of failure.

### *11.5.2* Cathode Strip Chambers

A very large number of neutrons ($\sim 10^{18}$) will be produced in and around the detector during the life of the experiment. They are produced at the production target, the transport solenoid collimators, the muon stopping target, and the muon beam dump. Neutron absorbers are needed to reduce the rate in the cosmic ray veto due to the relatively high detection efficiency of plastic scintillator. An alternative implementation of the cosmic ray veto would utilize gaseous detectors with $< 10^{-3}$ efficiency for detecting neutrons, making them essentially neutron-blind. An alternative cosmic ray veto design consisting of arrays of cathode strip chambers similar to those developed for the ATLAS and CMS endcap muon systems at the LHC have been evaluated [22].

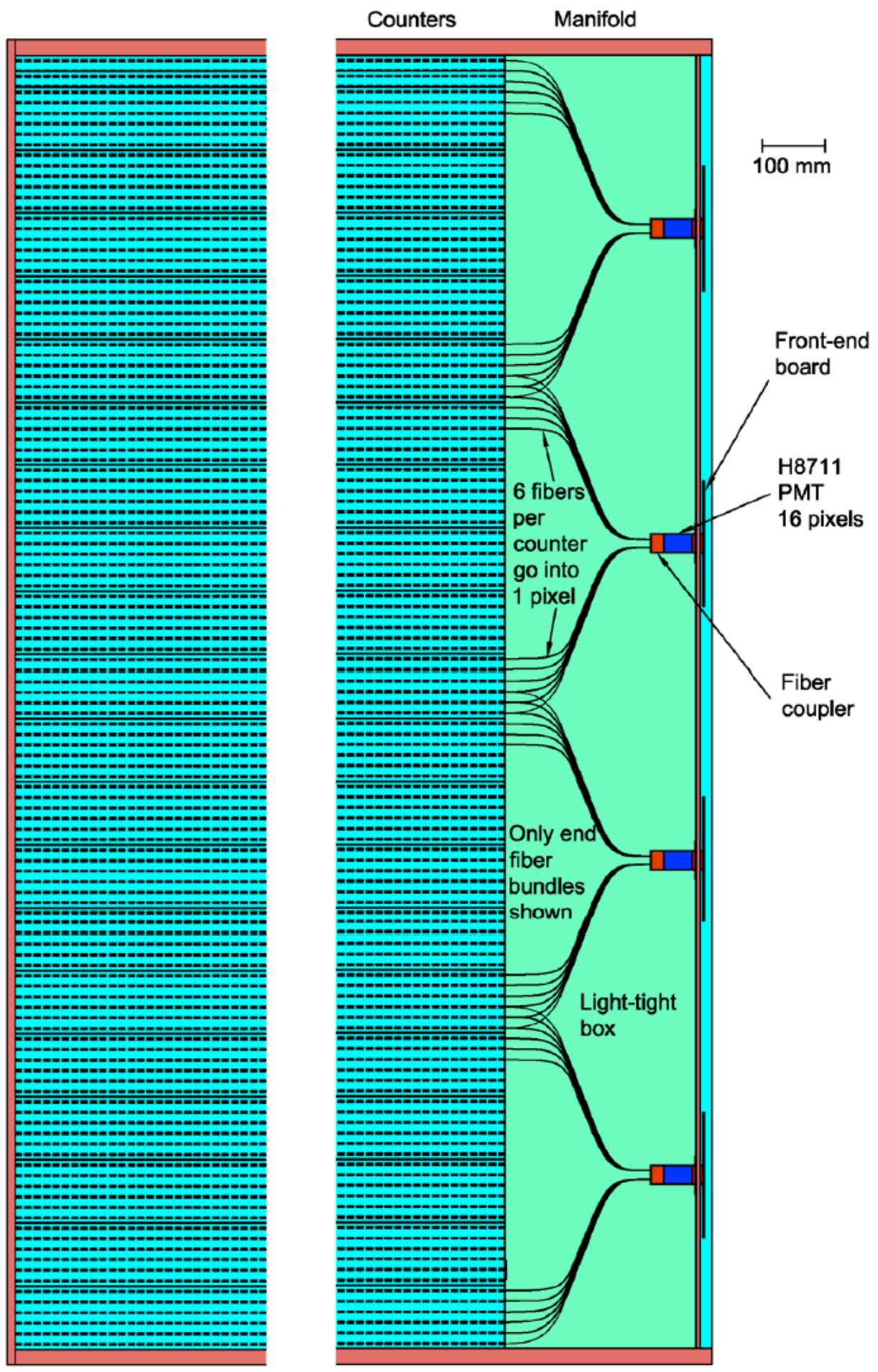

Figure 11.16. Front view of a module employing Hamamatsu H8711 photomultipliers. The routing of the fibers from the two counters furthest from each photomultiplier is shown.

Cathode strip chambers are multiwire proportional chambers developed for collider experiments requiring excellent spatial and time resolution and large area coverage [23]. They differ from conventional proportional chambers in two respects:

- The gas gap (or gaps in the case of multigap chambers) is formed by stiff yet lightweight honeycomb and copper-clad FR4 panels that can support the wire tension without heavy frames and use the copper-clad faces of the panels as the chamber's cathodes.
- The spatial and timing information is derived by reading the induced charge on one or both cathodes appropriately segmented in strips or pads according to the needs of the experiment.

Figure 11.17 shows a typical cross section of a four-layer cathode strip chambers. The absence of heavy frames results in compact lightweight chambers for easy installation over large areas in 4π detectors. By charge interpolation, a spatial resolution of ~50 μm can be obtained with a cathode strip pitch of 0.5 mm [24]. Such resolution is not needed for the Mu2e cosmic ray veto. The cathodes can be coarsely segmented to provide rough but adequate tracking with a minimum number of simple readout channels. The wires may or may not be read out for additional information. Timing resolution is limited by the ionization electrons' drift time. For a typical anode-cathode gap of 3 mm and a similar wire spacing it is of order 7-8 ns. Using the earliest arrival in a multi-layer chamber one can obtain an rms resolution of ~4 ns for a four-layer detector.

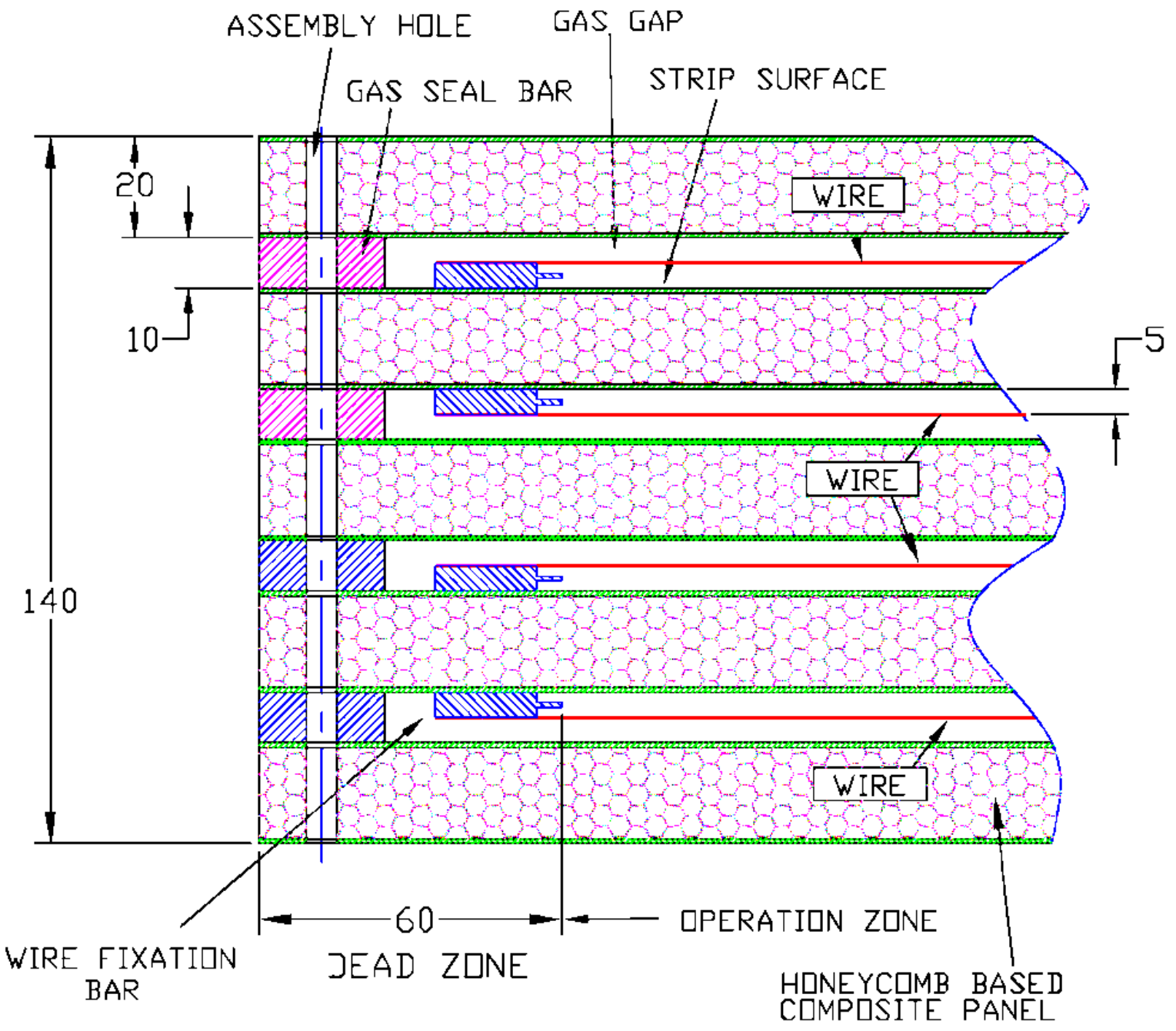

Figure 11.17. Cross sectional view of a four-gap cathode strip chamber. Dimensions are in mm.

The operating gas for a cathode strip chambers is typically a mixture of Ar-$CO_2$, sometimes with a fraction of $CF_4$ added if a faster gas for better timing is desired. $CF_4$ also provides additional quenching and some resistance to aging, which will not be a problem in Mu2e. In either case, the absence of hydrogen contributes to the very low neutron sensitivity of these detectors. In Figure 11.18 the efficiency of a cathode strip chambers for detecting neutrons in the energy range of 10 eV to 100 keV is shown. These measurements were obtained at the 2 GeV proton synchrotron of the St. Petersburg Nuclear Physics Institute using a time-of-flight method. Figure 11.19 shows the calculated efficiencies to neutrons for the four gaseous chamber technologies used in the ATLAS Muon Spectrometer. They cover the energy range 100 keV to 100 MeV. The differences among the various detectors are due primarily to gas composition and gas volume differences. For the cathode strip chambers and for the energy range of interest to the Mu2e experiment the neutron sensitivity is of the order of $10^{-4}$.

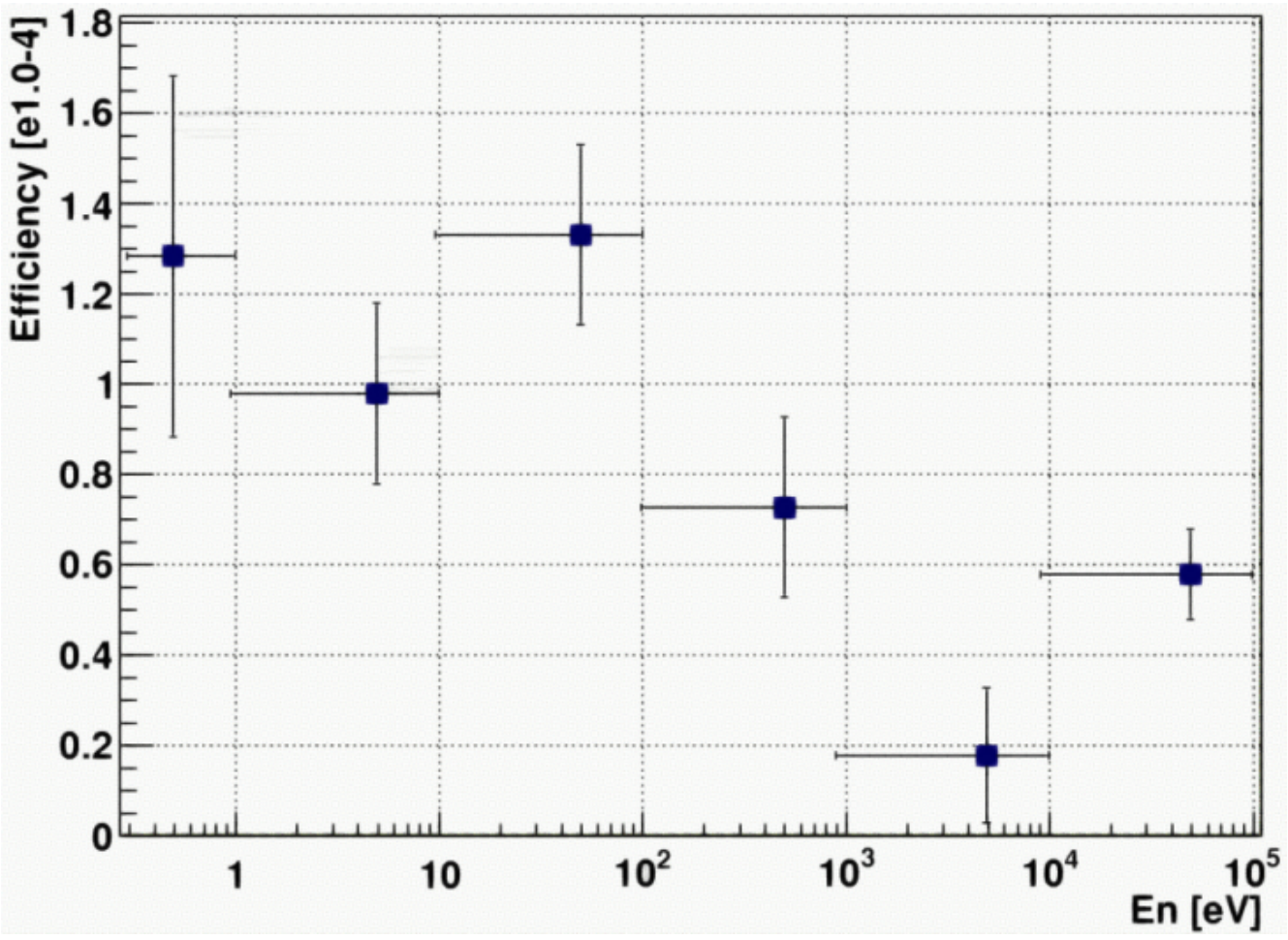

Figure 11.18. Measured neutron efficiency of a cathode strip chamber.

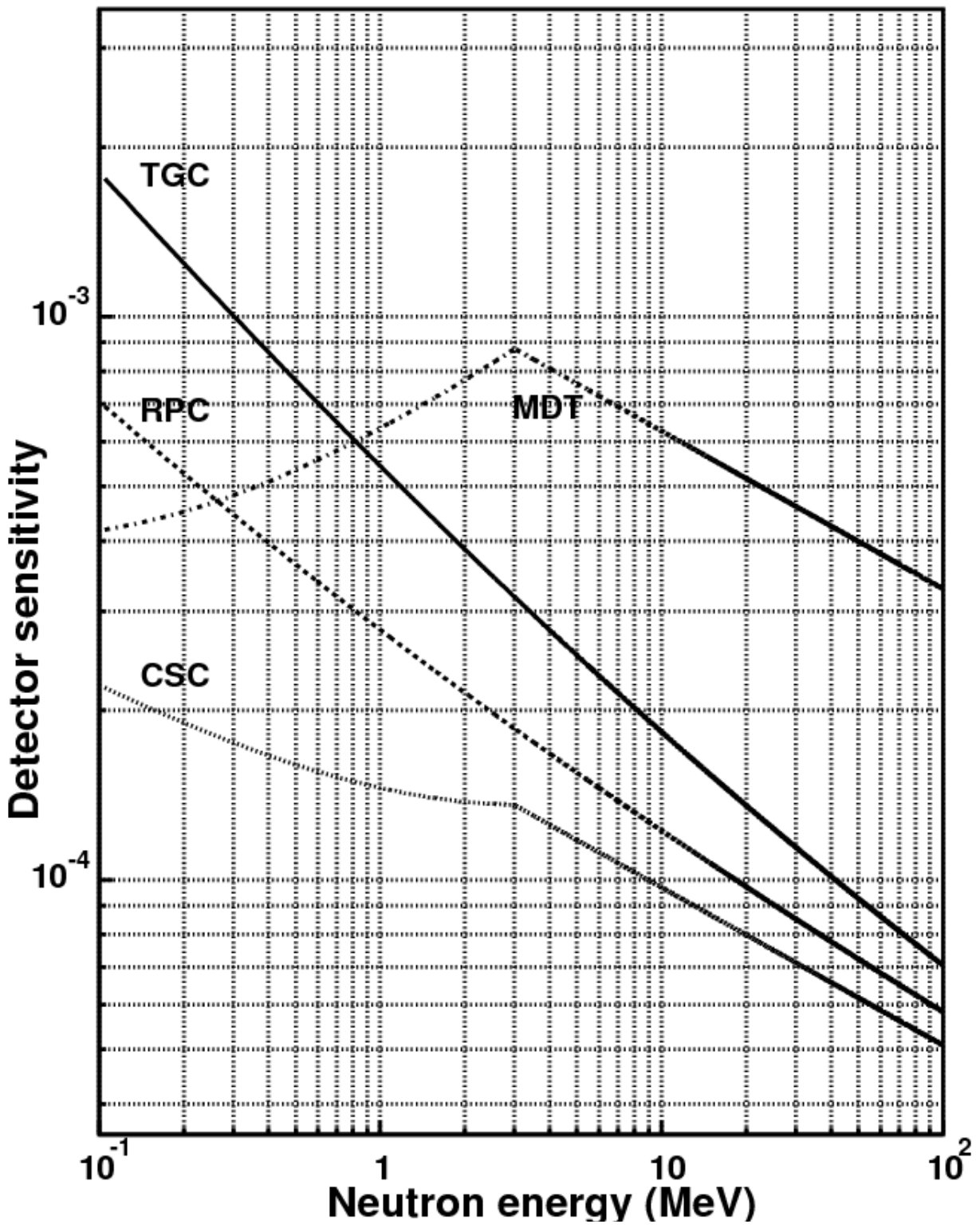

Figure 11.19. Calculated efficiencies to neutrons for the four gaseous chamber technologies used in the ATLAS Muon Spectrometer.

The maximum size of a cathode strip chamber is determined by the available material (FR4 panels), typically 8 × 4 ft$^2$ (2.44 × 1.22 m$^2$), although larger panels, 12 × 4 ft$^2$ (3.66 × 1.22 m$^2$), can also be obtained. The possibility of building slightly longer detectors, 4.7 m, in order to avoid overlapping in two dimensions will be investigated. This can, in principle, be accomplished by splicing together two panels. In this way, chambers of the same size as the baseline detector modules can be implemented as shown in Figure 11.20. Each chamber consists of four gas gaps. The efficiency of a single gap is better than 99% [22] so that the required veto efficiency of better than 99.99% can easily be achieved. As mentioned in the previous section, the cathodes can be segmented to satisfy the required readout granularity. For the layout shown in Figure 11.20, we assume a very large strip size to reduce the channel count, and hence cost. One of the cathodes in each gas gap is segmented in 32 strips that are 4 cm wide in the long direction and the other cathode is segmented into 16 short strips that are 30 cm wide. Table 11.4 shows the number of resulting readout channels. Further studies can refine this segmentation scheme, but it sets the scale of the readout system. The 4 cm wide strips would provide a ~1 cm position resolution for cosmic-ray tracking.

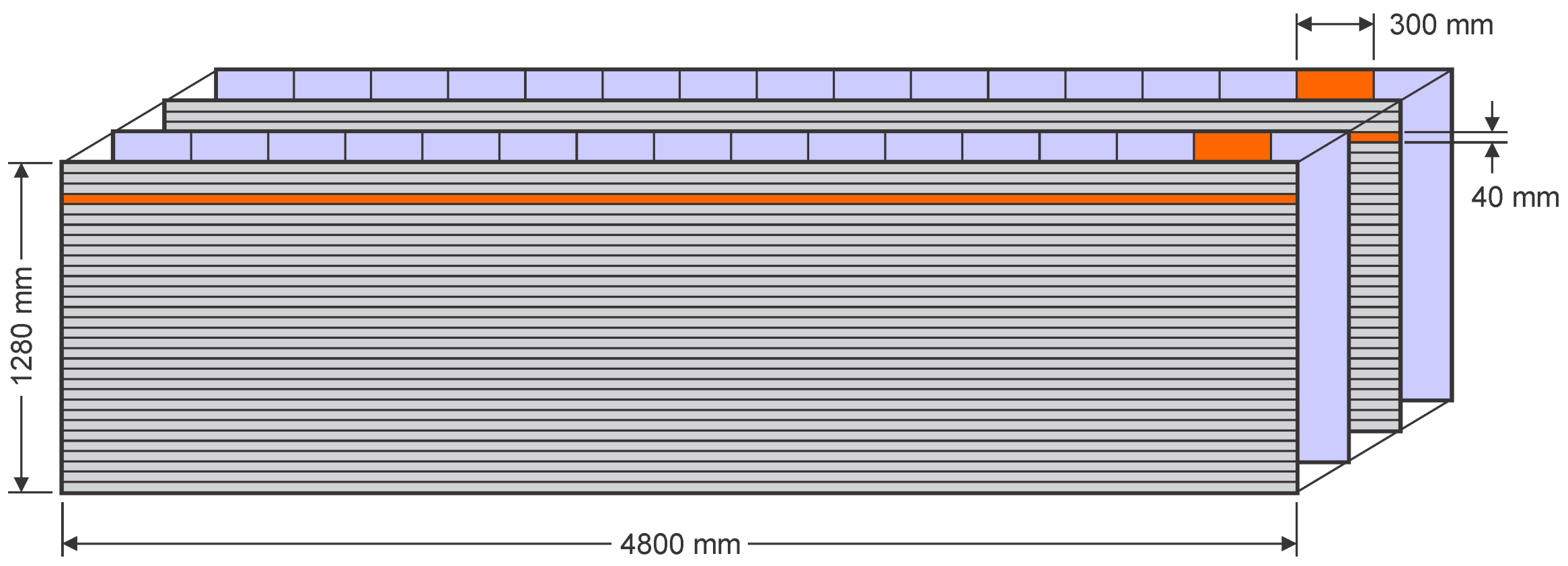

Figure 11.20. Conceptual layout of a cosmic ray veto cathode strip chamber showing the orientation of the strips.

| Strip Type | No. per Gap | No. per Chamber | Total |
|---|---|---|---|
| Long | 32 | 128 | 7,424 |
| Short | 16 | 64 | 3,712 |
| Total | 48 | 192 | 11,136 |

Table 11.4. Channel count for a cosmic ray veto constructed of 58 cathode strip chambers.

A readout system based on a front-end integrated circuit being developed for similar detectors to be used in the ATLAS experiment at CERN is envisioned. The key features of this device relevant to the cosmic ray veto are: a data driven architecture, a peak amplitude detector, a time stamp, and a built-in ADC.

## 11.6 ES&H

Polystyrene, the scintillator base and fiber core material, is classified according to DIN4102 as a "B3" product, meaning highly flammable or easily ignited. It burns and produces a dense black smoke. At temperatures above 300° C it releases combustible gases. This will have to be taken into account during the cosmic ray veto's production, assembly, storage and operation phases.

Small quantities of adhesive will be used in potting the fibers and for detector assembly. Ventilation appropriate for these quantities will be installed in the module production factory and personnel working with adhesives will wear the appropriate personal protective equipment.

The Fermilab-NICADD Extrusion Facility has a documented set of ES&H procedures that will be followed.

The size and weight of the modules require special precautions during handling. Explicit procedures for safely handling modules will be developed as part of a series of time-and-motion studies.

The photodetectors and electronics systems do not present any special safety issues. There will be no exposed low or high voltages and lockout/tag-out procedures will be used to ensure that systems are de-energized when they are being worked on. Everyone involved in the work on these systems will receive the required electrical safety training.

Each of the ES&H issues listed above are identified and discussed in the Mu2e Preliminary Hazard Analysis document [25].

## 11.7 Risks

The risks associated with the performance of the baseline design for the cosmic ray veto are few as the technology is mature and has been used successfully by several Fermilab experiments.

The most profound risk associated with the cosmic ray veto is that its efficiency for rejecting background is too low, either because the cosmic-ray-muon induced background rate is larger than expected or because the cosmic ray veto is less efficient than expected at identifying muons. The double-ended readout should give a photoelectron yield safety factor of two that should be adequate given the level of quality assurance that is planned. Prototypes will be fabricated and tested and the fiber diameter will be adjusted to meet specifications. A preproduction prototype will be built and tested to verify that it performs to specifications. Calculations of the cosmic-ray background are ongoing and every effort is being made to improve the simulation in both accuracy and statistics. We note that the Mu2e calculation described in Section 3.5.10 agrees with an earlier simulation done completely independently by MECO using a completely different code [26]. The cosmic-ray muon background will be measured as soon as the tracker has been installed in the Detector Solenoid.

Another risk is that the neutron rate in the cosmic ray veto is larger than anticipated, resulting in excessive deadtime in a detector that is sensitive to neutrons. It is unlikely that calculations of the neutron rate from the stopping target will be far off. The production mechanism for neutrons from captured muons is well measured and the simulations are straightforward and done by experts in the field. The discovery of a higher-than-anticipated neutron rate could be mitigated by placing additional neutron absorbers between the neutron shield and the Detector Solenoid. A study has been done showing that an inexpensive water tank could be placed around the detector solenoid to moderate and absorb neutrons (Section 8.3.8). Additionally, there is a significant neutron flux from the production target, and to a lesser extent, the Transport Solenoid. The cosmic ray veto is some distance away from these sources, so the rate is expected to be small compared to that from the stopping target. Calculating the rate from these sources, however, is more difficult. Again, should the rate prove to be too large it is possible to add additional neutron absorbers.

## 11.8 Quality Assurance

Quality assurance is vital to ensuring that the cosmic ray veto performs to specification and will be applied as part of the design, procurement, and fabrication processes. Detectors with similar components have been fabricated for MINOS, MINERνA, and NOνA. Lessons learned from those projects will be applied to Mu2e and we will make use of some of their test equipment to qualify components.

The Fermilab-NICADD Extrusion Facility has over the years developed an extensive set of quality assurance tests for their extrusions that includes validation of dimensional, bend, and twist tolerances as well as light yield measurements. Raw materials are obtained from vendors with proven track records and are validated prior to use. The extrusion machine variables and settings, including temperatures, operating speed, vacuum levels and water temperature are logged to help identify problems.

It is standard practice for fiber vendors to provide quality assurance data with each shipment. For the Mu2e waveshifting fibers the vendor will be asked to provide measurements of the fiber diameter every few centimeters and measurements of the fiber eccentricity, light yield and attenuation length every 1000 m. Upon delivery, measurements of the light yield and attenuation length will be made by Mu2e using an apparatus that was built by Michigan State for the NOvA waveshifting fiber, and used both at Kuraray and Michigan State.

After the fibers are installed in the counters at the module fabrication factory, each fiber will be tested to insure that they were not damaged in either the insertion process or in fly cutting. After each module is fabricated it will be moved to the cosmic-ray test stand where the light yield will be measured over the entire surface of every counter using the front-end electronics. Finally, after the modules have been shipped to Fermilab they will be tested before installation to insure no damage ensued in transit.

The SiPMs all need to be tested to insure they meet specifications. The nature of the SiPM production process ensures that certain characteristics are present in the fabricated devices. Specifically, the nature of photolithography and epitaxial growth process is such that the dimensions of the individual cells within the SiPM (and therefore the gain as a function of over-voltage) are very stable. This also applies to the variation of the breakdown voltage with temperature. Effectively, this means that the breakdown voltage, the gain of the device, and the rate of dark pulses as a function of overvoltage can be extracted with a DC measurement of current versus reverse bias voltage. The IV curve (current vs. voltage) can also be used to measure the relative response to light, if a stable source of light is used. This is the technique we propose to implement for quality assurance testing of all SiPMs in the cosmic ray veto. Such a campaign is much easier to carry out than one where parameters of the SiPM must be studied as a function of temperature, voltage, etc.

The quality assurance stand for production testing will consist of a large dark box where the temperature can be kept stable, a diffuse light source of approximately the right spectral distribution, and an array of small signal relays (these can be solid state opto-relays). A small carrier board, filled with a convenient number SiPMs will be placed into the test fixture and the relays will be operated by an appropriate (Kelvin type) connection to each SiPM in turn. The actual IV measurements will be carried out using standard semiconductor test instruments such as the Keithley 2400 series Sourcemeter. The carrier board, with a 12 × 12 array of SiPMs will be only 10 × 10 $cm^2$ and will be very inexpensive (approximately $50 each). The carrier boards will also be used to keep track of the individual SiPMs; we will ship the board to the vendor to insert the photodetectors. From that point on, the devices will always be kept on that board – through QA testing all the way to installation.

Keeping track of all quality assurance tests will be done using a standard database package.

## 11.9 Value Management

As estimates of the various costs become available it will be possible to tune the parameters that determine the detector efficiency to optimize the performance and the cost. The adjustable parameters include the scintillator thickness, the number of waveshifting fibers, the fiber diameter and the photodetector efficiency. It may also prove cost efficient to replace the AdvanSiD SiPMs with less expensive models from different vendors; the field has become very competitive. The dye concentration of the fiber can also be optimized. The Mu2e baseline design of 175 ppm of K27 dye is the same as MINOS even though the MINOS fibers are 8 m long. More K27 dye would produce more light at the cost of a reduced attenuation length. Some R&D is needed to find the right balance between light yield and attenuation length to optimize this parameter, which is not a cost driver. We are also evaluating wavelength shifting fiber from Bicron (Saint-Gobain) [28].

Some of this work has already led to cost savings. The use of SiPMs rather than photomultipliers, in addition to improving performance, has decreased the cosmic ray veto cost. The shorter length modules have also allowed the magnet height to be lowered, making the detector hall smaller.

We are aggressively pursuing quality control devices that have been used in previous experiments. A cosmic-ray muon test stand used for the calibration of the CsI crystals in the PIBETA experiment [27] has been shipped to the module fabrication factory at Virginia and will be used for R&D studies as well as testing modules after fabrication. We are negotiating the use the fiber test jig built by Michigan State and used to test all of the NOνA fibers. Several fly-cutters exist and we intend to use one of them in the module fabrication factory.

## 11.10 R&D Program

The design of the cosmic ray veto employs mature technology, with perhaps the exception of the photodetector. Hence the R&D is focused on tuning the technology to the requirements of the cosmic ray veto, validating the design, and developing quality assurance tools and procedures; in other words preparing for the detector construction. The R&D tasks are of three types: Monte Carlo simulations, counter performance measurements, and module fabrication studies. There are two simulation tasks: determining the required cosmic ray veto efficiency and determining the neutron background rate. The former has been done and the results are described in Section 3.5.10, although ongoing improvements remain in progress. The neutron background studies fall under the rubric of the muon beamline and are discussed therein. The counter performance measurements focus on measuring the photoelectron yield, the critical parameter needed to meet the efficiency requirement, and the neutron efficiency. Module fabrication studies will be done to validate our design. Below we describe the R&D tasks in more detail.

### *11.10.1* Required Veto Counter Efficiency

The simulation to estimate the rate of electrons from cosmic ray muons that are consistent with conversion events is described in Section 3.5.10. That estimate, accurate to about 25%, has been used to establish the overall efficiency required of the cosmic ray veto. The simulation is continuously being refined to describe changes in the building and detector design, incorporate improvements in the track-finding algorithms, and to speed up the simulation. Increasing the speed of the simulation is particularly important as we intend to determine the position dependence of the required cosmic ray veto efficiency, in particular, where coverage is vital and where it can be omitted. This work is being done by off-project physicists.

### *11.10.2* Counter Photoelectron Yield Measurements

It is vital that the counters meet the baseline design photoelectron yield requirements given in Section 11.4.2 In that section the results of a test-beam study of a prototype counter of an early design (very similar to that described in Section 11.4.1) are described. The performance of the prototype met requirements. We intend to fabricate prototype counters of the baseline design described above, including the readout electronics. Several working full-length counters will be used to make photoelectron yield measurements as a function of longitudinal and transverse positions, incident angle, and with unglued and glued fibers. The time resolution will also be measured. Although test beam studies are preferred, the convenient M-Test beam at Fermilab will be out of commission for a year starting in mid-2012, and hence cosmic-ray muons will be used. Two cosmic-ray test stands are being set up: one at the University of Virginia using drift chambers with sub-millimeter resolution [27], and one at Fermilab using cathode-strip chambers with sub-centimeter resolution fabricated for the ATLAS experiment [29]

Once these measurements have been made, the photoelectron yield can be tuned by changing the wavelength shifting fiber and SiPM diameters. As mentioned above, the cost associated with selecting a custom SiPM size is not large.

### *11.10.3* Neutron Efficiency Measurements

Several short counters consistent with the baseline design will be fabricated to make neutron efficiency measurements. We have identified three potential neutron facilities where well-characterized neutron beams exist to make the measurement: the NUMI neutrino beamline at Fermilab, the Fermilab Neutron Therapy Facility, and a research reactor at Dubna. The NUMI beam will be down for a year starting in mid-2012, but the Neutron Therapy Facility will continue to operate. Hence, our intention is to initially make the measurements at the Neutron Therapy Facility and, if needed, continue them at either NUMI or at Dubna. The measurements will determine the counter response to both high (> 0.1 MeV) and low (< 0.1 MeV) energy neutrons, and the efficiency of different absorber thicknesses and materials. The short counters will also be used for SiPM tests.

### *11.10.4* Extrusion Measurements

A new die is being fabricated for the Fermilab-NICADD Extrusion Facility. Prototype extrusions will be sent to the module fabrication factory to produce the counters needed for the photoelectron yield and neutron efficiency measurements, as well as for module fabrication studies. The die will be tuned to insure that the extrusions meet the flatness and size specifications, which have yet to be finalized, and to tune quality assurance procedures.

### *11.10.5* Fibers

Almost every experiment using scintillating fibers in recent years has chosen Kuraray fibers [7], which are known for excellent performance characteristics and good quality. Nevertheless, as part of our value management efforts we will procure and test Bicron fibers [28]. Quality assurance devices and procedures for testing the fibers upon arrival from the vendor and after counter fabrication will be obtained.

### *11.10.6* Photodetectors

SiPMs from different manufacturers will be acquired and tested, with the focus on measuring their relative quantum efficiency and noise rate as a function of bias, the operating bias range, and the temperature coefficient. Different mounting schemes, including surface mounting, will be pursued. As mentioned in Section 11.3.3, we will develop protocols by which the production testing will be done.

### *11.10.7* Module Fabrication

Initially short extrusions will be used to fabricate several non-working mockups in order to explore various fabrication options and to validate and tune the design. Adhesive studies will also be made using weighted counters. Once the design has been finalized a full-length module will be built, time-and-motion studies will be done, and the fabrication procedure will be tuned. In parallel, quality assurance jigs will be built and tested. Finally, a full-length working module will be fabricated and tested.

## 11.11 References

[1] E.C. Dukes, "Cosmic Ray veto Requirements," Mu2e-doc-944.

[2] V. Jorjadze, "Cosmic Ray Background Simulation for the Mu2e Experiment," Mu2e-doc-1312.

[3] E.C. Dukes, “Cosmic Ray Veto Parameters,” Mu2e-doc-862.
[4] D.G. Michael et al., Nucl. Instrum. Meth. **A596**, 190 (2008).
[5] D. Benznosko et al., FERMILAB-PUB-05-344 (2005).
[6] J.-C. Yun and A. Para, “Scintillator reflective layer coextrusion,” U.S. Patent 6,218,670, 2001.
[7] Kuraray America, Inc., 200 Park Ave., NY 10166 USA; 3-1-6, NIHONBASHI, CHUO-KU, TOKYO 103-8254, JAPAN.
[8] C. Piemonte et al., IEEE Trans. Nucl. Science, Vol. 54, NO. 1, (2007).
[9] Advanced Silicon Detectors, via Sommarive 18, 1-38123 Pova, Tento, Italy.
[10] Center of Perspective Technology and Apparatus, 1-ya Vladimirskaya St., 22-2 Moscow, Russian Federation 111141.
[11] Hamamatsu Corp., 360 Foothill Rd., Bridgewater, NJ 08807.
[12] M. Crisler, J. Hall, E. Ramberg, T. Kiper, FERMILAB-CONF-10-459-E (2010).
[13] V. Tumakov and W. Molzon, “Optimization of Counting Rate in Cosmic Ray Shield from Proton Secondaries,” Mu2e-docdb-239.
[14] J. Warchol, “CFT/PS status”, DØ Internal Note 111889, http://www-d0.hef.ru.nl//fullAgenda.php?ida=a111886.
[15] Y. Oksuzian, “CRV R&D Status Update,” Mu2e-doc-1155.
[16] Itasca Plastics, 3750 Ohio Ave., St. Charles, IL 60174.
[17] V. Biliyar, “Calculation of the Spectrum of Ejected Neutrons from Muon Capture,” Mu2e-doc-1619.
[18] V. Tumakov, “Simulation of Neutrons Down to Thermal Energies with GEANT3 Gcalor Code,” Mu2e-doc-236.
[19] N. Colonna and S. Altieri, Health Phys. **82**(6), 840 (2002).
[20] N.V. Mokhov and C.C. James, “The MARS Code System User’s Guide Version 15,” Fermilab (2010).
[21] MCNP 5.1.40, Los Alamos National Laboratory, LA-UR-05-8617 (2005).
[22] S. Chatrchyan et al., J. Instrum., **5** T03018 (2010).
[23] V. A. Polychronakos, “Interpolating Cathode Strip Chambers of the GEM Muon System Prototype Development and Performance,” SSC-GEM-TN-93-301 (1993).
[24] G. Bencze et al., Nucl. Instrum. Meth. **A357**, 40 (1995).
[25] R. Ray, “Mu2e Preliminary Hazard Analysis,” Mu2e-doc-675.
[26] M. Overlin, MECO-014.
[27] E. Frlež et al., Nucl. Instrum. Meth. **A440**, 57 (2000).
[28] Saint-Gobain Crystals, 17900 Great Lakes Pkwy, Hiram, OH 44234.
[29] T. Argyropoulos et al., IEEE Trans. Nucl. Science, Vol. 56, No. 3, 1568 (2009).

# 12 Trigger and Data Acquisition

## 12.1 Introduction

The Mu2e Trigger and Data Acquisition (DAQ) subsystem provides hardware and software for collecting digitized data from the Tracker, Calorimeter, Cosmic Ray Veto and Beam Monitoring systems, and delivering that data to online and offline processing for analysis. It is also responsible for detector synchronization and control.

## 12.2 Requirements

The Mu2e Collaboration has developed a complete set of requirements for the Trigger and Data Acquisition System [1]. The DAQ must monitor, select, and validate physics and calibration data from the Mu2e detector for final stewardship by the offline computing systems. If operated in a streaming mode, the off-detector bandwidth requirement for the DAQ is estimated to be approximately 30 GBytes/sec. The DAQ must combine information from ~275 detector data sources and apply filters (triggers) to reduce this rate by a factor of one thousand before the data can be transferred to offline storage.

The DAQ must provide precise control of the data sources and readout, including timestamp synchronization, live gate fraction, data compression, calibration and diagnostic modes, and channel specific settings (thresholds, gains, masks). The DAQ must provide sufficient bandwidth margin and readout frequency prescaling capability to accommodate acquisition modes with higher instantaneous rates. The DAQ must also provide a "slow control" network for operational control and monitoring of other Mu2e subsystems.

## 12.3 Proposed Design

Data transfer technologies and costs have improved to the point that it is now feasible to consider a continuous (streaming) readout system for Mu2e. The architecture described here (Figure 12.1) assumes that all data is zero-suppressed and moved off the detector prior to event selection. The data transfer, online processing and networking functions are all implemented using commercial components.

In a streaming DAQ system, an "event" is simply a fixed timeslice of data from all detector components. The timeslice size is optimized for network packet and server cache efficiency. A streaming architecture provides more flexibility in data analysis and can simplify design of the front-end electronics, but it results in a higher off-detector data rate.

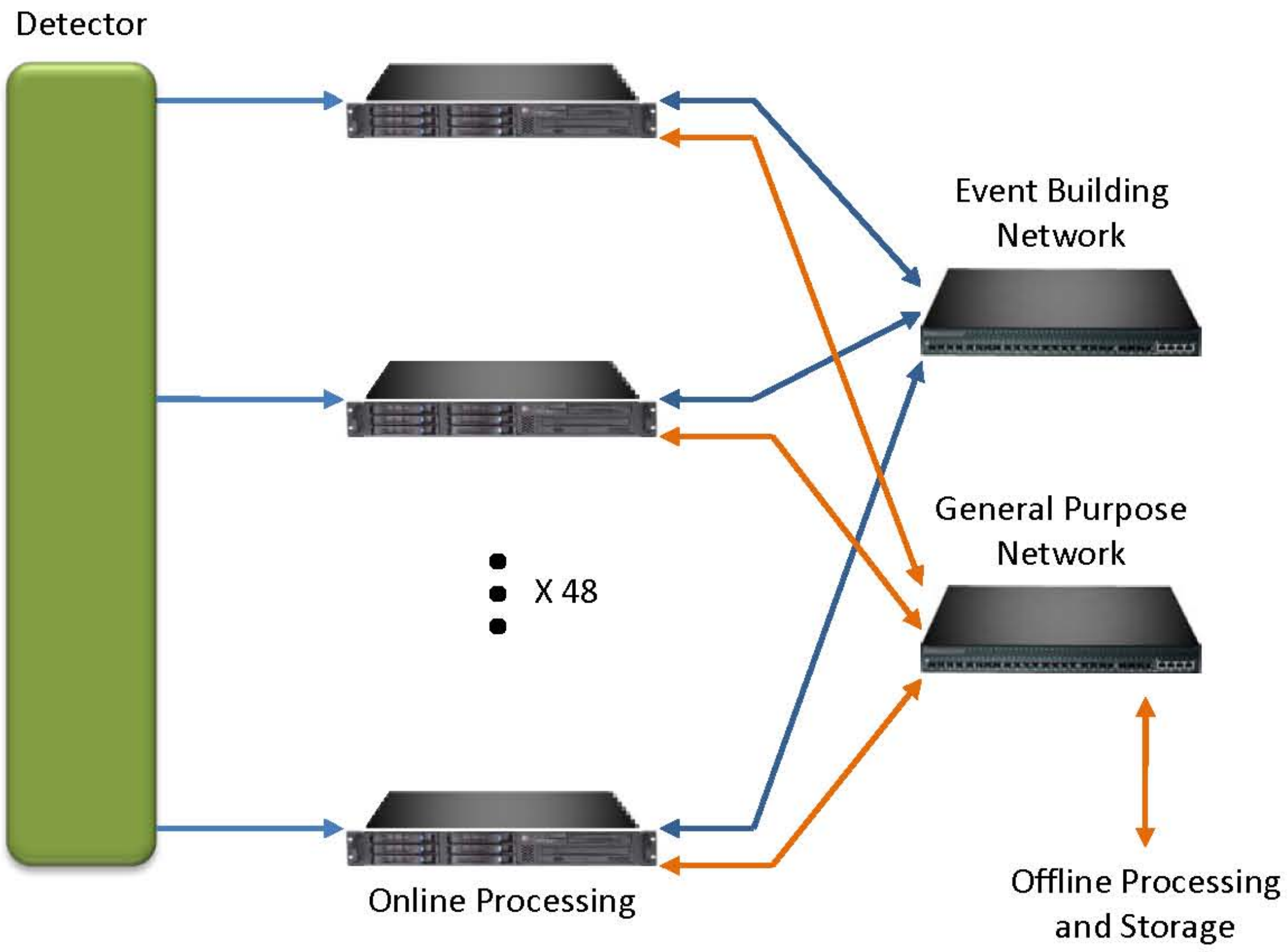

Figure 12.1. The basic DAQ architecture for Mu2e.

### *12.3.1* Data Sources

The DAQ system interfaces to the Data Source digitizing electronics. All Tracker and Calorimeter electronics are located inside the detector vacuum, so the DAQ boundary is defined as the external optical connectors of the feedthrough cables (Figure 12.2). The Cosmic Ray Veto digitizing electronics are located outside of the detector, but will use the same DAQ interface.

The distance from the detector to the control room where the DAQ electronics will be located is approximately 40 meters. We assume that the Data Source links operate at 2.5 Gbps, which is currently a cost-effective rate for FPGAs and optical transceivers, and that all detector subsystems use common 12 channel MTP/MPO optical connections to the DAQ. The total cross-section of the optical feedthrough cables is less than 5 $cm^2$. Data leaving a Data Source is sparsified (local timestamp / channel / data). Extended time and source information may be added at the receiving end.

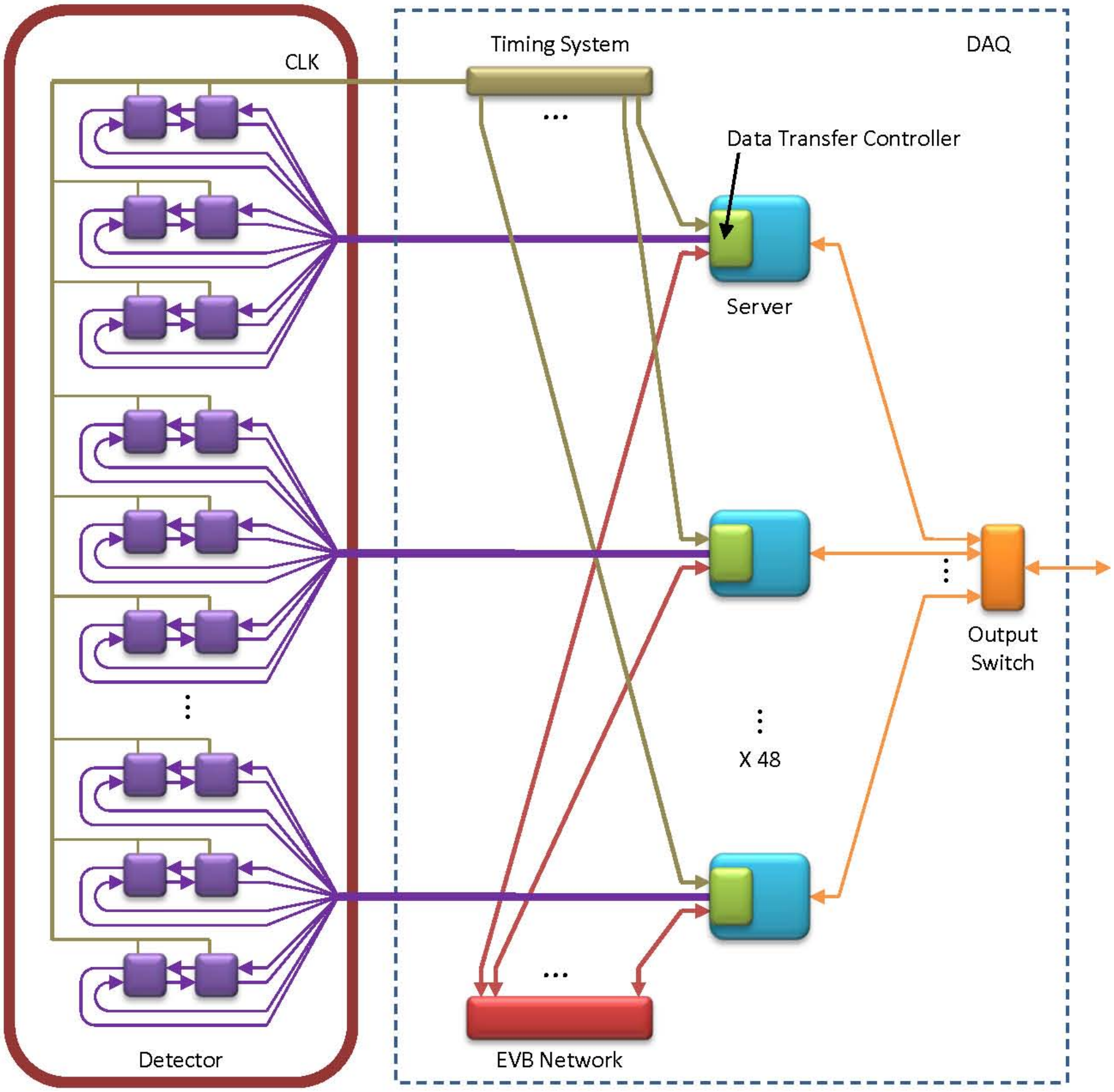

Figure 12.2. The interface between the various data sources and the DAQ.

The Tracker is the largest source of data in Mu2e. Each Tracker panel generates a maximum of 1 Gbps of zero suppressed data. A pair of counter-rotating control/data rings is used for redundancy and bandwidth expansion. Two Tracker panel controllers are connected to each pair of rings (Figure 12.3), and six rings are bundled into an MTP/MPO cable. For a tracker with 21,600 straws and 216 controllers, there are 36 optical cables connecting to the DAQ system.

The Calorimeter generates a similar "per channel" data rate. For a system with 1760 crystals, there would be a total of 88 controllers (22 per vane). Four or five Calorimeter controllers are placed in each ring to provide approximately the same number of channels and data rate per ring as in the Tracker. The Calorimeter will require 8 MTP/MPO cables connecting to the DAQ system.

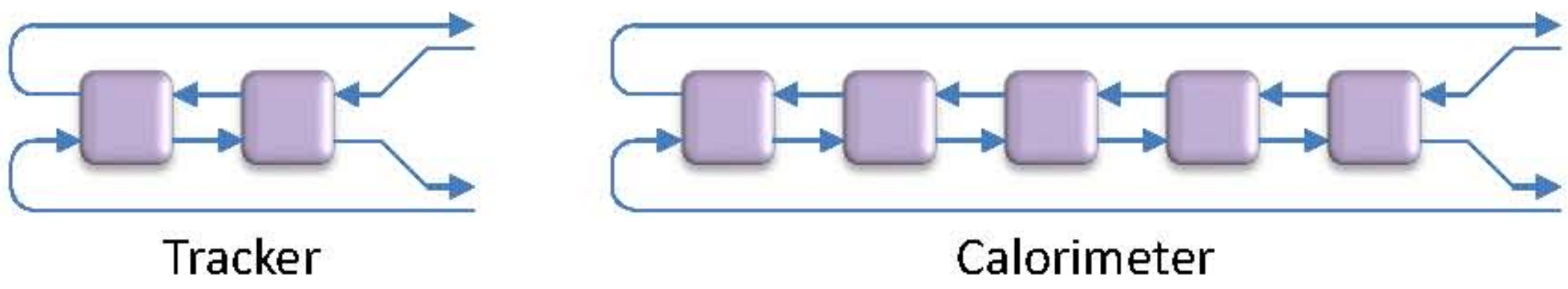

Figure 12.3. Dual ring interconnection for detector readout controllers.

Electronics for the Cosmic Ray Veto system, Stopping Target Monitor and Extinction Monitor are located outside the detector vacuum. The combined data rate for these sources is expected to be a few GBytes/sec, so the number of optical rings is determined by geography rather than bandwidth.

To make efficient use of optical link bandwidth, a Data Source readout controller should have enough buffer memory to hold zero-suppressed data from many timeslices (up to 1 second, or at least 256 Mbytes per controller).

### *12.3.2* Data Acquisition

A typical switch-based DAQ system is shown in Figure 12.4a. Data is collected by Data Transfer Controllers (DTCs) and transmitted unidirectionally through the Event building (EVB) Network to a farm of commodity processors. Final event assembly is done in the processors.

For Mu2e, the connections will be rearranged slightly as shown in Figure 12.4b. Data is collected by PCIe based DTCs in the processors, then exchanged bidirectionally between DTCs through the Event Builder Network before being forwarded to the processors. This has several advantages:

- Switch port utilization in the Event Builder Network is doubled, reducing the number of switches required.
- Event building can be done entirely in hardware, offloading this task from the processors.
- The Data Transfer Controller has access to complete events and can perform formatting and pre-processing (including L1 trigger selection) in the on-board FPGA, further offloading the processors.

The DTC (Figure 12.5) has three main ports; a Data Source port, an Event Builder Network port and a Processor port. The Data Source interface supports six 2.5 Gbps optical rings. The Event Builder Network interface is a single channel 10 Gbps Ethernet connection. The Processor interface is PCIe. There is an additional 2.5 Gbps port which receives control information from the timing system for distribution to the Data Sources. To reduce development cost, the DTC is based on a commercial PCIe card. The serial interfaces will likely be implemented on a FMC (VITA 57) daughtercard.

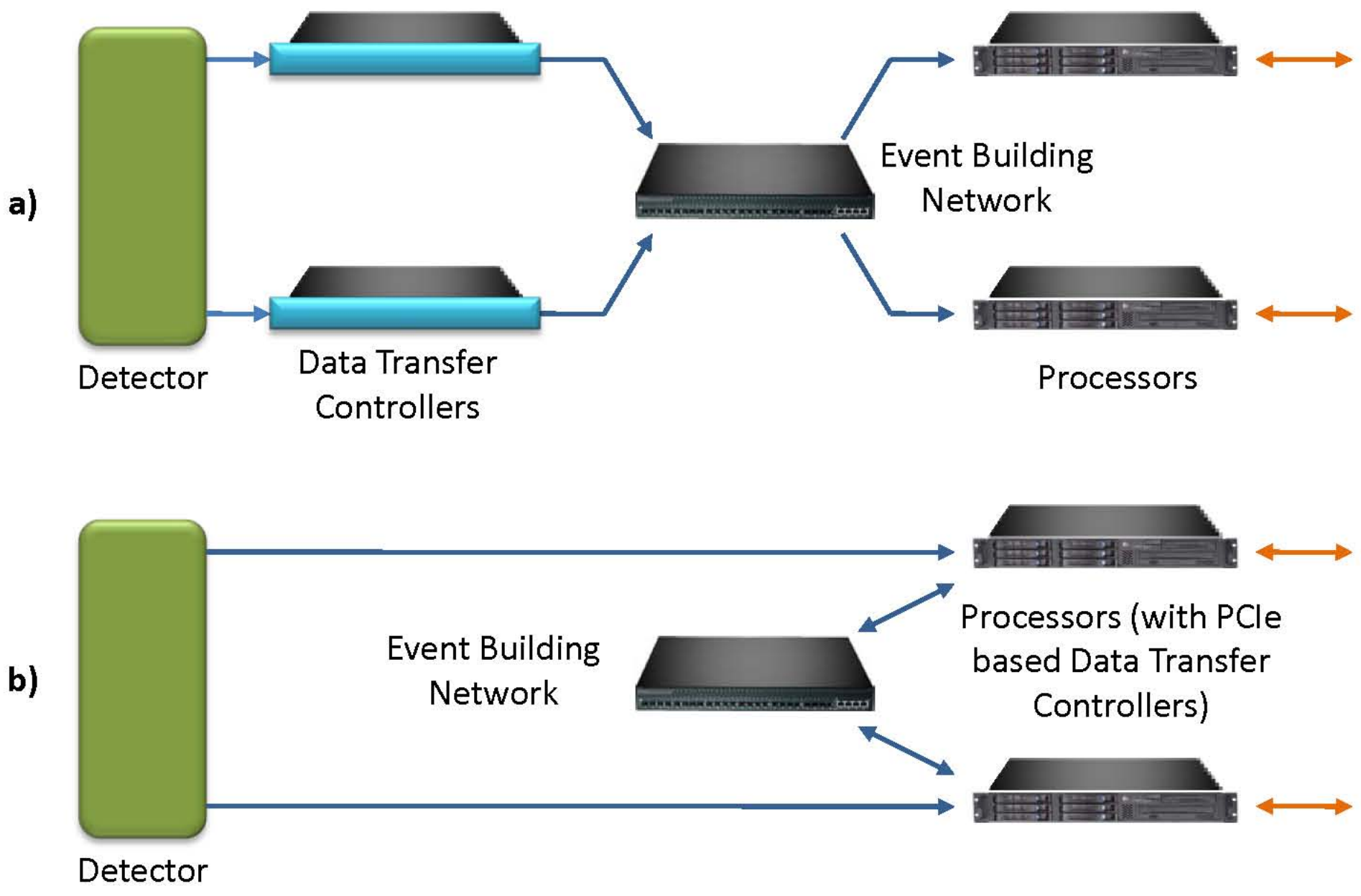

Figure 12.4. A typical switch-based DAQ system is shown in a). For Mu2e, the connections will be rearranged slightly as shown in b).

The DAQ system contains 48 Servers with integrated Data Transfer Controllers. Data is exchanged between DTCs using a single 48-port, 10 Gigabit Ethernet switch. This architecture allows a total of 288 data rings (48 DTCs X 6 rings). These would initially be assigned as follows; 216 for the Tracker, 48 for the Calorimeter, and 24 for all other subsystems (CRV, monitors, etc). Space requirements are estimated at five equipment rack (Figure 12.6).

The Timing System generates two signals:

1. System Clock - a 1695 ns (590KHz) low jitter, phase aligned clock, synchronized to the accelerator. The leading edge of this clock is the timing reference for all fast synchronous operations within a microbunch.

2. Control Link - a 2.5 Gbps packet based serial connection. Packets are framed by the System clock.

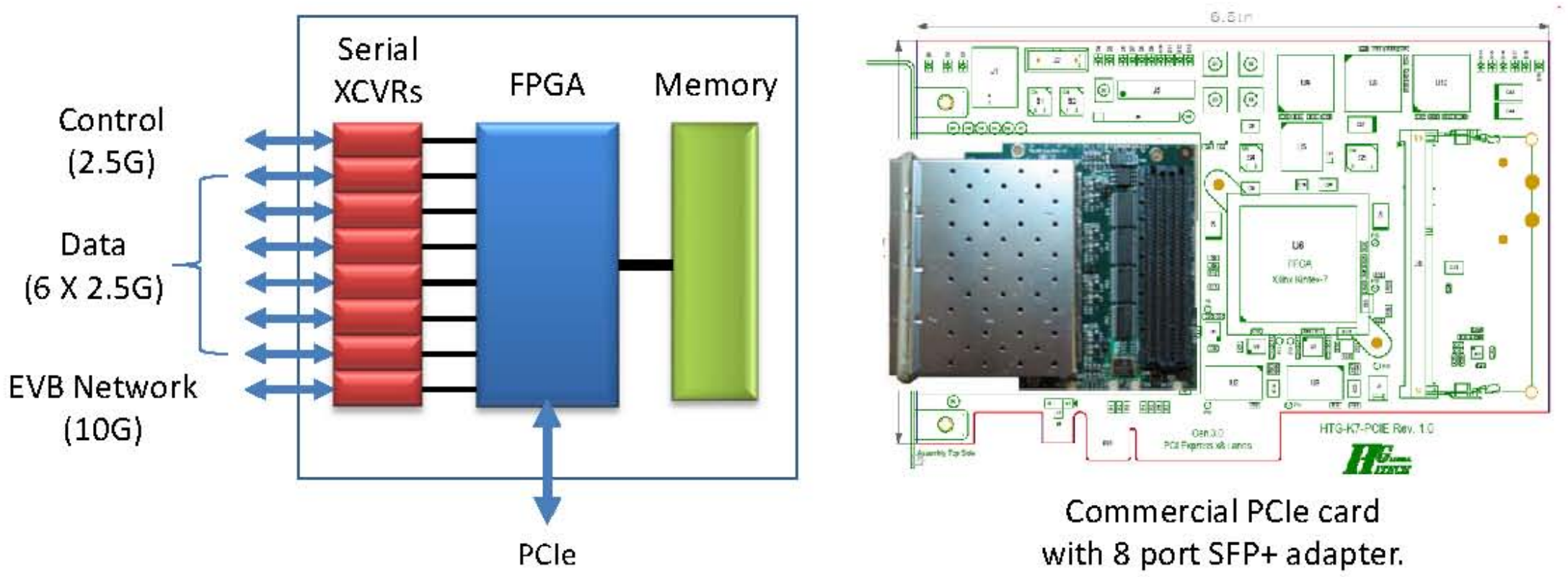

Figure 12.5. Implementation example of a Data Transfer Controller.

Data Sources (readout controllers) generate local acquisition clocks, phase locked to the system clock. Data Sources also provide the local timestamp counter, which is driven by the acquisition clock (Figure 12.7). Fanout and phase alignment of the system clock within the detector is the responsibility of the detector subsystem (using either hardware or software timing calibration).

Each Data Source connects to two 2.5 Gbps optical rings. Dual rings provide additional bandwidth for expansion, and redundancy against Data Source or interconnect failures. A ring carries both control and data packets. Control packets are prioritized, and allow the DAQ system to dynamically set Data Source operating parameters such as live gate fraction (acquisition start and stop times within a microbunch), data compression mode, and calibration or diagnostic mode. Control packets are also used to enable/disable data acquisition for each microbunch (to limit readout rate in high occupancy modes), and to set and check the local timestamp counters for timing synchronization. These commands are transmitted in the preceding System Clock frame (microbunch) and then synchronized by the Data Source to the leading edge or a specific timestamp in the next frame. Data Sources can insert control packets to indicate error conditions or buffer status.

The DAQ system will be designed to support partitioning. This is the ability for multiple operators to independently take data from non-overlapping fractions of the full detector. This feature allows commissioning, debugging, and beam-off calibration work to take place in parallel. In this model, data taking in one partition can be started, stopped, and otherwise controlled without affecting other partitions. It should be noted that partitioning has been done on several previous experiments (CDF, NOvA), so the needs and benefits of partitioning are well understood.

The partitioning scheme will support a pre-defined level of granularity, and a reasonable choice would be to specify that each of the 48 Data Transfer Controllers/Processors can be assigned to a particular partition. In this model, all of the Data Sources that are connected to these Data Transfer Controllers would be treated as an indivisible set. The DAQ applications that control data taking will ensure that the granularity choices are enforced and that DAQ components that cannot be shared will never be included in more than one partition. During production running of the experiment, the full detector will be controlled and read out in a single partition.

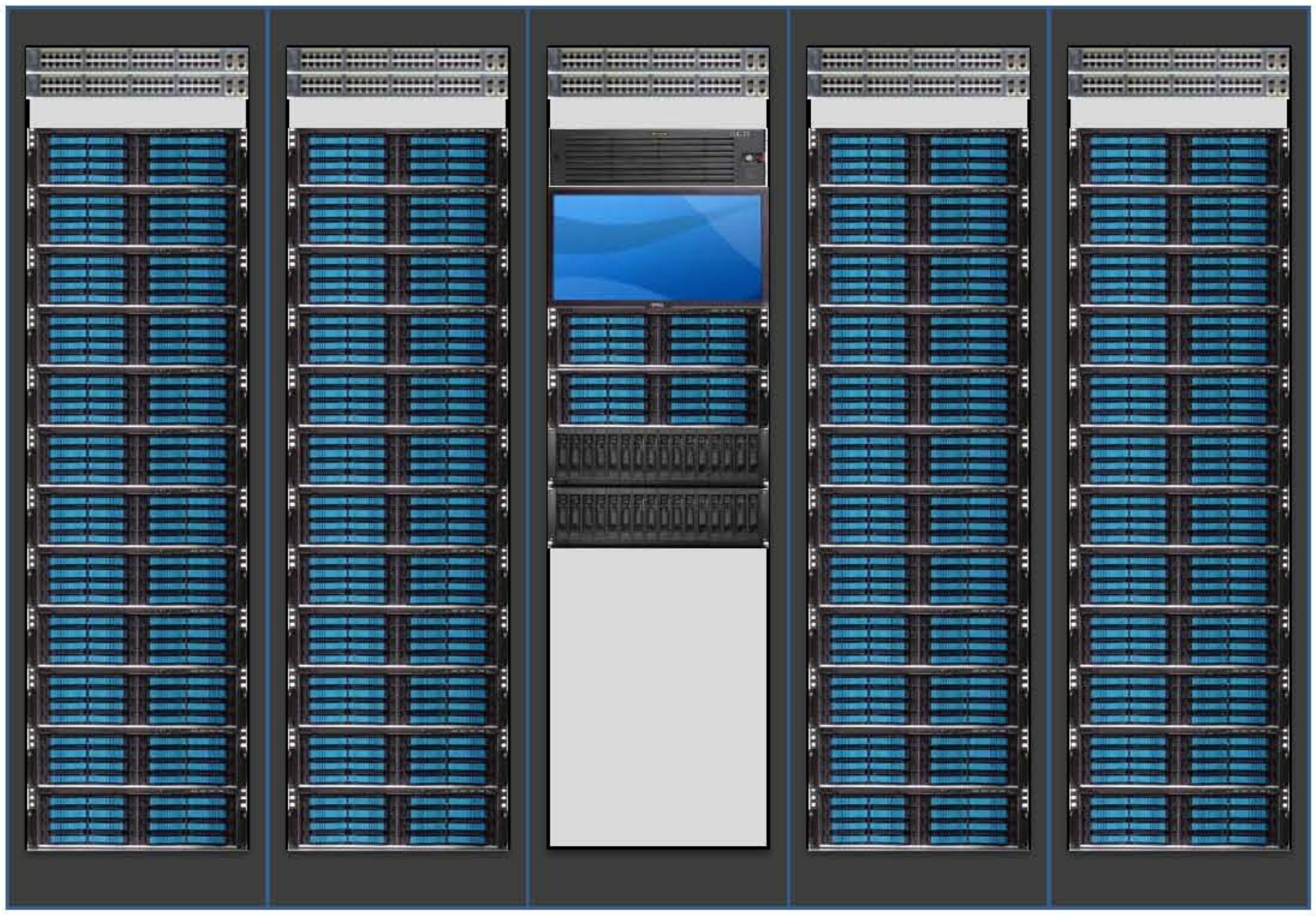

Figure 12.6. DAQ system rack space.

The data acquisition software components will manage the movement and processing of the data from the front-end crates to the storage system, and they will configure, control, and monitor the software and hardware components in the data path.

The specific software components include the following:

- Common utilities such as message passing and message logging.
- Data flow applications to handle movement of data between hosts, building of events, management of event data buffers in the processor farm, and logging of data to disk.
- Control applications to configure hardware and software components, start and stop data taking, and manage the assignment of resources to partitions.
- Applications to monitor data quality and DAQ system performance.

- Applications, libraries, and databases to store and retrieve run history and other archival information.
- Specialized kernels and device drivers that are needed for custom hardware components.

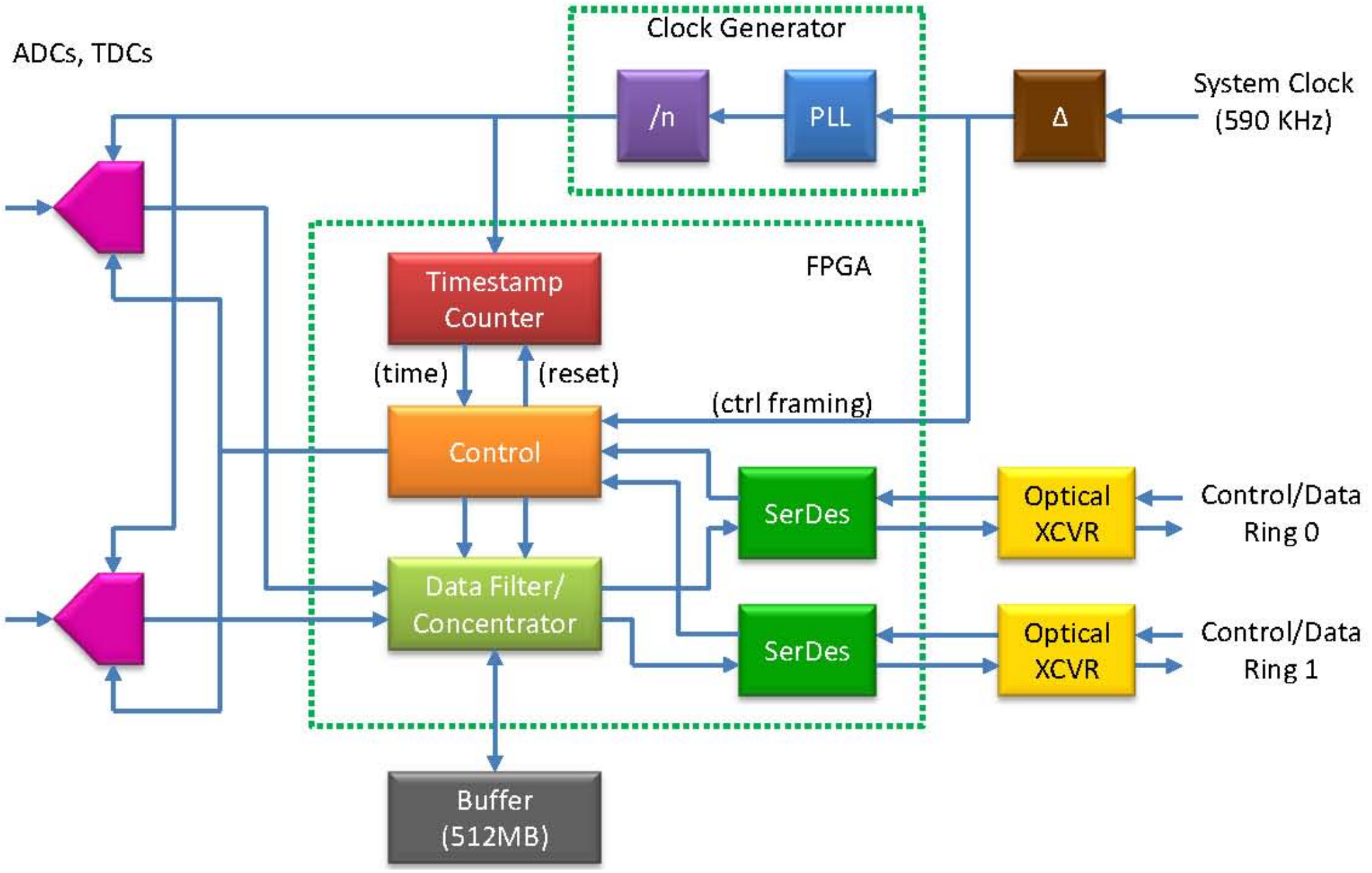

Figure 12.7. A typical Data Source interface in Mu2e, which is phased locked to the System Clock and provides a local timestamp.

In some cases, it may be possible to reuse common components from previous experiments (e.g. NOvA). In all cases, the selected components or underlying technologies will be matched to the rate requirements and specific needs of the experiment, as defined by the requirements gathered from the collaboration.

In addition to the DAQ software applications themselves, test environments will be constructed and maintained as part of this effort. This will include a scaled-down, portable version of the DAQ system that will allow work on individual components without requiring a full set of hardware. This version will support integrated component testing at an early stage in the development of the system. Additional test environments with real hardware components will allow for full integration testing as well as realistic stress testing. One or more of the planned test environments will support testing of individual hardware components.

### *12.3.3* Data Processing

A streaming system requires significant online processing resources. We assume one "Multi-Processing Unit" (MPU) per Data Transfer Controller. Each MPU contains multiple physical processors or cores. For high-performance cores, an MPU may be defined as a commercial server with 4 processors and 16 cores per processor. Of potentially greater interest is an emerging class of systems (Figure 12.8) using lower performance cores, but with many more cores per server (typically 256-512). Systems based on Intel MIC [2], Tilera GX [3]and ARM Cortex A-15 [4] devices are expected in 2011-13.

Benchmarks of the tracking code will determine the total processing requirements. The goal is to move as much of the Trigger as is economically practical to software. However, as described earlier, the proposed DAQ architecture also provides several TeraOPS of FPGA based processing capability with access to full event datasets. Cost estimates for the online processing can therefore be fixed, even if benchmarks are incomplete. A first level filter in the Data Transfer Controller FPGA is tuned to accept the maximum data rate that can be handled by the online processing system.

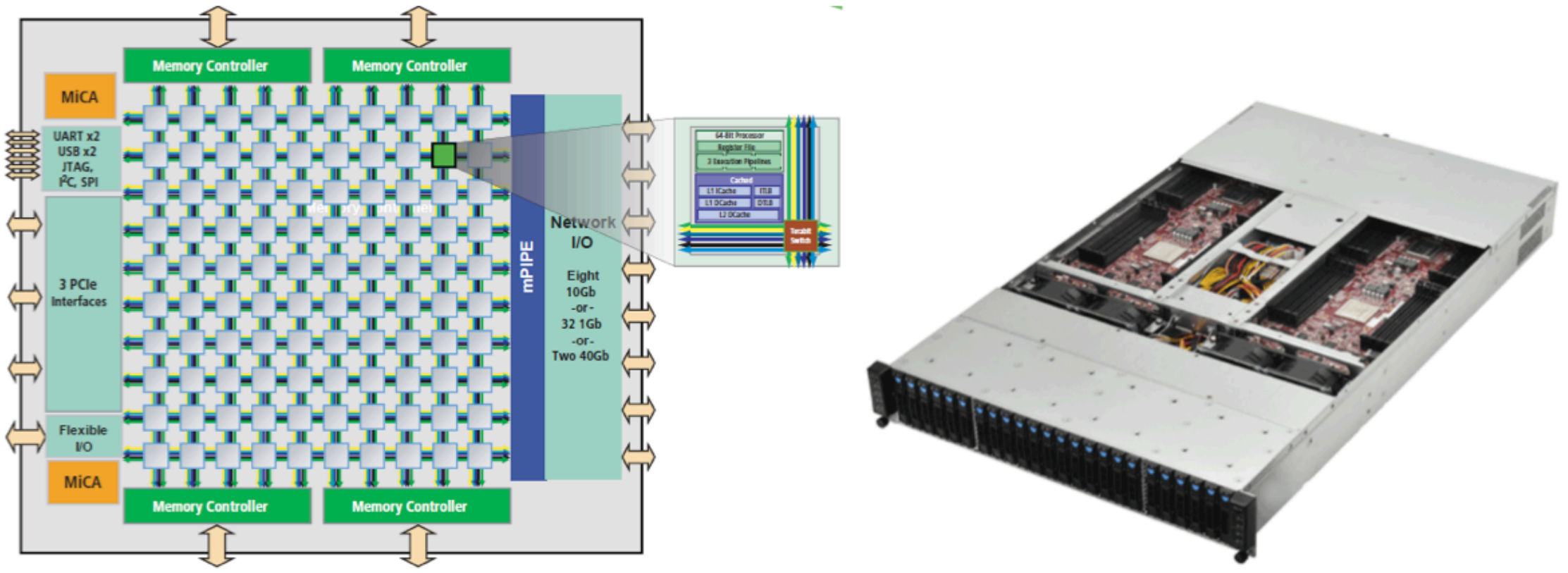

Figure 12.8. "Many-Core" Servers are becoming available that use lower performance cores but come packaged with many more cores per server (typically 256 – 512).

If a Level 1 trigger is not implemented in the Data Transfer Controller, the data rate between the DTC and MPU can still be reduced by incrementally transferring only the subset of data needed for each level of software trigger. For example, a Level 1 Calorimeter trigger running in software requires ~5% of the total event data, reducing the average MPU input rate to less than 50 MBytes/sec. The structure of the event data in the DTC memory will allow filtering of timeslices according to the dataset requirements of the software trigger(s).

Local data storage will be provided by disk drives incorporated either in the online processing farm nodes or in a centralized data logger. Offline storage of processed data will be provided by the Fermilab Computing Division shared storage facility. Offline storage requirements are estimated at 1 PetaByte per year.

The data processing software will contain all of the components necessary for performing the data filtering operations in the software trigger. This will include the event reconstruction and analysis framework, the data management application that will provide events to the reconstruction software and transfer accepted events downstream, and the infrastructure that is needed to configure the reconstruction code.

The offline event reconstruction and analysis framework will be used for the software trigger, with suitable changes to support the online environment. As part of the modified framework, sample reconstruction and analysis modules, data handling modules, and throughput monitoring tools will be provided. In addition, a portable simulation environment will be created so that analysis modules can be developed and tested without requiring the full DAQ system software chain.

### *12.3.4* Control and Networking

The network (Figure 12.9) provides connections to the Data Transfer Controllers, Timing System Controller, Data Sources, Online Processors and the Event Builder Network management ports. There is a distribution switch in each DAQ rack, and a central router for external connections (control room and data storage facility). A separate management/IPMI network handles operational monitoring of the servers and networking equipment.

The Mu2e operational control room will be located remotely (Figure 12.10). Control will be via network connection, with VPN access available from other locations. There is also a workstation in the Detector Hall Electronics Room with the same functionality as the remote control room stations.

The local electronics room houses the Online Processing farm, Data Transfer Controllers, and networking equipment. The total DAQ space requirement is estimated to be 5 racks. The total power and cooling requirement is estimated at 30KW.

The Slow Controls system will control and monitor voltages, temperatures, and other environmental parameters that are not part of the safety systems of the experiment. A combination of EPICS and LabView will be used for the Slow Controls infrastructure, similar to what is being done for the NOvA experiment. EPICS (the Experimental Physics and Industrial Control System) is an open-source set of software libraries, applications, and tools for soft real time control systems [5]. It provides standard patterns and applications for interfacing to hardware components and provides customizable displays for quickly creating monitor applications. LabView is a commercial application from National Instruments that provides a graphical interface for creating control and test applications. LabView will be used primarily in smaller test environments, while EPICS will be used for larger test environments and the full detector.

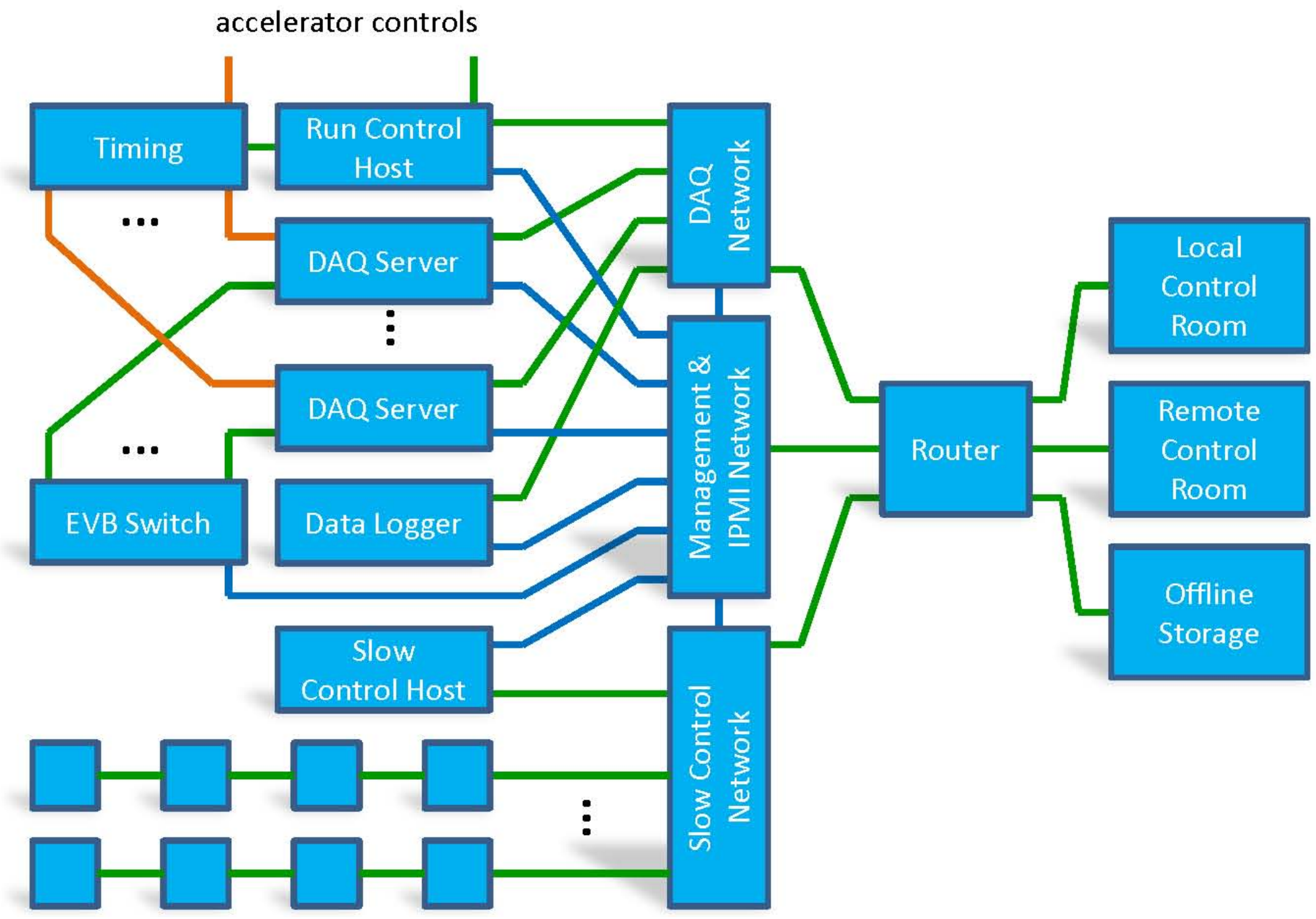

Figure 12.9. Mu2e networking and slow control connections

Specific components from the NOvA slow controls system will be considered for reuse, such as the EPICS interface to the power supplies and field point modules from National Instruments. Communication bridges between EPICS, LabView, and the DAQ messaging system will be provided to allow for convenient communication between the slow controls and DAQ systems.

Mu2e is on a private network, isolated from the main Fermilab network. Setup and management of the connecting router will be handled by the Fermilab Computing Division networking department, according to established security policies.

### *12.3.5* System Expansion

This architecture scales linearly to higher data rates. The Event Building network can be expanded to any size (Figure 12.11), and the number of DTCs/Servers increased accordingly. For higher front-end rates, either the number of Data Sources in a ring or the number of rings connected to a DTC can be reduced to maintain the same effective bandwidth per DTC.

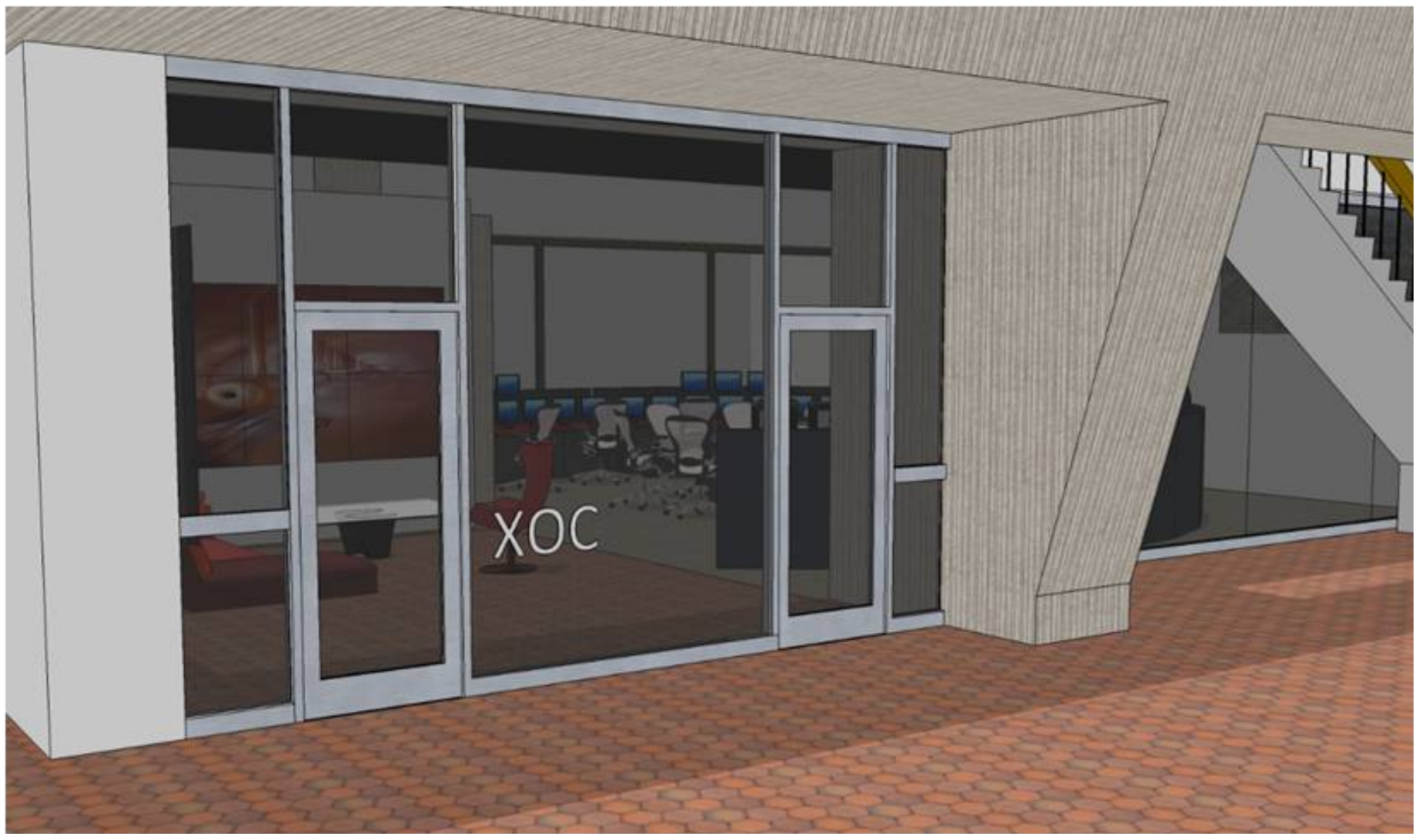

Figure 12.10. Remote control room in the Wilson Hall Experimental Operations Center (XOC)

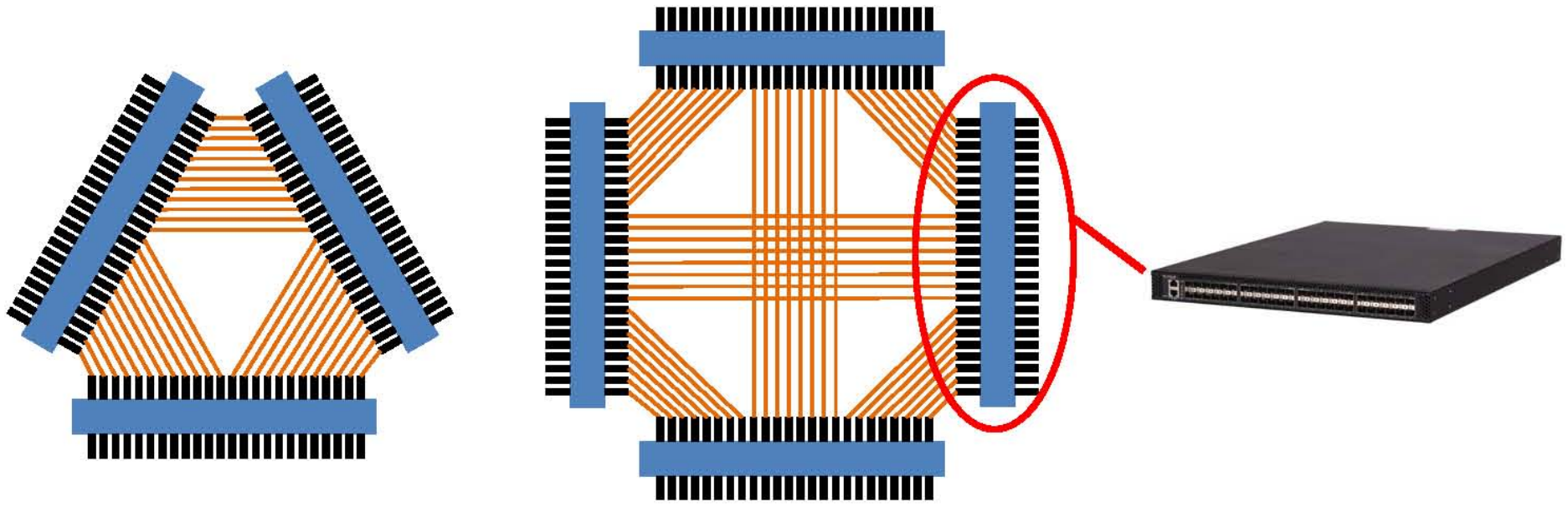

Figure 12.11. Event Building Network expansion (72 and 96 port)

## 12.4 Considered Alternatives to the Proposed Design

The primary alternative to the “streaming” DAQ architecture is a “triggered” system. In the triggered system, a smaller subset of data (Calorimeter sums in the case of Mu2e) is transmitted off the detector and the remaining data is buffered on the front-end ASICs or readout controllers. A fast decision is made by a hardware Level 1 Trigger and the accept signal is broadcast to all front-end modules. The buffered data is then transmitted or discarded. This reduces the off-detector data rate by a large factor, and in turn reduces the bandwidth requirements of the data acquisition and data processing systems.

The front-end electronics are more complicated in the triggered architecture, as there are additional levels of data multiplexing, along with the L1 trigger hardware itself and the L1 Accept timing distribution. But there may be an overall cost savings from the smaller off-detector DAQ.

The choice of DAQ architecture is also influenced by the front-end technology. One alternative proposal, for example, uses analog (switched capacitor) pipelines and waveform digitizers, which would require a triggered system.

The streaming system is generally more flexible, since the initial event selection criteria are not hardwired into the design.

## 12.5 ES&H

The Trigger and Data Acquisition subsystem presents no significant or unusual environmental, safety or health issues. DAQ System components all use 120 or 240V AC power. Power supplies are UL listed with modular AC inputs (no exposed AC connections). Internal circuitry operates at 12 V DC or lower.

Optical communication links use Class 1 (eye safe under all conditions of normal use) or Class 1M lasers (eye safe for all conditions of use except when passed through magnifying optics). Lasers are shielded when cables are attached.

## 12.6 Risks

Most of the risks associated with the DAQ system development are common to previous efforts. The subsystem risks that have been identified are described below.

- There is always some risk in estimating labor requirements for software and hardware design. Most of the hardware is commercially supplied which minimizes development cost and risk. The main risk in software development is in determining the right balance of costed/ uncosted labor.
- We are moving to a higher data rate technology (10Gbps) that has both risks and benefits, but is in common use.
- The required level of online processing is not yet well known, but preliminary benchmarks of the tracking filter indicate that it will work with existing server technology. This is not considered a significant risk since we have the option of implementing a prefilter or trigger in Data Transfer Controller hardware.

- The cost of the DAQ system scales with the data rate, and is dependent on reasonably accurate rate estimates. Current estimates are conservative and rates can be controlled by varying thresholds, etc.

None of these risks are considered to be high. They are all itemized in the Mu2e Risk Registry [6].

## 12.7 Quality Assurance and Quality Control

We will establish a system integration test environment for software development and hardware verification. Most of the DAQ hardware components are commercially sourced, and will undergo initial testing at the manufacturer, followed by "burn-in" testing in the integration test environment.

All components will be designed for in-situ and self-test, which reduces the need for dedicated test stands. The goal is to reduce or eliminate much of the difficult to maintain "teststand" infrastructure (backplanes, controllers, power supplies, etc) and allow development in a desktop/office environment.

All project engineering documentation and software versioning will be maintained in the standard Fermilab Teamcenter management system [7] as it becomes available.

## 12.8 Value Management

The DAQ system uses commodity hardware for the processing and networking components. This hardware tends to decrease in cost and increase in performance over time. Delaying purchases of these components can improve the cost/performance ratio significantly.

The cost/performance of FPGAs improves in the same way as other commodity components, but the cost at introduction and phase-out of specific devices can be much higher. Some degree of market timing in the selection and purchase of FPGA based components can be valuable.

The FPGAs in the Data Transfer Controllers have significant embedded processing capability. If utilized, this may reduce the cost of the online processing.

We plan to reuse software developed for NOvA and other recent DAQ systems wherever possible (especially for slow controls).

The choice of streaming vs. triggered DAQ architecture affects the cost of both DAQ and detector electronics. We have chosen a streaming architecture as the Mu2e baseline, but additional simulation could result in either higher requirements for digitization rate or lower requirements for trigger flexibility, which would favor the triggered architecture.

## 12.9 R&D Plan

R&D tasks are focused on evaluation and selection of components for the Data Transfer Controller, Timing Distribution and Online Processing subsystems. A small test system will be assembled using commercial modules (Figure 12.12).

This system will allow evaluation and follow-on development of;
1) DTC FPGA performance, firmware and interfaces,
2) timing system accuracy and stability,
3) 10G networking infrastructure,
4) online processing benchmarks,
5) architecture options (e.g., hardware vs software triggering and event building).

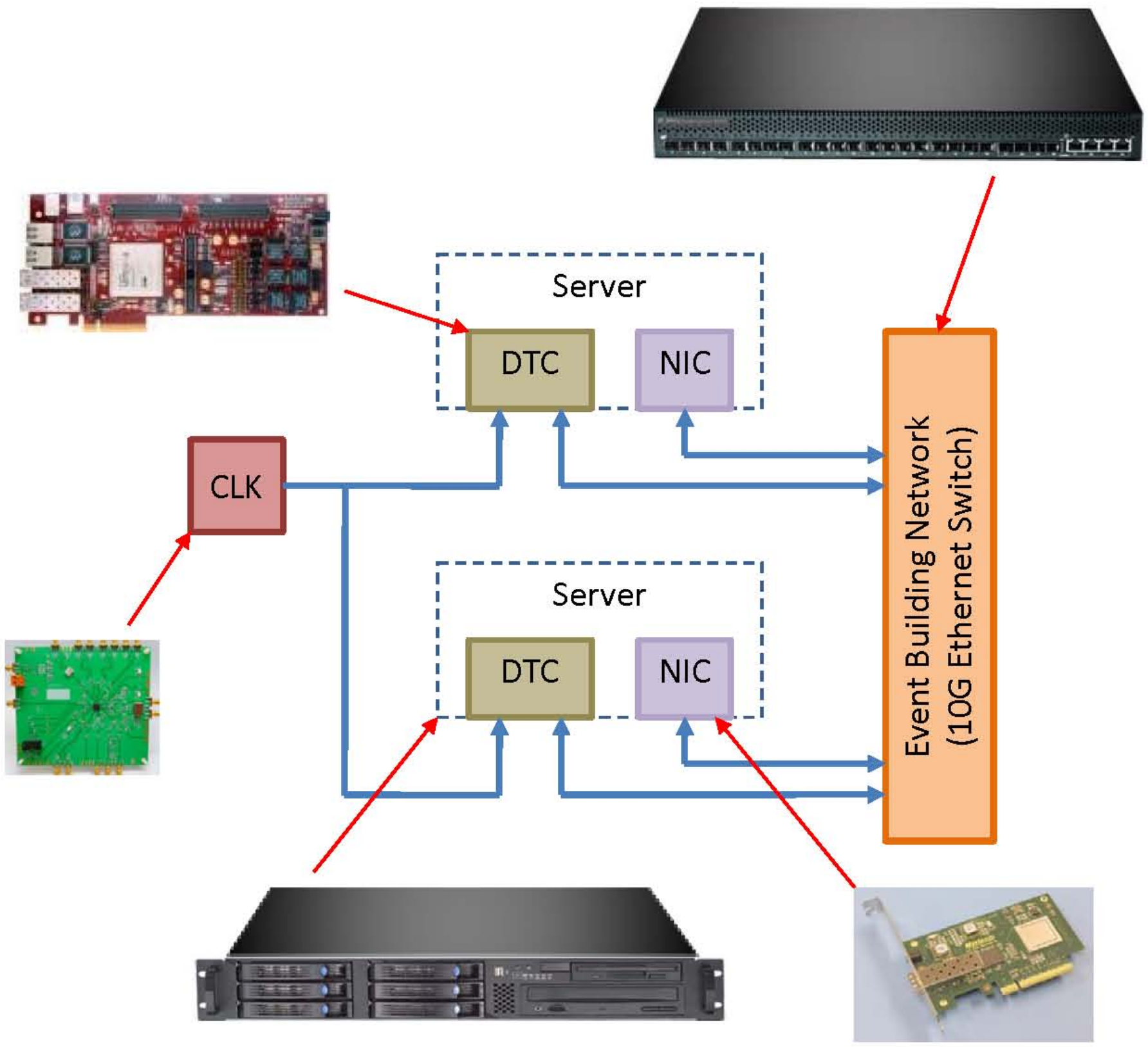

Figure 12.12. R&D system for DAQ component evaluation

## 12.10 References

[1] R. Tschirhart, "Requirements of the Mu2e Trigger and Data Acquisition System," Mu2e-doc-1150.

[2] http://newsroom.intel.com/servlet/JiveServlet/download/2152-17-5259/Intel_ISC_2011_MIC_Kirk_Skaugen_Presentation.pdf

[3] http://www.tilera.com/products/processors/TILE-Gx_Family

[4] http://www.arm.com/products/processors/cortex-a/cortex-a15.php

[5] http://www.aps.anl.gov/epics/

[6] M. Dinnon, "Mu2e Risk Registry," Mu2e-doc-1463.

[7] http://www.fnal.gov/faw/teamcenter/

# 13 Satisfying the Physics Requirements

The Mu2e experiment is aiming for sensitivity to the $\mu^- N \rightarrow e^- N$ conversion process that is several orders of magnitude better than anything that has come before it. This ambitious goal leads to a set of technical requirements, described in a number of Requirements Documents (Table 2.2) and a list of physics requirements, summarized in Section 3.6. The conceptual designs for the accelerator, the Mu2e solenoid systems, and the Mu2e detector sub-systems have been developed to satisfy these requirements and have been discussed in detail in Chapters 5 - 12. A description of how each Physics Requirement is satisfied by the current design appears below.

*Suppression of prompt backgrounds requires a pulsed proton beam where the extinction, the ratio of beam between pulses to the beam contained in a pulse, is less than $10^{-10}$.*

The accelerator scheme discussed in Chapter 5 will deliver protons to the Mu2e production target in pulses that are approximately 200 ns wide and spaced 1695 ns apart (center-to-center). In between pulses the number of protons reaching the production target is suppressed using a system of collimators in the Debuncher to remove momentum tails (cf. Section 5.5.1). The number of out-of-time protons is further suppressed by employing a fast AC dipole to sweep clean the inter-pulse beam. The AC dipole will reside in the beam line that brings the protons from the Debuncher to the Mu2e production target (cf. Section 5.8.1). Simulations estimate that this design achieves an extinction of $10^{-12}$.

*To suppress backgrounds from muon decays in orbit, the width of the conversion electron peak, including energy loss and resolution effects, should be on the order of 1 MeV FWHM or better with no significant high energy tails.*

The Tracker design discussed in Chapter 9 consists of 36 planes of straw tubes. Each plane consists of multiple straw layers in order to improve efficiency and to resolve left-right ambiguities in the drift distance. Every effort is made to minimize the amount of material in the fiducial volume of the Tracker. The straws will be kept thin (e.g. the total thickness of the straw walls is 15 μm) and the gas manifolds, front-end electronics, and associated cooling are placed at large radii, outside the active fiducial volume. The intrinsic high-side resolution (ignoring the effect of the target and all other material upstream of the tracker), which is the most important for distinguishing conversion electrons from backgrounds, has a core component sigma of 115 KeV/c, and a significant tail sigma of 176 KeV/c. The net resolution is significantly less than the estimated resolution due to energy loss in the upstream material.

Using a full tracker simulation that includes occupancy effects from background events, energy loss, straggling, detector resolution and a prototype pattern recognition and track fitting algorithm an overall resolution function of about 1 MeV FWHM is achieved (Figure 3.11).

*To suppress backgrounds from beam electrons the field in the upstream section of the Detector Solenoid must be graded. The graded field also increases the acceptance for conversion electrons.*

As discussed in Chapter 7, the field in the Detector Solenoid starts at 2.0 Tesla just after the transition from the Transport Solenoid. The field then decreases linearly down to 1.0 Tesla over roughly five meters. This graded field largely eliminates background from beam electrons that scatter in the last collimator or stopping target by pitching them forward, thus reducing their transverse momentum so that they fall out of the geometric acceptance of the Tracker. The stopping target is approximately centered in this graded region, which nearly doubles the geometric acceptance for electrons from the $\mu^- N \rightarrow e^- N$ conversion process.

*Suppression of backgrounds from cosmic rays requires a veto surrounding the detector. The cosmic ray veto should be nearly hermetic on the top and sides in the region of the collimator at the entrance to the Detector Solenoid, the muon stopping target, tracker, and calorimeter. The overall efficiency of the cosmic ray veto should be 0.9999 or better.*

Simulations show that cosmic ray induced background events are initiated along the tops and sides of the Detector Solenoid and also along the tops and sides of the Transport Solenoid straight section that interfaces with the Detector Solenoid. The veto system discussed in Chapter 11 provides hermetic coverage along the entire length of the Detector Solenoid and the adjacent half of the Transport Solenoid for the tops and sides. The current design employs three layers of scintillator and will employ a 2-out-of-3 veto logic. Test beam studies demonstrate that an efficiency of >99.99% can be achieved with the current design (cf. Section 11.4.1).

*Suppression of long transit time backgrounds places requirements on the magnetic field in the straight sections of the Transport Solenoid. The field gradient in the three straight sections of the Transport Solenoid must be continuously negative and the gradient must be relatively uniform.*

The conceptual design for the Transport Solenoid focused on satisfying these complex field requirements. A 3D model of the Transport Solenoid field was implemented using OPERA. The field in the straight sections is always negative and the gradient is within the specifications (cf. Sec 7.3.2 and Figs. 7.29-30).

*A thin window is required to absorb antiprotons.*

The Muon Beam Line includes a thin Kapton window placed in the middle of the collimator located in the central straight section of the Transport Solenoid. Simulations show that this window reduces background from antiprotons by over two orders of magnitude (Section 3.5.4) to a level that is negligible.

*The capacity to take data outside of the search window time interval must exist.*

The DAQ has been designed to handle the higher rates that are present at earlier times in the spill cycle. This capacity is required to determine backgrounds using data, in particular, electrons from radiative pion capture.

*The capacity to collect calibration electrons from $\pi^+ \rightarrow e^+\nu$ is required.*

The Muon Beamline, discussed in Chapter 8, is optimized to efficiently transport negatively charged low momentum muons from the production target to the stopping target. It can be made to efficiently transport positively charged particles by rotating the middle collimator. The design discussed in 8.3.2 has a rotating central collimator for the purpose of selecting and transporting positively charged pions to the stopping target. Stopped $\pi^+$ can decay to $\pi^+ \rightarrow e^+\nu$ providing a mono-energetic source of $e^+$ that can be used to calibrate the absolute momentum scale of the spectrometer. The Tracker is designed in a charge symmetric way (cf. Chapter 9) and the reconstruction efficiency and resulting momentum resolution for $e^-$ and $e^+$ are expected to be comparable. The Trigger and DAQ system can be configured to trigger on high energy $e^-$ or $e^+$ or both simultaneously (cf. Chapter 12).

The Mu2e conceptual design satisfies the full set of physics requirements and currently yields a single event sensitivity of $5.4 \times 10^{-17}$ using our current set of algorithms. This includes a preliminary track recognition package that met the interim goal of 50% of the eventual anticipated reconstruction efficiency. Obvious improvements, which have yet to be studied in detail, can improve the sensitivity to about $2.4 \times 10^{-17}$. This represents an improvement of about 4 orders of magnitude over the current world's best limit on the $\mu^- N \rightarrow e^- N$ conversion process and provides discovery sensitivity over a broad range of new physics models including Supersymmetry, Little Higgs, Extra Dimensions, generic 2HDM, and Fourth Generation models.

# 14 Environment, Safety, Health and Quality Assurance

## 14.1 ES&H

Fermi National Accelerator Laboratory is committed to the success of the mission objectives of Mu2e as well as to the safety of all users, staff and the public. The goal is to provide an injury free work environment. To attain this standard, safe working conditions and practices are a requirement for all staff and contractors. All design work on Mu2e shall be done with this in mind. ES&H must be fully integrated into the project in order to realize these goals. ES&H, as it applies to each subsystem, is discussed in Sections 5.10, 6.6, 7.6, 8.7, 9.6, 10.9, 11.6 and 12.5.

### *14.1.1* Integrated Safety Management

The philosophy of Integrated Safety Management (ISM) will be incorporated into all work on Mu2e, including any work done on the Fermilab site by subcontractors and sub-tier contractors. Integrated Safety Management is a system for performing work safely and in an environmentally responsible manner. The term “integrated” is used to indicate that the ES&H management systems are normal and natural elements of doing work. The intent is to integrate the management of ES&H with the management of the other primary elements of work: quality, cost and schedule. The seven principles of ISM are as follows:

1. Line Management Responsibility for Safety. Line management is responsible and accountable for the protection of the employees, the public and the environment.
2. Clear Roles and Responsibilities. The roles, responsibilities and authority at all levels of the organization, including potential sub-tier contractors, are clearly identified.
3. Competence Commensurate with Responsibility. Personnel possess the experience, knowledge, skills and abilities that are necessary to discharge their responsibilities.
4. Balanced Priorities. Resources are effectively allocated to address safety, programmatic and operational considerations. Protecting the public, the workers and the environment will be a priority whenever activities are planned and performed.
5. Identification of Safety Standards and Requirements. Before work is performed, the associated hazards are evaluated, and an agreed upon set of safety standards and requirements are established. These standards will provide adequate assurance that the public, the workers and the environment are protected from adverse consequences.

6. Hazard Controls Tailored to Work Being Performed. Administrative and engineering controls tailored to the work being performed are present to prevent and mitigate hazards.
7. Operations Authorization. The conditions and requirements to be satisfied for operations to be initiated and conducted are clearly established and understood by all.

The Mu2e ES&H program is intended to ensure that all relevant and necessary actions are taken to provide a safe working environment for the design, construction, installation, testing and operation of all components of the Mu2e Project.

### *14.1.2* Preliminary Hazard Analysis

A principle component of an effective ES&H program is to ensure that all hazards have been properly identified and controlled through design and procedure. To ensure that all potential hazards are identified during the conceptual design phase, a Preliminary Hazard Analysis (HA) has been performed to identify the hazards that could be encountered during the construction and operation of Mu2e [1].

The hazards and risks identified in the Preliminary HA are all well known to the accelerator and particle physics community. Years of experience with such facilities at Fermilab and within the DOE complex have generated well-defined design criteria and controls to eliminate and/or mitigate these risks. Table 14.1 summarizes the hazards that have been considered and the codes and standards that apply to the reduction of risk associated with each hazard. The Fermilab ES&H Manual (FESHM) has applicable standards for each of these potential hazards.

The Preliminary Hazard Analysis process began during the conceptual design phase to Mu2e to ensure that all significant hazards were identified and adequately addressed in early design work. Each of these issues will be followed as the design advances and as construction, installation and operations commence. The Preliminary HA divides the project into 8 zones that include the new Mu2e facility as well as the existing Recycler and Debuncher Rings where modifications for Mu2e will be made. A Baseline Hazards List was developed for each of these zones as the first step in identifying potential hazards. This list utilized the best available information, encompassing data from the Mu2e subproject managers, existing Fermilab safety documentation, subject matter expertise (with conventional facilities, accelerator systems, and ES&H) and lessons-learned from the DOE's accelerator and high energy physics community covering design criteria, regulatory requirements, and related occurrences. It also included a preliminary (pre-mitigation) risk assessment for hazards present in each zone. A list of design and operational mitigation strategies were developed for each hazard in each zone and the risk was re-analyzed taking into consideration only the passive design mitigations. Operational mitigations will further reduce risks.

| Hazard List | Applicable Regulations and Standards |
|---|---|
| **Mechanical Hazards**<br>Moving large, heavy equipment<br>Overhead cranes/hoists<br>Vacuum pumps<br>Power tools and equipment<br>Motor generator equipment<br>Compressed gases<br>Vacuum/pressure vessels<br>Open hatches | ANSI/ASME Standard B30.20 Overhead Cranes<br>FESHM 5021, 5023, 5024, 5025, 5031, 5033, 5034 |
| **Flammable Gas Hazards**<br>Flammable gases | FESM 6020.3 |
| **Electrical Hazards**<br>Stored energy exposure<br>High voltage exposure<br>Low voltage, high current exposure<br>Electrical faults<br>Battery bank and UPS equipment<br>Arc flash<br>Cable tray overloading/mixed utilities | NFPA 70 National Electrical Code<br>NFPA 70 E Standard for Electrical Safety in the Workplace<br>NFPA 70 B Recommended Practice for Electrical Equipment Maintenance<br>FESHM 5040, 5041, 5042, 5043, 5044, 5046 |
| **Fire Hazards**<br>Flammable/combustible materials<br>Wire and cable insulation and jackets<br>Electrical<br>Lighting | NFPA 101 Life Safety Code<br>NFPA 45 Fire Protection for Laboratories Using Chemicals<br>FESHM 6010, 6020.1, 6040 |
| **Radiation Hazards**<br>Calibration source exposure<br>Prompt radiation from beamline<br>Indirect radiation from beamline<br>Radioactive contamination<br>Activation<br>Creation of mixed waste<br>RF & microwave<br>Magnetic fields | FESHM 10010 |

| Hazard List | Applicable Regulations and Standards |
|---|---|
| **Toxic Material Hazards**<br>Chemical agents<br>Lead and other heavy metals | FESM 5052, 8040 |
| **Laser Hazards**<br>Lasers | ANSI Z136.1-2000 Safe Use of Lasers<br>FESHM 5062.1 |
| **Construction Hazards**<br>Site clearing<br>Excavation<br>Work at elevations<br>Material handling<br>Utility interfaces<br>Weather related conditions | 29 CFR 1926, Safety and Health Regulations for Construction.<br>FESHM 7010, 7011 |
| **Oxygen Deficiency Hazards**<br>Cryogenic spill<br>Cryogen leak<br>Ventilation failure<br>Sensor failure<br>Confined space | 29 CFR 1910.134, OSHA Respiratory Protection Standard<br>FESHM 5064, 5032 |
| **Cryogenic Hazards**<br>Oxygen deficiency<br>Cryogenic distribution system<br>Thermal<br>Pressure | FESHM 5032 |
| **Environmental Hazards**<br>Construction impacts<br>Storm water discharge<br>Soil activation<br>Air activation<br>Cooling water activation<br>Discharge/emission points | 40 CFR 61 - Subpart A, National Emissions Standards for Hazardous Air Pollutants (NESHAPS)<br>6 NYCRR 200 – 234 – NYSDEC Prevention and Control of Air contamination and Air Pollution |

Table 14.1 Hazards considered in the Mu2e Preliminary Hazard Analysis and applicable codes and standards.

All of the hazards identified in this process are typical of those encountered with other high energy physics projects at accelerator facilities across the DOE complex. The design and operational criteria to mitigate these hazards, resulting from many years of operational experience and lessons learned, will be applied to Mu2e. There are no unmitigated risks that are deemed to be *Critical* but a number of potentially *High* risks could be present in Mu2e in the absence of passive mitigation. When taking into account the planned passive mitigations there are no risks higher than *Moderate* and most are *Low* or *Minimal*. Active mitigation measures will reduce the risks even farther.

## 14.2 Quality Assurance

The Mu2e Project will design and build a facility and experimental apparatus that meets the Mu2e mission objectives. This will be accomplished with the assistance of a fully implemented Quality Assurance (QA) Program.

A QA Program Plan has been prepared by the Mu2e Project and approved by the Mu2e Project Manager [4]. This plan specifies the program requirements that apply to all Mu2e work. The primary objective of the QA program is to implement quality assurance criteria in a way that achieves adequate protection of the workers, the public, and the environment, taking into account the work to be performed and the associated hazards. The objectives include:

- Designing in quality and reliability.
- Assuring that all personnel involved in the project uphold the NSLS-II Quality Assurance Plan.
- Promoting early detection of problems to minimize failure costs and impact on schedule.
- Developing appropriate documentation to support construction and operational requirements.
- Assuring that personnel have the necessary training as needed before performing critical activities, especially activities that have environmental, safety, security, or health consequences.
- Defining the general requirements for design and readiness reviews, including environmental, safety, security, and health issues related to hardware, software, and processes.

Quality Assurance and Quality Control issues have been discussed for each of the L2 subsystems. Sections 6.7, 7.8, 8.5, 9.8, 10.11, 11.8 and 12.7 of this report can be consulted for those evaluations.

## 14.3 References

[1] Mu2e Preliminary Hazard Analysis, Mu2e-doc-675.
[2] Project Information Form for Mu2e, Mu2e-doc-1000.
[3] Mu2e Categorical Exclusion, Mu2e-doc-xxx.
[4] Mu2e Quality Management Plan, Mu2e-doc-677.

# 15 Risk Management

Risk management is based on a graded approach in which levels of risk are assessed for project activities and elements. This assessment is based upon the probability of occurrence and the impact should the risk materialize. Minimizing risk is a 5-step process:

1. Identifying potential project risk
2. Analyzing project risk
3. Planning risk abatement strategies
4. Executing risk abatement strategies
5. Monitoring the results of and revising risk abatement strategies.

Risk assessments are conducted throughout the project lifecycle. Risks identified include technical, cost, schedule and ES&H risks. The Mu2e Risk Management Plan [1] details the process for identifying, evaluating, mitigating, and managing risks in compliance with DOE Order 413.3a. These activities must be completed and fully implemented by CD-2 so that mitigation impacts can be fully captured in the Project's Performance Baseline.

For CD-1, a top-down study to identify significant risks that require mitigation as early as possible in the design process are identified and preliminary mitigation strategies are developed. These risks have been captured in the Primavera Risk Management Tool and are displayed in the Mu2e Risk Registry [2]. This will form the basis for the more thorough bottoms-up risk analysis that will be completed as Mu2e moves from the Conceptual Design towards a Preliminary Design. In some cases, mitigations have already been incorporated into the conceptual design.

The risk registry catalogs the potential problem (or opportunity), the potential cause and the consequence if the risk is realized. The probability, impact and severity of each risk are identified prior to mitigation and strategies for mitigating the risk are discussed. Finally, the risk registry is used to generate a cost and schedule range for the Project based on the cost to mitigate potential risks and the potential savings if opportunities can be taken advantage of.

Risk issues have been discussed for each of the L2 subsystems in Chapters 5 – 12. Sections 6.5, 7.7, 8.4, 9.7, 10.10, 11.7 and 12.6 of this report can be consulted for those evaluations.

## 15.1 Technical Risks

The technical risks facing the Mu2e Project are no greater than those facing other HEP projects. Risks that are identified will be managed as early as possible to assure that they do not derail the timely completion of the project or stress its budget in unexpected ways.

Many of the components required for the Mu2e project are similar to others recently built at Fermilab, for which there are numerous technical and cost benchmarks. Many of the hardware components will be reused or refurbished equipment already existing at Fermilab. The technical risks associated with this equipment are small. There are a few components for which the unmitigated technical risks are high, however. These include radiation shielding, interface issues and tracker failures. Mitigation strategies will reduce these risks significantly.

## 15.2 Cost and Schedule Risks

Use of fixed-price subcontracts and competition will be maximized to reduce cost risk. The Project is also making a major effort to understand how the cost of large ticket items that will not be delivered for several years are tied to exchange rate fluctuations, the producer price index or any other relevant index. This will allow a more precise determination of contingency on the elements that drive the cost of the project.

Schedule risk will be minimized by:

- R&D, including bench testing and time and motion studies
- Realistic planning
- Verification of subcontractor's credit and capacity during evaluation
- Close surveillance of subcontractor performance.

Incentive subcontracts, such as fixed-price with incentive, will be considered when a reasonably firm basis for pricing does not exist or the nature of the requirement is such that the subcontractor's assumption of a degree of cost risk will provide a positive profit incentive for effective cost and/or schedule control and performance.

Cost and schedule risks in Mu2e include the effect of continuing resolutions and funding delays, delays to the solenoids that define the critical path and resources that are not available when needed. Mitigation strategies, including a bottoms-up contingency analysis, will reduce these risks to manageable levels.

## 15.3 ES&H Risks

The management and mitigation of Environment, Safety and Health risks is of paramount importance. These risks are captured in the Mu2e Preliminary Safety Assessment Document, the Hazard Analysis and the Mu2e Environmental Assessment. They are managed through Fermilab's Integrated Safety Management program. ES&H has been incorporated into the planning, design and implementation of Mu2e from the beginning and has driven the design of various aspects of the Project. The Mu2e Risk Management Program evaluates potential ES&H risks and establishes strategies to mitigate those risks. ES&H risks include radiation exposure and construction accidents. Mitigation strategies will significantly reduce the probability that these incidents occur.

## 15.4 References

[1] Mu2e Risk Management Plan, Mu2e-doc-461.
[2] M. Dinnon, "Risk Registry," Mu2e-doc-1463.

# 16 Safeguards and Security

Neither Mu2e nor Fermilab are nuclear facilities and no part of the project is sensitive or classified. However, because of Fermilab's association with the U.S. Government the lab as a whole could be considered a potential target. This threat potential extends to the Mu2e facility since it resides on the Fermilab site. The probability of an attack is considered quite small and the existing Fermilab security envelope is more than adequate, but some common sense security precautions will be implemented. The most likely security concerns for Mu2e involves computer security, theft and vandalism. The later two are primarily a concern during the building construction phase.

During the R&D and construction phases of Mu2e, the Project ensures appropriate levels of protection by relying primarily on the existing security apparatus in place at the host institution. In particular, Mu2e activities fall under the security umbrella of Fermilab and the Laboratories and Universities of collaborators. This includes protection from unauthorized access, theft, destruction of DOE assets and other potential adverse impacts.

A Vulnerability Assessment [1] by the Project team has exposed one moderate security threat in the area of computer security. Unauthorized access and malicious attacks against computer systems are unfortunate facts of life for every organization. Fermilab's computing system is routinely attacked. Mu2e can expect a similarly hostile environment. The probability of attempted unauthorized accesses and malicious attacks against the Mu2e computing facilities is moderate. Appropriate security will be built into all computing systems as a requirement. As this is currently a rapidly changing and timely field with the onset of grid computing, we will wait as long as possible before committing to a particular protocol. We will follow the Fermilab Computing Division's lead on this issue and will be included within Fermilab's online security envelope.

The primary security threats at the various Mu2e sites where detector construction and assembly might take place are theft and vandalism. The Mu2e Project will have to ensure that adequate security is in place to protect its assets at all of the facilities where detector construction is taking place. Access restrictions, appropriate ES&H procedures and sufficient fire protection will be required at all locations. Explicit requirements will be spelled out in the Statement of Work issued to the institutions doing the work. All Mu2e activities will take place at National Laboratories, Universities or private industry. The probability of these threats being realized at any of these locations is small because of the existing adequate security that is already in place.

## 16.1 References

[1] Security Vulnerability Assessment Report for the Mu2e Project, Mu2e-doc-676.

# 17 Stakeholder Input

The primary stakeholder in the Mu2e project is the Mu2e Collaboration. The Collaboration has been involved in the design and planning of the Project, and it is expected that Mu2e will continue to attract university, national laboratory and international collaborators.

The design of Mu2e has been developed by the scientific and technical staff at Fermilab and the collaborating institutions. This combination is best equipped to define the optimal performance parameters for the facility and to carry out the science in a cost-effective manner.

Throughout the design and planning process for Mu2e, every effort has been made to maintain communications with DOE, with the Mu2e Collaboration and with the management of Fermilab. The Mu2e Project Manager has made several trips to Germantown to report on progress and interact with DOE management and bi-weekly reports on Mu2e progress are provided to the head of the DOE Office of High Energy Physics. Reports on the Mu2e Project are generally provided at various DOE Reviews of Fermilab. Regular Integrated Project Team meetings take place and monthly Working Group Meetings are held with DOE and Fermilab management.

# 18 Cost, Schedule and Scope Range

## 18.1 Project Deliverables, Test and Acceptance Criteria

### *18.1.1* Mu2e Detector Hall Facility

A new experimental hall facility must be constructed on the Fermilab site, north-west of the existing antiproton source near Kautz Road. The facility includes an underground enclosure for the Mu2e detector, an at-grade service building and a tunnel stub that connects to the Muon Campus External Beamline. The test and acceptance criteria for the facility are beneficial occupancy and completion of the final punch list of deficiencies.

### *18.1.2* Superconducting Solenoid System

A system of superconducting solenoids must be designed and constructed for Mu2e. The solenoid system consists of a *Production Solenoid* that contains the target for the primary proton beam, an S-shaped *Transport Solenoid* that serves as a magnetic channel for pions and muons of the correct charge and momentum range and a *Detector Solenoid* that houses the muon stopping target and the detector elements. The solenoid system shall be considered complete when each solenoid has been cooled down, powered and run at nominal field strength.

### *18.1.3* Mu2e Detector

The Mu2e Detector consists of a tracker, a calorimeter, a stopping target monitor, a cosmic ray veto, an extinction monitor and the electronics, trigger and data acquisition required to read out, select and store the data. The tracker accurately measures the trajectory of charged particles, the calorimeter provides independent measurements of energy, position and time, the stopping target monitor measures the characteristic X-ray spectrum from the formation of muonic atoms, the cosmic ray veto identifies cosmic ray muons traversing the detector region that can cause backgrounds and the extinction monitor detects scattered protons from the stopping target to determine the fraction of out-of-time beam. The detector will be considered complete when tracks from cosmic ray muons have been observed in the detector and recorded by the data acquisition system.

### *18.1.4* Accelerator Modifications

Parts of the Fermilab accelerator complex must be modified to transfer 8 GeV protons from the Fermilab Booster to the Mu2e detector while the 120 GeV neutrino program is operating. To accomplish this the existing Debuncher Ring will be modified to slow extract beam to the Mu2e detector through a new external beamline. The accelerator modifications shall be considered complete when the capability to extract 8 GeV protons from the Debuncher Ring and transport them through the Muon Campus External Beamline to the Mu2e Production Target is in place.

## 18.2 Cost Range

The Total Project Cost (TPC) range for Mu2e is $208 - $287M. The estimates are in Actual Year Dollars (AY$) and include contingency, overheads and escalation from the base year of 2012.

An estimate of the most likely Project cost was assembled and that cost is supported by BOEs for each Level 3 activity. The cost and schedule ranges are based on Project Risk and are detailed in the Mu2e Project Risk and Opportunity Register [1].

### *18.2.1* R&D Funding Requirements

R&D funds in the amount of $24M are needed during the period from FY10 – FY13. R&D funds are used to develop the Mu2e conceptual design, procure and test aluminum stabilized superconductor for solenoid R&D, fabricate prototype scintillator for the Cosmic Ray Veto, develop a prototype ASIC for tracker readout and a variety of other activities necessary as input to final designs.

### *18.2.2* PED Funding Requirements

PED funds in the amount of $49M are required during the period from FY12 to FY14 to develop preliminary and final designs for all aspects of the Project.

## 18.3 Schedule Range

The Key Performance Parameters will be satisfied by August 1 2019, corresponding to a construction schedule of 65 months that starts at CD-2/3a. Additional time is required for reviews and Project close-out. A schedule range of 56 - 80 months is estimated for the duration of the construction project, starting at CD-2/3a. This corresponds to a completion date ranging from November 2018 to November 2020. The Project duration is dominated by the Solenoid design, procurement and installation. Schedule variations in the other subsystems are unlikely to have an impact to the overall Project, so the schedule range is largely determined by the solenoids and funding risks/opportunities. The CD-4 date has 18 months of programmatic float added from the expected completion date, corresponding to the 2$^{nd}$ quarter of FY21.

## 18.4 Scope Range

Mu2e has little scope range. The only significant scope variation involves the calorimeter. In the baseline plan, 2/3 of the calorimeter cost is borne by INFN and provided in-kind. The Mu2e Project pays for the remaining 1/3. It's possible that the entire calorimeter could be provided by INFN. It's also possible, though considered unlikely, that INFN's circumstances might change and the full cost of the calorimeter would be borne by the Mu2e Project. This results in a calorimeter scope range that varies from 0 to 100%. This scope range is reflected in the cost range.

## 18.5 References

[1] M. Dinnon, “Mu2e Risk Registry,” Mu2e-doc-1463.